\documentclass[%
  manyauthors,
  nocleardouble,
  COMPASS,
  american,
  fleqn,  %
  ]{cernphprep}

\usepackage[absolute]{textpos}  %
\usepackage[nottoc]{tocbibind}  %
\usepackage{cite}               %
\usepackage{packages}
\makeatletter
\@ifpackageloaded{tikz}%
{}%
{\usepackage{tikz}}%
\makeatother

\hypersetup{%
  pdfborder={0 0 0.1}  %
}

\makeatletter
\g@addto@macro\bfseries{\boldmath}
\makeatother

\addto\captionsamerican{}  %
\crefname{section}{Sec.}{Secs.}                             %
\Crefname{section}{Section}{Sections}
\renewcommand{\thesection}{\Roman{section}}%
\renewcommand{\thesubsection}{\Alph{subsection}}%
\renewcommand{\thesubsubsection}{\arabic{subsubsection}}%
\makeatletter
\renewcommand{\p@section}{}%
\renewcommand{\p@subsection}{\thesection\,}%
\renewcommand{\p@subsubsection}{\thesection\,\thesubsection\,}%
\makeatother
\usepackage[titles,subfigure]{tocloft}

\makeatletter
\renewcommand{\appendix}{%
  \par
  \setcounter{section}\z@%
  \setcounter{subsection}\z@%
  \setcounter{subsubsection}\z@%
  \renewcommand{\thesection}{\Alph{section}}%
  \renewcommand{\theHsection}{Dummy.\Alph{section}}%
  \renewcommand{\thesubsection}{\arabic{subsection}}%
  \renewcommand{\thesubsubsection}{\alph{subsubsection}}%
  \renewcommand{\p@section}{}%
  \renewcommand{\p@subsection}{\thesection\,}%
  \renewcommand{\p@subsubsection}{\thesection\,\thesubsection\,}%
  \renewcommand{\theequation}{\thesection\arabic{equation}}%
  \addtocontents{toc}{\protect\appendix}%
}%
\makeatother

\graphicspath{%
  {./},%
  {../tikz/},%
  {../plots/mass_dependent_fit_14waves/},%
}

\usepackage{macros}

\usepackage{multirow}
\usepackage{wallnerstudymapping}

\newlength{\totalPlotWidth}
\newlength{\twoPlotWidth}
\newlength{\twoPlotSpacing}
\newlength{\threePlotWidth}
\newlength{\threePlotSpacing}
\newlength{\threePlotSmallWidth}
\newlength{\threePlotSmallSpacing}
\newlength{\fourPlotWidth}
\newlength{\fourPlotSpacing}
\begin{document}

%
%

%
%
%
%
%

\begin{titlepage}
  \PHnumber{2018--XXX}
  \PHdate{\today}
  %

  %
  %
  %
%
%

\title{Light isovector resonances in $\pi^-\, p \to \threePi\, p$ at~\SI{190}{\GeVc}}
   \makeatletter
  \ShortTitle{\@title}
  \makeatother
  \Collaboration{The COMPASS Collaboration}
  \ShortAuthor{}

  %
  %
  %
%
%

\begin{abstract}
  We have performed the most comprehensive resonance-model fit of
  \threePi states using the results of our previously published
  partial-wave analysis (PWA) of a large data set of
  diffractive-dissociation events from the reaction \reaction with a
  \SI{190}{\GeVc} pion beam.  The PWA results, which were obtained in
  100~bins of three-pion mass,
  \SIvalRange{0.5}{\mThreePi}{2.5}{\GeVcc}, and simultaneously in
  11~bins of the reduced four-momentum transfer squared,
  \SIvalRange{0.1}{\tpr}{1.0}{\GeVcsq}, are subjected to a
  resonance-model fit using Breit-Wigner amplitudes to simultaneously
  describe a subset of 14~selected waves using 11~isovector
  light-meson states with $\JPC = 0^{-+}$, $1^{++}$, $2^{++}$,
  $2^{-+}$, $4^{++}$, and spin-exotic $1^{-+}$ quantum numbers.  The
  model contains the well-known resonances \Ppi[1800], \PaOne, \PaTwo,
  \PpiTwo, \PpiTwo[1880], and \PaFour.  In addition, it includes the
  disputed \PpiOne[1600], the excited states \PaOne[1640],
  \PaTwo[1700], and \PpiTwo[2005], as well as the resonancelike
  \PaOne[1420].  We measure the resonance parameters mass and width of
  these objects by combining the information from the PWA results
  obtained in the 11 \tpr bins.  We extract the relative branching
  fractions of the $\Prho \pi$ and $\PfTwo \pi$ decays of \PaTwo and
  \PaFour, where the former one is measured for the first time.  In a
  novel approach, we extract the \tpr dependence of the intensity of
  the resonances and of their phases.  The \tpr dependence of the
  intensities of most resonances differs distinctly from the \tpr
  dependence of the nonresonant components.  For the first time, we
  determine the \tpr dependence of the phases of the production
  amplitudes and confirm that the production mechanism of the Pomeron
  exchange is common to all resonances.  We have performed extensive
  systematic studies on the model dependence and correlations of the
  measured physical parameters.
\end{abstract}

  %
  %
  \vspace*{20pt}
  \begin{flushleft}
    Keywords:
    %
%
experimental results, magnetic spectrometer;
hadron spectroscopy, meson, light;
CERN Lab;
CERN SPS;
COMPASS;
beam, pi-, 190 GeV/c;
pi-, hadroproduction, meson resonance;
pi-, diffraction, dissociation;
pi-, multiple production, (pi+ 2pi-);
target, hydrogen;
pi- p, inelastic scattering, exclusive reaction;
pi- p --> p pi+ 2pi-;
partial-wave analysis; isobar model; hadronic decay, amplitude analysis;
mass spectrum, (pi+ 2pi-);
spin, density matrix;
momentum transfer dependence, slope;
data analysis method;
scalar meson, isoscalar;
pseudoscalar meson, isovector;
vector meson, isovector;
axial-vector meson, isovector;
tensor meson;
f0(500); rho(770); f0(980); f2(1270); f0(1500); rho3(1690);
a1(1260); a2(1320); a1(1420); pi1(1600); a1(1640); pi2(1670); a2(1700); pi(1800); pi2(1880); pi2(2005); a4(2040)
   \end{flushleft}
  \Submitted{(submitted to Physical Review D)}
\end{titlepage}

{\pagestyle{empty}
%
%
\section*{The COMPASS Collaboration}
\label{app:collab}
\renewcommand\labelenumi{\textsuperscript{\theenumi}~}
\renewcommand\theenumi{\arabic{enumi}}
\begin{flushleft}
M.~Aghasyan\Irefn{triest_i},
M.G.~Alexeev\Irefn{turin_u},
G.D.~Alexeev\Irefn{dubna}, 
A.~Amoroso\Irefnn{turin_u}{turin_i},
V.~Andrieux\Irefnn{illinois}{saclay},
N.V.~Anfimov\Irefn{dubna}, 
V.~Anosov\Irefn{dubna}, 
A.~Antoshkin\Irefn{dubna}, 
K.~Augsten\Irefnn{dubna}{praguectu}, 
W.~Augustyniak\Irefn{warsaw}
% ...
\end{flushleft}
%
%
\begin{Authlist}
\item \Idef{aveiro}{University of Aveiro, Department of Physics, 3810-193 Aveiro, Portugal}
\item \Idef{bochum}{Universit\"at Bochum, Institut f\"ur Experimentalphysik, 44780 Bochum, Germany\Arefs{l}\Aref{s}}
\item \Idef{bonniskp}{Universit\"at Bonn, Helmholtz-Institut f\"ur  Strahlen- und Kernphysik, 53115 Bonn, Germany\Arefs{l}}
\item \Idef{bonnpi}{Universit\"at Bonn, Physikalisches Institut, 53115 Bonn, Germany\Arefs{l}}
\item \Idef{brno}{Institute of Scientific Instruments, AS CR, 61264 Brno, Czech Republic\Arefs{m}}
\item \Idef{calcutta}{Matrivani Institute of Experimental Research \& Education, Calcutta-700 030, India\Arefs{n}}
\item \Idef{dubna}{Joint Institute for Nuclear Research, 141980 Dubna, Moscow region, Russia}
\item \Idef{freiburg}{Universit\"at Freiburg, Physikalisches Institut, 79104 Freiburg, Germany\Arefs{l}\Aref{s}}
\item \Idef{cern}{CERN, 1211 Geneva 23, Switzerland}
\item \Idef{liberec}{Technical University in Liberec, 46117 Liberec, Czech Republic\Arefs{m}}
\item \Idef{lisbon}{LIP, 1000-149 Lisbon, Portugal\Arefs{p}}
\item \Idef{mainz}{Universit\"at Mainz, Institut f\"ur Kernphysik, 55099 Mainz, Germany\Arefs{l}}
\item \Idef{miyazaki}{University of Miyazaki, Miyazaki 889-2192, Japan\Arefs{q}}
\item \Idef{moscowlpi}{Lebedev Physical Institute, 119991 Moscow, Russia}
\item \Idef{munichtu}{Technische Universit\"at M\"unchen, Physik-Department, 85748 Garching, Germany\Arefs{l}\Aref{r}}
\item \Idef{nagoya}{Nagoya University, 464 Nagoya, Japan\Arefs{q}}
\item \Idef{praguecu}{Charles University in Prague, Faculty of Mathematics and Physics, 18000 Prague, Czech Republic\Arefs{m}}
\item \Idef{praguectu}{Czech Technical University in Prague, 16636 Prague, Czech Republic\Arefs{m}}
\item \Idef{protvino}{NRC \enquote{Kurchatov Institute}, IHEP, 142281 Protvino, Russia}
\item \Idef{saclay}{IRFU, CEA, Universit\'e Paris-Saclay, 91191 Gif-sur-Yvette, France\Arefs{s}}
\item \Idef{taipei}{Academia Sinica, Institute of Physics, Taipei 11529, Taiwan\Arefs{tw}}
\item \Idef{telaviv}{Tel Aviv University, School of Physics and Astronomy, 69978 Tel Aviv, Israel\Arefs{t}}
\item \Idef{triest_u}{University of Trieste, Department of Physics, 34127 Trieste, Italy}
\item \Idef{triest_i}{Trieste Section of INFN, 34127 Trieste, Italy}
\item \Idef{turin_u}{University of Turin, Department of Physics, 10125 Turin, Italy}
\item \Idef{turin_i}{Torino Section of INFN, 10125 Turin, Italy}
\item \Idef{tomsk}{Tomsk Polytechnic University, 634050 Tomsk, Russia\Arefs{nauka}}
\item \Idef{illinois}{University of Illinois at Urbana-Champaign, Department of Physics, Urbana, Illinois 61801-3080, USA\Arefs{nsf}}
\item \Idef{warsaw}{National Centre for Nuclear Research, 00-681 Warsaw, Poland\Arefs{u}}
\item \Idef{warsawu}{University of Warsaw, Faculty of Physics, 02-093 Warsaw, Poland\Arefs{u}}
\item \Idef{warsawtu}{Warsaw University of Technology, Institute of Radioelectronics, 00-665 Warsaw, Poland\Arefs{u} }
\item \Idef{yamagata}{Yamagata University, Yamagata 992-8510, Japan\Arefs{q} }
\end{Authlist}
%
%
\renewcommand\theenumi{\alph{enumi}}
\begin{Authlist}
\item [{\makebox[2mm][l]{\textsuperscript{\#}}}] Corresponding authors
\item [{\makebox[2mm][l]{\textsuperscript{*}}}] Deceased
\item \Adef{a}{Also at Instituto Superior T\'ecnico, Universidade de Lisboa, Lisbon, Portugal}
\item \Adef{b}{Also at Department of Physics, Pusan National University, Busan 609-735, Republic of Korea and at Physics Dept., Brookhaven National Laboratory, Upton, NY 11973, USA}
\item \Adef{ictp}{Also at Abdus Salam ICTP, 34151 Trieste, Italy}
\item \Adef{r}{Supported by the DFG cluster of excellence `Origin and Structure of the Universe' (www.universe-cluster.de) (Germany)}
\item \Adef{p2i}{Supported by the Laboratoire d'excellence P2IO (France)}
\item \Adef{c}{Supported by the DFG Research Training Group Programmes 1102 and 2044 (Germany)} 
\item \Adef{d}{Also at Chubu University, Kasugai, Aichi 487-8501, Japan\Arefs{q}}
\item \Adef{x}{Also at Department of Physics, National Central University, 300 Jhongda Road, Jhongli 32001, Taiwan}
\item \Adef{e}{Also at KEK, 1-1 Oho, Tsukuba, Ibaraki 305-0801, Japan}
\item \Adef{g}{Also at Moscow Institute of Physics and Technology, Moscow Region, 141700, Russia}
\item \Adef{yerevan}{Also at Yerevan Physics Institute, Alikhanian Br. Street, Yerevan, Armenia, 0036}
\item \Adef{y}{Also at Department of Physics, National Kaohsiung Normal University, Kaohsiung County 824, Taiwan}
\item \Adef{turin_p}{Also at University of Eastern Piedmont, 15100 Alessandria, Italy}
\item \Adef{h}{Present address: RWTH Aachen University, III.\ Physikalisches Institut, 52056 Aachen, Germany}
\item \Adef{i}{Present address: Uppsala University, Box 516, 75120 Uppsala, Sweden}
%
%
\item \Adef{l}{Supported by BMBF - Bundesministerium f\"ur Bildung und Forschung (Germany)}
\item \Adef{s}{Supported by FP7, HadronPhysics3, Grant 283286 (European Union)}
\item \Adef{m}{Supported by MEYS, Grant LG13031 (Czech Republic)}
\item \Adef{n}{Supported by SAIL (CSR) and B.Sen fund (India)}
\item \Adef{p}{\raggedright Supported by FCT - Funda\c{c}\~{a}o para a Ci\^{e}ncia e Tecnologia, COMPETE and QREN, Grants CERN/FP 116376/2010, 123600/2011 and CERN/FIS-NUC/0017/2015 (Portugal)}
\item \Adef{q}{Supported by MEXT and JSPS, Grants 18002006, 20540299, 18540281 and 26247032, the Daiko and Yamada Foundations (Japan)}
\item \Adef{tw}{Supported by the Ministry of Science and Technology (Taiwan)}
\item \Adef{t}{Supported by the Israel Academy of Sciences and Humanities (Israel)}
\item \Adef{nauka}{Supported by the Russian Federation  program ``Nauka'' (Contract No. 0.1764.GZB.2017) (Russia)}
\item \Adef{nsf}{Supported by the National Science Foundation, Grant no. PHY-1506416 (USA)}
\item \Adef{u}{Supported by NCN, Grant 2015/18/M/ST2/00550 (Poland)}
\end{Authlist}

   \clearpage
}

\setcounter{page}{1}
 \tableofcontents

\setlist{noitemsep}
\setlist{nolistsep}
\setlist[enumerate]{label=\textit{\roman*})}

\clearpage
%
%
%

\section{Introduction}
\label{sec:introduction}

The excitation spectrum of bound quark-antiquark states that are
composed of $u$, $d$, and $s$~quarks, \ie light-quark mesons, has
regained interest in recent years.  Excited light-quark mesons are
currently studied extensively in high-flux fixed-target experiments
with hadrons at CERN~\cite{Abbon:2014aex} and with photons at
Jefferson Lab~\cite{Battaglieri:2010zza,Ghoul:2015ifw}.  They are also
produced, for example, in multibody decays of heavy-quark mesons and
in $e^+ e^-$ collisions with initial-state radiation.  Both processes
are studied, for example, at BESIII~\cite{Asner:2008nq},
\babar~\cite{Bevan:2014iga}, and Belle~\cite{Bevan:2014iga}.
Recently, the formulation of QCD on the lattice has gained new
momentum because it now also addresses light-meson decays; see \eg\
\refsCite{Feng:2010es,Wilson:2015dqa,Prelovsek:2013ela,Wilson:2014cna}.
In the future, this will lead to more realistic predictions for masses
and widths of excited hadrons.  Hence obtaining more precise
experimental knowledge of the properties of light mesons has become
important.  Despite many decades of research, the spectroscopic
information coming from different experiments is sometimes
inconsistent or even controversial.  Extensive discussions of the
light-meson sector can be found
in~\refsCite{Patrignani:2016xqp,klempt:2007cp,Crede:2008vw,Ochs:2013gi,Meyer:2010ku,Meyer:2015eta,Brambilla:2014jmp}.

\begin{figure}[tbp]
  \centering
  \includegraphics[width=\linewidthOr{0.75\textwidth}]{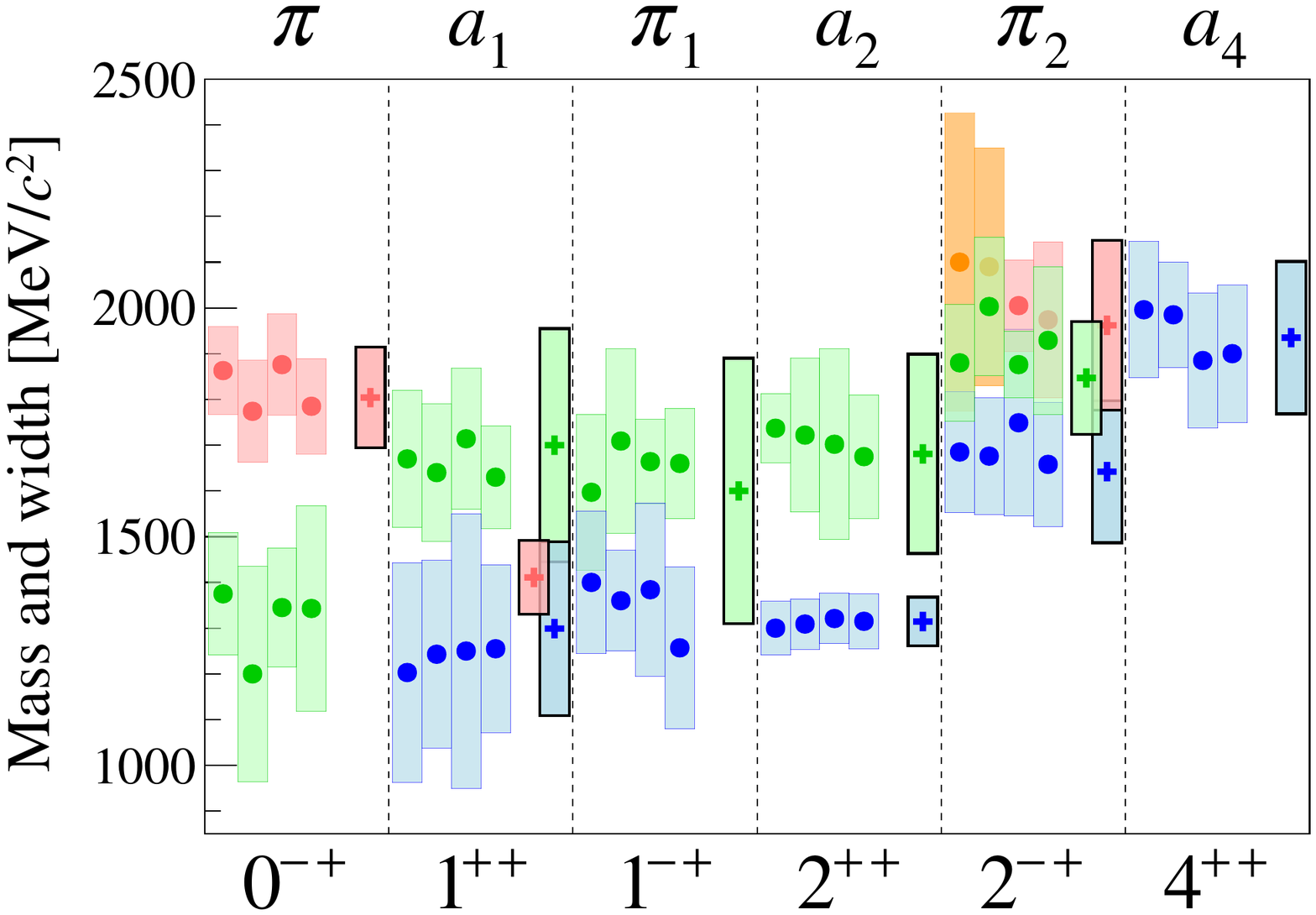}
  \caption{Masses and widths of light isovector mesons with positive
    $C$~parity and a $3\pi$ decay mode.  For each resonance, the four
    most recent measurements of masses (circles) and widths (vertical
    size of boxes), as listed by the PDG~\cite{Patrignani:2016xqp},
    are compared to the masses and widths obtained in this analysis
    (crosses and black-framed boxes, respectively).  The measurements
    are grouped according to the \JPC quantum numbers of the states.
    Higher excitations with the same \JPC are shown in different
    colors.}
  \label{fig:spectroscopy_recent_measurements}
\end{figure}

Light-meson states are characterized by spin~$J$, parity $P$, charge
conjugation~$C$,\footnote{Although the $C$~parity is not defined for
  the charged states considered here, it is customary to quote the
  \JPC quantum numbers of the corresponding neutral partner state in
  the isospin triplet.  The $C$~parity can be generalized to the
  $G$~parity, $G \equiv C\, e^{i \pi I_y}$, which is a multiplicative
  quantum number that is defined for the nonstrange states of a meson
  multiplet.  Here, $I_y$ is the $y$~component of the isospin.} and
isospin~$I$ quantum numbers.  The mesons are grouped into
SU(3)$_\text{flavor}$ multiplets that contain states with the same \JP
quantum numbers.  In this paper, we restrict ourselves to isovector
mesons with masses below about \SI{2.1}{\GeVcc}, which decay into
three charged pions and hence have negative $G$~parity.  The Particle
Data Group (PDG) provides a complete listing of the known
states~\cite{Patrignani:2016xqp}.
\Cref{fig:spectroscopy_recent_measurements} shows a summary of recent
measurements of masses and widths of these states grouped by their
\JPC quantum numbers.  For each resonance, the four most recent
entries from the PDG are confronted with the results that will be
presented in this work.  For some states, the variation of the
resonance parameters extracted from different experiments is by far
larger than the statistical uncertainties of the individual
measurements.  In many cases, these variations originate from
different analysis methods and model assumptions.  Substantial
differences among the measurements are found, for example, for the
parameters of the \PaOne* ground state, \PaOne, and the first excited
states of the \PaOne* and the \PaTwo*, \PaOne[1640] and \PaTwo[1700].
The situation is similar for the \PpiOne[1600], which has
\enquote{exotic} $\JPC = 1^{-+}$ quantum numbers that are forbidden
for ordinary \qqbarPrime quark-model states in the nonrelativistic
limit.  The resonance interpretation of the \PpiOne[1600] signal is
controversial, in particular in the $\Prho \pi$ decay mode that will
be addressed in this analysis.  For all states discussed here, we
exploit the observed dependence of the production amplitudes on the
squared four-momentum transfer in order to better separate resonant
and nonresonant contributions.  We also extract branching-fraction
ratios for the $\Prho \pi$ and $\PfTwo \pi$ decays of \PaTwo and
\PaFour.

The COMPASS Collaboration has already published properties of
isovector $3\pi$ resonances with masses in the range between
\SIlist{1.1;2.1}{\GeVcc}, produced in pion scattering off a solid-lead
target~\cite{alekseev:2009aa,adolph:2014mup}.  In particular, we
reported in \refCite{alekseev:2009aa} the observation of the
spin-exotic \PpiOne[1600] in the $\Prho \pi$ decay mode.  Our recent
observation of a new axial-vector resonancelike structure, the
\PaOne[1420], with the same quantum numbers as the elusive
\PaOne~\cite{Adolph:2015pws} has spurred much work on the
interpretation of states (including heavy-quark states), for which the
assignment to quark-model multiplets is unclear; see \eg\
\refsCite{wang:2014bua,Basdevant:2015ysa,Basdevant:2015wma,Ketzer:2015tqa,Wang:2015cis,Chen:2015fwa,Aceti:2016yeb,Gutsche:2017oro,Liu:2015taa,Guo:2017jvc}.
The present study uses the same data but yields more accurate
resonance parameters.

\begin{figure}[tbp]
  \centering
  \includegraphics[width=\linewidthOr{\twoPlotWidth}]{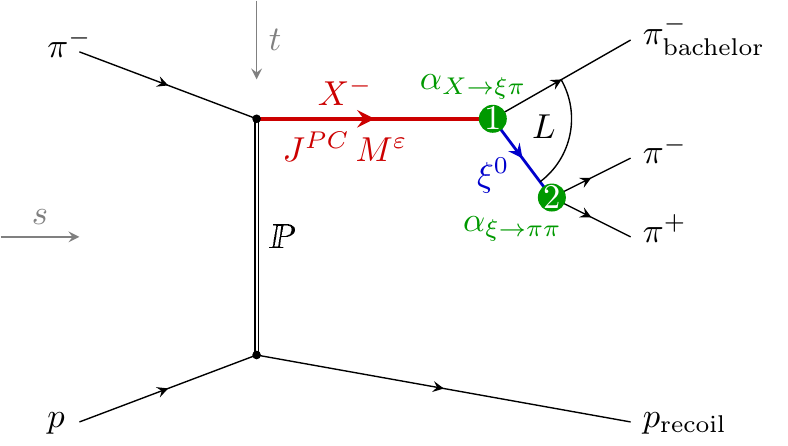}
  \caption{Diffractive dissociation of a beam pion on a target proton
    into the \threePi final state via an intermediate $3\pi$ state
    $X^-$.  The decay of $X^-$ is described using the isobar model,
    which assumes that the decay proceeds via an intermediate \twoPi
    state $\xi^0$, the so-called isobar.  At the two decay vertices,
    the couplings $\alpha_{X \to \xi \pi}$ (vertex~1) and
    $\alpha_{\xi \to \pi \pi}$ (vertex~2) appear, which are in general
    complex numbers.}
  \label{fig:3pi_reaction_isobar}
\end{figure}

This work is based on the world's largest data set to date on
diffractively produced mesons decaying into three charged pions.  The
data were obtained by the COMPASS experiment and were already
presented in detail in~\refCite{Adolph:2015tqa}.  They contain
exclusive events from the inelastic reaction
\begin{equation}
  \reaction,
  \label{eq:reaction}
\end{equation}
which was induced by a \SI{190}{\GeVc} $\pi^-$ beam impinging on a
liquid-hydrogen target.  The recoiling target proton is denoted by
$p_\text{recoil}$.  In such single-diffractive reactions, the target
particle stays intact and the beam pion is excited via the exchange of
a Pomeron with the target nucleon to a short-lived intermediate state
$X^-$ that then decays into \threePi as shown in
\cref{fig:3pi_reaction_isobar}.

Reaction~\eqref{eq:reaction} depends on two Mandelstam variables: the
squared $\pi^- p$ center-of-mass energy~$s$, which is fixed by the
beam energy, and the squared four-momentum~$t$ transferred by the
Pomeron.  It is convenient to define the \emph{reduced four-momentum
  transfer squared}
\begin{equation}
  \tpr \equiv \tabs - \tmin \geq 0,
  ~\text{where}~
  \tmin \approx \rBrk{\frac{\mThreePi^2 - m_\pi^2}{2 \Abs{\vec{p}_\text{beam}}}}^2
  \label{eq:tPrime}
\end{equation}
is the minimum absolute value of the four-momentum transfer
needed to excite the beam pion to a $3\pi$ state with invariant mass
\mThreePi.  The beam momentum $\vec{p}_\text{beam}$ is
defined in the laboratory frame.  The analysis is
limited to the kinematic range \SIvalRange{0.1}{\tpr}{1.0}{\GeVcsq}.
Typical values of \tmin are well below \SI{e-3}{\GeVcsq} for the
$3\pi$ mass range from \SIrange{0.5}{2.5}{\GeVcc} considered in this
analysis.

Since reaction~\eqref{eq:reaction} is dominated by Pomeron
exchange,\footnote{The Pomeron is a quasiparticle with vacuum quantum
  numbers and therefore has $\IG = 0^+$.} isospin and $G$ parity of
the beam pion are conserved so that the quantum numbers of the
intermediate state $X^-$ are restricted\footnote{We do not consider
  flavor-exotic states with isospin~2.} to $\IG = 1^-$.  This limits
the analysis to meson states that belong to the \piJ and \aJ
families.\footnote{Note that due to parity conservation, $a_0$ states
  cannot decay into \threePi.}  The $X^-$ decay is assumed to proceed
independently of the $X^-$ production; \ie the amplitude for the
process factorizes into production and decay amplitudes.

In our previous publication~\cite{Adolph:2015tqa}, the data were
subjected to a partial-wave analysis~(PWA) of the outgoing $3\pi$
system.  The employed PWA model relies on the isobar model, which
describes the $X^- \to \threePi$ decay as a sequence of two two-body
decays, $X^- \to \xi^0 \pi^-$ and $\xi^0 \to \twoPi$ via intermediate
\twoPi states $\xi^0$, the so-called \emph{isobars} (see
\cref{fig:3pi_reaction_isobar}).  Each isobar is characterized by its
\IGJPC quantum numbers and an assumed dependence of its decay
amplitude on the \twoPi invariant mass \mTwoPi, which in the simplest
case is a Breit-Wigner amplitude representing a $\pi\pi$ resonance.

The PWA model used in \refCite{Adolph:2015tqa} assumed that the data
are a mixture of interfering contributions of various partial waves
that are defined by the quantum numbers of the $X^-$ and their decay
modes.  This set of partial waves included six different isobars, and
we allowed for total spins $0 \leq J \leq 6$ and orbital angular
momenta $0 \leq L \leq 6$ between the isobars and the bachelor
$\pi^-$.  Independent fits of the set of partial-wave amplitudes to
the data were carried out in \num{1100} $(\mThreePi, \tpr)$ bins
without applying model assumptions about the resonance content of the
$3\pi$ system.  We refer to this first step that was performed prior
to the present analysis as \emph{mass-independent analysis}.  The
results of a PWA fit in a given $(\mThreePi, \tpr)$ bin were
represented in terms of a spin-density matrix that contains all
information about the partial-wave amplitudes and their mutual
interferences that can be extracted from the data.  This
mass-independent analysis is a prerequisite to searching for $3\pi$
resonances produced in reaction~\eqref{eq:reaction}, which can be
identified only if we combine the information contained in the
spin-density matrices over a wide range of \mThreePi.

In this paper, the results of the mass-independent analysis from
\refCite{Adolph:2015tqa} are used as input for a resonance-model fit,
which is also referred to as \emph{mass-dependent fit}.  In this
second analysis step, we search for $3\pi$ resonances that contribute
to the intermediate $X^-$ states by modeling the \mThreePi dependence
of the earlier extracted spin-density matrices over a wide range of
\mThreePi.  Resonances appear as characteristic structures in the
\mThreePi dependence not only of the moduli squared of the
partial-wave amplitudes, \ie in the \emph{partial-wave intensities},
but also of the mutual interference terms of the partial waves.  In
addition to the product of the moduli of the partial-wave amplitudes,
an interference term contains information about the relative phase
between a pair of waves.  The change of a relative phase with
increasing \mThreePi is called \emph{phase motion}.  The fit model
assumes that the partial-wave amplitudes can be described by a
coherent sum of Breit-Wigner amplitudes representing the resonances
and amplitudes that describe nonresonant components.  In a novel
approach, we extend this analysis technique that was used in most of
the previous analyses (see \eg\
\refsCite{daum:1980ay,Amelin:1995gt,Gunter:2000am,alekseev:2009aa,Salgado:2013dja})
by including for the first time to our knowledge the information on
the dependence of the partial-wave amplitudes on \tpr in the fit.  By
requiring that the shape parameters of the resonances are independent
of \tpr, a better separation of the resonant and nonresonant
components is achieved, which is a substantial improvement over
previous analyses.

Most of the details on the event selection and the mass-independent
analysis have already been presented in our previous
publication~\cite{Adolph:2015tqa}.  Therefore, we give in
\cref{sec:setup_and_event_selection} only a brief summary of the basic
features of the experimental setup and the event selection.
\Cref{sec:mass-independent_fit} contains a discussion of those details
of the mass-independent analysis from \refCite{Adolph:2015tqa} that
are relevant for the resonance-model fit.  In \cref{sec:method}, we
explain the fit model and the employed fitting method.  Because of the
large number of events, statistical uncertainties of the extracted
resonance parameters are negligible compared to systematic
uncertainties.  Hence we performed extensive systematic studies, which
are described in \cref{sec:systematics}.  The results of the
resonance-model fit are presented and discussed in \cref{sec:results}
grouped by the \JPC quantum numbers of the resonances.  This includes
a comparison of the obtained resonance parameters with world data and
a discussion of the extracted \tpr spectra of the resonant and
nonresonant components.  The \tpr dependence of the relative phases of
the wave components is discussed in \cref{sec:production_phases}.  In
\cref{sec:conclusions}, we summarize our findings.  The appendixes
contain the details about an alternative description of the
nonresonant contributions, about alternative formulations of the
\chisq~function that is minimized to determine the resonance
parameters, and about the systematic uncertainties of the extracted
resonance parameters.  The supplemental
material\ifMultiColumnLayout{~\cite{paper3_supplemental_material}}{ in
  \cref{sec:spin-dens_matrices,sec:phase-space_vol}} contains the
amplitude data that enter in the resonance-model fit, the full fit
result, and additional information required to perform the
resonance-model fit.  The data required to perform the resonance-model
fit are provided in computer-readable format at~\cite{paper3_hepdata}.
 %
%
%

\section{Experimental setup and event selection}
\label{sec:setup_and_event_selection}

The experimental setup and the data selection criteria are described
in detail in \refsCite{Adolph:2015tqa,haas:2014bzm}.  Here, we give
only a brief summary.

The COMPASS experiment~\cite{Abbon:2007pq,Abbon:2014aex} is located at
the M2~beam line of the CERN Super Proton Synchrotron.  The data used
for the analysis presented in this paper were recorded in the year
2008.  A beam of negatively charged hadrons with \SI{190}{\GeVc}
momentum and \SI{96.8}{\percent} $\pi^-$ content was incident on a
\SI{40}{\cm} long liquid-hydrogen target that was surrounded by a
recoil-proton detector (RPD).  Incoming pions were identified using a
pair of beam Cherenkov detectors (CEDARs) that were placed in the beam
line upstream of the target.  Outgoing charged particles were detected
by the tracking system, and their momenta were determined using two
large-aperture dipole magnets.  The large-acceptance high-precision
two-stage magnetic spectrometer was well suited for investigating
high-energy reactions at low to intermediate values of the reduced
four-momentum transfer squared \tpr.  For the present analysis, \tpr
was chosen to be in the range from \SIrange{0.1}{1.0}{\GeVcsq}, where
the lower bound is dictated by the acceptance of the RPD and the upper
bound by the decrease of the number of events with increasing \tpr.

Data were recorded using a trigger based on a recoil-proton signal in
the RPD in coincidence with an incoming beam particle and no signal in
the veto counters (see Sec.~II~B in \refCite{Adolph:2015tqa}).  In the
analysis, we require a production vertex located within the target
volume.  This vertex must have one incoming beam pion and three
outgoing charged particles.  The sum of the energies of the outgoing
particles, $E_\text{sum}$, is required to be equal to the average beam
energy within 2~standard deviations $\sigma_{E_\text{sum}}$, \ie
within $\pm \SI{3.78}{\GeV}$.  Contributions from double-diffractive
processes, in which also the target proton is excited, are suppressed
by the RPD and veto trigger signals and by requiring exactly one
recoil particle detected in the RPD that is back-to-back with the
outgoing \threePi system in the plane transverse to the beam
(transverse momentum balance; see Sec.~II~C in
\refCite{Adolph:2015tqa}).  Events are disregarded if the incoming
beam particle is identified by the CEDARs as a kaon.  If at least one
of the three forward-going particles is identified by the ring-imaging
Cherenkov detector (RICH) as not being a pion, the event is also
rejected.  In addition, we require Feynman-$x$ of the fastest
final-state $\pi^-$ to be below~0.9 for rapidity differences between
the fast $\pi^-$ and the slower \twoPi pair in the range from
\numrange{2.7}{4.5}.  This suppresses the small contamination by
centrally produced \twoPi final states in the analyzed mass range (see
Sec.~II~C in \refCite{Adolph:2015tqa}).  The selected kinematic region
of \SIvalRange{0.5}{\mThreePi}{2.5}{\GeVcc} and
\SIvalRange{0.1}{\tpr}{1.0}{\GeVcsq} contains a total of \num{46E6}
exclusive events that enter into the partial-wave analysis (see
\cref{sec:mass-independent_fit}).
 %
%
%

\section{Partial-wave decomposition}
\label{sec:mass-independent_fit}

We use a two-step procedure for the determination of the spectrum of
$3\pi$ resonances produced in the reaction \reaction.  In the first
analysis step published in \refCite{Adolph:2015tqa}, a partial-wave
decomposition was performed independently in 100~\mThreePi bins each
divided into 11~\tpr bins, which serves as input for the
resonance-model fit presented in this paper.  The PWA method and the
results are discussed in detail in \refCite{Adolph:2015tqa}.  Here, we
summarize the facts relevant for the resonance-model fit, which is
introduced in \cref{sec:method}.

Our basic assumption for the PWA model is that resonances dominate the
$3\pi$ intermediate states $X^-$ that are produced in the scattering
process.  We therefore describe the process as an inelastic two-body
scattering reaction $\pi^- + p \to X^- + p_{\text{recoil}}$ with
subsequent decay of $X^-$ into the three final-state pions,
$X^- \to \threePi$.

For fixed center-of-mass energy~$\sqrt{s}$, the kinematic distribution
of the final-state particles depends on \mThreePi, \tpr, and a set of
five additional phase-space variables represented by $\tau$.  The
latter fully describes the three-body decay.  The set of variables
used in our analysis is defined in Sec.~III~A of
\refCite{Adolph:2015tqa}.  For the reaction \reaction, a perfect
detector with unit acceptance would measure the intensity distribution
\begin{equation}
  \label{eq:intensity_def}
  \begin{splitOrNot}
  \mathcal{I}(\mThreePi, \tpr, \tau)
  \alignOrNot\equiv \frac{\dif{N}}{\dif{\mThreePi}\, \dif{\tpr}\, \dif{\varphi_3}(\mThreePi, \tau)} \newLineOrNot
  \alignOrNot\propto \frac{\dif{\sigma_{\reaction}}}{\dif{\mThreePi}\, \dif{\tpr}\, \dif{\varphi_3}(\mThreePi, \tau)} \newLineOrNot
  \alignOrNot\propto \mThreePi\, \Abs{\mathcal{M}_{fi}(\mThreePi, \tpr, \tau)}^2,
  \end{splitOrNot}
\end{equation}
where $N$ is the number of events, $\dif{\varphi_3}$ the
five-dimensional differential Lorentz-invariant three-body phase-space
element of the three outgoing pions, $\dif{\sigma_{\reaction}}$ the
differential cross section for the measured process, and
$\mathcal{M}_{fi}$ the transition matrix element from the initial to
the final state.\footnote{To simplify notation, the term
  $\Abs{\mathcal{M}_{fi}}^2$ is assumed to include incoherent sums,
  \eg over the helicities of the particles with nonzero spin [see
  \cref{eq:intensity_ansatz}].}  The right-hand side of
\cref{eq:intensity_def} is derived from Fermi's golden rule as given
\eg in \refCite{pdg_kinematics:2016}.  We factorize the phase space of
the four outgoing particles into the two-body phase space for $X^-$
and $p_{\text{recoil}}$ and the three-body phase space for the decay
$X^- \to \threePi$, which introduces the factor \mThreePi.  The
differential two-body phase space element is expressed in terms of
\tpr.  All constant factors have been dropped from the right-hand side
of \cref{eq:intensity_def}.  It is worth noting that, since
$\mathcal{I}$ is differential in the three-body phase-space element,
it is independent of the particular choice of the
variables~$\tau$.\footnote{The simplest parametrization of the
  differential three-body phase-space element is in terms of the
  energies of two of the final-state particles, \eg $E_1$ and $E_3$,
  and the Euler angles $(\alpha, \beta, \gamma)$ that define the
  spatial orientation of the plane that is formed by the daughter
  particles in the $X^-$~rest frame:
  \begin{equation*}
    \dif{\varphi_3}(\mThreePi, \underbrace{E_1, E_3, \alpha, \beta, \gamma}_{\equiv \tau})
    \propto \dif{E_1}\, \dif{E_3}\, \dif{\alpha}\, \dif{\cos \beta}\, \dif{\gamma}
  \end{equation*}
  For different choices of~$\tau$, the respective Jacobians have to be
  taken into account.}

Since we assume that the $3\pi$ intermediate state is dominated by
resonances, the production of $X^-$ can be treated independently of
its decay (see \cref{fig:3pi_reaction_isobar}).  The amplitude for a
particular intermediate state $X^-$ therefore factorizes into two
terms: \one~the \emph{transition amplitude}
$\mathcal{T}(\mThreePi, \tpr)$, which encodes the \mThreePi-dependent
strength and phase of the production of a state $X^-$ with specific
quantum numbers, and \two~the \emph{decay amplitude}
$\Psi(\mThreePi, \tau)$, which describes the decay of $X^-$ into a
particular \threePi final state.

As demonstrated in \refCite{Adolph:2015tqa}, we observe dominant
contributions of resonances in the \twoPi subsystem of the \threePi
final state.  Therefore, we factorize the three-body decay amplitude
into two two-body decay terms (see \cref{fig:3pi_reaction_isobar}).
This factorization is known as the \emph{isobar model}\footnote{An
  early detailed discussion can be found in \refCite{herndon:1973yn}.}
and the intermediate neutral \twoPi state $\xi^0$ is called the
\emph{isobar}.  In the first two-body decay, $X^- \to \xi^0 \pi^-$, a
relative orbital angular momentum $L$ appears.  The orbital angular
momentum in the isobar decay $\xi^0 \to \twoPi$ is equal to the spin
of the isobar.  For a given three-pion mass, the decay amplitude
accounts for the deviation of the kinematic distribution of the three
outgoing pions from the isotropic phase-space distribution and is
specified by the quantum numbers of $X^-$ (isospin $I$, $G$ parity,
spin $J$, parity $P$, $C$ parity, and the spin projection $M$) and its
decay mode ($\xi$, $L$).  For convenience, we introduce the
partial-wave index
\begin{equation}
  \label{eq:wave_index}
  a \equiv (\IG, \JPC, M, \xi, L).
\end{equation}
We describe the decay $X^- \to \xi^0 \pi^-$ in the Gottfried-Jackson
rest frame of the $X^-$ (see Sec.~III~A in \refCite{Adolph:2015tqa}),
where the quantization axis is chosen along the beam direction, and we
employ the \emph{reflectivity basis}, where positive and negative
values of the spin projection $M$ are combined to yield amplitudes
characterized by $M \geq 0$ and by the reflectivity quantum number
$\refl = \pm 1$~\cite{chung:1974fq}.  The reflectivity \refl is the
eigenvalue of the reflection through the $X^-$~production plane.  In
the high-energy limit, \refl corresponds to the naturality of the
exchange in the scattering process such that $\refl = +1$ corresponds
to natural spin parity of the exchanged Reggeon, \ie
$\JP = (\text{odd})^-$ or $(\text{even})^+$ transfer to the beam
particle.  Conversely, $\refl = -1$ corresponds to unnatural spin
parity of the exchanged Reggeon, \ie $\JP = (\text{even})^-$ or
$(\text{odd})^+$ transfer to the beam particle.

The isobar-model decay amplitudes are calculable using the helicity
formalism up to the unknown complex-valued couplings
$\alpha_{X \to \xi \pi}$ and $\alpha_{\xi \to \pi \pi}$, which appear
at each decay vertex (see \cref{fig:3pi_reaction_isobar}).  Assuming
that these couplings do not depend on the kinematics, they are moved
from the decay amplitudes into the transition amplitudes.  The
transition and decay amplitudes redefined in this way are represented
by $\Widebar{\mathcal{T}}_a(\mThreePi, \tpr)$ and
$\Widebar{\Psi}_a(\mThreePi, \tau)$.  It is worth noting that due to
this redefinition, the transition amplitudes $\Widebar{\mathcal{T}}_a$
depend not only on the $X^-$ quantum numbers but also on the $X^-$
decay mode.  Details are explained in Sec.~III~B of
\refCite{Adolph:2015tqa}.

We model the intensity distribution
$\mathcal{I}(\mThreePi, \tpr, \tau)$ of the final-state particles in
\cref{eq:intensity_def} as a truncated series of partial waves, which
are denoted by the index $a$ as defined in \cref{eq:wave_index}.  The
$N_\text{waves}^\refl$ partial-wave amplitudes for the contributing
intermediate $X^-$ states and their decays are summed coherently:
\begin{multlineOrEq}
  \label{eq:intensity_ansatz}
  \mathcal{I}(\mThreePi, \tpr, \tau) \newLineOrNot
  = \sum_{\refl = \pm1} \sum_{r = 1}^{N_r^\refl} \Abs[3]{\sum_a^{N_\text{waves}^\refl}
  \Widebar{\mathcal{T}}_a^{r \refl}(\mThreePi, \tpr)\,
  \Widebar{\Psi}_a^\refl(\mThreePi, \tau)}^2 \newLineOrNot
  + \Widebar{\mathcal{T}}_\text{flat}^2(\mThreePi, \tpr).
\end{multlineOrEq}
In the above formula,\footnote{\Cref{eq:intensity_ansatz} corresponds
  to Eq.~(17) in \refCite{Adolph:2015tqa}.  The explicit factor
  \mThreePi that appears on the right-hand side of
  \cref{eq:intensity_def} is absorbed into
  $\Widebar{\mathcal{T}}_a^{r \refl}(\mThreePi, \tpr)$.} the
contributions to the intensity distribution corresponding to
reflectivity \refl and \emph{rank index} $r$ (see next paragraph) are
summed incoherently.  The former is due to parity conservation that
forbids interference of states with different
reflectivities~\cite{chung:1974fq}.  We also introduced an additional
incoherently added wave that is isotropic in the three-body phase
space and is referred to as \emph{flat wave}.  The purpose of this
wave is to absorb intensity of events with three uncorrelated pions in
the final state, \eg nonexclusive background.  The corresponding
transition amplitude $\Widebar{\mathcal{T}}_\text{flat}$ is
real-valued.\footnote{The decay amplitude
  $\Widebar{\Psi}_\text{flat}(\mThreePi, \tau)$ of the flat wave is a
  constant and was set to unity.}

Several processes, \eg spin-flip and spin-nonflip processes or the
excitation of baryon resonances at the target vertex, may disturb the
coherence of the intermediate states.  Incoherence may also be
introduced by integrating over large ranges of \tpr, if intermediate
states are produced with different dependences on \tpr.  Incoherences
are incorporated by the additional rank index~$r$ for the transition
amplitudes, which is summed over incoherently [see
\cref{eq:intensity_ansatz}].  In general, the \emph{rank} $N_r$ may be
different in the two reflectivity sectors, \ie $N_r^\refl$.

The goal of the partial-wave analysis is to extract the unknown
transition amplitudes in \cref{eq:intensity_ansatz} from the data.
The $\Widebar{\mathcal{T}}_a^{r \refl}$ contain information about the
intermediate $3\pi$ resonances.  Since the \mThreePi dependence of the
transition amplitudes is unknown, the event sample is divided into
\mThreePi bins that are chosen to be much narrower than the width of
typical hadronic resonances.  The analyzed mass range
\SIvalRange{0.5}{\mThreePi}{2.5}{\GeVcc} is subdivided into
100~equidistant \mThreePi bins with a width of \SI{20}{\MeVcc}.
Within each mass bin, the \mThreePi dependence of the amplitudes is
assumed to be negligible, so that the transition amplitudes only
depend on \tpr.

We do not know \apriori the \tpr dependence of the transition
amplitudes.  In previous analyses, it was often assumed that the
\mThreePi and \tpr dependences are uncorrelated and the \tpr
dependence was modeled by real functions $g_a^\refl(\tpr)$.  These
functions were extracted from the analyzed data sample by integrating
over wide \mThreePi ranges, often only for groups of waves.  We have
shown in \refCite{Adolph:2015tqa} that for the process under study
this assumption is not valid. The \tpr dependence of the intensity of
individual waves depends on \mThreePi and may differ significantly
from wave to wave.  This agrees with previous studies of diffractive
dissociation of pions (see \eg\
\refsCite{alekseev:2009aa,dzierba:2005jg,kachaev:2001jj,daum:1980ay}),
which revealed contributions of nonresonant background processes such
as the Deck effect~\cite{deck:1964hm}.  The nonresonant processes
typically exhibit \mThreePi and \tpr dependences that are different
from those of resonances.  In particular, the analyses presented
in~\refsCite{dzierba:2005jg,daum:1980ay} showed the importance of the
kinematic variable \tpr in a partial-wave analysis of the
diffractively produced $3\pi$ system and illustrated the power of
accounting for the different \tpr dependences of the reaction
mechanisms and also of the different resonances.  Therefore, for each
\mThreePi bin the partial-wave decomposition was performed
independently in 11~nonequidistant \tpr slices of the analyzed range
\SIvalRange{0.1}{\tpr}{1.0}{\GeVcsq} as listed in \cref{tab:t-bins}.
Within each \tpr bin, we assumed the transition amplitudes to be
independent of \tpr.  In this work, we further develop this approach
to better disentangle resonant and nonresonant components (see
\cref{sec:method,sec:production_phases}).

\begin{table*}[tbp]
  \sisetup{%
    round-mode = places,
    round-precision = 3
  }
  \caption{Borders of the 11~nonequidistant \tpr bins, in which
    the partial-wave analysis is performed.  The intervals are chosen
    such that each bin contains approximately \num[round-mode =
    places, round-precision = 1]{4.6e6} events.  Only the last range
    from \SIrange{0.448588}{1.000000}{\GeVcsq} is subdivided further
    into two bins.}
  \label{tab:t-bins}
  \renewcommand{\arraystretch}{1.2}
  \newcolumntype{Z}{%
    >{\Makebox[0pt][c]\bgroup}%
    c%
    <{\egroup}%
  }
  \setlength{\tabcolsep}{0pt}  %
  \ifMultiColumnLayout{%
    \begin{tabular}{l@{\extracolsep{12pt}}c@{\extracolsep{6pt}}Z*{10}{cZ}c}
      \hline
      \hline
      Bin && 1 && 2 && 3 && 4 && 5 && 6 && 7 && 8 && 9 && 10 && 11 & \\
      \hline
      \tpr [\si{\GeVcsq}] &
      \num{0.100000} &&
      \num{0.112853} &&
      \num{0.127471} &&
      \num{0.144385} &&
      \num{0.164401} &&
      \num{0.188816} &&
      \num{0.219907} &&
      \num{0.262177} &&
      \num{0.326380} &&
      \num{0.448588} &&
      \num{0.724294} &&
      \num{1.000000} \\
      \hline
      \hline
    \end{tabular}%
  }{%
    \begin{tabular}{l@{\extracolsep{12pt}}c@{\extracolsep{6pt}}Z*{5}{cZ}c}
      \hline
      \hline
      Bin && 1 && 2 && 3 && 4 && 5 && 6 \\
      \hline
      \tpr [\si{\GeVcsq}] &
      \num{0.100000} &&
      \num{0.112853} &&
      \num{0.127471} &&
      \num{0.144385} &&
      \num{0.164401} &&
      \num{0.188816} &&
      \num{0.219907} \\
      \hline
      \hline
    \end{tabular}
    \\[3ex]
    \begin{tabular}{l@{\extracolsep{12pt}}c@{\extracolsep{6pt}}Z*{4}{cZ}c}
      \hline
      \hline
      Bin && 7 && 8 && 9 && 10 && 11 & \\
      \hline
      \tpr [\si{\GeVcsq}] &
      \num{0.219907} &&
      \num{0.262177} &&
      \num{0.326380} &&
      \num{0.448588} &&
      \num{0.724294} &&
      \num{1.000000} \\
      \hline
      \hline
    \end{tabular}%
  }
\end{table*}

In order to simplify notation, we consider the intensity in
\cref{eq:intensity_ansatz} in a particular $(\mThreePi, \tpr)$ bin.
Within this kinematic bin, \mThreePi and \tpr are considered to be
constant, and hence $\mathcal{I}$ is only a function of the set $\tau$
of phase-space variables.

In the resonance-model fit, special care has to be taken about the
normalization of the transition amplitudes.  A consistent
normalization that makes the transition amplitudes comparable across
different experiments is achieved by normalizing the decay amplitudes
to the integrals $I_{a a}^\refl$, which are the diagonal elements of
the integral matrix
\begin{equation}
  \label{eq:integral_matrix_def}
  I_{a b}^\refl(\mThreePi)
  \equiv \int\! \dif{\varphi_3(\tau; \mThreePi)}\,
  \Widebar{\Psi}_a^\refl(\tau; \mThreePi)\, \Widebar{\Psi}_b^{\refl \text{*}}(\tau; \mThreePi),
\end{equation}
where $a$~and~$b$ are wave indices as defined in \cref{eq:wave_index}.
We define\footnote{Since the decay amplitude
  $\Widebar{\Psi}_\text{flat}$ of the flat wave was set to unity, the
  corresponding normalized decay amplitude is given by
  \begin{equation}
    \label{eq:decay_amplitude_norm_flat}
    \Psi_\text{flat}(\tau; \mThreePi)
    \equiv \frac{1}{\sqrt{V_{\varphi_3}(\mThreePi)}}
  \end{equation}
  with
  \begin{equation}
    \label{eq:phase_space_vol}
    V_{\varphi_3}(\mThreePi)
    \equiv \int\! \dif{\varphi_3(\tau; \mThreePi)}.
  \end{equation}}
\begin{equation}
  \label{eq:decay_amplitude_norm}
  \Psi_a^\refl(\tau; \mThreePi)
  \equiv \frac{\Widebar{\Psi}_a^\refl(\tau; \mThreePi)}{\sqrt{I_{a
        a}^\refl(\mThreePi)}}.
\end{equation}

The normalization of the transition amplitudes is determined by the
expression for the number of events $N_{\text{pred}}$ predicted for
the $(\mThreePi, \tpr)$ bin by the model in
\cref{eq:intensity_ansatz}:
\begin{equation}
  \label{eq:expected_ev_nmb_corr}
  N_{\text{pred}}(\mThreePi, \tpr)
  = \int\! \dif{\varphi_3(\tau; \mThreePi)}\, \mathcal{I}(\tau; \mThreePi, \tpr).
\end{equation}
Based on \cref{eq:decay_amplitude_norm}, the transition amplitudes are
redefined according to\footnote{Similarly, the transition amplitude of
  the flat wave is redefined based on
  \cref{eq:decay_amplitude_norm_flat}:
  \begin{equation}
    \label{eq:flat_amplitude_norm}
    \mathcal{T}_\text{flat}(\mThreePi, \tpr)
    \equiv \Widebar{\mathcal{T}}_\text{flat}(\mThreePi, \tpr)\, \sqrt{\smash[b]{V_{\varphi_3}}(\mThreePi)}.
  \end{equation}}
\begin{equation}
  \label{eq:prod_amplitude_norm}
  \mathcal{T}_a^{r \refl}(\mThreePi, \tpr)
  \equiv \Widebar{\mathcal{T}}_a^{r \refl}(\mThreePi, \tpr)\, \sqrt{I_{a a}^\refl(\mThreePi)},
\end{equation}
so that $\mathcal{I}$ remains unchanged.  Using the fact that the
decay amplitudes $\Psi_a^\refl$ are normalized via
\cref{eq:decay_amplitude_norm,eq:decay_amplitude_norm_flat},
\cref{eq:expected_ev_nmb_corr} reads
\begin{multlineOrEq}
  \label{eq:expected_ev_nmb_corr_amp}
  N_{\text{pred}}
  = \sum_{\refl = \pm1} \Bigg\{ \sum_a^{N_\text{waves}^\refl}
    \sum_{r = 1}^{N_r^\refl} \Abs[1]{\mathcal{T}_a^{r \refl}}^2 \newLineOrNot
    \ifMultiColumnLayout{\mbox{}\hfill}{{}}
    + 2 \sum_{a < b}^{N_\text{waves}^\refl}
    \Re\!\sBrk[4]{\sum_{r = 1}^{N_r^\refl} \mathcal{T}_a^{r \refl}\, \mathcal{T}_b^{r \refl \text{*}}
      \frac{I_{a b}^\refl}{\sqrt{I_{a a}^\refl}\, \sqrt{I_{b b}^\refl}}} \Bigg\} \newLineOrNot
    + \mathcal{T}_\text{flat}^2.
\end{multlineOrEq}

We introduce the \emph{spin-density matrix} for the
$(\mThreePi, \tpr)$ bin,
\begin{equation}
  \label{eq:spin_density}
  \varrho_{a b}^\refl(\mThreePi, \tpr)
  \equiv \sum_{r = 1}^{N_r^\refl} \mathcal{T}_a^{r \refl}(\mThreePi, \tpr)\, \mathcal{T}_b^{r \refl \text{*}}(\mThreePi, \tpr),
\end{equation}
which represents the full information that can be obtained about the
$X^-$ states.  The parameter~$N_r^\refl$ is the \emph{rank} of the
spin-density matrix.  With the above,
\cref{eq:expected_ev_nmb_corr_amp} simplifies to
\begin{multlineOrEq}
  \label{eq:expected_ev_nmb_corr_rho}
  N_{\text{pred}}
  = \sum_{\refl = \pm1} \Bigg\{ \sum_a^{N_\text{waves}^\refl}
  \ifMultiColumnLayout{\overbrace}{\underbrace}{\varrho_{a a}^\refl
    \ifMultiColumnLayout{}{\vphantom{\sum_r}}}\ifMultiColumnLayout{^}{_}{%
    \mathclap{\displaystyle \text{\ifMultiColumnLayout{\hspace*{2.5em}}{}Intensities}}} \newLineOrNot
  \ifMultiColumnLayout{\mbox{}\hfill}{{}}
  + \sum_{a < b}^{N_\text{waves}^\refl}
  \underbrace{2 \Re\!\sBrk[4]{\varrho_{a b}^\refl
      \frac{I_{a b}^\refl}{\sqrt{I_{a a}^\refl}\, \sqrt{I_{b b}^\refl}}}}_{%
    \displaystyle \text{Overlaps}} \Bigg\} \newLineOrNot
  + \mathcal{T}_\text{flat}^2.
\end{multlineOrEq}
From this equation, we can derive an interpretation for the
spin-density matrix elements.  The diagonal elements
$\varrho_{a a}^\refl$ are the \emph{partial-wave intensities}, \ie the
expected number of events in wave~$a$.\footnote{For a real experiment,
  this corresponds to the acceptance-corrected number of events.}  The
off-diagonal elements $\varrho_{a b}^\refl$, which contain information
about the relative phase between waves~$a$ and~$b$, contribute to the
so-called \emph{overlaps}, which are the number of events originating
from the interference between waves~$a$ and~$b$.\footnote{For
  constructive interference, this number is positive; for destructive
  interference, it is negative.}  Limiting the summation in
\cref{eq:expected_ev_nmb_corr_rho} to a subset of partial waves yields
the expected number of events in these waves including all
interferences.  Such sums will be denoted as \emph{coherent sums} of
partial waves in the following text.

We used an extended maximum-likelihood approach~\cite{Barlow:1990vc}
to determine the unknown transition amplitudes
$\mathcal{T}_a^{r \refl}$ by fitting the model intensity
$\mathcal{I}(\tau)$ of \cref{eq:intensity_ansatz} to the measured
$\tau$ distribution, in narrow bins of \mThreePi and \tpr.  The
extended likelihood function for a $(\mThreePi, \tpr)$
bin,\footnote{For better readability, we do not explicitly write the
  \mThreePi and \tpr dependences.}
\begin{equation}
  \label{eq:likelihood_function_ansatz}
  \mathcal{L}
  = \underbrace{\frac{\Widebar{N}^{N}\, e^{-\Widebar{N}}}{N!\vphantom{\Widebar{N}}}}_{%
    \substack{\displaystyle{\text{Poisson\vphantom{y}}} \\ \displaystyle{\text{probability}}}}\,
  \prod_{i = 1}^{N}
  \underbrace{\frac{\mathcal{I}(\tau_i)}{\Widebar{N}}}_{%
    \mathclap{\substack{\displaystyle{\text{Probability}} \\ \displaystyle{\text{for event $i$}}}}},
\end{equation}
contains a Poisson term for the actually \emph{observed} number of
events $N(\mThreePi, \tpr)$ and the number of events
\begin{multlineOrEq}
  \label{eq:expected_ev_nmb}
  \Widebar{N}(\mThreePi, \tpr) \newLineOrNot
  = \int\! \dif{\varphi_3(\tau; \mThreePi)}\, \eta(\tau; \mThreePi, \tpr)\, \mathcal{I}(\tau; \mThreePi, \tpr)
\end{multlineOrEq}
that is \emph{expected} to be observed by the detector.  Via this
term, the detection efficiency $\eta(\tau; \mThreePi, \tpr)$ of the
experimental setup is taken into account by the PWA model.  In
addition, \cref{eq:expected_ev_nmb} together with
\cref{eq:decay_amplitude_norm,eq:decay_amplitude_norm_flat} ensures
the correct normalization of the transition amplitudes according to
\cref{eq:prod_amplitude_norm,eq:flat_amplitude_norm}.  This also fixes
the normalization of the diagonal elements of the spin-density matrix
in \cref{eq:spin_density} to the acceptance-corrected number of events
in the particular wave.

In principle, the partial-wave expansion in \cref{eq:intensity_ansatz}
includes an infinite number of waves.  In practice, the expansion
series has to be truncated.  We thus have to define a \emph{wave set}
describing the data sufficiently well, without too many free
parameters.  We included \pipiS, \Prho, \PfZero[980], \PfTwo,
\PfZero[1500], and \PrhoThree as isobars in the fit model, where
\pipiS represents a parametrization of the broad component of the
\pipiSW, which dominates the \mTwoPi spectrum from low to intermediate
two-pion masses and exhibits a slow phase motion (see Fig.~10 in
\refCite{Adolph:2015tqa}).  This selection of isobars is based on
features observed in the \twoPi invariant mass spectrum (see
\refCite{Adolph:2015tqa}) and on analyses of previous
experiments~\cite{amelin:1995gu,adams:1998ff,kachaev:2001jj,chung:2002pu,dzierba:2005jg,alekseev:2009aa}.
Based on the six isobars, we have constructed a set of 88~partial
waves, \ie 80~waves with reflectivity $\refl = +1$, seven waves with
$\refl = -1$, and a noninterfering flat wave representing three
uncorrelated pions (see Table~IX in Appendix~A of
\refCite{Adolph:2015tqa} for a complete list).  This wave set is the
largest used so far in a PWA of the \threePi final state.  It includes
partial waves with spin $J \leq 6$, orbital angular momentum
$L \leq 6$, and spin projection $M = \numlist{0;1;2}$.  The wave set
consists mainly of positive-reflectivity waves, which is expected due
to Pomeron dominance at high energies.  As discussed in
\refCite{Adolph:2015tqa}, it was found that the ranks
$N_r^{(\refl = +1)} = 1$ and $N_r^{(\refl = -1)} = 2$ describe the
data well.  In the reflectivity basis, partial waves are completely
defined by the wave index~$a$, as given in \cref{eq:wave_index}, and
the reflectivity \refl.  For the remaining text, we adopt the
\emph{partial-wave notation}
\wave{J}{PC}{M}{\refl}{[\text{isobar}]}{L}.

The \emph{total intensity} of all partial waves is defined as the
total number of acceptance-corrected events as given by
\cref{eq:expected_ev_nmb_corr}.  The \emph{relative intensity} of a
particular partial wave, as \eg listed in
\cref{tab:method:fitmodel:waveset} in \cref{sec:method}, is defined as
the ratio of its intensity integral over the analyzed range
\SIvalRange{0.5}{\mThreePi}{2.5}{\GeVcc} and the corresponding
integral of the total intensity.  Owing to interference effects
between the waves, \ie overlaps, this value is in general different
from the contribution of a wave to the total intensity.\footnote{The
  relative intensities include effects from interference due to Bose
  symmetrization of the two indistinguishable final-state $\pi^-$.}
Hence in our fit, the relative intensities of all 88~partial waves add
up to \SI{105.3}{\percent} instead of \SI{100}{\percent}.

As shown in \refCite{Adolph:2015tqa}, the waves with negative
reflectivity corresponding to unnatural-parity exchange processes
contribute only \SI{2.2}{\percent} to the total intensity and do not
interfere with the positive-reflectivity waves.  This dominance of
natural-parity exchange processes is consistent with the expected
dominance of the Pomeron contribution at COMPASS energies.  In this
paper, we only consider a selection of positive-reflectivity partial
waves.
 %
%
%

\section{Resonance-model fit}
\label{sec:method}

The goal of the analysis described in this paper is to extract $3\pi$
resonances contributing to the reaction \reaction and to determine
their quantum numbers and parameters, \ie masses and widths.  The
starting point of the analysis is the spin-density matrix
$\varrho_{a b}(\mThreePi, \tpr)$ as defined in \cref{eq:spin_density}.
It has been extracted from the data in the first step of the analysis
by performing a partial-wave decomposition independently in 100~bins
of \mThreePi and 11~bins of \tpr for each \mThreePi bin using a model
with 88~waves (see \refCite{Adolph:2015tqa} and
\cref{sec:mass-independent_fit}).

For the resonance extraction presented here, we select a subset of
waves that exhibit resonance signals in their intensity spectra and in
their phase motions.  Some waves contain well-known resonances that
are used as an interferometer to study the resonance content of more
interesting waves, such as the spin-exotic
\wave{1}{-+}{1}{+}{\Prho}{P} wave.  All selected waves have positive
reflectivity.  Since the spin-density submatrix of the $\refl = +1$
waves was chosen to have rank~1, we will drop reflectivity and rank
indices from \cref{eq:spin_density} and from all formulas that will
follow below.  We therefore write
\begin{equation}
  \label{eq:method:spindens}
  \varrho_{a b}(\mThreePi, \tpr)
  = \mathcal{T}_a(\mThreePi, \tpr)\, \mathcal{T}_b^\text{*}(\mThreePi, \tpr).
\end{equation}
For the selected waves, the \mThreePi and \tpr dependences of the
corresponding elements of the spin-density submatrix in
\cref{eq:method:spindens} are parametrized in terms of the transition
amplitudes.  The fit model must therefore reproduce not only the
measured partial-wave intensities but also their mutual interferences.
Performing the analysis on the amplitude level greatly improves the
sensitivity for potential resonance signals.  We employ a
parametrization similar to the ones used by previous analyses (see
\eg\
\refsCite{daum:1980ay,Amelin:1995gt,Chung:1999we,chung:2002pu,alekseev:2009aa,Salgado:2013dja}).
In the following, model quantities will be distinguished from the
corresponding measured quantities by a hat
(\enquote{\,$\widehat{\phantom{\rho}}$\;}).

We model the transition amplitudes $\mathcal{T}_a(\mThreePi, \tpr)$ as
the product of an amplitude $\mathcal{P}(\mThreePi, \tpr)$, which
accounts for the overall strength of the production of a $3\pi$ system
with mass \mThreePi at a given \tpr (see \cref{sec:method:fitmodel}),
and a term that coherently sums over possible resonance propagators
and nonresonant background contributions of the $3\pi$ system with
quantum numbers defined by the wave index~$a$ [see
\cref{eq:wave_index}].  The model $\widehat{\mathcal{T}}_a$ for the
measured transition amplitude $\mathcal{T}_a$ for wave~$a$ is
\begin{multlineOrEq}
  \label{eq:method:transitionampl}
  \widehat{\mathcal{T}}_a(\mThreePi, \tpr)
  = \sqrt{I_{a a}(\mThreePi)}\, \sqrt{\mThreePi\vphantom{I_a}}\;
  \mathcal{P}(\mThreePi, \tpr)\, \newLineTimesOrNot
  \sum_{\mathclap{j\; \in\; \mathbb{S}_a}} \mathcal{C}_a^j(\tpr)\, \mathcal{D}_j(\mThreePi, \tpr; \zeta_j).
\end{multlineOrEq}
Here, $I_{a a}$ is the \emph{decay phase-space volume} of wave~$a$ as
defined in \cref{eq:integral_matrix_def}.  This factor enters, because
the partial-wave intensities $\abs{\mathcal{T}_a}^2$ are normalized
via \cref{eq:prod_amplitude_norm} to represent the
acceptance-corrected number of events in wave~$a$.  The factor
$\sqrt{\mThreePi}$ results from the splitting of the four-body phase
space of the final-state particles in \cref{eq:intensity_def}.  The
functions $\mathcal{D}_j(\mThreePi, \tpr; \zeta_j)$ are the
\emph{dynamical amplitudes} that represent the resonant or nonresonant
wave components, which are enumerated by the index~$j$.  The coherent
sum runs over the subset $\mathbb{S}_a$ of the indices of those wave
components that we assume to appear in wave a.  The dynamical
amplitudes depend on the set $\zeta_j$ of \emph{shape parameters},
which are \eg the masses and widths in the case of resonance
components.  It should be stressed that if the same wave component
$\mathcal{D}_j(\mThreePi, \tpr; \zeta_j)$ appears in several partial
waves, which must have the same \JPC quantum numbers, it has the same
values of the shape parameters $\zeta_j$.  The coefficients
$\mathcal{C}_a^j(\tpr)$ in \cref{eq:method:transitionampl} are the
so-called \emph{coupling amplitudes}.  They collect the unknown parts
of the model, which are the \tpr dependences of the production
strengths and phases of the $X^-$ and the complex-valued couplings,
$\alpha_{X \to \xi \pi}$ and $\alpha_{\xi \to \pi \pi}$, which appear
at the two vertices in the isobar decay chain.

Based on \cref{eq:method:transitionampl}, we can formulate the model
for the spin-density submatrix of the selected waves
\begin{wideEqOrNot}%
  \begin{equation}
    \label{eq:method:param:spindens}
    \begin{aligned}
      \widehat{\varrho}_{a b}(\mThreePi, \tpr)
      &= \widehat{\mathcal{T}}_a(\mThreePi, \tpr)\, \widehat{\mathcal{T}}_b^\text{*}(\mThreePi, \tpr) \\
      &= \ifMultiColumnLayout{}{\aligned[t]}
        \ifMultiColumnLayout{}{&}\sqrt{I_{a a}(\mThreePi)}\, \sqrt{I_{b b}(\mThreePi)}\;
        \mThreePi\, \Abs[1]{\mathcal{P}(\mThreePi, \tpr)}^2 \ifMultiColumnLayout{}{\\}
        \ifMultiColumnLayout{}{&\times} \sBrk[4]{\,\smashoperator[r]{\sum_{j\; \in\; \mathbb{S}_a}}
          \mathcal{C}_a^j(\tpr)\, \mathcal{D}_j(\mThreePi, \tpr; \zeta_j)}
        \sBrk[4]{\,\smashoperator[r]{\sum_{k\; \in\; \mathbb{S}_b}}
          \mathcal{C}_b^k(\tpr)\, \mathcal{D}_k(\mThreePi, \tpr; \zeta_k)}^\text{*},
      \ifMultiColumnLayout{}{\endaligned}
    \end{aligned}
  \end{equation}
\end{wideEqOrNot}%
which describes the \mThreePi and \tpr dependences of the measured
spin-density matrix elements $\varrho_{a b}(\mThreePi, \tpr)$.  The
free parameters to be determined by the resonance-model fit are the
coupling amplitudes $\mathcal{C}_a^j(\tpr)$ and the shape parameters
$\zeta_j$.

In \cref{eq:method:param:spindens} we extended the commonly used
ansatz for the parametrization of the spin-density matrix to
explicitly include the \tpr dependence.  In particular, the coupling
amplitudes $\mathcal{C}_a^j(\tpr)$ are allowed to take different
values in each \tpr bin.  This novel approach allows us to perform for
the first time a \tpr-resolved resonance-model fit.  The \tpr
information that was extracted in the mass-independent analysis
performed in the first analysis step (see
\cref{sec:mass-independent_fit}) is exploited here to better separate
the resonant and nonresonant contributions by allowing them to have
different \tpr dependences.  The resonance-model fit yields as
additional results the \tpr dependence of the intensity and the
production phases of the wave components (see
\cref{sec:method:tp,sec:production_phases} ).

Assuming factorization of production and decay of the intermediate
$3\pi$ state $X^-$, the resonant amplitudes
$\mathcal{D}^\text{R}_j(\mThreePi; \zeta^\text{R}_j)$, which represent
the on-shell propagators of the produced $3\pi$ resonances, should be
independent of \tpr.  This is in particular true for the corresponding
shape parameters $\zeta^\text{R}_j$ of the resonant amplitudes, \ie
the masses and widths of the resonances.  This constraint is built
into the model by using the same shape parameters across all \tpr
bins.  Only the strengths and coupling phases of the resonant
components, which are represented by the $\mathcal{C}_a^j(\tpr)$, can
be chosen freely by the fit for each individual \tpr bin.  We exploit
the factorization of production and decay further for the case, where
a resonance appears in several partial waves, which have the same
\JPCMrefl quantum numbers.  These waves represent different decay
modes of the same $X^-$~state and differ only in the isobar~$\xi^0$ or
the orbital angular momentum~$L$.  The resonant amplitude is expected
to follow the same \tpr dependence in these partial waves.  This is
built into the model by fixing the \tpr dependence
$\mathcal{C}_b^j(\tpr)$ of a resonance~$j$ that appears in wave~$b$ to
the \tpr dependence $\mathcal{C}_a^j(\tpr)$ that this resonance has in
wave~$a$ via
\begin{equation}
  \label{eq:method:branchingdefinition}
  \mathcal{C}_b^j(\tpr)
  = \prescript{}{b}{\mathcal{B}}_a^j\, \mathcal{C}_a^j(\tpr).
\end{equation}
This replaces the set of independent coupling amplitudes
$\mathcal{C}_b^j(\tpr)$ for wave~$b$ by a single \tpr-independent
complex-valued \emph{branching amplitude}
$\prescript{}{b}{\mathcal{B}}_a^j$ as a free fit parameter.  This
quantity represents the relative strength and phase of the two decay
modes of resonance~$j$.  The constraint expressed by
\cref{eq:method:branchingdefinition} significantly reduces the number
of free parameters and was also found to stabilize the fit (see
\cref{sec:systematics,sec:results}).

In general, the above assumptions do not hold for the nonresonant
amplitudes
$\mathcal{D}^\text{NR}_j(\mThreePi, \tpr; \zeta^\text{NR}_j)$.  The
shape of their \mThreePi distribution may vary with \tpr and may also
depend on the $X^-$ quantum numbers and decay mode.  Therefore, for
each wave in the fit, a separate nonresonant component is added to the
model.  Although the nonresonant amplitudes may have an explicit \tpr
dependence, the shape parameters $\zeta^\text{NR}_j$ are kept the same
across all \tpr bins.

\subsection{Fit model}
\label{sec:method:fitmodel}

Ideally, the resonance model would describe the \mThreePi dependence
of the full $88 \times 88$ spin-density matrix obtained from the PWA
fit in the first analysis step.  However, in practice such a fit would
require very large computing resources owing to the large number of
free parameters.  In addition, some partial waves, which mostly have
small relative intensities, are affected by imperfections in the PWA
model.  These imperfections may cause artifacts at the stage of the
mass-independent analysis that the physical model is not able to
describe.  Thus the resonance-model fit is commonly performed using
only a selected submatrix of the spin-density matrix.  For the present
analysis, we selected a subset of 14~waves that are listed in
\cref{tab:method:fitmodel:waveset} out of the 88~waves used in the
partial-wave decomposition (see Table~IX in Appendix~A of
\refCite{Adolph:2015tqa}).  Compared to previous analyses of the
$3\pi$ final state this constitutes the so far largest wave set
included in a resonance-model fit.  The sum of the relative
intensities (see definition in \cref{sec:mass-independent_fit}) of the
14~waves is \SI{56.8}{\percent}, whereas the coherent sum of these
waves amounts to \SI{57.9}{\percent}.  The intensity distributions of
the waves are discussed in detail in \refCite{Adolph:2015tqa} with the
exception of the spin-exotic \wave{1}{-+}{1}{+}{\Prho}{P} wave.  The
waves contain signals of the well-known resonances \PaOne, \PaTwo,
\PpiTwo, \Ppi[1800], \PpiTwo[1880], and \PaFour, which appear as peaks
in the intensity distributions of the partial waves with the
corresponding quantum numbers.  In addition, the set of selected waves
includes a clear signal of the novel resonancelike \PaOne[1420], which
was first reported in \refCite{Adolph:2015pws}, and potential signals
of the less well-known or disputed states \PpiOne[1600], \PaOne[1640],
and \PaTwo[1700].  In the development of the analysis model it was
found that a third $\JPC = 2^{-+}$ resonance, the \PpiTwo[2005], is
required to describe the data.

\subsubsection{Parametrization of the dynamical amplitudes for resonances}
\label{sec:method:fitmodel:resonances}

The selected 14~waves are described using the resonance model of
\cref{eq:method:param:spindens} with six \aJ-like and five \piJ-like
resonances.  The resonances are parametrized using relativistic
Breit-Wigner amplitudes~\cite{breit:1936zzb},
\begin{equation}
  \label{eq:BreitWigner}
  \mathcal{D}^\text{R}_j(\mThreePi; \underbrace{m_j, \Gamma_j}_{\displaystyle{\equiv \zeta^\text{R}_j}})
  = \frac{m_j\, \Gamma_j}{m_j^2 - \mThreePi^2 - i\, m_j\, \Gamma_{j, \text{tot}}(\mThreePi)},
\end{equation}
with the mass-dependent total width
$\Gamma_{j, \text{tot}}(\mThreePi)$.  The shape parameters to be
determined by the fit are mass $m_j$ and width $\Gamma_j$ of the
resonance~$j$.  For most resonances, the decay modes and relative
branching fractions are not or only poorly known.  In these cases, we
approximate the mass-dependent width by a constant:
\begin{equation}
  \label{eq:method:fixedwidth}
  \Gamma_{j, \text{tot}}(\mThreePi) \approx \Gamma_j.
\end{equation}

Only for \PaOne and \PaTwo are different parametrizations used.  Due
to the large width of the \PaOne, we use the Bowler parametrization
[\namecref{eq:method:bowlerG}~(9) in \refCite{Bowler:1987bj}] to
account for the variation of the decay phase space across the
resonance width:
\begin{multlineOrEq}
  \label{eq:method:bowlerG}
  \Gamma_{\PaOne, \text{tot}}(\mThreePi)
  \newLineOrNot
  = \Gamma_{\PaOne}\, \frac{I_{aa}(\mThreePi)}{I_{aa}(m_{\PaOne})}\, \frac{m_{\PaOne}}{\mThreePi}
\end{multlineOrEq}
with $a = \wave{1}{++}{0}{+}{\Prho}{S}$.  Here, $I_{aa}$ is the decay
phase-space volume of the \wave{1}{++}{0}{+}{\Prho}{S} wave calculated
according to \cref{eq:integral_matrix_def}, which takes into account
the finite width of the \Prho, the angular-momentum barrier factor in
the \Prho decay, and the Bose symmetrization of the decay amplitude.

For the \PaTwo, we approximate the total width by assuming that it is
saturated by the two dominant decay modes, $\Prho \pi$ and
$\Peta* \pi$, both in a
$D$~wave~\cite{beladidze:1993km,Adolph:2014rpp},\footnote{We neglect
  the additional mass dependence of the \PaTwo width that would be
  induced by the $\omega\pi\pi$ and \KKbar decay modes, which have
  branching fractions of \SI{10.6(32)}{\percent} and
  \SI{4.9(8)}{\percent}, respectively~\cite{Patrignani:2016xqp}.}
\begin{wideEqOrNot}%
  \begin{multline}
    \label{eq:method:a2dynamicwidth}
    \Gamma_{\PaTwo, \text{tot}}(\mThreePi)
    = \Gamma_{\PaTwo}\, \frac{m_{\PaTwo}}{\mThreePi}\,
    \Bigg[ (1 - x)\, \frac{q_{\Prho* \pi}(\mThreePi)}{q_{\Prho* \pi}(m_{\PaTwo})}\,
    \frac{F_2^2\rBrk[1]{q_{\Prho* \pi}(\mThreePi)}}{F_2^2\rBrk[1]{q_{\Prho* \pi}(m_{\PaTwo})}} \\
    {} + x\,  \frac{q_{\Peta* \pi}(\mThreePi)}{q_{\Peta* \pi}(m_{\PaTwo})}\,
    \frac{F_2^2\rBrk[1]{q_{\Peta* \pi}(\mThreePi)}}{F_2^2\rBrk[1]{q_{\Peta* \pi}(m_{\PaTwo})}} \Bigg].
  \end{multline}
\end{wideEqOrNot}%
In \cref{eq:method:a2dynamicwidth}, we neglect the width of the \Prho
and use the quasi-two-body approximation, where $q_{\xi \pi}$ is the
two-body breakup momentum in the decay $X^- \to \xi^0 \pi^-$.  It is
given by
\begin{multlineOrEq}
  \label{eq:method:breakupmomentum}
  q_{\xi \pi}^2(\mThreePi)
  \newLineOrNot
  = \frac{\sBrk{\mThreePi^2 - (m_\pi + m_\xi)^2} \sBrk{\mThreePi^2 - (m_\pi - m_\xi)^2}}{4\mThreePi^2}
\end{multlineOrEq}
with $m_\xi$ being the mass of the isobar~$\xi^0$.\footnote{For the
  \PaTwo, the lower bound of the fitted \mThreePi range was chosen
  such that $q_{\Prho* \pi}^2 > 0$.}  The $F_\ell(q_{\xi \pi})$ terms
in \cref{eq:method:a2dynamicwidth} are the Blatt-Weisskopf
angular-momentum barrier factors~\cite{blatt:1952}, which take into
account the centrifugal-barrier effect caused by the orbital angular
momentum $\ell = 2$ between the bachelor $\pi^-$ and the \Prho or the
\Peta*.  We use the parametrization of von~Hippel and
Quigg~\cite{VonHippel:1972fg} as given in Sec.~IV~A of
\refCite{Adolph:2015tqa} with a range parameter of
$q_R = \SI{200}{\MeVc}$.\footnote{This corresponds to an assumed
  strong-interaction range of \SI{1}{\femto\meter}.}  We approximate
the relative branching fraction between both \PaTwo decay modes by
setting $x = 0.2$.\footnote{The masses of $\pi$, \Peta*, and \Prho in
  \cref{eq:method:breakupmomentum} are set to
  $m_\pi = \SI{139}{\MeVcc}$, $m_{\Peta*} = \SI{547}{\MeVcc}$, and
  $m_{\Prho*} = \SI{770}{\MeVcc}$.}

\subsubsection{Parametrization of the dynamical amplitudes for nonresonant components}
\label{sec:method:fitmodel:nonres}

For each of the 14~selected partial waves, a separate nonresonant
component is included in the fit model.  We adopt a phenomenological
parametrization for the nonresonant amplitude in the form of a
Gaussian in the two-body breakup momentum~$q$ of the decay that was
inspired by \refCite{tornqvist:1995kr}.  We extend this
parametrization to have a more flexible threshold behavior and to
include an explicit empirical \tpr dependence:
\begin{multlineOrEq}
  \label{eq:method:nonresterm}
  \mathcal{D}^\text{NR}_j(\mThreePi, \tpr;
  \ifMultiColumnLayout{\overbrace}{\underbrace}{b, c_0, c_1, c_2}\ifMultiColumnLayout{^}{_}{%
    \displaystyle{\equiv \zeta^\text{NR}_j}}) \newLineOrNot
  = \sBrk{\frac{\mThreePi - m_\text{thr}}{m_\text{norm}}}^b\,
  e^{-(c_0 + c_1 \tpr + c_2\tpr^2)\, \tilde{q}_{\xi \pi}^2(\mThreePi)}.
\end{multlineOrEq}
Here, $b$ and the $c_i$ are the free shape parameters for the
nonresonant component~$j$.\footnote{In order to simplify notation, we
  omit the subscript~$j$ for these parameters.}  The parameters
$m_\text{norm}$ and $m_\text{thr}$ are the same for all nonresonant
components and are empirically fixed to \SI{1}{\GeVcc} and
\SI{0.5}{\GeVcc}, respectively.  The quasi-two-body breakup momentum
for the decay $X^- \to \xi^0 \pi^-$ is represented by
$\tilde{q}_{\xi \pi}(\mThreePi)$.  However, we cannot use
\cref{eq:method:breakupmomentum} to calculate this quantity because
$q_{\xi \pi}(\mThreePi)$ becomes imaginary for
$\mThreePi < m_\pi + m_\xi$.  We therefore construct an approximation,
$\tilde{q}_{\xi \pi}(\mThreePi)$, to the two-body breakup momentum,
which is valid also below the quasi-two-body threshold and takes into
account the finite width of the isobar~$\xi^0$,\footnote{We start from
  the ansatz that the two-body phase-space volume
  $\varphi_2 \propto q_{\xi \pi} / \mThreePi$ approximates the
  three-body phase-space volume $I_{aa}$ well at large values of
  \mThreePi because the effects from the finite width of the $\xi^0$
  and from the barrier factors become negligible.  For lower values of
  \mThreePi, these effects are taken into account by defining an
  \enquote{effective} two-body breakup momentum via
  $I_{aa} \propto \tilde{q}_{\xi \pi} / \mThreePi$.}
\begin{equation}
  \label{eq:method:nonresterm:qtilde}
  \tilde{q}_{\xi \pi}(\mThreePi)
  \equiv q_{\xi \pi}(m_\text{norm})\, \frac{I_{aa}(\mThreePi)}{I_{aa}(m_\text{norm})}\,
  \frac{\mThreePi}{m_\text{norm}}.
\end{equation}
Here, $\tilde{q}_{\xi \pi}$ is normalized such that it is equal to the
value of $q_{\xi \pi}$ at
$m_\text{norm} = \SI{2.4}{\GeVcc}$.\footnote{The value of
  $m_\text{norm}$ was somewhat arbitrarily chosen to lie above the
  maximum of the fit range of \SI{2.3}{\GeVcc} (see
  \cref{tab:method:fitmodel:waveset}) and low enough so that the decay
  phase-space volume $I_{aa}(m_\text{norm})$ can be calculated
  reliably.}  The decay phase-space volume $I_{aa}$ of wave~$a$ is
calculated according to \cref{eq:integral_matrix_def}.

For partial waves with small relative intensities
$\leq \SI{2.4}{\percent}$, we simplify the parametrization in
\cref{eq:method:nonresterm} to
\begin{multlineOrEq}
  \label{eq:method:nonrestermsmall}
  \mathcal{D}^\text{NR}_j(\mThreePi; b = 0, c_0, c_1 = 0, c_2 = 0)
  \newLineOrNot
  =  e^{-c_0\, \tilde{q}_{\xi \pi}^2(\mThreePi)}.
\end{multlineOrEq}
This reduces the number of free parameters and increases the fit
stability. The only exception is the spin-exotic
\wave{1}{-+}{1}{+}{\Prho}{P} wave because of its dominant nonresonant
contribution.

\subsubsection{Parametrization of the production probability}
\label{sec:method:fitmodel:prod}

At high energies, hadronic scattering reactions are dominated by
$t$-channel Pomeron (\Ppom) exchange.  In earlier measurements of
inclusive diffractive reactions of the type $p + p \to X^+ + p$ at the
CERN ISR~\cite{Albrow:1976sv}, the differential cross section
$\dif[2]{\sigma} / \dif{m_X^2} \dif{t}$ was observed to fall
approximately as $s / m_X^2$, with $\sqrt{s}$ being the center-of-mass
energy of the reaction and $m_X$ the invariant mass of the produced
system~$X^+$.  This behavior is described by Regge
theory~\cite{Collins:1977jy,Kaidalov:1979jz},
\begin{equation}
  \label{eq:method:param:regge_cross_sect}
  \frac{\dif[2]{\sigma}}{\dif{m_X^2}\, \dif{t}}
  = g^\Ppom_{pp}(t)\, \sigma_{\Ppom p}^\text{tot}(m_X^2, t)\, \sBrk[3]{\frac{s}{m_X^2}}^{2 \alpha_\Ppom(t) - 1},
\end{equation}
where $g^\Ppom_{pp}$ is the $t$-dependent proton-proton-Pomeron
coupling and $\sigma_{\Ppom p}^\text{tot}(m_X^2, t)$ is the total
Pomeron-proton cross section.  The Regge trajectory of the Pomeron is
$\alpha_\Ppom(t) = \alpha_0 + \alpha'\, t$, which yields the
$\dif[2]{\sigma} / \dif{m_X^2} \dif{t} \propto s / m_X^2$ behavior for
$\alpha_0 = 1$ and $\alpha' = 0$.

In \refCite{Ataian:1991gn}, a phenomenological Regge framework was
developed to describe exclusive central-production reactions of the
type $p + p \to p + X^0 + p$ in terms of double-Pomeron exchange.  In
these calculations, the cross section is proportional to the so-called
\enquote{Pomeron flux} factor
\begin{equation}
  \label{eq:method:param:pomeron_flux}
  F_{\Ppom p}(x_\Ppom, t)
  \propto \frac{e^{-b_\Ppom\, \abs{t}}}{x_\Ppom^{2 \alpha_\Ppom(t) - 1}}
\end{equation}
using the approximate relation $m_X^2 / s \approx x_\Ppom$ with
$x_\Ppom$ being the longitudinal proton-momentum fraction carried by
the Pomeron in the center-of-mass frame of the reaction.  The slope
parameter of the Pomeron exchange is $b_\Ppom$.
\Cref{eq:method:param:pomeron_flux} can be interpreted as the
probability for Pomeron emission by the proton, which in the limit of
$\alpha_0 = 1$ and $\alpha' = 0$ is proportional to $1 / x_\Ppom$ and
therefore similar to the probability of photon emission in the case of
bremsstrahlung.  Assuming that \cref{eq:method:param:pomeron_flux} is
universal, it can be used to model various diffractive processes in
terms of single-Pomeron exchange~\cite{Cox:2000jt}.  We follow this
approach and have chosen the $3\pi$ production probability in
\cref{eq:method:param:spindens} to be proportional to the probability
of Pomeron emission by the target proton:
\begin{equation}
  \label{eq:method:param:prods}
  \Abs[1]{\mathcal{P}(\mThreePi, \tpr)}^2
  \equiv \frac{1}{x_\Ppom^{2 \alpha_\Ppom(\tpr) - 1}}
  = \sBrk[3]{\frac{s}{\mThreePi^2}}^{2 \alpha_\Ppom(\tpr) - 1}.
\end{equation}
Here, \mThreePi takes the role of $m_X$ and we have made the
approximation $\tpr \approx -t$ thereby neglecting \tmin, so that
$\alpha_\Ppom(\tpr) = \alpha_0 - \alpha' \tpr$.  The normalization and
the explicitly \tpr-dependent factor $e^{-b_\Ppom\, \tpr}$ in
\cref{eq:method:param:pomeron_flux} are both absorbed into the
coupling amplitudes $\mathcal{C}_a^j(\tpr)$ in
\cref{eq:method:param:spindens}.  We use a value of $\alpha_0 = 1.2$,
based on an analysis of data from the H1~experiment at
HERA~\cite{Adloff:1997sc}, while for the shrinkage parameter we use a
value of $\alpha' = \SI{0.26}{\perGeVcsq}$, which was obtained from a
simultaneous fit to CDF (Fermilab) and ISR (CERN)
data~\cite{Abe:1993xx}.\footnote{The result for $\alpha_0$ in
  \refCite{Adloff:1997sc} is based on the $\alpha'$ value from
  \refCite{Abe:1993xx}.  The results of our resonance-model fit are
  not sensitive to the particular choice of the values for $\alpha_0$
  and $\alpha'$.}  \Cref{fig:method:param:prods} shows the deviation
of \cref{eq:method:param:prods} from the $s / \mThreePi^2$ dependence
in the analyzed kinematic range.

\begin{figure}[tbp]
  \centering
  \includegraphics[width=\twoPlotWidth]{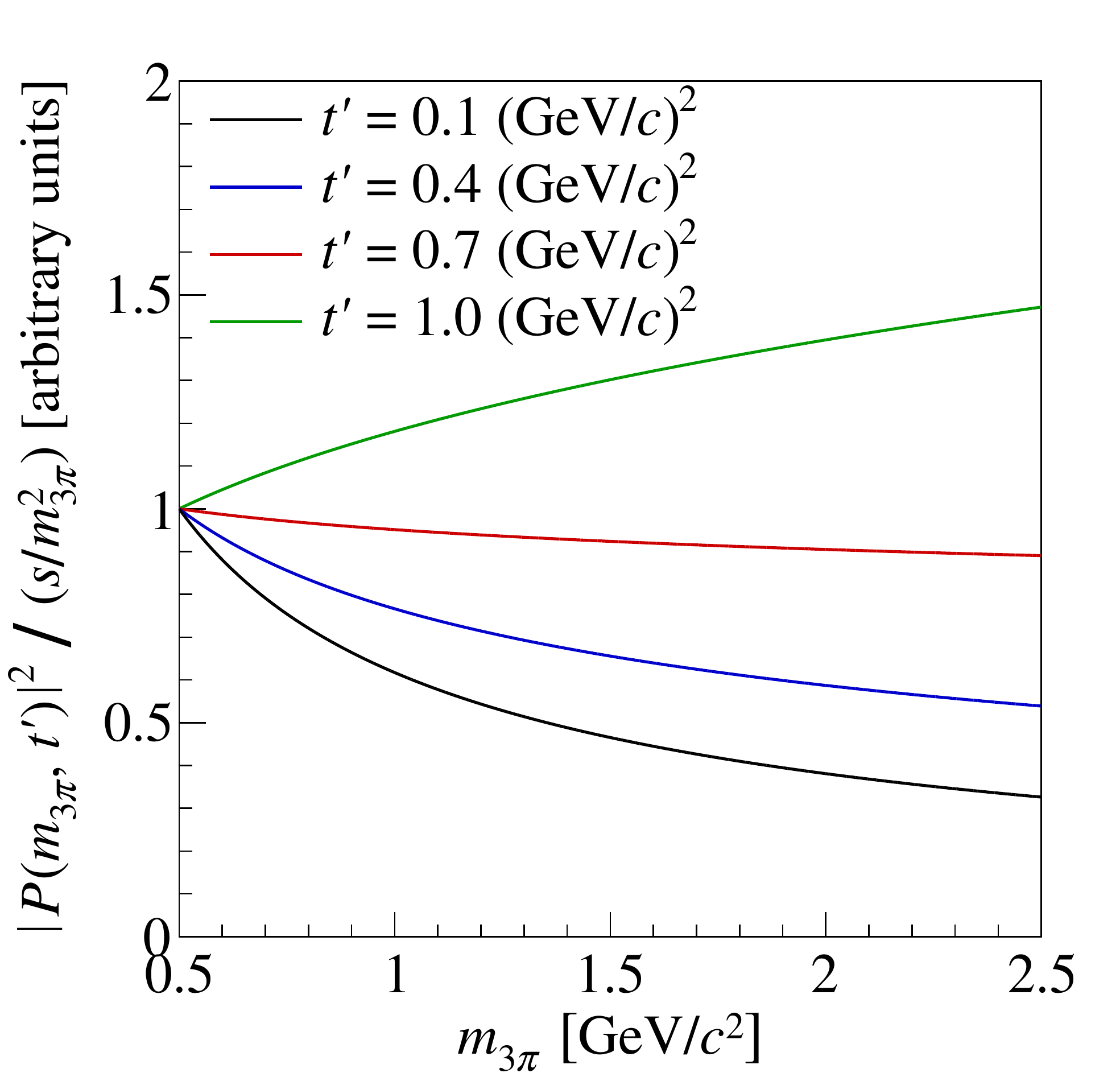}
  \caption{Deviation of the \mThreePi dependence of the $3\pi$
    production probability $\Abs[1]{\mathcal{P}(\mThreePi, \tpr)}^2$,
    as given by \cref{eq:method:param:prods}, from the
    $s / \mThreePi^2$ dependence for various \tpr values.  The curves
    are normalized to~1 at $\mThreePi = \SI{0.5}{\GeVcsq}$.}
  \label{fig:method:param:prods}
\end{figure}

\subsubsection{Discussion of the fit model}
\label{sec:method:fitmodel:discussion}

Our analysis focuses on $3\pi$ resonances with masses up to about
\SI{2}{\GeVcc}.  The goal was to parametrize the data with a minimum
number of resonances while at the same time covering an \mThreePi
range as large as possible.  The employed \mThreePi fit ranges are
listed in \cref{tab:method:fitmodel:waveset}.  For most waves, the
lower bound of the fit range is determined either by thresholds
applied in the PWA (see Table~IX in Appendix~A of
\refCite{Adolph:2015tqa}) or by the phase-space opening.  For some
waves, the reduced phase-space volume at low \mThreePi causes
ambiguities in the solutions of the mass-independent analysis leading
to unphysical structures.  Such regions are excluded.\footnote{By
  limiting the fit ranges, \SI{4.2}{\percent} of the summed
  intensities of all 14~waves are excluded from the fit.}  Seven of
the 14~waves are described by the model up to masses of
\SI{2.3}{\GeVcc}.  For the other waves, the model departs from the
data already at lower masses.  This could be due to higher-lying
excited states above \SI{2}{\GeVcc} or due to increased nonresonant
contributions.  Motivations for the particular choice of the fit
ranges will be discussed in more detail in \cref{sec:results}.

We summarize in \cref{tab:method:fitmodel:waveset} the 14-wave fit
model.  In total, the model has 722~free real-valued parameters, to be
determined by the fit: 22~resonance shape parameters, 29~shape
parameters for the nonresonant components, 22~real-valued parameters
for the branching amplitudes $\prescript{}{b}{\mathcal{B}}_a^j$ [see
\cref{eq:method:branchingdefinition}], and 649~real-valued parameters
for the coupling amplitudes.  The coupling amplitudes for the \PaOne
in the \wave{1}{++}{0}{+}{\Prho}{S} wave are chosen to be real.

\begin{wideTableOrNot}[tbp]
  \renewcommand{\arraystretch}{1.2}
  \centering
  \caption{Fit model with 11~resonances to describe the elements of
    the spin-density matrix of the selected 14~partial waves from six
    \JPC sectors using \cref{eq:method:param:spindens}.  The relative
    intensities listed in the second column are evaluated as a sum
    over the 11~\tpr bins and are normalized to the total number of
    acceptance-corrected events~\cite{Adolph:2015tqa}.  The relative
    intensities do not include interference effects between the waves.
    The third column lists the resonances used to describe the waves.
    For most resonances, the total width is approximated by a constant
    [see \cref{eq:method:fixedwidth}].  For the other resonances, the
    width parametrization is given in square brackets.  The fourth
    column lists the parametrizations used for the nonresonant
    components, the last column the fit ranges (see
    \cref{sec:method:fitmethod} for details).}
  \label{tab:method:fitmodel:waveset}
  \begin{tabular}{lrp{\ifMultiColumnLayout{0.25\textwidth}{0.33\textwidth}}cc}
    \hline
    \hline
    Partial wave                   & {Relative}          & Resonances                                   & Nonresonant                       & \mThreePi fit range             \\
                                   & {intensity}         &                                        & component Eq.                          & [\si\GeVcc]           \\
    \hline
    \wave{0}{-+}{0}{+}{\PfZero}{S} &  \SI{2.4}{\percent} & \Ppi[1800]                                & \eqref{eq:method:nonrestermsmall} & \numrange{1.20}{2.30} \\[1.2ex]

    \wave{1}{++}{0}{+}{\Prho}{S}   & \SI{32.7}{\percent} & \PaOne~[\cref{eq:method:bowlerG}],
                                                           \PaOne[1640]                                 & \eqref{eq:method:nonresterm}      & \numrange{0.90}{2.30} \\
    \wave{1}{++}{0}{+}{\PfZero}{P} &  \SI{0.3}{\percent} & \PaOne[1420]                                 & \eqref{eq:method:nonrestermsmall} & \numrange{1.30}{1.60} \\
    \wave{1}{++}{0}{+}{\PfTwo}{P}  &  \SI{0.4}{\percent} & \PaOne~[\cref{eq:method:bowlerG}],
                                                           \PaOne[1640]                                 & \eqref{eq:method:nonrestermsmall} & \numrange{1.40}{2.10} \\[1.2ex]

    \wave{1}{-+}{1}{+}{\Prho}{P}   &  \SI{0.8}{\percent} & \PpiOne[1600]                                & \eqref{eq:method:nonresterm}      & \numrange{0.90}{2.00} \\[1.2ex]

    \wave{2}{++}{1}{+}{\Prho}{D}   &  \SI{7.7}{\percent} & \rdelim\}{3}{\linewidth}[~\mbox{\PaTwo~[\cref{eq:method:a2dynamicwidth}],~\PaTwo[1700]}]
                                                                                                        & \eqref{eq:method:nonresterm}      & \numrange{0.90}{2.00} \\
    \wave{2}{++}{2}{+}{\Prho}{D}   &  \SI{0.3}{\percent} &                                              & \eqref{eq:method:nonrestermsmall} & \numrange{1.00}{2.00} \\
    \wave{2}{++}{1}{+}{\PfTwo}{P}  &  \SI{0.5}{\percent} &                                              & \eqref{eq:method:nonrestermsmall} & \numrange{1.00}{2.00} \\[1.2ex]

    \wave{2}{-+}{0}{+}{\Prho}{F}   &  \SI{2.2}{\percent} & \rdelim\}{4}{\linewidth}[~\mbox{\PpiTwo, \PpiTwo[1880], \PpiTwo[2005]}]
                                                                                                        & \eqref{eq:method:nonresterm}      & \numrange{1.20}{2.10} \\
    \wave{2}{-+}{0}{+}{\PfTwo}{S}  &  \SI{6.7}{\percent} &                                              & \eqref{eq:method:nonresterm}      & \numrange{1.40}{2.30} \\
    \wave{2}{-+}{1}{+}{\PfTwo}{S}  &  \SI{0.9}{\percent} &                                              & \eqref{eq:method:nonrestermsmall} & \numrange{1.40}{2.30} \\
    \wave{2}{-+}{0}{+}{\PfTwo}{D}  &  \SI{0.9}{\percent} &                                              & \eqref{eq:method:nonrestermsmall} & \numrange{1.60}{2.30} \\[1.2ex]

    \wave{4}{++}{1}{+}{\Prho}{G}   &  \SI{0.8}{\percent} & \rdelim\}{2}{\linewidth}[~\mbox{\PaFour}]
                                                                                                        & \eqref{eq:method:nonrestermsmall} & \numrange{1.25}{2.30} \\
    \wave{4}{++}{1}{+}{\PfTwo}{F}  &  \SI{0.2}{\percent} &                                              & \eqref{eq:method:nonrestermsmall} & \numrange{1.40}{2.30} \\[1.2ex]

    Intensity sum                  & \SI{56.8}{\percent} &                                              &                                   &                       \\
    \hline
    \hline
  \end{tabular}
\end{wideTableOrNot}

In the partial-wave decomposition (see
\cref{sec:mass-independent_fit}), resolution effects of the
spectrometer in \mThreePi and \tpr are not corrected, because the
analysis is performed independently in $(\mThreePi, \tpr)$ bins.
Since the estimated resolution effects are small,\footnote{The $3\pi$
  mass resolution varies between \SI{5.4}{\MeVcc} at small \mThreePi
  (in the range from \SIrange{0.5}{1.0}{\GeVcc}) and \SI{15.5}{\MeVcc}
  at large \mThreePi (in the range from \SIrange{2.0}{2.5}{\GeVcc}).
  The \tpr resolution as obtained from the reconstructed $3\pi$ final
  state ranges between \SIlist{7e-3;20e-3}{\GeVcsq} depending on the
  \mThreePi and \tpr region. See \refCite{Adolph:2015tqa} for
  details.} they are neglected in the resonance-model fit.

Although the fit model describes the data rather well (see
\cref{sec:results}), it has a number of potential caveats and
limitations that are mainly rooted in its
simplicity~\cite{pdg_resonances:2016}.  Breit-Wigner amplitudes are in
general good approximations only for single narrow resonances.  When
using a constant-width parametrization [\cref{eq:method:fixedwidth}],
the resonance in addition has to be far above thresholds.  The
description of a set of resonances with the same quantum numbers as a
sum of Breit-Wigner amplitudes may violate unitarity and is a good
approximation only for well-separated resonances with little overlap.
In particular for the $\JPC = 2^{-+}$ resonances, this condition is
not well fulfilled.  Also coupled-channel effects are not taken into
account.  All the above effects render the extracted Breit-Wigner
parameters model and process dependent.  An additional process and
model dependence is introduced by the decomposition of the
partial-wave amplitudes into resonant and nonresonant components,
which is not unique.  However, our results can be compared directly to
previous analyses of diffractive three-pion production (see \eg\
\refCite{daum:1980ay,adams:1998ff,chung:2002pu,alekseev:2009aa}). The
model assumption that the phase of the nonresonant amplitudes does not
depend on \mThreePi may not be well justified for cases where these
amplitudes exhibit pronounced peaks in their intensity distribution.
One may also remark that singularities in the scattering matrix that
are not related to resonances might mimic Breit-Wigner resonances.  A
possible example is the \PaOne[1420]~\cite{Adolph:2015pws}, which
could be the singularity of a triangle
diagram~\cite{Ketzer:2015tqa,Aceti:2016yeb} (see also
\cref{sec:onePP}).

Some of the potential issues mentioned above are expected to be
mitigated by the fact that in our model most of the resonances are
fitted in at least two decay modes.  In addition, we combine in the
fit the information of 11~\tpr bins while forcing the resonances to
appear with the same parameters in each \tpr bin.  By performing such
a \tpr-resolved analysis, resonance parameters are constrained by the
various production processes that may contribute with different
strengths and phases to the reaction under study depending on the \tpr
region.

Instead of Breit-Wigner parameters, one could attempt to extract the
poles on the second Riemann sheet of the scattering amplitude, which
correspond to resonances.  The location of a resonance pole in the
complex energy plane and its residue represent the universal resonance
properties.  However, the construction of coupled-channel models for
the reaction \reaction that are consistent with the fundamental
principles of unitarity and analyticity is a formidable task.  In the
past, quasi-two-body $K$-matrix approaches were applied to analyze
$3\pi$ resonances in diffractive production (see \eg\
\refCite{daum:1980ay,amelin:1995gu}).  The extraction of resonance
pole positions using an analytical model based on the principles of
the relativistic $S$-matrix is currently under
development~\cite{Jackura:2016llm,Mikhasenko:2017jtg}.  A first
successful application of this model to the $\eta \pi$ $D$-wave
extracted from COMPASS data yielded pole positions for the \PaTwo and
\PaTwo[1700]~\cite{Jackura:2017amb}.  In \cref{sec:twoPP_discussion}
we compare those results to the ones from our analysis.

\subsection{Fit method}
\label{sec:method:fitmethod}

The free parameters of the model in \cref{eq:method:param:spindens},
\ie the set of coupling amplitudes $\mathcal{C}_a^j(\tpr)$ and the set
of shape parameters $\zeta_j$ of the wave components, are extracted by
a fit to the spin-density matrix $\varrho_{a b}(\mThreePi, \tpr)$ that
was extracted in the mass-independent analysis (see
\cref{sec:mass-independent_fit}).  In the resonance-model fit, the
information of the Hermitian spin-density matrix is represented by a
real-valued matrix $\Lambda_{a b}(\mThreePi, \tpr)$ of the same
dimension.  The elements of this matrix are defined by the upper
triangular part of $\varrho_{a b}$:
\begin{equation}
  \label{eq:method:fitmethod:rhoredef}
  \Lambda_{a b}(\mThreePi, \tpr) =
  \begin{cases}
    \Re\sBrk{\varrho_{a b}(\mThreePi, \tpr)} & \text{for}~a < b, \\[1.1ex]
    \Im\sBrk{\varrho_{b a}(\mThreePi, \tpr)} & \text{for}~a > b, \\[1.1ex]
    \varrho_{a a}(\mThreePi, \tpr) = \Abs{\mathcal{T}_a(\mThreePi, \tpr)}^2 & \text{for}~a = b.
  \end{cases}
\end{equation}
Hence the diagonal elements of $\Lambda_{a b}(\mThreePi, \tpr)$ are
the partial-wave intensities, the upper off-diagonal elements are the
real parts of the interference terms, and the lower off-diagonal
elements are the corresponding imaginary parts.

The deviation of the resonance model $\widehat{\Lambda}_{a b}$ from
the matrix $\Lambda_{a b}$, which is extracted from data, is measured
by summing up the squared Pearson's residuals~\cite{Pearson:1900} of
all matrix elements for all \mThreePi and \tpr
bins~\cite{adams:1998ff}:
\begin{wideEqOrNot}%
  \begin{equation}
    \label{eq:method:fitmethod:chi2}
    \chisq
    = \sum_{a, b}^{N_{\text{waves}}\vphantom{)_{a b}}}\;
    \sum^{\text{\tpr bins}\vphantom{N_{\text{w}})_{a b}}}\;
    \sum^{(\text{\mThreePi bins})_{a b}\vphantom{N_{\text{w}}}}
    \sBrk{\frac{\Lambda_{a b}(\mThreePi, \tpr)
        - \widehat{\Lambda}_{a b}(\mThreePi, \tpr)}{\sigma_{a b}(\mThreePi, \tpr)}}^2.
  \end{equation}
\end{wideEqOrNot}%
Here, $N_{\text{waves}}$ is the number of partial waves included in
the fit model and $\sigma_{a b}(\mThreePi, \tpr)$ is the statistical
uncertainty of $\Lambda_{a b}(\mThreePi, \tpr)$ as determined by the
mass-independent analysis.  The sum in \cref{eq:method:fitmethod:chi2}
runs over all 11~\tpr bins and those \mThreePi bins that lie within
the fit ranges.  The fit ranges for the intensity terms
$\Lambda_{a a}$ are listed in \cref{tab:method:fitmodel:waveset}.  The
fit ranges for the off-diagonal interference terms $\Lambda_{a b}$ are
defined by the intersections of the fit ranges for the intensities of
waves~$a$ and~$b$.  The values of the model parameters are determined
by minimizing the \chisq function using the \textsc{Migrad} algorithm
of the \textsc{Minuit} program~\cite{James:1975dr}.

Although we use the notation \chisq in \cref{eq:method:fitmethod:chi2}
for the quantity that is minimized in the resonance-model fit, it is
important to note that the minimum of \cref{eq:method:fitmethod:chi2}
does not follow a \chisq~distribution.  Therefore, the expectation
value of~\chisq is neither the number of degrees of freedom~(n.d.f.)
nor is its deviation from the n.d.f. an absolute measure for the
goodness of the fit.  The reason for this is that
\cref{eq:method:fitmethod:chi2} does not take into account
correlations among the spin-density matrix elements.  Although the
spin-density matrix elements from different \mThreePi or \tpr bins are
independent from each other, within an $(\mThreePi, \tpr)$ bin, two
kinds of correlations appear: \one~statistical correlations of the
spin-density matrix elements and \two~mathematical dependences caused
by using a rank-1 spin-density matrix for the positive-reflectivity
waves in the partial-wave decomposition (see
\cref{sec:mass-independent_fit}).  The result of the mass-independent
analysis in principle includes the covariance matrix of the extracted
transition amplitudes $\mathcal{T}_a$.  However, the propagation of
this information to the covariance matrix for $\Lambda_{a b}$ is not
well-defined because the spin-density matrix has more free real-valued
parameters than the set of transition amplitudes.\footnote{In each
  $(\mThreePi, \tpr)$ bin, the resonance-model fit minimizes the
  distance to $N_{\text{waves}}^2$ data points, which are the elements
  of $\Lambda_{a b}(\mThreePi, \tpr)$.  However, the transition
  amplitudes extracted in the mass-independent analysis with rank-1
  spin-density matrix represent only $(2 N_{\text{waves}} - 1)$ data
  points.}  The rank-1 condition leads to analytical relations among
the spin-density matrix elements for waves~$a$, $b$, $c$, and~$d$ of
the form
\begin{equation}
  \label{eq:spin_dens_corr}
  \varrho_{a b}\, \varrho_{c d}
  = \varrho_{a d}\, \varrho_{c b}.
\end{equation}

We have performed studies using alternative formulations of~\chisq
that take into account the statistical correlations and
\cref{eq:spin_dens_corr} (see \cref{sec:alt_chi_2}).  For most
parameters, the obtained results are similar to those obtained with
\cref{eq:method:fitmethod:chi2} and the systematic effects are smaller
than those from the other systematic studies (see
\cref{sec:systematics}).  Exceptions are discussed in
\cref{sec:results,sec:syst_uncert}.  Given the limitations of our
model in describing details of the data, the \chisq~formulation in
\cref{eq:method:fitmethod:chi2} has practical advantages.  The
information from the 14~waves enters symmetrically; \ie
\cref{eq:method:fitmethod:chi2} does not require one to choose a
reference wave as it is the case in the alternative
\chisq~formulations.  In addition, compared to the alternative
\chisq~formulations, \cref{eq:method:fitmethod:chi2} effectively
assigns more weight to the interference terms, which contain the phase
information.  This tends to improve the fit stability as imperfections
in the description of the intensity distributions of some waves have
less influence.  A possible issue of neglecting the correlations of
the spin-density matrix elements in \cref{eq:method:fitmethod:chi2} is
that it may lead to biased estimates for the statistical uncertainties
of the fit parameters.  However, in our analysis this effect can be
safely ignored because, due to the large data set, all uncertainties
on physical parameters are dominated by systematic effects outweighing
the statistical ones.  For the above reasons, we use the
\chisq~definition of \cref{eq:method:fitmethod:chi2} to determine the
physical parameters.

The extraction of the resonance parameters using the fit model
described in \cref{sec:method:fitmodel} is based on highly precise
physical information obtained from the mass-independent analysis.  The
722~free parameters of the model are constrained by the matrix
$\Lambda_{a b}(\mThreePi, \tpr)$, which has $14 \times 14$~elements
for each of the 100~\mThreePi and 11~\tpr bins.  Taking into account
the chosen \mThreePi fit ranges (see
\cref{tab:method:fitmodel:waveset}), this yields a total number of
\num{76505}~data points that enter into the fit.

The fit model described in \cref{sec:method:fitmodel} is highly
nonlinear in the shape parameters $\zeta_j$ of the wave components.
Some of the model parameters are also strongly correlated.  In
addition, the employed parametrizations are only approximations or in
the case of the nonresonant components purely empirical.  Hence they
often do not describe all details of our high-precision data.  The
resulting deviations between model and data lead to a multimodal
behavior of the minimized \chisq~function.  Therefore, the fit result
may depend on the start values for the fit parameters.  To avoid the
fit being trapped in local \chisq~minima, we perform numerous fit
attempts using different sets of start values for the shape
parameters, which are randomly picked from uniform distributions.  For
the resonance parameters, conservatively wide ranges are chosen for
these distributions based on previous
measurements~\cite{Patrignani:2016xqp}.  The ranges are shown as
dotted rectangles in \cref{fig:method:fitmethod:startparameters}.  For
the shape parameters of the nonresonant components, we use wide
uniform distributions to pick the start values as there is no prior
knowledge.  Details are discussed in \refCite{msc_thesis_wallner}.
The central values for the fit parameters are estimated by performing
fits with 250~different sets of start values, which are shown as dots
in \cref{fig:method:fitmethod:startparameters}.  For the systematic
studies discussed in \cref{sec:systematics}, we typically use
50~random sets of start values.

\begin{figure}[tbp]
  \centering
  \includegraphics[width=\linewidthOr{0.8\linewidth}]{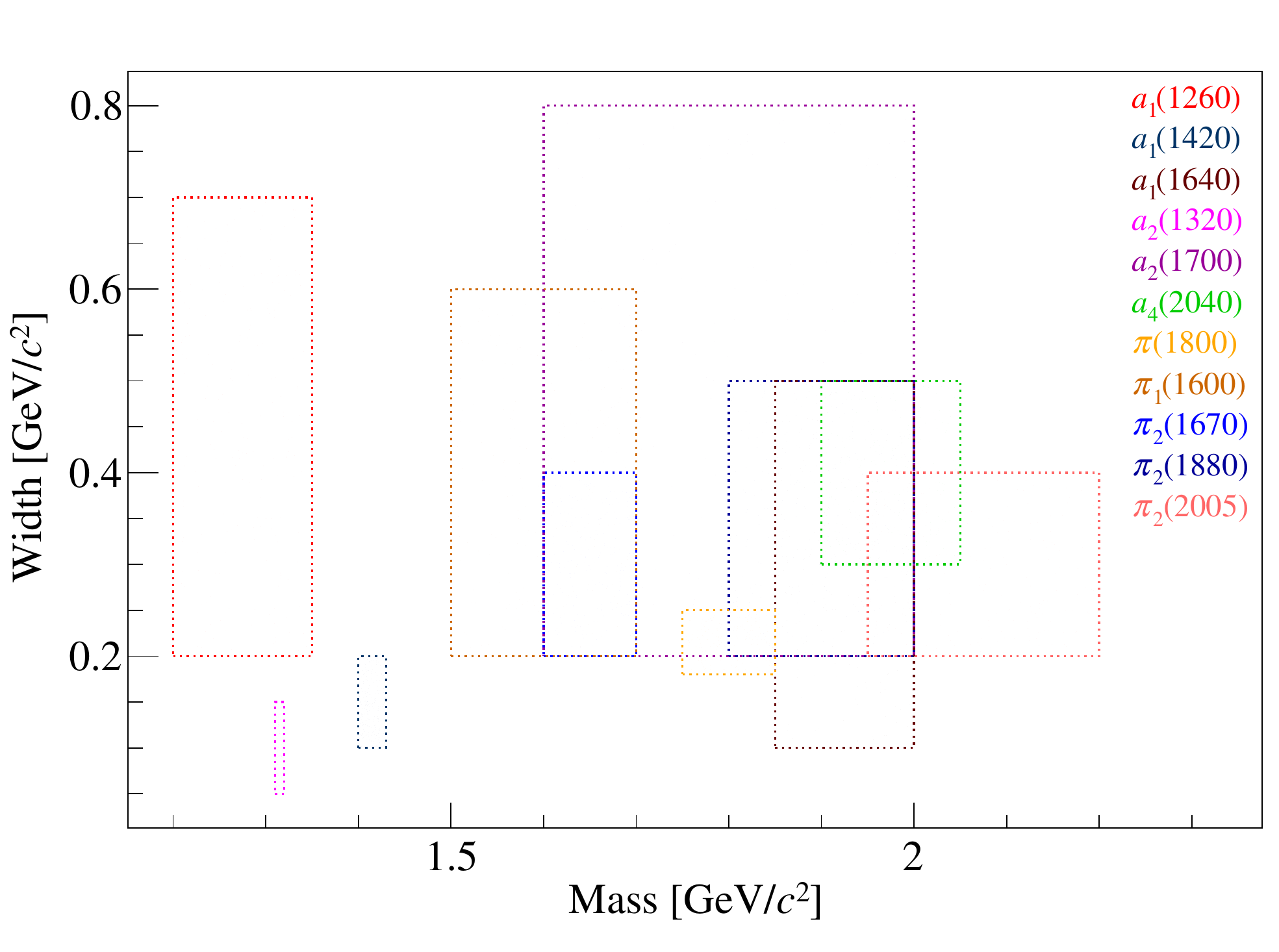}
  \caption{The dotted boxes indicate the ranges of the uniform
    distributions that were used to randomly generate start values for
    the mass and width parameters of the resonances included in the
    fit model.  Different colors encode different resonances.}
  \label{fig:method:fitmethod:startparameters}
\end{figure}

With the randomly chosen start values for the fit parameters, it is in
general not possible to fit all 722~free parameters at once.
Therefore, a multistaged approach is used, where first only a subset
of the parameters is left free, while the others are kept fixed.  The
parameter values found in this first stage are then used as start
values for the next fit stages, in which in addition some of the
previously fixed parameters are freed.  In the last fit stage, all
722~model parameters are left free.  Since also the order, in which
the parameters are released during the fit, may influence the fit
result, we perform for each set of start values four different schemes
of releasing the fit parameters (see \refCite{msc_thesis_wallner} for
details).  Using this procedure, the central values of the model
parameters are estimated based on a total of \num{1000}~fit attempts
performed using the 250~independent randomly chosen sets of start
values.

\Cref{fig:method:fitmethod:chi2:beforecuts} shows the frequency
distribution of the \chisq~values from the \num{1000} fit attempts in
narrow bins of \SI{0.1}{units} of~\chisq.  We assume that fits falling
into the same \chisq~bin correspond to identical solutions.  In order
to remove unphysical solutions from this set of solutions, we apply a
series of selection criteria.  Most of these criteria aim at rejecting
solutions, where components of the resonance model are misused to
compensate for imperfections in the model.  The fit ranges listed in
\cref{tab:method:fitmodel:waveset} were chosen such that they cover
the peak regions of the resonances included in our model.  Therefore,
solutions are rejected if the mass value of any of the resonance
components lies outside of the respective fit ranges\footnote{An
  exception is made for the \PaOne component in the
  \wave{1}{++}{0}{+}{\PfTwo}{P} wave.} (see
\cref{tab:method:fitmodel:waveset}).  Solutions are also rejected if
any of the resonance width values lie at the border of the allowed
parameter range from \SIrange{40}{1000}{\MeVcc}.  Furthermore,
solutions are rejected if a component that represents an excited
resonance is misused by the fit to describe a lower-lying state and
vice versa.  Such solutions are clearly unphysical.  For example, in
some unphysical solutions the \PaOne and \PaOne[1640] components
become wide and have nearly identical masses to better describe the
dominant peak in the intensity distribution of the
\wave{1}{++}{0}{+}{\Prho}{S} wave.  The above condition removes in
particular all 17~solutions, which have a lower~\chisq than the
selected physical solution [the latter one is shown in red in
\cref{fig:method:fitmethod:chi2:beforecuts}].  In the last step, we
remove solutions that are found only once.\footnote{With this step, we
  remove in particular solutions, where the fitting algorithm was
  trapped in shallow local minima.  It is worth stressing that all
  solutions removed by this criterion have a larger \chisq than the
  selected physical solution.}  More details can be found in
\refsCite{msc_thesis_wallner,msc_thesis_schmeing}.

The fit method described above is computationally expensive, but it
avoids constraining the range of parameter values in the fit, while at
the same time it allows us to use wide ranges for the random choice of
the start values.

\begin{figure}
  \centering
  \subfloat[][]{%
    \includegraphics[width=\twoPlotWidth]{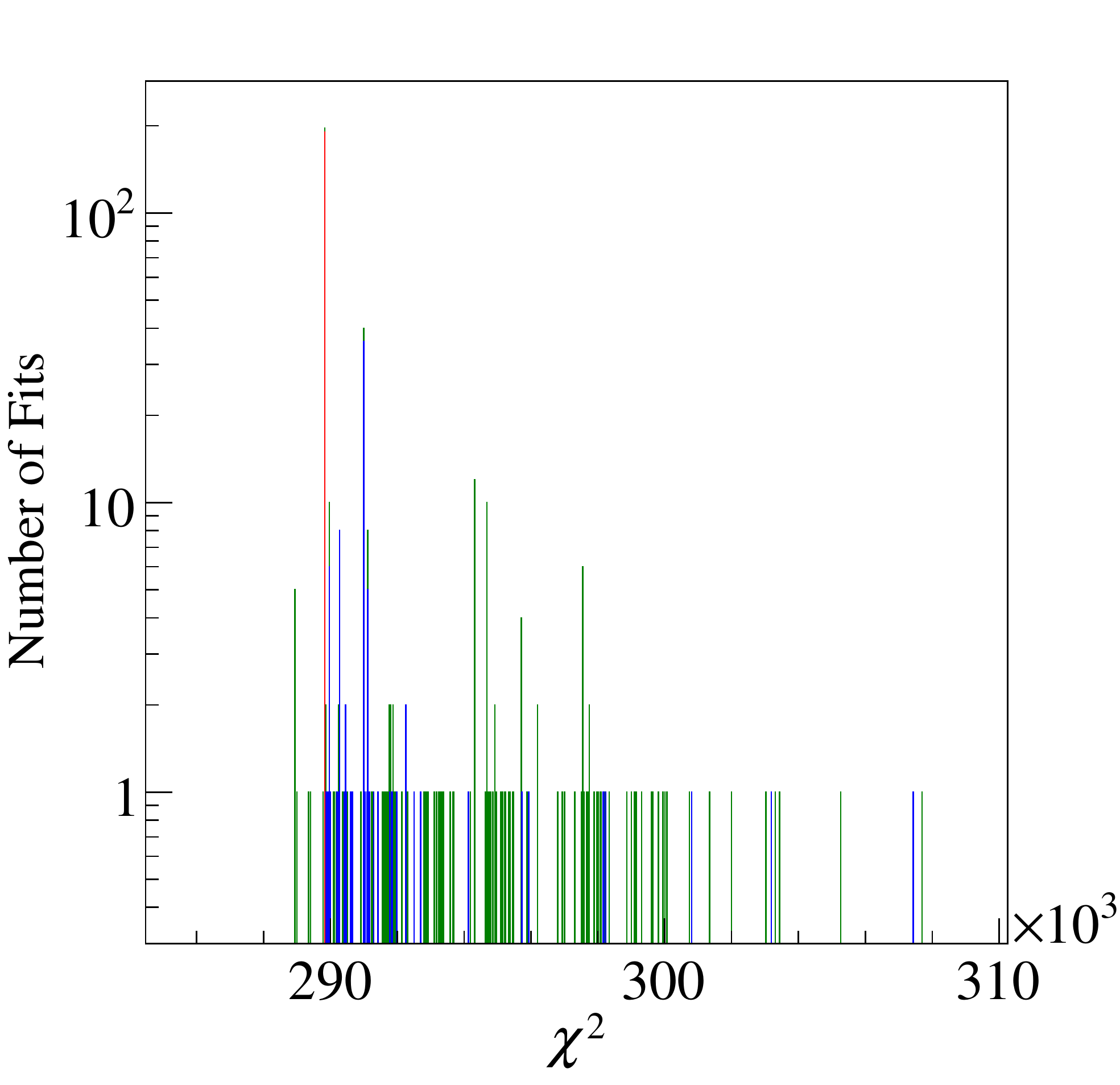}%
    \label{fig:method:fitmethod:chi2:beforecuts}%
  }%
  \newLineOrHspace{\twoPlotSpacing}%
  \subfloat[][]{%
    \includegraphics[width=\twoPlotWidth]{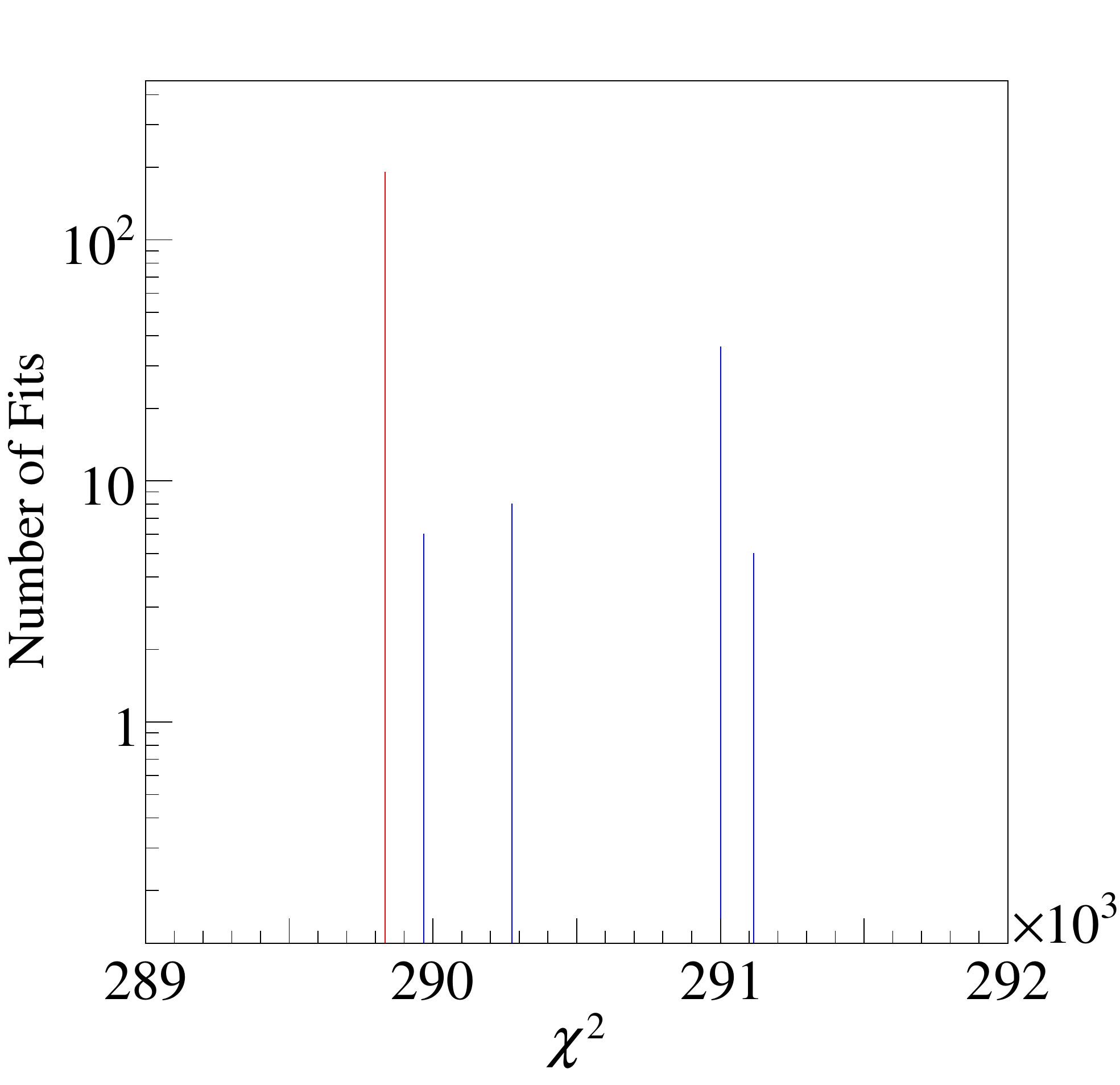}%
    \label{fig:method:fitmethod:chi2:aftercuts}%
  }%
  \caption{\subfloatLabel{fig:method:fitmethod:chi2:beforecuts}~Distribution
    of the \chisq~values of the 832~fits that converged out of the
    \num{1000}~fit attempts.  The selected physical solution (see
    text) is shown in red.  Additional solutions that are considered
    physical are shown in blue and unphysical solutions in green.
    \subfloatLabel{fig:method:fitmethod:chi2:aftercuts}~Corresponding
    distribution after removing all unphysical solutions. Note the
    narrower \chisq~range.}
\end{figure}

For~252 out of the total of \num{1000}~fit attempts, the \chisq
minimization procedure converged and the resulting solution passed the
selection criteria.  The \chisq~distribution of those solutions is
shown in \cref{fig:method:fitmethod:chi2:aftercuts}.  The solution
with the lowest~\chisq of \num{289834} is shown in red and is found
190~times.  In addition, \cref{fig:method:fitmethod:chi2:aftercuts}
shows four physical solutions with slightly larger \chisq~values.  For
all four solutions, the parameter values lie within the estimated
systematic uncertainties (see \cref{sec:systematics}).  The solution
with the lowest \chisq, which is also the most frequently found
solution, is called \emph{main solution} in the remaining text.  It is
interesting to disentangle the contributions from the intensities and
interference terms to the \chisq in \cref{eq:method:fitmethod:chi2}.
This is visualized for the main solution in \cref{fig:chi2Matrix} in
the form of a matrix, which shows the \chisq contributions (summed
over the \mThreePi and \tpr bins) from the elements of the matrix
$\Lambda_{a b}(\mThreePi, \tpr)$ defined in
\cref{eq:method:fitmethod:rhoredef}.  The diagonal elements in
\cref{fig:chi2Matrix} show the \chisq contributions from the intensity
distributions of each partial wave, the off-diagonal elements the
\chisq contributions from the real (upper triangle) and imaginary
parts (lower triangle) of the interference terms between the waves.
The intensity distribution of the \wave{1}{++}{0}{+}{\Prho}{S} wave
gives by far the largest contribution to the \chisq.  Also the \chisq
contributions of some of its interference terms are large.  The reason
for this is that the model is not able to describe all details of this
partial-wave amplitude within the extremely small statistical
uncertainties, which are a consequence of the large relative intensity
of the \wave{1}{++}{0}{+}{\Prho}{S} wave of \SI{32.7}{\percent} and
the large data set.  Due to the dominant contribution of the
\wave{1}{++}{0}{+}{\Prho}{S} amplitude to the \chisq, the parameters
of resonances in other waves are sensitive to the parametrizations
used for the $1^{++}$ waves (see
\cref{sec:systematics,sec:syst_uncert}).

\begin{figure}[tbp]
  \centering
  \includegraphics[width=\linewidthOr{\twoPlotWidth}]{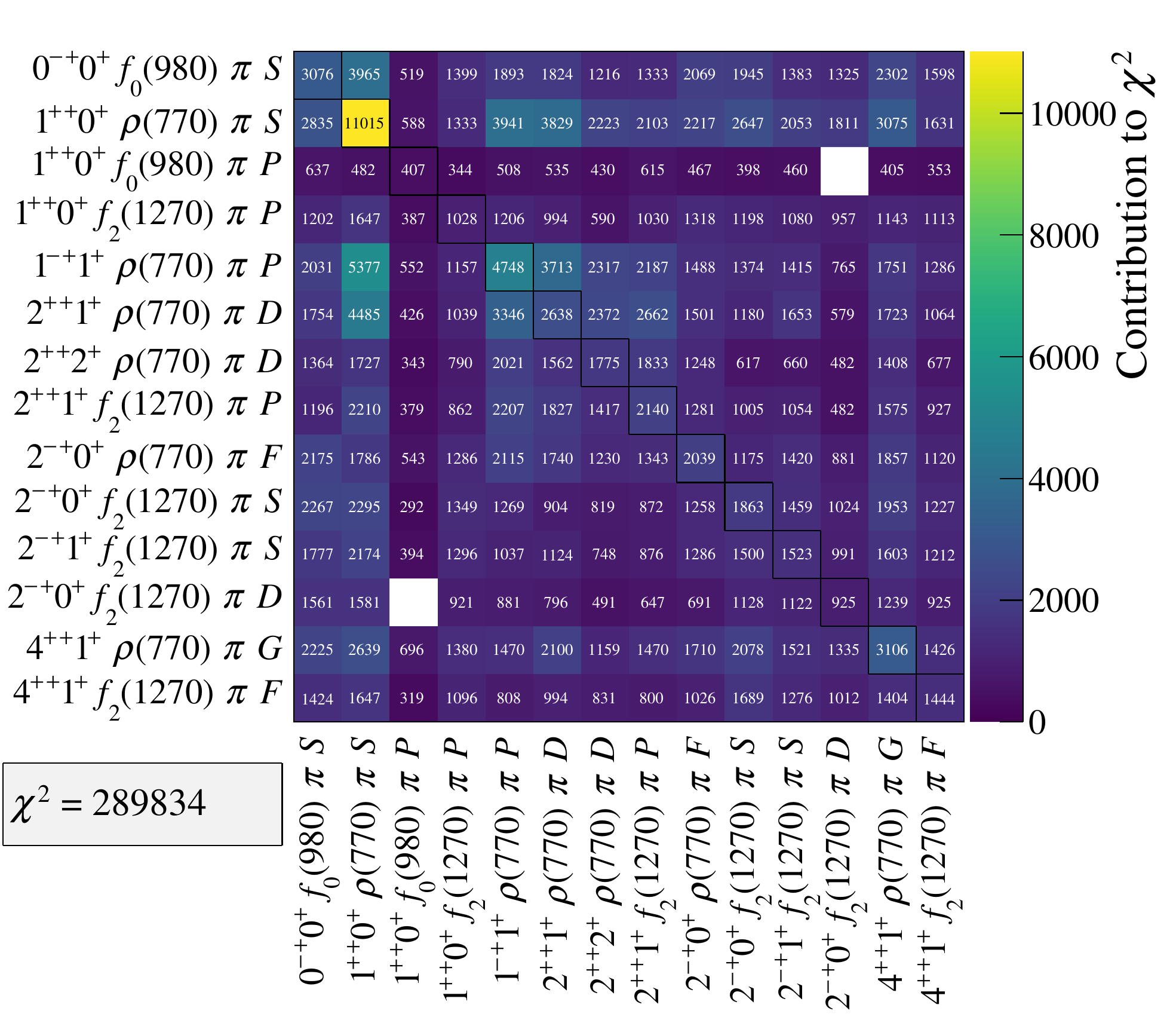}
  \caption{Contributions from the intensities and interference terms
    to the \chisq in \cref{eq:method:fitmethod:chi2} summed over the
    \mThreePi and \tpr bins.  The two cells for the interference term
    of the \wave{1}{++}{0}{+}{\PfZero[980]}{P} and
    \wave{2}{-+}{0}{+}{\PfTwo}{D} waves are empty because the fit
    ranges for these two waves do not overlap (see
    \cref{tab:method:fitmodel:waveset}).}
  \label{fig:chi2Matrix}
\end{figure}

\subsection{Extraction of \tpr spectra of wave components}
\label{sec:method:tp}

Performing the partial-wave analysis in bins of \tpr not only helps to
better disentangle resonant and nonresonant contributions via their
different \tpr dependences but also allows us to determine the \tpr
dependence of each wave component in the resonance model.  Since the
analysis is performed on the amplitude level, we can extract the \tpr
dependence of the intensity, \ie the \tpr spectrum, of each wave
component and the \tpr dependence of the relative phases of the
coupling amplitudes $\mathcal{C}_a^j(\tpr)$ of the components.  The
latter is discussed in more detail in \cref{sec:production_phases}.

Starting from \cref{eq:method:transitionampl}, we can write the model
$\widehat{\mathcal{T}}_a$ for the transition amplitude of wave~$a$ as
\begin{equation}
  \widehat{\mathcal{T}}_a(\mThreePi, \tpr)
  = \smashoperator[r]{\sum_{j\; \in\; \mathbb{S}_a}} \widehat{\mathcal{T}}_a^j(\mThreePi, \tpr).
\end{equation}
Here, $\widehat{\mathcal{T}}_a^j$ is the transition amplitude for
component~$j$ in this wave and given by
\begin{multlineOrEq}
  \label{eq:method:transitionamplcomp}
  \widehat{\mathcal{T}}_a^j(\mThreePi, \tpr)
  \equiv \sqrt{I_{aa}(\mThreePi)}\, \sqrt{\mThreePi\vphantom{I_a}}\;
  \mathcal{P}(\mThreePi, \tpr)\,
  \newLineTimesOrNot
  \mathcal{C}_a^j(\tpr)\, \mathcal{D}_j(\mThreePi, \tpr; \zeta_j).
\end{multlineOrEq}
With the above, the partial-wave intensity reads
\begin{wideEqOrNot}%
  \begin{multline}
    \label{eq:method:intensitywave}
    \Abs[1]{\widehat{\mathcal{T}}_a(\mThreePi, \tpr)}^2
    = \smashoperator[r]{\sum_{j\; \in\; \mathbb{S}_a}}
    \Overbrace{I_{aa}(\mThreePi)\, \mThreePi\, \Abs[1]{\mathcal{P}(\mThreePi, \tpr)}^2\,
      \Abs[1]{\mathcal{C}_a^j(\tpr)}^2\, \Abs[1]{\mathcal{D}_j(\mThreePi, \tpr; \zeta_j)}^2}{%
      \text{Intensity of wave component $j$}} \\
    + \sum_{\mathclap{j\, <\, k\; \in\; \mathbb{S}_a}} I_{aa}(\mThreePi)\, \mThreePi\, \Abs[1]{\mathcal{P}(\mThreePi, \tpr)}^2\,
    \Underbrace{2\Re\sBrk{\mathcal{C}_a^j(\tpr)\, \mathcal{D}_j(\mThreePi, \tpr; \zeta_j)\,
        \mathcal{C}_a^{k \text{*}}(\tpr)\, \mathcal{D}_k^\text{*}(\mThreePi, \tpr; \zeta_k)}}{%
      \text{Overlap of wave components $j$ and $k$}}.
  \end{multline}
\end{wideEqOrNot}%
Due to the chosen normalization of the transition amplitudes via
\cref{eq:prod_amplitude_norm}, the partial-wave intensity in
\cref{eq:method:intensitywave} corresponds to the expected number of
events in wave~$a$.  Using the same reasoning as for
\cref{eq:expected_ev_nmb_corr_rho}, we interpret the terms
\begin{equation}
  \label{eq:method:intensitycomp}
  \begin{splitOrNot}
    \Abs[1]{\widehat{\mathcal{T}}_a^j(\mThreePi, \tpr)}^2
    \alignOrNot= \begin{multlinedOrNot} I_{aa}(\mThreePi)\, \mThreePi\, \Abs[1]{\mathcal{P}(\mThreePi, \tpr)}^2\,
    \newLineTimesOrNot
    \Abs[1]{\mathcal{C}_a^j(\tpr)}^2\, \Abs[1]{\mathcal{D}_j(\mThreePi, \tpr; \zeta_j)}^2 \end{multlinedOrNot}
    \newLineOrNot
    \alignOrNot\equiv \frac{\dif{N_a^j}}{\dif{\mThreePi}\, \dif{\tpr}}
  \end{splitOrNot}
\end{equation}
as the expected number of events $N_a^j$ in component~$j$ in wave~$a$
in the $(\mThreePi, \tpr)$ bin.  Integrating
\cref{eq:method:intensitycomp} over \mThreePi gives the \tpr-dependent
yield, \ie the \tpr spectrum
$\mathcal{I}_a^j(\tpr) \equiv \dif{N_a^j} / \dif{\tpr}$ of wave
component~$j$ in wave~$a$.  To account for the nonequidistant \tpr
binning, we normalize in each \tpr bin the intensity to the respective
bin width $\Delta \tpr$:
\begin{multlineOrEq}
  \label{eq:tprim-dependence}
  \mathcal{I}_a^j(\tpr)
  = \frac{1}{\Delta \tpr}\, \Abs[1]{\mathcal{C}_a^j(\tpr)}^2
  \smashoperator[r]{\int_{m_\text{min}}^{m_\text{max}}}\! \dif{\mThreePi}\,
  I_{aa}(\mThreePi)\, \mThreePi\,
  \newLineTimesOrNot
  \Abs[1]{\mathcal{P}(\mThreePi, \tpr)}^2\, \Abs[1]{\mathcal{D}_j(\mThreePi, \tpr; \zeta_j)}^2.
\end{multlineOrEq}
The model for the nonresonant amplitudes is valid only within the
applied fit ranges in \mThreePi.  Therefore, we use the fit ranges
from \cref{tab:method:fitmodel:waveset} as the \mThreePi integration
range in \cref{eq:tprim-dependence} for all wave components.

As an example, we show in \cref{fig:method:tp:examplespectrum:m0} the
\tpr spectrum of the \PpiTwo[1880] component in the
\wave{2}{-+}{0}{+}{\Prho}{F} wave.  In each \tpr bin, the black
horizontal line indicates the central value of the intensity
$\mathcal{I}_a^j(\tpr)$ of the wave component as determined by
\cref{eq:tprim-dependence}.  The horizontal extent of the line
indicates the width of the \tpr bin.  The statistical uncertainty is
represented by the height of the gray box around the central value.
It is calculated from the statistical uncertainties of the
resonance-model parameters using Monte Carlo error propagation.  For
many wave components, the statistical uncertainties are very small and
barely visible in the diagrams.

\begin{figure}[tbp]
  \centering
  \subfloat[][]{%
    \includegraphics[width=\twoPlotWidth]{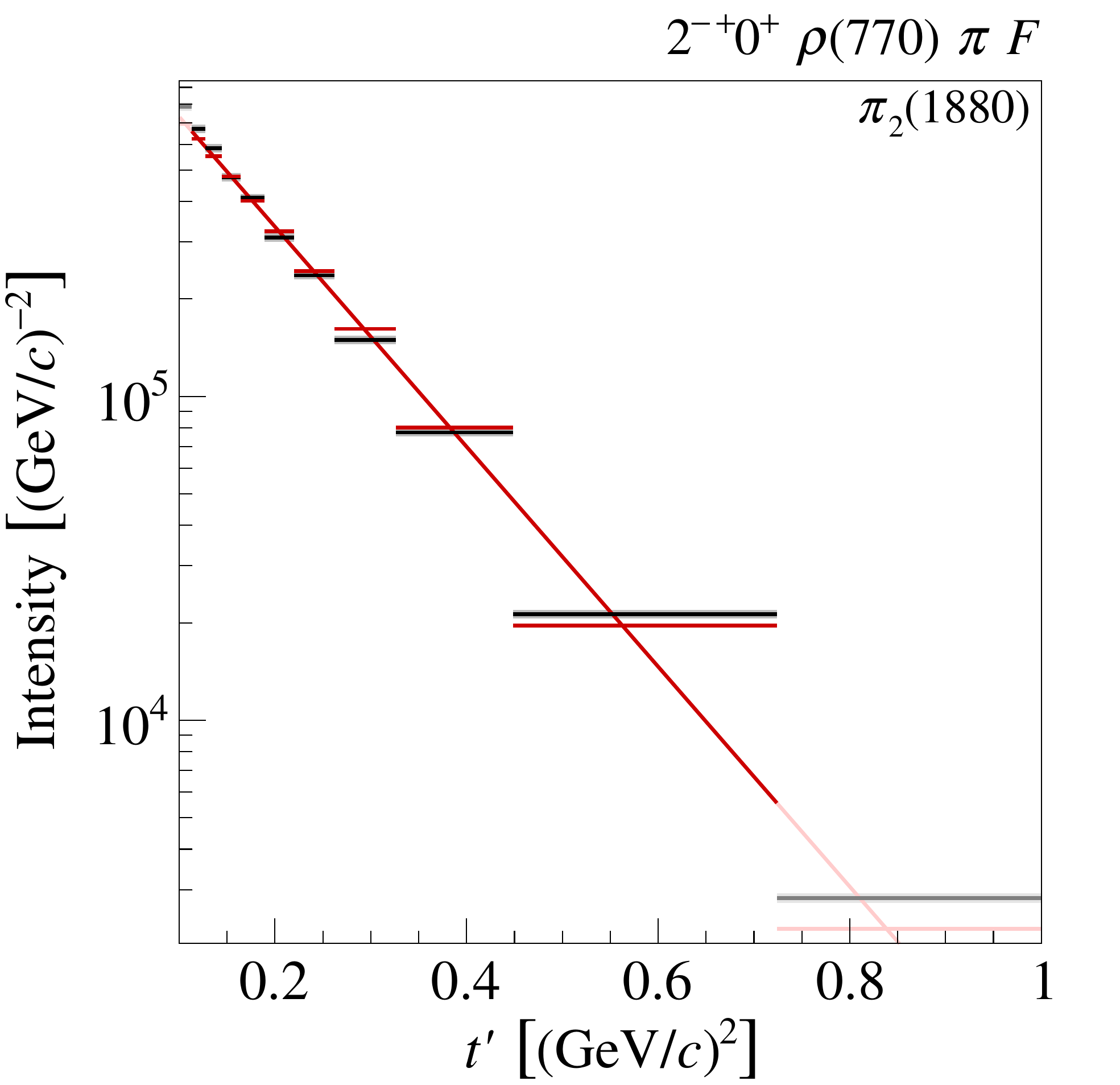}%
    \label{fig:method:tp:examplespectrum:m0}%
  }%
  \newLineOrHspace{\twoPlotSpacing}%
  \subfloat[][]{%
    \includegraphics[width=\twoPlotWidth]{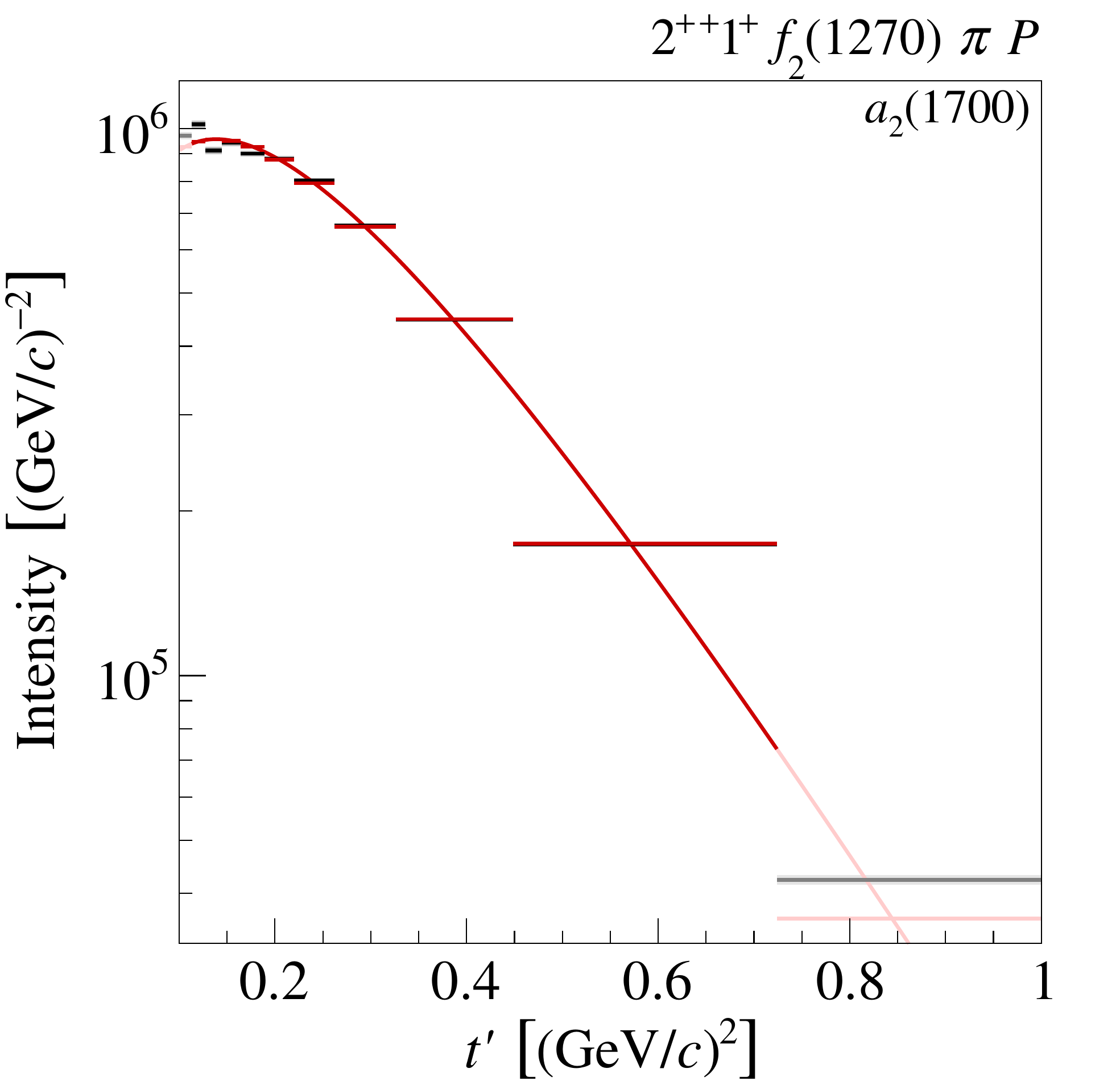}%
    \label{fig:method:tp:examplespectrum:m1}%
  }%
  \caption{Examples for \tpr spectra of wave components extracted
    according to \cref{eq:tprim-dependence}. The black horizontal
    lines indicate the central values, and the gray boxes the
    statistical uncertainties (see text for details).
    \subfloatLabel{fig:method:tp:examplespectrum:m0}~\PpiTwo[1880]
    component in the \wave{2}{-+}{0}{+}{\Prho}{F} wave;
    \subfloatLabel{fig:method:tp:examplespectrum:m1}~\PaTwo[1700]
    component in the \wave{2}{++}{1}{+}{\PfTwo}{P} wave.  The red
    curves and lines show the result of a fit of
    \cref{eq:slope-parametrization} to the data (see text for
    details).}
  \label{fig:method:tp:examplespectrum}
\end{figure}

The intensities of most wave components fall approximately
exponentially with increasing \tpr.  This is consistent with Regge
theory, which at high energies describes the scattering process as
Pomeron exchange between the beam pion and target proton.  For waves
with spin projection $M \neq 0$, the exponential behavior is modified
by an additional $\rbrk{\tpr}^{\abs{M}}$ factor, which is given by the
forward limit of the Wigner $D$-functions~\cite{perl:1974high} and
suppresses the intensity at small~\tpr [see for example
\cref{fig:method:tp:examplespectrum:m1}].  We therefore parametrize
the \tpr spectra by the model
\begin{equation}
  \label{eq:slope-parametrization}
  \widehat{\mathcal{I}}_a^j(\tpr)
  = \dod{\widehat{N}_a^j}{\tpr}
  = A_a^j \cdot \rBrk{\tpr}^{\abs{M}} \cdot e^{-b_a^j\, \tpr}
\end{equation}
with the real-valued amplitude parameter $A_a^j$ and the slope
parameter $b_a^j$ for component~$j$ in wave~$a$ as free parameters.
The red curves in \cref{fig:method:tp:examplespectrum} show the result
of a \chisq fit of \cref{eq:slope-parametrization} to the data.  In
the formulation of the \chisq, the model function is integrated over
each \tpr bin (red horizontal lines) and compared to the data (black
horizontal lines).  For most wave components, the simple model in
\cref{eq:slope-parametrization} holds only approximately and in a
limited \tpr range.  Therefore, we exclude the two extremal \tpr bins
and fit the data in the reduced range
\SIvalRange{0.113}{\tpr}{0.724}{\GeVcsq}.  For some wave components,
narrower fit ranges are used (see \cref{tab:slopes} in
\cref{sec:results}).  The \tpr bins excluded from the fit and the
extrapolations of the model curve are shown in lighter colors.

Special cases are resonance components, for which the coupling
amplitudes in different waves are constrained via
\cref{eq:method:branchingdefinition}.  This constrains the \tpr
dependence of the coupling amplitudes $\mathcal{C}_a^j(\tpr)$ in the
different waves to be the same up to complex-valued proportionality
constants, \ie the branching amplitudes
$\prescript{}{b}{\mathcal{B}}_a^j$.  Although the dynamic amplitude
$\mathcal{D}_j$ for a resonance component is independent of \tpr, the
\tpr spectra of the resonance component in the different waves can be
slightly different even in this case.  This is caused by the
$I_{aa}(\mThreePi)\, \Abs[1]{\mathcal{P}(\mThreePi, \tpr)}^2$ term in
the integrand in \cref{eq:tprim-dependence} as the function
$I_{aa}(\mThreePi)$ is different for different waves.  In addition,
the statistical uncertainties of the extracted intensities
$\mathcal{I}_a^j(\tpr)$ are different in the different waves.
Therefore, the slope parameters of resonances in different waves,
which are extracted using \cref{eq:slope-parametrization}, may be
slightly different even though the coupling amplitudes are related by
\cref{eq:method:branchingdefinition}.

\subsection{Extraction of branching-fraction ratios}
\label{sec:method:br}

In order to extract the branching-fraction ratios of resonances that
appear in more than one decay channel, we calculate the yields
$\mathcal{N}_a^j(\tpr)$ of resonance component~$j$ in the
corresponding waves.  To this end, we integrate the resonance
intensity in a given \tpr bin over \mThreePi:
\begin{multlineOrEq}
  \label{eq:resonance_yield}
  \mathcal{N}_a^j(\tpr)
  = \Abs[1]{\mathcal{C}_a^j(\tpr)}^2
  \smashoperator[r]{\int_{m_\text{min}}^{m_\text{max}}}\! \dif{\mThreePi}\,
  I_{aa}(\mThreePi)\, \mThreePi\,
  \newLineTimesOrNot
  \Abs[1]{\mathcal{D}_j^\text{R}(\mThreePi; \zeta_j^\text{R})}^2.
\end{multlineOrEq}
This expression corresponds to \cref{eq:tprim-dependence} with the
production probability $\Abs[1]{\mathcal{P}(\mThreePi, \tpr)}^2$ set
to unity\footnote{\Cref{eq:resonance_yield} does not include the
  production probability because the branching-fraction ratio is a
  property of the resonance decay only.  Therefore, the yields have
  arbitrary units and are not normalized to number of events.}  and
without the division by the \tpr bin width.  The branching-fraction
ratio for resonance component~$j$ is defined as the ratio of the
\tpr-summed yields in the two waves~$a$ and~$b$:
\begin{equation}
  \label{eq:branch_fract_ratio}
  B_{ab}^j
  \equiv \frac{\dsum^{\tpr~\text{bins}} \mathcal{N}_a^j(\tpr)}{\dsum^{\tpr~\text{bins}} \mathcal{N}_b^j(\tpr)}.
\end{equation}
It is important to note that due to the phase-space factor
$I_{aa}(\mThreePi)$ in \cref{eq:resonance_yield},
$\mathcal{N}_{a, b}^j(\tpr)$ and therefore also $B_{ab}^j$ depend on
the chosen \mThreePi integration limits.  We use
$m_\text{min} = \SI{0.5}{\GeVcc}$ and
$m_\text{max} = \SI{2.5}{\GeVcc}$ for all resonances in all waves.
This mass range is much wider than the width of any of the resonances.
 %
%
%

\section{Systematic studies}
\label{sec:systematics}

The physical parameters obtained from the resonance-model fit, \ie the
resonance parameters, the branching-fraction ratios, and the \tpr
slope parameters of the wave components, are subject to systematic
uncertainties related to our fit model and fitting method (see
\cref{sec:method}).  In order to estimate these uncertainties, we
performed a large variety of studies.  In each study, an aspect of the
analysis is modified and the result is compared to our main result.
In addition to studies that test the stability of the fit result, we
performed studies to evaluate the evidence for selected resonance
signals.  These studies are discussed in
\cref{sec:results,sec:syst_uncert}.

Due to the multimodal nature of the \chisq~function (see
\cref{sec:method:fitmethod}), the effects observed in the various
systematic studies are statistically not always independent of one
another.  In fact, for some studies the systematic effects are
correlated in a highly nonlinear way.  Because of the complexity of
the resonance-model fits and their high computational cost, it is not
possible to estimate the correlations between the various systematic
studies.  We therefore estimate the systematic uncertainty intervals
using the minimum and maximum values of the physical parameters
observed in the performed studies.  The uncertainties estimated with
this approach do in general not represent Gaussian uncertainties.
Unless stated otherwise in \cref{sec:results,sec:syst_uncert}, all
systematic studies discussed below are included in the estimation of
the uncertainty intervals for the extracted parameters.  The obtained
systematic uncertainties are found to be at least 1~order of magnitude
larger than the statistical uncertainties.  Hence we quote in
\cref{sec:results} only the systematic uncertainties and omit
statistical uncertainties.

In this section, we describe only the most important studies that
either define the systematic uncertainties of some resonance
parameters or illustrate interesting effects.  We will discuss in
\cref{sec:results,sec:syst_uncert} the effects of these studies on the
resonance and \tpr slope parameters in detail.  For easier reference,
the studies are labeled by uppercase letters.

\paragraph*{\StudyK:}
In this study, the influence of background contaminations from kaon
diffraction, kaon pairs in the final state, central-production
reactions, and nonexclusive events in the selected data sample on the
fit result is studied.  To this end, the analysis is performed on a
data sample, in which \one~the information from the
particle-identification detectors for the beam (CEDARs) and the
final-state particles (RICH) was not used, \two~the rejection of
central-production events was not applied, and \three~the requirements
of exactly one recoil proton detected in the RPD and of transverse
momentum balance were not applied in the event selection (see
\cref{sec:setup_and_event_selection}).  Possible background
contributions are expected to be enhanced in this data sample, which
is \SI{76.2}{\percent} larger than that used for the main analysis.

\paragraph*{Studies~\studyA through~\studyD:}
The selection of the 14~waves that enter the resonance-model fit (see
\cref{tab:method:fitmodel:waveset}) is to some extent subjective.  In
addition, the fit model has difficulties describing details of some
partial-wave amplitudes.  This in particular is true for the intensity
distribution of the \wave{1}{++}{0}{+}{\Prho}{S} wave, which is the
most dominant wave in the data.  We therefore investigate in
Studies~\studyA through~\studyD how various waves influence the fit
result, by omitting single waves or combinations of waves from the
fit.  The various studies are listed in
\cref{tab:syst_studies:wave_set}.

\begin{table}[tbp]
  \renewcommand{\arraystretch}{1.2}
  \centering
  \caption{List of studies performed on smaller wave sets, in which some
    of the 14~waves that are used in the main fit (see
    \cref{tab:method:fitmodel:waveset}) are omitted.}
  \label{tab:syst_studies:wave_set}
  \begin{tabular}{cl}
    \hline
    \hline
    Study & Omitted waves \\
    \hline
    \studyA & All four $2^{-+}$ waves \\
    \studyB & \wave{1}{++}{0}{+}{\Prho}{S} and \wave{1}{++}{0}{+}{\PfTwo}{P} \\
    \studyC & All two $4^{++}$ waves \\
    \studyH & \wave{2}{++}{1}{+}{\Prho}{D} \\
    \studyI & \wave{2}{++}{2}{+}{\Prho}{D} \\
    \studyJ & \wave{2}{++}{1}{+}{\PfTwo}{P} \\
    \studyE & \wave{2}{++}{2}{+}{\Prho}{D} and \wave{2}{++}{1}{+}{\PfTwo}{P} \\
    \studyF & \wave{2}{++}{1}{+}{\Prho}{D} and \wave{2}{++}{1}{+}{\PfTwo}{P} \\
    \studyG & \wave{2}{++}{1}{+}{\Prho}{D} and \wave{2}{++}{2}{+}{\Prho}{D} \\
    \studyD & All three $2^{++}$ waves \\
    \hline
    \hline
  \end{tabular}
\end{table}

\paragraph*{\StudyL:}
We investigate the impact of the \tpr binning by applying a coarser
\tpr binning to the data using only eight bins, which are given in
\cref{tab:t-bins_8}.

\begin{table*}[tbp]
  \sisetup{%
    round-mode = places,
    round-precision = 3
  }
  \caption{Borders of the eight nonequidistant \tpr bins used for
    \StudyL.  The intervals are chosen such that each bin contains
    approximately \num[round-mode = places, round-precision =
    1]{5.8e6} events.}
  \label{tab:t-bins_8}
  \renewcommand{\arraystretch}{1.2}
  \newcolumntype{Z}{%
    >{\Makebox[0pt][c]\bgroup}%
    c%
    <{\egroup}%
  }
  \setlength{\tabcolsep}{0pt}  %
  \begin{tabular}{l@{\extracolsep{12pt}}c@{\extracolsep{6pt}}Z*{7}{cZ}c}
    \hline
    \hline
    Bin && 1 && 2 && 3 && 4 && 5 && 6 && 7 && 8 & \\
    \hline
    \tpr [\si{\GeVcsq}] &
    \num{0.100} &&
    \num{0.116} &&
    \num{0.136} &&
    \num{0.159} &&
    \num{0.188} &&
    \num{0.227} &&
    \num{0.285} &&
    \num{0.395} &&
    \num{1.000} \\
    \hline
    \hline
  \end{tabular}%
\end{table*}

\paragraph*{\StudyT:}
The impact of the assumption that the \tpr dependence of resonance
amplitudes is the same in partial waves with the same \JPCMrefl
quantum numbers but different decay modes is investigated in this
study.  To this end, we performed a resonance-model fit without the
constraint in \cref{eq:method:branchingdefinition}, so that the \tpr
dependence of the resonance amplitudes can be chosen freely by the fit
in all partial waves.  This model has 942~free parameters in
comparison to the 722~free parameters of the main fit.  Despite the
largely increased number of free parameters, the minimum \chisq~value
decreases only by a factor of \num{0.93} \wrt the main fit.  This
shows that for many resonances the constraint in
\cref{eq:method:branchingdefinition} is consistent with the data.
\StudyT plays a special role in the determination of the systematic
uncertainties of the branching-fraction ratios that are calculated
using \cref{eq:branch_fract_ratio}.  For a true resonance, the
branching-fraction ratio is expected to be independent of \tpr.  We
include the values found in the individual \tpr bins in the estimation
of the uncertainty intervals for the branching-fraction ratios.

\paragraph*{\StudyO:}
As described in \cref{sec:method:fitmodel}, we use a purely
phenomenological parametrization for the nonresonant contributions
[see \cref{eq:method:nonresterm}].  The choice of this parametrization
may impact the fit result, in particular for waves with significant
nonresonant contributions.  Although, we cannot uniquely identify the
underlying physics processes, Deck-like processes~\cite{deck:1964hm}
are believed to play a major role.  Several models exist for the Deck
process.  An example is shown in \cref{fig:deck_model}.  Using the
Deck model in \cref{eq:deck_ampl}, which is discussed in
\cref{sec:deck_model}, we generated \num{e8}~Monte Carlo events and
performed a mass-independent analysis using the same model with
88~waves as for the real data.  In \StudyO, we replace the
parametrizations of the nonresonant amplitudes [see
\cref{eq:method:nonresterm,eq:method:nonrestermsmall}] by the square
root of the intensity distributions of the Deck Monte Carlo data in
each partial wave.  As in the main fit, the phases of these
partial-wave projections of the Deck amplitude are assumed to be
independent of \mThreePi.  In \StudyO, the fit model has 693~free
parameters in comparison to the 722~free parameters of the main
fit. \Wrt the main fit, the minimum \chisq~value increases by a factor
of~\num{1.42}.  In order to find out which partial-wave amplitudes are
described differently, we decompose the \chisq~difference between the
study and the main fit into contributions from the elements of the
matrix $\Lambda_{a b}(\mThreePi, \tpr)$ defined in
\cref{eq:method:fitmethod:rhoredef}.  This is visualized in
\cref{fig:DeckMC_chi2difference} in the same way as in
\cref{fig:chi2Matrix}.  The diagonal elements show the contributions
to the \chisq~difference from the intensity distributions of each
partial wave, the off-diagonal elements the contributions from the
real (upper triangle) and imaginary parts (lower triangle) of the
interference terms between the waves.
\Cref{fig:DeckMC_chi2difference} shows that the largest contribution
to the \chisq~increase in \StudyO comes from the
\wave{1}{++}{0}{+}{\Prho}{S} amplitude.  \StudyO is particularly
relevant for the interpretation of the resonance signals in the
$1^{++}$ and $1^{-+}$ waves (see
\cref{sec:onePP_results,sec:oneMP_results}).

\begin{figure}[tbp]
  \centering
  \ifMultiColumnLayout{\includegraphics[width=\linewidth]{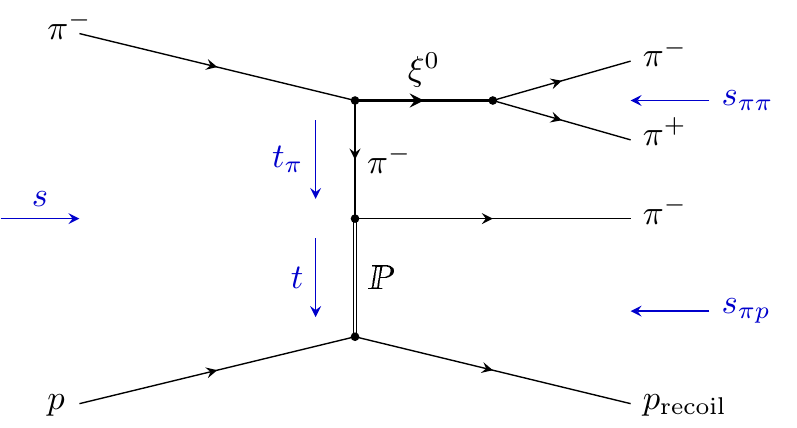}}{\includegraphics[scale=1]{fig8}}
  \caption{Example for a nonresonant production process for the $3\pi$
    final state as proposed by Deck~\cite{deck:1964hm}.  In this
    process, the beam pion dissociates into the isobar~$\xi^0$ and the
    bachelor~$\pi^-$, followed by diffractive scattering of one of
    these beam fragments (typically the~$\pi^-$, as shown here) off
    the target proton.}
  \label{fig:deck_model}
\end{figure}

\begin{figure}[tbp]
  \centering
  \includegraphics[width=\linewidthOr{\twoPlotWidth}]{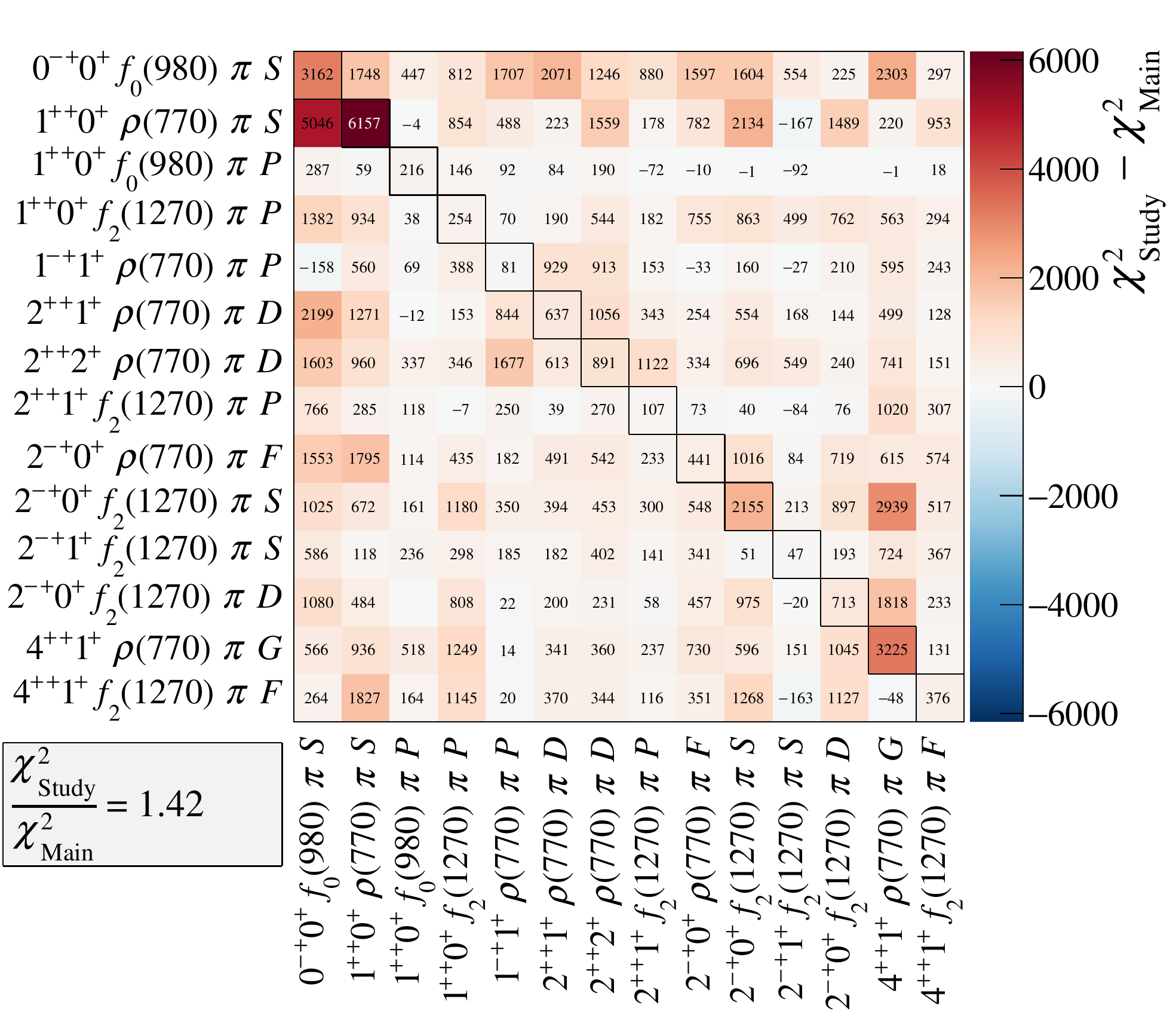}
  \caption{Decomposition of the \chisq~difference between the main fit
    and the fit, in which the parametrization of the nonresonant
    amplitude was replaced by the square root of the intensity
    distribution of the partial-wave projections of Deck Monte Carlo
    data [\StudyO].  The \chisq~difference is visualized in the form
    of a matrix, which shows the contributions (summed over the
    \mThreePi and \tpr bins) from the intensities and interference
    terms to the \chisq~difference.  Positive values (red colors)
    indicate that the data are described less well in the study.  The
    rare negative values (blue colors) indicate that the data are
    described better in the study.}
  \label{fig:DeckMC_chi2difference}
\end{figure}

\paragraph*{Studies~\studyS and~\studyR:}
As explained in \cref{sec:method:fitmethod}, the minimum value of the
\chisq~function that is determined by the resonance-model fit does not
follow a \chisq~distribution because \cref{eq:method:fitmethod:chi2}
does not take into account the correlations of the spin-density matrix
elements.  In order to test the potential bias introduced by this, we
constructed two possible \chisq~functions that take into account these
correlations (see \cref{sec:alt_chi_2}).  In \StudyS, we use the
\chisq~formulation in \cref{eq:systematics:chi2_alt1} with
\cref{eq:systematics:deviation1,eq:systematics:deviation2,eq:systematics:deviation3},
which is based on a single row of the spin-density matrix.  In
\StudyR, we use \cref{eq:systematics:chi2_alt1} with
\cref{eq:systematics:substitution}, which directly compares the
modeled and measured transition amplitudes.  The differences between
the resonance parameters estimated in Studies~\studyS and~\studyR are
small compared to the systematic uncertainties.  Comparing the two
studies with the main solution, large effects are only observed for
the resonances in the $1^{++}$ and $1^{-+}$ waves.  They are discussed
in \cref{sec:onePP_results,sec:oneMP_results}.

\paragraph*{\StudyAF:}
The model we employ for the diffractive-production probability
$\Abs[0]{\mathcal{P}(\mThreePi, \tpr)}^2$ in
\cref{eq:method:param:spindens,eq:method:param:prods} also influences
the fit result.  In order to estimate the systematic effect, we
performed a study, in which this factor was set to unity.  \Wrt the
main fit, the minimum \chisq~value increased by a factor of \num{1.01}
while the number of free parameters remained unchanged.  This shows
that both models describe the data on average equally well.  For most
of the resonance parameters the effects observed in \StudyAF are
small.  Exceptions are the \PaOne[1640] (see
\cref{sec:syst_uncert_onePP}), the \Ppi[1800] (see
\cref{sec:syst_uncert_zeroMP}), and the \PpiTwo* resonances (see
\cref{sec:syst_uncert_twoMP}).

\paragraph*{Studies~\studyM and~\studyN:}
We also studied the effect of the range parameter~$q_R$ of the
Blatt-Weisskopf factors in the decay $X^- \to \xi^0 \pi^-$ (vertex~1
in \cref{fig:3pi_reaction_isobar}).  These factors appear explicitly
in \cref{eq:method:a2dynamicwidth} and implicitly in the phase-space
integrals~$I_{aa}$ in
\cref{eq:method:transitionampl,eq:method:nonresterm:qtilde}.  In
\StudyM we set~$q_R$ to \SI{267}{\MeVc} and in \StudyN to
\SI{155}{\MeVc} corresponding to assumed strong-interaction ranges of
\SI{0.75}{\femto\meter} and \SI{1.29}{\femto\meter}, respectively.
Most resonance parameters change only slightly in both studies.
Exceptions are the \PaTwo* resonances (see
\cref{sec:syst_uncert_twoPP}), the \PpiOne[1600] (see
\cref{sec:syst_uncert_oneMP}), and the \PaFour (see
\cref{sec:syst_uncert_fourPP}).

Integrating the model function in \cref{eq:method:fitmethod:chi2} over
the \mThreePi bins instead of taking the function values at the mass
bin centers does not significantly influence the resonance parameters.
 %
%
%

\section{Results on resonance parameters and \tpr spectra of wave components}
\label{sec:results}

In this section, we describe and discuss the results of the
resonance-model fit grouped by the \JPC quantum numbers of the
resonances.  The subsections are ordered by increasing complexity of
the results.  We start with the \JPC sectors that contain the clearest
resonance signals that are well described by our model and later
discuss the more complicated cases, where several resonances with the
same \JPC quantum numbers appear.  In the last \cref{sec:oneMP}, we
discuss the resonance content of the spin-exotic $\JPC = 1^{-+}$ wave.
The extracted Breit-Wigner resonance parameters and their systematic
uncertainties are listed in \cref{tab:parameters} and are compared to
the PDG averages as listed in \refCite{Patrignani:2016xqp}.  The
positions of the resonance poles of the Breit-Wigner amplitudes in the
complex energy plane are discussed in \cref{sec:pole_positions}.  The
\tpr slope parameters of the resonant and nonresonant wave components,
determined by fitting the extracted \tpr spectra using
\cref{eq:slope-parametrization} (see \cref{sec:method:tp}), are listed
in \cref{tab:slopes}.  In the presentation of the results, we restrict
ourselves to figures that illustrate the typical quality of the fit or
certain aspects of the analysis.  The full fit result can be found in
the supplemental
material\ifMultiColumnLayout{~\cite{paper3_supplemental_material}}{ in
  \cref{sec:spin-dens_matrices,sec:phase-space_vol}} together with
additional information required to perform the resonance-model fit.
The data required to perform the resonance-model fit are provided in
computer-readable format at~\cite{paper3_hepdata}.

%
%
%

\begin{wideTableOrNot}[tbh]
  \renewcommand{\arraystretch}{\ifMultiColumnLayout{1.0}{1.2}}
  \centering
  \captionsetup[subtable]{position=top}
  \caption{Resonance parameters with systematic uncertainties as
    extracted in this analysis.  The statistical uncertainties are at
    least an order of magnitude smaller than the systematic ones and
    are hence omitted.  For comparison, the PDG averages are
    listed~\cite{Patrignani:2016xqp}.  For the \PaTwo, we quote the
    PDG average for the $3\pi$ decay mode. For the two entries marked
    with a \enquote{\,*\,} no PDG average exists.  The \PaOne[1420] is
    listed as \enquote{omitted from summary table} and the quoted mass
    and width values were estimated in an earlier COMPASS analysis
    based on the same data set that is used here but with only three
    waves in the resonance-model fit~\cite{Adolph:2015pws}.  The
    \PpiTwo[2005] is listed as a \enquote{further state} and we quote
    for comparison the parameters measured by the BNL E852
    experiment~\cite{Lu:2004yn} with the statistical and systematic
    uncertainties added in quadrature.}
  \label{tab:parameters}
  \let\parColWidth\relax
  \newlength{\parColWidth}
  \ifMultiColumnLayout{\setlength{\parColWidth}{0.095\linewidth}}{\setlength{\parColWidth}{0.111\linewidth}}
  \subfloat[$a_{J}$-like resonances]{%
    \label{tab:parameters:a}%
    \begin{tabular}{llp{\parColWidth}p{\parColWidth}p{\parColWidth}p{\parColWidth}p{\parColWidth}p{\parColWidth}}
      \hline
      \hline
      & &
      \multicolumn{1}{c}{\PaOne} &
      \multicolumn{1}{c}{\PaOne[1420]} &
      \multicolumn{1}{c}{\PaOne[1640]} &
      \multicolumn{1}{c}{\PaTwo} &
      \multicolumn{1}{c}{\PaTwo[1700]} &
      \multicolumn{1}{c}{\PaFour} \\
      & &
      \multicolumn{3}{c}{(\cref{sec:onePP})} &
      \multicolumn{2}{c}{(\cref{sec:twoPP})} &
      \multicolumn{1}{c}{(\cref{sec:fourPP})} \\
      \hline
      \rule{0pt}{1.1\normalbaselineskip}
      \multirow{4}{*}{\rotatebox{90}{\hspace{3pt} COMPASS}} & Mass &
      \multirow{2}{*}{$1299\,^{+12}_{-28}$} &
      \multirow{2}{*}{$1411\,^{+4}_{-5}$} &
      \multirow{2}{*}{$1700\,^{\phantom{1}+35}_{-130}$} &
      \multirow{2}{*}{$1314.5\,^{+4.0}_{-3.3}$} &
      \multirow{2}{*}{$1681\,^{+22}_{-35}$} &
      \multirow{2}{*}{$1935\,^{+11}_{-13}$} \\[\ifMultiColumnLayout{0ex}{-1ex}]
      & {\small [\si{\MeVcc}]} & & & & & & \\
      & Width &
      \multirow{2}{*}{$\phantom{1}380 \pm 80$} &
      \multirow{2}{*}{$\phantom{1}161\,^{+11}_{-14}$} &
      \multirow{2}{*}{$\phantom{1}510\,^{+170}_{\phantom{1}-90}$} &
      \multirow{2}{*}{$\phantom{1}106.6\,^{+3.4}_{-7.0}$} &
      \multirow{2}{*}{$\phantom{1}436\,^{+20}_{-16}$}  &
      \multirow{2}{*}{$\phantom{1}333\,^{+16}_{-21}$} \\[\ifMultiColumnLayout{0ex}{-1ex}]
      & {\small [\si{\MeVcc}]} & & & & & & \\[0.5ex]
      \hline
      \rule{0pt}{1.1\normalbaselineskip}
      \multirow{4}{*}{\rotatebox{90}{\centering PDG}} & Mass &
      \multirow{2}{*}{$1230 \pm 40$} &
      \multirow{2}{*}{$1414\,^{+15}_{-13}$} &
      \multirow{2}{*}{$1647 \pm 22$} &
      \multirow{2}{*}{$1319.0\,^{+1.0}_{-1.3}$} &
      \multirow{2}{*}{$1732 \pm 16$} &
      \multirow{2}{*}{$1995\,^{+10}_{\phantom{1}-8}$} \\[\ifMultiColumnLayout{0ex}{-1ex}]
      & {\small [\si{\MeVcc}]} & & & & & & \\
      & Width &
      \multirow{2}{*}{$250$ to $600$} &
      \multirow{2}{*}{$\phantom{1}153\,^{\phantom{1}+8}_{-23}$} &
      \multirow{2}{*}{$\phantom{1}254 \pm 27$} &
      \multirow{2}{*}{$\phantom{1}105.0\,^{+1.6}_{-1.9}$} &
      \multirow{2}{*}{$\phantom{1}194 \pm 40$} &
      \multirow{2}{*}{$\phantom{1}257\,^{+25}_{-23}$} \\[\ifMultiColumnLayout{0ex}{-1ex}]
      & {\small [\si{\MeVcc}]} & & & & & & \\[\ifMultiColumnLayout{-3ex}{-1ex}]
      & & & \multicolumn{1}{c}{*} & & & & \\[\ifMultiColumnLayout{-2ex}{-1ex}]
      \hline
      \hline
    \end{tabular}%
  }%
  \\
  \subfloat[$\pi_{J}$-like resonances]{%
    \label{tab:parameters:p}%
    \begin{tabular}{llp{\parColWidth}p{\parColWidth}p{\parColWidth}p{\parColWidth}p{\parColWidth}}
      \hline
      \hline
      & &
      \multicolumn{1}{c}{\Ppi[1800]} &
      \multicolumn{1}{c}{\PpiOne[1600]} &
      \multicolumn{1}{c}{\PpiTwo} &
      \multicolumn{1}{c}{\PpiTwo[1880]} &
      \multicolumn{1}{c}{\PpiTwo[2005]} \\
      & &
      \multicolumn{1}{c}{(\cref{sec:zeroMP})} &
      \multicolumn{1}{c}{(\cref{sec:oneMP})} &
      \multicolumn{3}{c}{(\cref{sec:twoMP})} \\
      \hline
      \rule{0pt}{1.1\normalbaselineskip}
      \multirow{4}{*}{\rotatebox{90}{\hspace{3pt}  COMPASS}} & Mass &
      \multirow{2}{*}{$1804\,^{+6}_{-9}$} &
      \multirow{2}{*}{$1600\,^{+110}_{\phantom{1}-60}$} &
      \multirow{2}{*}{$1642\,^{+12}_{\phantom{1}-1}$} &
      \multirow{2}{*}{$1847\,^{+20}_{\phantom{1}-3}$} &
      \multirow{2}{*}{$1962\,^{+17}_{-29}$} \\[\ifMultiColumnLayout{0ex}{-1ex}]
      & {\small [\si{\MeVcc}]} & & & & & \\
      & Width &
      \multirow{2}{*}{$\phantom{1}220\,^{\phantom{1}+8}_{-11}$} &
      \multirow{2}{*}{$\phantom{1}580\,^{+100}_{-230}$} &
      \multirow{2}{*}{$\phantom{1}311\,^{+12}_{-23}$} &
      \multirow{2}{*}{$\phantom{1}246\,^{+33}_{-28}$} &
      \multirow{2}{*}{$\phantom{1}371\,^{\phantom{1}+16}_{-120}$} \\[\ifMultiColumnLayout{0ex}{-1ex}]
      & {\small [\si{\MeVcc}]} & & & & & \\[0.5ex]
      \hline
      \rule{0pt}{1.1\normalbaselineskip}
      \multirow{4}{*}{\rotatebox{90}{\centering PDG}} & Mass &
      \multirow{2}{*}{$1812 \pm 12$} &
      \multirow{2}{*}{$1662\,^{+8}_{-9}$} &
      \multirow{2}{*}{$1672.2 \pm 3.0$} &
      \multirow{2}{*}{$1895 \pm 16$} &
      \multirow{2}{*}{$1974 \pm 84$} \\[\ifMultiColumnLayout{0ex}{-1ex}]
      & {\small [\si{\MeVcc}]} & & & & & \\
      & Width &
      \multirow{2}{*}{$\phantom{1}208 \pm 12$} &
      \multirow{2}{*}{$\phantom{1}241 \pm 40$} &
      \multirow{2}{*}{$\phantom{1}260\phantom{.1} \pm 9$} &
      \multirow{2}{*}{$\phantom{1}235 \pm 34$} &
      \multirow{2}{*}{$\phantom{1}341 \pm 152$} \\[\ifMultiColumnLayout{0ex}{-1ex}]
      & {\small [\si{\MeVcc}]} & & & & & \\[\ifMultiColumnLayout{-3ex}{-1ex}]
      & & & & & & \multicolumn{1}{c}{*} \\[\ifMultiColumnLayout{-2ex}{-1ex}]
      \hline
      \hline
    \end{tabular}%
  }%
\end{wideTableOrNot}
 
%
%
%

%
%
%
%
%
\ifMultiColumnLayout{%
  \begin{table*}[p]
    \renewcommand{\thempfootnote}{[\arabic{mpfootnote}]}  %
}{%
  \begin{table}[h!]
    \begin{minipage}{\textwidth}
      \renewcommand\footnoterule{}}
  \renewcommand{\arraystretch}{1.4}
  \captionsetup[subtable]{position=top}
  \caption{The \tpr slope parameters~$b_a^j$ in units of
    \si{\perGeVcsq} extracted by fitting
    \cref{eq:slope-parametrization} to the \tpr spectra of the wave
    components.  The quoted uncertainties are of systematic origin.
    Statistical uncertainties are at least an order of magnitude
    smaller than the systematic ones and are hence omitted.  For most
    wave components, the fits are performed in the range
    \SIvalRange{0.113}{\tpr}{0.724}{\GeVcsq}.  Reduced fit ranges are
    given in the footnotes.  Cases where the model is not able to
    describe the \tpr spectrum are marked by a dagger
    (\enquote{$\dagger$}).  Partial waves, for which the \tpr
    dependence of the resonance amplitudes is connected via
    \cref{eq:method:branchingdefinition}, are marked with a star
    (\enquote{\,*\,}).  Slight differences of the extracted
    slope-parameter values for the resonances in these waves originate
    from differences in the decay phase-space volumes and in the
    statistical uncertainties (see \cref{sec:method:tp}).}
  \label{tab:slopes}
  \let\parColWidth\relax
  \newlength{\parColWidth}
  \ifMultiColumnLayout{\setlength{\parColWidth}{0.08\linewidth}}{\setlength{\parColWidth}{0.14\linewidth}}
  \let\parColWidthA\relax
  \newlength{\parColWidthA}
  \ifMultiColumnLayout{\setlength{\parColWidthA}{0.1475\linewidth}}{\setlength{\parColWidthA}{0.2\linewidth}}
  \ifMultiColumnLayout{}{\captionsetup[subtable]{%
      singlelinecheck=false,
      justification=raggedright,
      margin=6em,  %
    }}
  \subfloat[$0^{-+}$ Waves (\cref{sec:zeroMP})]{%
    \label{tab:slopes:0mp}%
    \small%
    \begin{tabular}{p{\parColWidthA}cp{\parColWidth}p{\parColWidth}}
      \hline
      \hline
      \multicolumn{1}{c}{Wave}            & \phantom{*} & \multicolumn{1}{c}{\Ppi[1800]}      & \multicolumn{1}{c}{Nonresonant} \\
      \hline
      \wave{0}{-+}{0}{+}{\PfZero[980]}{S} &             & $\phantom{1}8.8\,^{+0.7}_{-0.3}$ & $26\,^{+6}_{-5}$\stepcounter{footnote}\footnotemark[\value{footnote}]           \\
      \hline
      \hline
    \end{tabular}%
  }%
  \\
  \subfloat[$1^{++}$ Waves (\cref{sec:onePP})]{%
    \label{tab:slopes:1pp}%
    \small%
    \begin{tabular}{p{\parColWidthA}cp{\parColWidth}p{\parColWidth}p{\parColWidth}p{\parColWidth}}
    	\hline
      \hline
    	\multicolumn{1}{c}{Wave}            & \phantom{*} & \multicolumn{1}{c}{\PaOne} & \multicolumn{1}{c}{\PaOne[1420]}    & \multicolumn{1}{c}{\PaOne[1640]} & \multicolumn{1}{c}{Nonresonant} \\
      \hline
    	\wave{1}{++}{0}{+}{\Prho}{S}        & *           & $11.8\,^{+0.9}_{-4.2}$  & \multicolumn{1}{c}{---}             & $\phantom{1}7.7\,^{+6.2}_{-0.4}$ & $12.5\,^{+2.1}_{-1.5}$           \\
    	\wave{1}{++}{0}{+}{\PfZero[980]}{P} &             & \multicolumn{1}{c}{---}    & $\phantom{1}9.5\,^{+0.6}_{-1.0}$ & \multicolumn{1}{c}{---}          & $11.8\,^{+0.8}_{-1.2}$           \\
    	\wave{1}{++}{0}{+}{\PfTwo}{P}       & *           & $11 \pm 4$     & \multicolumn{1}{c}{---}             & $\phantom{1}7.6\,^{+1.6}_{-0.5}$ & $11.2\,^{+2.7}_{-2.2}$           \\
      \hline
      \hline
    \end{tabular}%
  }%
  \\
  \subfloat[$1^{-+}$ Waves (\cref{sec:oneMP})]{%
    \label{tab:slopes:1mp}%
    \small%
    \begin{tabular}{p{\parColWidthA}cp{\parColWidth}p{\parColWidth}}
    	\hline
      \hline
    	\multicolumn{1}{c}{Wave}     & \phantom{*} & \multicolumn{1}{c}{\PpiOne[1600]} & \multicolumn{1}{c}{Nonresonant} \\
      \hline
    	\wave{1}{-+}{1}{+}{\Prho}{P} &             & \multicolumn{1}{c}{$\dagger$}  & $19.1\,^{+1.4}_{-4.7}$\stepcounter{footnote}\footnotemark[\value{footnote}]           \\
      \hline
      \hline
    \end{tabular}%
  }%
  \\
  \subfloat[$2^{++}$ Waves (\cref{sec:twoPP})]{%
    \label{tab:slopes:2pp}%
    \small%
    \begin{tabular}{p{\parColWidthA}cp{\parColWidth}p{\parColWidth}p{\parColWidth}}
    	\hline
      \hline
    	\multicolumn{1}{c}{Wave}      & \phantom{*} & \multicolumn{1}{c}{\PaTwo}          & \multicolumn{1}{c}{\PaTwo[1700]} & \multicolumn{1}{c}{Nonresonant} \\
      \hline
    	\wave{2}{++}{1}{+}{\Prho}{D}  & *           & $\phantom{1}7.9 \pm 0.5$       & $\phantom{1}7.3\,^{+2.4}_{-0.9}$ & $13.6\,^{+0.4}_{-1.8}$           \\
    	\wave{2}{++}{2}{+}{\Prho}{D}  &             & $\phantom{1}9.0\,^{+1.2}_{-0.7}$ & \multicolumn{1}{c}{$\dagger$}                 & $\phantom{1}8.1\,^{+1.6}_{-0.5}$ \\
    	\wave{2}{++}{1}{+}{\PfTwo}{P} & *           & $\phantom{1}7.8\,^{+0.6}_{-0.5}$ & $\phantom{1}7.2\,^{+1.1}_{-0.8}$ & \multicolumn{1}{c}{$\dagger$} \\
      \hline
      \hline
    \end{tabular}%
  }%
  \\
  \subfloat[$2^{-+}$ Waves (\cref{sec:twoMP})]{%
    \label{tab:slopes:2mp}%
    \small%
    \begin{tabular}{p{\parColWidthA}cp{\parColWidth}p{\parColWidth}p{\parColWidth}p{\parColWidth}}
      \hline
      \hline
      \multicolumn{1}{c}{Wave}      & \phantom{*} & \multicolumn{1}{c}{\PpiTwo}         & \multicolumn{1}{c}{\PpiTwo[1880]} & \multicolumn{1}{c}{\PpiTwo[2005]}   & \multicolumn{1}{c}{Nonresonant} \\
      \hline
      \wave{2}{-+}{0}{+}{\Prho}{F}  & *           & $\phantom{1}8.5\,^{+0.9}_{-0.5}$ & $\phantom{1}7.8\,^{+0.5}_{-0.9}$  & $\phantom{1}6.8\,^{+0.4}_{-3.9}$ & \multicolumn{1}{c}{$\dagger$} \\
      \wave{2}{-+}{0}{+}{\PfTwo}{S} & *           & $\phantom{1}8.5\,^{+0.9}_{-0.5}$ & $\phantom{1}7.8\,^{+7.5}_{-0.9}$  & $\phantom{1}6.7\,^{+0.4}_{-1.3}$ & \multicolumn{1}{c}{$\dagger$} \\
      \wave{2}{-+}{1}{+}{\PfTwo}{S} &             & \multicolumn{1}{c}{$\dagger$} & \multicolumn{1}{c}{$\dagger$}  & $\phantom{1}7.1\,^{+3.5}_{-2.6}$ & $\phantom{1}6.9\,^{+1.1}_{-1.9}$ \\
      \wave{2}{-+}{0}{+}{\PfTwo}{D} & *           & $\phantom{1}8.4\,^{+0.8}_{-1.7}$ & $\phantom{1}7.8\,^{+0.5}_{-0.9}$  & $\phantom{1}6.7\,^{+0.4}_{-1.3}$ & $12\,^{+6}_{-2}$           \\
      \hline
      \hline
    \end{tabular}%
  }%
  \\
  \subfloat[$4^{++}$ Waves (\cref{sec:fourPP})]{%
    \label{tab:slopes:4pp}%
    \small%
    \begin{tabular}{p{\parColWidthA}cp{\parColWidth}p{\parColWidth}}
      \hline
      \hline
      \multicolumn{1}{c}{Wave}      & \phantom{*} & \multicolumn{1}{c}{\PaFour}         & \multicolumn{1}{c}{Nonresonant} \\
      \hline
      \wave{4}{++}{1}{+}{\Prho}{G}  & *           & $\phantom{1}9.2\,^{+0.8}_{-0.5}$ & $14 \pm 4$                 \\
      \wave{4}{++}{1}{+}{\PfTwo}{F} & *           & $\phantom{1}9.2\,^{+0.8}_{-0.5}$ & $14.5\,^{+1.8}_{-3.7}$           \\
      \hline
      \hline
    \end{tabular}%
  }%
  \ifMultiColumnLayout{\renewcommand{\thempfootnote}{\arabic{mpfootnote}}}{}  %
  \addtocounter{footnote}{-1}%
  \footnotetext[\value{footnote}]{Fit range \SIvalRange{0.113}{\tpr}{0.326}{\GeVcsq}}
  \stepcounter{footnote}%
  \footnotetext[\value{footnote}]{Fit range \SIvalRange{0.113}{\tpr}{0.449}{\GeVcsq}.}
\ifMultiColumnLayout{\end{table*}}{\end{minipage}\end{table}}
 \clearpage
%
%
%

\subsection{$\JPC = 0^{-+}$ resonances}
\label{sec:zeroMP}

\subsubsection{Results on $0^{-+}$ resonances}
\label{sec:zeroMP_results}

The only $\JPC = 0^{-+}$ wave included in the resonance-model fit is
the \wave{0}{-+}{0}{+}{\PfZero[980]}{S} wave.  It contributes
\SI{2.4}{\percent} to the total intensity in the mass range from
\SIrange{0.5}{2.5}{\GeVcc}.  The intensity distribution of this wave
is shown in
\cref{fig:intensity_0mp_tbin1,fig:intensity_0mp_tbin10,fig:intensity_0mp_tbin11}
for three \tpr bins.  Except in the highest \tpr bin, the intensity
distributions exhibit a clear peak of the \Ppi[1800] resonance at
$\mThreePi \approx \SI{1.8}{\GeVcc}$ with a shoulder toward lower
masses.  The picture changes dramatically in the highest \tpr bin,
where the intensity at the \Ppi[1800] peak position is close to zero
and hence the low-mass shoulder dominates the spectrum.
\Cref{fig:intensity_phases_0mp} also shows, as an example, the
\mThreePi dependence of the relative phases of the $0^{-+}$ wave \wrt
the \wave{1}{++}{0}{+}{\Prho}{S} and the \wave{2}{-+}{1}{+}{\PfTwo}{S}
waves.  Clearly rising phase motions are observed in the
\SI{1.8}{\GeVcc} mass region.

\ifMultiColumnLayout{\begin{figure*}[p]}{\begin{figure}[tbp]}
  \centering
  \subfloat[][]{%
    \includegraphics[width=\threePlotWidth]{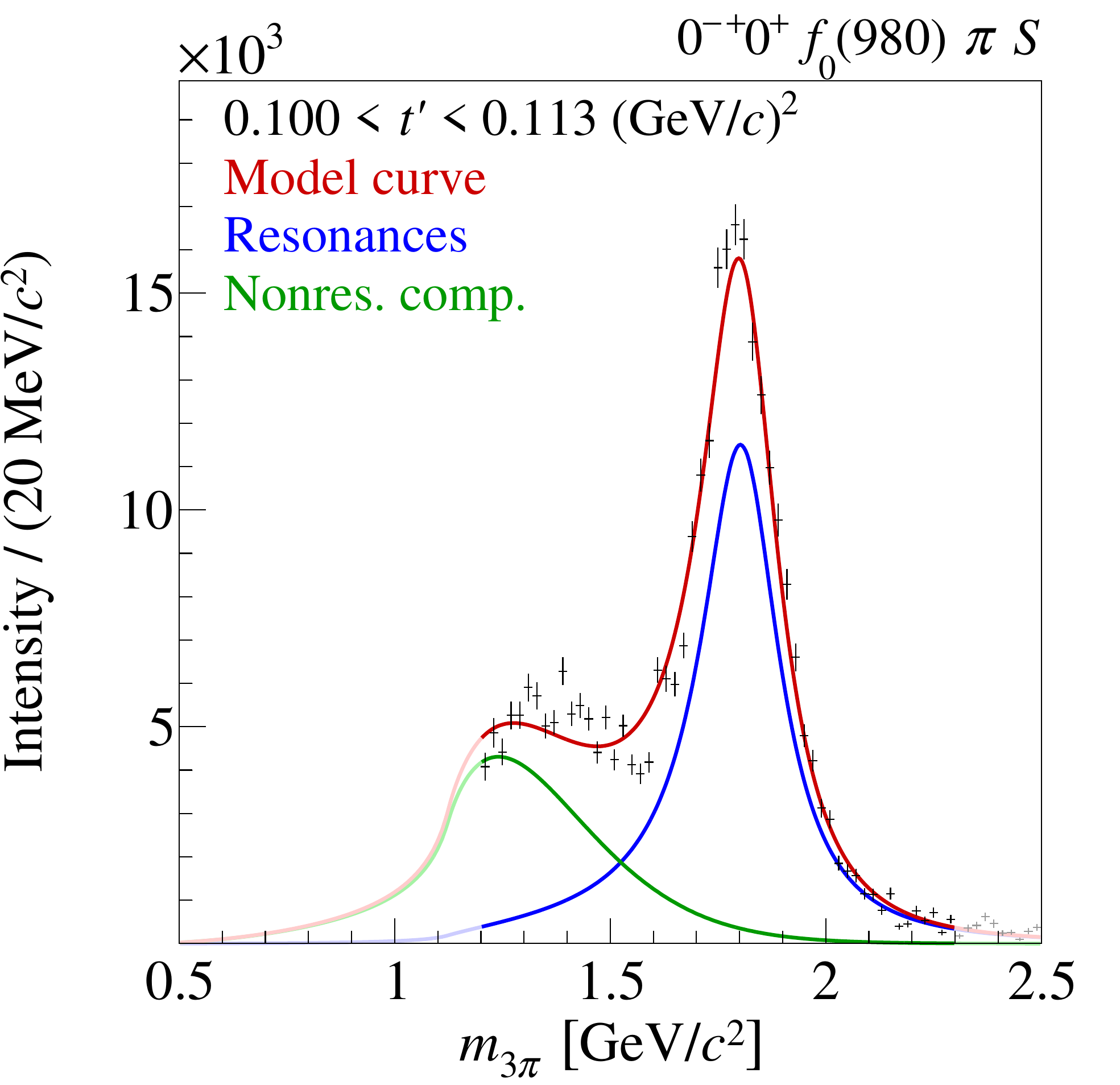}%
    \label{fig:intensity_0mp_tbin1}%
  }%
  \hspace*{\threePlotSpacing}%
  \subfloat[][]{%
    \includegraphics[width=\threePlotWidth]{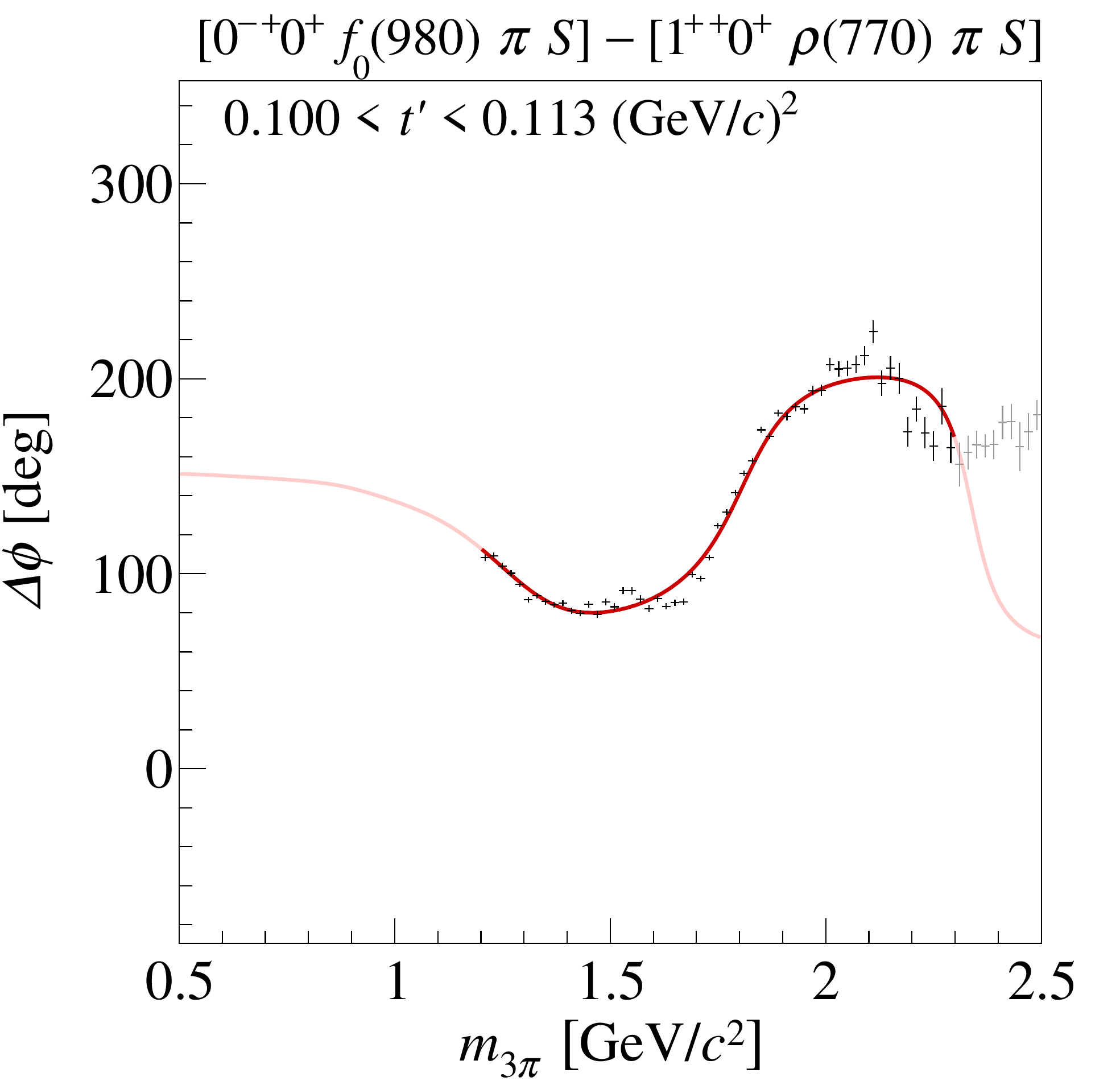}%
    \label{fig:phase_0mp_1pp_rho_tbin1}%
  }%
  \hspace*{\threePlotSpacing}%
  \subfloat[][]{%
    \includegraphics[width=\threePlotWidth]{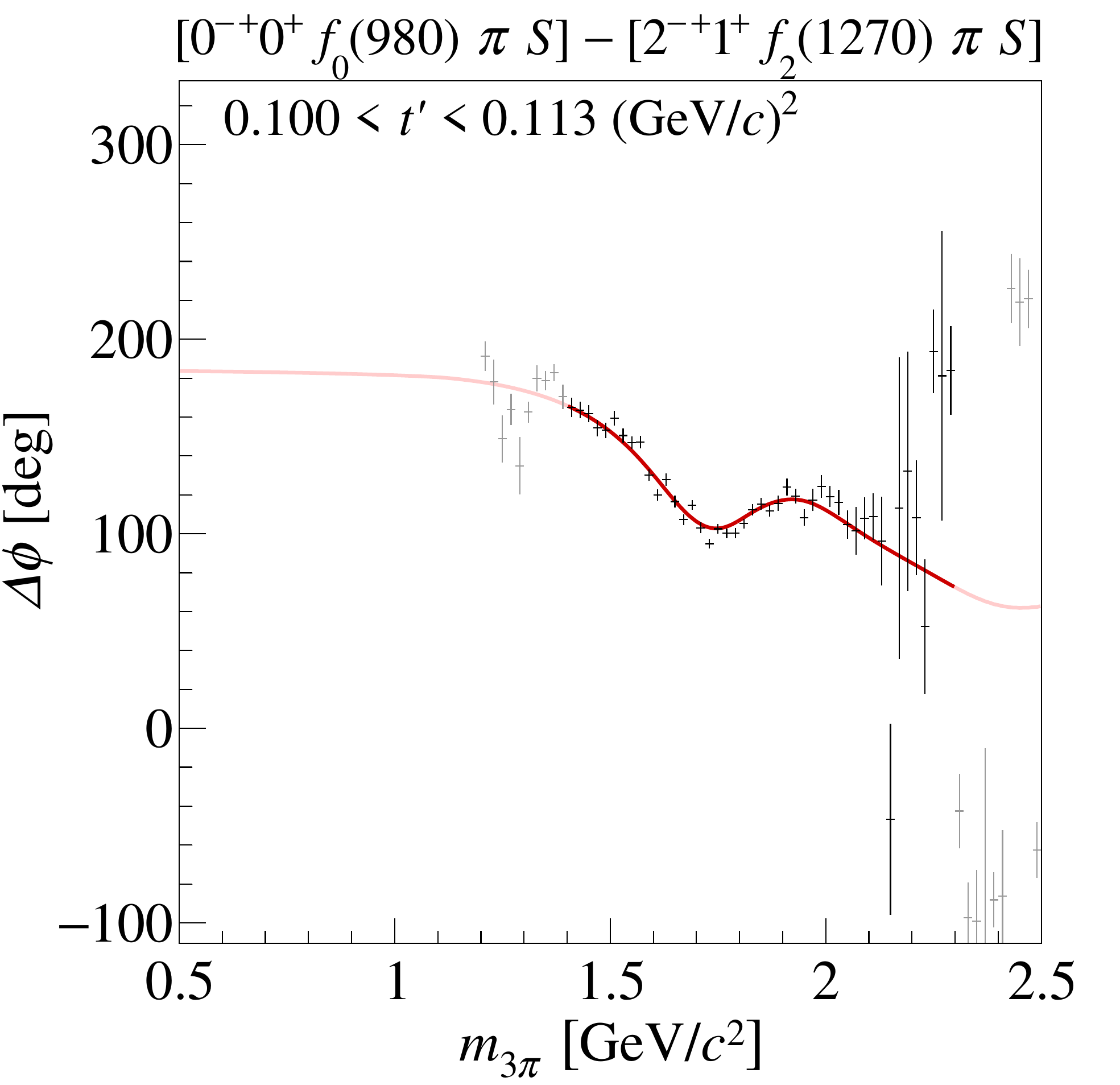}%
    \label{fig:phase_0mp_2mp_f2_tbin1}%
  }%
  \\
  \subfloat[][]{%
    \includegraphics[width=\threePlotWidth]{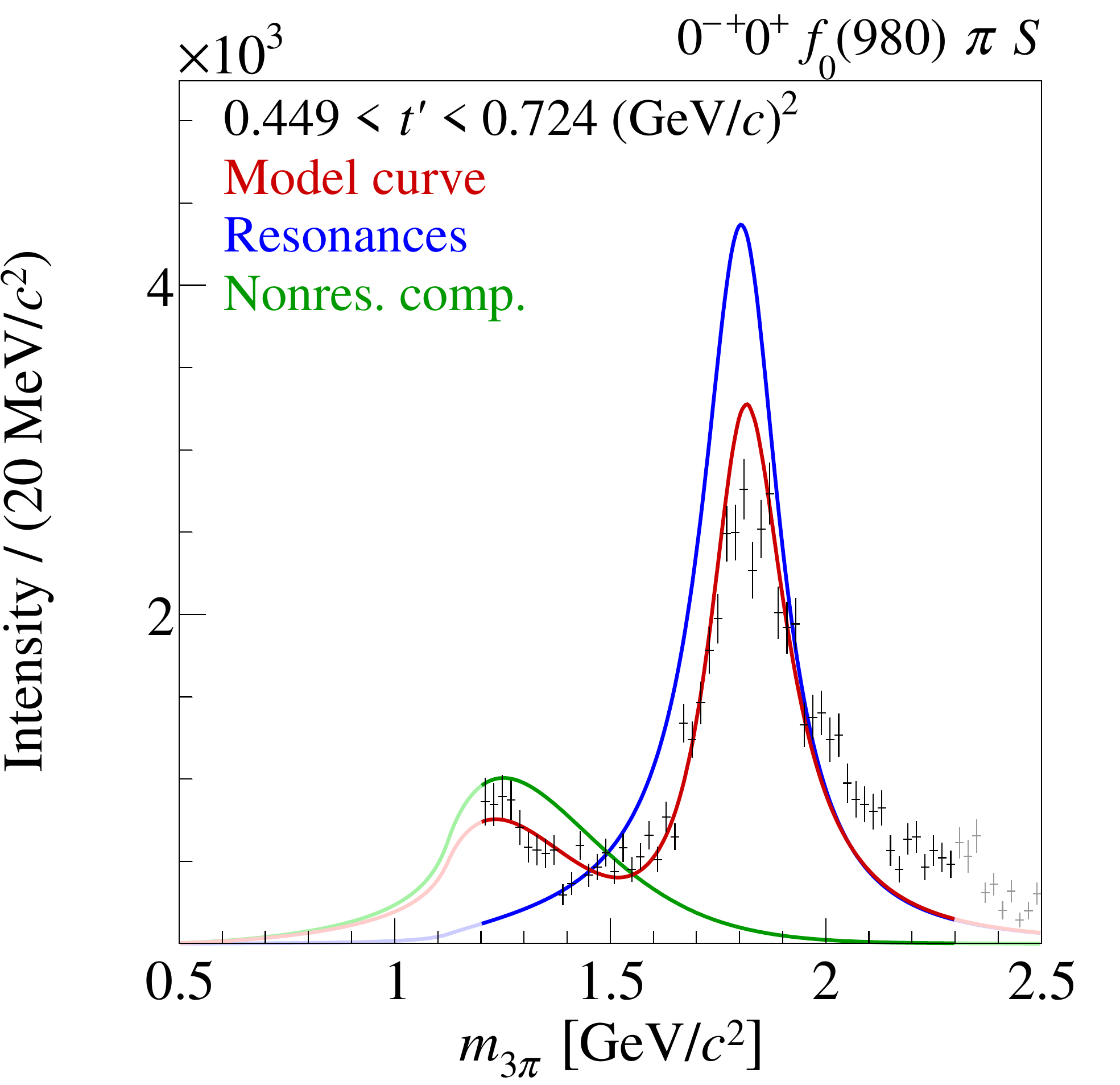}%
    \label{fig:intensity_0mp_tbin10}%
  }%
  \hspace*{\threePlotSpacing}%
  \subfloat[][]{%
    \includegraphics[width=\threePlotWidth]{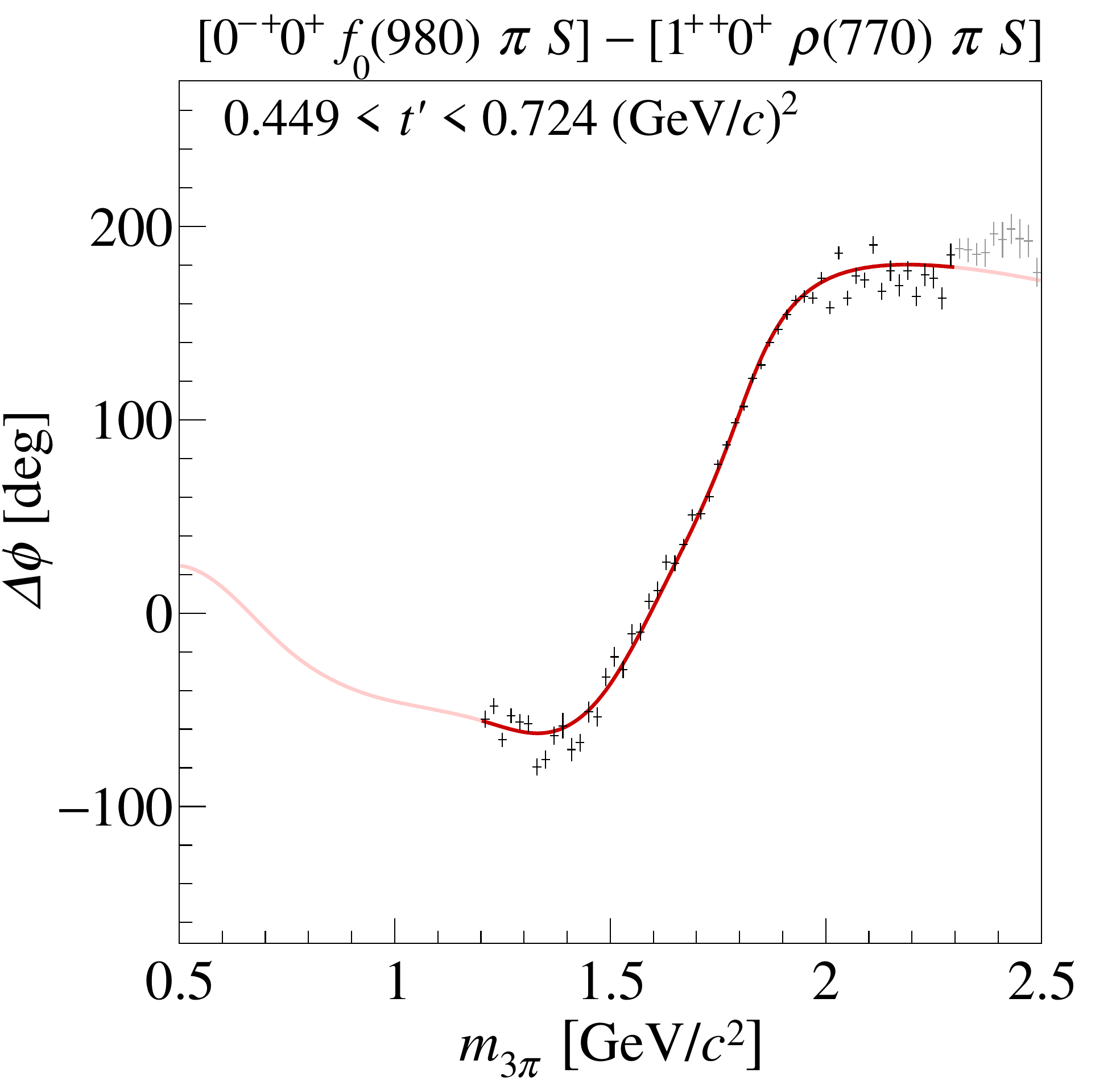}%
    \label{fig:phase_0mp_1pp_rho_tbin10}%
  }%
  \hspace*{\threePlotSpacing}%
  \subfloat[][]{%
    \includegraphics[width=\threePlotWidth]{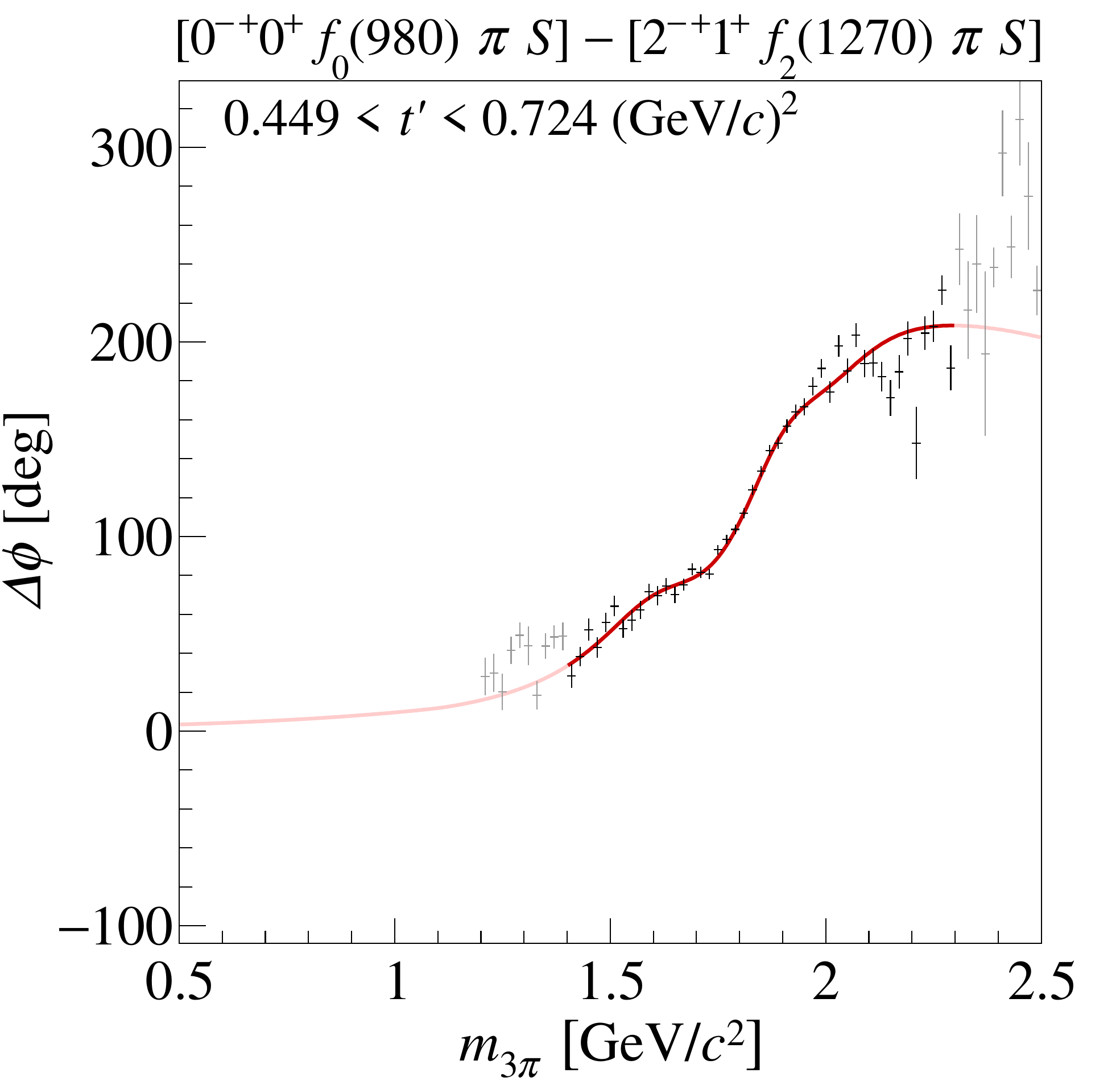}%
    \label{fig:phase_0mp_2mp_f2_tbin10}%
  }%
  \\
  \subfloat[][]{%
    \includegraphics[width=\threePlotWidth]{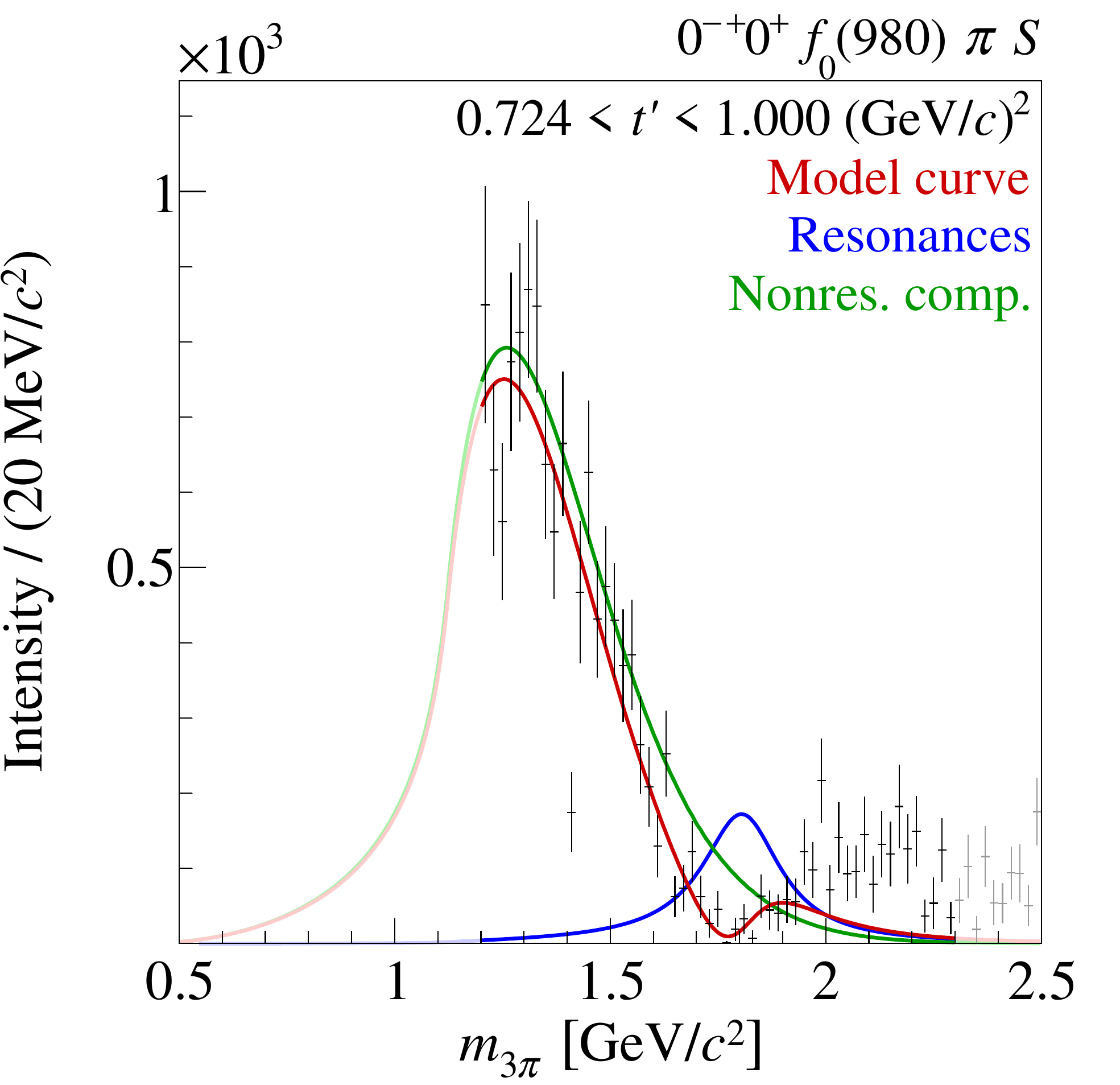}%
    \label{fig:intensity_0mp_tbin11}%
  }%
  \hspace*{\threePlotSpacing}%
  \subfloat[][]{%
    \includegraphics[width=\threePlotWidth]{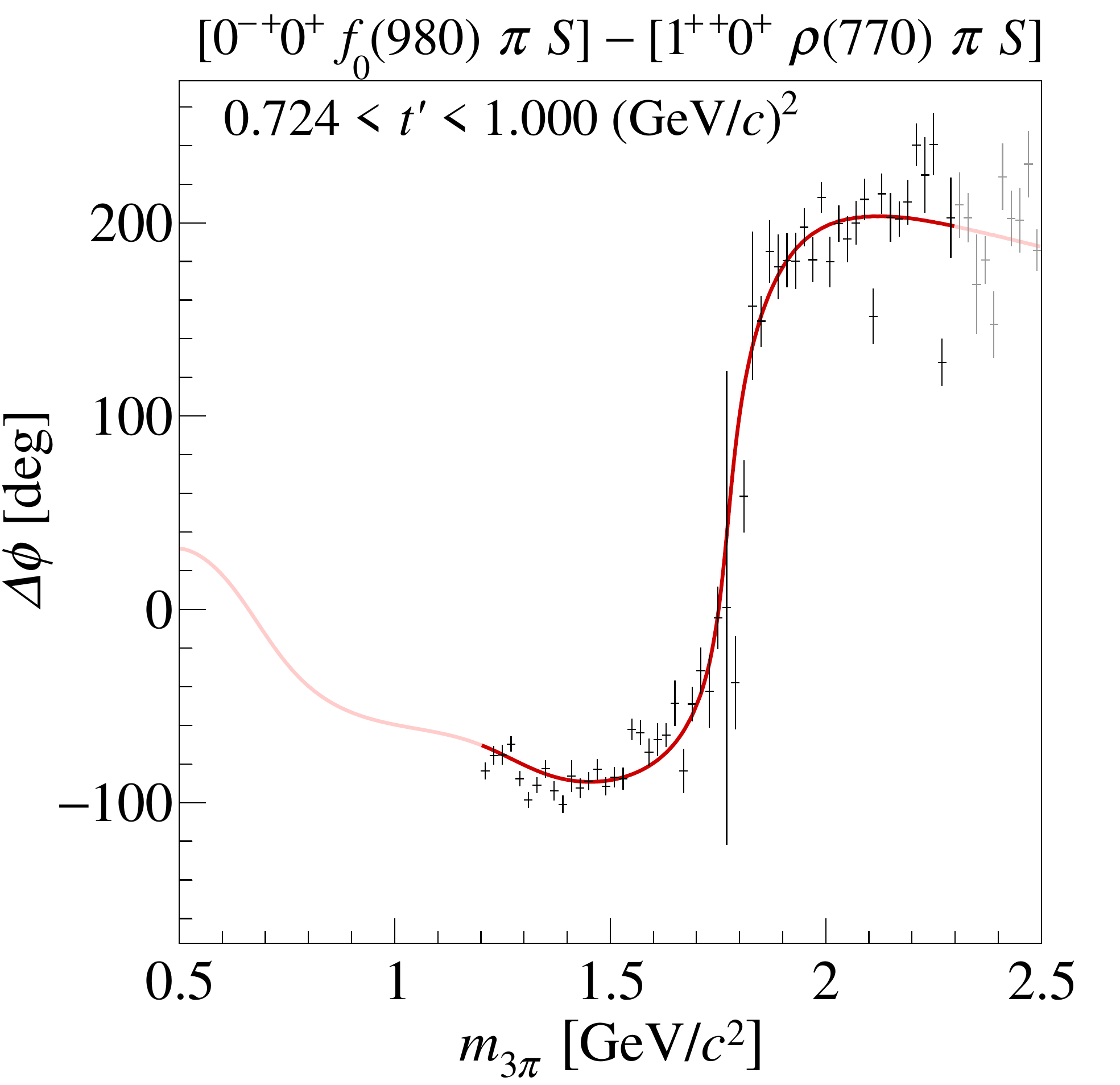}%
    \label{fig:phase_0mp_1pp_rho_tbin11}%
  }%
  \hspace*{\threePlotSpacing}%
  \subfloat[][]{%
    \includegraphics[width=\threePlotWidth]{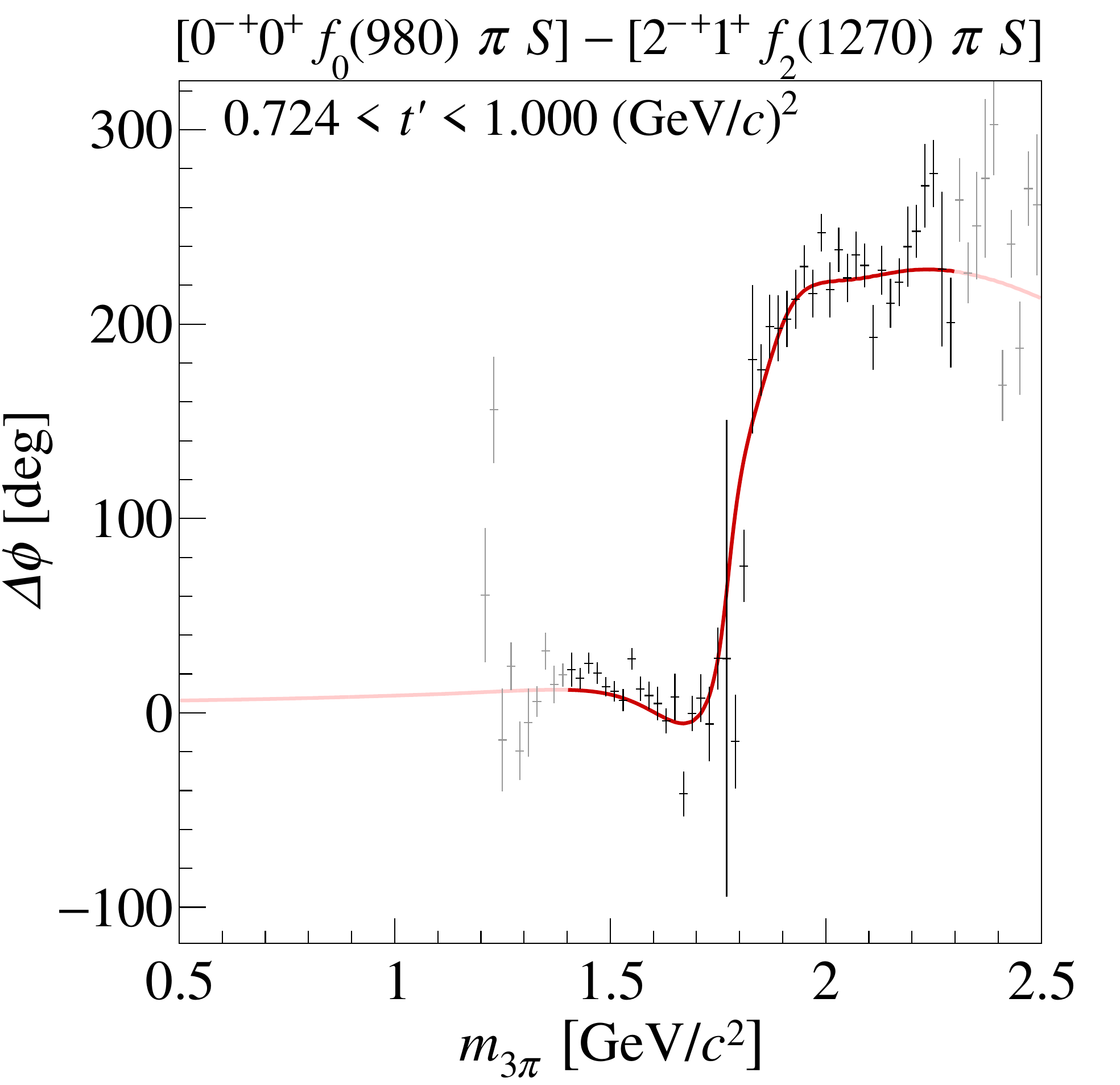}%
    \label{fig:phase_0mp_2mp_f2_tbin11}%
  }%
  \caption{The \wave{0}{-+}{0}{+}{\PfZero[980]}{S} partial-wave
    amplitude in three \tpr bins (rows): (left column) intensity
    distributions; (central column) phase motions \wrt the
    \wave{1}{++}{0}{+}{\Prho}{S} wave; (right column) phase motions
    \wrt the \wave{2}{-+}{1}{+}{\PfTwo}{S} wave.  The data points are
    taken from \refCite{Adolph:2015tqa} and represent the so-called
    mass-independent analysis (see \cref{sec:mass-independent_fit}).
    The red curve represents the full model (see
    \cref{tab:method:fitmodel:waveset}), which is the coherent sum of
    the wave components.  The other curves represent the wave
    components: \Ppi[1800] resonance (blue curves), nonresonant
    contribution (green curves).  The extrapolations of the model and
    the wave components beyond the fit range are shown in lighter
    colors.}
  \label{fig:intensity_phases_0mp}
\ifMultiColumnLayout{\end{figure*}}{\end{figure}}

The data are well described by the fit model (red curves in
\cref{fig:intensity_phases_0mp}), which contains two $0^{-+}$
components: a Breit-Wigner resonance for the \Ppi[1800] (blue curves)
and a nonresonant component (green curves).  The extrapolations of
these curves below and above the fitted mass range of
\SIvalRange{1.2}{\mThreePi}{2.3}{\GeVcc} are shown in lighter colors
in \cref{fig:intensity_phases_0mp}.  The \Ppi[1800] is parametrized
using \cref{eq:BreitWigner,eq:method:fixedwidth}, the nonresonant
component using \cref{eq:method:nonrestermsmall} (see
\cref{tab:method:fitmodel:waveset}).  In our fit model, the
nonresonant contribution is attributed to the low-mass shoulder.  At
low values of \tpr, it interferes constructively with the resonance at
the \Ppi[1800] peak position.  At higher values of \tpr, the
interference of the two components is destructive at the peak position
due to a sign flip of the coupling amplitude of the nonresonant
component at $\tpr \approx \SI{0.3}{\GeVcsq}$ (see discussion in
\cref{sec:production_phases}).  In the highest \tpr bin, the
destructive interference of the two components is complete and leads
to a dip in the intensity distribution around the \Ppi[1800] mass.
The remaining low-mass shoulder is completely described by the
nonresonant component.  In the intensity distributions, the model
exhibits some disagreement with the observed peak shape and does not
reproduce the high-mass shoulder in the two highest \tpr bins.

The strong variation of the intensity of the $0^{-+}$ wave with \tpr
originates from the very different \tpr dependences of the amplitudes
of the two $0^{-+}$ wave components.  \Cref{fig:tprim_0mp} shows the
\tpr spectra for both components as determined using
\cref{eq:tprim-dependence} together with the results of fits using
\cref{eq:slope-parametrization}.  While the intensity of the
\Ppi[1800] component exhibits an approximately exponential behavior
with slope parameter \SIaerrSys{8.8}{0.7}{0.3}{\perGeVcsq}, the
intensity of the nonresonant component first drops steeply with
$b = \SIaerrSys{26}{6}{5}{\perGeVcsq}$ at low values of \tpr, before
it starts to rise again with \tpr, forming a dip at
$\tpr \approx \SI{0.3}{\GeVcsq}$.

\begin{figure}[tbp]
  \centering
  \includegraphics[width=\twoPlotWidth]{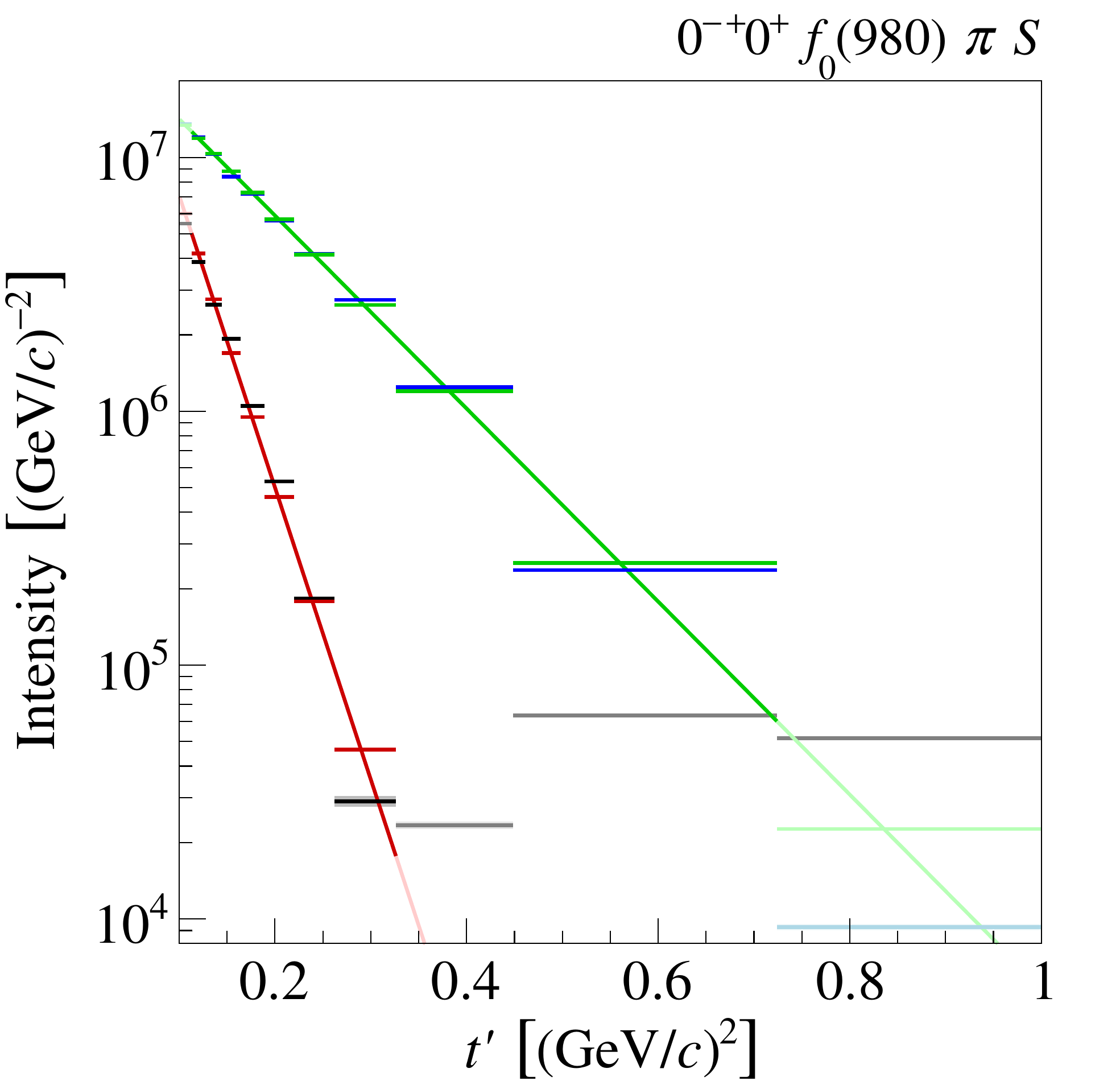}
  \caption{Similar to \cref{fig:method:tp:examplespectrum}, but
    showing the \tpr spectra of the two $\JPC = 0^{-+}$ wave
    components as given by \cref{eq:tprim-dependence}: the \Ppi[1800]
    component is shown as blue lines (central values) and light blue
    boxes (statistical uncertainties; not visible for most bins), the
    nonresonant component is shown as black lines and gray boxes as in
    \cref{fig:method:tp:examplespectrum}.  The red and green curves
    and horizontal lines represent fits using
    \cref{eq:slope-parametrization}.}
  \label{fig:tprim_0mp}
\end{figure}

The $0^{-+}$ wave exhibits clearly rising phases \wrt the
\wave{1}{++}{0}{+}{\Prho}{S} wave in the \Ppi[1800] region (see
central column of \cref{fig:intensity_phases_0mp}).  At low \tpr, the
relative phase decreases at about \SI{1.3}{\GeVcc} due to the \PaOne
and rises at about \SI{1.8}{\GeVcc} due to the \Ppi[1800].  At higher
values of \tpr, the decrease is less pronounced and the relative phase
rises steeply starting at \SI{1.5}{\GeVcc}.  This is explained in our
fit model by a sign change of the coupling amplitude of the
nonresonant $0^{-+}$ component, which dominates the low-mass region,
leading to an additional rise of the total phase of the $0^{-+}$
amplitude.  The extremely rapid phase motion at \SI{1.8}{\GeVcc} in
the highest \tpr bin is a direct consequence of the nearly vanishing
intensity at this mass.  Since the phase is defined \wrt the origin of
the complex plane, its value changes rapidly by \SI{\pm 180}{\degree}
if the amplitude passes close to the origin~\cite{Grigorenko:1999}.
We observe a similar \tpr dependence of the phase motions \wrt other
waves.  As an example, we show in the right column of
\cref{fig:intensity_phases_0mp} the phase motions \wrt the
\wave{2}{-+}{1}{+}{\PfTwo}{S} wave.  Here, the phase drop in the
lowest \tpr bin, which appears at about \SI{1.6}{\GeVcc}, is caused by
the \PpiTwo.  Within the fit ranges, the model describes all relative
phases of the $0^{-+}$ wave well in all \tpr bins.

From the fit, we obtain the Breit-Wigner resonance parameters
$m_{\Ppi[1800]} = \SIaerrSys{1804}{6}{9}{\MeVcc}$ and
$\Gamma_{\Ppi[1800]} = \SIaerrSys{220}{8}{11}{\MeVcc}$.  The
\Ppi[1800] resonance parameters are rather insensitive to changes of
the fit model discussed in \cref{sec:systematics}.  The estimated
systematic uncertainties are therefore the smallest of all \piJ-like
resonances in the model.  More details on the results of the
systematic studies are discussed in \cref{sec:syst_uncert_zeroMP}.
It is worth mentioning that in the study, in which the fit range was
narrowed to \SIvalRange{1.6}{\mThreePi}{2.3}{\GeVcc}, the nonresonant
component is practically vanishing.  This demonstrates that indeed
most of the peak structure arises from the \Ppi[1800].

\subsubsection{Discussion of results on $0^{-+}$ resonances}
\label{sec:zeroMP_discussion}

Although the \Ppi[1800] in principle has been well known for more than
three decades, its resonance parameters are not well determined.  In
particular the \Ppi[1800] mass values extracted by previous
experiments show a large spread and fall into two
clusters~\cite{Patrignani:2016xqp}: one with central values around
\SI{1780}{\MeVcc} and the other around \SI{1860}{\MeVcc}.  Our result
for the \Ppi[1800] mass of
$m_{\Ppi[1800]} = \SIaerrSys{1804}{6}{9}{\MeVcc}$ falls between these
two clusters and is in good agreement with the PDG world average of
$m_{\Ppi[1800]} = \SI{1812(12)}{\MeVcc}$~\cite{Patrignani:2016xqp}.
This is also true for the \Ppi[1800] width, for which the PDG average
is $\Gamma_{\Ppi[1800]} = \SI{208(12)}{\MeVcc}$ compared to our value
of $\Gamma_{\Ppi[1800]} = \SIaerrSys{220}{8}{11}{\MeVcc}$.  Our
measurement of the \Ppi[1800] parameters is the most precise and
accurate so far.  It is also consistent within uncertainties with a
previous COMPASS measurement using a lead
target~\cite{alekseev:2009aa}.

In the \wave{0}{-+}{0}{+}{\pipiS}{S} wave, we observe a peak that is
similar in shape and position to the \Ppi[1800] peak in the
\wave{0}{-+}{0}{+}{\PfZero[980]}{S} wave [see Figs.~24 and 25(b) in
\refCite{Adolph:2015tqa}].  Although the $\pipiS \pi$ wave was not
included in the resonance model fit for reasons discussed below, the
observed similarity of the peaks suggests that the \Ppi[1800]
resonance parameters would be similar in this wave.

The \Ppi[1800] is the second radial excitation of the pion.  Its
lighter partner state is the \Ppi[1300].  This state has been observed
in the $\Prho \pi$ and $\pipiS \pi$ final states, as well as in
$\gamma \gamma$ production~\cite{Patrignani:2016xqp}.  The parameters
of the \Ppi[1300] are only poorly known.  The world averages estimated
by the PDG are $m_{\Ppi[1300]} = \SI{1300(100)}{\MeVcc}$ and
$\Gamma_{\Ppi[1300]} =
\SIrange{200}{600}{\MeVcc}$~\cite{Patrignani:2016xqp}.  Also the
coupling of the \Ppi[1300] to the $\pipiS \pi$ final state is
controversial.  The Obelix Collaboration claims that the coupling is
\num{2.2(4)}~times stronger than for the $\Prho \pi$ final state and
extracts a resonance mass of
$m_{\Ppi[1300]}=\SI{1200(40)}{\MeVcc}$~\cite{salvini:2004gz}.  The
Crystal Barrel Collaboration, however, sets an upper limit for the
coupling to the $\pipiS \pi$ decay channel of 0.15~times the coupling
to $\Prho \pi$~\cite{Abele:2001js} and quotes a mass of
$m_{\Ppi[1300]}=\SI{1375(40)}{\MeVcc}$.  The two experiments also
disagree on the value of the \Ppi[1300] width.  In our data, we
observe an unusually strong \tpr dependence of the intensity of the
\wave{0}{-+}{0}{+}{\pipiS}{S} wave in the \SI{1.3}{\GeVcc} region [see
Figs.~24 and~35(a) in \refCite{Adolph:2015tqa}], which is similar to
that of the nonresonant component in the
\wave{0}{-+}{0}{+}{\PfZero[980]}{S} wave (see \cref{fig:tprim_0mp}).
In addition, the intensity in this mass region is strongly dependent
on the PWA model employed for the mass-independent analysis.  We
therefore did not include the \wave{0}{-+}{0}{+}{\pipiS}{S} wave in
the resonance-model fit.

Since the intensity spectra of the \wave{0}{-+}{0}{+}{\PfZero[980]}{S}
wave show a significant shoulder at
$\mThreePi \approx \SI{1.3}{\GeVcc}$, we tried an alternative
description of this partial-wave amplitude using a \Ppi[1300]
resonance component instead of the nonresonant component.  This model
describes the data less well than the main fit and does not yield
meaningful \Ppi[1300] resonance parameters.\footnote{In the solutions
  with the lowest \chisq~values, the \Ppi[1300] mass is found at the
  lower parameter limit of \SI{1}{\GeVcc}.}
If we include in another study a nonresonant component, the minimum
\chisq~value decreases by a factor of~\num{0.97} \wrt the main
fit.\footnote{Compared to the \num{722} free parameters of the main
  fit, this fit has \num{746} free parameters.}  In the solution with
the lowest~\chisq, the \Ppi[1300] is found with a mass and width of
about \SI{1630}{\MeVcc} and \SI{380}{\MeVcc}, respectively.  While the
width value is compatible with previous measurements, the mass value
is clearly not.\footnote{Only local minima with significantly larger
  \chisq~values yield \Ppi[1300] masses of about \SI{1270}{\MeVcc}
  that are compatible with previous measurements.}  Moreover, the
\wave{0}{-+}{0}{+}{\PfZero[980]}{S} wave does not show any phase rise
in the \SI{1.3}{\GeVcc} mass region.  Within our model, we therefore
conclude that the data do not support a \Ppi[1300] signal in the
$\PfZero[980] \pi$ decay mode.  This conclusion is consistent with the
fact that so far no observation of such a \Ppi[1300] decay has been
claimed.

Heavier excited pion states with masses around
\SIlist{2070;2360}{\MeVcc} were reported by the authors of
\refCite{anisovich:2001pn}.  We do not see clear resonance signals of
heavy pions in the mass range from \SIrange{2000}{2500}{\MeVcc} in the
\wave{0}{-+}{0}{+}{\PfZero[980]}{S} wave.
 %
%
%

\subsection{$\JPC = 4^{++}$ resonances}
\label{sec:fourPP}

\subsubsection{Results on $4^{++}$ resonances}
\label{sec:fourPP_results}

We include two $\JPC = 4^{++}$ waves, \wave{4}{++}{1}{+}{\Prho}{G} and
\wave{4}{++}{1}{+}{\PfTwo}{F}, in the resonance-model fit.  Both have
small intensities and contribute \SI{0.8}{\percent} and
\SI{0.2}{\percent}, respectively, to the total intensity in the mass
range from \SIrange{0.5}{2.5}{\GeVcc}.  The intensity distributions of
the two waves are shown for the lowest \tpr bin in
\cref{fig:intensity_4pp_f2_tbin1,fig:intensity_4pp_rho_tbin1}.  In
both waves, a clear peak around \SI{1.9}{\GeVcc} is observed.  The
shape of the intensity distributions depends only weakly on \tpr.  The
\wave{4}{++}{1}{+}{\PfTwo}{F} wave exhibits a slight shoulder at high
masses.  In the \wave{4}{++}{1}{+}{\Prho}{G} wave, this shoulder is
more pronounced, and in addition a low-mass shoulder is visible.  In
both waves, these features are most pronounced at low \tpr and vanish
in the highest \tpr bin.  \Cref{fig:intensity_phases_4pp} also shows,
as an example, the \mThreePi dependence of the relative phases of the
$4^{++}$ waves \wrt the \wave{1}{++}{0}{+}{\Prho}{S} and
\wave{1}{++}{0}{+}{\PfTwo}{P} waves in the lowest \tpr bin.  Clearly
rising phases are observed in the \SI{1.9}{\GeVcc} mass region.  In
addition, \cref{fig:phase_4pp_rho_4pp_f2_tbin1} shows the relative
phase between the two $4^{++}$ waves.  The approximately constant
phase indicates that there is a common dominant resonance in the two
waves.

\begin{wideFigureOrNot}[tbp]
  \centering
  \subfloat[][]{%
    \includegraphics[width=\fourPlotWidth]{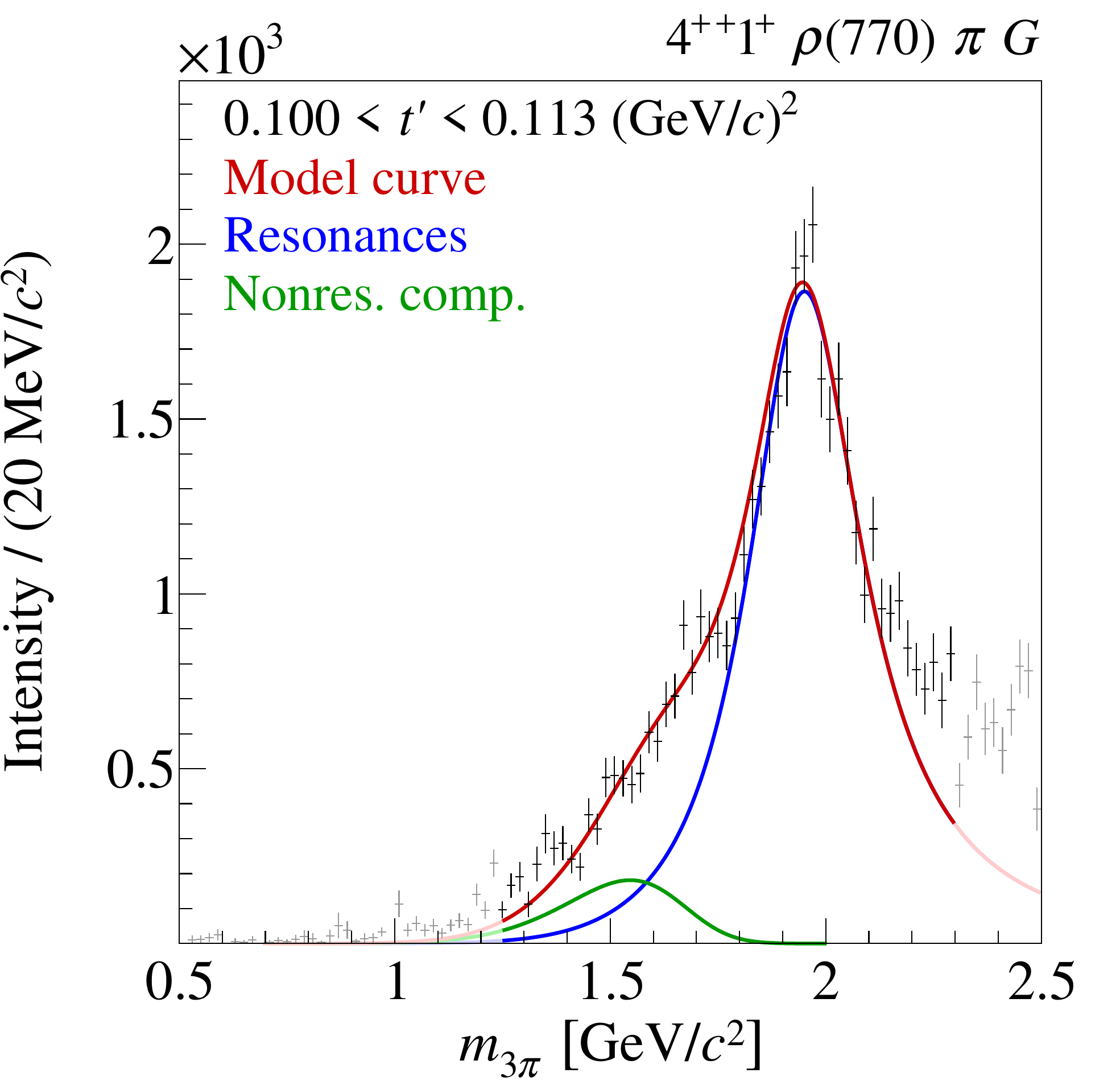}%
    \label{fig:intensity_4pp_rho_tbin1}%
  }%
  \hspace*{\fourPlotSpacing}%
  \subfloat[][]{%
    \includegraphics[width=\fourPlotWidth]{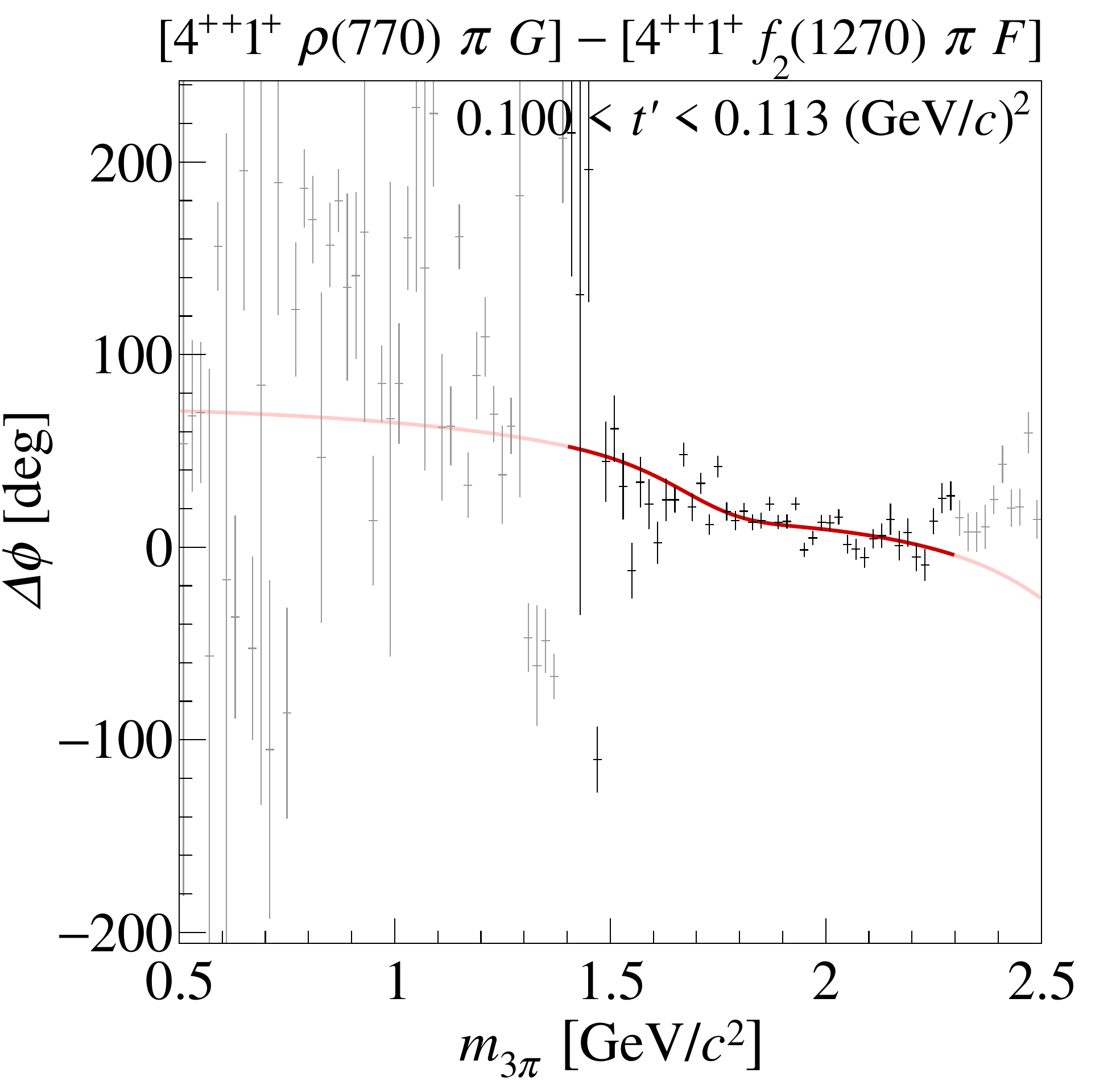}%
    \label{fig:phase_4pp_rho_4pp_f2_tbin1}%
  }%
  \hspace*{\fourPlotSpacing}%
  \subfloat[][]{%
    \includegraphics[width=\fourPlotWidth]{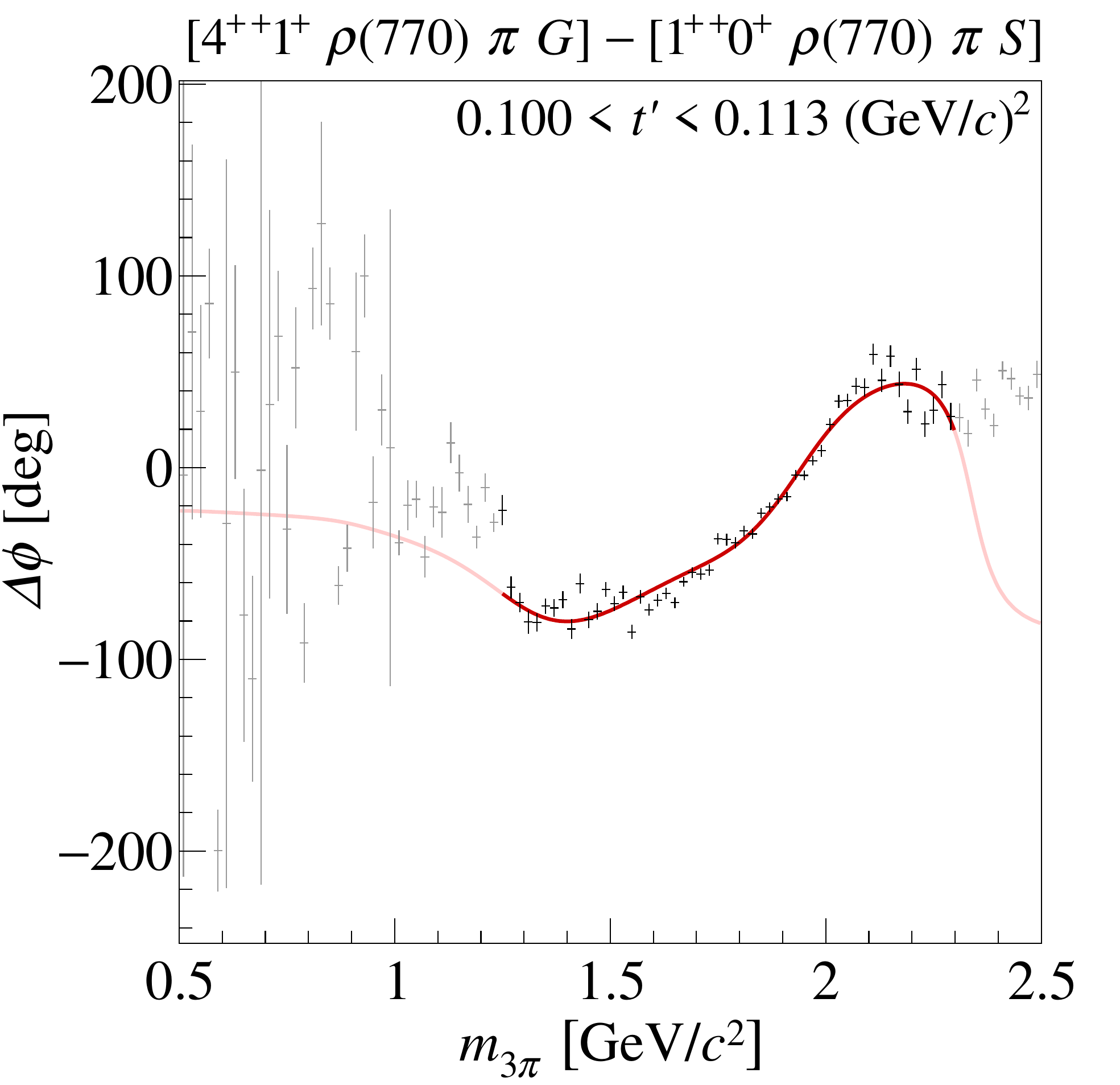}%
    \label{fig:phase_4pp_rho_1pp_rho_tbin1}%
  }%
  \hspace*{\fourPlotSpacing}%
  \subfloat[][]{%
    \includegraphics[width=\fourPlotWidth]{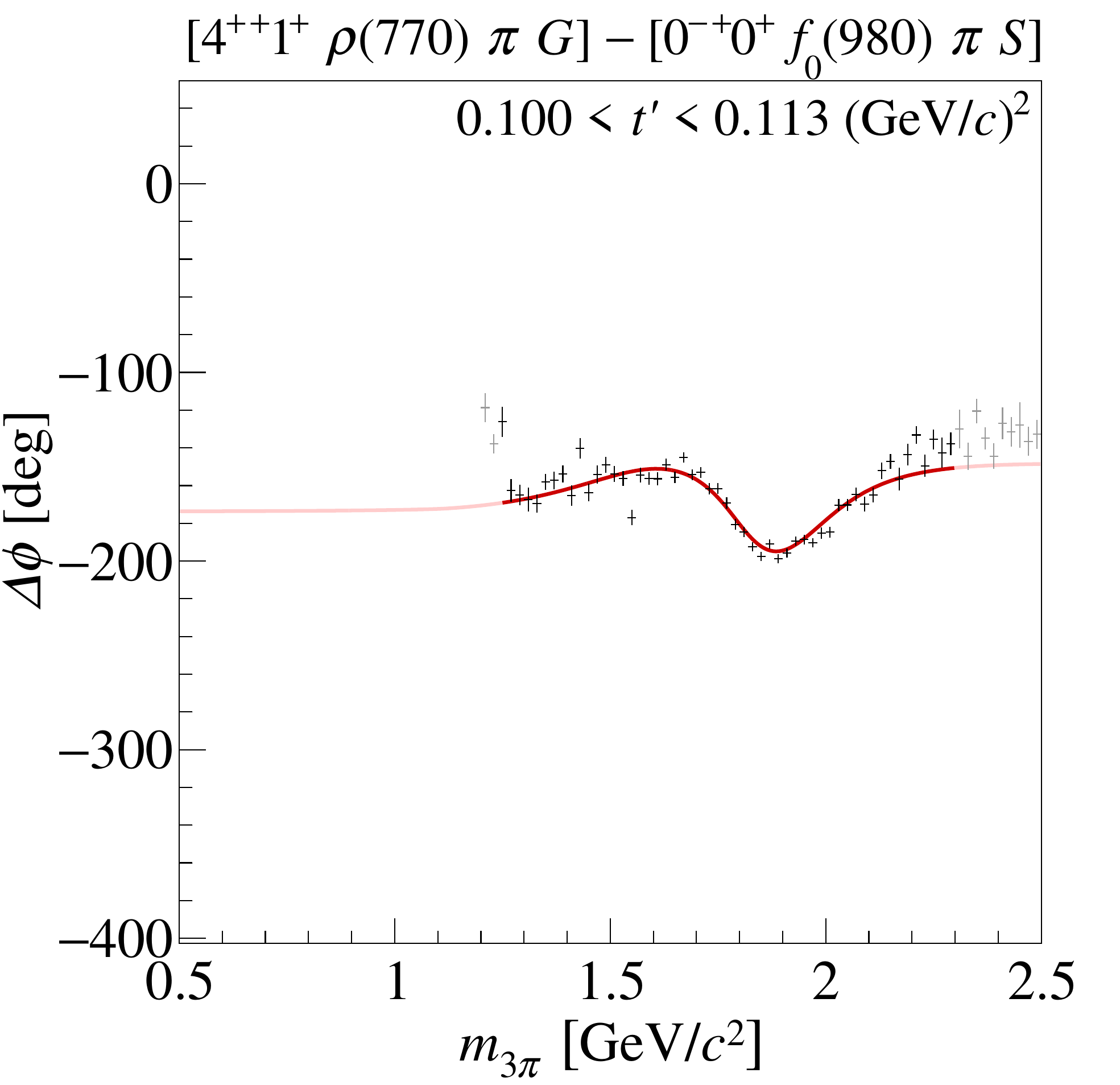}%
    \label{fig:phase_4pp_rho_0mp_tbin1}%
  }%
  \\
  \hspace*{\fourPlotWidth}%
  \hspace*{\fourPlotSpacing}%
  \subfloat[][]{%
    \includegraphics[width=\fourPlotWidth]{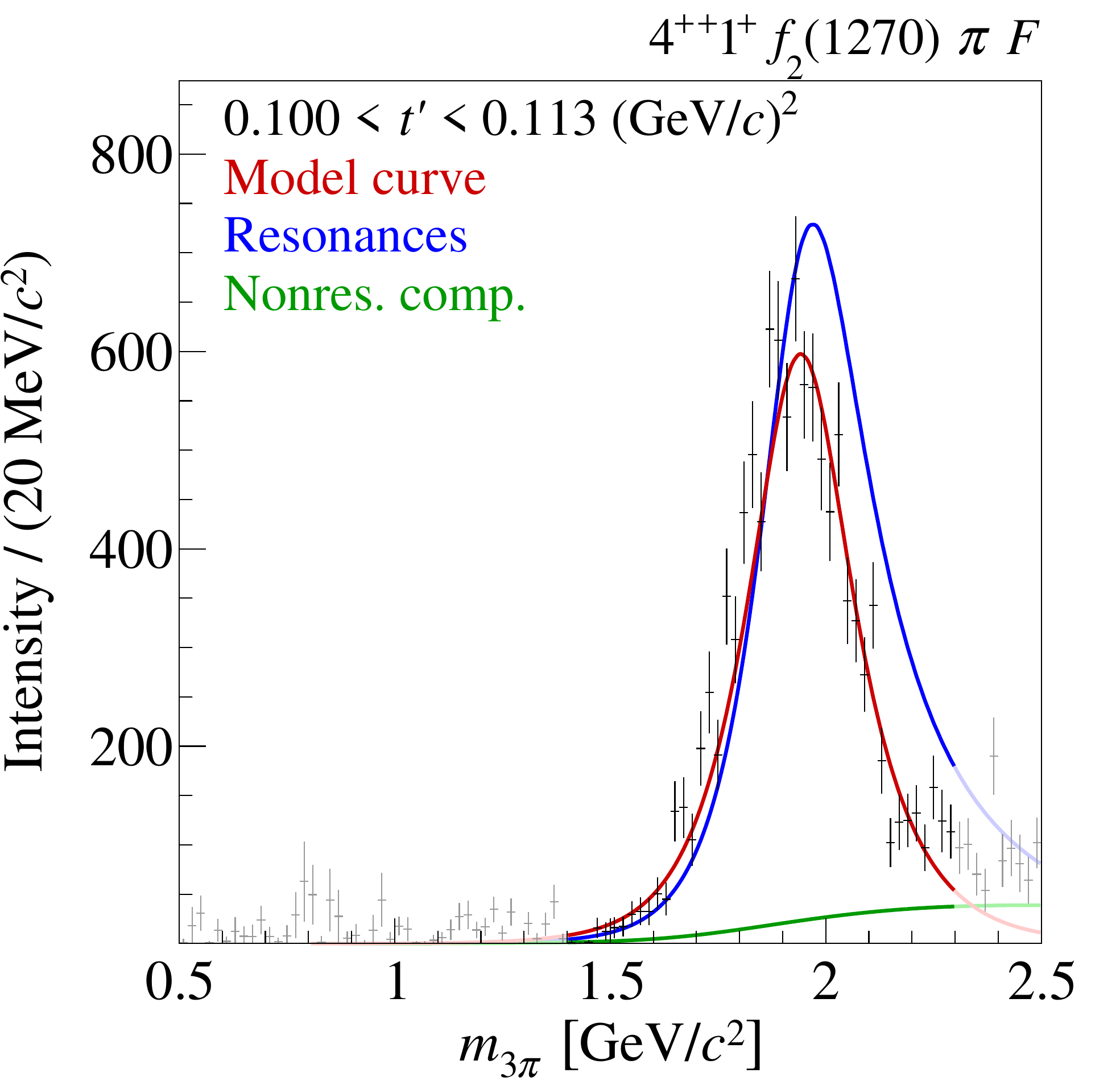}%
    \label{fig:intensity_4pp_f2_tbin1}%
  }%
  \hspace*{\fourPlotSpacing}%
  \subfloat[][]{%
    \includegraphics[width=\fourPlotWidth]{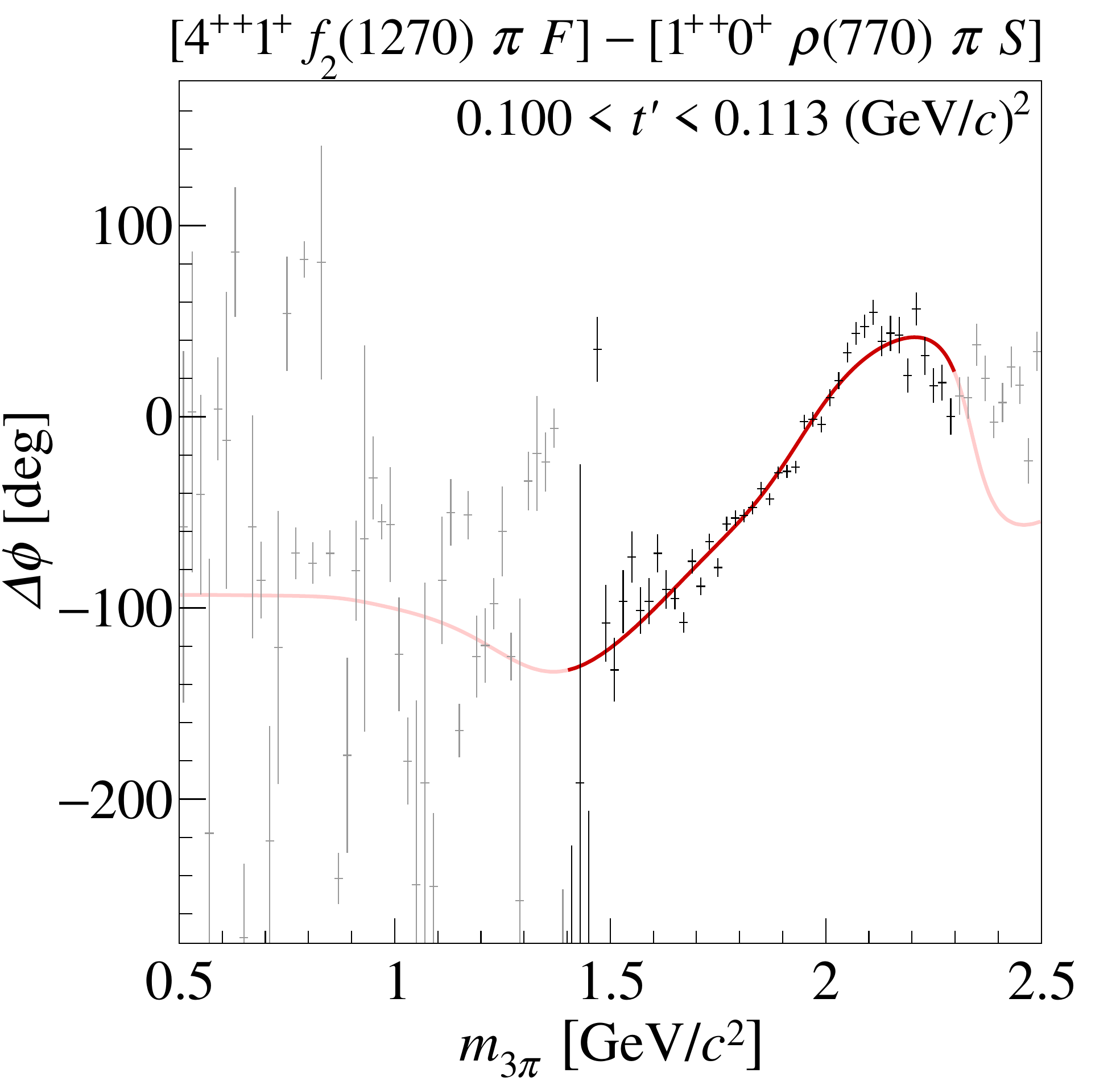}%
    \label{fig:phase_4pp_f2_1pp_rho_tbin1}%
  }%
  \hspace*{\fourPlotSpacing}%
  \subfloat[][]{%
    \includegraphics[width=\fourPlotWidth]{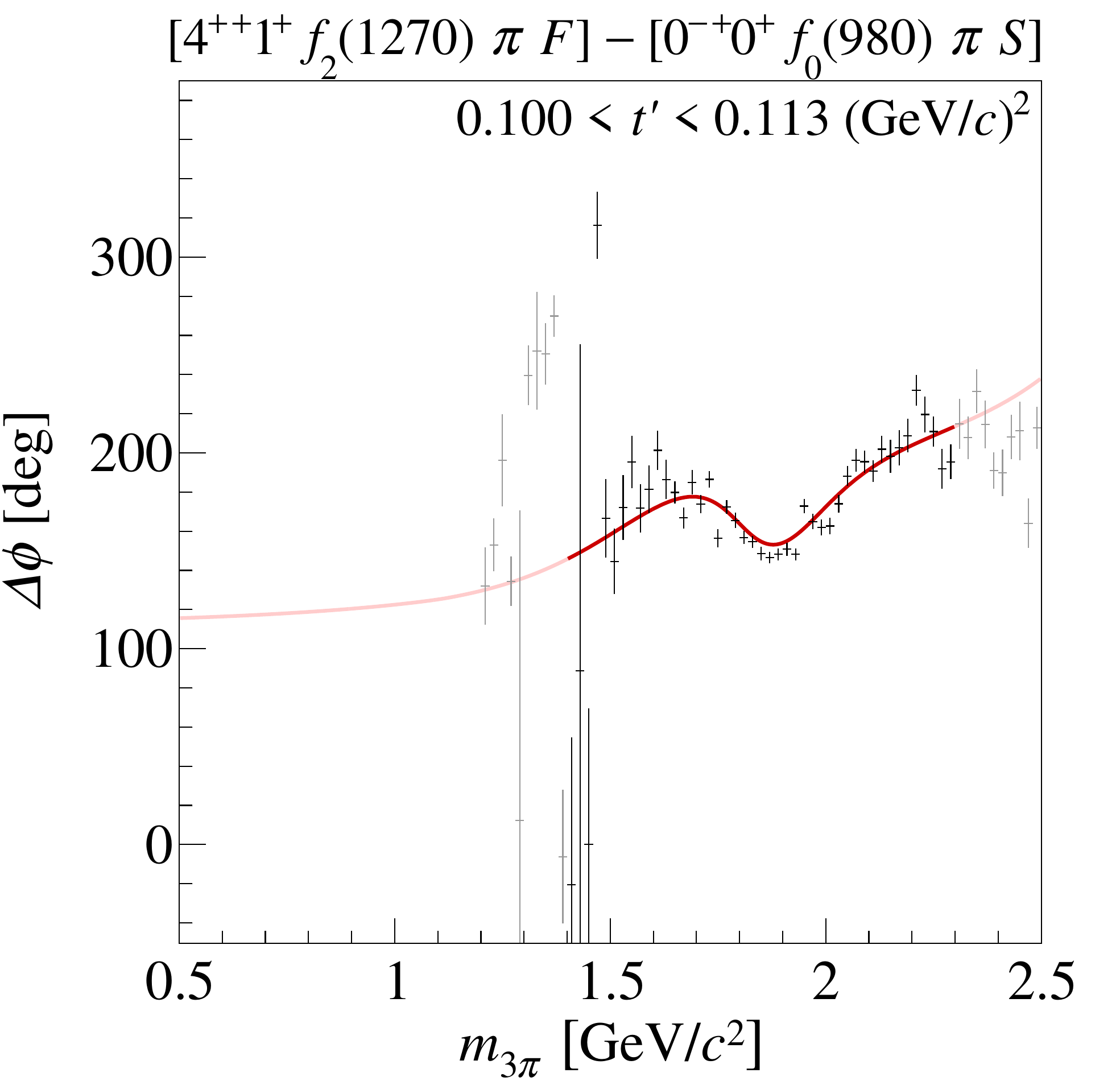}%
    \label{fig:phase_4pp_f2_0mp_tbin1}%
  }%
  \caption{Amplitudes of the two $\JPC = 4^{++}$ waves in the lowest
    \tpr bin.
    \subfloatLabel{fig:intensity_4pp_rho_tbin1}~through~\subfloatLabel{fig:phase_4pp_rho_0mp_tbin1}:
    intensity distribution and relative phases for the
    \wave{4}{++}{1}{+}{\Prho}{G} wave.
    \subfloatLabel{fig:intensity_4pp_f2_tbin1}~through~\subfloatLabel{fig:phase_4pp_f2_0mp_tbin1}:
    intensity distribution and relative phases for the
    \wave{4}{++}{1}{+}{\PfTwo}{F} wave. The model and the wave
    components are represented as in \cref{fig:intensity_phases_0mp},
    except that here the blue curve represents the \PaFour.}
  \label{fig:intensity_phases_4pp}
\end{wideFigureOrNot}

Our model contains one $\JPC = 4^{++}$ resonance, the \PaFour, which
is the only confirmed isovector state with these quantum
numbers~\cite{Patrignani:2016xqp}.  The \PaFour is parametrized using
\cref{eq:BreitWigner,eq:method:fixedwidth}, the nonresonant components
using \cref{eq:method:nonrestermsmall} (see
\cref{tab:method:fitmodel:waveset}).  The data are well described
within the fit range, which for the $\PfTwo \pi F$ wave is
\SIvalRange{1.4}{\mThreePi}{2.3}{\GeVcc}.  The low-mass tail of the
$\Prho \pi G$ wave allows us to extend the fit range for this wave
down to \SI{1.25}{\GeVcc}.  In our fit model, the nonresonant
components are small in both $4^{++}$ waves.  Their contribution
decreases with increasing \tpr and almost vanishes at higher values of
\tpr.  The nonresonant components interfere destructively with the
high-mass tail of the \PaFour in the $\PfTwo \pi F$ wave and
constructively with the low-mass tail of the \PaFour in the
$\Prho \pi G$ wave.  The model is not able to reproduce the high-mass
shoulder in the intensity distributions of the $\Prho \pi G$ wave at
low \tpr.

The two $4^{++}$ waves exhibit clearly rising phases in the
\SI{1.9}{\GeVcc} mass region, \eg\ \wrt the
\wave{1}{++}{0}{+}{\Prho}{S} wave, as shown in
\cref{fig:phase_4pp_f2_1pp_rho_tbin1,fig:phase_4pp_rho_1pp_rho_tbin1}.
This rise is observed for all \tpr bins.  Its magnitude is slightly
smaller for the \wave{4}{++}{1}{+}{\Prho}{G} wave.  The phase
variations \wrt the \wave{0}{-+}{0}{+}{\PfZero[980]}{S} wave exhibit a
more complex pattern.  The phase drop around \SI{1.8}{\GeVcc} due to a
stronger \Ppi[1800] signal in the \wave{0}{-+}{0}{+}{\PfZero[980]}{S}
wave is compensated by the phase motion of the \PaFour leading to a
rising phase around \SI{1.9}{\GeVcc} [see
\cref{fig:phase_4pp_f2_0mp_tbin1,fig:phase_4pp_rho_0mp_tbin1}].  The
magnitude of these phase motions decreases with \tpr.

The relative phase between the two $4^{++}$ partial-wave amplitudes
shows only little variation over the fitted mass region.  Together
with the phase motions discussed in the previous paragraph, this
demonstrates that the two waves are dominated by resonances and that
they have the same resonance content.  The residual slight rise of the
phase between the $\PfTwo \pi F$ and $\Prho \pi G$ waves is caused by
differences in the small nonresonant components.

We extract the Breit-Wigner parameters of the \PaFour and find
$m_{\PaFour} = \SIaerrSys{1935}{11}{13}{\MeVcc}$ and
$\Gamma_{\PaFour} = \SIaerrSys{333}{16}{21}{\MeVcc}$.  The \PaFour
resonance parameters are rather insensitive to the systematic studies
(see \cref{sec:systematics,sec:syst_uncert_fourPP}).

The \tpr spectra of the $4^{++}$ wave components are shown in
\cref{fig:tprim_4pp} together with the results of fits using
\cref{eq:slope-parametrization}.  In our model, the \tpr dependence of
the amplitudes of the \PaFour components in the two waves is
constrained by \cref{eq:method:branchingdefinition}.  The fit finds a
relative phase of the branching amplitudes close to \SI{0}{\degree}
for the \PaFour components in the two waves (see
\cref{sec:production_phases}).  The slope parameters of the \tpr
spectra of the \PaFour component in the two waves have practically
identical values of \SIaerrSys{9.2}{0.8}{0.5}{\perGeVcsq}.  For both
\tpr spectra, the model curve undershoots the data at small values of
\tpr.  This could indicate that in this \tpr range our resonance model
overestimates the \PaFour yields in both waves.  The nonresonant
contributions have steeper falling \tpr spectra with almost identical
slope parameters of \SIerrSys{14}{4}{\perGeVcsq} for the $\Prho \pi G$
wave and \SIaerrSys{14.5}{1.8}{3.7}{\perGeVcsq} for the $\PfTwo \pi F$
wave.  It is worth noting that, if we do not constrain the \PaFour
coupling amplitudes via \cref{eq:method:branchingdefinition} and thus
allow them to have different \tpr dependence [\StudyT; see
\cref{sec:systematics}], we obtain \PaFour resonance and slope
parameters that are consistent within the systematic uncertainties.
Also the relative phase of approximately \SI{0}{\degree} between the
$\Prho \pi G$ and $\PfTwo \pi F$ decay modes is recovered.  This
confirms the assumptions contained in
\cref{eq:method:branchingdefinition}.

\begin{figure}[tbp]
  \centering
  \subfloat[][]{%
    \includegraphics[width=\twoPlotWidth]{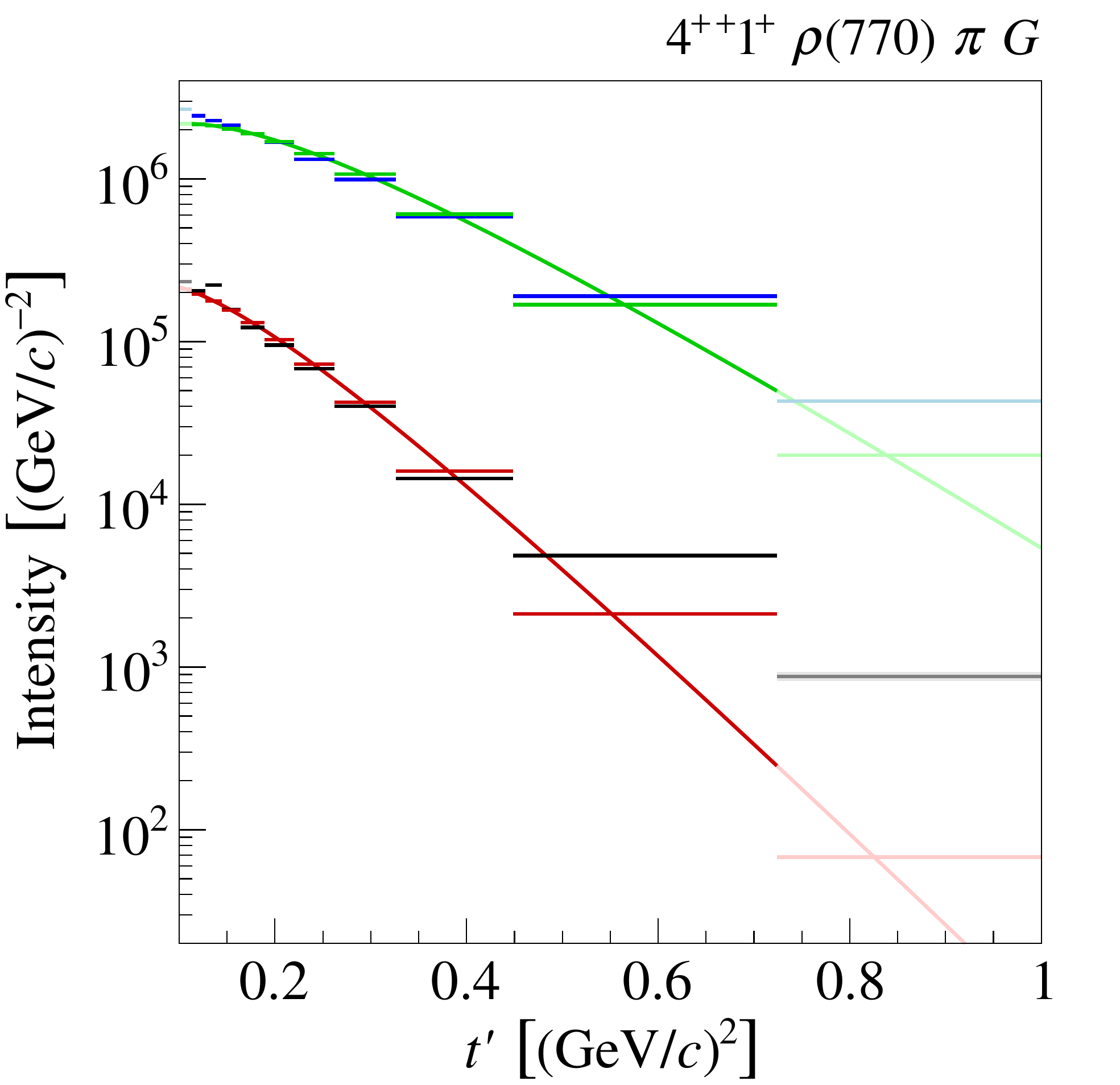}%
    \label{fig:tprim_4pp_rho}%
  }%
  \newLineOrHspace{\twoPlotSpacing}%
  \subfloat[][]{%
    \includegraphics[width=\twoPlotWidth]{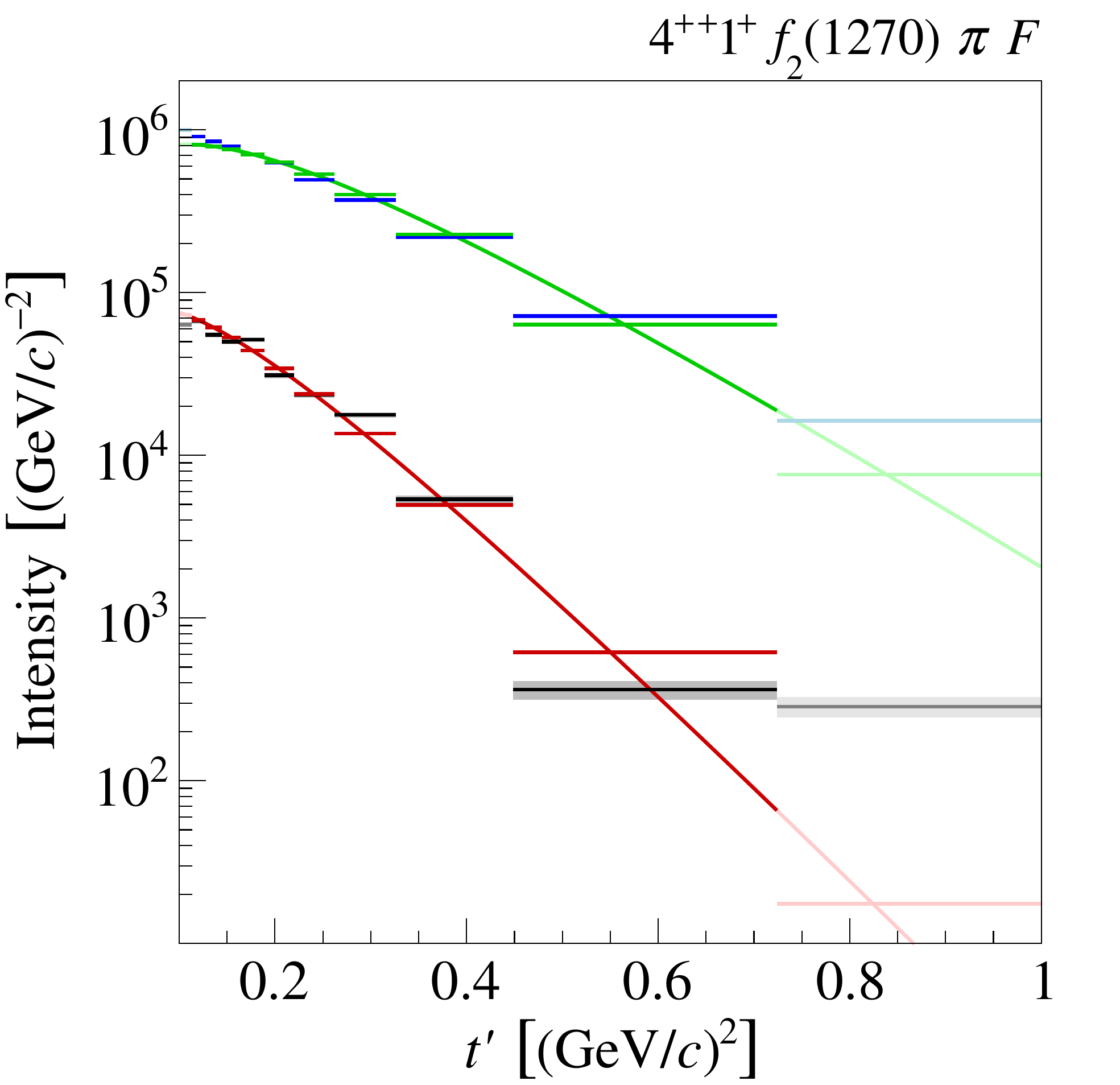}%
    \label{fig:tprim_4pp_f2}%
  }%
  \caption{Similar to \cref{{fig:tprim_0mp}}, but showing the \tpr
    spectra of \subfloatLabel{fig:tprim_4pp_rho}~the
    \wave{4}{++}{1}{+}{\Prho}{G} and
    \subfloatLabel{fig:tprim_4pp_f2}~the \wave{4}{++}{1}{+}{\PfTwo}{F}
    wave components as given by \cref{eq:tprim-dependence}: the
    \PaFour component is shown as blue lines and light blue boxes, and
    the nonresonant components as black lines and gray boxes.  The red
    and green curves and horizontal lines represent fits using
    \cref{eq:slope-parametrization}.}
  \label{fig:tprim_4pp}
\end{figure}

From the \PaFour yields in the two analyzed decay branches we derive
the ratio of branching fractions according to
\cref{eq:branch_fract_ratio}:
\begin{multlineOrEq}
  \label{eq:branch_fract_ratio_a4}
  B_{\Prho* \pi G, \PfTwo* \pi F}^{\PaFour*}
  \newLineOrNot
  \begin{alignedOrNot}
    \alignOrNot= \frac{\text{BF}\!\sBrk{\PaFour^- \to \Prho^0 \pi^- \to \threePi}\hfill}{\text{BF}\!\sBrk{\PaFour^- \to \PfTwo \pi^- \to \threePi}}
    \newLineOrNot
    \alignOrNot= \numaerrSys{2.5}{0.5}{0.3}.
  \end{alignedOrNot}
\end{multlineOrEq}
Taking into account the unobserved decays
$\PaFour*^- \to \Prho*^- \pi^0$ and $\PaFour*^- \to \PfTwo* \pi^-$ to
the \threePiN final state and assuming isospin symmetry, this value
increases by a factor of $4/3$:
\begin{equation}
  \label{eq:branch_fract_ratio_a4_iso}
  \begin{splitOrNot}
    B_{\Prho* \pi G, \PfTwo* \pi F}^{\PaFour*, \text{iso}}
    \alignOrNot= \frac{\text{BF}\!\sBrk{\PaFour^- \to \Prho \pi \to 3\pi}\hfill}{\text{BF}\!\sBrk{\PaFour^- \to \PfTwo \pi \to 3\pi}}
    \newLineOrNot
    \alignOrNot= \numaerrSys{3.3}{0.7}{0.4}.
  \end{splitOrNot}
\end{equation}
The isospin factor needs to be corrected for self-interference
effects.  Unlike the $\Prho \pi$ channel, the $\PfTwo \pi$ channel is
affected by different Bose symmetrizations in the \threePi and
\threePiN final states.  In addition, the branching fraction of the
\PfTwo into $2\pi$ of
\SIaerr{84.2}{2.9}{0.9}{\percent}~\cite{Patrignani:2016xqp} needs to
be included.  Taking both effects into account, the isospin factor
$4/3$ should be replaced by \numaerr{1.19}{0.04}{0.02}\footnote{We
  only take into account the uncertainty of the $\PfTwo \to 2\pi$
  branching fraction.} leading to the corrected ratio
\begin{equation}
  \label{eq:branch_fract_ratio_a4_corr}
  \begin{splitOrNot}
    B_{\Prho* \pi G, \PfTwo* \pi F}^{\PaFour*, \text{corr}}
    \alignOrNot= \frac{\text{BF}\!\sBrk{\PaFour^- \to \Prho \pi}\hfill}{\text{BF}\!\sBrk{\PaFour^- \to \PfTwo \pi}}
    \newLineOrNot
    \alignOrNot= \numaerr{2.9}{0.6}{0.4}.
  \end{splitOrNot}
\end{equation}

\subsubsection{Discussion of results on $4^{++}$ resonances}
\label{sec:fourPP_discussion}

The PDG world averages for mass and width of the \PaFour are
$m_{\PaFour} = \SIaerr{1995}{10}{8}{\MeVcc}$ and
$\Gamma_{\PaFour} =
\SIaerr{257}{25}{23}{\MeVcc}$~\cite{Patrignani:2016xqp}.  Our
measurement of the \PaFour parameters of
$m_{\PaFour} = \SIaerrSys{1935}{11}{13}{\MeVcc}$ and
$\Gamma_{\PaFour} = \SIaerrSys{333}{16}{21}{\MeVcc}$ is the most
accurate and precise so far, but we find the \PaFour mass to be
\SI{60}{\MeVcc} smaller and the width \SI{76}{\MeVcc} larger than the
world average.  We agree with our two previous analyses: the one based
on the measurement of the \threePi final state diffractively produced
on a solid lead target~\cite{alekseev:2009aa}, and the other based on
the measurement of the $\eta \pi$ and $\eta' \pi$ final states
diffractively produced on a liquid-hydrogen
target~\cite{Adolph:2014rpp}.  Also, the results on diffractively
produced \threePi by the BNL E852 experiment~\cite{chung:2002pu} and
$\omega \pi^- \pi^0$ by the VES experiment~\cite{Amelin:1999gk} are in
good agreement with our results.

Our measurement of the \PaFour width is especially at variance with
the value of $\Gamma_{\PaFour} = \SI{180(30)}{\MeVcc}$ obtained by the
authors of \refCite{anisovich:2001pn}.  They analyzed $3\pi^0$,
$\eta \pi^0$, and $\eta' \pi^0$ final states produced in \pbarp
annihilations.  They used a model with two $4^{++}$ resonances below
\SI{2.5}{\GeVcc} and claimed an excited \PaFour* state with a mass of
\SI{2255(40)}{\MeVcc} and a width of \SIaerr{330}{110}{50}{\MeVcc}.
In the two analyzed waves, we do not see clear resonance signals of
heavier \PaFour* resonances in the mass range from
\SIrange{2000}{2500}{\MeVcc}.

The measured value of \numaerrSys{2.5}{0.5}{0.3} of the
branching-fraction ratio $B_{\Prho* \pi G, \PfTwo* \pi F}^{\PaFour*}$
in \cref{eq:branch_fract_ratio_a4} is larger than the value
\numerrs{1.1}{0.2}{0.2} that was reported by the BNL E852 experiment
in a study of the same channel at \SI{18}{\GeVc} beam
momentum~\cite{chung:2002pu}.  Taking into account the unobserved
\threePiN decay mode and the \PfTwo branching fraction into $2\pi$,
the present result of \numaerr{2.9}{0.6}{0.4} for
$B_{\Prho* \pi G, \PfTwo* \pi F}^{\PaFour*, \text{corr}}$ in
\cref{eq:branch_fract_ratio_a4_corr} agrees with the value of
\num{3.3} predicted by the ${}^3P_0$ decay model~\cite{barnes:1996ff}.
In this model, the strong decay of a \qqbar state to the
$(q\widebar{q}\,')\, (q'\widebar{q})$ exit channel proceeds via
production of a $q'\widebar{q}\,'$ pair with vacuum quantum numbers,
$\JPC = 0^{++}$.  Note that the \PaFour width predicted by this model
is a factor of~2 smaller than our measured value of
$\Gamma_{\PaFour}$.
 %
%
%

\subsection{$\JPC = 2^{++}$ resonances}
\label{sec:twoPP}

\subsubsection{Results on $2^{++}$ resonances}
\label{sec:twoPP_results}

We include three $\JPC = 2^{++}$ waves in the resonance-model fit:
\wave{2}{++}{1}{+}{\Prho}{D}, \wave{2}{++}{2}{+}{\Prho}{D}, and
\wave{2}{++}{1}{+}{\PfTwo}{P}.  The \wave{2}{++}{1}{+}{\Prho}{D} wave
has the third largest intensity of the 88 waves in the
mass-independent analysis (see \cref{sec:mass-independent_fit}) and
contributes \SI{7.7}{\percent} to the total intensity in the mass
range from \SIrange{0.5}{2.5}{\GeVcc}.  The two other $2^{++}$ waves
contribute \SI{0.3}{\percent} and \SI{0.8}{\percent} to the total
intensity, respectively.  The intensity distributions of the three
waves are shown in
\cref{fig:intensity_2pp_m1_rho_tbin1,fig:intensity_2pp_m2_rho_tbin1,fig:intensity_2pp_f2_tbin1}
for the lowest \tpr bin and in
\cref{fig:intensity_2pp_m1_rho_tbin11,fig:intensity_2pp_m2_rho_tbin11,fig:intensity_2pp_f2_tbin11}
for the highest \tpr bin.

\ifMultiColumnLayout{\begin{figure*}[t]}{\begin{figure}[tbp]}
  \centering
  \subfloat[][]{%
    \includegraphics[width=\fourPlotWidth]{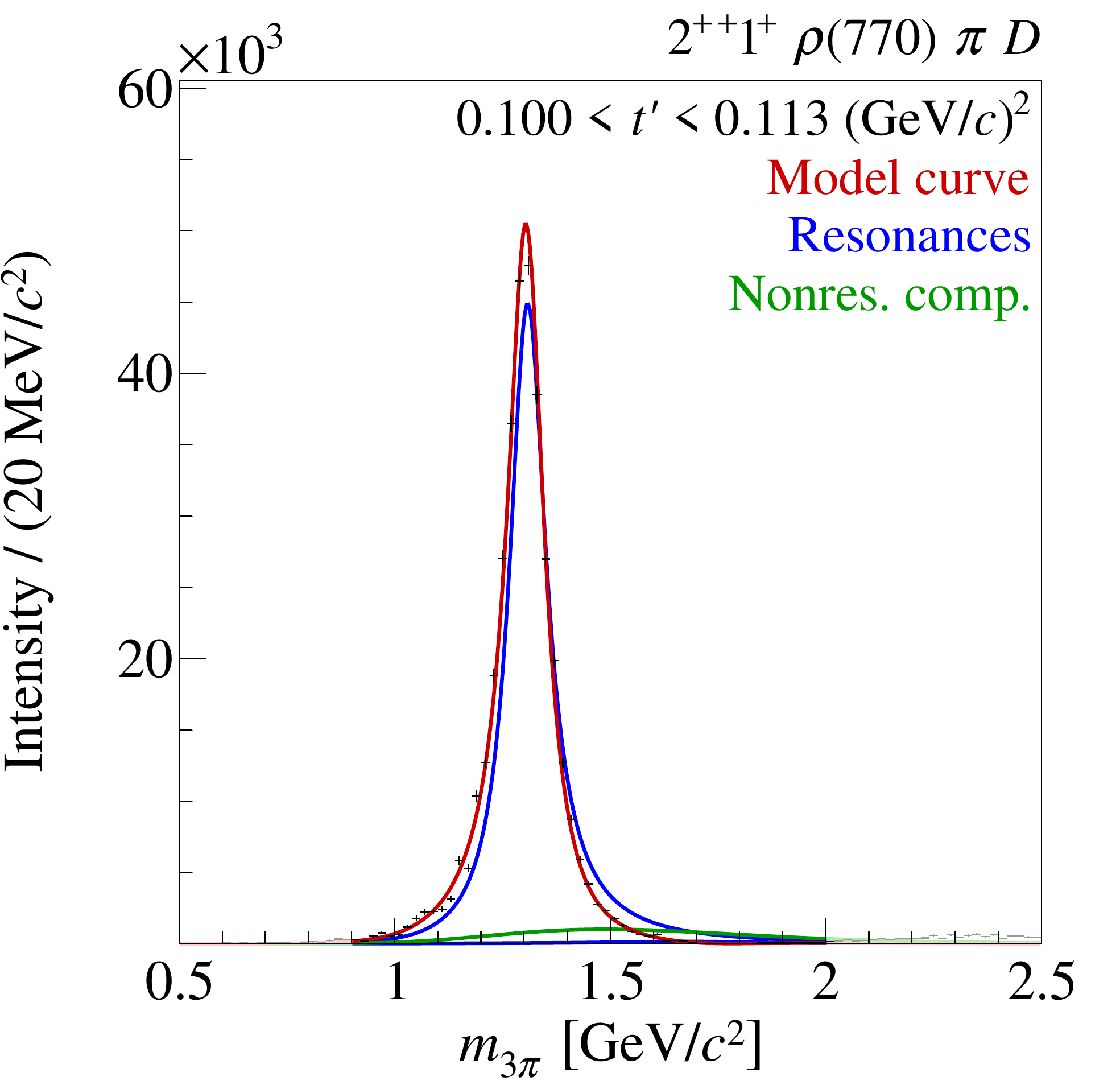}%
    \label{fig:intensity_2pp_m1_rho_tbin1}%
  }%
  \hspace*{\fourPlotSpacing}%
  \subfloat[][]{%
    \includegraphics[width=\fourPlotWidth]{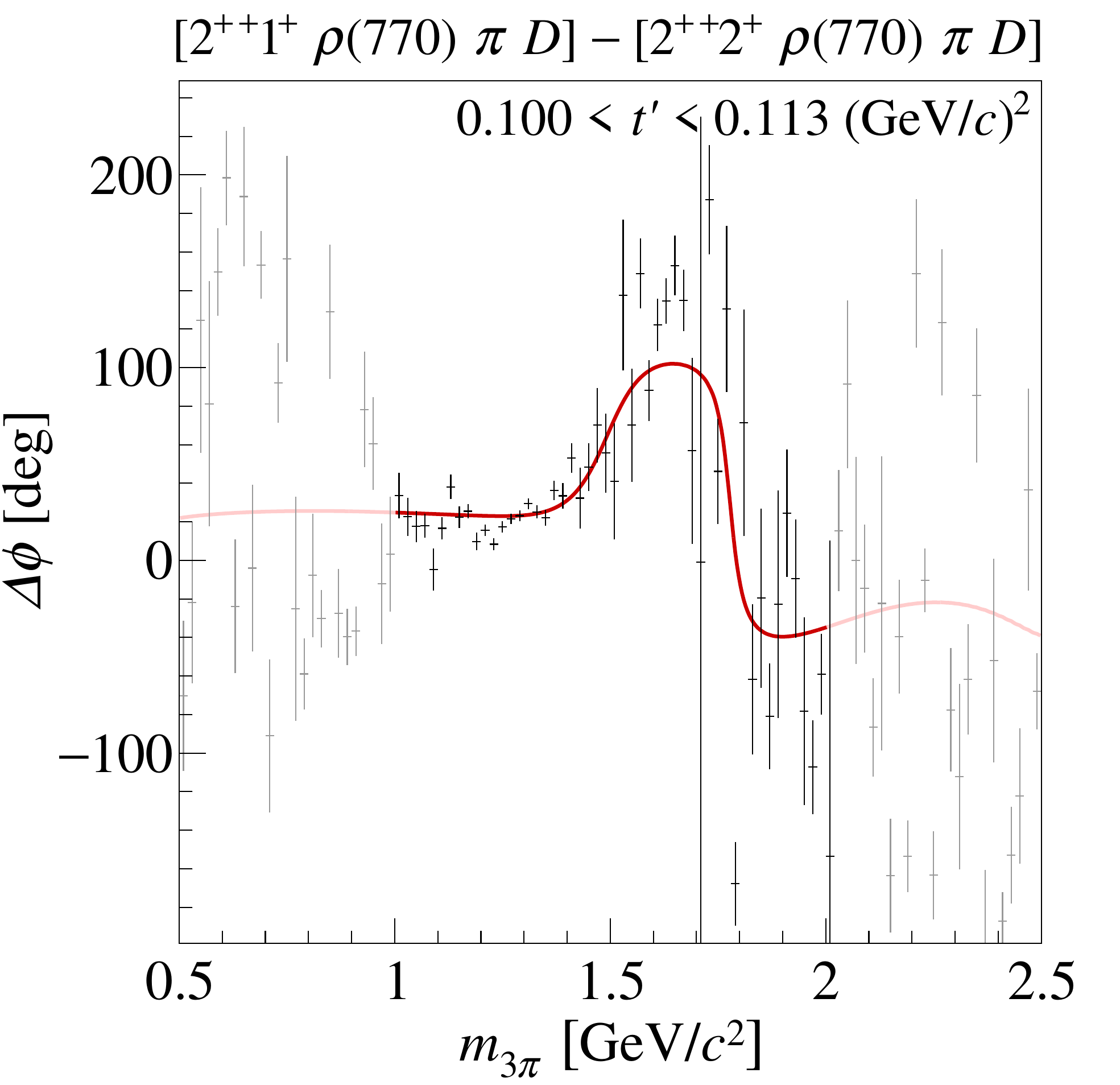}%
    \label{fig:phase_2pp_m1_rho_2pp_m2_rho_tbin1}%
  }%
  \hspace*{\fourPlotSpacing}%
  \subfloat[][]{%
    \includegraphics[width=\fourPlotWidth]{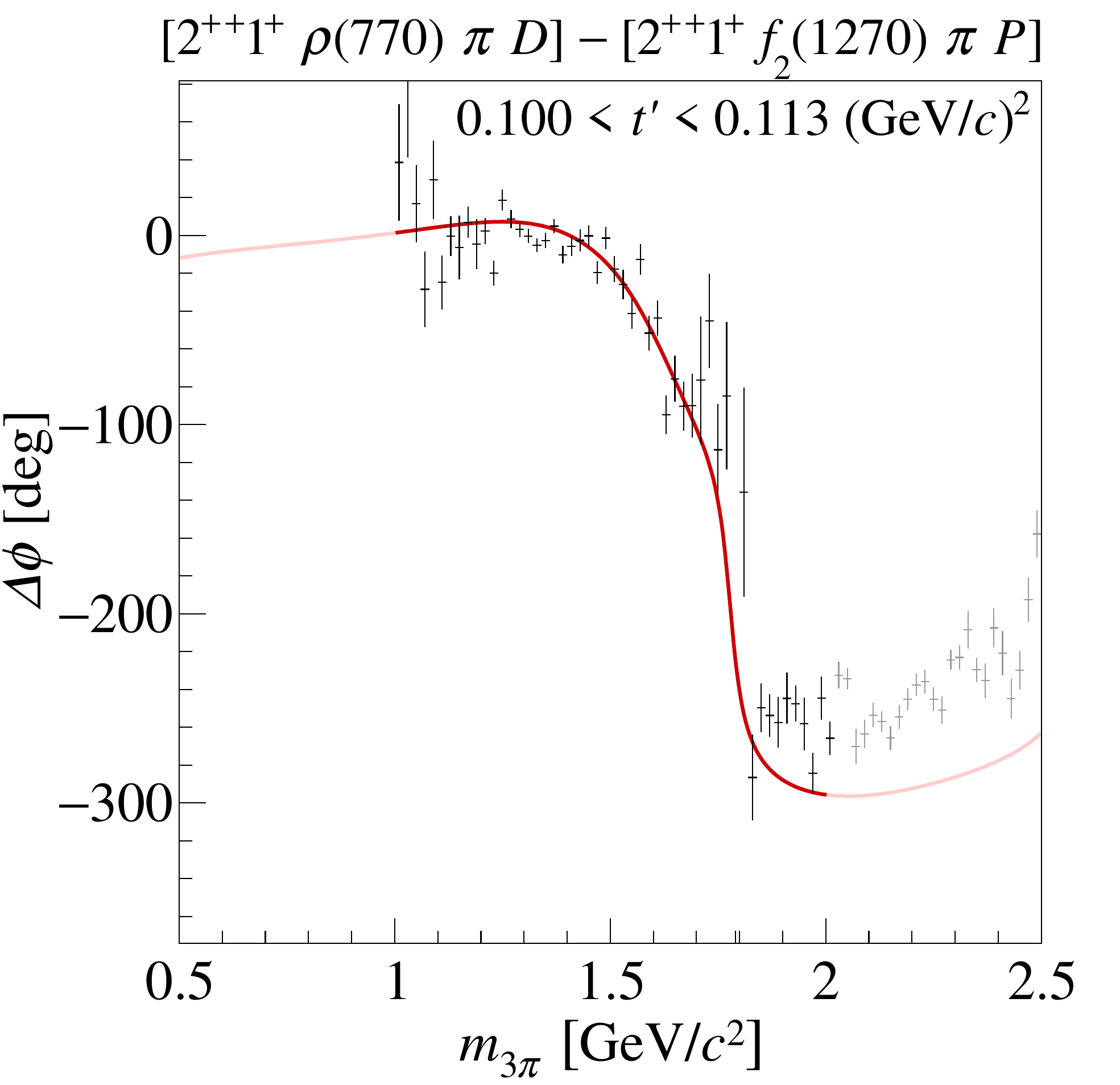}%
    \label{fig:phase_2pp_m1_rho_2pp_f2_tbin1}%
  }%
  \hspace*{\fourPlotSpacing}%
  \subfloat[][]{%
    \includegraphics[width=\fourPlotWidth]{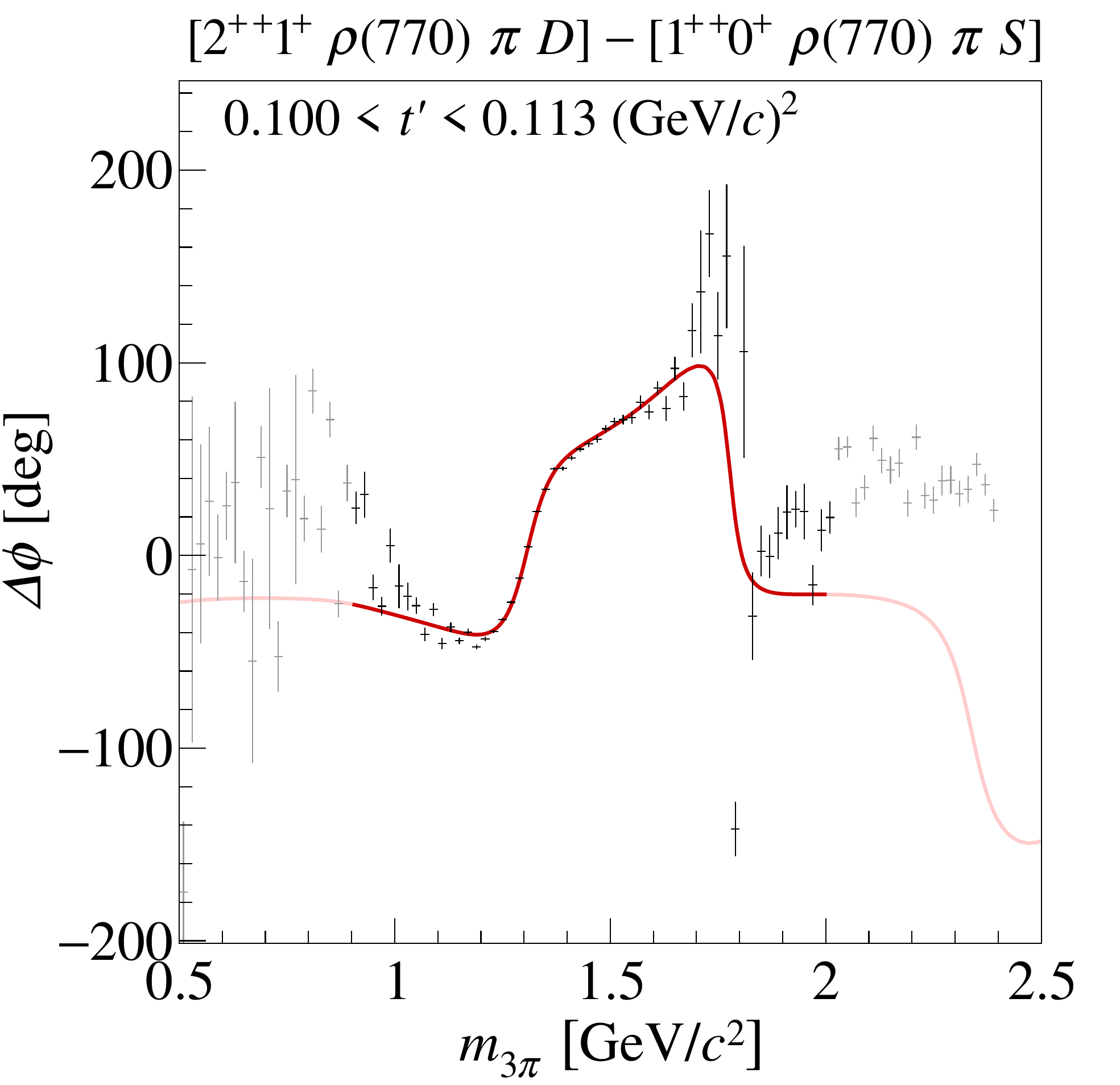}%
    \label{fig:phase_2pp_m1_rho_1pp_rho_tbin1}%
  }%
  \\
  \hspace*{\fourPlotWidth}%
  \hspace*{\fourPlotSpacing}%
  \subfloat[][]{%
    \includegraphics[width=\fourPlotWidth]{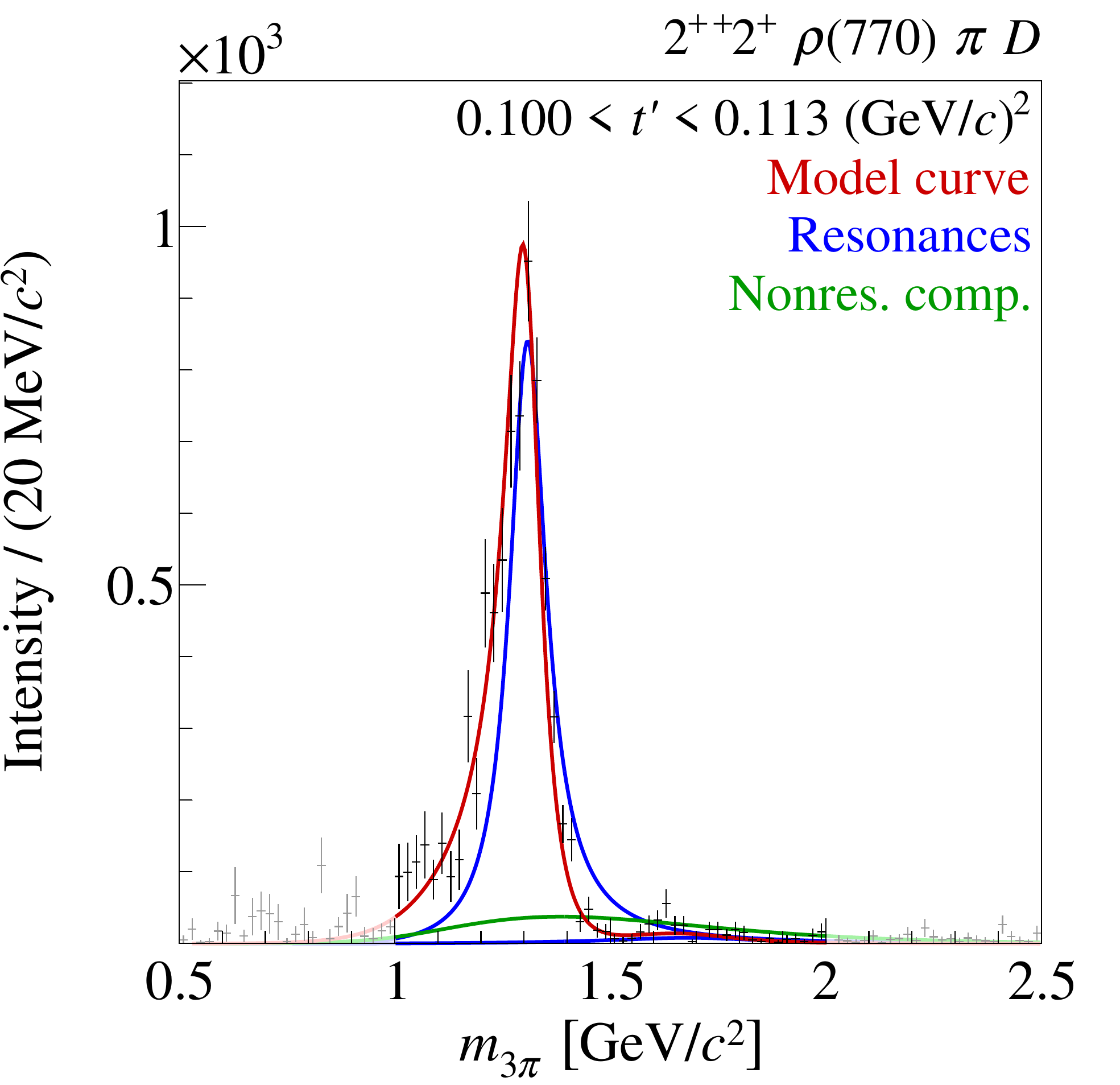}%
    \label{fig:intensity_2pp_m2_rho_tbin1}%
  }%
  \hspace*{\fourPlotSpacing}%
  \subfloat[][]{%
    \includegraphics[width=\fourPlotWidth]{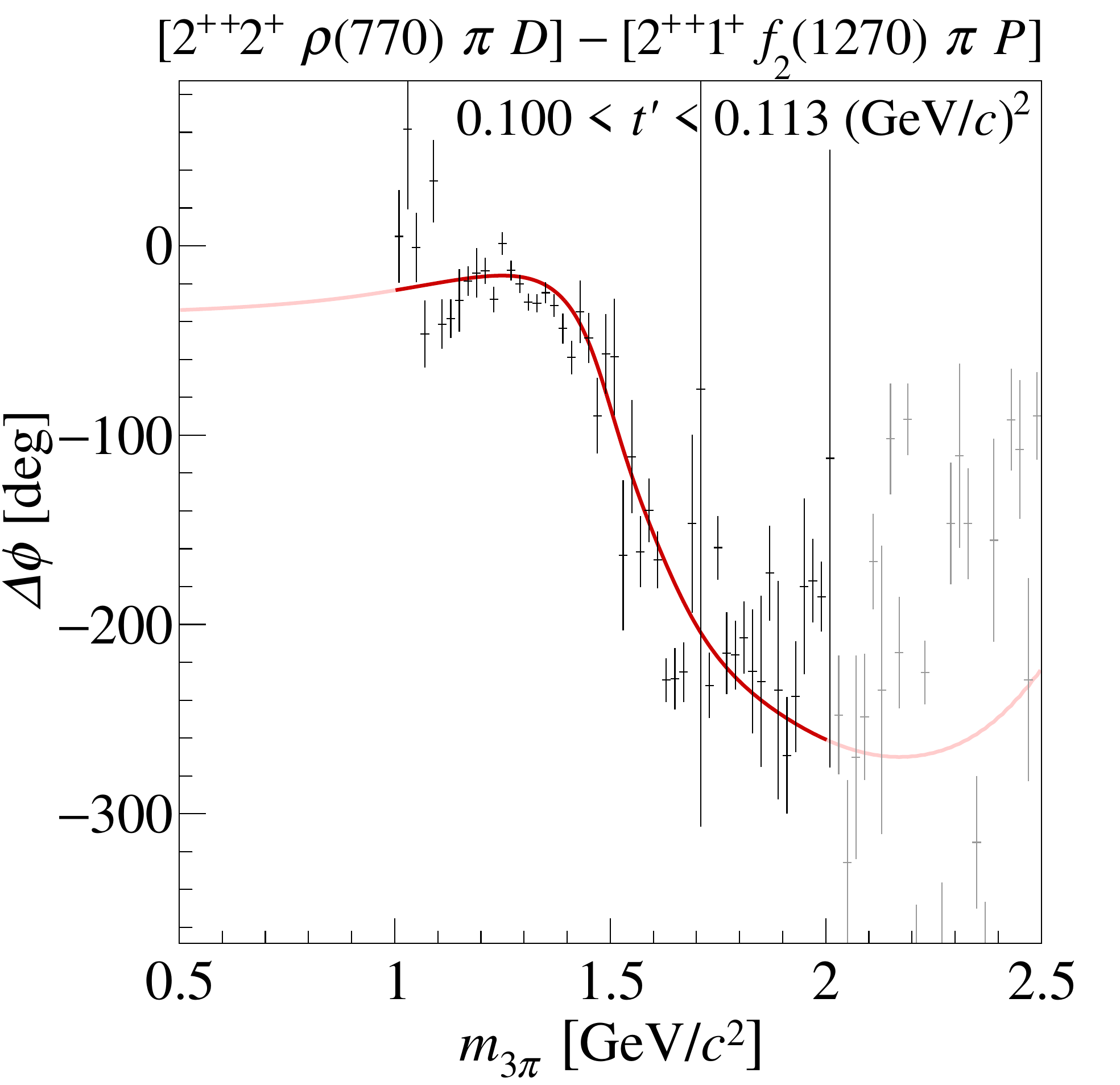}%
    \label{fig:phase_2pp_m2_rho_2pp_f2_tbin1}%
  }%
  \hspace*{\fourPlotSpacing}%
  \subfloat[][]{%
    \includegraphics[width=\fourPlotWidth]{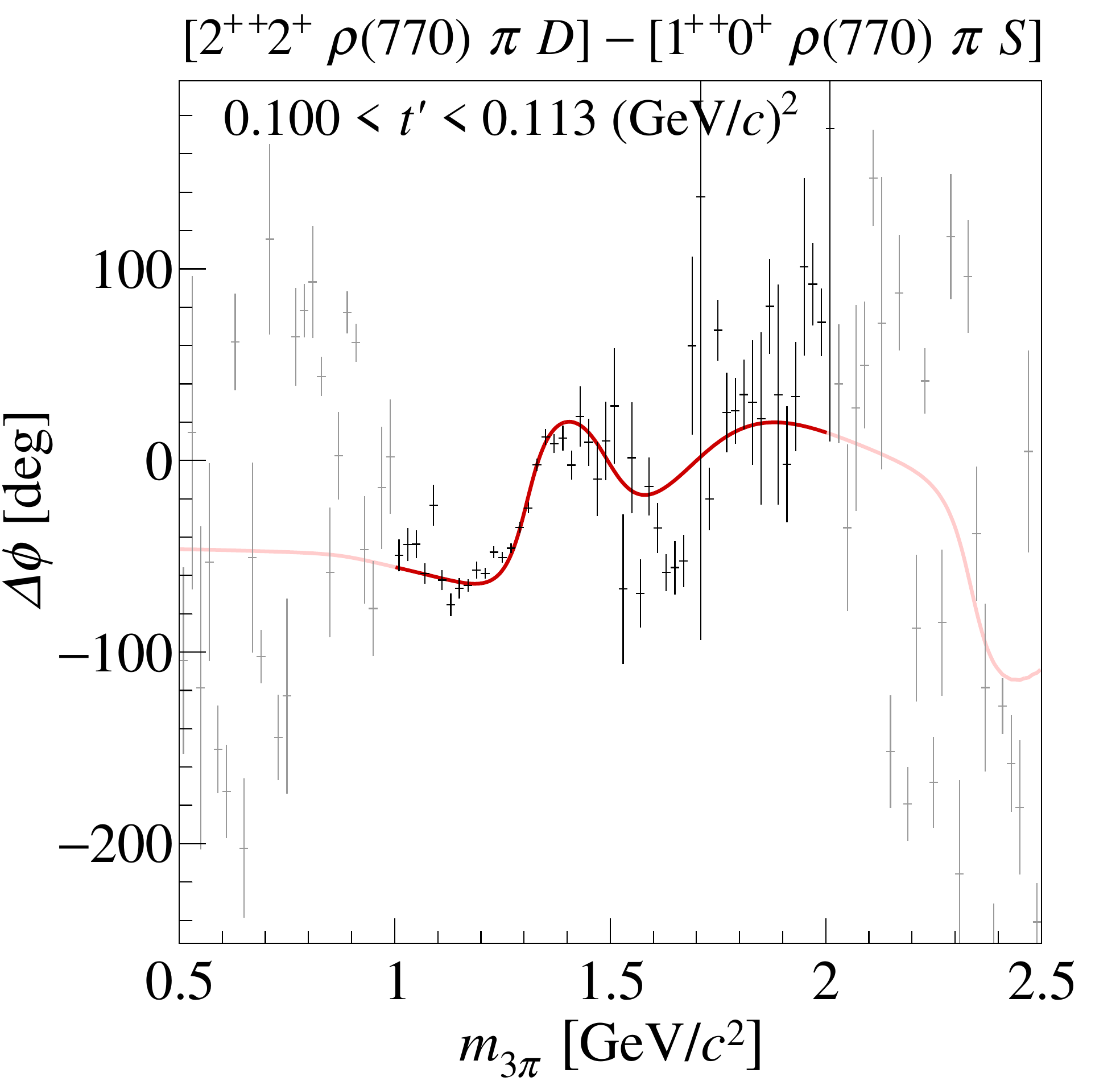}%
    \label{fig:phase_2pp_m2_rho_1pp_rho_tbin1}%
  }%
  \\
  \hspace*{\fourPlotWidth}%
  \hspace*{\fourPlotSpacing}%
  \hspace*{\fourPlotWidth}%
  \hspace*{\fourPlotSpacing}%
  \subfloat[][]{%
    \includegraphics[width=\fourPlotWidth]{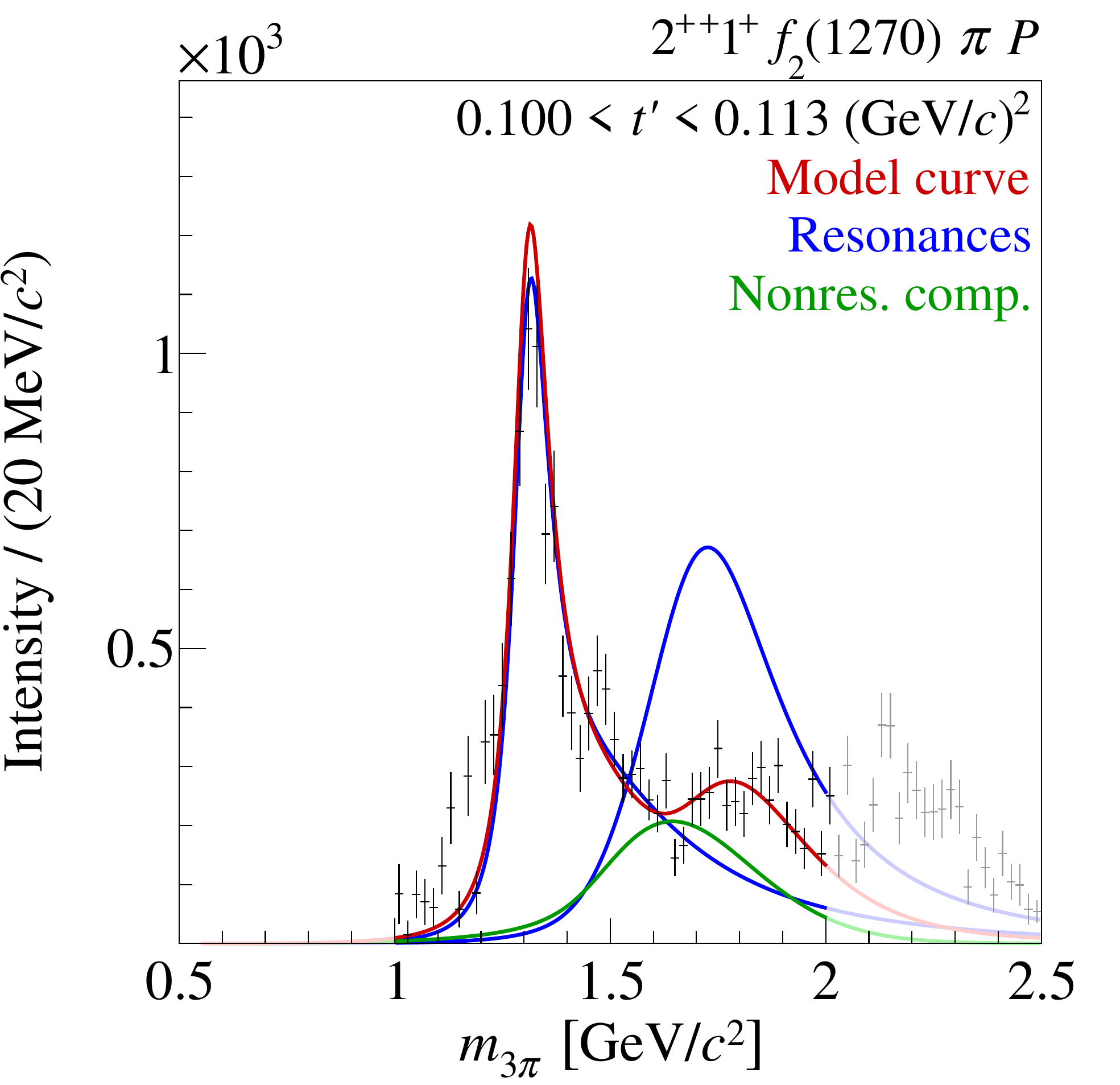}%
    \label{fig:intensity_2pp_f2_tbin1}%
  }%
  \hspace*{\fourPlotSpacing}%
  \subfloat[][]{%
    \includegraphics[width=\fourPlotWidth]{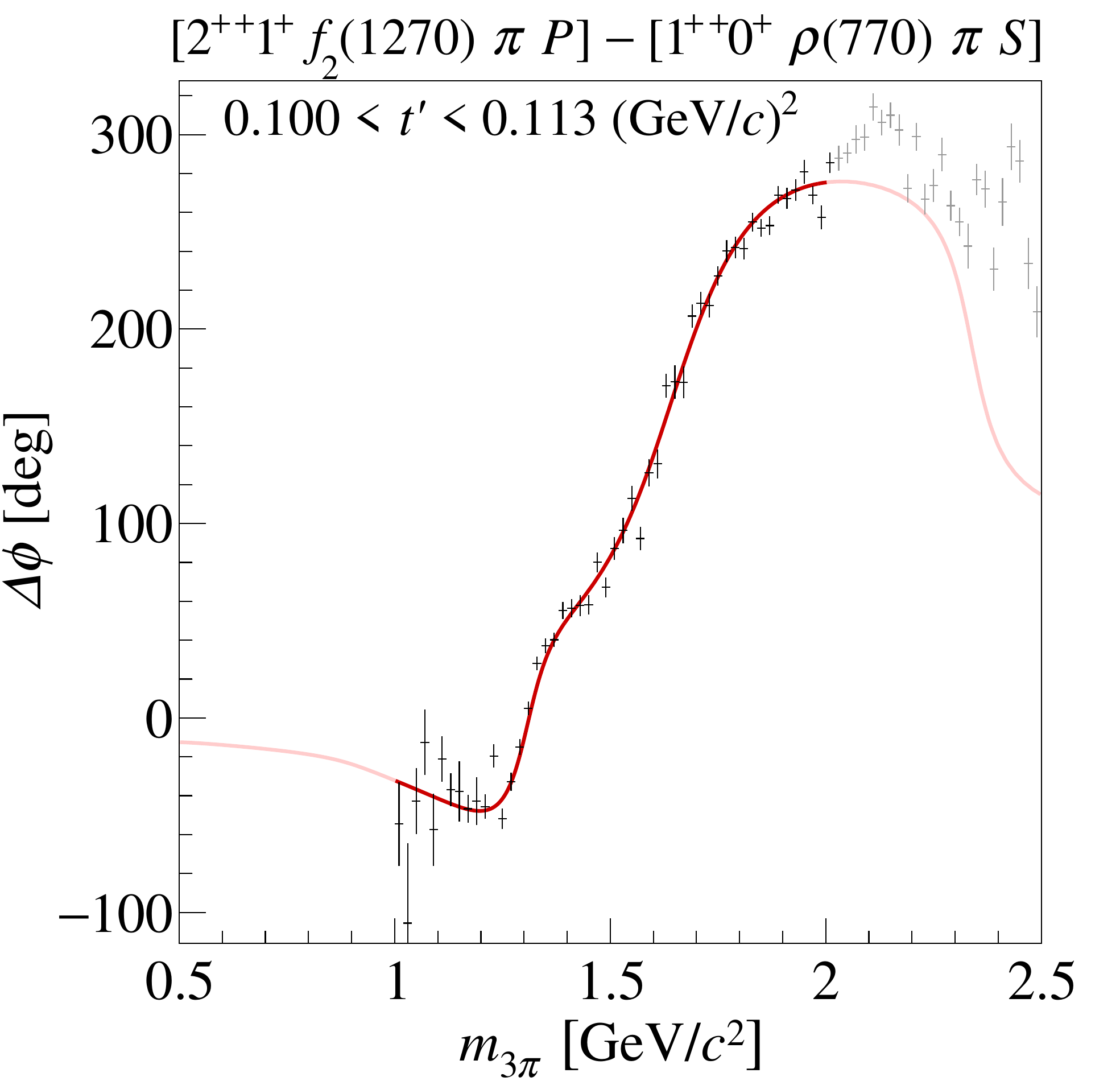}%
    \label{fig:phase_2pp_f2_1pp_rho_tbin1}%
  }%
  \caption{Amplitudes of the three $\JPC = 2^{++}$ waves in the lowest
    \tpr bin.
    \subfloatLabel{fig:intensity_2pp_m1_rho_tbin1}~through~\subfloatLabel{fig:phase_2pp_m1_rho_1pp_rho_tbin1}:
    Intensity distribution and relative phases for the
    \wave{2}{++}{1}{+}{\Prho}{D}
    wave. \subfloatLabel{fig:intensity_2pp_m2_rho_tbin1}~through~\subfloatLabel{fig:phase_2pp_m2_rho_1pp_rho_tbin1}:
    Intensity distribution and relative phases for the
    \wave{2}{++}{2}{+}{\Prho}{D} wave.
    \subfloatLabel{fig:intensity_2pp_f2_tbin1}~and~\subfloatLabel{fig:phase_2pp_f2_1pp_rho_tbin1}:
    Intensity distribution and relative phase for the
    \wave{2}{++}{1}{+}{\PfTwo}{P} wave.  The model and the wave
    components are represented as in \cref{fig:intensity_phases_0mp},
    except that here the blue curves represent the \PaTwo and the
    \PaTwo[1700].}
  \label{fig:intensity_phases_2pp_tbin1}
\ifMultiColumnLayout{\end{figure*}}{\end{figure}}

\ifMultiColumnLayout{\begin{figure*}[t]}{\begin{figure}[tbp]}
  \centering
  \subfloat[][]{%
    \includegraphics[width=\fourPlotWidth]{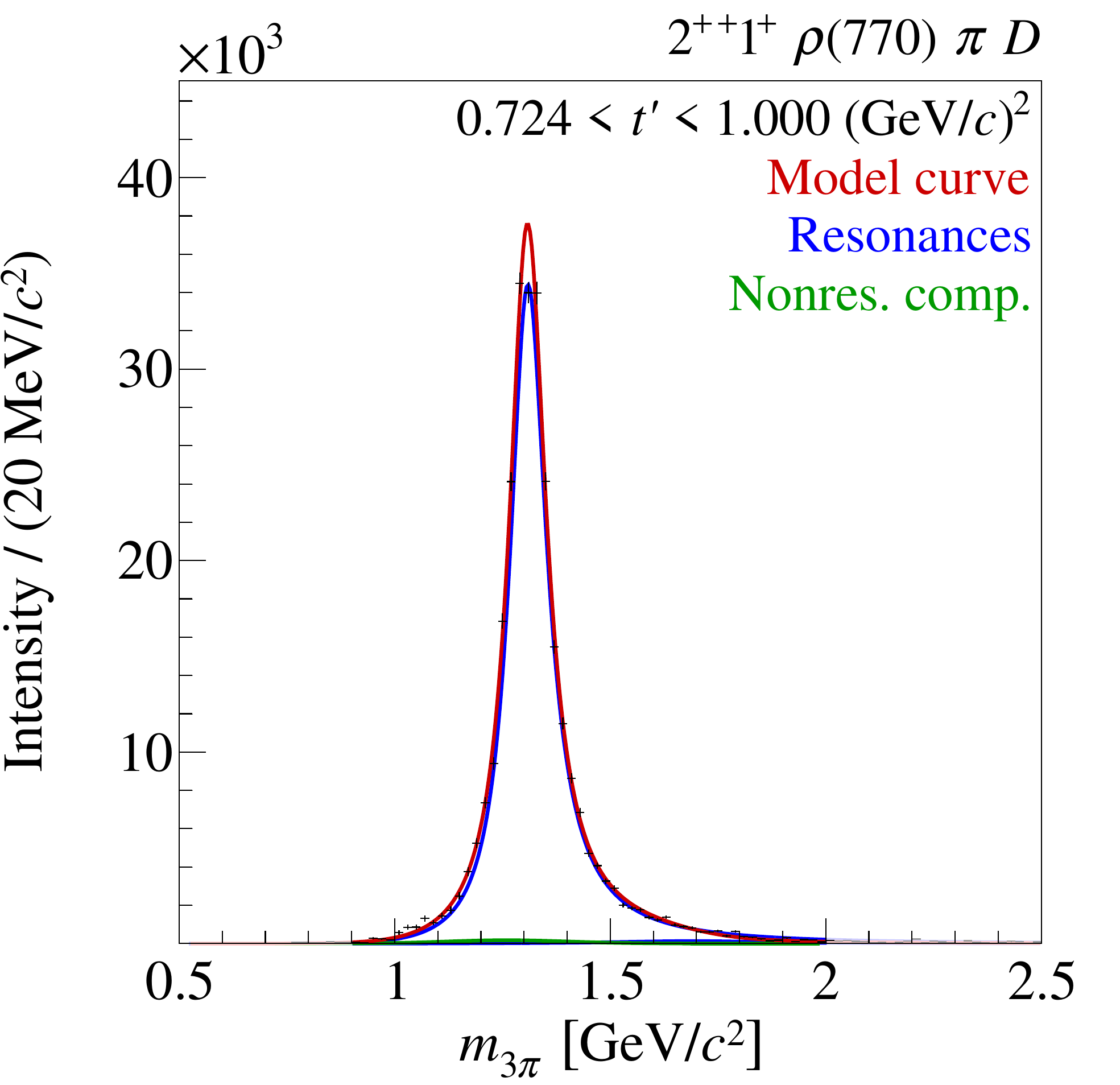}%
    \label{fig:intensity_2pp_m1_rho_tbin11}%
  }%
  \hspace*{\fourPlotSpacing}%
  \subfloat[][]{%
    \includegraphics[width=\fourPlotWidth]{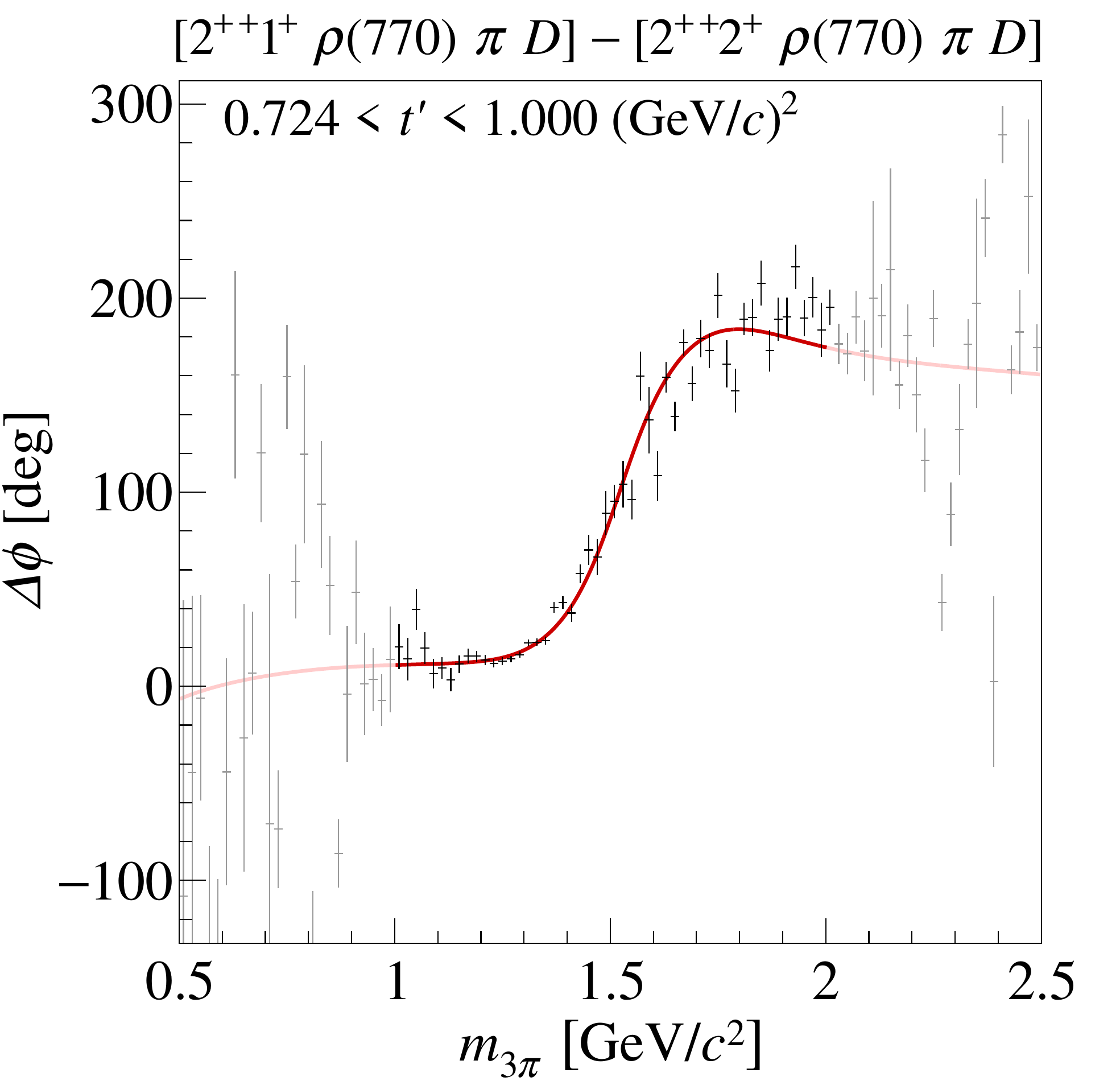}%
    \label{fig:phase_2pp_m1_rho_2pp_m2_rho_tbin11}%
  }%
  \hspace*{\fourPlotSpacing}%
  \subfloat[][]{%
    \includegraphics[width=\fourPlotWidth]{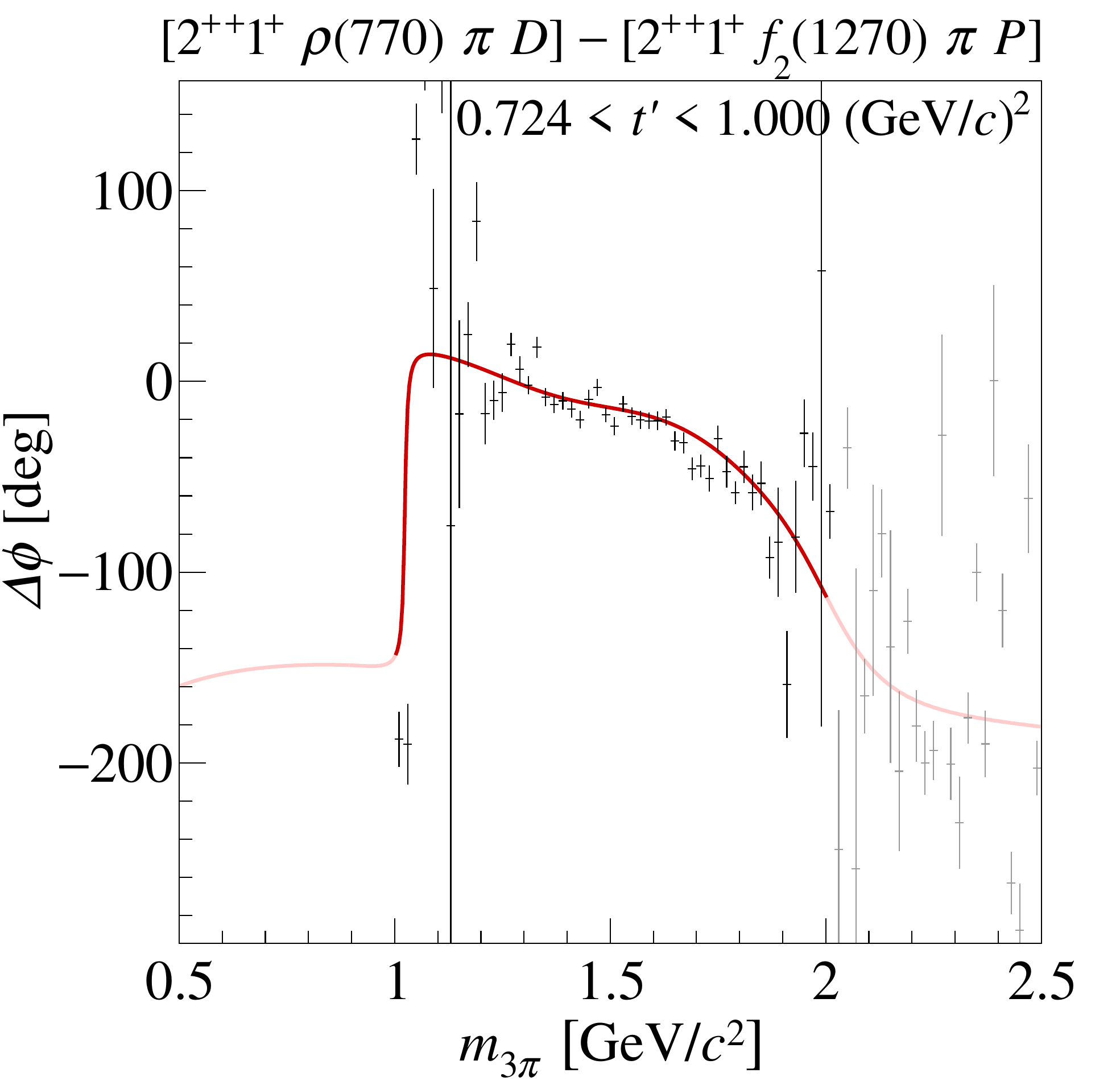}%
    \label{fig:phase_2pp_m1_rho_2pp_f2_tbin11}%
  }%
  \hspace*{\fourPlotSpacing}%
  \subfloat[][]{%
    \includegraphics[width=\fourPlotWidth]{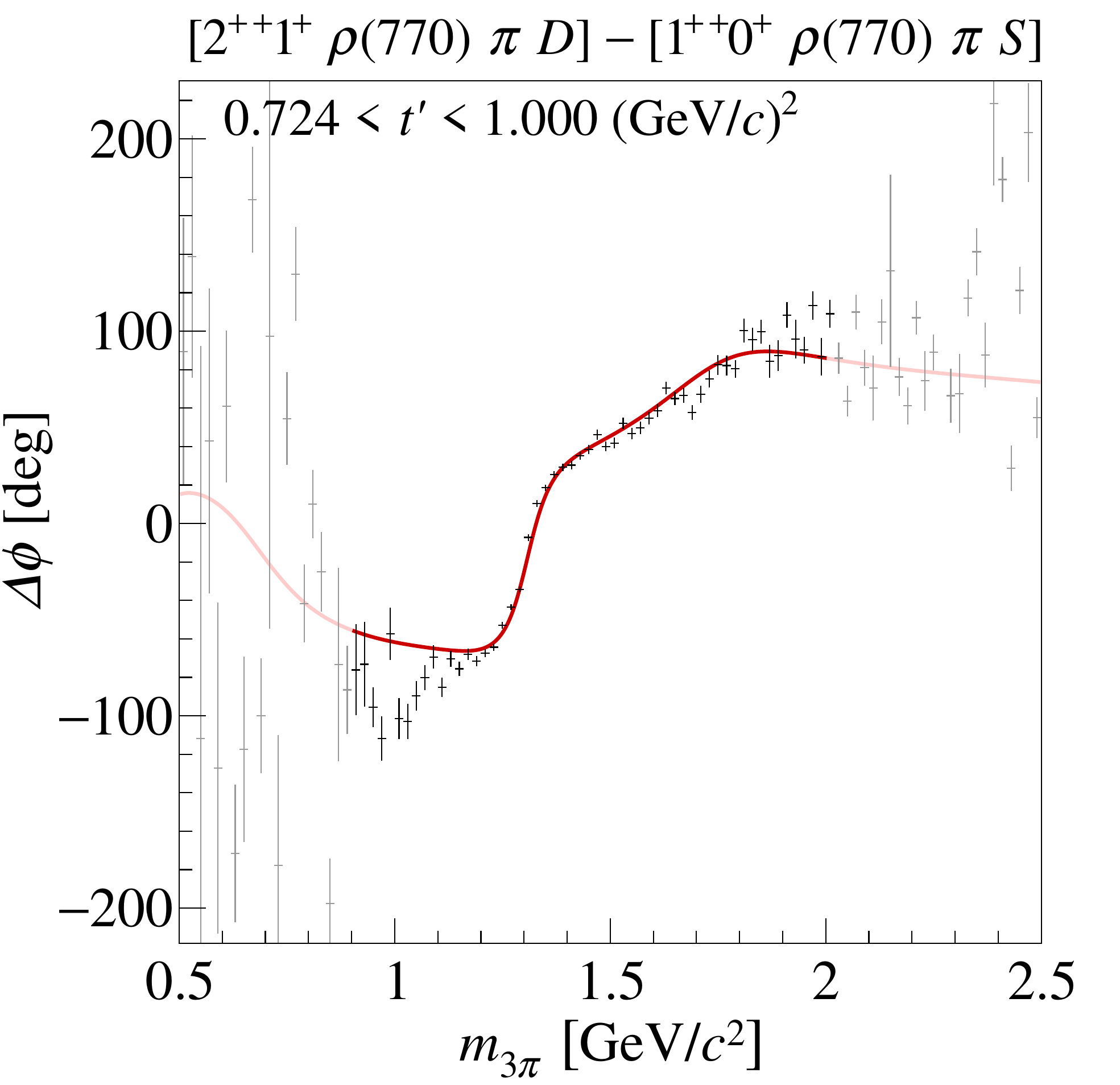}%
    \label{fig:phase_2pp_m1_rho_1pp_rho_tbin11}%
  }%
  \\
  \hspace*{\fourPlotWidth}%
  \hspace*{\fourPlotSpacing}%
  \subfloat[][]{%
    \includegraphics[width=\fourPlotWidth]{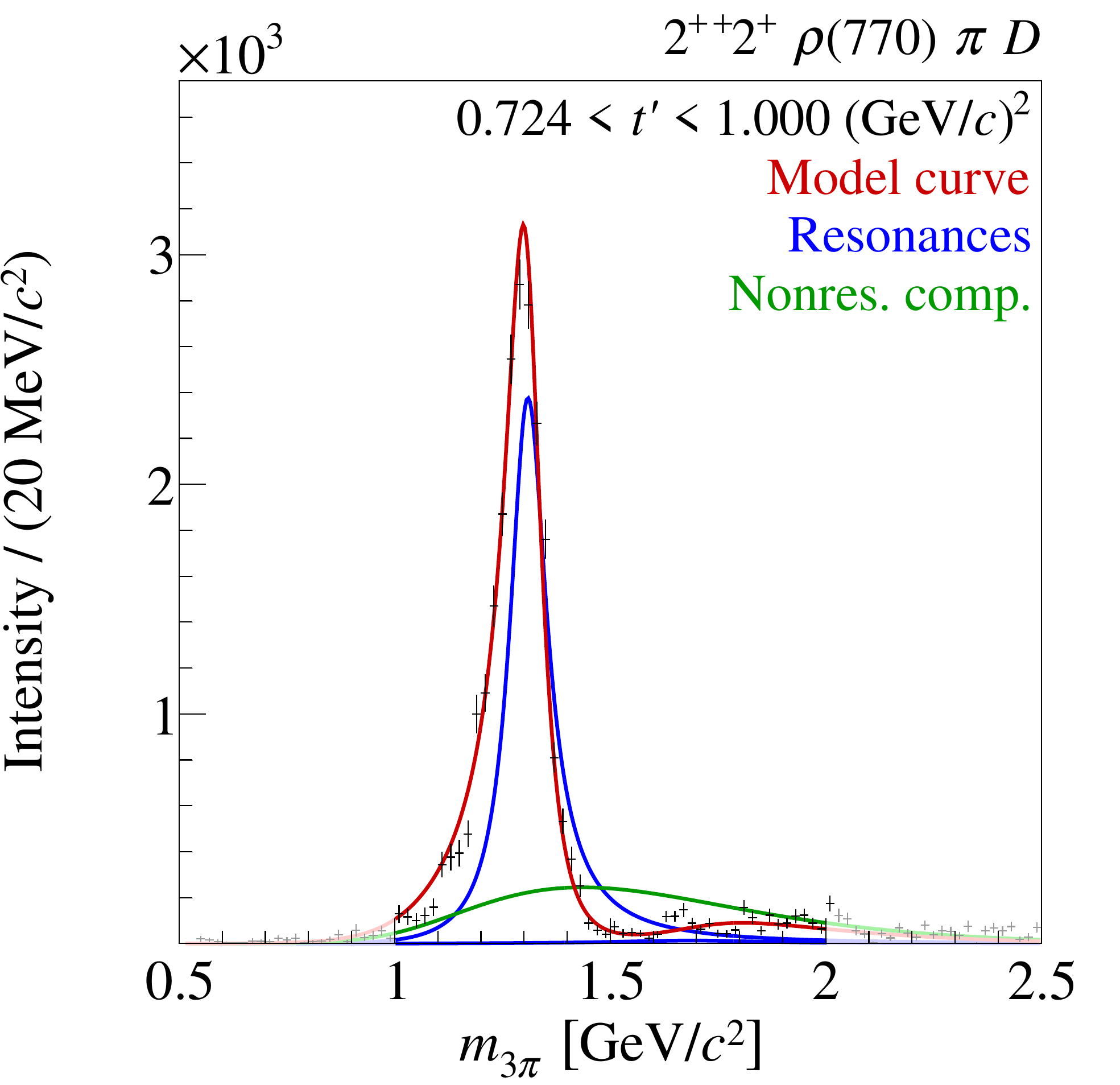}%
    \label{fig:intensity_2pp_m2_rho_tbin11}%
  }%
  \hspace*{\fourPlotSpacing}%
  \subfloat[][]{%
    \includegraphics[width=\fourPlotWidth]{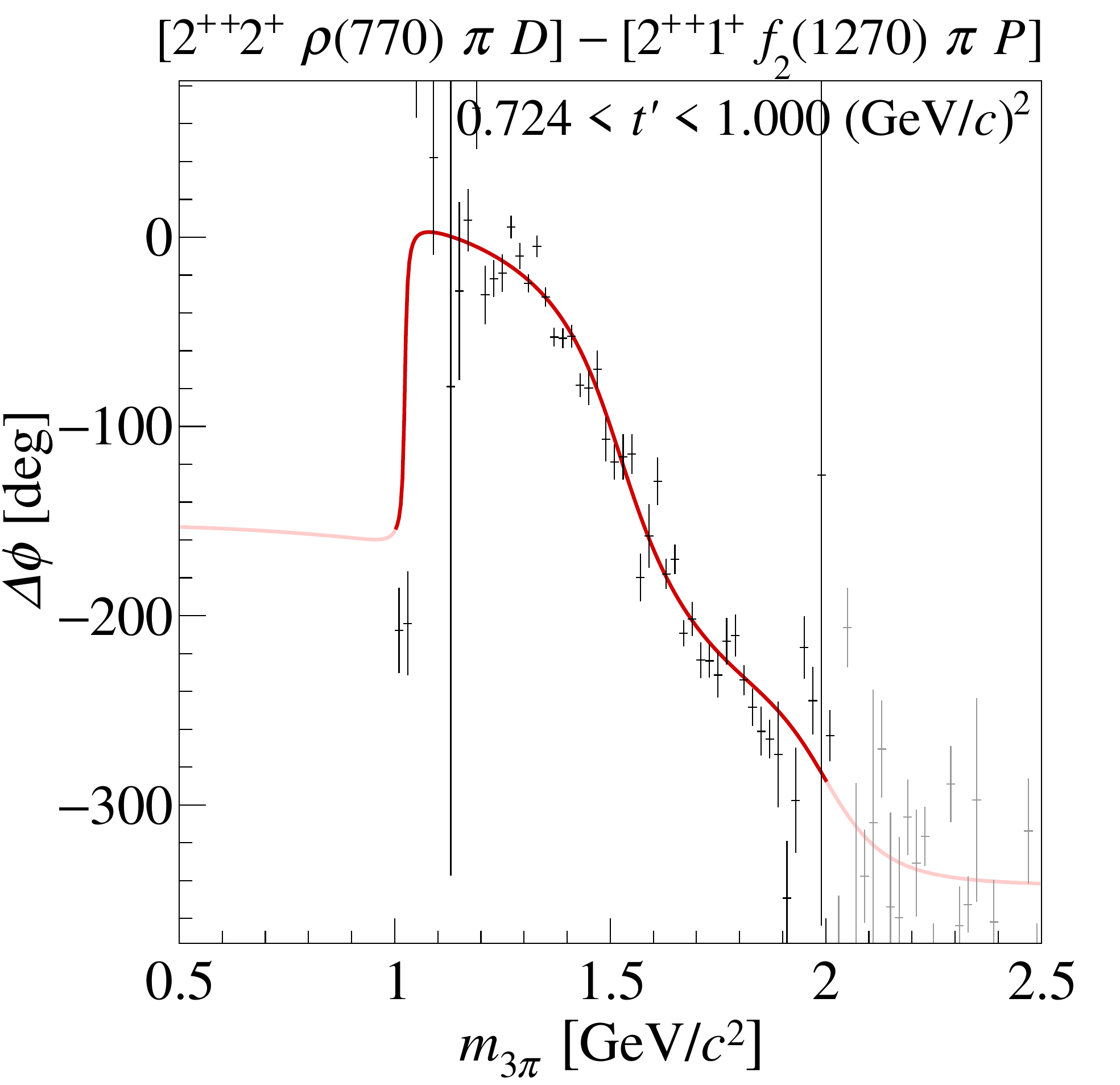}%
    \label{fig:phase_2pp_m2_rho_2pp_f2_tbin11}%
  }%
  \hspace*{\fourPlotSpacing}%
  \subfloat[][]{%
    \includegraphics[width=\fourPlotWidth]{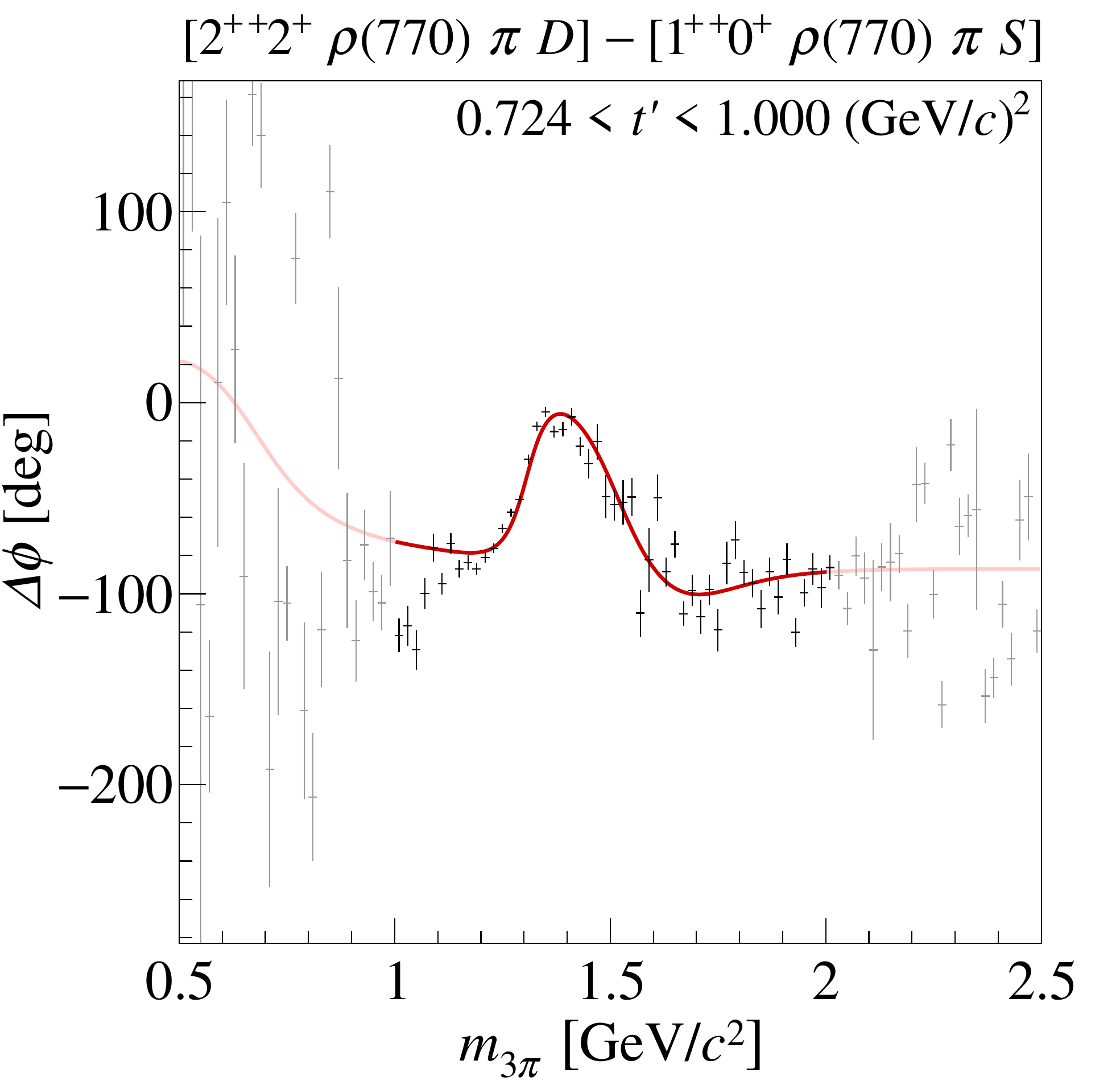}%
    \label{fig:phase_2pp_m2_rho_1pp_rho_tbin11}%
  }%
  \\
  \hspace*{\fourPlotWidth}%
  \hspace*{\fourPlotSpacing}%
  \hspace*{\fourPlotWidth}%
  \hspace*{\fourPlotSpacing}%
  \subfloat[][]{%
    \includegraphics[width=\fourPlotWidth]{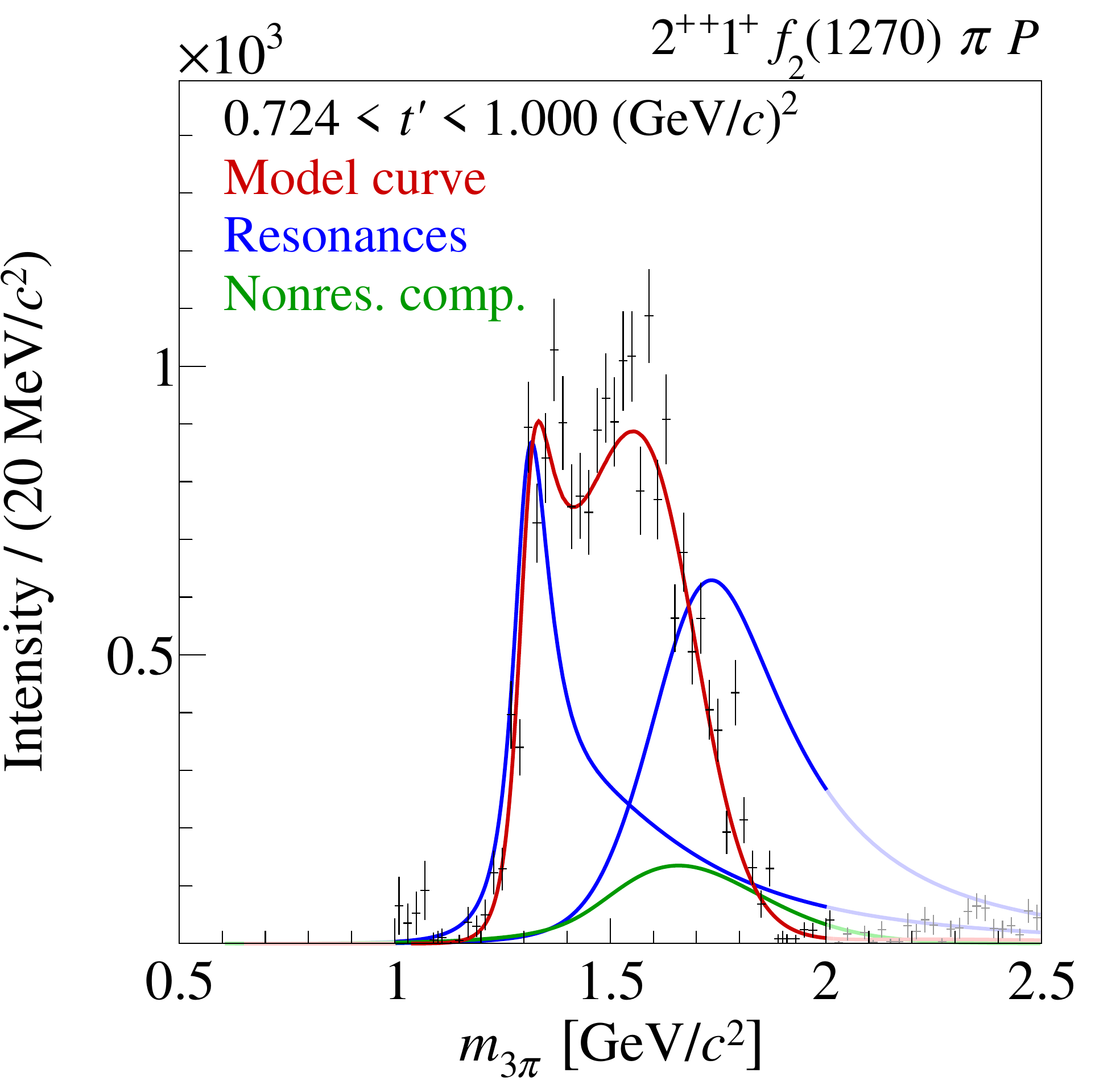}%
    \label{fig:intensity_2pp_f2_tbin11}%
  }%
  \hspace*{\fourPlotSpacing}%
  \subfloat[][]{%
    \includegraphics[width=\fourPlotWidth]{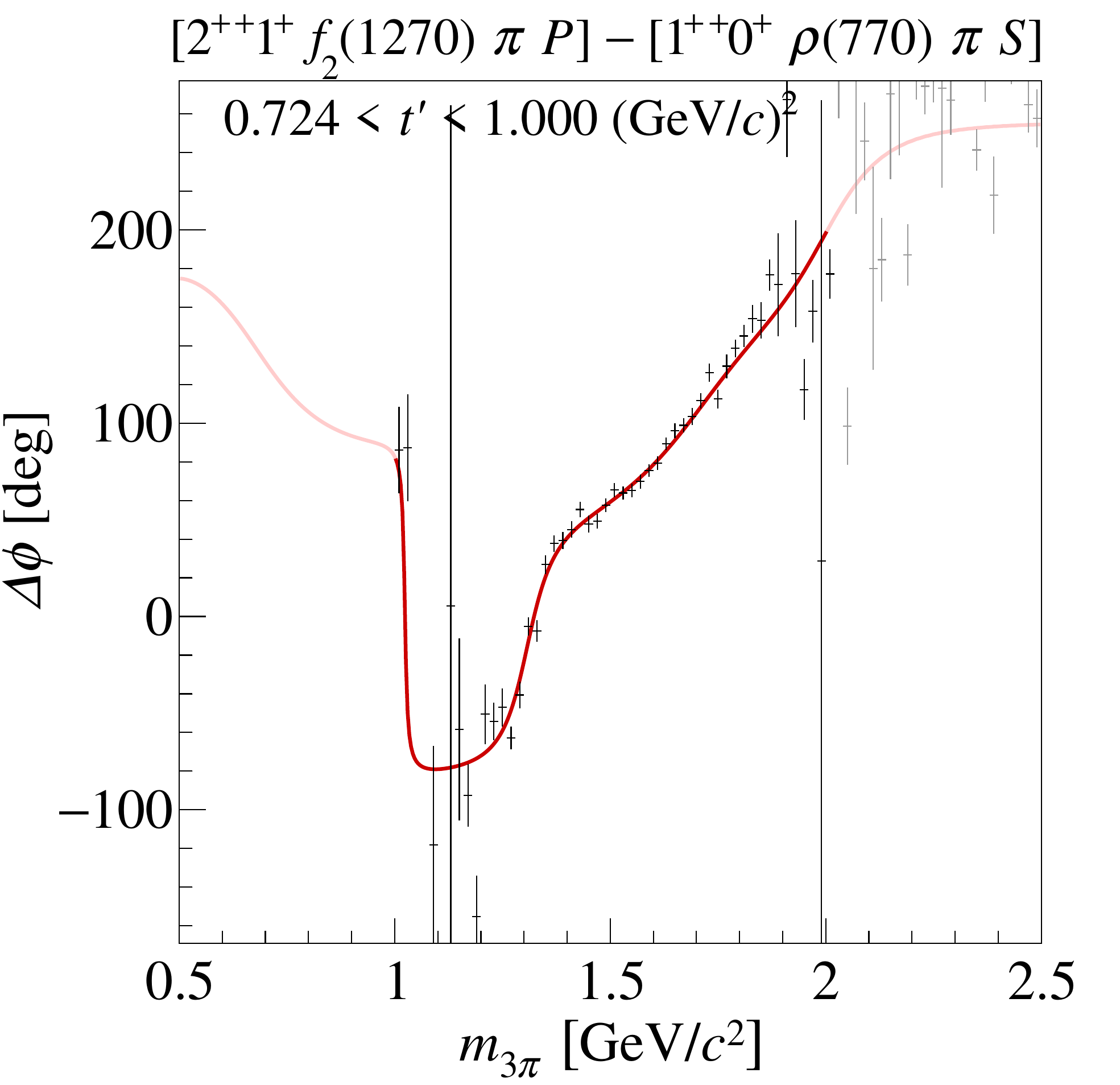}%
    \label{fig:phase_2pp_f2_1pp_rho_tbin11}%
  }%
  \caption{Similar to \cref{fig:intensity_phases_2pp_tbin1} but for
    the highest \tpr bin.}
  \label{fig:intensity_phases_2pp_tbin11}
\ifMultiColumnLayout{\end{figure*}}{\end{figure}}

All three waves exhibit a clear peak around \SI{1.3}{\GeVcc}.  The
intensity distributions of the two $\Prho \pi D$ waves are dominated
by this peak.  The peak shape is nearly independent of \tpr.  At low
\tpr, the $\Prho \pi D$ wave with $M = 1$ exhibits a dip in the
intensity distribution at about \SI{1.8}{\GeVcc} [see
\cref{fig:intensity_2pp_m1_rho_tbin1_log}].  With increasing \tpr,
this dip moves toward higher masses and becomes shallower until it
disappears in the two highest \tpr bins [see
\cref{fig:intensity_2pp_m1_rho_tbin8_log,fig:intensity_2pp_m1_rho_tbin11_log}].
A much stronger variation of the shape of the intensity distribution
with increasing \tpr is observed for the $\PfTwo \pi P$ wave.  In
addition to the peak at \SI{1.3}{\GeVcc}, this wave exhibits a
shoulder at about \SI{1.6}{\GeVcc}, which is absent at low \tpr and
increases with increasing \tpr, and a high-mass tail that becomes
weaker with increasing \tpr.

\begin{wideFigureOrNot}[tbp]
  \centering
  \subfloat[][]{%
    \includegraphics[width=\threePlotWidth]{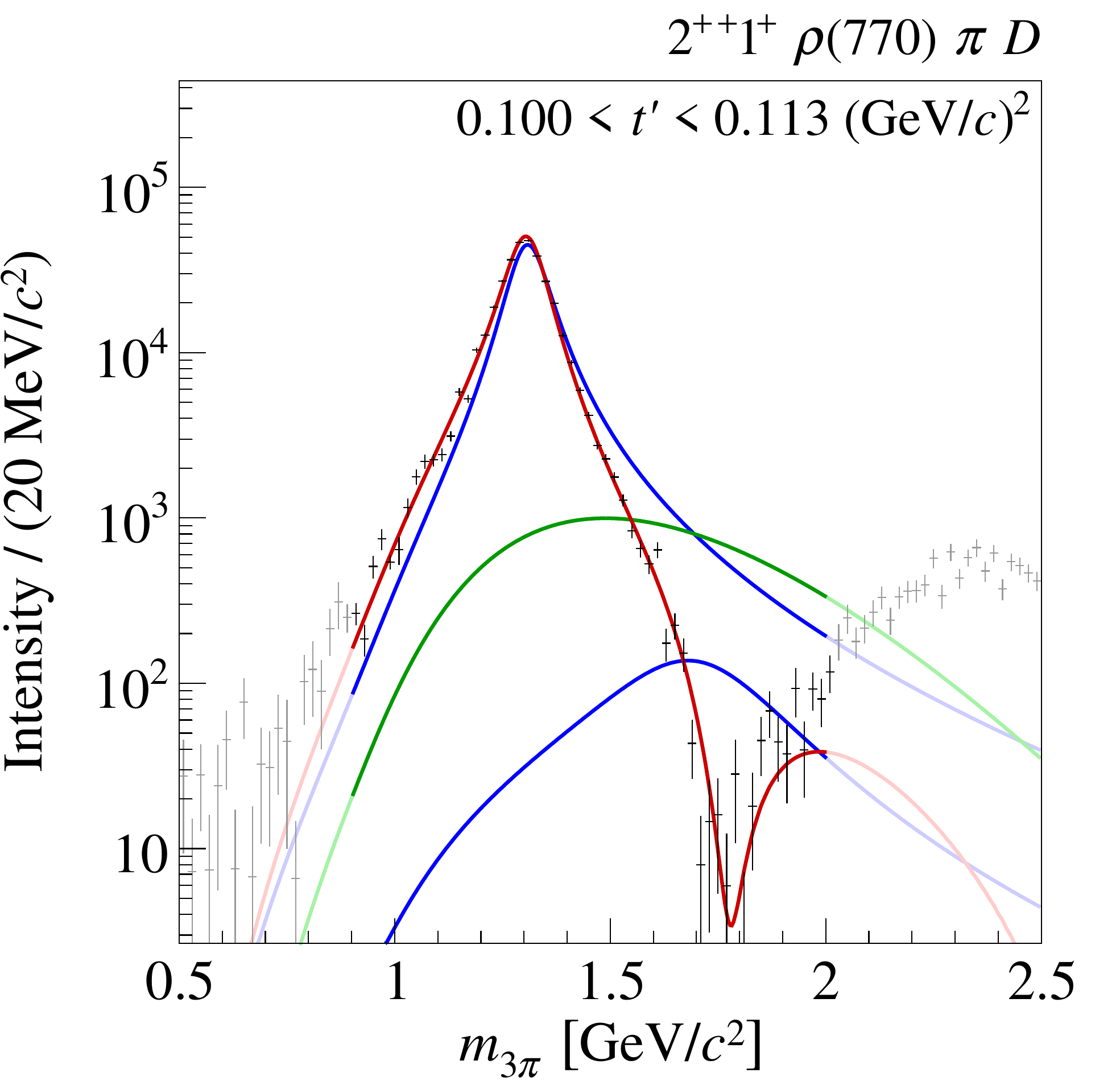}%
    \label{fig:intensity_2pp_m1_rho_tbin1_log}%
  }%
  \hspace*{\threePlotSpacing}%
  \subfloat[][]{%
    \includegraphics[width=\threePlotWidth]{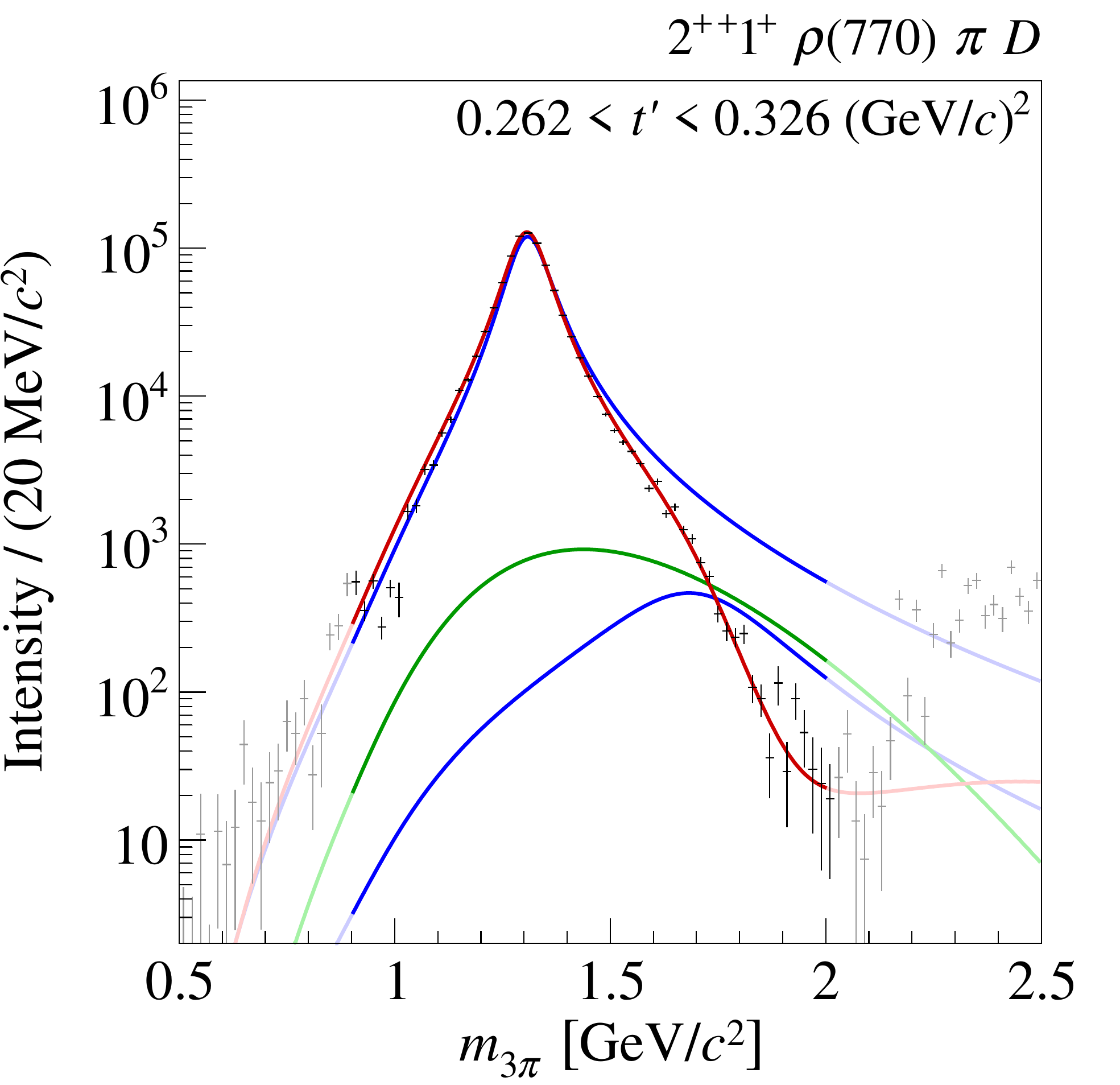}%
    \label{fig:intensity_2pp_m1_rho_tbin8_log}%
  }%
  \hspace*{\threePlotSpacing}%
  \subfloat[][]{%
    \includegraphics[width=\threePlotWidth]{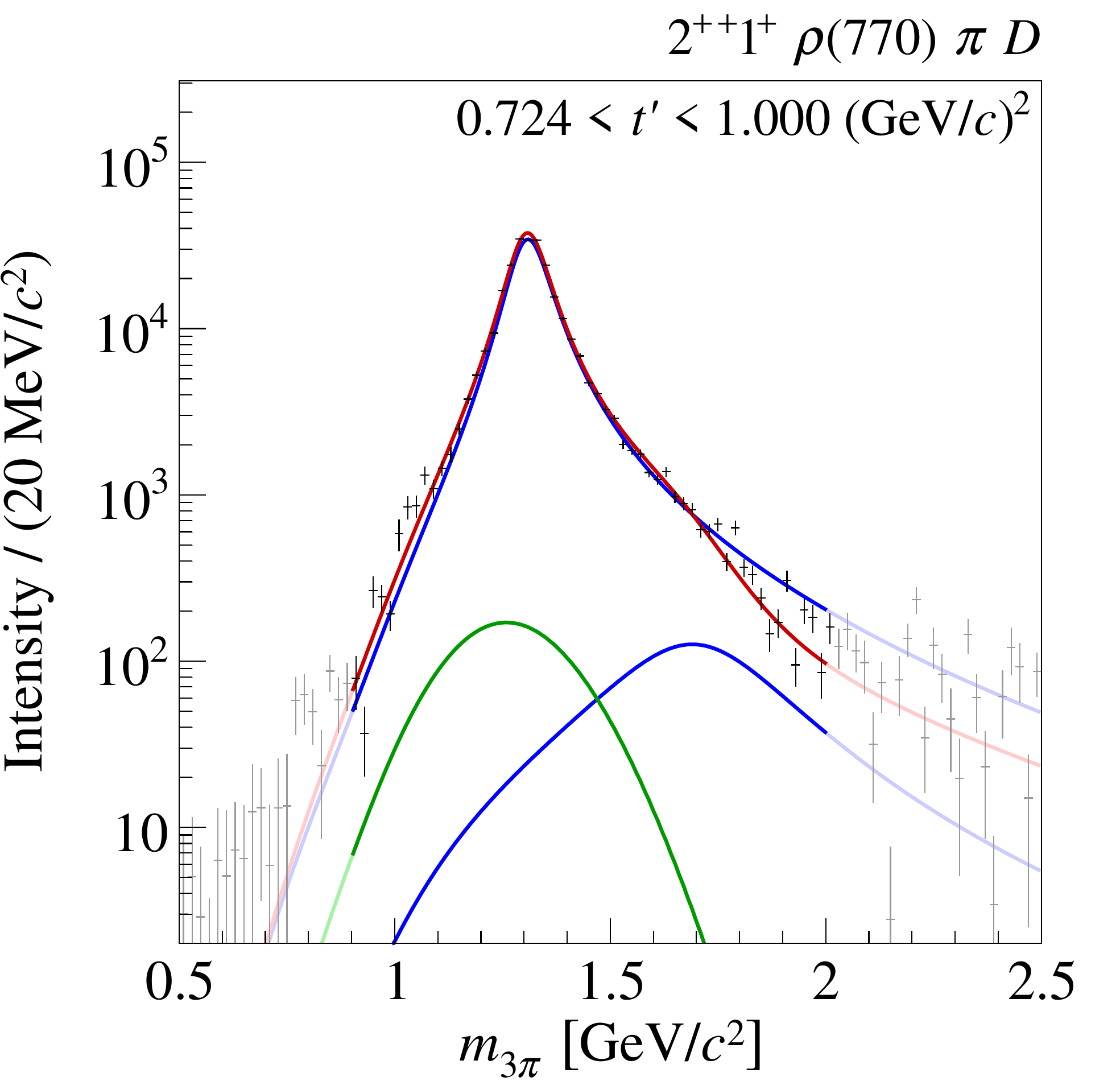}%
    \label{fig:intensity_2pp_m1_rho_tbin11_log}%
  }%
  \caption{Intensity distributions of the \wave{2}{++}{1}{+}{\Prho}{D}
    wave for three \tpr bins shown in logarithmic scale.  The model
    and the wave components are represented as in
    \cref{fig:intensity_phases_2pp_tbin1}.}
  \label{fig:intensity_2pp_log}
\end{wideFigureOrNot}

The right columns of
\cref{fig:intensity_phases_2pp_tbin1,fig:intensity_phases_2pp_tbin11}
show the \mThreePi dependence of the relative phases of the $2^{++}$
waves \wrt the \wave{1}{++}{0}{+}{\Prho}{S} wave.  Clearly rising
phases are observed in the \SI{1.3}{\GeVcc} mass region in all \tpr
bins.
\Cref{fig:intensity_phases_2pp_tbin1,fig:intensity_phases_2pp_tbin11}
also show the relative phases between the three $2^{++}$ waves.  Here,
a more complex pattern is observed that points to different relative
contributions of the components in these waves.

In our model, the three $\JPC = 2^{++}$ waves are described using two
resonances, \PaTwo and \PaTwo[1700].  The \PaTwo is parametrized using
\cref{eq:BreitWigner,eq:method:a2dynamicwidth}, the \PaTwo[1700] using
\cref{eq:BreitWigner,eq:method:fixedwidth}, and the nonresonant
components using \cref{eq:method:nonresterm} for the
\wave{2}{++}{1}{+}{\Prho}{D} wave and \cref{eq:method:nonrestermsmall}
for the other two $2^{++}$ waves (see
\cref{tab:method:fitmodel:waveset}).

Taking into account the high precision of the data in particular for
the \wave{2}{++}{1}{+}{\Prho}{D} wave, the model describes the data
well within the fit range, which is
\SIvalRange{0.9}{\mThreePi}{2.0}{\GeVcc} for the
\wave{2}{++}{1}{+}{\Prho}{D} wave and
\SIvalRange{1.0}{\mThreePi}{2.0}{\GeVcc} for the other two waves.  The
two $\Prho \pi D$ waves are dominated by the \PaTwo with only small
contributions from the \PaTwo[1700].  This is strikingly different in
the $\PfTwo \pi P$ wave, in which the \PaTwo[1700] has an intensity
comparable to that of the \PaTwo and the relative \PaTwo[1700]
intensity grows with increasing \tpr.  In our fit model, the
nonresonant components behave differently in the three $2^{++}$ waves.
Compared to the dominant \PaTwo peak, the nonresonant component in the
\wave{2}{++}{1}{+}{\Prho}{D} wave is small and vanishes nearly
completely in the highest \tpr bin.  The corresponding wave with
$M = 2$ exhibits a larger nonresonant contribution relative to the
\PaTwo, which slightly increases with increasing \tpr.  We find the
largest nonresonant contribution \wrt the \PaTwo peak in the
\wave{2}{++}{1}{+}{\PfTwo}{P} wave.  The relative nonresonant
intensity, which grows slightly with increasing \tpr, is concentrated
mostly in the \PaTwo[1700] region and---as in the other two $2^{++}$
waves---is small in the \PaTwo peak region.

In the \PaTwo region, interference effects of the wave components are
small in all three waves.  The largest effect is a slight asymmetric
distortion of the \PaTwo peak in the \wave{2}{++}{2}{+}{\Prho}{D} wave
due to interference of the \PaTwo with the nonresonant component.
This is different for the \PaTwo[1700] region.  In the
\wave{2}{++}{1}{+}{\Prho}{D} wave, a complicated interplay between
\PaTwo, \PaTwo[1700], and the nonresonant contribution becomes
apparent.  At low \tpr, destructive interference causes the intensity
to drop by 4~orders of magnitude from the \PaTwo peak down to the dip
at about \SI{1.8}{\GeVcc}.  In the two highest \tpr bins, the
nonresonant contribution practically vanishes in the \PaTwo[1700]
region and the interference pattern changes so that the dip in the
high-mass region disappears.  This distinct interference pattern helps
the fit to separate the small \PaTwo[1700] contribution despite the
presence of the dominant \PaTwo.  In the \wave{2}{++}{1}{+}{\PfTwo}{P}
wave, the high-mass shoulder is described by a relatively large
\PaTwo[1700] contribution.  At high \tpr, the rather sharp drop of
this shoulder around \SI{1.8}{\GeVcc} [see
\cref{fig:intensity_2pp_f2_tbin11}] is described by the interference
of all three wave components.

Although the fit model describes the intensity distributions in
general well, it falls short in some regions.  In the dominant
\wave{2}{++}{1}{+}{\Prho}{D} wave, it does not reproduce well the
high-mass tail, which is most pronounced at low \tpr [see \eg
\cref{fig:intensity_2pp_m1_rho_tbin1_log}].  Also, the extrapolation
of the fit model above \SI{2.0}{\GeVcc}, which is the upper limit of
the fit range, disagrees with the data.  We observe a similar behavior
also in the \wave{2}{++}{1}{+}{\PfTwo}{P} wave [see \eg
\cref{fig:intensity_2pp_f2_tbin1}].  In this wave, the model in
addition undershoots the low-mass tail below \SI{1.2}{\GeVcc}, which
is, however, mainly defined by the opening of the $\PfTwo \pi$ phase
space.

The interpretation of the structures in the intensity distributions in
terms of resonances is supported by the relative phases \wrt selected
waves.  The \wave{2}{++}{1}{+}{\Prho}{D} wave exhibits rapidly rising
phases \wrt the \wave{1}{++}{0}{+}{\Prho}{S} wave in the
\SI{1.3}{\GeVcc} region, which are caused by the \PaTwo, and slower
rising phases in the \SI{1.6}{\GeVcc} region [see
\cref{fig:phase_2pp_m1_rho_1pp_rho_tbin1,fig:phase_2pp_m1_rho_1pp_rho_tbin11}].
Both features depend only weakly on \tpr.  The dominant \PaTwo leads
to approximately constant phases relative to the other $2^{++}$ waves
in the region between \SIlist{1.0;1.4}{\GeVcc} [see
\cref{fig:phase_2pp_m1_rho_2pp_m2_rho_tbin1,fig:phase_2pp_m1_rho_2pp_f2_tbin1,fig:phase_2pp_m1_rho_2pp_m2_rho_tbin11,fig:phase_2pp_m1_rho_2pp_f2_tbin11}].
The extremely rapidly decreasing phases around \SI{1.8}{\GeVcc} in
\cref{fig:phase_2pp_m1_rho_2pp_m2_rho_tbin1,fig:phase_2pp_m1_rho_2pp_f2_tbin1,fig:phase_2pp_m1_rho_1pp_rho_tbin1}
are connected to the dip in the intensity distribution of the
\wave{2}{++}{1}{+}{\Prho}{D} wave.  At this dip, the partial-wave
amplitude becomes nearly zero due to destructive interference.  This
behavior of the phases is analogous to the one observed in the
\wave{0}{-+}{0}{+}{\PfZero[980]}{S} wave [see
\cref{sec:zeroMP_results,fig:phase_0mp_1pp_rho_tbin11,fig:phase_0mp_2mp_f2_tbin11}].
As the dip in the intensity distributions, the phase drop disappears
toward higher \tpr.  At large values of \tpr, the phase of the
\wave{2}{++}{1}{+}{\Prho}{D} wave \wrt the
\wave{2}{++}{1}{+}{\PfTwo}{P} wave becomes approximately constant [see
\cref{fig:phase_2pp_m1_rho_2pp_f2_tbin11}], consistent with the
\PaTwo[1700] appearing in both waves.  The phases of the
\wave{2}{++}{2}{+}{\Prho}{D} wave \wrt the
\wave{1}{++}{0}{+}{\Prho}{S} wave also exhibit the rapid rise in the
\SI{1.3}{\GeVcc} region due to the \PaTwo.  It is followed by a drop
of the phase toward the \SI{1.7}{\GeVcc} region.  The missing rising
phase from the \PaTwo[1700] is consistent with the small intensity of
this component in this wave.  The phase motion changes only slightly
with \tpr.  The phase \wrt the other two $2^{++}$ waves are
approximately constant around the \PaTwo.  The phase relative to the
\wave{2}{++}{1}{+}{\PfTwo}{P} wave falls by more than
\SI{180}{\degree} above about \SI{1.4}{\GeVcc}.  This drop is
approximately independent of \tpr and covers the mass region of the
\PaTwo[1700].\footnote{A similar behavior is observed for the
  \PaOne[1640] in the \wave{1}{++}{0}{+}{\PfTwo}{P} wave (see
  \cref{sec:onePP_discussion}).}  The phases of the
\wave{2}{++}{1}{+}{\PfTwo}{P} wave \wrt the
\wave{1}{++}{0}{+}{\Prho}{S} wave exhibit two consecutive phase rises
due to \PaTwo and \PaTwo[1700].  Unlike the intensity distributions of
this wave, the phase motions do not change drastically with \tpr.

From the fit, we extract the following \PaTwo Breit-Wigner resonance
parameters: $m_{\PaTwo} = \SIaerrSys{1314.5}{4.0}{3.3}{\MeVcc}$ and
$\Gamma_{\PaTwo} = \SIaerrSys{106.6}{3.4}{7.0}{\MeVcc}$.  Due to the
large intensity of the \PaTwo, its small width, and the small
contributions from the nonresonant components in the \SI{1.3}{\GeVcc}
region, the systematic uncertainties of the \PaTwo resonance
parameters are the smallest of all resonances in the model (see
\cref{sec:syst_uncert_twoPP}).

The extracted Breit-Wigner resonance parameters for the \PaTwo[1700]
are $m_{\PaTwo[1700]} = \SIaerrSys{1681}{22}{35}{\MeVcc}$ and
$\Gamma_{\PaTwo[1700]} = \SIaerrSys{436}{20}{16}{\MeVcc}$.  They are
mainly determined by the \wave{2}{++}{1}{+}{\PfTwo}{P} wave.  Since
the \PaTwo[1700] signal is much smaller than that of the \PaTwo, the
\PaTwo[1700] parameters have much larger systematic uncertainties.
The \PaTwo[1700] parameters are sensitive to the parametrization of
the nonresonant component and to the value of the range parameter
$q_R$ in the Blatt-Weisskopf factors (see
\cref{sec:syst_uncert_twoPP}).

The \tpr dependence of the intensities of the resonant and nonresonant
$2^{++}$ wave components is shown in
\cref{fig:tprim_2pp,fig:tprim_2pp_f2_nonres} together with the results
of fits using \cref{eq:slope-parametrization}.  The coupling
amplitudes of the resonance components in the two $2^{++}$ waves with
$\Mrefl = 1^+$ are constrained by
\cref{eq:method:branchingdefinition}.  Therefore, the extracted values
of the slope parameters are nearly identical: the \PaTwo slope
parameter has a value of \SIerrSys{7.9}{0.5}{\perGeVcsq} in the
$\Prho \pi D$ wave with $M = 1$ and of
\SIaerrSys{7.8}{0.6}{0.5}{\perGeVcsq} in the $\PfTwo \pi P$ wave (see
\cref{tab:slopes}).  Similar to the $1^{++}$ and $2^{-+}$ sectors (see
\cref{sec:onePP_results,sec:twoMP_results}, respectively), the slope
parameter of the higher-mass state, here the \PaTwo[1700], is smaller.
It has a value of \SIaerrSys{7.3}{2.4}{0.9}{\perGeVcsq} in the
$\Prho \pi D$ wave with $M = 1$ and
\SIaerrSys{7.2}{1.1}{0.8}{\perGeVcsq} in the $\PfTwo \pi P$ wave.  If
we do not constrain the coupling amplitudes of the resonance
components via \cref{eq:method:branchingdefinition} [\StudyT; see
\cref{sec:systematics}], the above slope values remain essentially
unchanged.  The only exception is the \PaTwo[1700] in the
$\Prho \pi P$ wave, the slope of which becomes about
\SI{2}{\perGeVcsq} steeper.

\begin{wideFigureOrNot}[tbp]
  \centering
  \subfloat[][]{%
    \includegraphics[width=\threePlotWidth]{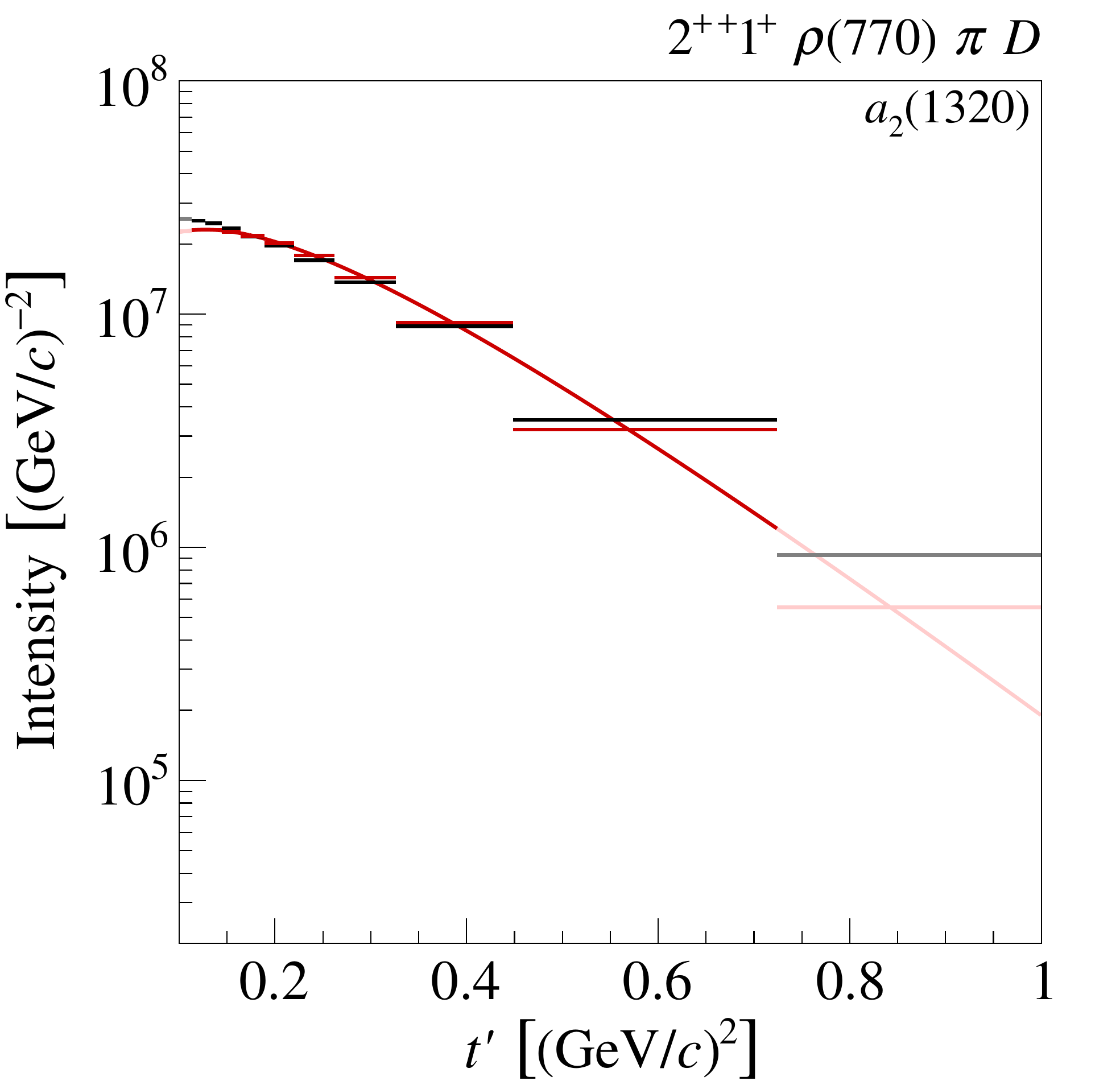}%
  }%
  \hspace*{\threePlotSpacing}%
  \subfloat[][]{%
    \includegraphics[width=\threePlotWidth]{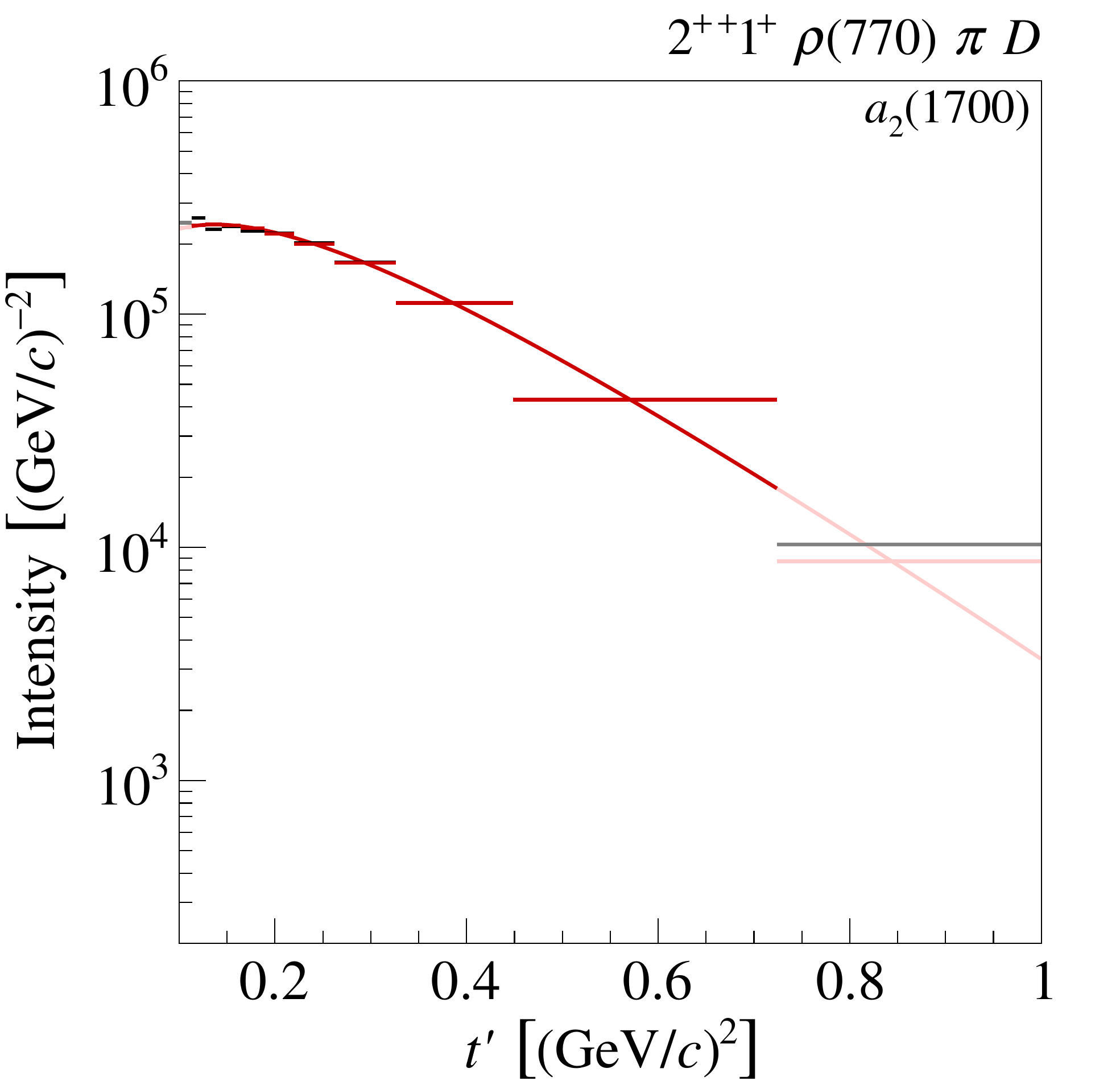}%
  }%
  \hspace*{\threePlotSpacing}%
  \subfloat[][]{%
    \includegraphics[width=\threePlotWidth]{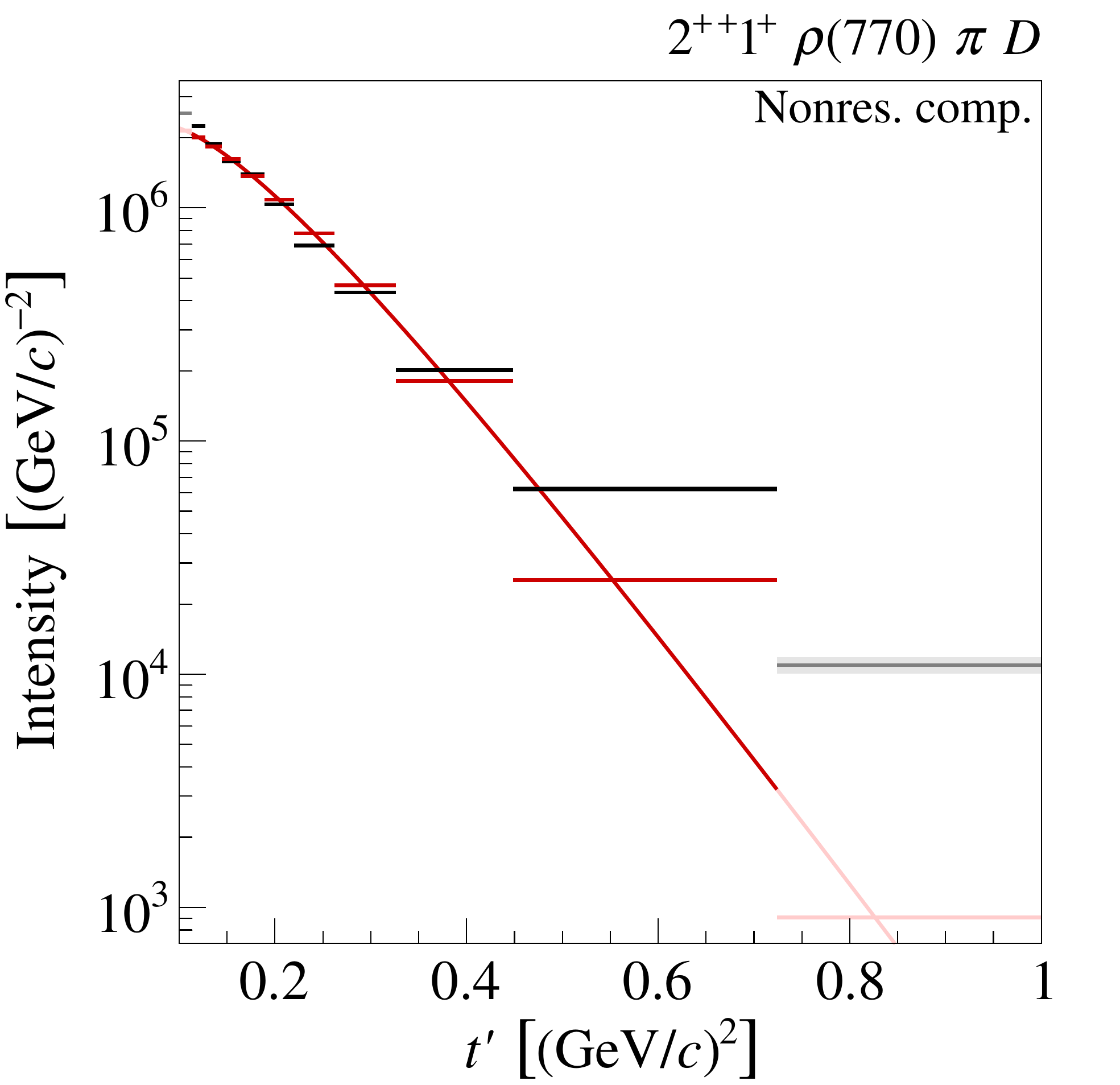}%
  }%
  \\
  \subfloat[][]{%
    \includegraphics[width=\threePlotWidth]{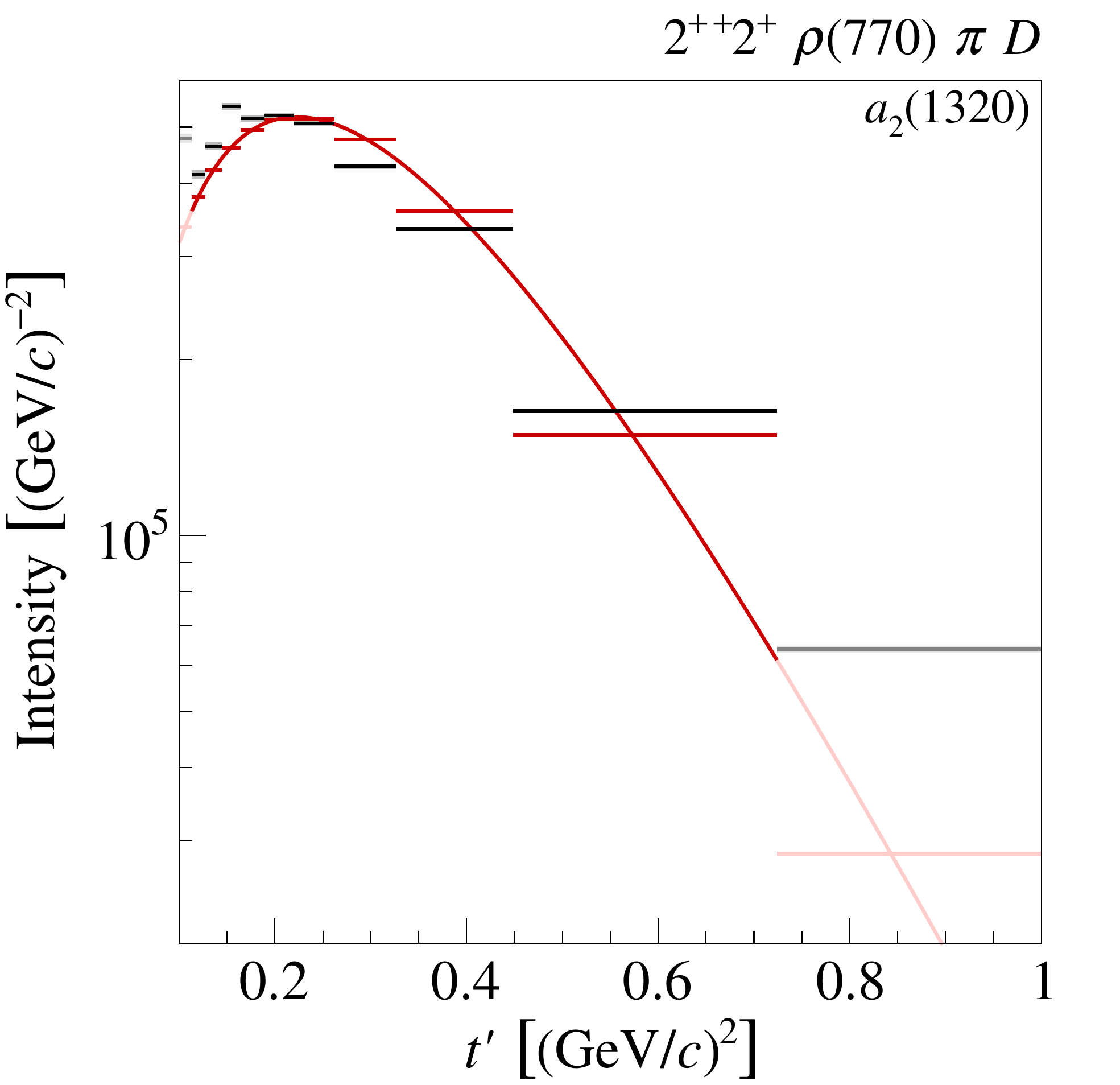}%
  }%
  \hspace*{\threePlotSpacing}%
  \subfloat[][]{%
    \includegraphics[width=\threePlotWidth]{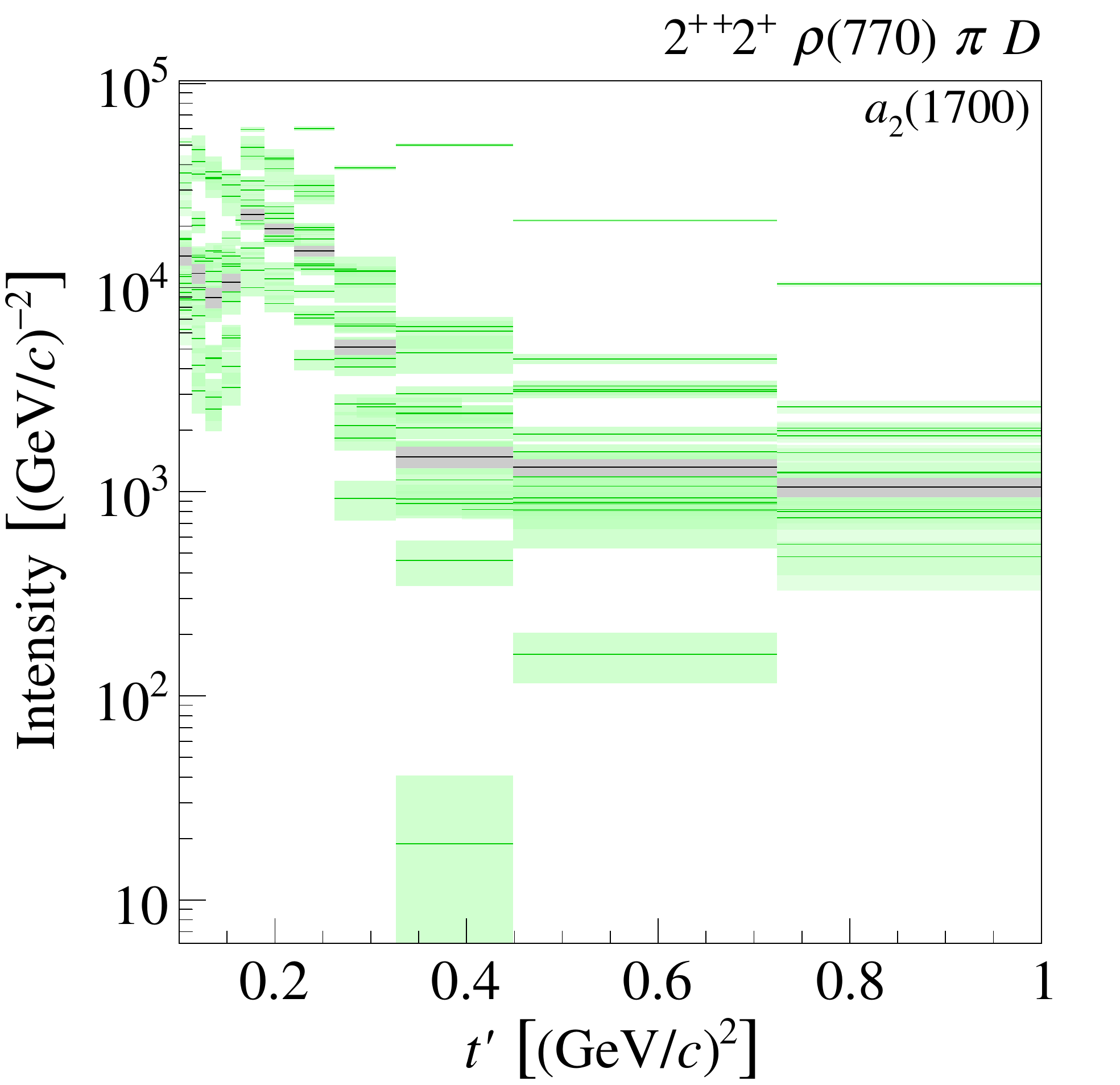}%
    \label{fig:tprim_2pp_m2}%
  }%
  \hspace*{\threePlotSpacing}%
  \subfloat[][]{%
    \includegraphics[width=\threePlotWidth]{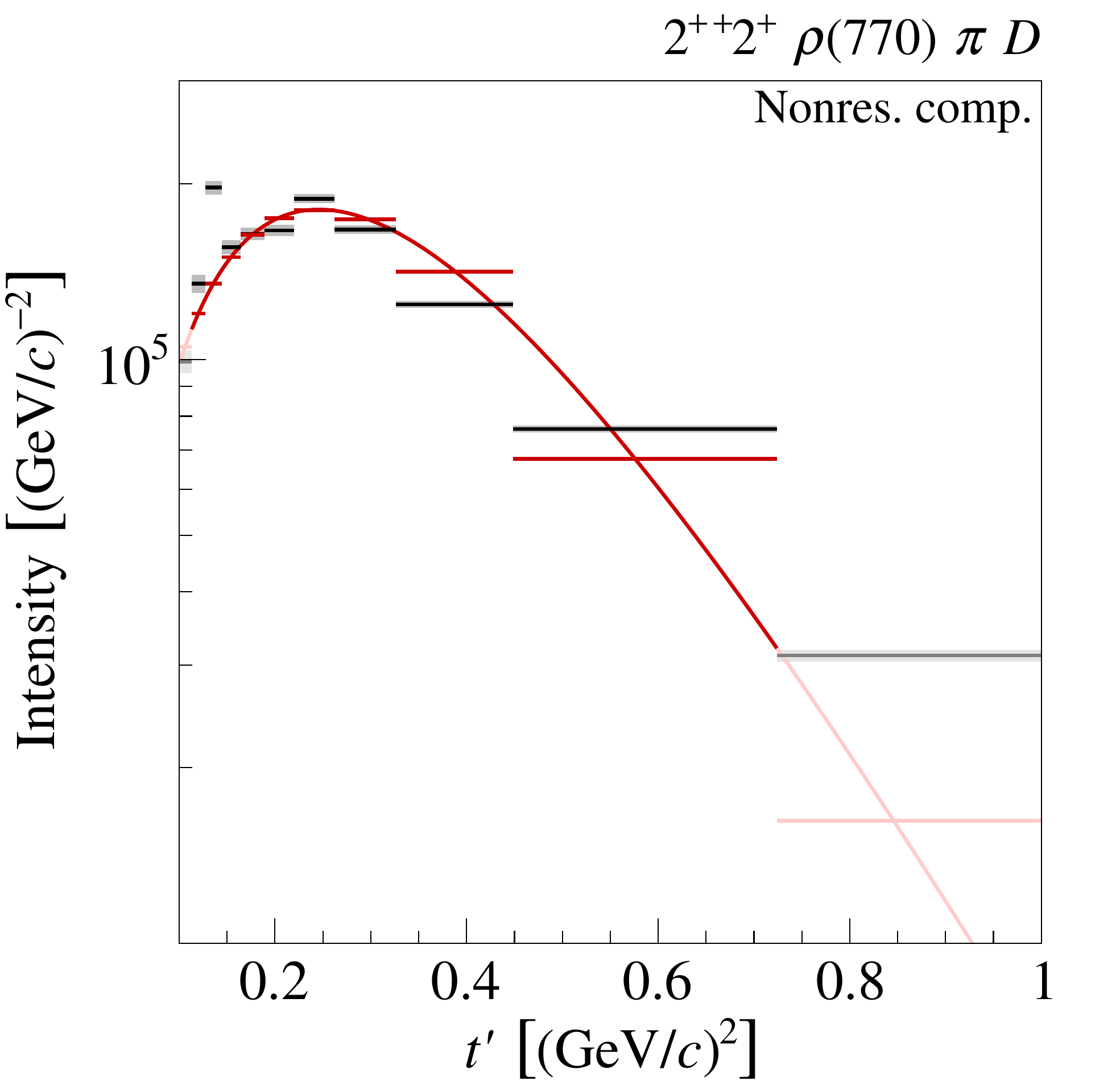}%
  }%
  \caption{Similar to \cref{fig:method:tp:examplespectrum} but showing
    the \tpr spectra of the components in the
    \wave{2}{++}{}{}{\Prho}{D} waves with (upper row) $\Mrefl = 1^+$
    and (lower row) $\Mrefl = 2^+$ as given by
    \cref{eq:tprim-dependence}: (left) \PaTwo component, (center)
    \PaTwo[1700] component, and (right) nonresonant components.  The
    red curves and horizontal lines represent fits using
    \cref{eq:slope-parametrization}.  \subfloatLabel{fig:tprim_2pp_m2}
    shows in addition to the \PaTwo[1700] \tpr spectrum from the main
    fit (black/gray) the \tpr spectra obtained in the various
    systematic studies (central values shown in green, statistical
    uncertainties in light green).}
  \label{fig:tprim_2pp}
\end{wideFigureOrNot}

\begin{figure}[tbp]
  \centering
  \includegraphics[width=\twoPlotWidth]{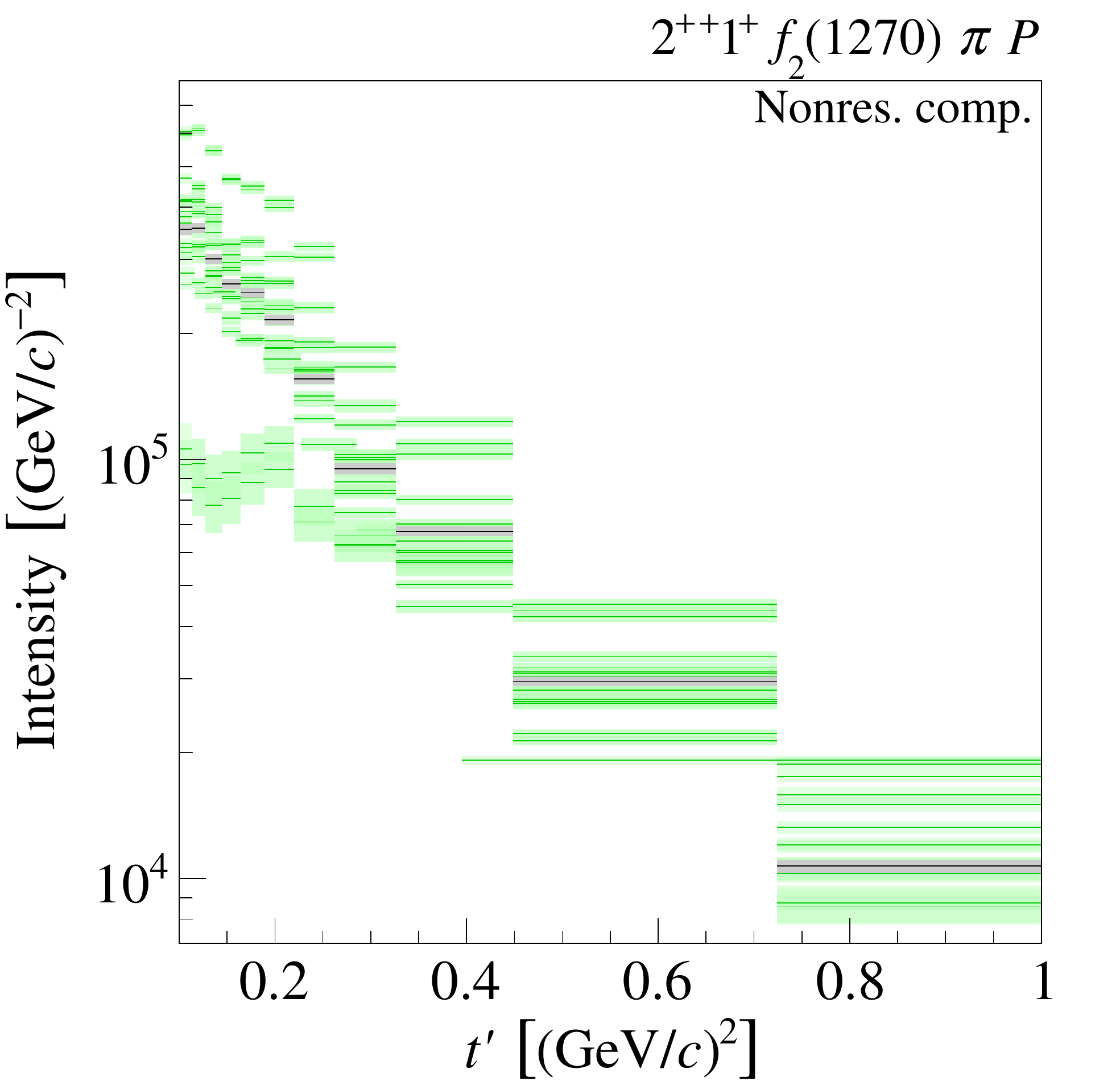}
  \caption{Similar to \cref{fig:tprim_2pp_m2}, but showing the \tpr
    spectrum of the nonresonant component in the
    \wave{2}{++}{1}{+}{\PfTwo}{P} wave as given by
    \cref{eq:tprim-dependence}.}
  \label{fig:tprim_2pp_f2_nonres}
\end{figure}

As in other waves, we observe that the \tpr spectra of the nonresonant
components in the \wave{2}{++}{1}{+}{\Prho}{D} and
\wave{2}{++}{1}{+}{\PfTwo}{P} waves are distinctly different from
those of the resonances.  In particular in the $\Prho \pi D$ wave with
$M = 1$, the nonresonant component exhibits a much steeper \tpr
spectrum with a slope parameter value of
\SIaerrSys{13.6}{0.4}{1.8}{\perGeVcsq}. The nonresonant \tpr spectrum
in the $\PfTwo \pi P$ wave is sensitive to changes of the fit model
discussed in \cref{sec:systematics} (see
\cref{fig:tprim_2pp_f2_nonres}).  It is not well described by the
model, \cref{eq:slope-parametrization}.  The $\rbrk{\tpr}^{\abs{M}}$
factor in the model induces a downturn toward lower \tpr, which is
inconsistent with the data.  From the above, we conclude that the
nonresonant component in this wave seems to have too much freedom.  We
also cannot exclude that it is distorted by leakage into the small
$\PfTwo \pi P$ wave at the stage of the mass-independent analysis.

In the fit model, the \tpr dependence of the coupling amplitudes of
the resonant components in the \wave{2}{++}{2}{+}{\Prho}{D} wave is
not constrained by \cref{eq:method:branchingdefinition} and is
therefore determined independently of the other two $2^{++}$ waves.
In the $M = 2$ wave, we observe slope parameters for the \PaTwo and
the nonresonant contribution of \SIaerrSys{9.0}{1.2}{0.7}{\perGeVcsq}
and \SIaerrSys{8.1}{1.6}{0.5}{\perGeVcsq}, respectively.  The value
for the \PaTwo is slightly larger than in the other two $2^{++}$
waves, while the one for the nonresonant component is significantly
smaller.  Both effects are not understood at present and illustrate
the limitations of our model.  The \tpr spectrum of the \PaTwo[1700]
in the $M = 2$ wave differs strongly from the \tpr spectra in the
other two waves [see \cref{fig:tprim_2pp_m2}].  It has a rather
peculiar shape: after an initial rise with increasing \tpr, the
intensity drops sharply with \tpr until about \SI{0.3}{\GeVcsq} and
then levels off.  The fit function in \cref{eq:slope-parametrization}
is not able to describe these data. The \tpr spectrum is sensitive to
changes of the fit model discussed in \cref{sec:systematics}.  We
therefore conclude that with our model the \PaTwo[1700] signal in the
$M = 2$ wave is too small in order to reliably extract \PaTwo[1700]
yields, although it helps to constrain the \PaTwo[1700] parameters.

We extract the branching-fraction ratio for the decays of the \PaTwo
into the $\Prho \pi D$ and $\PfTwo \pi P$ decay modes with $M = 1$,
where the latter one is a subthreshold decay.  Using
\cref{eq:branch_fract_ratio} we get
\begin{multlineOrEq}
  \label{eq:branch_fract_ratio_a2}
  B_{\Prho* \pi D, \PfTwo* \pi P}^{\PaTwo*}
  \newLineOrNot
  \begin{alignedOrNot}
    \alignOrNot= \frac{\text{BF}\!\sBrk{\PaTwo^- \to \Prho^0 \pi^- \to \threePi}\hfill}{\text{BF}\!\sBrk{\PaTwo^- \to \PfTwo \pi^- \to \threePi}}
    \newLineOrNot
    \alignOrNot= \numaerrSys{17.6}{1.1}{2.6}.
  \end{alignedOrNot}
\end{multlineOrEq}
This is the first measurement of this quantity.  As for the \PaFour
(see \cref{sec:fourPP_results}), this ratio increases by a factor of
$4/3$ when we take into account the unobserved decays
$\PaTwo*^- \to \Prho*^- \pi^0$ and $\PaTwo*^- \to \PfTwo* \pi^-$ to
the \threePiN final state and assume isospin symmetry.  Hence
\begin{equation}
  \label{eq:branch_fract_ratio_a2_iso}
  \begin{splitOrNot}
    B_{\Prho* \pi D, \PfTwo* \pi P}^{\PaTwo*, \text{iso}}
    \alignOrNot= \frac{\text{BF}\!\sBrk{\PaTwo^- \to \Prho \pi \to 3\pi}\hfill}{\text{BF}\!\sBrk{\PaTwo^- \to \PfTwo \pi \to 3\pi}}
    \newLineOrNot
    \alignOrNot= \numaerrSys{23.5}{1.5}{3.5}.
  \end{splitOrNot}
\end{equation}
Taking into account the branching fraction of the \PfTwo to $2\pi$ and
the effect of the different Bose symmetrizations in the \threePi and
\threePiN final states, the isospin factor $4/3$ should be replaced by
\numaerr{0.936}{0.032}{0.010}\footnote{We only take into account the
  uncertainty of the $\PfTwo \to 2\pi$ branching fraction.} yielding
the corrected branching-fraction ratio
\begin{equation}
  \label{eq:branch_fract_ratio_a2_corr}
  \begin{splitOrNot}
    B_{\Prho* \pi D, \PfTwo* \pi P}^{\PaTwo*, \text{corr}}
    \alignOrNot= \frac{\text{BF}\!\sBrk{\PaTwo^- \to \Prho \pi}\hfill}{\text{BF}\!\sBrk{\PaTwo^- \to \PfTwo \pi}}
    \newLineOrNot
    \alignOrNot= \numaerr{16.5}{1.2}{2.4}.
  \end{splitOrNot}
\end{equation}

\subsubsection{Discussion of results on $2^{++}$ resonances}
\label{sec:twoPP_discussion}

From our analysis, we conclude that we observe two resonances with
$\JPC = 2^{++}$.  The \PaTwo appears as a clear peak in all three
$2^{++}$ waves, whereas the \PaTwo[1700] shows up most prominently in
the $\PfTwo \pi P$ wave and is seen to couple only weakly to
$\Prho \pi D$.  In order to study the significance of the extracted
\PaTwo[1700] signal, we performed a fit, in which the \PaTwo[1700]
component was removed from the fit model.  Compared to the main fit,
this fit has a minimum \chisq~value that is larger by a factor
of~\num{1.48}.\footnote{Compared to the \num{722} free parameters of
  the main fit, this fit has \num{674} free parameters.}
\Cref{fig:no-a2(1700)_chi2difference} shows the contributions from the
spin-density matrix elements to the \chisq~difference between this and
the main fit.  As expected, the largest contribution to the observed
\chisq~increase comes from the \wave{2}{++}{1}{+}{\PfTwo}{P} wave
intensity and from its interferences.  This shows that most of the
support for the \PaTwo[1700] component comes from the
\wave{2}{++}{1}{+}{\PfTwo}{P} wave, which is consistent with the
observation that the \PaTwo[1700] signal is small in the two
$\Prho \pi D$ waves.  \Cref{fig:no-a2(1700)_2pp_f2_intensity} shows
that the \wave{2}{++}{1}{+}{\PfTwo}{P} wave cannot be described
without the \PaTwo[1700].  The model without the \PaTwo[1700] that is
represented by the dashed red curve is in particular unable to
describe the shoulder at about \SI{1.6}{\GeVcc} in the intensity
distribution.

\begin{figure}[tbp]
  \centering
  \includegraphics[width=\linewidthOr{\twoPlotWidth}]{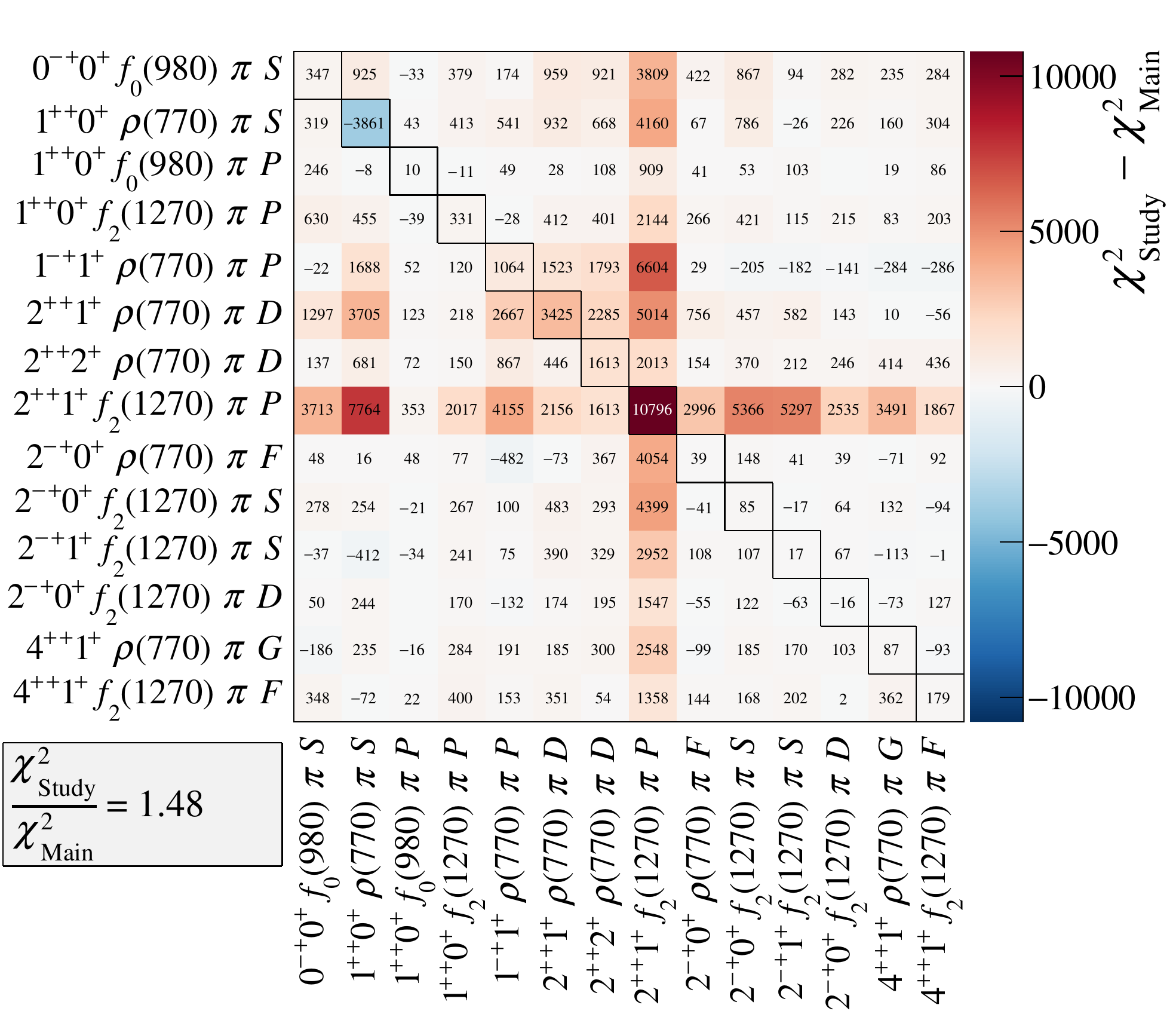}
  \caption{Similar to \cref{fig:DeckMC_chi2difference}, but for the
    study, in which the \PaTwo[1700] component was omitted from the
    fit model.}
  \label{fig:no-a2(1700)_chi2difference}
\end{figure}

\begin{figure}[tbp]
  \centering
  \includegraphics[width=\twoPlotWidth]{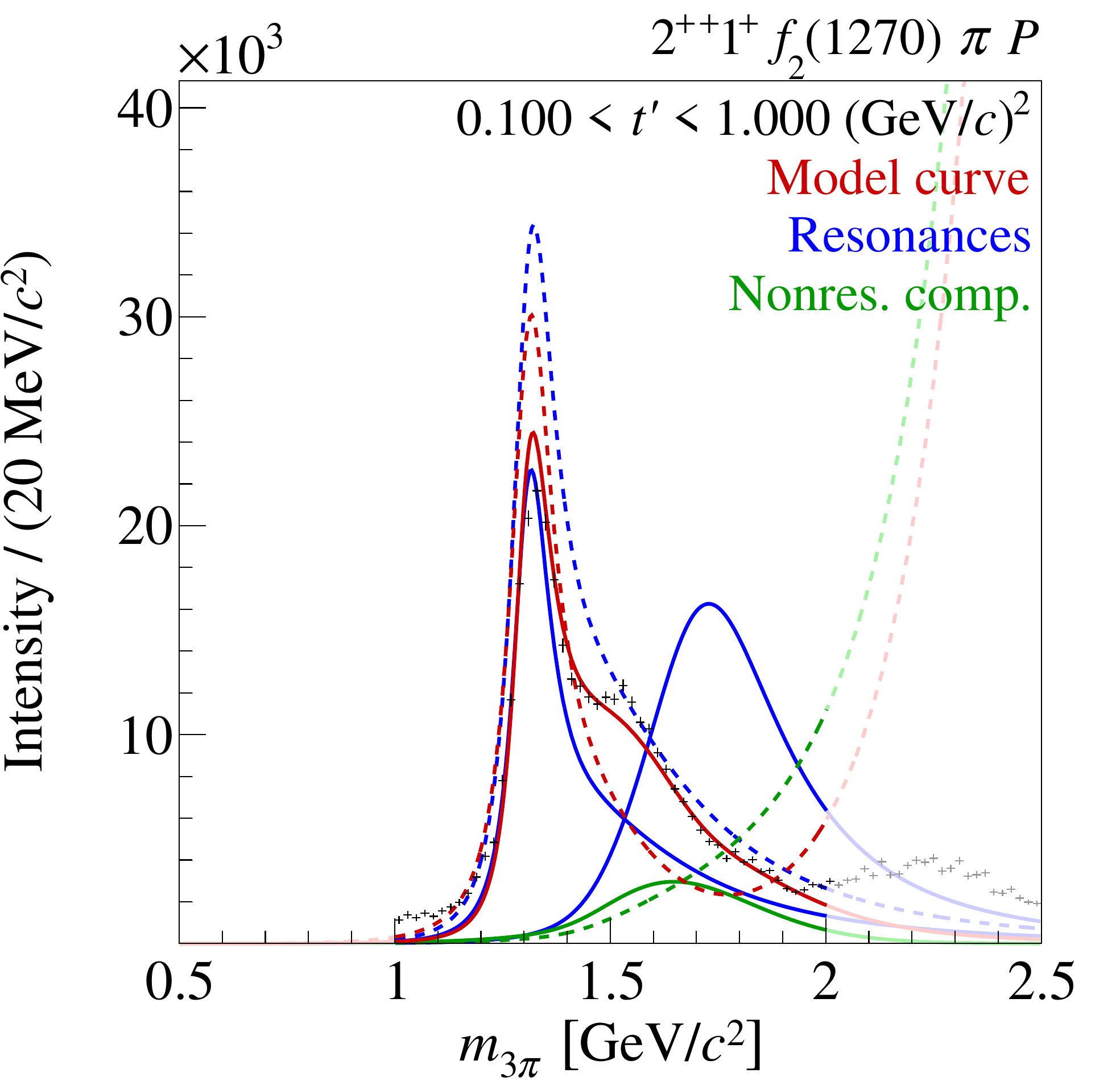}
  \caption{\tpr-summed intensity of the \wave{2}{++}{1}{+}{\PfTwo}{P}
    wave with the result of the main fit (continuous curves) and of
    the fit, in which the \PaTwo[1700] component was removed from the
    fit model (dashed curves).  The model and the wave components are
    represented as in \cref{fig:intensity_phases_2pp_tbin1}.}
  \label{fig:no-a2(1700)_2pp_f2_intensity}
\end{figure}

We clearly observe the production of \PaTwo with $M = 2$.  This is
consistent with the peak observed in the $M = 2$ $D$-wave of the
$\pi^-\eta$ final state [see Fig.~3(g) in \refCite{Adolph:2014rpp}].
Also the intensity ratio of the $M = 1$ and $M = 2$ waves at the
\PaTwo peak position is similar for the two final states.  In the
present analysis, we have studied in detail the \tpr dependence of the
\PaTwo component in the $\Prho \pi D$ waves with $M = 1$ and $M = 2$.
Despite the different functional dependence due to the
$\rbrk{\tpr}^{\abs{M}}$ factor in \cref{eq:slope-parametrization}, the
extracted slope parameters have similar values.  In addition, the
relative phase of the coupling amplitudes of the \PaTwo in the two
waves exhibits only a weak \tpr dependence and departs from zero by no
more than \SI{20}{\degree} (see \cref{sec:production_phases}).  All
this points to the same production mechanism and shows that Pomeron
exchange can transfer helicity~2 to the produced state.

The PDG quotes world averages for the \PaTwo parameters of
$m_{\PaTwo} = \SIaerr{1319.0}{1.0}{1.3}{\MeVcc}$ and
$\Gamma_{\PaTwo} =
\SIaerr{105}{1.6}{1.9}{\MeVcc}$~\cite{Patrignani:2016xqp} for the
$3\pi$ decay mode.  While our estimate of
$m_{\PaTwo} = \SIaerrSys{1314.5}{4.0}{3.3}{\MeVcc}$ is
\SI{4.5}{\MeVcc} lower, our width value of
$\Gamma_{\PaTwo} = \SIaerrSys{106.6}{3.4}{7.0}{\MeVcc}$ agrees well
with the PDG average.  Our present \PaTwo parameters agree with the
results of our two previous analyses: the one based on the measurement
of the \threePi final state diffractively produced on a solid lead
target~\cite{alekseev:2009aa}, and the other based on the measurement
of the $\eta \pi$ and $\eta' \pi$ final states diffractively produced
on a liquid-hydrogen target~\cite{Adolph:2014rpp}.  The finite
resolution in \mThreePi, which is neglected in our analysis, is
estimated to affect the width by less than \SI{1}{\MeVcc}.  Our values
for the slope parameter of the \PaTwo in the $\Prho \pi D$ and
$\PfTwo \pi P$ waves with $M = 1$ are in good agreement with the value
of \SI{7.3(1)}{\perGeVcsq} measured by ACCMOR~\cite{daum:1980ay}.

The \PaTwo[1700] is listed by the PDG as \enquote{omitted from summary
  table} with world averages for mass and width of
$m_{\PaTwo[1700]} = \SI{1732(16)}{\MeVcc}$ and
$\Gamma_{\PaTwo[1700]} =
\SI{194(40)}{\MeVcc}$~\cite{Patrignani:2016xqp}.  Our result of
$m_{\PaTwo[1700]} = \SIaerrSys{1681}{22}{35}{\MeVcc}$ is consistent
with the world average, but our width value of
$\Gamma_{\PaTwo[1700]} = \SIaerrSys{436}{20}{16}{\MeVcc}$ is
\SI{242}{\MeVcc} larger.  Our width estimate is especially in
disagreement with the result of the Belle experiment, which measured
an enhancement in the invariant mass spectrum of $K^+ K^-$ pairs
produced in two-photon collisions~\cite{Abe:2003vn} with a width of
only \SIerrs{151}{22}{24}{\MeVcc}.  The PDG assigns this measurement
to the \PaTwo[1700] and includes it in the world average.  It is
interesting to compare our results with an analysis of the $\eta \pi$
$D$-wave intensity using an analytical model based on the principles
of the relativistic $S$-matrix~\cite{Jackura:2017amb}.  The analysis
is based on the partial-wave decomposition of COMPASS data from
\refCite{Adolph:2014rpp}.  The extracted \PaTwo pole parameters from
\refCite{Jackura:2017amb} are consistent with the values of our
Breit-Wigner parameters.  The same is true for the \PaTwo[1700] mass,
but the \PaTwo[1700] width of \SIerrs{280}{10}{70}{\MeVcc} that is
found in \refCite{Jackura:2017amb} appears to be lower than our value.
This is a hint that our simplifying model assumptions may cause an
overestimation of the \PaTwo[1700] width.

We observe that the \PaTwo[1700] predominantly decays into
$\PfTwo \pi P$ and less into $\Prho \pi D$.  This finding is difficult
to reconcile with the dominance of the $\Prho \pi$ over the
$\PfTwo \pi$ decay mode observed by the L3~experiment in an analysis
of the $\pi^+\pi^-\pi^0$ final state produced in two-photon
collisions~\cite{Shchegelsky:2006es}.  At the current stage of the
analysis we do not make a quantitative statement on the \PaTwo[1700]
branching fractions because the \PaTwo[1700] region in the two
$\Prho \pi D$ waves is dominated by the \PaTwo high-mass tail and the
nonresonant components.

A number of observations of potential higher excited \PaTwo* states
are listed by the PDG as \enquote{further
  states}~\cite{Patrignani:2016xqp}:
\PaTwo[1950]~\cite{anisovich:2001pn},
\PaTwo[1990]~\cite{Shchegelsky:2006es,Lu:2004yn},
\PaTwo[2030]~\cite{anisovich:2001pn},
\PaTwo[2175]~\cite{anisovich:2001pn}, and
\PaTwo[2255]~\cite{anisovich:2001pp}.  We do not see clear resonance
signals of heavy \PaTwo* states above the \PaTwo[1700] in the analyzed
waves.
 %
%
%

\subsection{$\JPC = 2^{-+}$ resonances}
\label{sec:twoMP}

\subsubsection{Results on $2^{-+}$ resonances}
\label{sec:twoMP_results}

We include four waves with $\JPC = 2^{-+}$ in the resonance-model fit.
The \wave{2}{-+}{0}{+}{\Prho}{F} and \wave{2}{-+}{0}{+}{\PfTwo}{S}
waves have relatively large intensities and contribute
\SI{2.2}{\percent} and \SI{6.7}{\percent} to the total intensity,
respectively.  The \wave{2}{-+}{1}{+}{\PfTwo}{S} and
\wave{2}{-+}{0}{+}{\PfTwo}{D} waves have smaller intensities and each
contributes \SI{0.9}{\percent} to the total intensity.
\Cref{fig:intensity_phases_2mp} shows the intensity distributions of
the four waves for the lowest and the highest \tpr bins (first and
third rows, respectively).

\ifMultiColumnLayout{\begin{figure*}[p]}{\begin{figure}[tbp]}
  \centering
  \subfloat[][]{%
    \includegraphics[width=\fourPlotWidth]{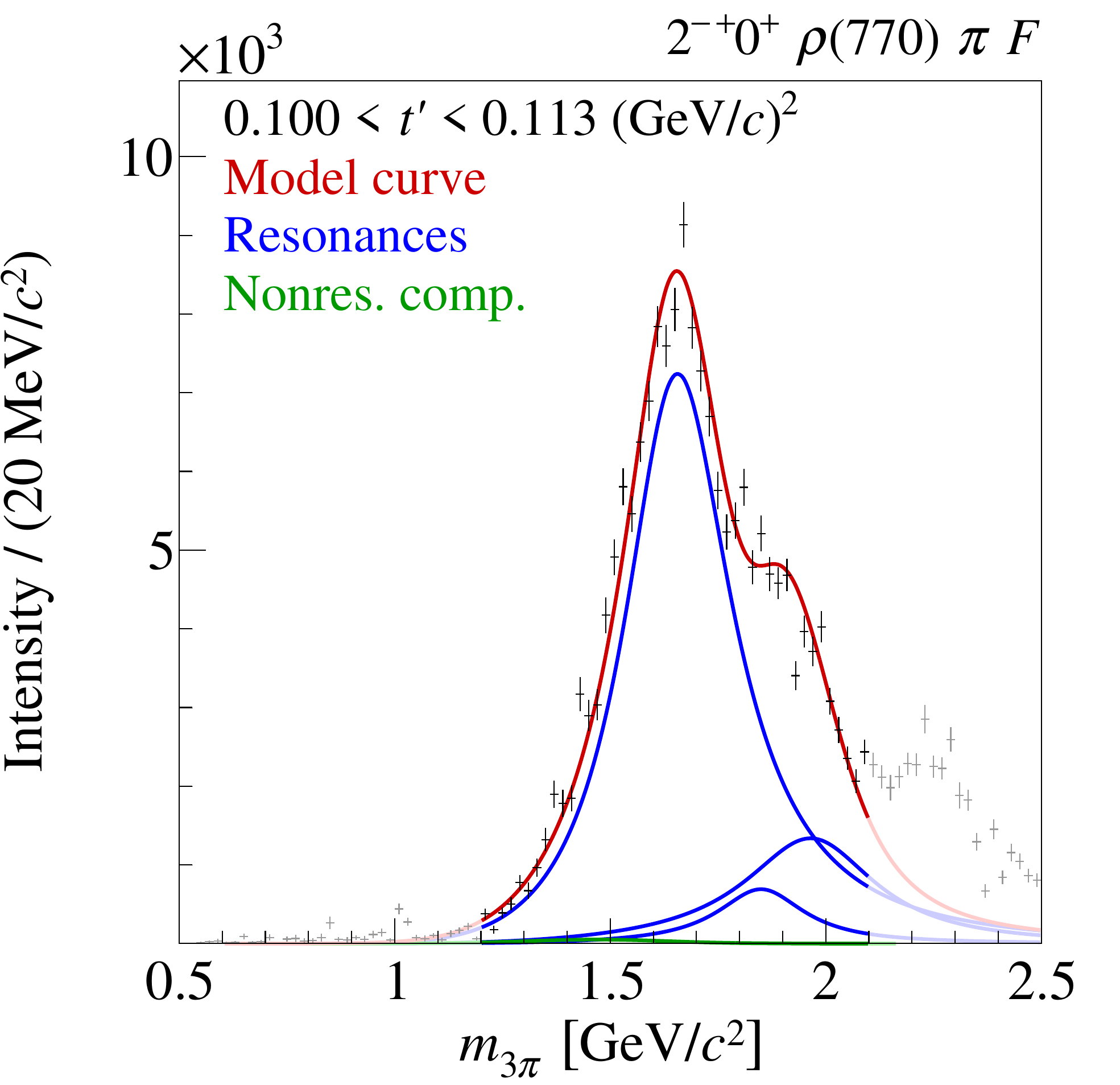}%
    \label{fig:intensity_2mp_rho_tbin1}%
  }%
  \hspace*{\fourPlotSpacing}%
  \subfloat[][]{%
    \includegraphics[width=\fourPlotWidth]{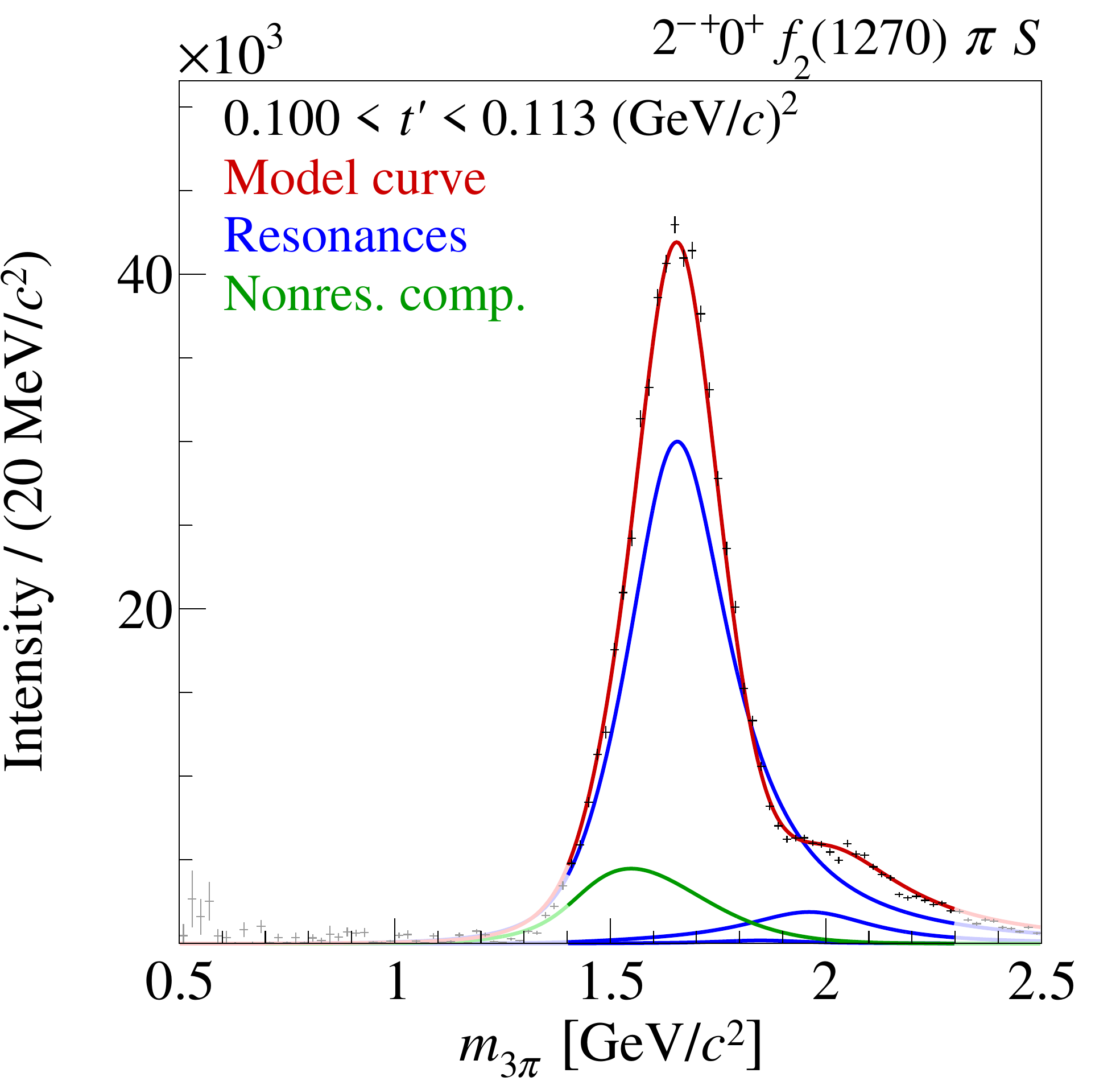}%
    \label{fig:intensity_2mp_m0_f2_S_tbin1}%
  }%
  \hspace*{\fourPlotSpacing}%
  \subfloat[][]{%
    \includegraphics[width=\fourPlotWidth]{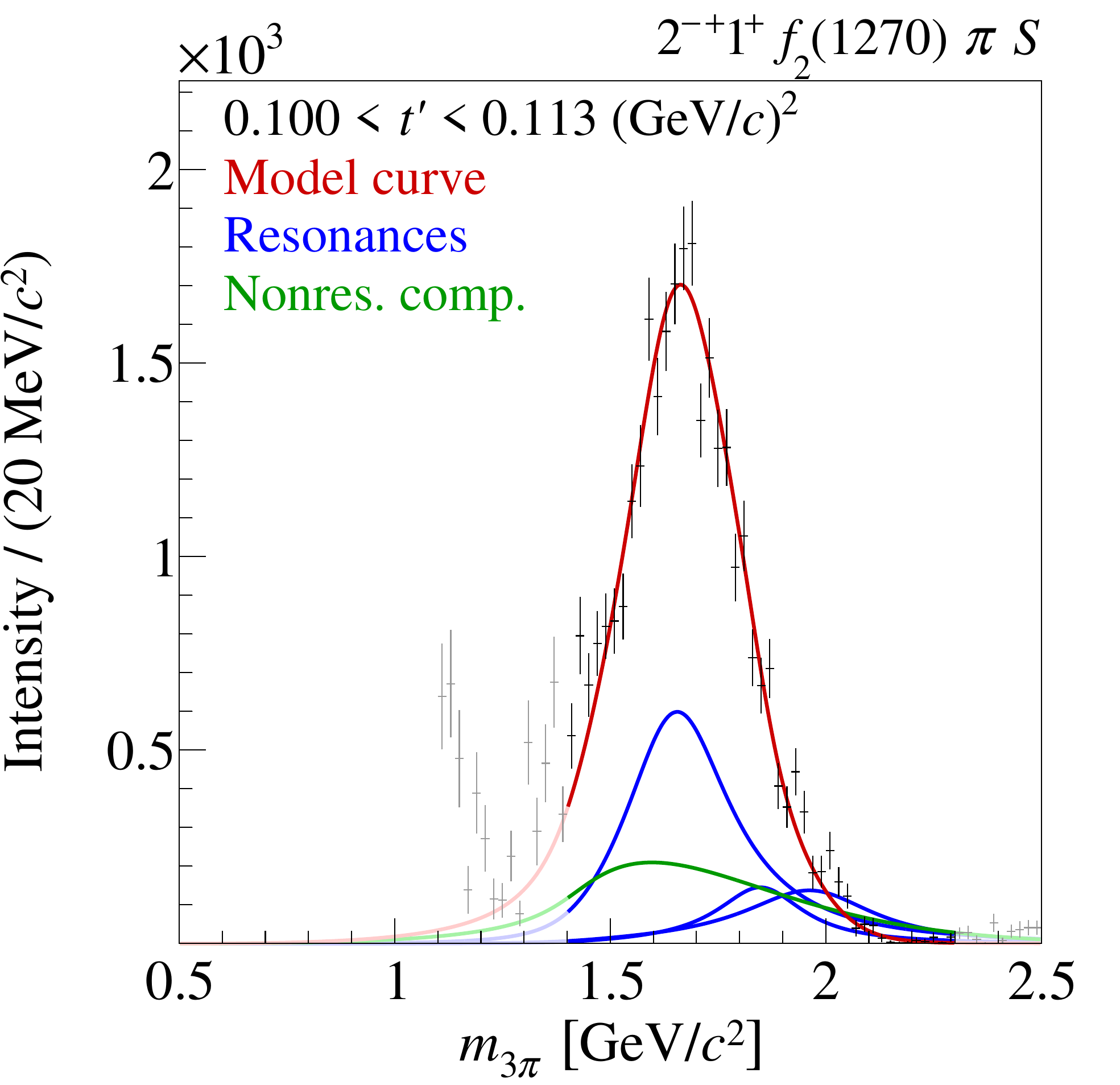}%
    \label{fig:intensity_2mp_m1_f2_S_tbin1}%
  }%
  \hspace*{\fourPlotSpacing}%
  \subfloat[][]{%
    \includegraphics[width=\fourPlotWidth]{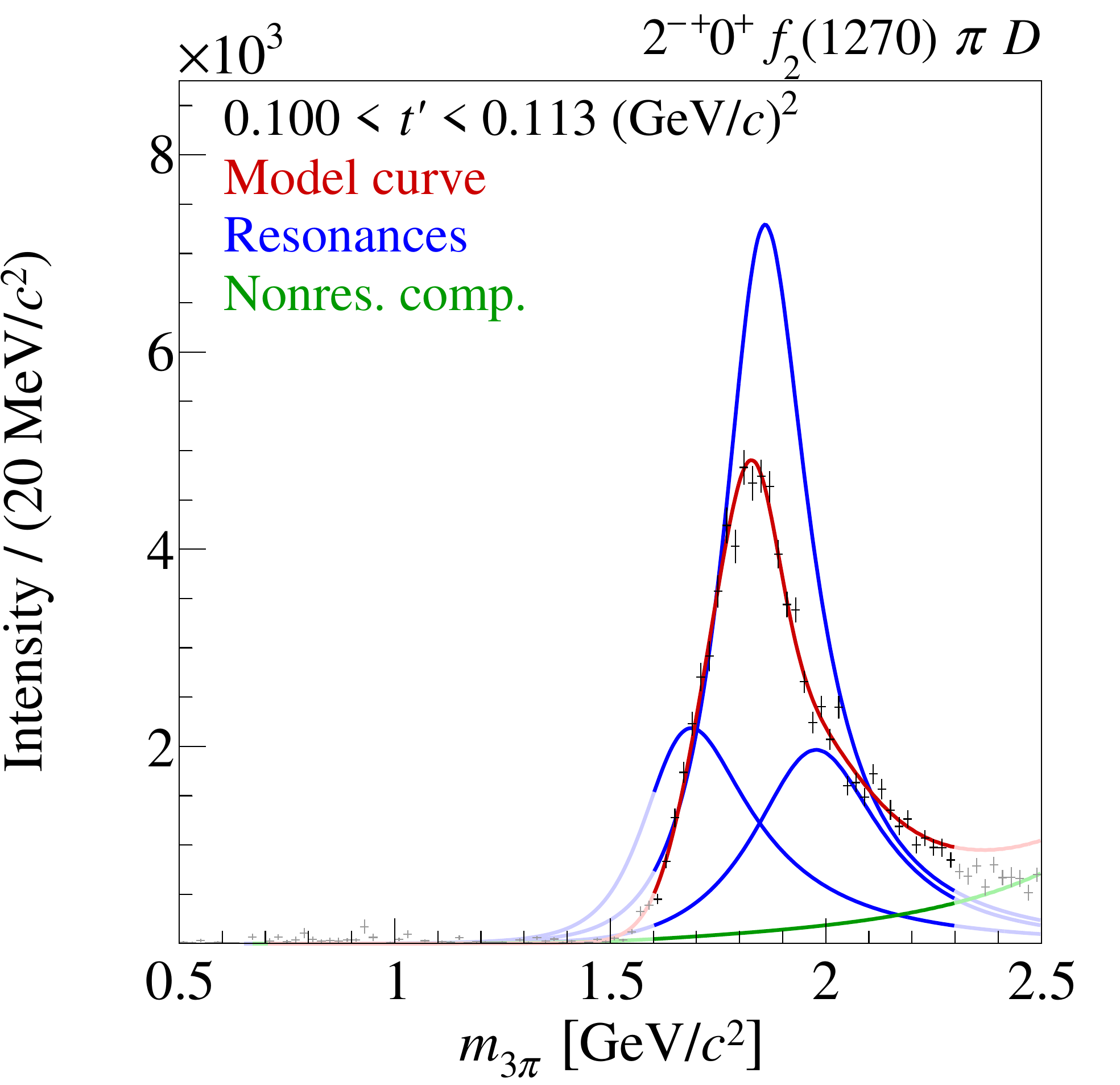}%
    \label{fig:intensity_2mp_f2_D_tbin1}%
  }%
  \\
  \subfloat[][]{%
    \includegraphics[width=\fourPlotWidth]{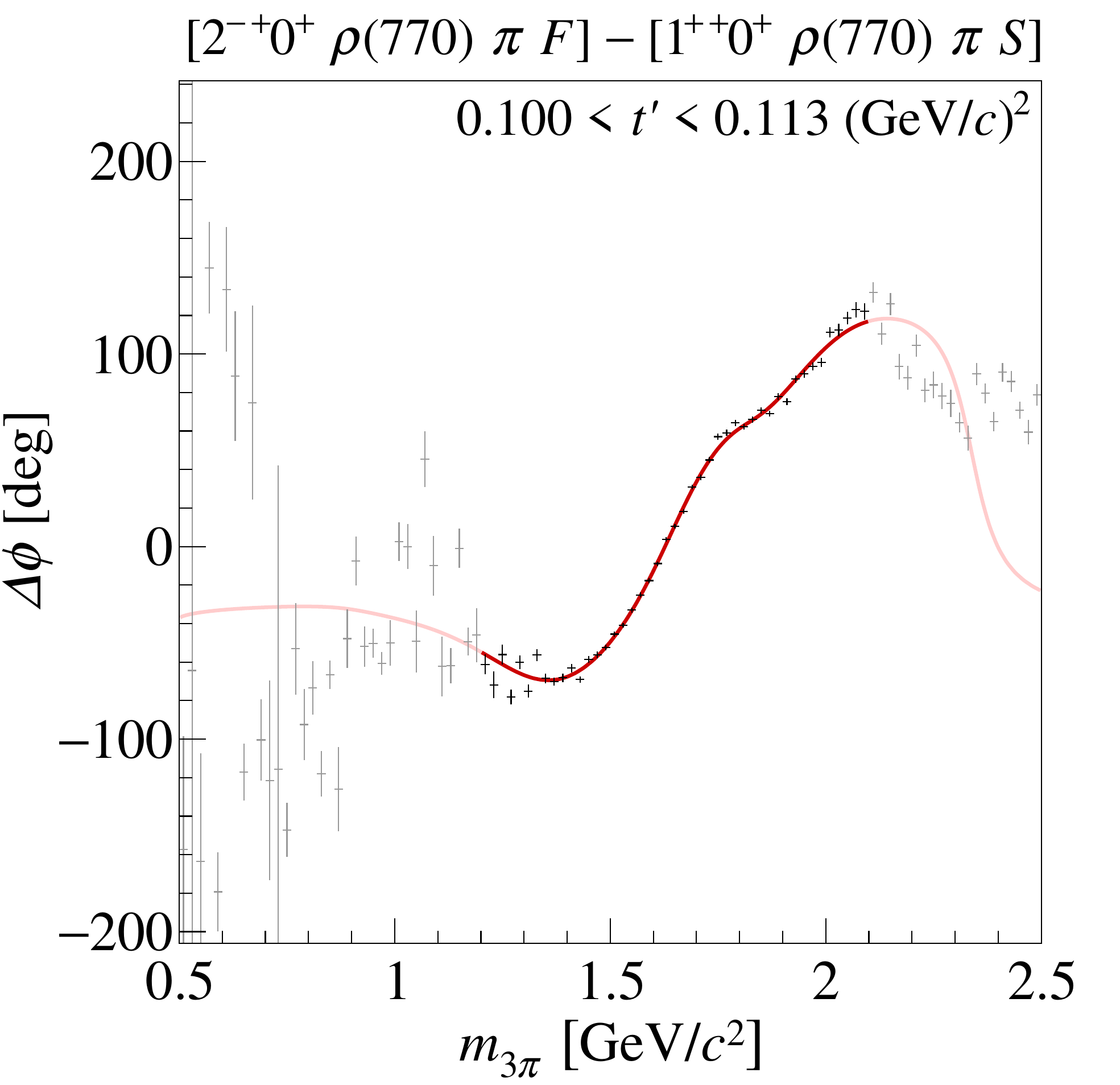}%
    \label{fig:phase_2mp_rho_1pp_rho_tbin1}%
  }%
  \hspace*{\fourPlotSpacing}%
  \subfloat[][]{%
    \includegraphics[width=\fourPlotWidth]{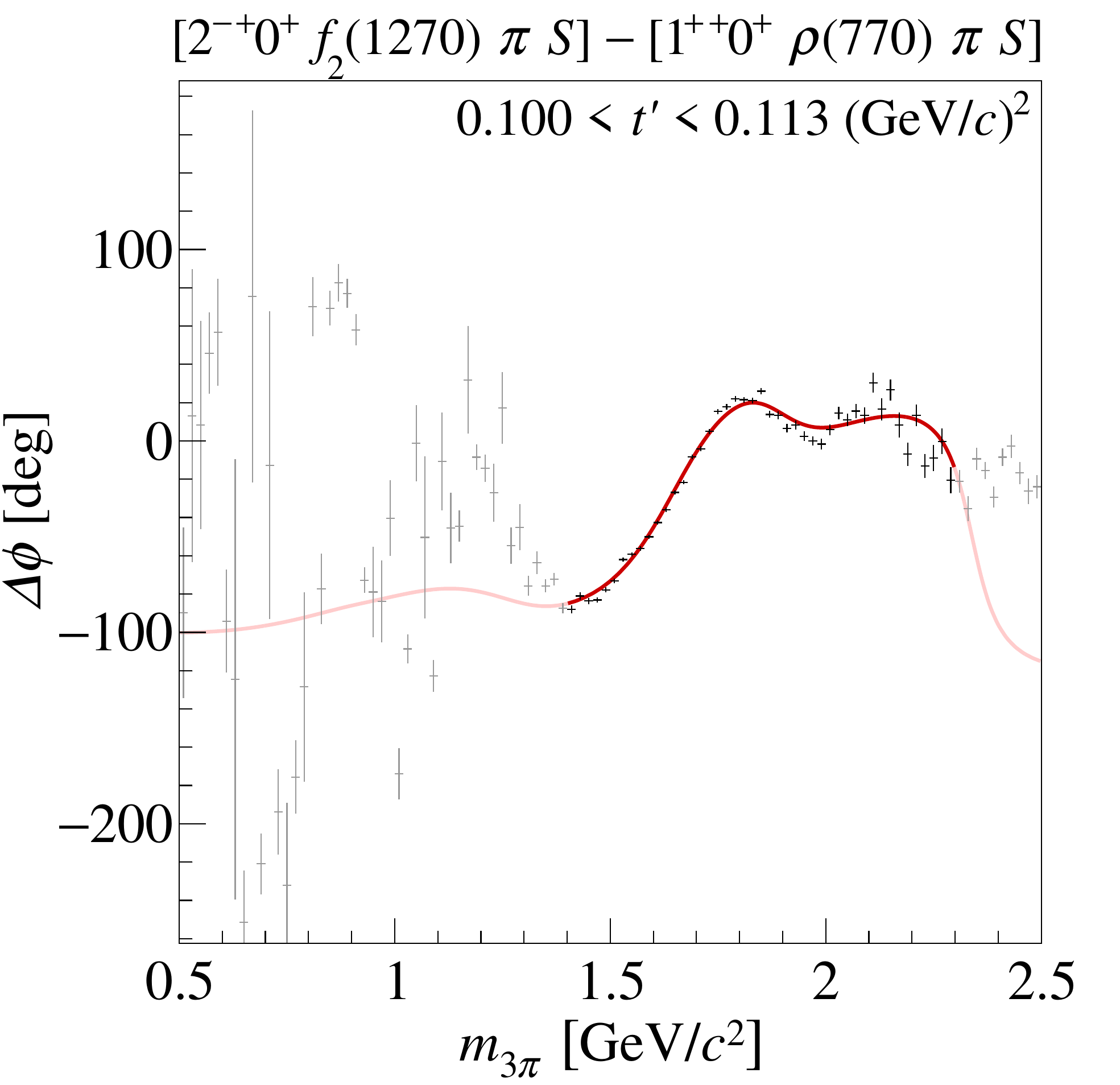}%
    \label{fig:phase_2mp_m0_f2_S_1pp_rho_tbin1}%
  }%
  \hspace*{\fourPlotSpacing}%
  \subfloat[][]{%
    \includegraphics[width=\fourPlotWidth]{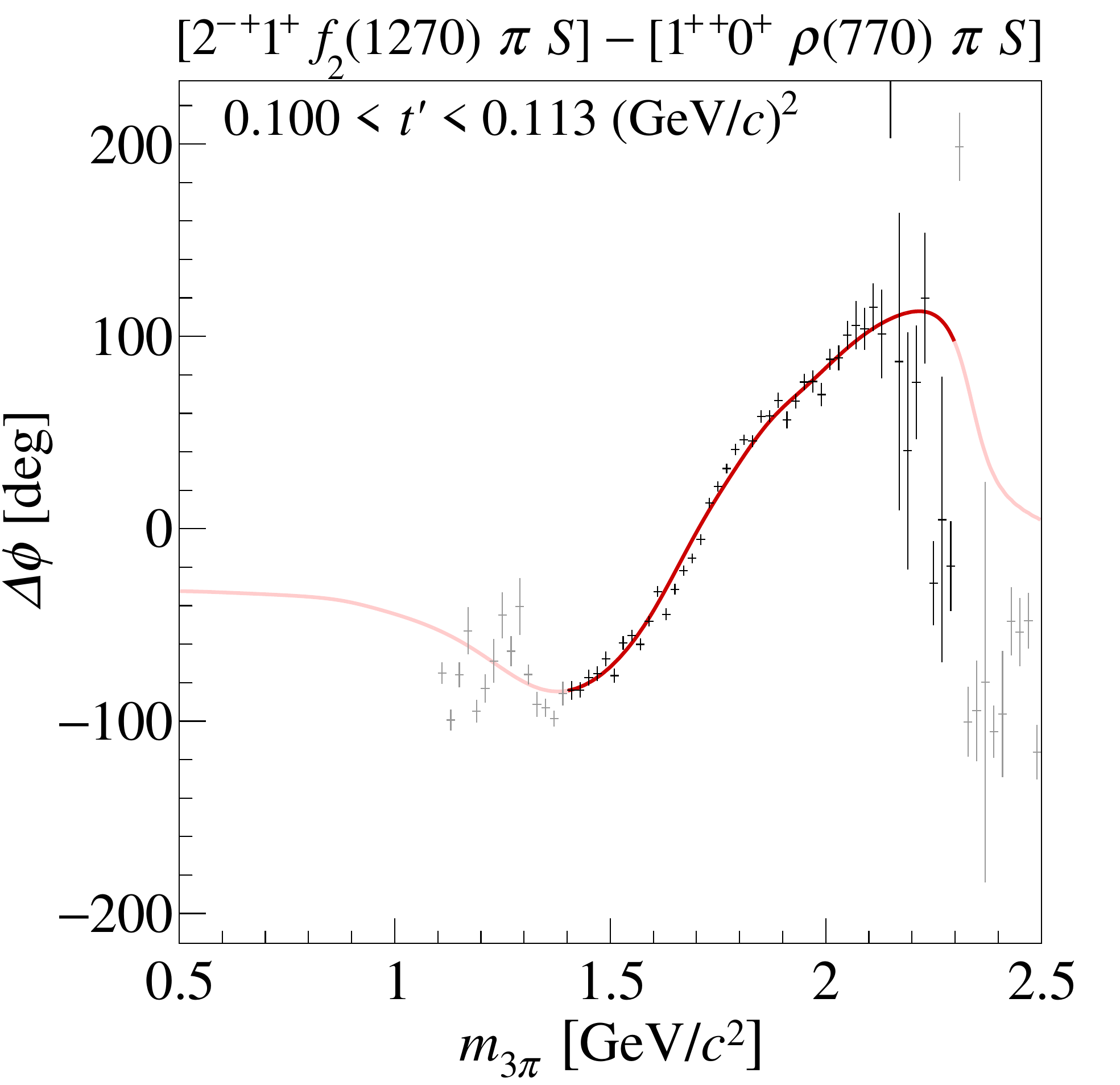}%
    \label{fig:phase_2mp_m1_f2_S_1pp_rho_tbin1}%
  }%
  \hspace*{\fourPlotSpacing}%
  \subfloat[][]{%
    \includegraphics[width=\fourPlotWidth]{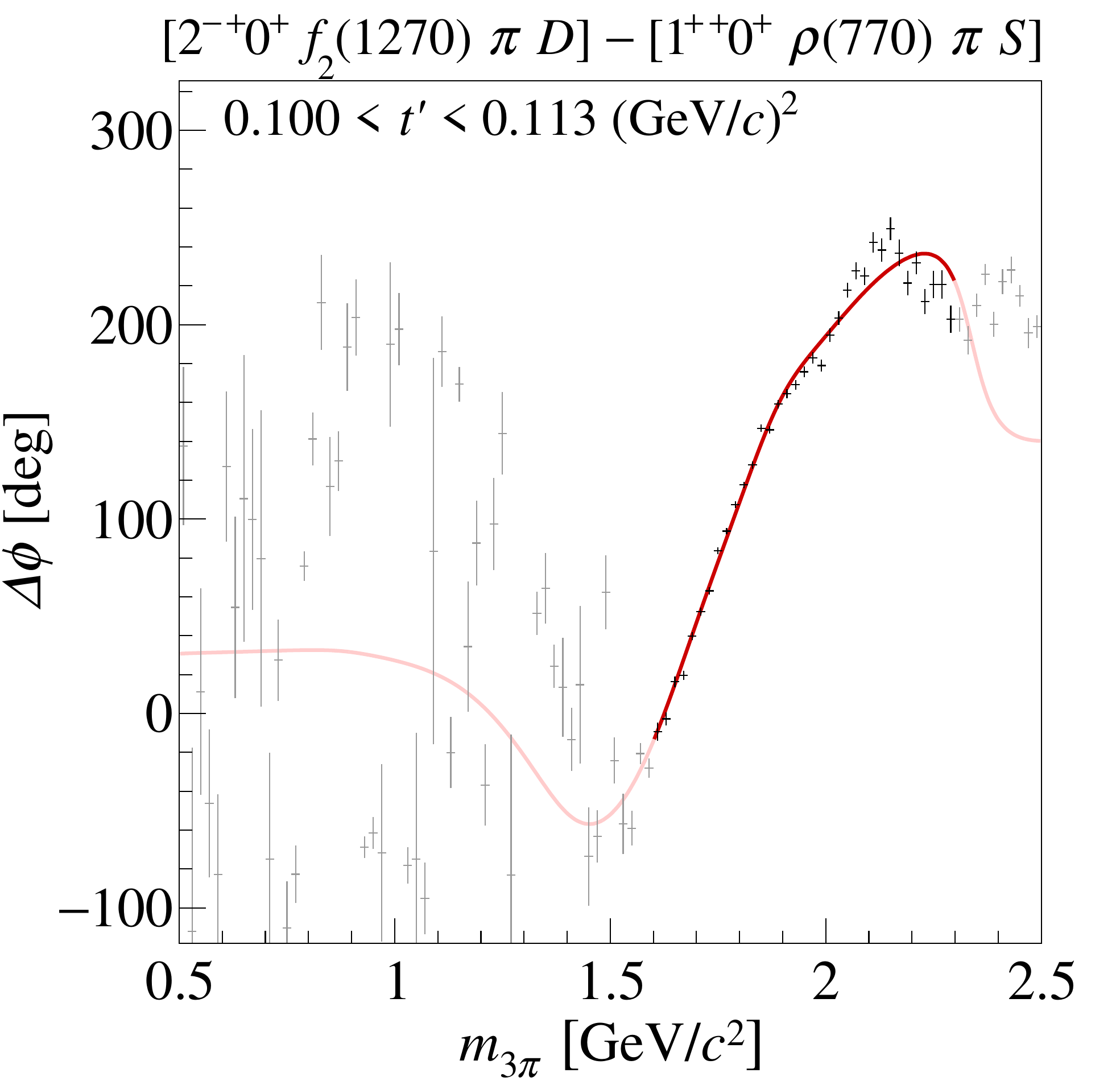}%
    \label{fig:phase_2mp_f2_D_1pp_rho_tbin1}%
  }%
  \\
  \subfloat[][]{%
    \includegraphics[width=\fourPlotWidth]{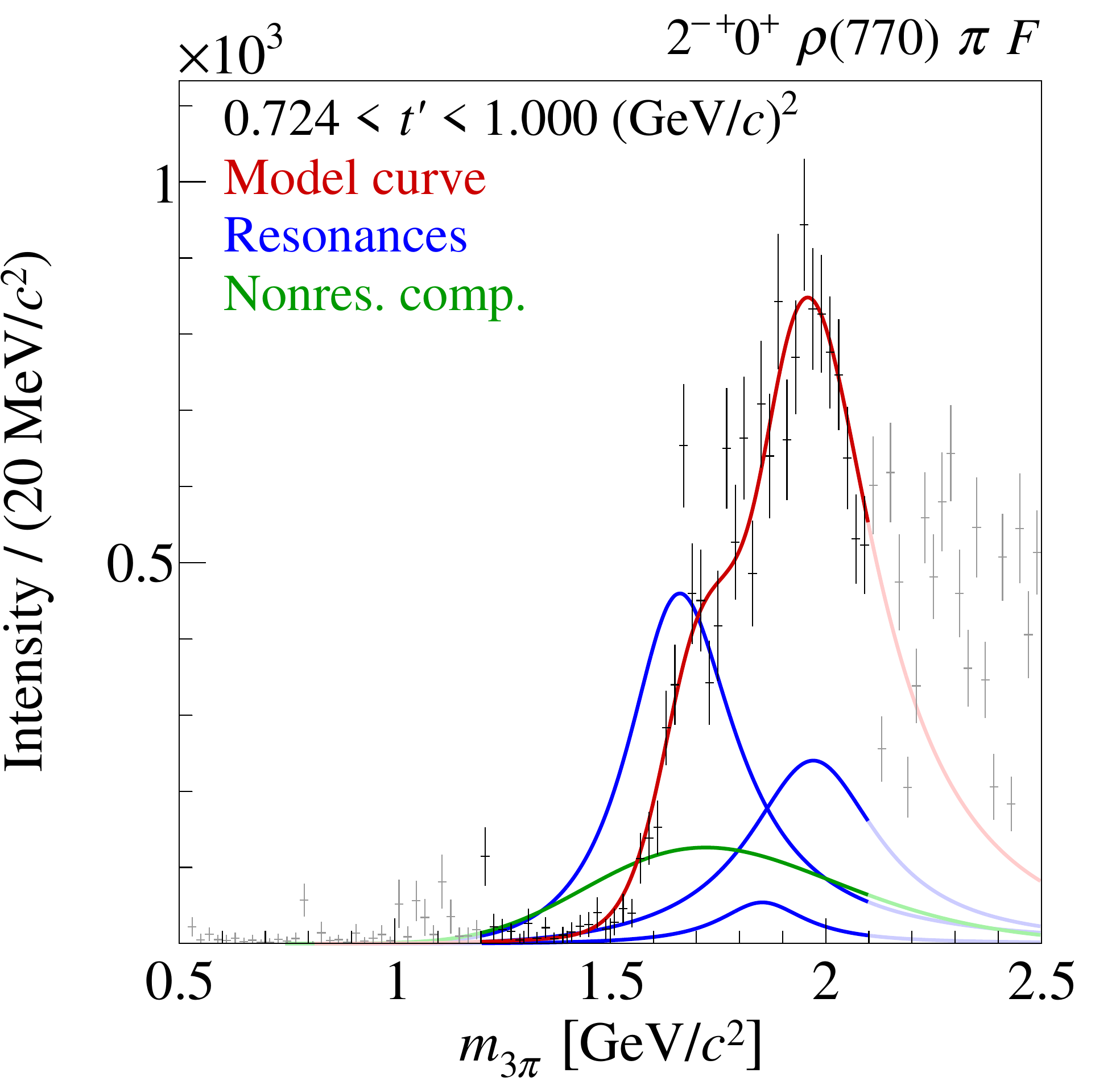}%
    \label{fig:intensity_2mp_rho_tbin11}%
  }%
  \hspace*{\fourPlotSpacing}%
  \subfloat[][]{%
    \includegraphics[width=\fourPlotWidth]{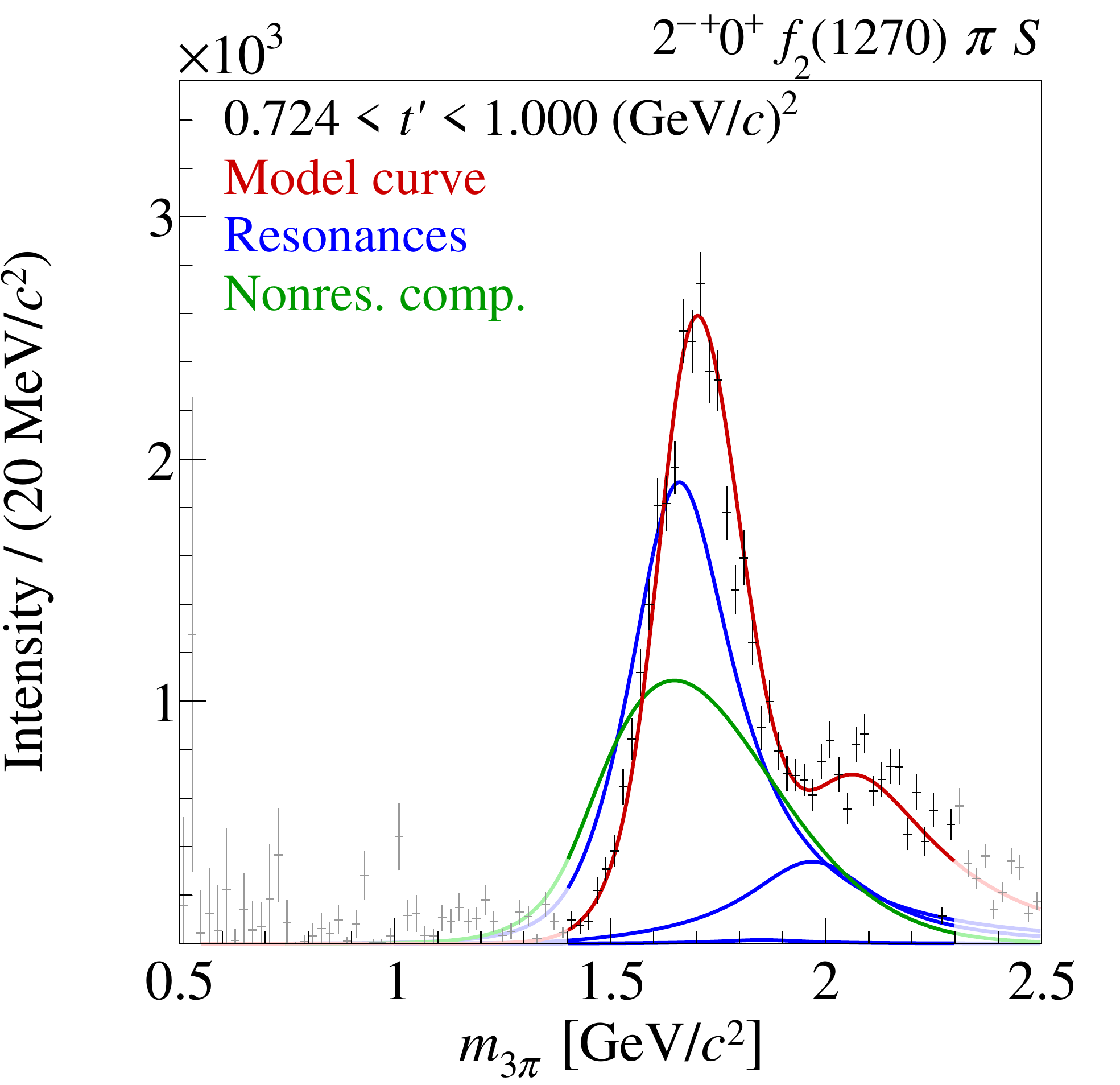}%
    \label{fig:intensity_2mp_m0_f2_S_tbin11}%
  }%
  \hspace*{\fourPlotSpacing}%
  \subfloat[][]{%
    \includegraphics[width=\fourPlotWidth]{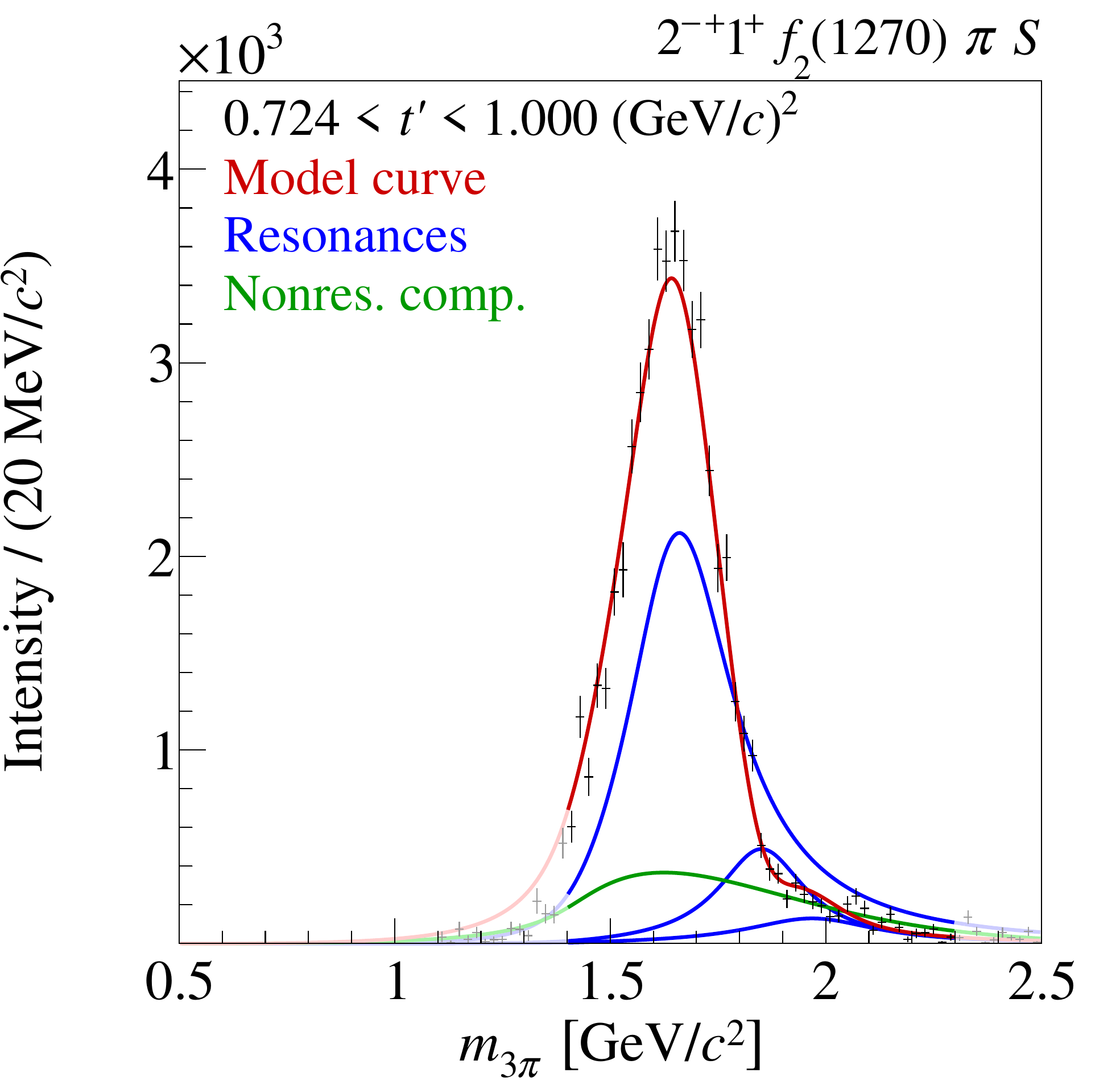}%
    \label{fig:intensity_2mp_m1_f2_S_tbin11}%
  }%
  \hspace*{\fourPlotSpacing}%
  \subfloat[][]{%
    \includegraphics[width=\fourPlotWidth]{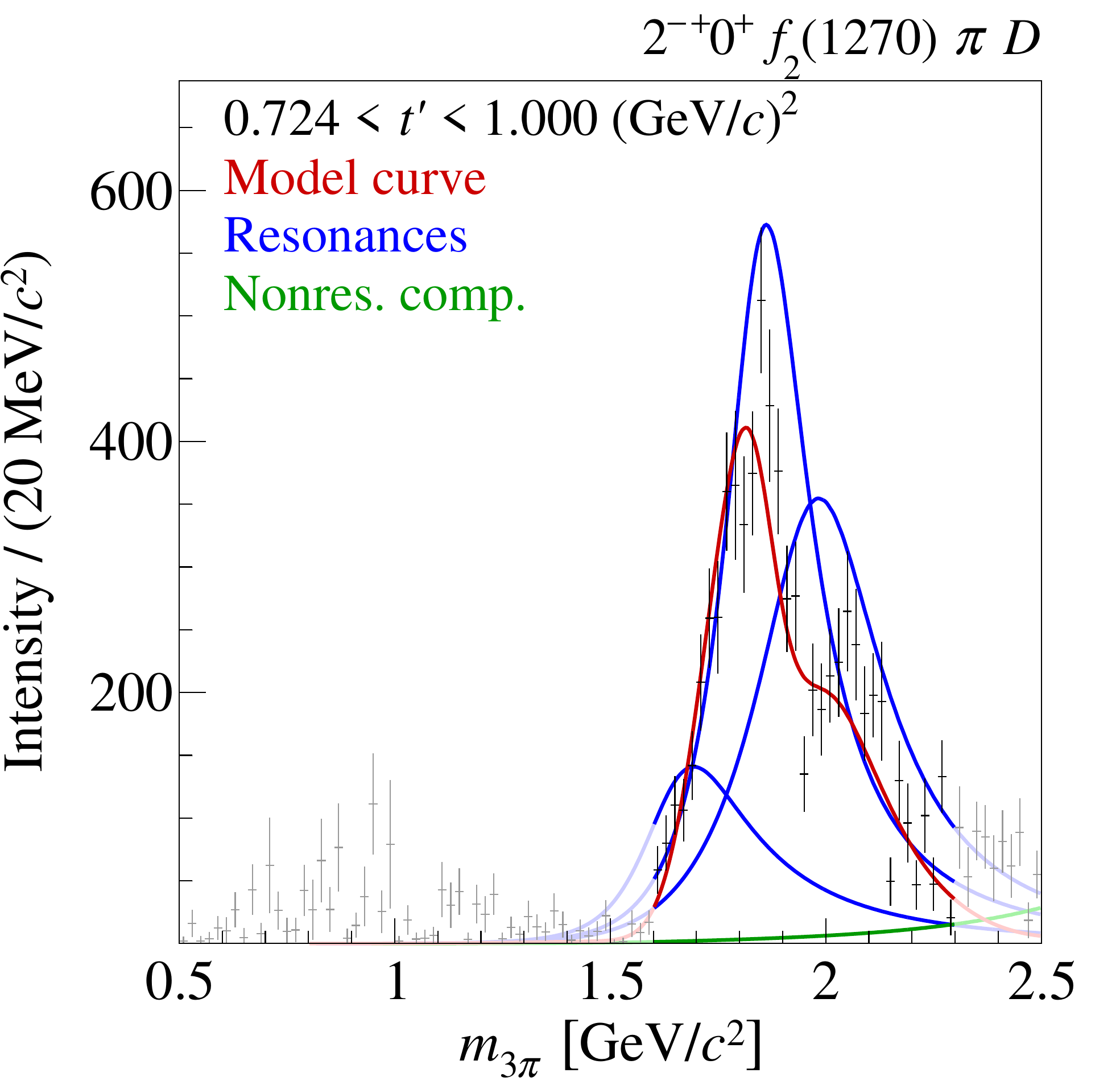}%
    \label{fig:intensity_2mp_f2_D_tbin11}%
  }%
  \\
  \subfloat[][]{%
    \includegraphics[width=\fourPlotWidth]{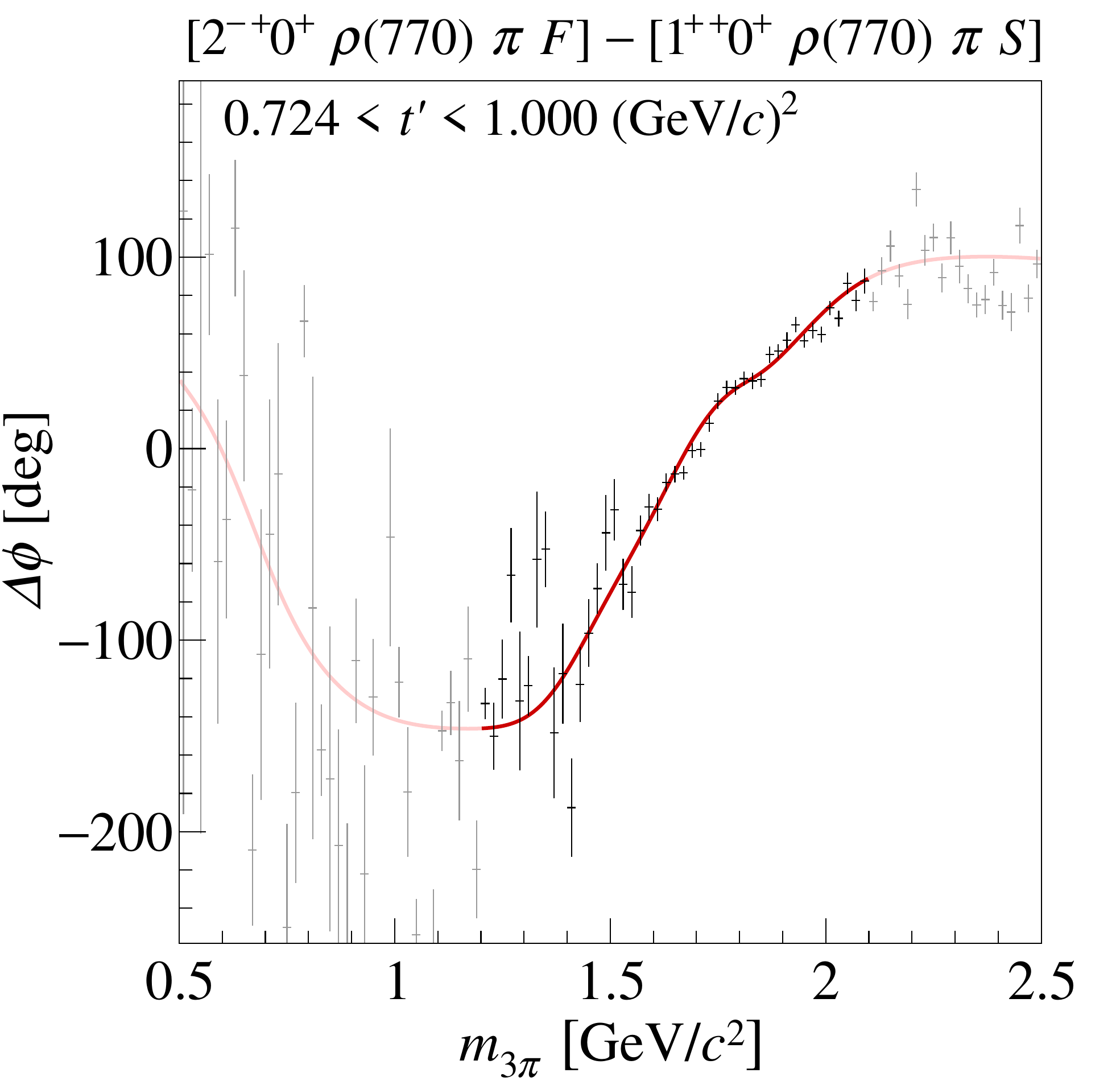}%
    \label{fig:phase_2mp_rho_1pp_rho_tbin11}%
  }%
  \hspace*{\fourPlotSpacing}%
  \subfloat[][]{%
    \includegraphics[width=\fourPlotWidth]{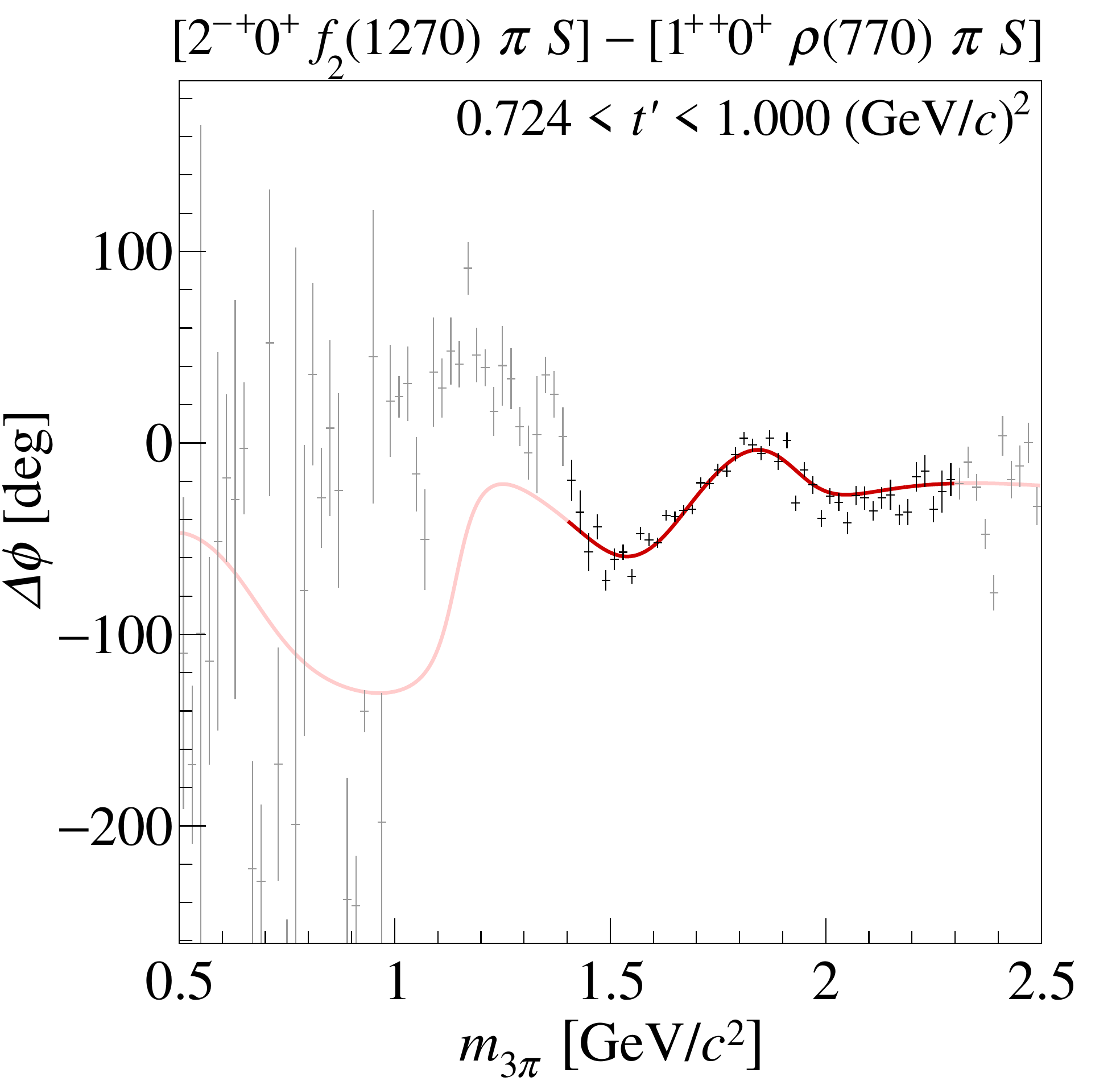}%
    \label{fig:phase_2mp_m0_f2_S_1pp_rho_tbin11}%
  }%
  \hspace*{\fourPlotSpacing}%
  \subfloat[][]{%
    \includegraphics[width=\fourPlotWidth]{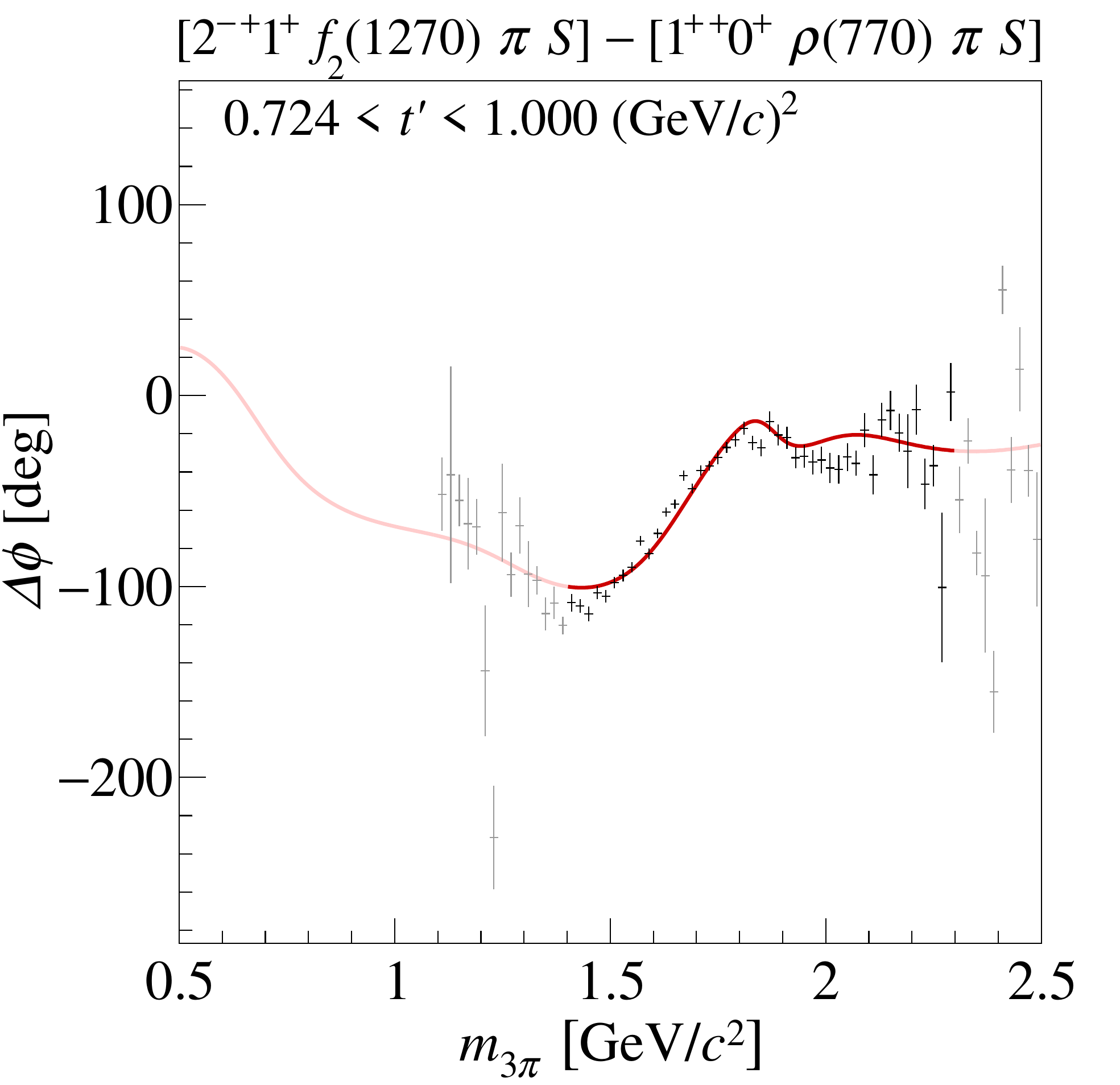}%
    \label{fig:phase_2mp_m1_f2_S_1pp_rho_tbin11}%
  }%
  \hspace*{\fourPlotSpacing}%
  \subfloat[][]{%
    \includegraphics[width=\fourPlotWidth]{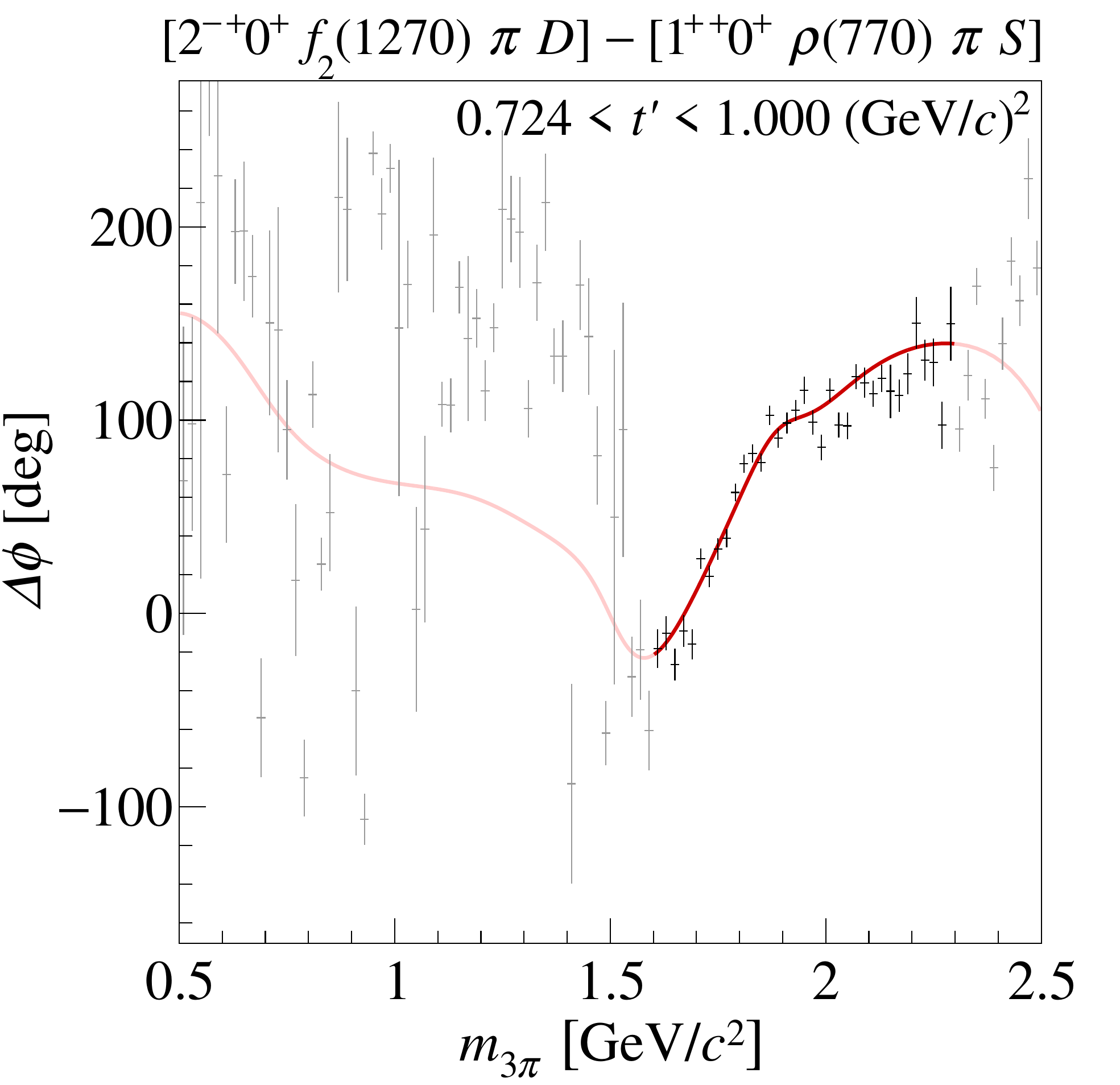}%
    \label{fig:phase_2mp_f2_D_1pp_rho_tbin11}%
  }%
  \caption{Amplitudes of the four $\JPC = 2^{-+}$ waves: (first
    column) \wave{2}{-+}{0}{+}{\Prho}{F} wave, (second column)
    \wave{2}{-+}{0}{+}{\PfTwo}{S} wave, (third column)
    \wave{2}{-+}{1}{+}{\PfTwo}{S} wave, and (fourth column)
    \wave{2}{-+}{0}{+}{\PfTwo}{D} wave.  (first and third rows)
    Intensity distributions in the lowest and highest \tpr bins,
    respectively.  (second and fourth rows) Phases relative to the
    \wave{1}{++}{0}{+}{\Prho}{S} wave in the lowest and highest \tpr
    bins, respectively.  The model and the wave components are
    represented as in \cref{fig:intensity_phases_0mp}, except that
    here the blue curves represent the \PpiTwo, the \PpiTwo[1880], and
    the \PpiTwo[2005].}
  \label{fig:intensity_phases_2mp}
\ifMultiColumnLayout{\end{figure*}}{\end{figure}}

The intensities of the $\Prho \pi F$ wave and of the two
$\PfTwo \pi S$ waves exhibit a clear peak at \SI{1.65}{\GeVcc}, which
dominates in particular the $\PfTwo \pi S$ waves.  The position of
this peak does not depend strongly on \tpr.  The $\Prho \pi F$ wave
has an additional high-mass shoulder at \SI{1.9}{\GeVcc}, which
becomes a dominant peak in the highest \tpr bin.  The $\PfTwo \pi S$
wave with $M = 0$ has a smaller high-mass shoulder at about
\SI{2.05}{\GeVcc}, which also grows relative to the \SI{1.65}{\GeVcc}
peak with increasing \tpr.  This shoulder is absent in the
$\PfTwo \pi S$ wave with $M = 1$.  The $\PfTwo \pi D$ wave has no
structure at \SI{1.65}{\GeVcc}.  Instead, it exhibits a dominant peak
at \SI{1.8}{\GeVcc} and a slight high-mass shoulder at
\SI{2.05}{\GeVcc}, which becomes more pronounced toward higher \tpr.
The position of the peak is independent of \tpr.

The \wave{2}{-+}{0}{+}{\Prho}{F} wave and the two
$2^{-+} \PfTwo \pi S$ waves exhibit clearly rising phases \wrt the
\wave{1}{++}{0}{+}{\Prho}{S} wave in the region of the
\SI{1.65}{\GeVcc} peak (see second and fourth rows in
\cref{fig:intensity_phases_2mp}).  At low \tpr, the phases of the
\wave{2}{-+}{0}{+}{\Prho}{F} and \wave{2}{-+}{1}{+}{\PfTwo}{S} waves
continue to rise in the \SI{1.9}{\GeVcc} region [see
\cref{fig:phase_2mp_rho_1pp_rho_tbin1,fig:phase_2mp_m1_f2_S_1pp_rho_tbin1}].
The phase motion of the \wave{2}{-+}{0}{+}{\Prho}{F} wave is
approximately independent of \tpr, whereas the phase of the
\wave{2}{-+}{1}{+}{\PfTwo}{S} wave flattens out at about
\SI{1.9}{\GeVcc} at higher \tpr, making the phase motion of this wave
similar to that of the corresponding $M = 0$ wave.  The phase motion
of the \wave{2}{-+}{0}{+}{\PfTwo}{D} wave \wrt the
\wave{1}{++}{0}{+}{\Prho}{S} wave exhibits a rapid rise in the region
of the \SI{1.8}{\GeVcc} peak and a slower rise in the region of the
\SI{2.05}{\GeVcc} shoulder.  The amplitude of the phase motion
decreases with increasing \tpr.

The fit model contains three resonances, \PpiTwo, \PpiTwo[1880], and
\PpiTwo[2005], to describe the four $\JPC = 2^{-+}$ waves.  The
resonances are parametrized using
\cref{eq:BreitWigner,eq:method:fixedwidth}, the nonresonant components
using \cref{eq:method:nonresterm} for the
\wave{2}{-+}{0}{+}{\PfTwo}{S} and \wave{2}{-+}{0}{+}{\Prho}{F} waves
and \cref{eq:method:nonrestermsmall} for the other two $2^{-+}$ waves
(see \cref{tab:method:fitmodel:waveset}).  The $\Prho \pi F$ wave is
fit in the range from \SIrange{1.2}{2.1}{\GeVcc}, the two
$\PfTwo \pi S$ waves from \SIrange{1.4}{2.3}{\GeVcc}, and the
$\PfTwo \pi D$ wave from \SIrange{1.6}{2.3}{\GeVcc}.

The $\Prho \pi F$ wave and the two $\PfTwo \pi S$ waves are dominated
by the \PpiTwo.  In the $\Prho \pi F$ wave, the nonresonant component
is small compared to the \PpiTwo component.  Only in the two highest
\tpr bins does it have a larger intensity.  The contributions from the
nonresonant components are larger in the two $\PfTwo \pi S$ waves.
These waves also show a stronger interference of the wave components
in the \PpiTwo region, in particular at lower \tpr.  In the
$\Prho \pi F$ wave and the two $\PfTwo \pi S$ waves, the intensities
of the two excited \PpiTwo* components are comparable to those of the
nonresonant components or even smaller.  In the $\PfTwo \pi S$ wave
with $M = 0$, the \PpiTwo[1880] component is practically vanishing.
The excited \PpiTwo* components show different interference patterns.
In the $\Prho \pi F$ wave, significant constructive interference of
the wave components describes the high-mass shoulder at
\SI{1.9}{\GeVcc}.  In the $\PfTwo \pi S$ wave with $M = 0$ these
interference effects are much smaller, whereas in the $\PfTwo \pi S$
wave with $M = 1$ the components interfere destructively leading to a
steeper drop of the intensity in the \SI{1.8}{\GeVcc} region at larger
\tpr.

The composition of the $\PfTwo \pi D$ wave is strikingly different.
In this wave, all three resonance components play a significant role,
with the \PpiTwo[1880] being the dominant one that destructively
interferes with the other components.  At lower values of \tpr, the
\PpiTwo and the \PpiTwo[2005] appear with similar intensities.  In the
two highest \tpr bins, the \PpiTwo[2005] component becomes larger.
The contribution from the nonresonant component is small.

Within the fit ranges, the fit model describes the intensity
distributions in general well.  This is in particular true for the two
$\PfTwo \pi S$ waves.  The fit model does not reproduce the details of
the high-mass shoulder at \SI{2.05}{\GeVcc} in the $\PfTwo \pi D$
wave.  In this wave, also the extrapolation of the fit model above the
fit range of \SI{2.3}{\GeVcc} deviates from the data, in particular at
lower \tpr.  In the $\Prho \pi F$ wave, the fit model does not
reproduce details of the peak at \SI{1.65}{\GeVcc} and of the shoulder
at \SI{1.9}{\GeVcc}.  The extrapolation of the fit model above the fit
range of \SI{2.1}{\GeVcc} deviates from the data.

The dominance of the \PpiTwo in the $\Prho \pi F$ wave and in the two
$\PfTwo \pi S$ waves is supported by the clearly rising phases of
these waves \wrt the \wave{1}{++}{0}{+}{\Prho}{S} wave (see
\cref{fig:intensity_phases_2mp}).  It is also consistent with the
approximately constant relative phases among these three $2^{-+}$
waves in the \SI{1.6}{\GeVcc} region at low \tpr (see
\cref{fig:intensity_phases_2mp_tbin1}).  Above the \PpiTwo region, the
similar relative strengths of \PpiTwo[1880] and \PpiTwo[2005] in the
\wave{2}{-+}{0}{+}{\Prho}{F} and \wave{2}{-+}{1}{+}{\PfTwo}{S} waves
lead to only small variations of their relative phase.  The
\wave{2}{-+}{0}{+}{\Prho}{F} and \wave{2}{-+}{1}{+}{\PfTwo}{S} waves
exhibit more pronounced phase motions in the \SI{1.9}{\GeVcc} region
\wrt the \wave{2}{-+}{0}{+}{\PfTwo}{S} wave because of the vanishing
\PpiTwo[1880] component in the latter wave.  The interference pattern
of the three $2^{-+}$ waves changes toward higher \tpr mainly because
of the changing composition of the \wave{2}{-+}{0}{+}{\Prho}{F} wave
(see \cref{fig:intensity_phases_2mp_tbin11}).  The phase of the
\wave{2}{-+}{0}{+}{\PfTwo}{D} wave \wrt the
\wave{1}{++}{0}{+}{\Prho}{S} wave rises in the \SI{1.8}{\GeVcc} region
and, less rapidly, in the \SI{2.0}{\GeVcc} region.  This phase motion
is caused by \PpiTwo[1880] and \PpiTwo[2005] and is connected to the
phases of the coupling amplitudes of the two heavier \PpiTwo*, which
are close to \SI{180}{\degree} relative to the \PpiTwo in this wave
(see \cref{sec:production_phases}).  \Wrt the other three $2^{-+}$
waves, the $\PfTwo \pi D$ wave shows similar phase motions.  This is
consistent with the large contributions from \PpiTwo[1880] and
\PpiTwo[2005] in this wave compared to the \PpiTwo component.

\ifMultiColumnLayout{\begin{figure*}[t]}{\begin{figure}[p]}
  \centering
  \subfloat[][]{%
    \includegraphics[width=\fourPlotWidth]{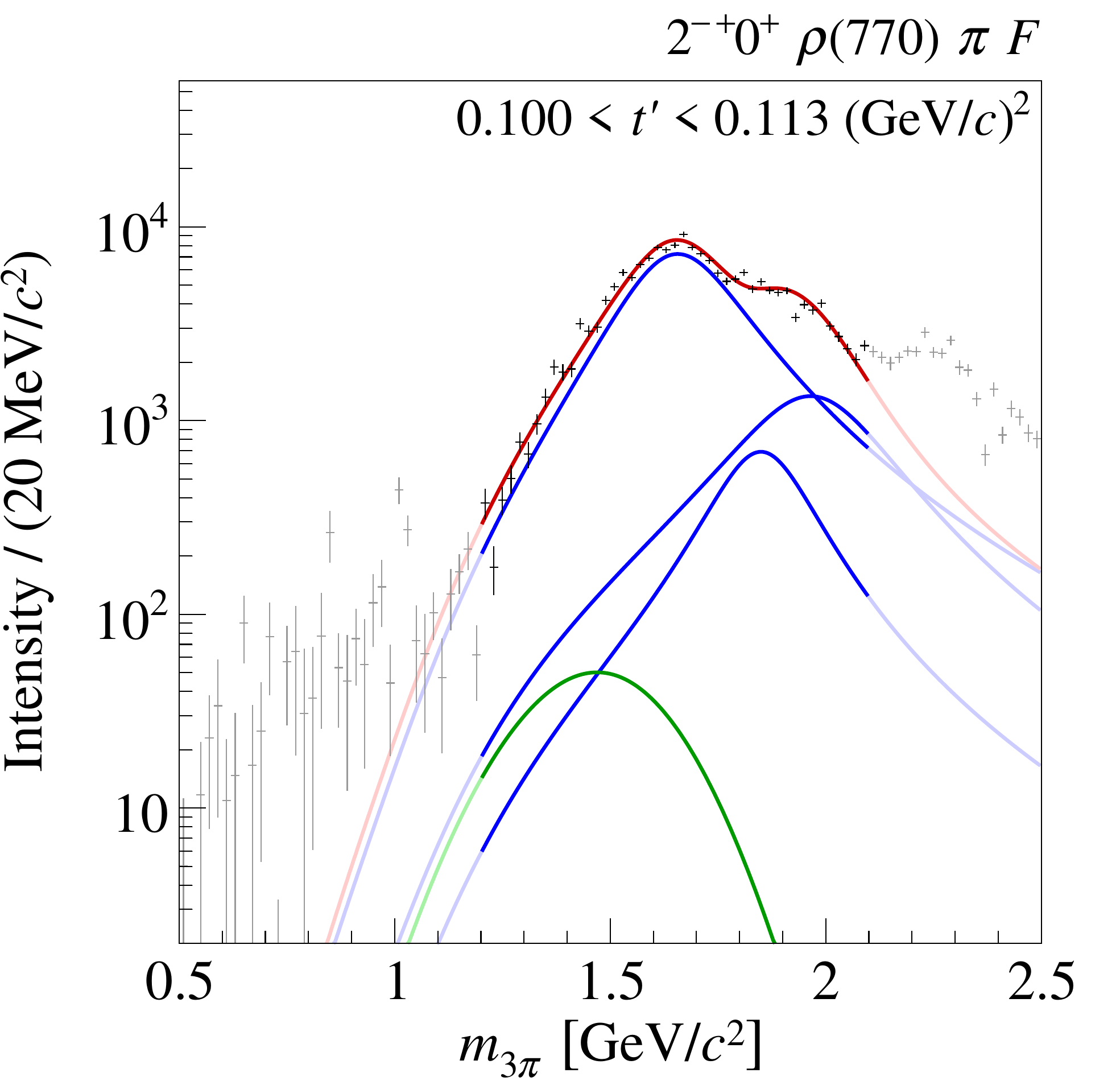}%
    \label{fig:intensity_2mp_rho_tbin1_log}%
  }%
  \hspace*{\fourPlotSpacing}%
  \subfloat[][]{%
    \includegraphics[width=\fourPlotWidth]{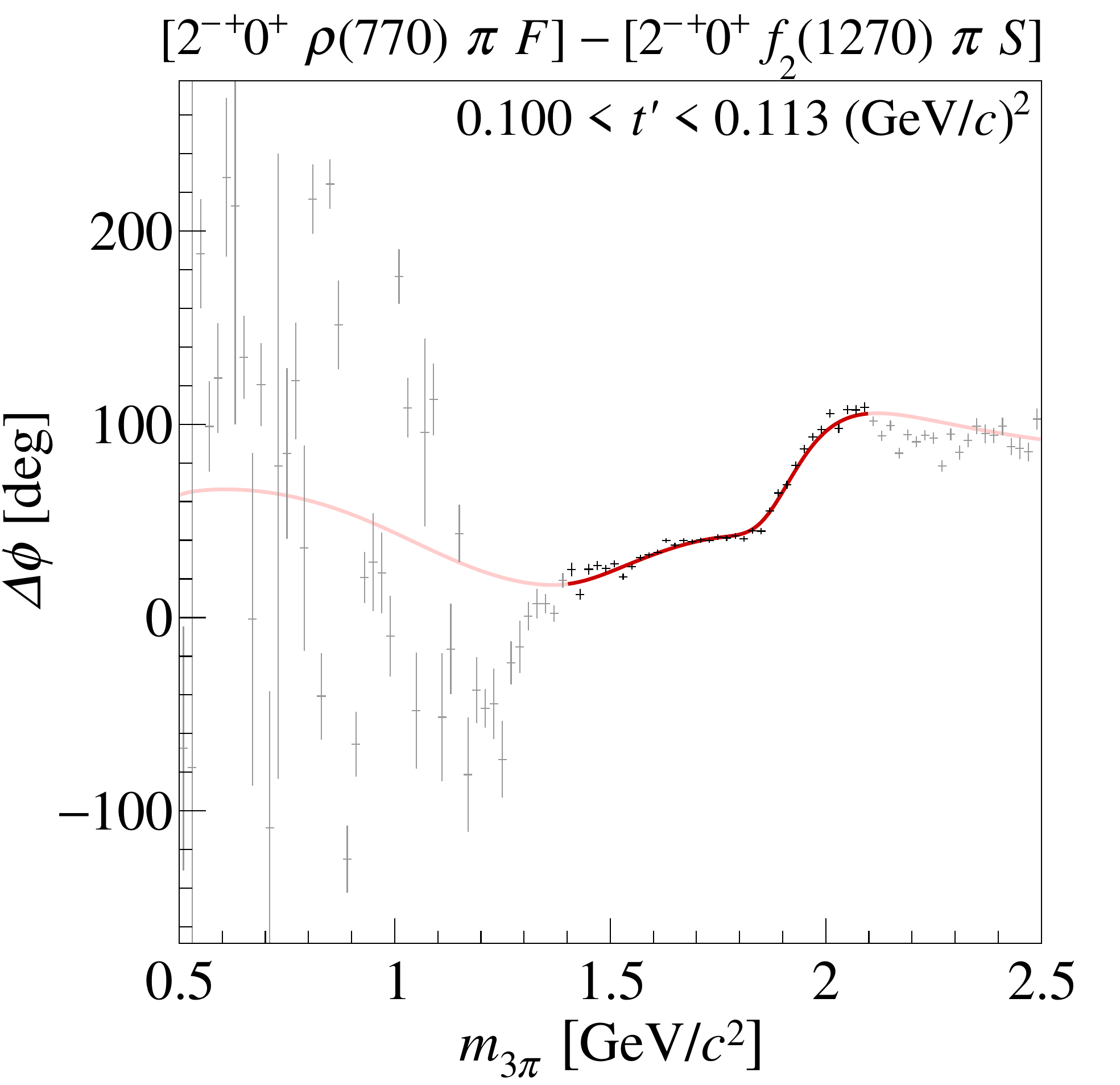}%
    \label{fig:phase_2mp_rho_2mp_m0_f2_S_tbin1}%
  }%
  \hspace*{\fourPlotSpacing}%
  \subfloat[][]{%
    \includegraphics[width=\fourPlotWidth]{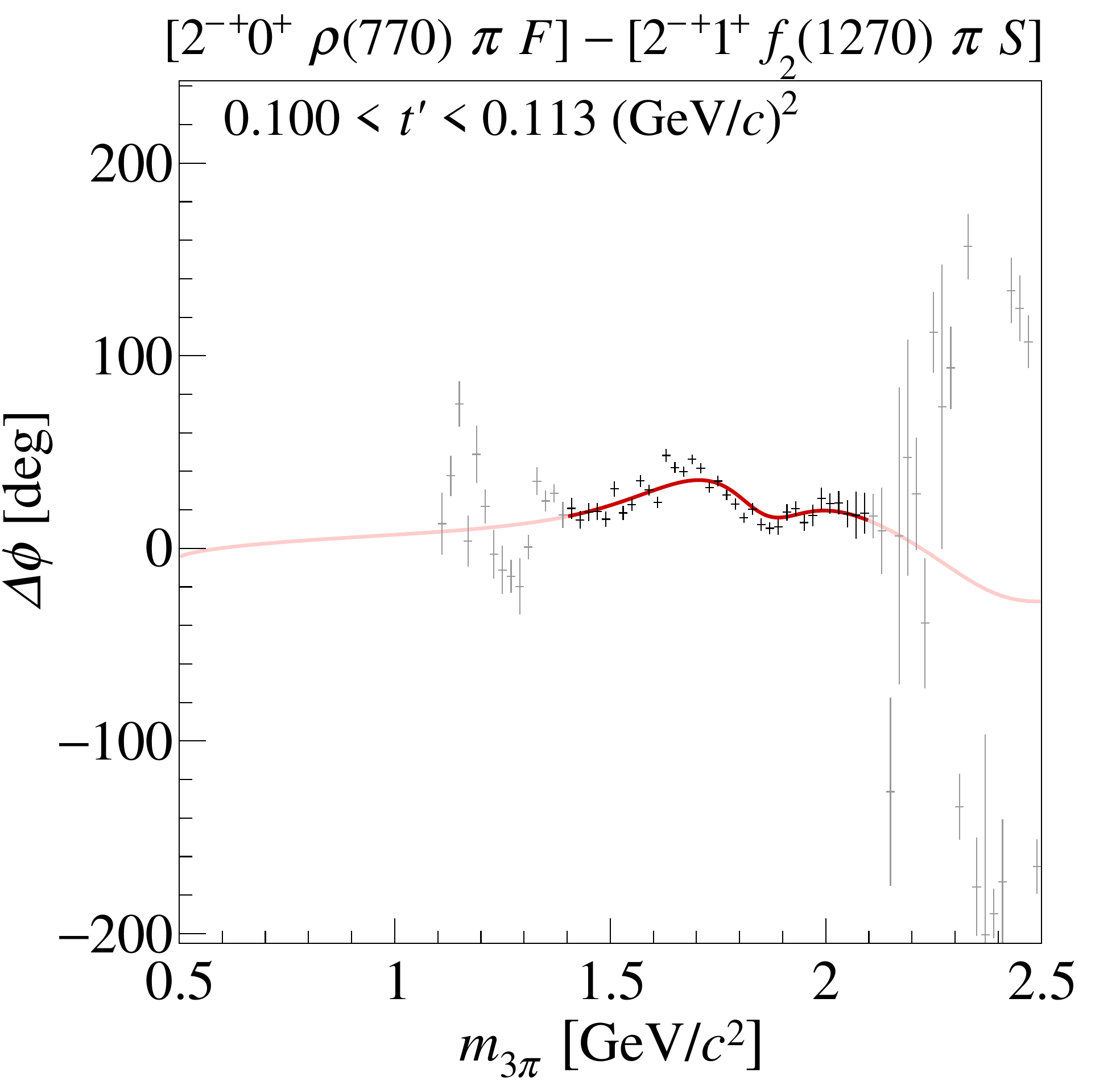}%
    \label{fig:phase_2mp_rho_2mp_m1_f2_S_tbin1}%
  }%
  \hspace*{\fourPlotSpacing}%
  \subfloat[][]{%
    \includegraphics[width=\fourPlotWidth]{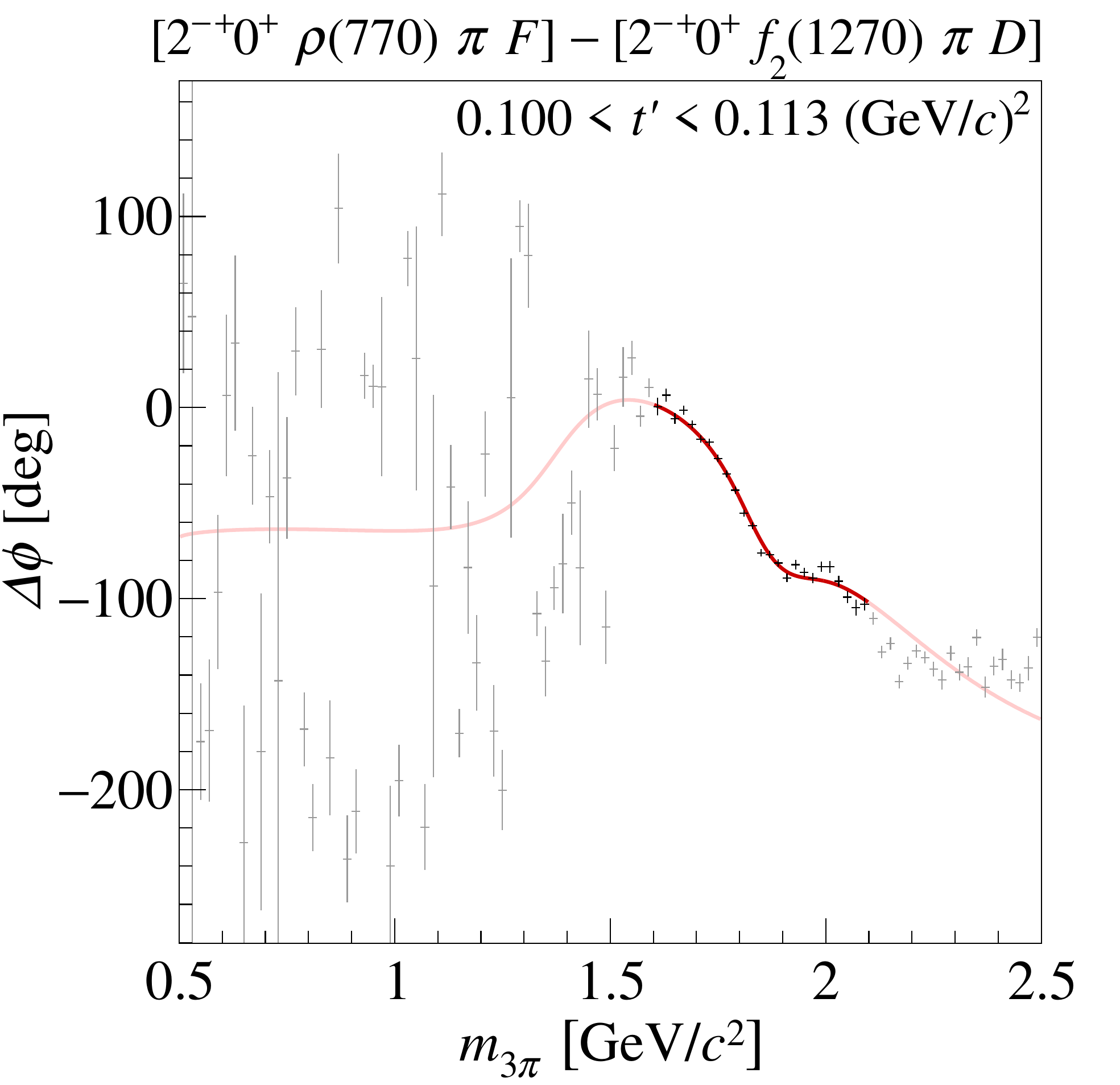}%
    \label{fig:phase_2mp_rho_2mp_f2_D_tbin1}%
  }%
  \\
  \hspace*{\fourPlotWidth}%
  \hspace*{\fourPlotSpacing}%
  \subfloat[][]{%
    \includegraphics[width=\fourPlotWidth]{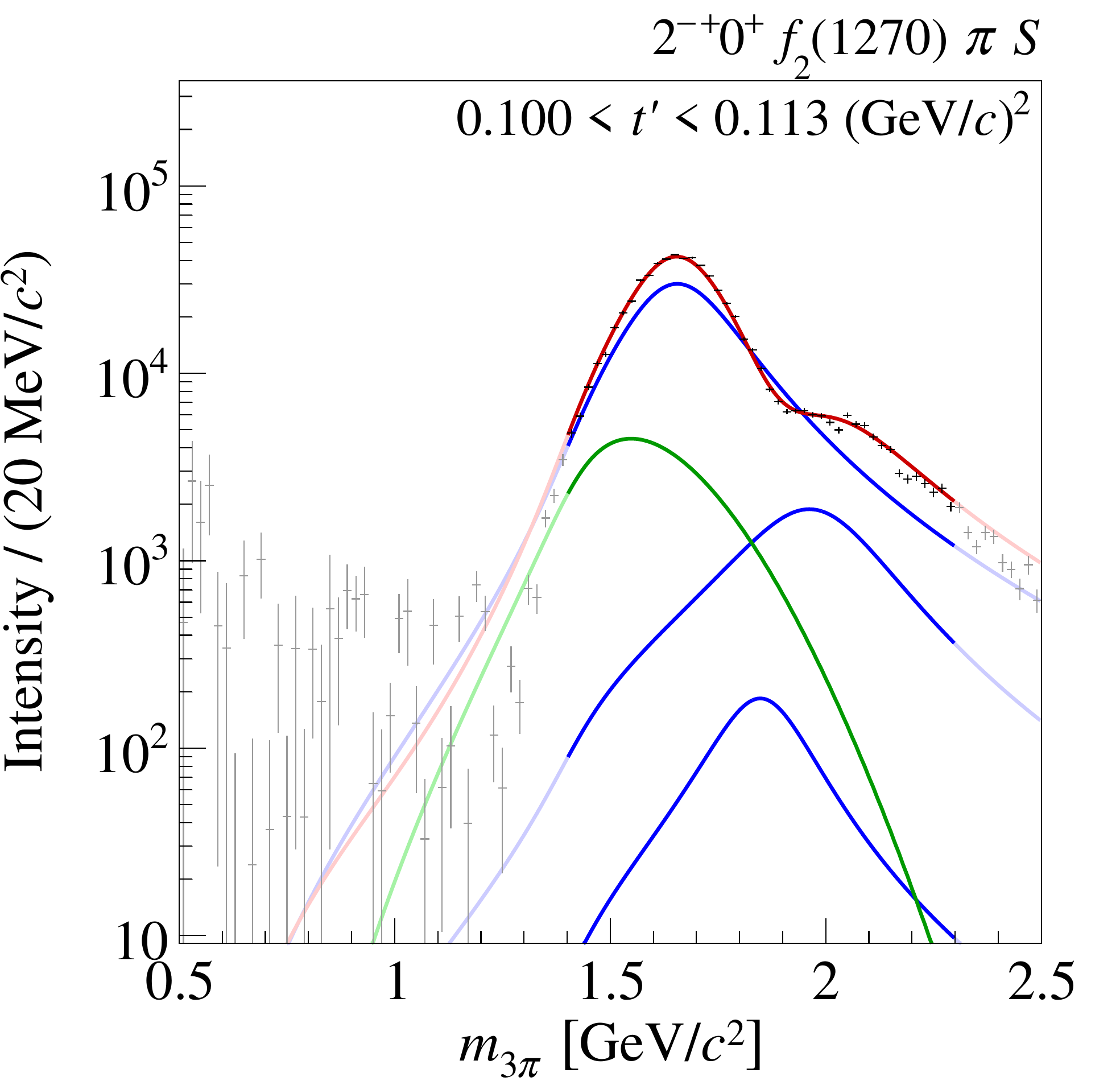}%
    \label{fig:intensity_2mp_m0_f2_S_tbin1_log}%
  }%
  \hspace*{\fourPlotSpacing}%
  \subfloat[][]{%
    \includegraphics[width=\fourPlotWidth]{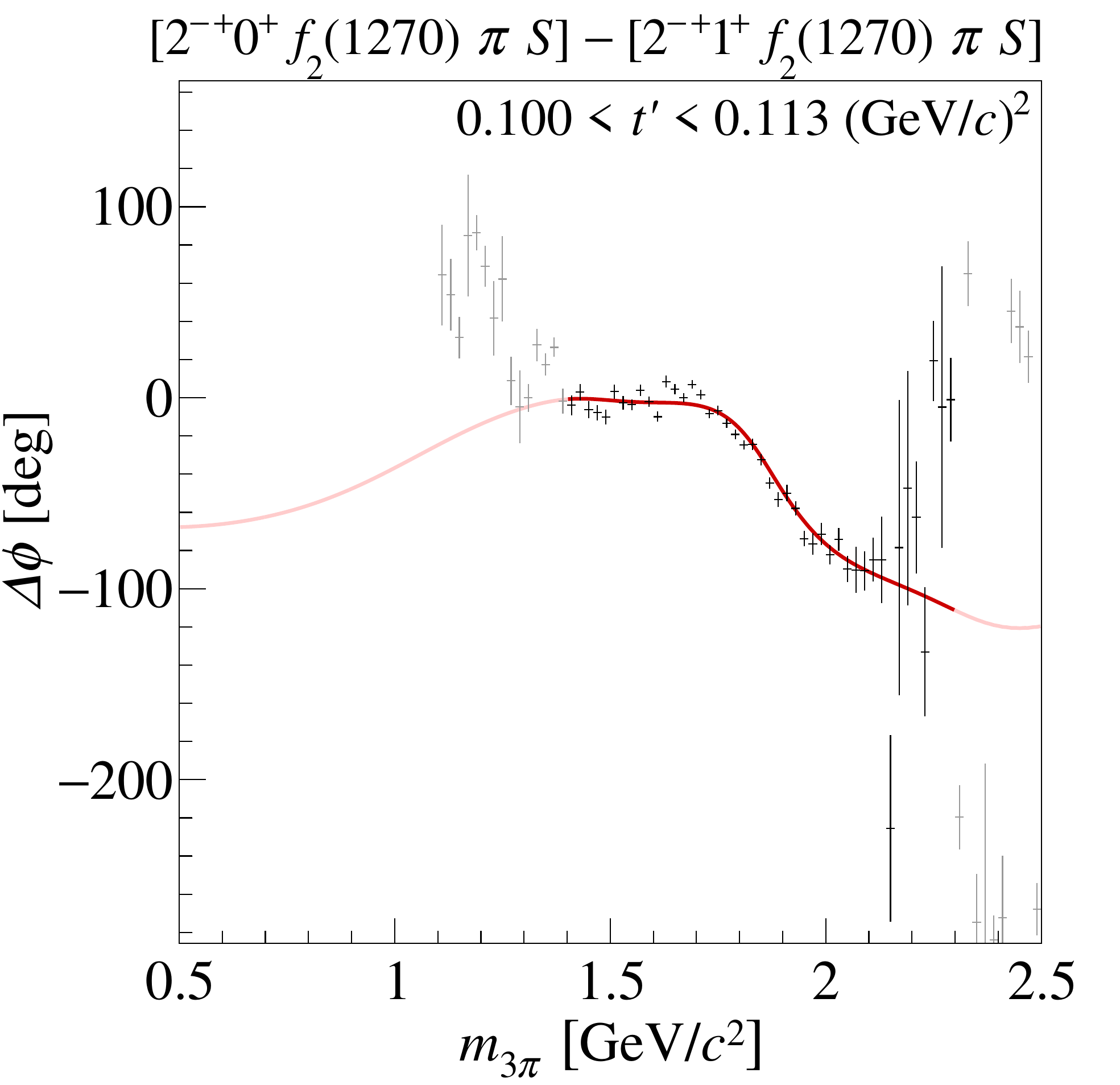}%
    \label{fig:phase_2mp_m0_f2_S_2mp_m1_f2_S_tbin1}%
  }%
  \hspace*{\fourPlotSpacing}%
  \subfloat[][]{%
    \includegraphics[width=\fourPlotWidth]{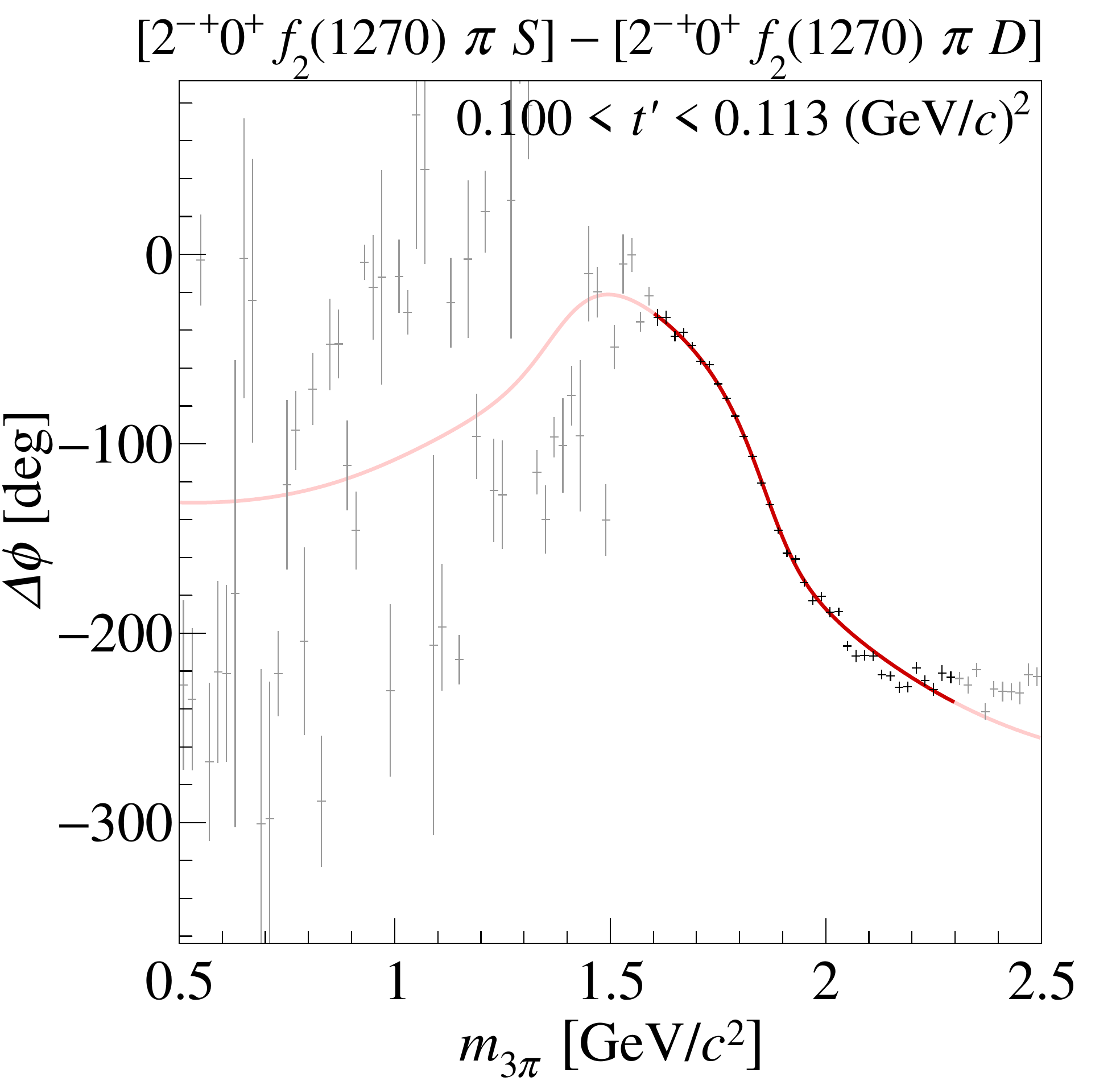}%
    \label{fig:phase_2mp_m0_f2_S_2mp_f2_D_tbin1}%
  }%
  \\
  \hspace*{\fourPlotWidth}%
  \hspace*{\fourPlotSpacing}%
  \hspace*{\fourPlotWidth}%
  \hspace*{\fourPlotSpacing}%
  \subfloat[][]{%
    \includegraphics[width=\fourPlotWidth]{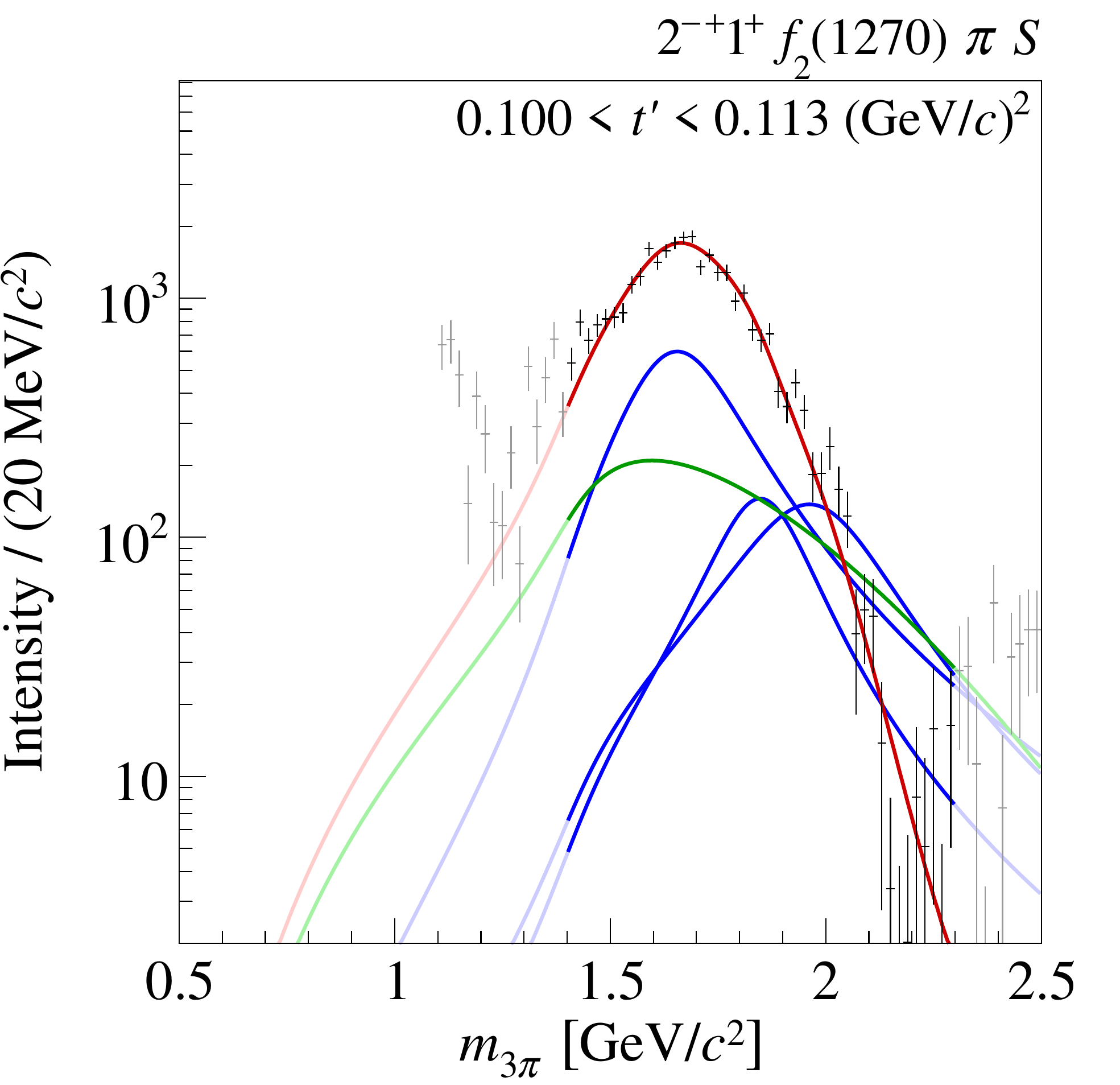}%
    \label{fig:intensity_2mp_m1_f2_S_tbin1_log}%
  }%
  \hspace*{\fourPlotSpacing}%
  \subfloat[][]{%
    \includegraphics[width=\fourPlotWidth]{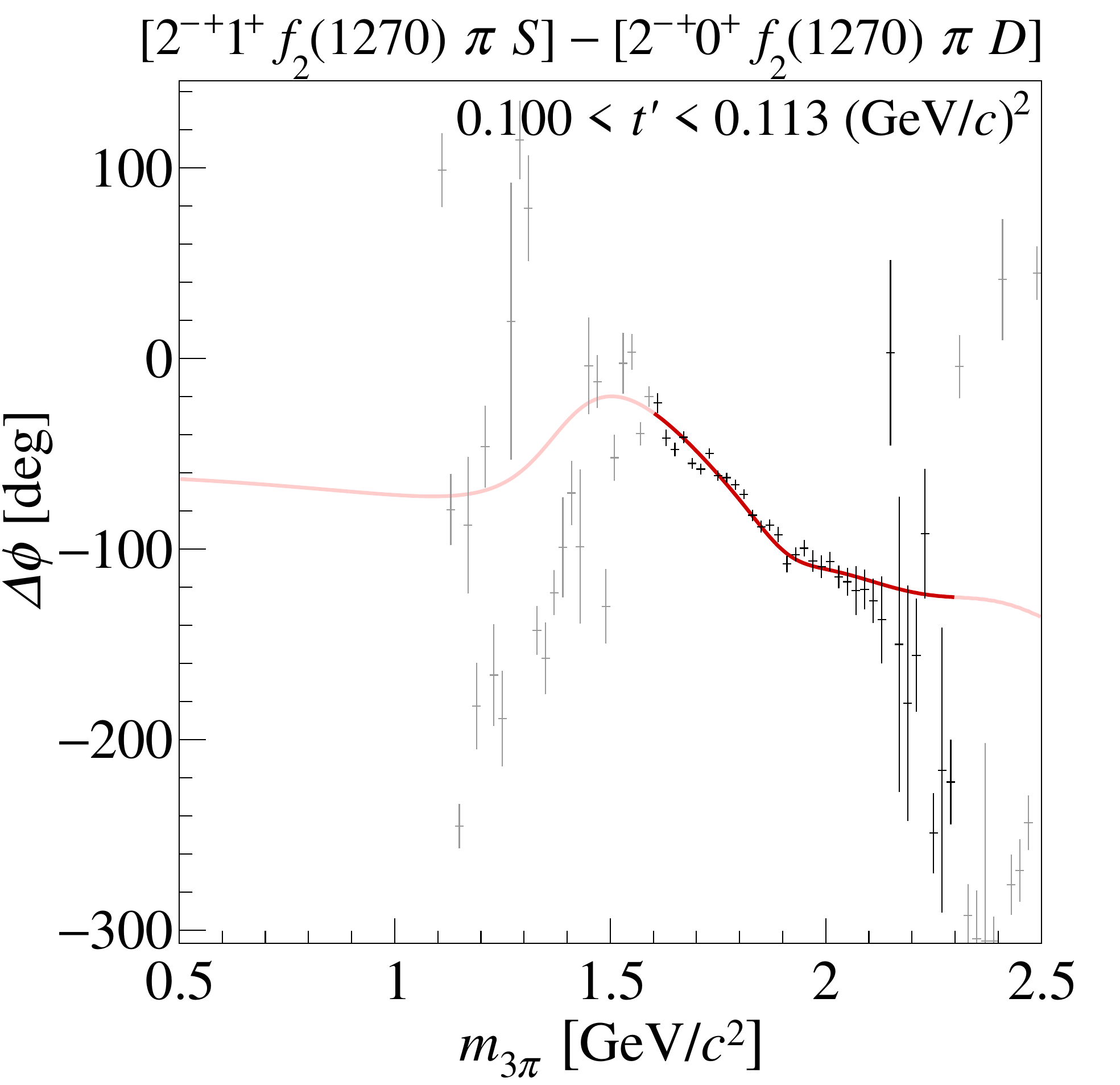}%
    \label{fig:phase_2mp_m1_f2_S_2mp_f2_D_tbin1}%
  }%
  \\
  \hspace*{\fourPlotWidth}%
  \hspace*{\fourPlotSpacing}%
  \hspace*{\fourPlotWidth}%
  \hspace*{\fourPlotSpacing}%
  \hspace*{\fourPlotWidth}%
  \hspace*{\fourPlotSpacing}%
  \subfloat[][]{%
    \includegraphics[width=\fourPlotWidth]{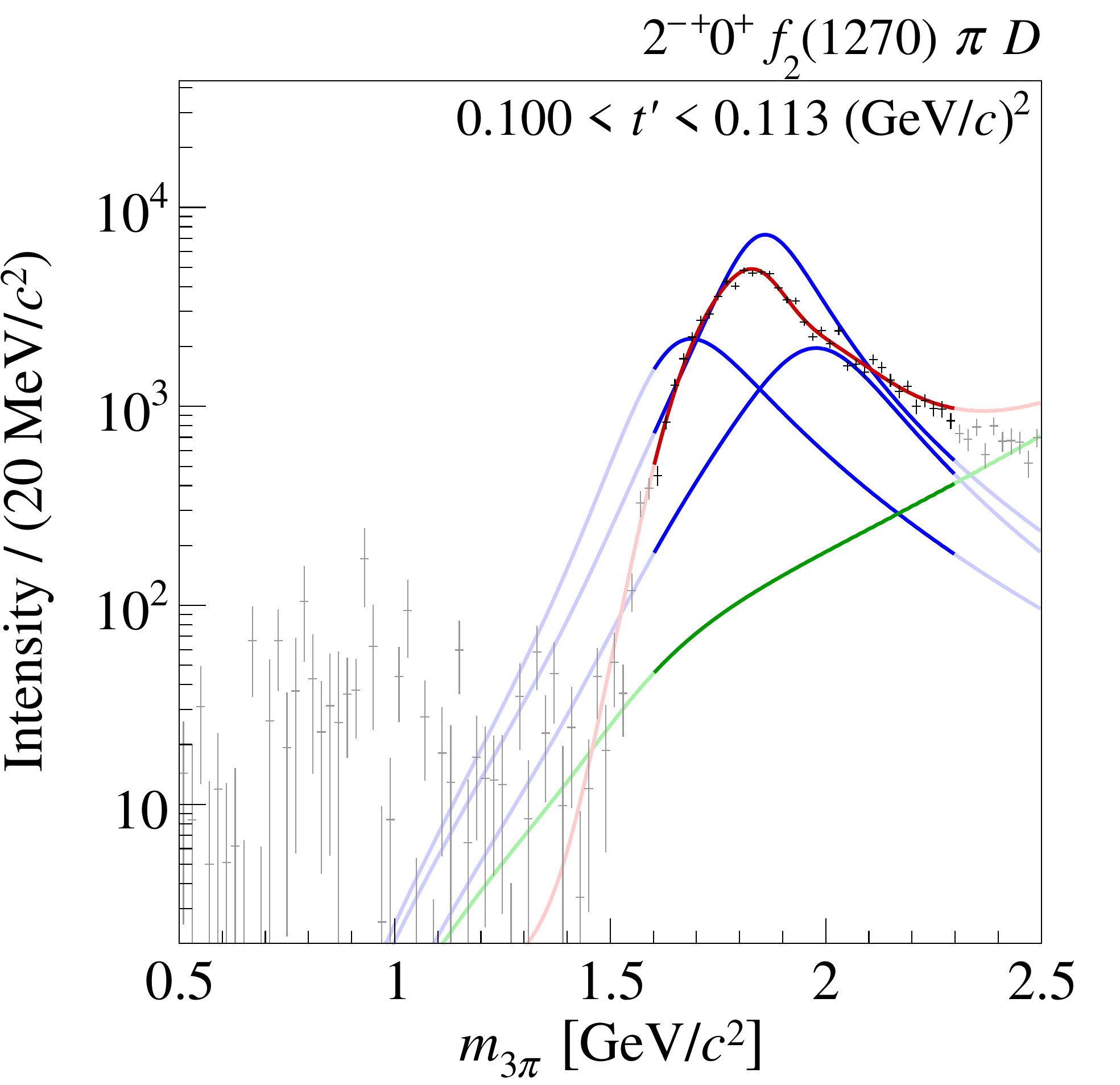}%
    \label{fig:intensity_2mp_f2_D_tbin1_log}%
  }%
  \caption{Amplitudes of the four $\JPC = 2^{-+}$ waves in the lowest
    \tpr bin.
    \subfloatLabel{fig:intensity_2mp_rho_tbin1_log}~through~\subfloatLabel{fig:phase_2mp_rho_2mp_f2_D_tbin1}:
    Intensity distribution and relative phases for the
    \wave{2}{-+}{0}{+}{\Prho}{F}
    wave. \subfloatLabel{fig:intensity_2mp_m0_f2_S_tbin1_log}~through~\subfloatLabel{fig:phase_2mp_m0_f2_S_2mp_f2_D_tbin1}:
    Intensity distribution and relative phases for the
    \wave{2}{-+}{0}{+}{\PfTwo}{S} wave.
    \subfloatLabel{fig:intensity_2mp_m1_f2_S_tbin1_log}~and~\subfloatLabel{fig:phase_2mp_m1_f2_S_2mp_f2_D_tbin1}:
    Intensity distribution and relative phase for the
    \wave{2}{-+}{1}{+}{\PfTwo}{S} wave.
    \subfloatLabel{fig:intensity_2mp_f2_D_tbin1_log}: Intensity
    distribution for the \wave{2}{-+}{0}{+}{\PfTwo}{D} wave.  The
    model and the wave components are represented as in
    \cref{fig:intensity_phases_2mp}.}
  \label{fig:intensity_phases_2mp_tbin1}
\ifMultiColumnLayout{\end{figure*}}{\end{figure}}

\ifMultiColumnLayout{\begin{figure*}[t]}{\begin{figure}[p]}
  \centering
  \subfloat[][]{%
    \includegraphics[width=\fourPlotWidth]{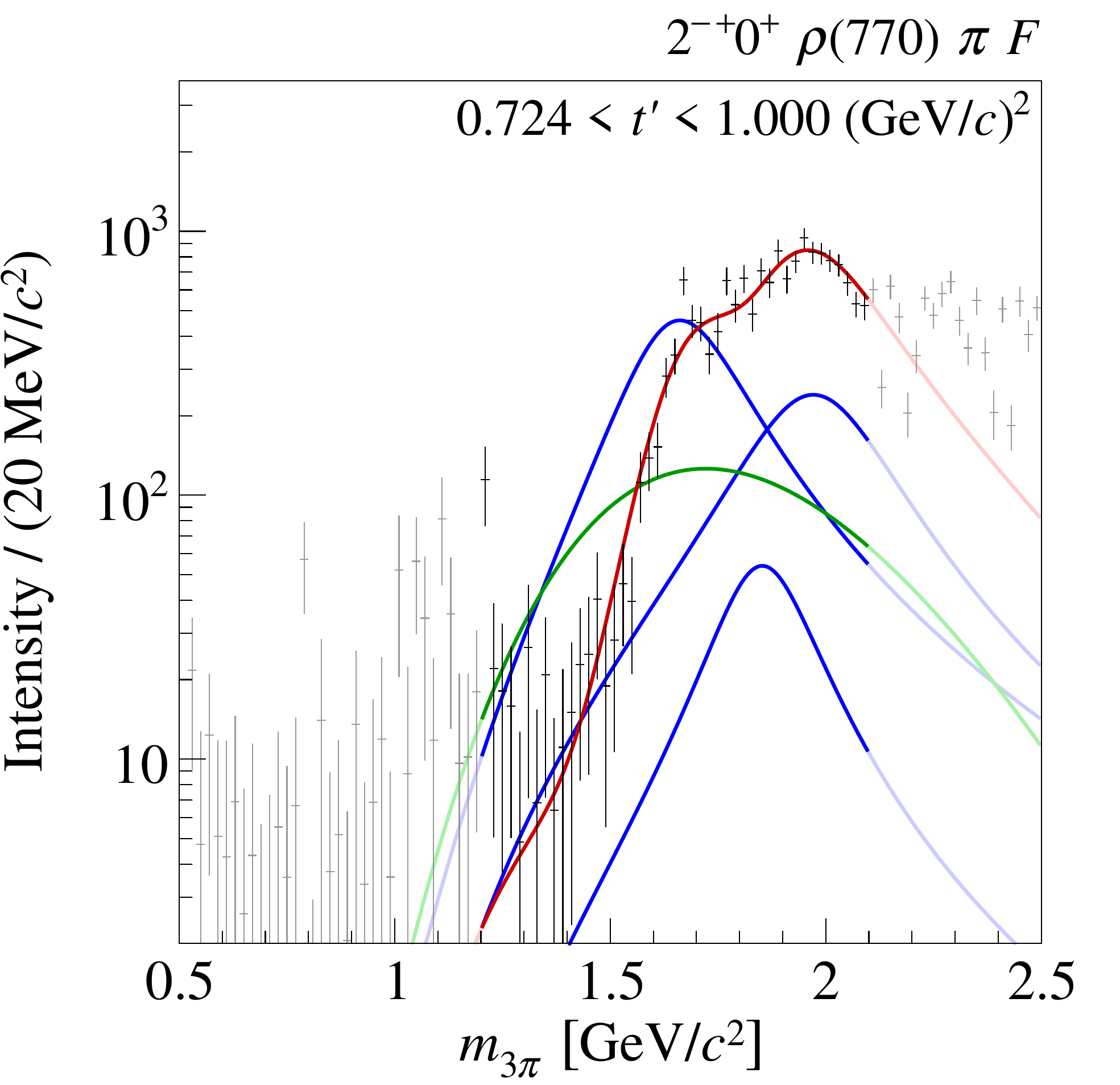}%
    \label{fig:intensity_2mp_rho_tbin11_log}%
  }%
  \hspace*{\fourPlotSpacing}%
  \subfloat[][]{%
    \includegraphics[width=\fourPlotWidth]{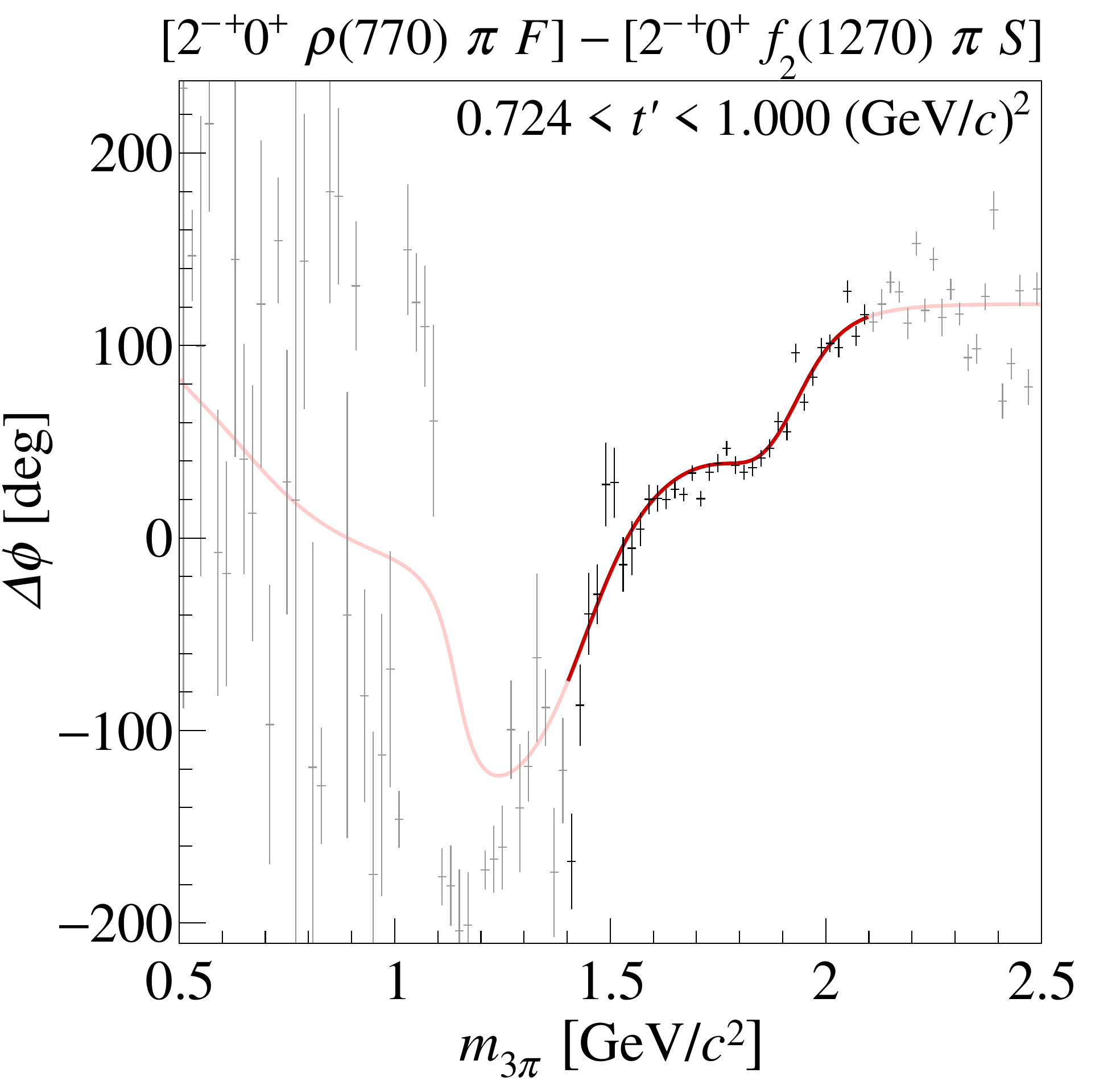}%
    \label{fig:phase_2mp_rho_2mp_m0_f2_S_tbin11}%
  }%
  \hspace*{\fourPlotSpacing}%
  \subfloat[][]{%
    \includegraphics[width=\fourPlotWidth]{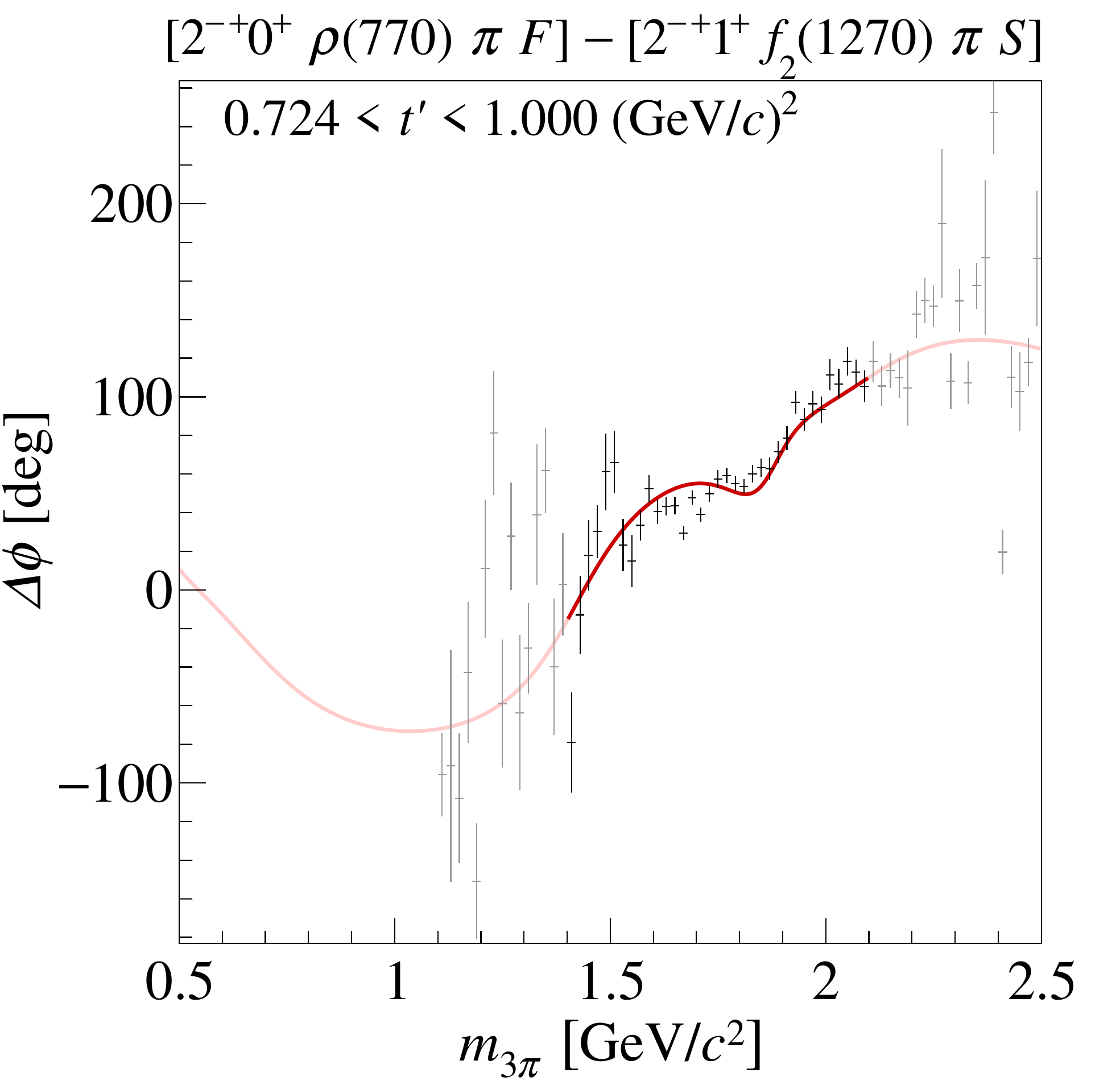}%
    \label{fig:phase_2mp_rho_2mp_m1_f2_S_tbin11}%
  }%
  \hspace*{\fourPlotSpacing}%
  \subfloat[][]{%
    \includegraphics[width=\fourPlotWidth]{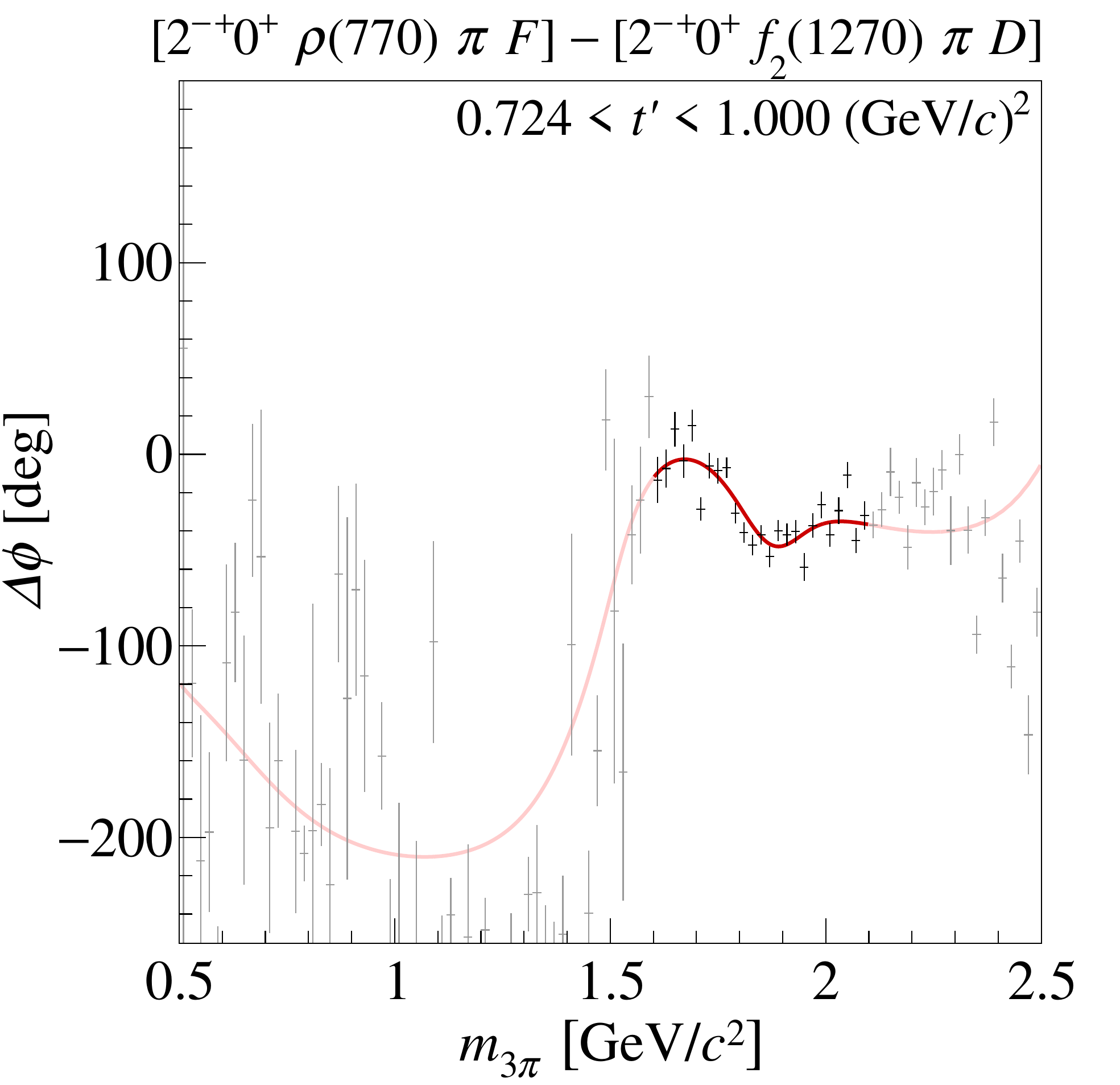}%
    \label{fig:phase_2mp_rho_2mp_f2_D_tbin11}%
  }%
  \\
  \hspace*{\fourPlotWidth}%
  \hspace*{\fourPlotSpacing}%
  \subfloat[][]{%
    \includegraphics[width=\fourPlotWidth]{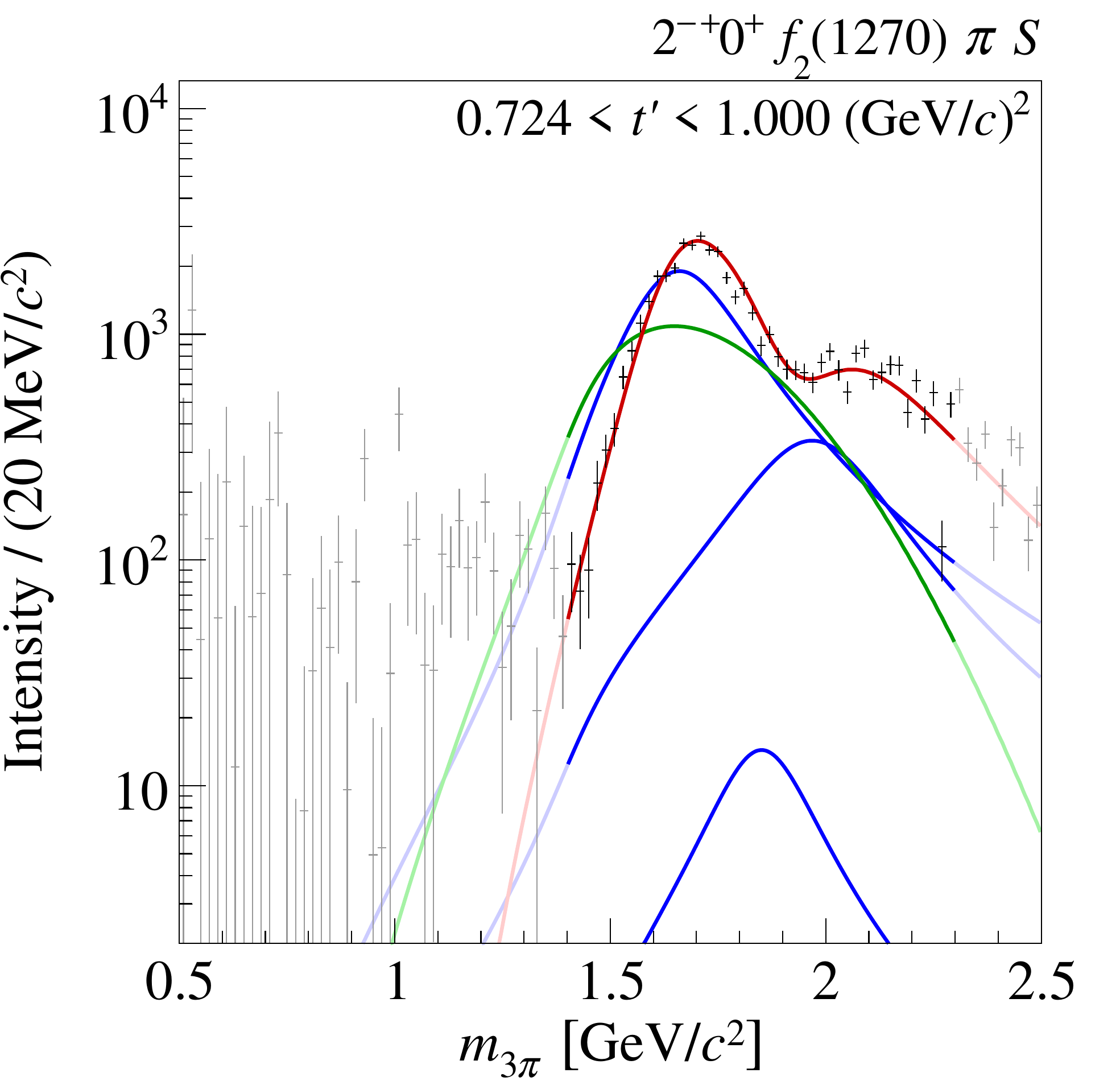}%
    \label{fig:intensity_2mp_m0_f2_S_tbin11_log}%
  }%
  \hspace*{\fourPlotSpacing}%
  \subfloat[][]{%
    \includegraphics[width=\fourPlotWidth]{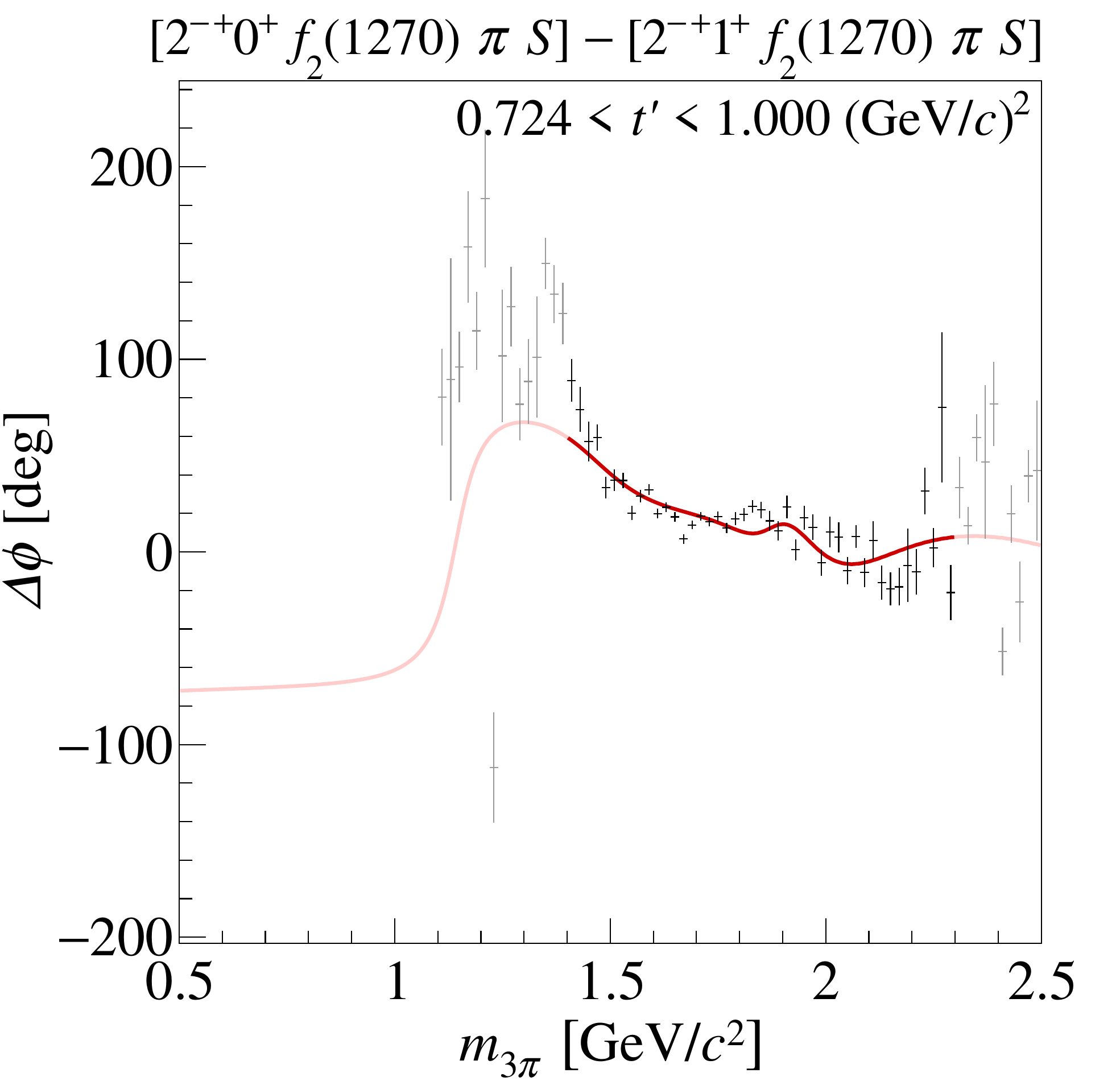}%
    \label{fig:phase_2mp_m0_f2_S_2mp_m1_f2_S_tbin11}%
  }%
  \hspace*{\fourPlotSpacing}%
  \subfloat[][]{%
    \includegraphics[width=\fourPlotWidth]{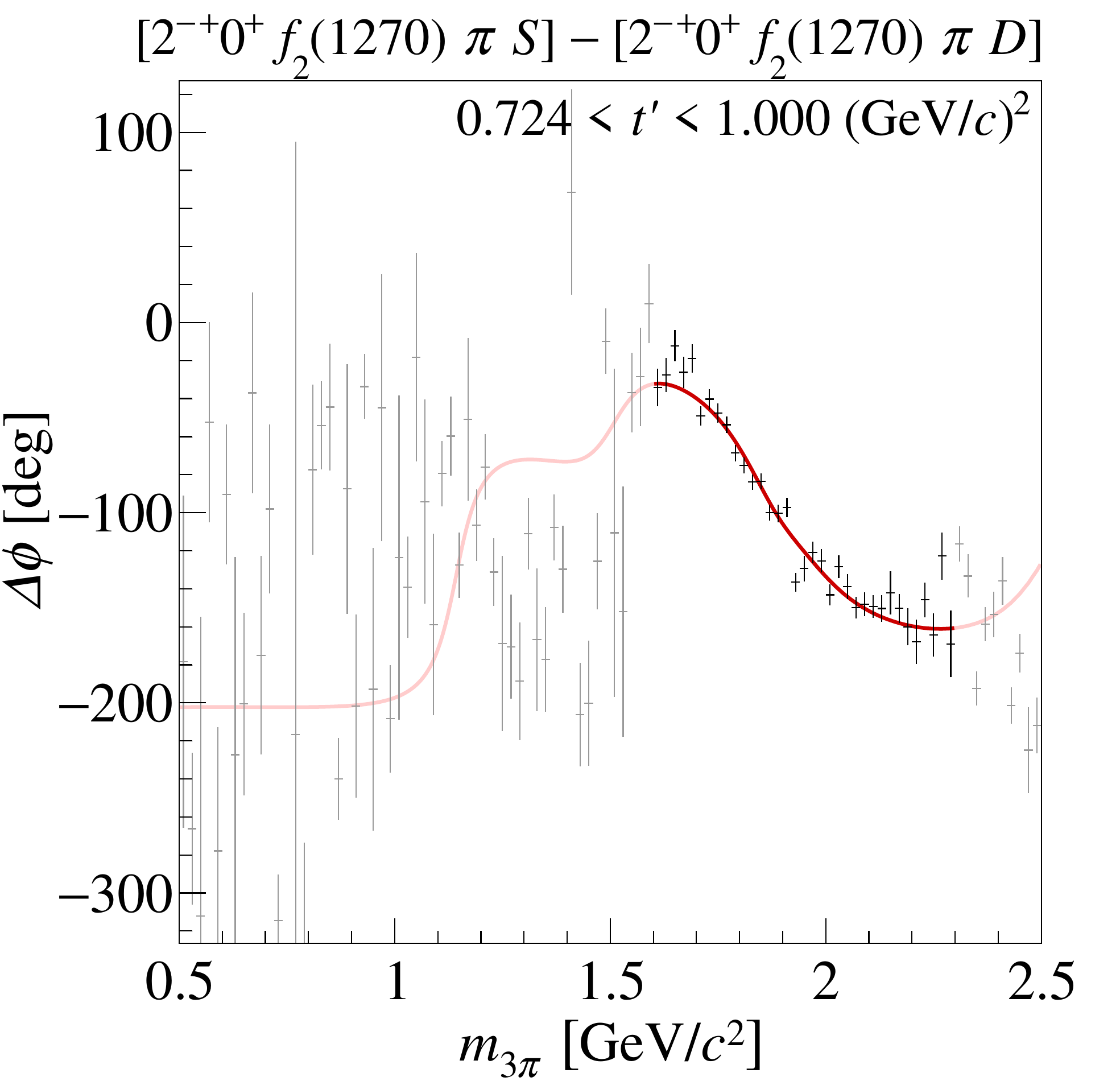}%
    \label{fig:phase_2mp_m0_f2_S_2mp_f2_D_tbin11}%
  }%
  \\
  \hspace*{\fourPlotWidth}%
  \hspace*{\fourPlotSpacing}%
  \hspace*{\fourPlotWidth}%
  \hspace*{\fourPlotSpacing}%
  \subfloat[][]{%
    \includegraphics[width=\fourPlotWidth]{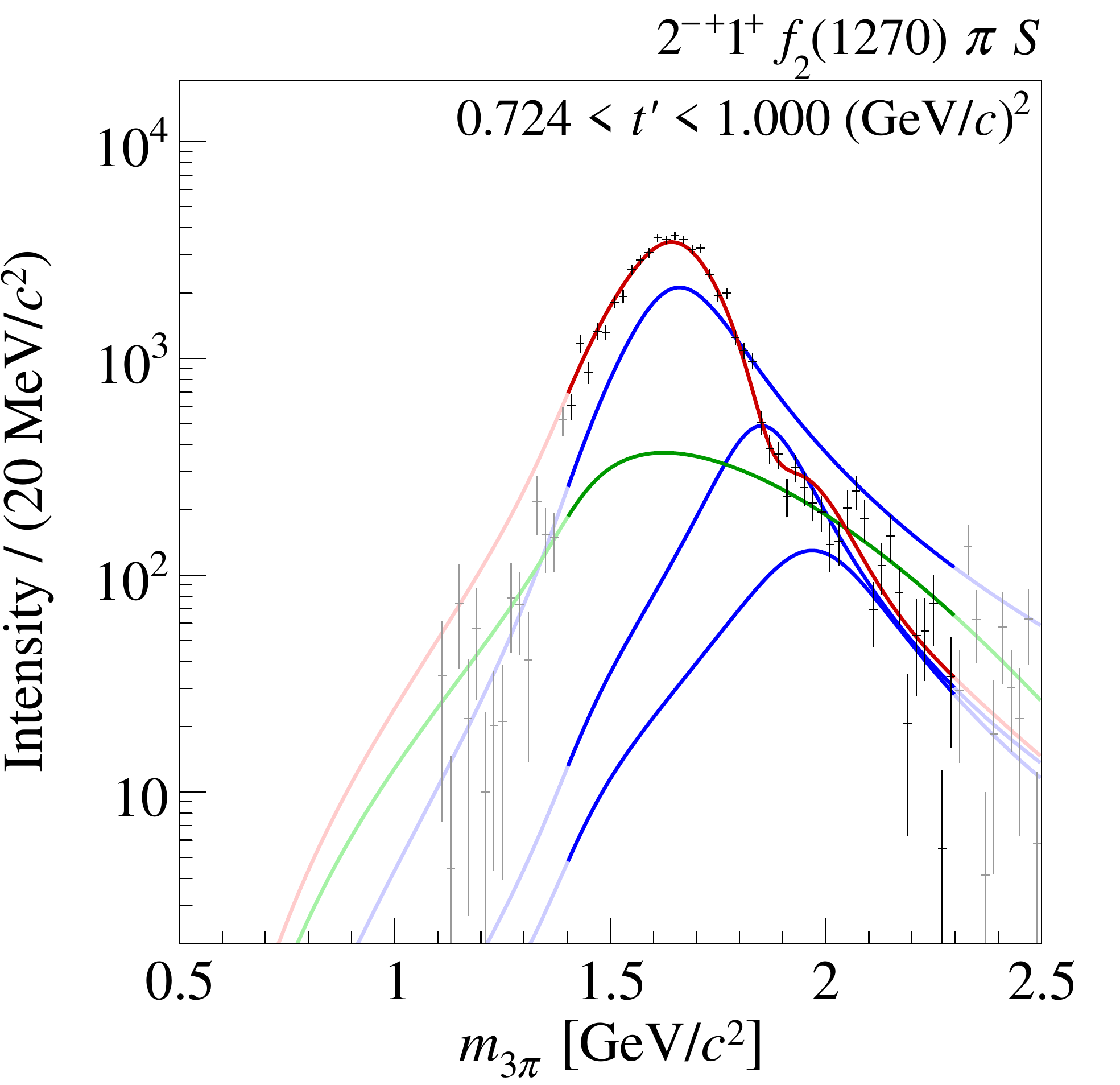}%
    \label{fig:intensity_2mp_m1_f2_S_tbin11_log}%
  }%
  \hspace*{\fourPlotSpacing}%
  \subfloat[][]{%
    \includegraphics[width=\fourPlotWidth]{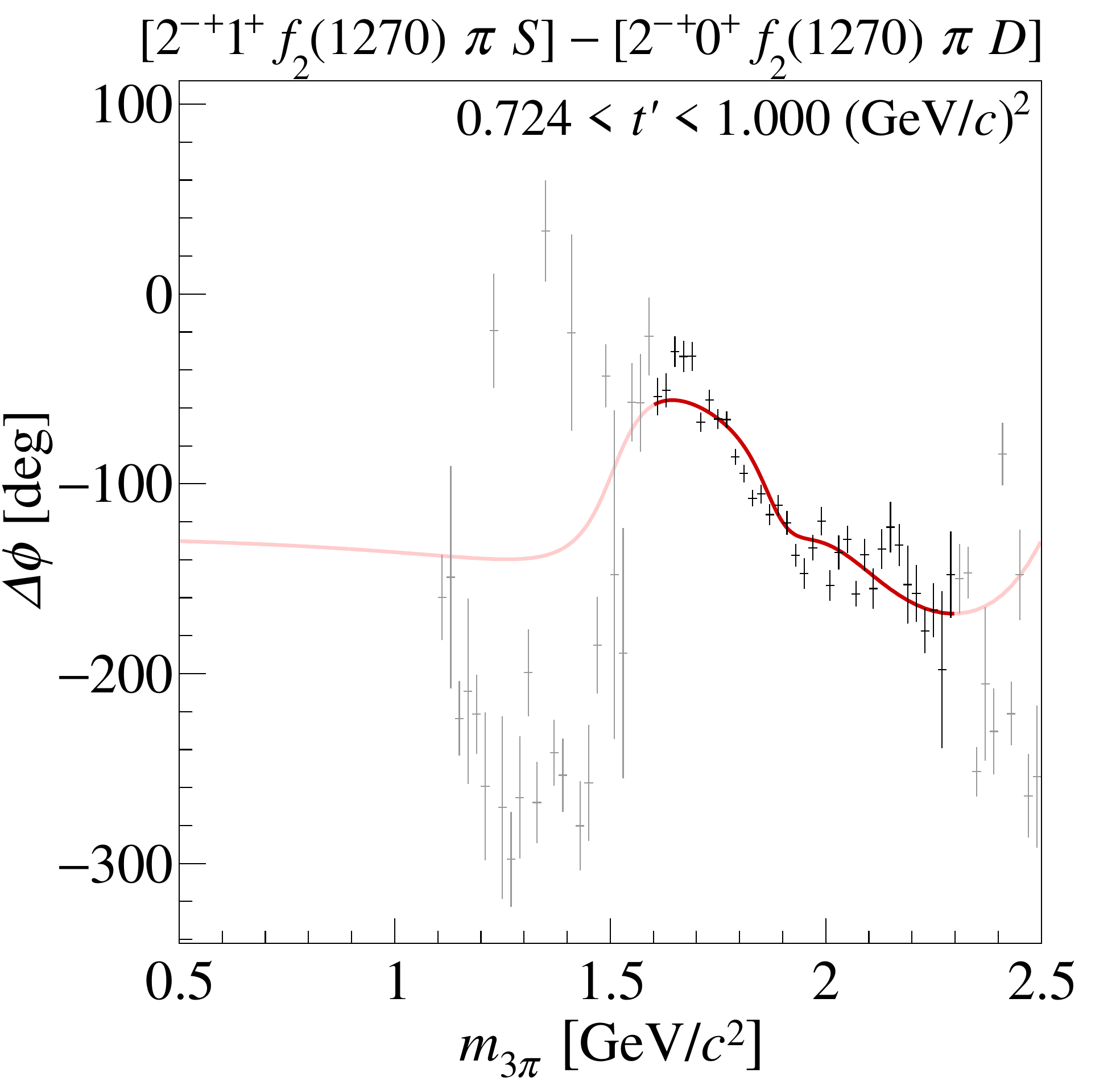}%
    \label{fig:phase_2mp_m1_f2_S_2mp_f2_D_tbin11}%
  }%
  \\
  \hspace*{\fourPlotWidth}%
  \hspace*{\fourPlotSpacing}%
  \hspace*{\fourPlotWidth}%
  \hspace*{\fourPlotSpacing}%
  \hspace*{\fourPlotWidth}%
  \hspace*{\fourPlotSpacing}%
  \subfloat[][]{%
    \includegraphics[width=\fourPlotWidth]{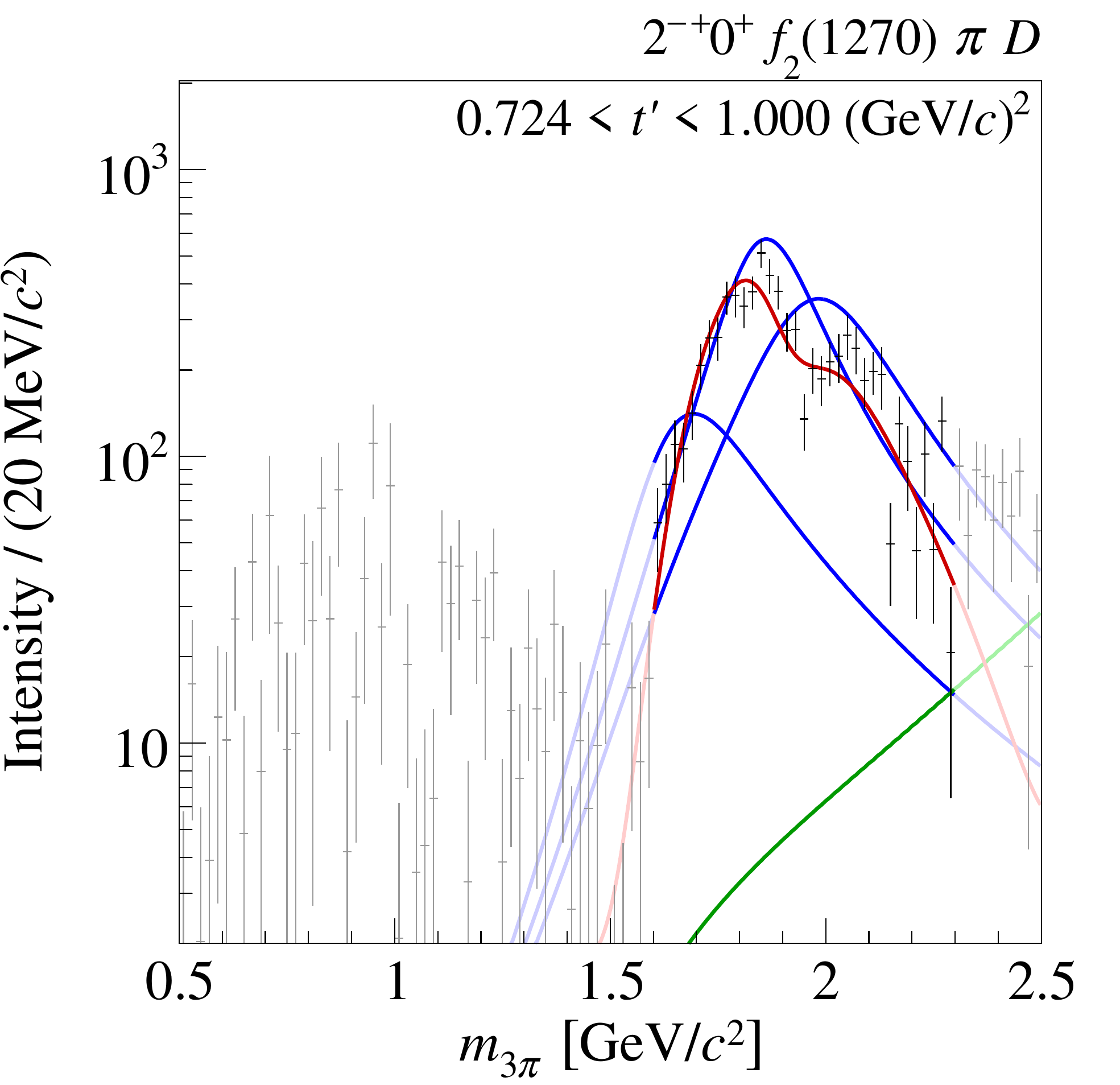}%
    \label{fig:intensity_2mp_f2_D_tbin11_log}%
  }%
  \caption{Similar to \cref{fig:intensity_phases_2mp_tbin1} but for
    the highest \tpr bin.}
  \label{fig:intensity_phases_2mp_tbin11}
\ifMultiColumnLayout{\end{figure*}}{\end{figure}}

Within the fit range, the model is able to describe well most of the
phase motions.  Some details in the high-mass regions are not
reproduced.  Often, the high-mass extrapolations of the fit model
deviate from the data (see \eg \cref{fig:intensity_phases_2mp_tbin1}
and the second row of \cref{fig:intensity_phases_2mp}).  In some
cases, this is also true for the extrapolations below the low-mass
limits of the fit ranges (see \eg
\cref{fig:intensity_phases_2mp_tbin1}).  However, in many of these
cases the intensities of the waves are small.

The extracted resonance parameters for \PpiTwo, \PpiTwo[1880], and
\PpiTwo[2005] are
\begin{align*}
  m_{\PpiTwo} &= \SIaerrSys{1642}{12}{1}{\MeVcc}\text{,} \\
  \Gamma_{\PpiTwo} &= \SIaerrSys{311}{12}{23}{\MeVcc}\text{,} \\
  m_{\PpiTwo[1880]} &= \SIaerrSys{1847}{20}{3}{\MeVcc}\text{,} \\
  \Gamma_{\PpiTwo[1880]} &= \SIaerrSys{246}{33}{28}{\MeVcc}\text{,} \\
  m_{\PpiTwo[2005]} &= \SIaerrSys{1962}{17}{29}{\MeVcc}\text{, and} \\
  \Gamma_{\PpiTwo[2005]} &= \SIaerrSys{371}{16}{120}{\MeVcc}.
\end{align*}
Hence in the $2^{-+}$ sector, the model assumption of well-separated
resonances with little overlap is not well fulfilled.  Although
constrained by the amplitudes of four waves, the $2^{-+}$ resonance
parameters exhibit a larger sensitivity to changes of the fit model
discussed in \cref{sec:systematics}.  They therefore have larger
systematic uncertainties than, for example, the parameters of the
\Ppi[1800].  In addition, some of the systematic uncertainty intervals
are highly asymmetric.  The parameters of the three $2^{-+}$
resonances are correlated in a complicated way and depend, among other
things, on the set of waves included in the fit.  Also the number of
background events in the selected data sample influences the resonance
parameters.  The parameters of \PpiTwo[1880] and \PpiTwo[2005] are in
addition sensitive to the number of \tpr bins.  This underlines the
importance of using a fine-grained \tpr binning in order to capture
the evolution of the $2^{-+}$ amplitudes with \tpr.  The $2^{-+}$
resonance parameters exhibit an exceptionally large sensitivity to the
\mThreePi and \tpr dependences of the production probability
$\Abs[0]{\mathcal{P}(\mThreePi, \tpr)}^2$ in
\cref{eq:method:param:spindens,eq:method:param:prods}.
The widths of \PpiTwo and \PpiTwo[1880] are also affected by the
interference of the $2^{-+}$ waves with the low-mass part of the
\wave{0}{-+}{0}{+}{\PfZero[980]}{S} wave.  More details on the results
of the systematic studies can be found in
\cref{sec:syst_uncert_twoMP}.

\Cref{fig:tprim_2mp_1,fig:tprim_2mp_2} show the \tpr dependence of the
intensities of the resonant and nonresonant $2^{-+}$ wave components
together with the results of fits using
\cref{eq:slope-parametrization}.  In our fit model, the coupling
amplitudes of the resonance components in the three $2^{-+}$ waves
with $M = 0$ are constrained by \cref{eq:method:branchingdefinition}.
The \tpr spectra of the resonance components are well described by the
exponential model in \cref{eq:slope-parametrization}.  The extracted
values of the slope parameters for \PpiTwo, \PpiTwo[1880], and
\PpiTwo[2005] are approximately \SI{8.5}{\perGeVcsq},
\SI{7.8}{\perGeVcsq}, and \SI{6.7}{\perGeVcsq}, respectively [see
\cref{tab:slopes} for details], which are typical values for
resonances.  As for the $1^{++}$ and $2^{++}$ resonances, the slope
parameter decreases with increasing mass of the resonance.  This
flattening of the \tpr slope with increasing \mThreePi was also
observed in the \tpr spectra before partial-wave decomposition (see
\eg Fig.~31 in \refCite{Adolph:2015tqa}).  The three-component Deck
model~\cite{CohenTannoudji:1976tj,Kaidalov:1979jz,Antunes:1984cy} may
explain this behavior.  The relative enhancement of higher-mass states
at larger values of \tpr helps to better disentangle the various
resonance components.

In the fit model, the \tpr dependence of the coupling amplitudes of
the resonant components in the \wave{2}{-+}{1}{+}{\PfTwo}{S} wave is
not constrained by \cref{eq:method:branchingdefinition}.  Due to the
relative smallness of this wave, the intensities of the wave
components are extracted less
reliably. \Cref{eq:slope-parametrization} does not describe well the
\tpr spectra of the wave components. This is in particular true for
the \PpiTwo and \PpiTwo[1880].  Hence only a rough comparison of the
slope parameters is possible.  The slope parameter values for
\PpiTwo[1880] and \PpiTwo[2005] are compatible with those found in the
other three $2^{-+}$ waves.  However, the slope of the \PpiTwo\ \tpr
spectrum is significantly smaller with $b = \SI{5.0}{\perGeVcsq}$.
This effect is not understood but it is consistent with the shallower
\tpr slope of the intensity of this wave in the \PpiTwo mass region
(see Table~VI in \refCite{Adolph:2015tqa}).

Compared to the other \JPC sectors, where we observe in general a
steeper \tpr slope for the nonresonant components than for the
resonances, the nonresonant components in the $2^{-+}$ sector behave
somewhat irregularly.  The only exception is the nonresonant component
in the $\PfTwo \pi D$ wave.  Its \tpr spectrum is well described by
the exponential in \cref{eq:slope-parametrization} and has a slope of
\SIaerrSys{12}{6}{2}{\perGeVcsq}, which is considerably steeper than
the slopes of the \PpiTwo* resonances [see
\cref{fig:tprim_2mp_f2_D_nonres}].  The \tpr spectrum of the
nonresonant component in the $\PfTwo \pi S$ wave with $M = 1$ has a
shallower slope of \SIaerrSys{6.9}{1.1}{1.9}{\perGeVcsq} that is
comparable to those of the \PpiTwo* resonances [see
\cref{fig:tprim_2mp_m1_f2_S_nonres}].  However, at low \tpr the data
deviate from the fit model.  Also for the nonresonant component in the
$\PfTwo \pi S$ wave with $M = 0$, the model deviates from the data at
low \tpr [see \cref{fig:tprim_2mp_m0_f2_S_nonres}].
\Cref{eq:slope-parametrization} cannot reproduce the step at
$\tpr \approx \SI{0.16}{\GeVcsq}$.  The extracted value of
\SI{5.1}{\perGeVcsq} for the slope parameter is smaller than that for
the \PpiTwo* resonances, but is not well defined.  The nonresonant
component in the $\Prho \pi F$ wave exhibits a complicated \tpr
spectrum [see \cref{fig:tprim_2mp_rho_nonres}].  It has a narrow dip
at about \SI{0.16}{\GeVcsq} at the same location where we observe a
step in the \tpr spectrum of the nonresonant component in the
\wave{2}{-+}{0}{+}{\PfTwo}{S} wave.  \Cref{eq:slope-parametrization}
cannot describe such a distribution.  The complicated shape of the
\tpr spectrum may be an artifact caused by forcing the same \tpr
dependence of the resonances in the $M = 0$ waves via
\cref{eq:method:branchingdefinition}.  However, if we leave the \tpr
dependence of all resonance components free [see discussion of \StudyT
below], the dip at low \tpr remains.  Since at low \tpr the
nonresonant component is much smaller than any of the three resonance
components, its intensity is less well determined and more sensitive
to systematic effects.  Monte Carlo studies of a model for the Deck
effect (see \cref{sec:deck_model}) have shown that the projection of
this nonresonant amplitude into the $\Prho \pi F$ wave is vanishingly
small so that the observed nonresonant intensity is presumably of
different origin.

\begin{wideFigureOrNot}[tbp]
  \centering
  \subfloat[][]{%
    \includegraphics[width=\fourPlotWidth]{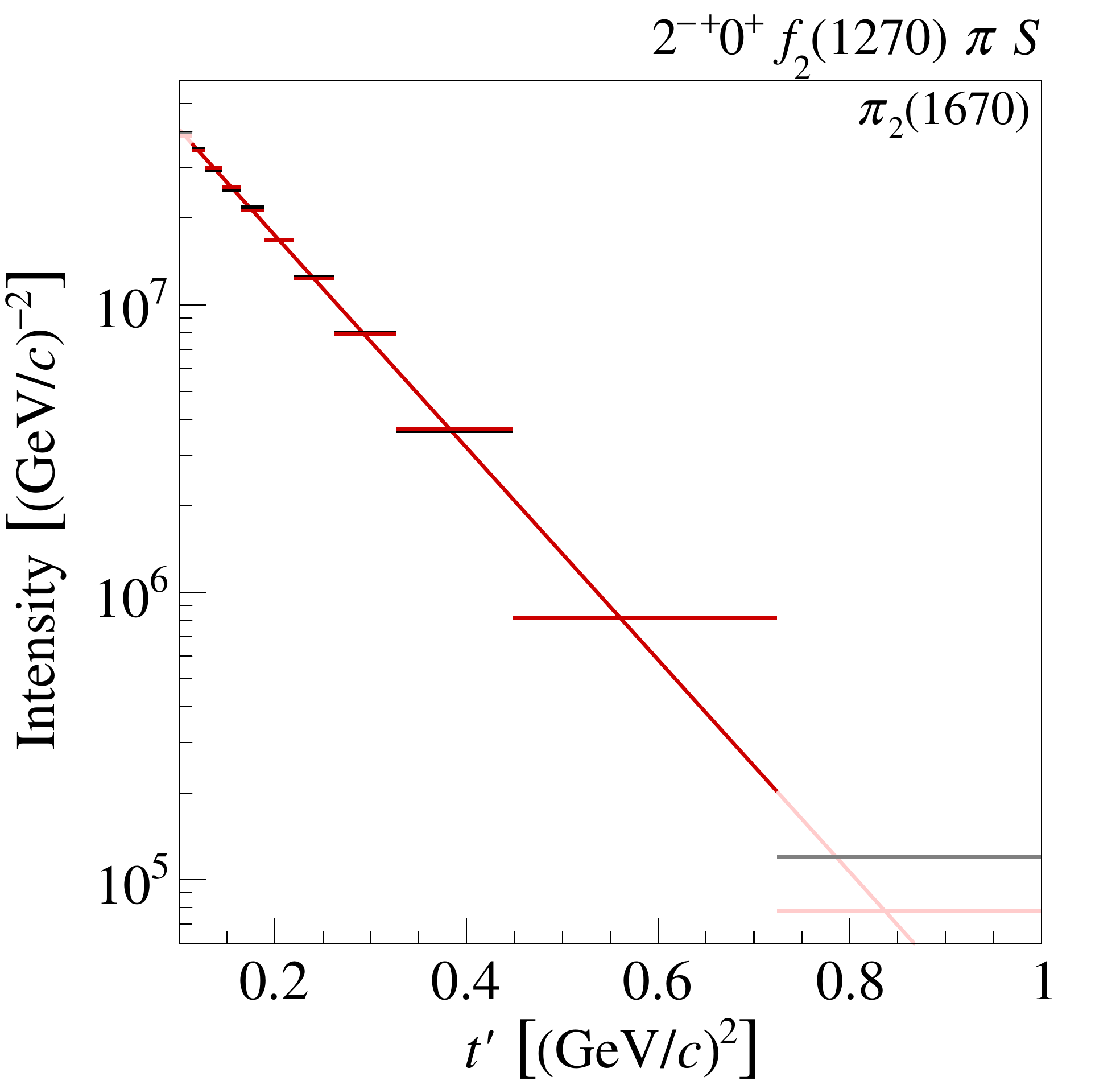}%
  }%
  \hspace*{\fourPlotSpacing}%
  \subfloat[][]{%
    \includegraphics[width=\fourPlotWidth]{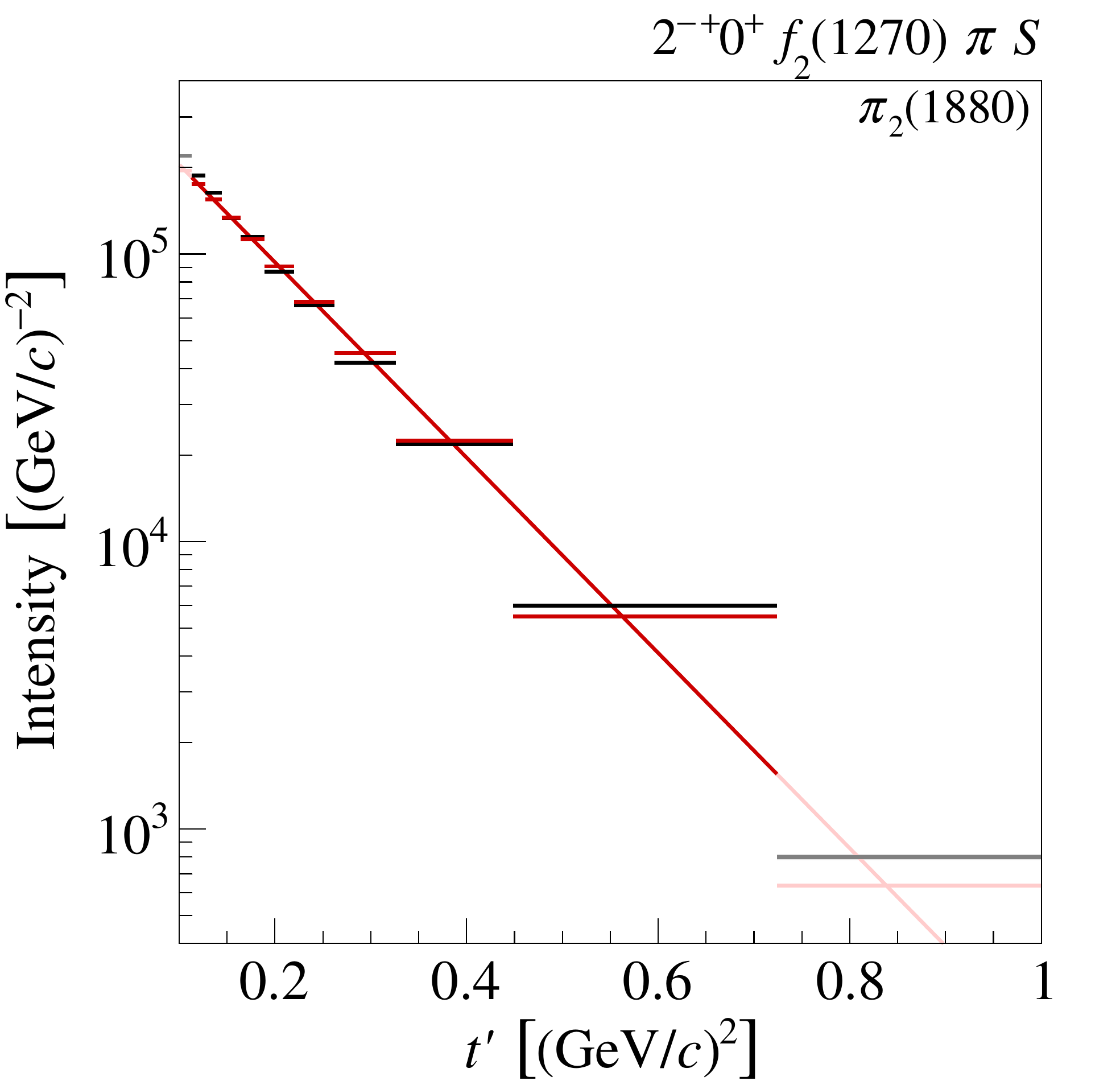}%
    \label{fig:tprim_2mp_m0_f2_S_pi2_1880}%
  }%
  \hspace*{\fourPlotSpacing}%
  \subfloat[][]{%
    \includegraphics[width=\fourPlotWidth]{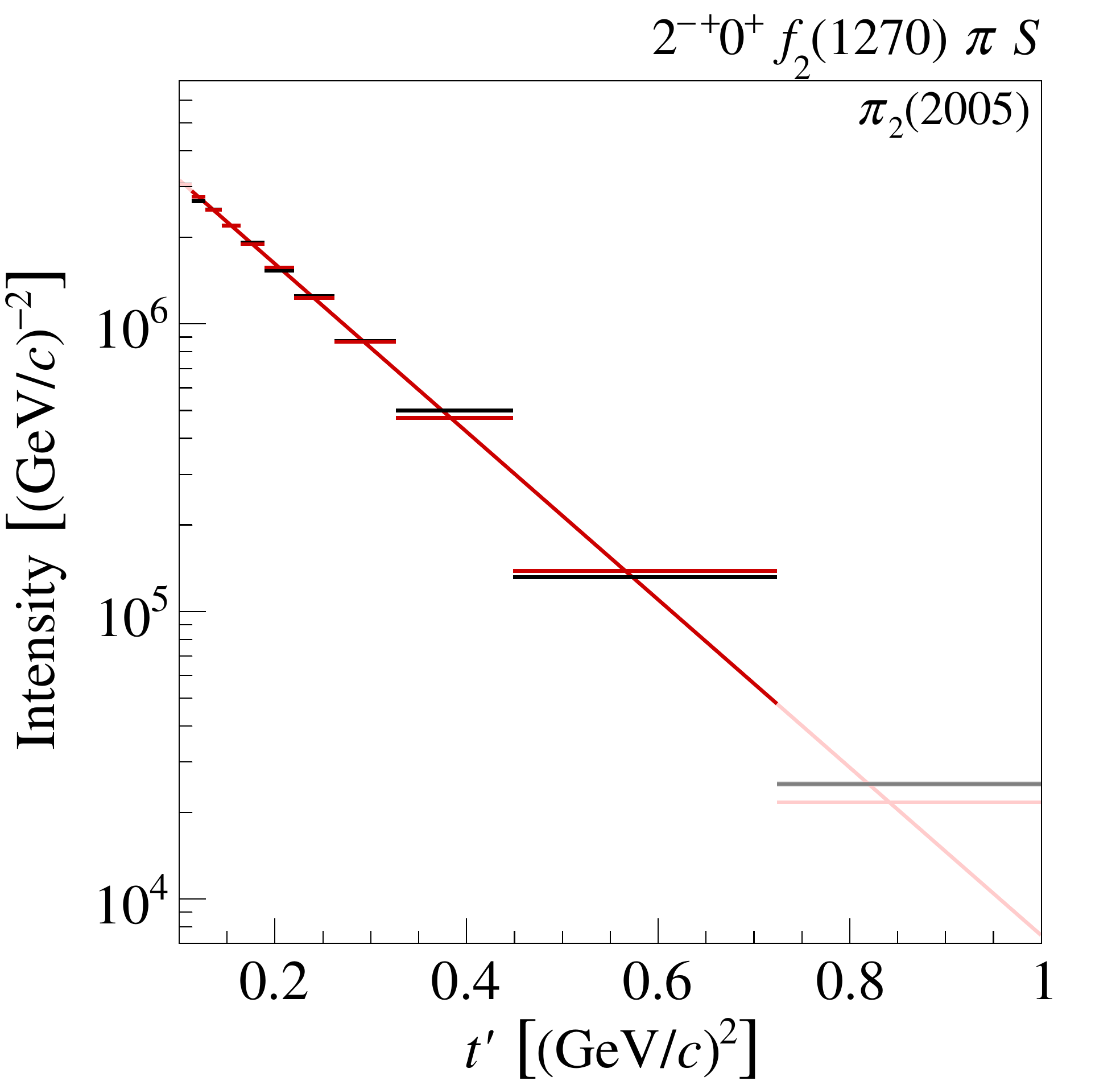}%
  }%
  \hspace*{\fourPlotSpacing}%
  \subfloat[][]{%
    \includegraphics[width=\fourPlotWidth]{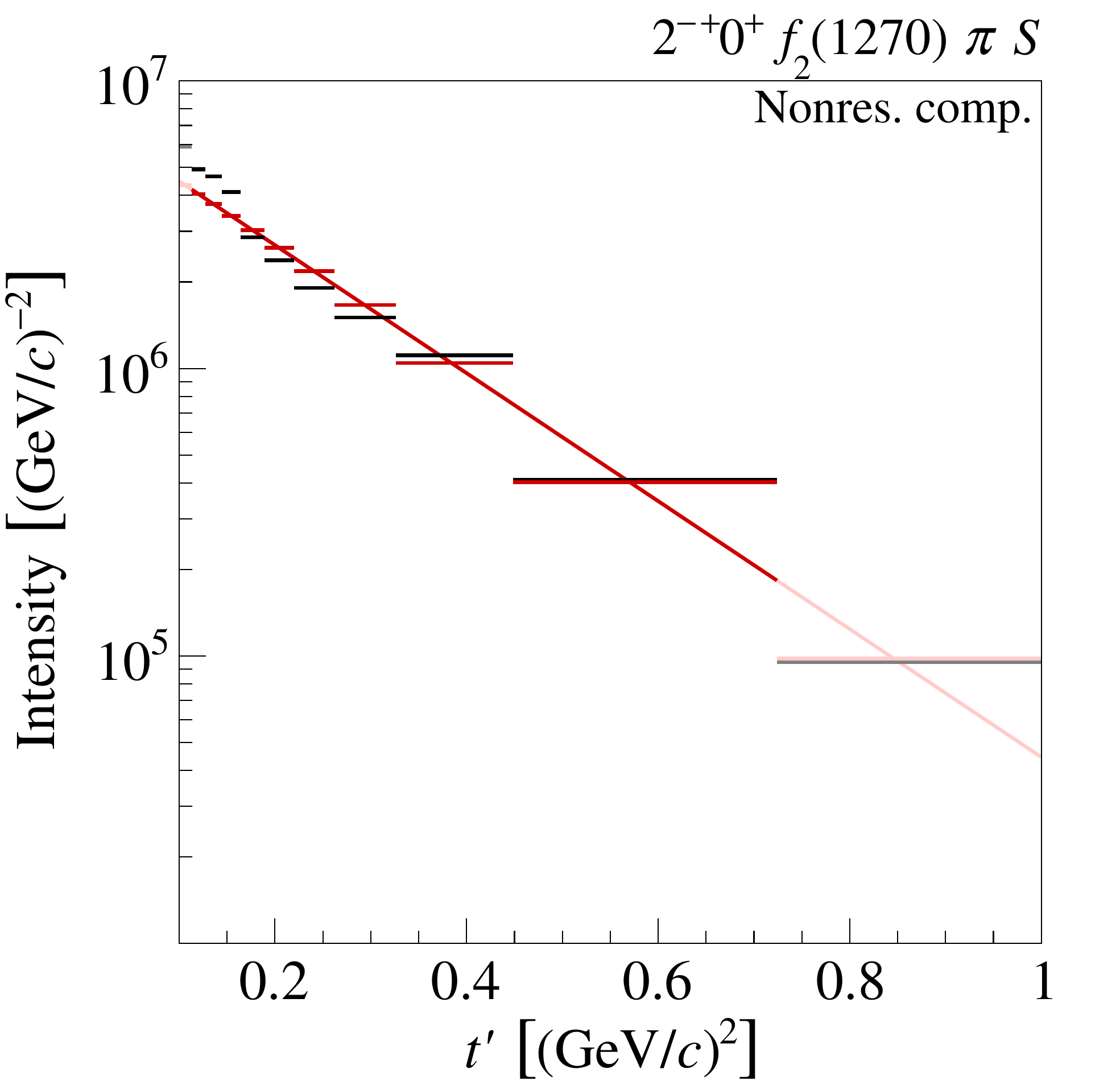}%
    \label{fig:tprim_2mp_m0_f2_S_nonres}%
  }%
  \\
  \subfloat[][]{%
    \includegraphics[width=\fourPlotWidth]{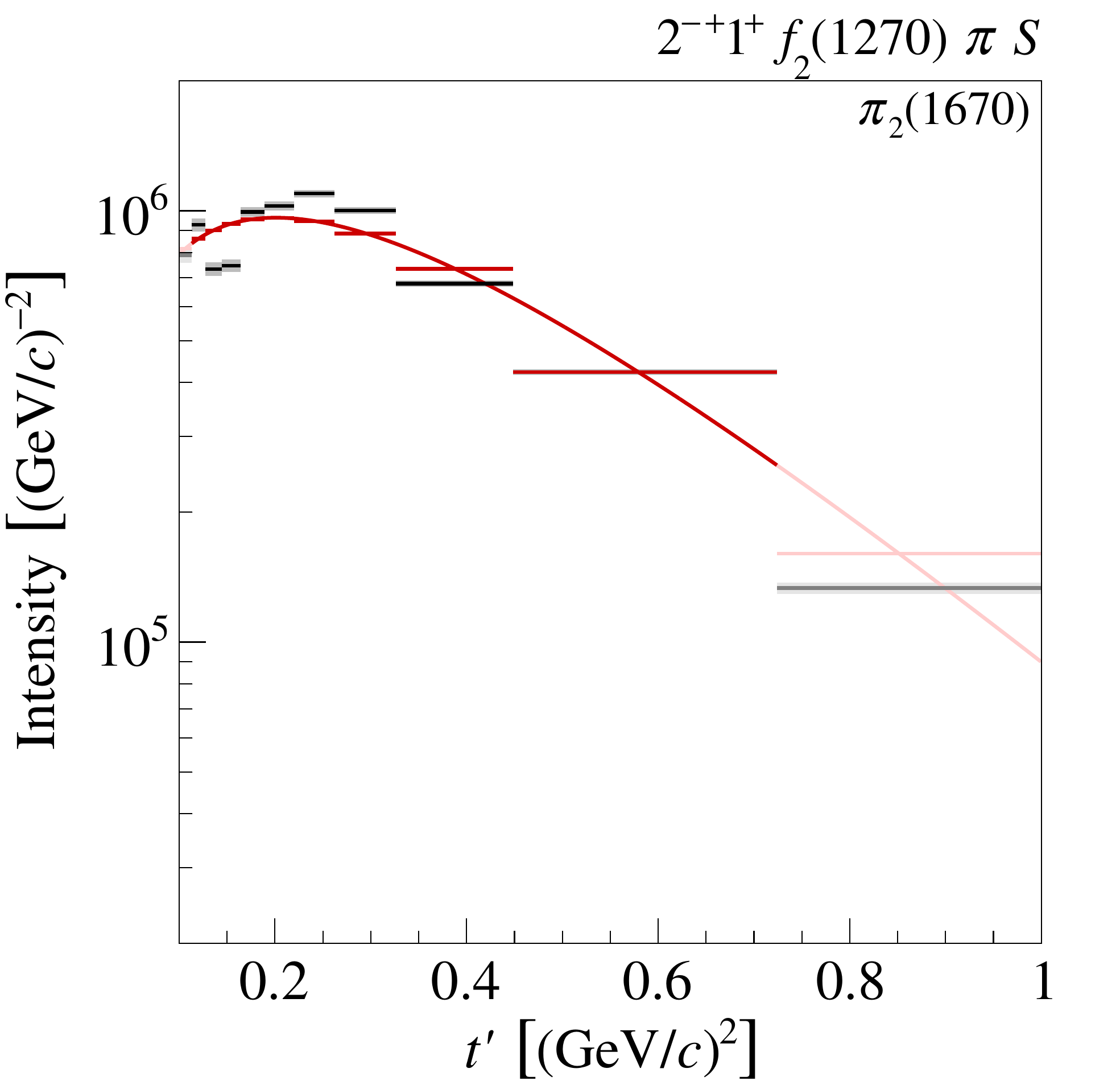}%
  }%
  \hspace*{\fourPlotSpacing}%
  \subfloat[][]{%
    \includegraphics[width=\fourPlotWidth]{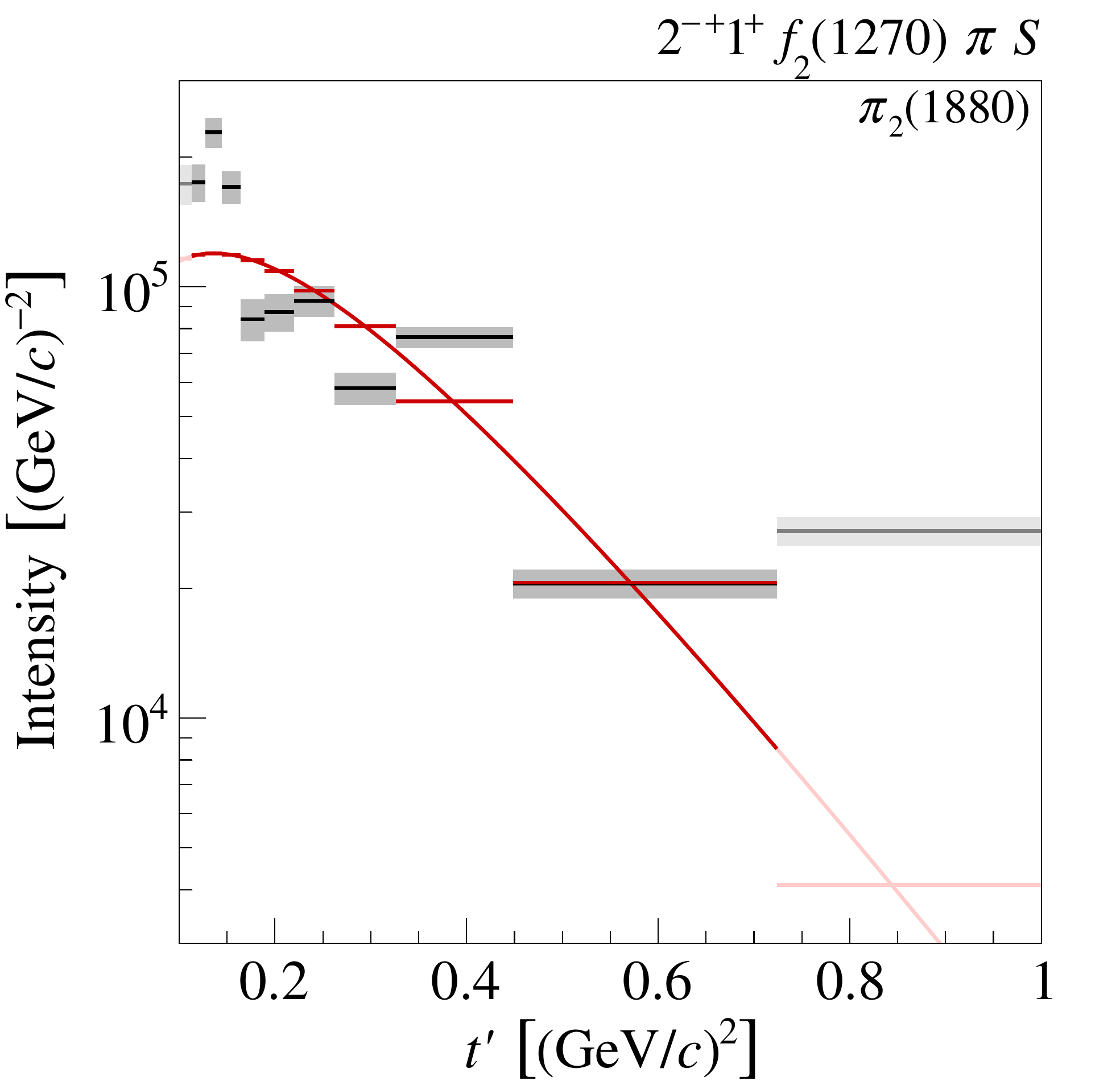}%
    \label{fig:tprim_2mp_m1_f2_S_pi2_1880}%
  }%
  \hspace*{\fourPlotSpacing}%
  \subfloat[][]{%
    \includegraphics[width=\fourPlotWidth]{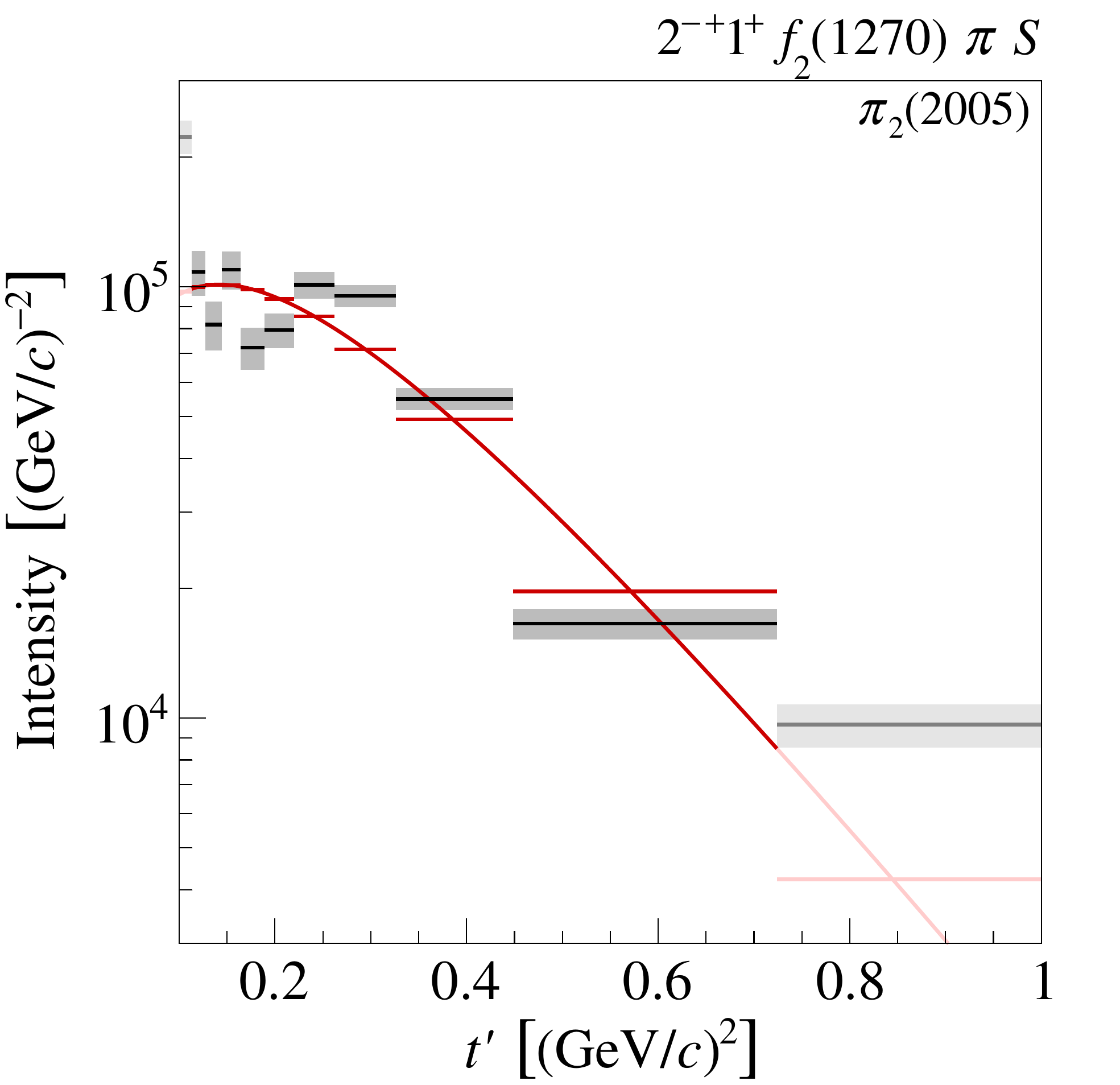}%
  }%
  \hspace*{\fourPlotSpacing}%
  \subfloat[][]{%
    \includegraphics[width=\fourPlotWidth]{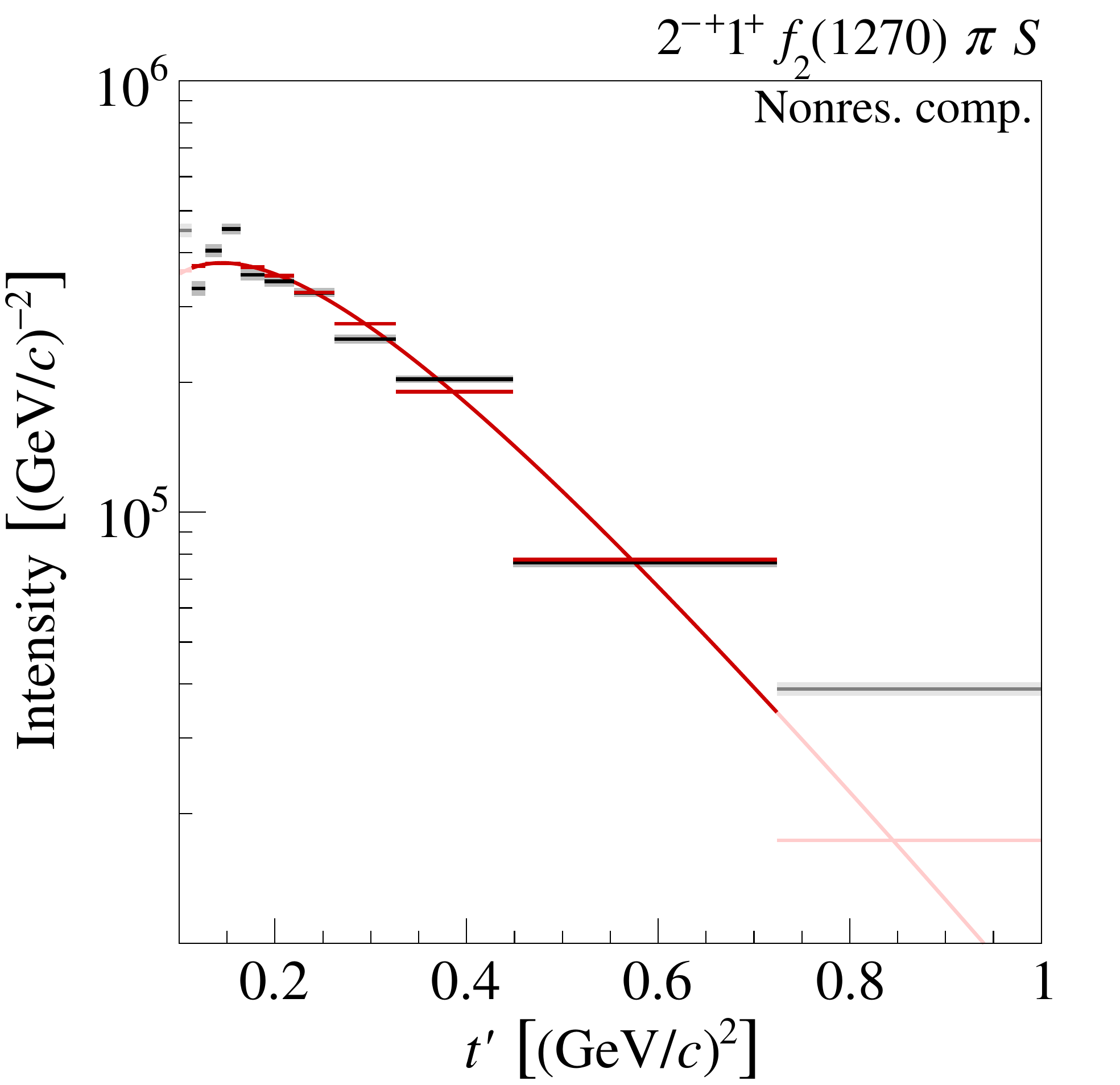}%
    \label{fig:tprim_2mp_m1_f2_S_nonres}%
  }%
  \caption{Similar to \cref{fig:method:tp:examplespectrum}, but
    showing the \tpr spectra of the components in the two
    $2^{-+} \PfTwo \pi S$ waves as given by
    \cref{eq:tprim-dependence}: (top row) $\Mrefl = 0^+$ wave and
    (bottom row) $\Mrefl = 1^+$ wave; (first column) \PpiTwo
    component, (second column) \PpiTwo[1880] component, (third column)
    \PpiTwo[2005] component, and (fourth column) nonresonant
    components.  The red curves and horizontal lines represent fits
    using \cref{eq:slope-parametrization}.}
  \label{fig:tprim_2mp_1}
\end{wideFigureOrNot}

\begin{figure}[tbp]
  \centering
  \subfloat[][]{%
    \includegraphics[width=\twoPlotWidth]{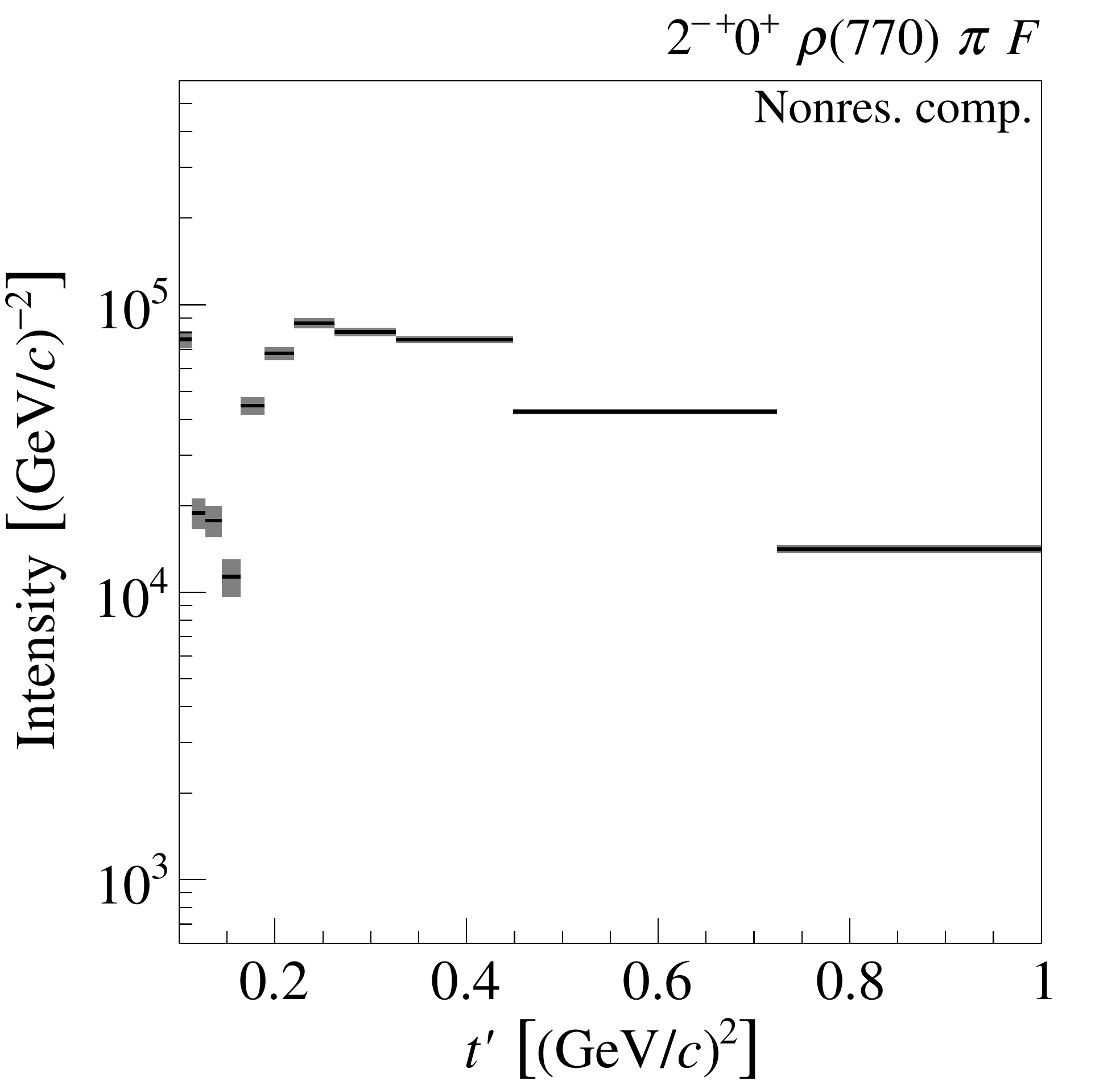}%
    \label{fig:tprim_2mp_rho_nonres}%
  }%
  \newLineOrHspace{\twoPlotSpacing}%
  \subfloat[][]{%
    \includegraphics[width=\twoPlotWidth]{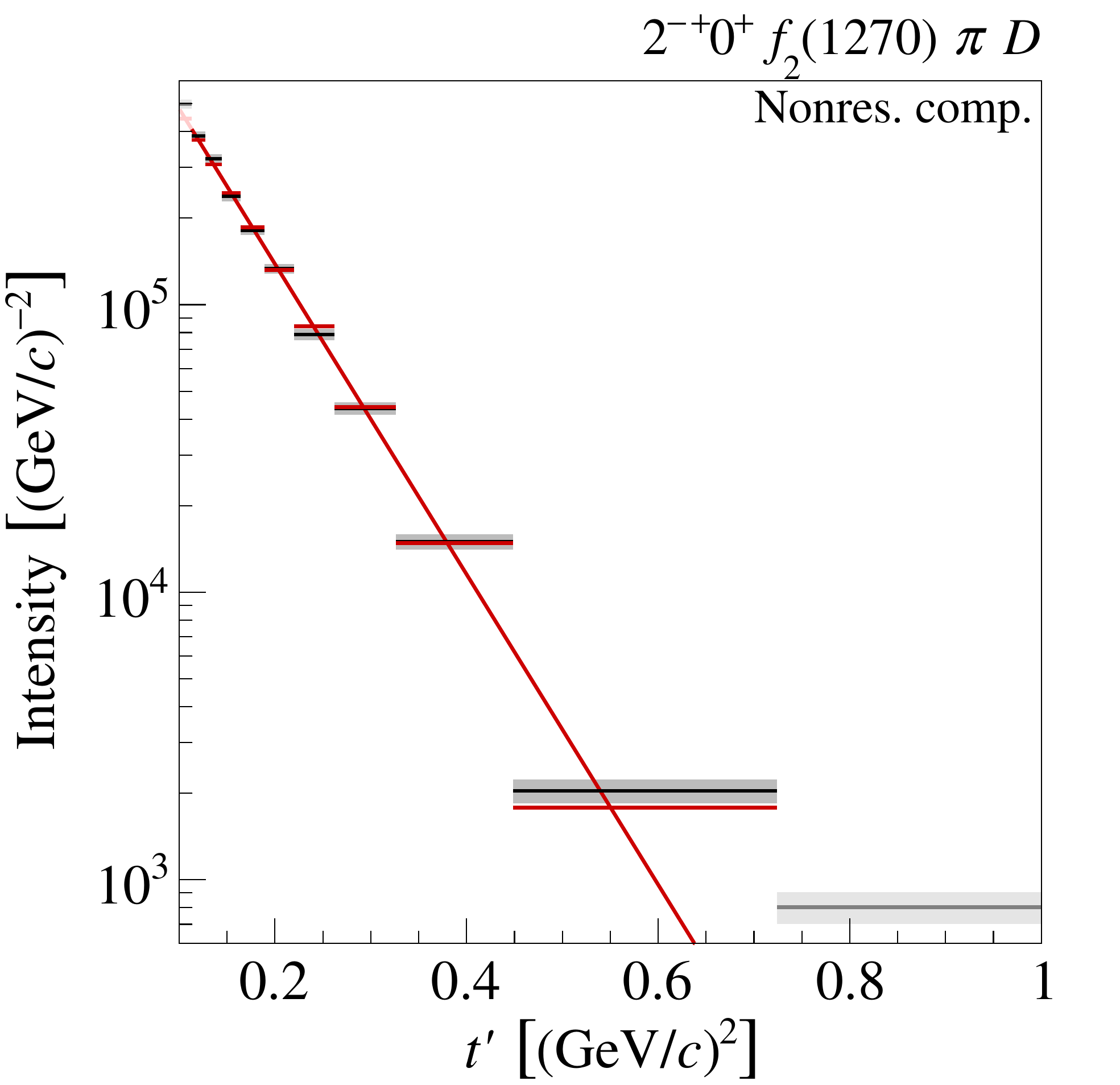}%
    \label{fig:tprim_2mp_f2_D_nonres}%
  }%
  \caption{Similar to \cref{fig:method:tp:examplespectrum}, but
    showing the \tpr spectra of the nonresonant components
    \subfloatLabel{fig:tprim_2mp_rho_nonres}~in the
    \wave{2}{-+}{0}{+}{\Prho}{F} wave and
    \subfloatLabel{fig:tprim_2mp_f2_D_nonres}~in the
    \wave{2}{-+}{0}{+}{\PfTwo}{D} wave.  The red curve and horizontal
    lines in \subfloatLabel{fig:tprim_2mp_f2_D_nonres} represent a fit
    using \cref{eq:slope-parametrization}.}
  \label{fig:tprim_2mp_2}
\end{figure}

If we do not constrain the coupling amplitudes via
\cref{eq:method:branchingdefinition} and thus allow the resonance
components to have different \tpr dependences [\StudyT; see
\cref{sec:systematics}], the extracted \tpr spectra agree in general
less with the simple model of \cref{eq:slope-parametrization}.  The
components of the \wave{2}{-+}{1}{+}{\PfTwo}{S} and
\wave{2}{-+}{0}{+}{\PfTwo}{D} waves show similar \tpr spectra with
slope parameters that deviate by at most \SI{2}{\perGeVcsq} from those
of the main fit.  This is also true for the \PpiTwo component in the
other two $2^{-+}$ waves, the \PpiTwo[1880] component in the
\wave{2}{-+}{0}{+}{\Prho}{F} wave, and the \PpiTwo[2005] component in
the \wave{2}{-+}{0}{+}{\PfTwo}{S} wave.  However, in the latter wave,
the slope parameter of the \PpiTwo[1880] becomes almost twice as large
and thus inconsistent with the \PpiTwo[1880] slope parameters in the
other three waves.  In the $\Prho \pi F$ wave, the \PpiTwo[2005] \tpr
spectrum changes drastically and becomes similar to the \tpr spectrum
of the nonresonant component in the main fit.  In turn, the \tpr
spectrum of the nonresonant component becomes steeper.  In addition to
the \tpr spectra, also the resonance parameters of \PpiTwo[1880] and
\PpiTwo[2005] change in \StudyT.  The \PpiTwo[1880] becomes
\SI{29}{\MeVcc} wider, whereas the \PpiTwo[2005] becomes
\SI{75}{\MeVcc} narrower.  The results of this study indicate that
without the constraint of \cref{eq:method:branchingdefinition}, the
relative intensities of the three \PpiTwo* states and the nonresonant
components are not well constrained by the data.  A possible reason
for this behavior is that our approach to model the partial-wave
amplitudes as a sum of Breit-Wigner amplitudes might not be a good
approximation anymore because of the considerable overlap of the three
\PpiTwo* resonances.  Applying more advanced models is the topic of
future research~\cite{Jackura:2016llm,Mikhasenko:2017jtg}.

\subsubsection{Discussion of results on $2^{-+}$ resonances}
\label{sec:twoMP_discussion}

We observe three distinct resonances with $\JPC = 2^{-+}$ in our data
set, which are clearly identified owing to their different production
characteristic and decay paths.  The \PpiTwo appears as a dominant
peak with associated phase motion in the $\Prho \pi F$ and the two
$\PfTwo \pi S$ waves with $\Mrefl = 0^+$ and $1^+$.  The strongest
signal for the \PpiTwo[1880] appears in the $\PfTwo \pi D$ wave in the
form of a dominant peak with associated phase motion.  The relative
intensity of the \PpiTwo[1880] in the other three $2^{-+}$ waves is
small, which is in particular true for the
\wave{2}{-+}{0}{+}{\PfTwo}{S} wave.  The \PpiTwo[2005] appears as
high-mass shoulders in the \wave{2}{-+}{0}{+}{\Prho}{F},
\wave{2}{-+}{0}{+}{\PfTwo}{S}, and \wave{2}{-+}{0}{+}{\PfTwo}{D}
waves, which due to the shallower \tpr slope of the \PpiTwo[2005] are
more pronounced in the highest \tpr bin.  In the $\Prho \pi F$ wave,
this shoulder even turns into a clear peak at large \tpr.  The
\PpiTwo[2005] contribution is significantly larger than that of the
\PpiTwo[1880] in the $\Prho \pi F$ wave and in the $\PfTwo \pi S$ wave
with $M = 0$.  In the $\PfTwo \pi S$ wave with $M = 1$, the two
contributions are of comparable strength.

The parameters of the \PpiTwo are well known.  The PDG quotes world
averages for its mass and width of
$m_{\PpiTwo} = \SI{1672.2(30)}{\MeVcc}$ and
$\Gamma_{\PpiTwo} = \SI{260(9)}{\MeVcc}$,
respectively~\cite{Patrignani:2016xqp}.  We find a mass of
$m_{\PpiTwo} = \SIaerrSys{1642}{12}{1}{\MeVcc}$, which is smaller by
\SI{30}{\MeVcc}, and a width of
$\Gamma_{\PpiTwo} = \SIaerrSys{311}{12}{23}{\MeVcc}$, which is larger
by \SI{51}{\MeVcc}.  However, within uncertainties our result is
consistent with our previous measurement of the \threePi final state
diffractively produced on a solid lead target~\cite{alekseev:2009aa}.
It is interesting to note that a study with a reduced set of only
11~waves, from which all $2^{-+}$ waves but the
\wave{2}{-+}{0}{+}{\PfTwo}{S} wave have been removed, yields \PpiTwo
resonance parameters of $m_{\PpiTwo} = \SI{1663}{\MeVcc}$ and
$\Gamma_{\PpiTwo} = \SI{256}{\MeVcc}$, which are close to the world
average.  In that fit, the \wave{2}{-+}{0}{+}{\PfTwo}{S} amplitude was
described in a smaller mass range from \SIrange{1.4}{1.9}{\GeVcc}
using the \PpiTwo as the only $2^{-+}$ resonance component.

The \PpiTwo[1880] appears to be experimentally well established
according to the PDG, although its measured mass and width values vary
considerably.  The PDG lists no observation for the decay
$\PpiTwo[1880] \to 3\pi$.  The PDG world averages of the \PpiTwo[1880]
parameters are $m_{\PpiTwo[1880]} = \SI{1895(16)}{\MeVcc}$ and
$\Gamma_{\PpiTwo[1880]} =
\SI{235(34)}{\MeVcc}$~\cite{Patrignani:2016xqp}.  While we find a
value for the \PpiTwo[1880] width of
$\Gamma_{\PpiTwo[1880]} = \SIaerrSys{246}{33}{28}{\MeVcc}$ that is
compatible with the world average, our mass value of
$m_{\PpiTwo[1880]} = \SIaerrSys{1847}{20}{3}{\MeVcc}$ is
\SI{48}{\MeVcc} smaller.  The four measurements listed by the PDG fall
into two subsets.  The first consists of two measurements with lower
masses $m_{\PpiTwo[1880]} \leq \SI{1880}{\MeVcc}$ and smaller widths
$\Gamma_{\PpiTwo[1880]} \leq
\SI{255}{\MeVcc}$~\cite{anisovich:2001hj,Lu:2004yn}.  Our estimate of
the \PpiTwo[1880] parameters is within uncertainties compatible with
these two measurements, although there is some disagreement with the
extremely small width estimate of \SIerrs{146}{17}{62}{\MeVcc} from
\refCite{Lu:2004yn}.  The other two measurements with larger masses
$m_{\PpiTwo[1880]} \geq \SI{1929}{\MeVcc}$ and larger widths
$\Gamma_{\PpiTwo[1880]} \geq
\SI{306}{\MeVcc}$~\cite{kuhn:2004en,eugenio:2008zza} are better
compatible with our estimates for the \PpiTwo[2005] parameters.

The \PpiTwo[2005] is listed by the PDG only as a \enquote{further
  state} with two observations~\cite{Patrignani:2016xqp}.  It was
claimed in an analysis by the BNL~E852 experiment of the
$\omega \pi^0 \pi^-$ final state diffractively produced on a proton
target~\cite{Lu:2004yn} and in two analyses based on \ppbar
annihilation data from the Crystal Barrel experiment: a combined
analysis of $3\pi^0$, $\pi^0 \eta$, and $\pi^0 \eta'$ final
states~\cite{anisovich:2001pn} and an analysis of
$\eta \eta \pi^0$~\cite{anisovich:2001pp}.  The mass range explored in
\ppbar annihilations in flight starts only around \SI{1.95}{\GeVcc}
and thus covers only the high-mass part of the \PpiTwo[2005]
resonance.  Within uncertainties, our estimate for the \PpiTwo[2005]
parameters, $m_{\PpiTwo[2005]} = \SIaerrSys{1962}{17}{29}{\MeVcc}$ and
$\Gamma_{\PpiTwo[2005]} = \SIaerrSys{371}{16}{120}{\MeVcc}$, is
compatible with either measurement.

In order to study the significance of the \PpiTwo[2005] signal in our
data, we have performed a systematic study, in which we omitted the
\PpiTwo[2005] from the fit model.  The minimum \chisq~value found in
this fit is \num{1.07}~times larger than the one of the main
fit.\footnote{Compared to the \num{722} free parameters of the main
  fit, this fit has \num{672} free parameters.}
\Cref{fig:no-pi2(2005)_chi2difference} shows the contributions from
the spin-density matrix elements to the \chisq~difference between this
and the main fit.  Without the \PpiTwo[2005], the model describes the
$2^{-+}$ intensity distributions and interference terms less well, in
particular for the \wave{2}{-+}{0}{+}{\PfTwo}{S} and
\wave{2}{-+}{0}{+}{\PfTwo}{D} waves.
\Cref{fig:no-pi2(2005)_intensities} shows that the high-mass shoulders
cannot be reproduced well.  Omitting the \PpiTwo[2005] component also
shifts some of the resonance parameters.  On the one hand, the
\PpiTwo[1880] becomes \SI{20}{\MeVcc} lighter and \SI{100}{\MeVcc}
wider, which would be contradictory to all previous measurements.  On
the other hand, the \PpiTwo parameters move closer to the PDG world
average.\footnote{The \PpiTwo becomes \SI{17}{\MeVcc} heavier and
  \SI{20}{\MeVcc} narrower.  Large changes are also observed for the
  \PaOne[1640], which becomes \SI{42}{\MeVcc} lighter and
  \SI{82}{\MeVcc} wider, and for the \PpiOne[1600], which becomes
  \SI{51}{\MeVcc} narrower.}

\begin{figure}[tbp]
  \centering
  \includegraphics[width=\linewidthOr{\twoPlotWidth}]{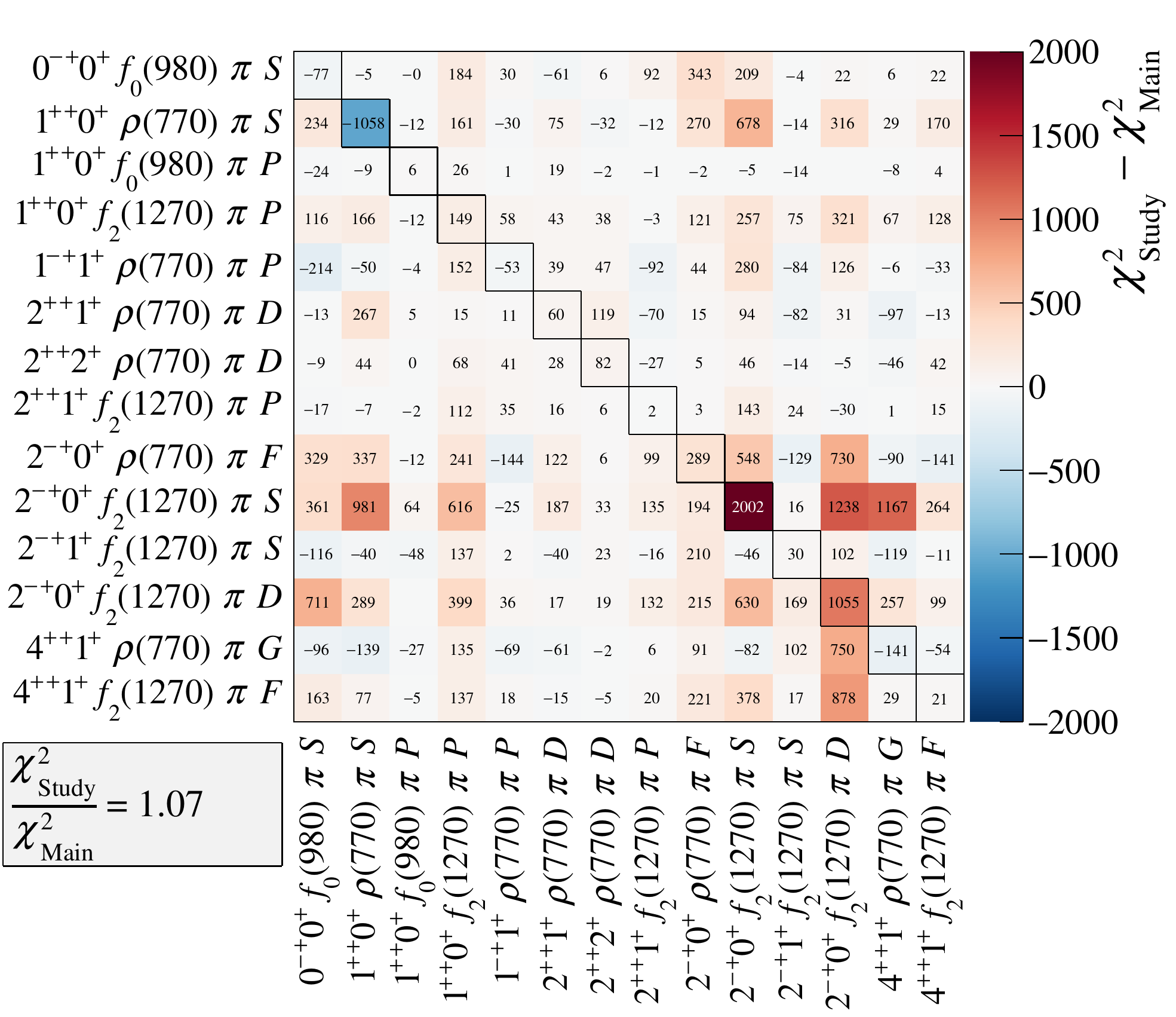}
  \caption{Similar to \cref{fig:DeckMC_chi2difference}, but for the
    study in which the \PpiTwo[2005] resonance was omitted from the
    fit model.}
  \label{fig:no-pi2(2005)_chi2difference}
\end{figure}

\begin{wideFigureOrNot}[tbp]
  \centering
  \subfloat[][]{%
    \includegraphics[width=\threePlotWidth]{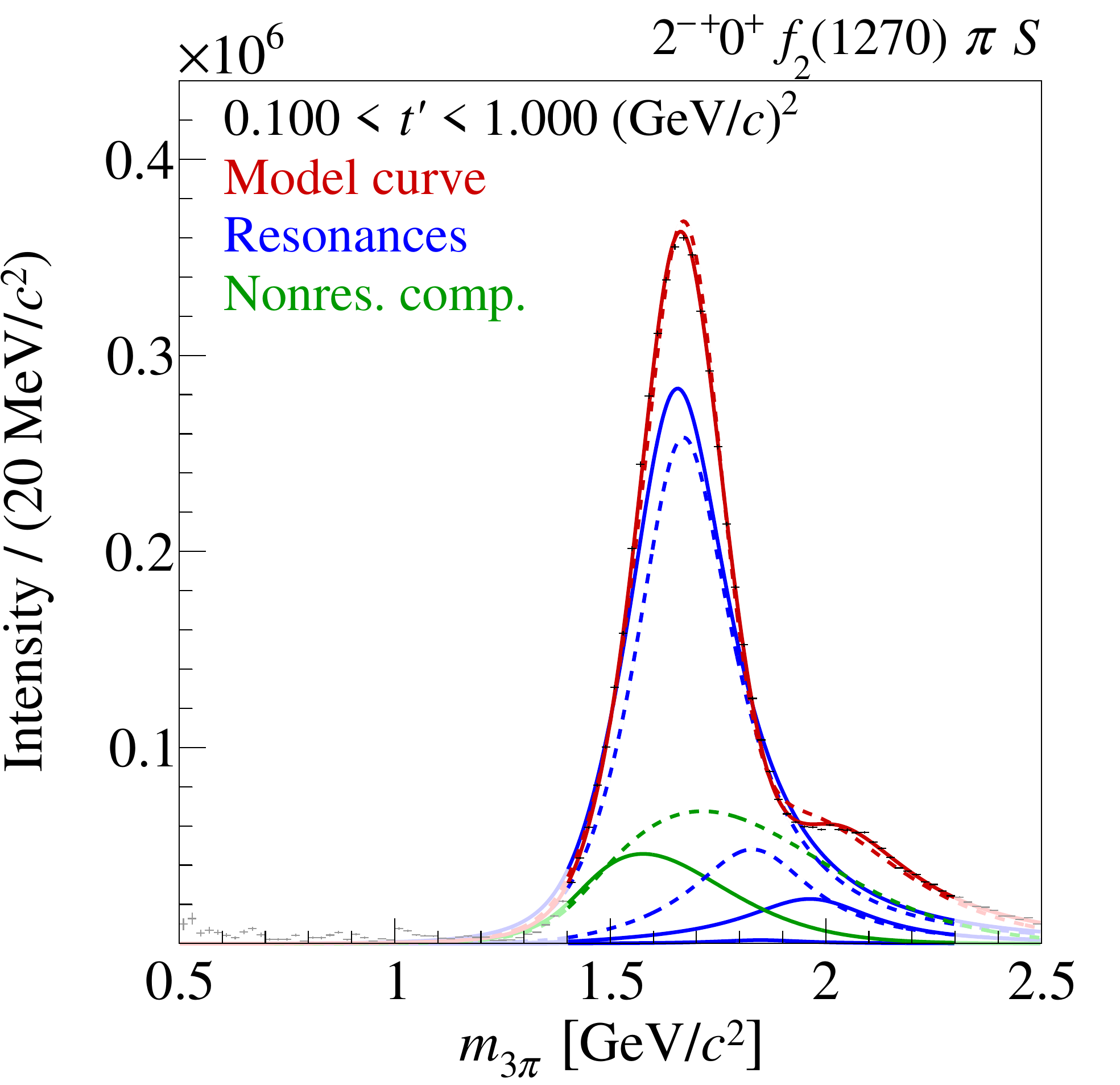}%
    \label{fig:no-pi2(2005)_intensity_2mp_m0_f2_S}%
  }%
  \hspace*{\threePlotSpacing}%
  \subfloat[][]{%
    \includegraphics[width=\threePlotWidth]{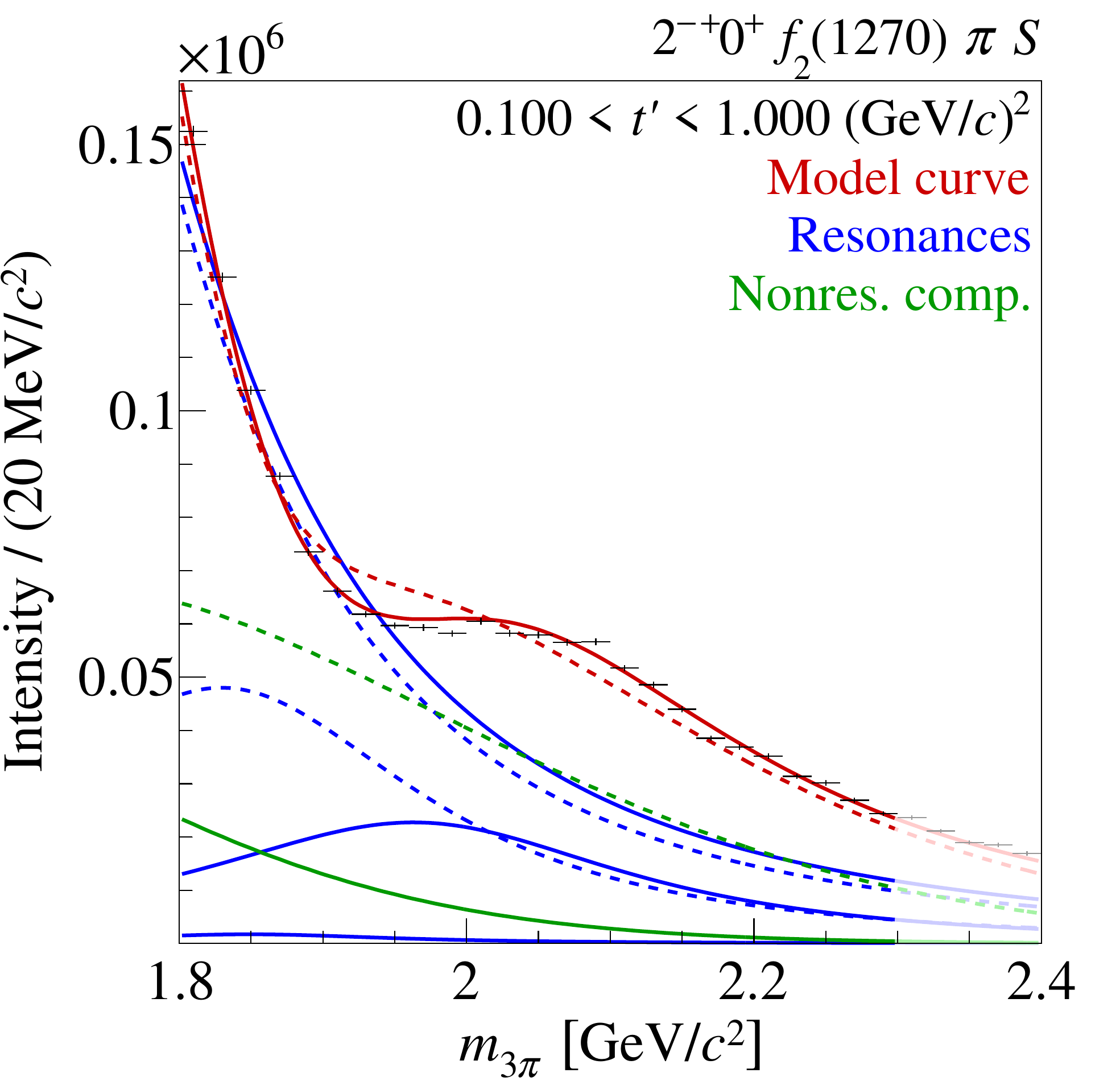}%
    \label{fig:no-pi2(2005)_intensity_2mp_m0_f2_S_zoom}%
  }%
  \hspace*{\threePlotSpacing}%
  \subfloat[][]{%
    \includegraphics[width=\threePlotWidth]{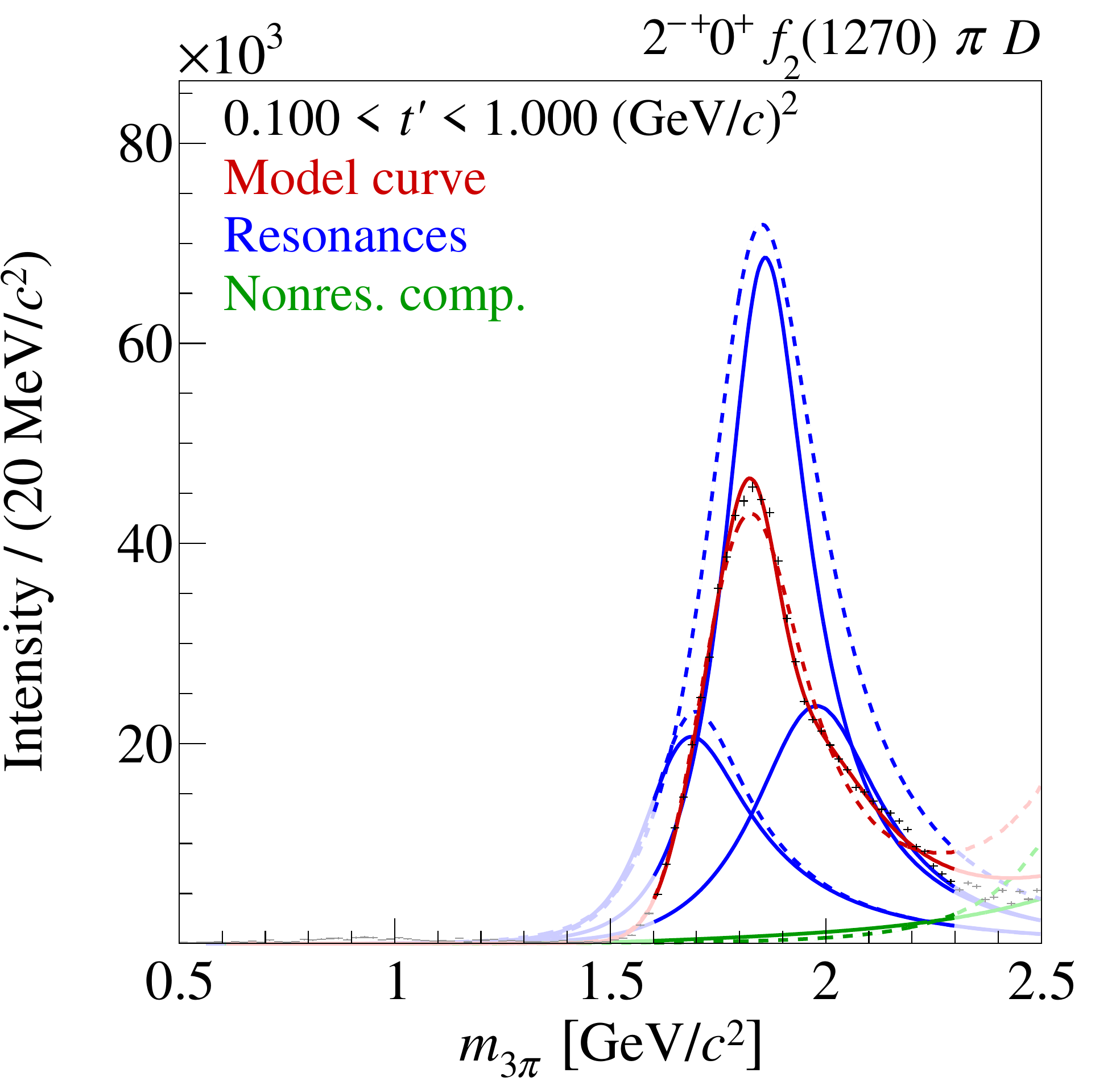}%
    \label{fig:no-pi2(2005)_intensity_2mp_f2_D}%
  }%
  \caption{\tpr-summed intensities of
    \subfloatLabel{fig:no-pi2(2005)_intensity_2mp_m0_f2_S}~the
    \wave{2}{-+}{0}{+}{\PfTwo}{S} wave and
    \subfloatLabel{fig:no-pi2(2005)_intensity_2mp_f2_D}~the
    \wave{2}{-+}{0}{+}{\PfTwo}{D} wave with the result of the main fit
    (continuous curves) and of the fit in which the \PpiTwo[2005]
    resonance was omitted from the fit model (dashed curves).  The
    model and the wave components are represented as in
    \cref{fig:intensity_phases_2mp}. In~\subfloatLabel{fig:no-pi2(2005)_intensity_2mp_m0_f2_S_zoom},
    a zoomed view of the high-mass region in
    \subfloatLabel{fig:no-pi2(2005)_intensity_2mp_m0_f2_S} is shown.}
  \label{fig:no-pi2(2005)_intensities}
\end{wideFigureOrNot}

In addition to \PpiTwo, \PpiTwo[1880], and \PpiTwo[2005], the PDG
lists the \PpiTwo[2100] as \enquote{omitted from summary
  table}~\cite{Patrignani:2016xqp}.  The PDG entry is based on two
observations reported by the ACCMOR~\cite{daum:1980ay} and the VES
experiments~\cite{amelin:1995gu} in the diffractively produced
\threePi final state.  The \PpiTwo[2100] thus requires further
experimental confirmation.  It is close in mass to the \PpiTwo[2005],
but has a much larger width of \SI{625(50)}{\MeVcc}.  In the ACCMOR
analysis, the intensity distributions of the
\wave{2}{-+}{0}{+}{\pipiS}{D}, \wave{2}{-+}{0}{+}{\Prho}{P},
\wave{2}{-+}{0}{+}{\PfTwo}{S}, and \wave{2}{-+}{0}{+}{\PfTwo}{D} waves
were fit together with selected relative phases of these waves using a
model with two $2^{-+}$ resonances, \PpiTwo and \PpiTwo[2100], which
was based on the $K$-matrix approach~\cite{daum:1980ay}.  In this
model, the dominant peak at \SI{1.8}{\GeVcc} in the $\PfTwo \pi D$
wave is explained as a constructive interference of the two resonance
components.  The VES analysis is similar and confirms this
finding~\cite{amelin:1995gu}.  It is worth noting that in both
analyses rather high \PpiTwo masses $\geq \SI{1710}{\MeVcc}$ are
found.  Our data exhibit similar features as the ACCMOR and VES data.
In particular, considering the uncertainties it is likely that our
\PpiTwo[2005] signal corresponds to the \PpiTwo[2100] measurements
discussed above, although the width estimates differ significantly.
The main difference of our analysis is that we include different waves
in the resonance-model fit.  We did not include the $\Prho \pi P$ wave
because it exhibits a sizable and not well understood low-mass
enhancement below the \PpiTwo region [see Fig.~57(e) in
\refCite{Adolph:2015tqa_suppl}].  The $\pipiS \pi D$ and
$\PfZero[980] \pi D$ waves have complicated intensity distributions
[see Figs.~25(c) and 25(d) in \refCite{Adolph:2015tqa}].  At low
\mThreePi, both partial-wave intensities are sensitive to the wave set
that is used in the mass-independent analysis.  They also may be
affected by the particular parametrizations chosen for the \pipiS and
\PfZero[980] isobar amplitudes.  A less model-dependent analysis, in
which the amplitude of the \twoPi $S$-wave subsystem was extracted
from the data instead of using a parametrization with fixed functional
form, shows a clear correlation of a peak in the region of
$\mThreePi = \SI{1.9}{\GeVcc}$, which is presumably the \PpiTwo[1880],
with a peak in the \PfZero[980] region in the \twoPi mass spectrum
[see Figs.~40 and 43(c) in \refCite{Adolph:2015tqa}].  However, shape,
position, and strength of the observed peak structure in the \PpiTwo
region depend strongly on \tpr, which hints at large contributions
from nonresonant components.

As discussed above, the four $2^{-+}$ waves selected for the
resonance-model fit are not well described if we include only two
$2^{-+}$ Breit-Wigner resonances in the model.  In particular, we do
not observe solutions similar to those found by ACCMOR or VES with a
second resonance in the \SI{2.1}{\GeVcc} region.  It is therefore
unlikely that the \PpiTwo[1880] signal is caused by a constructive
interference of the other two resonances.  It is also unlikely that
the \PpiTwo[1880] signal arises from an interference with a
nonresonant component since the \tpr spectrum of the \PpiTwo[1880]
exhibits a resonancelike behavior [see
\cref{fig:tprim_2mp_m0_f2_S_pi2_1880,fig:tprim_2mp_m1_f2_S_pi2_1880}].

The PDG lists another potential higher excited \PpiTwo* state, the
\PpiTwo[2285], as a \enquote{further state}~\cite{Patrignani:2016xqp}.
It was reported with the parameters
$m_{\PpiTwo[2285]} = \SIerrs{2285}{20}{25}{\MeVcc}$ and
$\Gamma_{\PpiTwo[2285]} = \SIerrs{250}{20}{25}{\MeVcc}$ by the authors
of \refCite{Anisovich:2010nh} in an analysis of the
$\eta \pi^0 \pi^0 \pi^0$ final state produced in \ppbar annihilations
in flight, which was based on data from the Crystal Barrel experiment.
Although we do not see clear resonance signals of heavy \PpiTwo*
states in the mass range from \SIrange{2200}{2500}{\MeVcc} in the
analyzed waves, we cannot exclude that the observed deviations of the
model from the data at high masses, in particular in the
\wave{2}{-+}{0}{+}{\Prho}{F} wave, are due to additional excited
\PpiTwo* states.

The mass of the \PpiTwo agrees well with the quark-model prediction
for the \PpiTwo* ground state by Godfrey and
Isgur~\cite{Godfrey:1985xj}.  The mass of the \PpiTwo[2005] agrees
with the prediction for the first radial excitation of the \PpiTwo*.
However, the \PpiTwo[1880] does not fit into this picture.  The
interpretations of the \PpiTwo[1880] are manifold.  It has been
interpreted as a supernumerous exotic meson with conventional quantum
numbers.  It has in particular been considered as a good candidate for
a hybrid meson by the authors of
\refsCite{anisovich:2001hj,klempt:2007cp}.  In contrast, Li and Zhou
argue in \refCite{Li:2008xy} that the observed decay width of
approximately \SI{235}{\MeVcc} is too large for a pure hybrid state,
for which a smaller width of rather \SI{100}{\MeVcc} would be
expected.  In addition, the dominant decay into the $\PfTwo \pi D$
wave and the small coupling to the $\PfTwo \pi S$ wave that we observe
in our data contradict the hybrid-meson interpretation based on model
calculations for the decay of such objects performed by Page, Swanson,
and Szczepaniak in \refCite{page:1998gz}, which predict the opposite
behavior for a hybrid resonance.  Li and Zhou argue that the
\PpiTwo[1880] decay pattern is more similar to model predictions for
the first radial excitation of the conventional
\PpiTwo*~\cite{Li:2008xy}.  However, they do not exclude a possible
small admixture of a hybrid state.

In an alternative approach, Dudek and Szczepaniak have proposed in
\refCite{Dudek:2006ud} that the \SI{1.65}{\GeVcc} peak in the
$\PfTwo \pi S$ wave and the \SI{1.8}{\GeVcc} peak in the
$\PfTwo \pi D$ wave are caused by the same \PpiTwo* ground-state
resonance.  The seemingly different structures are caused by
interference of this resonance with a type of nonresonant background
originally proposed by Deck~\cite{deck:1964hm}, which is much stronger
in the $\PfTwo \pi S$ wave.  In order to explain the phase motions,
this model requires a second \PpiTwo* resonance at a higher mass as in
the ACCMOR and VES analyses discussed above.  Their hypothesis may be
tested by including the \tpr dependence and the population of the $M$
substates of the Deck amplitude.
 %
%
%

\subsection{$\JPC = 1^{++}$ resonances}
\label{sec:onePP}

\subsubsection{Results on $1^{++}$ resonances}
\label{sec:onePP_results}

The resonance-model fit includes three waves with $\JPC = 1^{++}$.
The \wave{1}{++}{0}{+}{\Prho}{S} wave is the most dominant wave in the
88-wave set with a relative intensity of \SI{32.7}{\percent}.  The
\wave{1}{++}{0}{+}{\PfZero[980]}{P} and \wave{1}{++}{0}{+}{\PfTwo}{P}
waves are about 2~orders of magnitude less intense with relative
intensities of \SI{0.4}{\percent} and \SI{0.3}{\percent},
respectively.  The intensity distributions of the three $1^{++}$
waves, as shown in
\cref{fig:intensity_1pp_rho_tbin1_log,fig:intensity_1pp_f0_tbin1,fig:intensity_1pp_f2_tbin1}
for the lowest \tpr bin and in
\cref{fig:intensity_1pp_rho_tbin11_log,fig:intensity_1pp_f0_tbin11,fig:intensity_1pp_f2_tbin11}
for the highest \tpr bin, are surprisingly different.

\ifMultiColumnLayout{\begin{figure*}[t]}{\begin{figure}[p]}
  \centering
  \subfloat[][]{%
    \includegraphics[width=\fourPlotWidth]{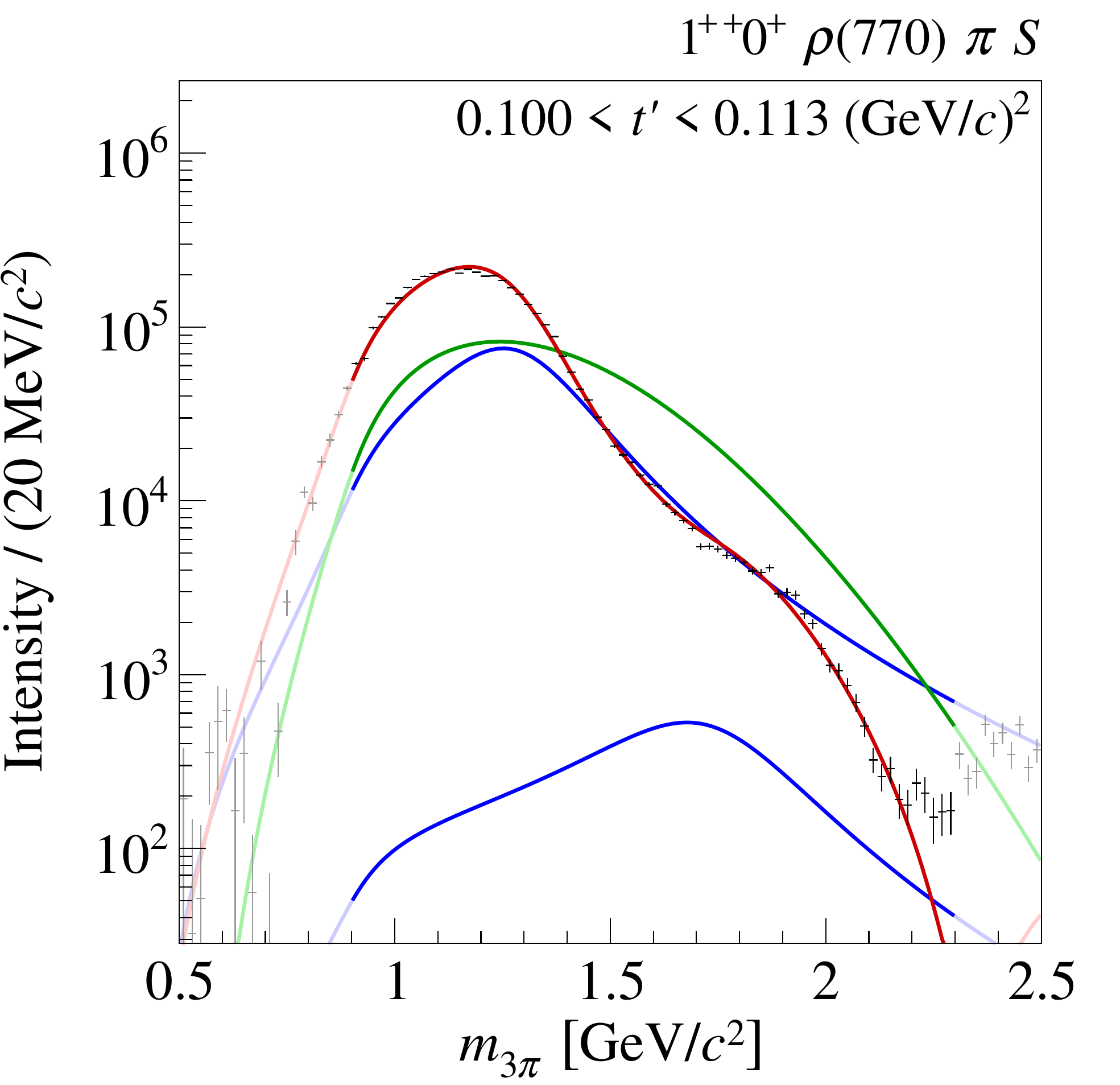}%
    \label{fig:intensity_1pp_rho_tbin1_log}%
  }%
  \hspace*{\fourPlotSpacing}%
  \subfloat[][]{%
    \includegraphics[width=\fourPlotWidth]{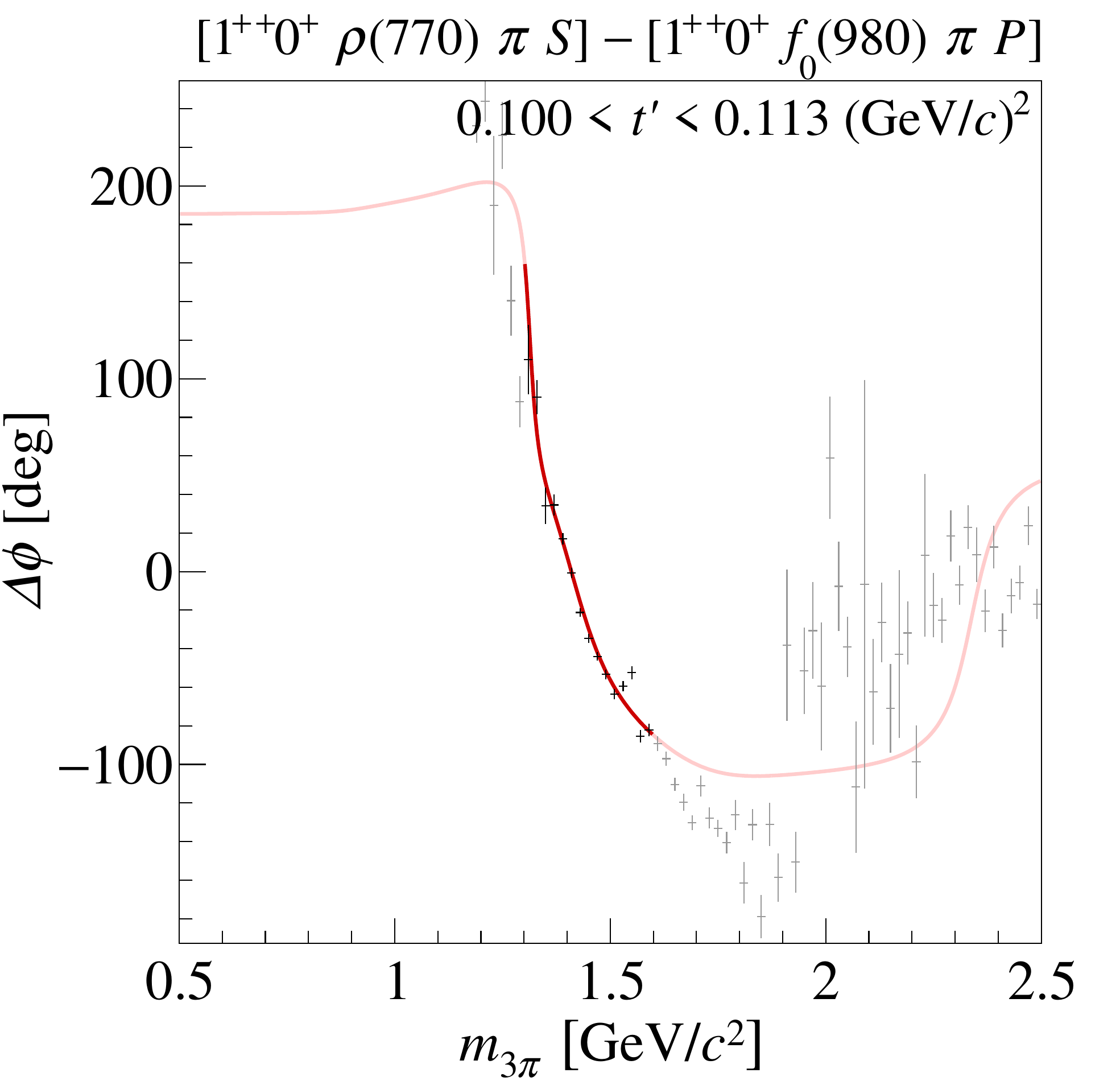}%
    \label{fig:phase_1pp_rho_1pp_f0_tbin1}%
  }%
  \hspace*{\fourPlotSpacing}%
  \subfloat[][]{%
    \includegraphics[width=\fourPlotWidth]{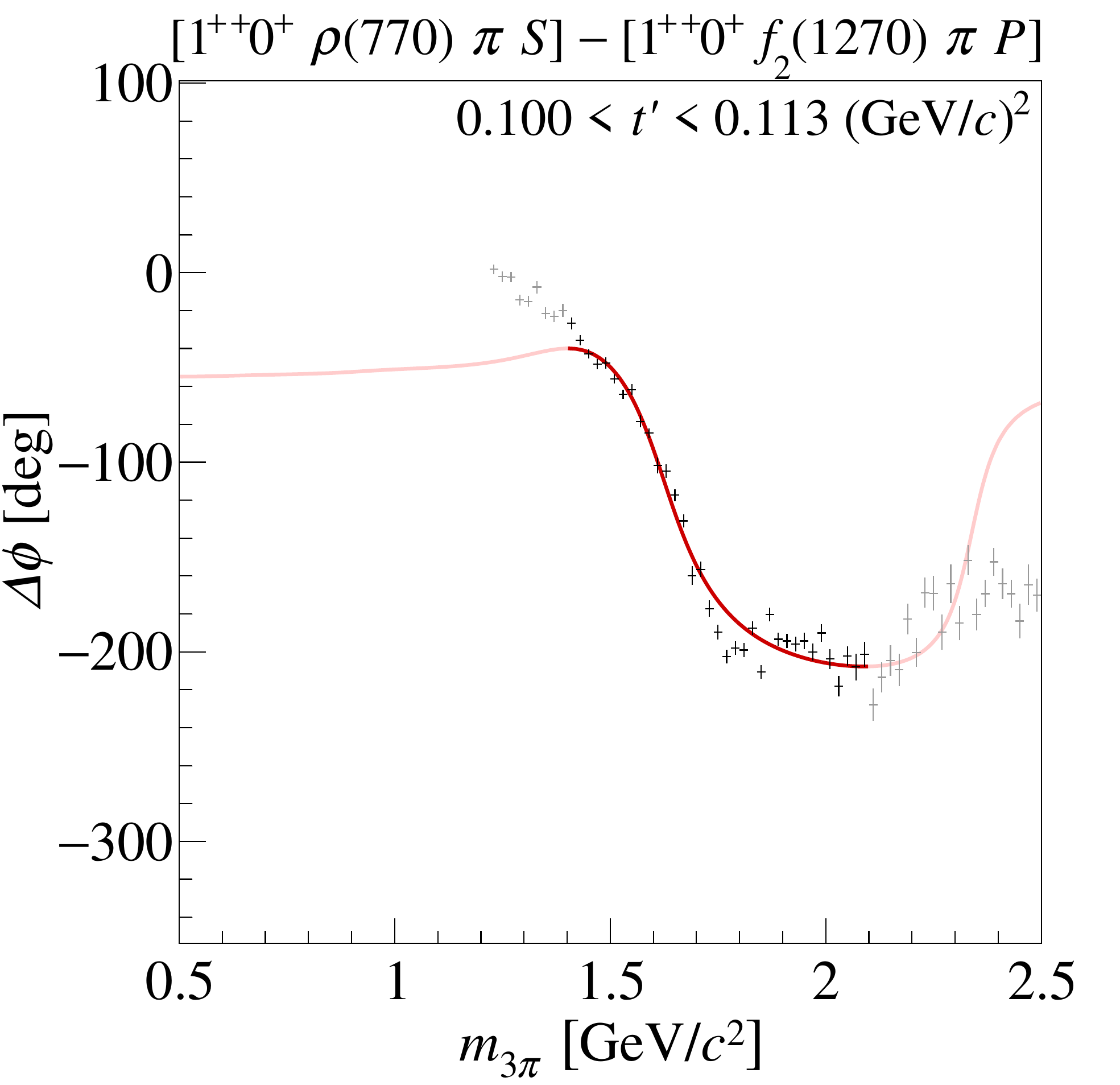}%
    \label{fig:phase_1pp_rho_1pp_f2_tbin1}%
  }%
  \hspace*{\fourPlotSpacing}%
  \subfloat[][]{%
    \includegraphics[width=\fourPlotWidth]{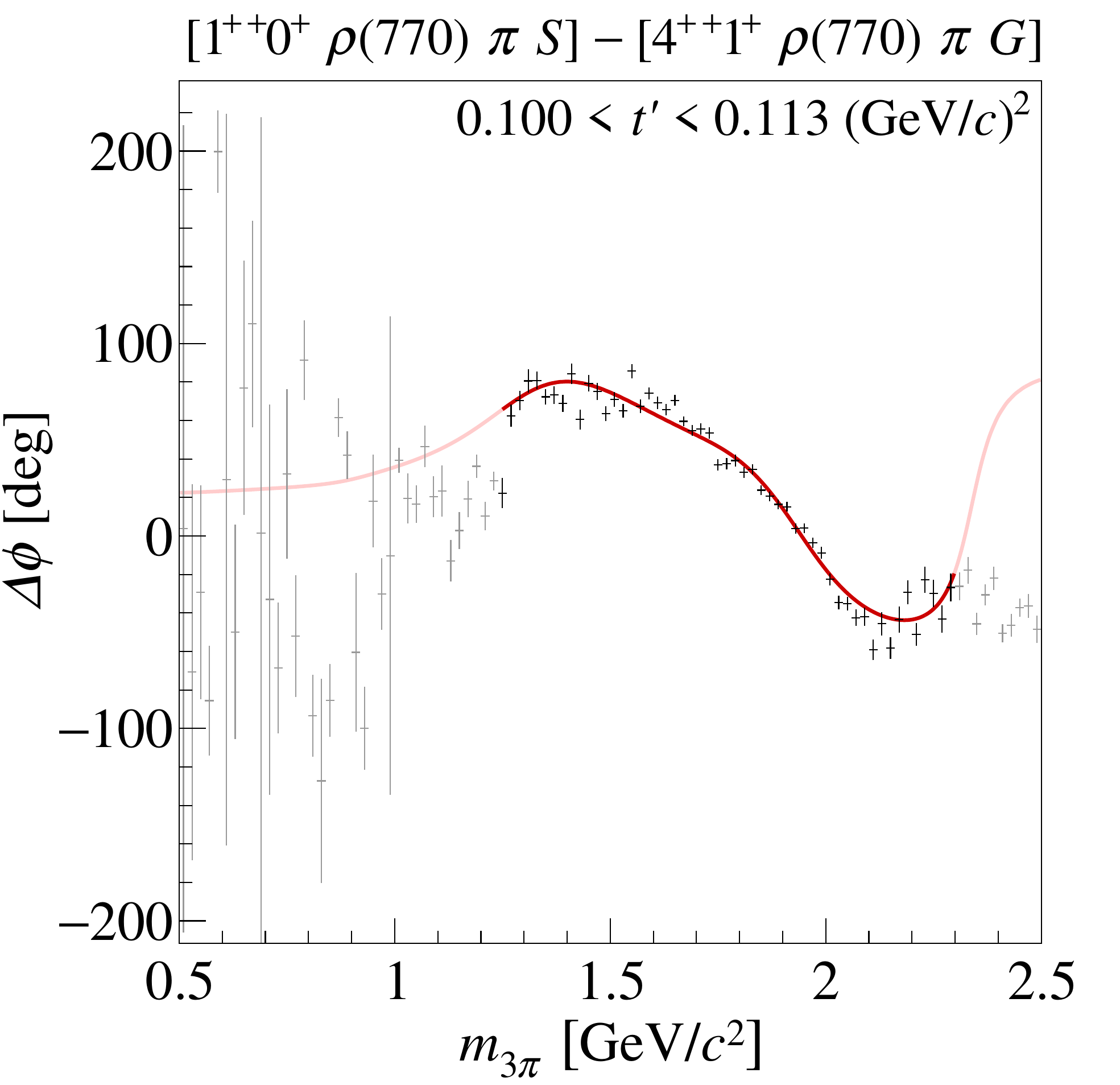}%
    \label{fig:phase_1pp_rho_4pp_rho_tbin1}%
  }%
  \\
  \hspace*{\fourPlotWidth}%
  \hspace*{\fourPlotSpacing}%
  \subfloat[][]{%
    \includegraphics[width=\fourPlotWidth]{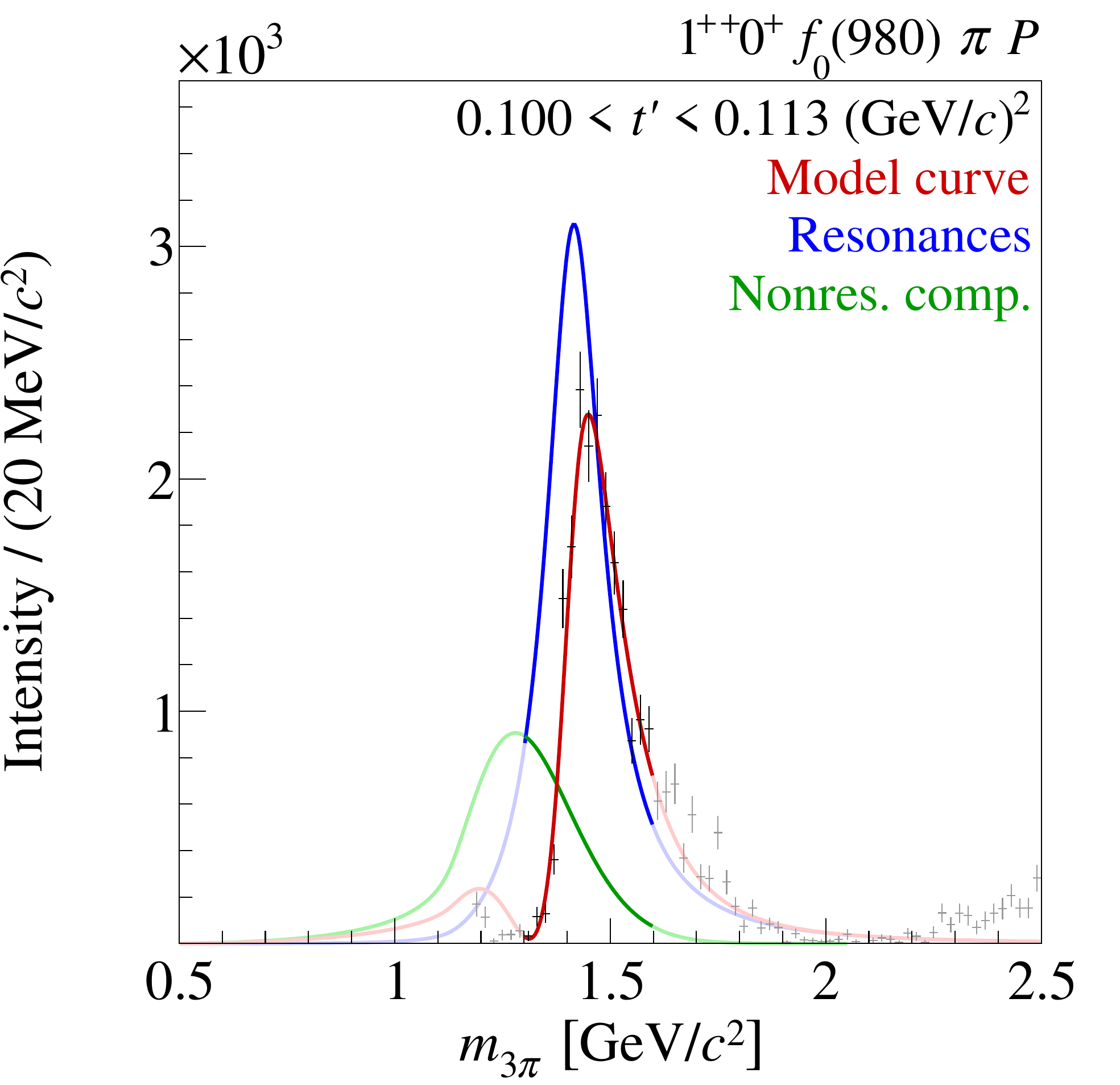}%
    \label{fig:intensity_1pp_f0_tbin1}%
  }%
  \hspace*{\fourPlotSpacing}%
  \subfloat[][]{%
    \includegraphics[width=\fourPlotWidth]{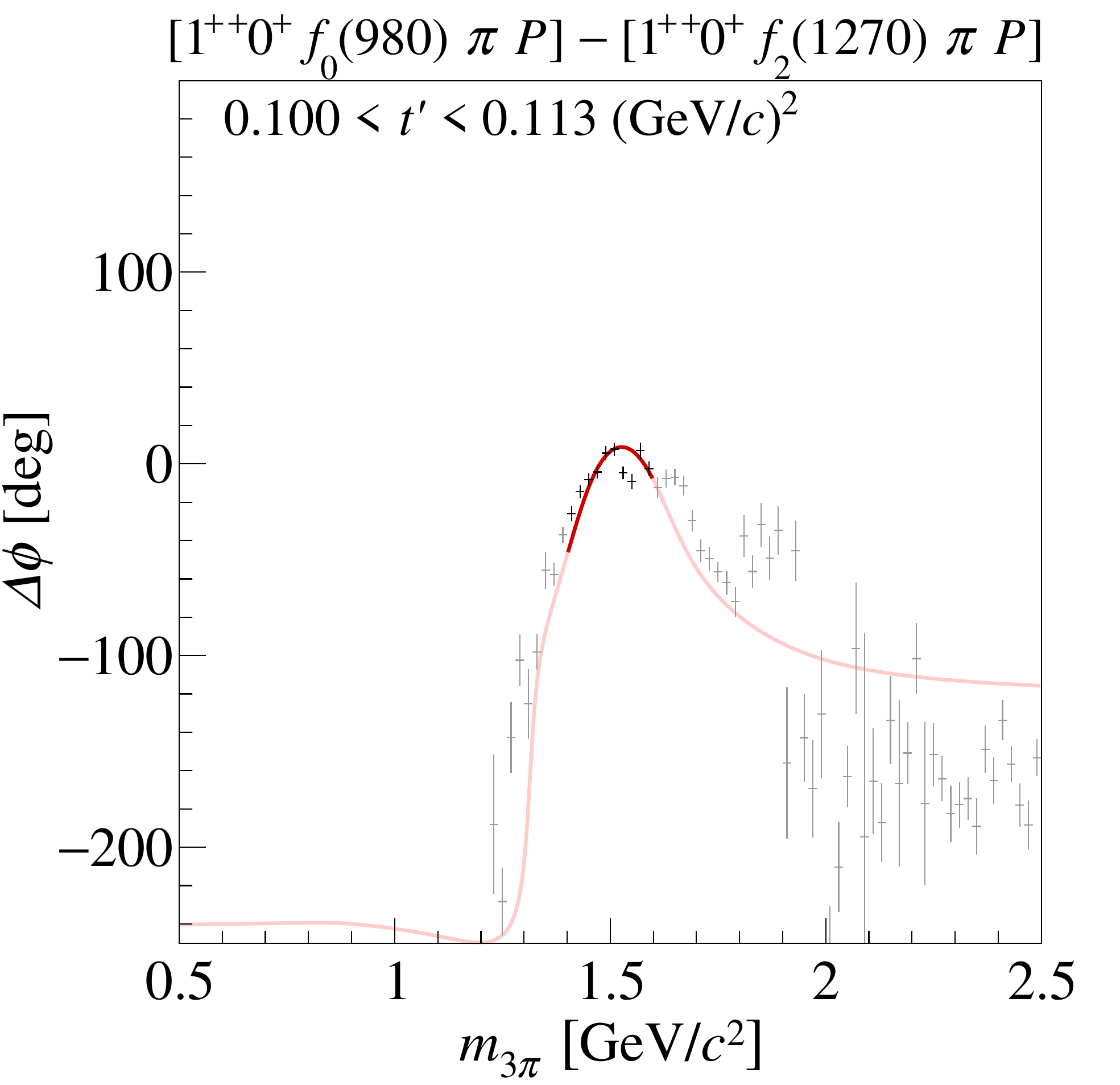}%
    \label{fig:phase_1pp_f0_1pp_f2_tbin1}%
  }%
  \hspace*{\fourPlotSpacing}%
  \subfloat[][]{%
    \includegraphics[width=\fourPlotWidth]{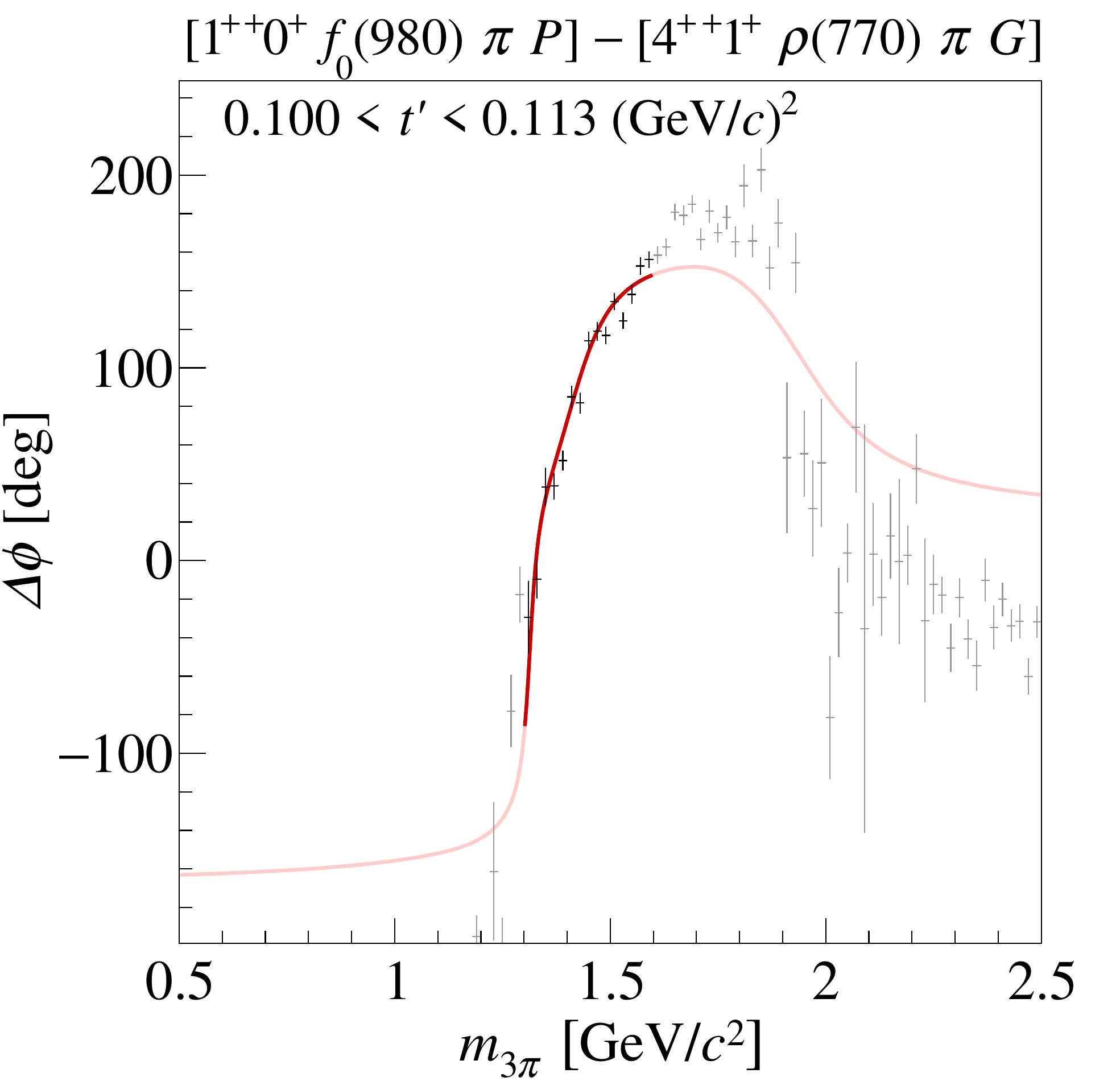}%
    \label{fig:phase_1pp_f0_4pp_rho_tbin1}%
  }%
  \\
  \hspace*{\fourPlotWidth}%
  \hspace*{\fourPlotSpacing}%
  \hspace*{\fourPlotWidth}%
  \hspace*{\fourPlotSpacing}%
  \subfloat[][]{%
    \includegraphics[width=\fourPlotWidth]{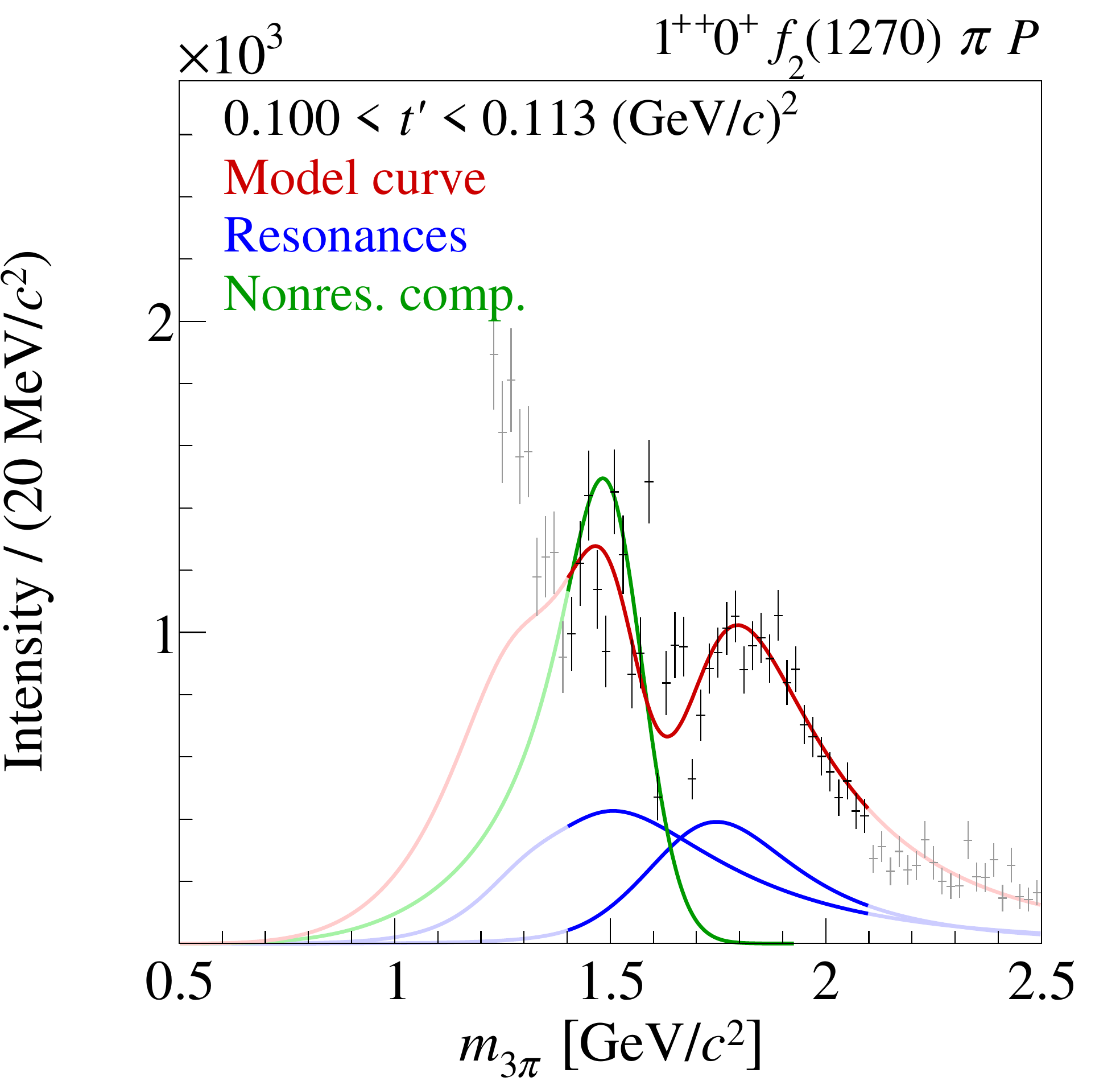}%
    \label{fig:intensity_1pp_f2_tbin1}%
  }%
  \hspace*{\fourPlotSpacing}%
  \subfloat[][]{%
    \includegraphics[width=\fourPlotWidth]{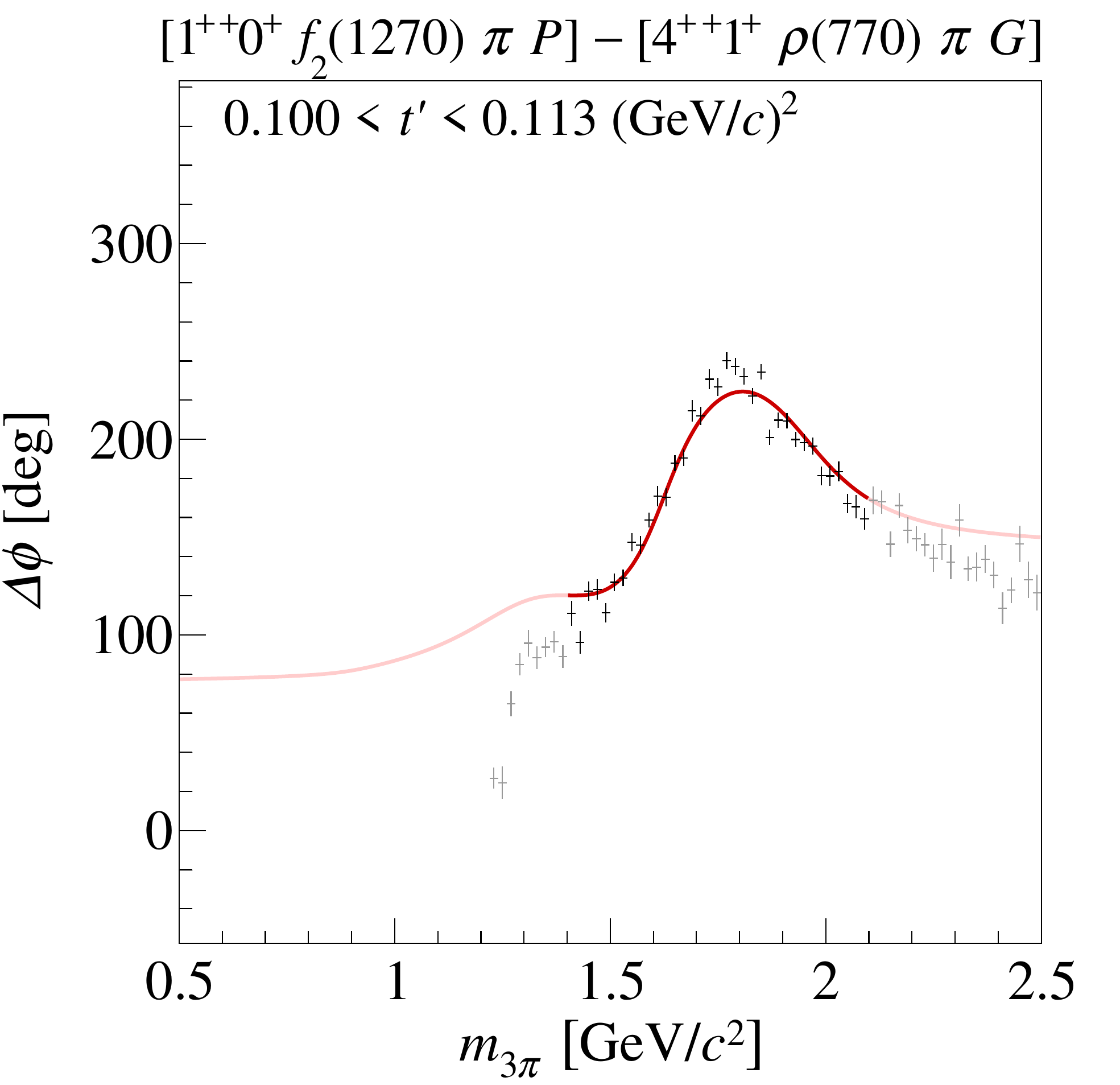}%
    \label{fig:phase_1pp_f2_4pp_rho_tbin1}%
  }%
  \caption{Amplitudes of the three $\JPC = 1^{++}$ waves in the lowest
    \tpr bin.
    \subfloatLabel{fig:intensity_1pp_rho_tbin1_log}~through~\subfloatLabel{fig:phase_1pp_rho_4pp_rho_tbin1}:
    Intensity distribution and relative phases for the
    \wave{1}{++}{0}{+}{\Prho}{S} wave.  Note that the intensity
    distribution in~\subfloatLabel{fig:intensity_1pp_rho_tbin1_log} is
    shown in logarithmic scale.
    \subfloatLabel{fig:phase_1pp_rho_4pp_rho_tbin1} Corresponds to
    \cref{fig:phase_4pp_rho_1pp_rho_tbin1}. \subfloatLabel{fig:intensity_1pp_f0_tbin1}~through~\subfloatLabel{fig:phase_1pp_f0_4pp_rho_tbin1}:
    Intensity distribution and relative phases for the
    \wave{1}{++}{0}{+}{\PfZero[980]}{P} wave.
    \subfloatLabel{fig:intensity_1pp_f2_tbin1}~and~\subfloatLabel{fig:phase_1pp_f2_4pp_rho_tbin1}:
    Intensity distribution and relative phase for the
    \wave{1}{++}{0}{+}{\PfTwo}{P} wave.  The model and the wave
    components are represented as in \cref{fig:intensity_phases_0mp},
    except that in
    \subfloatLabel{fig:intensity_1pp_rho_tbin1_log}~and~\subfloatLabel{fig:intensity_1pp_f2_tbin1}
    the blue curves represent the \PaOne and the \PaOne[1640], whereas
    in~\subfloatLabel{fig:intensity_1pp_f0_tbin1} the blue curve
    represents the \PaOne[1420].}
  \label{fig:intensity_phases_1pp_tbin1}
\ifMultiColumnLayout{\end{figure*}}{\end{figure}}

\ifMultiColumnLayout{\begin{figure*}[t]}{\begin{figure}[p]}
  \centering
  \subfloat[][]{%
    \includegraphics[width=\fourPlotWidth]{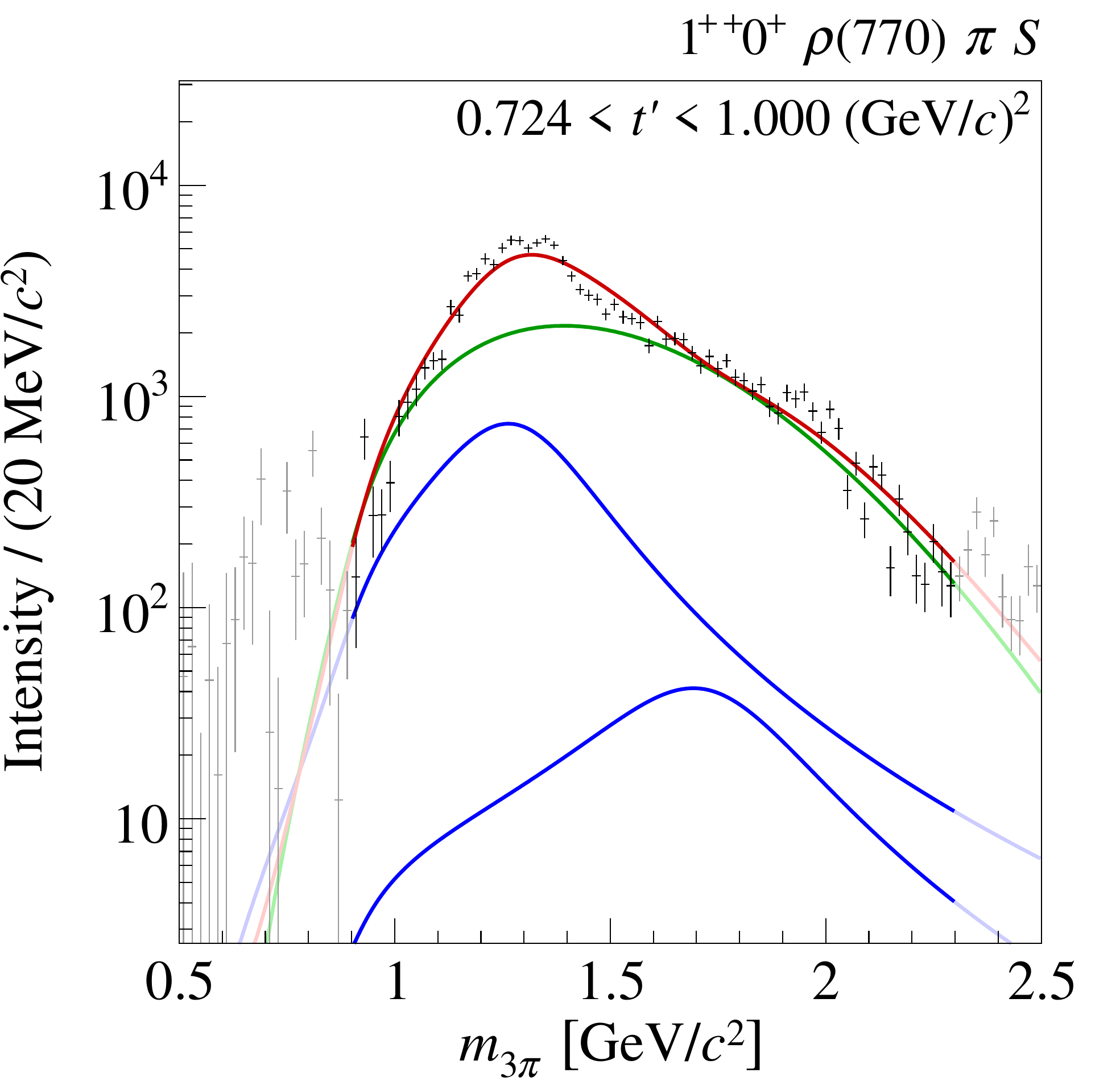}%
    \label{fig:intensity_1pp_rho_tbin11_log}%
  }%
  \hspace*{\fourPlotSpacing}%
  \subfloat[][]{%
    \includegraphics[width=\fourPlotWidth]{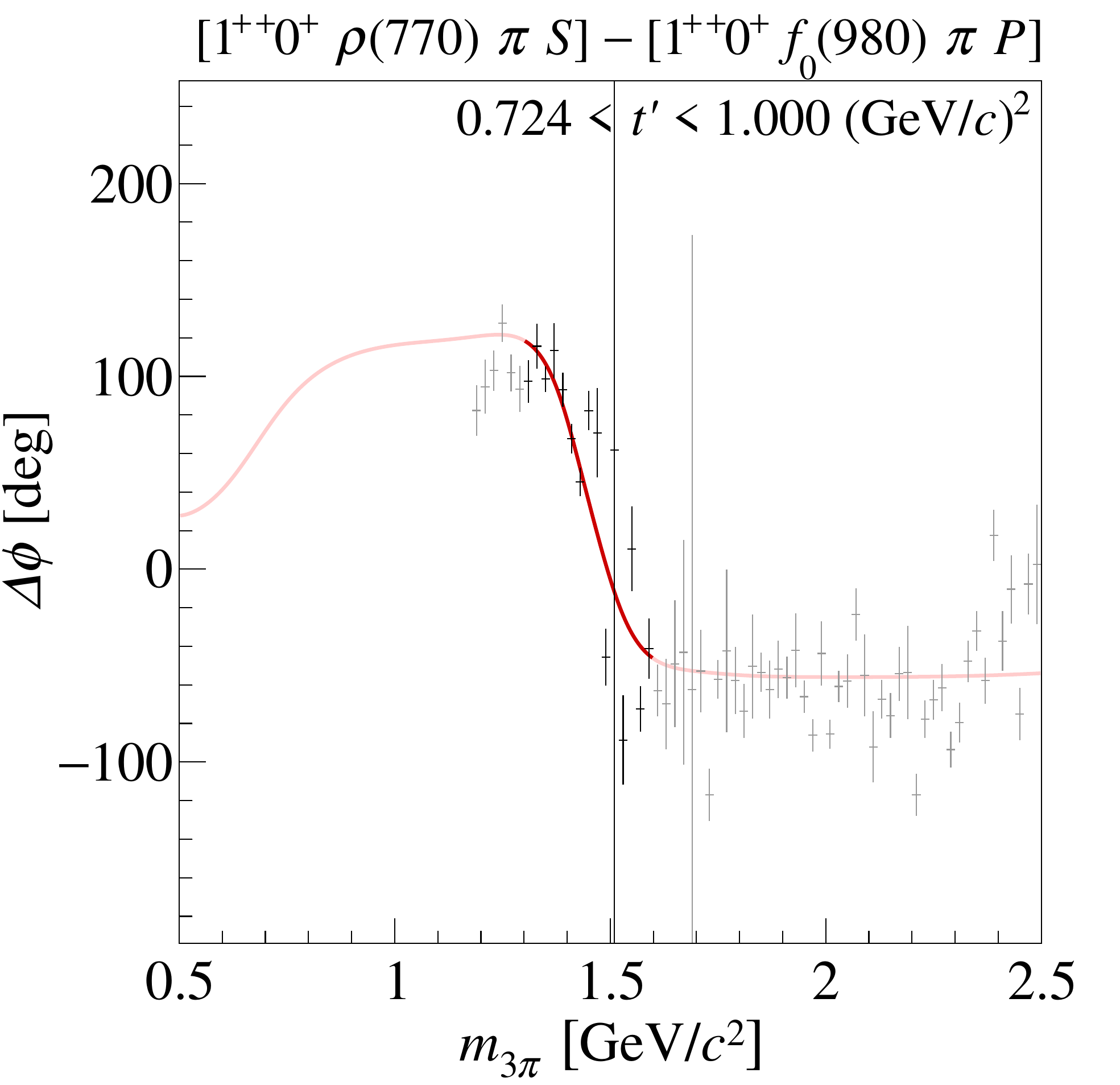}%
    \label{fig:phase_1pp_rho_1pp_f0_tbin11}%
  }%
  \hspace*{\fourPlotSpacing}%
  \subfloat[][]{%
    \includegraphics[width=\fourPlotWidth]{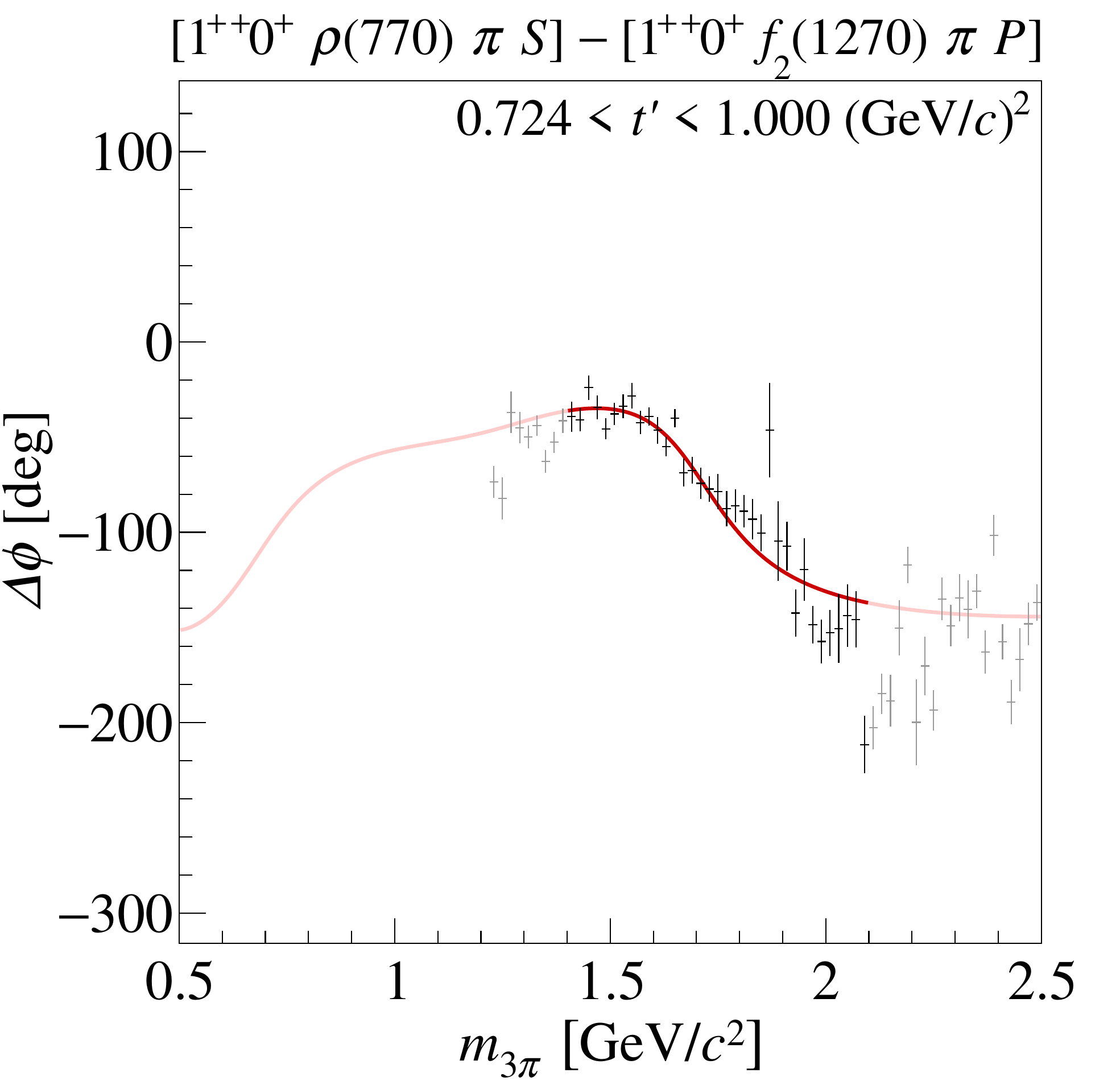}%
    \label{fig:phase_1pp_rho_1pp_f2_tbin11}%
  }%
  \hspace*{\fourPlotSpacing}%
  \subfloat[][]{%
    \includegraphics[width=\fourPlotWidth]{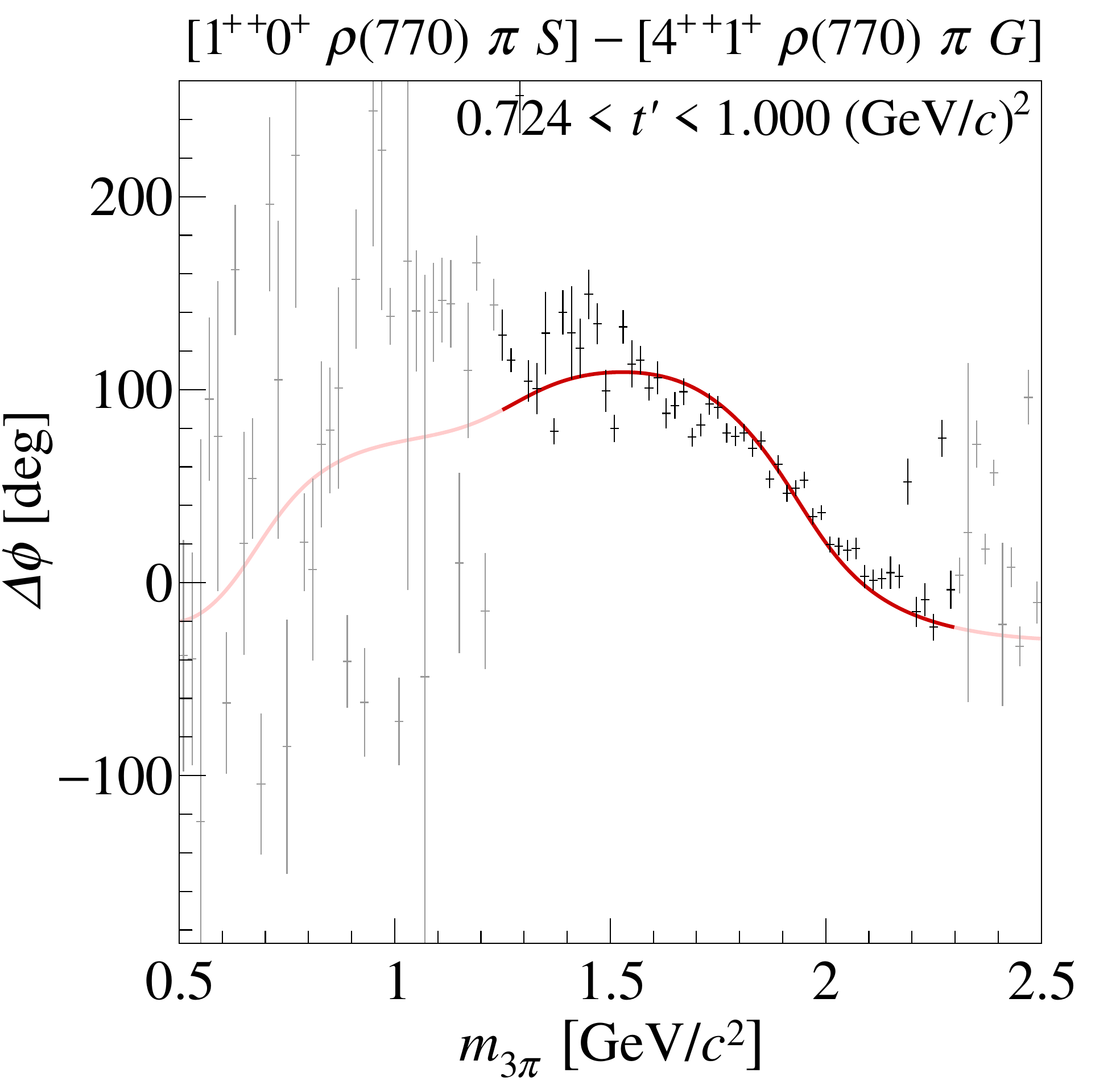}%
    \label{fig:phase_1pp_rho_4pp_rho_tbin11}%
  }%
  \\
  \hspace*{\fourPlotWidth}%
  \hspace*{\fourPlotSpacing}%
  \subfloat[][]{%
    \includegraphics[width=\fourPlotWidth]{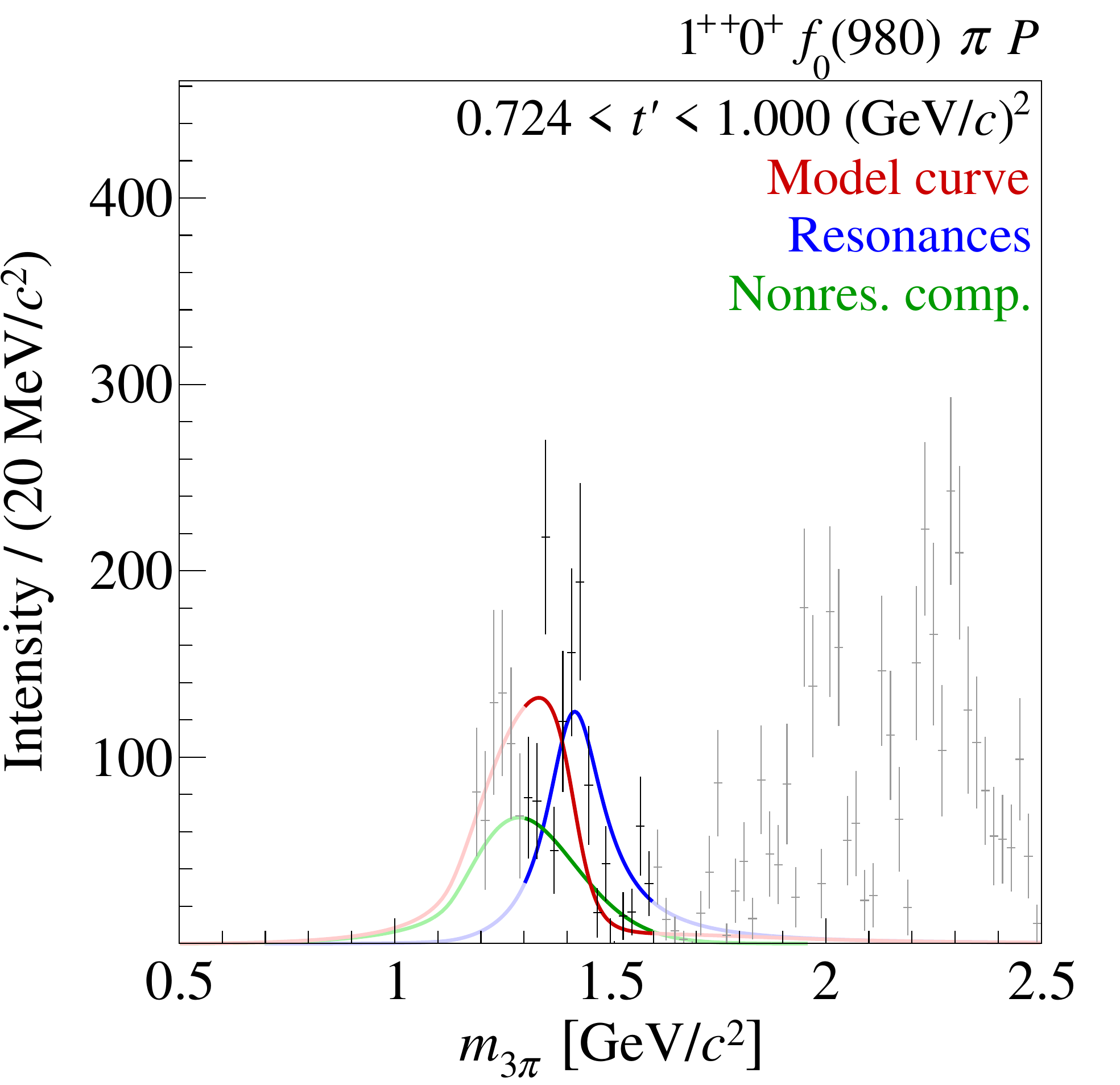}%
    \label{fig:intensity_1pp_f0_tbin11}%
  }%
  \hspace*{\fourPlotSpacing}%
  \subfloat[][]{%
    \includegraphics[width=\fourPlotWidth]{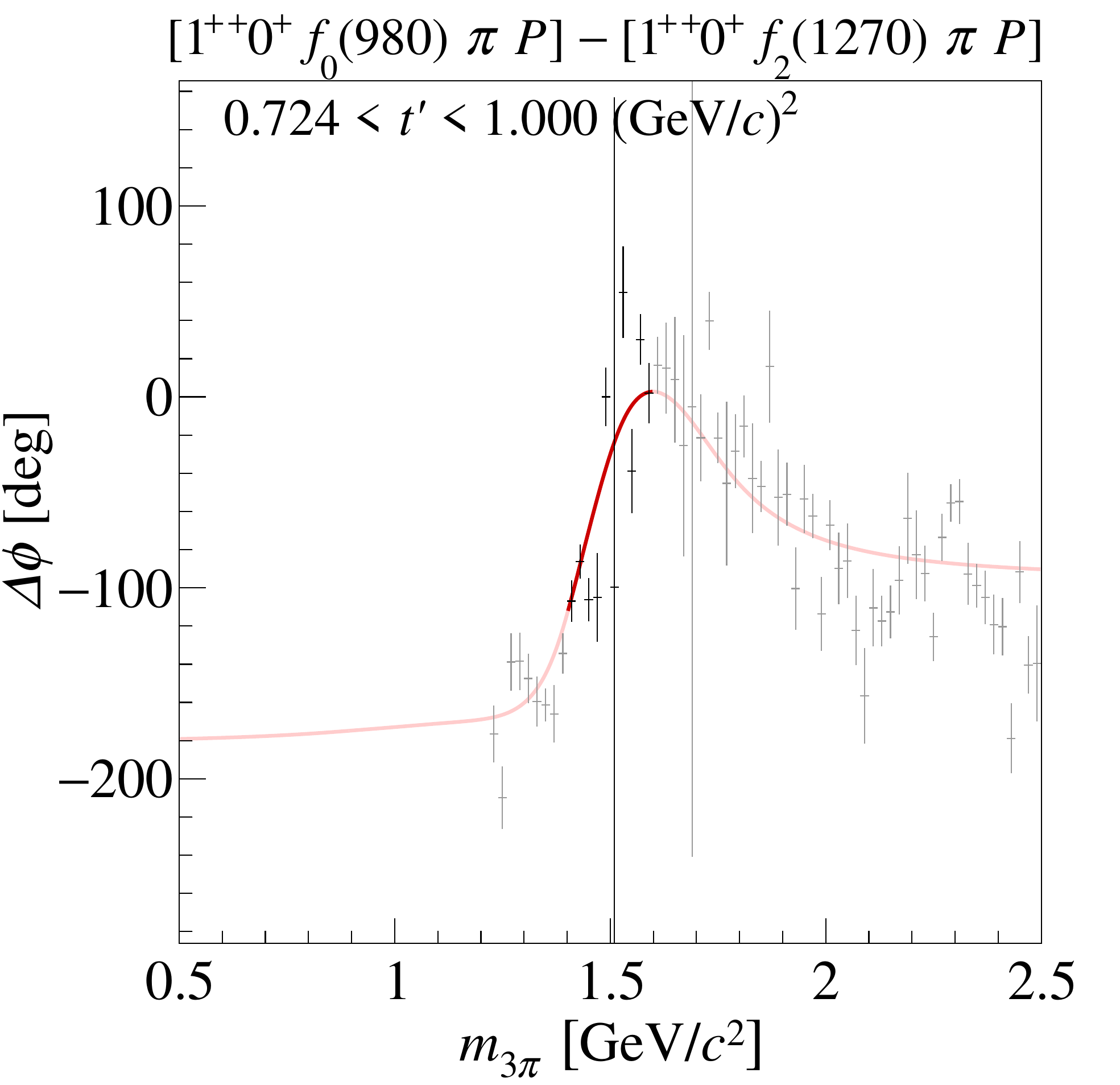}%
    \label{fig:phase_1pp_f0_1pp_f2_tbin11}%
  }%
  \hspace*{\fourPlotSpacing}%
  \subfloat[][]{%
    \includegraphics[width=\fourPlotWidth]{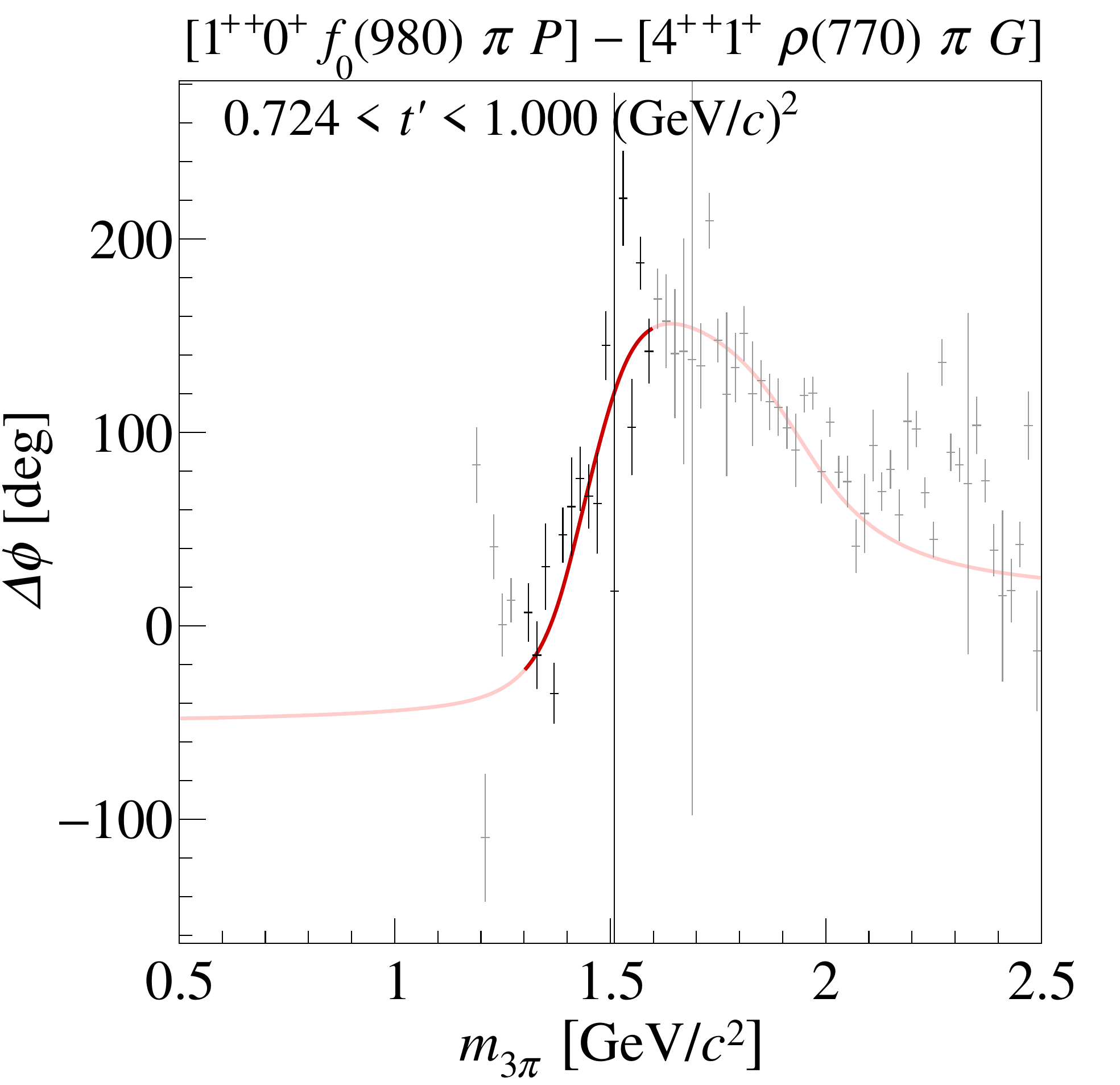}%
    \label{fig:phase_1pp_f0_4pp_rho_tbin11}%
  }%
  \\
  \hspace*{\fourPlotWidth}%
  \hspace*{\fourPlotSpacing}%
  \hspace*{\fourPlotWidth}%
  \hspace*{\fourPlotSpacing}%
  \subfloat[][]{%
    \includegraphics[width=\fourPlotWidth]{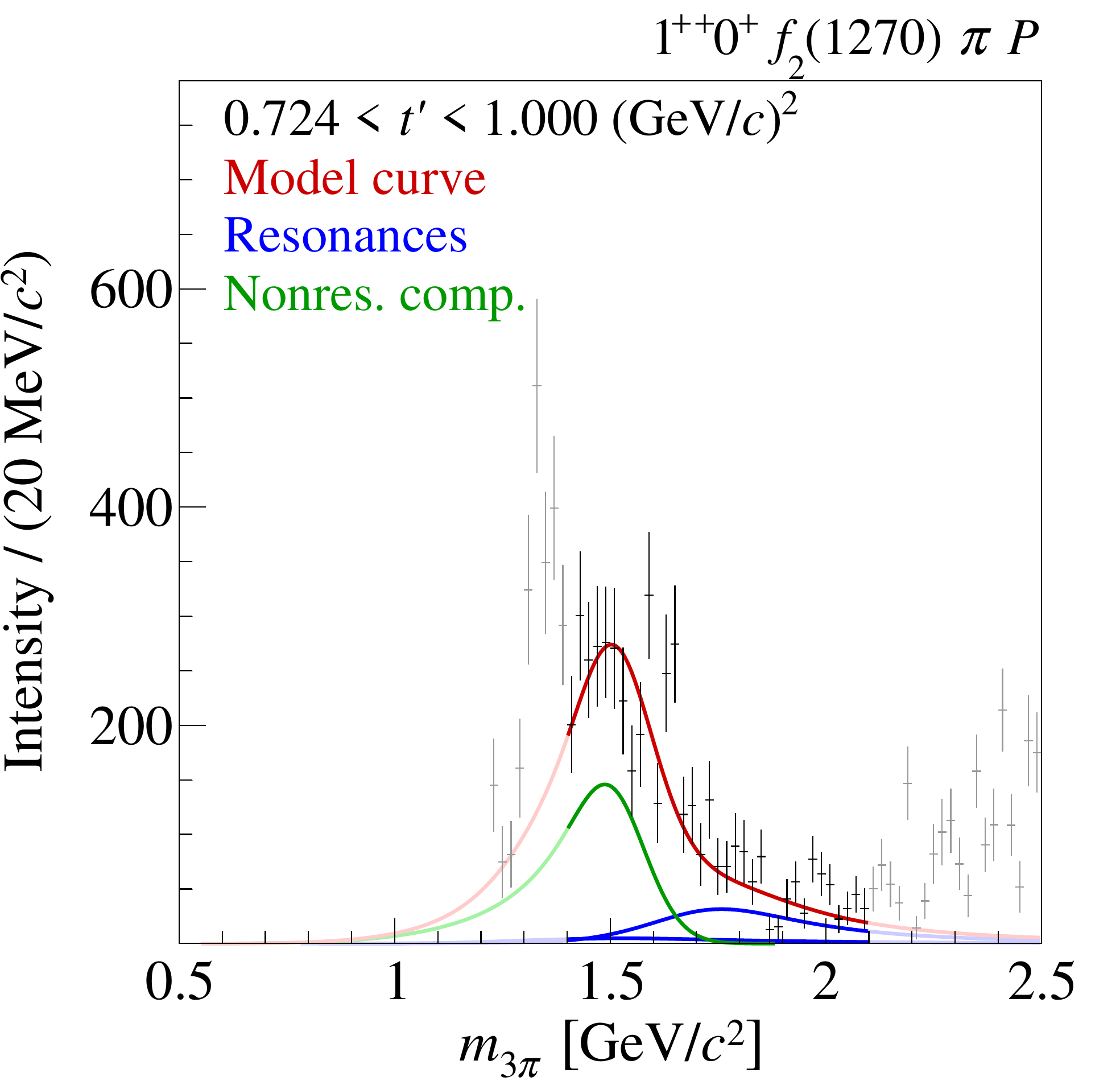}%
    \label{fig:intensity_1pp_f2_tbin11}%
  }%
  \hspace*{\fourPlotSpacing}%
  \subfloat[][]{%
    \includegraphics[width=\fourPlotWidth]{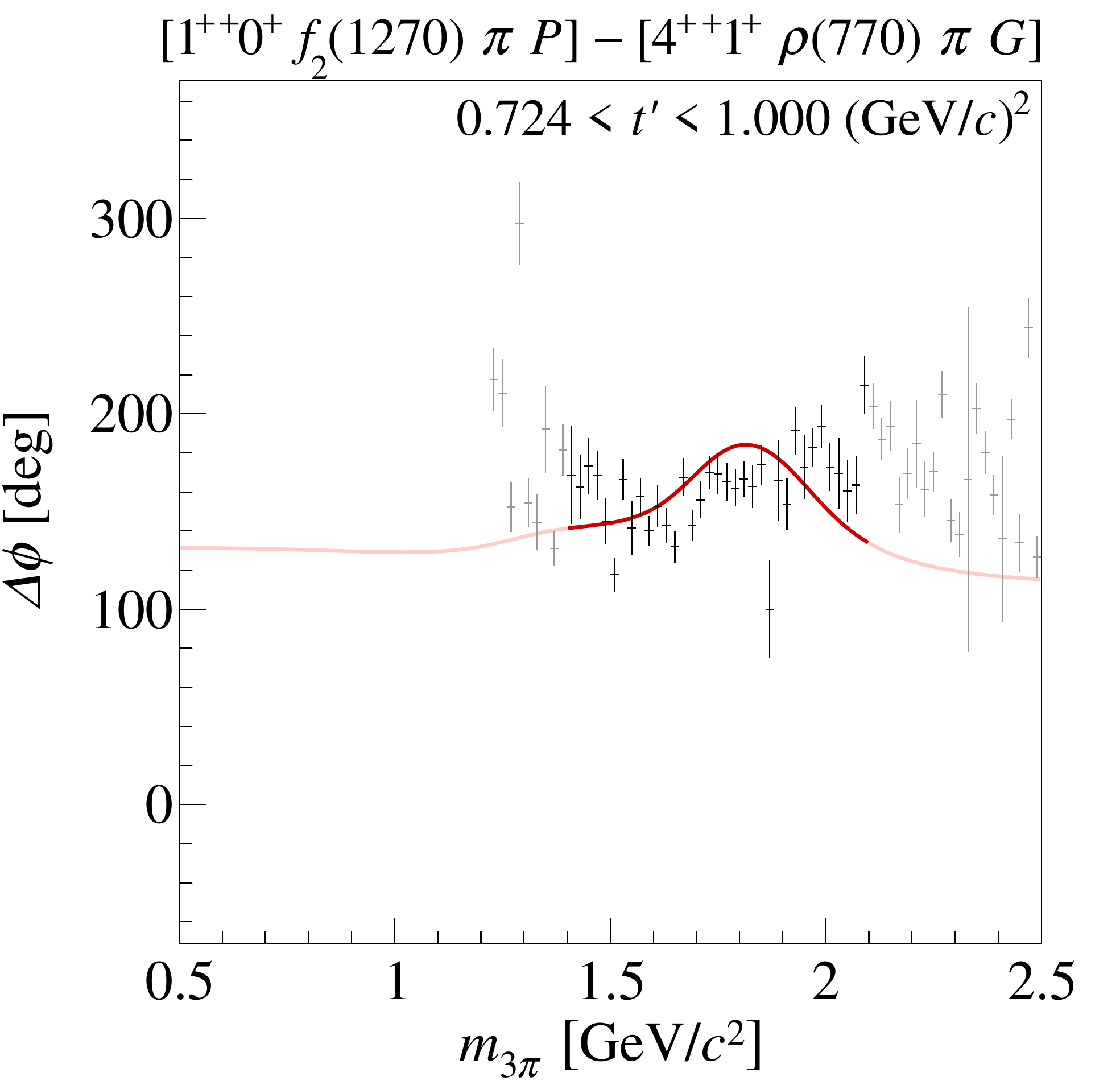}%
    \label{fig:phase_1pp_f2_4pp_rho_tbin11}%
  }%
  \caption{Similar to \cref{fig:intensity_phases_1pp_tbin1} but for
    the highest \tpr bin.}
  \label{fig:intensity_phases_1pp_tbin11}
\ifMultiColumnLayout{\end{figure*}}{\end{figure}}

The \wave{1}{++}{0}{+}{\Prho}{S} intensity exhibits a broad peak
around \SI{1.2}{\GeVcc}, which changes its shape and shifts by about
\SI{140}{\MeVcc} toward higher masses with increasing \tpr (see
\cref{fig:intensity_1pp_rho_pi_S}).  This behavior suggests large
contributions from nonresonant components in addition to the expected
\PaOne signal and underlines the importance of a \tpr-resolved
analysis to better disentangle these components.

\ifMultiColumnLayout{\begin{figure*}[t]}{\begin{figure}[p]}
  \centering
  \subfloat[][]{%
    \includegraphics[width=\threePlotSmallWidth]{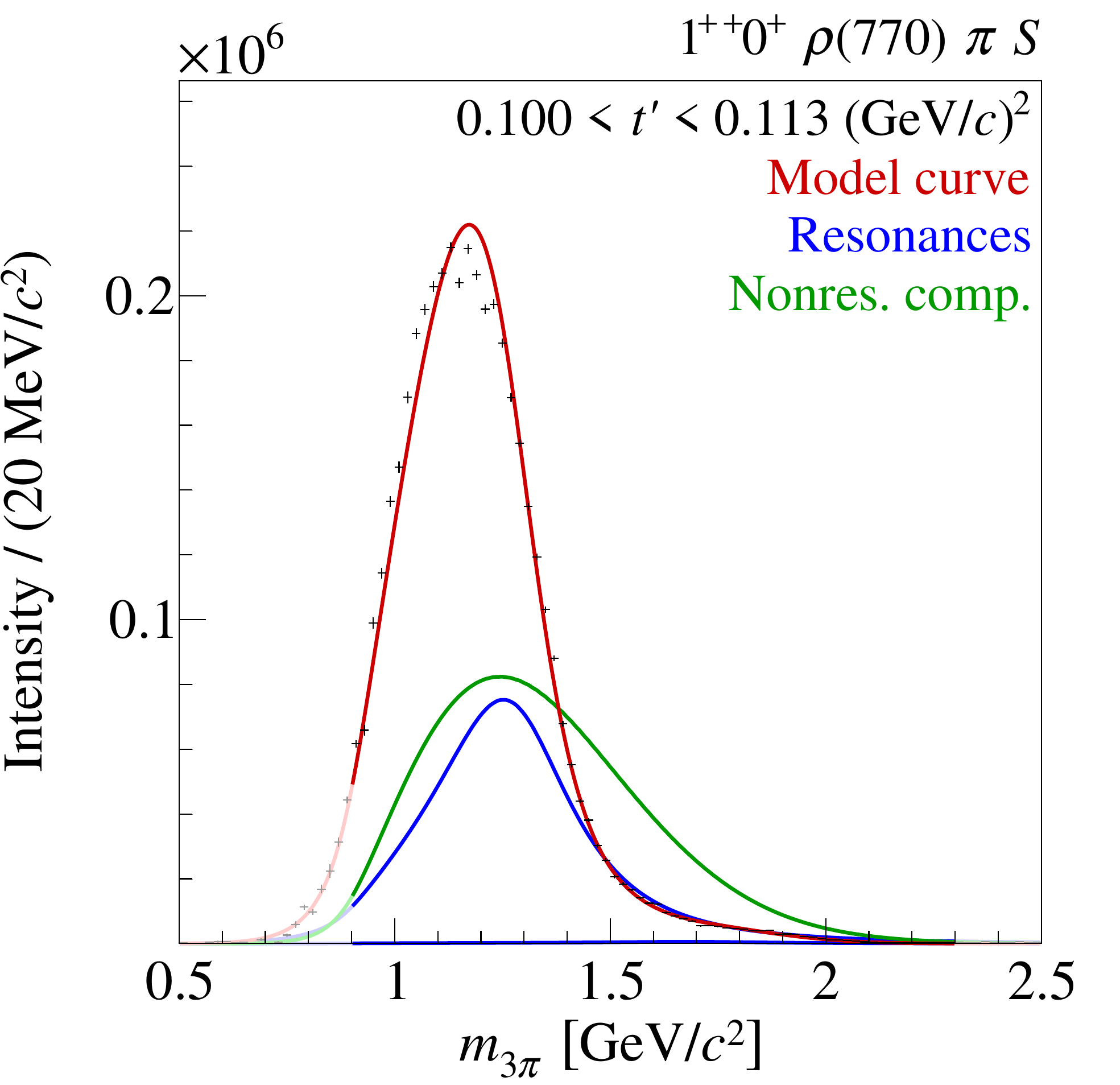}%
    \label{fig:intensity_1pp_rho_pi_S_tbin1}%
  }%
  \hspace*{\threePlotSmallSpacing}%
  \subfloat[][]{%
    \includegraphics[width=\threePlotSmallWidth]{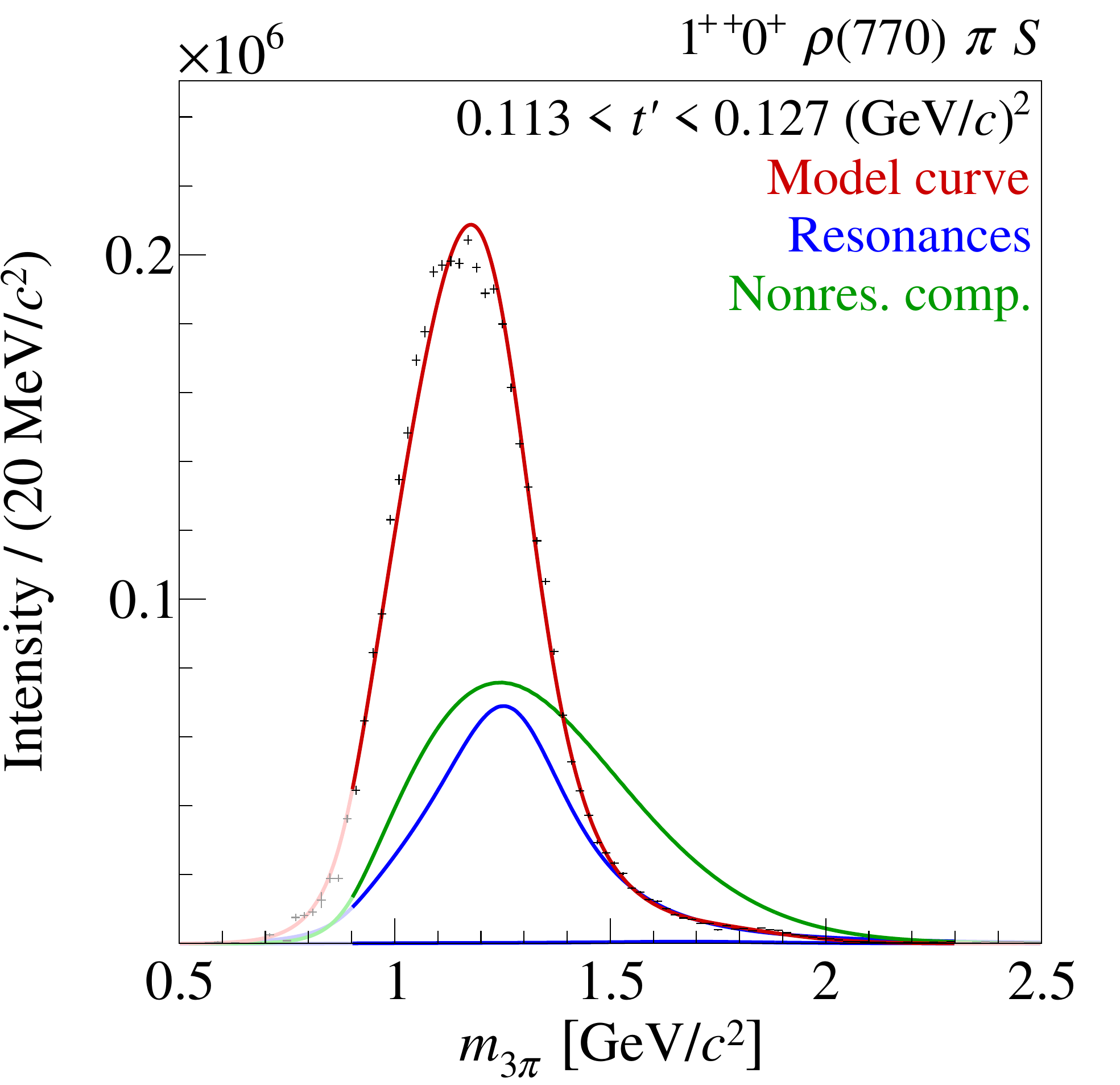}%
  }%
  \hspace*{\threePlotSmallSpacing}%
  \subfloat[][]{%
    \includegraphics[width=\threePlotSmallWidth]{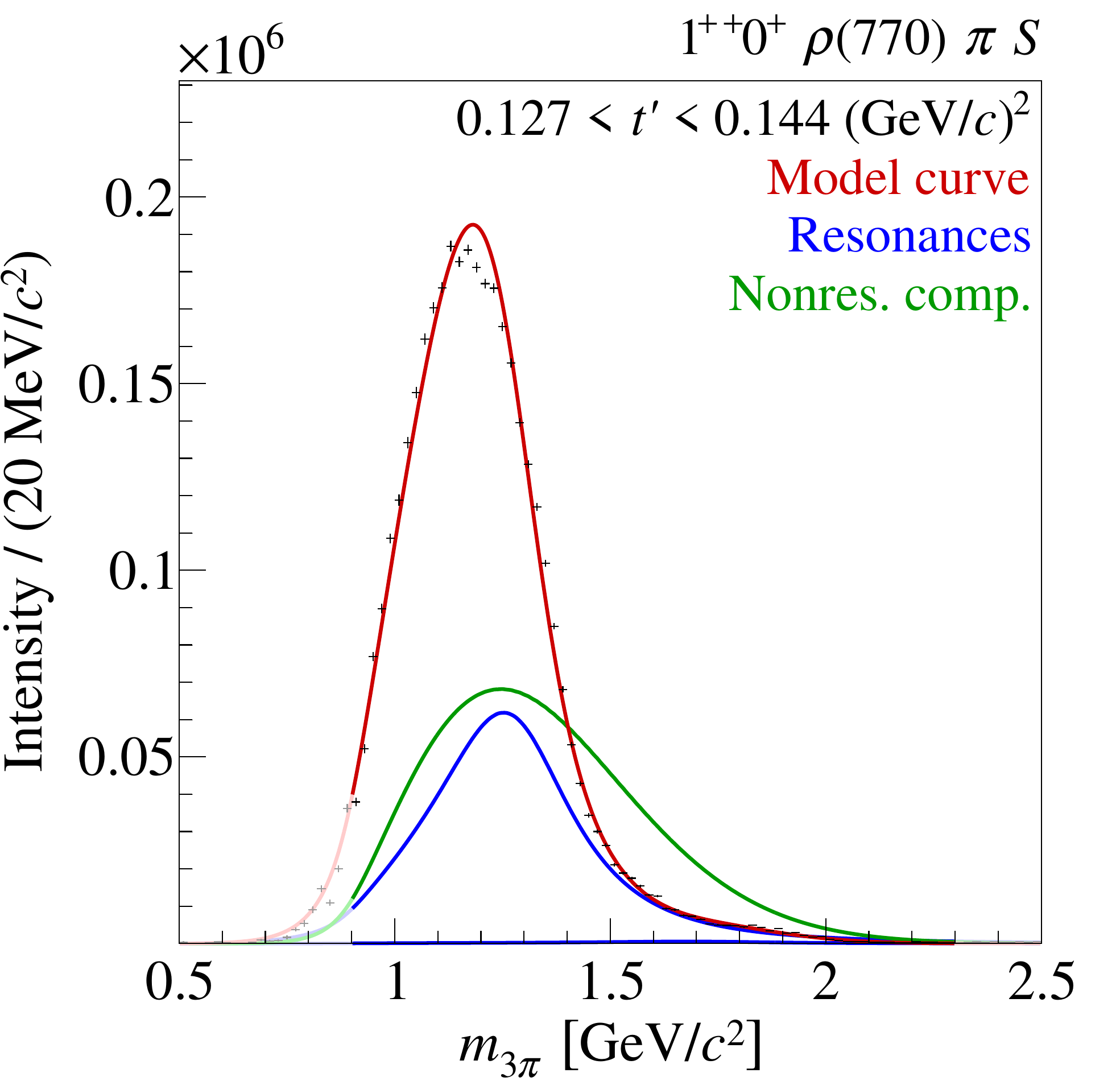}%
  }%
  \\
  \subfloat[][]{%
    \includegraphics[width=\threePlotSmallWidth]{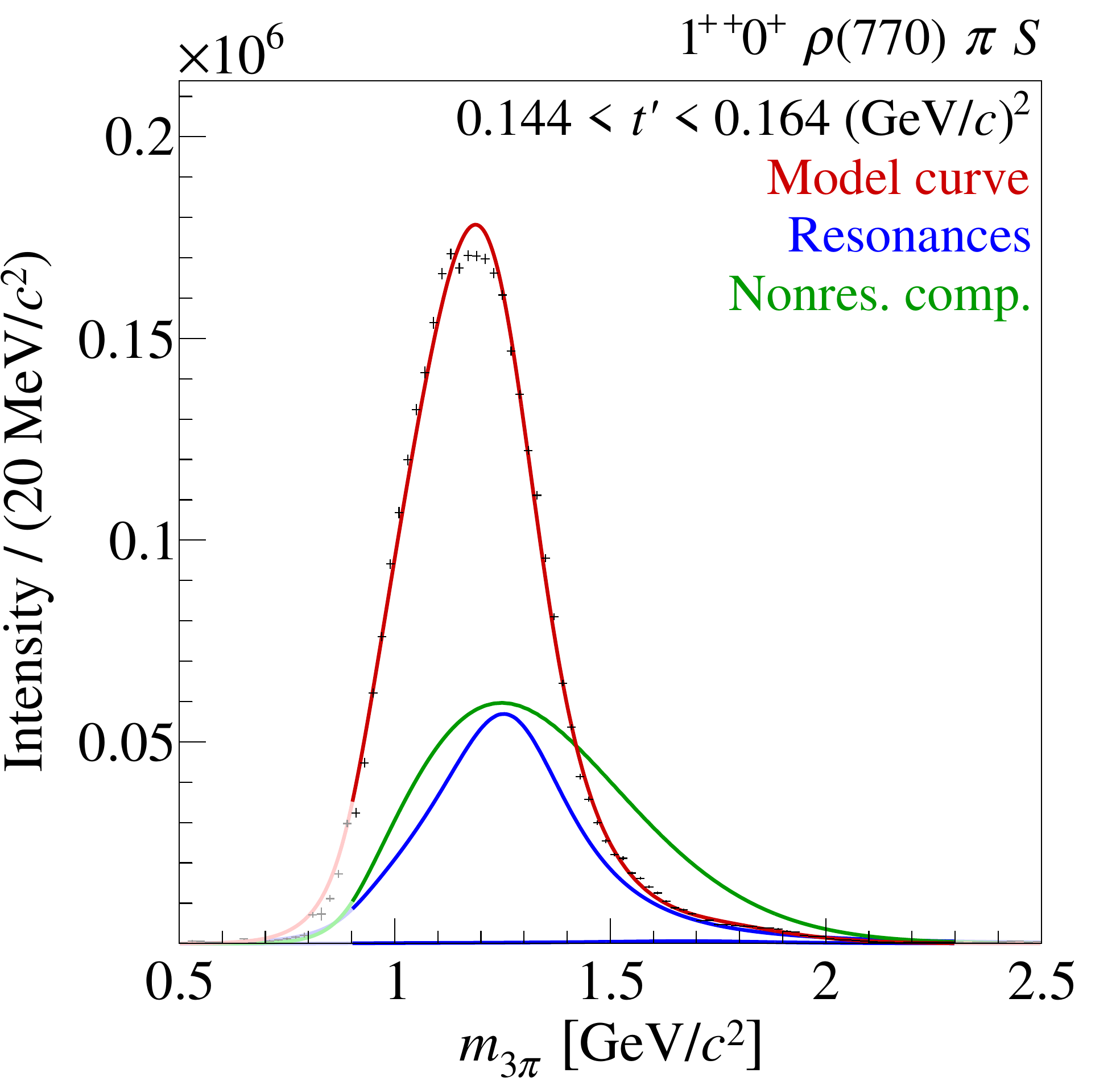}%
  }%
  \hspace*{\threePlotSmallSpacing}%
  \subfloat[][]{%
    \includegraphics[width=\threePlotSmallWidth]{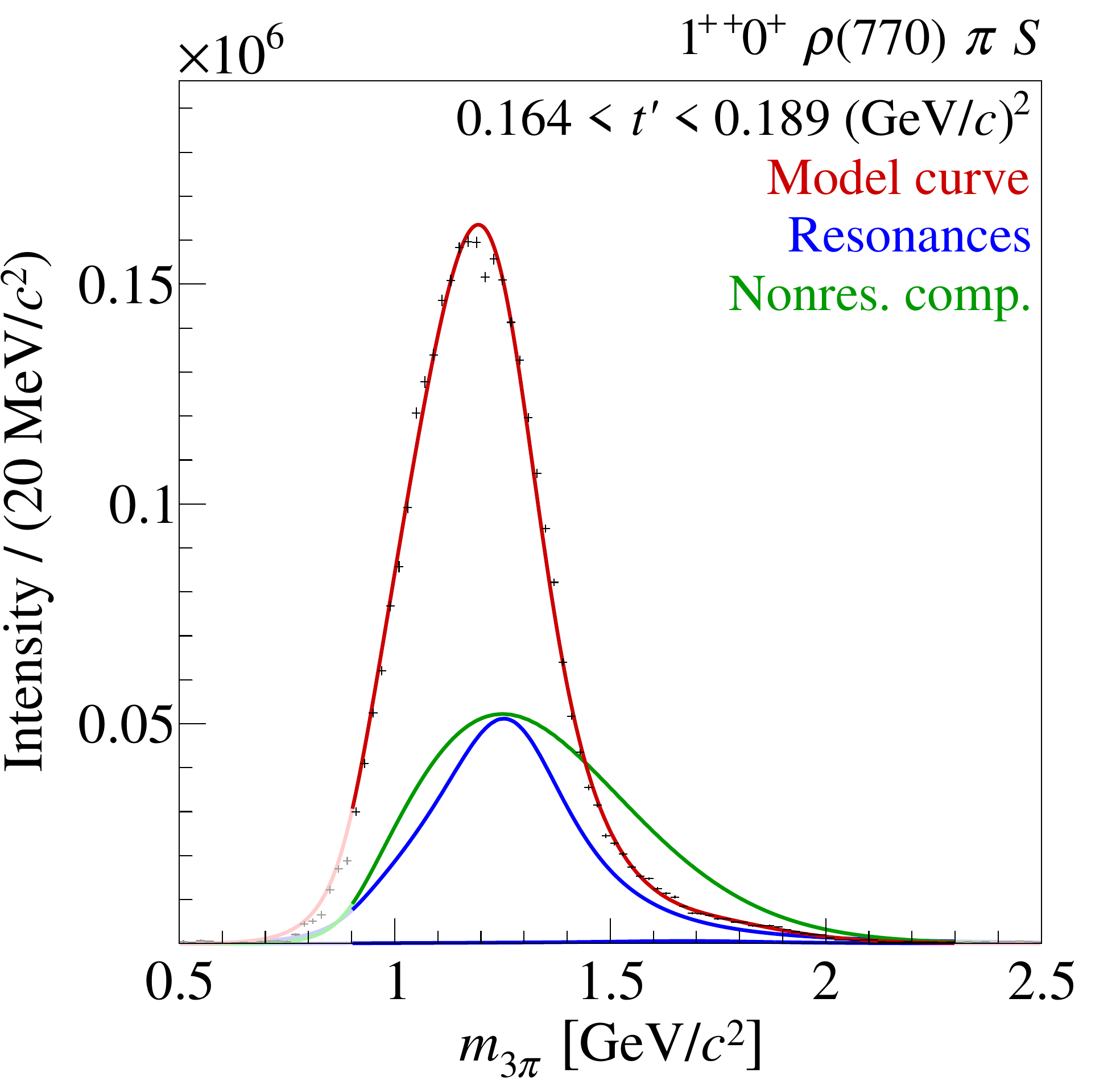}%
  }%
  \hspace*{\threePlotSmallSpacing}%
  \subfloat[][]{%
    \includegraphics[width=\threePlotSmallWidth]{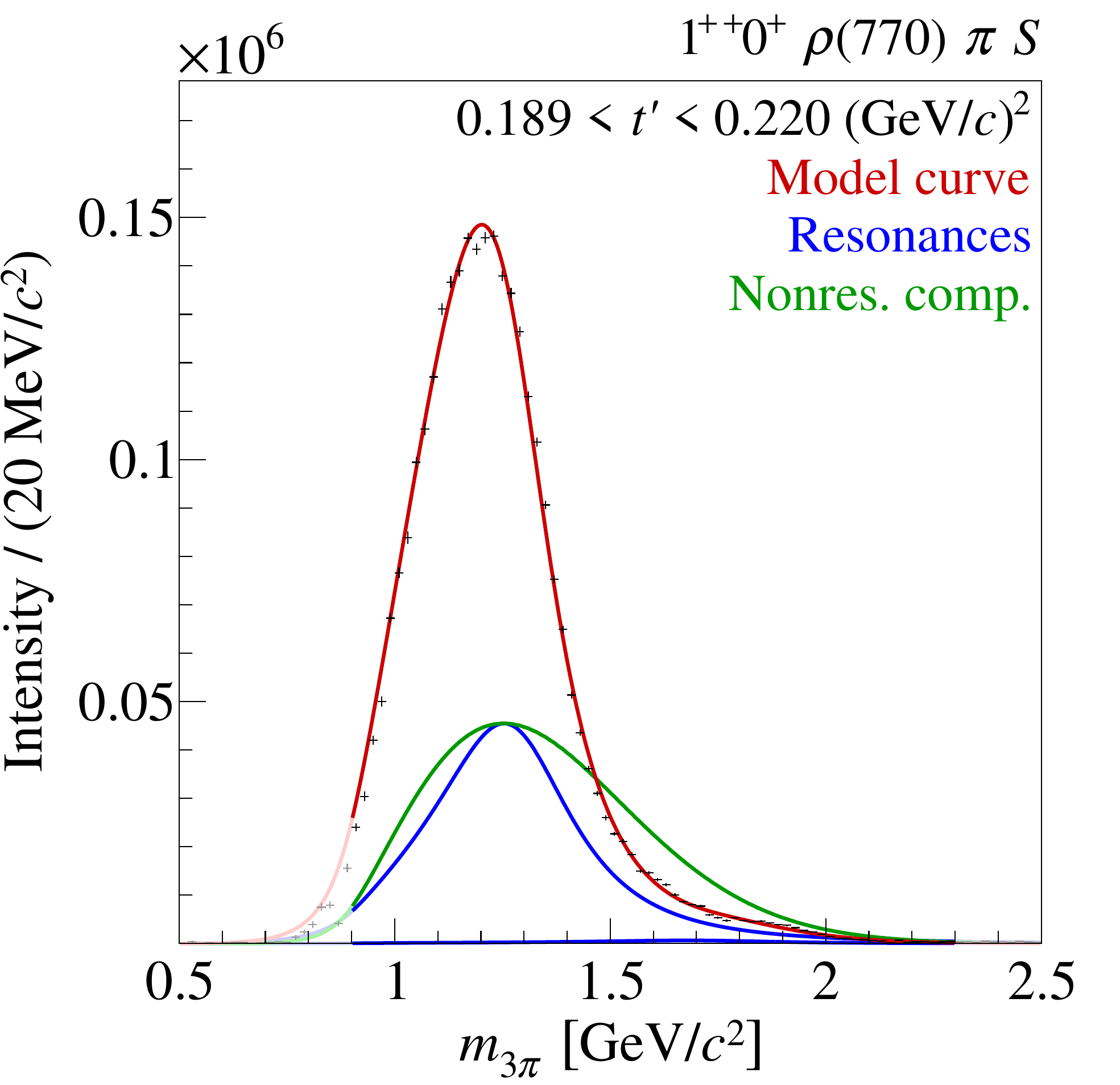}%
  }%
  \\
  \subfloat[][]{%
    \includegraphics[width=\threePlotSmallWidth]{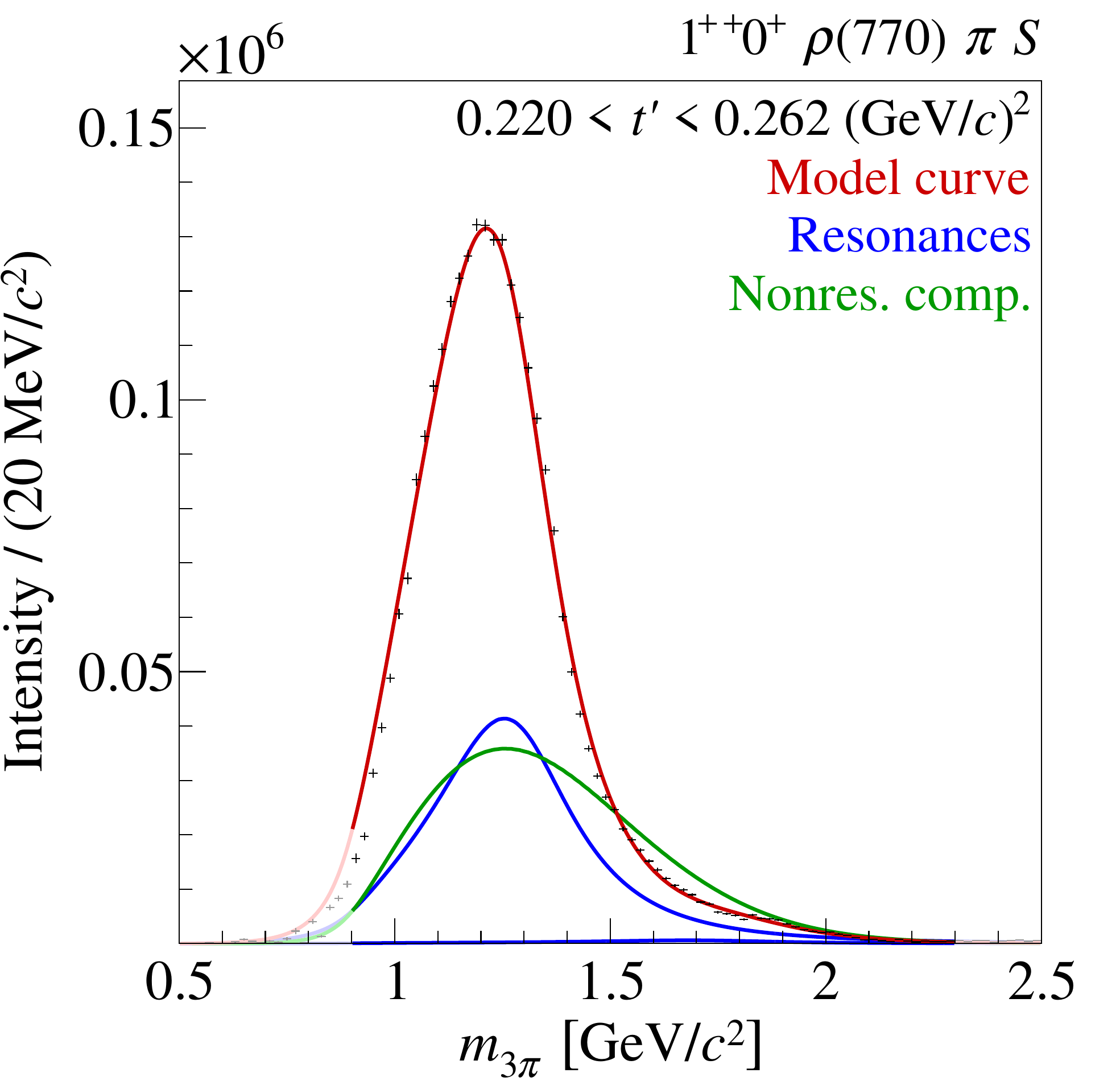}%
  }%
  \hspace*{\threePlotSmallSpacing}%
  \subfloat[][]{%
    \includegraphics[width=\threePlotSmallWidth]{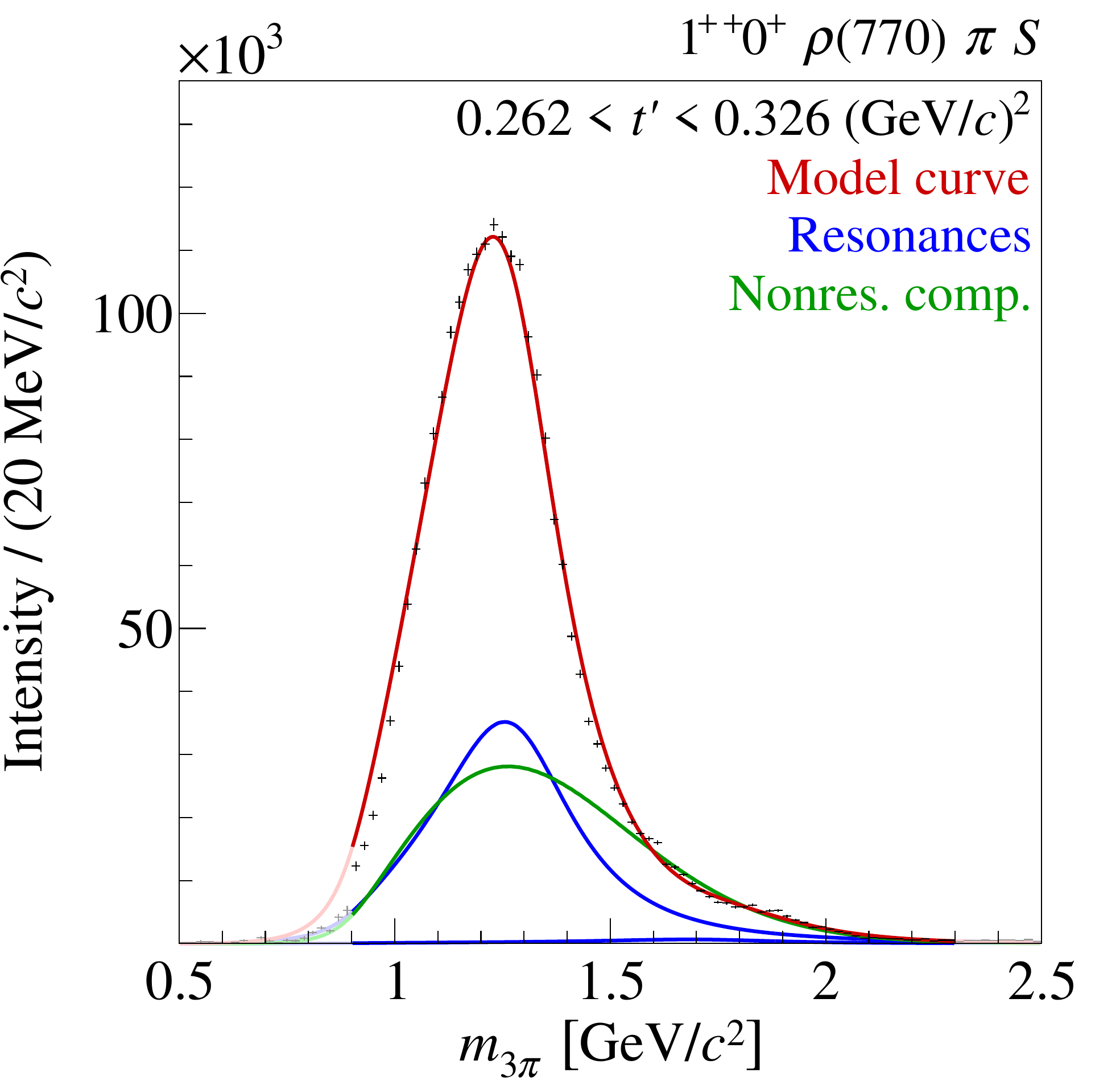}%
  }%
  \hspace*{\threePlotSmallSpacing}%
  \subfloat[][]{%
    \includegraphics[width=\threePlotSmallWidth]{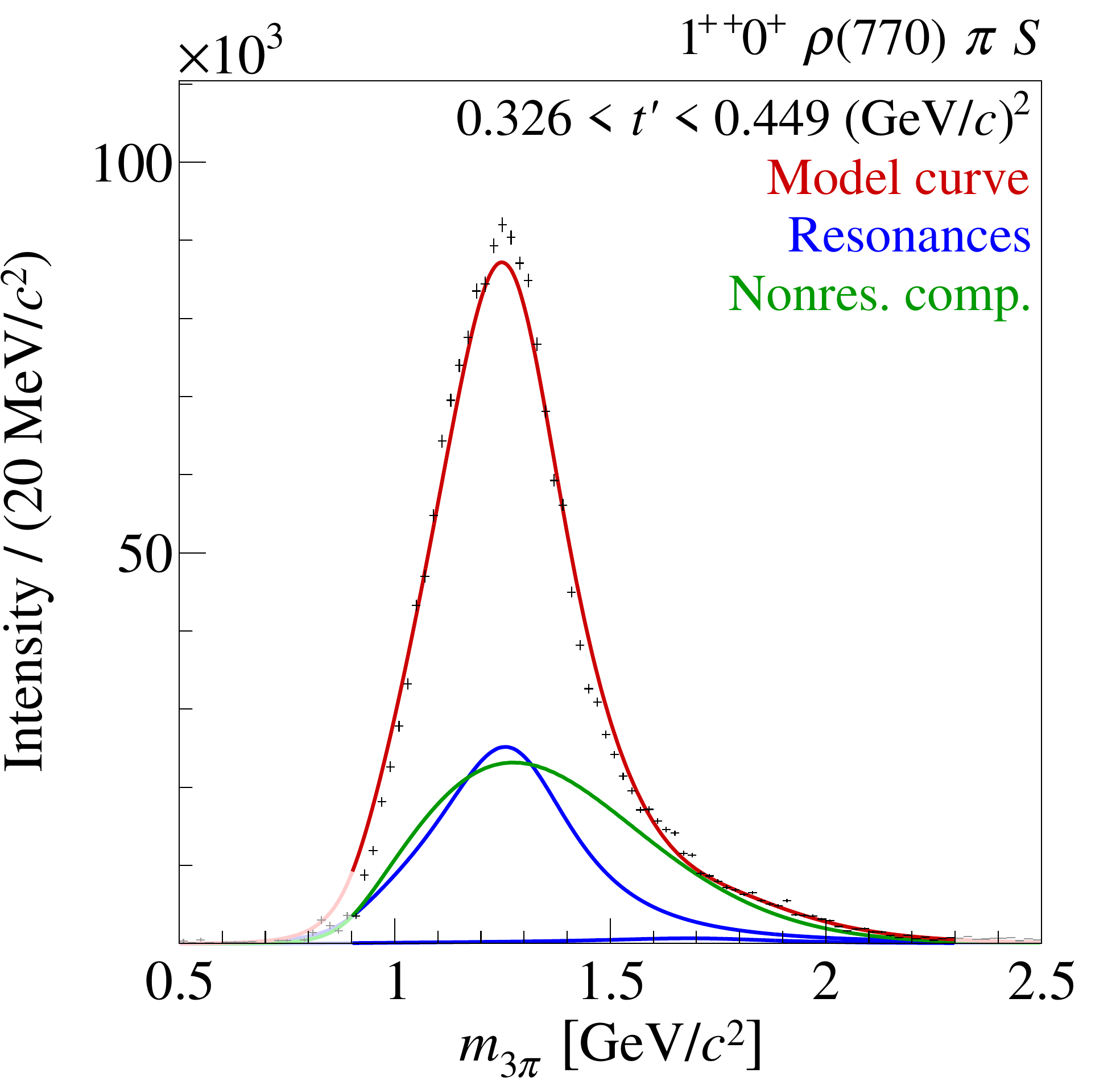}%
  }%
  \\
  \subfloat[][]{%
    \includegraphics[width=\threePlotSmallWidth]{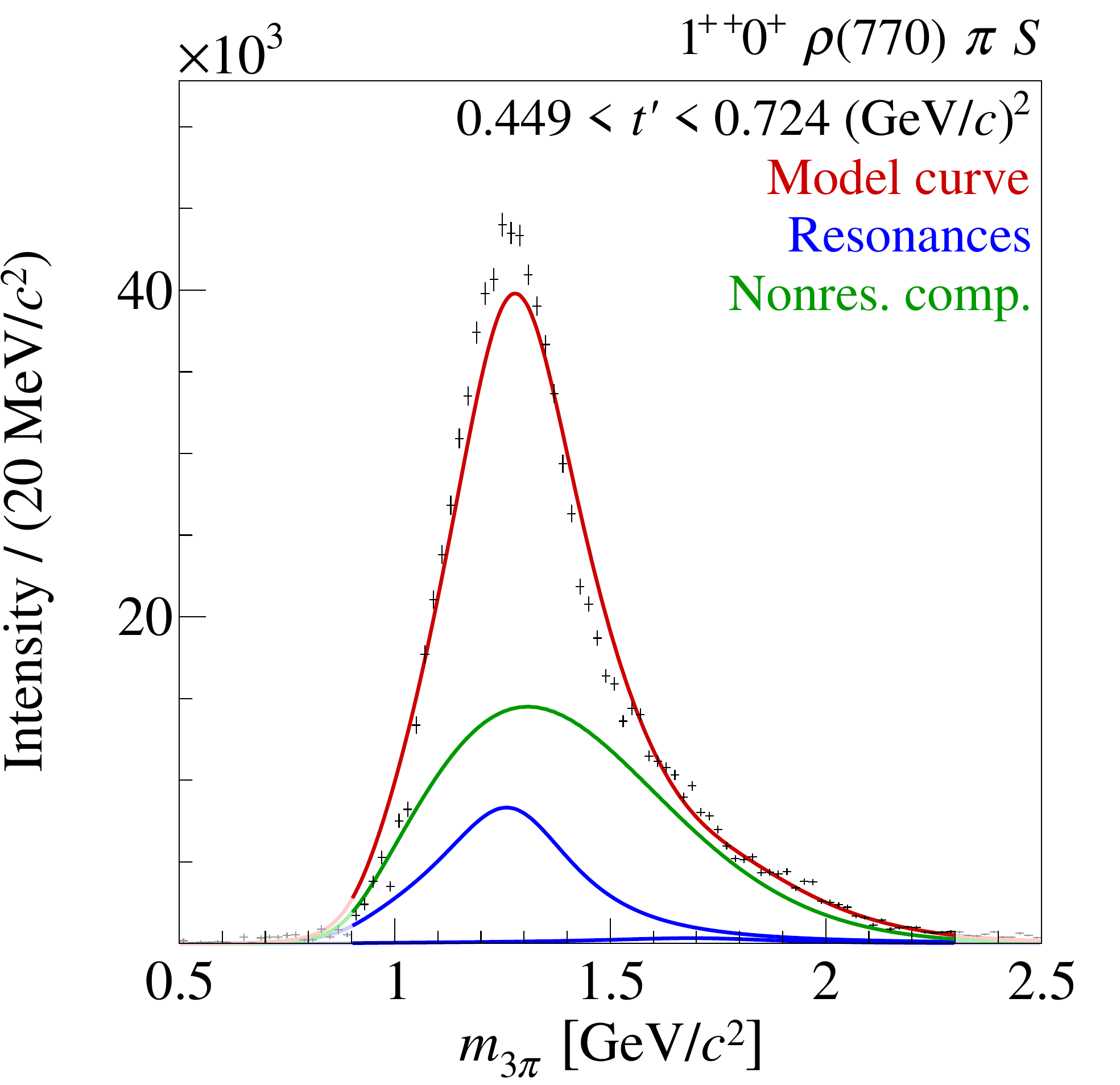}%
    \label{fig:intensity_1pp_rho_pi_S_tbin10}%
  }%
  \hspace*{\threePlotSmallSpacing}%
  \subfloat[][]{%
    \includegraphics[width=\threePlotSmallWidth]{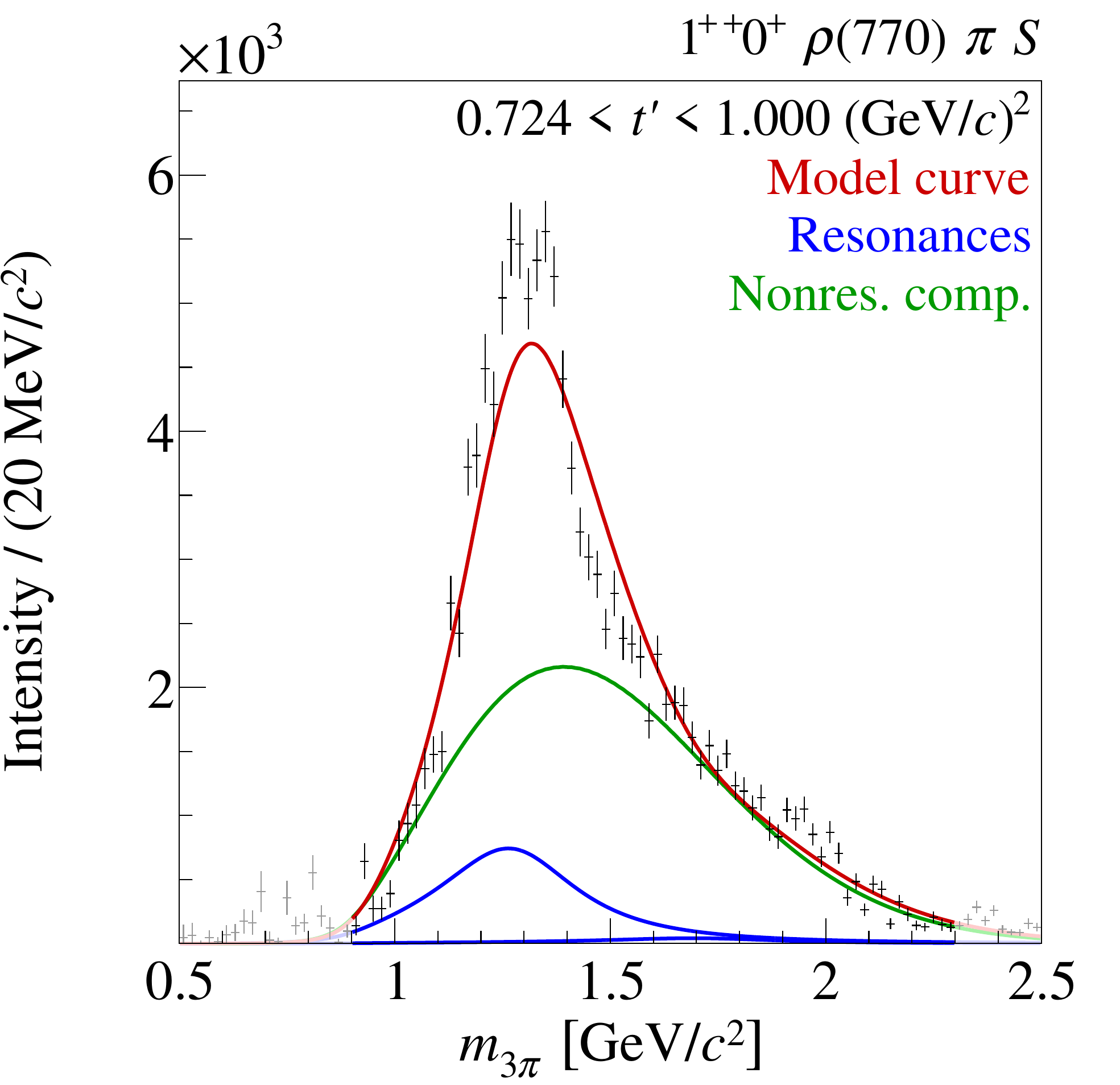}%
    \label{fig:intensity_1pp_rho_pi_S_tbin11}%
  }%
  \hspace*{\threePlotSmallSpacing}%
  \subfloat[][]{%
    \includegraphics[width=\threePlotSmallWidth]{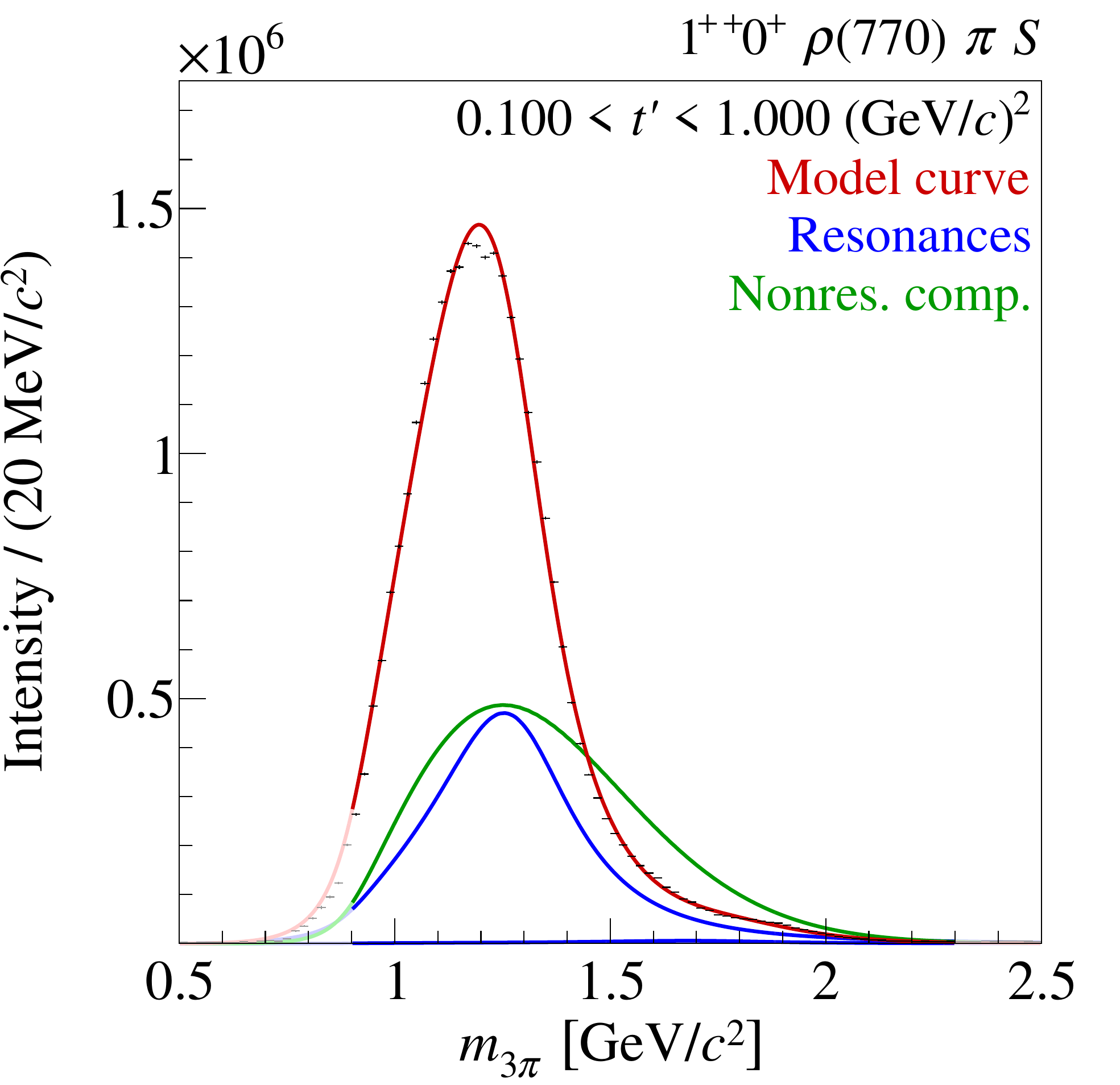}%
    \label{fig:intensity_1pp_rho_pi_S_tsum}%
  }%
  \caption{\subfloatLabel{fig:intensity_1pp_rho_pi_S_tbin1}~to~\subfloatLabel{fig:intensity_1pp_rho_pi_S_tbin11}:
    Intensity distributions of the \wave{1}{++}{0}{+}{\Prho}{S} wave
    in the 11~\tpr bins.
    \subfloatLabel{fig:intensity_1pp_rho_pi_S_tsum}~The \tpr-summed
    intensity.  The model and the wave components are represented as
    in \cref{fig:intensity_phases_1pp_tbin1}.  The contribution of the
    \PaOne[1640] component is so small that it is barely visible in
    linear scale.}
  \label{fig:intensity_1pp_rho_pi_S}
\ifMultiColumnLayout{\end{figure*}}{\end{figure}}

The \wave{1}{++}{0}{+}{\PfTwo}{P} intensity distribution exhibits a
low-mass enhancement below threshold and a broad peak structure at
about \SI{1.8}{\GeVcc} that disappears in the two highest \tpr bins.
In addition, a weaker enhancement appears around \SI{1.5}{\GeVcc} at
lower \tpr.  A portion of the low-mass enhancement might originate
from leakage within the $1^{++}$ sector at the stage of the
partial-wave decomposition.\footnote{This is supported by our finding
  that the low-mass enhancement in the intensity distribution of the
  \wave{1}{++}{0}{+}{\PfTwo}{P} wave changes significantly if a
  reduced set of 53~waves is used for the partial-wave decomposition
  (see Sec.~IV~F in \refCite{Adolph:2015tqa}).}  This leakage is
presumably induced by Deck-like nonresonant contributions.  Monte
Carlo simulations of a model for the Deck amplitude (see
\cref{sec:deck_model}) have shown that at low \tpr, the shapes of the
isobars are distorted, especially that of the \Prho.  This might cause
leakage into the $\PfTwo \pi P$ wave, which has an intensity that is
2~orders of magnitude smaller than that of the $\Prho \pi S$ wave.

The most peculiar intensity distribution is observed for the
\wave{1}{++}{0}{+}{\PfZero[980]}{P} wave.  It has a dominant narrow
peak at approximately \SI{1.45}{\GeVcc} that disappears in the highest
\tpr bin.  In this mass region, large and rapid phase motions of the
$\PfZero \pi P$ wave are observed relative to the other two $1^{++}$
waves in all \tpr bins [see
\cref{fig:phase_1pp_rho_1pp_f0_tbin1,fig:phase_1pp_f0_1pp_f2_tbin1,fig:phase_1pp_rho_1pp_f0_tbin11,fig:phase_1pp_f0_1pp_f2_tbin11}].
This suggests that the $\PfZero \pi P$ wave has a different resonance
content.  Similar phase motions are also observed \wrt other waves.
As an example,
\cref{fig:phase_1pp_f0_4pp_rho_tbin1,fig:phase_1pp_f0_4pp_rho_tbin11}
show the phases relative to the \wave{4}{++}{1}{+}{\Prho}{G} wave,
where the latter was discussed in \cref{sec:fourPP_results}.

Also the relative phase between the \wave{1}{++}{0}{+}{\Prho}{S} and
\wave{1}{++}{0}{+}{\PfTwo}{P} waves changes substantially with
\mThreePi [see
\cref{fig:phase_1pp_rho_1pp_f2_tbin1,fig:phase_1pp_rho_1pp_f2_tbin11}],
which suggests that the wave components contribute with different
strengths to these two waves.  At high \tpr, this phase becomes
approximately constant in the \PaOne region and the phase motion in
the \SI{1.6}{\GeVcc} region becomes shallower.  In general, the
\wave{1}{++}{0}{+}{\Prho}{S} wave shows only slowly changing or
approximately constant phases \wrt other waves in the \PaOne mass
region.  As an example,
\cref{fig:phase_0mp_1pp_rho_tbin1,fig:phase_0mp_1pp_rho_tbin10,fig:phase_0mp_1pp_rho_tbin11}
in \cref{sec:zeroMP_results} show the phase \wrt the
\wave{0}{-+}{0}{+}{\PfZero}{S} wave.  The dominant feature is a rising
phase in the \SI{1.8}{\GeVcc} region due to the \Ppi[1800].  In a
similar way, the phase \wrt the \wave{4}{++}{1}{+}{\Prho}{G} wave is
dominated by the \PaFour\ [see
\cref{fig:phase_1pp_rho_4pp_rho_tbin1,fig:phase_1pp_rho_4pp_rho_tbin11}].
The \wave{1}{++}{0}{+}{\PfTwo}{P} wave shows phase motions in the
\SI{1.65}{\GeVcc} region, for example \wrt the other two $1^{++}$
waves and the \wave{4}{++}{1}{+}{\Prho}{G} wave [see
\cref{fig:phase_1pp_rho_1pp_f2_tbin1,fig:phase_1pp_f0_1pp_f2_tbin1,fig:phase_1pp_f2_4pp_rho_tbin1,fig:phase_1pp_rho_1pp_f2_tbin11,fig:phase_1pp_f0_1pp_f2_tbin11}].
In the highest \tpr bin, the phase \wrt the
\wave{4}{++}{1}{+}{\Prho}{G} wave becomes constant [see
\cref{fig:phase_1pp_f2_4pp_rho_tbin11}].

We model the three $1^{++}$ waves using three resonance components,
\PaOne, \PaOne[1420], and \PaOne[1640].  The \PaOne and \PaOne[1640]
appear in both the \wave{1}{++}{0}{+}{\Prho}{S} and
\wave{1}{++}{0}{+}{\PfTwo}{P} waves, whereas the
\wave{1}{++}{0}{+}{\PfZero[980]}{P} wave is described using the
\PaOne[1420] as the only resonance component (see
\cref{tab:method:fitmodel:waveset}).  The \PaOne is parametrized by
\cref{eq:BreitWigner,eq:method:bowlerG}, and the \PaOne[1420] and
\PaOne[1640] by \cref{eq:BreitWigner,eq:method:fixedwidth}.  For the
nonresonant component in the $\Prho \pi S$ wave we use
\cref{eq:method:nonresterm}, for those in the other two waves
\cref{eq:method:nonrestermsmall}.  The $\Prho \pi S$ wave is fit in
the mass range from \SIrange{0.9}{2.3}{\GeVcc} and the $\PfTwo \pi P$
wave from \SIrange{1.4}{2.1}{\GeVcc}.  For the $\PfZero[980] \pi P$
wave, a narrower fit range from \SIrange{1.3}{1.6}{\GeVcc} was
chosen.\footnote{Therefore, this wave has no overlap with the fit
  range of the \wave{2}{-+}{0}{+}{\PfTwo}{D} wave, which starts only
  at \SI{1.6}{\GeVcc}.}

The employed model is in fair agreement with the data.  In particular
it is able to describe the change of the $\Prho \pi S$ intensity with
\tpr in terms of a \tpr-dependent interference between the \PaOne and
the nonresonant component (see \cref{fig:intensity_1pp_rho_pi_S}).
The relative phase of the coupling amplitudes of the nonresonant
component \wrt the \PaOne changes from approximately \SI{0}{\degree}
at low \tpr to \SI{+100}{\degree} at high \tpr (see
\cref{fig:tprim_phase_1pp_rho} in \cref{sec:production_phases}).
Although the model reproduces the main features of the data, the
extremely small statistical uncertainties of the $\Prho \pi S$ data
points lead to significant disagreement of the model with the data in
the \PaOne region.  The intensity distributions of the $\Prho \pi S$
wave and the real and imaginary parts of its interference terms in the
11~\tpr bins contribute together already
about \SI{25}{\percent} to the total~\chisq of the model [see
\cref{eq:method:fitmethod:chi2}].  The model systematically deviates
from the $\Prho \pi S$ intensity in the low- and high-mass flanks of
the peak and also cannot well describe the tip of the peak (see
\cref{fig:intensity_1pp_rho_pi_S_zoom}).  Some of the discontinuities
in this mass region might be induced by the thresholds applied to some
of the 88~waves used in the partial-wave decomposition (see Table~IX
in Appendix~A of \refCite{Adolph:2015tqa}).  The deviations of the
model from the $\Prho \pi S$ intensity increase with \tpr.  In the two
highest \tpr bins, the peak becomes significantly narrower, which the
model is not able to reproduce [see
\cref{fig:intensity_1pp_rho_pi_S_tbin10,fig:intensity_1pp_rho_pi_S_tbin11}].
The model also does not reproduce smaller details in the high-mass
region.  The \PaOne and the nonresonant component contribute with
similar intensities to the $\Prho \pi S$ wave and interfere
constructively in the \PaOne region.  In the low-\tpr region, the two
components interfere destructively at higher masses.  The contribution
of the \PaOne[1640] component to the $\Prho \pi S$ wave is
approximately 2~orders of magnitude smaller than that of the \PaOne.
It accounts for the small shoulder at \SI{1.8}{\GeVcc}.

\begin{figure}[tbp]
  \centering
  \includegraphics[width=\twoPlotWidth]{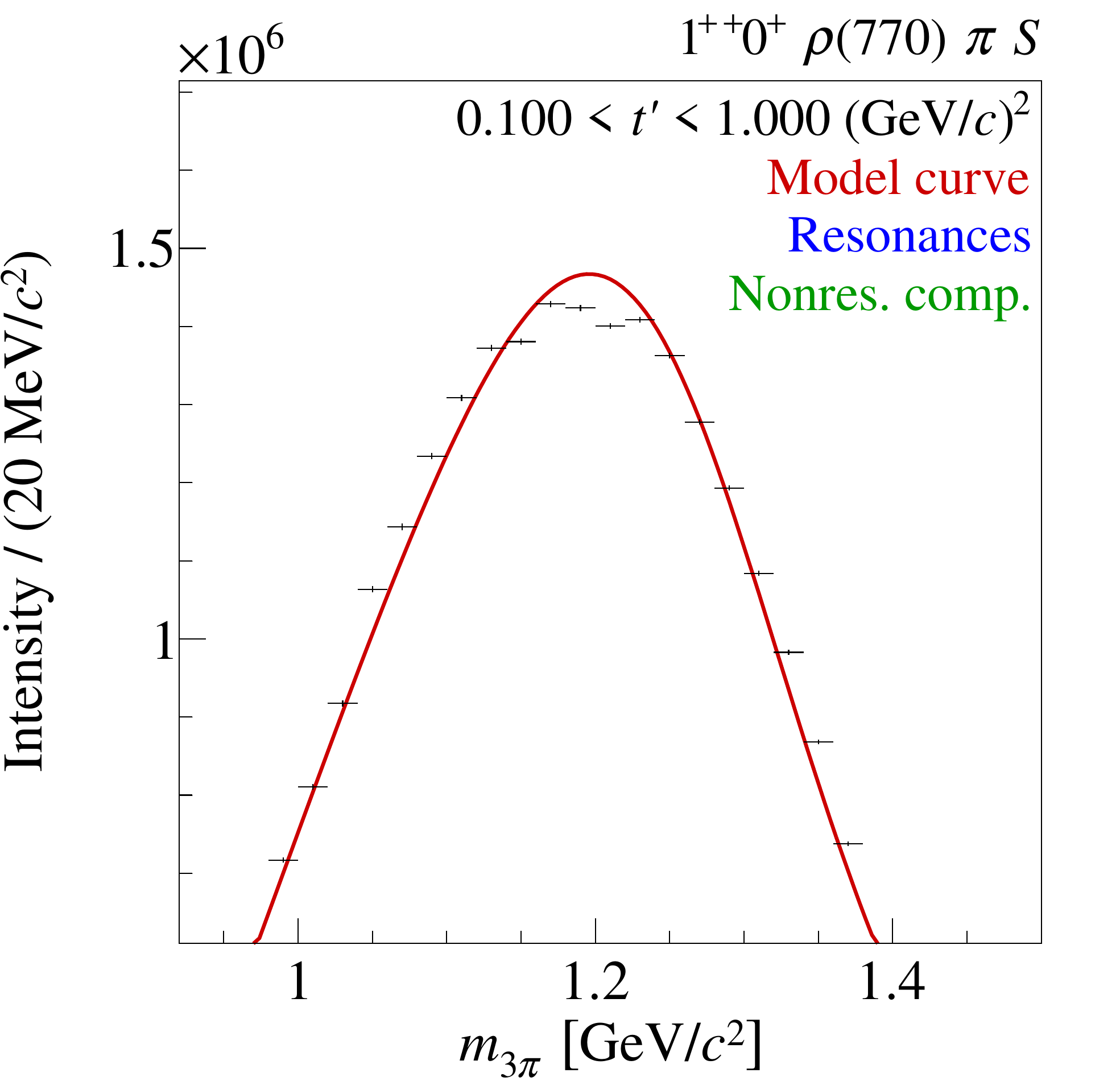}%
  \caption{Zoomed view of the \tpr-summed \wave{1}{++}{0}{+}{\Prho}{S}
    intensity distribution in \cref{fig:intensity_1pp_rho_pi_S_tsum}.}
  \label{fig:intensity_1pp_rho_pi_S_zoom}
\end{figure}

The \PaOne[1640] parameters are mainly determined by the
$\PfTwo \pi P$ wave.  The model describes the low-mass enhancement of
the $\PfTwo \pi P$ intensity by a dominant nonresonant component that
is sharply peaked in the \PaOne region and a comparatively small
\PaOne component.  The high-mass region of the $\PfTwo \pi P$
intensity is dominated by a peak at about \SI{1.8}{\GeVcc} that is
described well as the constructive interference of the \PaOne and
\PaOne[1640] components.  The peak disappears toward
$\tpr = \SI{1.0}{\GeVcsq}$ and so do the resonance components.  The
extrapolations of the model below and above the fit range undershoot
the $\PfTwo \pi P$ intensity at low and high \mThreePi [see
\cref{fig:intensity_1pp_f2_tbin1,fig:intensity_1pp_f2_tbin11}].

The intensity distribution of the $\PfZero[980] \pi P$ wave is
peculiar in that it shows a peak slightly above the \PaOne but
significantly narrower.  The peak is well described by the third
$1^{++}$ resonance in our model, the \PaOne[1420].  The \PaOne[1420]
interferes destructively with a smaller nonresonant component that
peaks at about \SI{1.3}{\GeVcc}.  The model is not able to describe
the high-mass tail, which grows with increasing \tpr.  This is why the
fit range was limited to below \SI{1.6}{\GeVcc}.

Within the fit ranges, the model describes the relative phases of the
$1^{++}$ waves better than the intensity distributions discussed
above.  In particular the rapid phase motion of the
\wave{1}{++}{0}{+}{\PfZero[980]}{P} \wrt other waves is well
reproduced.  This is also true for the phase motions of the
\wave{1}{++}{0}{+}{\PfTwo}{P} wave in the \SI{1.6}{\GeVcc} region,
which are caused by the \PaOne[1640].  A significant \PaOne[1640]
component in this wave is also consistent with the phase relative to
the \wave{2}{++}{1}{+}{\PfTwo}{P} wave (see \cref{fig:phases_1pp_f2}).
The \PaTwo causes a decreasing phase in the \SI{1.3}{\GeVcc} region.
At higher masses, the relative phase varies only slightly due to a
compensation of the phase motions of \PaTwo[1700] and \PaOne[1640].
The phase of the \wave{1}{++}{0}{+}{\Prho}{S} wave shows a completely
different behavior [see
\cref{fig:phase_2pp_f2_1pp_rho_tbin1,fig:phase_2pp_f2_1pp_rho_tbin11}].
In addition to the rapid phase motion caused by the \PaTwo, also the
\PaTwo[1700] creates a clear phase motion that is not canceled by the
\PaOne[1640].  Hence neither \PaOne nor \PaOne[1640] causes strong
phase motions of the \wave{1}{++}{0}{+}{\Prho}{S} wave.  This is also
true for the phases of this wave \wrt other waves [see \eg
\cref{fig:phase_1pp_rho_1pp_f0_tbin1,fig:phase_1pp_rho_1pp_f2_tbin1,fig:phase_1pp_rho_4pp_rho_tbin1,fig:phase_1pp_rho_1pp_f0_tbin11,fig:phase_1pp_rho_1pp_f2_tbin11,fig:phase_1pp_rho_4pp_rho_tbin11}].
The behavior of the phases is consistent with the large nonresonant
component over the full mass range and the weak signal of the
\PaOne[1640] compared to the \PaOne ground state in the $\Prho \pi S$
wave.  For many phases, the extrapolations of the model below and
above the fit range follow approximately the data.  Deviations appear
in particular at low \tpr.  For the phases of the
\wave{1}{++}{0}{+}{\Prho}{S} wave, the model extrapolations deviate
from the data in the region above \SI{2.3}{\GeVcc} [see \eg
\cref{fig:phase_1pp_rho_1pp_f2_tbin1,fig:phase_1pp_rho_4pp_rho_tbin1}].
For the phases of the \wave{1}{++}{0}{+}{\PfTwo}{P} wave, the model
deviates typically at low masses [see \eg
\cref{fig:phase_1pp_rho_1pp_f2_tbin1,fig:phase_1pp_f2_4pp_rho_tbin1}]
where also the intensity distribution is not well reproduced.

\begin{figure}[tbp]
  \centering
  \subfloat[][]{%
    \includegraphics[width=\twoPlotWidth]{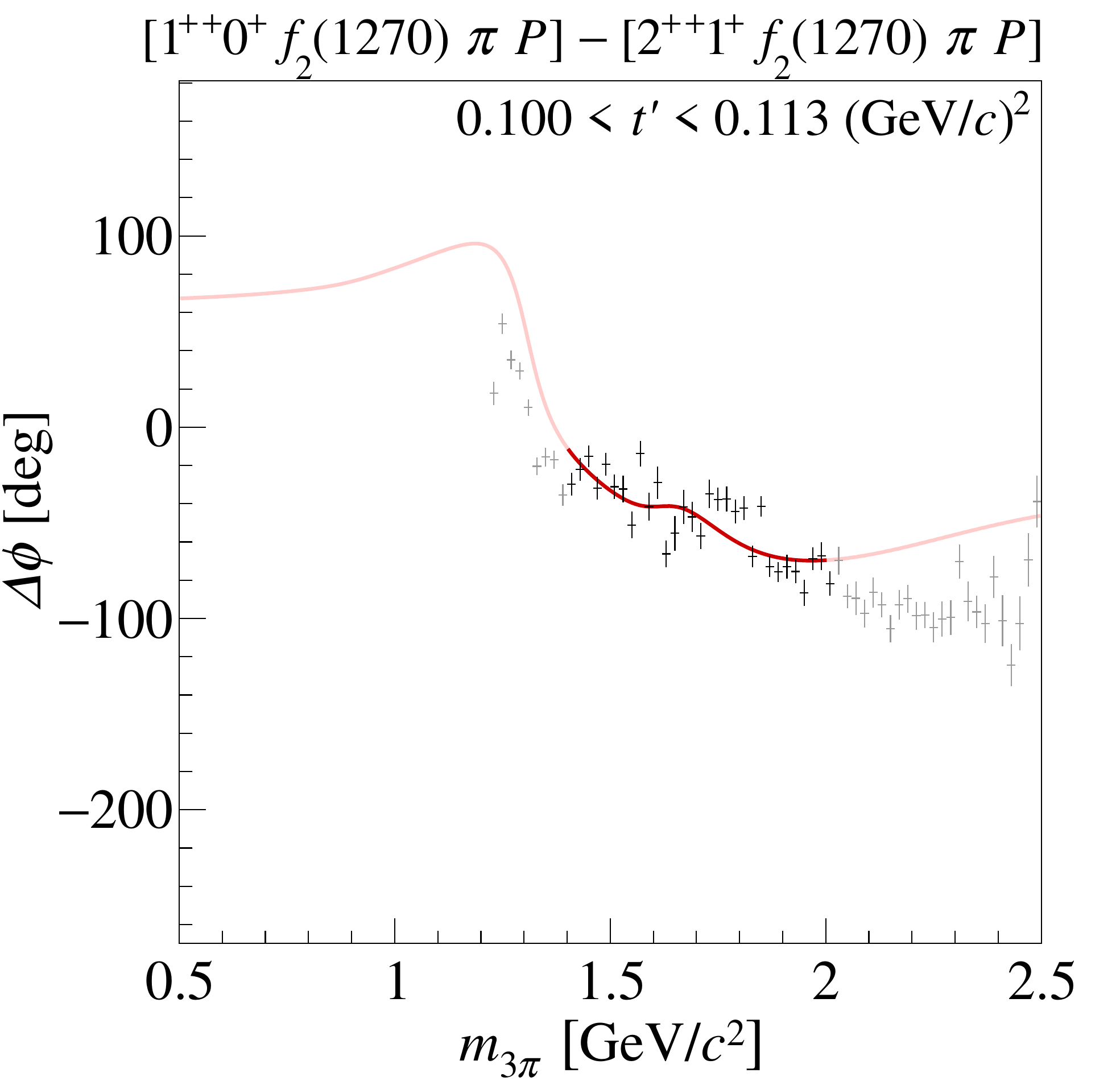}%
    \label{fig:phase_1pp_f2_2pp_f2_tbin1}%
  }%
  \newLineOrHspace{\twoPlotSpacing}%
  \subfloat[][]{%
    \includegraphics[width=\twoPlotWidth]{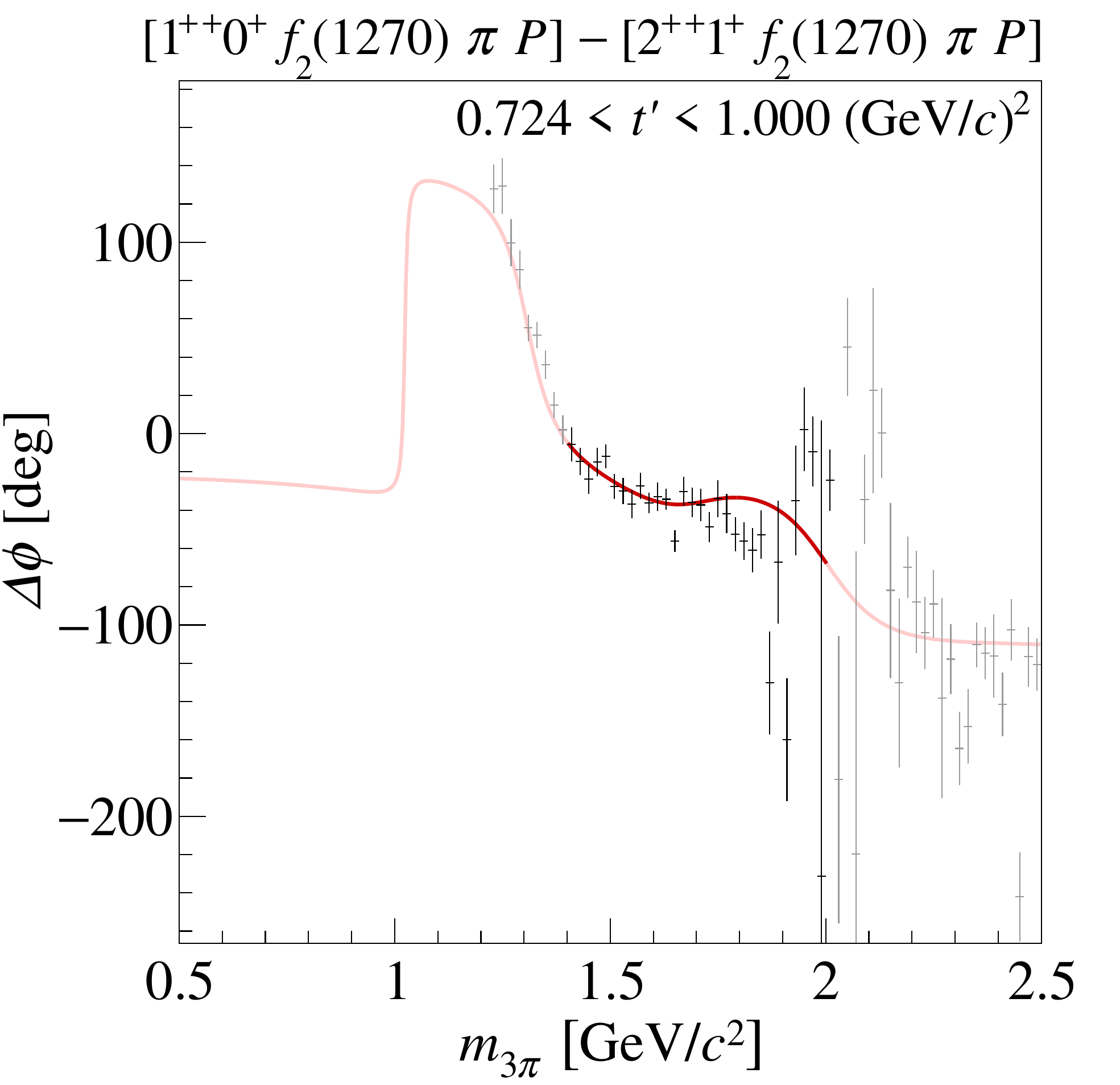}%
    \label{fig:phase_1pp_f2_2pp_f2_tbin11}%
  }%
  \caption{Phase of the \wave{1}{++}{0}{+}{\PfTwo}{P} wave relative to
    the \wave{2}{++}{1}{+}{\PfTwo}{P} wave,
    \subfloatLabel{fig:phase_1pp_f2_2pp_f2_tbin1}~for the lowest \tpr
    bin and \subfloatLabel{fig:phase_1pp_f2_2pp_f2_tbin11}~for the
    highest \tpr bin.  The model is represented as in
    \cref{fig:intensity_phases_1pp_tbin1}.}
  \label{fig:phases_1pp_f2}
\end{figure}

\Cref{fig:tprim_1pp} shows the \tpr spectra of the $1^{++}$ wave
components together with the results of fits using
\cref{eq:slope-parametrization}.  The \tpr dependence of the
amplitudes of \PaOne and \PaOne[1640] in the
\wave{1}{++}{0}{+}{\Prho}{S} and \wave{1}{++}{0}{+}{\PfTwo}{P} waves
is constrained via \cref{eq:method:branchingdefinition}.  The \tpr
dependence of the \PaOne[1420] amplitude in the
\wave{1}{++}{0}{+}{\PfZero[980]}{P} wave is independently determined
by the fit.  The simple exponential model in
\cref{eq:slope-parametrization} is in fair agreement with the \tpr
spectra of all $1^{++}$ wave components.  The extracted
slope-parameter values for the \PaOne are
\SIaerrSys{11.8}{0.9}{4.2}{\perGeVcsq} in the $\Prho \pi S$ wave and
\SIerrSys{11}{4}{\perGeVcsq} in the $\PfTwo \pi P$ wave.  The \PaOne
has the steepest \tpr spectrum of all resonances in the model (see
\cref{tab:slopes}) although the uncertainty toward smaller slope
values is considerable.  The \PaOne slope values agree within
uncertainties with the slope values of the nonresonant components in
all three $1^{++}$ waves.  This is in contrast to most other waves,
for which we typically observe steeper \tpr spectra for the
nonresonant components.  This might be a hint that the model is not
able to completely separate the \PaOne from the nonresonant
components.  As expected, the \PaOne[1640] has a shallower \tpr
spectrum with slope-parameter values close to \SI{8}{\perGeVcsq}.
This value is similar to those of other resonances.  In particular, it
agrees with the slopes of the \PaTwo[1700], which has similar
resonance parameters.

\begin{wideFigureOrNot}[tbp]
  \centering
  \subfloat[][]{%
    \includegraphics[width=\threePlotWidth]{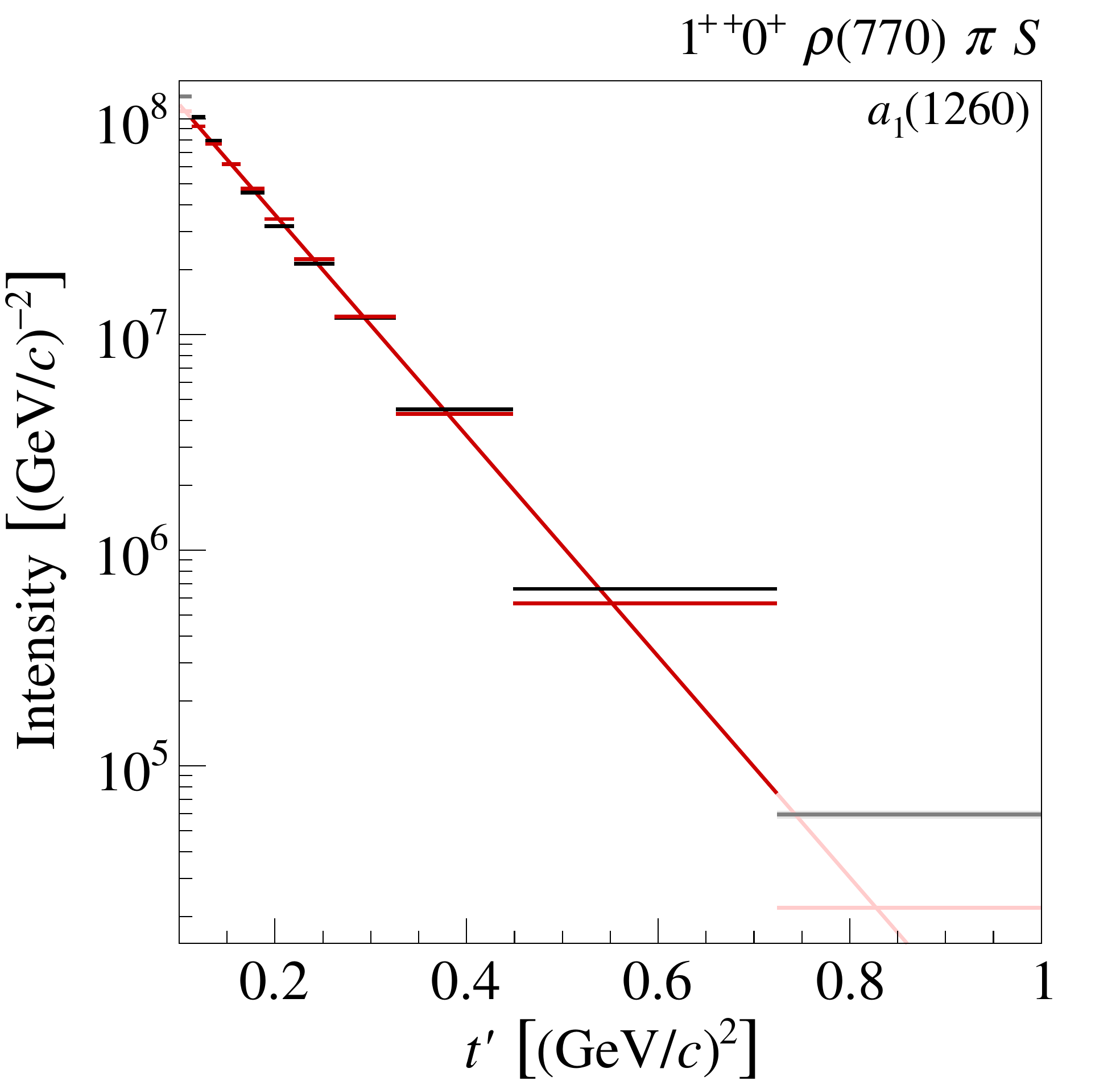}%
  }%
  \hspace*{\threePlotSpacing}%
  \subfloat[][]{%
    \includegraphics[width=\threePlotWidth]{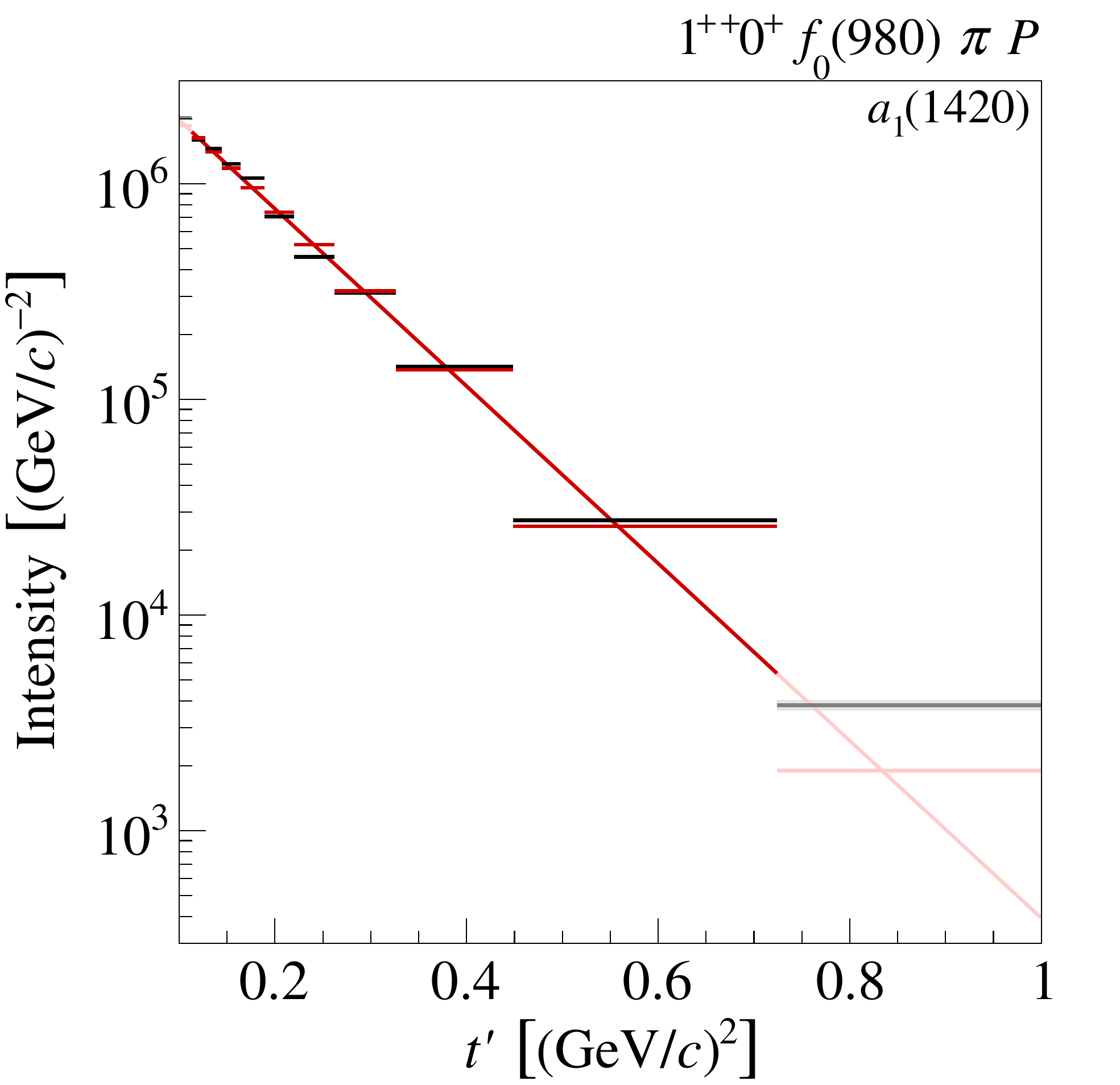}%
  }%
  \hspace*{\threePlotSpacing}%
  \subfloat[][]{%
    \includegraphics[width=\threePlotWidth]{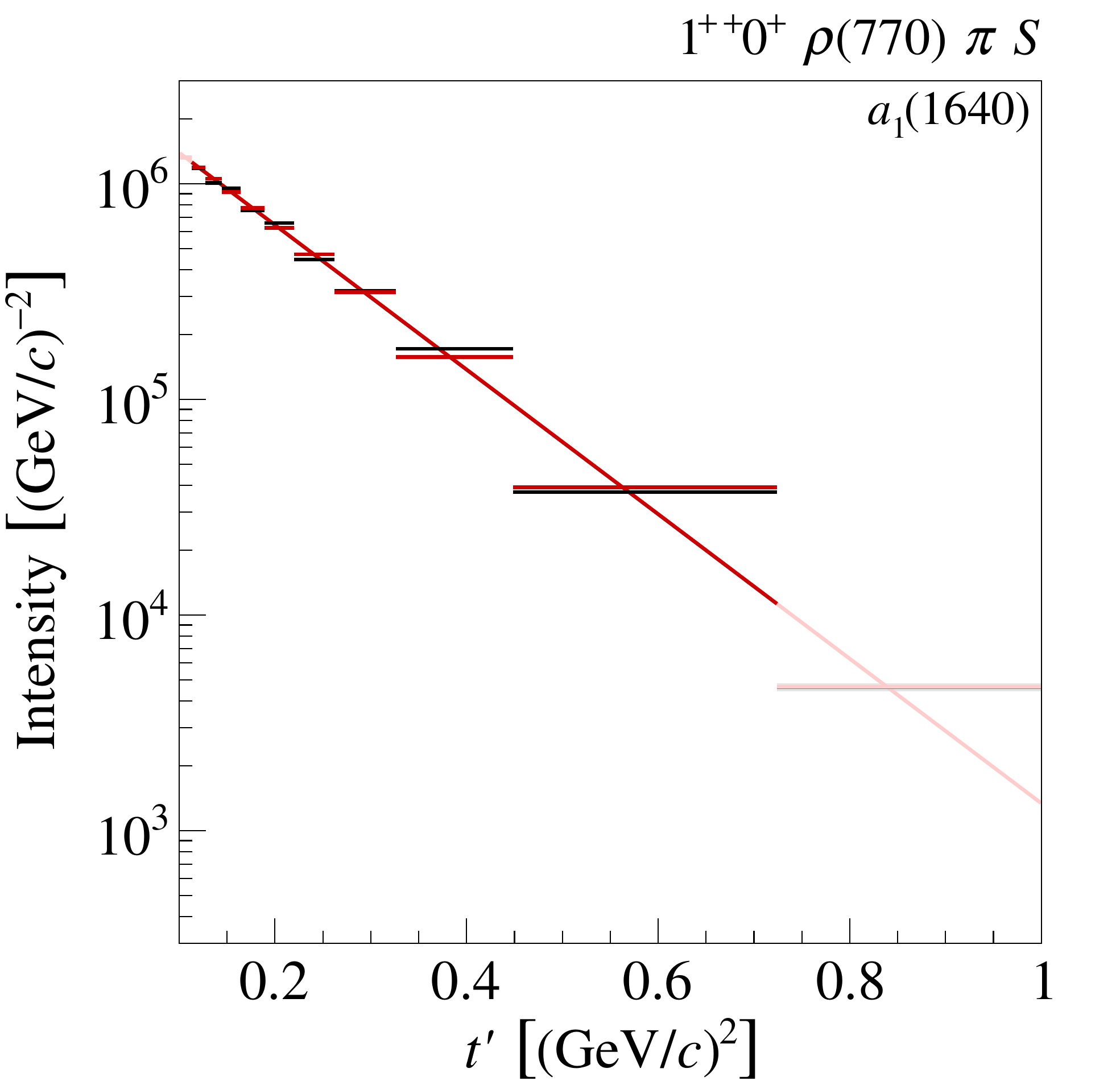}%
  }%
  \\
  \subfloat[][]{%
    \includegraphics[width=\threePlotWidth]{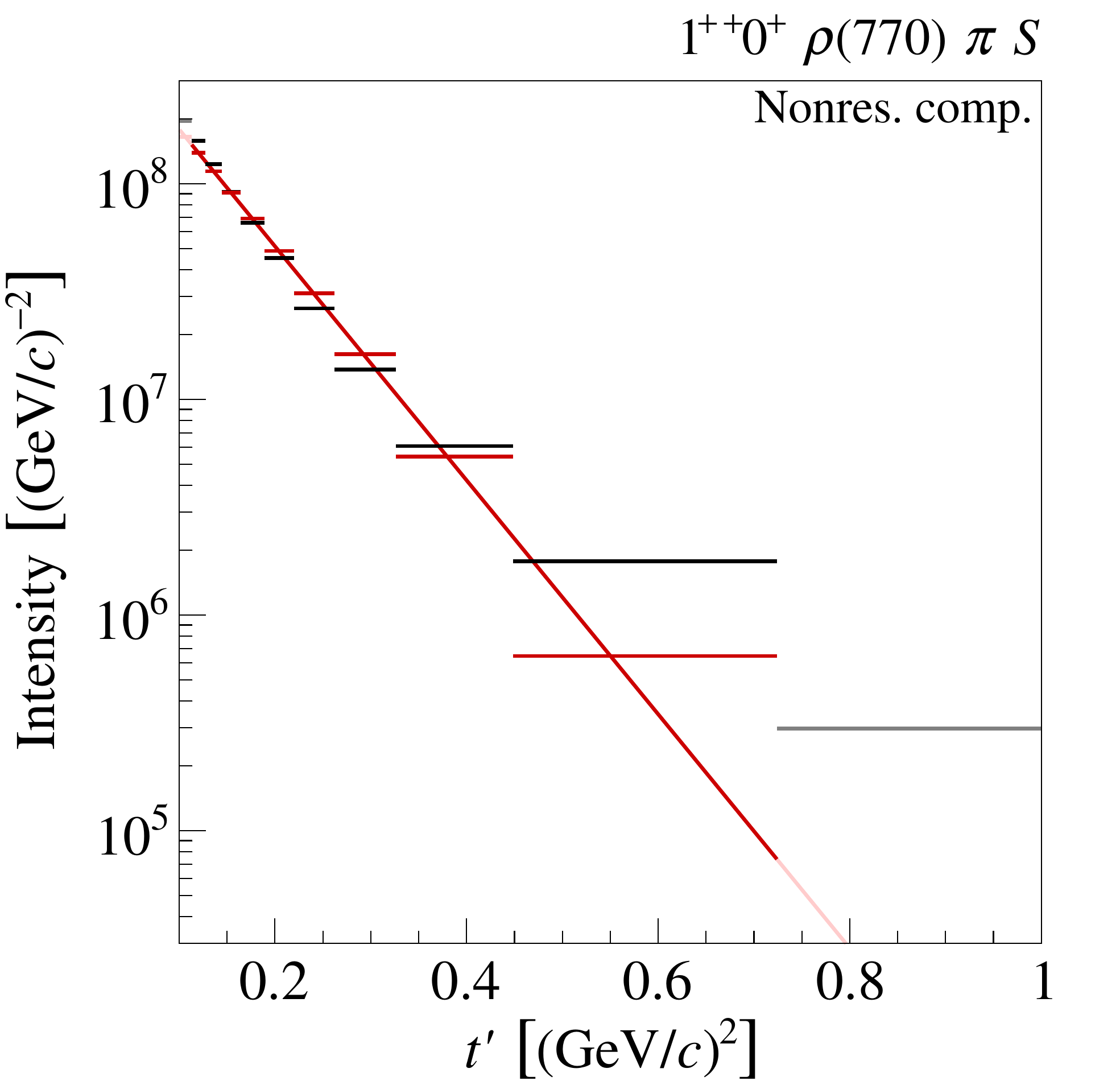}%
  }%
  \hspace*{\threePlotSpacing}%
  \subfloat[][]{%
    \includegraphics[width=\threePlotWidth]{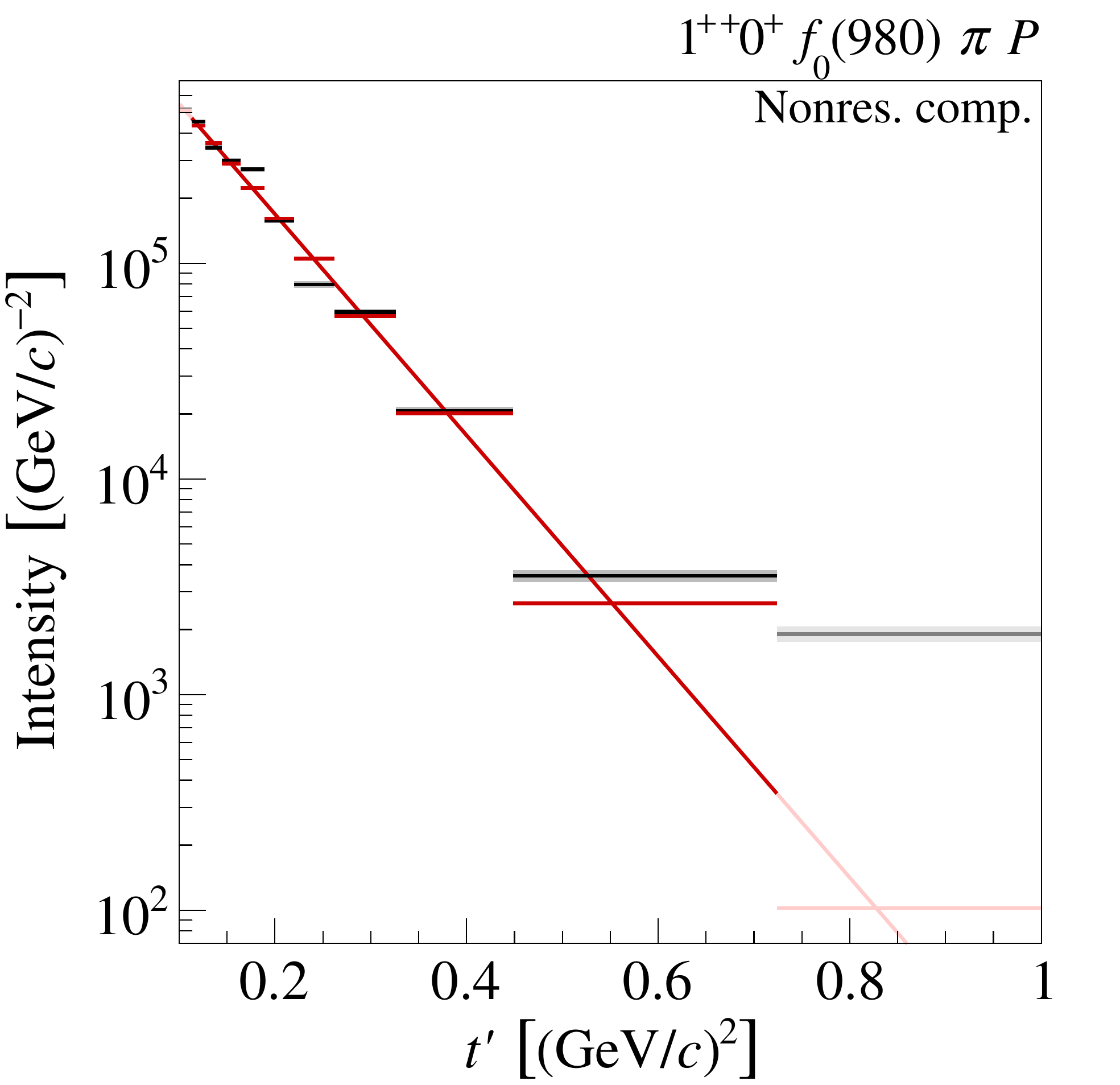}%
  }%
  \hspace*{\threePlotSpacing}%
  \subfloat[][]{%
    \includegraphics[width=\threePlotWidth]{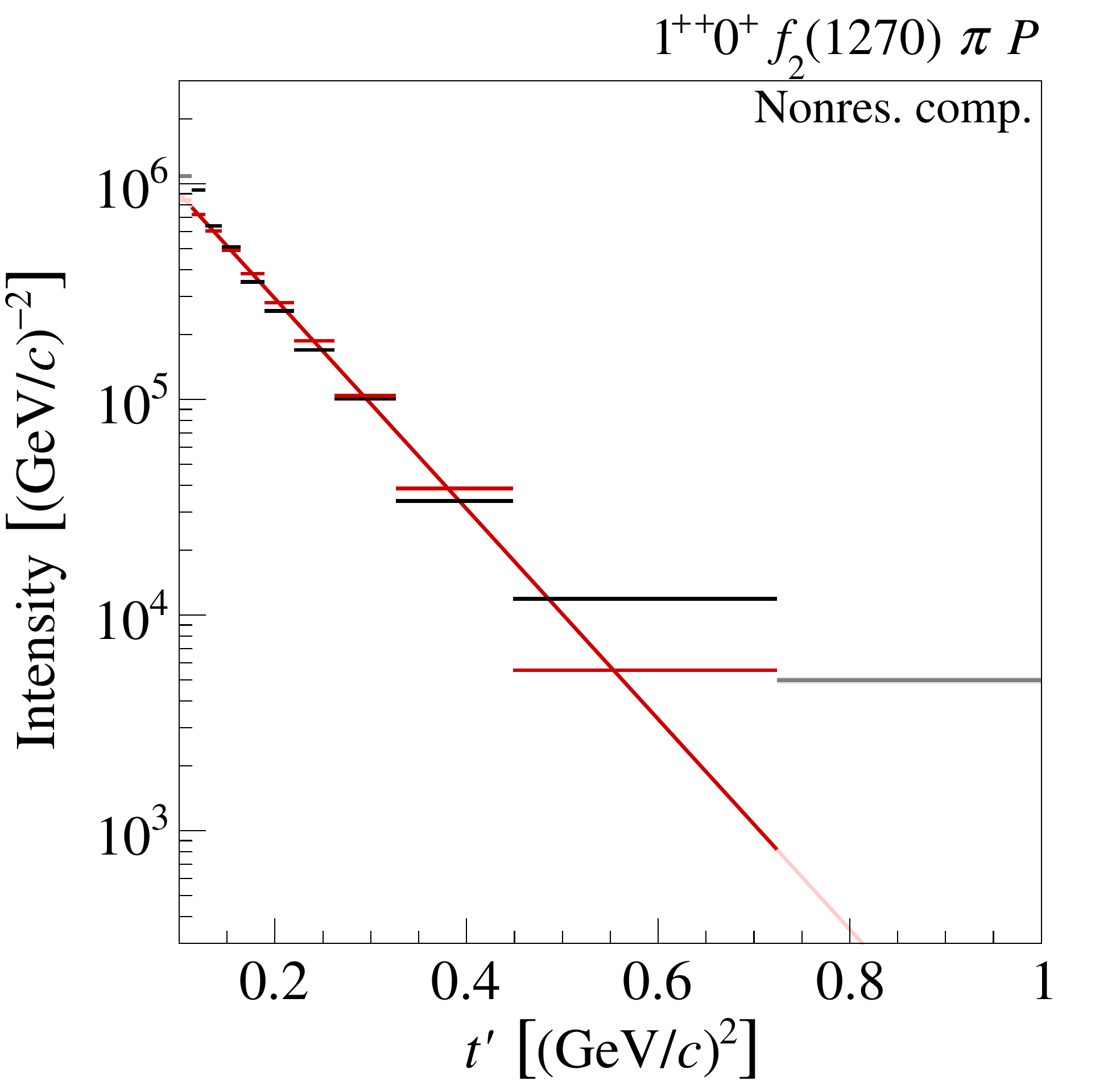}%
  }%
  \caption{Similar to \cref{fig:method:tp:examplespectrum}, but
    showing the \tpr spectra of some of the components in the three
    $1^{++}$ waves as given by \cref{eq:tprim-dependence}.  The red
    curves and horizontal lines represent fits using
    \cref{eq:slope-parametrization}.}
  \label{fig:tprim_1pp}
\end{wideFigureOrNot}

The \tpr spectrum of the \PaOne[1420] in the $\PfZero[980] \pi P$ wave
is consistent with the resonance interpretation of this signal.  The
\PaOne[1420] slope parameter has a value of
\SIaerrSys{9.5}{0.6}{1.0}{\perGeVcsq}, which confirms the tendency
that slopes decrease with increasing mass.

If none of the coupling amplitudes of the resonance components is
constrained via \cref{eq:method:branchingdefinition} [\StudyT; see
\cref{sec:systematics}], the model has more freedom and can better
describe the intensity distribution of the $\Prho \pi S$ wave at high
\tpr [see \cref{fig:intensity_1pp_rho_tbin11_noCoupling}].  The \PaOne
resonance parameters change only slightly.  However, the extracted
\PaOne slope parameters become inconsistent: \SI{9.0}{\perGeVcsq} in
the $\Prho \pi S$ and \SI{15}{\perGeVcsq} in the $\PfTwo \pi P$ wave.
The slope of the \PaOne[1640] increases to \SI{14}{\perGeVcsq} in the
$\Prho \pi S$ wave but remains practically unchanged in the
$\PfTwo \pi P$ wave.  This confirms that the \PaOne[1640] resonance is
well determined by the $\PfTwo \pi P$ wave.  The results of \StudyT
also indicate that without the constraint of
\cref{eq:method:branchingdefinition} the relative intensities of the
two \PaOne* states and the nonresonant components are not well
constrained by the data.

\begin{figure}[tbp]
  \centering
  \subfloat[][]{%
    \includegraphics[width=\threePlotWidth]{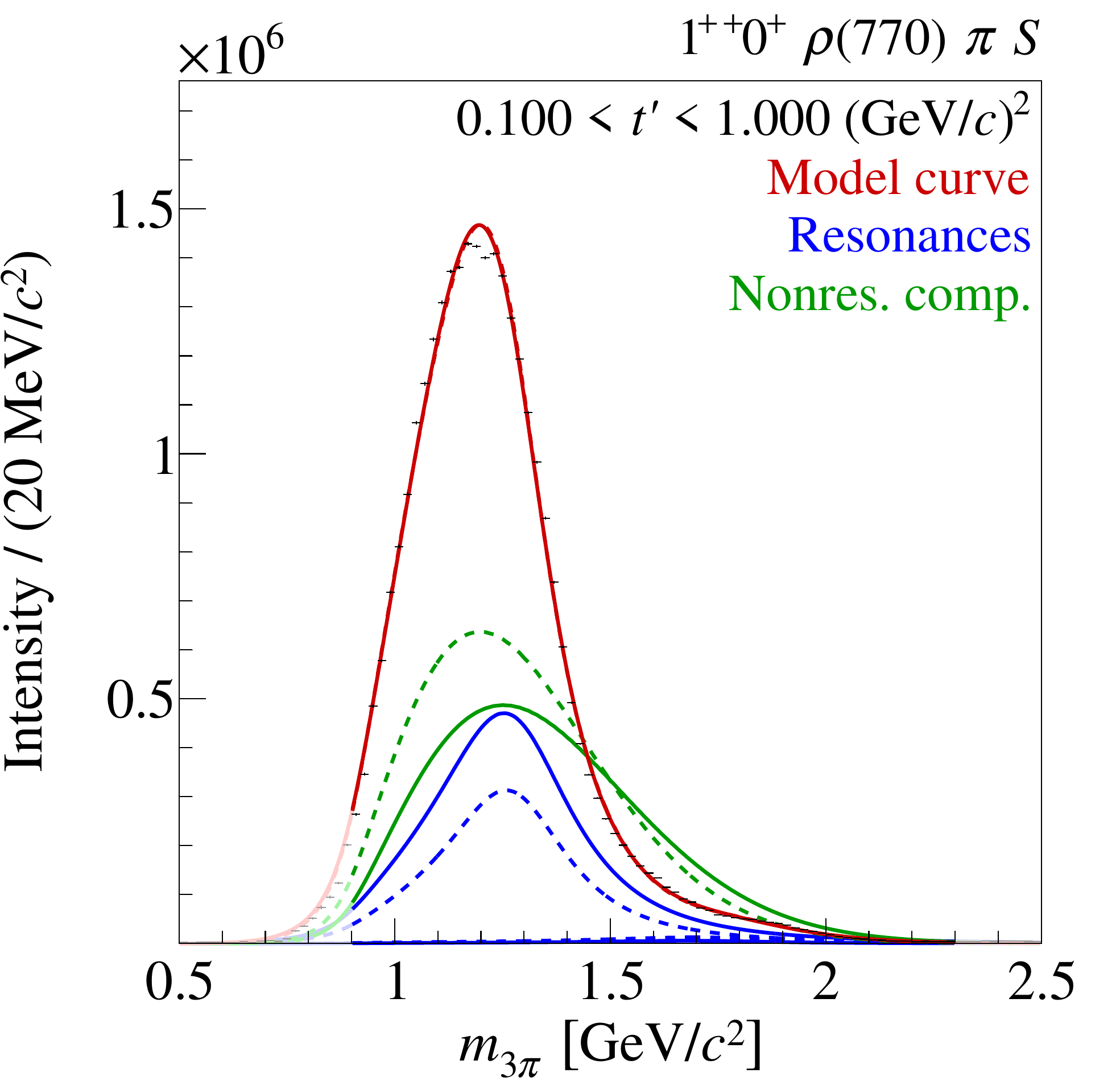}%
    \label{fig:intensity_1pp_rho_noCoupling}%
  }%
  \newLineOrHspace{\threePlotSpacing}%
  \subfloat[][]{%
    \includegraphics[width=\threePlotWidth]{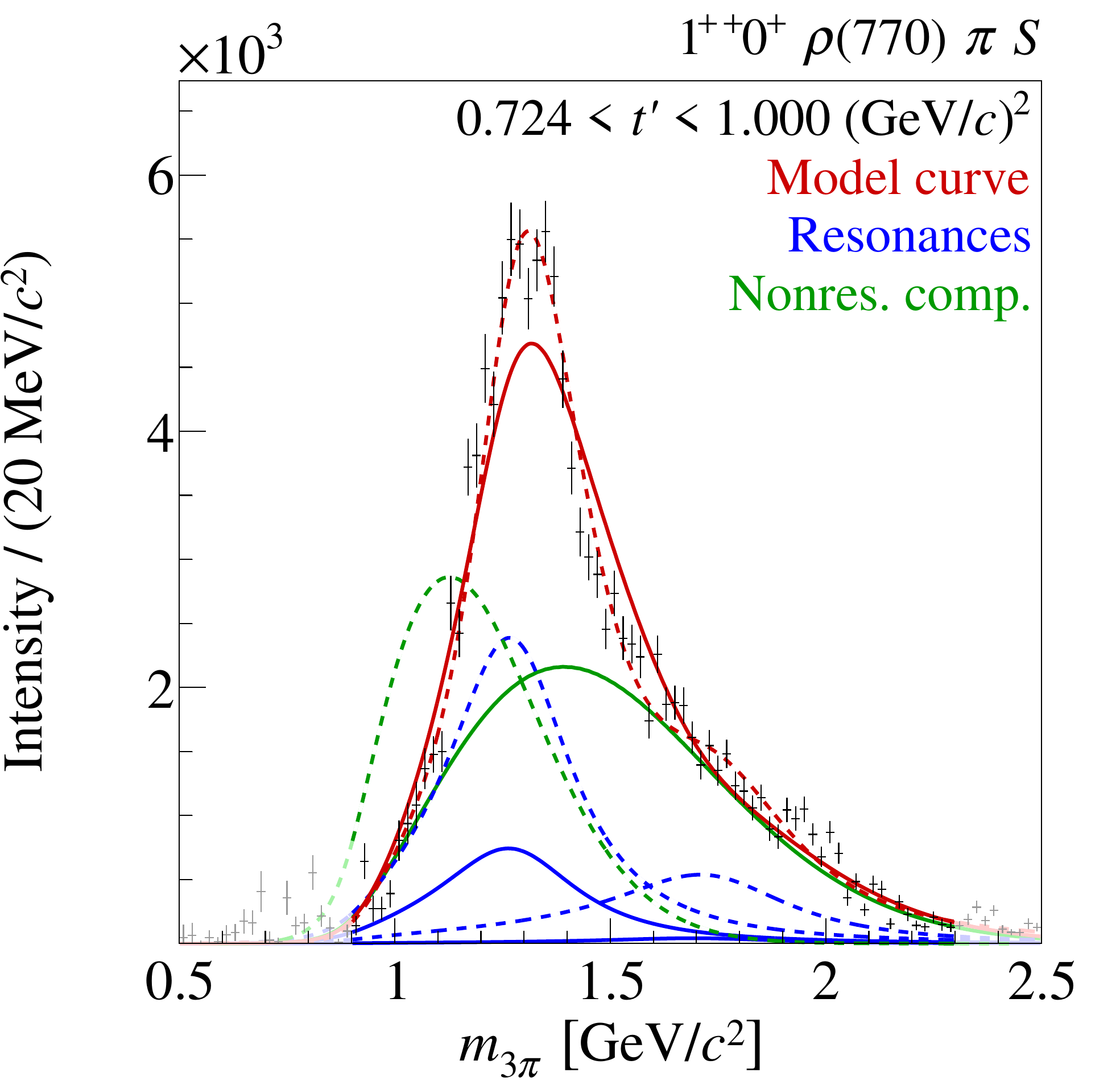}%
    \label{fig:intensity_1pp_rho_tbin11_noCoupling}%
  }%
  \newLineOrHspace{\threePlotSpacing}%
  \subfloat[][]{%
    \includegraphics[width=\threePlotWidth]{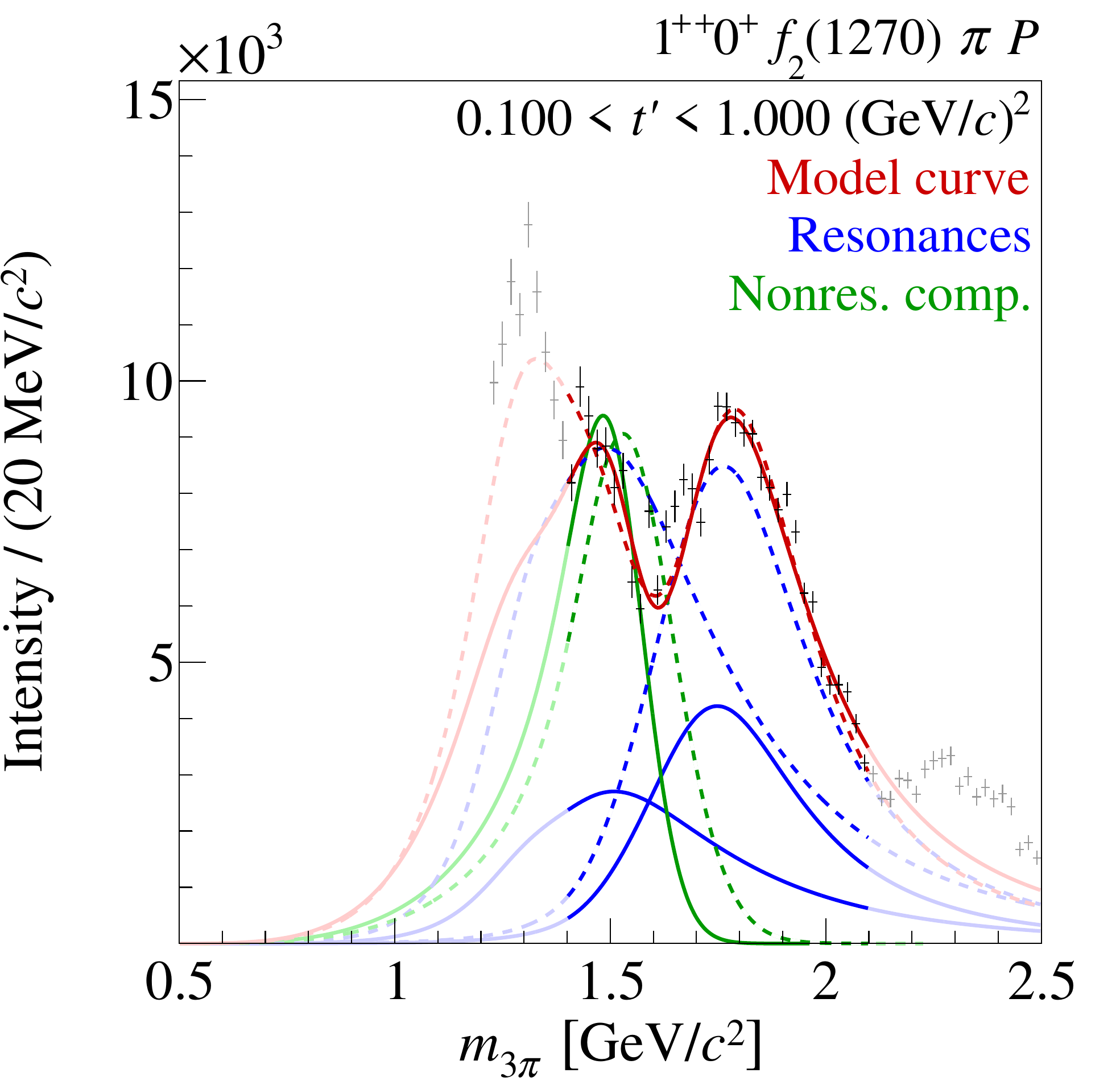}%
    \label{fig:intensity_1pp_f2_noCoupling}%
  }%
  \caption{\subfloatLabel{fig:intensity_1pp_rho_noCoupling}~\tpr-summed
    intensity of the \wave{1}{++}{0}{+}{\Prho}{S} wave.
    \subfloatLabel{fig:intensity_1pp_rho_tbin11_noCoupling} Intensity
    of this wave in the highest \tpr bin.
    \subfloatLabel{fig:intensity_1pp_f2_noCoupling}~\tpr-summed
    intensity of the \wave{1}{++}{0}{+}{\PfTwo}{P} wave.  The result
    of the main fit is represented by the continuous curves.  The fit,
    in which none of the coupling amplitudes of the resonance
    components was constrained via
    \cref{eq:method:branchingdefinition} [\StudyT; see
    \cref{sec:systematics}], is represented by the dashed curves.  The
    model and the wave components are represented as in
    \cref{fig:intensity_phases_1pp_tbin1}.}
  \label{fig:intensities_1pp_noCoupling}
\end{figure}

For the \PaOne, we extract the resonance parameters
$m_{\PaOne} = \SIaerrSys{1299}{12}{28}{\MeVcc}$ and
$\Gamma_{\PaOne} = \SIerrSys{380}{80}{\MeVcc}$.  The extracted
resonance parameters for the \PaOne[1640] are
$m_{\PaOne[1640]} = \SIaerrSys{1700}{35}{130}{\MeVcc}$ and
$\Gamma_{\PaOne[1640]} = \SIaerrSys{510}{170}{90}{\MeVcc}$.  Due to
the dominance of the \PaOne signal, the parameters of the \PaOne[1640]
are correlated with those of the \PaOne.  The fit model does not
describe well the $\Prho \pi S$ and $\PfTwo \pi P$ intensities in some
mass regions.  This leads to a bimodal behavior of the fit with a
second solution with a narrower \PaOne and a wider and heavier
\PaOne[1640].  In the main fit, this solution has a larger~\chisq but
in some of the systematic studies (see \cref{sec:systematics}), the
solution with the narrow \PaOne is preferred.
The parameters of \PaOne and \PaOne[1640] depend strongly on the
interference of the $1^{++}$ and $2^{++}$ waves and therefore on the
set of $2^{++}$ waves included in the fit.
We also observe a large dependence of the parameters of \PaOne and
\PaOne[1640] on the number of background events in the selected data
sample.
Studies~\studyS and~\studyR with alternative \chisq~formulations (see
\cref{sec:alt_chi_2}) indicate that the model deviates more from the
measured intensity distributions than from the phases of the $1^{++}$
waves.
The results from the above mentioned systematic studies are discussed
in more detail in \cref{sec:syst_uncert_onePP}.

Since the \wave{1}{++}{0}{+}{\Prho}{S} wave has a large nonresonant
component, the fit result depends on the choice of the parametrization
used for the nonresonant component.  Also the strongly peaked shape of
the nonresonant component at about \SI{1.5}{\GeVcc} in the
\wave{1}{++}{0}{+}{\PfTwo}{P} wave seems rather implausible.  We
therefore studied the dependence of the fit result on the
parametrization used for the nonresonant component.  In \StudyO, we
replace the parametrization of the nonresonant amplitude by the square
root of the intensity distribution of the partial-wave decomposition
of Deck Monte Carlo data that were generated according to the model
described in \cref{sec:deck_model}.  This model describes the measured
\wave{1}{++}{0}{+}{\Prho}{S} amplitude well [see
\cref{fig:DeckMC_chi2difference,fig:intensity_1pp_rho_DeckMC,fig:intensity_1pp_rho_tbin11_DeckMC}].
The $\Prho \pi S$ intensity distribution in the high-\tpr region is
described even better than in the main fit.  The shape of the
nonresonant component from the Deck model in the $\Prho \pi S$ wave is
qualitatively similar to that obtained in the main fit.  The \PaOne
parameters change only slightly but the yield of the \PaOne component
becomes larger and that of the nonresonant component smaller in
particular at high \tpr.  The model is also in fair agreement with the
\wave{1}{++}{0}{+}{\PfTwo}{P} intensity distribution [see
\cref{fig:intensity_1pp_f2_DeckMC}], although the shape of the
nonresonant component from the Deck model is drastically different
from that used in the main fit.  The \PaOne[1640] width increases by
\SI{126}{\MeVcc} in \StudyO.

\begin{wideFigureOrNot}[tbp]
  \centering
  \subfloat[][]{%
    \includegraphics[width=\twoPlotWidth]{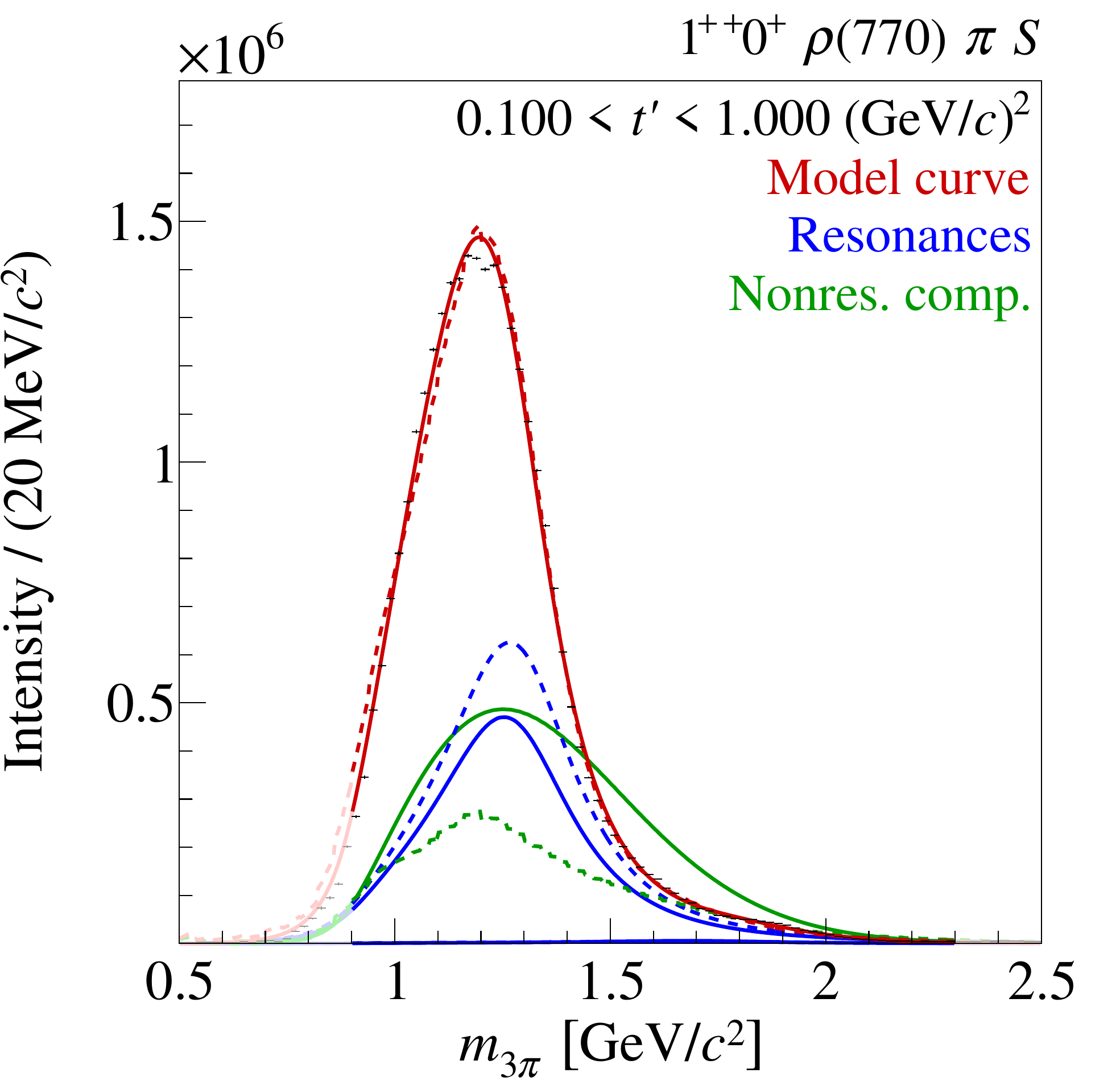}%
    \label{fig:intensity_1pp_rho_DeckMC}%
  }%
  \hspace*{\twoPlotSpacing}%
  \subfloat[][]{%
    \includegraphics[width=\twoPlotWidth]{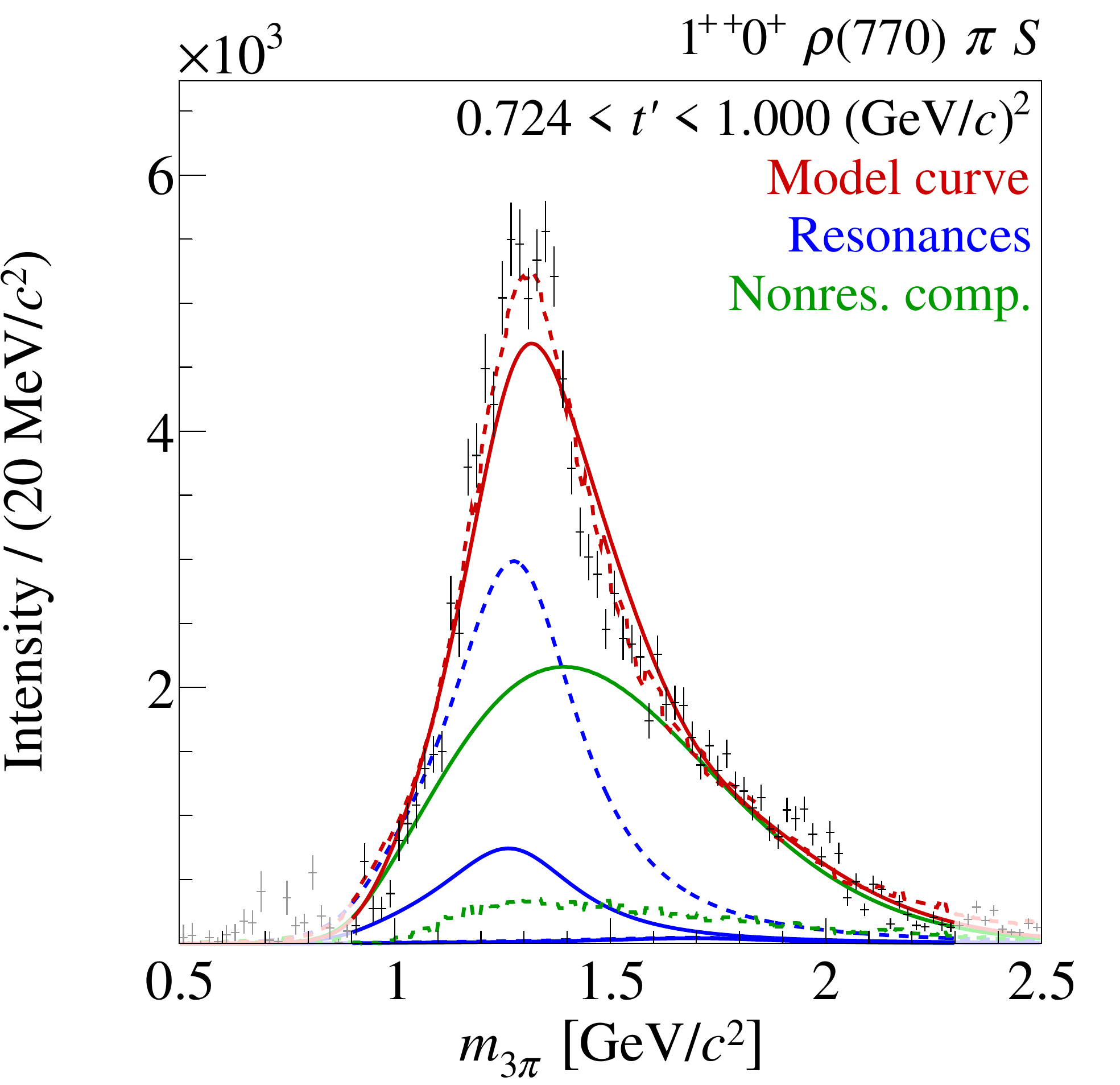}%
    \label{fig:intensity_1pp_rho_tbin11_DeckMC}%
  }%
  \\
  \subfloat[][]{%
    \includegraphics[width=\twoPlotWidth]{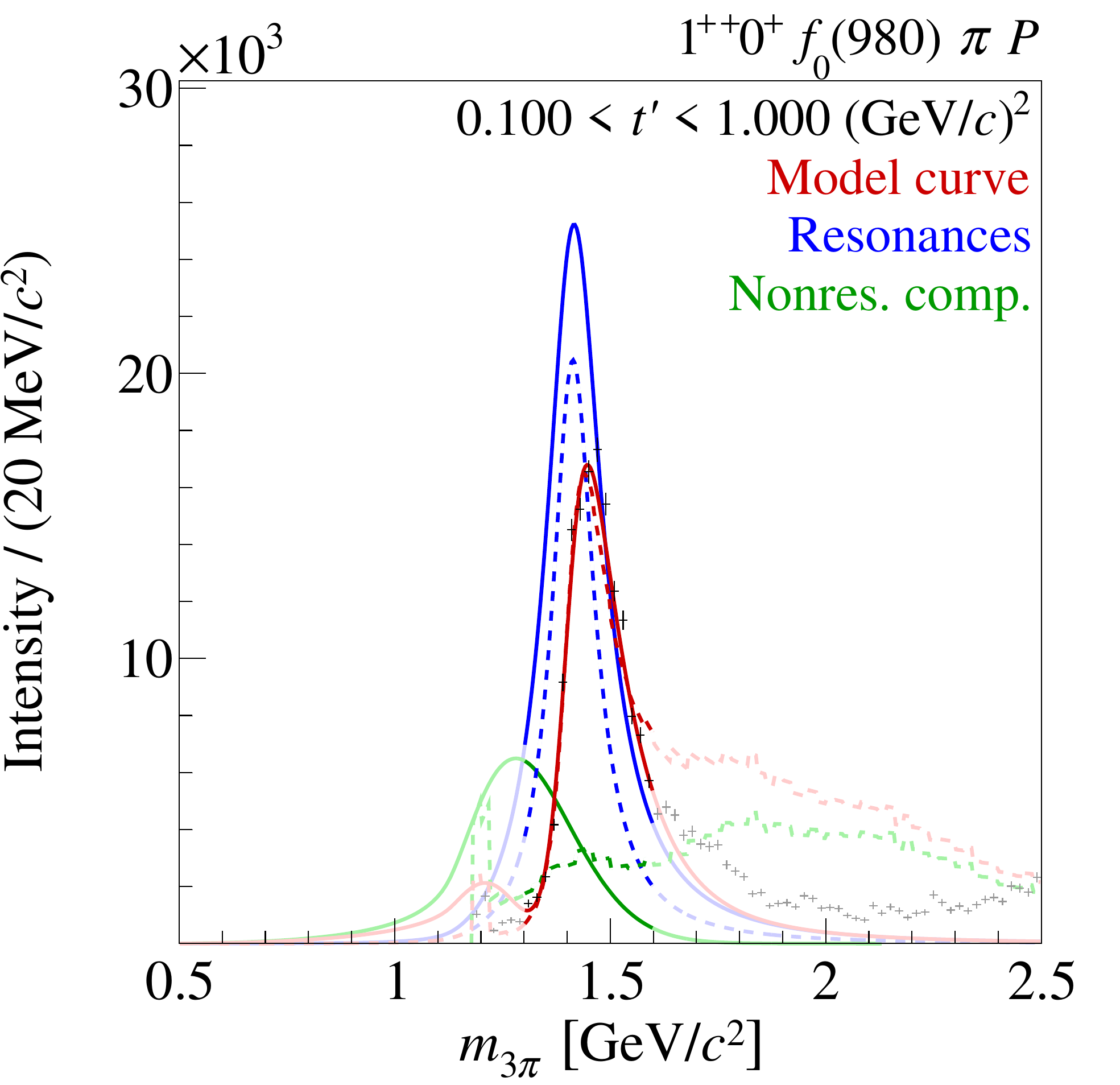}%
    \label{fig:intensity_1pp_f0_DeckMC}%
  }%
  \hspace*{\twoPlotSpacing}%
  \subfloat[][]{%
    \includegraphics[width=\twoPlotWidth]{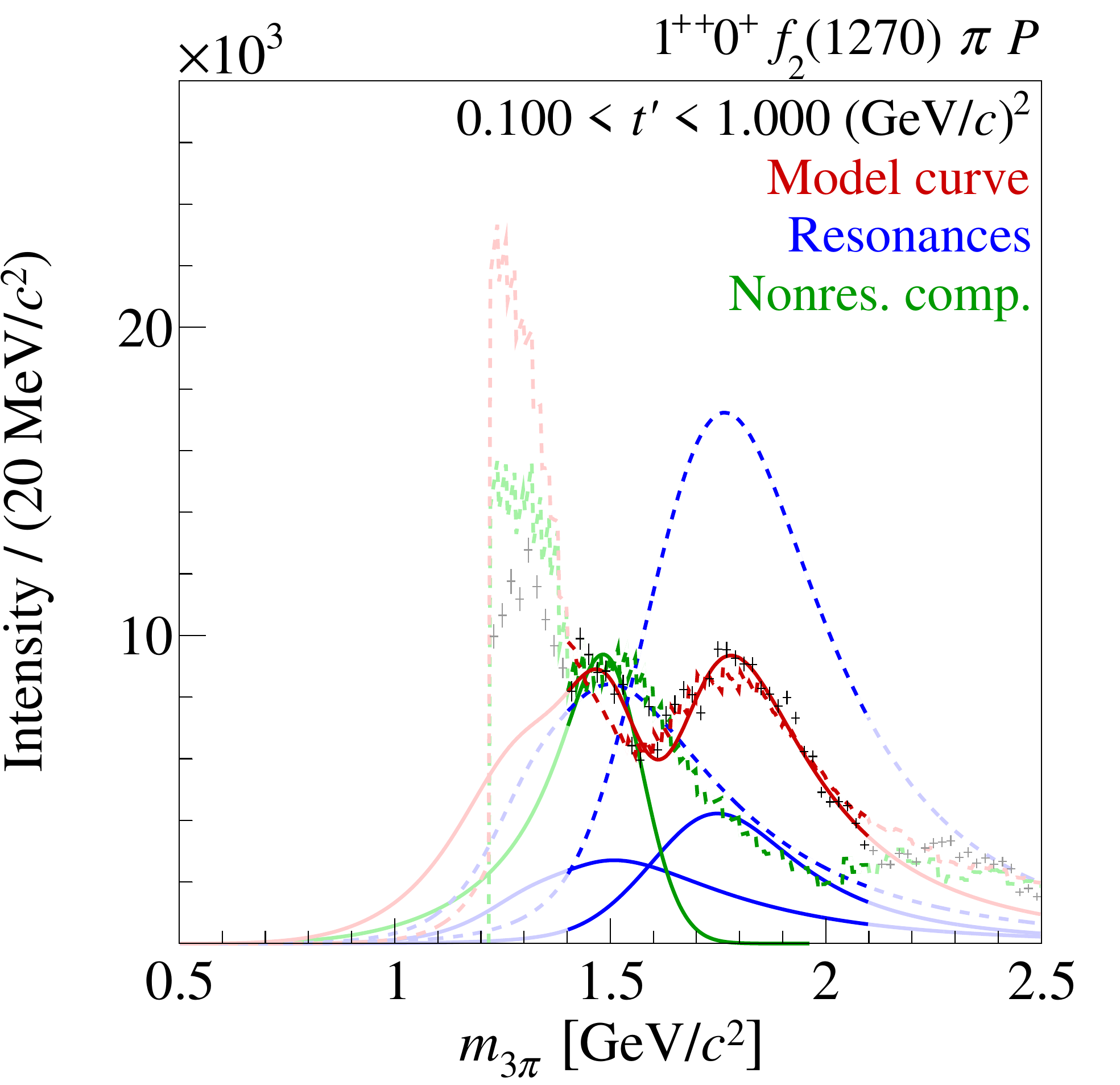}%
    \label{fig:intensity_1pp_f2_DeckMC}%
  }%
  \caption{\subfloatLabel{fig:intensity_1pp_rho_DeckMC}~\tpr-summed
    intensity of the \wave{1}{++}{0}{+}{\Prho}{S} wave.
    \subfloatLabel{fig:intensity_1pp_rho_tbin11_DeckMC} Intensity of
    this wave in the highest \tpr bin.
    \subfloatLabel{fig:intensity_1pp_f0_DeckMC}~and~\subfloatLabel{fig:intensity_1pp_f2_DeckMC}:
    The \tpr-summed intensity of the
    \wave{1}{++}{0}{+}{\PfZero[980]}{P} and the
    \wave{1}{++}{0}{+}{\PfTwo}{P} wave, respectively.  The result of
    the main fit is represented by the continuous curves.  The fit, in
    which the parametrization of the nonresonant amplitude was
    replaced by the square root of the intensity distribution of the
    partial-wave decomposition of Deck Monte Carlo data [\StudyO; see
    \cref{sec:systematics}], is represented by the dashed curves.  The
    model and the wave components are represented as in
    \cref{fig:intensity_phases_1pp_tbin1}.}
  \label{fig:intensities_1pp_DeckMC}
\end{wideFigureOrNot}

Compared to the studies discussed above, the \PaOne and \PaOne[1640]
parameters depend only weakly on the particular choice of the \tpr
binning.  This was verified in \StudyL, in which the analysis was
performed using only 8~\tpr bins.

From the fit, we extract \PaOne[1420] resonance parameters of
$m_{\PaOne[1420]} = \SIaerrSys{1411}{4}{5}{\MeVcc}$ and
$\Gamma_{\PaOne[1420]} = \SIaerrSys{161}{11}{14}{\MeVcc}$.  In spite
of the smallness of the \PaOne[1420] signal, its resonance parameters
are found to be remarkably stable in the systematic studies described
above, which results in small systematic uncertainties\footnote{We
  excluded \StudyO in the determination of the systematic uncertainty
  of the \PaOne[1420] parameters, because the shape of the intensity
  distribution of the Deck model in the
  \wave{1}{++}{0}{+}{\PfZero[980]}{P} wave contradicts the data [see
  \cref{fig:intensity_1pp_f0_DeckMC}].} (see
\cref{sec:syst_uncert_onePP} for details). This result supersedes our
previous measurement of the \PaOne[1420] parameters reported in
\refCite{Adolph:2015pws}, which was obtained using the same data set
and the same analysis technique but with only three waves included in
the resonance-model fit.

\subsubsection{Discussion of results on $1^{++}$ resonances}
\label{sec:onePP_discussion}

We observe three $\JPC = 1^{++}$ resonances in our analysis.  The
\PaOne appears in the \wave{1}{++}{0}{+}{\Prho}{S} wave, which is the
most dominant wave, together with a large contribution of the
nonresonant component.  The contribution of the \PaOne to the
\wave{1}{++}{0}{+}{\PfTwo}{P} wave is not well determined, since the
model does not describe well the data in the region below
\SI{1.5}{\GeVcc} because of the apparent leakage as pointed out in
\cref{sec:onePP_results} above.  The \PaOne[1640] appears clearly as a
peak in the $\PfTwo \pi P$ wave with associated phase motion but has
only a small relative contribution to the $\Prho \pi S$ wave.  In
general, the description of the $\Prho \pi S$ and $\PfTwo \pi P$
intensities appears to be difficult.  The disagreement of the model
with the data induces large systematic uncertainties.  The
\PaOne[1420] is observed as a clear peak in the
\wave{1}{++}{0}{+}{\PfZero[980]}{P} wave with associated phase motion
(see
\cref{fig:intensity_phases_1pp_tbin1,fig:intensity_phases_1pp_tbin11}).
There is no clear signature for the presence of the \PaOne[1420] in
the other two $1^{++}$ waves.

In order to study the significance of the \PaOne[1420] resonance we
have removed it from the fit model, so that the
\wave{1}{++}{0}{+}{\PfZero[980]}{P} wave is described by the
nonresonant component only.  This fit has a minimum \chisq~value that
is \num{1.44} times larger than that of the main
fit.\footnote{Compared to the \num{722} free parameters of the main
  fit, this fit has \num{698} free parameters.}  Without the
\PaOne[1420], the model is not able to describe the
$\PfZero[980] \pi P$ intensity and relative phases (red dashed curve
in \cref{fig:intensity_phase_1pp_f0_no_a1_1420}).

\begin{figure}[tbp]
  \centering
  \subfloat[][]{%
    \includegraphics[width=\twoPlotWidth]{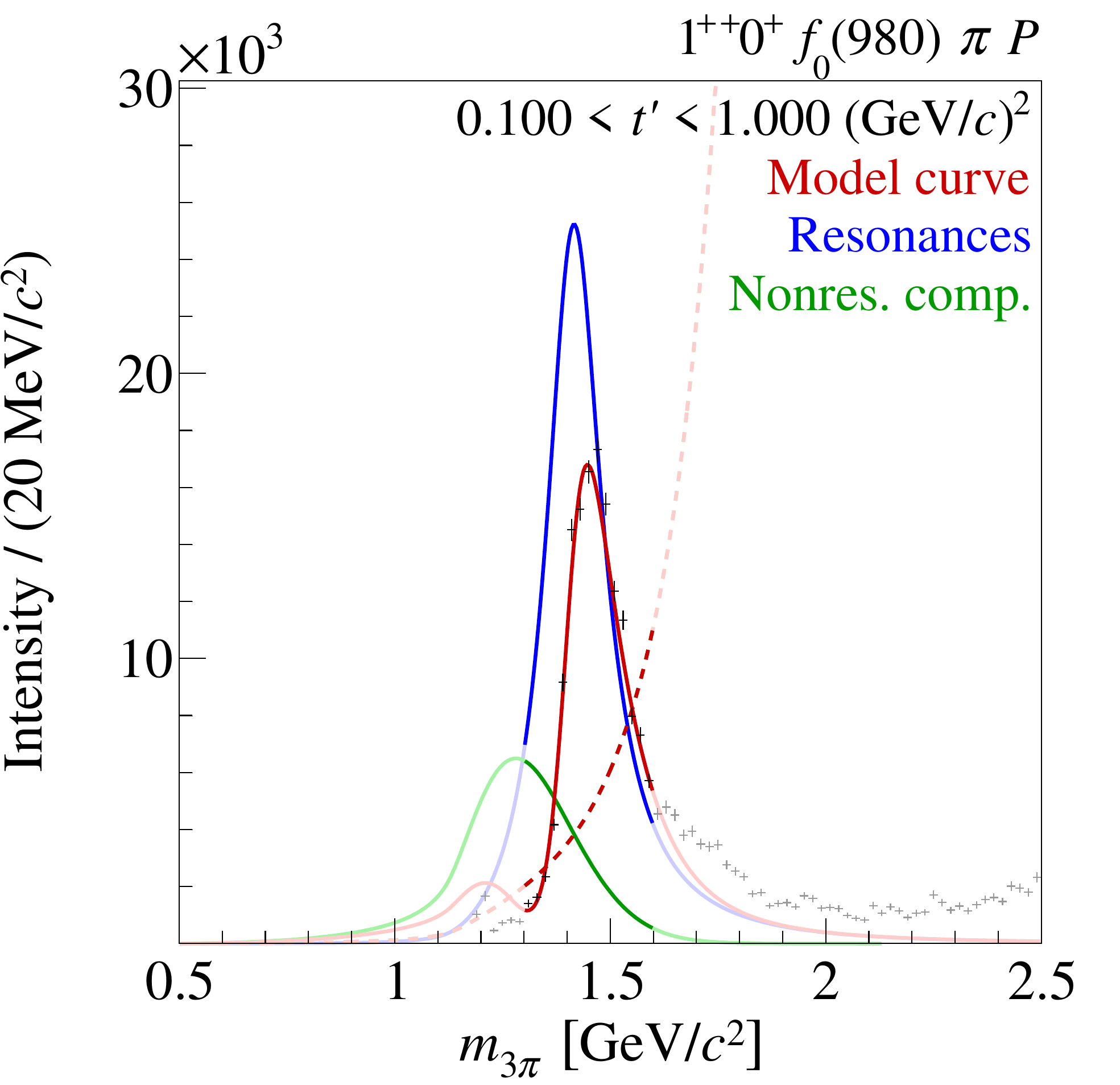}%
    \label{fig:intensity_1pp_f0_no_a1_1420}%
  }%
  \newLineOrHspace{\twoPlotSpacing}%
  \subfloat[][]{%
    \includegraphics[width=\twoPlotWidth]{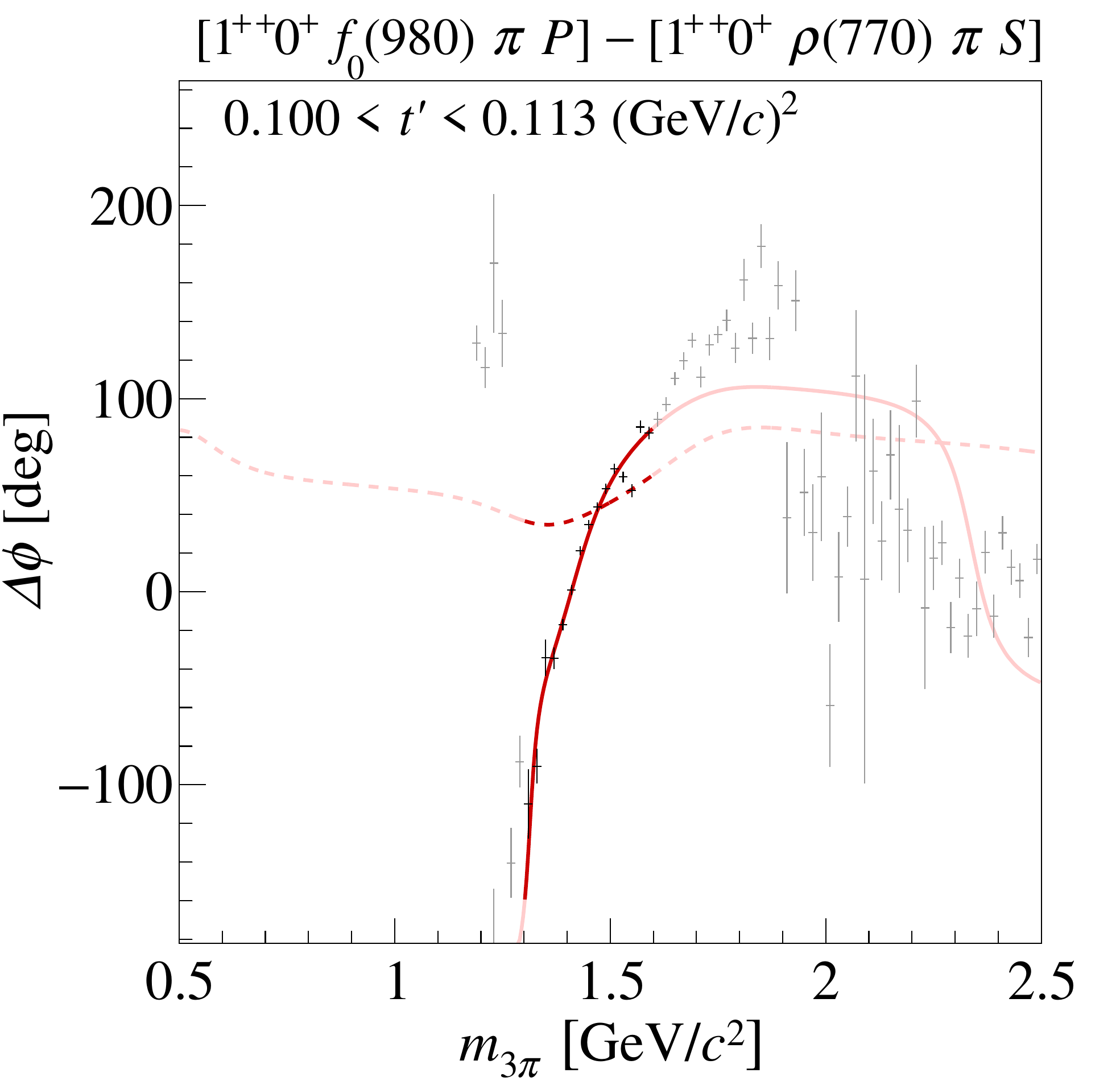}%
    \label{fig:phase_1pp_f0_1pp_rho_tbin1_no_a1_1420}%
  }%
  \caption{\subfloatLabel{fig:intensity_1pp_f0_no_a1_1420}~\tpr-summed
    intensity of the \wave{1}{++}{0}{+}{\PfZero[980]}{P} wave and
    \subfloatLabel{fig:phase_1pp_f0_1pp_rho_tbin1_no_a1_1420}~phase of
    this wave \wrt the \wave{1}{++}{0}{+}{\Prho}{S} wave in the lowest
    \tpr bin.  The result of the main fit is represented by the
    continuous curves.  The fit, in which the \PaOne[1420] component
    was removed from the model, is represented by the dashed curves.
    These curves correspond to the nonresonant component.  The model
    and the wave components are represented as in
    \cref{fig:intensity_phases_1pp_tbin1}.}
  \label{fig:intensity_phase_1pp_f0_no_a1_1420}
\end{figure}

In order to check if the peak in the $\PfZero[980] \pi P$ wave could
be a threshold effect of the \PaOne, we performed a fit, in which the
\PaOne[1420] component is replaced by the \PaOne component, so that
the latter appears in all three $1^{++}$ waves.  The minimum
\chisq~value of this fit is \num{1.09} times larger than that of the
main fit.\footnote{Compared to the \num{722} free parameters of the
  main fit, this fit has \num{700} free parameters.}
\Cref{fig:a1(1260)_in_1pp_f0_chi2difference} shows the contributions
from the spin-density matrix elements to the \chisq~difference between
this and the main fit.  The model with the \PaOne in the
$\PfZero[980] \pi P$ wave describes the peak in this wave less well
[see \cref{fig:intensity_1pp_f0_a1(1260)_in_1pp_f0}].  The model
requires a larger nonresonant component and a more destructive
interference.  While the description of the peak in the $\Prho \pi S$
wave is slightly improved [see
\cref{fig:intensity_1pp_rho_a1(1260)_in_1pp_f0}], the interference
term of this wave with the \wave{2}{++}{1}{+}{\Prho}{D} wave is
described less well [see
\cref{fig:a1(1260)_in_1pp_f0_chi2difference}].  The \PaOne resonance
parameters and the decomposition of the \wave{1}{++}{0}{+}{\Prho}{S}
wave in terms of its components change drastically.  The \PaOne
becomes \SI{85}{\MeVcc} heavier and \SI{188}{\MeVcc} narrower so that
its resonance parameters actually become close to those of the
\PaOne[1420] in the main fit [cf.\ continuous and dashed blue curves
in \cref{fig:intensity_1pp_f0_a1(1260)_in_1pp_f0}].\footnote{Also the
  parameters of the \PaOne[1640] change.  It becomes \SI{85}{\MeVcc}
  heavier and \SI{20}{\MeVcc} wider.}  The $\Prho \pi S$ intensity is
described nearly completely by the nonresonant component with only a
small contribution from the \PaOne\ [see
\cref{fig:intensity_1pp_rho_a1(1260)_in_1pp_f0}].  This interpretation
of the $\Prho \pi S$ intensity seems implausible and would disagree
with previous results on the \PaOne (see discussion below).  We
therefore conclude that the peak in the $\PfZero[980] \pi P$ wave
requires a resonance in our model, which is not the \PaOne.

\begin{figure}[tbp]
  \centering
  \includegraphics[width=\linewidthOr{\twoPlotWidth}]{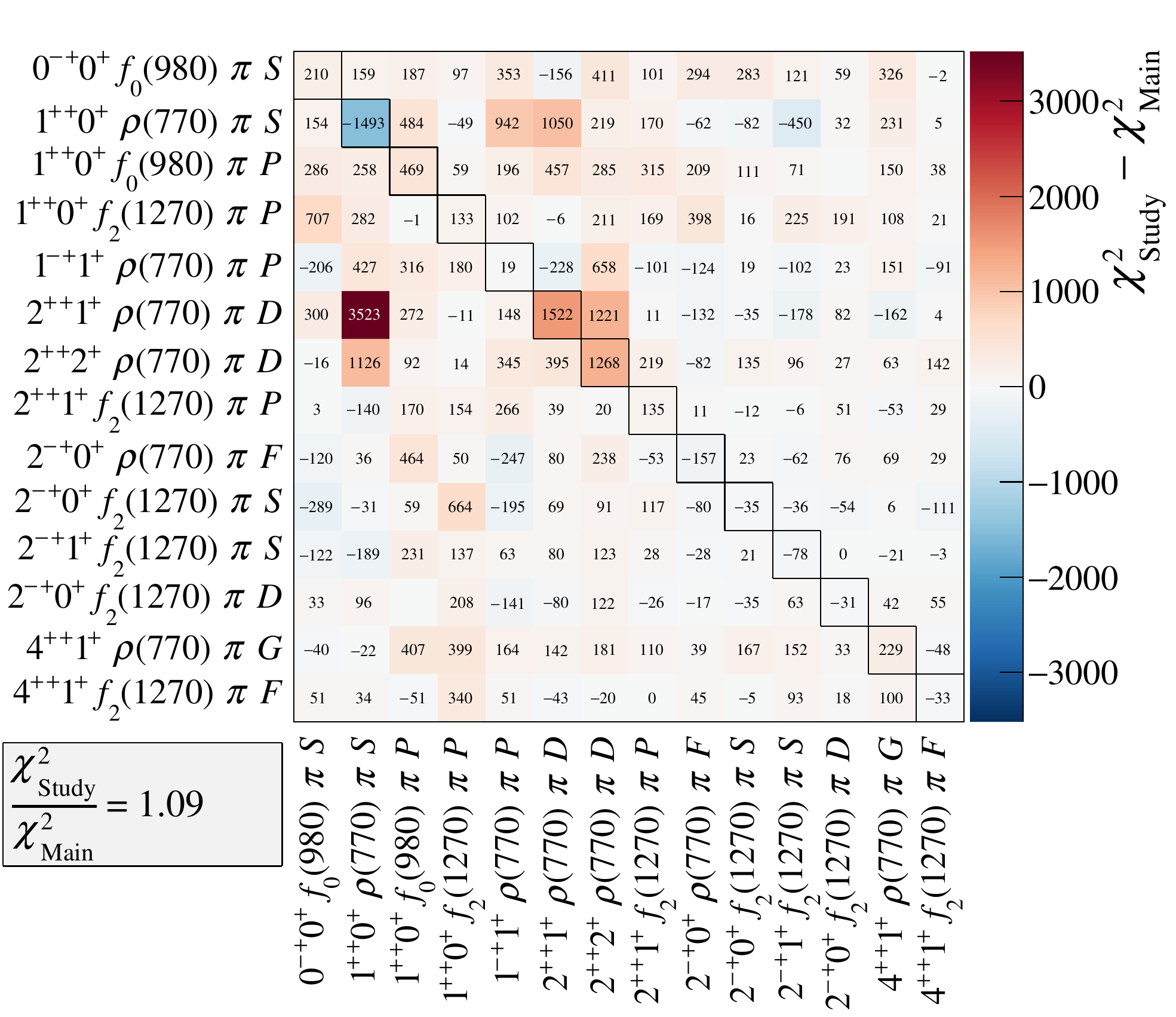}
  \caption{Similar to \cref{fig:DeckMC_chi2difference}, but for the
    study, in which the \PaOne[1420] resonance in the
    \wave{1}{++}{0}{+}{\PfZero[980]}{P} wave was replaced by the
    \PaOne.}
  \label{fig:a1(1260)_in_1pp_f0_chi2difference}
\end{figure}

\begin{figure}[tbp]
  \centering
  \subfloat[][]{%
    \includegraphics[width=\twoPlotWidth]{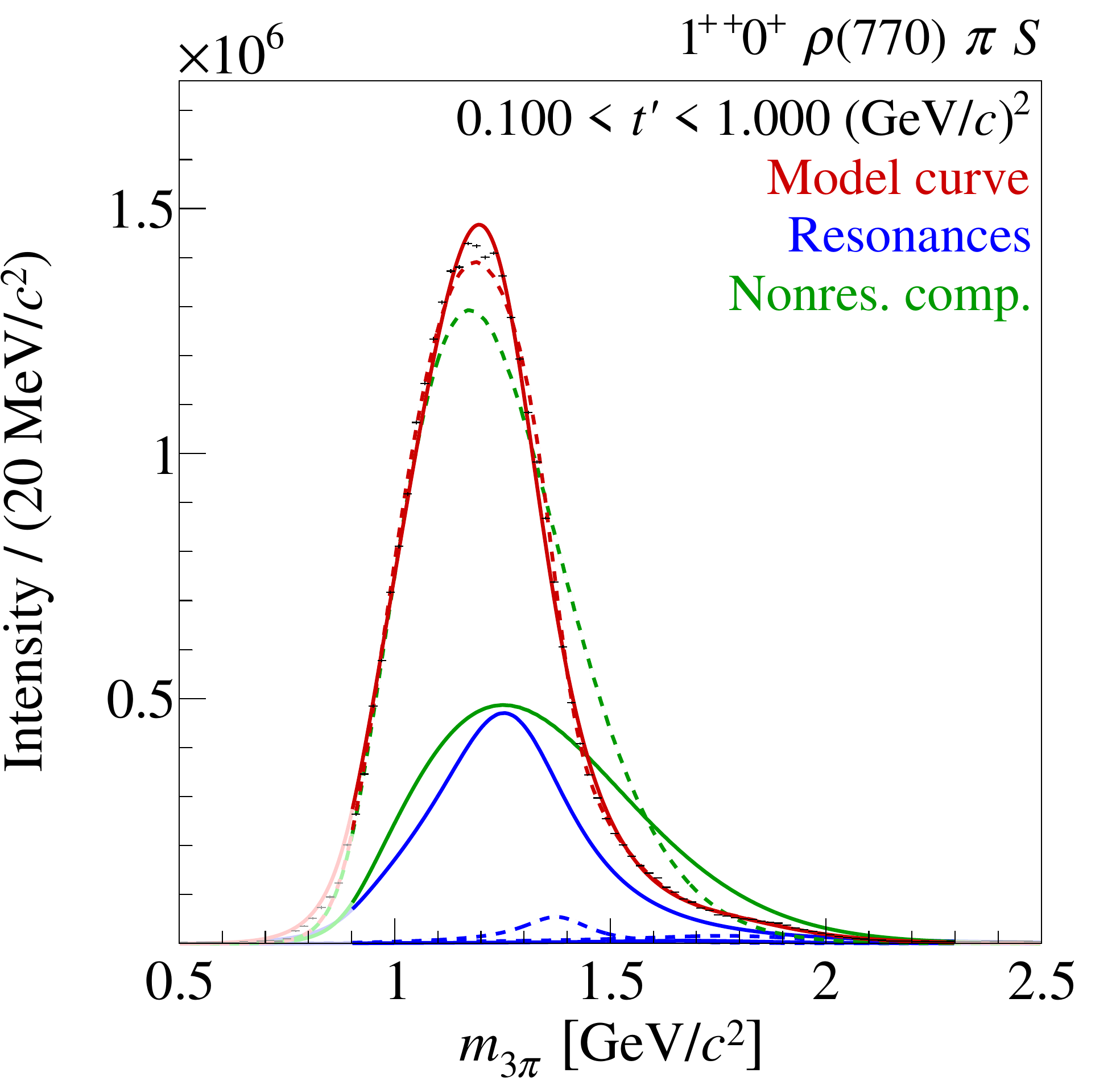}%
    \label{fig:intensity_1pp_rho_a1(1260)_in_1pp_f0}%
  }%
  \newLineOrHspace{\twoPlotSpacing}%
  \subfloat[][]{%
    \includegraphics[width=\twoPlotWidth]{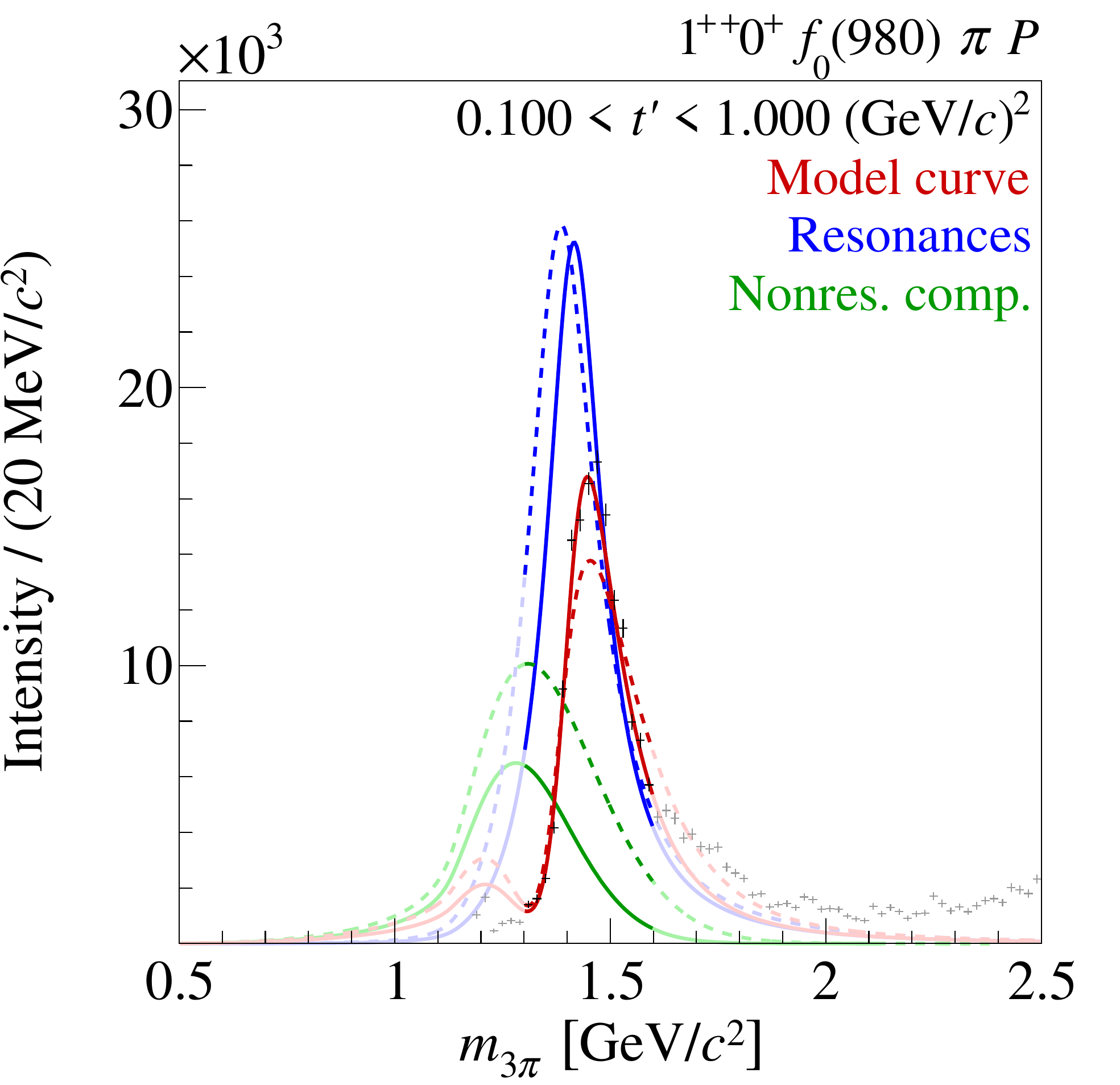}%
    \label{fig:intensity_1pp_f0_a1(1260)_in_1pp_f0}%
  }%
  \caption{\tpr-summed intensities of the \wave{1}{++}{0}{+}{\Prho}{S}
    and the \wave{1}{++}{0}{+}{\PfZero[980]}{P} wave.  The result of
    the main fit is represented by the continuous curves. The fit, in
    which the \PaOne[1420] resonance in the
    \wave{1}{++}{0}{+}{\PfZero[980]}{P} wave was replaced by the
    \PaOne, is represented by the dashed curves.  The model and the
    wave components are represented as in
    \cref{fig:intensity_phases_1pp_tbin1}.}
  \label{fig:intensities_1pp_a1(1260)_in_1pp_f0}
\end{figure}

We estimate the strength of a possible \PaOne[1420] component in the
other two $1^{++}$ waves by adding the \PaOne[1420] component to the
amplitudes of these waves.  The minimum~\chisq of this fit is
\num{0.96} times smaller than that of the main fit.\footnote{Compared
  to the \num{722} free parameters of the main fit, this fit has
  \num{766} free parameters.}  The largest contribution to this
improvement in the description of the data comes from the intensity of
the $\Prho \pi S$ wave (see
\cref{fig:a1(1420)_in_all_1pp_chi2difference}).  Adding the
\PaOne[1420] component to this wave improves the description of the
peak in the \PaOne region [see
\cref{fig:intensity_1pp_rho_a1(1420)_in_all_1pp}].  Within the fit
range, the description of the $\PfTwo \pi P$ wave changes only
slightly [see \cref{fig:intensity_1pp_f2_a1(1420)_in_all_1pp}].
However, the extrapolation of the model toward lower masses disagrees
even more strongly with the data than in the main fit.  The
description of the $\PfZero[980] \pi P$ wave remains practically
unchanged.\footnote{The \PaOne[1420] parameters change only slightly.
  Its mass increases by \SI{4}{\MeVcc} and its width by
  \SI{11}{\MeVcc}.  In contrast, the parameters of the \PaOne and
  \PaOne[1640] change substantially.  The \PaOne becomes
  \SI{27}{\MeVcc} lighter and \SI{75}{\MeVcc} wider as compared to the
  main fit.  The width of the \PaOne[1640] decreases by
  \SI{93}{\MeVcc}.}  The relative contributions of the \PaOne[1420] to
the $\PfTwo \pi P$ and in particular to the $\Prho \pi S$ wave are
small.  The coupling amplitudes of the \PaOne[1420] in the three waves
are not constrained by \cref{eq:method:branchingdefinition} and are
therefore freely determined by the fit.  The values of the
\PaOne[1420] slope parameters in the three waves differ significantly:
in the $\Prho \pi S$ wave the slope is \SI{6.7}{\perGeVcsq}, in the
$\PfTwo \pi P$ wave it is \SI{17.5}{\perGeVcsq}, and in the
$\PfZero[980] \pi P$ wave it is \SI{9.5}{\perGeVcsq}.  The latter
value is identical to the one from the main fit (see
\cref{tab:slopes}).  The phase of the \PaOne[1420] coupling amplitude
in the $\Prho \pi S$ and the $\PfTwo \pi P$ waves exhibits a stronger
dependence on \tpr than that in the $\PfZero[980] \pi P$ wave.  From
the above, we conclude that we do not see convincing evidence for an
\PaOne[1420] component in the $\Prho \pi S$ or the $\PfTwo \pi P$
wave, although we cannot rule out a small contribution.

\begin{figure}[tbp]
  \centering
  \includegraphics[width=\linewidthOr{\twoPlotWidth}]{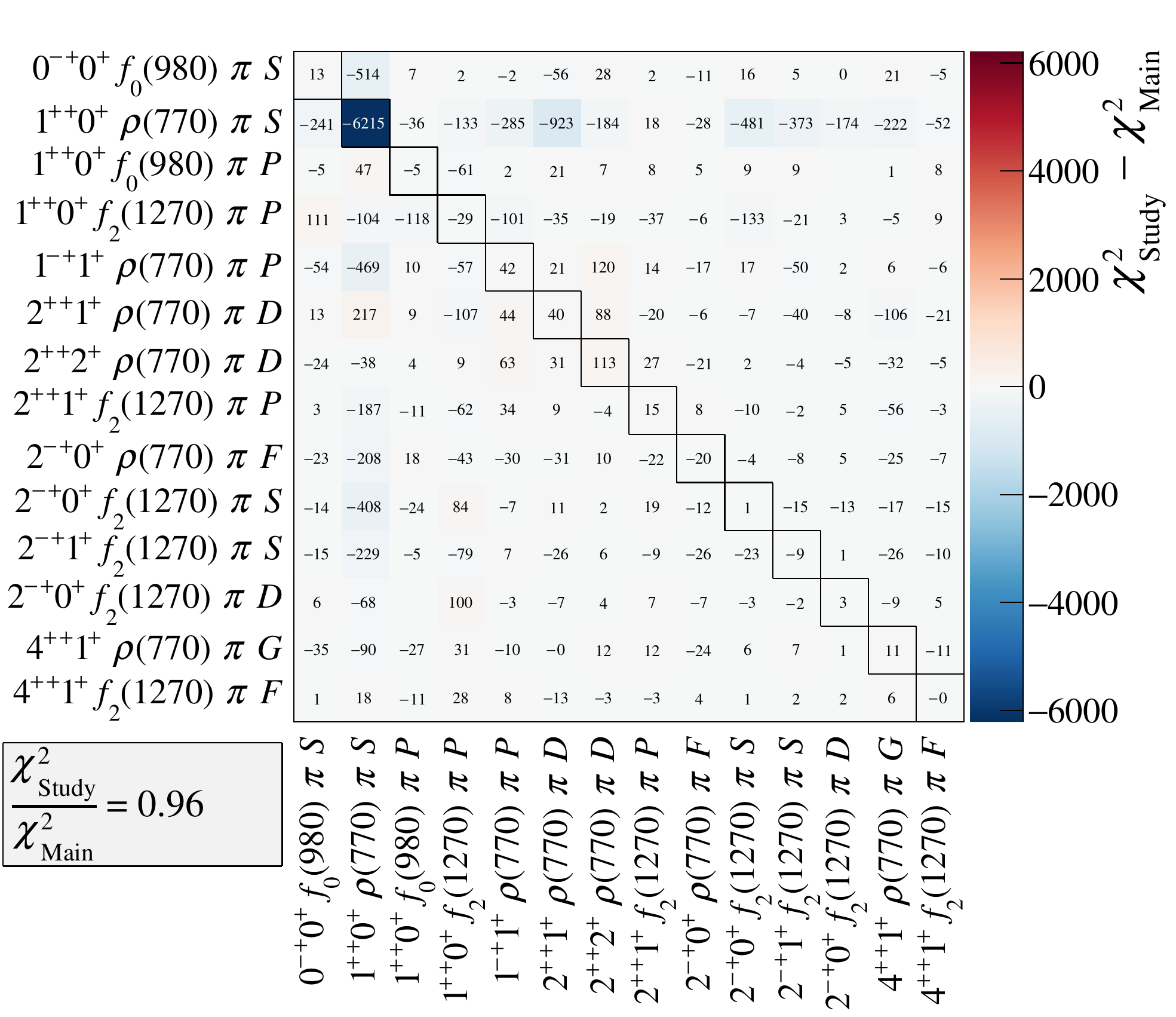}
  \caption{Similar to \cref{fig:DeckMC_chi2difference}, but for the
    study, in which the \PaOne[1420] resonance was also included in
    the \wave{1}{++}{0}{+}{\Prho}{S} and \wave{1}{++}{0}{+}{\PfTwo}{P}
    waves.}
  \label{fig:a1(1420)_in_all_1pp_chi2difference}
\end{figure}

\begin{figure}[tbp]
  \centering
  \subfloat[][]{%
    \includegraphics[width=\twoPlotWidth]{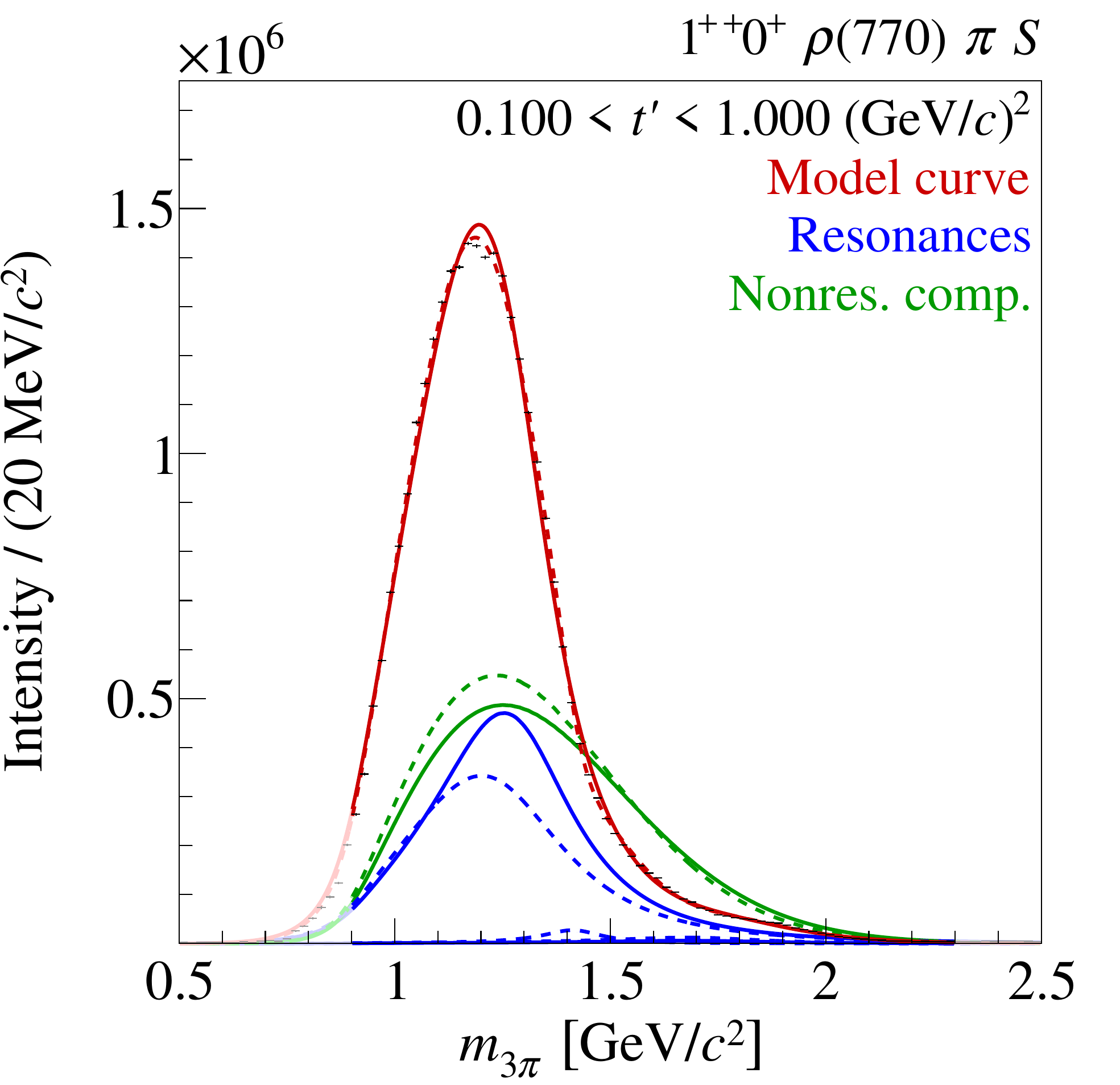}%
    \label{fig:intensity_1pp_rho_a1(1420)_in_all_1pp}%
  }%
  \newLineOrHspace{\twoPlotSpacing}%
  \subfloat[][]{%
    \includegraphics[width=\twoPlotWidth]{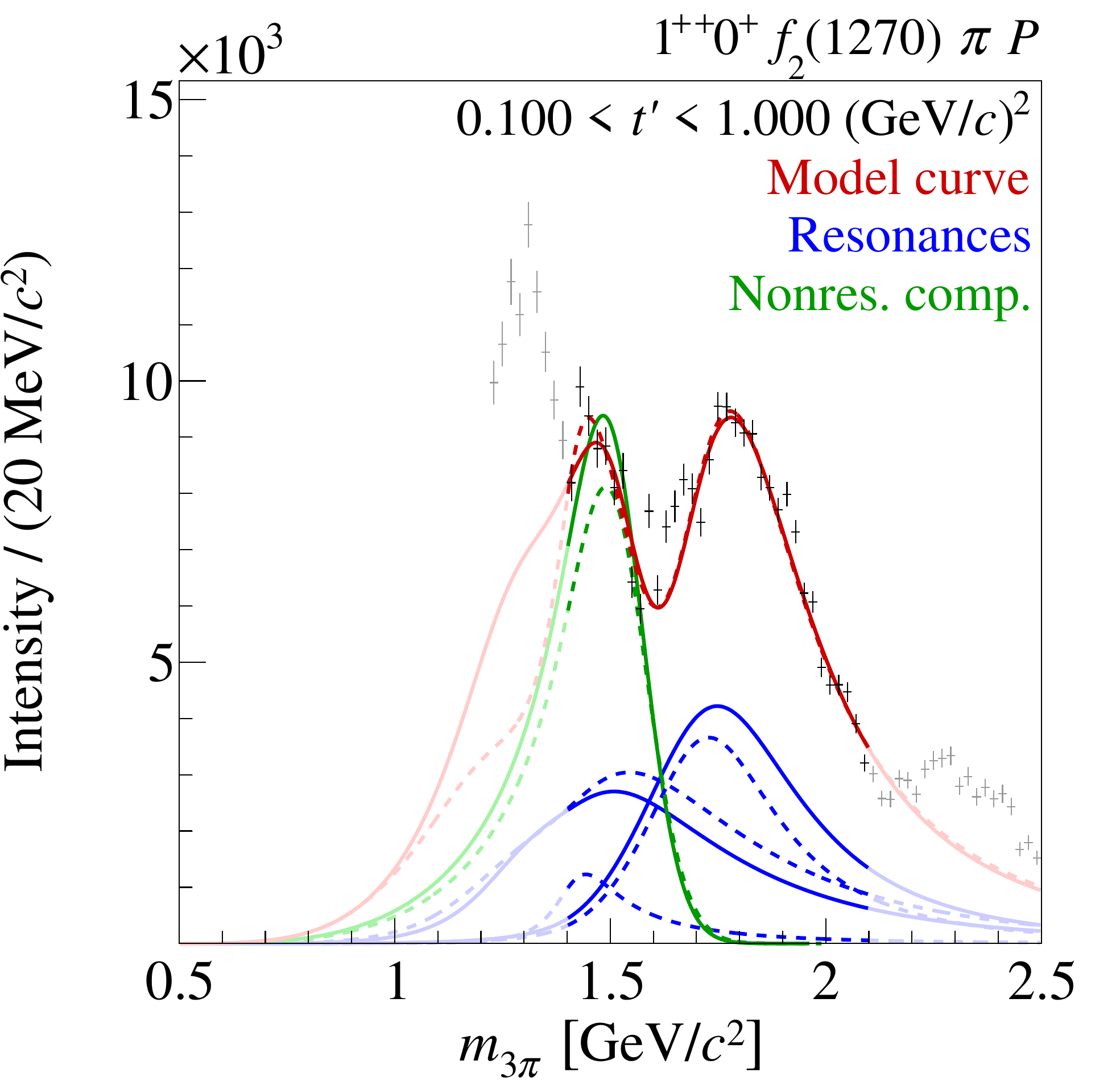}%
    \label{fig:intensity_1pp_f2_a1(1420)_in_all_1pp}%
  }%
  \caption{\tpr-summed intensities of
    \subfloatLabel{fig:intensity_1pp_rho_a1(1420)_in_all_1pp}~the
    \wave{1}{++}{0}{+}{\Prho}{S} wave and
    \subfloatLabel{fig:intensity_1pp_f2_a1(1420)_in_all_1pp}~the
    \wave{1}{++}{0}{+}{\PfTwo}{P} wave.  The result of the main fit is
    represented by the continuous curves.  The fit, in which the
    \PaOne[1420] resonance is also included in the
    \wave{1}{++}{0}{+}{\Prho}{S} and \wave{1}{++}{0}{+}{\PfTwo}{P}
    waves, is represented by the dashed curves.  The model and the
    wave components are represented as in
    \cref{fig:intensity_phases_1pp_tbin1}.}
  \label{fig:intensities_1pp_a1(1420)_in_all_1pp}
\end{figure}

In order to study the significance of the \PaOne[1640] component, we
performed a fit, in which we omitted the \PaOne[1640] resonance from
the fit model.  The minimum \chisq~value of this fit is \num{1.13}
times larger than that of the main fit.\footnote{Compared to the
  \num{722} free parameters of the main fit, this fit has \num{696}
  free parameters.} \Cref{fig:no-a1(1640)_chi2difference} shows the
contributions from the spin-density matrix elements to the
\chisq~difference between this and the main fit.  Without the
\PaOne[1640], the model describes less well in particular the
intensity distributions of the \wave{1}{++}{0}{+}{\Prho}{S} and
\wave{1}{++}{0}{+}{\PfTwo}{P} waves (see
\cref{fig:intensities_1pp_no-a1(1640)}).  The width of the \PaOne
becomes \SI{47}{\MeVcc} larger.  From the above, we conclude that the
\PaOne[1640] component is necessary to describe the data but its
parameters are not well determined.

\begin{figure}[tbp]
  \centering
  \includegraphics[width=\linewidthOr{\twoPlotWidth}]{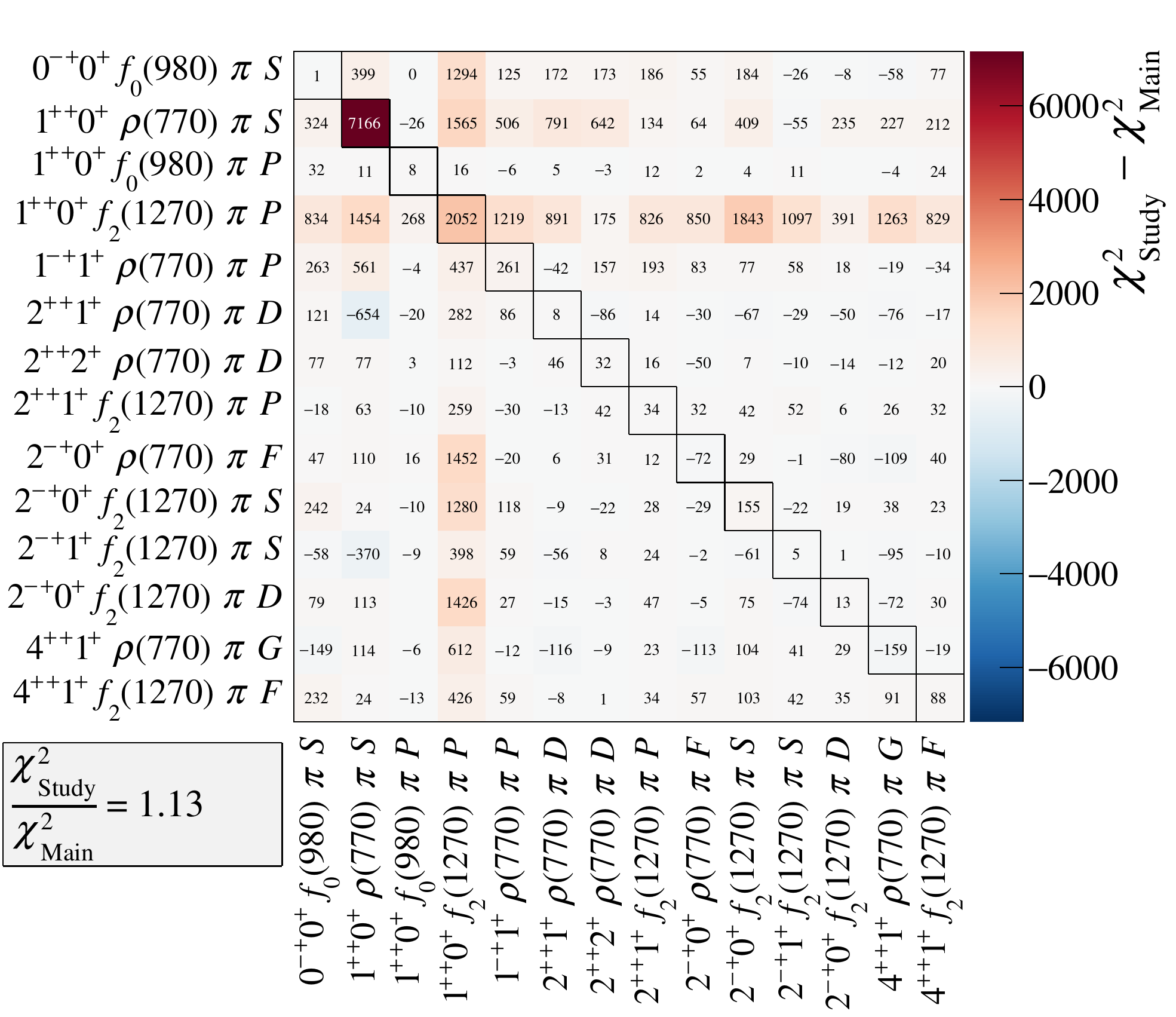}
  \caption{Similar to \cref{fig:DeckMC_chi2difference}, but for the
    study, in which the \PaOne[1640] resonance was omitted from the
    fit model.}
  \label{fig:no-a1(1640)_chi2difference}
\end{figure}

\begin{wideFigureOrNot}[tbp]
  \centering
  \subfloat[][]{%
    \includegraphics[width=\threePlotWidth]{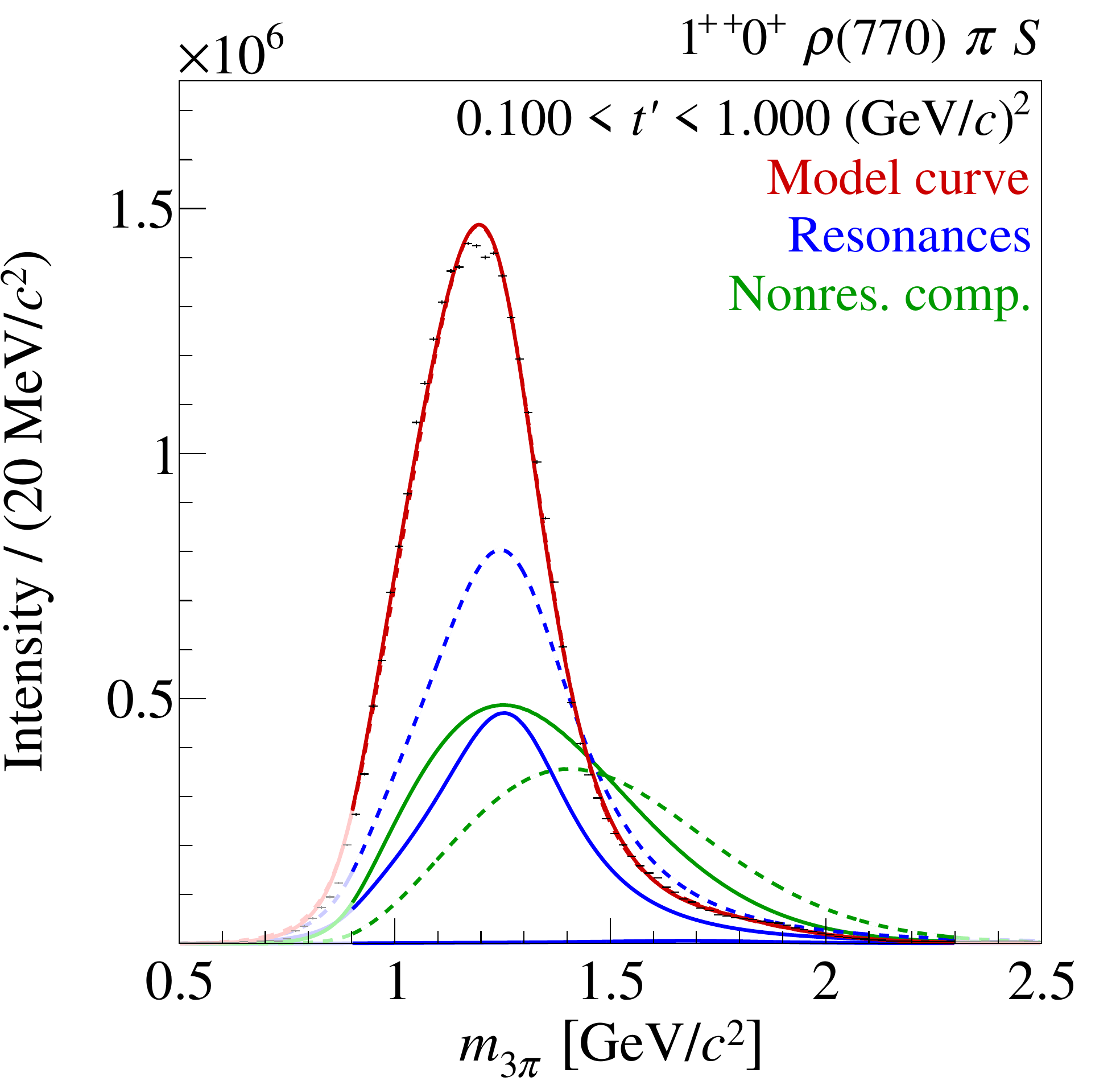}%
    \label{fig:intensity_1pp_rho_no-a1(1640)}%
  }%
  \hspace*{\threePlotSpacing}%
  \subfloat[][]{%
    \includegraphics[width=\threePlotWidth]{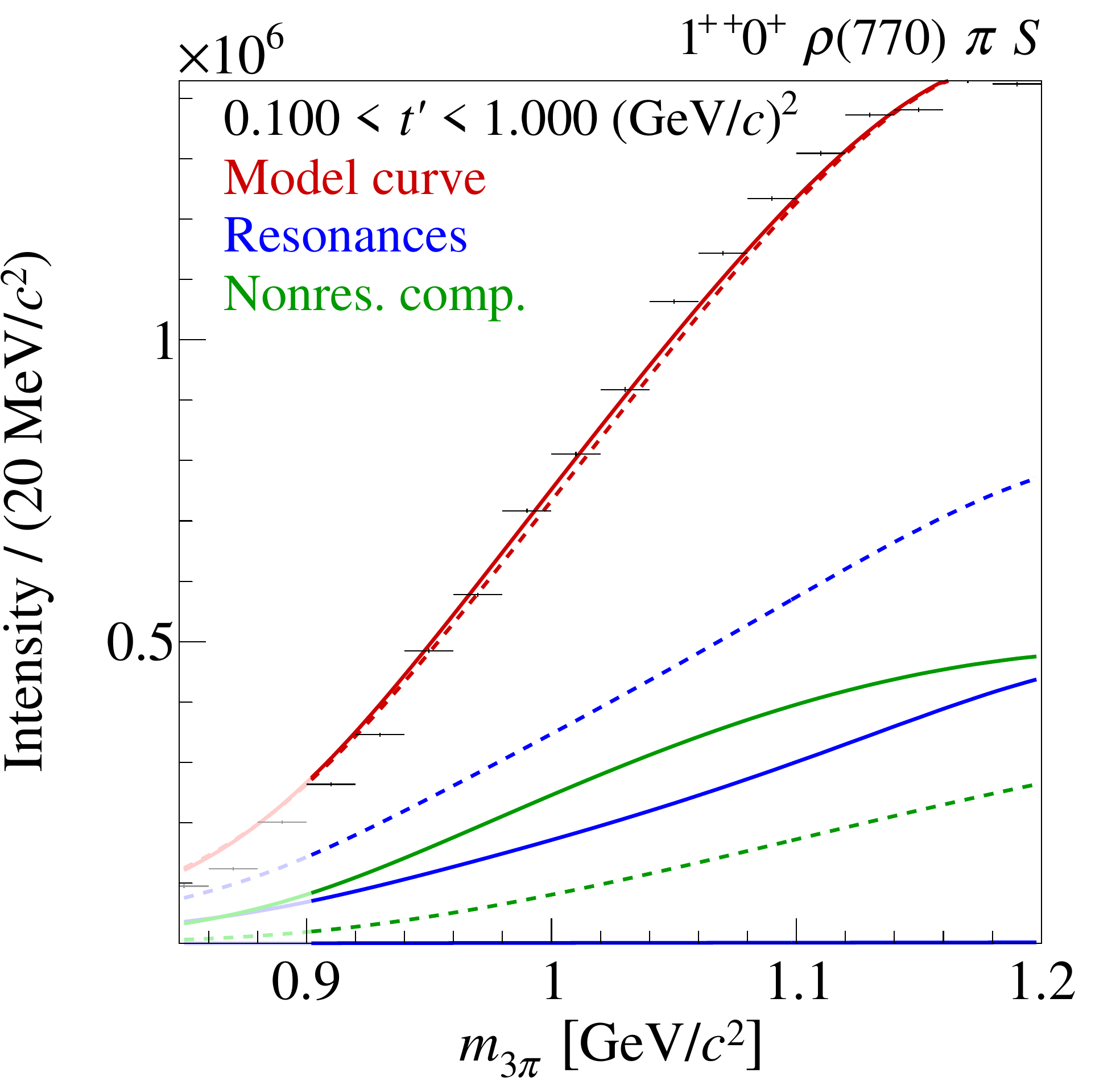}%
    \label{fig:intensity_1pp_rho_no-a1(1640)_zoom}%
  }%
  \hspace*{\threePlotSpacing}%
  \subfloat[][]{%
    \includegraphics[width=\threePlotWidth]{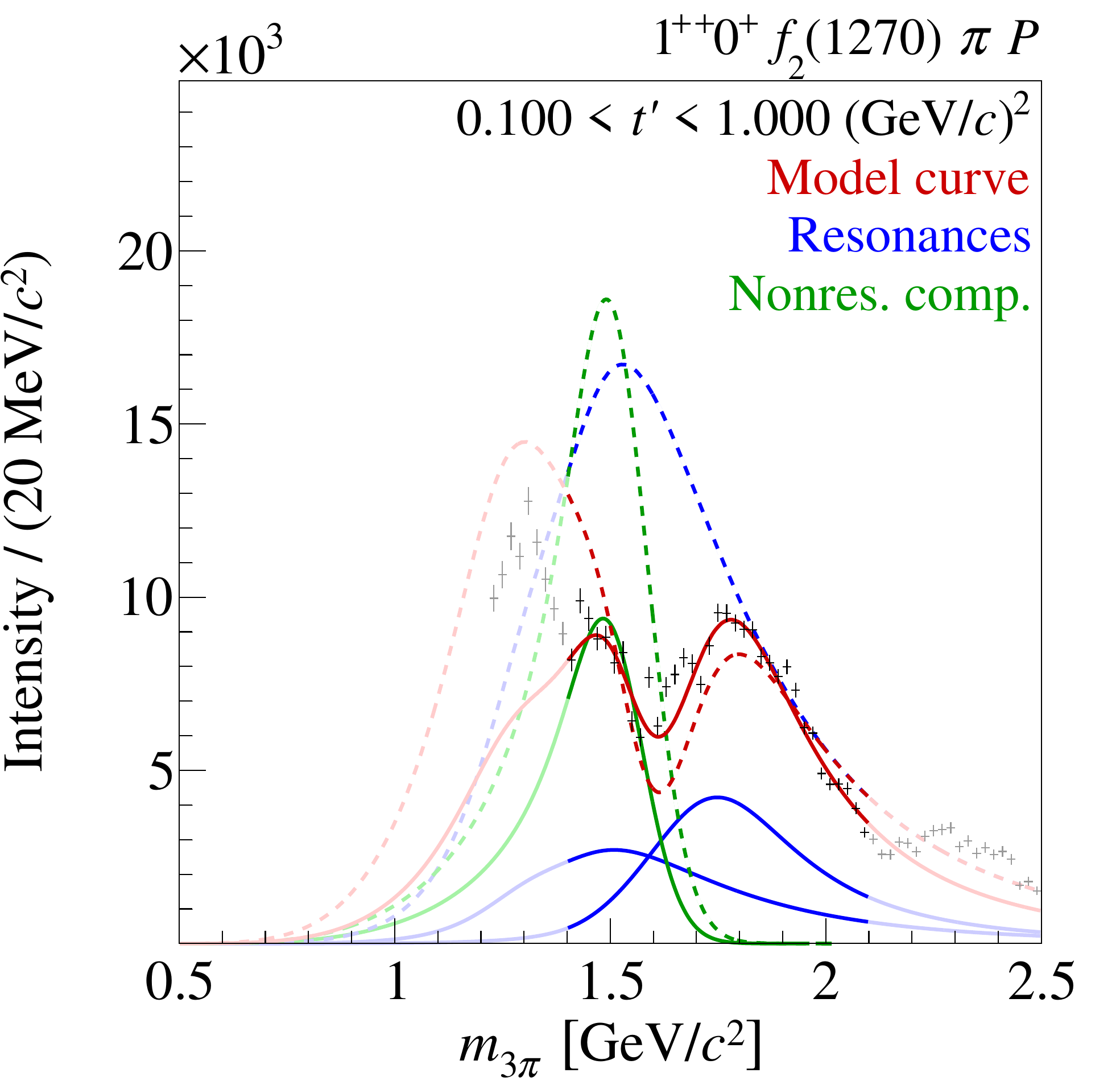}%
    \label{fig:intensity_1pp_f2_no-a1(1640)}%
  }%
  \caption{\tpr-summed intensities of
    \subfloatLabel{fig:intensity_1pp_rho_no-a1(1640)}~the
    \wave{1}{++}{0}{+}{\Prho}{S} wave and
    \subfloatLabel{fig:intensity_1pp_f2_no-a1(1640)}~the
    \wave{1}{++}{0}{+}{\PfTwo}{P} wave.
    In~\subfloatLabel{fig:intensity_1pp_rho_no-a1(1640)_zoom}, a
    zoomed view of~\subfloatLabel{fig:intensity_1pp_rho_no-a1(1640)}
    is shown.  The result of the main fit is represented by the
    continuous curves. The fit, in which the \PaOne[1640] component
    was removed from the fit model, is represented by the dashed
    curves.  The model and the wave components are represented as in
    \cref{fig:intensity_phases_1pp_tbin1}.}
  \label{fig:intensities_1pp_no-a1(1640)}
\end{wideFigureOrNot}

Although the \PaOne is a well-established resonance that has been
observed in many experiments, its parameters are not well determined.
Depending on the analyzed process and the employed parametrizations,
the values of the \PaOne parameters differ
substantially~\cite{pdg_a1_1260:2006}.  The measurements listed by the
PDG cover a wide range of mass values from
\SI{1041(13)}{\MeVcc}~\cite{Gavillet:1977kx} up to
\SIerrs{1331}{10}{3}{\MeVcc}~\cite{Asner:1999kj} and width values from
\SI{230(50)}{\MeVcc}~\cite{Gavillet:1977kx} up to
\SIerrs{814}{36}{13}{\MeVcc}~\cite{Asner:1999kj}.  Due to the large
spread of the measured parameter values, the PDG does not perform an
average but provides only an estimate of
$m_{\PaOne} = \SI{1230(40)}{\MeVcc}$ and
$\Gamma_{\PaOne} =
\SIrange{250}{600}{\MeVcc}$~\cite{Patrignani:2016xqp}.  Our measured
\PaOne mass of $m_{\PaOne} = \SIaerrSys{1299}{12}{28}{\MeVcc}$ is
larger than the PDG estimate but compatible within our large
uncertainties.  Our measured width of
$\Gamma_{\PaOne} = \SIerrSys{380}{80}{\MeVcc}$ has large uncertainties
and is close to the center of the range estimated by the PDG.
Compared to our previous measurement of the \threePi final state
diffractively produced on a solid lead target~\cite{alekseev:2009aa},
the width agrees well but we obtain a larger mass that is in slight
disagreement.  However, since the lead-target data sample is
approximately 2~orders of magnitude smaller, the analysis in
\refCite{alekseev:2009aa} was performed by integrating over the \tpr
range from \SIrange{0.1}{1.0}{\GeVcsq} and assuming a model for the
\tpr dependence of the partial-wave amplitudes.  Considering the
unexpected \tpr dependence of the shape of the
\wave{1}{++}{0}{+}{\Prho}{S} intensity distribution as observed in
\cref{fig:intensity_1pp_rho_pi_S}, this might have been an inadequate
approximation in former analyses and might explain the mass
difference.

As already discussed in \refCite{Adolph:2015pws}, the nature of the
peculiar resonancelike \PaOne[1420] signal, which is listed by the PDG
as \enquote{omitted from summary table}~\cite{Patrignani:2016xqp}, is
still unclear and several interpretations were proposed.  In
\refCite{Adolph:2015pws} and in this analysis we have shown that it is
consistent with a Breit-Wigner amplitude.  Hence it could be the
isospin partner to the $f_1(1420)$.  Isovector
$[n\, n]\,[\overline{n}\,\overline{n}]$ and
$[n\, s]\,[\overline{n}\,\overline{s}]$ states with $n = u$ or $d$
were predicted in the \SI{1.4}{\GeVcc} mass range in quark-model
calculations that included tetraquark states~\cite{Vijande:2005jd}.
The \PaOne[1420] signal was also described as a two-quark-tetraquark
mixed state~\cite{wang:2014bua} and as a tetraquark with mixed flavor
symmetry~\cite{Chen:2015fwa}.  In addition, calculations based on a
soft-wall AdS/QCD approach predict a
$[n\, \overline{s}]\,[s\, \overline{n}]$ tetraquark with a mass of
\SI{1414}{\MeVcc}~\cite{Gutsche:2017oro}.  The authors of
\refCite{Gutsche:2017twh} studied the two-body decay rates for the
modes $\PaOne[1420] \to \PfZero[980] \pi$ and
$\PaOne[1420] \to K \PKbar^*(892)$ for four-quark configurations using
the covariant confined quark model.  They found that a molecular
configuration is preferred over a compact diquark-antidiquark state.
However, other models were proposed that do not require an additional
resonance.  Basdevant and Berger proposed resonant rescattering
corrections in the Deck process as an
explanation~\cite{Basdevant:2015ysa,Basdevant:2015wma}, whereas the
authors of \refCite{Ketzer:2015tqa} suggested an anomalous triangle
singularity in the rescattering diagram for
$\PaOne \to K \PKbar^*(892) \to K \PKbar \pi \to \PfZero[980] \pi$.
The results of the latter calculation were confirmed in
\refCite{Aceti:2016yeb}.  Preliminary studies show that the amplitude
for the triangle diagram describes the data equally well as the
Breit-Wigner model.  In
the case of a triangle singularity, the production rates of the
\PaOne[1420] would be completely determined by those of the \PaOne.
Therefore, the slope parameters of the two peaks would be equal.
Unfortunately, in our analysis the systematic uncertainties of the
slope parameters are too large in order to draw any conclusion (see
\cref{tab:slopes}).  Hence more detailed studies are still needed in
order to distinguish between different models for the \PaOne[1420].

The \PaOne[1640] is listed by the PDG as \enquote{omitted from summary
  table} based on four
measurements~\cite{Bellini:1984fz,Baker:1999fc,chung:2002pu,baker:2003jh}.
This state therefore requires further confirmation.  The PDG world
averages for the \PaOne[1640] parameters are
$m_{\PaOne[1640]} = \SI{1647(22)}{\MeVcc}$ and
$\Gamma_{\PaOne[1640]} =
\SI{254(27)}{\MeVcc}$~\cite{Patrignani:2016xqp}.  Compared to other
waves, the agreement of our model with the
\wave{1}{++}{0}{+}{\Prho}{S} and \wave{1}{++}{0}{+}{\PfTwo}{P}
intensities is worse and thus our measured \PaOne[1640] parameters,
$m_{\PaOne[1640]} = \SIaerrSys{1700}{35}{130}{\MeVcc}$ and
$\Gamma_{\PaOne[1640]} = \SIaerrSys{510}{170}{90}{\MeVcc}$, have large
systematic uncertainties.  Our \PaOne[1640] mass value is larger but
within uncertainties compatible with the world average.  However, our
width value is significantly larger.  As the study with the
\PaOne[1420] component in all three $1^{++}$ waves suggests (see
discussion above), this discrepancy might be due to the disagreement
between model and data in the mass region between \PaOne and
\PaOne[1640] in the $\Prho \pi S$ and $\PfTwo \pi P$ intensities.  It
might also be a consequence of not including any higher-lying \PaOne*
states in the fit model.

The PDG~\cite{Patrignani:2016xqp} lists three further \PaOne* states:
\PaOne[1930]~\cite{anisovich:2001pn}, \PaOne[2095]~\cite{kuhn:2004en},
and \PaOne[2270]~\cite{anisovich:2001pn}.  Although we do not see
clear resonance signals of heavy \PaOne* states in the mass range from
\SIrange{1900}{2500}{\MeVcc} in the analyzed waves, we cannot exclude
that some of the observed deviations of the model from the data at
high masses are due to additional excited \PaOne* states that we do
not take into account.
 %
%
%

\subsection{$\JPC = 1^{-+}$ resonances}
\label{sec:oneMP}

\subsubsection{Results on $1^{-+}$ resonances}
\label{sec:oneMP_results}

In addition to waves with ordinary \qqbar quantum numbers, our
analysis also includes the \wave{1}{-+}{1}{+}{\Prho}{P} wave with an
exotic \JPC\ combination.  This wave contributes \SI{0.8}{\percent} to
the total intensity.  \Cref{fig:intensities_1mp} shows the intensity
distributions for all 11~\tpr bins.  The shapes of these distributions
exhibit a surprisingly strong dependence on \tpr.  At low \tpr, the
intensity distribution is dominated by a broad structure that extends
from about \SIrange{1.0}{1.7}{\GeVcc} with a maximum at approximately
\SI{1.2}{\GeVcc}.  With increasing \tpr, the structure becomes
narrower and the maximum moves to about \SI{1.6}{\GeVcc}.  This
behavior suggests large contributions from nonresonant processes in
this wave.  In the highest \tpr bin, a dip appears at
\SI{1.25}{\GeVcc} where the intensity nearly vanishes.  At low \tpr, a
narrow enhancement appears at \SI{1.1}{\GeVcc} on top of the broad
structure.  This enhancement is sensitive to details of the wave set
that is used in the partial-wave decomposition and we therefore
suspect it to be an artifact induced by imperfections in the PWA
model.

\Cref{fig:phases_1mp} shows selected phases of the
\wave{1}{-+}{1}{+}{\Prho}{P} wave \wrt other waves in the lowest and
the highest \tpr bins (top and bottom rows, respectively).  At low
\tpr, decreasing phases appear at masses that correspond to resonances
in the other waves.\footnote{The slightly decreasing phase \wrt the
  \wave{1}{++}{0}{+}{\Prho}{S} wave around \SI{1.2}{\GeVcc} is caused
  by the \PaOne\ [see
  \cref{fig:phase_1mp_1pp_rho_tbin1,sec:onePP_results}].  The rapidly
  decreasing phase \wrt the \wave{2}{++}{1}{+}{\Prho}{D} wave around
  \SI{1.3}{\GeVcc} is caused by the \PaTwo\ [see
  \cref{fig:phase_1mp_2pp_m1_rho_tbin1,sec:twoPP_results}].  The
  slightly decreasing phase \wrt the \wave{2}{-+}{0}{+}{\PfTwo}{S}
  wave around \SI{1.7}{\GeVcc} is caused by the \PpiTwo\ [see
  \cref{fig:phase_1mp_2mp_m0_f2_tbin1,sec:twoMP_results}].  The
  decreasing phase \wrt the \wave{4}{++}{1}{+}{\Prho}{G} wave around
  \SI{1.9}{\GeVcc} is caused by the \PaFour\ [see
  \cref{fig:phase_1mp_4pp_rho_tbin1,sec:fourPP_results}].}  In
\cref{fig:phase_1mp_1pp_rho_tbin1,fig:phase_1mp_4pp_rho_tbin1}
slightly rising phases are observed in the \SI{1.6}{\GeVcc} region.
The phase \wrt the \wave{2}{++}{1}{+}{\Prho}{D} wave is approximately
constant between \SIlist{1.4;1.6}{\GeVcc}.  Its rapid rise at
\SI{1.7}{\GeVcc} [see \cref{fig:phase_1mp_2pp_m1_rho_tbin1}] is
induced by the nearly vanishing intensity of the
\wave{2}{++}{1}{+}{\Prho}{D} wave (see \cref{sec:twoPP_results}).
Compared to the intensity of the $1^{-+}$ wave, its phase motions \wrt
most waves show less dependence on \tpr in the \SI{1.6}{\GeVcc} region
(see bottom row of \cref{fig:phases_1mp}).  At high \tpr, rapidly
decreasing phases appear at \SI{1.25}{\GeVcc} because of the nearly
vanishing intensity of the $1^{-+}$ wave.\footnote{This is the same
  effect as seen in the \wave{0}{-+}{0}{+}{\PfZero[980]}{S} wave; see
  \cref{sec:zeroMP_results,fig:phase_0mp_1pp_rho_tbin11,fig:phase_0mp_2mp_f2_tbin11}.}
It is worth noting that we do not observe any phase motions in the
\SI{1.1}{\GeVcc} region, where the narrow enhancement is observed in
the intensity distribution.  This supports interpretation of this
structure as a model artifact.

\ifMultiColumnLayout{\begin{figure*}[t]}{\begin{figure}[p]}
  \centering
  \subfloat[][]{%
    \includegraphics[width=\threePlotSmallWidth]{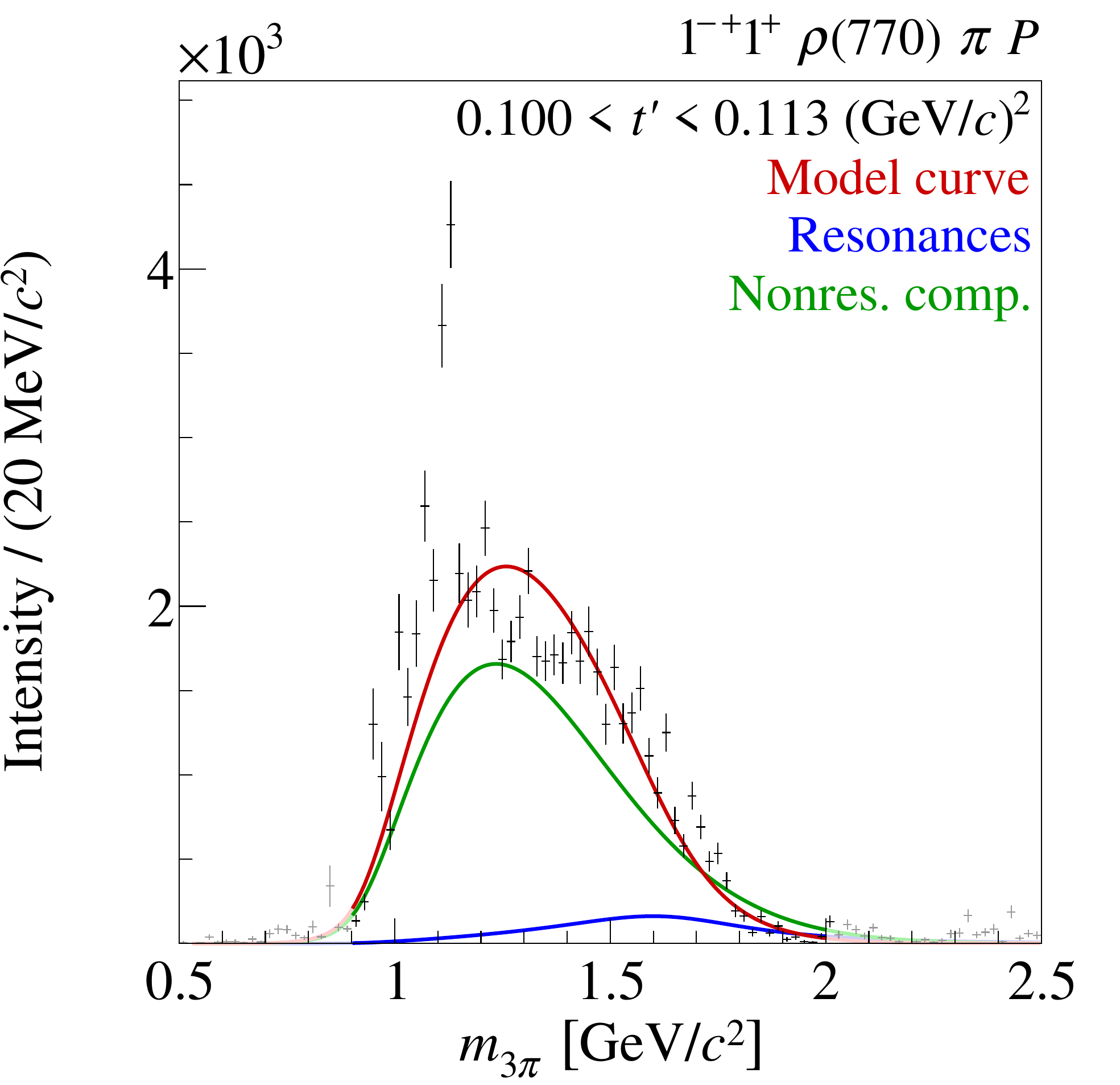}%
    \label{fig:intensity_1mp_tbin1}%
  }%
  \hspace*{\threePlotSmallSpacing}%
  \subfloat[][]{%
    \includegraphics[width=\threePlotSmallWidth]{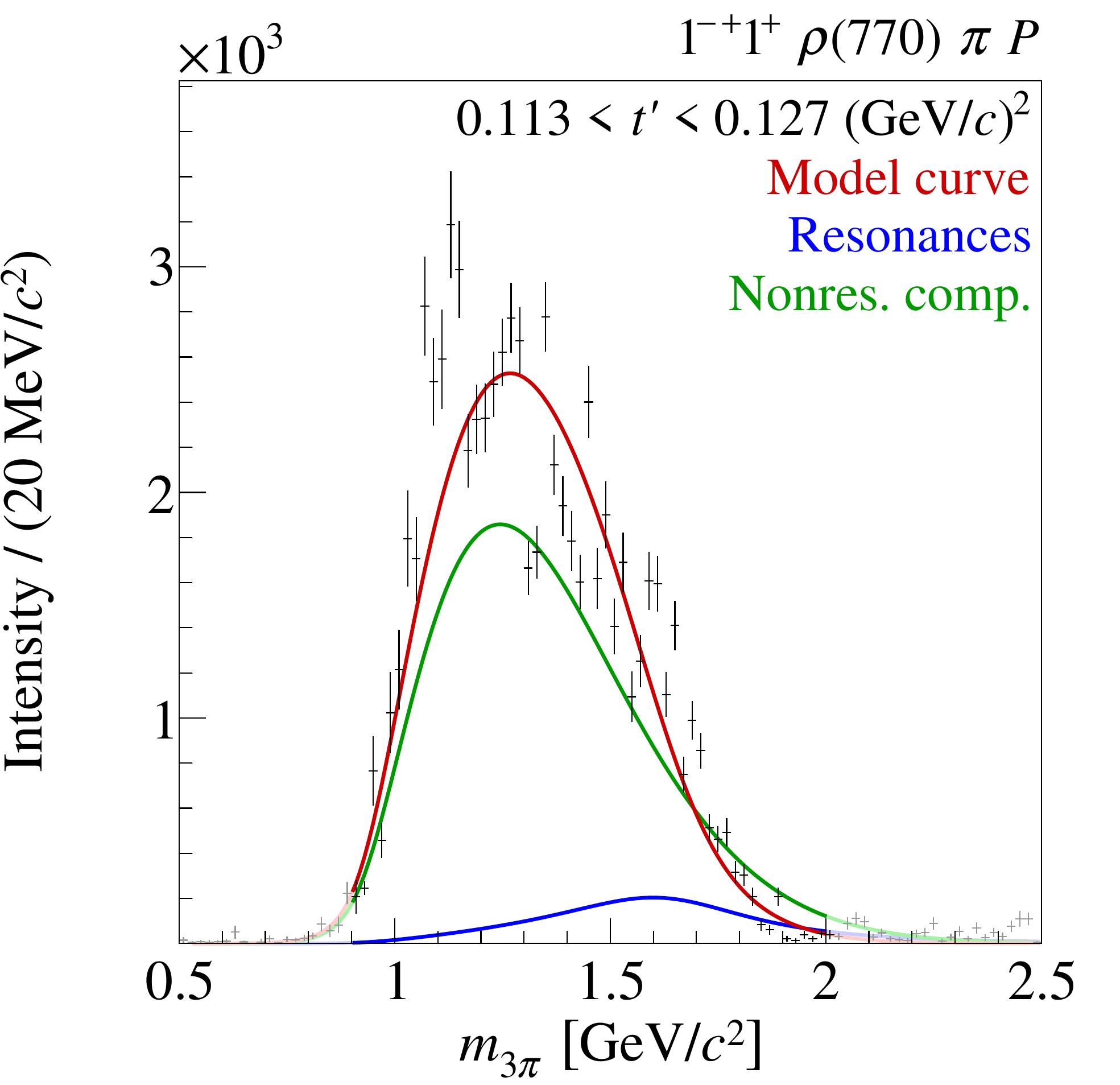}%
  }%
  \hspace*{\threePlotSmallSpacing}%
  \subfloat[][]{%
    \includegraphics[width=\threePlotSmallWidth]{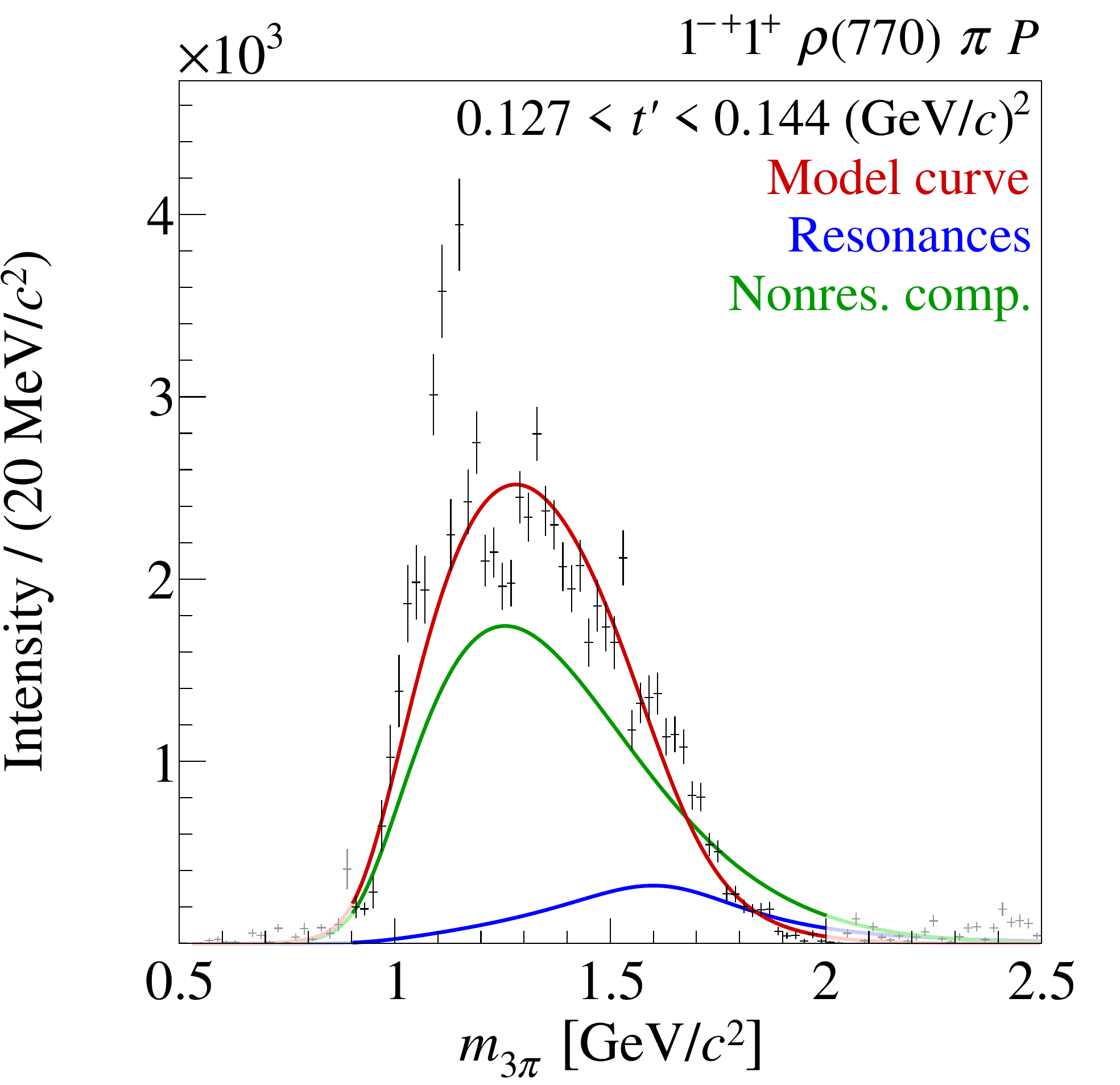}%
  }%
  \\
  \subfloat[][]{%
    \includegraphics[width=\threePlotSmallWidth]{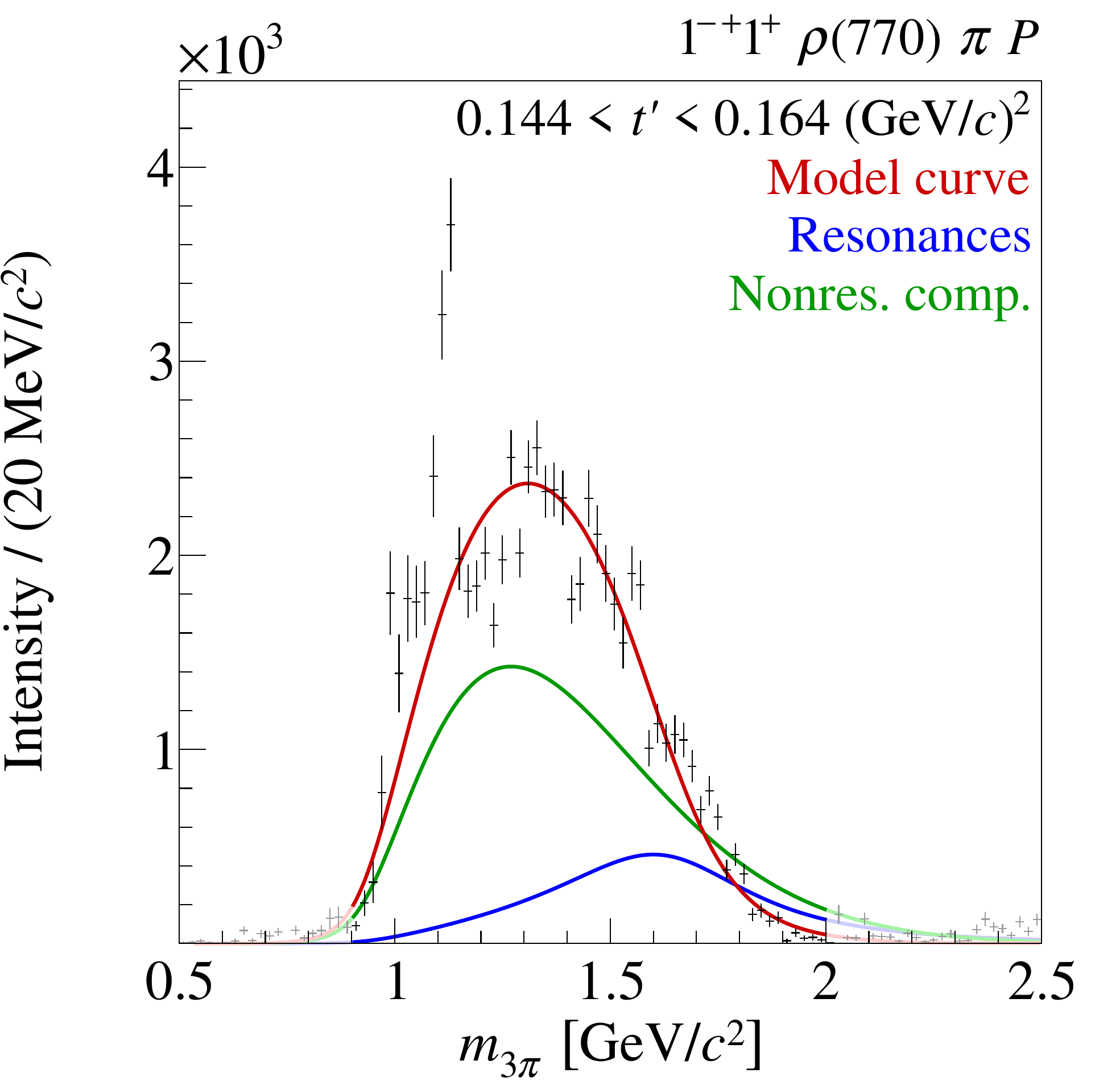}%
  }%
  \hspace*{\threePlotSmallSpacing}%
  \subfloat[][]{%
    \includegraphics[width=\threePlotSmallWidth]{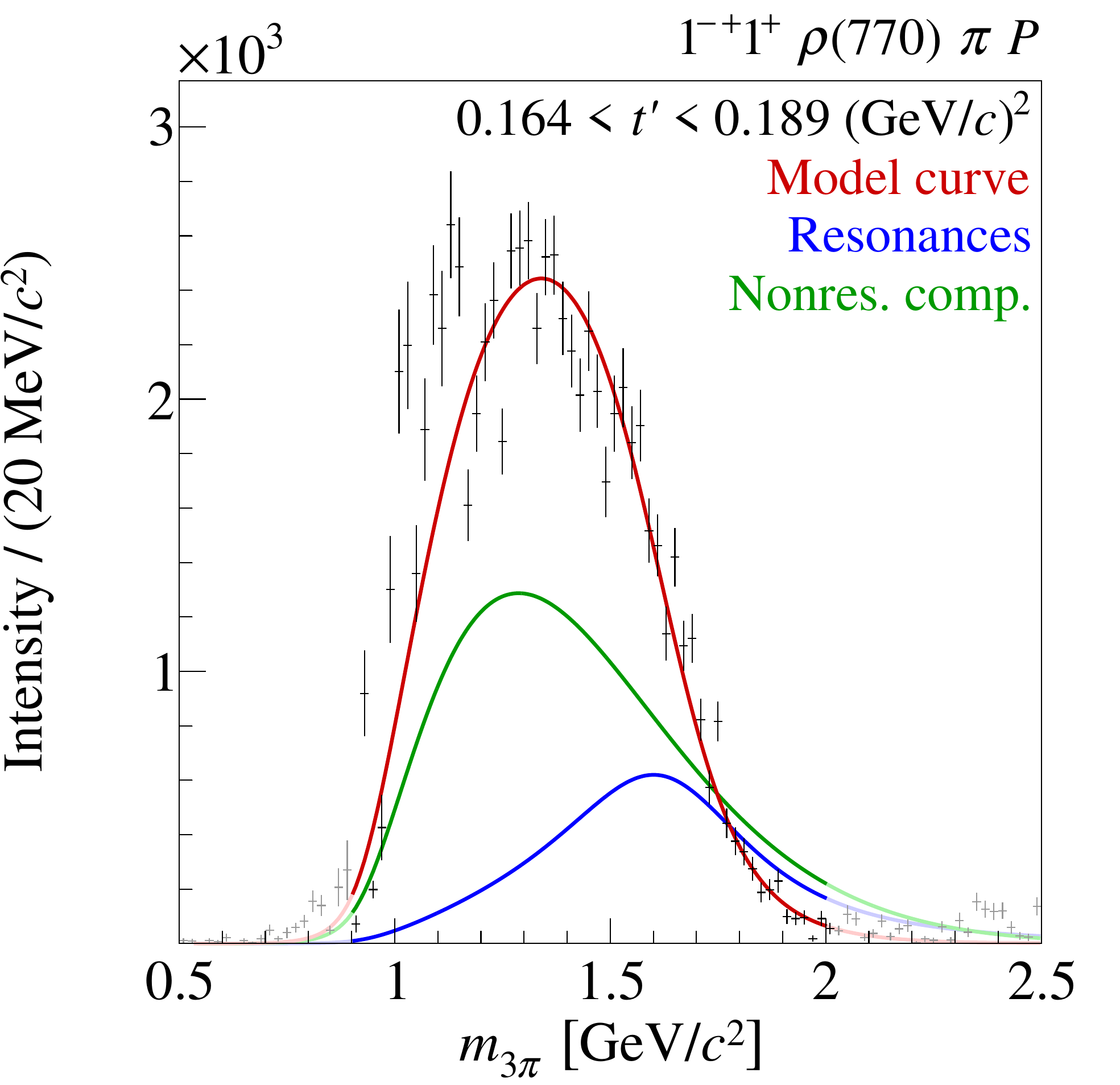}%
  }%
  \hspace*{\threePlotSmallSpacing}%
  \subfloat[][]{%
    \includegraphics[width=\threePlotSmallWidth]{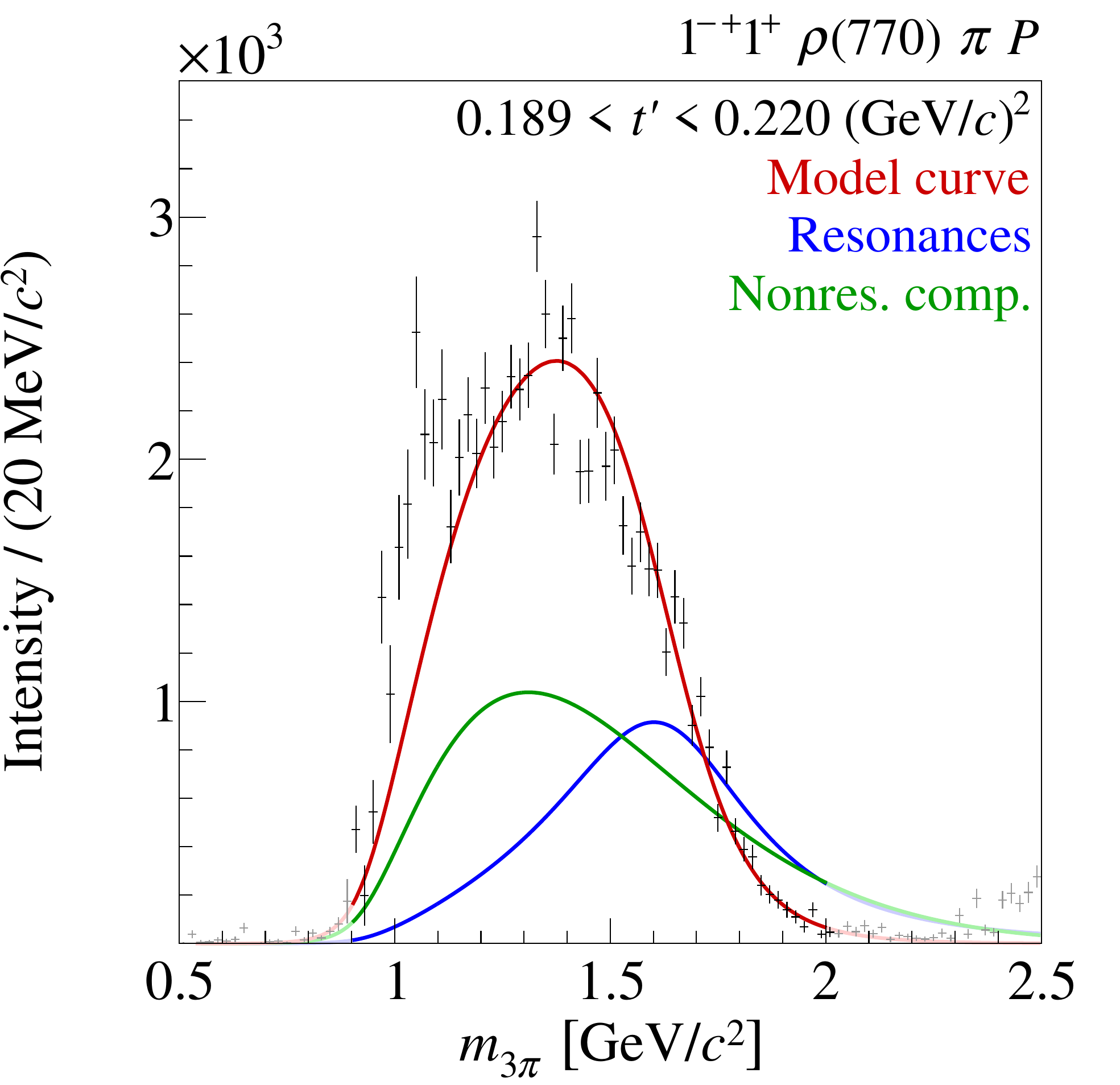}%
  }%
  \\
  \subfloat[][]{%
    \includegraphics[width=\threePlotSmallWidth]{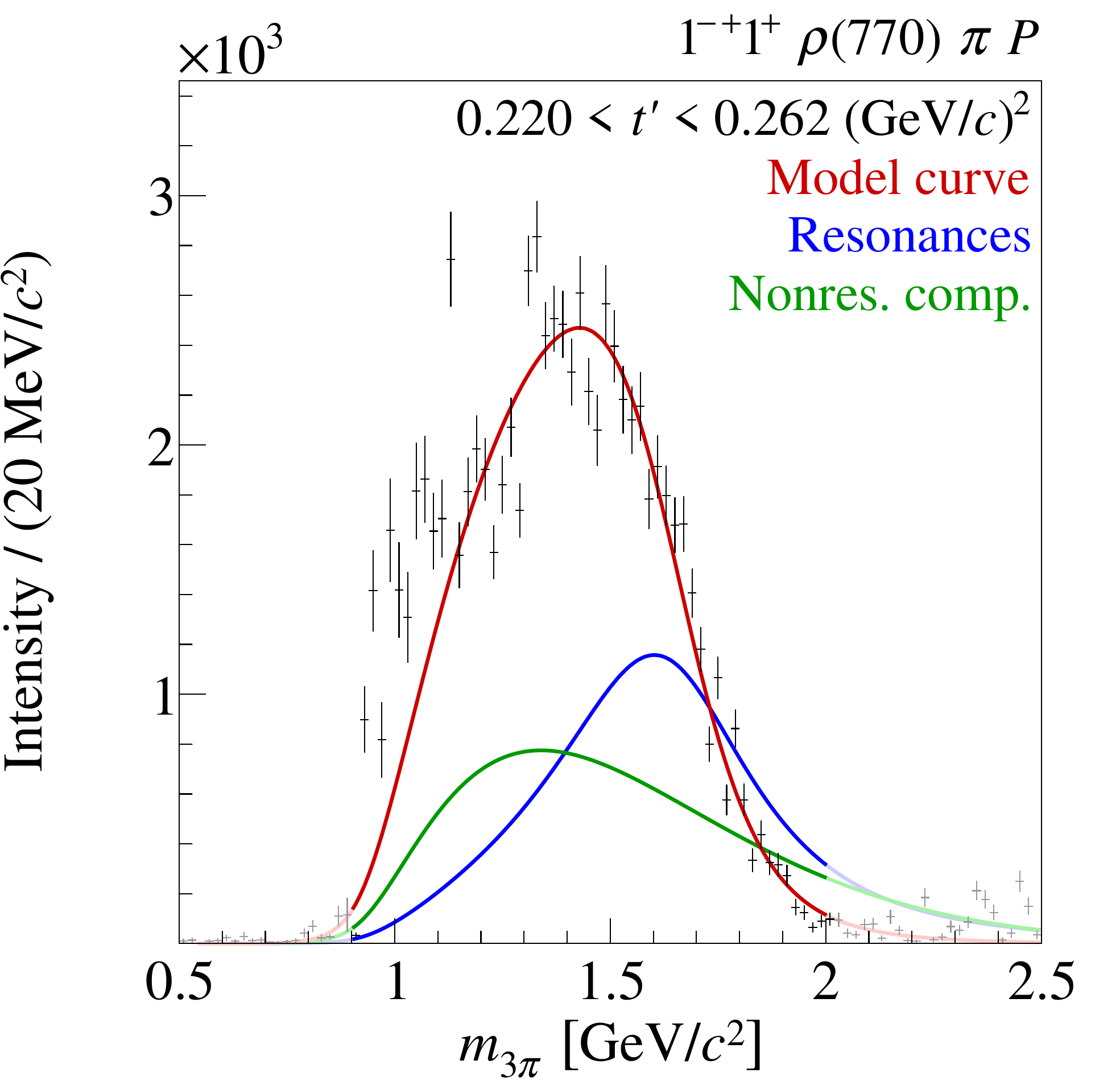}%
  }%
  \hspace*{\threePlotSmallSpacing}%
  \subfloat[][]{%
    \includegraphics[width=\threePlotSmallWidth]{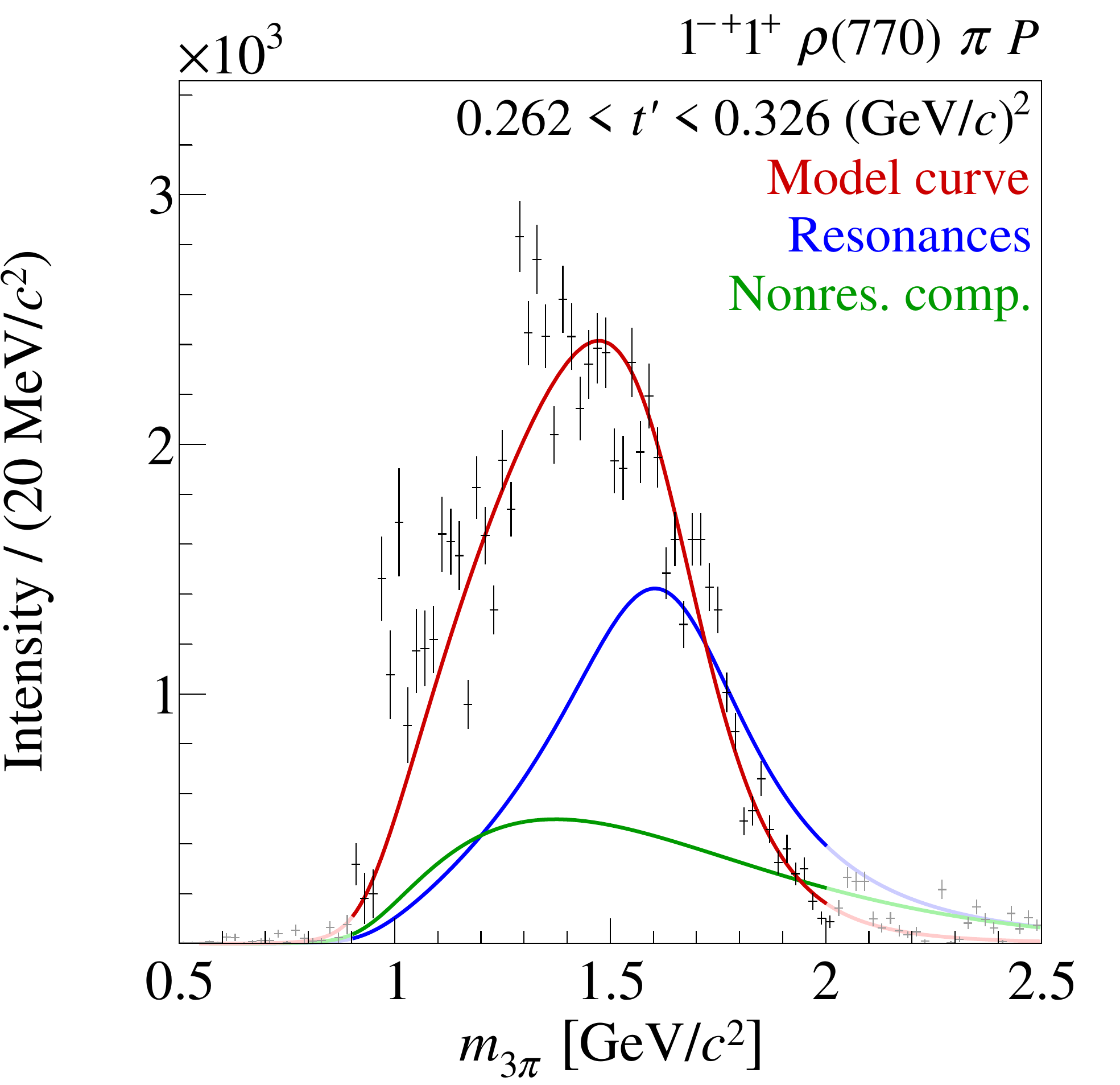}%
  }%
  \hspace*{\threePlotSmallSpacing}%
  \subfloat[][]{%
    \includegraphics[width=\threePlotSmallWidth]{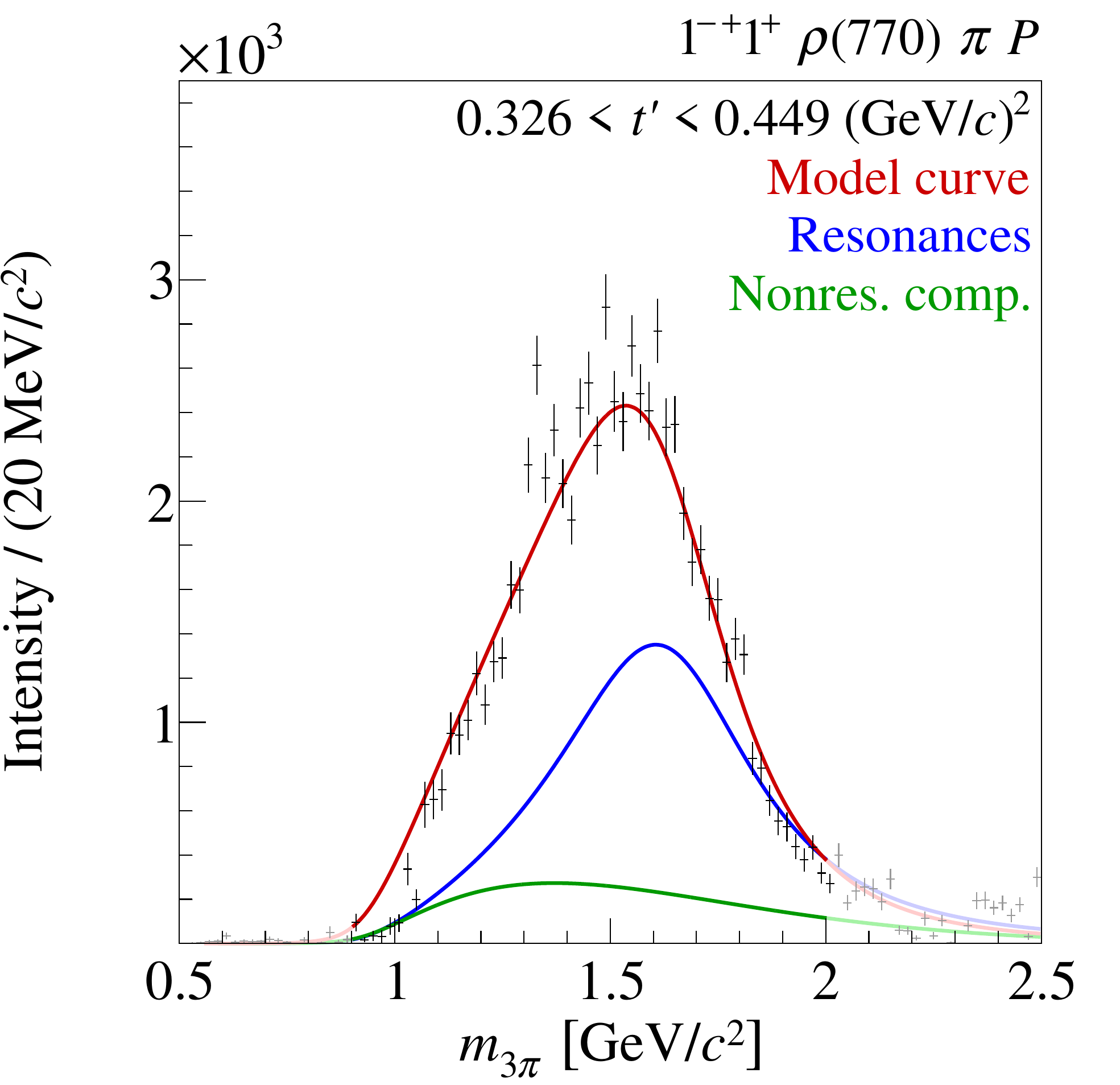}%
  }%
  \\
  \subfloat[][]{%
    \includegraphics[width=\threePlotSmallWidth]{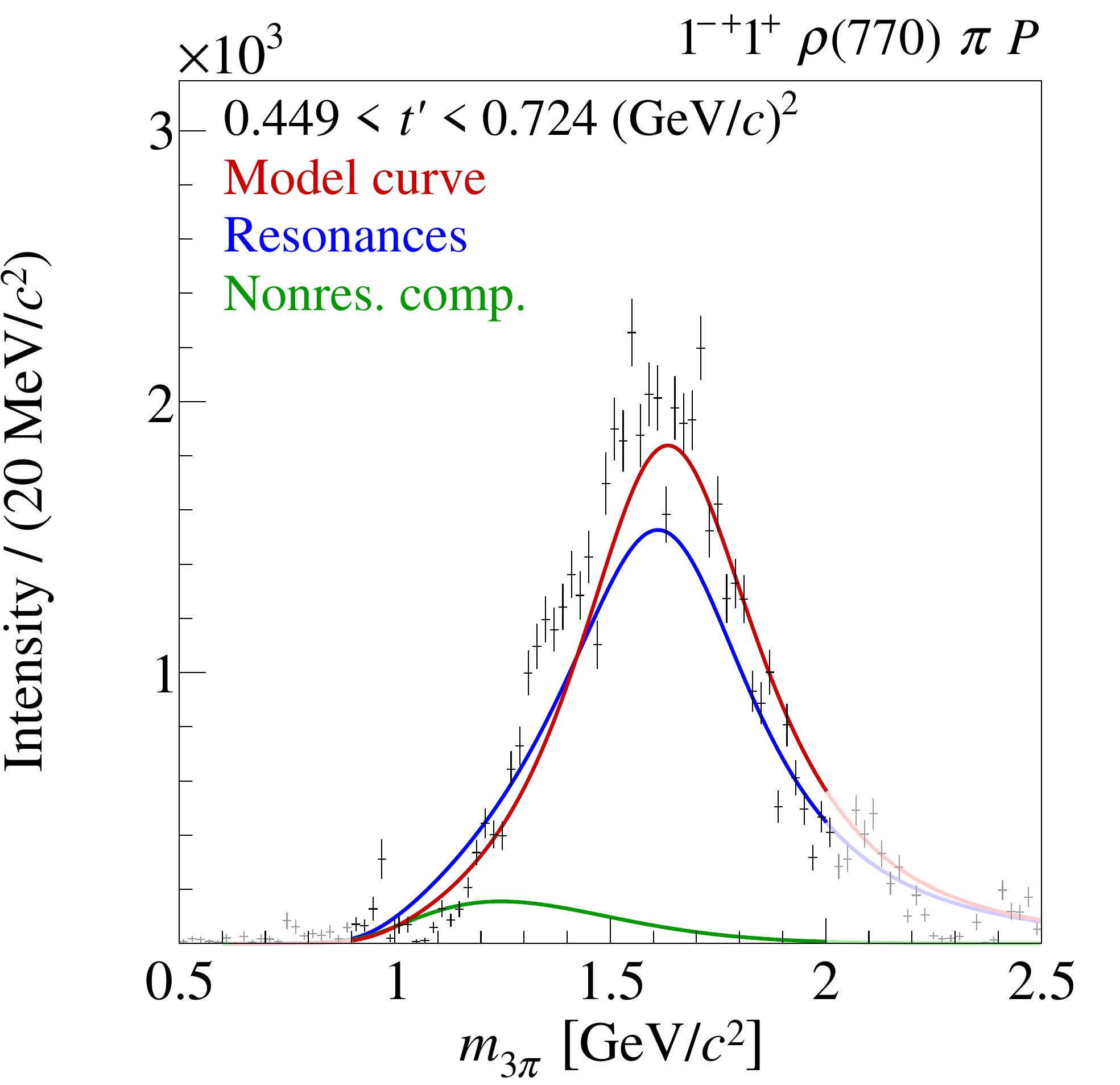}%
    \label{fig:intensity_1mp_tbin10}%
  }%
  \hspace*{\threePlotSmallSpacing}%
  \subfloat[][]{%
    \includegraphics[width=\threePlotSmallWidth]{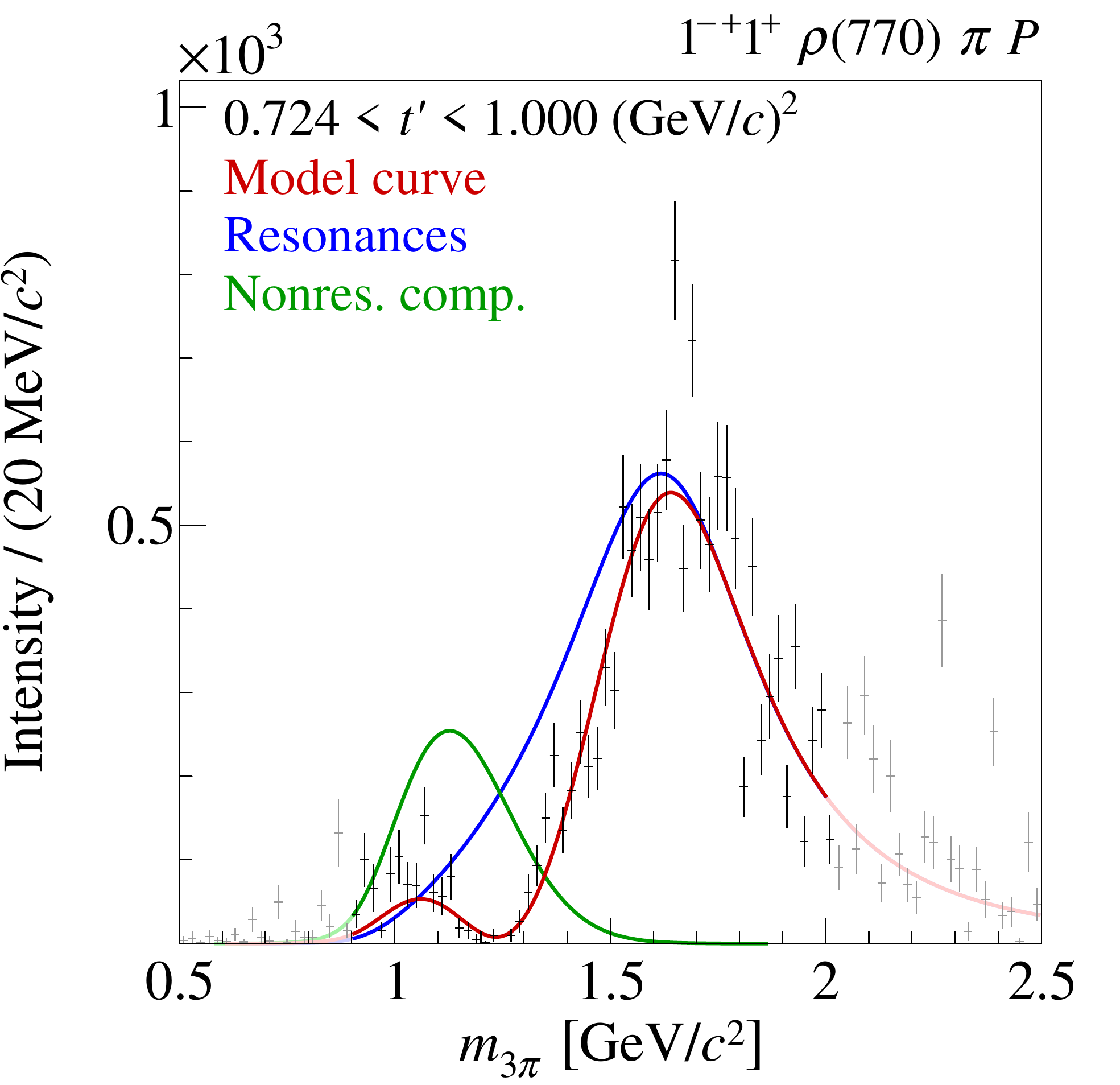}%
    \label{fig:intensity_1mp_tbin11}%
  }%
  \hspace*{\threePlotSmallSpacing}%
  \subfloat[][]{%
    \includegraphics[width=\threePlotSmallWidth]{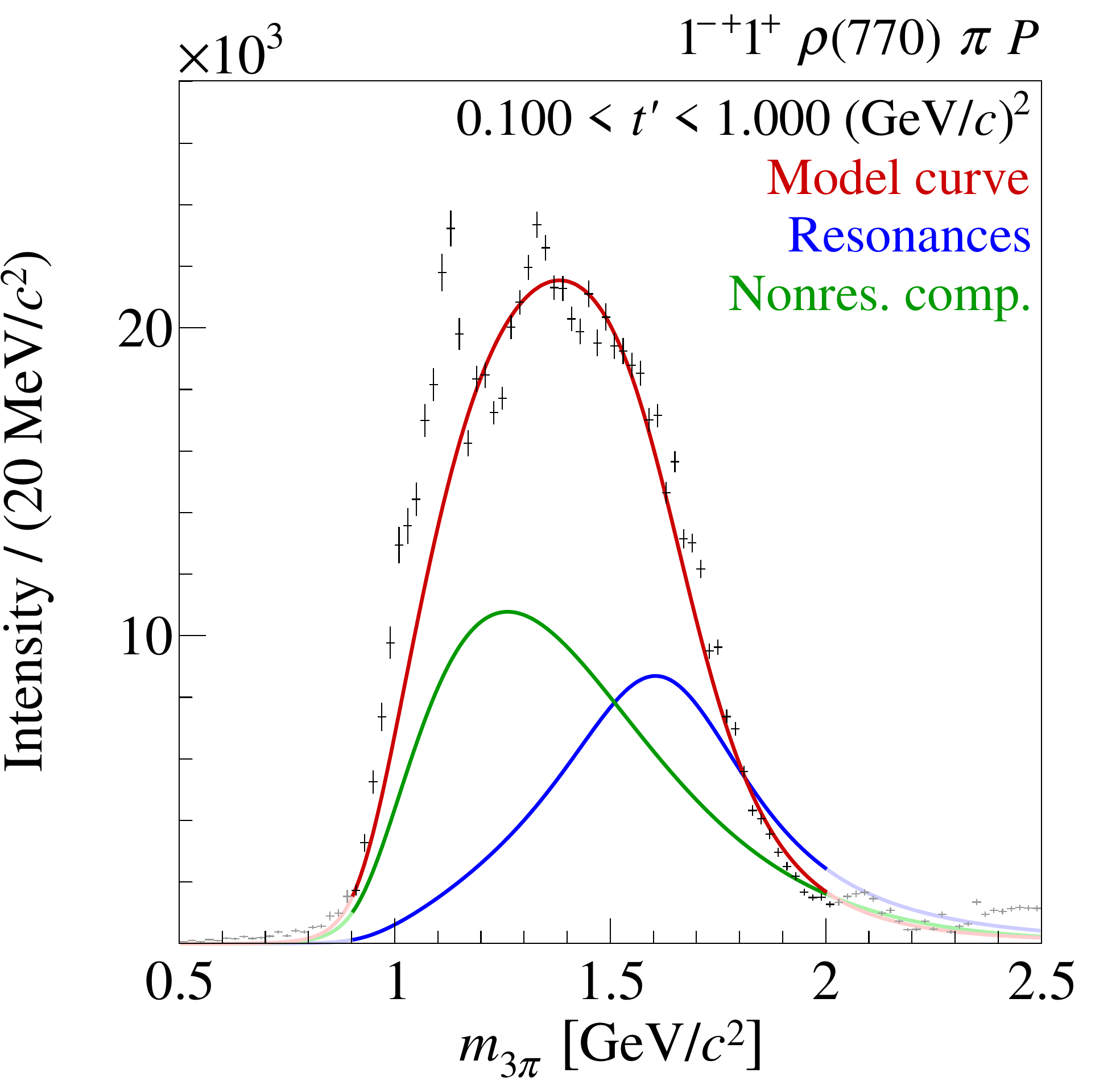}%
    \label{fig:intensity_1mp_tsum}%
  }%
  \caption{\subfloatLabel{fig:intensity_1mp_tbin1}~to~\subfloatLabel{fig:intensity_1mp_tbin11}:
    Intensity distribution of the \wave{1}{-+}{1}{+}{\Prho}{P} wave in
    the 11~\tpr bins.  \subfloatLabel{fig:intensity_1mp_tsum}~The
    \tpr-summed intensity.  The model and the wave components are
    represented as in \cref{fig:intensity_phases_0mp}, except that the
    blue curve represents the \PpiOne[1600].}
  \label{fig:intensities_1mp}
\ifMultiColumnLayout{\end{figure*}}{\end{figure}}

\begin{wideFigureOrNot}[tbp]
  \centering
  \subfloat[][]{%
    \includegraphics[width=\fourPlotWidth]{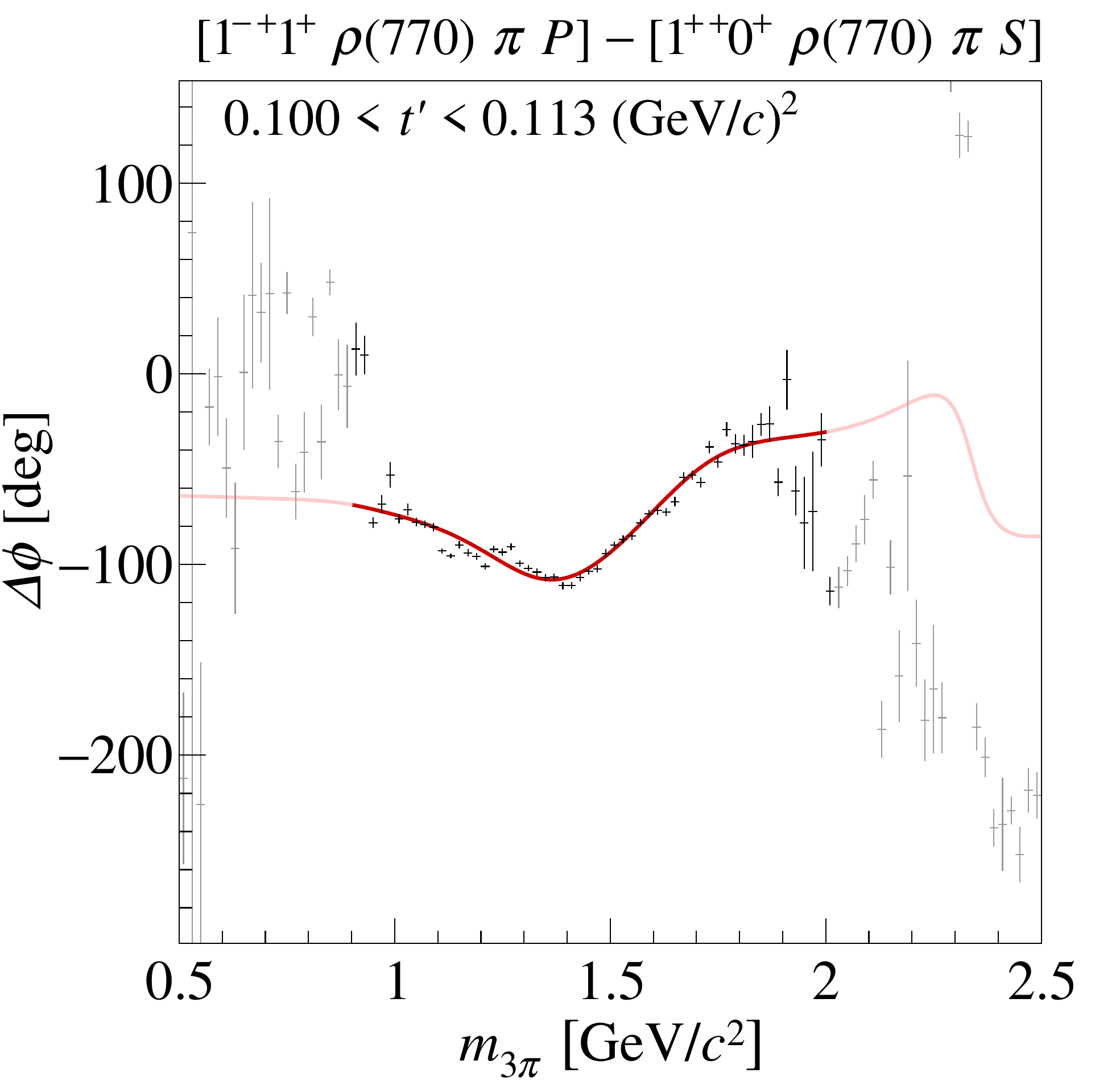}%
    \label{fig:phase_1mp_1pp_rho_tbin1}%
  }%
  \hspace*{\fourPlotSpacing}%
  \subfloat[][]{%
    \includegraphics[width=\fourPlotWidth]{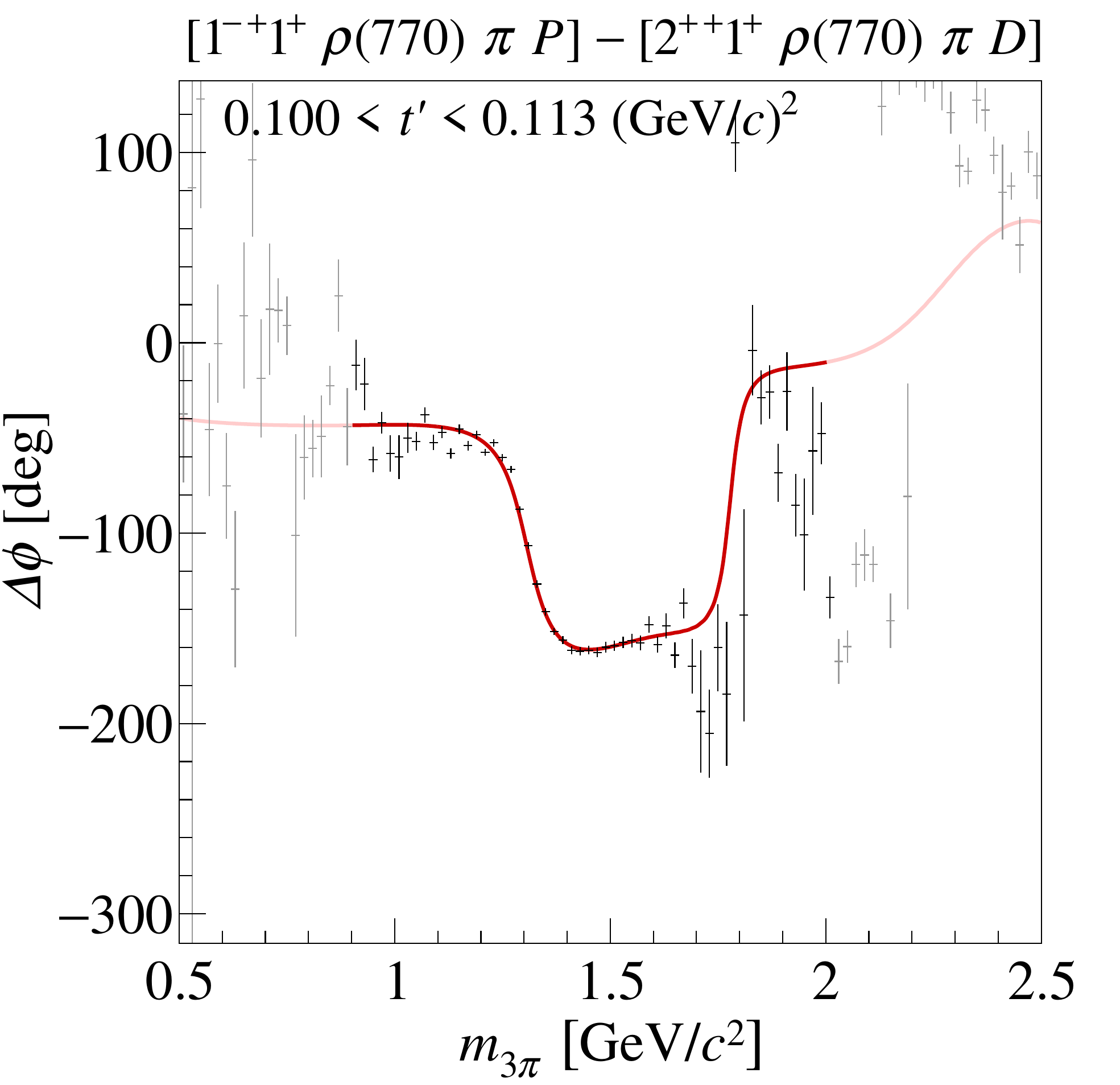}%
    \label{fig:phase_1mp_2pp_m1_rho_tbin1}%
  }%
  \hspace*{\fourPlotSpacing}%
  \subfloat[][]{%
    \includegraphics[width=\fourPlotWidth]{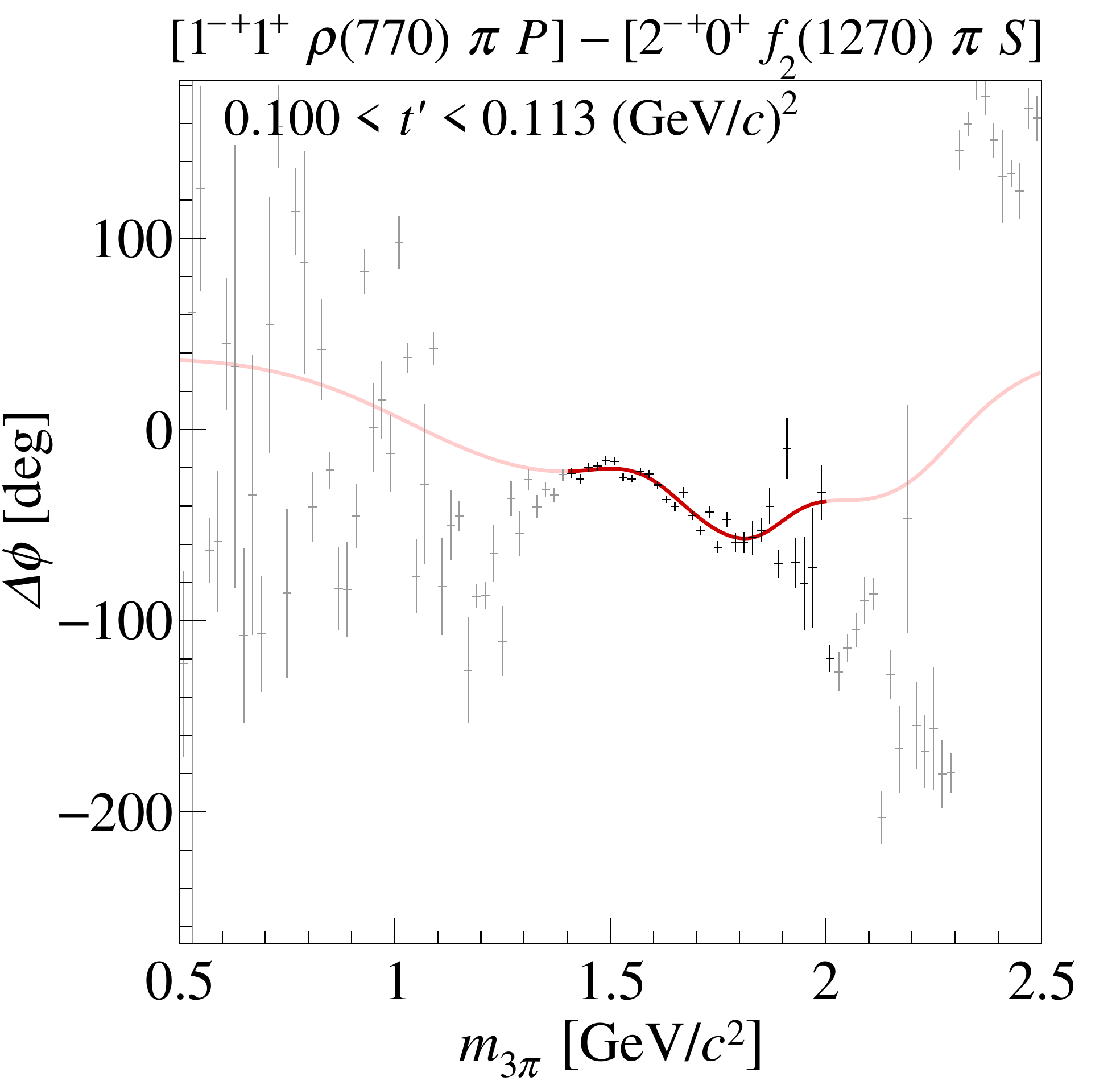}%
    \label{fig:phase_1mp_2mp_m0_f2_tbin1}%
  }%
  \hspace*{\fourPlotSpacing}%
  \subfloat[][]{%
    \includegraphics[width=\fourPlotWidth]{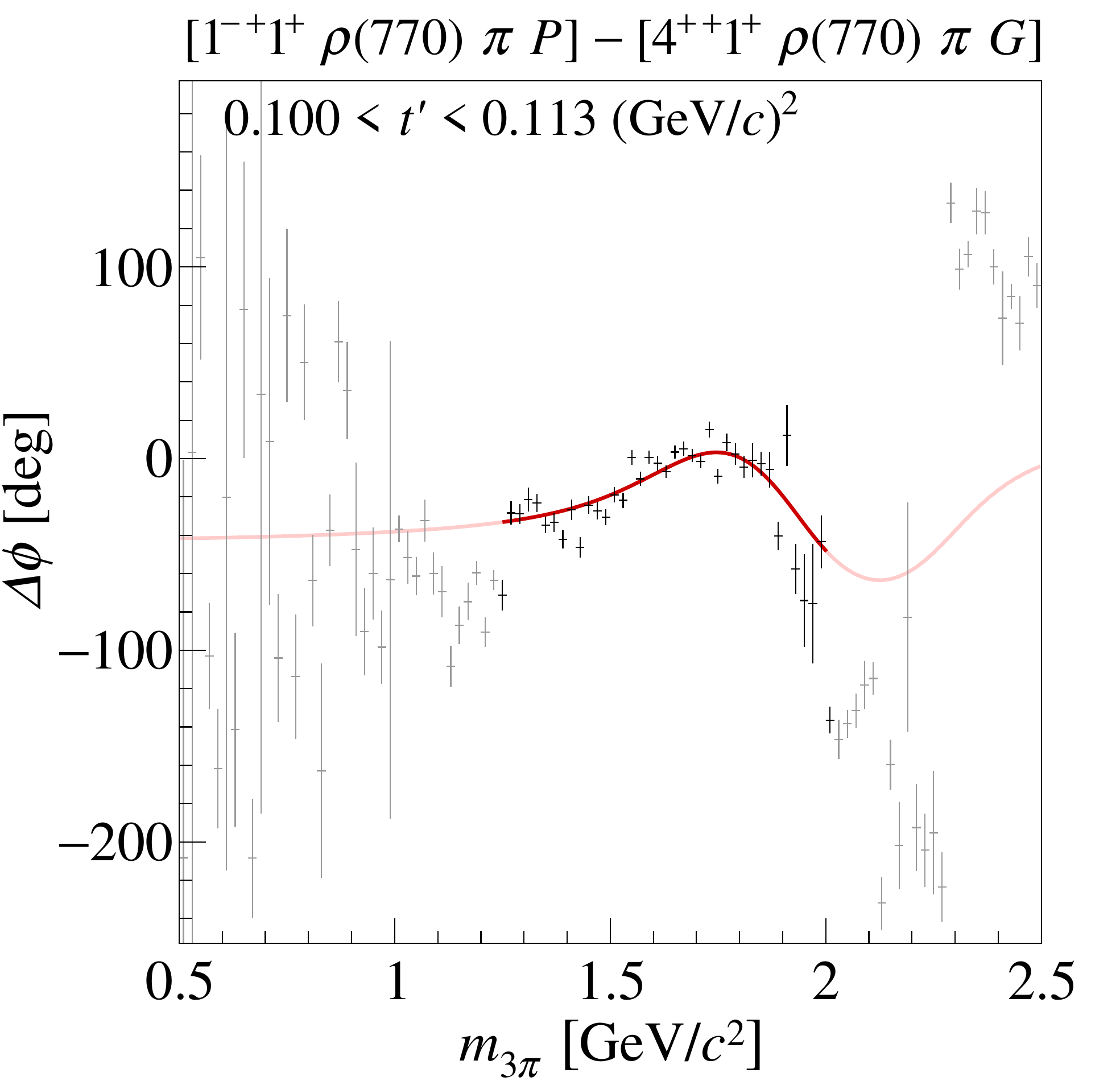}%
    \label{fig:phase_1mp_4pp_rho_tbin1}%
  }%
  \\
  \subfloat[][]{%
    \includegraphics[width=\fourPlotWidth]{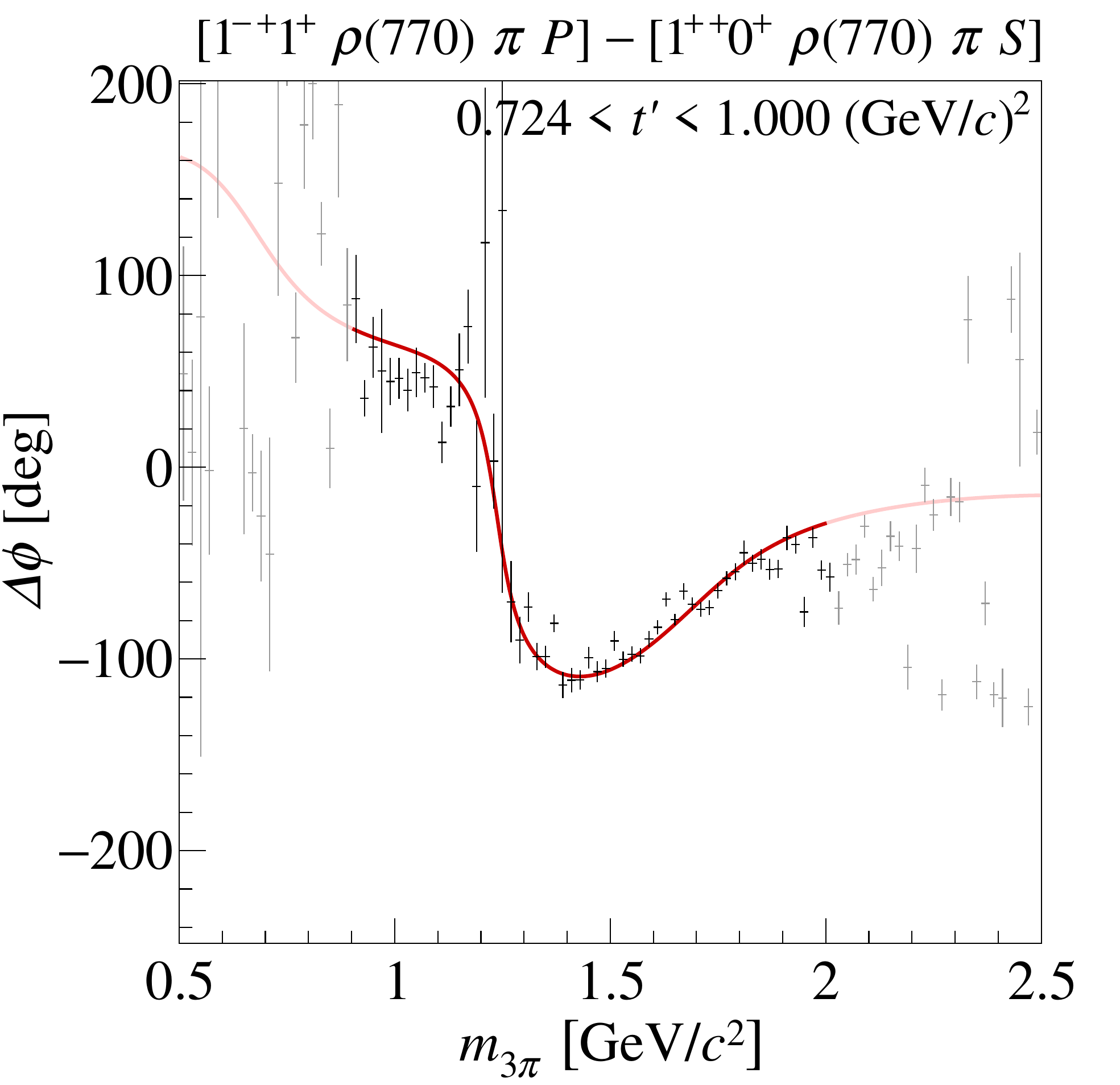}%
    \label{fig:phase_1mp_1pp_rho_tbin11}%
  }%
  \hspace*{\fourPlotSpacing}%
  \subfloat[][]{%
    \includegraphics[width=\fourPlotWidth]{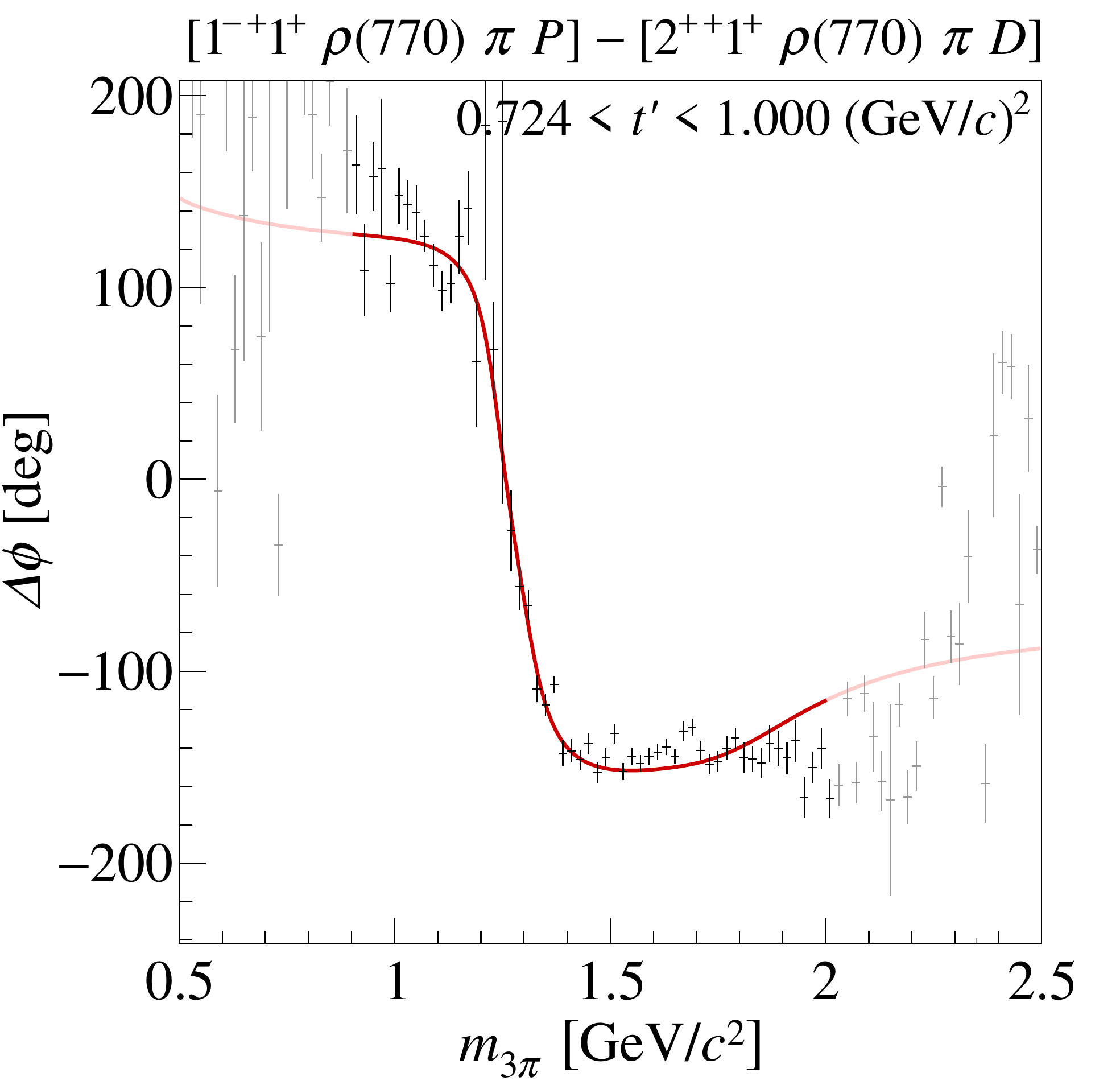}%
    \label{fig:phase_1mp_2pp_m1_rho_tbin11}%
  }%
  \hspace*{\fourPlotSpacing}%
  \subfloat[][]{%
    \includegraphics[width=\fourPlotWidth]{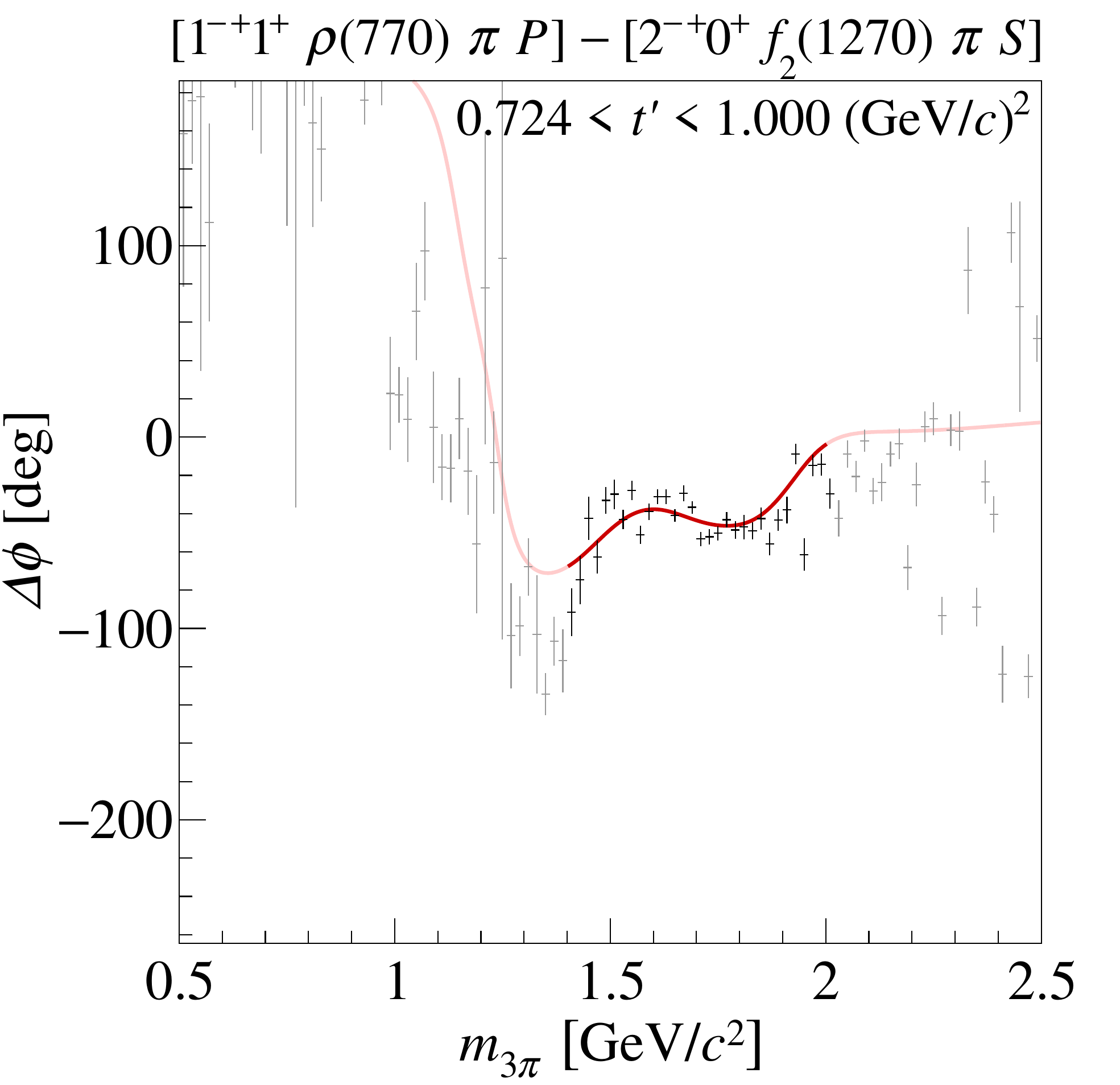}%
    \label{fig:phase_1mp_2mp_m0_f2_tbin11}%
  }%
  \hspace*{\fourPlotSpacing}%
  \subfloat[][]{%
    \includegraphics[width=\fourPlotWidth]{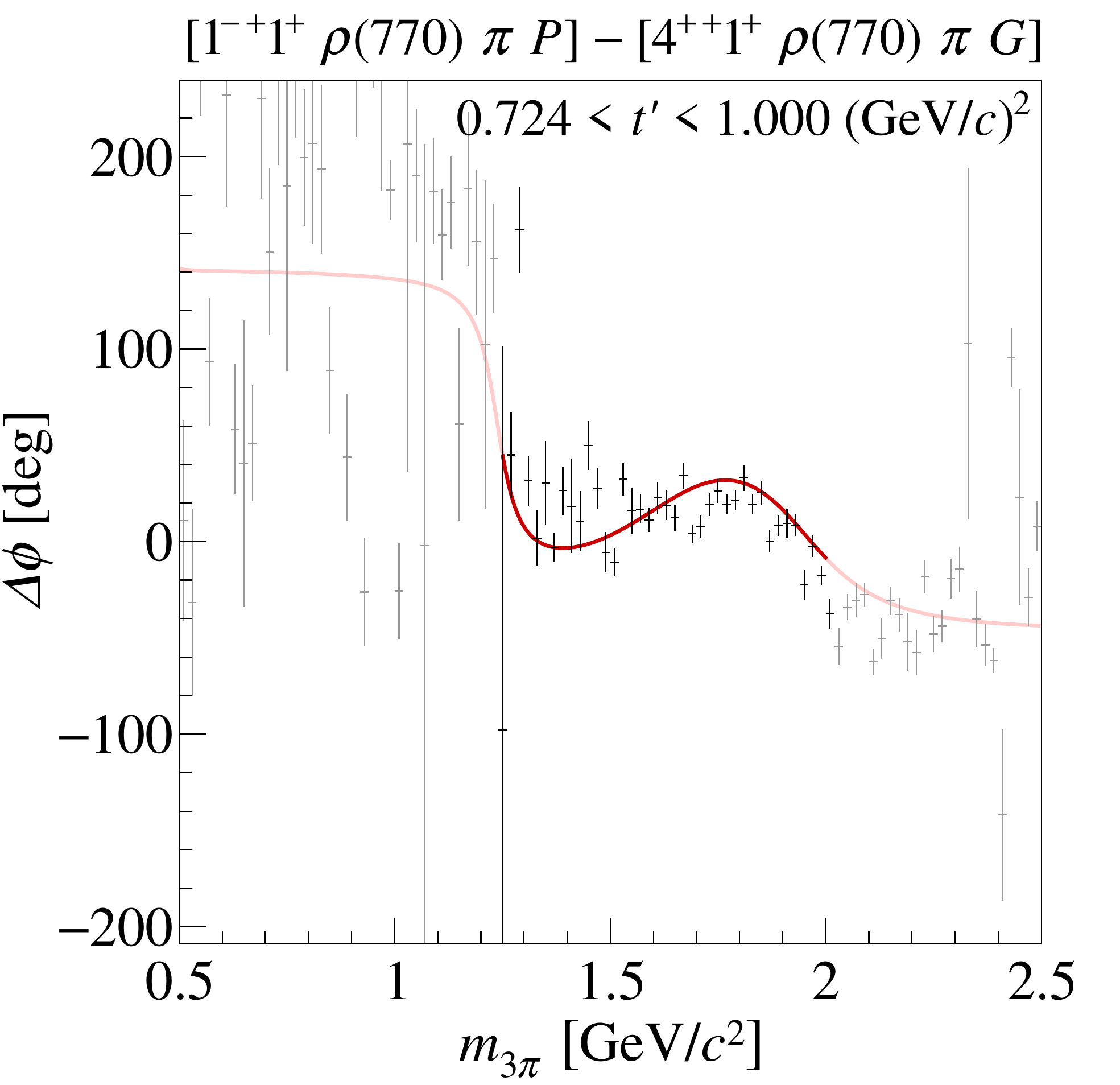}%
    \label{fig:phase_1mp_4pp_rho_tbin11}%
  }%
  \caption{Phase of the \wave{1}{-+}{1}{+}{\Prho}{P} wave relative to
    \subfloatLabel{fig:phase_1mp_1pp_rho_tbin1}~the
    \wave{1}{++}{0}{+}{\Prho}{S},
    \subfloatLabel{fig:phase_1mp_2pp_m1_rho_tbin1}~the
    \wave{2}{++}{1}{+}{\Prho}{D},
    \subfloatLabel{fig:phase_1mp_2mp_m0_f2_tbin1}~the
    \wave{2}{-+}{0}{+}{\PfTwo}{S}, and
    \subfloatLabel{fig:phase_1mp_4pp_rho_tbin1}~the
    \wave{4}{++}{1}{+}{\Prho}{G} wave for the lowest \tpr bin.
    \subfloatLabel{fig:phase_1mp_1pp_rho_tbin11}~through~\subfloatLabel{fig:phase_1mp_4pp_rho_tbin11}:
    The phases for the highest \tpr bin.  The model is represented as
    in \cref{fig:intensities_1mp}.}
  \label{fig:phases_1mp}
\end{wideFigureOrNot}

We describe the \wave{1}{-+}{1}{+}{\Prho}{P} amplitude by a
spin-exotic $\JPC = 1^{-+}$ resonance, the \PpiOne[1600], and a
nonresonant component.  The \PpiOne[1600] is parametrized by
\cref{eq:BreitWigner,eq:method:fixedwidth}, the nonresonant component
using \cref{eq:method:nonresterm} (see
\cref{tab:method:fitmodel:waveset}).  The $1^{-+}$ wave is fit in the
mass range from \SIrange{0.9}{2.0}{\GeVcc}.

The model is in fair agreement with the intensity distributions.  It
reproduces in particular the strong \tpr dependence of the shape of
the intensity distribution by a \tpr-dependent interference of the
\PpiOne[1600] with the nonresonant component.  The latter strongly
changes shape, strength, and phase with \tpr.  At low \tpr, the
intensity is dominated by the large nonresonant component, which
interferes constructively with the \PpiOne[1600] at low masses.  With
increasing \tpr, the strength of the nonresonant component decreases
quickly so that the \PpiOne[1600] becomes the dominant component.  In
the two highest \tpr bins, the nonresonant component is small or even
vanishes in the \SI{1.6}{\GeVcc} region and the broad peak in the data
is nearly entirely described by the \PpiOne[1600].  The intensity dip
at \SI{1.25}{\GeVcc} in the highest \tpr bin is reproduced by a
destructive interference of the \PpiOne[1600] and the nonresonant
component.  However, the shape of the nonresonant component in the
highest \tpr bin seems implausible since it is inconsistent with the
continuous evolution with increasing \tpr (see the discussion of the
Deck model below).  At low \tpr, the model does not describe well the
low-mass part of the intensity distribution.  In particular, the model
cannot reproduce the presumably artificial narrow enhancement at
\SI{1.1}{\GeVcc}.

The model describes the phases of the $1^{-+}$ wave well within the
fit range.  The \PpiOne[1600] component causes only slight phase
motions.  This becomes particularly obvious in the nearly constant
phase \wrt the \wave{4}{++}{1}{+}{\Prho}{G} wave in the
\SI{1.6}{\GeVcc} region [see
\cref{fig:phase_1mp_4pp_rho_tbin1,fig:phase_1mp_4pp_rho_tbin11}].  The
$4^{++}$ wave contains no resonance in this mass range.  For some
waves, the model extrapolations to low or high masses deviate from the
data [see \eg
\cref{fig:phase_1mp_1pp_rho_tbin1,fig:phase_1mp_2mp_m0_f2_tbin1}].

The strong \tpr dependence of the relative strength of the nonresonant
and the \PpiOne[1600] components is shown in
\cref{fig:tprim_1mp_main}.  For $\tpr \gtrsim \SI{0.3}{\GeVcsq}$, the
\PpiOne[1600] contribution dominates, whereas in the lowest \tpr bin
the intensity of the nonresonant component, integrated over the fit
range, is nearly an order of magnitude larger.  The \tpr spectrum of
the \PpiOne[1600] is not well described by the parametrization in
\cref{eq:slope-parametrization}.  The model is not able to reproduce
the downturn toward low \tpr.  This may be a hint that, at low \tpr,
the fit is not able to separate the small \PpiOne[1600] component from
the dominant nonresonant component due to an inappropriate description
of the shape of the latter.  This hypothesis is supported by the
result of a study, in which the shape of the nonresonant component was
determined from a Deck model (see discussion below).  Limiting the fit
range to the region \SIvalRange{0.189}{\tpr}{0.724}{\GeVcsq}, where
the model is able to describe the data, yields a \PpiOne[1600] slope
parameter of \SI{7.3}{\perGeVcsq}.  This value lies in the range that
is typical for resonances and is clearly much smaller than the slope
value of the nonresonant component. The model in
\cref{eq:slope-parametrization} is in fair agreement with the \tpr
spectrum of the nonresonant component, which has a slope parameter
value of \SIaerrSys{19.1}{1.4}{4.7}{\perGeVcsq}.  This is the second
largest slope value of all wave components in the fit.\footnote{Only
  the nonresonant component in the \wave{0}{-+}{0}{+}{\PfZero[980]}{S}
  wave has an even steeper slope (see
  \cref{tab:slopes,sec:zeroMP_results}).}

\begin{wideFigureOrNot}[tbp]
  \centering
  \subfloat[][]{%
    \includegraphics[width=\twoPlotWidth]{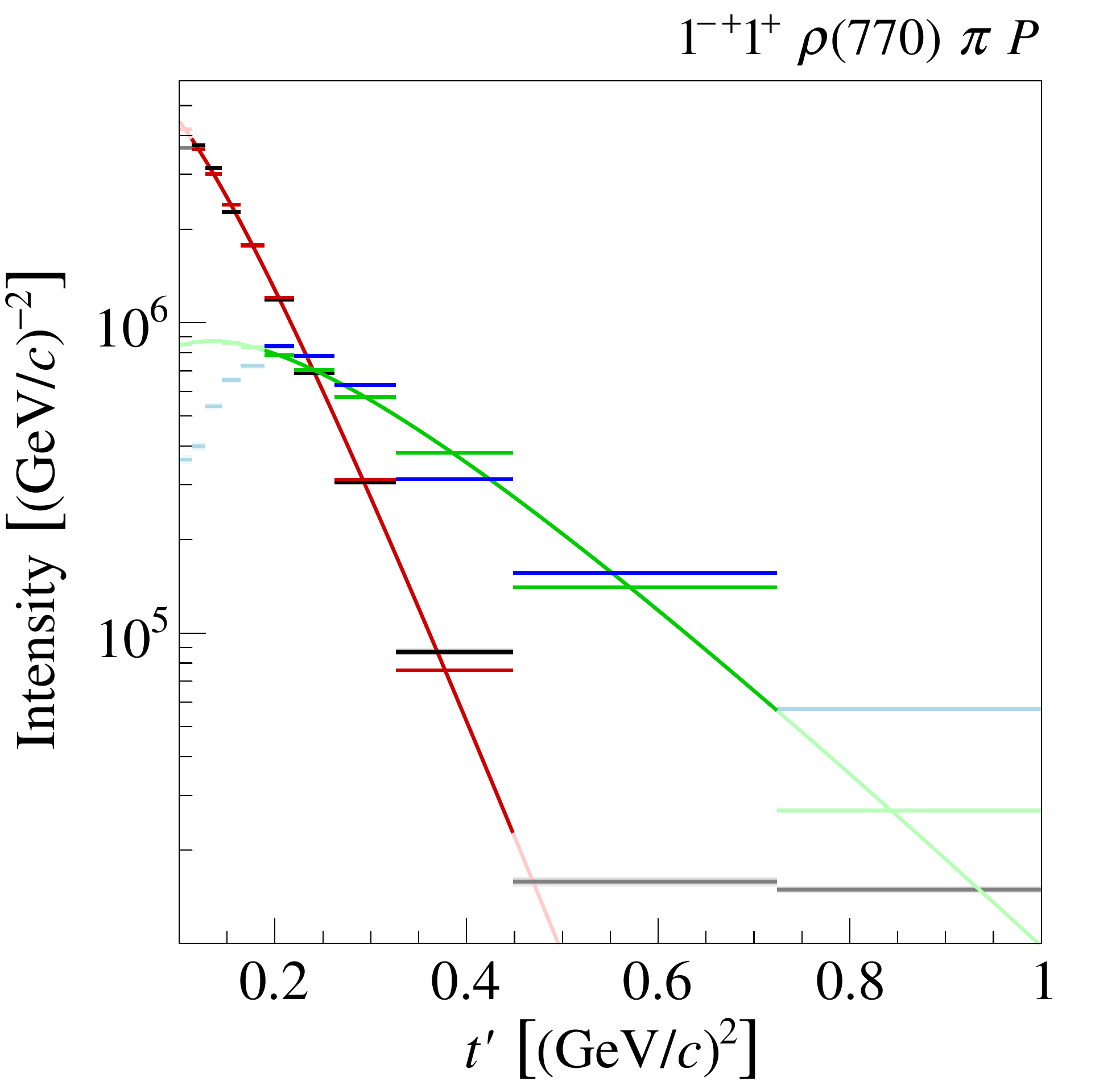}%
    \label{fig:tprim_1mp_main}%
  }%
  \hspace*{\twoPlotSpacing}%
  \subfloat[][]{%
    \includegraphics[width=\twoPlotWidth]{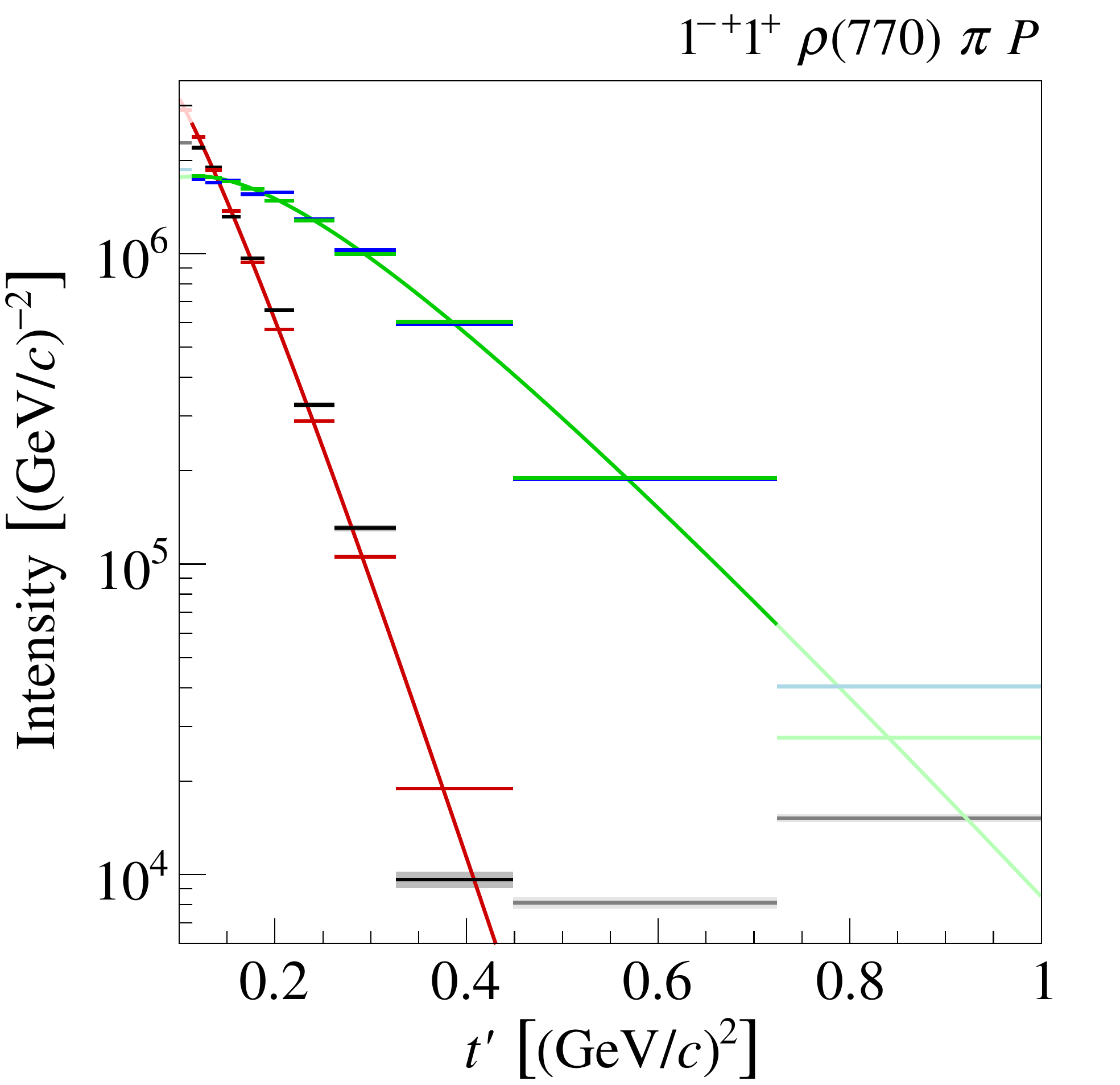}%
    \label{fig:tprim_1mp_Deck}%
  }%
  \caption{Similar to \cref{{fig:tprim_0mp}}, but showing the \tpr
    spectra of the two $\JPC = 1^{-+}$ wave components as given by
    \cref{eq:tprim-dependence}: the \PpiOne[1600] component is shown
    as blue lines and light blue boxes, and the nonresonant component
    as black lines and gray boxes.  The red and green curves and
    horizontal lines represent fits using
    \cref{eq:slope-parametrization}. \subfloatLabel{fig:tprim_1mp_main}~The
    result of the main fit.  \subfloatLabel{fig:tprim_1mp_Deck}~The
    result of a fit, in which the parametrization of the nonresonant
    amplitude was replaced by the square root of the intensity
    distribution of the partial-wave decomposition of Deck Monte Carlo
    data [\StudyO; see
    \cref{sec:systematics,fig:intensities_phase_1mp_DeckMC}].}
  \label{fig:tprim_1mp}
\end{wideFigureOrNot}

From the fit, we obtain the Breit-Wigner resonance parameters
$m_{\PpiOne[1600]} = \SIaerrSys{1600}{110}{60}{\MeVcc}$ and
$\Gamma_{\PpiOne[1600]} = \SIaerrSys{580}{100}{230}{\MeVcc}$.  Since
the $1^{-+}$ wave has a small intensity and is dominated by
nonresonant contributions, the \PpiOne[1600] resonance parameters are
sensitive to changes of the fit model discussed in
\cref{sec:systematics} and hence have large systematic uncertainties.
In the systematic studies, we observe a correlation of the
\PpiOne[1600] parameters with the \PaOne, \PaOne[1640], and
\PaTwo[1700] parameters.
We also observe that the \PpiOne[1600] parameters depend on the choice
of the waves included in the fit.
Studies~\studyS and~\studyR with alternative \chisq~formulations (see
\cref{sec:alt_chi_2}) indicate that larger width values are preferred
when less weight is given to the phase information in the
\chisq~function.
The \PpiOne[1600] parameters are also sensitive to the range
parameter~$q_R$ in the Blatt-Weisskopf factors.  More details on the
results of these systematic studies are discussed in
\cref{sec:syst_uncert_oneMP}.

Since the $1^{-+}$ wave is dominated by the nonresonant component, the
fit result depends on the choice of the parametrization for the
nonresonant component.  In order to estimate this dependence, we
performed \StudyO, in which the parametrization of the nonresonant
amplitude was replaced by the square root of the intensity
distribution of the partial-wave decomposition of Deck Monte Carlo
data generated according to the model described in
\cref{sec:deck_model}.  This fit describes the $1^{-+}$ amplitude
fairly well (see
\cref{fig:DeckMC_chi2difference,fig:intensities_phase_1mp_DeckMC}).
The Deck model behaves qualitatively similar to the empirical
parametrization used in the main fit, except in the highest \tpr bin,
where the Deck model has a more plausible shape.  The main difference
\wrt the main fit is a larger \PpiOne[1600] yield at low \tpr.  The
resulting \tpr spectrum for the \PpiOne[1600] [see
\cref{fig:tprim_1mp_Deck}] is much better described by the
parametrization in \cref{eq:slope-parametrization} than the \tpr
spectrum of the main fit [see \cref{fig:tprim_1mp_main}].  The slope
value of \SI{8.5}{\perGeVcsq} that is extracted using a fit range of
\SIvalRange{0.113}{\tpr}{0.742}{\GeVcsq} is in the range typical for
resonances. Mass and width of the \PpiOne[1600] resonance decrease by
\SI{60}{\MeVcc}.  \StudyO defines the lower boundary of the
uncertainty interval for the \PpiOne[1600] mass.

\begin{wideFigureOrNot}[tbp]
  \centering
  \subfloat[][]{%
    \includegraphics[width=\twoPlotWidth]{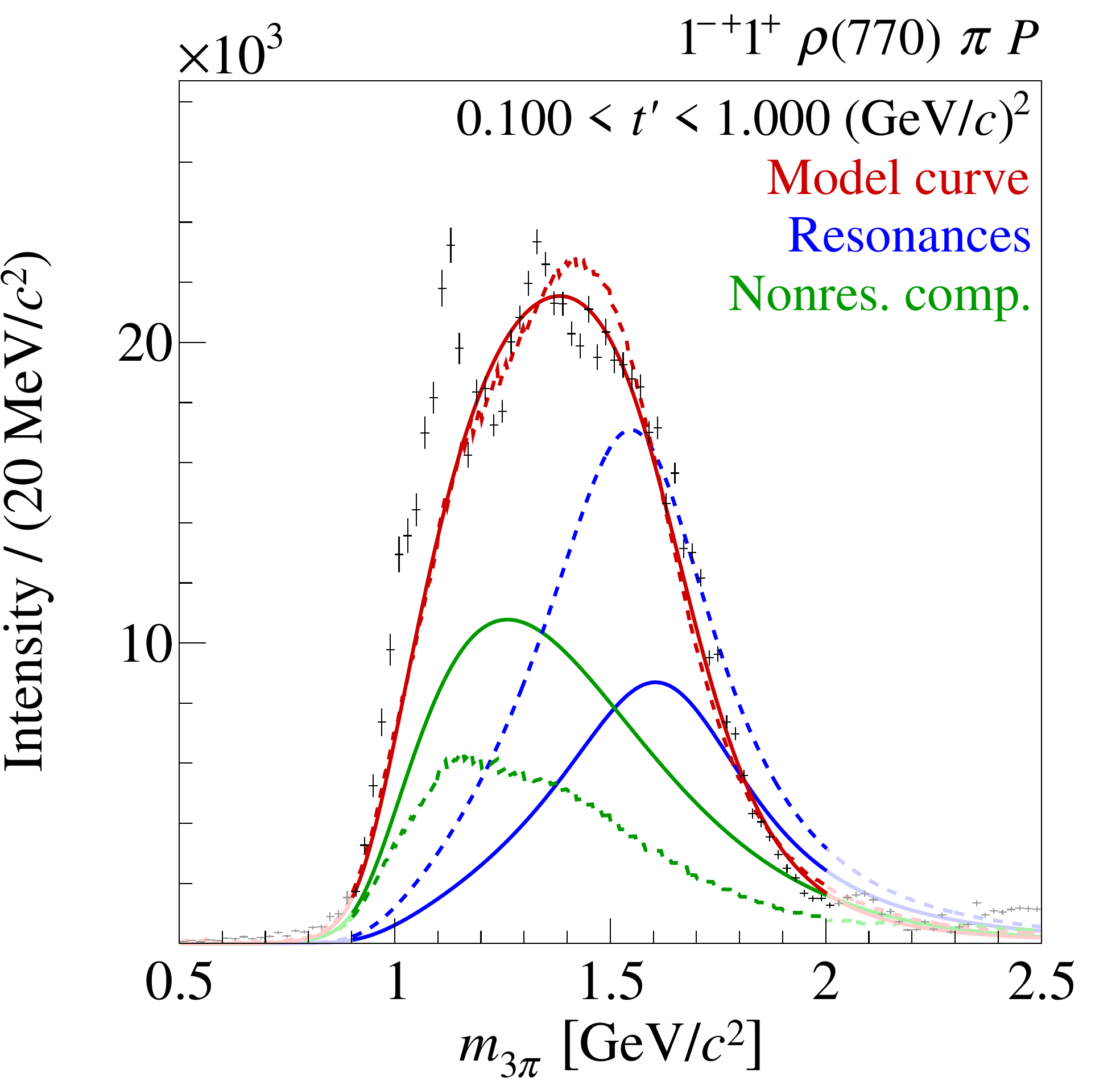}%
    \label{fig:intensity_1mp_DeckMC}%
  }%
  \hspace*{\twoPlotSpacing}%
  \subfloat[][]{%
    \includegraphics[width=\twoPlotWidth]{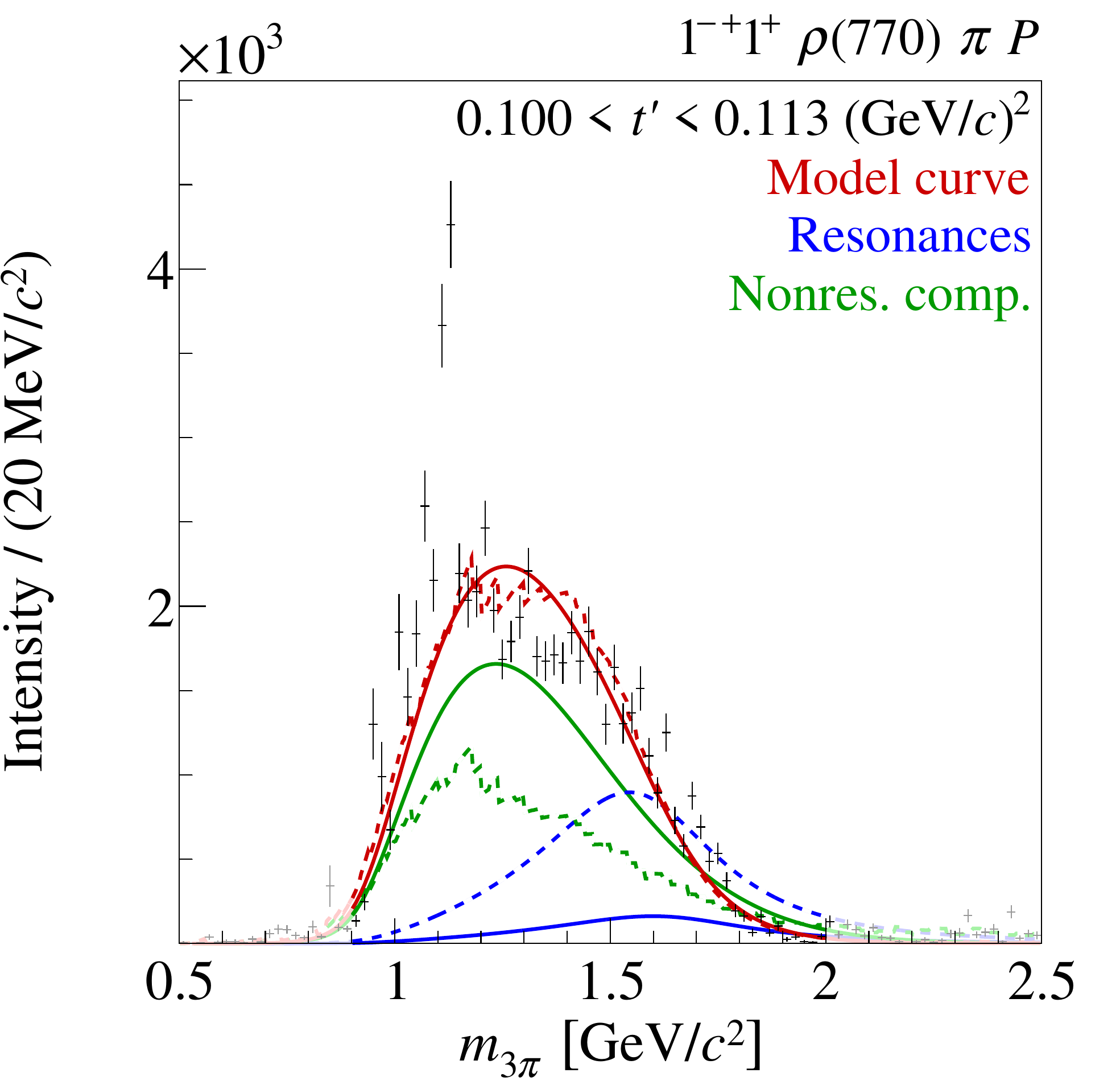}%
    \label{fig:intensity_1mp_tbin1_DeckMC}%
  }%
  \\
  \subfloat[][]{%
    \includegraphics[width=\twoPlotWidth]{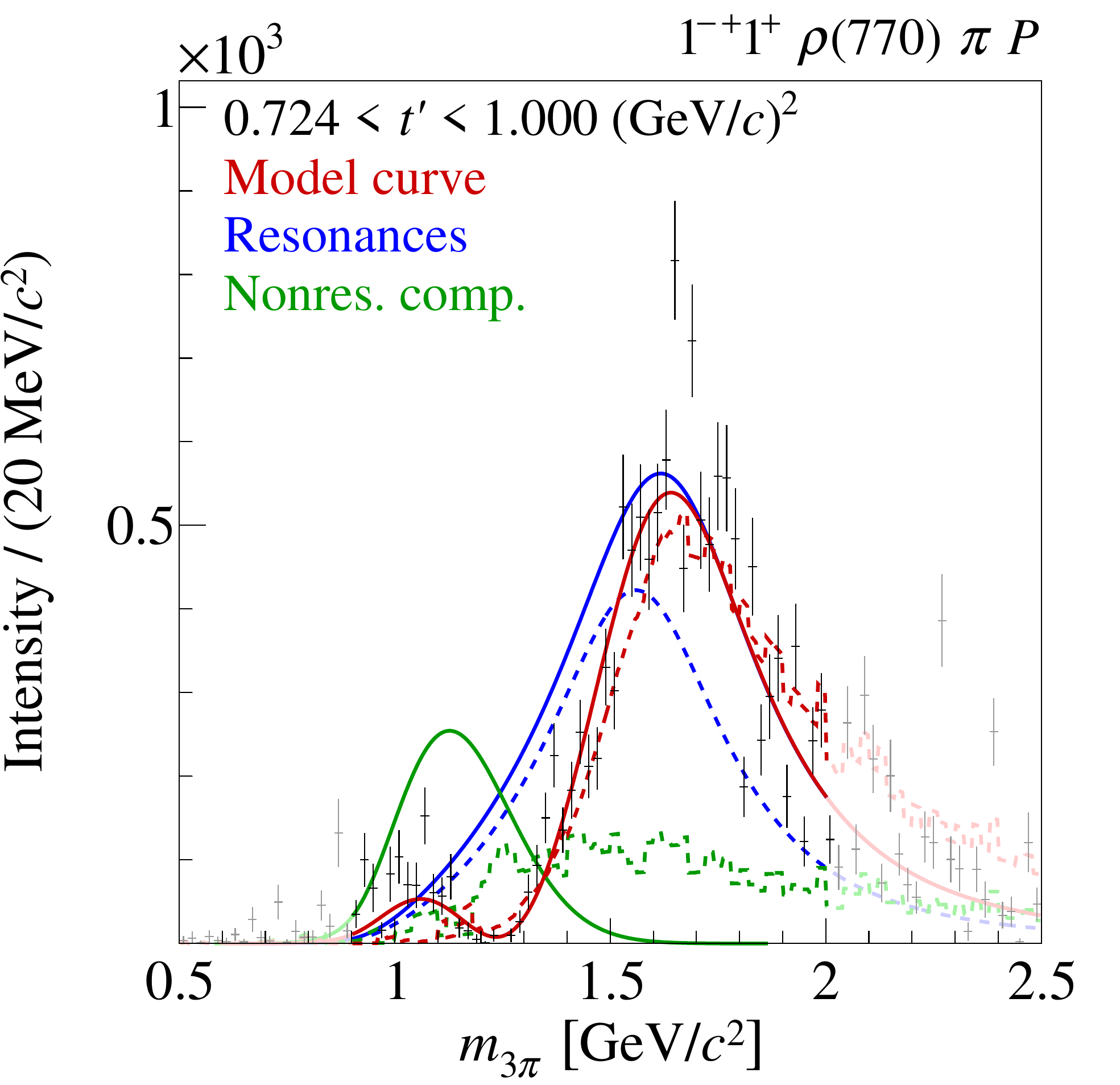}%
    \label{fig:intensity_1mp_tbin11_DeckMC}%
  }%
  \hspace*{\twoPlotSpacing}%
  \subfloat[][]{%
    \includegraphics[width=\twoPlotWidth]{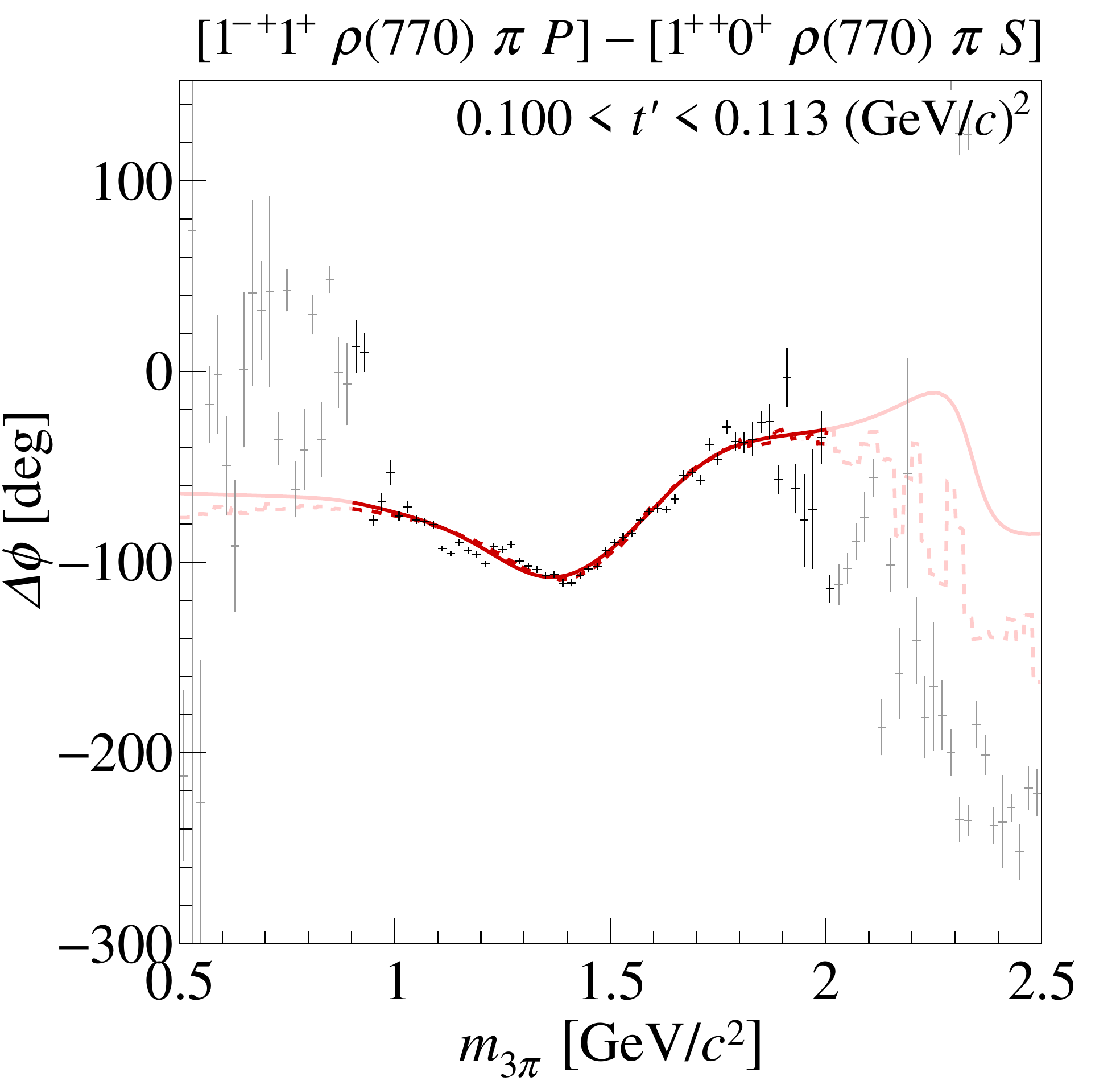}%
    \label{fig:phase_1mp_1pp_rho_tbin11_DeckMC}%
  }%
  \caption{\subfloatLabel{fig:intensity_1mp_DeckMC}~\tpr-summed
    intensity of the \wave{1}{-+}{1}{+}{\Prho}{P} wave.
    \subfloatLabel{fig:intensity_1mp_tbin1_DeckMC}~and~\subfloatLabel{fig:intensity_1mp_tbin11_DeckMC}:
    Intensity of this wave in the lowest and highest \tpr bins,
    respectively.
    \subfloatLabel{fig:phase_1mp_1pp_rho_tbin11_DeckMC}~Phase of the
    $1^{-+}$ wave relative to the \wave{1}{++}{0}{+}{\Prho}{S} wave in
    the highest \tpr bin.  The result of the main fit is represented
    by the continuous curves.  The fit, in which the parametrization
    of the nonresonant amplitude was replaced by the square root of
    the intensity distribution of the partial-wave decomposition of
    Deck Monte Carlo data [\StudyO; see \cref{sec:systematics}], is
    represented by the dashed curves.  The model and the wave
    components are represented as in \cref{fig:intensities_1mp}.}
  \label{fig:intensities_phase_1mp_DeckMC}
\end{wideFigureOrNot}

In the related \StudyK, we estimate the effect of an increased
background contamination on the fit result by using weaker
event-selection criteria.  The \tpr-summed \PpiOne[1600] yield remains
approximately unchanged while the strength of the nonresonant
component increases.\footnote{However, the \PpiOne[1600] parameters
  change.  It becomes \SI{46}{\MeVcc} heavier and \SI{130}{\MeVcc}
  narrower.}

Since at low \tpr the intensity distribution of the $1^{-+}$ wave
exhibits presumably artificial structures in the low-mass region, we
performed a study, in which the fit range for the $1^{-+}$ wave was
limited to \SIvalRange{1.4}{\mThreePi}{2.0}{\GeVcc}. In this study,
the mass of the \PpiOne[1600] increases by \SI{60}{\MeVcc} but remains
within the systematic uncertainty while the width remains unchanged.
A similar result is obtained in \StudyL, in which the analysis was
performed using only eight~\tpr bins, so that the subdivision of the
analyzed \tpr range into 11~bins seems to be sufficient to capture the
rapid change of the shape of the intensity distribution of the
$1^{-+}$ wave with \tpr.

We obtain slightly changed values for mass and width, \ie
$m_{\PpiOne[1600]} = \SI{1650}{\MeVcc}$ and
$\Gamma_{\PpiOne[1600]} = \SI{560}{\MeVcc}$, if we use a
mass-dependent width for the parametrization of the \PpiOne[1600]
analogous to \cref{eq:method:a2dynamicwidth} and assume that this
width is saturated by the $\Prho \pi P$-wave decay mode.

\subsubsection{Discussion of results on $1^{-+}$ resonances}
\label{sec:oneMP_discussion}

The results of previous experiments on the existence of a
\PpiOne[1600] signal in the $3\pi$ final state are contradictory.  On
the one hand, the BNL~E852 experiment, which analyzed pion diffraction
at \SI{18}{\GeVc} beam momentum, claimed a \PpiOne[1600] signal in the
$\Prho \pi$ decay mode~\cite{adams:1998ff,chung:2002pu}.  On the other
hand, the authors of \refCite{dzierba:2005jg} concluded that the peak
structure in the \wave{1}{-+}{1}{+}{\Prho}{P} wave that was reported
in \refsCite{adams:1998ff,chung:2002pu} was due to leakage caused by a
too small wave set and that they do not observe a significant
\PpiOne[1600] signal in the $\Prho \pi$ channel.  This conclusion was
based on a partial-wave analysis of a much larger $3\pi$ data set also
from the BNL~E852 experiment in the kinematic range
\SIvalRange{0.08}{\tpr}{0.53}{\GeVcsq} using an extended wave set.
However, a \PpiOne[1600] signal was observed in a combined analysis of
$\eta' \pi$, $\PbOne \pi$, and $\Prho \pi$ final states from pion
diffraction at \SI{36.6}{\GeVc} beam momentum by the
VES~experiment~\cite{zaitsev:2000rc,Khokhlov:2000tk}.  No
\PpiOne[1600] signal was found by the CLAS experiment in \threePi
photoproduction~\cite{Nozar:2008aa,Eugenio:2013xua}.

We have studied the significance of the \PpiOne[1600] signal by
performing a fit, in which we omitted the \PpiOne[1600] component from
the model.  Hence in this fit, the \wave{1}{-+}{1}{+}{\Prho}{P} wave
is described solely by the nonresonant component.  The minimum
\chisq~value of this fit is \num{1.17} times larger than that of the
main fit.\footnote{Compared to the \num{722} free parameters of the
  main fit, this fit has \num{698} free parameters.}
\Cref{fig:no-pi1(1600)_chi2difference} shows the contributions from
the spin-density matrix elements to the \chisq~difference between this
and the main fit.  In particular, the intensity of the $1^{-+}$ wave
and its phase relative to the \wave{1}{++}{0}{+}{\Prho}{S} wave are
described less well by the model without the \PpiOne[1600] (see
\cref{fig:intensities_phase_no-pi1(1600)}).  The disagreement is
largest in the two highest \tpr bins where the model cannot describe
the data.  However, at lower \tpr the nonresonant component is
sufficient to describe the basic features of the data.\footnote{The
  omission of the \PpiOne[1600] also affects some of the resonance
  parameters in the fit.  Most striking is the impact on the
  description of the intensity distribution of the
  \wave{1}{++}{0}{+}{\Prho}{S} wave.  Although this wave has a
  relative intensity that is about 40~times larger than that of the
  $1^{-+}$ wave and although the \PaOne and the \PpiOne[1600] have a
  mass difference of about \SI{300}{\MeVcc}, the \PaOne becomes
  \SI{56}{\MeVcc} heavier and \SI{78}{\MeVcc} narrower if the
  \PpiOne[1600] is omitted from the model.  Also the \PaOne[1640] and
  \PaTwo[1700] parameters change substantially.  The \PaOne[1640]
  becomes \SI{92}{\MeVcc} heavier and \SI{26}{\MeVcc} wider; the
  \PaTwo[1700] becomes \SI{28}{\MeVcc} heavier and \SI{60}{\MeVcc}
  wider.}  Furthermore, we performed a fit with a model that describes
the $1^{-+}$ amplitude using two independent coherent nonresonant
contributions but no \PpiOne[1600].  Also this fit does not yield a
satisfactory description of the data.  Based upon the items discussed
above, we conclude that the significance of the \PpiOne[1600] signal
is strongly \tpr dependent.  At \tpr below about \SI{0.5}{\GeVcsq},
there is only weak evidence for the \PpiOne[1600].  This is consistent
with the nonobservation of the \PpiOne[1600] in the BNL~E852 data in
the kinematic range $\tpr < \SI{0.53}{\GeVcsq}$~\cite{dzierba:2005jg},
as discussed above.  However, our data show that a resonancelike
signal is required to describe the data in the \tpr region above about
\SI{0.5}{\GeVcsq}, which was not analyzed in \refCite{dzierba:2005jg}.

\begin{figure}[tbp]
  \centering
  \includegraphics[width=\linewidthOr{\twoPlotWidth}]{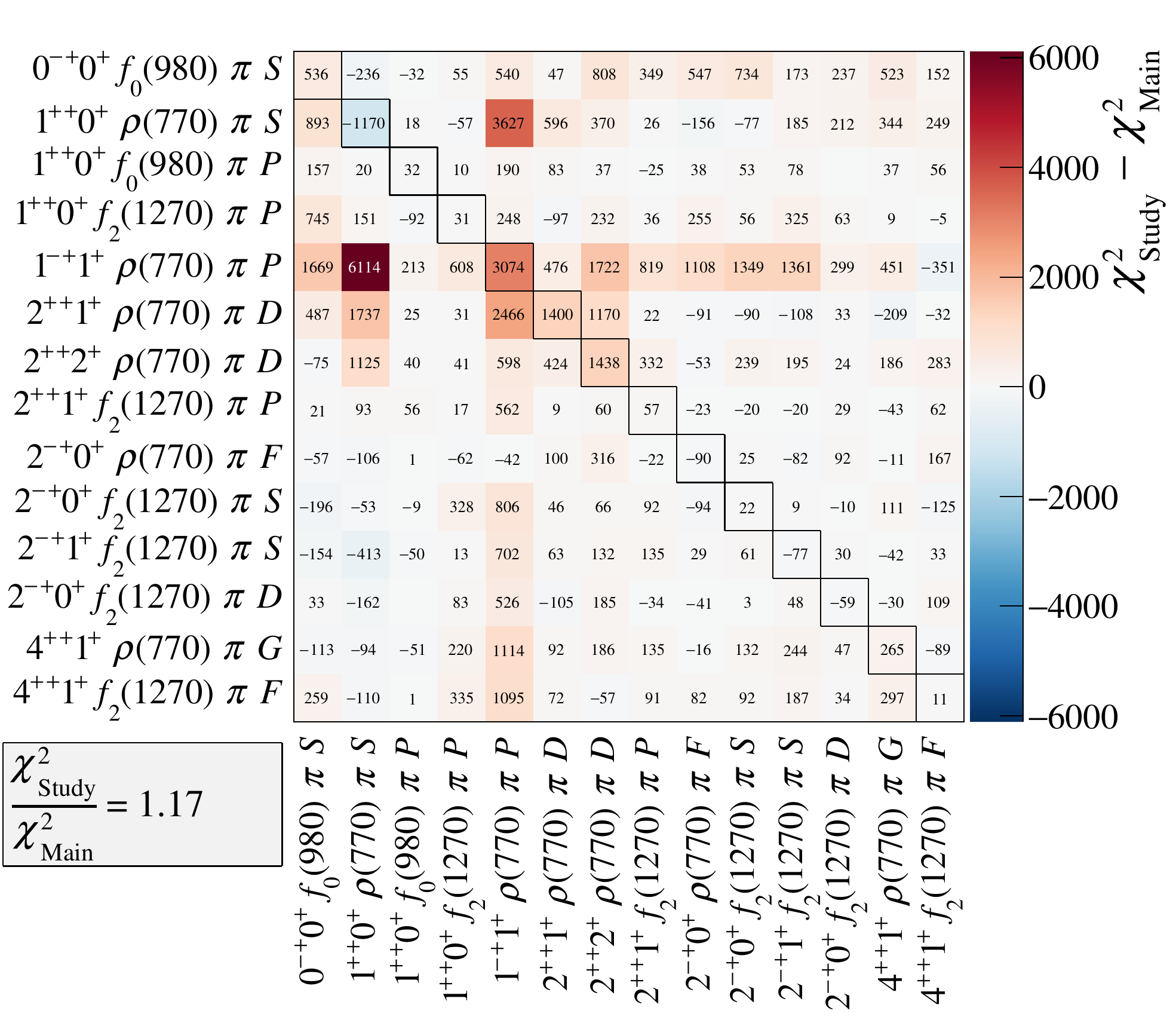}
  \caption{Similar to \cref{fig:DeckMC_chi2difference}, but for the
    study, in which the \PpiOne[1600] resonance was omitted from the
    fit model.}
  \label{fig:no-pi1(1600)_chi2difference}
\end{figure}

\begin{wideFigureOrNot}[tbp]
  \centering
  \subfloat[][]{%
    \includegraphics[width=\twoPlotWidth]{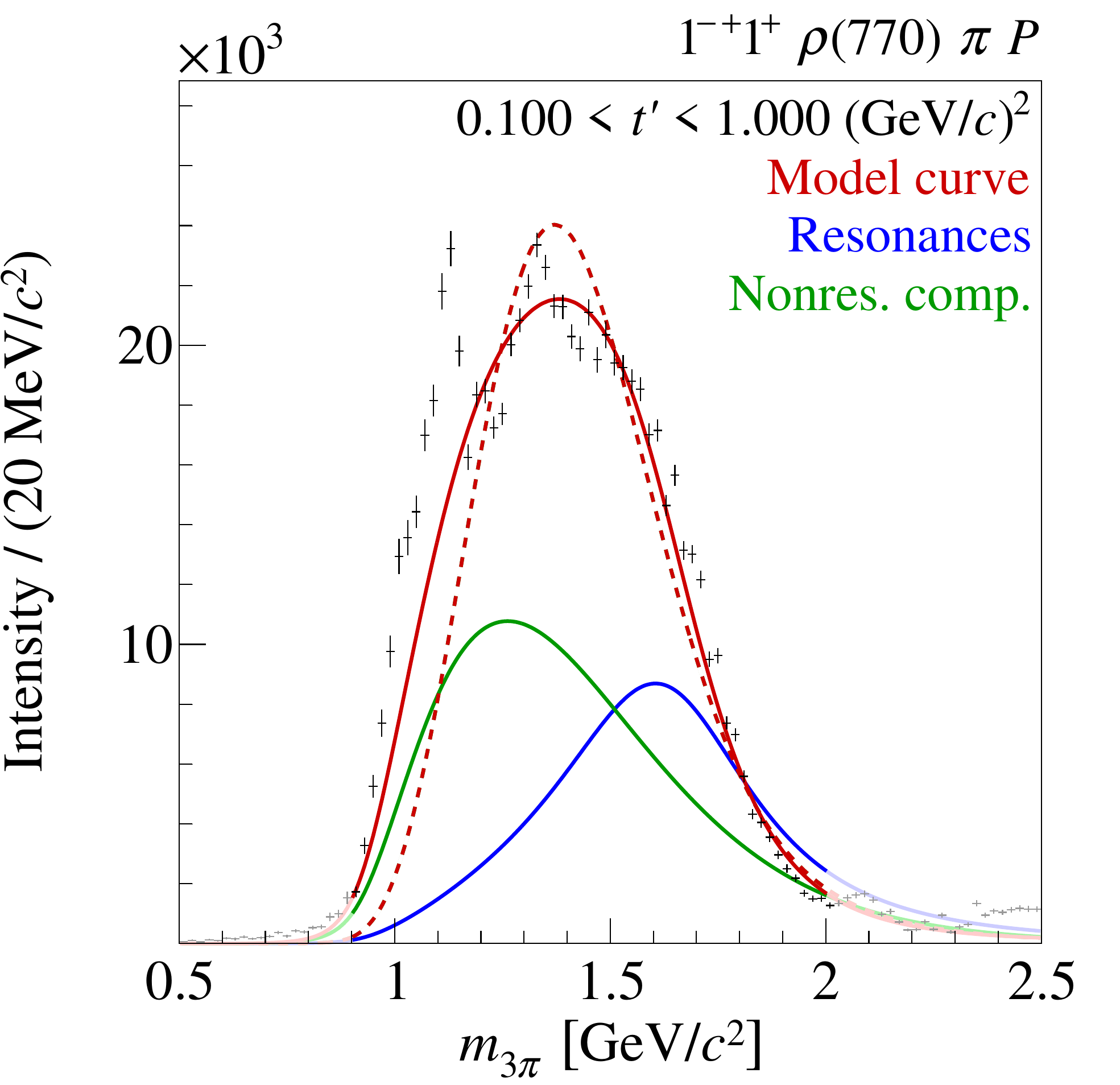}%
    \label{fig:intensity_1mp_no-pi1(1600)}%
  }%
  \hspace*{\twoPlotSpacing}%
  \subfloat[][]{%
    \includegraphics[width=\twoPlotWidth]{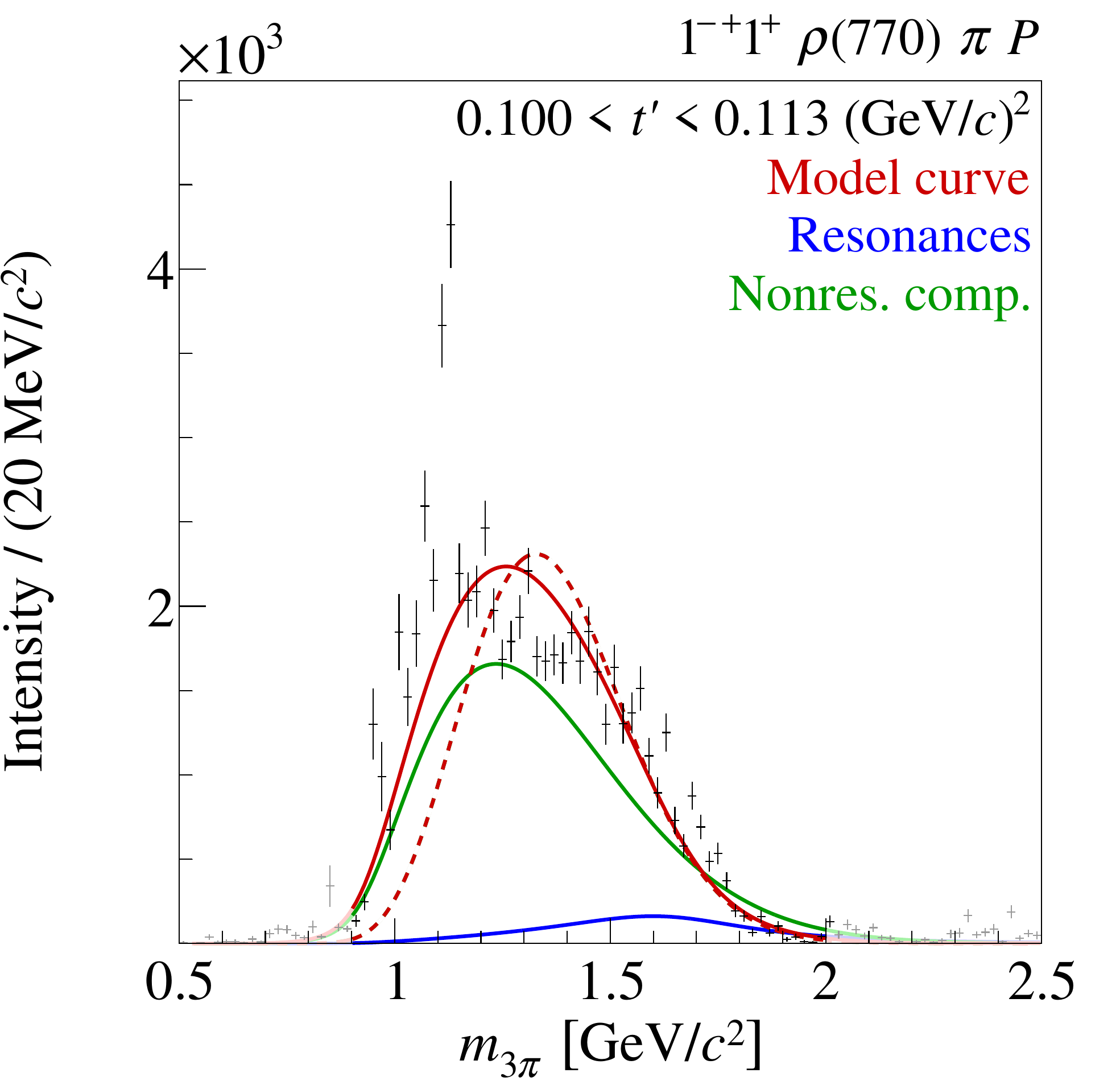}%
    \label{fig:intensity_1mp_tbin1_no-pi1(1600)}%
  }%
  \\
  \subfloat[][]{%
    \includegraphics[width=\twoPlotWidth]{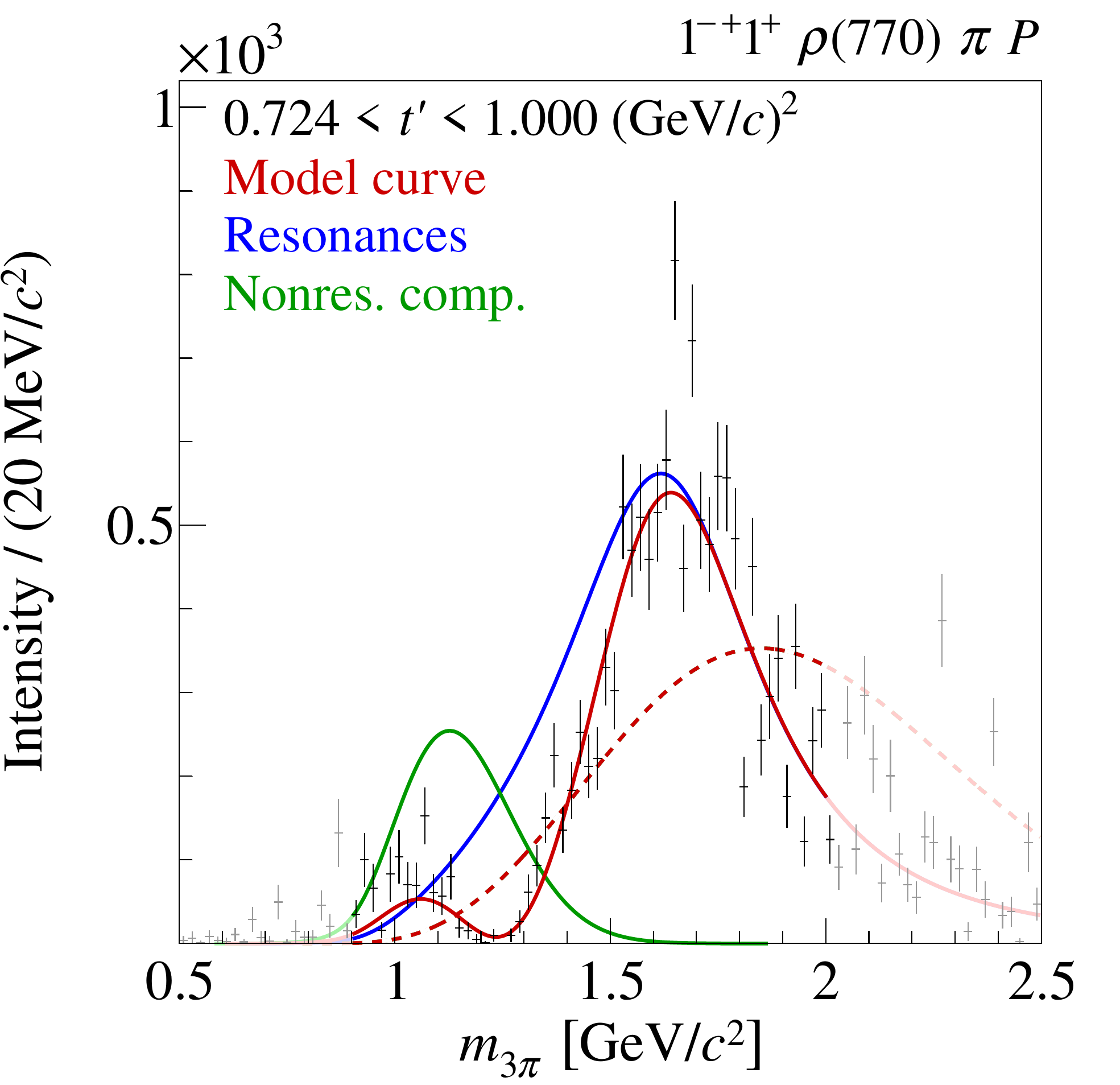}%
    \label{fig:intensity_1mp_tbin11_no-pi1(1600)}%
  }%
  \hspace*{\twoPlotSpacing}%
  \subfloat[][]{%
    \includegraphics[width=\twoPlotWidth]{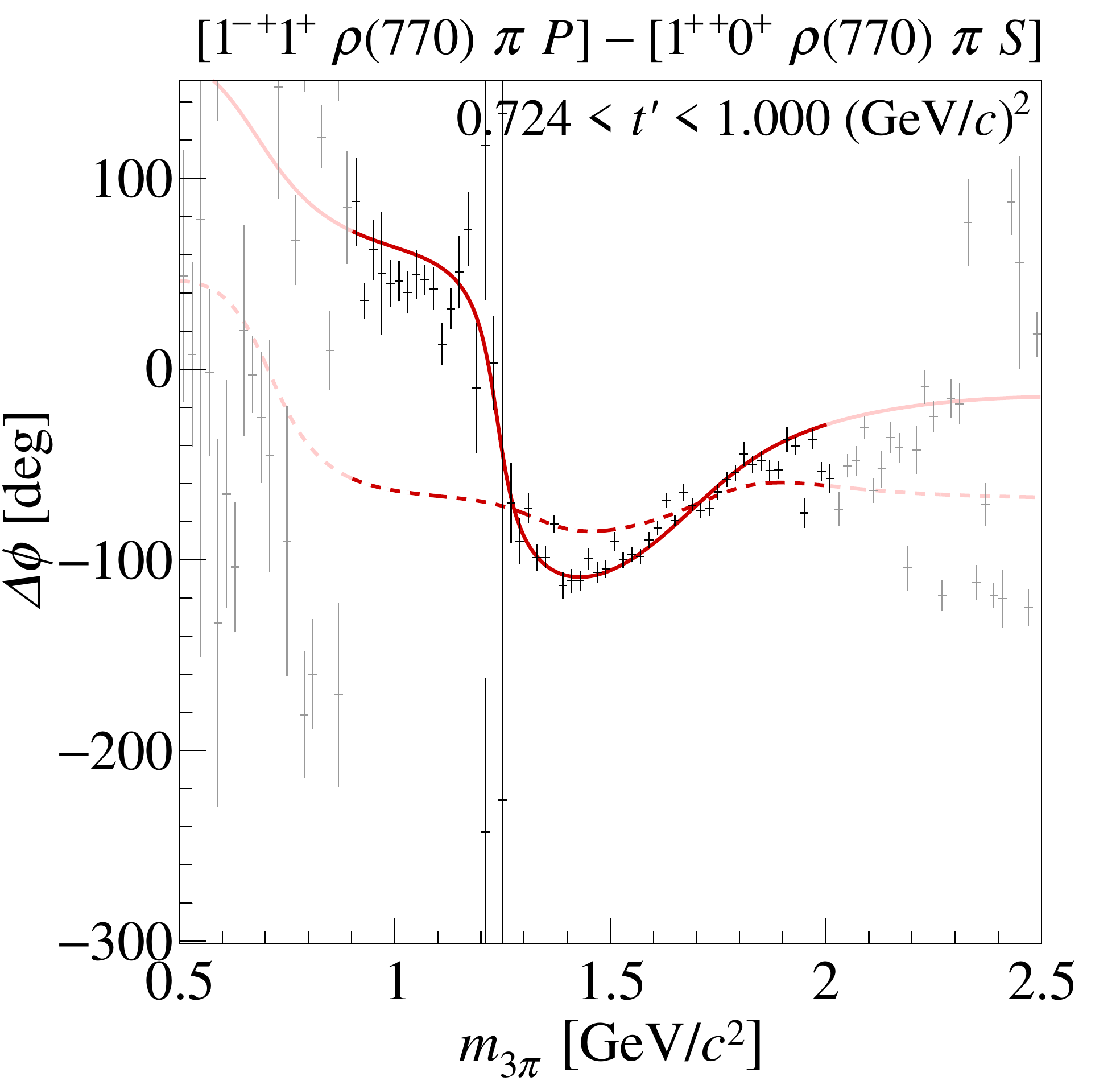}%
    \label{fig:phase_1mp_1pp_rho_tbin11_no-pi1(1600)}%
  }%
  \caption{\subfloatLabel{fig:intensity_1mp_no-pi1(1600)}~\tpr-summed
    Intensity of the \wave{1}{-+}{1}{+}{\Prho}{P} wave.
    \subfloatLabel{fig:intensity_1mp_tbin1_no-pi1(1600)}~and~\subfloatLabel{fig:intensity_1mp_tbin11_no-pi1(1600)}:
    intensity of this wave in the lowest and highest \tpr bins,
    respectively.
    \subfloatLabel{fig:phase_1mp_1pp_rho_tbin11_no-pi1(1600)}~Phase of
    the $1^{-+}$ wave relative to the \wave{1}{++}{0}{+}{\Prho}{S}
    wave in the highest \tpr bin.  The result of the main fit is
    represented by the continuous curves.  The fit, in which the
    \PpiOne[1600] resonance was omitted from the fit model, is
    represented by the dashed curves.  These curves correspond to the
    nonresonant component.  The model and the wave components are
    represented as in \cref{fig:intensities_1mp}.}
  \label{fig:intensities_phase_no-pi1(1600)}
\end{wideFigureOrNot}

The \PpiOne[1600] is considered by the PDG to be an established state.
It was seen by the BNL~E852 and VES experiments in diffractively
produced $\eta' \pi$~\cite{beladidze:1993km,Ivanov:2001rv},
$\eta \pi^+ \pi^- \pi^-$~\cite{kuhn:2004en,amelin:2005ry}, and
$\omega \pi^- \pi^0$~\cite{Lu:2004yn,amelin:2005ry} final states.
Evidence for the \PpiOne[1600] was also found in an analysis of
$\ppbar \to \omega \pi^+ \pi^- \pi^0$ Crystal Barrel
data~\cite{baker:2003jh} and in $\chi_{c1} \to \eta' \pi^+ \pi^-$
decays by the CLEO-c experiment~\cite{Adams:2011sq}.

The PDG world averages for mass and width of the \PpiOne[1600] are
$m_{\PpiOne[1600]} = \SIaerr{1662}{8}{9}{\MeVcc}$ and
$\Gamma_{\PpiOne[1600]} = \SI{241(40)}{\MeVcc}$,
respectively~\cite{Patrignani:2016xqp}.  Our measured \PpiOne[1600]
mass of $m_{\PpiOne[1600]} = \SIaerrSys{1600}{110}{60}{\MeVcc}$ is
consistent with the world average within the large systematic
uncertainties; however, our measured \PpiOne[1600] width of
$\Gamma_{\PpiOne[1600]} = \SIaerrSys{580}{100}{230}{\MeVcc}$ is
larger.  This discrepancy is mainly due to the extremely small width
value of $\Gamma_{\PpiOne[1600]} = \SIerrs{185}{25}{28}{\MeVcc}$
quoted by the BNL~E852 experiment for the $\omega \pi^- \pi^0$ final
state~\cite{Lu:2004yn}.  The present width is also larger than our
previously published one from an analysis of the same process on a
solid-lead target~\cite{alekseev:2009aa}.  Due to the approximately
2~orders of magnitude smaller data sample, the analysis in
\refCite{alekseev:2009aa} was performed by integrating over the \tpr
range from \SIrange{0.1}{1.0}{\GeVcsq} and by assuming a model for the
\tpr dependence of the partial-wave amplitudes.  Therefore, the \tpr
dependence of the shape of the \wave{1}{-+}{1}{+}{\Prho}{P} amplitude
was not taken into account.  It is remarkable that in the lead-target
data, the contribution of the nonresonant component is much smaller
than that in the proton-target data so that the \tpr-integrated
lead-target data resemble the high-\tpr region of the proton-target
data.

The PDG summary table lists the \PpiOne[1400] as an additional
$\JPC = 1^{-+}$ resonance.  This state was observed by several
experiments in the $\eta \pi$ final
state~\cite{Alde:1988bv,Aoyagi:1993kn,Thompson:1997bs,Chung:1999we,Adams:2006sa,Dorofeev:2001xu,Abele:1998gn,Abele:1999tf}.
In the $\Prho \pi$ channel, it was only observed by the Obelix
experiment~\cite{salvini:2004gz}.  We do not see any clear resonance
signal below \SI{1.5}{\GeVcc} in the \wave{1}{-+}{1}{+}{\Prho}{P}
wave.  Aside from the presumably artificial narrow structure at
\SI{1.1}{\GeVcc}, the description of the intensities and phases by our
model leaves little room for a possible \PpiOne[1400] component in the
$\Prho \pi P$ wave.

The BNL~E852 experiment also reported a heavy spin-exotic state, \ie
the \PpiOne[2015], in the $\PfOne \pi$~\cite{kuhn:2004en} and
$\PbOne \pi$~\cite{Lu:2004yn} decay modes.  We do not see any clear
resonance signal of a heavy \PpiOne* state in the mass range from
\SIrange{1900}{2500}{\MeVcc} in the $\Prho \pi P$ wave.  However, we
cannot exclude that some of the observed deviations of the model from
the data at high masses are due to an additional excited \PpiOne*
state.
 %
%
%

\section{Results on \tpr dependence of relative phases of coupling amplitudes}
\label{sec:production_phases}

As discussed in \cref{sec:method}, our fit model in
\cref{eq:method:param:spindens} contains coupling amplitudes
$\mathcal{C}_a^j(\tpr)$ for each wave component~$j$ in partial
wave~$a$, in addition to the shape parameters of the resonant and
nonresonant components.  The coupling amplitudes in the 11~\tpr bins
are independent parameters of the model, which are determined by the
fit.  In order to reduce the number of these fit parameters, the
coupling amplitudes of resonance components that appear in waves with
the same \JPCMrefl quantum numbers but different decay modes are
constrained to have the same \tpr dependence via
\cref{eq:method:branchingdefinition}.  In \cref{sec:results}, we
already discussed the \tpr-dependent yields of the resonant and
nonresonant components as given by \cref{eq:tprim-dependence}.  Most
of these \tpr spectra approximately follow the simple model in
\cref{eq:slope-parametrization}.

In this section, we discuss the \tpr dependence of the relative phases
between the coupling amplitudes of wave component~$j$ in wave~$a$ and
of wave component~$k$ in wave~$b$,
\begin{equation}
  \label{eq:coupling_phase}
  \Delta \phi_\text{coupl.}^{j, a; k, b}(\tpr)
  \equiv \arg\!\sBrk{\mathcal{C}_a^j(\tpr)\, \mathcal{C}_b^{k \text{*}}(\tpr)}.
\end{equation}
In the text below, we refer to these relative phases as \emph{coupling
  phases}.  Coupling amplitudes of the same resonance in different
decay channels, which are constrained via
\cref{eq:method:branchingdefinition}, have \tpr-independent relative
coupling phases that correspond to
$\arg \sBrk[1]{\prescript{}{b}{\mathcal{B}}_a^j}$.

As the coupling amplitude of a particular wave component is the
product of the actual production amplitude of this wave component and
the complex-valued couplings, $\alpha_{X \to \xi \pi}$ and
$\alpha_{\xi \to \pi \pi}$, which appear in its decay via the
isobar~$\xi^0$ (see \cref{sec:method}), the physical interpretation of
the coupling phase is not straightforward.  Assuming that a single
production mechanism dominates, we would expect the coupling phases of
resonances to be approximately independent of \tpr.  These phases may
be altered by effects from final-state interactions.

Our fit model assumes that resonances are described by Breit-Wigner
amplitudes and that they have the same masses and widths in different
waves and in all \tpr bins.  In contrast, the shape of the nonresonant
components can be adapted individually for each wave by the fit.  For
some waves, we allow the shape of the nonresonant component to change
with \tpr (see \cref{tab:method:fitmodel:waveset}).  In addition, the
fit has the freedom to choose the relative strengths and phases for
the different components within a single wave and the relative
strengths and phases between different waves.  The imperfections in
our model, in particular concerning the parametrization of the
nonresonant components (see \cref{sec:results}), might cause offsets
in the relative phases, which may even be uncorrelated across \tpr
bins.  Considering these possible artifacts, we consider small phase
differences up to \SI{20}{\degree} as insignificant for the physical
interpretation.

The discussion of coupling phases will focus mostly on the resonance
components.  In \cref{fig:tprim_phase_all_res}, we show the \tpr
dependence of the coupling phases of the 11~resonance components in
the dominant wave of the respective \JPC sector relative to the
\PpiTwo in the \wave{2}{-+}{0}{+}{\PfTwo}{S} wave.  The dominant waves
are characterized by a large contribution from the respective
ground-state resonance, while the contributions from higher excited
states are substantially smaller.  Since the \PaOne[1420] does not
appear in the dominant $1^{++}$ wave, its coupling phase is shown for
the \wave{1}{++}{0}{+}{\PfZero[980]}{P} wave.  In
\cref{fig:tprim_phase_all_res}, we have chosen the \PpiTwo as the
reference component because it turned out to be relatively stable in
our systematic studies.  The coupling phases of all resonances show a
smooth variation as a function of \tpr.  Since the model does not
contain any assumptions on the \tpr behavior of the coupling phases,
the observed continuous behavior is a nontrivial result, which
supports our analysis model.  We observe a similar behavior for the
coupling phases of the nonresonant components, although the variation
with \tpr is typically larger (see
\crefrange{fig:tprim_phase_2mp}{fig:tprim_phase_4pp} in
\cref{sec:phases_piJ,sec:phases_aJ} below).

\begin{figure}[tbp]
  \centering
  \includegraphics[width=\linewidthOr{1.22125\twoPlotWidth}]{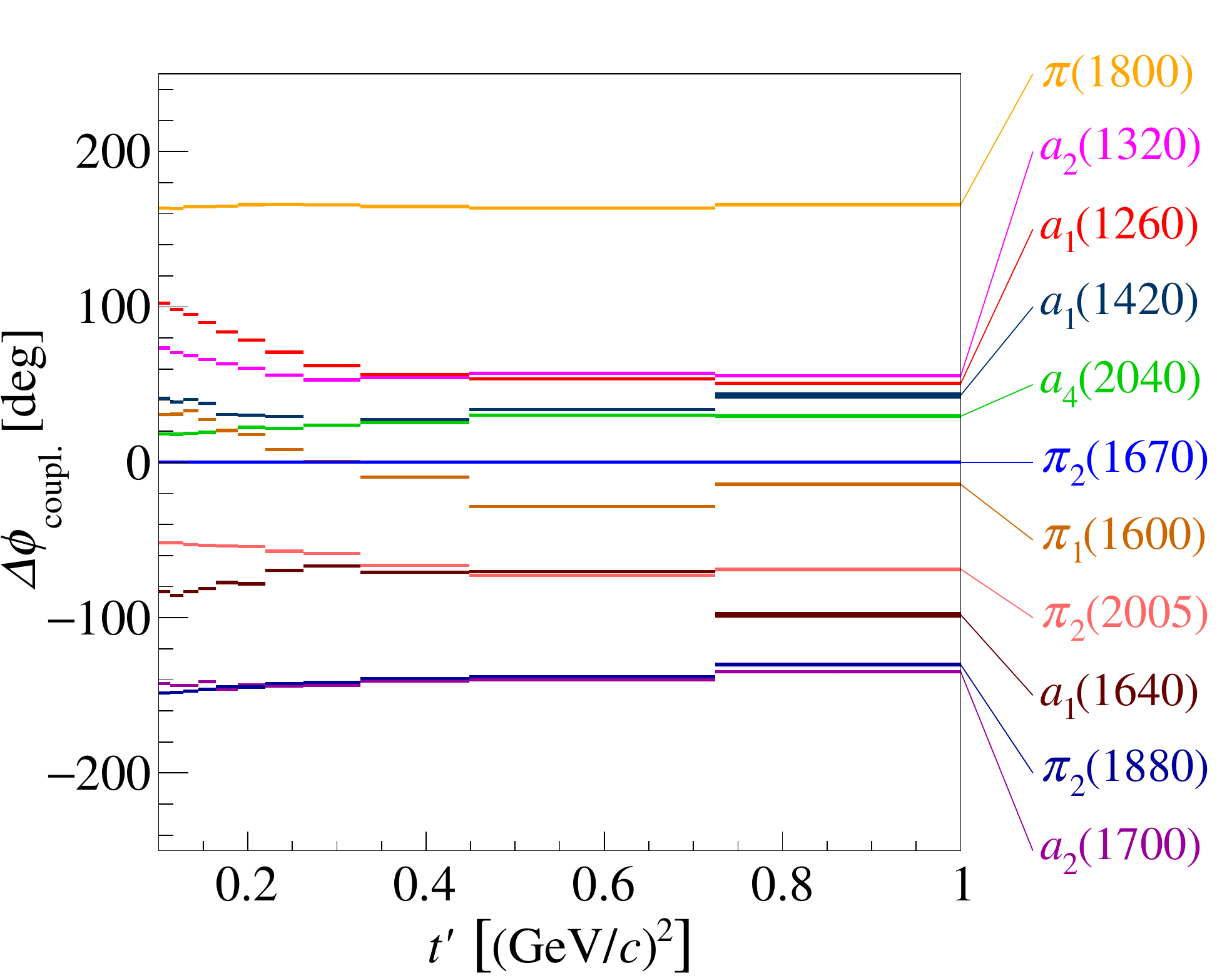}
  \caption{\tpr dependence of the relative phases
    $\Delta \phi_\text{coupl.}$ of the coupling amplitudes [see
    \cref{eq:coupling_phase}] of the 11~resonance components in the
    fit model \wrt the \PpiTwo.  The coupling phases are shown for the
    dominant wave of the respective \JPC sector:
    \wave{0}{-+}{0}{+}{\PfZero}{S}, \wave{1}{++}{0}{+}{\Prho}{S},
    \wave{1}{-+}{1}{+}{\Prho}{P}, \wave{2}{++}{1}{+}{\Prho}{D},
    \wave{2}{-+}{0}{+}{\PfTwo}{S}, and \wave{4}{++}{1}{+}{\Prho}{G}.
    The only exception is the \PaOne[1420], which appears only in the
    \wave{1}{++}{0}{+}{\PfZero}{P} wave.  The width of the horizontal
    lines represents the statistical uncertainty. The systematic
    uncertainty is not shown.}
  \label{fig:tprim_phase_all_res}
\end{figure}

The coupling phases of the resonance components exhibit three striking
features in their \tpr dependence: \one~for most resonances, we find
for $\tpr \lesssim \SI{0.3}{\GeVcsq}$ a slow change of the coupling
phases with \tpr, whereas for $\tpr \gtrsim \SI{0.3}{\GeVcsq}$ the
phases level off; \two~with the exception of the \PaOne[1420], the
coupling phases of different states with the same \JPC show large
relative offsets in the highest \tpr bin; and \three~the coupling
phases of the ground-state resonances do not deviate by more than
\SI{\pm 60}{\degree} from the phase of the \PpiTwo in the highest \tpr
bin.  In particular the nearly constant phases of all resonances for
$\tpr \gtrsim \SI{0.3}{\GeVcsq}$ are remarkable and appear to be
characteristic of resonances.  This behavior is consistent with a
common production mechanism for the resonances.

\subsection{Relative phases of the coupling amplitudes of the \piJ resonances}
\label{sec:phases_piJ}

The \tpr dependence of the coupling phases of the $2^{-+}$ wave
components is shown in \cref{fig:tprim_phase_2mp} relative to the
\PpiTwo in the \wave{2}{-+}{0}{+}{\PfTwo}{S} wave.  The \tpr
dependence of the coupling phases of the resonance components in the
three $2^{-+}$ waves with $M = 0$ is constrained via
\cref{eq:method:branchingdefinition}.  Therefore, in these waves the
coupling phases of the resonances follow the same \tpr dependence but
may have relative offsets, which correspond to the phase of the
branching amplitudes $\prescript{}{b}{\mathcal{B}}_a^j$.  As for the
resonance parameters and the \tpr spectra that were discussed in
\crefrange{sec:zeroMP}{sec:oneMP}, the uncertainties of the coupling
phases are dominated by systematic effects; \ie statistical
uncertainties are negligible in comparison.  The \tpr dependence of
the coupling phases differs in the various systematic studies (see
\cref{sec:systematics}).  In order to illustrate the magnitude of the
systematic effects at least qualitatively, we show in
\cref{fig:tprim_phase_2mp} for each wave component in addition to the
continuous lines, which represent the result of the main fit, two sets
of dashed lines.  They represent the results of the two systematic
studies that in the highest \tpr bin have the largest deviation from
the coupling phase of the main fit.  In order to guide the eye, the
region between the two sets of dashed lines is shaded.\footnote{Note
  that the shaded areas defined in this way cannot be interpreted as
  systematic uncertainties.}

\begin{wideFigureOrNot}[tbp]
  \centering
  \subfloat[][]{%
    \includegraphics[width=\twoPlotWidth]{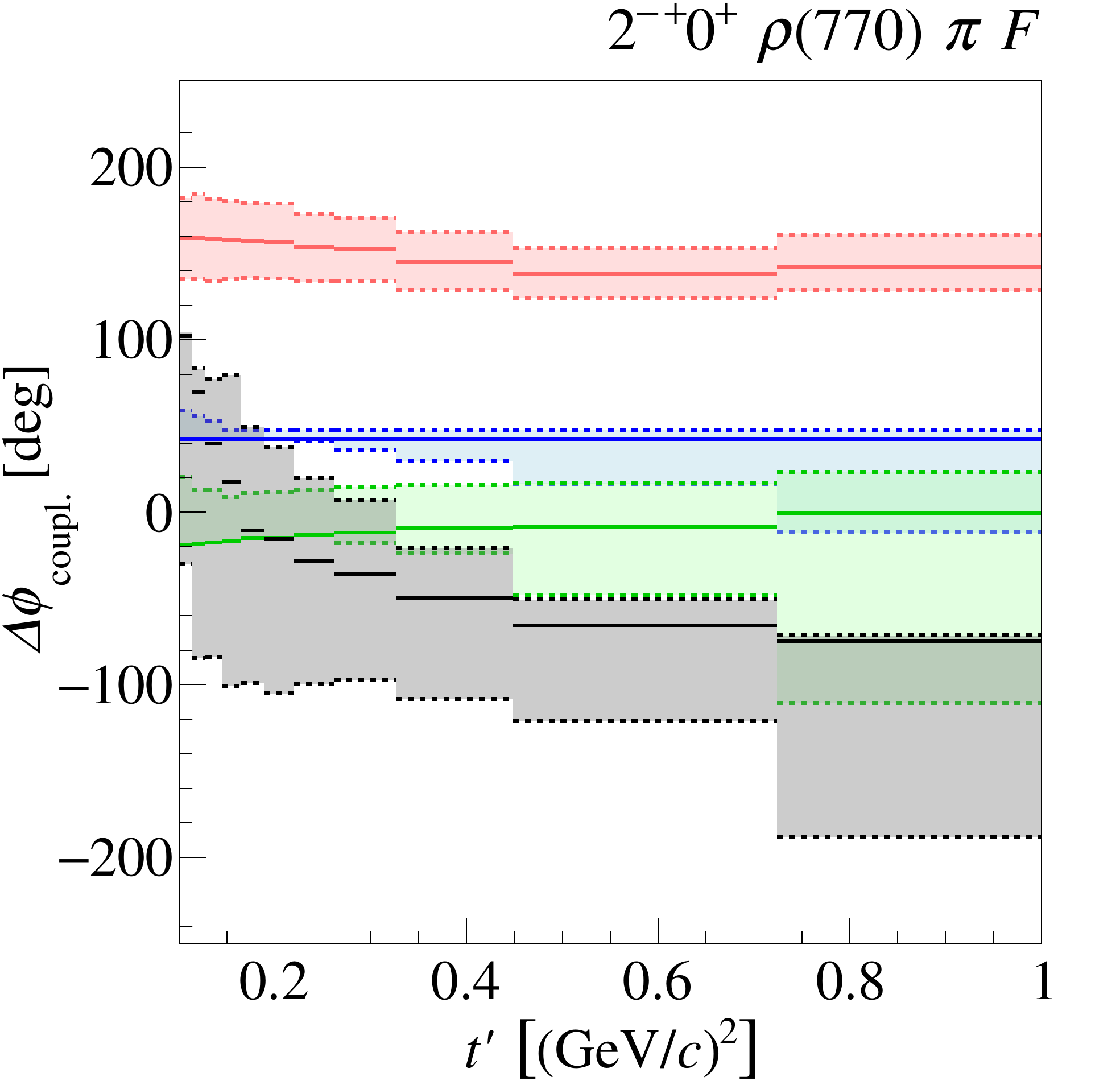}%
    \label{fig:tprim_phase_2mp_rho}%
  }%
  \hspace*{\twoPlotSpacing}%
  \subfloat[][]{%
    \includegraphics[width=\twoPlotWidth]{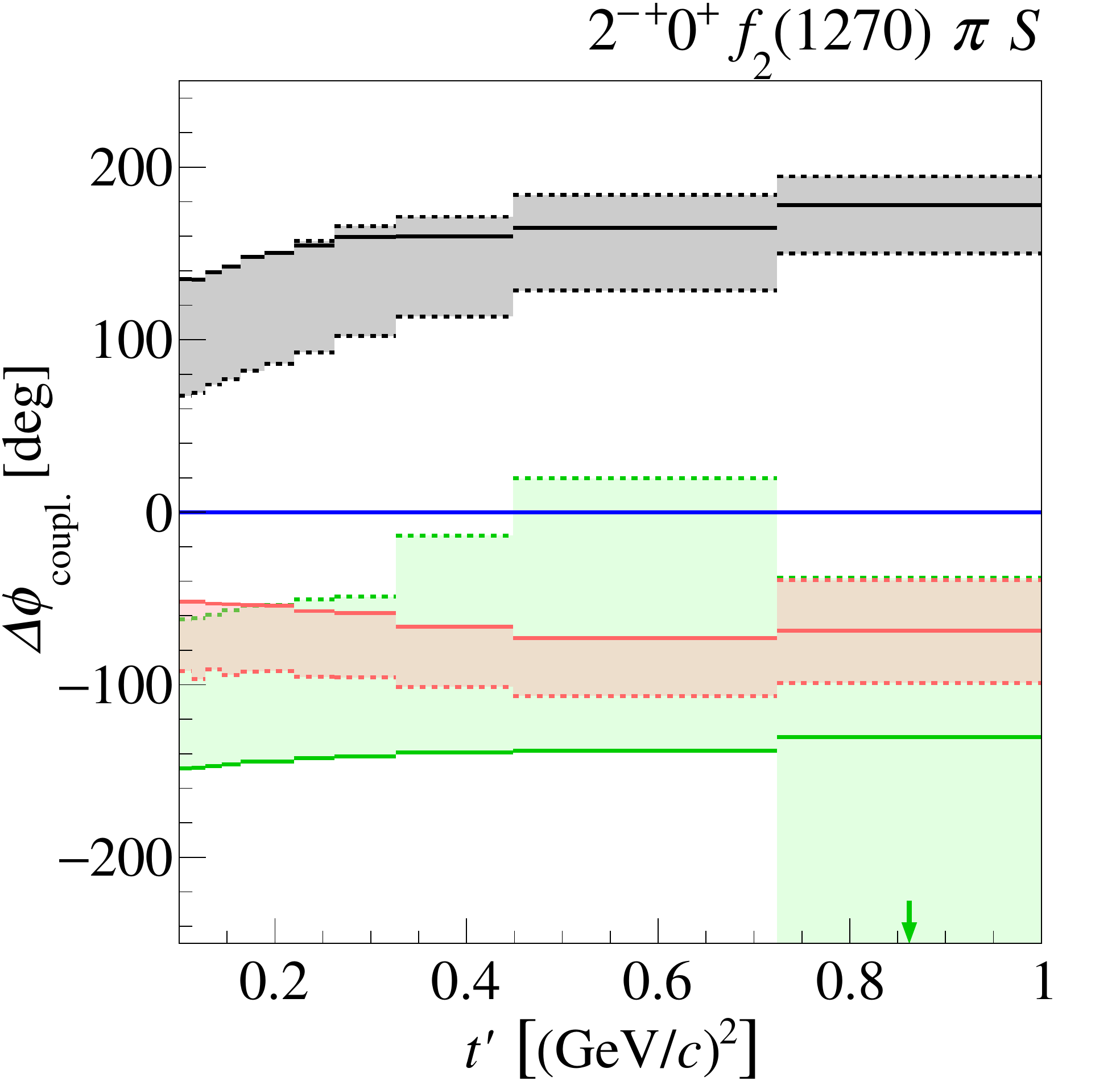}%
    \label{fig:tprim_phase_2mp_m0_f2_S}%
  }%
  \\
  \subfloat[][]{%
    \includegraphics[width=\twoPlotWidth]{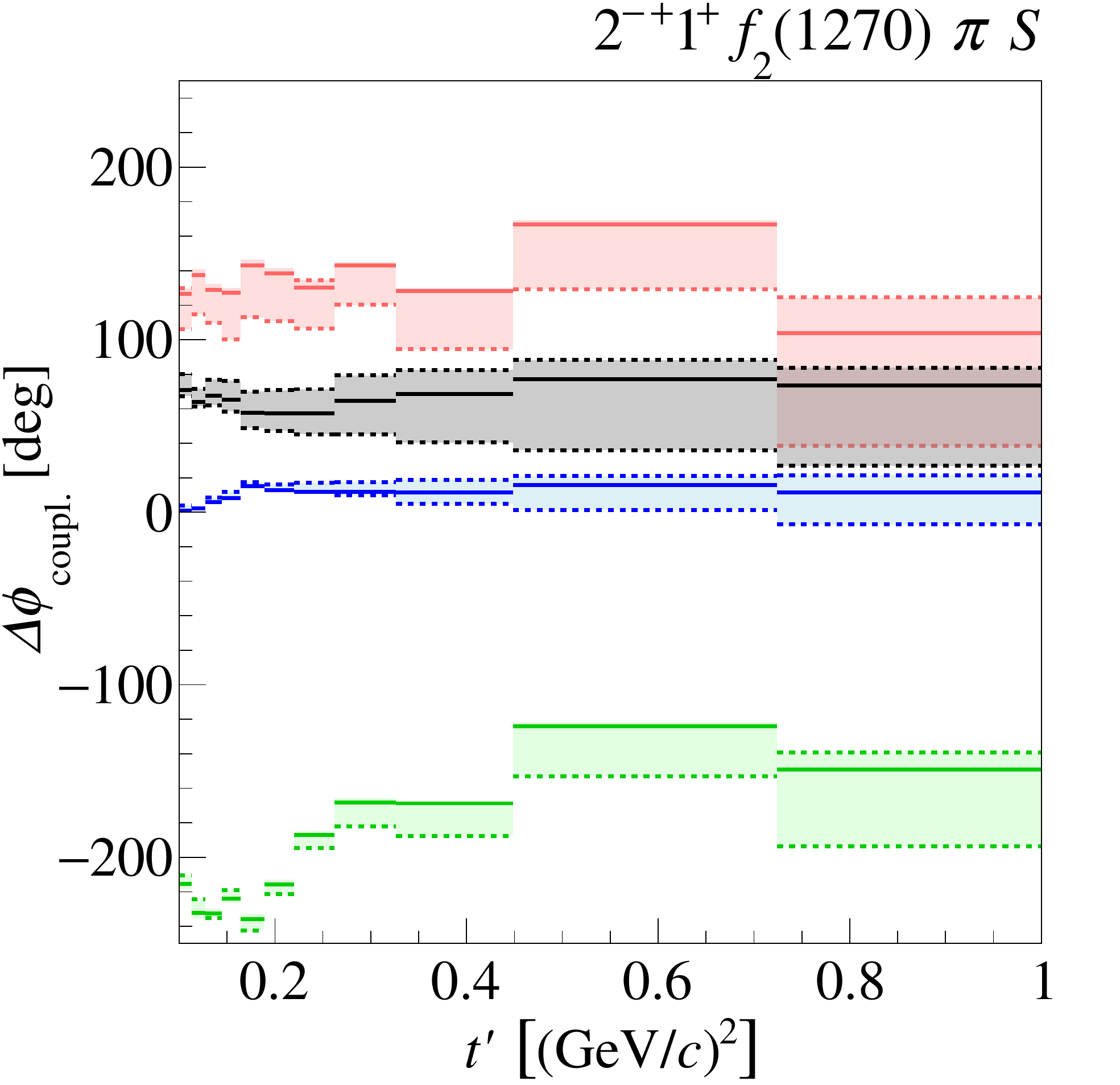}%
    \label{fig:tprim_phase_2mp_m1_f2_S}%
  }%
  \hspace*{\twoPlotSpacing}%
  \subfloat[][]{%
    \includegraphics[width=\twoPlotWidth]{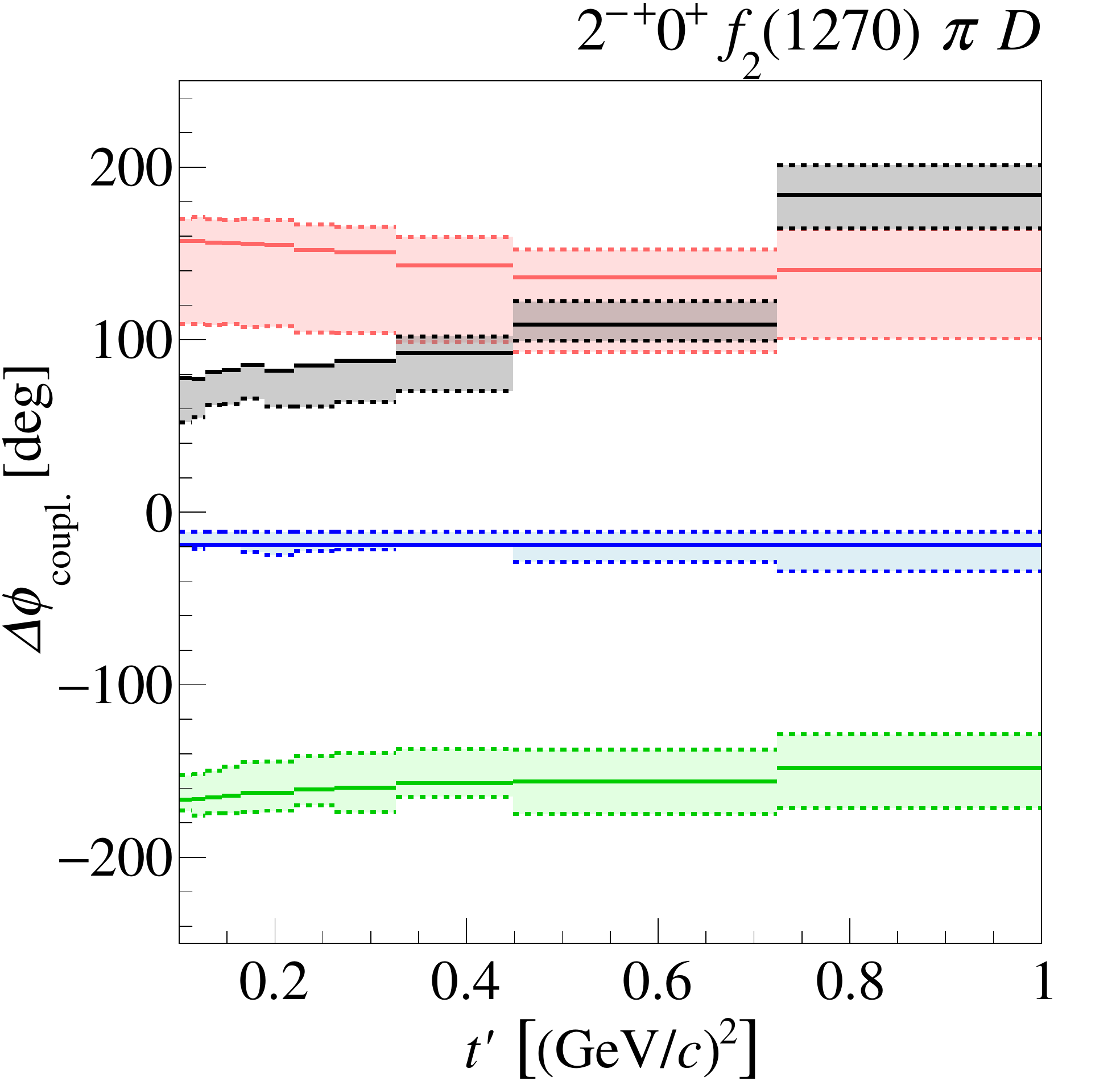}%
    \label{fig:tprim_phase_2mp_f2_D}%
  }%
  \caption{\tpr dependence of the coupling phases of the wave
    components in \subfloatLabel{fig:tprim_phase_2mp_rho}~the
    \wave{2}{-+}{0}{+}{\Prho}{F},
    \subfloatLabel{fig:tprim_phase_2mp_m0_f2_S}~the
    \wave{2}{-+}{0}{+}{\PfTwo}{S},
    \subfloatLabel{fig:tprim_phase_2mp_m1_f2_S}~the
    \wave{2}{-+}{1}{+}{\PfTwo}{S}, and
    \subfloatLabel{fig:tprim_phase_2mp_f2_D}~the
    \wave{2}{-+}{0}{+}{\PfTwo}{D} wave.  The coupling phases of the
    \PpiTwo (blue lines), the \PpiTwo[1880] (green lines), the
    \PpiTwo[2005] (red lines), and the nonresonant components (black
    lines) are shown relative to the \PpiTwo in the
    \wave{2}{-+}{0}{+}{\PfTwo}{S} wave.  For each wave component, the
    magnitude of the effects observed in the systematic studies (see
    \cref{sec:systematics}) is illustrated qualitatively by two sets
    of dashed lines with shaded area in between (see text).}
  \label{fig:tprim_phase_2mp}
\end{wideFigureOrNot}

The coupling phase of the \PpiTwo in the $\PfTwo \pi S$ wave with
$M = 0$ is zero by definition.  The coupling phases of the \PpiTwo in
the $\PfTwo \pi D$ wave and in the $\PfTwo \pi S$ wave with $M = 1$
are similar and offset by less than \SI{\pm 20}{\degree} (see
\cref{fig:tprim_phase_2mp}).  The latter observation is remarkable
because the coupling phase of the \PpiTwo in the $\PfTwo \pi S$ wave
with $M = 1$ is not constrained via
\cref{eq:method:branchingdefinition}.  In the $\Prho \pi F$ wave, the
\PpiTwo coupling amplitude shows a larger offset of about
\SI{+50}{\degree}.

The \PpiTwo[1880] shows a coupling phase offset of about
\minusDeg{180}\footnote{This is, of course, mathematically equivalent
  to \plusDeg{180}.} in the $\PfTwo \pi D$ wave (see
\cref{fig:tprim_phase_2mp}).  In this wave, the \PpiTwo[1880] is the
dominant component.  Therefore, the corresponding coupling phase is
relatively stable \wrt the systematic studies.  In contrast, the
relative contribution of the \PpiTwo[1880] to the other $2^{-+}$ waves
is much smaller, which leads to larger variations of these coupling
phases in the systematic studies.  In the two $\PfTwo \pi S$ waves,
the coupling phase remains at an offset of about \minusDeg{180}.
However, in the $\Prho \pi F$ wave the \PpiTwo[1880] has a coupling
phase of about \SI{0}{\degree}.

The \PpiTwo[2005] is best determined by the $\Prho \pi F$ wave and
shows a phase offset of about \SI{+150}{\degree} (see
\cref{fig:tprim_phase_2mp}).  Similar offsets, although with larger
systematic variations, are also observed in the $\PfTwo \pi D$ wave
and in the $\PfTwo \pi S$ wave with $M = 1$.  In contrast, the
coupling phase in the $\PfTwo \pi S$ wave with $M = 0$ is about
\SI{-90}{\degree}.

The \tpr dependence of the coupling phase of the \Ppi[1800] follows
that of the \PpiTwo with an offset close to \plusDeg{180} (see
\cref{fig:tprim_phase_all_res}).  The black lines in
\cref{fig:tprim_phase_0mp} show the coupling phase of the nonresonant
component in the \wave{0}{-+}{0}{+}{\PfZero[980]}{S} wave relative to
the \Ppi[1800].  At low \tpr, the nonresonant coupling phase is offset
by about \SI{+100}{\degree}.  It then jumps by about
\SI{+150}{\degree} at $\tpr \approx \SI{0.3}{\GeVcsq}$, thereby
changing the sign of the coupling amplitude \wrt the \Ppi[1800].  At
the same \tpr value, we observe a dip in the \tpr spectrum of the
nonresonant component (see \cref{fig:tprim_0mp}).

\begin{figure}[tbp]
  \centering
  \includegraphics[width=\twoPlotWidth]{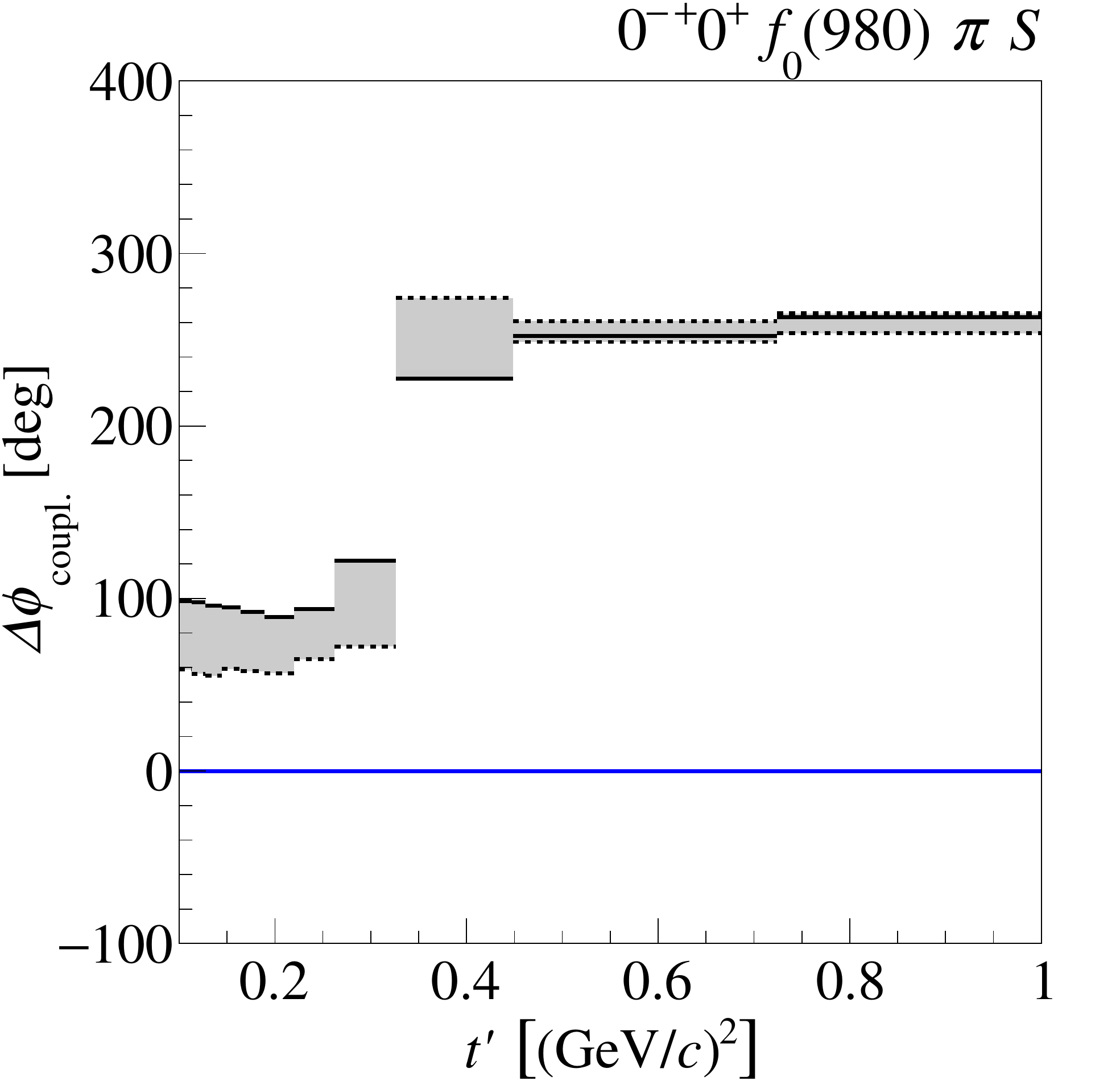}
  \caption{Similar to \cref{fig:tprim_phase_2mp} but for the wave
    components in the \wave{0}{-+}{0}{+}{\PfZero[980]}{S} wave.  The
    coupling phase of the nonresonant component (black lines) is shown
    relative to the \Ppi[1800] (blue line).}
  \label{fig:tprim_phase_0mp}
\end{figure}

The coupling phase of the \PpiOne[1600] relative to the \PpiTwo shows
the most pronounced \tpr dependence of all resonances in
\cref{fig:tprim_phase_all_res} but stays within about \SI{\pm
  30}{\degree} of the \PpiTwo coupling phase.  Qualitatively, the
\PpiOne[1600] coupling phase behaves similar to that of the
ground-state resonances.  The coupling phase of the nonresonant
component in the \wave{1}{-+}{1}{+}{\Prho}{P} wave relative to the
\PpiOne[1600] shows a strong \tpr dependence (see
\cref{fig:tprim_phase_1mp}).  Below $\tpr \approx \SI{0.3}{\GeVcsq}$,
the coupling phase of the nonresonant component is approximately
similar to the coupling phase of the \PpiOne[1600] with a negligible
offset.  In this \tpr region, the $1^{-+}$ wave is dominated by the
nonresonant component (see \cref{sec:oneMP_results}).  Therefore, the
\PpiOne[1600] is not well separated from the nonresonant component.
Above $\tpr \approx \SI{0.3}{\GeVcsq}$, the coupling phase rises
rapidly to about \plusDeg{180}.  This rapid change of the interference
pattern between the \PpiOne[1600] and the nonresonant component at
high \tpr is needed for the model to describe the changing shape of
the $1^{-+}$ intensity distribution.  However, the variation of the
coupling phase in the systematic studies is large as in the case of
the \PpiOne[1600] resonance parameters.

\begin{figure}[tbp]
  \centering
  \includegraphics[width=\twoPlotWidth]{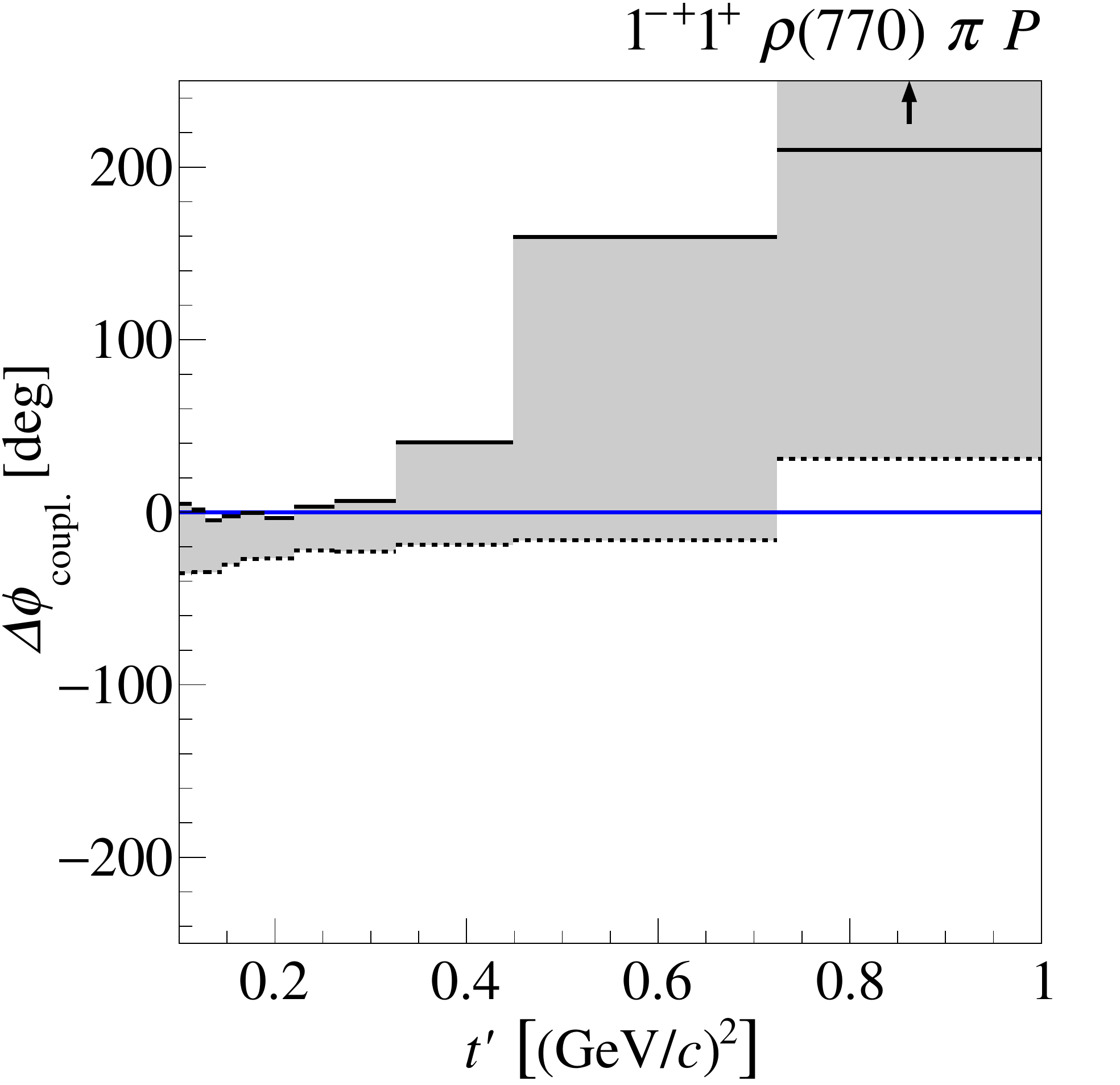}
  \caption{Similar to \cref{fig:tprim_phase_2mp} but for the wave
    components in the \wave{1}{-+}{1}{+}{\Prho}{P} wave.  The coupling
    phase of the nonresonant component (black lines) is shown relative
    to the \PpiOne[1600] (blue line).}
  \label{fig:tprim_phase_1mp}
\end{figure}

\subsection{Relative phases of the coupling amplitudes of the \aJ resonances}
\label{sec:phases_aJ}

The coupling phase of the \PaOne relative to the \PpiTwo shows the
largest variation with \tpr of all ground-state resonances in
\cref{fig:tprim_phase_all_res}.  It starts at \SI{+100}{\degree} at
$\tpr = \SI{0.1}{\GeVcsq}$ and falls until
$\tpr \approx \SI{0.3}{\GeVcsq}$, after which it levels off at about
\SI{+50}{\degree}.  \Cref{fig:tprim_phase_1pp} shows the coupling
phases of the $1^{++}$ wave components relative to the \PaOne in the
$\Prho \pi S$ wave.  The \tpr dependence of the coupling phases of the
resonance components in the $\Prho \pi S$ and $\PfTwo \pi P$ waves are
constrained via \cref{eq:method:branchingdefinition}.  The phase
offset between the coupling phases of the \PaOne in these two waves is
about \SI{+50}{\degree}.  However, the variation of the \PaOne
coupling phase in the $\PfTwo \pi P$ wave in the systematic studies is
large.  In the $\Prho \pi S$ wave, the coupling phase of the
nonresonant component rises by about \SI{100}{\degree} \wrt the \PaOne
over the analyzed \tpr range.  This change of the interference pattern
is needed for the model to describe the movement of the peak in the
intensity distribution of the $\Prho \pi S$ wave with \tpr.

\begin{wideFigureOrNot}[tbp]
  \centering
  \subfloat[][]{%
    \includegraphics[width=\threePlotWidth]{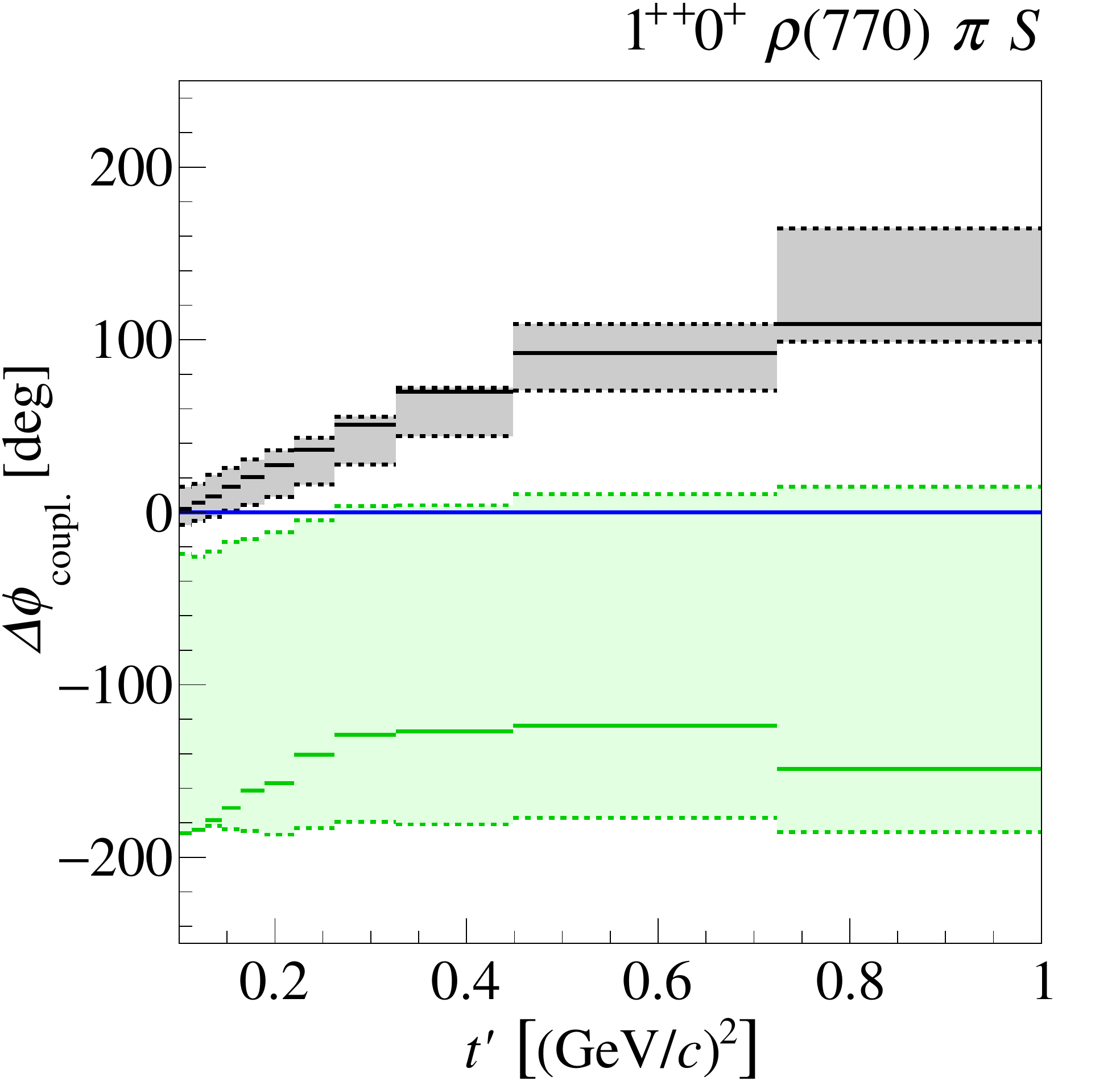}%
    \label{fig:tprim_phase_1pp_rho}%
  }%
  \hspace*{\threePlotSpacing}%
  \subfloat[][]{%
    \includegraphics[width=\threePlotWidth]{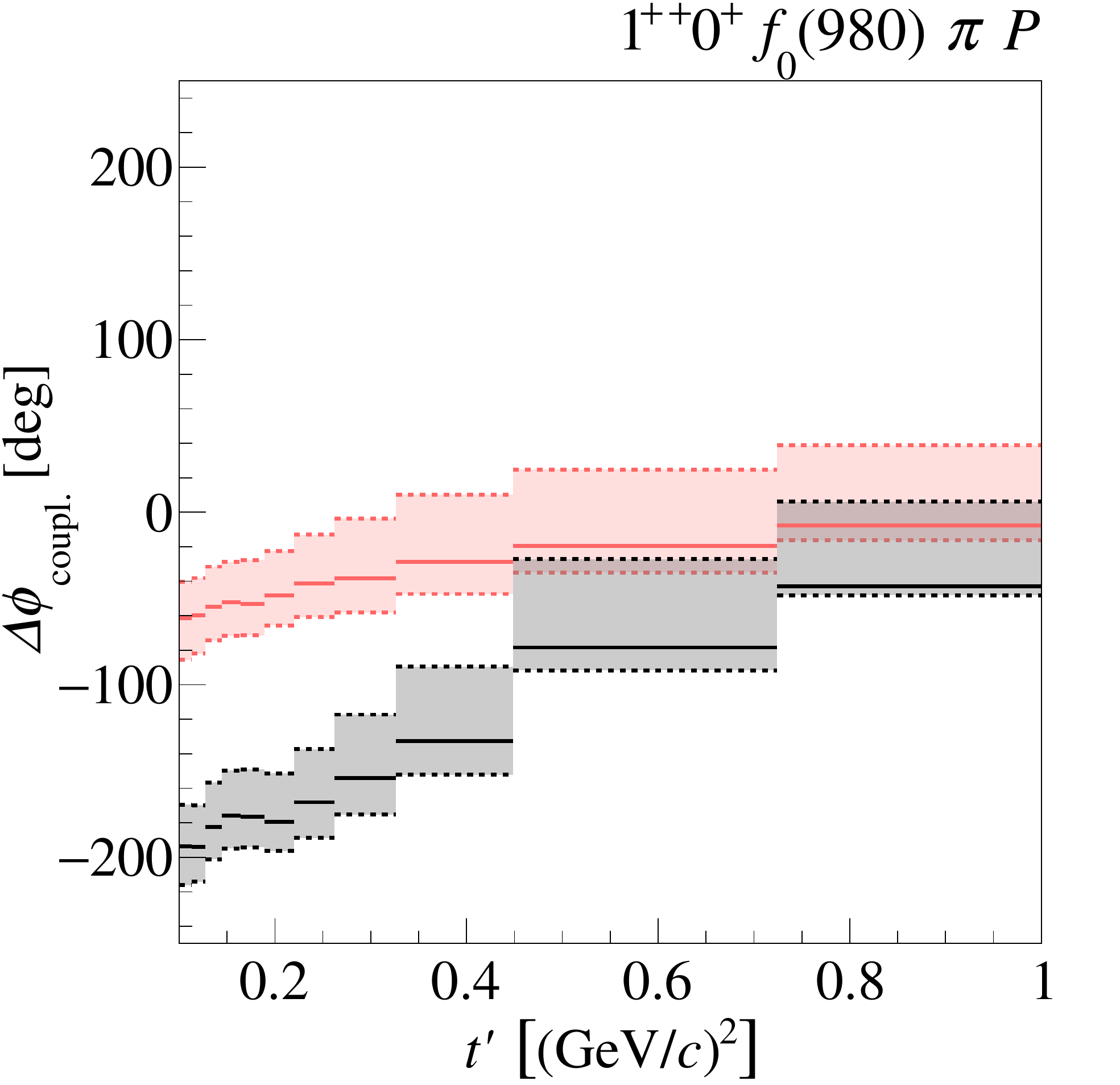}%
    \label{fig:tprim_phase_1pp_f0}%
  }%
  \hspace*{\threePlotSpacing}%
  \subfloat[][]{%
    \includegraphics[width=\threePlotWidth]{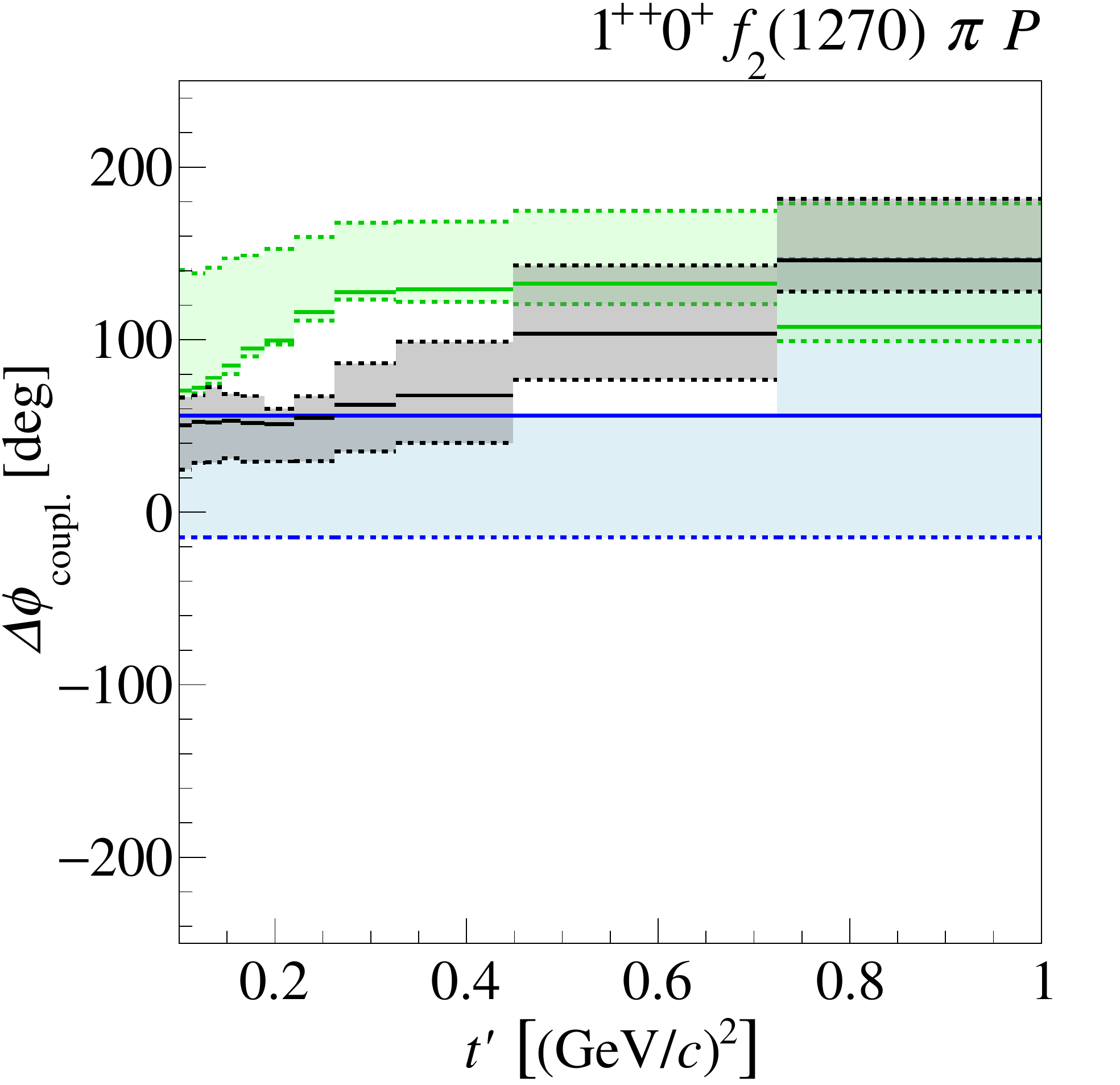}%
    \label{fig:tprim_phase_1pp_f2}%
  }%
  \caption{Similar to \cref{fig:tprim_phase_2mp} but for the wave
    components in \subfloatLabel{fig:tprim_phase_1pp_rho}~the
    \wave{1}{++}{0}{+}{\Prho}{S},
    \subfloatLabel{fig:tprim_phase_1pp_f0}~the
    \wave{1}{++}{0}{+}{\PfZero[980]}{P}, and
    \subfloatLabel{fig:tprim_phase_1pp_f2}~the
    \wave{1}{++}{0}{+}{\PfTwo}{P} wave.  The coupling phases of the
    \PaOne (blue lines), the \PaOne[1420] (red lines), the
    \PaOne[1640] (green lines), and the nonresonant components (black
    lines) are shown relative to the \PaOne in the
    \wave{1}{++}{0}{+}{\Prho}{S} wave.}
  \label{fig:tprim_phase_1pp}
\end{wideFigureOrNot}

The \PaOne[1640] coupling phase \wrt the \PpiTwo is approximately
independent of \tpr with an offset of about \SI{-70}{\degree} (see
\cref{fig:tprim_phase_all_res}).  Relative to the \PaOne, the coupling
phases of the \PaOne[1640] rise up to $\tpr \approx \SI{0.3}{\GeVcsq}$
and then level off (see \cref{fig:tprim_phase_1pp}).  As discussed in
\cref{sec:onePP_results}, the parameters of the \PaOne[1640] are
mainly determined by the $\PfTwo \pi P$ wave.  This is also true for
its coupling phase, which has a much smaller systematic variation in
the $\PfTwo \pi P$ wave.  In this wave, the \PaOne[1640] has a phase
offset \wrt the \PaOne of about \SI{+130}{\degree} at high \tpr,
whereas in the $\Prho \pi S$ wave, the phase offset is about
\SI{-130}{\degree}.  However, the variation of the latter coupling
phase in the systematic studies is large because the \PaOne[1640] is
only a small signal in the tail of the dominant \PaOne.

The \PaOne[1420] has a nearly constant coupling phase relative to the
\PpiTwo with an offset of about \SI{+40}{\degree} (see
\cref{fig:tprim_phase_all_res}).  It therefore behaves qualitatively
similar to the ground-state resonances.  In our model, the
\PaOne[1420] appears only in the $\PfZero[980] \pi P$ wave.  Its
coupling phase relative to the \PaOne is shown as red lines in
\cref{fig:tprim_phase_1pp_f0}.  This phase rises from about
\SI{-60}{\degree} at low \tpr to about \SI{0}{\degree} at high \tpr
and thus changes more strongly than the one \wrt the \PpiTwo.

The \PaTwo is the narrowest resonance in our analysis. In the two
\wave{2}{++}{}{}{\Prho}{D} waves, all other wave components are very
small in the \SI{1.3}{\GeVcc} mass range.  The coupling phase of the
\PaTwo in the $\Prho \pi D$ wave with $M = 1$ relative to the \PpiTwo
shows a weak dependence on \tpr with an offset of about
\SI{+60}{\degree} (see \cref{fig:tprim_phase_all_res}).
\Cref{fig:tprim_phase_2pp} shows the coupling phases of the $2^{++}$
wave components relative to the \PaTwo in the $\Prho \pi D$ wave with
$M = 1$.  The \tpr dependence of the coupling phases of the resonance
components in the $\Prho \pi D$ wave with $M = 1$ and in the
$\PfTwo \pi P$ wave are constrained via
\cref{eq:method:branchingdefinition}.  The phase offset of the \PaTwo
in these two waves is close to zero, which confirms that we indeed see
the $\PfTwo \pi$ decay mode of the \PaTwo.  The coupling phases of the
\PaTwo in the $\Prho \pi D$ wave with $M = 2$ is practically identical
to that in the $\Prho \pi D$ wave with $M = 1$.  This result is
particularly remarkable since the $\Prho \pi D$ wave with $M = 2$ has
a small relative intensity and the coupling phase of the \PaTwo
component in this wave is not constrained via
\cref{eq:method:branchingdefinition}.

\begin{wideFigureOrNot}[tbp]
  \centering
  \subfloat[][]{%
    \includegraphics[width=\threePlotWidth]{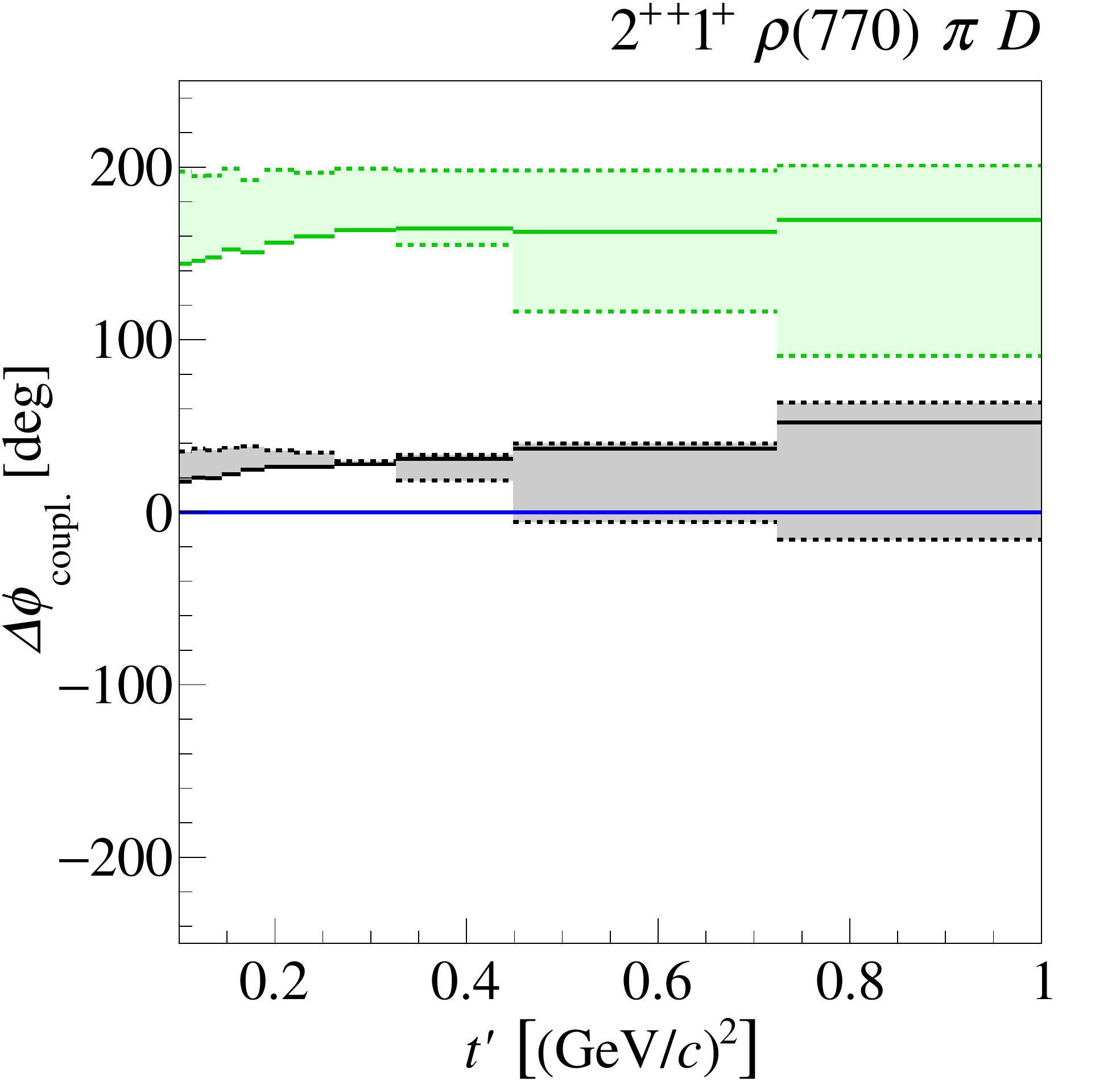}%
    \label{fig:tprim_phase_2pp_m1_rho}%
  }%
  \hspace*{\threePlotSpacing}%
  \subfloat[][]{%
    \includegraphics[width=\threePlotWidth]{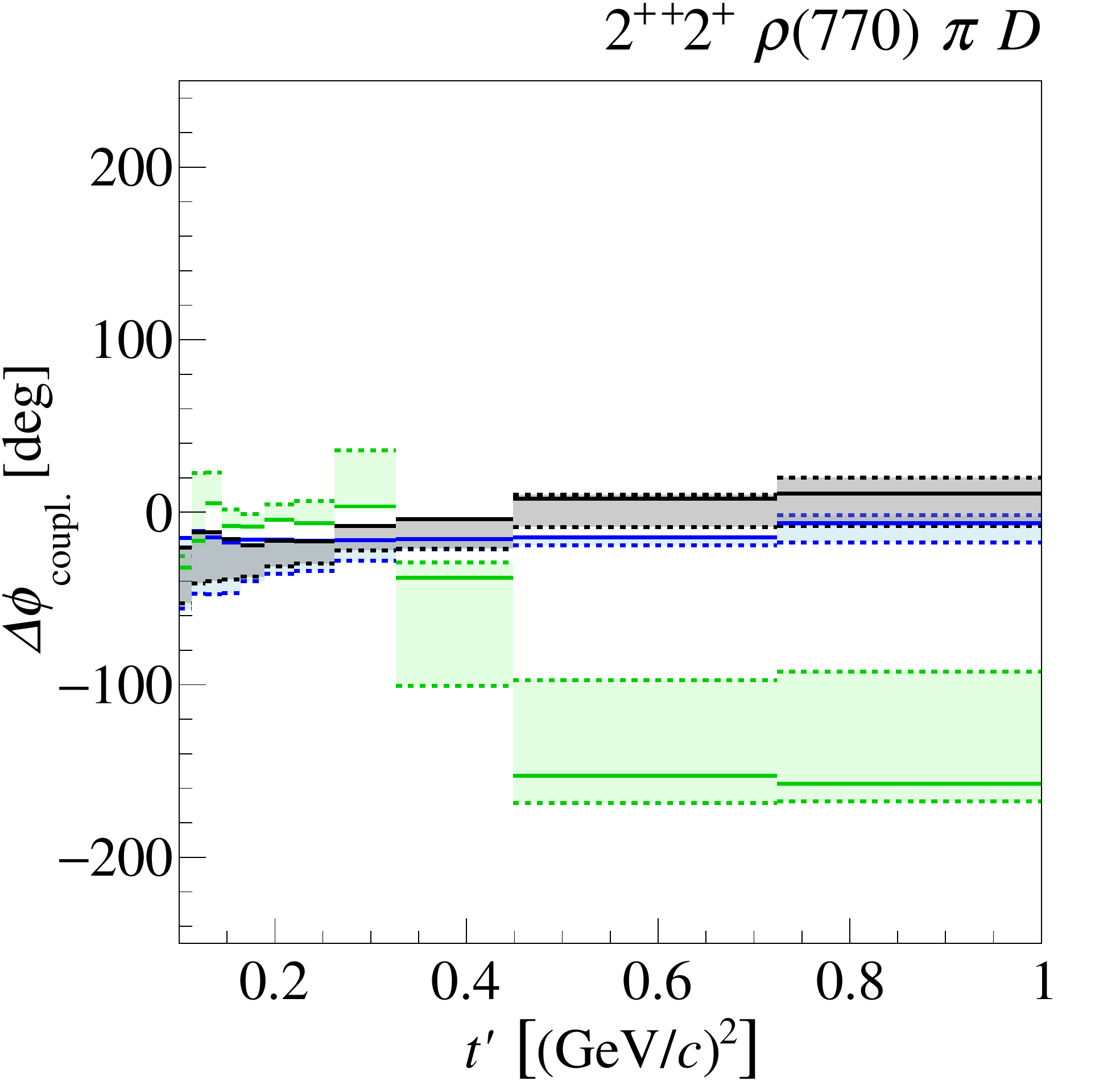}%
    \label{fig:tprim_phase_2pp_m2_rho}%
  }%
  \hspace*{\threePlotSpacing}%
  \subfloat[][]{%
    \includegraphics[width=\threePlotWidth]{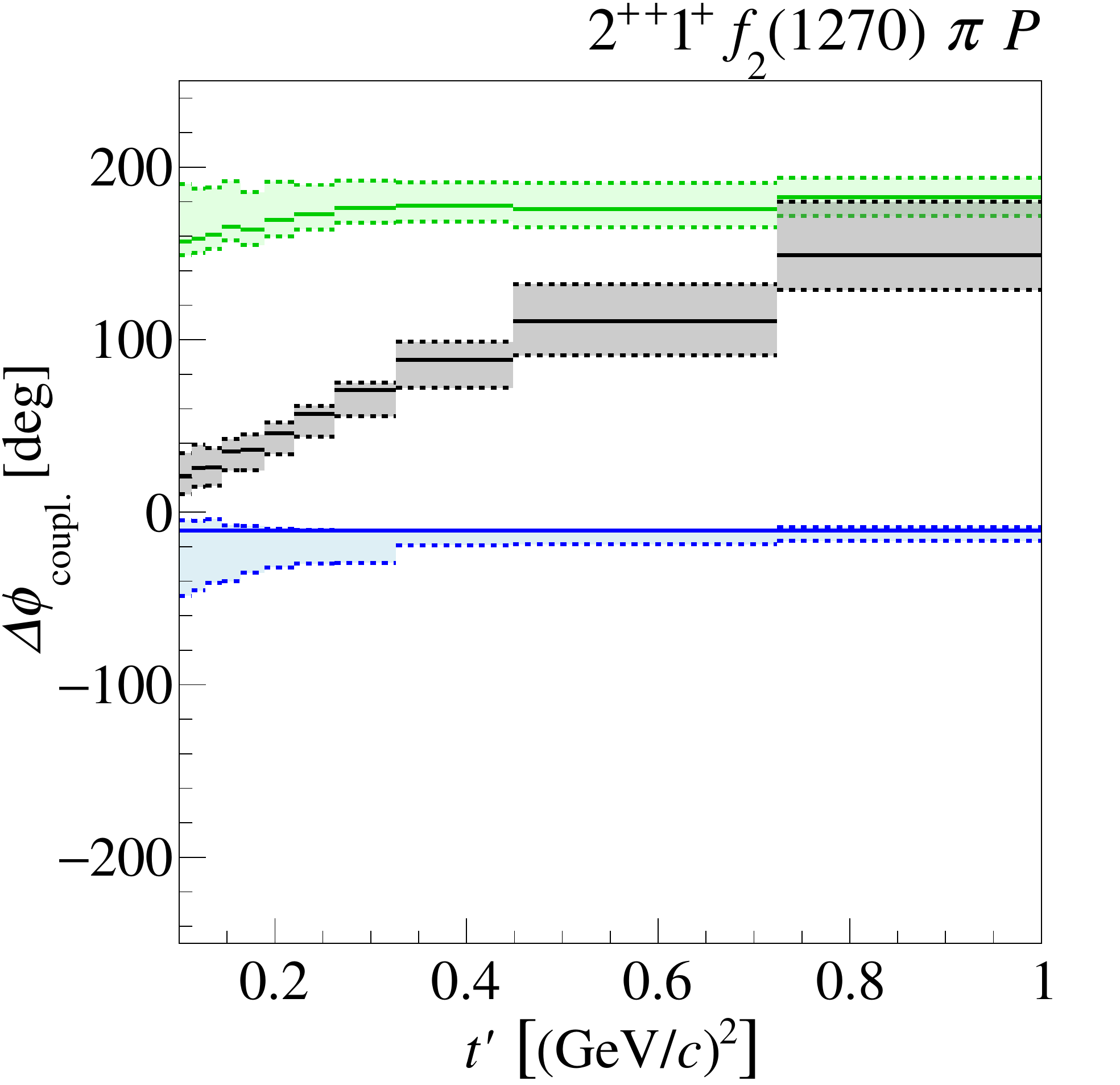}%
    \label{fig:tprim_phase_2pp_f2}%
  }%
  \caption{Similar to \cref{fig:tprim_phase_2mp} but for the wave
    components in \subfloatLabel{fig:tprim_phase_2pp_m1_rho}~the
    \wave{2}{++}{1}{+}{\Prho}{D},
    \subfloatLabel{fig:tprim_phase_2pp_m2_rho}~the
    \wave{2}{++}{2}{+}{\Prho}{D}, and
    \subfloatLabel{fig:tprim_phase_2pp_f2}~the
    \wave{2}{++}{1}{+}{\PfTwo}{P} wave.  The coupling phases of the
    \PaTwo (blue lines), the \PaTwo[1700] (green lines), and the
    nonresonant component (black lines) are shown relative to the
    \PaTwo in the \wave{2}{++}{1}{+}{\Prho}{D} wave.}
  \label{fig:tprim_phase_2pp}
\end{wideFigureOrNot}

The coupling phase of the \PaTwo[1700] in the $\Prho \pi D$ wave with
$M = 1$ has a nearly constant offset of \SI{-140}{\degree} \wrt the
\PpiTwo\ (see \cref{fig:tprim_phase_all_res}).  Relative to the
\PaTwo, the coupling phase of the \PaTwo[1700] shows a similar
behavior in the $\Prho \pi D$ wave with $M = 1$ and in the
$\PfTwo \pi P$ wave with a nearly constant offset of about
\plusDeg{180} (see \cref{fig:tprim_phase_2pp}).  In the
\wave{2}{++}{2}{+}{\Prho}{D} wave, the coupling phase starts at
\SI{0}{\degree} at low \tpr and decreases to \minusDeg{180} at high
\tpr.  However, the \PaTwo[1700] signal is very small in this wave and
therefore not extracted reliably (see \cref{sec:twoPP_results}).

Compared to the other ground-state resonances in
\cref{fig:tprim_phase_all_res}, the coupling phase of the \PaFour is
closest to that of the \PpiTwo with an offset of about
\SI{+30}{\degree}.  \Cref{fig:tprim_phase_4pp} shows the coupling
phases of the $4^{++}$ wave components relative to the \PaFour in the
$\Prho \pi G$ wave.  The \tpr dependence of the coupling phases of the
\PaFour in the $\Prho \pi G$ and $\PfTwo \pi F$ waves are constrained
via \cref{eq:method:branchingdefinition}.  The coupling phase offset
of the \PaFour in the $\PfTwo \pi F$ wave is close to \SI{0}{\degree}.

\begin{wideFigureOrNot}[tbp]
  \centering
  \subfloat[][]{%
    \includegraphics[width=\twoPlotWidth]{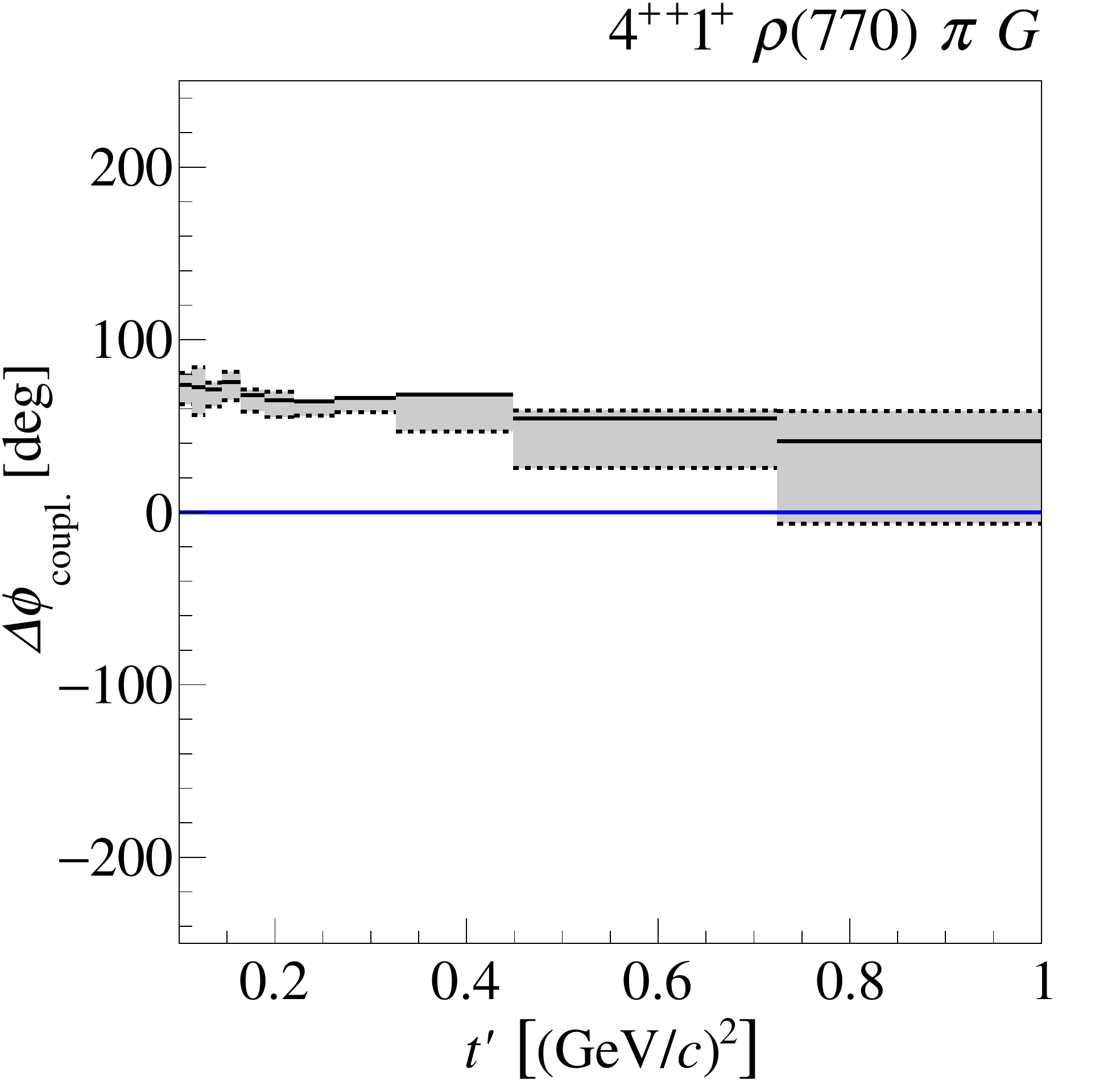}%
    \label{fig:tprim_phase_4pp_rho}%
  }%
  \hspace*{\twoPlotSpacing}%
  \subfloat[][]{%
    \includegraphics[width=\twoPlotWidth]{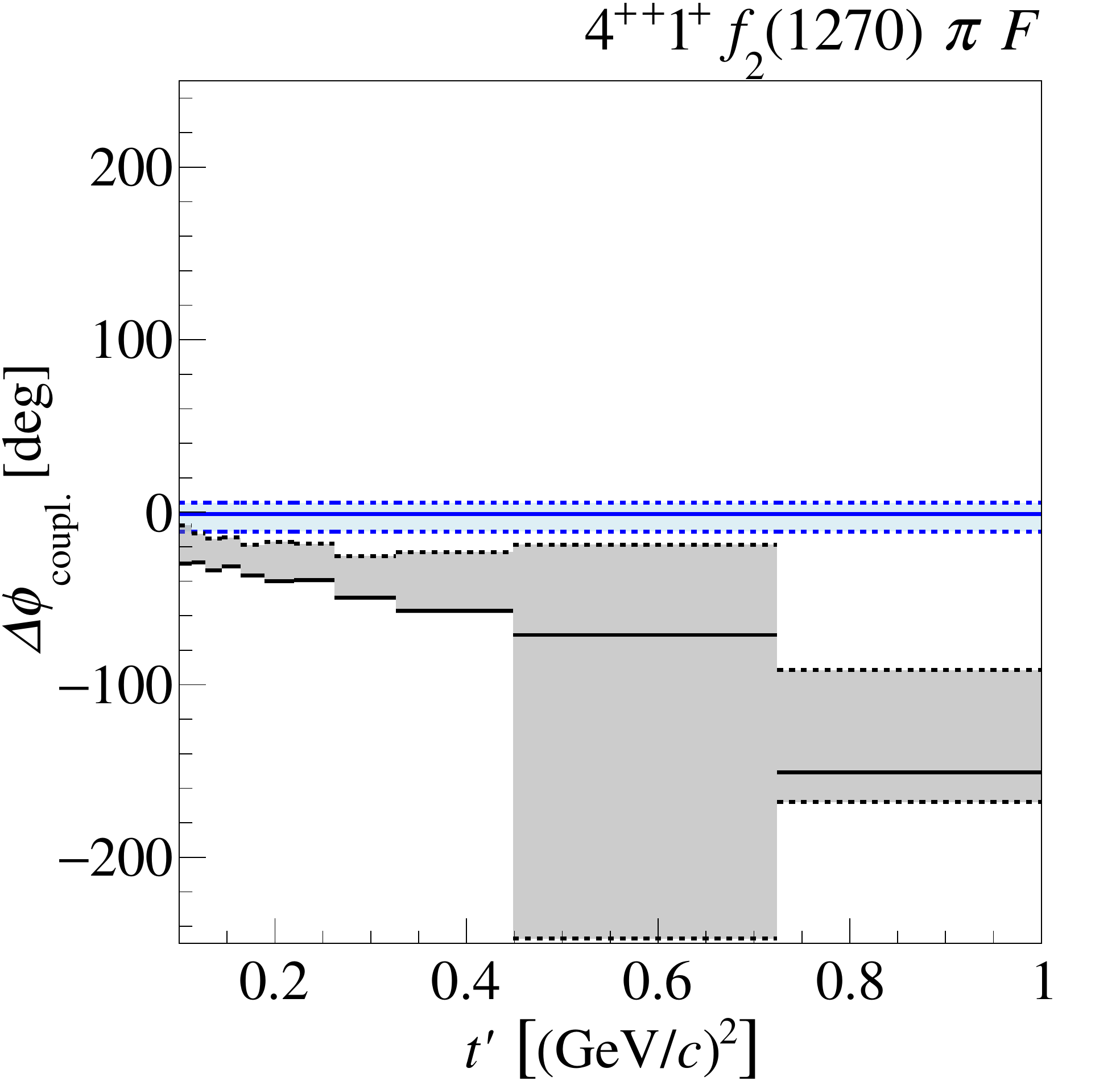}%
    \label{fig:tprim_phase_4pp_f2}%
  }%
  \caption{Similar to \cref{fig:tprim_phase_2mp} but for the wave
    components in \subfloatLabel{fig:tprim_phase_4pp_rho}~the
    \wave{4}{++}{1}{+}{\Prho}{G} and
    \subfloatLabel{fig:tprim_phase_4pp_f2}~the
    \wave{4}{++}{1}{+}{\PfTwo}{F} wave.  The coupling phases of the
    \PaFour (blue lines) and the nonresonant component (black lines)
    are shown relative to the \PaFour in the
    \wave{4}{++}{1}{+}{\Prho}{G} wave.}
  \label{fig:tprim_phase_4pp}
\end{wideFigureOrNot}
 %
%
%

\section{Summary and conclusions}
\label{sec:conclusions}

In this paper, we have presented the results of a fit of a
Breit-Wigner resonance model to 14~selected partial-wave amplitudes
with $\JPC = 0^{-+}$, $1^{++}$, $2^{++}$, $2^{-+}$, $4^{++}$, and
spin-exotic $1^{-+}$ quantum numbers. The amplitudes result from a
partial-wave analysis of \num{46e6}~exclusive events of the
diffractive reaction \reaction using a model with 88 partial
waves~\cite{Adolph:2015tqa}.

We have measured the masses and widths of the \aJ-like resonances:
\PaOne, \PaOne[1640], \PaTwo, \PaTwo[1700], \PaFour, and of the
resonancelike \PaOne[1420] [see \cref{fig:summary:a,tab:parameters}];
and those of the \piJ-like resonances: \Ppi[1800], \PpiTwo,
\PpiTwo[1880], \PpiTwo[2005], and the spin-exotic \PpiOne[1600] [see
\cref{fig:summary:p,tab:parameters}].

\begin{wideFigureOrNot}[tbp]
  \centering
  \subfloat[][]{%
    \includegraphics[width=0.5\textwidth]{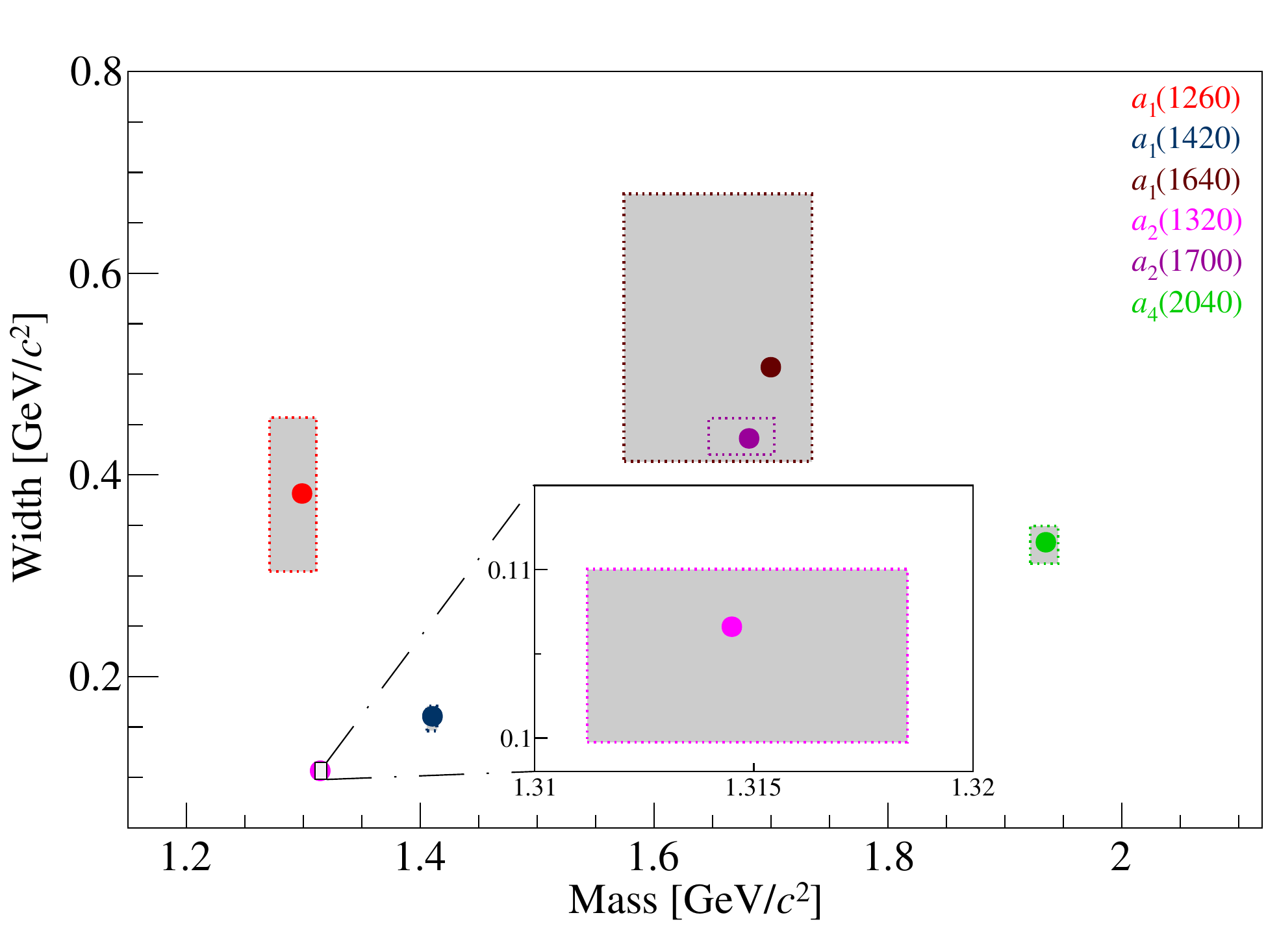}%
    \label{fig:summary:a}%
  }%
  \subfloat[][]{%
    \includegraphics[width=0.5\textwidth]{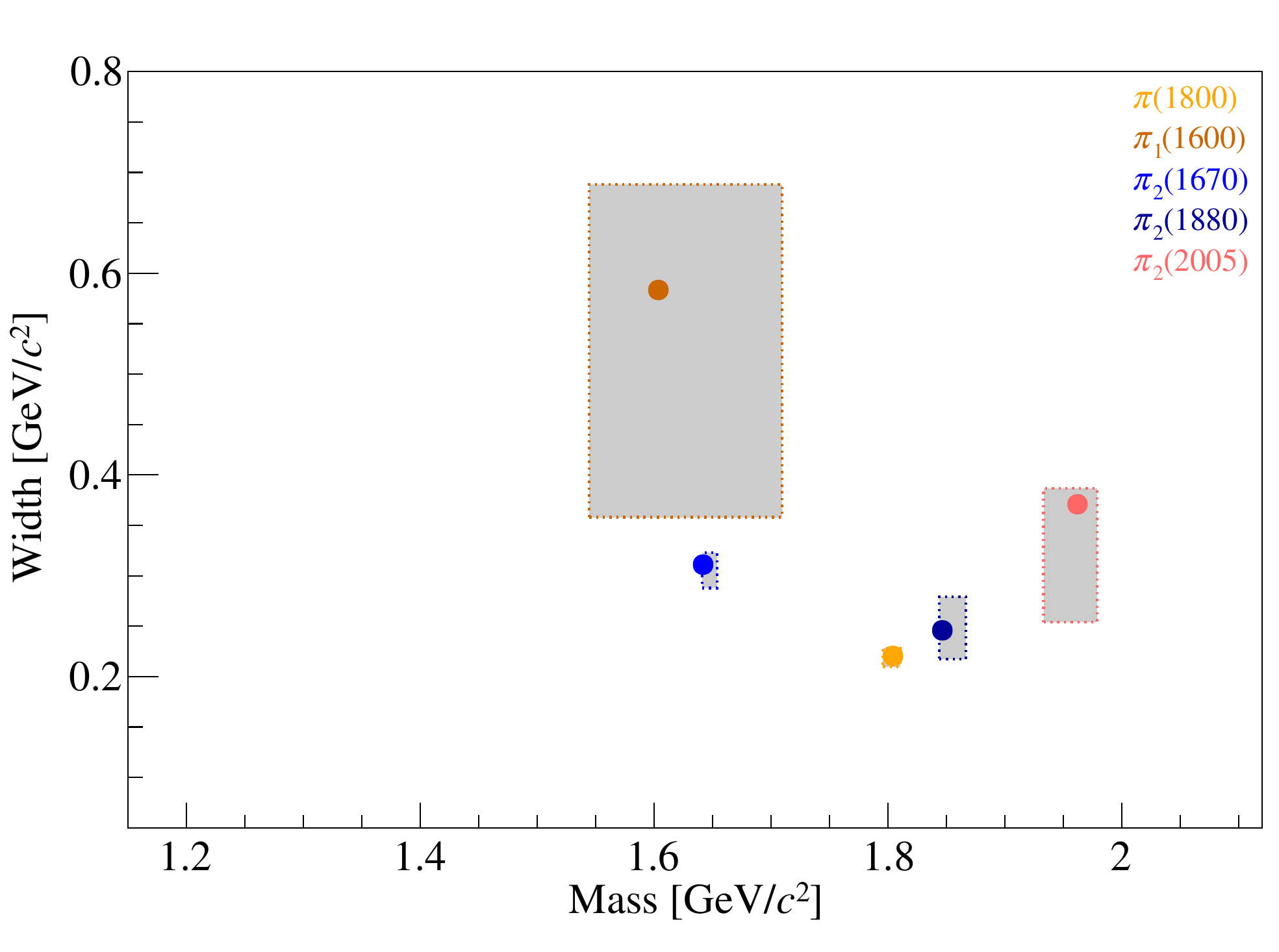}%
    \label{fig:summary:p}%
  }%
  \caption{Masses and widths of \subfloatLabel{fig:summary:a}~\aJ-like
    and \subfloatLabel{fig:summary:p}~\piJ-like resonances extracted
    in this analysis (points).  The systematic uncertainties are
    represented by the boxes.  The statistical uncertainties are at
    least an order of magnitude smaller than the systematic ones and
    are hence omitted.  Different colors encode different resonances.}
  \label{fig:summary}
\end{wideFigureOrNot}

The parameters of \PaOne[1420], \PaTwo, \PaFour, \Ppi[1800], and
\PpiTwo are reliably extracted with comparatively small uncertainties.
The consistency of the \PaOne[1420] signal with a Breit-Wigner
amplitude is confirmed.  The \PaOne[1420] parameter values are
consistent with those from a simpler analysis of the same data
in~\refCite{Adolph:2015pws}, but have smaller uncertainties.  The
\PaTwo and \Ppi[1800] parameter values are consistent with previous
measurements. The measured values of the \PaFour mass and width are
the most accurate so far.  We find a lower \PaFour mass and a larger
width than some of the previous experiments.

We observe production of the \PaTwo with spin projection $M = 2$ along
the beam axis.  In order to describe the $2^{++}$ partial-wave
amplitudes, the inclusion of an excited \PaTwo[1700] is necessary.  It
appears most strongly in the \wave{2}{++}{1}{+}{\PfTwo}{P} wave.  The
\PaTwo[1700] mass is consistent with previous measurements while the
width is larger.

In order to describe the four $2^{-+}$ partial-wave amplitudes that
are included in the fit, three resonances are needed, the \PpiTwo, the
\PpiTwo[1880], and the \PpiTwo[2005].  The latter one is not an
established state.  The measured \PpiTwo[2005] parameter values are
consistent with the two measurements by previous experiments.  We find
the \PpiTwo to be lighter and broader than the world average.  The
$\PpiTwo[1880] \to 3\pi$ decay is observed for the first time.  The
measured \PpiTwo[1880] width is consistent with the world average, and
the mass is found to be smaller.

The \wave{1}{++}{0}{+}{\Prho}{S} and \wave{1}{-+}{1}{+}{\Prho}{P}
partial-wave amplitudes are dominated by the nonresonant components
and are difficult to describe.  This is a main source of systematic
uncertainty.  The shape of the intensity distributions of both waves
depends strongly on \tpr.  By fitting the resonance model
simultaneously in 11~\tpr~bins, we achieve a better separation of the
resonant and nonresonant components in these waves compared to
previous analyses of diffractive-dissociation reactions.  In both
waves, the intensity of the nonresonant components behaves similar to
a model for the Deck effect.  The resonance model is not able to
describe all details of the \PaOne peak in the
\wave{1}{++}{0}{+}{\Prho}{S} wave, which leads to comparatively large
uncertainties for the \PaOne parameters.  The data require an excited
\PaOne[1640] state.  However, because of the dominant \PaOne, the
\PaOne[1640] parameters are not well determined.  The data also
require a spin-exotic resonance, the \PpiOne[1600], in the
\wave{1}{-+}{1}{+}{\Prho}{P} wave.  The \tpr-resolved analysis allows
us to establish for the first time that a significant \PpiOne[1600]
signal appears only for $\tpr \gtrsim \SI{0.5}{\GeVcsq}$, whereas at
low \tpr the intensity of the spin-exotic wave is saturated by
Deck-like nonresonant contributions.  The \PpiOne parameters have
large uncertainties.  The measured width is significantly larger than
that observed in previous experiments including our own result from
the data taken with a lead target, but it has a large systematic
uncertainty toward smaller values.

The resonance yields are found to be much more sensitive to model
assumptions than the resonance parameters.  For the \PaTwo and \PaFour
the systematic uncertainties are small enough to extract their
branching-fraction ratios for the decays into $\Prho \pi$ and
$\PfTwo \pi$.  The branching-fraction ratio for the \PaFour was
measured to be
$B_{\Prho* \pi G, \PfTwo* \pi F}^{\PaFour*, \text{corr}} =
\numaerr{2.9}{0.6}{0.4}$.  This value is corrected for the unobserved
\threePiN decay mode, the effects from self-interference, and the
branching fraction of the \PfTwo into $2\pi$.  The measured value is
in good agreement with predictions by the ${}^3P_0$ decay model.  The
corresponding branching-fraction ratio
$B_{\Prho* \pi D, \PfTwo* \pi P}^{\PaTwo*, \text{corr}} =
\numaerr{16.5}{1.2}{2.4}$ for the \PaTwo was measured for the first
time to our knowledge.

Since the resonance-model fit is performed simultaneously in 11~bins
of \tpr, the \tpr dependence of the amplitudes of the resonant and
nonresonant wave components has been studied in unprecedented detail.
The \tpr dependence of the intensities of most of the resonance
signals follows approximately the expected exponential behavior with
slope parameters between about \SIrange{7}{9}{\perGeVcsq} (see
\cref{tab:slopes}).  This is in particular true for the \PaOne[1420].
The \PpiOne[1600] exhibits an exponential \tpr spectrum only, if the
Deck model is used to describe the nonresonant components.  The slope
parameters of the higher-mass states are found to be smaller than
those of the ground states.  In many waves, the slope of the
nonresonant component is steeper than that of the resonances.

The \tpr dependence of the relative phases of the wave components was
studied for the first time to our knowledge.  Most resonances,
including the \PaOne[1420], are produced with a phase that is
approximately independent of \tpr, which is expected if the production
mechanism is the same over the analyzed \tpr range.  The production
phase of the \PpiOne[1600] exhibits a stronger dependence on \tpr.  In
many waves the production phase of the nonresonant component exhibits
a strong \tpr dependence, which is a hint that more than one
production mechanism contributes.

%
%
%
\appendix
\makeatletter
\@addtoreset{equation}{section}  %
\makeatother
%
%
%

\section{Pole positions}
\label{sec:pole_positions}

For those resonances that are described by the simple relativistic
Breit-Wigner amplitude,
\begin{equation*}
  \mathcal{D}^\text{R}_j(\mThreePi; m_j, \Gamma_j)
  = \frac{m_j\, \Gamma_j}{m_j^2 - \mThreePi^2 - i\, m_j\, \Gamma_j}
\end{equation*}
[see \cref{eq:BreitWigner,eq:method:fixedwidth} in
\cref{sec:method:fitmodel:resonances}], we can calculate the pole
positions in the complex energy plane.  The pole position $s_{\text{R}, j}$
of the Breit-Wigner amplitude for resonance~$j$ is given by
\begin{equation}
  \label{eq:BreitWignerPole}
  s_{\text{R}, j}
  = m_j^2 - i\, m_j\, \Gamma_j.
\end{equation}
Traditionally, the pole position is related to the resonance mass
$m_{\text{R}, j}$ and the total width $\Gamma_{\text{R}, j}$
by~\cite{pdg_resonances:2016}
\begin{equation}
  \label{eq:poleResPar}
  \sqrt{s_{\text{R}, j}}
  = m_{\text{R}, j} - i\, \frac{\Gamma_{\text{R}, j}}{2}.
\end{equation}
In \cref{tab:poleParameters}, we compare the Breit-Wigner parameters
$m_j$ and $\Gamma_j$ from \cref{tab:parameters} with the pole
parameters $m_{\text{R}, j}$ and $\Gamma_{\text{R}, j}$.  Except for
the \PaTwo, the listed pole parameters are estimated using
\cref{eq:BreitWignerPole,eq:poleResPar}.  The parametrization for the
\PaTwo uses the mass-dependent width in
\cref{eq:method:a2dynamicwidth}.  Therefore, \cref{eq:BreitWignerPole}
does not hold and the \PaTwo pole position was estimated by numerical
methods.\footnote{The \PaTwo amplitude has additional poles that are
  caused by the phase-space terms in \cref{eq:method:a2dynamicwidth}.
  However, these poles lie below \SI{1}{\GeVcc}, far away from the
  \PaTwo pole.}  The \PaOne is an even more complicated case.  In
order to calculate its pole position, one would need to analytically
continue the phase-space integral $I_{aa}(\mThreePi)$ in
\cref{eq:method:bowlerG} into the complex plane.  We therefore omit
the \PaOne in \cref{tab:poleParameters}.

\begin{wideTableOrNot}[tbp]
  \renewcommand{\arraystretch}{\ifMultiColumnLayout{1.0}{1.2}}
  \centering
  \captionsetup[subtable]{position=top}
  \caption{Breit-Wigner resonance parameters from
    \cref{tab:parameters} compared to pole parameters defined in
    \cref{eq:poleResPar}.}
  \label{tab:poleParameters}
  \let\parColWidth\relax
  \newlength{\parColWidth}
  \ifMultiColumnLayout{\setlength{\parColWidth}{0.095\linewidth}}{\setlength{\parColWidth}{0.111\linewidth}}
  \subfloat[$a_{J}$-like resonances]{%
    \label{tab:poleParameters:a}%
    \begin{tabular}{llp{\parColWidth}p{\parColWidth}p{\parColWidth}p{\parColWidth}p{\parColWidth}}
      \hline
      \hline
      & &
      \multicolumn{1}{c}{\PaOne[1420]} &
      \multicolumn{1}{c}{\PaOne[1640]} &
      \multicolumn{1}{c}{\PaTwo} &
      \multicolumn{1}{c}{\PaTwo[1700]} &
      \multicolumn{1}{c}{\PaFour} \\
      & &
      \multicolumn{2}{c}{(\cref{sec:onePP})} &
      \multicolumn{2}{c}{(\cref{sec:twoPP})} &
      \multicolumn{1}{c}{(\cref{sec:fourPP})} \\
      \hline
      \rule{0pt}{1.1\normalbaselineskip}
      \multirow{4}{*}{\rotatebox{90}{\centering BW}} & Mass &
      \multirow{2}{*}{$1411\,^{+4}_{-5}$} &
      \multirow{2}{*}{$1700\,^{\phantom{1}+35}_{-130}$} &
      \multirow{2}{*}{$1314.5\,^{+4.0}_{-3.3}$} &
      \multirow{2}{*}{$1681\,^{+22}_{-35}$} &
      \multirow{2}{*}{$1935\,^{+11}_{-13}$} \\[\ifMultiColumnLayout{0ex}{-1ex}]
      & {\small [\si{\MeVcc}]} & & & & & \\
      & Width &
      \multirow{2}{*}{$\phantom{1}161\,^{+11}_{-14}$} &
      \multirow{2}{*}{$\phantom{1}510\,^{+170}_{\phantom{1}-90}$} &
      \multirow{2}{*}{$\phantom{1}106.6\,^{+3.4}_{-7.0}$} &
      \multirow{2}{*}{$\phantom{1}436\,^{+20}_{-16}$}  &
      \multirow{2}{*}{$\phantom{1}333\,^{+16}_{-21}$} \\[\ifMultiColumnLayout{0ex}{-1ex}]
      & {\small [\si{\MeVcc}]} & & & & & \\[\ifMultiColumnLayout{-0.5ex}{0.5ex}]
      \hline
      \rule{0pt}{1.1\normalbaselineskip}
      \multirow{4}{*}{\rotatebox{90}{\centering Pole}} & Mass &
      \multirow{2}{*}{$1413$} &
      \multirow{2}{*}{$1718$} &
      \multirow{2}{*}{$1306.8$} &
      \multirow{2}{*}{$1695$} &
      \multirow{2}{*}{$1942$} \\[\ifMultiColumnLayout{0ex}{-1ex}]
      & {\small [\si{\MeVcc}]} & & & & & \\
      & Width &
      \multirow{2}{*}{$\phantom{1}160$} &
      \multirow{2}{*}{$\phantom{1}501$} &
      \multirow{2}{*}{$\phantom{1}105.2$} &
      \multirow{2}{*}{$\phantom{1}433$} &
      \multirow{2}{*}{$\phantom{1}332$} \\[\ifMultiColumnLayout{0ex}{-1ex}]
      & {\small [\si{\MeVcc}]} & & & & & \\[\ifMultiColumnLayout{-1ex}{0ex}]
      \hline
      \hline
    \end{tabular}%
  }%
  \\
  \subfloat[$\pi_{J}$-like resonances]{%
    \label{tab:poleParameters:p}%
    \begin{tabular}{llp{\parColWidth}p{\parColWidth}p{\parColWidth}p{\parColWidth}p{\parColWidth}}
      \hline
      \hline
      & &
      \multicolumn{1}{c}{\Ppi[1800]} &
      \multicolumn{1}{c}{\PpiOne[1600]} &
      \multicolumn{1}{c}{\PpiTwo} &
      \multicolumn{1}{c}{\PpiTwo[1880]} &
      \multicolumn{1}{c}{\PpiTwo[2005]} \\
      & &
      \multicolumn{1}{c}{(\cref{sec:zeroMP})} &
      \multicolumn{1}{c}{(\cref{sec:oneMP})} &
      \multicolumn{3}{c}{(\cref{sec:twoMP})} \\
      \hline
      \rule{0pt}{1.1\normalbaselineskip}
      \multirow{4}{*}{\rotatebox{90}{\centering BW}} & Mass &
      \multirow{2}{*}{$1804\,^{+6}_{-9}$} &
      \multirow{2}{*}{$1600\,^{+110}_{\phantom{1}-60}$} &
      \multirow{2}{*}{$1642\,^{+12}_{\phantom{1}-1}$} &
      \multirow{2}{*}{$1847\,^{+20}_{\phantom{1}-3}$} &
      \multirow{2}{*}{$1962\,^{+17}_{-29}$} \\[\ifMultiColumnLayout{0ex}{-1ex}]
      & {\small [\si{\MeVcc}]} & & & & & \\
      & Width &
      \multirow{2}{*}{$\phantom{1}220\,^{\phantom{1}+8}_{-11}$} &
      \multirow{2}{*}{$\phantom{1}580\,^{+100}_{-230}$} &
      \multirow{2}{*}{$\phantom{1}311\,^{+12}_{-23}$} &
      \multirow{2}{*}{$\phantom{1}246\,^{+33}_{-28}$} &
      \multirow{2}{*}{$\phantom{1}371\,^{\phantom{1}+16}_{-120}$} \\[\ifMultiColumnLayout{0ex}{-1ex}]
      & {\small [\si{\MeVcc}]} & & & & & \\[\ifMultiColumnLayout{-0.5ex}{0.5ex}]
      \hline
      \rule{0pt}{1.1\normalbaselineskip}
      \multirow{4}{*}{\rotatebox{90}{\centering Pole}} & Mass &
      \multirow{2}{*}{$1808$} &
      \multirow{2}{*}{$1629$} &
      \multirow{2}{*}{$1649$} &
      \multirow{2}{*}{$1851$} &
      \multirow{2}{*}{$1971$} \\[\ifMultiColumnLayout{0ex}{-1ex}]
      & {\small [\si{\MeVcc}]} & & & & & \\
      & Width &
      \multirow{2}{*}{$\phantom{1}220$} &
      \multirow{2}{*}{$\phantom{1}574$} &
      \multirow{2}{*}{$\phantom{1}310$} &
      \multirow{2}{*}{$\phantom{1}245$} &
      \multirow{2}{*}{$\phantom{1}369$} \\[\ifMultiColumnLayout{0ex}{-1ex}]
      & {\small [\si{\MeVcc}]} & & & & & \\[\ifMultiColumnLayout{-1ex}{0ex}]
      \hline
      \hline
    \end{tabular}%
  }%
\end{wideTableOrNot}

The width values of the pole positions are nearly identical to the
Breit-Wigner width values.  For some resonances, the pole masses
differ slightly from the Breit-Wigner masses.  Interestingly the pole
masses for the \PaTwo and \PaTwo[1700] are closer to the pole masses
of \SIerrs{1307}{1}{6}{\MeVcc} and \SIerrs{1720}{10}{60}{\MeVcc},
respectively, which were obtained in an analysis of the $\eta \pi$
$D$-wave intensity using an analytical model based on the principles
of the relativistic $S$-matrix~\cite{Jackura:2017amb}.  However, the
discrepancy in the \PaTwo[1700] width remains (see
\cref{sec:twoPP_discussion}).  The caveats of our simple Breit-Wigner
model, which are discussed in \cref{sec:method:fitmodel:discussion},
also apply to the extracted pole parameters and may be the reason for
this discrepancy.

\section{Deck model}
\label{sec:deck_model}

To construct a model for the Deck process~\cite{deck:1964hm} (see also
\cref{fig:deck_model}), we follow \refCite{daum:1980ay}, where the
Deck amplitude is described as a product of two vertex amplitudes and
a pion propagator in the $t$~channel:
\begin{multlineOrEq}
  \label{eq:deck_ampl}
  \mathcal{A}(s_{\pi\pi}, s_{\pi p}, t_\pi, t)
  \newLineOrNot
  = \mathcal{A}_{\pi\pi}(s_{\pi\pi})\, \mathcal{A}_{\pi p}(s_{\pi p}, t)\,
  \frac{e^{-b_2\, (m_{\pi}^2 - t_{\pi})}}{m_\pi^2 - t_\pi}.
\end{multlineOrEq}
Here, $t_\pi$ is the squared four-momentum of the exchanged pion.  The
amplitude $\mathcal{A}_{\pi\pi}$, which depends on the squared
center-of-mass energy $s_{\pi\pi}$ of the \twoPi system, describes
production, propagation, and decay of the isobar $\xi^0$.  As a
parametrization of $\mathcal{A}_{\pi\pi}$, we use the elastic \pipi
scattering amplitude from \refCite{hyams:1973zf}, which includes the
dominant isobars used in our PWA model: \pipiS, \Prho, \PfZero,
\PfTwo, and \PrhoThree.  The amplitude
\begin{equation}
  \label{eq:pi_p_scattering_ampl}
  \mathcal{A}_{\pi p}(s_{\pi p}, t) = i\, s_{\pi p}\, \sigma_{\pi p \to \pi p}\, e^{b_1\, t}
\end{equation}
describes the elastic scattering of pion and proton and depends on the
squared center-of-mass energy $s_{\pi p}$ of the bachelor pion and the
recoil proton and on the squared four-momentum~$t$ transferred to the
target nucleon.  We use a value of
$\sigma_{\pi p \to \pi p} = \SI{64}{\per\GeV\squared} =
\SI{25}{\milli\barn}$ for the total $\pi^- p$ elastic scattering cross
section and choose the slope parameter to be
$b_1 = \SI{8}{\perGeVcsq}$.  The description of the observed
$t$~dependence around $\mThreePi = \SI{1}{\GeVcc}$ requires the
additional exponential factor in \cref{eq:deck_ampl} with
$b_2 = \SI{0.45}{\perGeVcsq}$.

\section{Alternative \chisq formulations}
\label{sec:alt_chi_2}

The elements of a rank-1 spin-density matrix $\varrho_{a b}$ are
related by \cref{eq:spin_dens_corr}.  Therefore, the full information
from the mass-independent analysis is already contained in a single
row (or column) of $\varrho_{a b}$.  For a chosen reference wave with
index~$r$, the elements of the corresponding row vector
$\varrho_{r a} = \mathcal{T}_r\, \mathcal{T}_a^{\text{*}}$ represent
in total $(2 N_{\text{wave}} - 1)$ independent real values.  This
corresponds to the number of independent real values of the
$N_{\text{wave}}$ transition amplitudes.  The deviation of the model
from the data is measured by the quantities
\begin{equation}
  \label{eq:systematics:deviation1}
  \Delta_a^{\Re} = \Re[\varrho_{r a}] - \Re[\widehat{\varrho}_{r a}]
\end{equation}
and
\begin{equation}
  \label{eq:systematics:deviation2}
  \Delta_a^{\Im} = \Im[\varrho_{r a}] - \Im[\widehat{\varrho}_{r a}]
\end{equation}
for $a \neq r$, and
\begin{equation}
  \label{eq:systematics:deviation3}
  \Delta_r = \varrho_{r r} - \widehat{\varrho}_{r r}
\end{equation}
for $a = r$.  These deviations are collected into the
$(2 N_{\text{wave}} - 1)$-dimensional vector\footnote{Note that here
  the wave indices~$a$ and~$r$ represent both the quantum numbers of
  the waves as defined in \cref{eq:wave_index} and the numerical index
  in the list of $N_{\text{wave}}$ waves included in the
  resonance-model fit.}
\begin{multlineOrEq}
  \mathbold{\Delta} \equiv \big(\Delta_1^{\Re}, \Delta_1^{\Im}, \Delta_2^{\Re}, \Delta_2^{\Im}, \ldots,
  \newLineOrNot
  \Delta_{r - 1}^{\Im}, \Delta_r, \Delta_{r + 1}^{\Re}, \ldots,
  \Delta_{N_{\text{wave}}}^{\Re}, \Delta_{N_{\text{wave}}}^{\Im} \big).
\end{multlineOrEq}
The total deviation of the model from the data is given by the sum of
the squared Mahalanobis distances~\cite{Mahalanobis:1936} over all
\mThreePi and \tpr bins:
\begin{wideEqOrNot}%
  \begin{equation}
    \label{eq:systematics:chi2_alt1}
    \chisq
    = \sum_{i, j}^{2 N_{\text{wave}} - 1\vphantom{)_{i j}}}\;
    \sum^{\text{\tpr bins}\vphantom{N_{\text{w}})_{i j}}}\;
    \sum^{(\text{\mThreePi bins})_{i j}\vphantom{N_{\text{w}}}}
    \mathbold{\Delta}_i(\mThreePi, \tpr)\, V_{ij}^{-1}(\mThreePi, \tpr)\, \mathbold{\Delta}_j(\mThreePi, \tpr).
  \end{equation}
\end{wideEqOrNot}%
Here, $i$~and~$j$ are the indices of the elements of
$\mathbold{\Delta}$ and $V_{ij}$ is the covariance matrix of the
corresponding terms that appear in $\mathbold{\Delta}$.  The matrix
$V_{ij}$ is calculated from the covariance matrix of the transition
amplitudes using Gaussian error propagation.

In contrast to \cref{eq:method:fitmethod:chi2}, the \chisq~formulation
in \cref{eq:systematics:chi2_alt1} requires choosing a reference wave.
This wave needs to have significant intensity over the full analyzed
mass range, which extends from \SIrange{0.9}{2.3}{\GeVcc}.  Also the
model has to describe this wave over this mass range.  In our
analysis, only the \wave{1}{++}{0}{+}{\Prho}{S} wave fulfills these
criteria.  In addition, \cref{eq:systematics:chi2_alt1} is asymmetric
\wrt the way the information of the partial waves enters.  The
transition amplitude of the reference wave enters in every term of the
sum, whereas the transition amplitudes of the other waves enter each
only in two interference terms per $(\mThreePi, \tpr)$ bin.
Furthermore, the transition amplitudes of the reference wave appear
with a maximum power of~4, whereas the transition amplitudes of the
other waves have a maximum power of~2.  This is in contrast to
\cref{eq:method:fitmethod:chi2}, where the transition amplitudes of
all waves enter in a symmetric way.

Another possible approach is to construct the \chisq~function from the
differences of the modeled transition amplitudes and those obtained
from real data.  However, in order to fix the immeasurable global
phase, this approach also requires a reference wave.  The deviation of
the model from the data is measured in terms of the rotated transition
amplitudes
\begin{equation}
  e^{i \varphi_r}\, \mathcal{T}_a^{\text{*}}
  = \frac{\mathcal{T}_r}{\abs{\mathcal{T}_r}}\, \mathcal{T}_a^{\text{*}}
\end{equation}
The corresponding \chisq~function can be derived from
\cref{eq:systematics:deviation1,eq:systematics:deviation2} using the
substitution
\begin{equation}
  \label{eq:systematics:substitution}
  \varrho_{r a}
  = \mathcal{T}_r\, \mathcal{T}_a^{\text{*}}
  \to \frac{\mathcal{T}_r}{\abs{\mathcal{T}_r}}\, \mathcal{T}_a^{\text{*}}
  = \frac{\varrho_{r a}}{\abs{\mathcal{T}_r}}
\end{equation}
The resulting \chisq~function is similar to
\cref{eq:systematics:chi2_alt1}.  The only difference is that each
term of the sum now contains the phase of the reference wave instead
of the full transition amplitude.

\section{Systematic uncertainties of resonance parameters}
\label{sec:syst_uncert}

In this section, we discuss the results of selected systematic
studies, in addition to the studies already covered in
\cref{sec:results}.  We focus in particular on studies that yield the
largest deviations of resonance parameters from those of the main fit
and therefore define the systematic uncertainties.  The systematic
studies are explained in \cref{sec:systematics}.

\subsection{Systematic uncertainties of the \Ppi[1800]
  parameters}
\label{sec:syst_uncert_zeroMP}

The \Ppi[1800] parameters vary only slightly among the systematic
studies.  They are in particular only weakly sensitive to how well the
nonresonant component describes the low-mass shoulder.  Using the mass
shapes of the nonresonant components from the partial-wave
decomposition of a model for the Deck amplitude in \StudyO, the fit is
not able to reproduce the enhancement at \SI{1.3}{\GeVcc} in the
intensity distributions.  Nevertheless, the \Ppi[1800] width remains
practically unchanged and the mass increases only slightly by
\SI{6}{\MeVcc}, which defines the upper limit of the uncertainty
interval for the \Ppi[1800] mass.

A similar result is obtained in a study, in which the fit range for
the $0^{-+}$ wave is narrowed to the \Ppi[1800] peak region of
\SIvalRange{1.6}{\mThreePi}{2.3}{\GeVcc}.  In this study, the
nonresonant component nearly vanishes and the width of the \Ppi[1800]
increases by only \SI{8}{\MeVcc}, which defines the upper limit of the
uncertainty interval for the \Ppi[1800] width.

The \Ppi[1800] parameters also depend on the model for the production
probability $\Abs[0]{\mathcal{P}(\mThreePi, \tpr)}^2$ in
\cref{eq:method:param:spindens,eq:method:param:prods}.  In \StudyAF,
in which $\Abs[0]{\mathcal{P}(\mThreePi, \tpr)}^2$ is set to unity,
the \Ppi[1800] mass decreases by \SI{9}{\MeVcc} and the width by
\SI{11}{\MeVcc}, which both define the lower limits of the respective
uncertainty intervals.

\subsection{Systematic uncertainties of parameters of the $\JPC = 1^{++}$ resonances}
\label{sec:syst_uncert_onePP}

The parameters of \PaOne and \PaOne[1640] depend strongly on the
interference of the $1^{++}$ and $2^{++}$ waves.  In Studies~\studyH
through~\studyD (see \cref{tab:syst_studies:wave_set}), the solution
with the narrow \PaOne (see discussion in \cref{sec:onePP_results})
has the lowest~\chisq.  \StudyF defines the upper limit of the
uncertainty interval for the \PaOne mass and the lower limit of the
uncertainty interval for the \PaOne width.  \StudyG defines the upper
limit of the uncertainty interval for the \PaOne[1640] mass.

A strong dependence of the parameters of \PaOne and \PaOne[1640] on
the number of background events in the selected data sample is
observed in \StudyK.  In this study, weaker event-selection criteria
lead to an increased background.  \StudyK defines the lower limit of
the uncertainty interval for the \PaOne mass.  This study also defines
the lower limit of the uncertainty interval for the \PaOne[1640] mass
and the upper limit of the uncertainty interval for the \PaOne[1640]
width.

The upper limit of the uncertainty interval for the \PaOne width and
the lower limit for the \PaOne[1640] width are defined by the study,
which included the \PaOne[1420] resonance also in the
\wave{1}{++}{0}{+}{\Prho}{S} and \wave{1}{++}{0}{+}{\PfTwo}{P} waves
(see discussion in \cref{sec:onePP_discussion}).  This study and
\StudyK discussed above are the only two studies that yield a
significantly broader \PaOne.

In Studies~\studyS and~\studyR, alternative \chisq~formulations (see
\cref{sec:alt_chi_2}) are used that, compared to the main fit, give
more relative weight to the intensity distributions than to the
phases.  As discussed in \cref{sec:onePP_results}, the model is not
able to describe all details of the \wave{1}{++}{0}{+}{\Prho}{S}
intensity distributions and the resulting deviations of the model from
the data give a large contribution to the~\chisq.  In both studies,
the fit tries to compensate the deviations by using unphysical values
for the \PaOne and \PaOne[1640] parameters\footnote{Both resonances
  become approximately \SI{600}{\MeVcc} wide and have nearly identical
  masses around \SI{1.35}{\GeVcc}.}~\cite{msc_thesis_wallner}.
Therefore, the results of Studies~\studyS and~\studyR are not
considered for the systematic uncertainties of the \PaOne and
\PaOne[1640] parameters.

In \StudyAF, in which the production probability
$\Abs[0]{\mathcal{P}(\mThreePi, \tpr)}^2$ in
\cref{eq:method:param:spindens,eq:method:param:prods} is set to unity,
the \PaOne parameters are only slightly affected but the \PaOne[1640]
width increases by \SI{96}{\MeVcc}.

The parameters of the \PaOne[1420] have significantly smaller
systematic uncertainties than the other two $1^{++}$ resonances.  The
upper limits of the uncertainty intervals for the \PaOne[1420] mass
and width are defined by the study, in which the \PaOne[1420]
resonance is also included in the \wave{1}{++}{0}{+}{\Prho}{S} and
\wave{1}{++}{0}{+}{\PfTwo}{P} waves (see discussion in
\cref{sec:onePP_discussion}).  \StudyK defines the lower limit of the
uncertainty interval for the \PaOne[1420] mass, and \StudyR the one
for the \PaOne[1420] width.

\subsection{Systematic uncertainties of the \PpiOne[1600]
  parameters}
\label{sec:syst_uncert_oneMP}

As discussed in \cref{sec:oneMP_results}, the \PpiOne[1600] parameters
depend on the description used for the nonresonant component.  The
lower limit of the uncertainty interval for the \PpiOne[1600] mass is
defined by \StudyO, in which a model for the Deck amplitude is used to
determine the shape of the nonresonant contribution.

The \PpiOne[1600] parameters are also sensitive to the range
parameter~$q_R$ in the Blatt-Weisskopf factors.  In \StudyM, in which
$q_R$~was set to \SI{267}{\MeVc} corresponding to an assumed
strong-interaction range of \SI{0.75}{\femto\meter}, the \PpiOne[1600]
mass increases by \SI{110}{\MeVcc} and the width decreases by
\SI{90}{\MeVcc}.  This study defines the upper limit of the
uncertainty interval for the \PpiOne[1600] mass.  It is worth noting
that increasing the interaction radius in \StudyN to
\SI{1.29}{\femto\meter}, which corresponds to $q_R = \SI{155}{\MeVc}$,
leaves the \PpiOne[1600] parameters practically unchanged.

A particularly large effect on the \PpiOne[1600] parameters is
observed if the two \wave{2}{++}{}{}{\Prho}{D} waves are omitted from
the fit [\StudyG].  In this study, the \PpiOne[1600] mass increases by
\SI{80}{\MeVcc} and the width decreases by
\SI{230}{\MeVcc}.\footnote{In \StudyG, also the \PaOne becomes
  narrower and the \PaOne[1640] heavier and wider (see
  \cref{sec:onePP_results}).}  The latter defines the lower limit of
the uncertainty interval for the \PpiOne[1600] width.

Studies~\studyS and~\studyR with alternative \chisq~formulations (see
\cref{sec:alt_chi_2}) also influence the \PpiOne[1600] parameters.
The mass decreases by \SI{30}{\MeVcc}, and the width increases by
\SI{100}{\MeVcc}.  The latter defines the upper limit of the
uncertainty interval for the \PpiOne[1600] width.  These studies show
that larger width values are preferred when less weight is given to
the phase information in the \chisq~function.\footnote{In
  Studies~\studyS and~\studyR also the parameters of the \PaOne and
  the \PaOne[1640] change significantly (see
  \cref{sec:syst_uncert_onePP}) and the parameters of the
  \PpiOne[1600] are sensitive to these \PaOne* parameters.}

\subsection{Systematic uncertainties of parameters of the $\JPC = 2^{++}$ resonances}
\label{sec:syst_uncert_twoPP}

As mentioned in \cref{sec:twoPP_results}, the $2^{++}$ resonance
parameters are sensitive to the parametrization of the nonresonant
components.  We investigated this, by determining the mass shape of
the nonresonant component from the partial-wave decomposition of a
model for the Deck amplitude [\StudyO; see \cref{sec:systematics}].
In all three $2^{++}$ waves, the shape of the Deck intensity is
distinctly different from that of the nonresonant components
determined from data in the main fit.  \StudyO defines the lower
limits of the uncertainty intervals for the masses of \PaTwo and
\PaTwo[1700].

The $2^{++}$ resonance parameters also depend on the choice of the
wave set included in the fit.  The \PaTwo parameters change only
slightly if we omit the two dominant $1^{++}$ waves [\StudyB], the
four $2^{-+}$ waves [\StudyA], or the two $4^{++}$ waves [\StudyC].
However, \StudyF, in which only the low-intensity
\wave{2}{++}{2}{+}{\Prho}{D} wave was included in the fit, defines the
upper limit of the uncertainty interval for the \PaTwo mass and also
the lower limit for the \PaTwo width.  The \PaTwo[1700] parameters do
not depend strongly on the wave set used in the fit.  The only
exceptions are Studies~\studyJ, \studyE, and \studyF, in which the
\wave{2}{++}{1}{+}{\PfTwo}{P} wave is omitted from the fit.  If, for
example, only the two \wave{2}{++}{}{}{\Prho}{D} waves are included in
the fit [\StudyJ], we observe a strong increase of the \PaTwo[1700]
mass by \SI{150}{\MeVcc} and of the width by \SI{41}{\MeVcc}.
However, the two $\Prho \pi D$ waves are dominated by the \PaTwo and
contain only very weak \PaTwo[1700] signals.  Therefore, the
\PaTwo[1700] parameters are not reliably determined in these three
studies and they have been omitted from the determination of the
systematic uncertainties.

Also the value of the range parameter~$q_R$ in the Blatt-Weisskopf
factors influences the $2^{++}$ resonance parameters.  \StudyM, in
which $q_R$ was set to \SI{267}{\MeVc} corresponding to an assumed
strong-interaction range of \SI{0.75}{\femto\meter}, defines the upper
limits of the uncertainty intervals for the \PaTwo width and the
\PaTwo[1700] mass.  The lower limit of the uncertainty interval for
the \PaTwo[1700] width is defined by \StudyN, in which $q_R$ was set
to \SI{155}{\MeVc}, which corresponds to a range of
\SI{1.29}{\femto\meter}.

The upper limit of the uncertainty interval for the \PaTwo[1700] width
is defined by \StudyR, in which an alternative \chisq~formulation (see
\cref{sec:alt_chi_2}) was used.

\subsection{Systematic uncertainties of parameters of the $\JPC = 2^{-+}$ resonances}
\label{sec:syst_uncert_twoMP}

As discussed in \cref{sec:twoMP_results}, the parameters of the
\PpiTwo* resonances depend on the wave set.  \StudyC, in which the two
$4^{++}$ waves are omitted from the fit, defines the lower limits of
the uncertainty intervals for the masses of \PpiTwo and \PpiTwo[1880].
The omission of the \wave{1}{++}{0}{+}{\Prho}{S} wave from the fit in
\StudyH leads to the largest \PpiTwo[2005] width.

\StudyK, in which weaker event-selection criteria lead to an increased
background, defines the lower limits of the uncertainty intervals for
the widths of \PpiTwo and \PpiTwo[1880].  The parameters of the
\PpiTwo[2005] are only weakly affected.

The parameters of \PpiTwo[1880] and \PpiTwo[2005] also depend on the
number of \tpr bins.  \StudyL, in which the analysis was performed
using only eight~\tpr bins, defines the upper limit of the uncertainty
interval for the \PpiTwo[1880] mass and the lower limit for the
\PpiTwo[2005] mass.  The parameters of the \PpiTwo change only
slightly.

The \PpiTwo* resonance parameters are exceptionally sensitive to the
\mThreePi and \tpr dependences of the production probability
$\Abs[0]{\mathcal{P}(\mThreePi, \tpr)}^2$ in
\cref{eq:method:param:spindens,eq:method:param:prods}.  \StudyAF, in
which this factor was set to unity, defines the upper limits of the
systematic uncertainty intervals for the \PpiTwo mass and the
\PpiTwo[1880] width.  It also defines the lower limit for the
\PpiTwo[2005] width.

Studies~\studyS and~\studyR with alternative \chisq~formulations (see
\cref{sec:alt_chi_2}) leave the \PpiTwo parameters virtually
unchanged.  The \PpiTwo[1880] width increases by about
\SI{20}{\MeVcc}.  The strongest effect is observed for the
\PpiTwo[2005] parameters in \StudyR, where the \PpiTwo[2005] mass
increases by \SI{17}{\MeVcc}, and the width decreases by
\SI{63}{\MeVcc}.  The former value defines the upper limit of the
systematic uncertainty interval for the \PpiTwo[2005] mass.

The interference of the $2^{-+}$ wave with the
\wave{0}{-+}{0}{+}{\PfZero[980]}{S} wave affects the widths of the
\PpiTwo and the \PpiTwo[1880].  If the lower limit of the fit range in
the $0^{-+}$ wave is increased from \SI{1.2}{\GeVcc} to
\SI{1.6}{\GeVcc}, the width of the \PpiTwo increases by
\SI{12}{\MeVcc} and that of the \PpiTwo[1880] by \SI{32}{\MeVcc}.  The
former value defines the upper limit of the systematic uncertainty
interval for the \PpiTwo width.  The latter value is close to the
upper limit for the \PpiTwo[1880] width.  The width of the
\PpiTwo[2005] decreases by \SI{90}{\MeVcc}.

When we use the mass shapes of the nonresonant components from the
partial-wave decomposition of a model for the Deck amplitude in
\StudyO, the intensities and interference terms of all four $2^{-+}$
waves are described less well by the model (see
\cref{fig:DeckMC_chi2difference}).  In this study, the fit finds
smaller intensities for the nonresonant components.  In contrast, the
resonance components have larger intensities and exhibit a sizable
destructive interference.  We therefore conclude that the used Deck
model does not describe well the nonresonant components in the
$2^{-+}$ waves.

\subsection{Systematic uncertainties of the \PaFour
  parameters}
\label{sec:syst_uncert_fourPP}

The \PaFour resonance parameters depend only weakly on the set of
waves included in the fit.  This is in particular true for
Studies~\studyD and~\studyA, in which we omitted the $2^{++}$ and
$2^{-+}$ waves from the fit, respectively.  In \StudyB, in which the
\wave{1}{++}{0}{+}{\Prho}{S} and \wave{1}{++}{0}{+}{\PfTwo}{P} waves
are omitted, the \PaFour width increases by \SI{13}{\MeVcc}.

Also the value of the range parameter $q_R$ in the Blatt-Weisskopf
factors influences the \PaFour parameters.  \StudyM, in which $q_R$
was set to \SI{267}{\MeVc} corresponding to an assumed
strong-interaction range of \SI{0.75}{\femto\meter}, defines the lower
limits of the uncertainty intervals for the \PaFour mass and width.

The upper limit of the uncertainty interval for the \PaFour mass is
defined by~\StudyK, in which weaker event-selection criteria lead to
an increased background.

\StudyO, in which the parametrization of the nonresonant amplitude was
replaced by the square root of the intensity distribution of the
partial-wave decomposition of Deck Monte Carlo data generated
according to the model described in \cref{sec:deck_model}, defines the
upper limit of the uncertainty interval for the \PaFour width.  While
the shape of the Deck intensity in the $\PfTwo \pi F$ wave is similar
to that of the nonresonant component found in the main fit, it
deviates in the $\Prho \pi G$ wave leading to a worse description of
the data (see \cref{fig:DeckMC_chi2difference}).
 %
%
%

\section*{Acknowledgements}
\label{sec:acknowledgements}

We have received many suggestions and input during a series of PWA
workshops: a joint COMPASS-JLab-GSI Workshop on Physics and Methods in
Meson Spectroscopy (Garching/2008), Workshops on Spectroscopy at
COMPASS held 2009 and 2011 in Garching, and in the context of the
ATHOS workshop series (Camogli/2012, Kloster Seeon/2013, Ashburn/2015,
and Bad Honnef/2017).  We are especially indebted to V.~Mathieu,
W.~Ochs, J.~Pelaez, M.~Pennington, and A.~Szczepaniak for their help
and suggestions.  S.U.~Chung would like to thank the IAS at the TU
M\"unchen and together with D.~Ryabchikov the Excellence Cluster
\enquote{Universe} for supporting many visits to Munich during the
past years.

We gratefully acknowledge the support of the CERN management and staff
as well as the skills and efforts of the technicians of the
collaborating institutions.  This work is supported by MEYS (Czech
Republic); \enquote{HadronPhysics3} Integrating Activity in FP7
(European Union); CEA, Laboratoire d'Excellence P2IO and ANR (France);
BMBF, DFG cluster of excellence \enquote{Origin and Structure of the
  Universe}, the DFG Collaborative Research Centre/Transregio~110, the
computing facilities of the Computational Center for Particle and
Astrophysics (C2PAP), IAS-TUM, and Humboldt Foundation (Germany); SAIL
(CSR) (India); ISF (Israel); INFN (Italy); MEXT, JSPS, Daiko, and
Yamada Foundations (Japan); NRF (Republic of Korea); NCN (Poland); FCT
(Portugal).

%
%
%
\bibliographystyle{utphys_bgrube}
\providecommand{\href}[2]{#2}\begingroup\raggedright\endgroup
\clearpage
\section*{\textsc{Supplemental Material}}
\addtocontents{toc}{\protect\contentsline{section}{\textsc{Supplemental Material}}{}{}}
\clearpage
\clearpage{}%
%
%

In this supplemental material\ifMultiColumnLayout{ to
  \refCite{paper3}}{}, we provide additional information necessary to
repeat the analysis.  In \cref{sec:spin-dens_matrices}, we present the
full data set together with the result of the resonance-model fit.  In
\cref{sec:phase-space_vol}, we provide the decay phase-space integrals
$I_{a a}$ that enter \ifMultiColumnLayout{Eqs.~(2) and~(3) in Sec.~IV
  of
  \refCite{paper3}}{\cref{eq:method:transitionampl,eq:method:param:spindens}}.
The data required to perform the resonance-model fit are also provided
in computer-readable format at~\cite{paper3_hepdata}.

\section{Spin-density matrices in \tpr bins}
\label{sec:spin-dens_matrices}

In this section, we present the used data from the partial-wave
analysis presented in \refCite{Adolph:2015tqa} together with the
result of the resonance-model fit.  The measured spin-density matrix
elements of the 14~selected waves and the fit model are presented in
terms of the partial-wave intensities and the relative phases between
the partial waves, which are visualized in the form of a
$14 \times 14$ upper-triangular matrix of graphs.  The \mThreePi
dependence of the intensities is shown as diagonal elements.  The
\mThreePi dependence of the relative phases is shown in the
off-diagonal elements.  A relative phase $\Delta \phi_{ab}$ between
two waves is defined as the phase difference of partial wave~$a$ in
the row and partial wave~$b$ in the column of the matrix:
$\Delta \phi_{ab} \equiv \phi_a - \phi_b$.  In order to be able to
show all phases with a common axis, we plot instead
$\Delta \phi_{ab} - \delta \phi_{ab}$.  The phase offset
$\delta \phi_{ab}$ is calculated as the arithmetic average of the
minimum and maximum value of $\Delta \phi_{ab}$ in the respective fit
range.  For each phase motion, the value of $\delta \phi_{ab}$ is
given in the corresponding graph in the matrix.

In each graph of the matrix, the data from the partial-wave analysis
are shown by black crosses with horizontal lines that indicate the
\mThreePi bin width and vertical lines that indicate the statistical
uncertainties.  The data are overlaid by the red model curve.  In each
intensity distribution, also the resonances (blue curves) and the
nonresonant component (green curve) are shown.  All wave components
interfere among each other so that in general the intensities of the
wave components do not add up to the model curve.  In each graph of
the matrix, points outside the fit range are shown in gray.  The
extrapolations of the model curve and of the curves of the wave
components outside the fit range are shown in lighter colors.

Due to the large size of the matrix, it is broken down into
10~submatrices labeled~A through~J.  This is illustrated in
\cref{tab:spin-dens_matrix_overview}.  Each of the following
\crefrange{sec:spin-dens_submatrix_1}{sec:spin-dens_submatrix_10}
shows the corresponding submatrix in the 11~bins of the analyzed \tpr range
from \SIrange{0.1}{1.0}{\GeVcsq}.

\newcommand{\mathResult}[1]{\pgfmathparse{#1}\edef\tempMathResult{\pgfmathresult}}
\newcommand*{\sdmLabels}{{"A","B","C","D","E","F","G","H","I","J"}}
\newcommand*{\sdmColors}{{1,0,1,0,1,0,1,1,0,1}}
\DeclareRobustCommand{\sdmLabel}[1]{%
  \mathResult{\sdmLabels[#1-1]}%
  \tempMathResult%
}
\newcommand*{\sdmColor}[1]{%
  \mathResult{\sdmColors[#1-1]}%
  \ifthenelse{\isodd{\tempMathResult}}%
  {\cellcolor{blue!15}}%
  {\cellcolor{green!15}}%
}

\ifMultiColumnLayout{\begin{table}[!b]}{\begin{table}[tbp]}
  \newcommand*{\rotHd}[1]{\multicolumn{1}{l}{\rlap{\rotatebox{60}{#1}~}}}
  \newcommand*{\sdm}[1]{%
    \sdmColor{#1}%
    \hyperref[sec:spin-dens_submatrix_#1]{\sdmLabel{#1}}%
  }
  \renewcommand{\arraystretch}{1.2}
  \centering
  \caption{Subdivision scheme for the $14 \times 14$ matrix of graphs
    that represents the spin-density matrix.  The \mThreePi dependence
    of the partial-wave intensities are shown as
    diagonal elements, the \mThreePi dependence of the relative phases
    as off-diagonal elements.  The matrix is
    subdivided into 10~submatrices labeled A~through~J.  Each
    submatrix is shown in all 11~\tpr bins in the corresponding
    \crefrange{sec:spin-dens_submatrix_1}{sec:spin-dens_submatrix_10}.}
  \label{tab:spin-dens_matrix_overview}
  \begin{tabular}{ccccccccccccccl}

    \rotHd{\wave{0}{-+}{0}{+}{\PfZero}{S}} &
    \rotHd{\wave{1}{++}{0}{+}{\Prho}{S}}   &
    \rotHd{\wave{1}{++}{0}{+}{\PfZero}{P}} &
    \rotHd{\wave{1}{++}{0}{+}{\PfTwo}{P}}  &

    \rotHd{\wave{1}{-+}{1}{+}{\Prho}{P}}   &
    \rotHd{\wave{2}{++}{1}{+}{\Prho}{D}}   &
    \rotHd{\wave{2}{++}{2}{+}{\Prho}{D}}   &
    \rotHd{\wave{2}{++}{1}{+}{\PfTwo}{P}}  &

    \rotHd{\wave{2}{-+}{0}{+}{\Prho}{F}}   &
    \rotHd{\wave{2}{-+}{0}{+}{\PfTwo}{S}}  &
    \rotHd{\wave{2}{-+}{1}{+}{\PfTwo}{S}}  &
    \rotHd{\wave{2}{-+}{0}{+}{\PfTwo}{D}}  &

    \rotHd{\wave{4}{++}{1}{+}{\Prho}{G}}   &
    \rotHd{\wave{4}{++}{1}{+}{\PfTwo}{F}}  &
    \\

  	\sdm{1} & \sdm{1} & \sdm{1} & \sdm{1} & \sdm{2} & \sdm{2} & \sdm{2} & \sdm{2} & \sdm{3} & \sdm{3} & \sdm{3} & \sdm{3} & \sdm{4}  & \sdm{4}  & \wave{0}{-+}{0}{+}{\PfZero}{S} \\
  	        & \sdm{1} & \sdm{1} & \sdm{1} & \sdm{2} & \sdm{2} & \sdm{2} & \sdm{2} & \sdm{3} & \sdm{3} & \sdm{3} & \sdm{3} & \sdm{4}  & \sdm{4}  & \wave{1}{++}{0}{+}{\Prho}{S}   \\
  	        &         & \sdm{1} & \sdm{1} & \sdm{2} & \sdm{2} & \sdm{2} & \sdm{2} & \sdm{3} & \sdm{3} & \sdm{3} & \sdm{3} & \sdm{4}  & \sdm{4}  & \wave{1}{++}{0}{+}{\PfZero}{P} \\
  	        &         &         & \sdm{1} & \sdm{2} & \sdm{2} & \sdm{2} & \sdm{2} & \sdm{3} & \sdm{3} & \sdm{3} & \sdm{3} & \sdm{4}  & \sdm{4}  & \wave{1}{++}{0}{+}{\PfTwo}{P}  \\

  	        &         &         &         & \sdm{5} & \sdm{5} & \sdm{5} & \sdm{5} & \sdm{6} & \sdm{6} & \sdm{6} & \sdm{6} & \sdm{7}  & \sdm{7}  & \wave{1}{-+}{1}{+}{\Prho}{P}   \\
  	        &         &         &         &         & \sdm{5} & \sdm{5} & \sdm{5} & \sdm{6} & \sdm{6} & \sdm{6} & \sdm{6} & \sdm{7}  & \sdm{7}  & \wave{2}{++}{1}{+}{\Prho}{D}   \\
  	        &         &         &         &         &         & \sdm{5} & \sdm{5} & \sdm{6} & \sdm{6} & \sdm{6} & \sdm{6} & \sdm{7}  & \sdm{7}  & \wave{2}{++}{2}{+}{\Prho}{D}   \\
  	        &         &         &         &         &         &         & \sdm{5} & \sdm{6} & \sdm{6} & \sdm{6} & \sdm{6} & \sdm{7}  & \sdm{7}  & \wave{2}{++}{1}{+}{\PfTwo}{P}  \\

  	        &         &         &         &         &         &         &         & \sdm{8} & \sdm{8} & \sdm{8} & \sdm{8} & \sdm{9}  & \sdm{9}  & \wave{2}{-+}{0}{+}{\Prho}{F}   \\
  	        &         &         &         &         &         &         &         &         & \sdm{8} & \sdm{8} & \sdm{8} & \sdm{9}  & \sdm{9}  & \wave{2}{-+}{0}{+}{\PfTwo}{S}  \\
  	        &         &         &         &         &         &         &         &         &         & \sdm{8} & \sdm{8} & \sdm{9}  & \sdm{9}  & \wave{2}{-+}{1}{+}{\PfTwo}{S}  \\
  	        &         &         &         &         &         &         &         &         &         &         & \sdm{8} & \sdm{9}  & \sdm{9}  & \wave{2}{-+}{0}{+}{\PfTwo}{D}  \\

  	        &         &         &         &         &         &         &         &         &         &         &         & \sdm{10} & \sdm{10} & \wave{4}{++}{1}{+}{\Prho}{G}   \\
  	        &         &         &         &         &         &         &         &         &         &         &         &          & \sdm{10} & \wave{4}{++}{1}{+}{\PfTwo}{F}  \\

  \end{tabular}
\end{table}

\ifMultiColumnLayout{\onecolumngrid}{}
\clearpage
\newlength{\blockDistanceToTop}
\ifMultiColumnLayout{\setlength{\blockDistanceToTop}{0.1\paperheight}}{\setlength{\blockDistanceToTop}{0.15\paperheight}}
\newlength{\matrixHeight}
\ifMultiColumnLayout{\setlength{\matrixHeight}{0.65\paperheight}}{\setlength{\matrixHeight}{0.6\paperheight}}
\newlength{\matrixHeightHalf}
\setlength{\matrixHeightHalf}{0.5\matrixHeight}
\subsection{Submatrix A}
\label{sec:spin-dens_submatrix_1}

\begin{textblock*}{\textwidth}[0.5,0](0.5\paperwidth,\blockDistanceToTop)
 \begin{minipage}{\textwidth}
   \makeatletter
   \def\@captype{figure}
   \makeatother
   \centering
   \includegraphics[height=\matrixHeight]{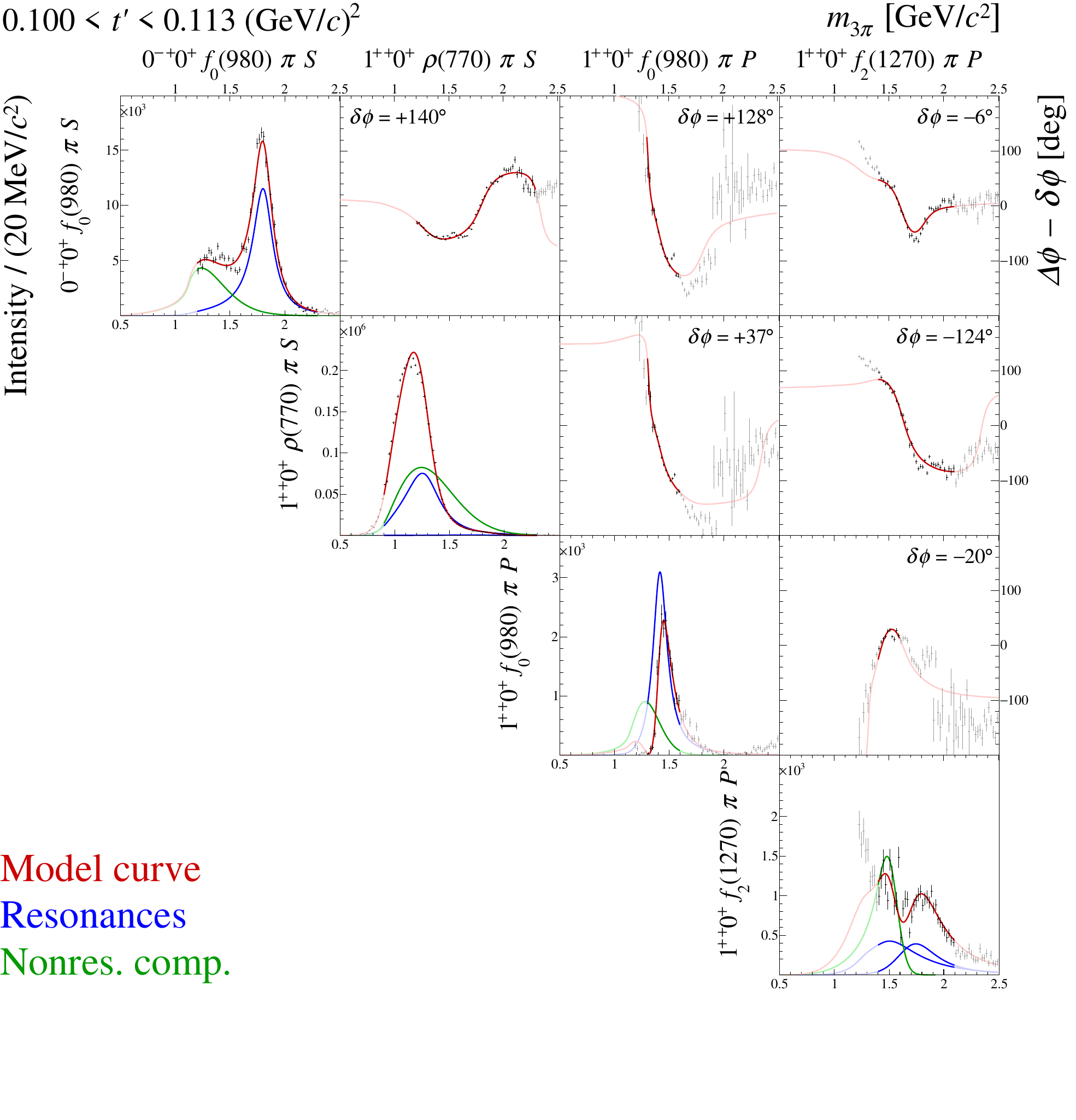}%
   \caption{Submatrix~A of the $14 \times 14$ matrix of graphs that
     represents the spin-density matrix (see
     \cref{tab:spin-dens_matrix_overview}).}
   \label{fig:spin-dens_submatrix_1_tbin_1}
 \end{minipage}
\end{textblock*}

\newpage\null
\begin{textblock*}{\textwidth}[0.5,0](0.5\paperwidth,\blockDistanceToTop)
 \begin{minipage}{\textwidth}
   \makeatletter
   \def\@captype{figure}
   \makeatother
   \centering
   \includegraphics[height=\matrixHeight]{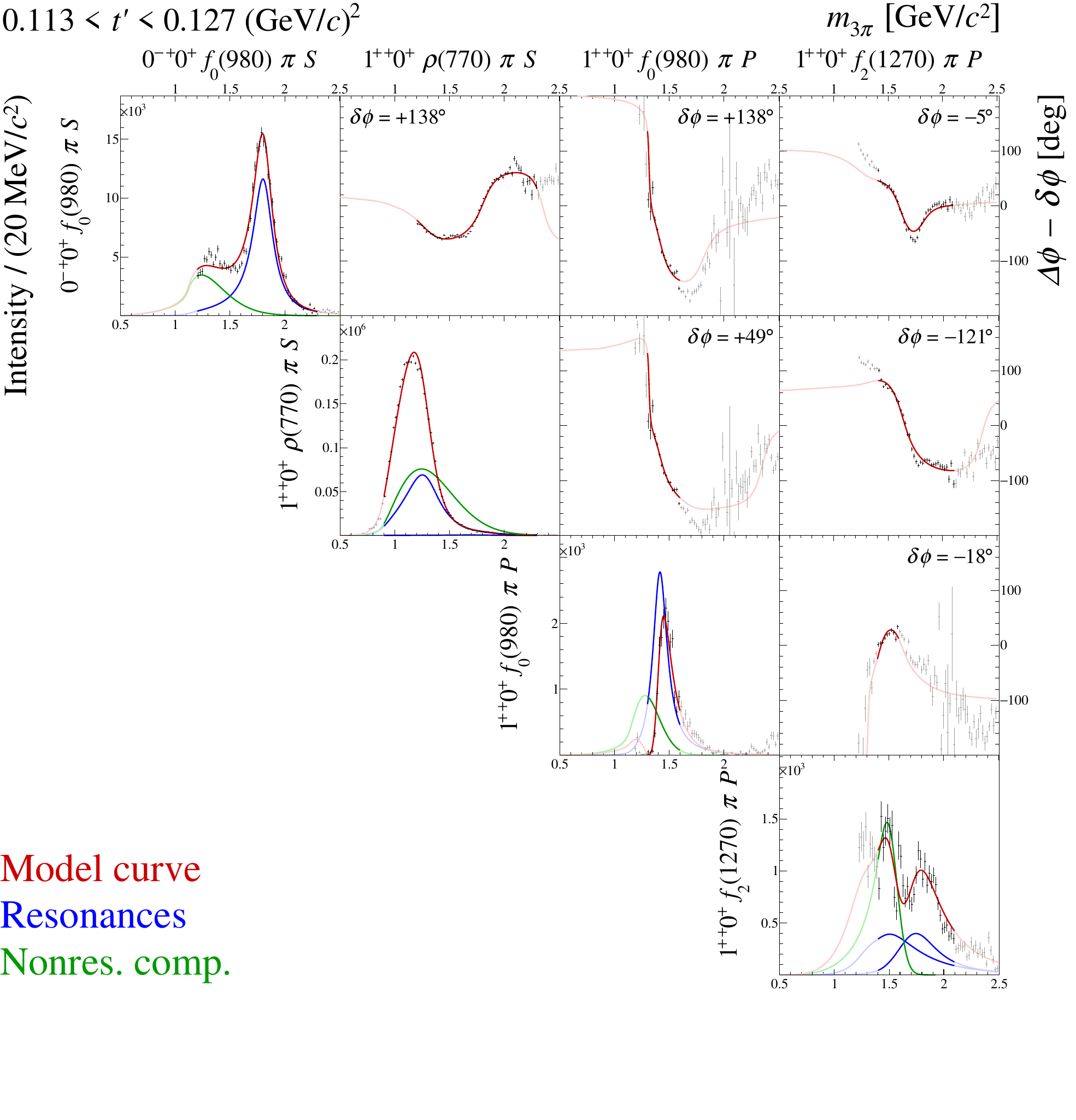}%
   \caption{Submatrix~A of the $14 \times 14$ matrix of graphs that
     represents the spin-density matrix (see
     \cref{tab:spin-dens_matrix_overview}).}
   \label{fig:spin-dens_submatrix_1_tbin_2}
 \end{minipage}
\end{textblock*}

\newpage\null
\begin{textblock*}{\textwidth}[0.5,0](0.5\paperwidth,\blockDistanceToTop)
 \begin{minipage}{\textwidth}
   \makeatletter
   \def\@captype{figure}
   \makeatother
   \centering
   \includegraphics[height=\matrixHeight]{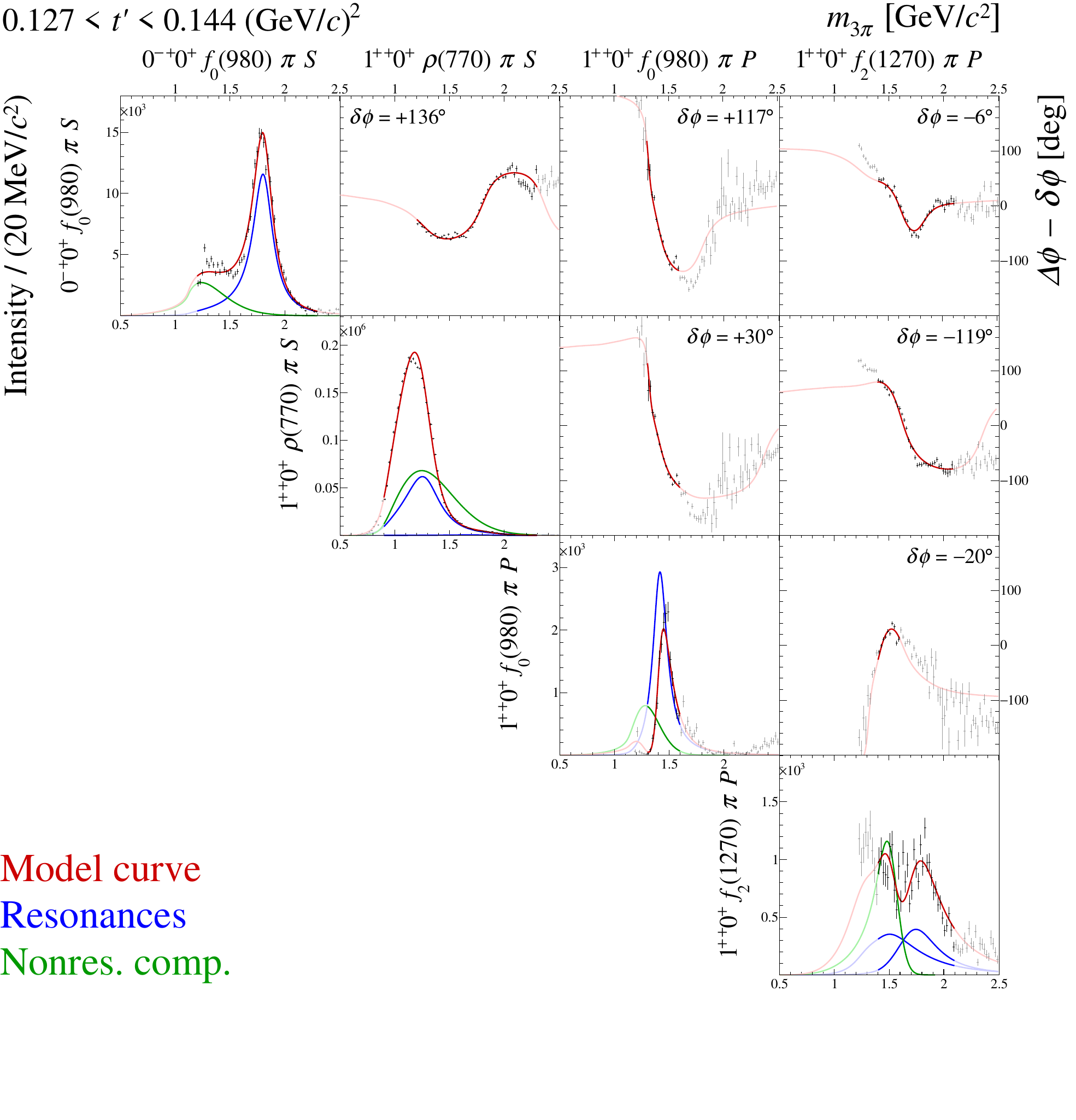}%
   \caption{Submatrix~A of the $14 \times 14$ matrix of graphs that
     represents the spin-density matrix (see
     \cref{tab:spin-dens_matrix_overview}).}
   \label{fig:spin-dens_submatrix_1_tbin_3}
 \end{minipage}
\end{textblock*}

\newpage\null
\begin{textblock*}{\textwidth}[0.5,0](0.5\paperwidth,\blockDistanceToTop)
 \begin{minipage}{\textwidth}
   \makeatletter
   \def\@captype{figure}
   \makeatother
   \centering
   \includegraphics[height=\matrixHeight]{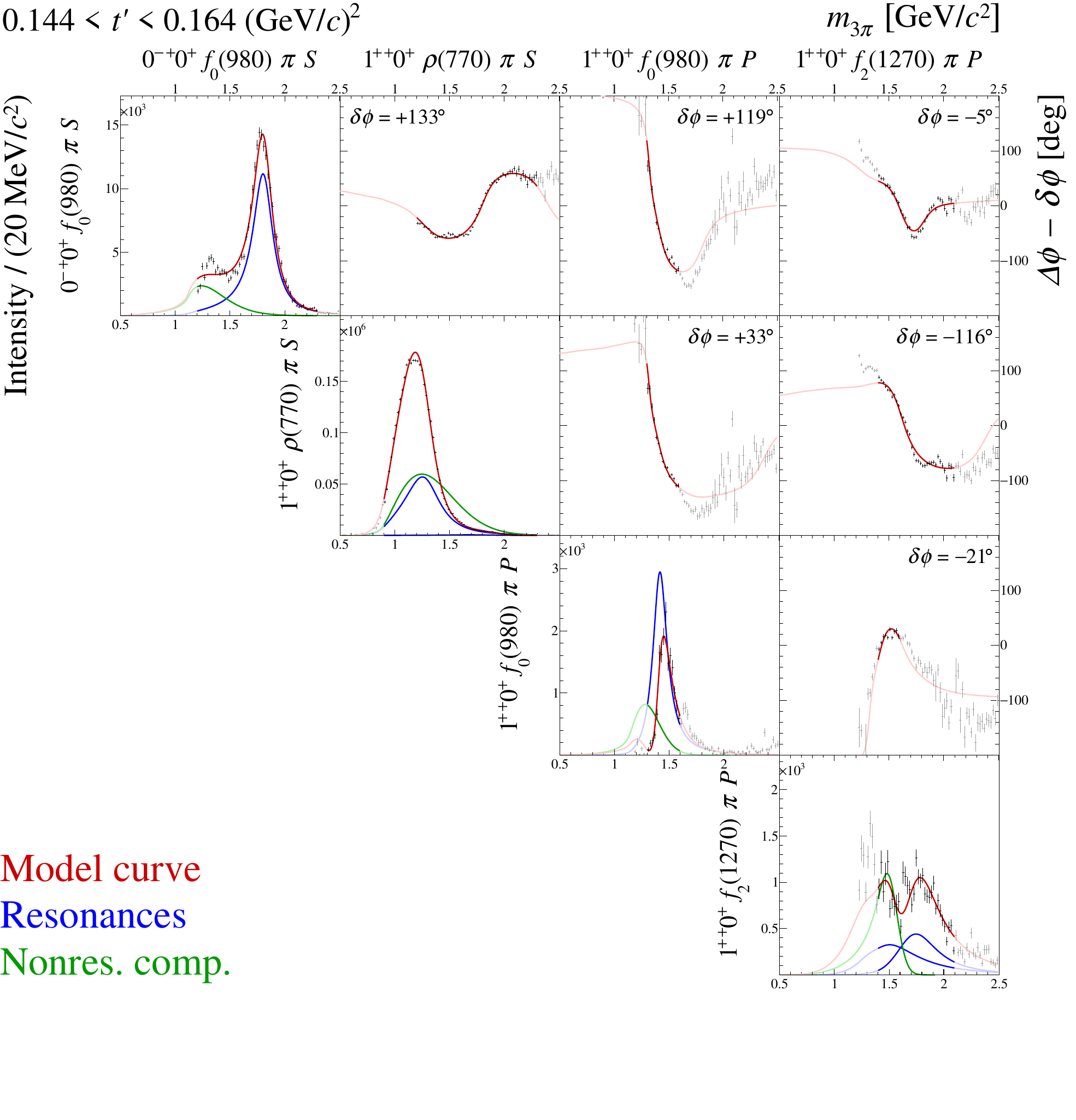}%
   \caption{Submatrix~A of the $14 \times 14$ matrix of graphs that
     represents the spin-density matrix (see
     \cref{tab:spin-dens_matrix_overview}).}
   \label{fig:spin-dens_submatrix_1_tbin_4}
 \end{minipage}
\end{textblock*}

\newpage\null
\begin{textblock*}{\textwidth}[0.5,0](0.5\paperwidth,\blockDistanceToTop)
 \begin{minipage}{\textwidth}
   \makeatletter
   \def\@captype{figure}
   \makeatother
   \centering
   \includegraphics[height=\matrixHeight]{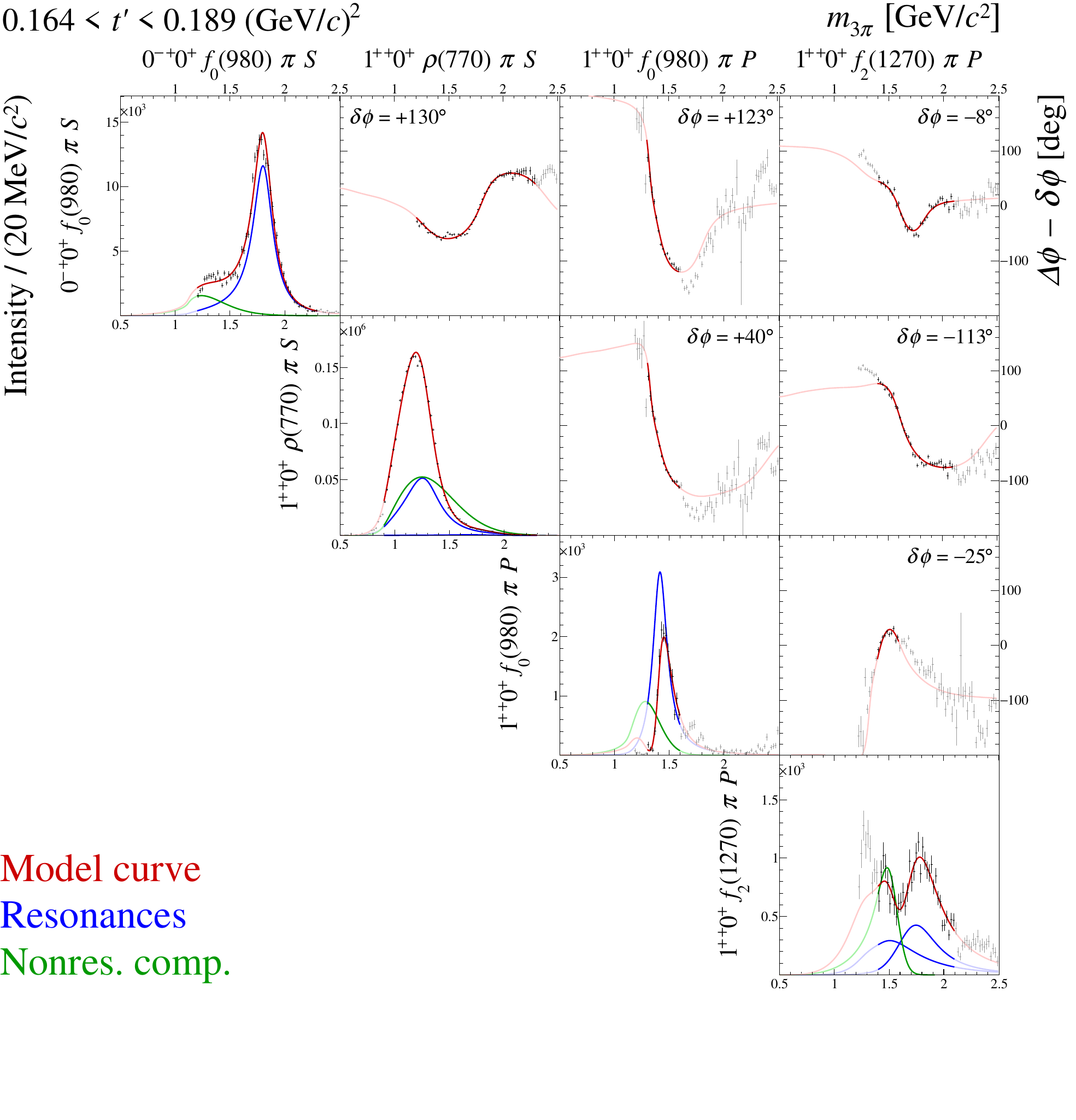}%
   \caption{Submatrix~A of the $14 \times 14$ matrix of graphs that
     represents the spin-density matrix (see
     \cref{tab:spin-dens_matrix_overview}).}
   \label{fig:spin-dens_submatrix_1_tbin_5}
 \end{minipage}
\end{textblock*}

\newpage\null
\begin{textblock*}{\textwidth}[0.5,0](0.5\paperwidth,\blockDistanceToTop)
 \begin{minipage}{\textwidth}
   \makeatletter
   \def\@captype{figure}
   \makeatother
   \centering
   \includegraphics[height=\matrixHeight]{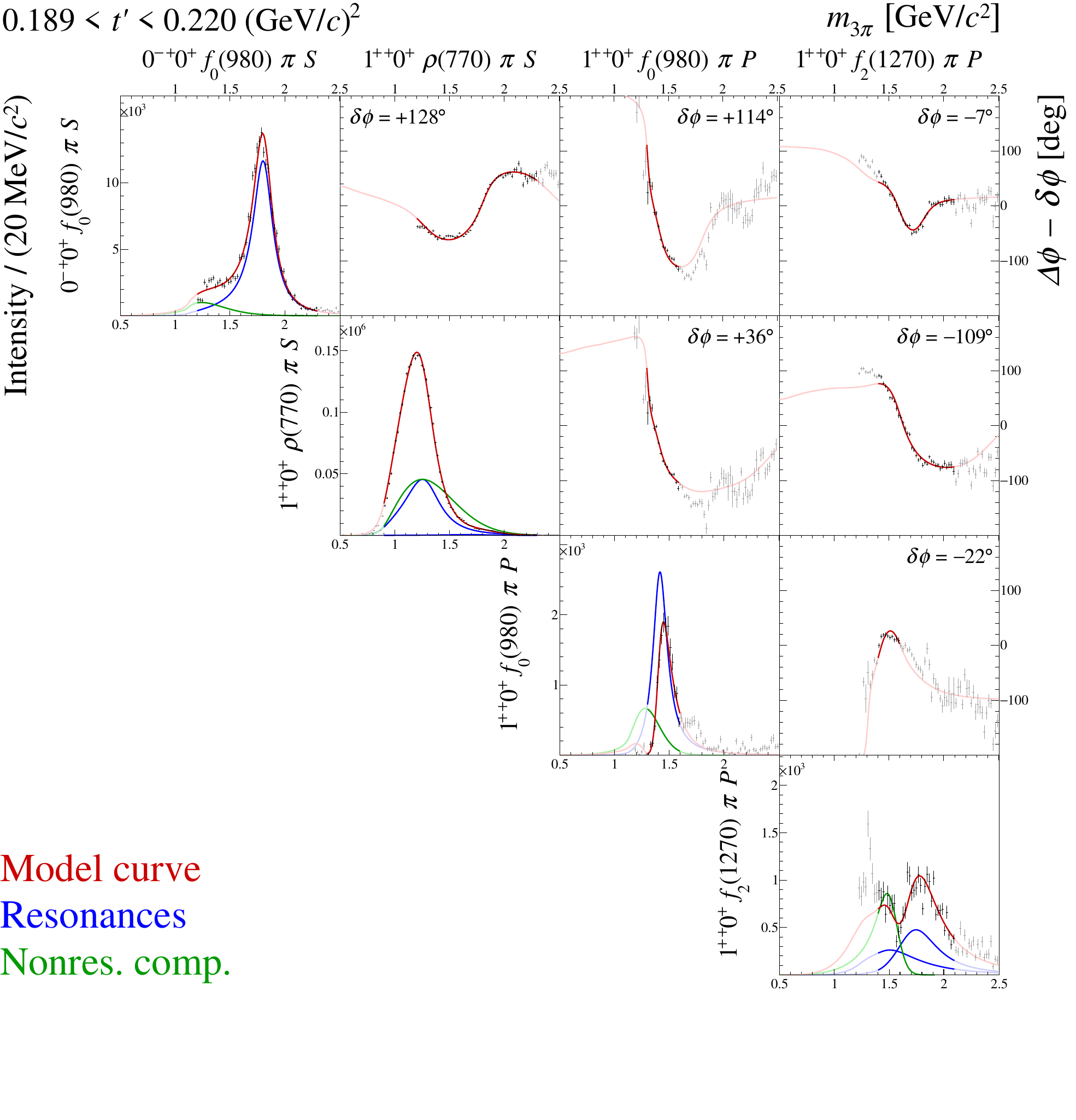}%
   \caption{Submatrix~A of the $14 \times 14$ matrix of graphs that
     represents the spin-density matrix (see
     \cref{tab:spin-dens_matrix_overview}).}
   \label{fig:spin-dens_submatrix_1_tbin_6}
 \end{minipage}
\end{textblock*}

\newpage\null
\begin{textblock*}{\textwidth}[0.5,0](0.5\paperwidth,\blockDistanceToTop)
 \begin{minipage}{\textwidth}
   \makeatletter
   \def\@captype{figure}
   \makeatother
   \centering
   \includegraphics[height=\matrixHeight]{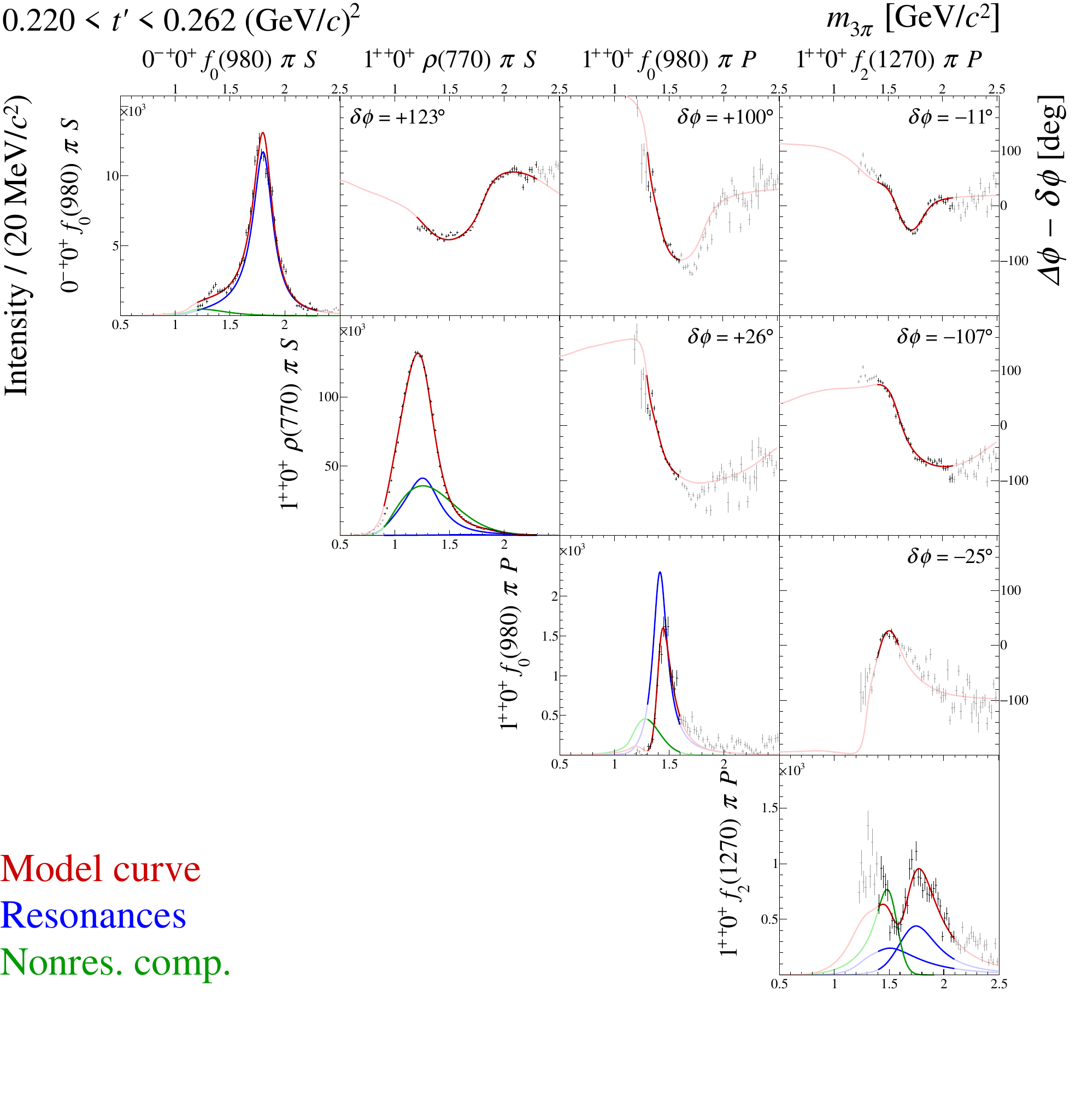}%
   \caption{Submatrix~A of the $14 \times 14$ matrix of graphs that
     represents the spin-density matrix (see
     \cref{tab:spin-dens_matrix_overview}).}
   \label{fig:spin-dens_submatrix_1_tbin_7}
 \end{minipage}
\end{textblock*}

\newpage\null
\begin{textblock*}{\textwidth}[0.5,0](0.5\paperwidth,\blockDistanceToTop)
 \begin{minipage}{\textwidth}
   \makeatletter
   \def\@captype{figure}
   \makeatother
   \centering
   \includegraphics[height=\matrixHeight]{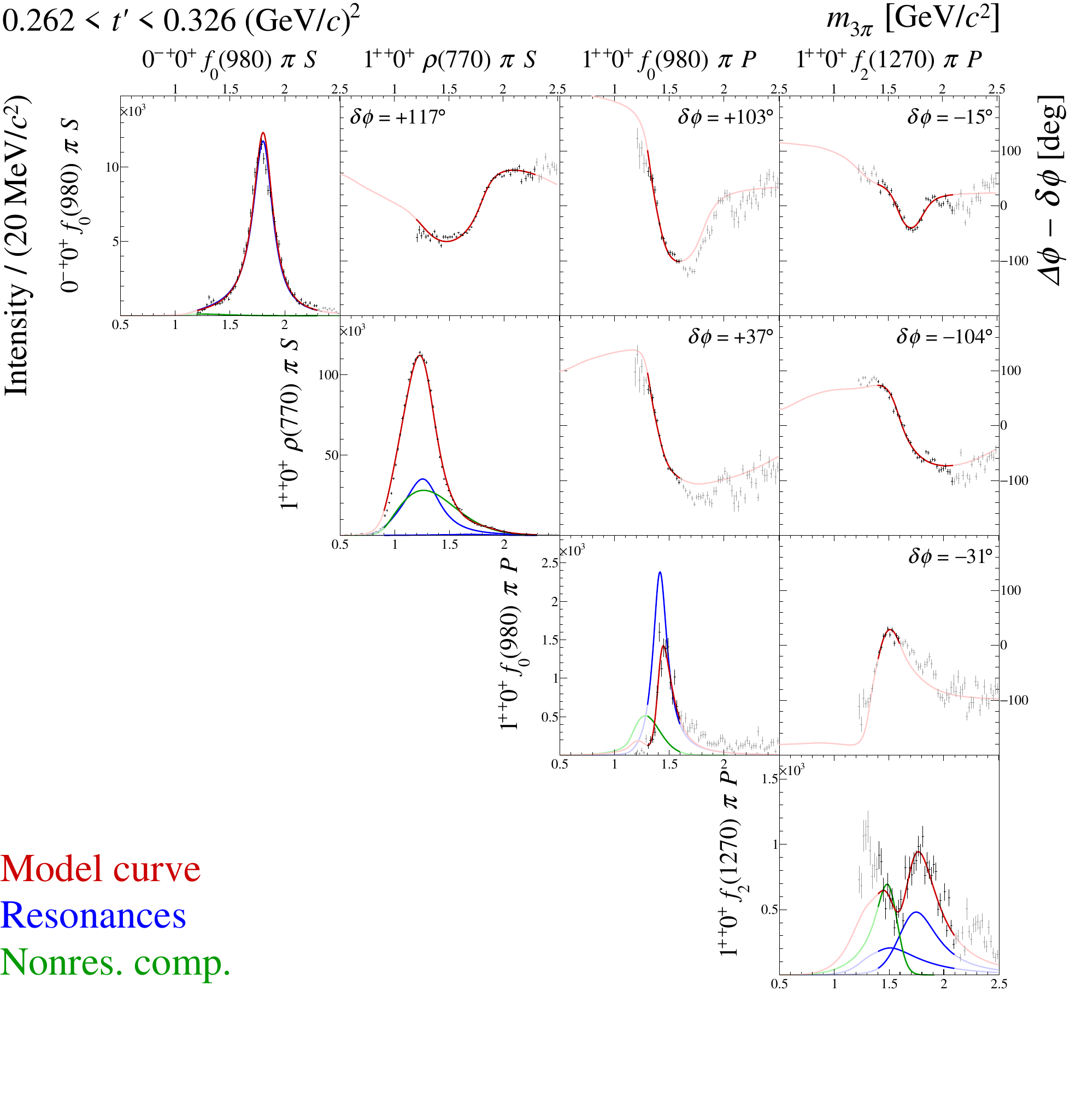}%
   \caption{Submatrix~A of the $14 \times 14$ matrix of graphs that
     represents the spin-density matrix (see
     \cref{tab:spin-dens_matrix_overview}).}
   \label{fig:spin-dens_submatrix_1_tbin_8}
 \end{minipage}
\end{textblock*}

\newpage\null
\begin{textblock*}{\textwidth}[0.5,0](0.5\paperwidth,\blockDistanceToTop)
 \begin{minipage}{\textwidth}
   \makeatletter
   \def\@captype{figure}
   \makeatother
   \centering
   \includegraphics[height=\matrixHeight]{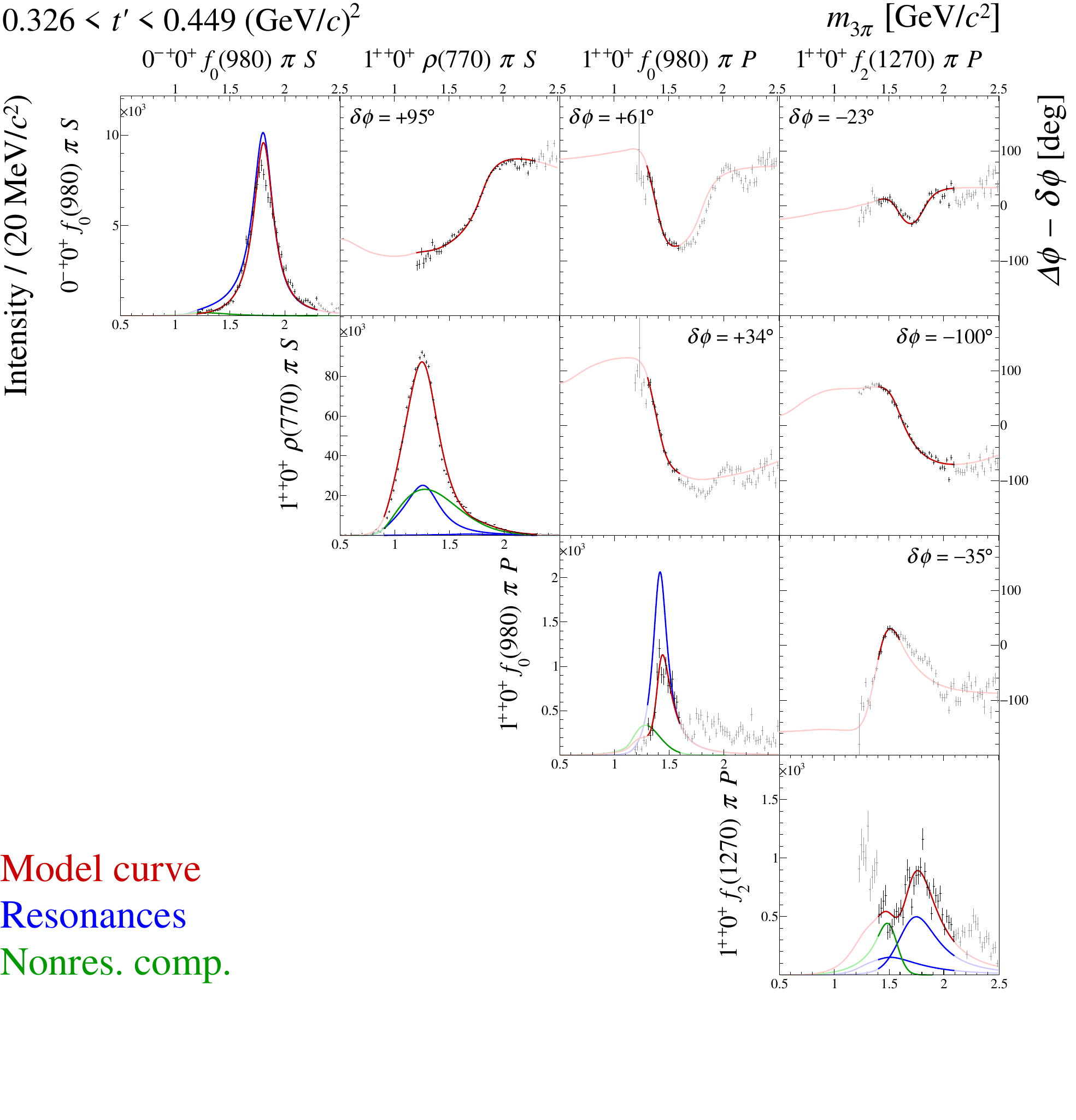}%
   \caption{Submatrix~A of the $14 \times 14$ matrix of graphs that
     represents the spin-density matrix (see
     \cref{tab:spin-dens_matrix_overview}).}
   \label{fig:spin-dens_submatrix_1_tbin_9}
 \end{minipage}
\end{textblock*}

\newpage\null
\begin{textblock*}{\textwidth}[0.5,0](0.5\paperwidth,\blockDistanceToTop)
 \begin{minipage}{\textwidth}
   \makeatletter
   \def\@captype{figure}
   \makeatother
   \centering
   \includegraphics[height=\matrixHeight]{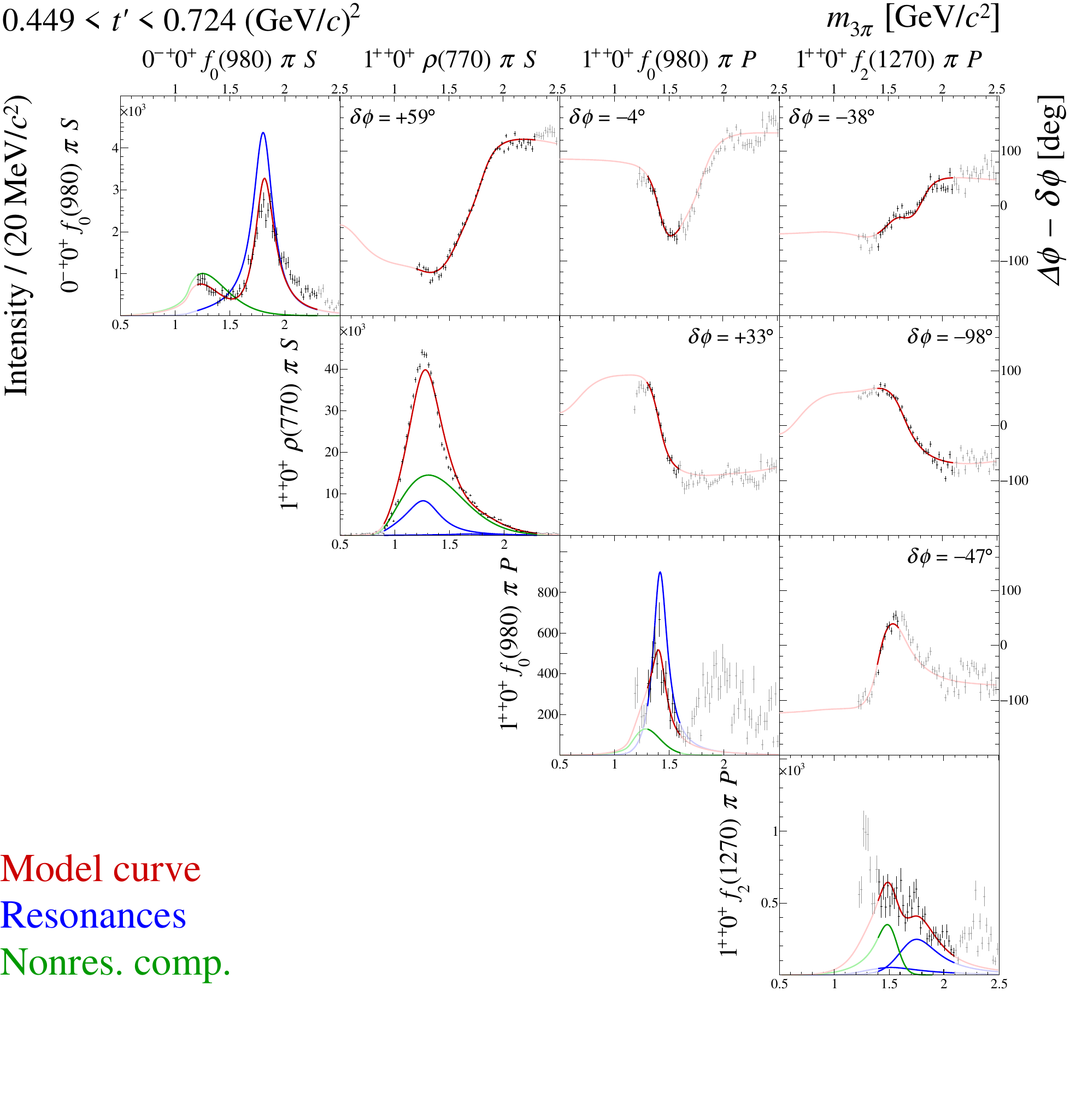}%
   \caption{Submatrix~A of the $14 \times 14$ matrix of graphs that
     represents the spin-density matrix (see
     \cref{tab:spin-dens_matrix_overview}).}
   \label{fig:spin-dens_submatrix_1_tbin_10}
 \end{minipage}
\end{textblock*}

\newpage\null
\begin{textblock*}{\textwidth}[0.5,0](0.5\paperwidth,\blockDistanceToTop)
 \begin{minipage}{\textwidth}
   \makeatletter
   \def\@captype{figure}
   \makeatother
   \centering
   \includegraphics[height=\matrixHeight]{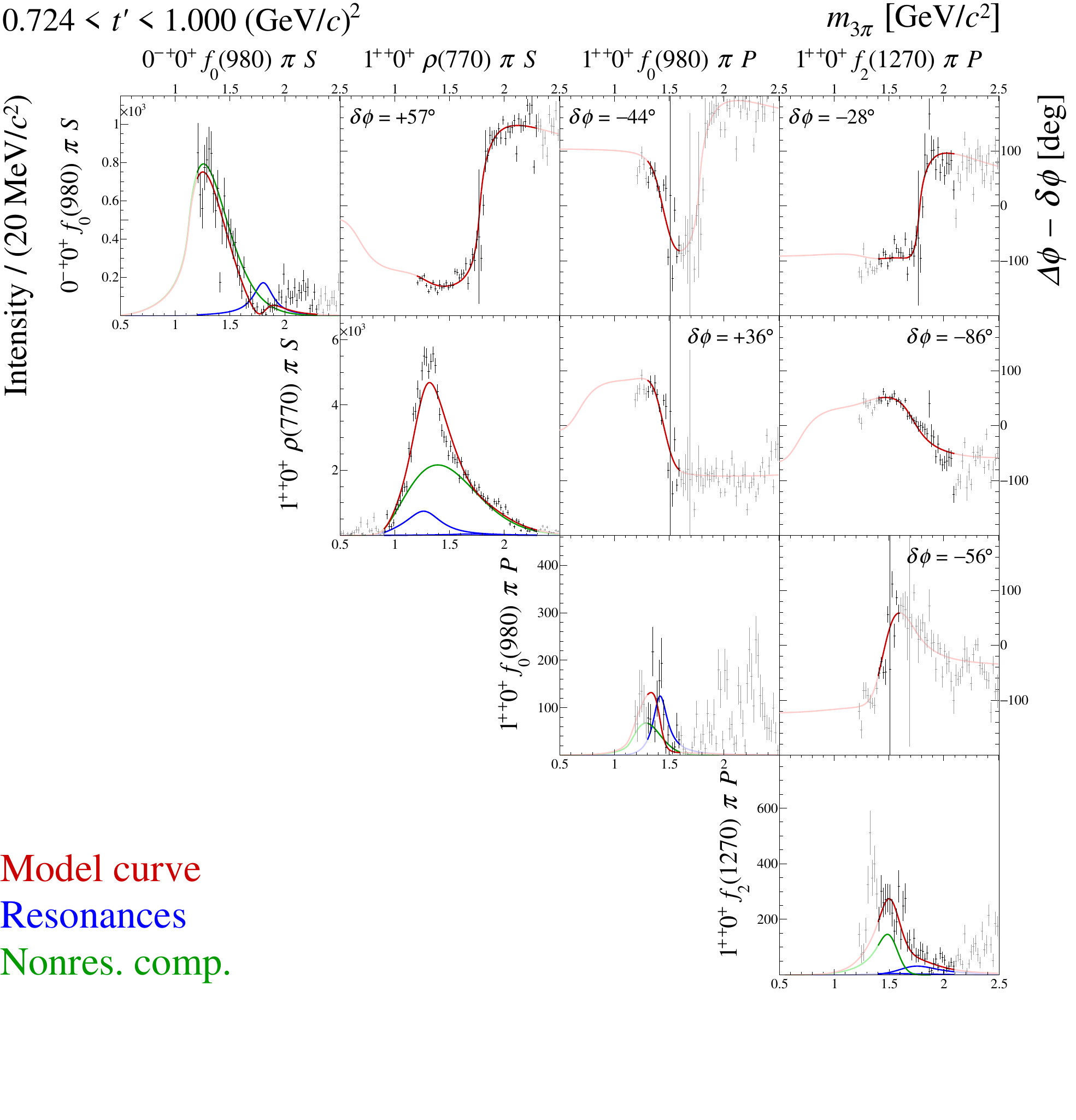}%
   \caption{Submatrix~A of the $14 \times 14$ matrix of graphs that
     represents the spin-density matrix (see
     \cref{tab:spin-dens_matrix_overview}).}
   \label{fig:spin-dens_submatrix_1_tbin_11}
 \end{minipage}
\end{textblock*}

\clearpage
\subsection{Submatrix B}
\label{sec:spin-dens_submatrix_2}

\begin{textblock*}{\textwidth}[0.5,0](0.5\paperwidth,\blockDistanceToTop)
 \begin{minipage}{\textwidth}
   \makeatletter
   \def\@captype{figure}
   \makeatother
   \centering
   \includegraphics[height=\matrixHeight]{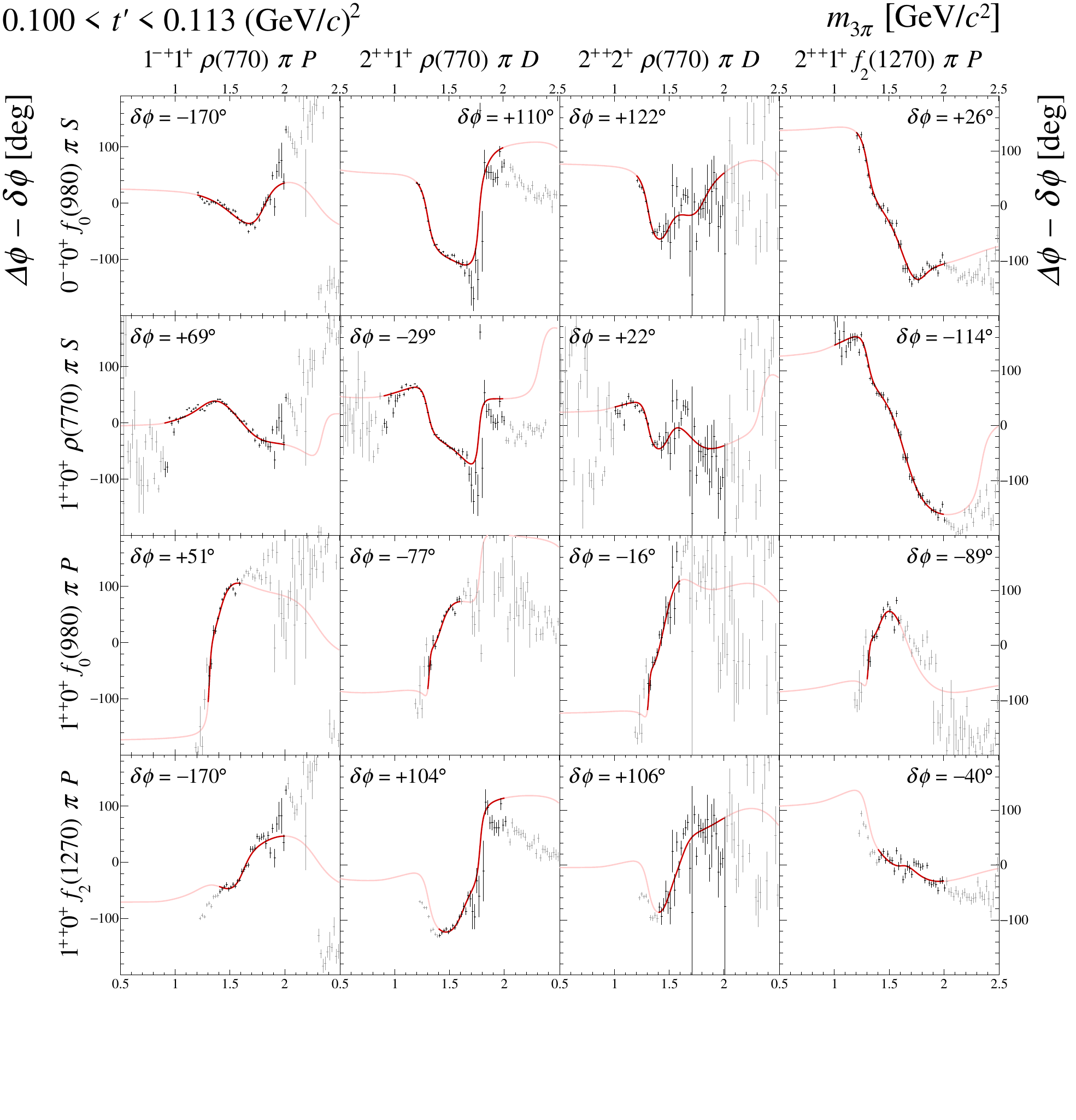}%
   \caption{Submatrix~B of the $14 \times 14$ matrix of graphs that
     represents the spin-density matrix (see
     \cref{tab:spin-dens_matrix_overview}).}
   \label{fig:spin-dens_submatrix_2_tbin_1}
 \end{minipage}
\end{textblock*}

\newpage\null
\begin{textblock*}{\textwidth}[0.5,0](0.5\paperwidth,\blockDistanceToTop)
 \begin{minipage}{\textwidth}
   \makeatletter
   \def\@captype{figure}
   \makeatother
   \centering
   \includegraphics[height=\matrixHeight]{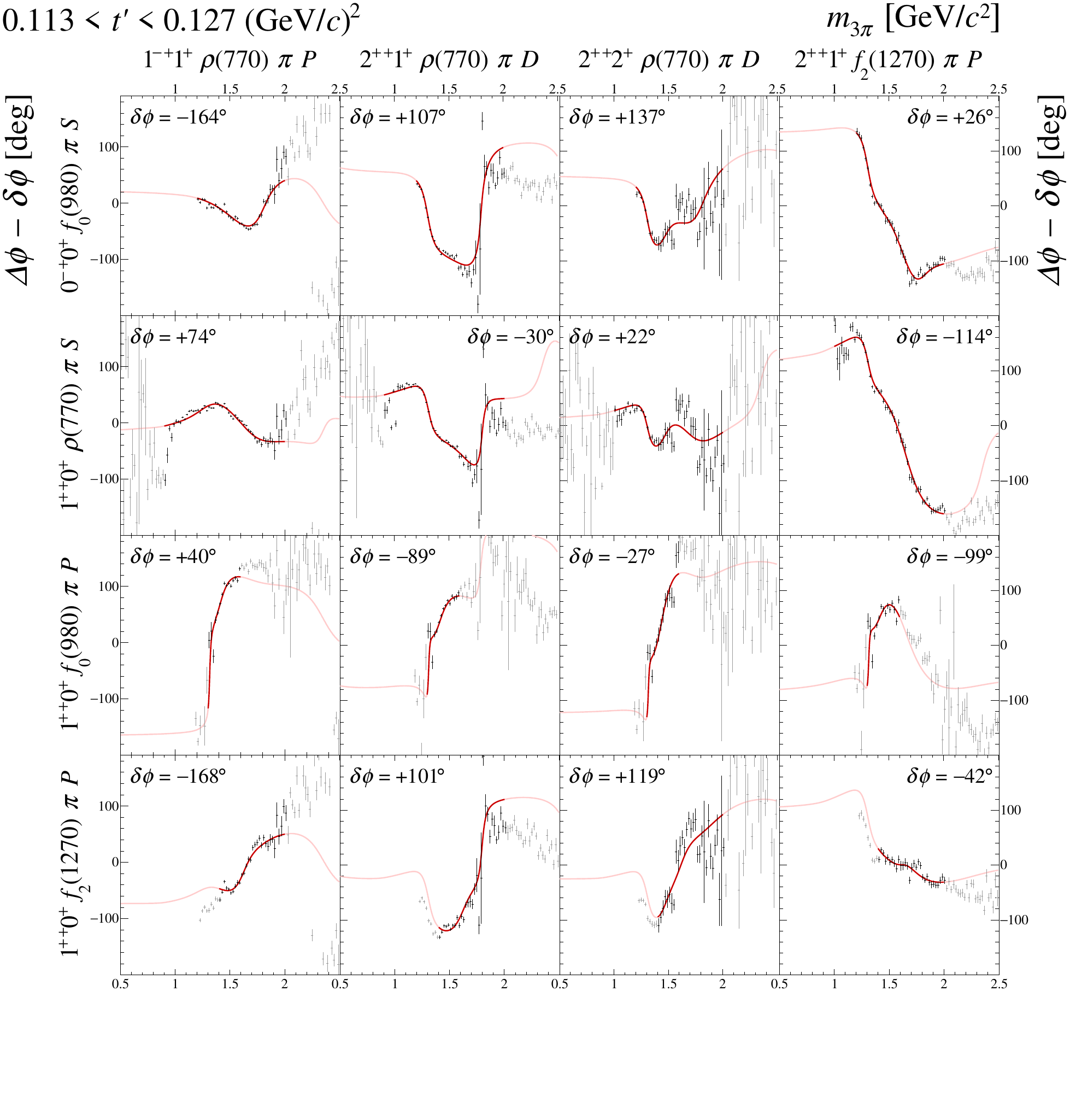}%
   \caption{Submatrix~B of the $14 \times 14$ matrix of graphs that
     represents the spin-density matrix (see
     \cref{tab:spin-dens_matrix_overview}).}
   \label{fig:spin-dens_submatrix_2_tbin_2}
 \end{minipage}
\end{textblock*}

\newpage\null
\begin{textblock*}{\textwidth}[0.5,0](0.5\paperwidth,\blockDistanceToTop)
 \begin{minipage}{\textwidth}
   \makeatletter
   \def\@captype{figure}
   \makeatother
   \centering
   \includegraphics[height=\matrixHeight]{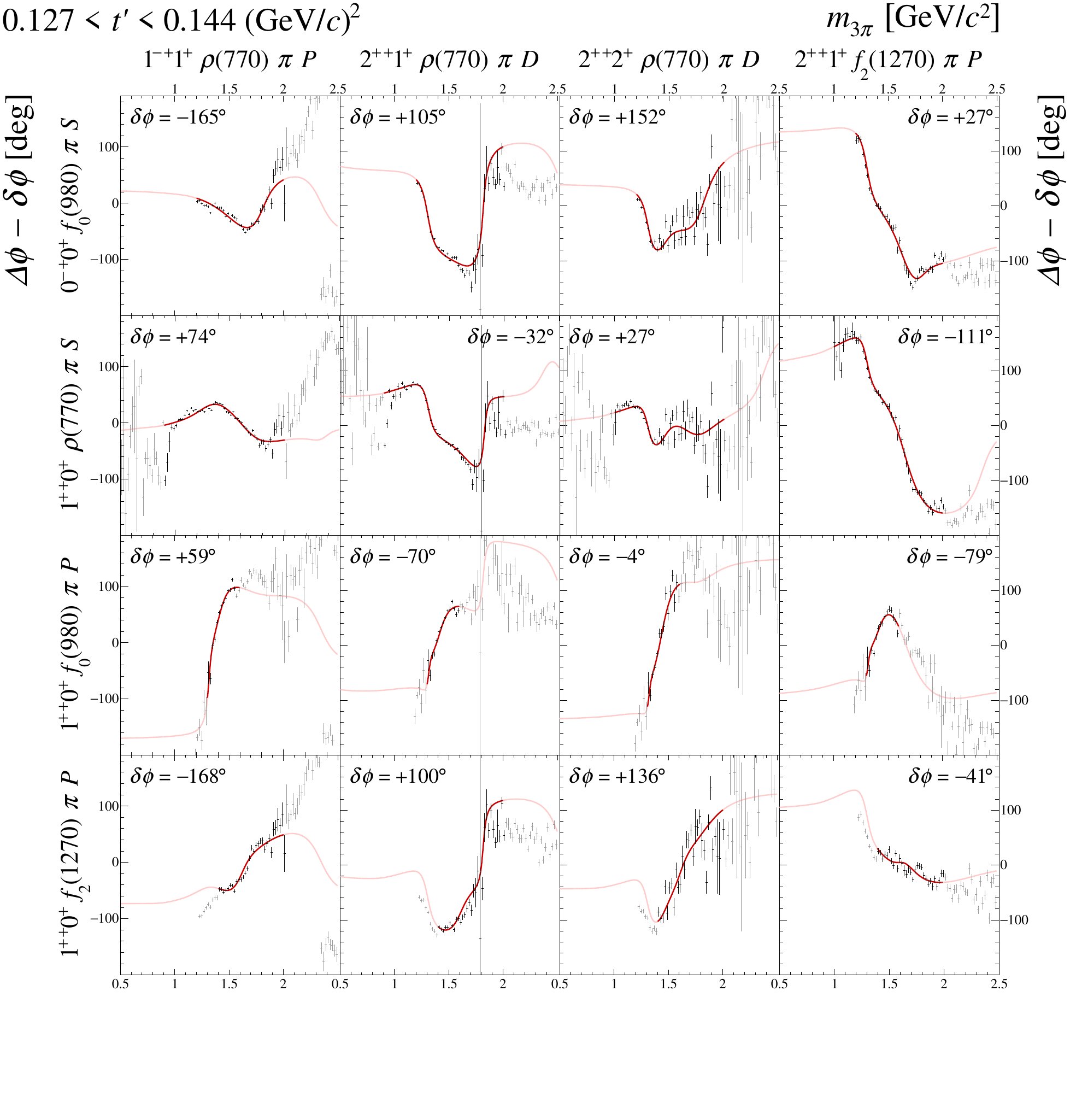}%
   \caption{Submatrix~B of the $14 \times 14$ matrix of graphs that
     represents the spin-density matrix (see
     \cref{tab:spin-dens_matrix_overview}).}
   \label{fig:spin-dens_submatrix_2_tbin_3}
 \end{minipage}
\end{textblock*}

\newpage\null
\begin{textblock*}{\textwidth}[0.5,0](0.5\paperwidth,\blockDistanceToTop)
 \begin{minipage}{\textwidth}
   \makeatletter
   \def\@captype{figure}
   \makeatother
   \centering
   \includegraphics[height=\matrixHeight]{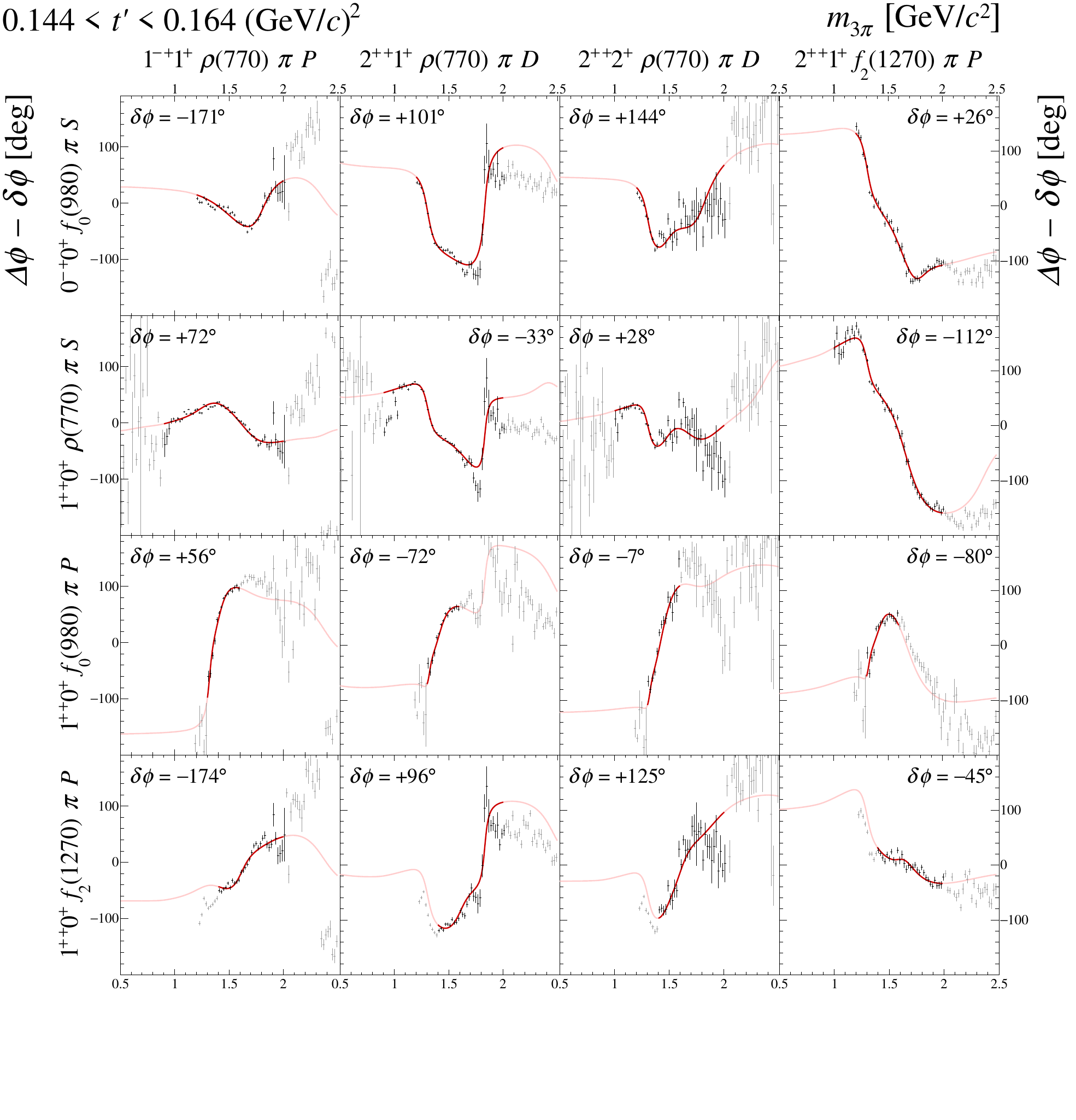}%
   \caption{Submatrix~B of the $14 \times 14$ matrix of graphs that
     represents the spin-density matrix (see
     \cref{tab:spin-dens_matrix_overview}).}
   \label{fig:spin-dens_submatrix_2_tbin_4}
 \end{minipage}
\end{textblock*}

\newpage\null
\begin{textblock*}{\textwidth}[0.5,0](0.5\paperwidth,\blockDistanceToTop)
 \begin{minipage}{\textwidth}
   \makeatletter
   \def\@captype{figure}
   \makeatother
   \centering
   \includegraphics[height=\matrixHeight]{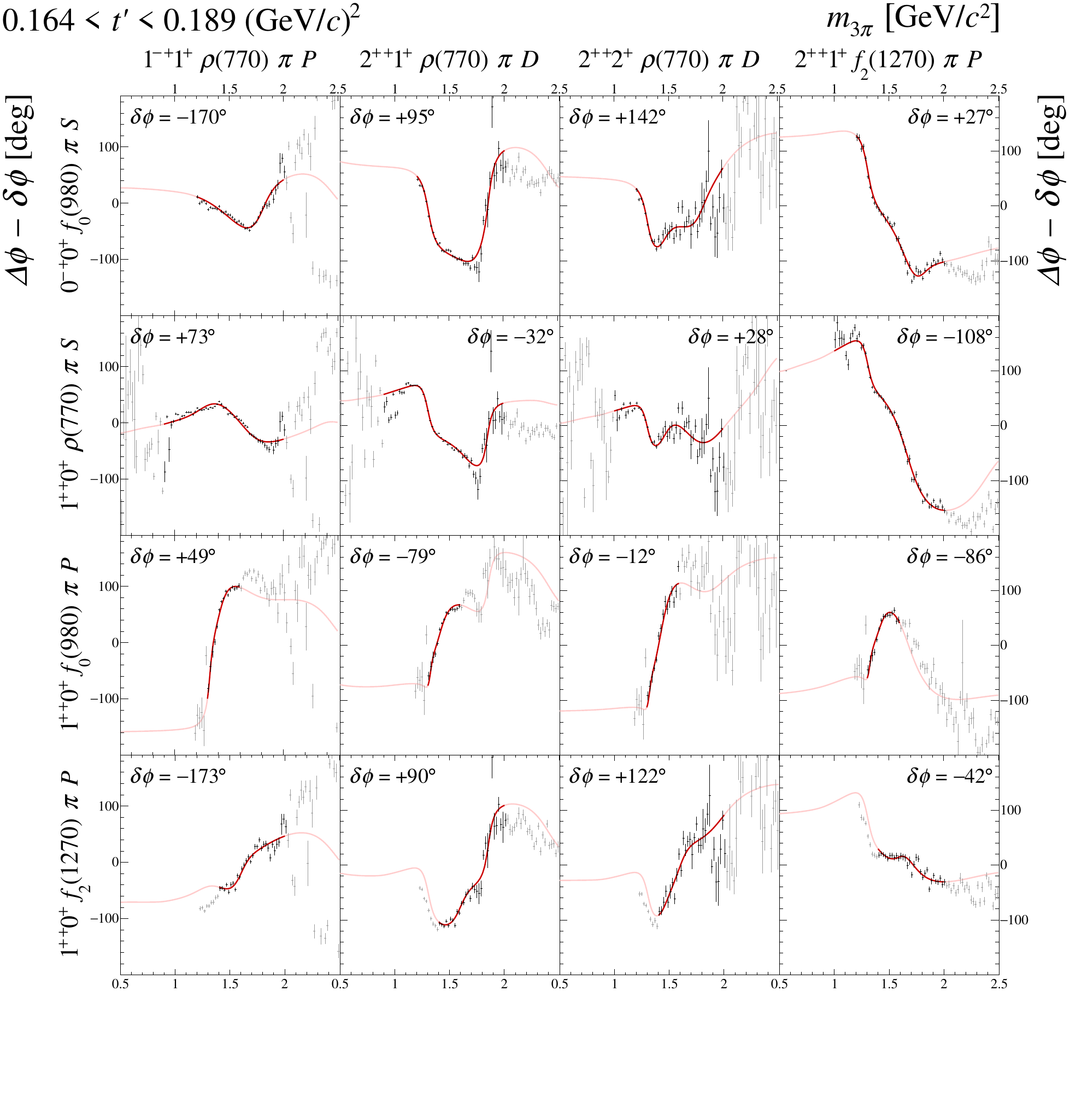}%
   \caption{Submatrix~B of the $14 \times 14$ matrix of graphs that
     represents the spin-density matrix (see
     \cref{tab:spin-dens_matrix_overview}).}
   \label{fig:spin-dens_submatrix_2_tbin_5}
 \end{minipage}
\end{textblock*}

\newpage\null
\begin{textblock*}{\textwidth}[0.5,0](0.5\paperwidth,\blockDistanceToTop)
 \begin{minipage}{\textwidth}
   \makeatletter
   \def\@captype{figure}
   \makeatother
   \centering
   \includegraphics[height=\matrixHeight]{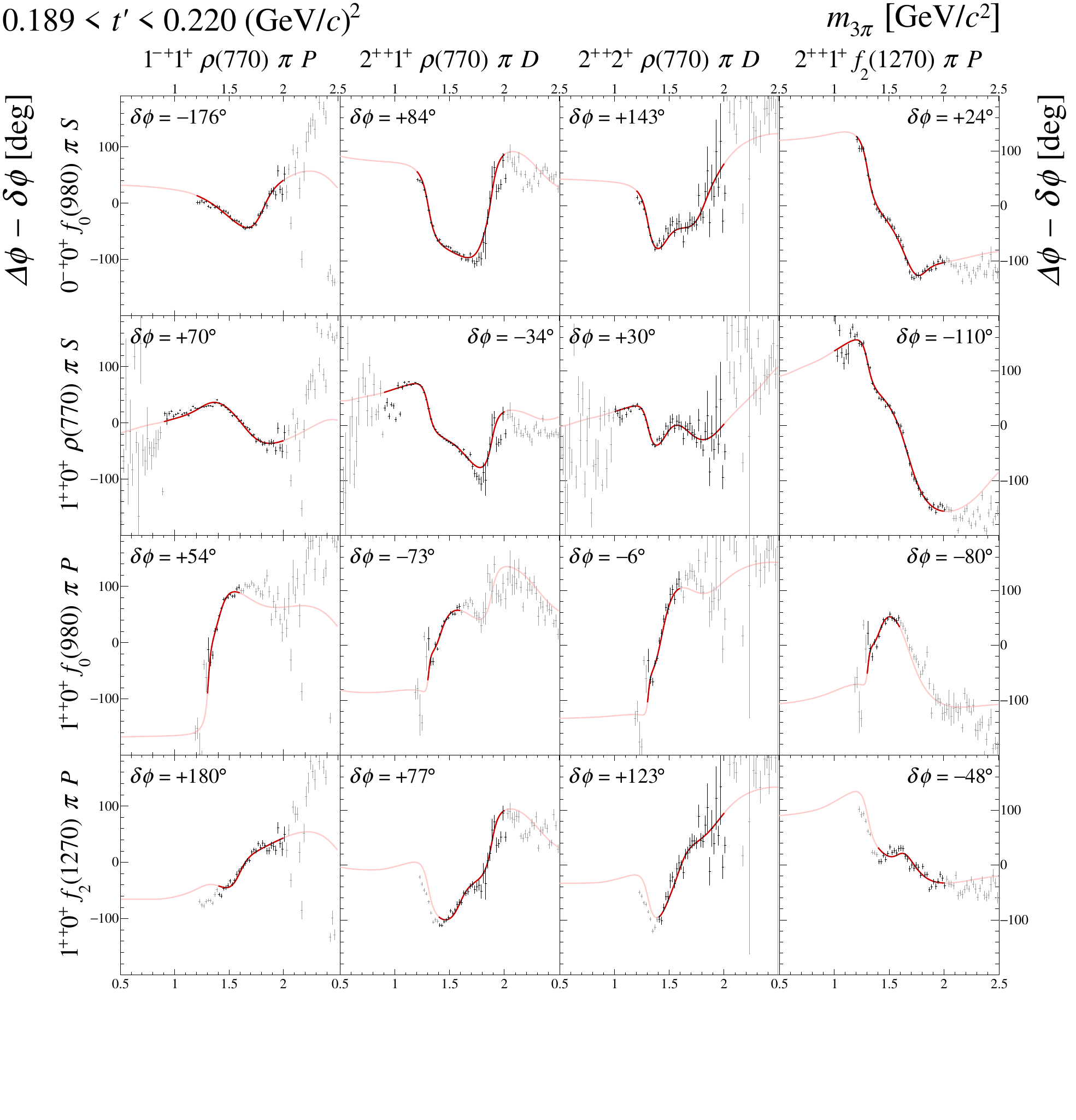}%
   \caption{Submatrix~B of the $14 \times 14$ matrix of graphs that
     represents the spin-density matrix (see
     \cref{tab:spin-dens_matrix_overview}).}
   \label{fig:spin-dens_submatrix_2_tbin_6}
 \end{minipage}
\end{textblock*}

\newpage\null
\begin{textblock*}{\textwidth}[0.5,0](0.5\paperwidth,\blockDistanceToTop)
 \begin{minipage}{\textwidth}
   \makeatletter
   \def\@captype{figure}
   \makeatother
   \centering
   \includegraphics[height=\matrixHeight]{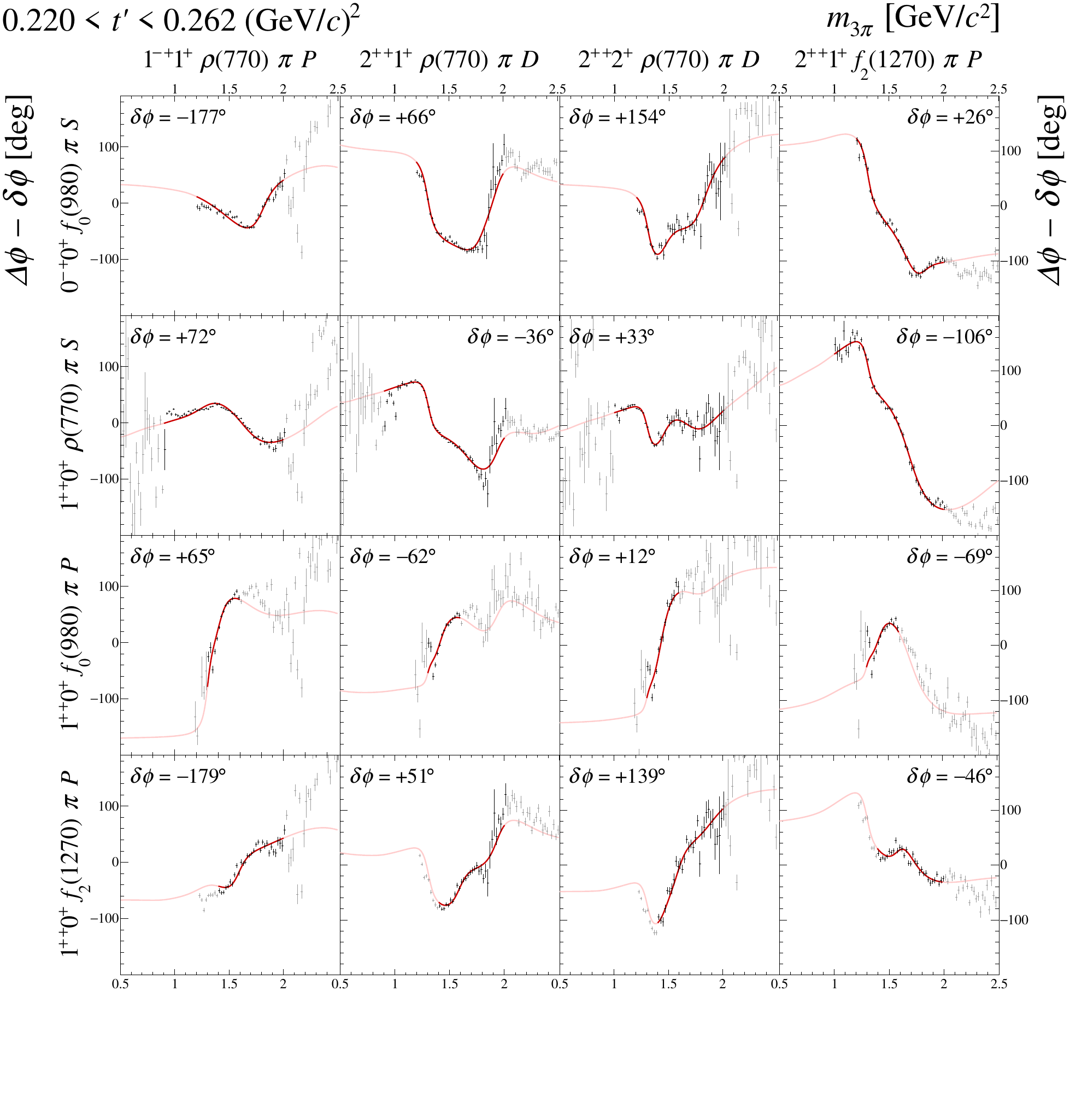}%
   \caption{Submatrix~B of the $14 \times 14$ matrix of graphs that
     represents the spin-density matrix (see
     \cref{tab:spin-dens_matrix_overview}).}
   \label{fig:spin-dens_submatrix_2_tbin_7}
 \end{minipage}
\end{textblock*}

\newpage\null
\begin{textblock*}{\textwidth}[0.5,0](0.5\paperwidth,\blockDistanceToTop)
 \begin{minipage}{\textwidth}
   \makeatletter
   \def\@captype{figure}
   \makeatother
   \centering
   \includegraphics[height=\matrixHeight]{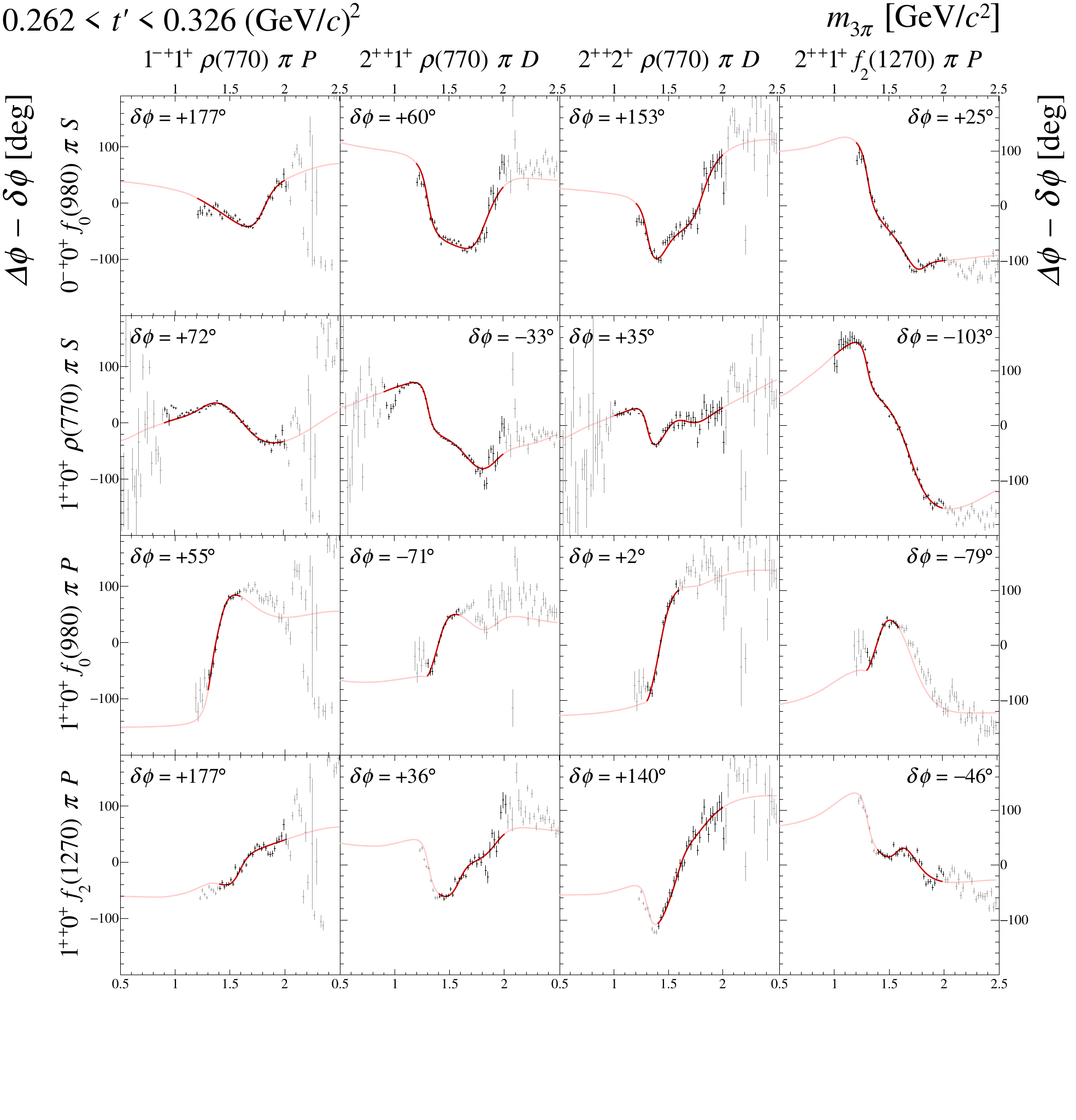}%
   \caption{Submatrix~B of the $14 \times 14$ matrix of graphs that
     represents the spin-density matrix (see
     \cref{tab:spin-dens_matrix_overview}).}
   \label{fig:spin-dens_submatrix_2_tbin_8}
 \end{minipage}
\end{textblock*}

\newpage\null
\begin{textblock*}{\textwidth}[0.5,0](0.5\paperwidth,\blockDistanceToTop)
 \begin{minipage}{\textwidth}
   \makeatletter
   \def\@captype{figure}
   \makeatother
   \centering
   \includegraphics[height=\matrixHeight]{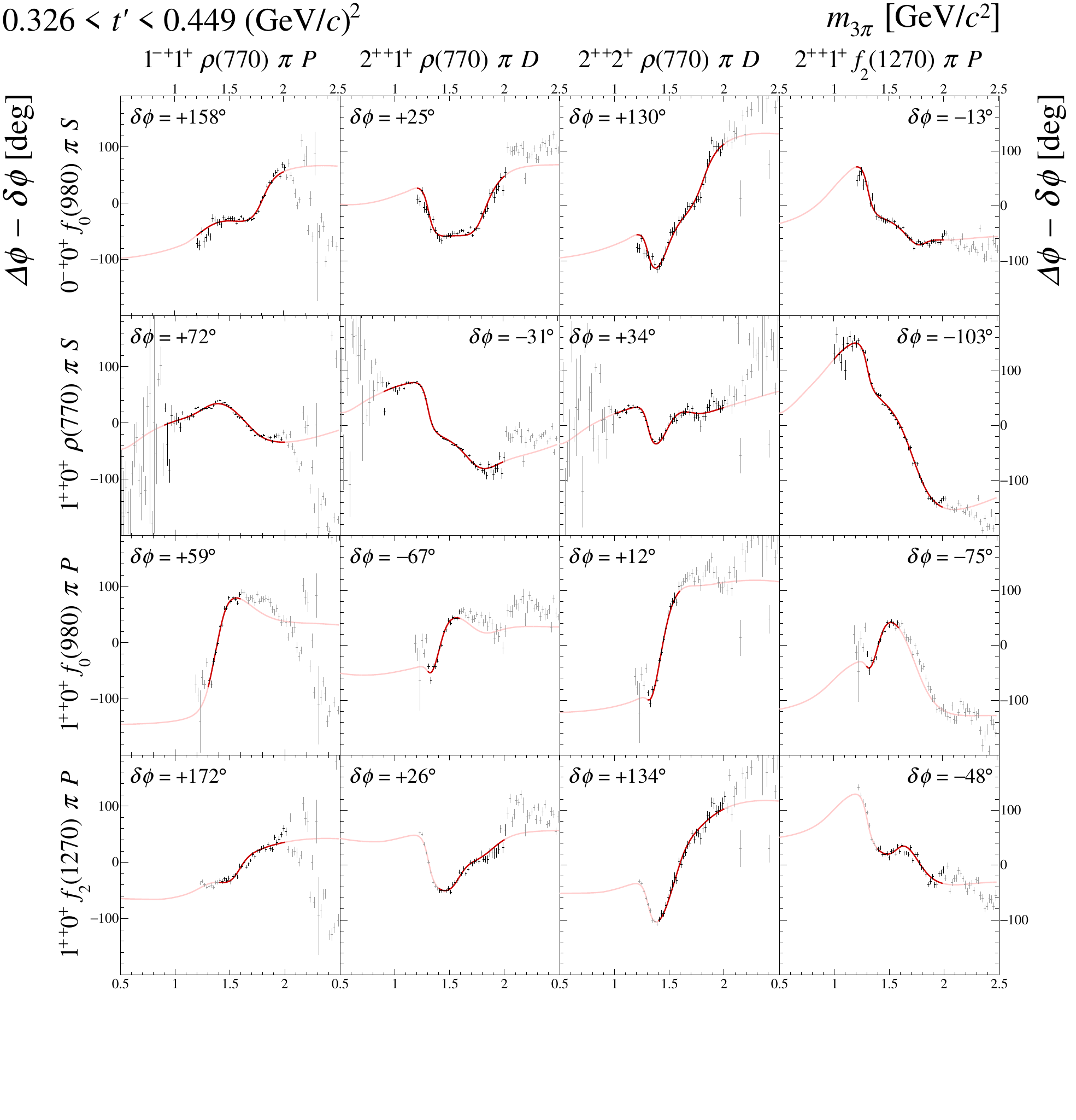}%
   \caption{Submatrix~B of the $14 \times 14$ matrix of graphs that
     represents the spin-density matrix (see
     \cref{tab:spin-dens_matrix_overview}).}
   \label{fig:spin-dens_submatrix_2_tbin_9}
 \end{minipage}
\end{textblock*}

\newpage\null
\begin{textblock*}{\textwidth}[0.5,0](0.5\paperwidth,\blockDistanceToTop)
 \begin{minipage}{\textwidth}
   \makeatletter
   \def\@captype{figure}
   \makeatother
   \centering
   \includegraphics[height=\matrixHeight]{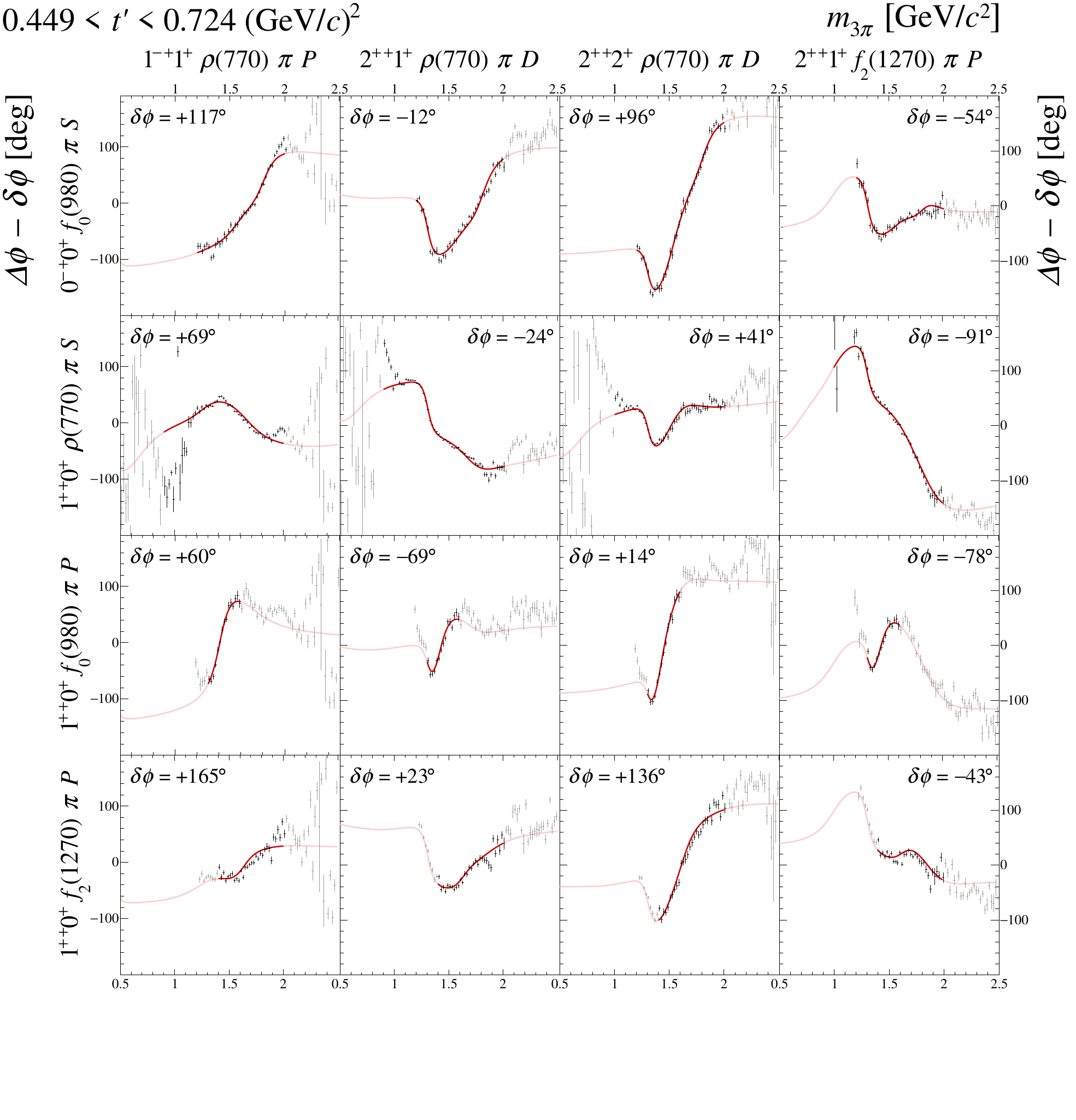}%
   \caption{Submatrix~B of the $14 \times 14$ matrix of graphs that
     represents the spin-density matrix (see
     \cref{tab:spin-dens_matrix_overview}).}
   \label{fig:spin-dens_submatrix_2_tbin_10}
 \end{minipage}
\end{textblock*}

\newpage\null
\begin{textblock*}{\textwidth}[0.5,0](0.5\paperwidth,\blockDistanceToTop)
 \begin{minipage}{\textwidth}
   \makeatletter
   \def\@captype{figure}
   \makeatother
   \centering
   \includegraphics[height=\matrixHeight]{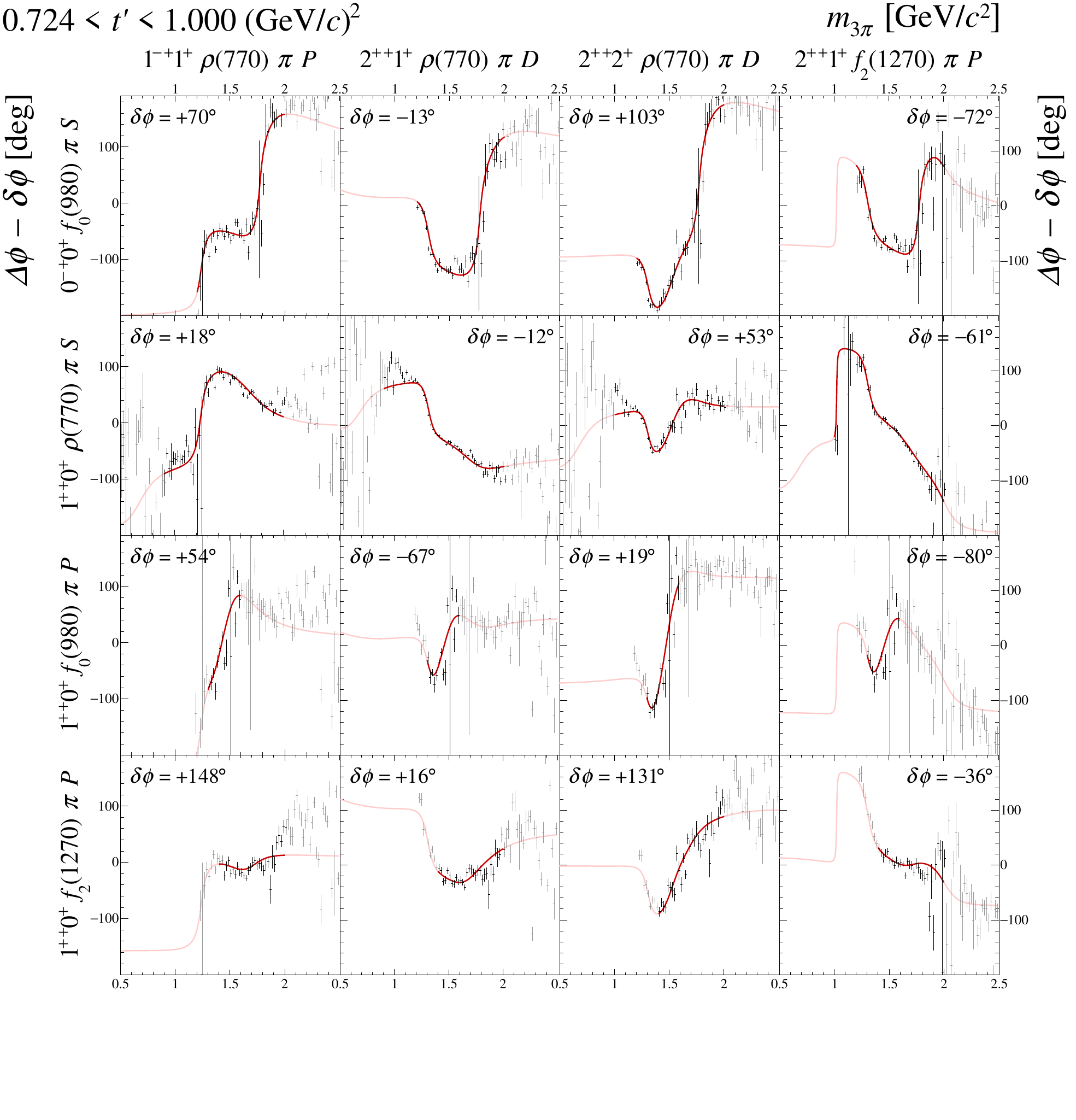}%
   \caption{Submatrix~B of the $14 \times 14$ matrix of graphs that
     represents the spin-density matrix (see
     \cref{tab:spin-dens_matrix_overview}).}
   \label{fig:spin-dens_submatrix_2_tbin_11}
 \end{minipage}
\end{textblock*}

\clearpage
\subsection{Submatrix C}
\label{sec:spin-dens_submatrix_3}

\begin{textblock*}{\textwidth}[0.5,0](0.5\paperwidth,\blockDistanceToTop)
 \begin{minipage}{\textwidth}
   \makeatletter
   \def\@captype{figure}
   \makeatother
   \centering
   \includegraphics[height=\matrixHeight]{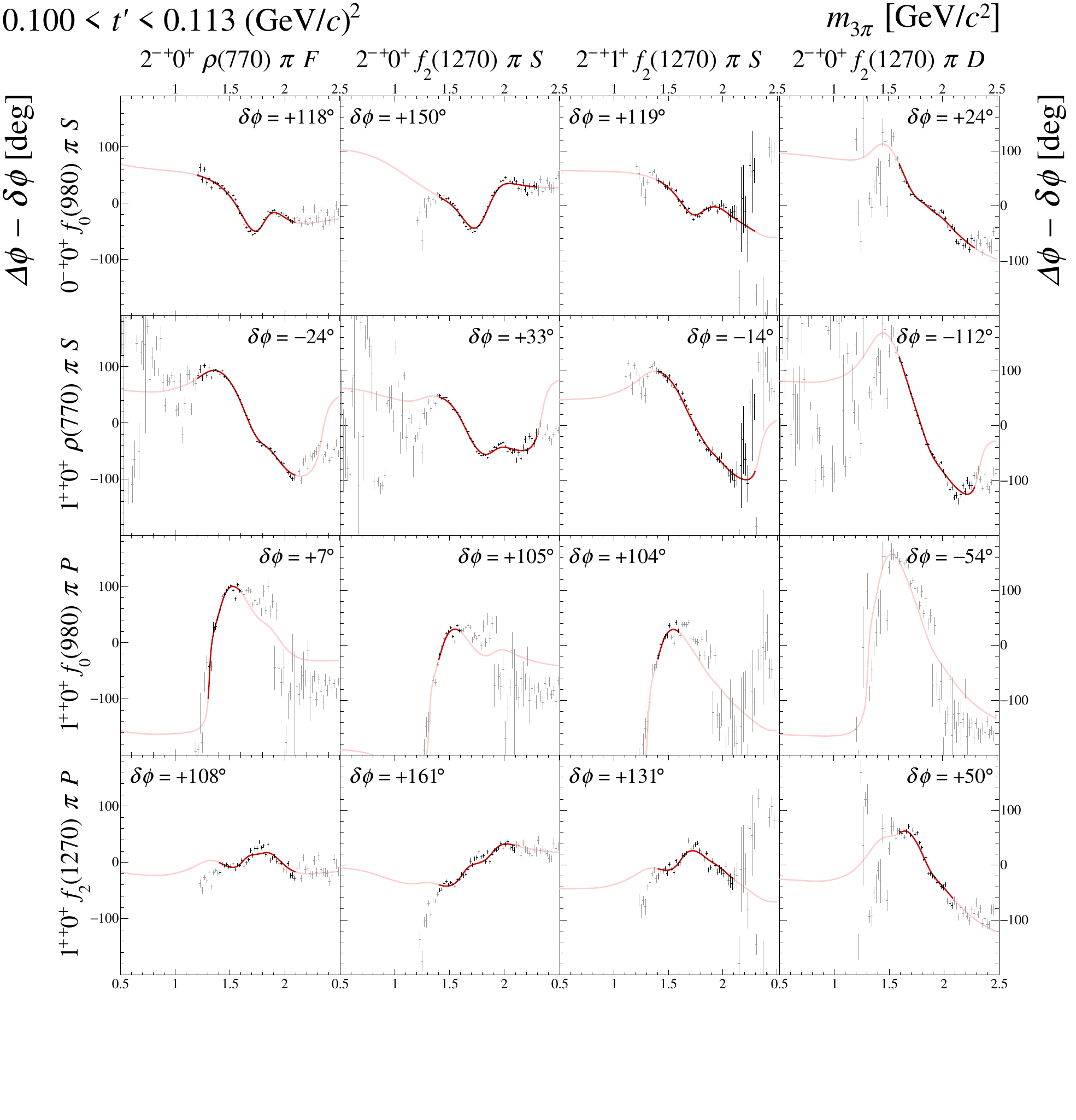}%
   \caption{Submatrix~C of the $14 \times 14$ matrix of graphs that
     represents the spin-density matrix (see
     \cref{tab:spin-dens_matrix_overview}).}
   \label{fig:spin-dens_submatrix_3_tbin_1}
 \end{minipage}
\end{textblock*}

\newpage\null
\begin{textblock*}{\textwidth}[0.5,0](0.5\paperwidth,\blockDistanceToTop)
 \begin{minipage}{\textwidth}
   \makeatletter
   \def\@captype{figure}
   \makeatother
   \centering
   \includegraphics[height=\matrixHeight]{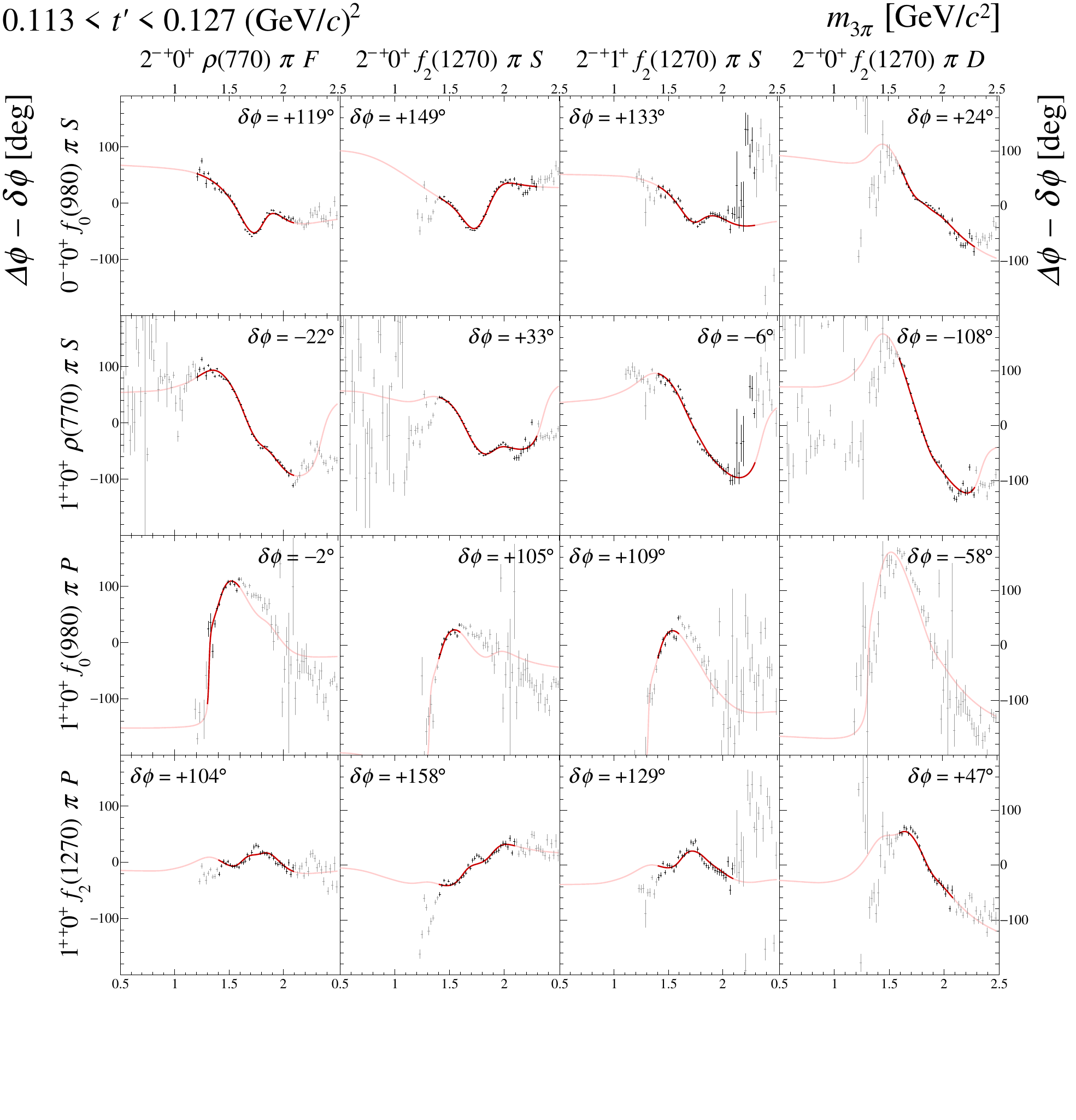}%
   \caption{Submatrix~C of the $14 \times 14$ matrix of graphs that
     represents the spin-density matrix (see
     \cref{tab:spin-dens_matrix_overview}).}
   \label{fig:spin-dens_submatrix_3_tbin_2}
 \end{minipage}
\end{textblock*}

\newpage\null
\begin{textblock*}{\textwidth}[0.5,0](0.5\paperwidth,\blockDistanceToTop)
 \begin{minipage}{\textwidth}
   \makeatletter
   \def\@captype{figure}
   \makeatother
   \centering
   \includegraphics[height=\matrixHeight]{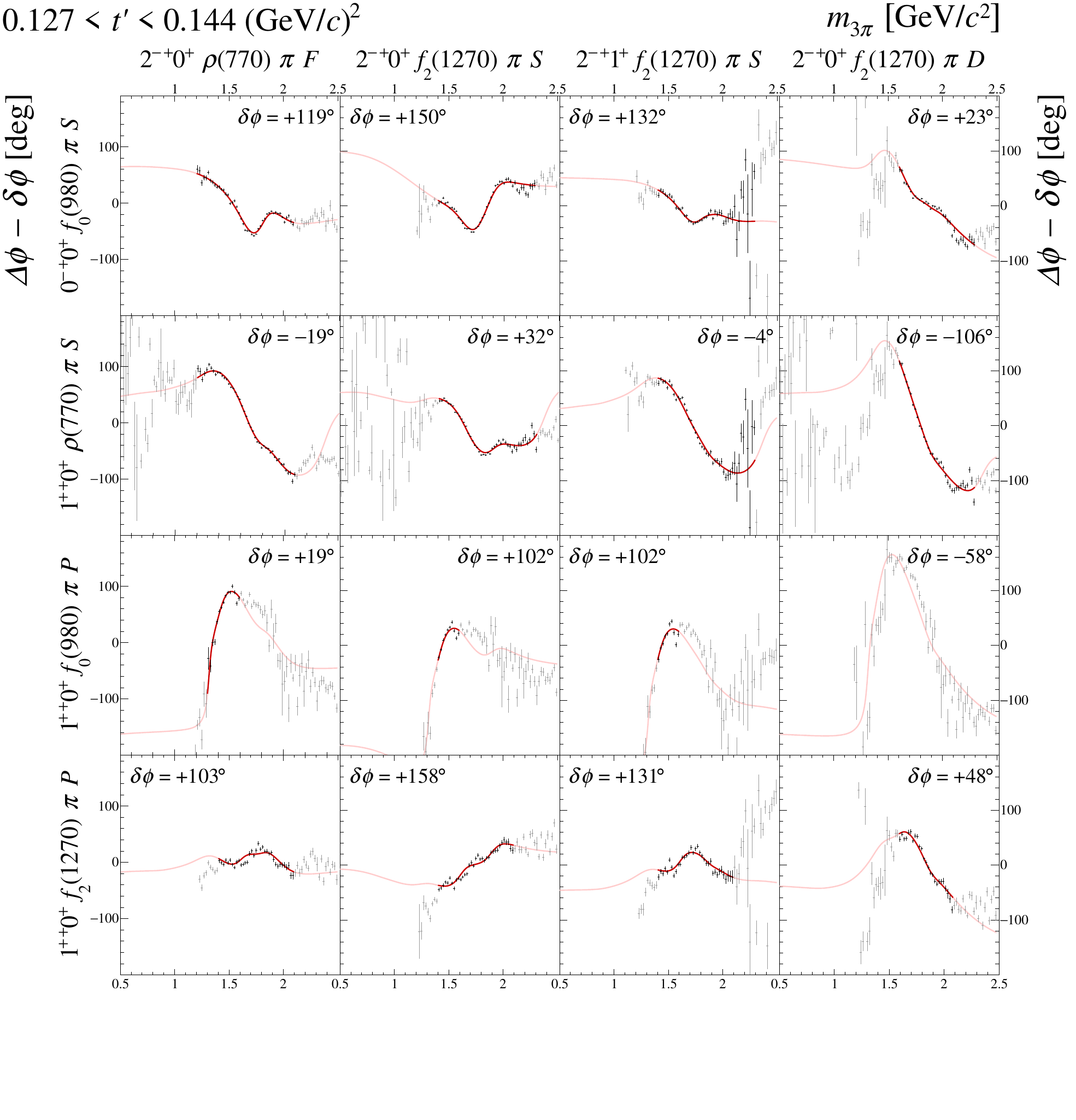}%
   \caption{Submatrix~C of the $14 \times 14$ matrix of graphs that
     represents the spin-density matrix (see
     \cref{tab:spin-dens_matrix_overview}).}
   \label{fig:spin-dens_submatrix_3_tbin_3}
 \end{minipage}
\end{textblock*}

\newpage\null
\begin{textblock*}{\textwidth}[0.5,0](0.5\paperwidth,\blockDistanceToTop)
 \begin{minipage}{\textwidth}
   \makeatletter
   \def\@captype{figure}
   \makeatother
   \centering
   \includegraphics[height=\matrixHeight]{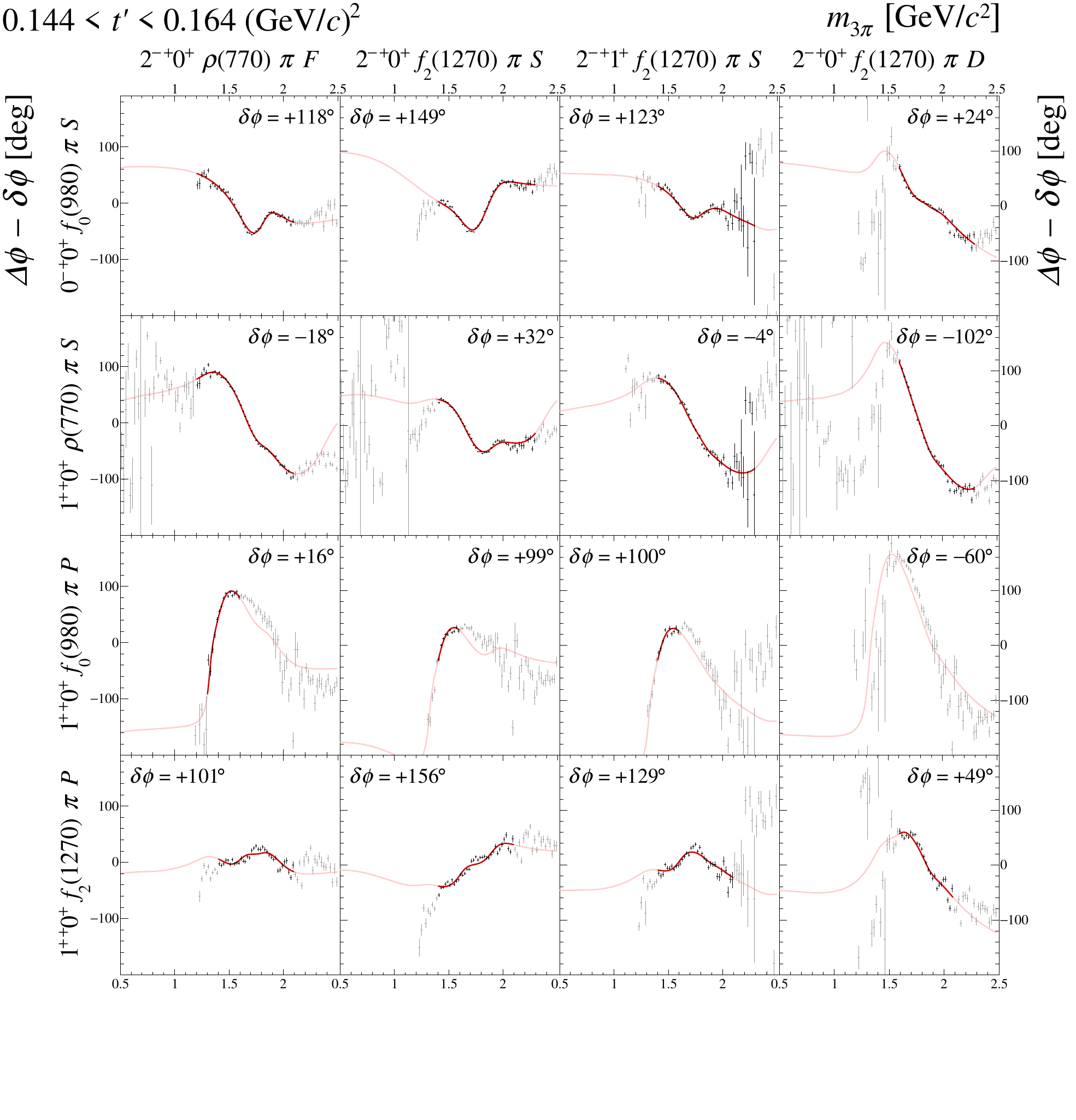}%
   \caption{Submatrix~C of the $14 \times 14$ matrix of graphs that
     represents the spin-density matrix (see
     \cref{tab:spin-dens_matrix_overview}).}
   \label{fig:spin-dens_submatrix_3_tbin_4}
 \end{minipage}
\end{textblock*}

\newpage\null
\begin{textblock*}{\textwidth}[0.5,0](0.5\paperwidth,\blockDistanceToTop)
 \begin{minipage}{\textwidth}
   \makeatletter
   \def\@captype{figure}
   \makeatother
   \centering
   \includegraphics[height=\matrixHeight]{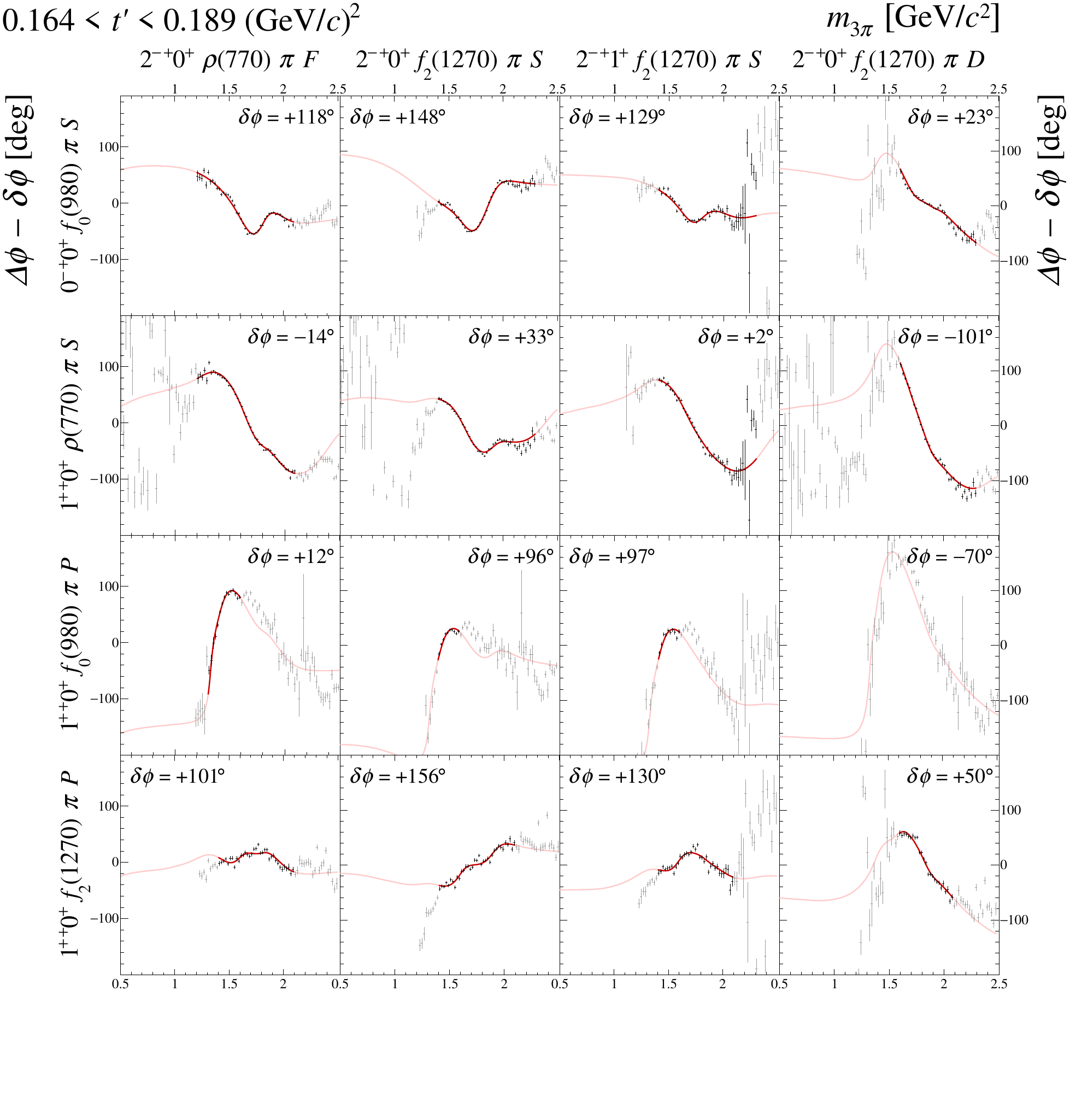}%
   \caption{Submatrix~C of the $14 \times 14$ matrix of graphs that
     represents the spin-density matrix (see
     \cref{tab:spin-dens_matrix_overview}).}
   \label{fig:spin-dens_submatrix_3_tbin_5}
 \end{minipage}
\end{textblock*}

\newpage\null
\begin{textblock*}{\textwidth}[0.5,0](0.5\paperwidth,\blockDistanceToTop)
 \begin{minipage}{\textwidth}
   \makeatletter
   \def\@captype{figure}
   \makeatother
   \centering
   \includegraphics[height=\matrixHeight]{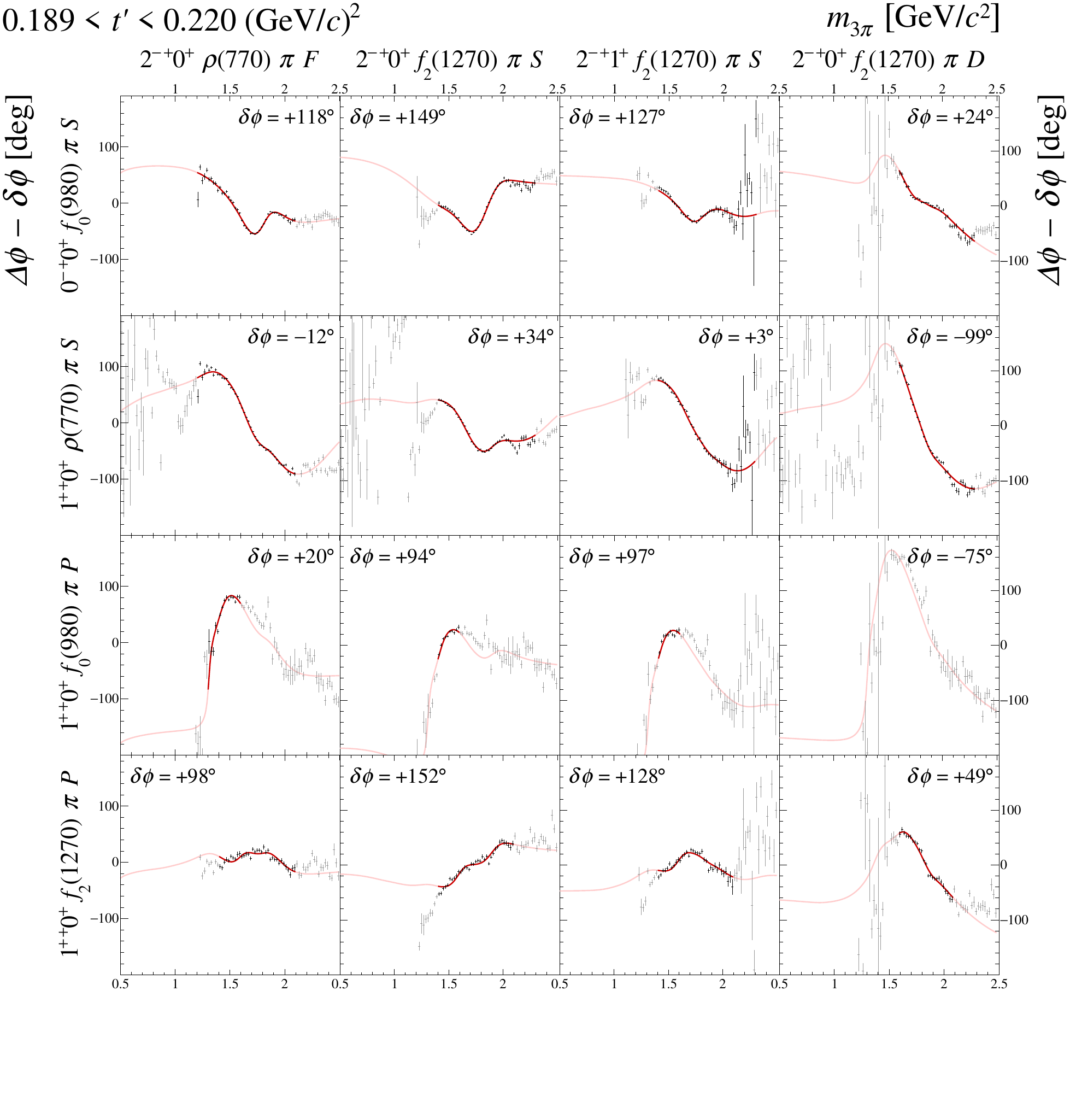}%
   \caption{Submatrix~C of the $14 \times 14$ matrix of graphs that
     represents the spin-density matrix (see
     \cref{tab:spin-dens_matrix_overview}).}
   \label{fig:spin-dens_submatrix_3_tbin_6}
 \end{minipage}
\end{textblock*}

\newpage\null
\begin{textblock*}{\textwidth}[0.5,0](0.5\paperwidth,\blockDistanceToTop)
 \begin{minipage}{\textwidth}
   \makeatletter
   \def\@captype{figure}
   \makeatother
   \centering
   \includegraphics[height=\matrixHeight]{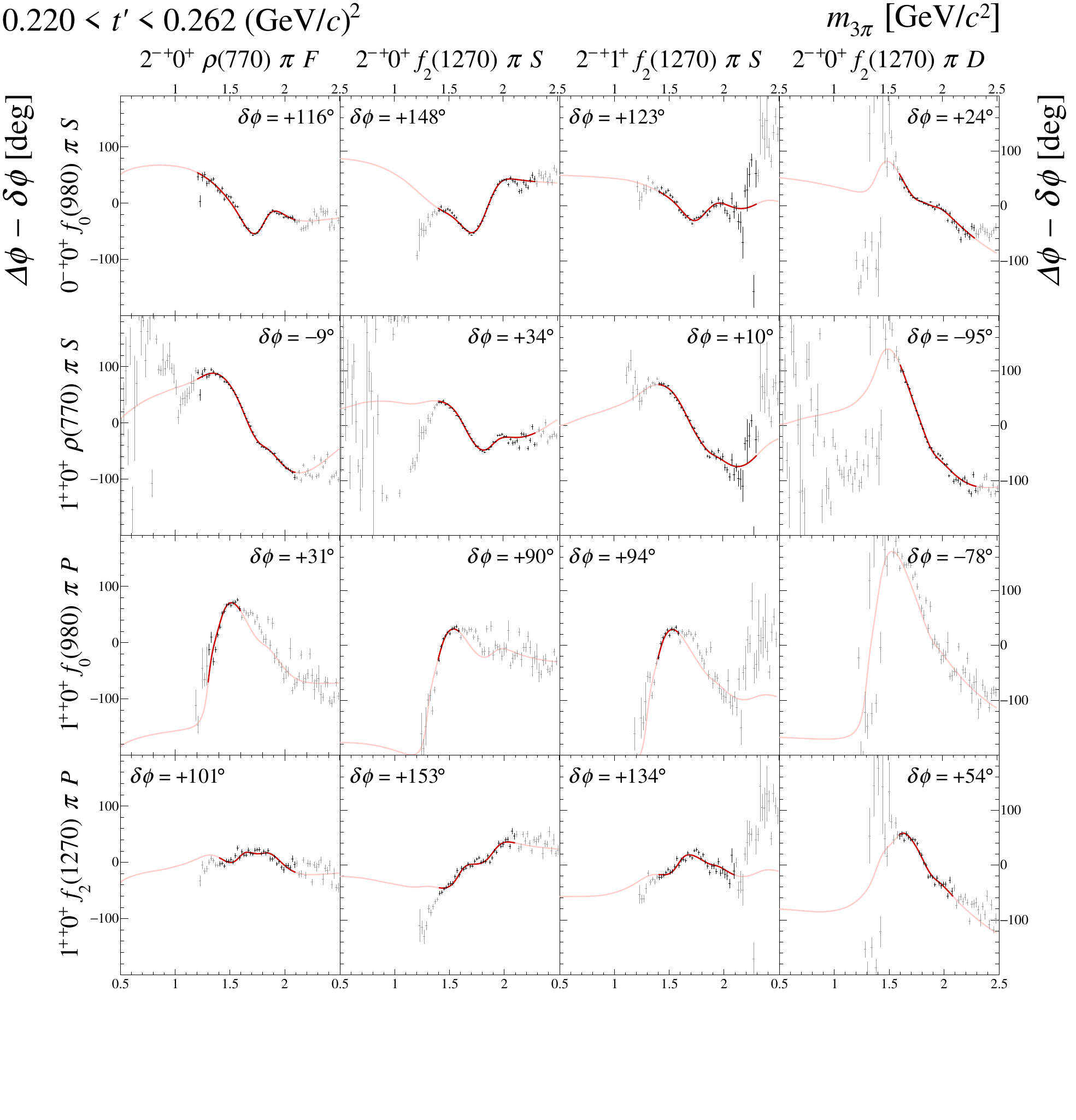}%
   \caption{Submatrix~C of the $14 \times 14$ matrix of graphs that
     represents the spin-density matrix (see
     \cref{tab:spin-dens_matrix_overview}).}
   \label{fig:spin-dens_submatrix_3_tbin_7}
 \end{minipage}
\end{textblock*}

\newpage\null
\begin{textblock*}{\textwidth}[0.5,0](0.5\paperwidth,\blockDistanceToTop)
 \begin{minipage}{\textwidth}
   \makeatletter
   \def\@captype{figure}
   \makeatother
   \centering
   \includegraphics[height=\matrixHeight]{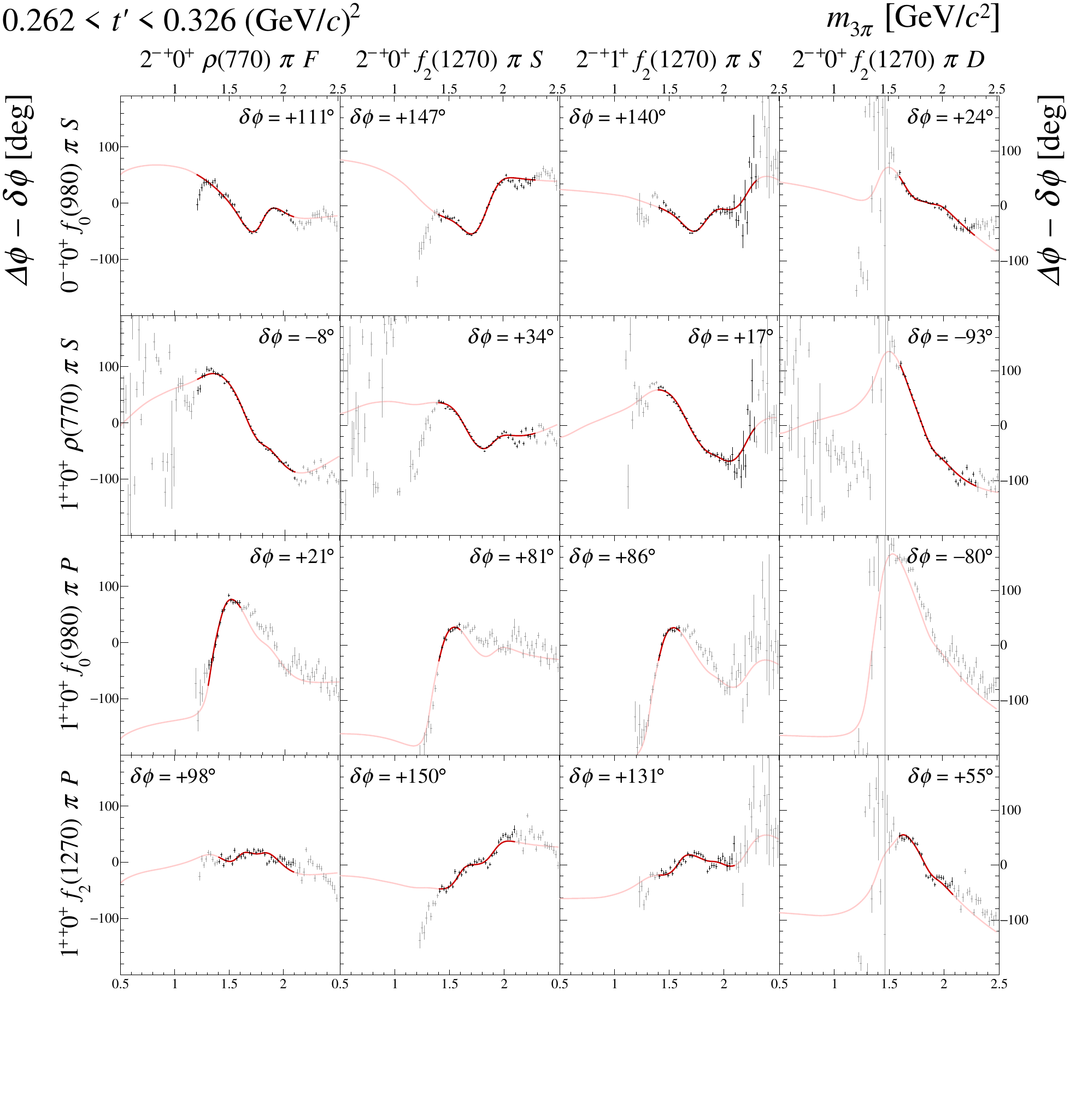}%
   \caption{Submatrix~C of the $14 \times 14$ matrix of graphs that
     represents the spin-density matrix (see
     \cref{tab:spin-dens_matrix_overview}).}
   \label{fig:spin-dens_submatrix_3_tbin_8}
 \end{minipage}
\end{textblock*}

\newpage\null
\begin{textblock*}{\textwidth}[0.5,0](0.5\paperwidth,\blockDistanceToTop)
 \begin{minipage}{\textwidth}
   \makeatletter
   \def\@captype{figure}
   \makeatother
   \centering
   \includegraphics[height=\matrixHeight]{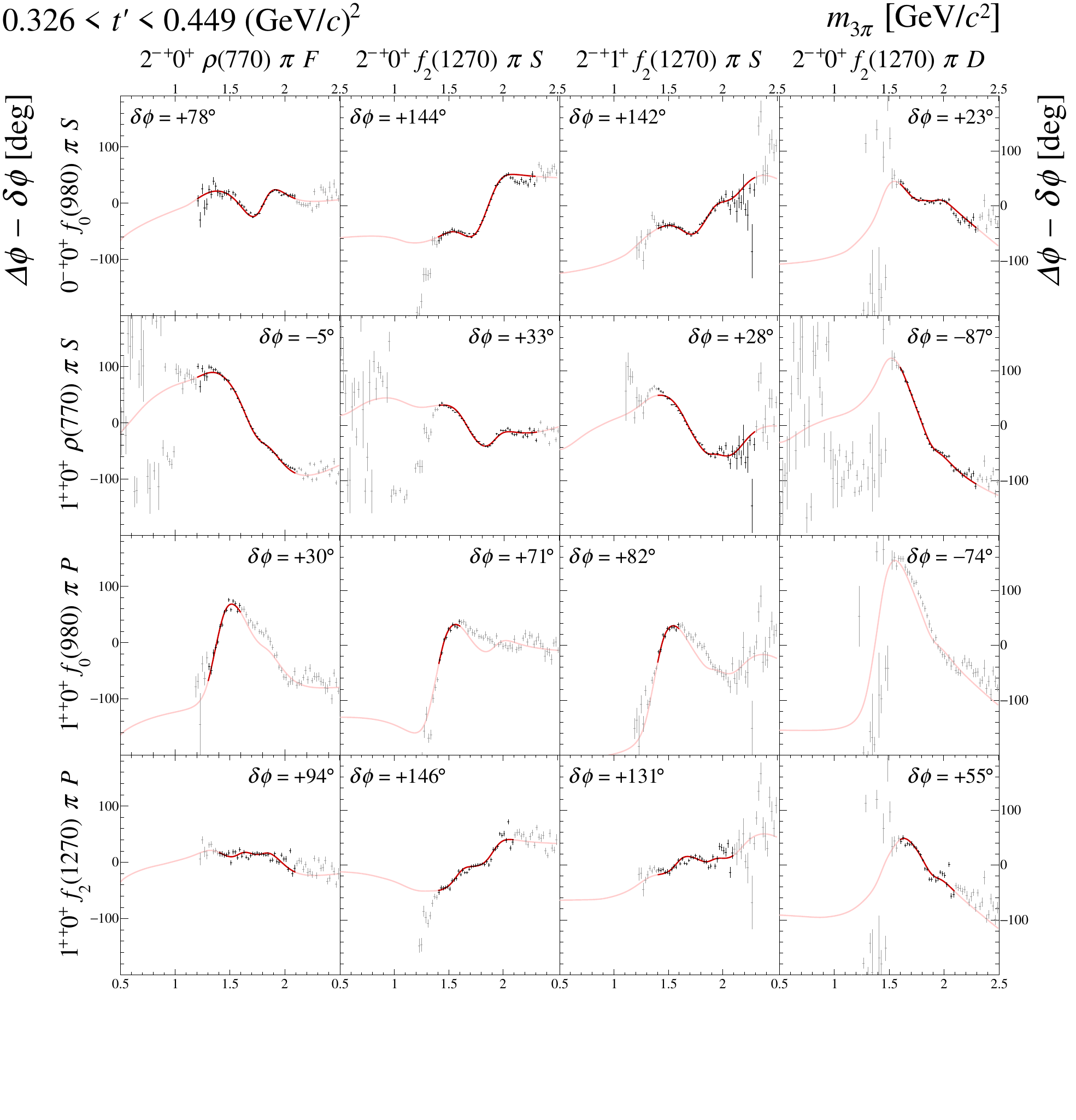}%
   \caption{Submatrix~C of the $14 \times 14$ matrix of graphs that
     represents the spin-density matrix (see
     \cref{tab:spin-dens_matrix_overview}).}
   \label{fig:spin-dens_submatrix_3_tbin_9}
 \end{minipage}
\end{textblock*}

\newpage\null
\begin{textblock*}{\textwidth}[0.5,0](0.5\paperwidth,\blockDistanceToTop)
 \begin{minipage}{\textwidth}
   \makeatletter
   \def\@captype{figure}
   \makeatother
   \centering
   \includegraphics[height=\matrixHeight]{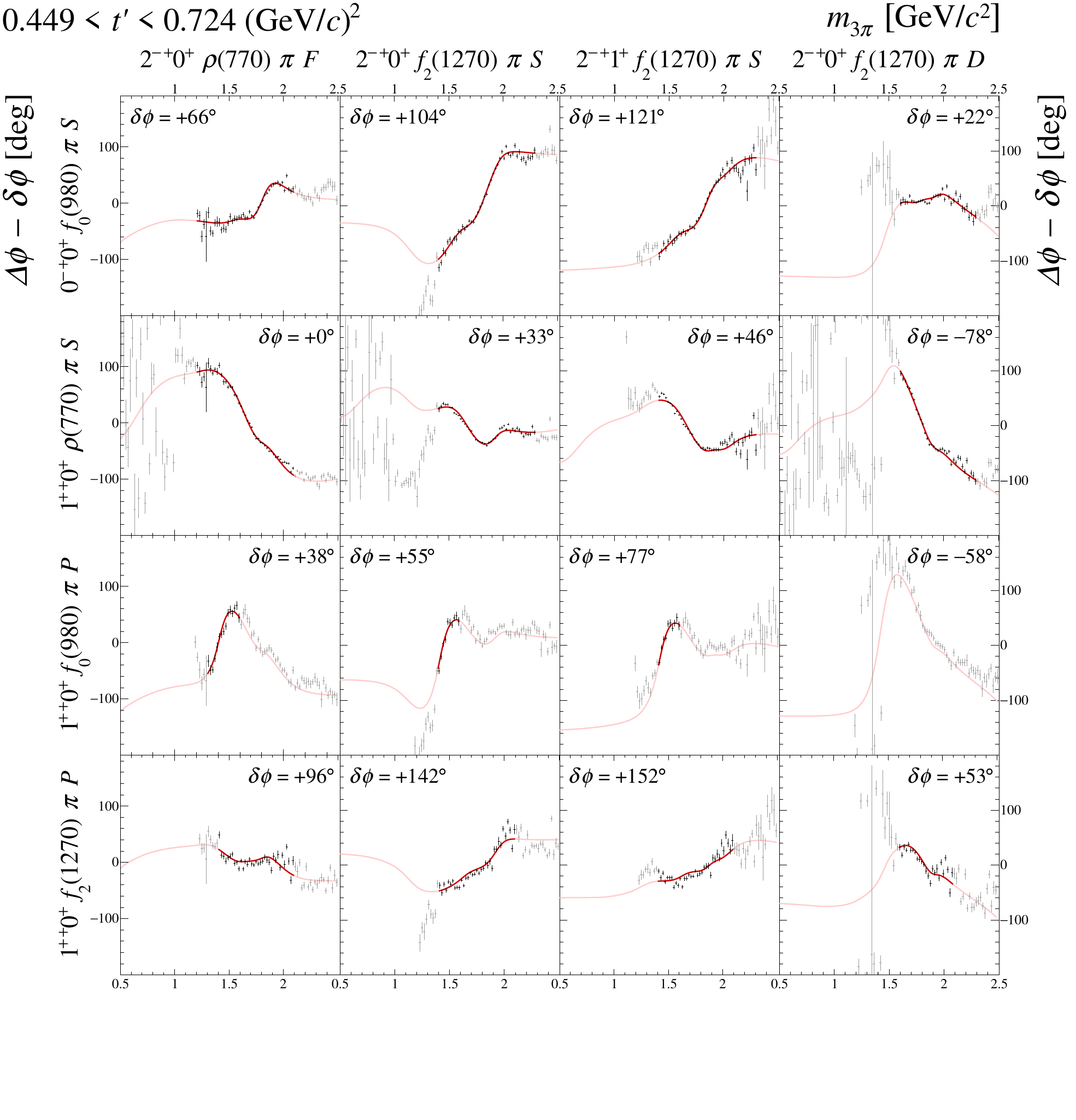}%
   \caption{Submatrix~C of the $14 \times 14$ matrix of graphs that
     represents the spin-density matrix (see
     \cref{tab:spin-dens_matrix_overview}).}
   \label{fig:spin-dens_submatrix_3_tbin_10}
 \end{minipage}
\end{textblock*}

\newpage\null
\begin{textblock*}{\textwidth}[0.5,0](0.5\paperwidth,\blockDistanceToTop)
 \begin{minipage}{\textwidth}
   \makeatletter
   \def\@captype{figure}
   \makeatother
   \centering
   \includegraphics[height=\matrixHeight]{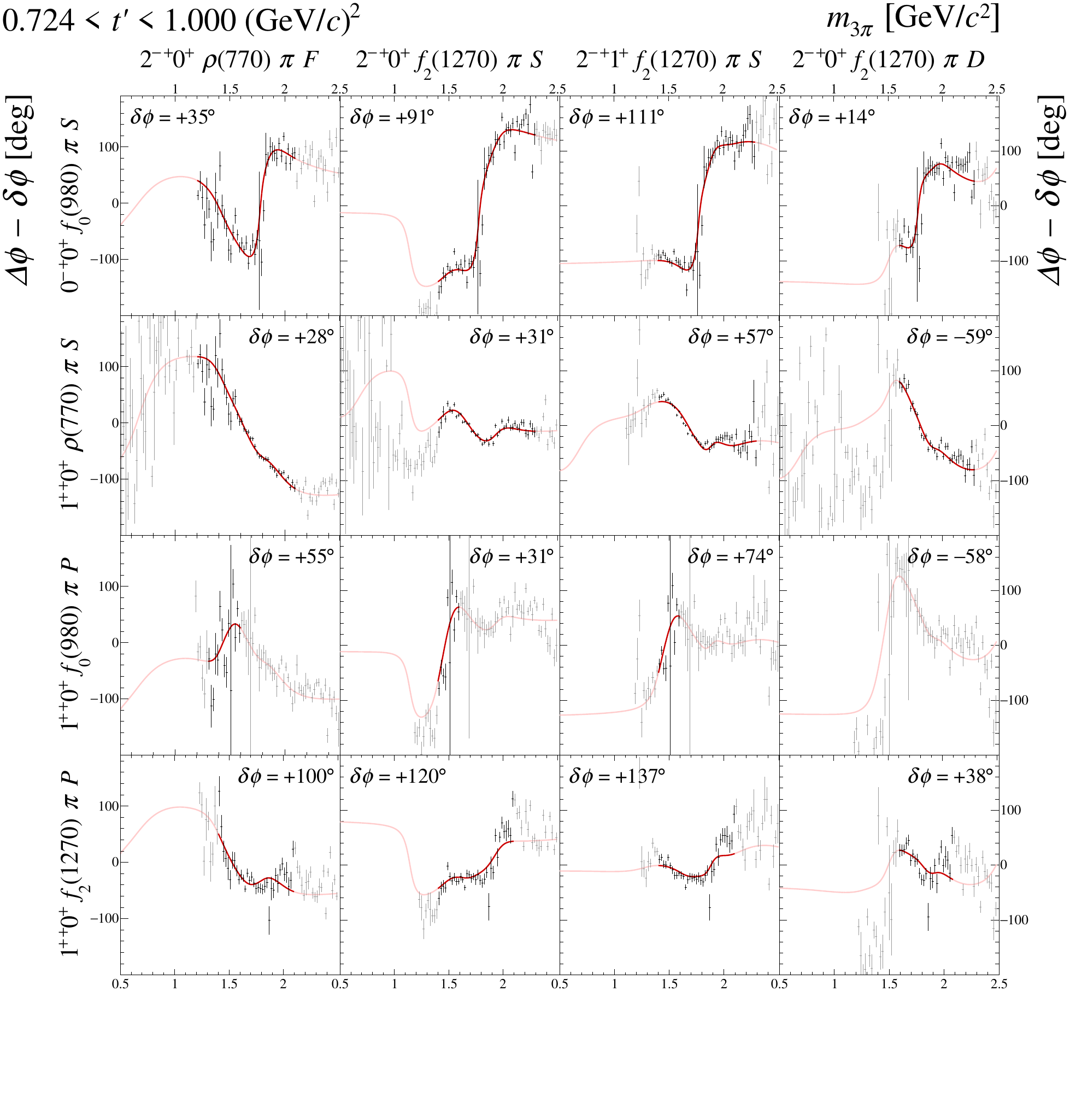}%
   \caption{Submatrix~C of the $14 \times 14$ matrix of graphs that
     represents the spin-density matrix (see
     \cref{tab:spin-dens_matrix_overview}).}
   \label{fig:spin-dens_submatrix_3_tbin_11}
 \end{minipage}
\end{textblock*}

\clearpage
\subsection{Submatrix D}
\label{sec:spin-dens_submatrix_4}

\begin{textblock*}{\textwidth}[0.5,0](0.5\paperwidth,\blockDistanceToTop)
 \begin{minipage}{\textwidth}
   \makeatletter
   \def\@captype{figure}
   \makeatother
   \centering
   \includegraphics[height=\matrixHeight]{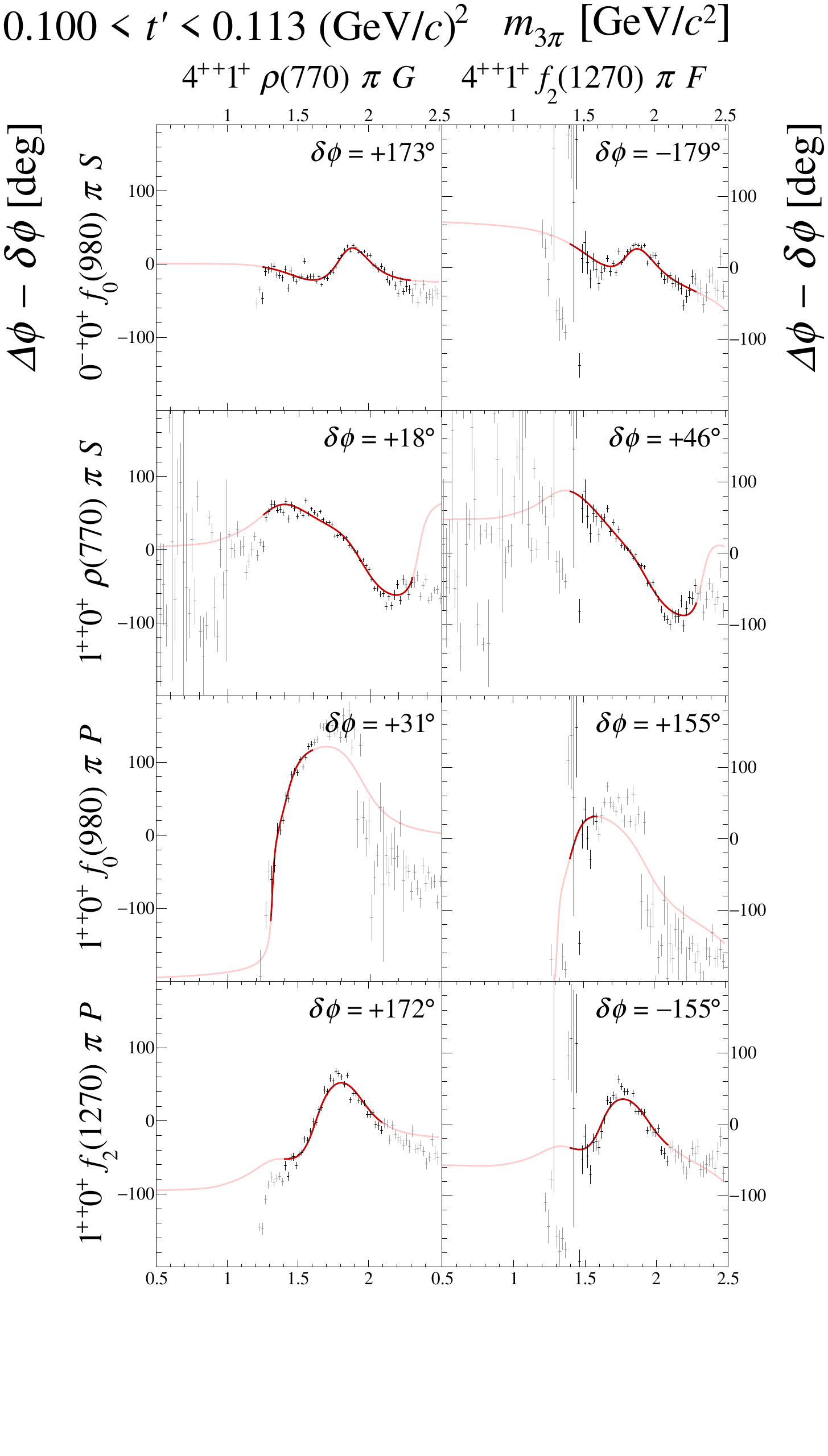}%
   \caption{Submatrix~D of the $14 \times 14$ matrix of graphs that
     represents the spin-density matrix (see
     \cref{tab:spin-dens_matrix_overview}).}
   \label{fig:spin-dens_submatrix_4_tbin_1}
 \end{minipage}
\end{textblock*}

\newpage\null
\begin{textblock*}{\textwidth}[0.5,0](0.5\paperwidth,\blockDistanceToTop)
 \begin{minipage}{\textwidth}
   \makeatletter
   \def\@captype{figure}
   \makeatother
   \centering
   \includegraphics[height=\matrixHeight]{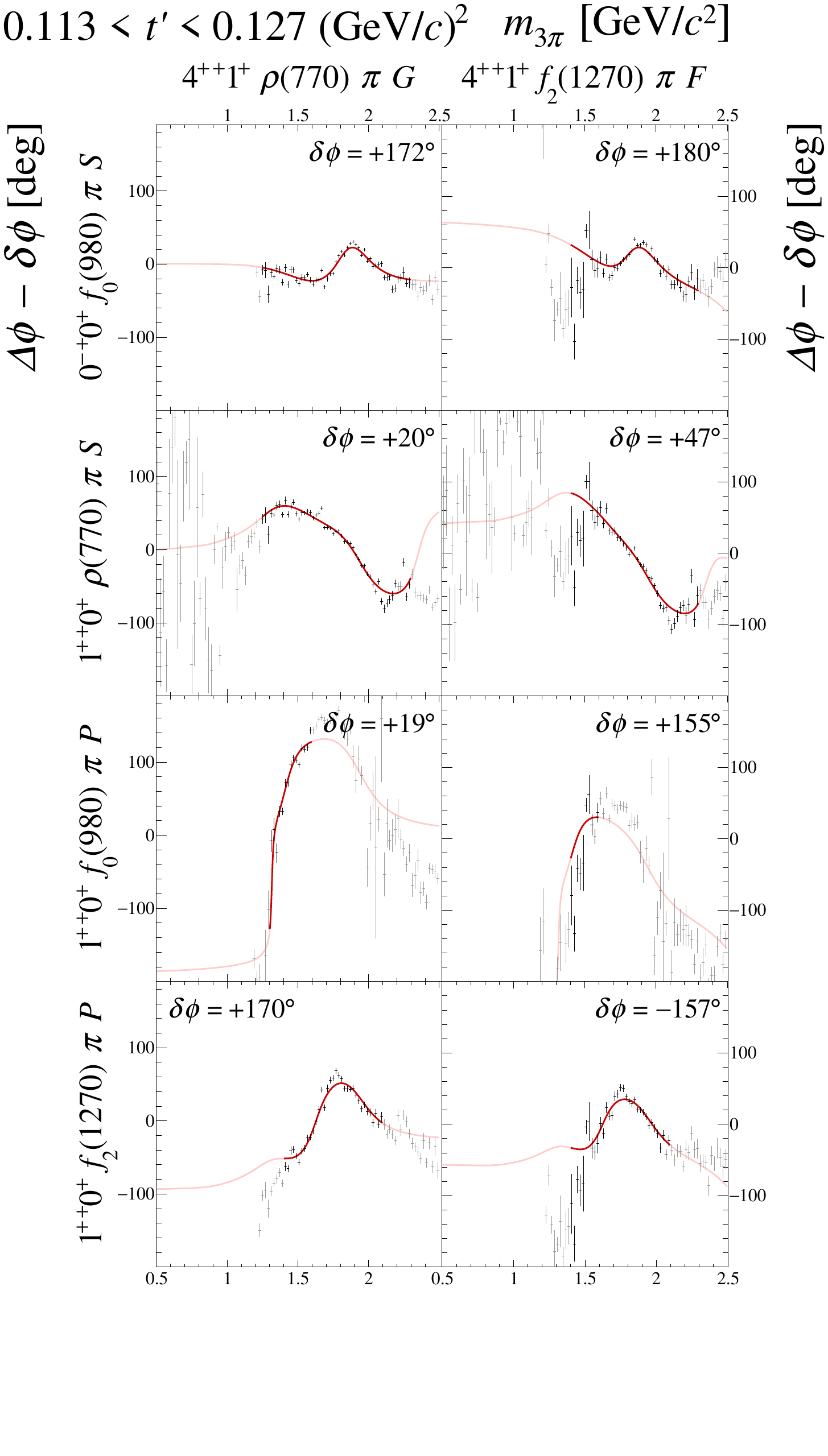}%
   \caption{Submatrix~D of the $14 \times 14$ matrix of graphs that
     represents the spin-density matrix (see
     \cref{tab:spin-dens_matrix_overview}).}
   \label{fig:spin-dens_submatrix_4_tbin_2}
 \end{minipage}
\end{textblock*}

\newpage\null
\begin{textblock*}{\textwidth}[0.5,0](0.5\paperwidth,\blockDistanceToTop)
 \begin{minipage}{\textwidth}
   \makeatletter
   \def\@captype{figure}
   \makeatother
   \centering
   \includegraphics[height=\matrixHeight]{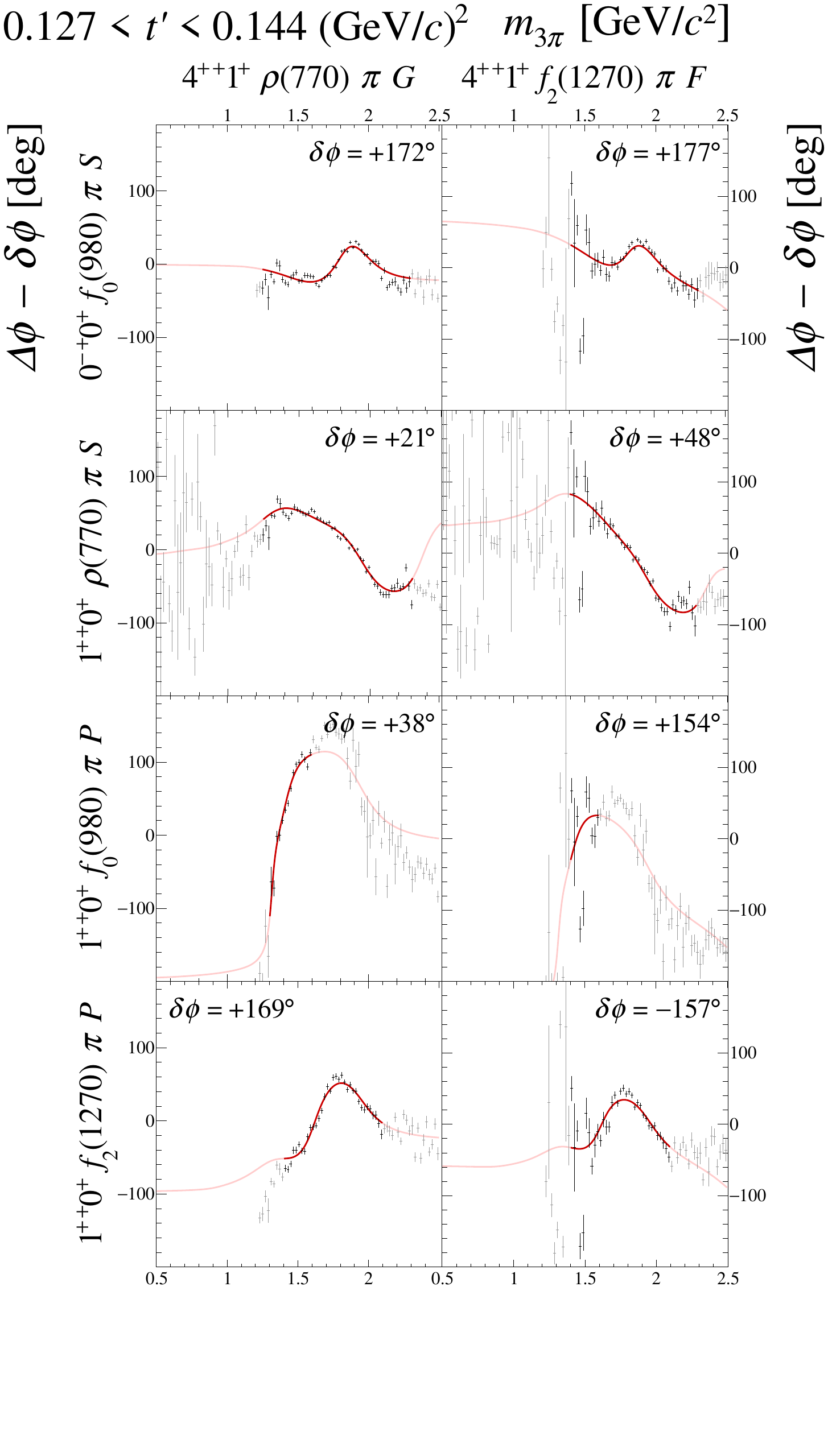}%
   \caption{Submatrix~D of the $14 \times 14$ matrix of graphs that
     represents the spin-density matrix (see
     \cref{tab:spin-dens_matrix_overview}).}
   \label{fig:spin-dens_submatrix_4_tbin_3}
 \end{minipage}
\end{textblock*}

\newpage\null
\begin{textblock*}{\textwidth}[0.5,0](0.5\paperwidth,\blockDistanceToTop)
 \begin{minipage}{\textwidth}
   \makeatletter
   \def\@captype{figure}
   \makeatother
   \centering
   \includegraphics[height=\matrixHeight]{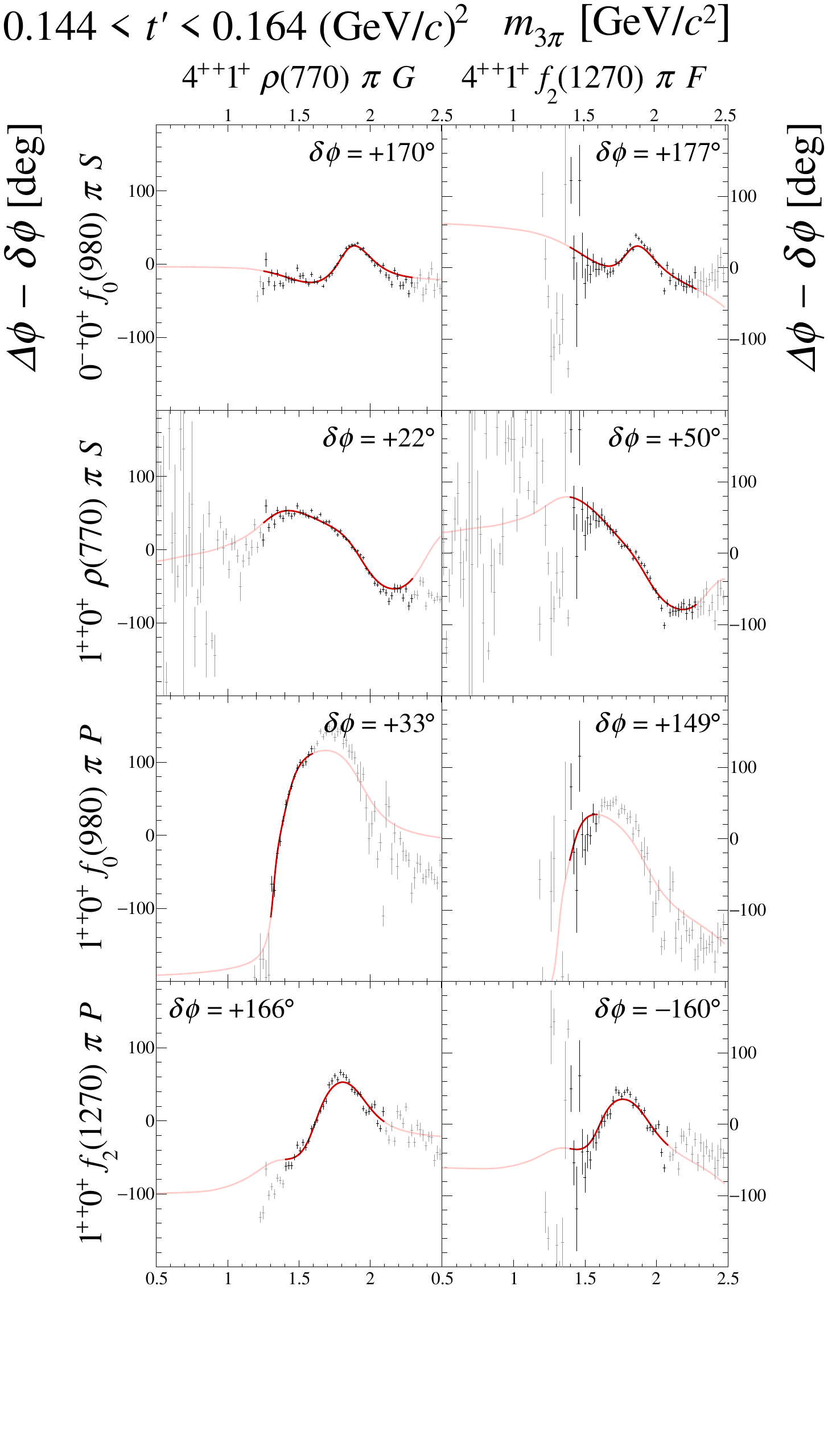}%
   \caption{Submatrix~D of the $14 \times 14$ matrix of graphs that
     represents the spin-density matrix (see
     \cref{tab:spin-dens_matrix_overview}).}
   \label{fig:spin-dens_submatrix_4_tbin_4}
 \end{minipage}
\end{textblock*}

\newpage\null
\begin{textblock*}{\textwidth}[0.5,0](0.5\paperwidth,\blockDistanceToTop)
 \begin{minipage}{\textwidth}
   \makeatletter
   \def\@captype{figure}
   \makeatother
   \centering
   \includegraphics[height=\matrixHeight]{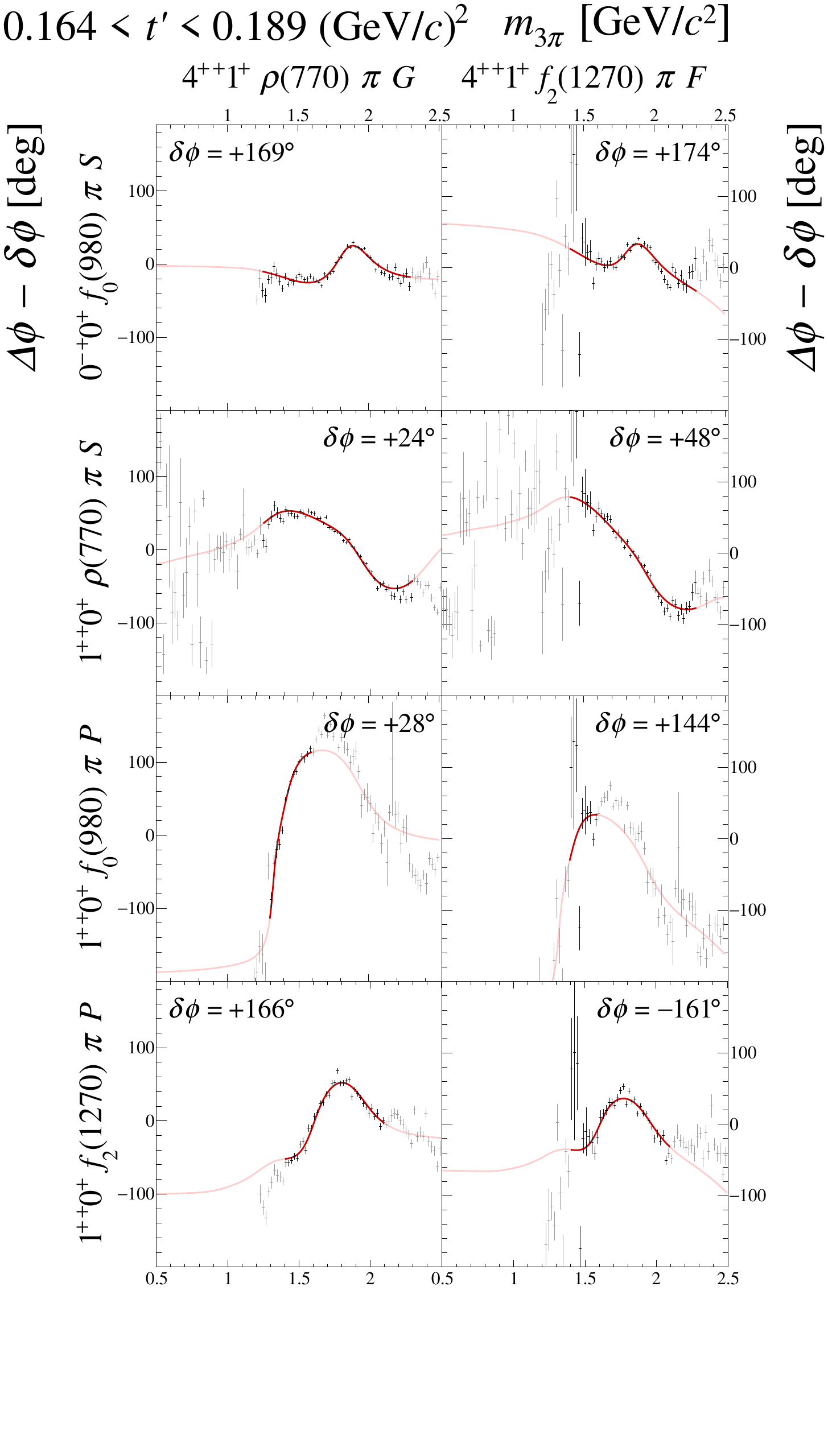}%
   \caption{Submatrix~D of the $14 \times 14$ matrix of graphs that
     represents the spin-density matrix (see
     \cref{tab:spin-dens_matrix_overview}).}
   \label{fig:spin-dens_submatrix_4_tbin_5}
 \end{minipage}
\end{textblock*}

\newpage\null
\begin{textblock*}{\textwidth}[0.5,0](0.5\paperwidth,\blockDistanceToTop)
 \begin{minipage}{\textwidth}
   \makeatletter
   \def\@captype{figure}
   \makeatother
   \centering
   \includegraphics[height=\matrixHeight]{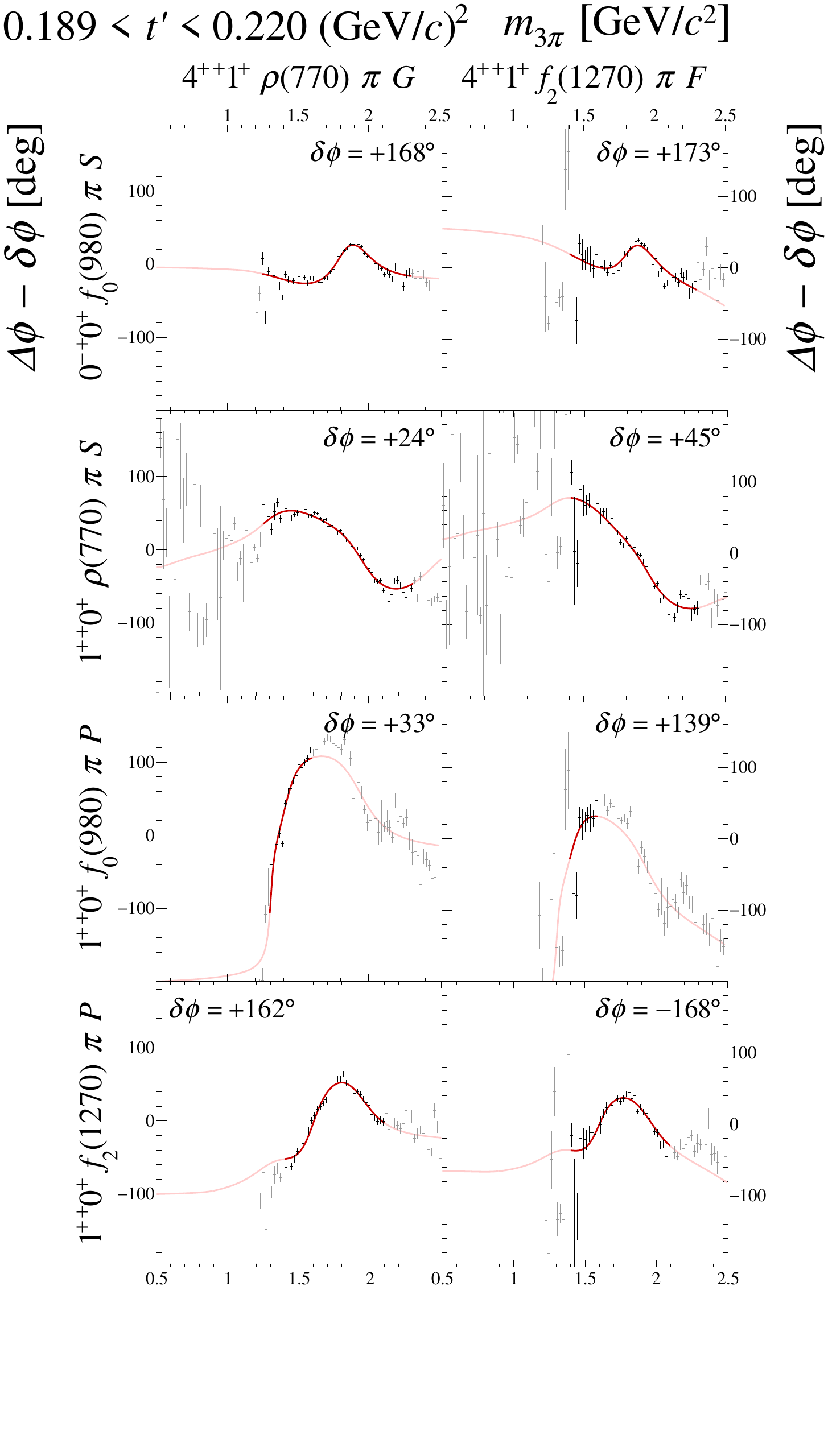}%
   \caption{Submatrix~D of the $14 \times 14$ matrix of graphs that
     represents the spin-density matrix (see
     \cref{tab:spin-dens_matrix_overview}).}
   \label{fig:spin-dens_submatrix_4_tbin_6}
 \end{minipage}
\end{textblock*}

\newpage\null
\begin{textblock*}{\textwidth}[0.5,0](0.5\paperwidth,\blockDistanceToTop)
 \begin{minipage}{\textwidth}
   \makeatletter
   \def\@captype{figure}
   \makeatother
   \centering
   \includegraphics[height=\matrixHeight]{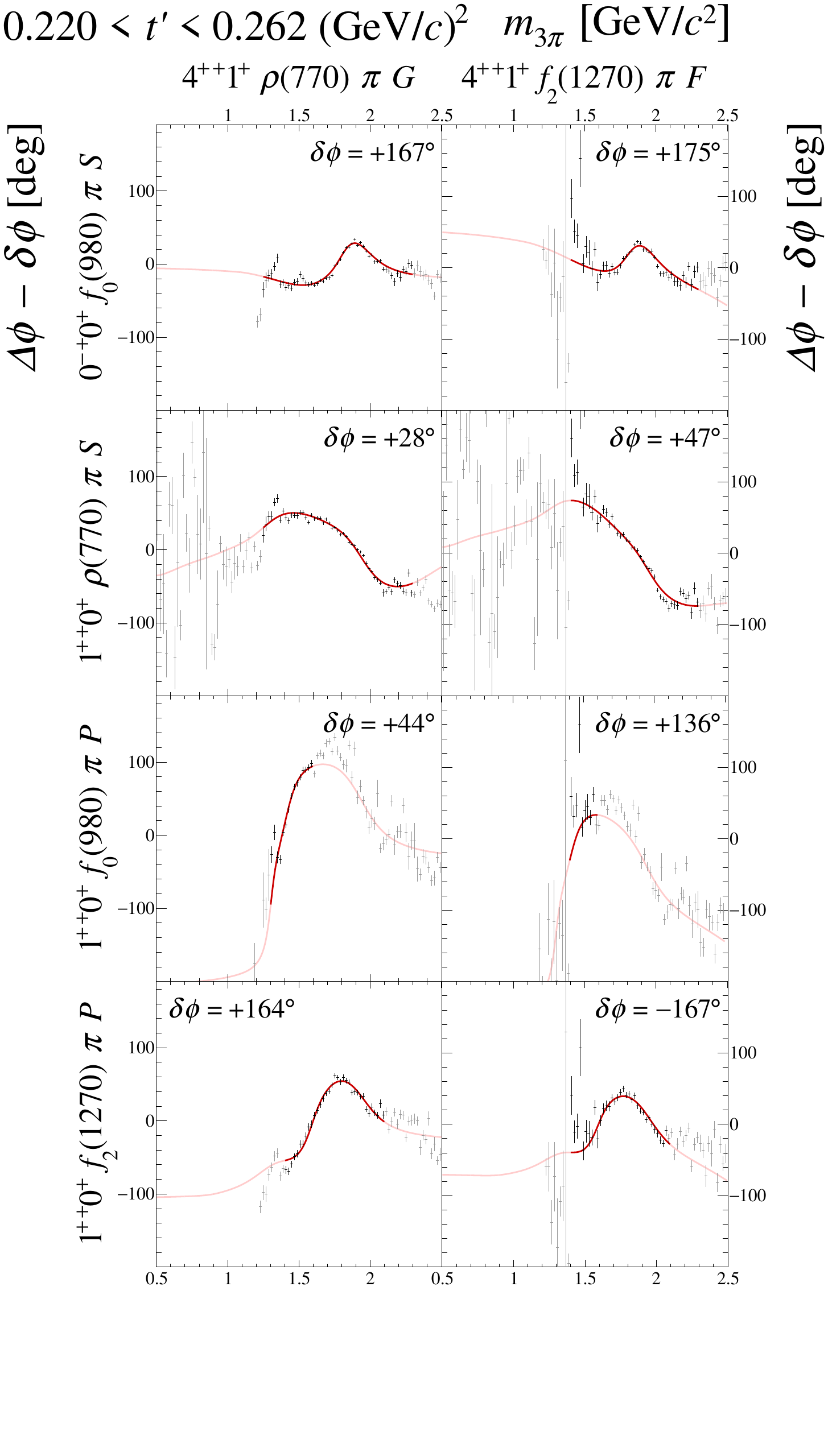}%
   \caption{Submatrix~D of the $14 \times 14$ matrix of graphs that
     represents the spin-density matrix (see
     \cref{tab:spin-dens_matrix_overview}).}
   \label{fig:spin-dens_submatrix_4_tbin_7}
 \end{minipage}
\end{textblock*}

\newpage\null
\begin{textblock*}{\textwidth}[0.5,0](0.5\paperwidth,\blockDistanceToTop)
 \begin{minipage}{\textwidth}
   \makeatletter
   \def\@captype{figure}
   \makeatother
   \centering
   \includegraphics[height=\matrixHeight]{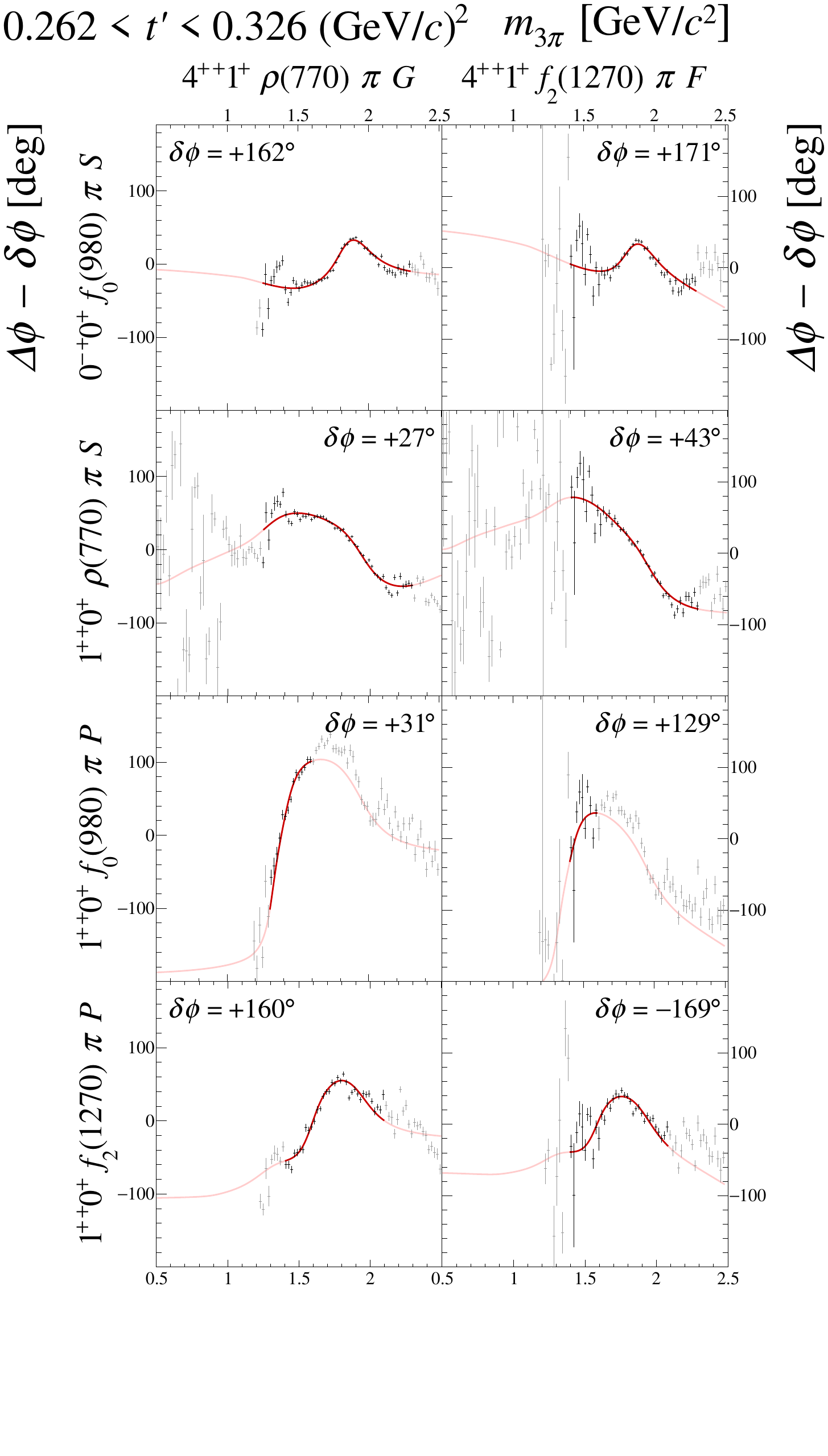}%
   \caption{Submatrix~D of the $14 \times 14$ matrix of graphs that
     represents the spin-density matrix (see
     \cref{tab:spin-dens_matrix_overview}).}
   \label{fig:spin-dens_submatrix_4_tbin_8}
 \end{minipage}
\end{textblock*}

\newpage\null
\begin{textblock*}{\textwidth}[0.5,0](0.5\paperwidth,\blockDistanceToTop)
 \begin{minipage}{\textwidth}
   \makeatletter
   \def\@captype{figure}
   \makeatother
   \centering
   \includegraphics[height=\matrixHeight]{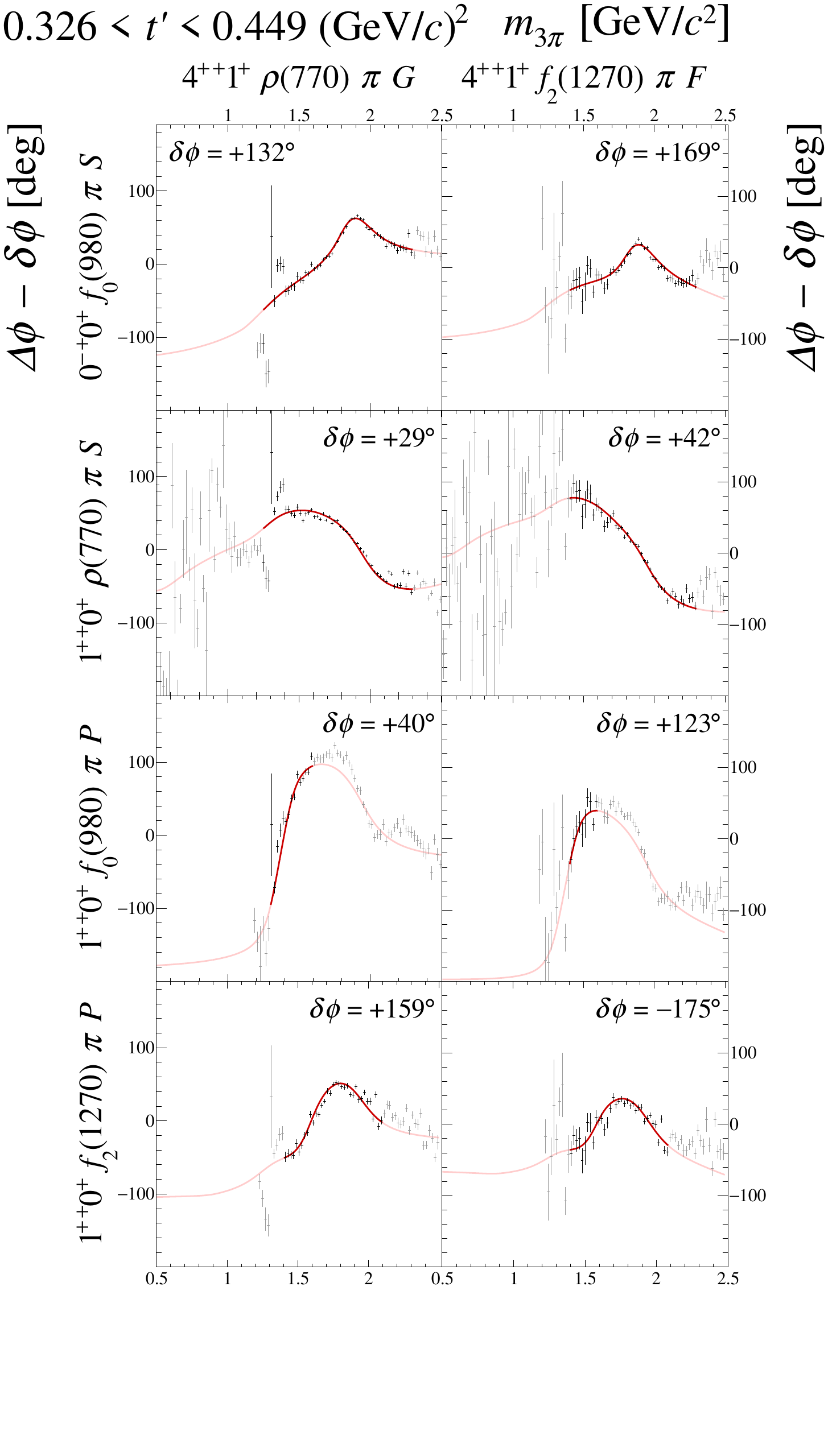}%
   \caption{Submatrix~D of the $14 \times 14$ matrix of graphs that
     represents the spin-density matrix (see
     \cref{tab:spin-dens_matrix_overview}).}
   \label{fig:spin-dens_submatrix_4_tbin_9}
 \end{minipage}
\end{textblock*}

\newpage\null
\begin{textblock*}{\textwidth}[0.5,0](0.5\paperwidth,\blockDistanceToTop)
 \begin{minipage}{\textwidth}
   \makeatletter
   \def\@captype{figure}
   \makeatother
   \centering
   \includegraphics[height=\matrixHeight]{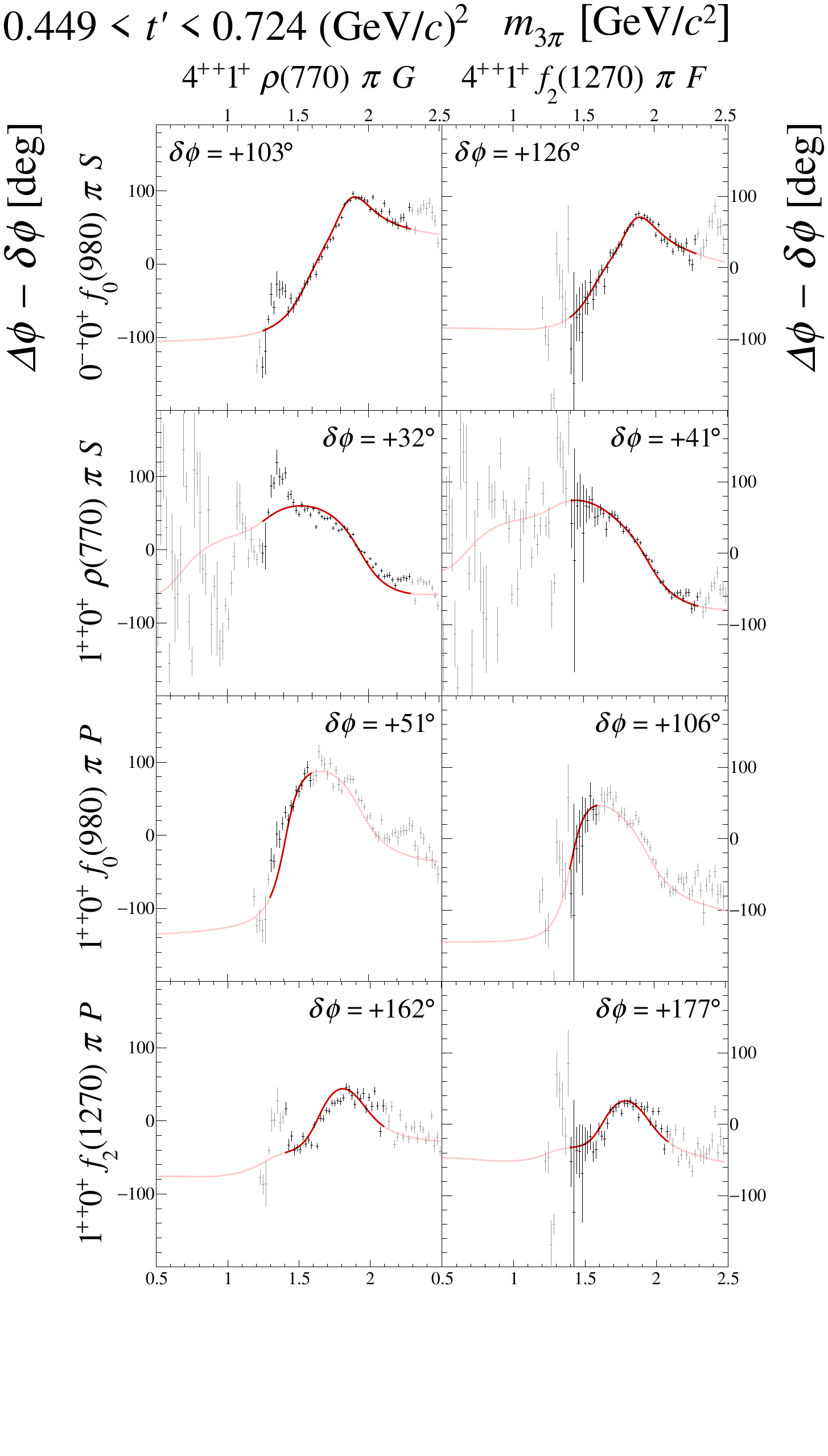}%
   \caption{Submatrix~D of the $14 \times 14$ matrix of graphs that
     represents the spin-density matrix (see
     \cref{tab:spin-dens_matrix_overview}).}
   \label{fig:spin-dens_submatrix_4_tbin_10}
 \end{minipage}
\end{textblock*}

\newpage\null
\begin{textblock*}{\textwidth}[0.5,0](0.5\paperwidth,\blockDistanceToTop)
 \begin{minipage}{\textwidth}
   \makeatletter
   \def\@captype{figure}
   \makeatother
   \centering
   \includegraphics[height=\matrixHeight]{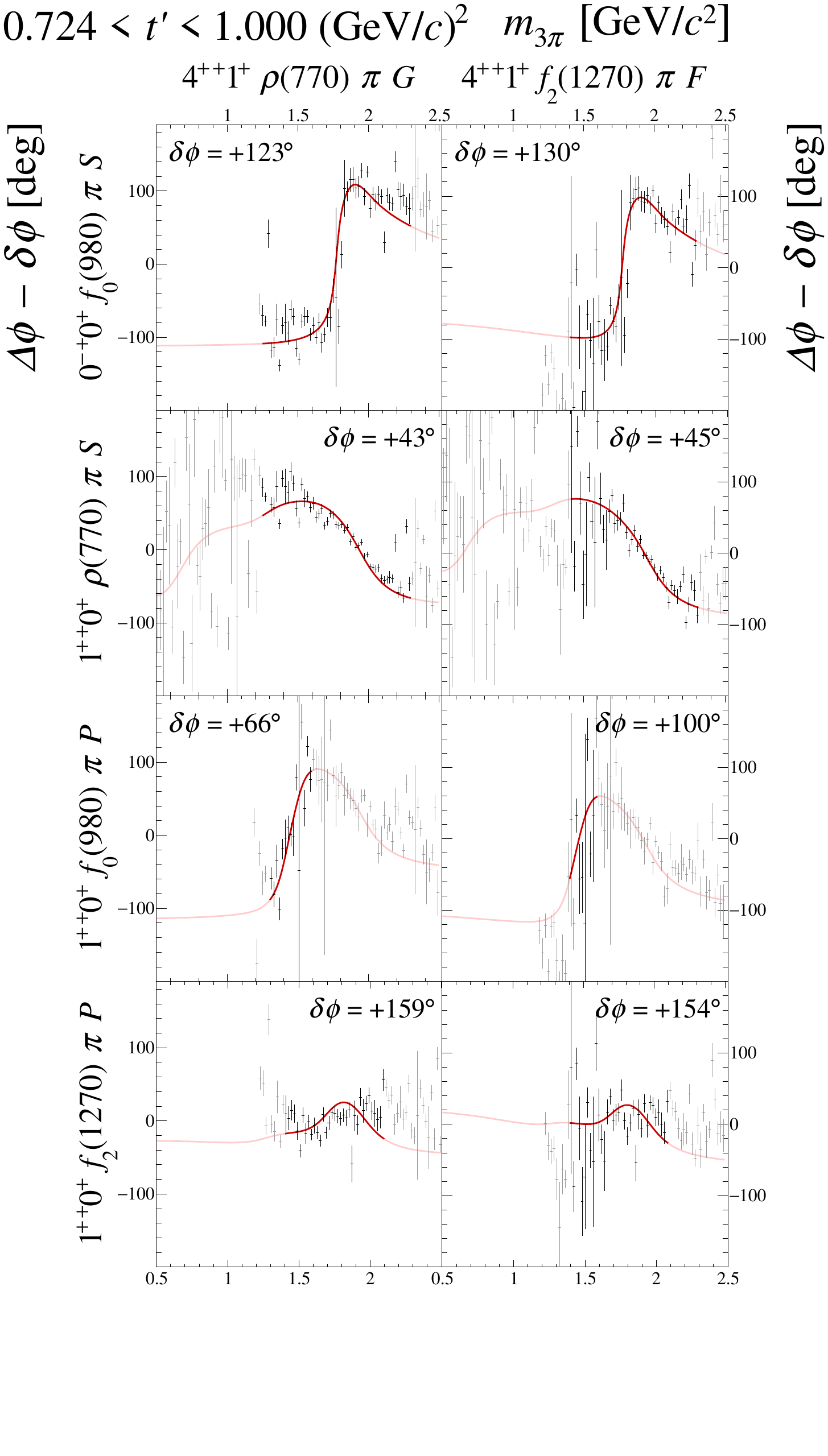}%
   \caption{Submatrix~D of the $14 \times 14$ matrix of graphs that
     represents the spin-density matrix (see
     \cref{tab:spin-dens_matrix_overview}).}
   \label{fig:spin-dens_submatrix_4_tbin_11}
 \end{minipage}
\end{textblock*}

\clearpage
\subsection{Submatrix E}
\label{sec:spin-dens_submatrix_5}

\begin{textblock*}{\textwidth}[0.5,0](0.5\paperwidth,\blockDistanceToTop)
 \begin{minipage}{\textwidth}
   \makeatletter
   \def\@captype{figure}
   \makeatother
   \centering
   \includegraphics[height=\matrixHeight]{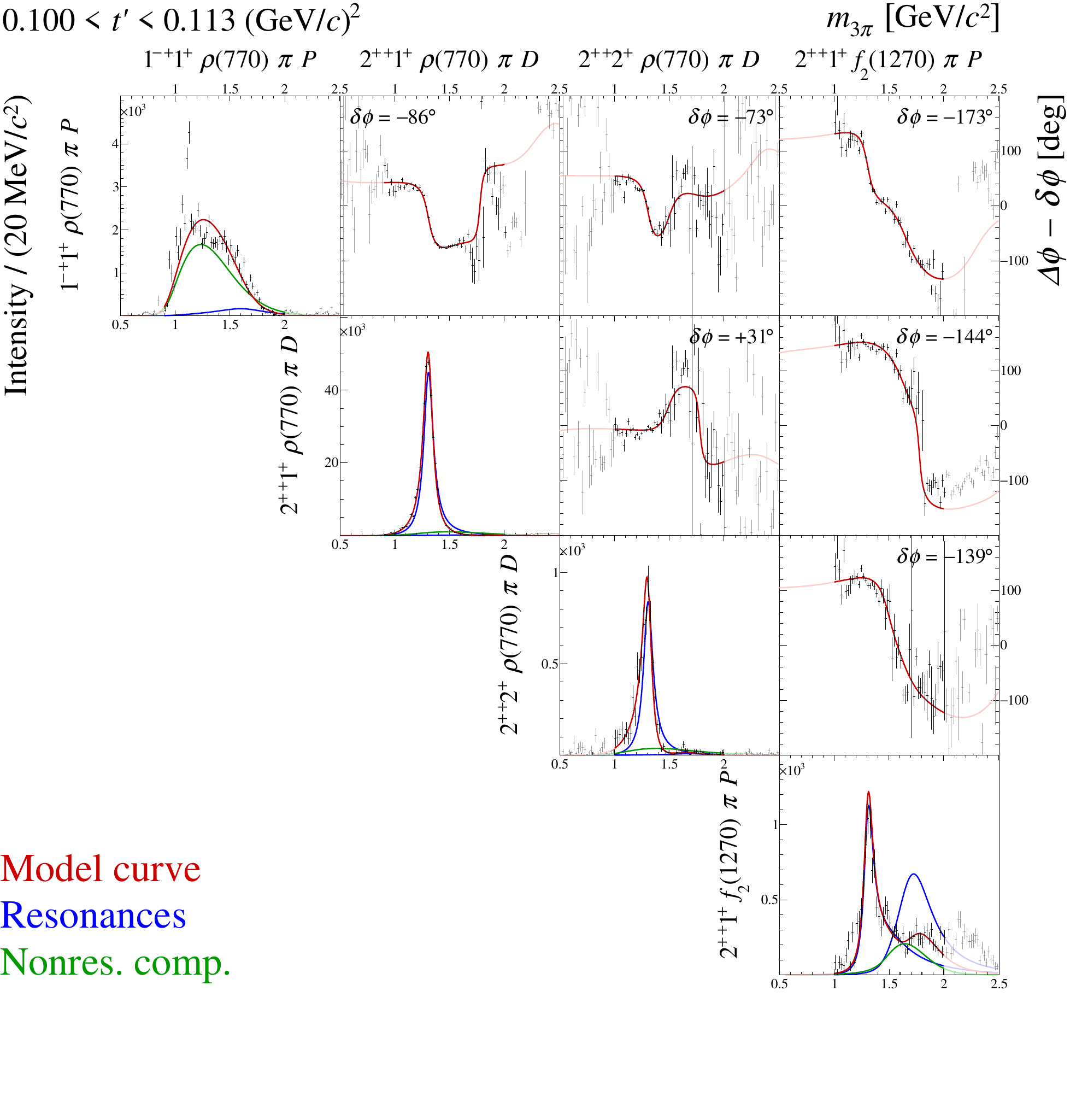}%
   \caption{Submatrix~E of the $14 \times 14$ matrix of graphs that
     represents the spin-density matrix (see
     \cref{tab:spin-dens_matrix_overview}).}
   \label{fig:spin-dens_submatrix_5_tbin_1}
 \end{minipage}
\end{textblock*}

\newpage\null
\begin{textblock*}{\textwidth}[0.5,0](0.5\paperwidth,\blockDistanceToTop)
 \begin{minipage}{\textwidth}
   \makeatletter
   \def\@captype{figure}
   \makeatother
   \centering
   \includegraphics[height=\matrixHeight]{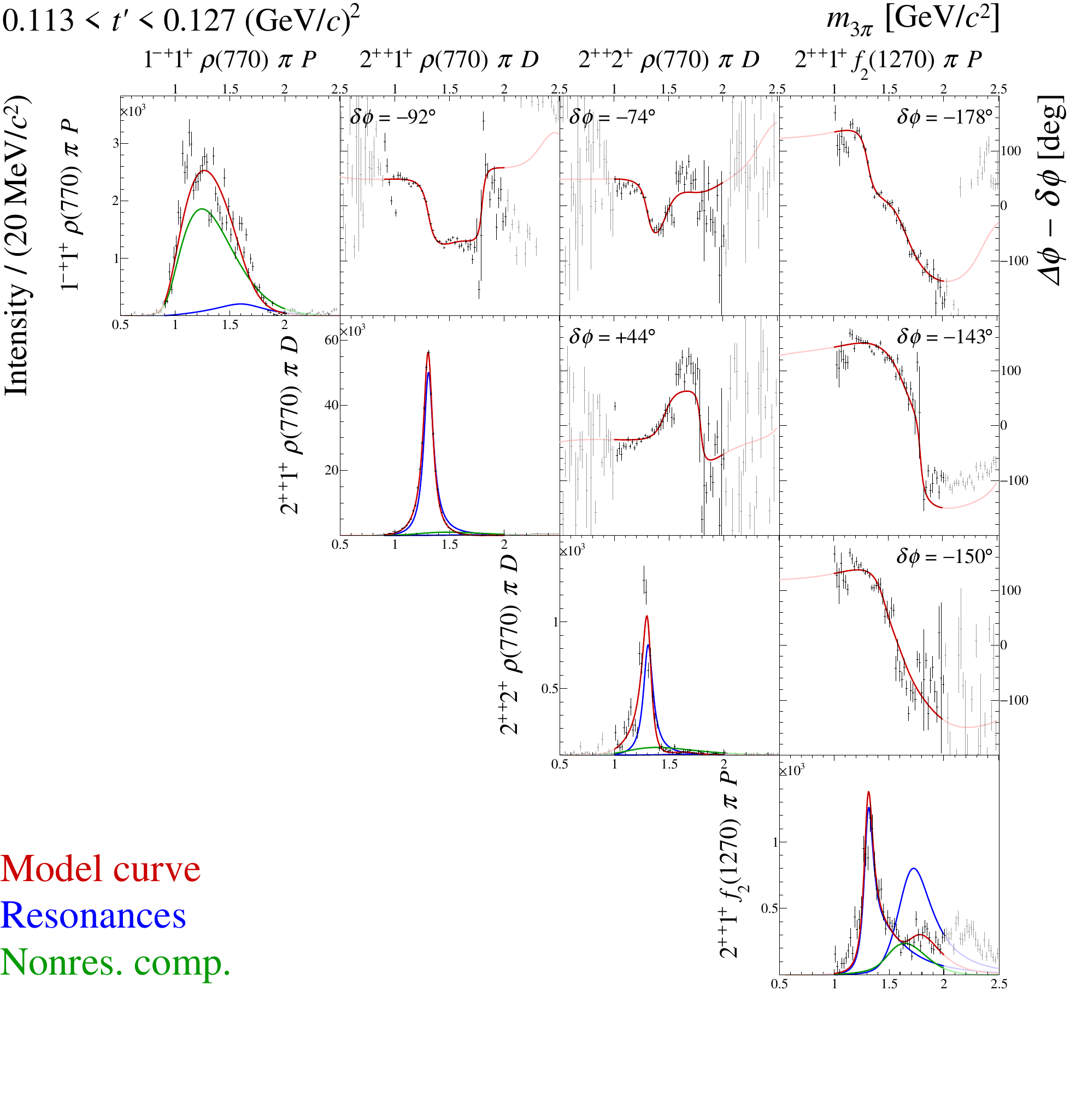}%
   \caption{Submatrix~E of the $14 \times 14$ matrix of graphs that
     represents the spin-density matrix (see
     \cref{tab:spin-dens_matrix_overview}).}
   \label{fig:spin-dens_submatrix_5_tbin_2}
 \end{minipage}
\end{textblock*}

\newpage\null
\begin{textblock*}{\textwidth}[0.5,0](0.5\paperwidth,\blockDistanceToTop)
 \begin{minipage}{\textwidth}
   \makeatletter
   \def\@captype{figure}
   \makeatother
   \centering
   \includegraphics[height=\matrixHeight]{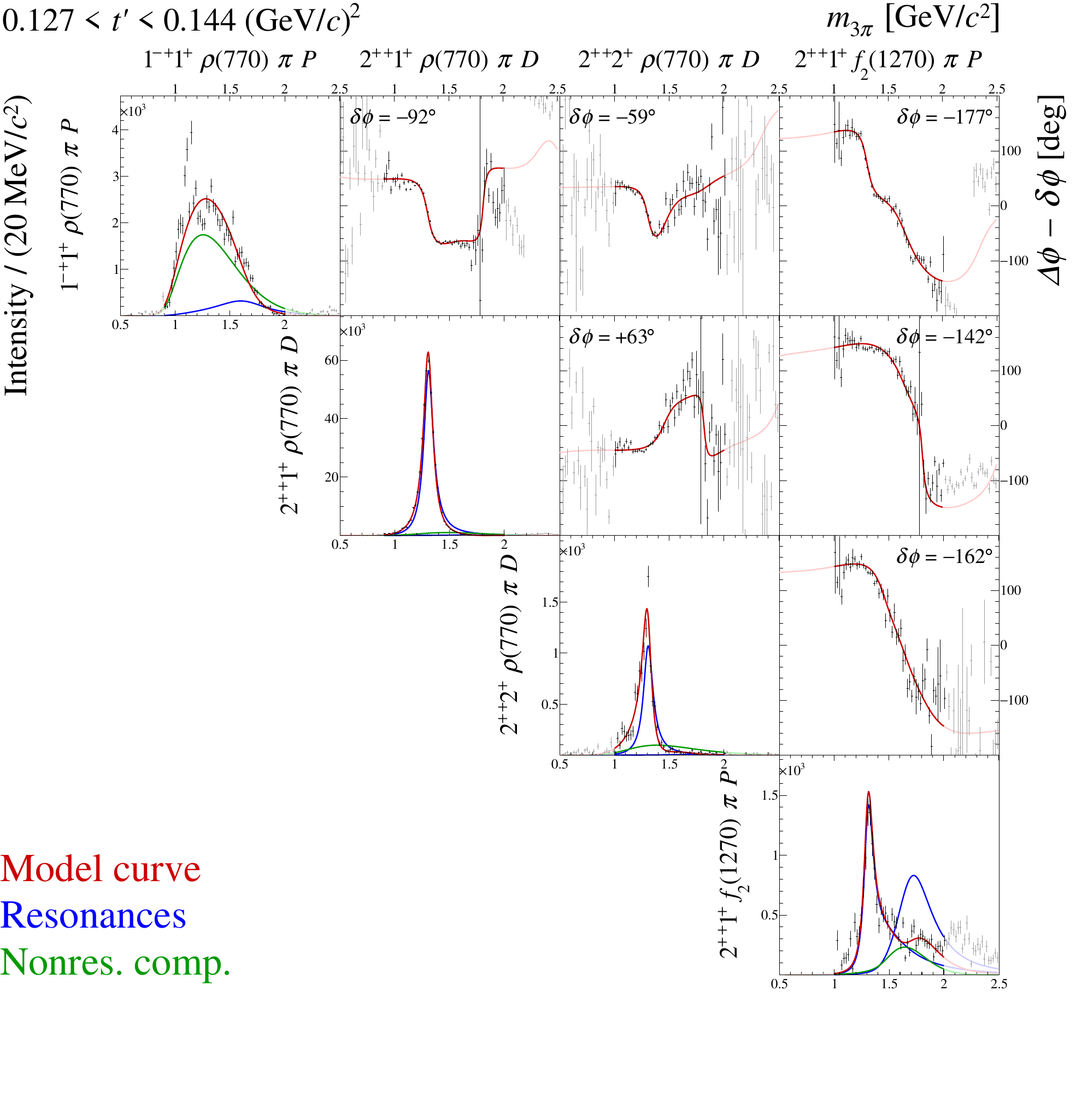}%
   \caption{Submatrix~E of the $14 \times 14$ matrix of graphs that
     represents the spin-density matrix (see
     \cref{tab:spin-dens_matrix_overview}).}
   \label{fig:spin-dens_submatrix_5_tbin_3}
 \end{minipage}
\end{textblock*}

\newpage\null
\begin{textblock*}{\textwidth}[0.5,0](0.5\paperwidth,\blockDistanceToTop)
 \begin{minipage}{\textwidth}
   \makeatletter
   \def\@captype{figure}
   \makeatother
   \centering
   \includegraphics[height=\matrixHeight]{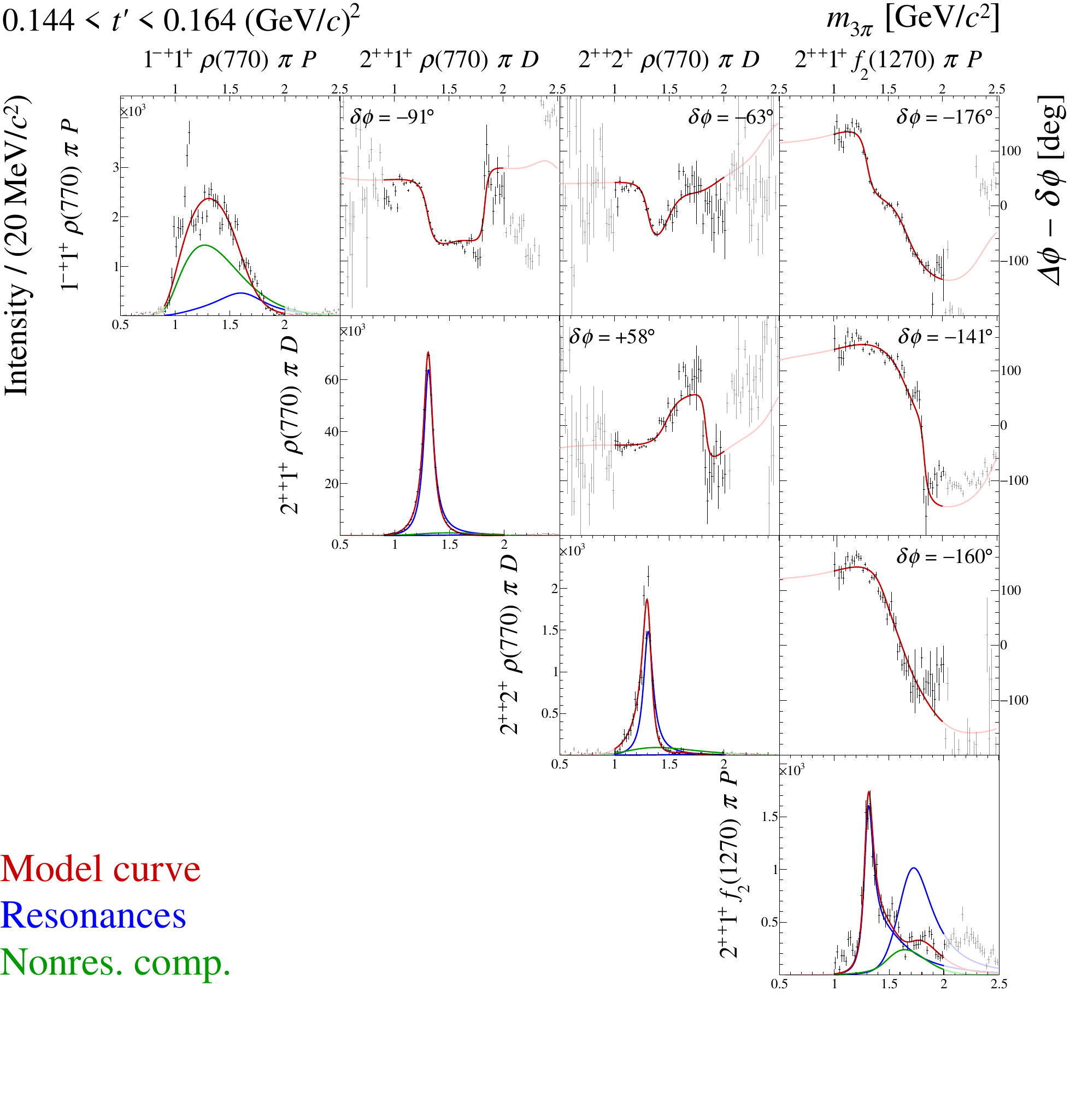}%
   \caption{Submatrix~E of the $14 \times 14$ matrix of graphs that
     represents the spin-density matrix (see
     \cref{tab:spin-dens_matrix_overview}).}
   \label{fig:spin-dens_submatrix_5_tbin_4}
 \end{minipage}
\end{textblock*}

\newpage\null
\begin{textblock*}{\textwidth}[0.5,0](0.5\paperwidth,\blockDistanceToTop)
 \begin{minipage}{\textwidth}
   \makeatletter
   \def\@captype{figure}
   \makeatother
   \centering
   \includegraphics[height=\matrixHeight]{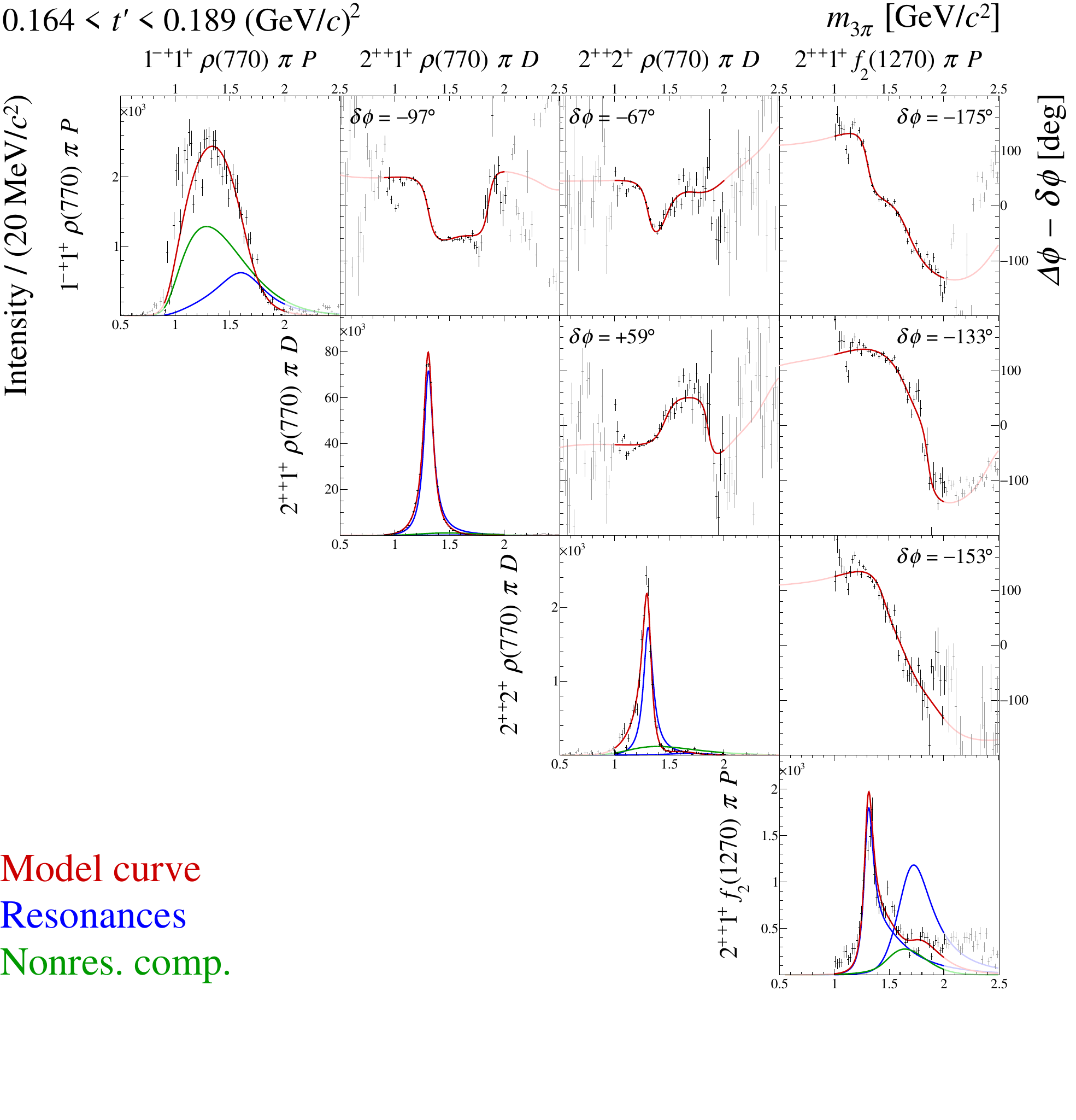}%
   \caption{Submatrix~E of the $14 \times 14$ matrix of graphs that
     represents the spin-density matrix (see
     \cref{tab:spin-dens_matrix_overview}).}
   \label{fig:spin-dens_submatrix_5_tbin_5}
 \end{minipage}
\end{textblock*}

\newpage\null
\begin{textblock*}{\textwidth}[0.5,0](0.5\paperwidth,\blockDistanceToTop)
 \begin{minipage}{\textwidth}
   \makeatletter
   \def\@captype{figure}
   \makeatother
   \centering
   \includegraphics[height=\matrixHeight]{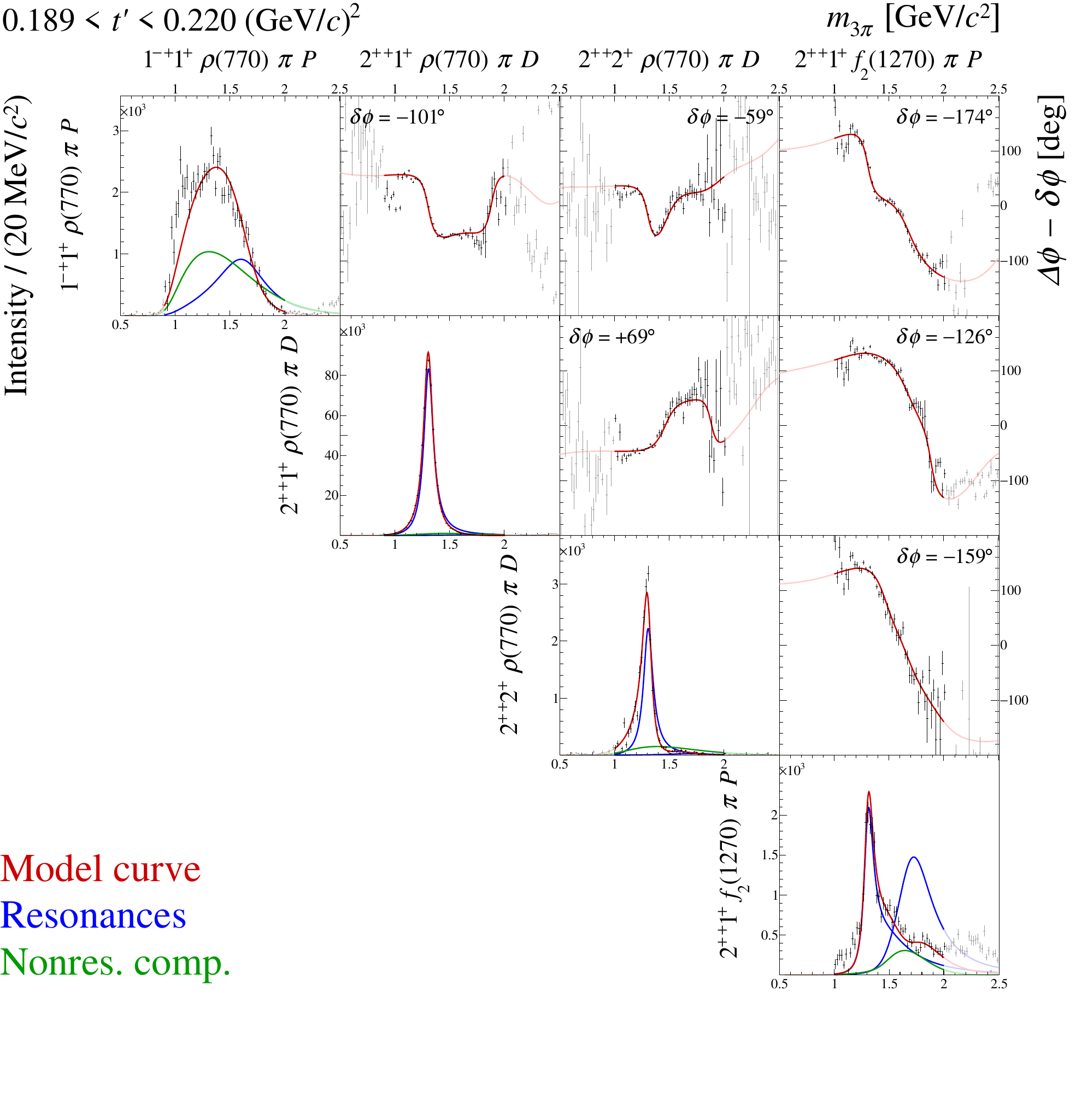}%
   \caption{Submatrix~E of the $14 \times 14$ matrix of graphs that
     represents the spin-density matrix (see
     \cref{tab:spin-dens_matrix_overview}).}
   \label{fig:spin-dens_submatrix_5_tbin_6}
 \end{minipage}
\end{textblock*}

\newpage\null
\begin{textblock*}{\textwidth}[0.5,0](0.5\paperwidth,\blockDistanceToTop)
 \begin{minipage}{\textwidth}
   \makeatletter
   \def\@captype{figure}
   \makeatother
   \centering
   \includegraphics[height=\matrixHeight]{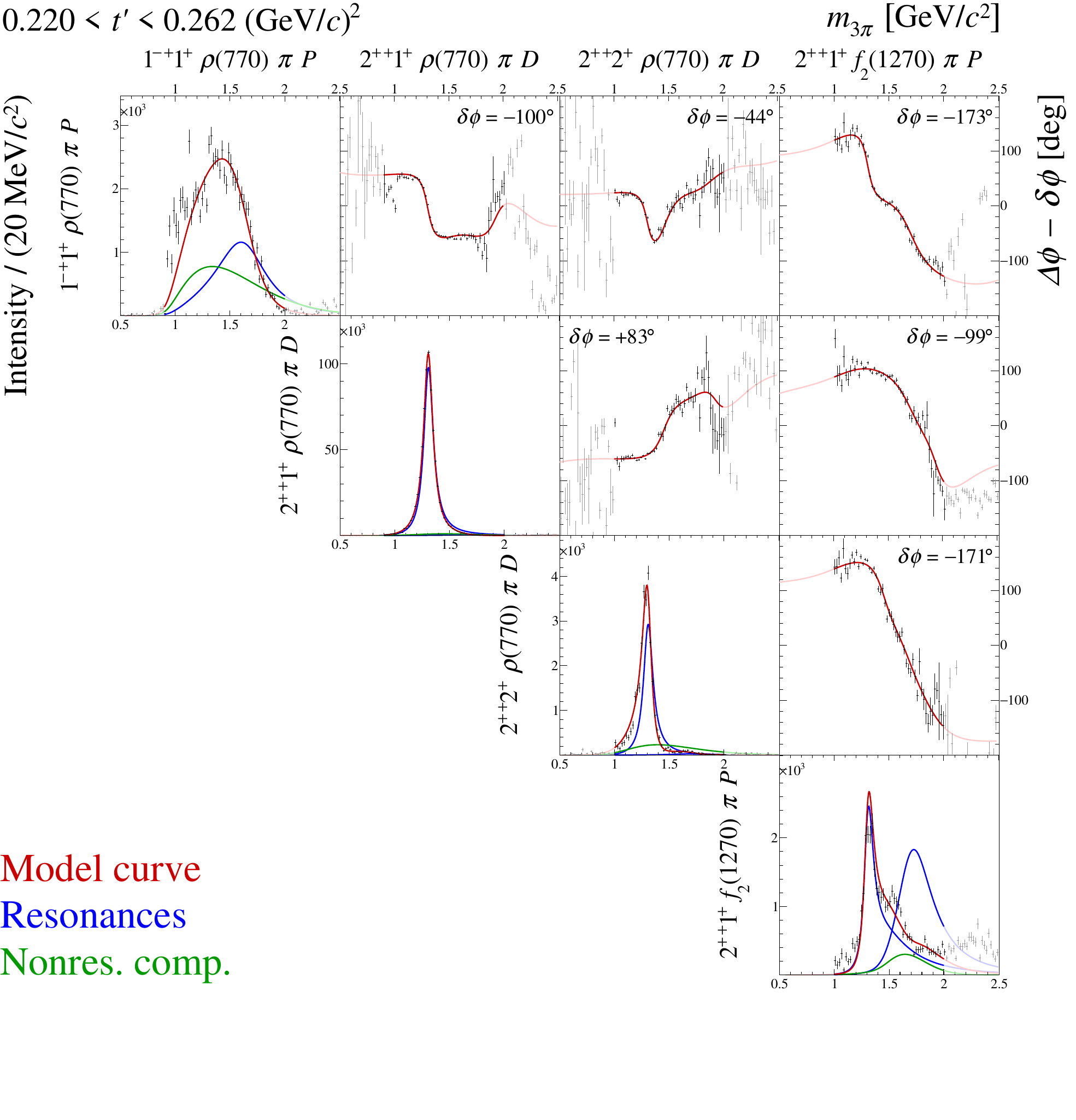}%
   \caption{Submatrix~E of the $14 \times 14$ matrix of graphs that
     represents the spin-density matrix (see
     \cref{tab:spin-dens_matrix_overview}).}
   \label{fig:spin-dens_submatrix_5_tbin_7}
 \end{minipage}
\end{textblock*}

\newpage\null
\begin{textblock*}{\textwidth}[0.5,0](0.5\paperwidth,\blockDistanceToTop)
 \begin{minipage}{\textwidth}
   \makeatletter
   \def\@captype{figure}
   \makeatother
   \centering
   \includegraphics[height=\matrixHeight]{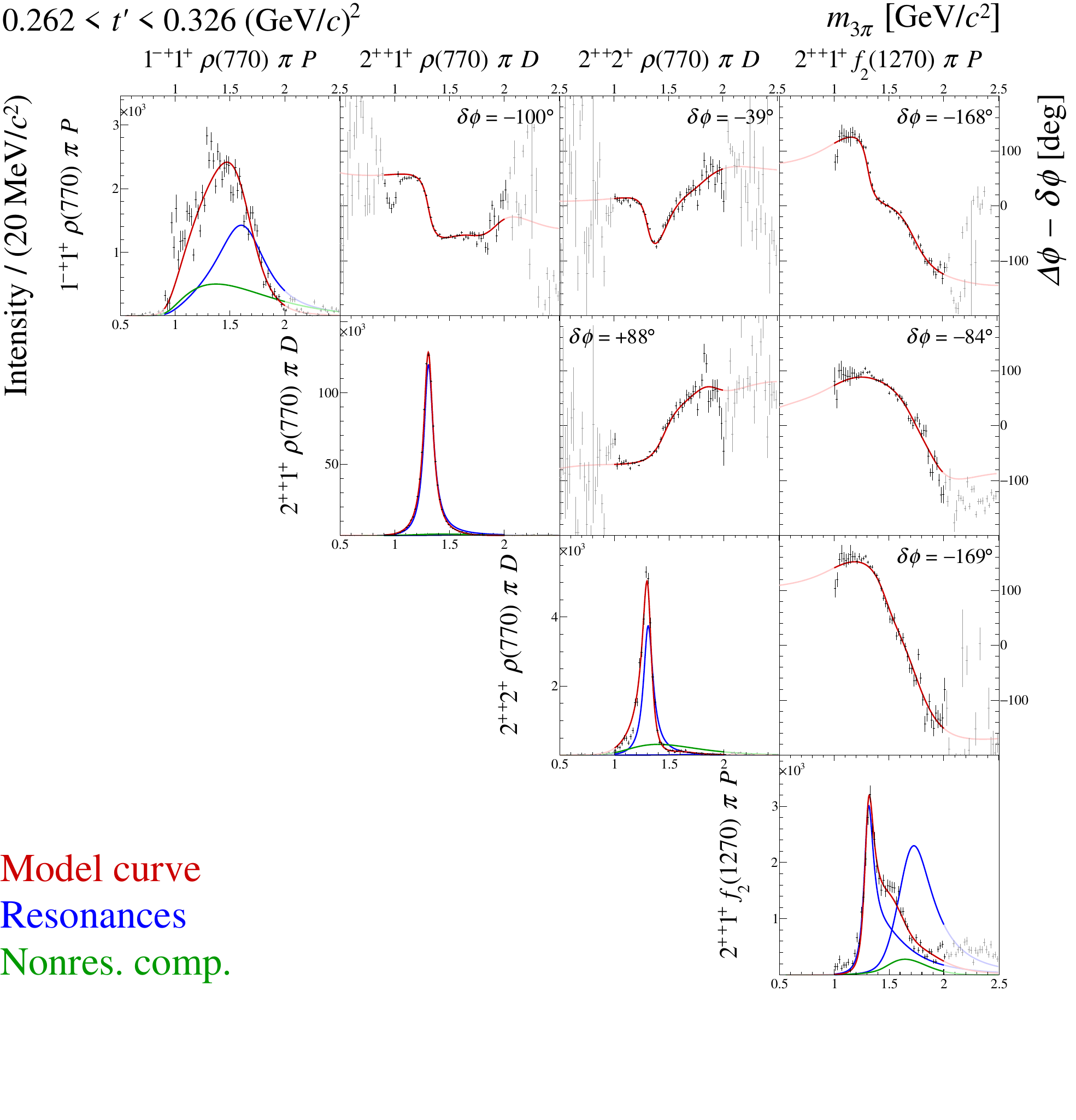}%
   \caption{Submatrix~E of the $14 \times 14$ matrix of graphs that
     represents the spin-density matrix (see
     \cref{tab:spin-dens_matrix_overview}).}
   \label{fig:spin-dens_submatrix_5_tbin_8}
 \end{minipage}
\end{textblock*}

\newpage\null
\begin{textblock*}{\textwidth}[0.5,0](0.5\paperwidth,\blockDistanceToTop)
 \begin{minipage}{\textwidth}
   \makeatletter
   \def\@captype{figure}
   \makeatother
   \centering
   \includegraphics[height=\matrixHeight]{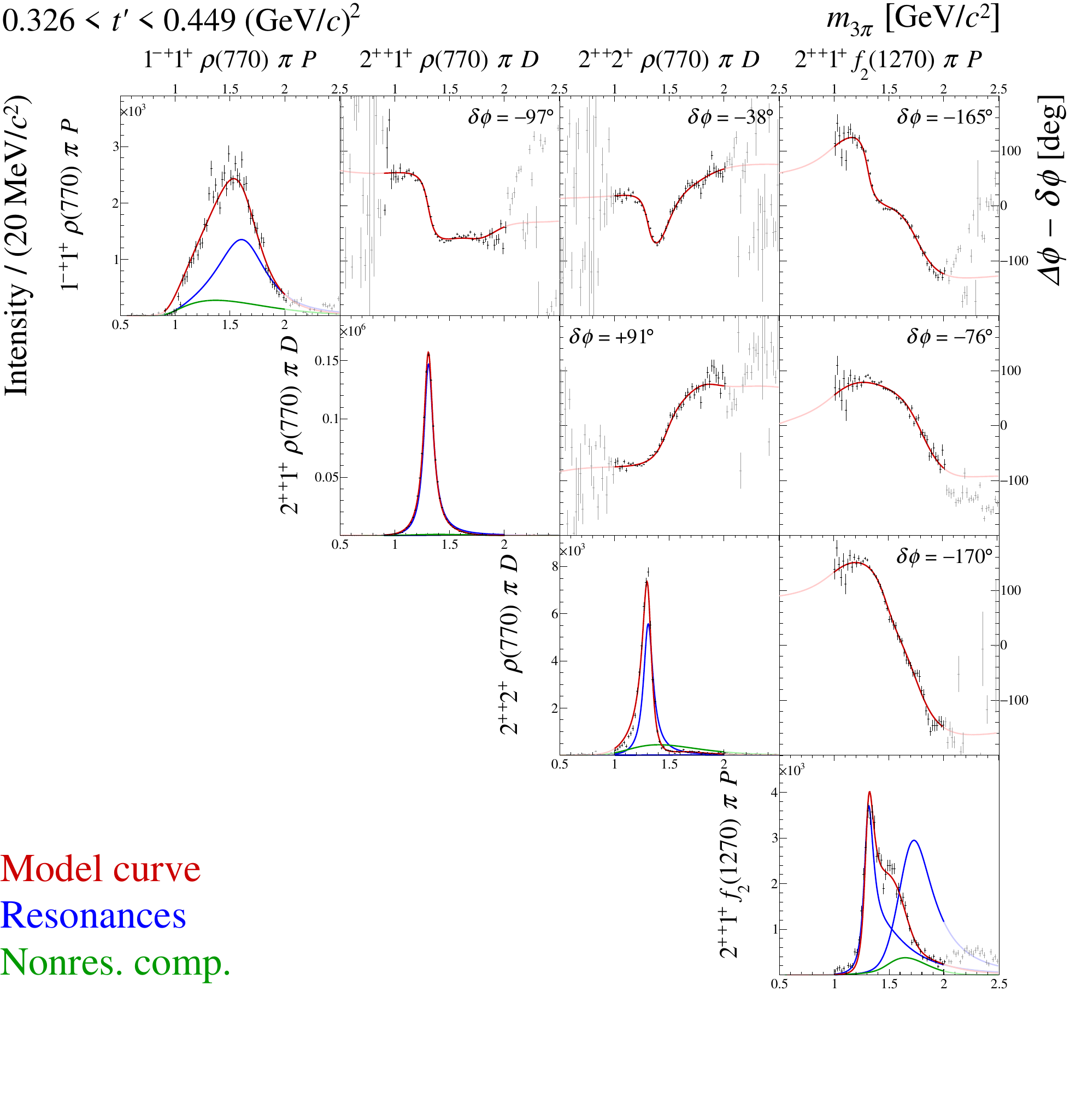}%
   \caption{Submatrix~E of the $14 \times 14$ matrix of graphs that
     represents the spin-density matrix (see
     \cref{tab:spin-dens_matrix_overview}).}
   \label{fig:spin-dens_submatrix_5_tbin_9}
 \end{minipage}
\end{textblock*}

\newpage\null
\begin{textblock*}{\textwidth}[0.5,0](0.5\paperwidth,\blockDistanceToTop)
 \begin{minipage}{\textwidth}
   \makeatletter
   \def\@captype{figure}
   \makeatother
   \centering
   \includegraphics[height=\matrixHeight]{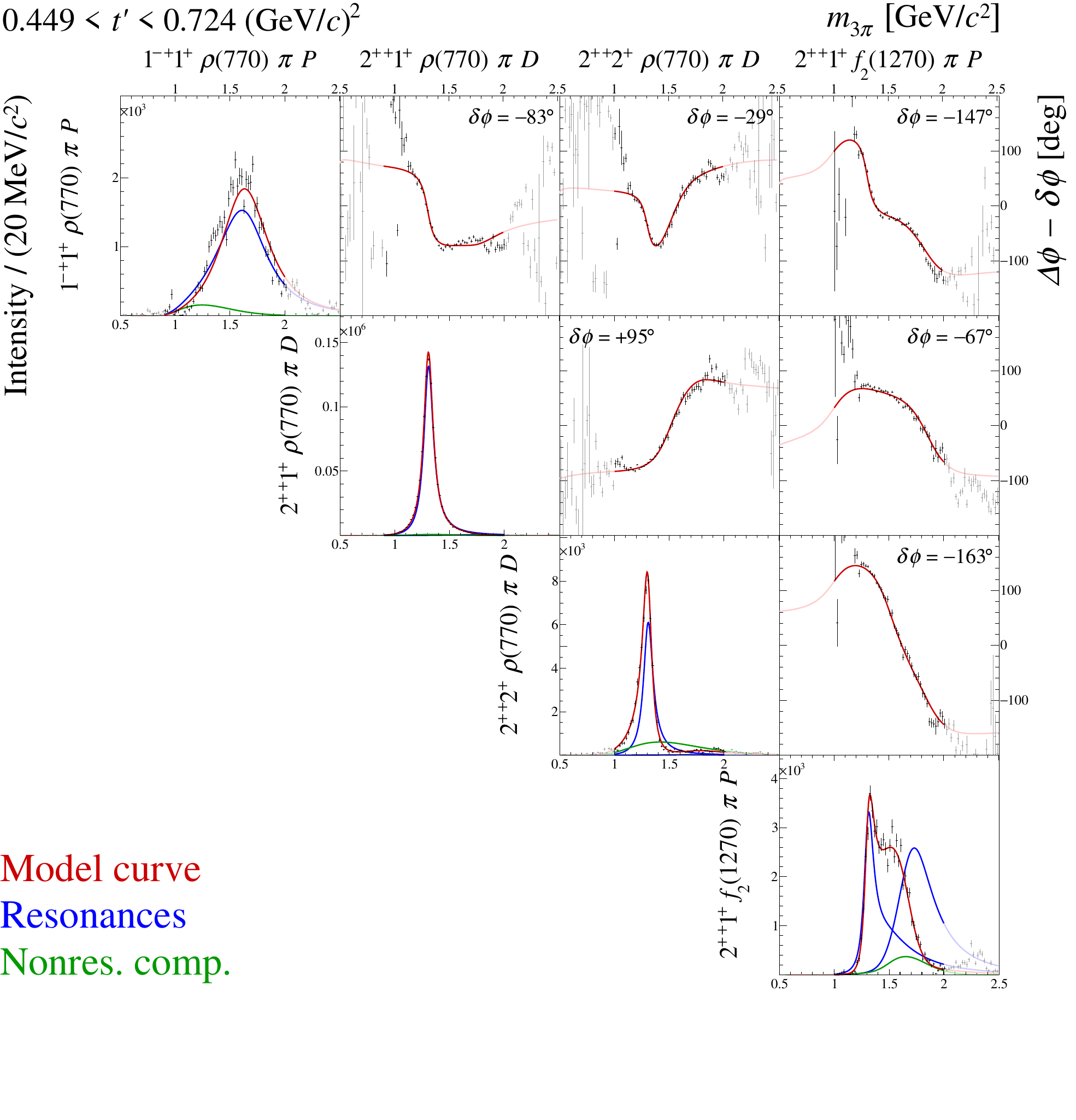}%
   \caption{Submatrix~E of the $14 \times 14$ matrix of graphs that
     represents the spin-density matrix (see
     \cref{tab:spin-dens_matrix_overview}).}
   \label{fig:spin-dens_submatrix_5_tbin_10}
 \end{minipage}
\end{textblock*}

\newpage\null
\begin{textblock*}{\textwidth}[0.5,0](0.5\paperwidth,\blockDistanceToTop)
 \begin{minipage}{\textwidth}
   \makeatletter
   \def\@captype{figure}
   \makeatother
   \centering
   \includegraphics[height=\matrixHeight]{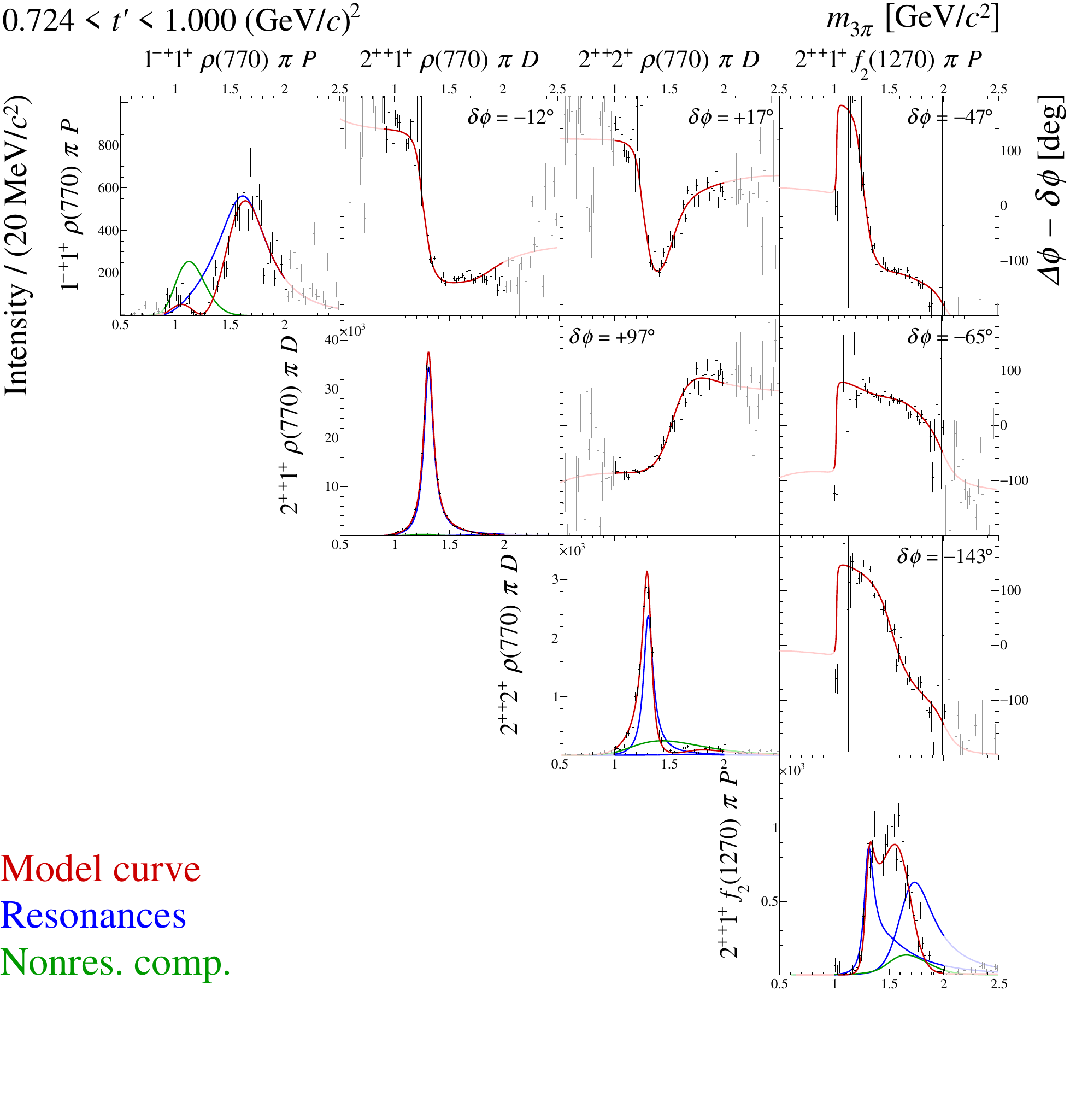}%
   \caption{Submatrix~E of the $14 \times 14$ matrix of graphs that
     represents the spin-density matrix (see
     \cref{tab:spin-dens_matrix_overview}).}
   \label{fig:spin-dens_submatrix_5_tbin_11}
 \end{minipage}
\end{textblock*}

\clearpage
\subsection{Submatrix F}
\label{sec:spin-dens_submatrix_6}

\begin{textblock*}{\textwidth}[0.5,0](0.5\paperwidth,\blockDistanceToTop)
 \begin{minipage}{\textwidth}
   \makeatletter
   \def\@captype{figure}
   \makeatother
   \centering
   \includegraphics[height=\matrixHeight]{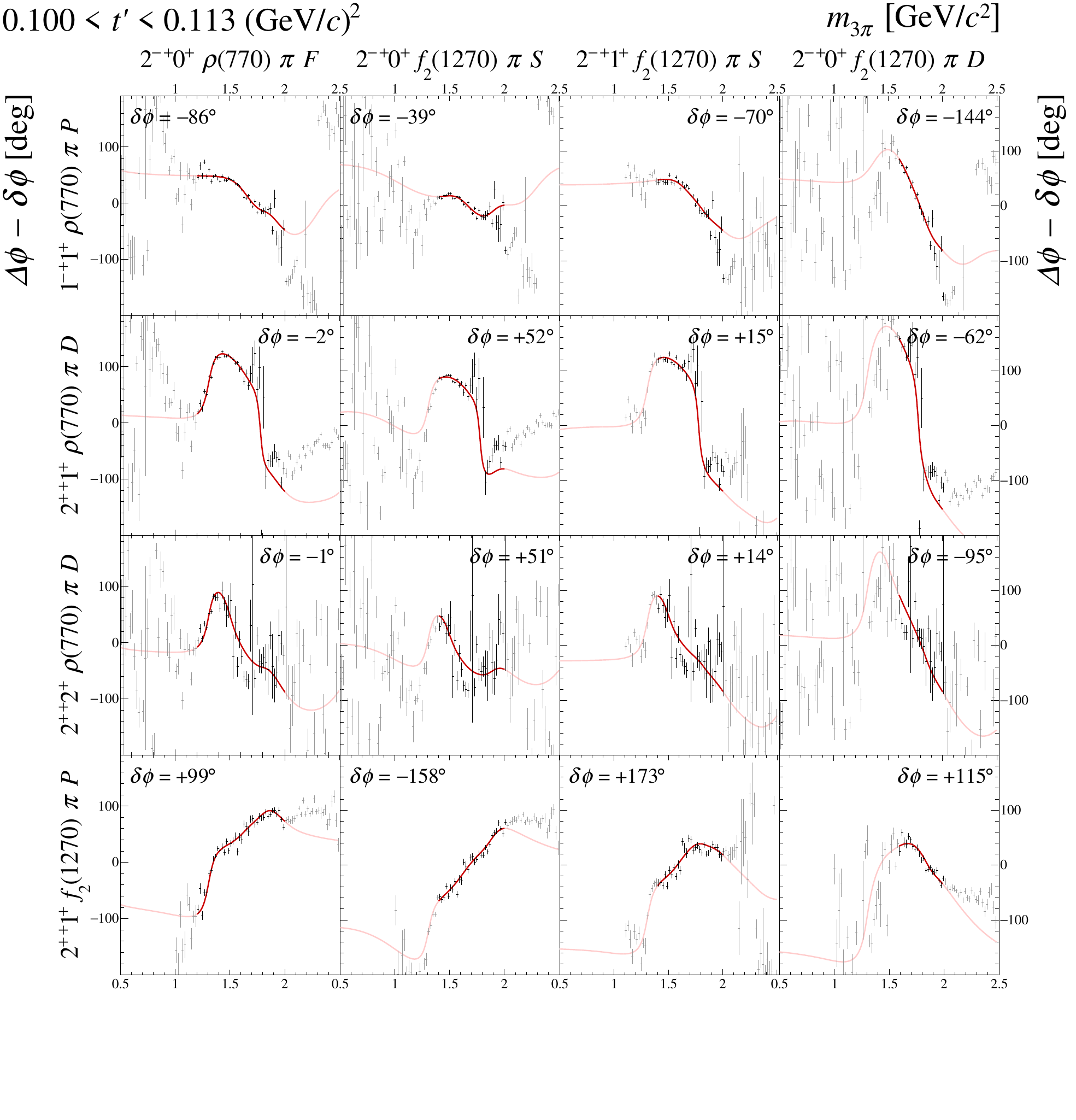}%
   \caption{Submatrix~F of the $14 \times 14$ matrix of graphs that
     represents the spin-density matrix (see
     \cref{tab:spin-dens_matrix_overview}).}
   \label{fig:spin-dens_submatrix_6_tbin_1}
 \end{minipage}
\end{textblock*}

\newpage\null
\begin{textblock*}{\textwidth}[0.5,0](0.5\paperwidth,\blockDistanceToTop)
 \begin{minipage}{\textwidth}
   \makeatletter
   \def\@captype{figure}
   \makeatother
   \centering
   \includegraphics[height=\matrixHeight]{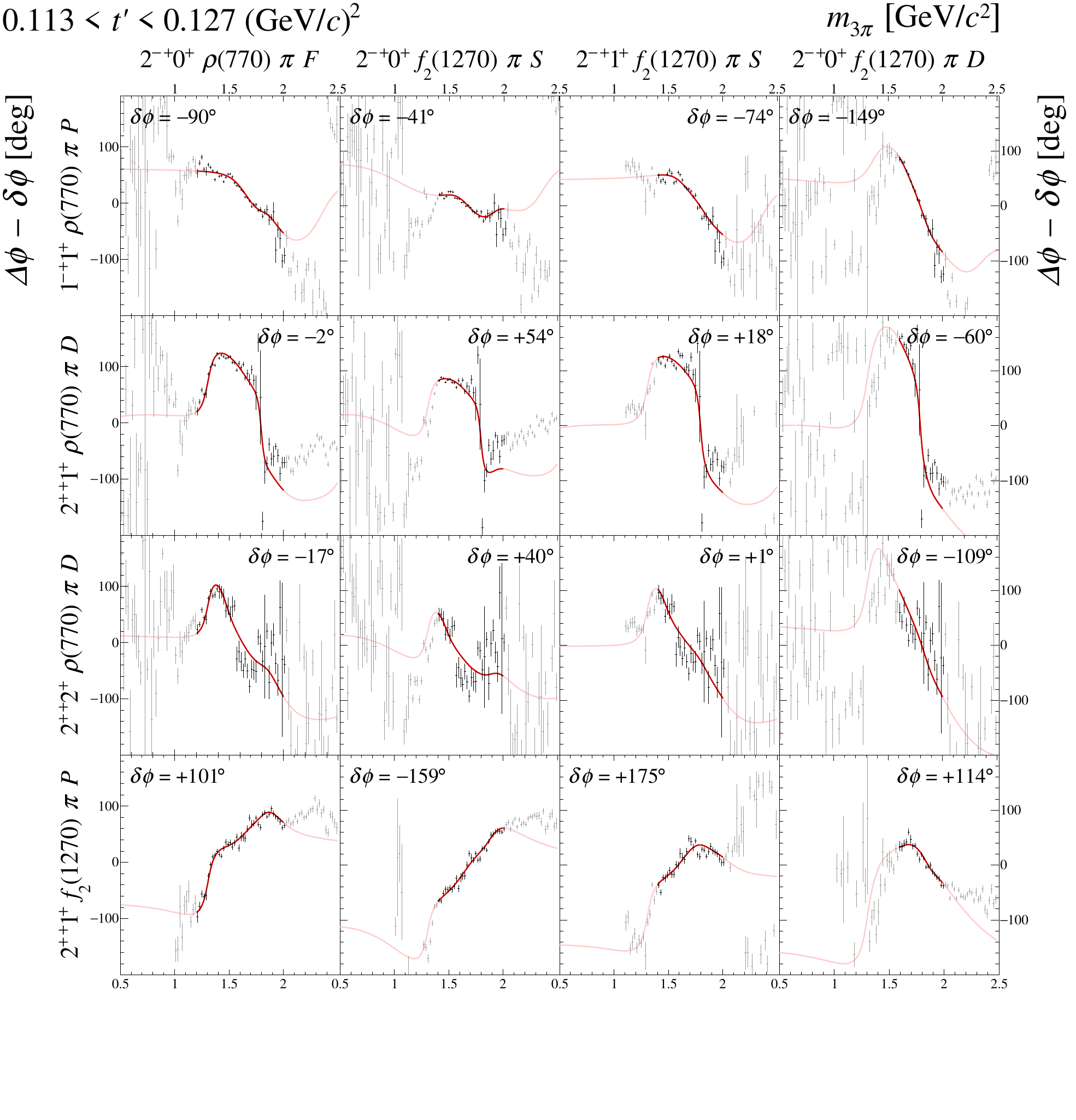}%
   \caption{Submatrix~F of the $14 \times 14$ matrix of graphs that
     represents the spin-density matrix (see
     \cref{tab:spin-dens_matrix_overview}).}
   \label{fig:spin-dens_submatrix_6_tbin_2}
 \end{minipage}
\end{textblock*}

\newpage\null
\begin{textblock*}{\textwidth}[0.5,0](0.5\paperwidth,\blockDistanceToTop)
 \begin{minipage}{\textwidth}
   \makeatletter
   \def\@captype{figure}
   \makeatother
   \centering
   \includegraphics[height=\matrixHeight]{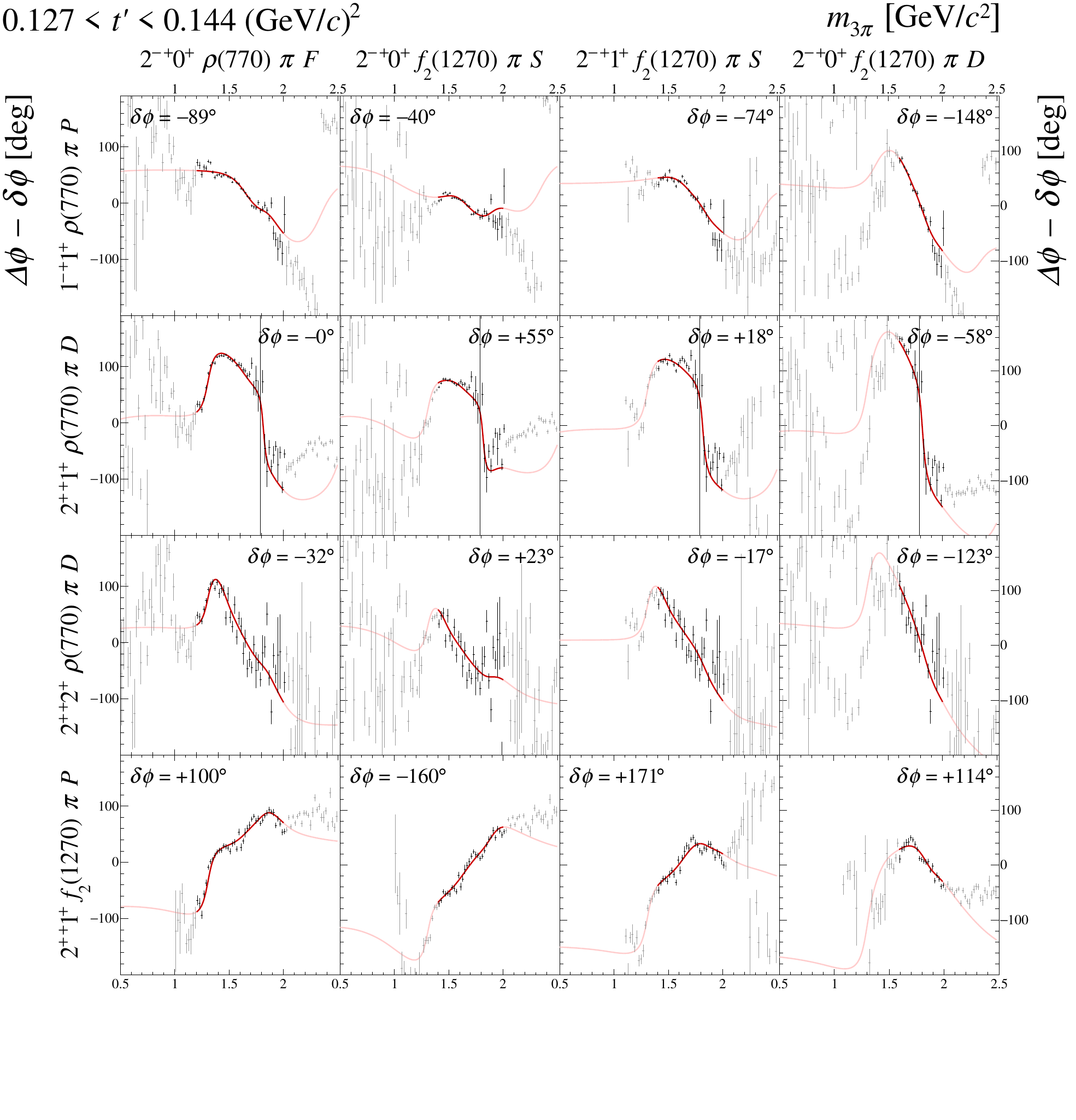}%
   \caption{Submatrix~F of the $14 \times 14$ matrix of graphs that
     represents the spin-density matrix (see
     \cref{tab:spin-dens_matrix_overview}).}
   \label{fig:spin-dens_submatrix_6_tbin_3}
 \end{minipage}
\end{textblock*}

\newpage\null
\begin{textblock*}{\textwidth}[0.5,0](0.5\paperwidth,\blockDistanceToTop)
 \begin{minipage}{\textwidth}
   \makeatletter
   \def\@captype{figure}
   \makeatother
   \centering
   \includegraphics[height=\matrixHeight]{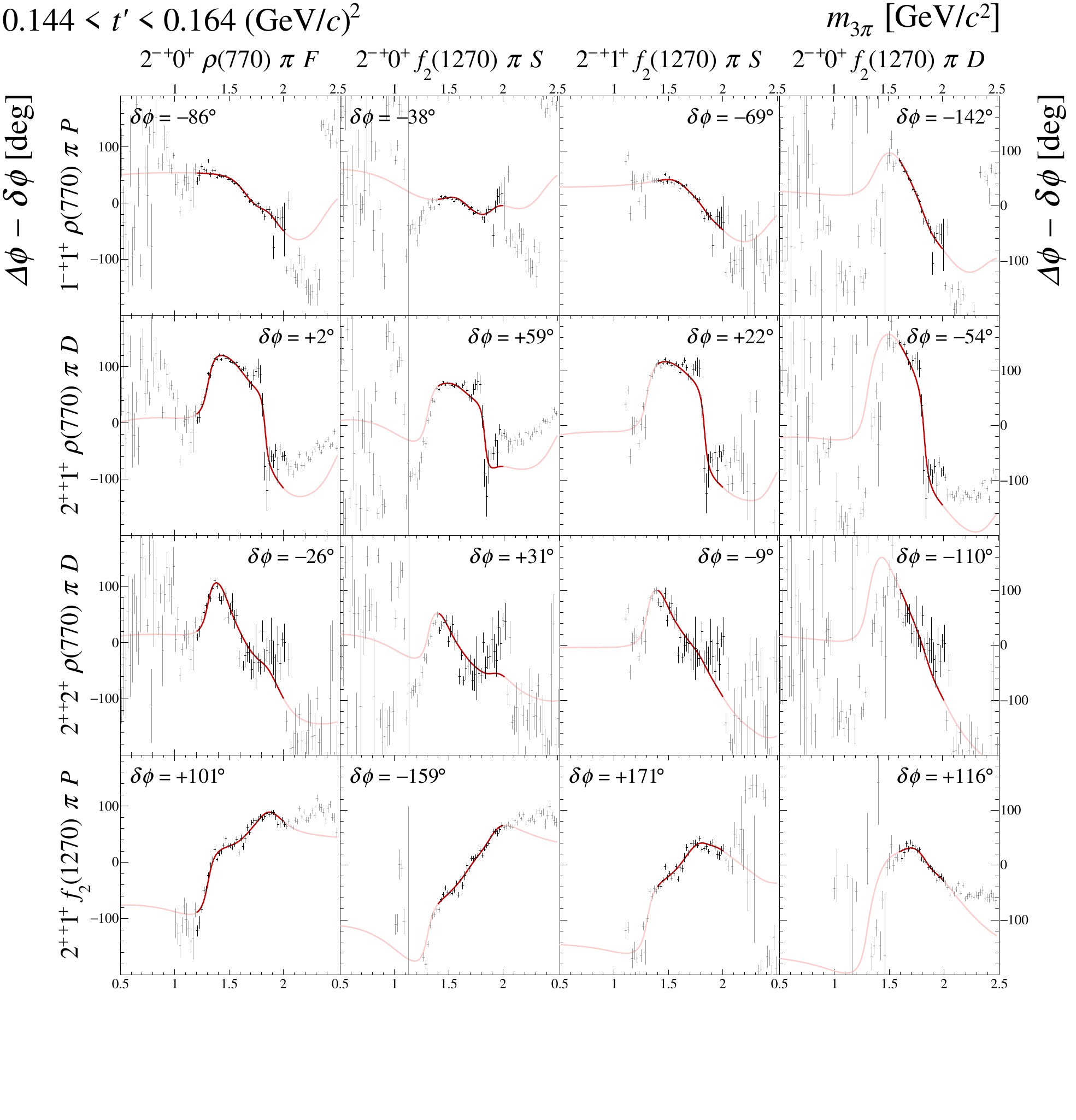}%
   \caption{Submatrix~F of the $14 \times 14$ matrix of graphs that
     represents the spin-density matrix (see
     \cref{tab:spin-dens_matrix_overview}).}
   \label{fig:spin-dens_submatrix_6_tbin_4}
 \end{minipage}
\end{textblock*}

\newpage\null
\begin{textblock*}{\textwidth}[0.5,0](0.5\paperwidth,\blockDistanceToTop)
 \begin{minipage}{\textwidth}
   \makeatletter
   \def\@captype{figure}
   \makeatother
   \centering
   \includegraphics[height=\matrixHeight]{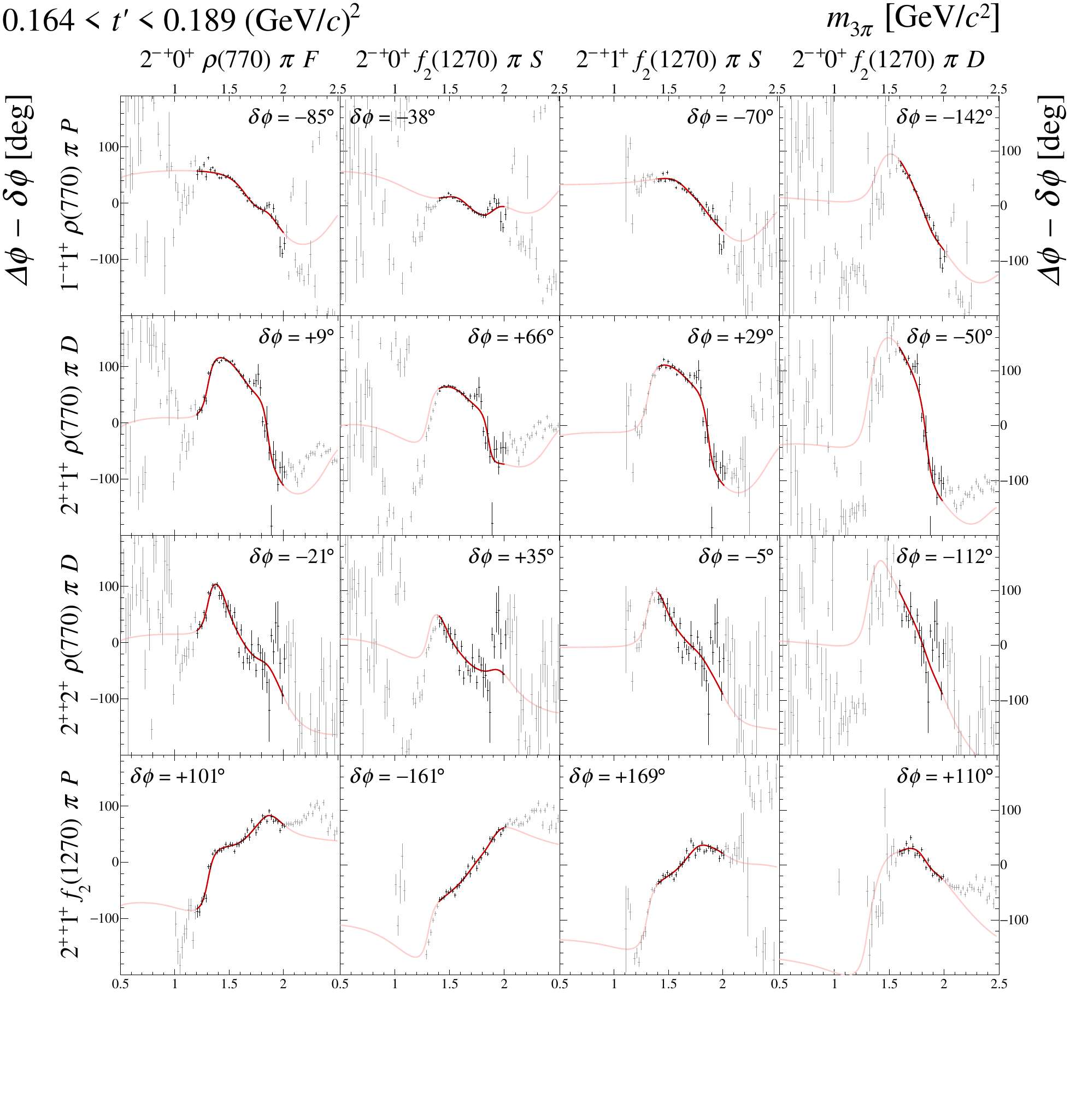}%
   \caption{Submatrix~F of the $14 \times 14$ matrix of graphs that
     represents the spin-density matrix (see
     \cref{tab:spin-dens_matrix_overview}).}
   \label{fig:spin-dens_submatrix_6_tbin_5}
 \end{minipage}
\end{textblock*}

\newpage\null
\begin{textblock*}{\textwidth}[0.5,0](0.5\paperwidth,\blockDistanceToTop)
 \begin{minipage}{\textwidth}
   \makeatletter
   \def\@captype{figure}
   \makeatother
   \centering
   \includegraphics[height=\matrixHeight]{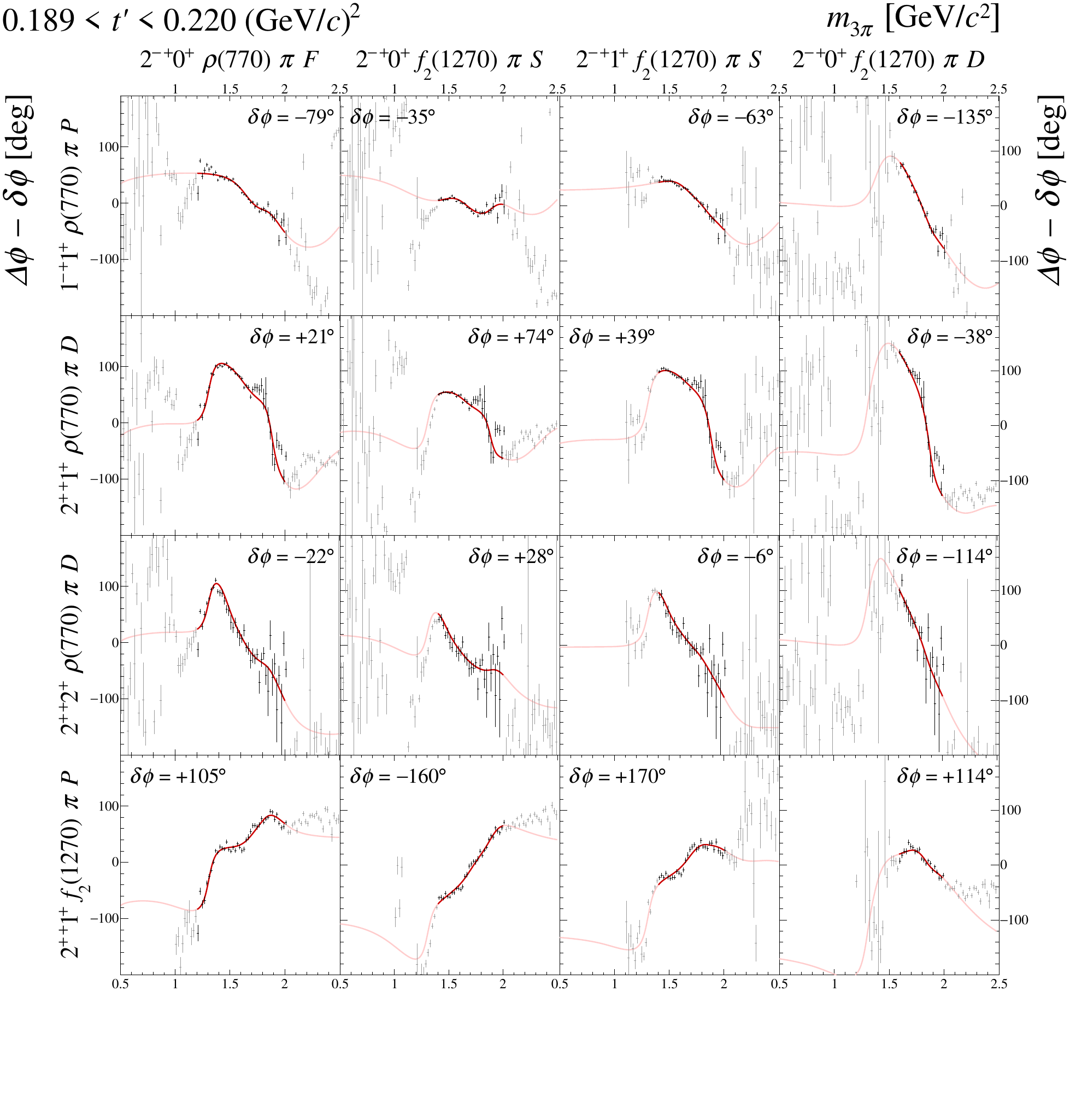}%
   \caption{Submatrix~F of the $14 \times 14$ matrix of graphs that
     represents the spin-density matrix (see
     \cref{tab:spin-dens_matrix_overview}).}
   \label{fig:spin-dens_submatrix_6_tbin_6}
 \end{minipage}
\end{textblock*}

\newpage\null
\begin{textblock*}{\textwidth}[0.5,0](0.5\paperwidth,\blockDistanceToTop)
 \begin{minipage}{\textwidth}
   \makeatletter
   \def\@captype{figure}
   \makeatother
   \centering
   \includegraphics[height=\matrixHeight]{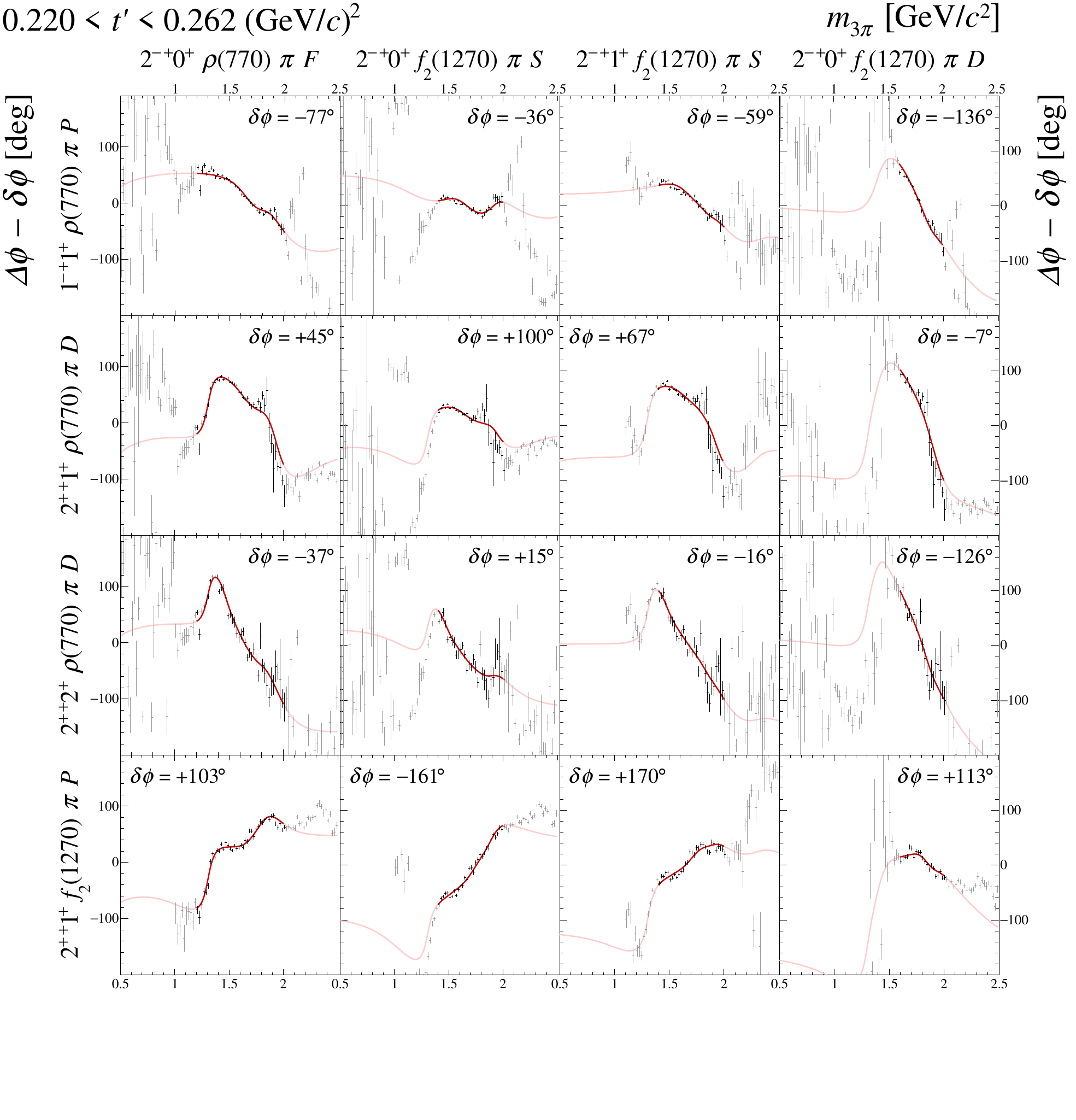}%
   \caption{Submatrix~F of the $14 \times 14$ matrix of graphs that
     represents the spin-density matrix (see
     \cref{tab:spin-dens_matrix_overview}).}
   \label{fig:spin-dens_submatrix_6_tbin_7}
 \end{minipage}
\end{textblock*}

\newpage\null
\begin{textblock*}{\textwidth}[0.5,0](0.5\paperwidth,\blockDistanceToTop)
 \begin{minipage}{\textwidth}
   \makeatletter
   \def\@captype{figure}
   \makeatother
   \centering
   \includegraphics[height=\matrixHeight]{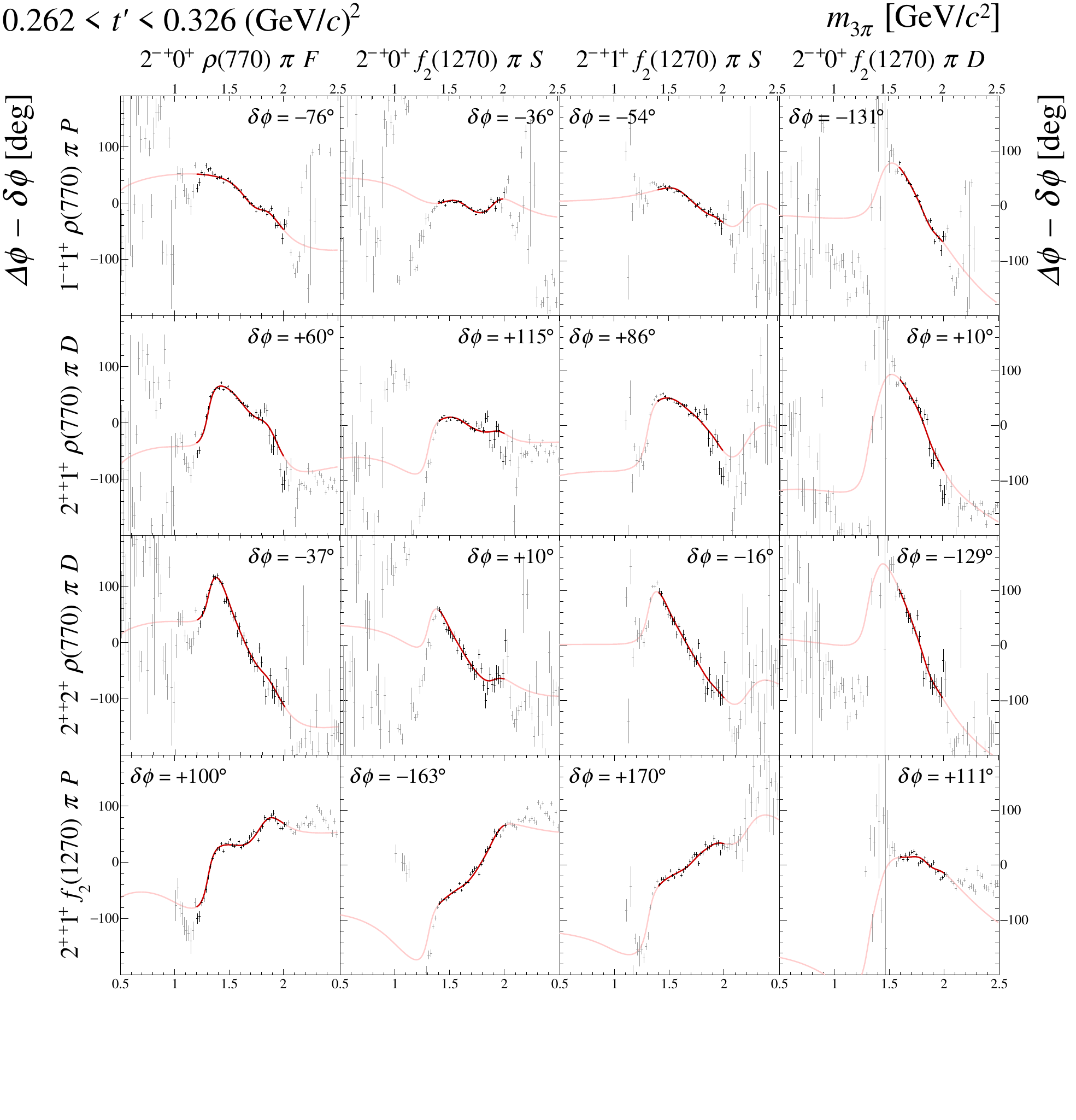}%
   \caption{Submatrix~F of the $14 \times 14$ matrix of graphs that
     represents the spin-density matrix (see
     \cref{tab:spin-dens_matrix_overview}).}
   \label{fig:spin-dens_submatrix_6_tbin_8}
 \end{minipage}
\end{textblock*}

\newpage\null
\begin{textblock*}{\textwidth}[0.5,0](0.5\paperwidth,\blockDistanceToTop)
 \begin{minipage}{\textwidth}
   \makeatletter
   \def\@captype{figure}
   \makeatother
   \centering
   \includegraphics[height=\matrixHeight]{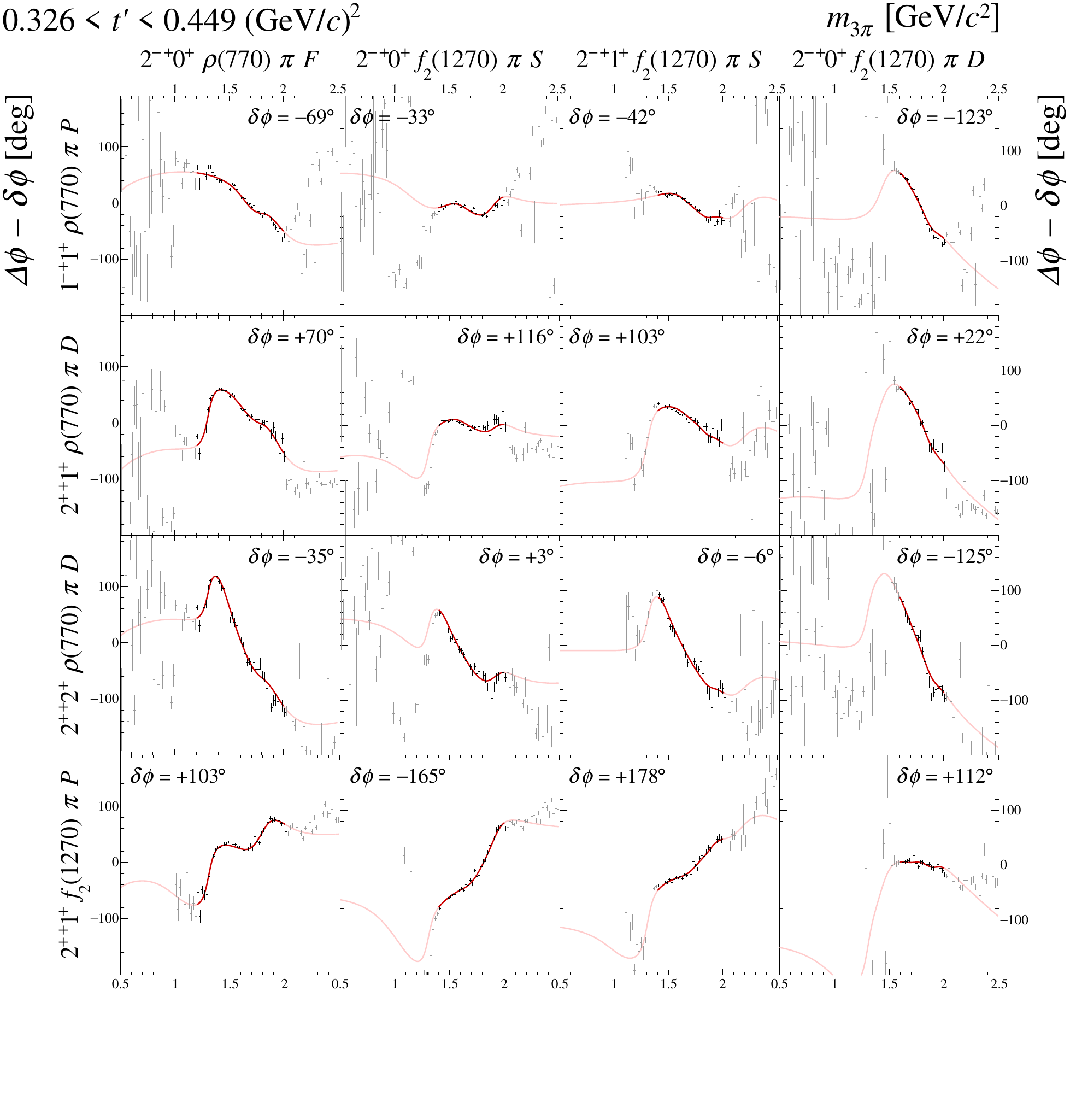}%
   \caption{Submatrix~F of the $14 \times 14$ matrix of graphs that
     represents the spin-density matrix (see
     \cref{tab:spin-dens_matrix_overview}).}
   \label{fig:spin-dens_submatrix_6_tbin_9}
 \end{minipage}
\end{textblock*}

\newpage\null
\begin{textblock*}{\textwidth}[0.5,0](0.5\paperwidth,\blockDistanceToTop)
 \begin{minipage}{\textwidth}
   \makeatletter
   \def\@captype{figure}
   \makeatother
   \centering
   \includegraphics[height=\matrixHeight]{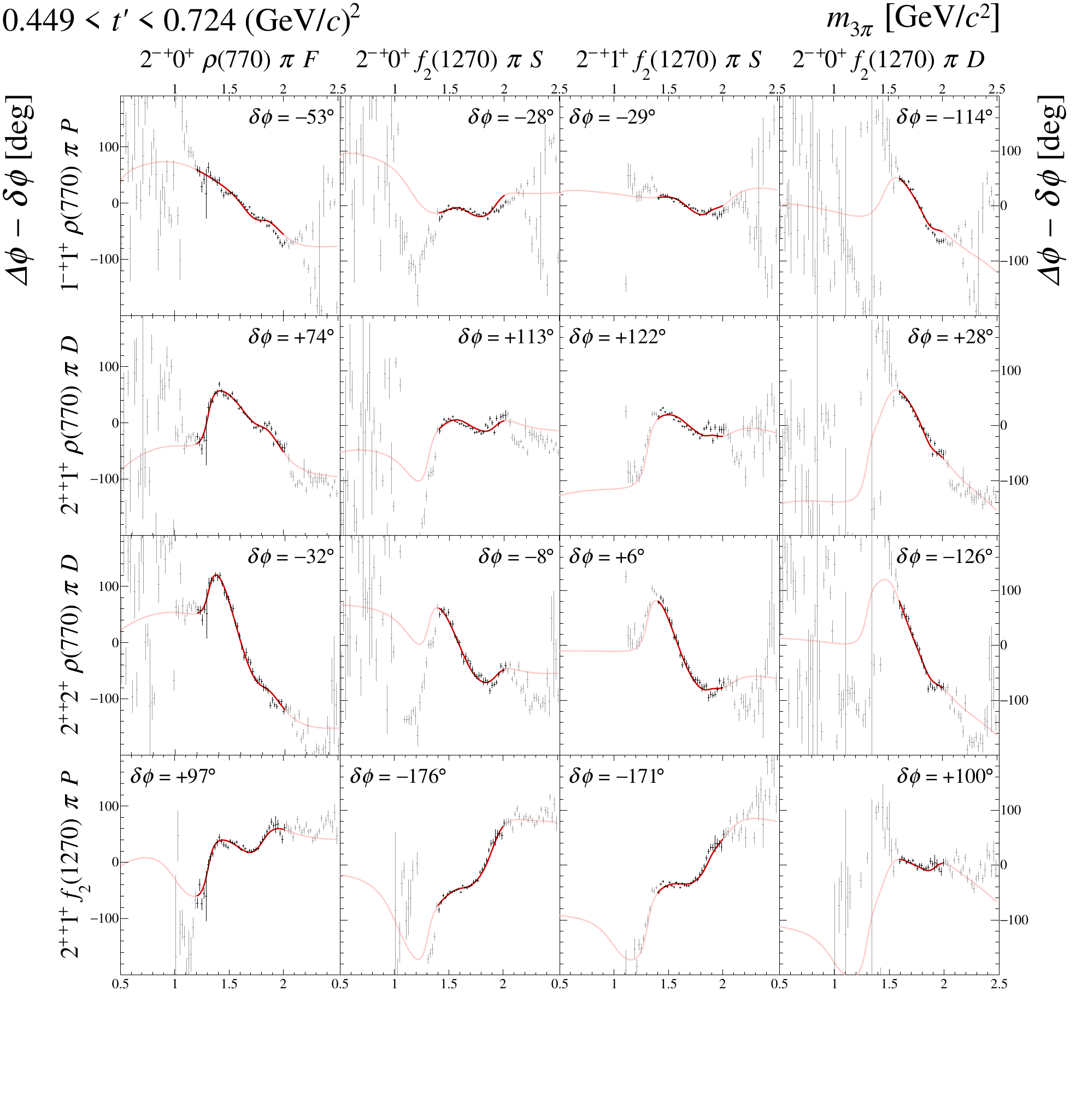}%
   \caption{Submatrix~F of the $14 \times 14$ matrix of graphs that
     represents the spin-density matrix (see
     \cref{tab:spin-dens_matrix_overview}).}
   \label{fig:spin-dens_submatrix_6_tbin_10}
 \end{minipage}
\end{textblock*}

\newpage\null
\begin{textblock*}{\textwidth}[0.5,0](0.5\paperwidth,\blockDistanceToTop)
 \begin{minipage}{\textwidth}
   \makeatletter
   \def\@captype{figure}
   \makeatother
   \centering
   \includegraphics[height=\matrixHeight]{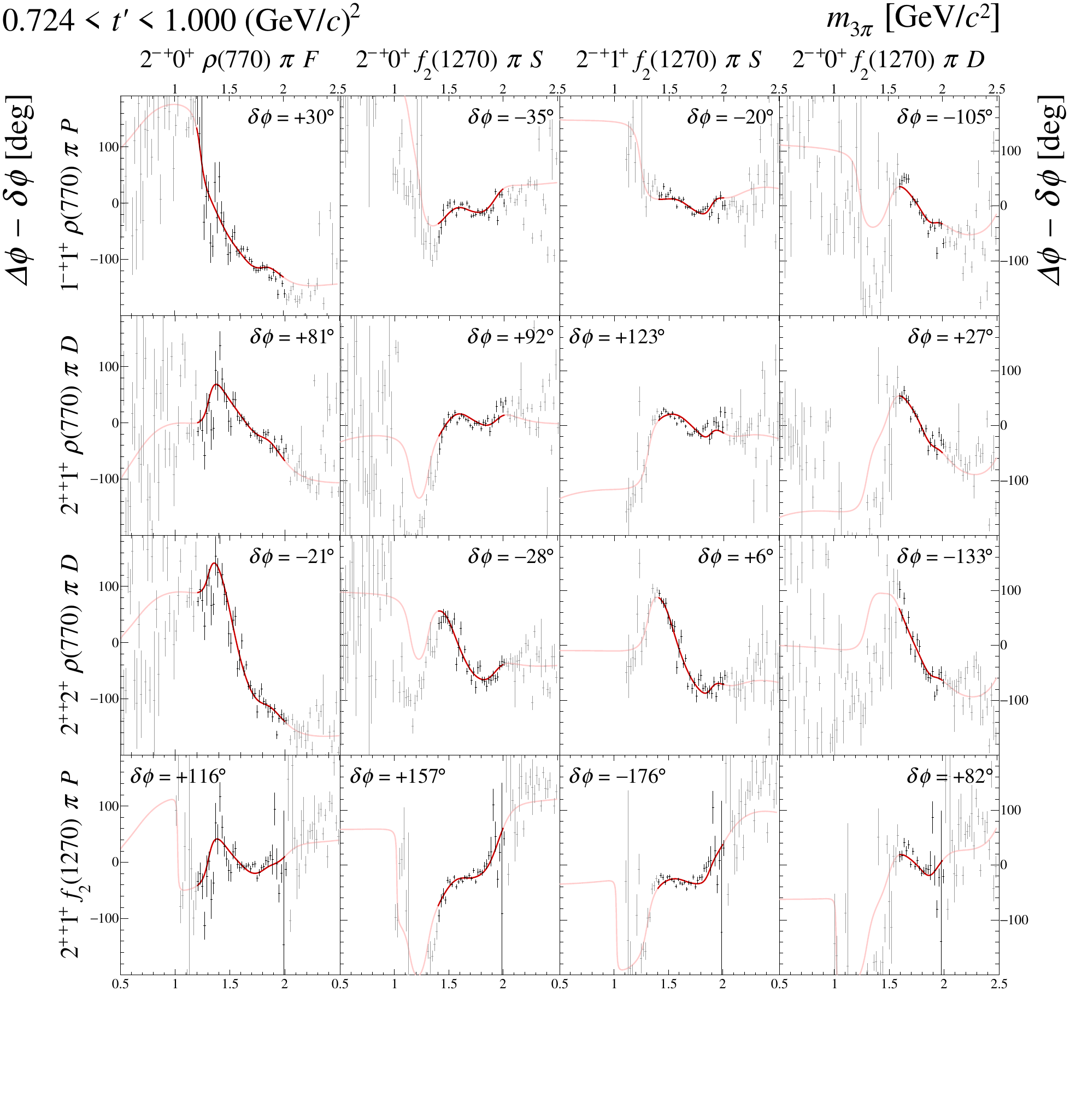}%
   \caption{Submatrix~F of the $14 \times 14$ matrix of graphs that
     represents the spin-density matrix (see
     \cref{tab:spin-dens_matrix_overview}).}
   \label{fig:spin-dens_submatrix_6_tbin_11}
 \end{minipage}
\end{textblock*}

\clearpage
\subsection{Submatrix G}
\label{sec:spin-dens_submatrix_7}

\begin{textblock*}{\textwidth}[0.5,0](0.5\paperwidth,\blockDistanceToTop)
 \begin{minipage}{\textwidth}
   \makeatletter
   \def\@captype{figure}
   \makeatother
   \centering
   \includegraphics[height=\matrixHeight]{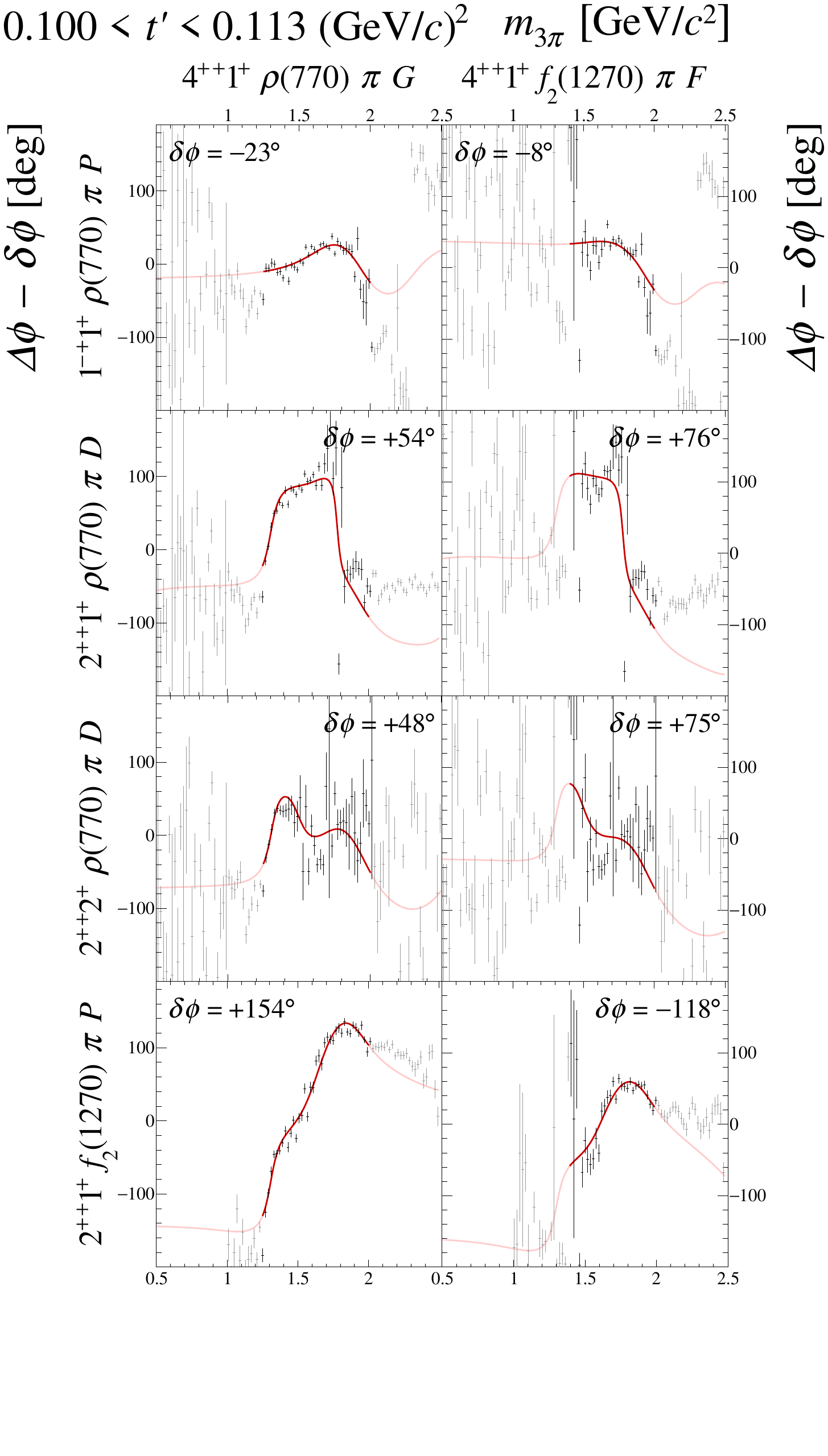}%
   \caption{Submatrix~G of the $14 \times 14$ matrix of graphs that
     represents the spin-density matrix (see
     \cref{tab:spin-dens_matrix_overview}).}
   \label{fig:spin-dens_submatrix_7_tbin_1}
 \end{minipage}
\end{textblock*}

\newpage\null
\begin{textblock*}{\textwidth}[0.5,0](0.5\paperwidth,\blockDistanceToTop)
 \begin{minipage}{\textwidth}
   \makeatletter
   \def\@captype{figure}
   \makeatother
   \centering
   \includegraphics[height=\matrixHeight]{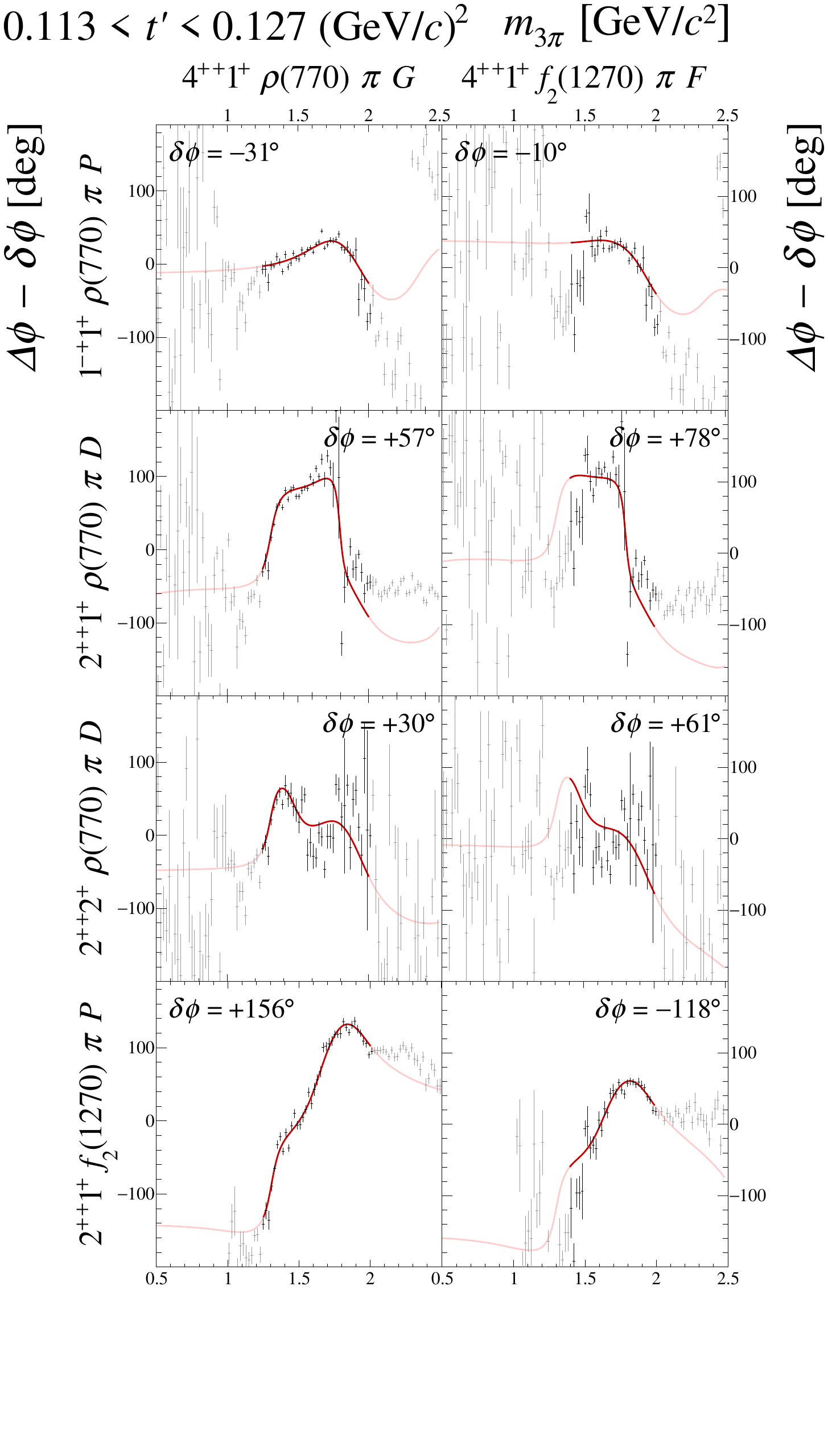}%
   \caption{Submatrix~G of the $14 \times 14$ matrix of graphs that
     represents the spin-density matrix (see
     \cref{tab:spin-dens_matrix_overview}).}
   \label{fig:spin-dens_submatrix_7_tbin_2}
 \end{minipage}
\end{textblock*}

\newpage\null
\begin{textblock*}{\textwidth}[0.5,0](0.5\paperwidth,\blockDistanceToTop)
 \begin{minipage}{\textwidth}
   \makeatletter
   \def\@captype{figure}
   \makeatother
   \centering
   \includegraphics[height=\matrixHeight]{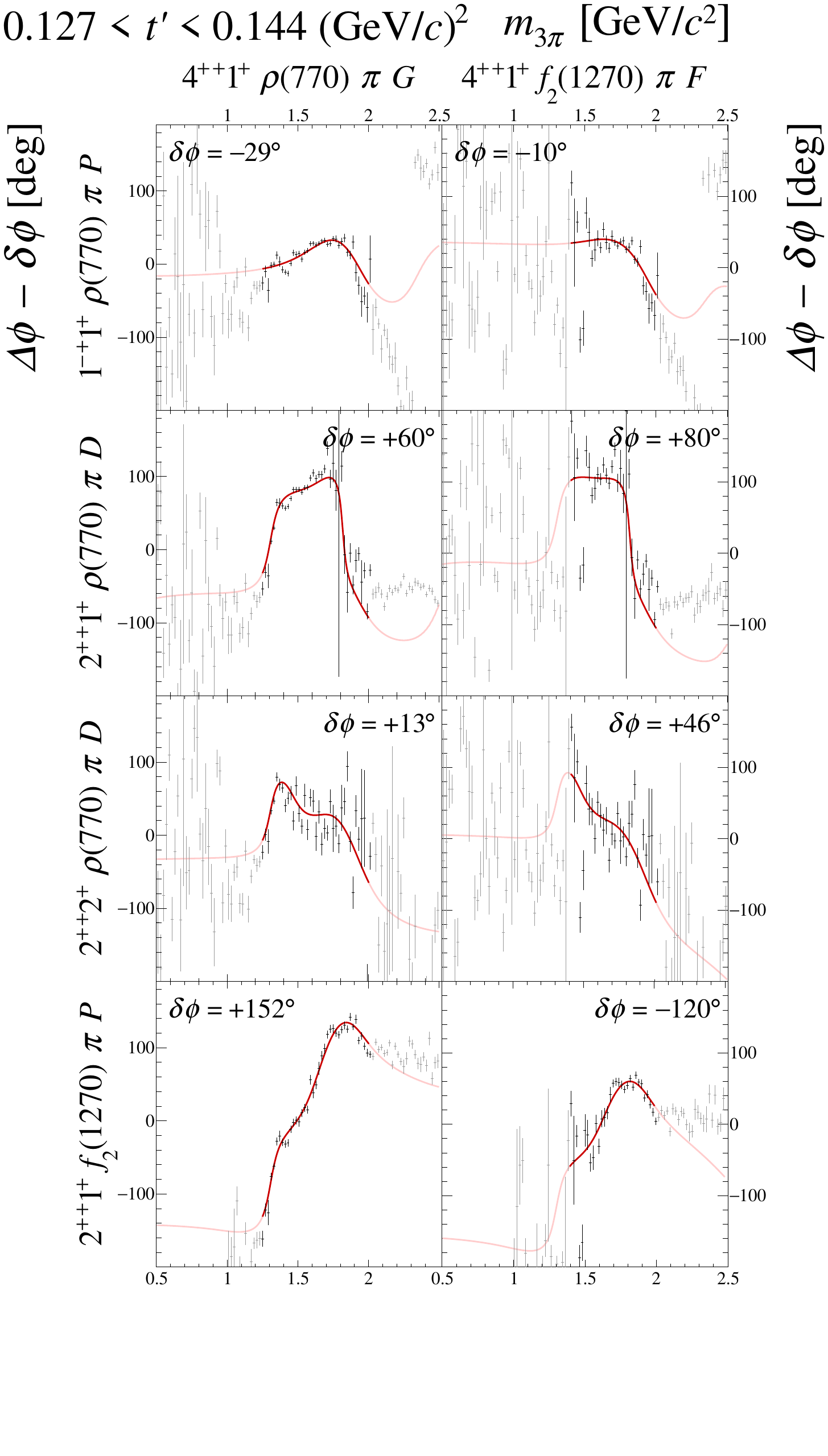}%
   \caption{Submatrix~G of the $14 \times 14$ matrix of graphs that
     represents the spin-density matrix (see
     \cref{tab:spin-dens_matrix_overview}).}
   \label{fig:spin-dens_submatrix_7_tbin_3}
 \end{minipage}
\end{textblock*}

\newpage\null
\begin{textblock*}{\textwidth}[0.5,0](0.5\paperwidth,\blockDistanceToTop)
 \begin{minipage}{\textwidth}
   \makeatletter
   \def\@captype{figure}
   \makeatother
   \centering
   \includegraphics[height=\matrixHeight]{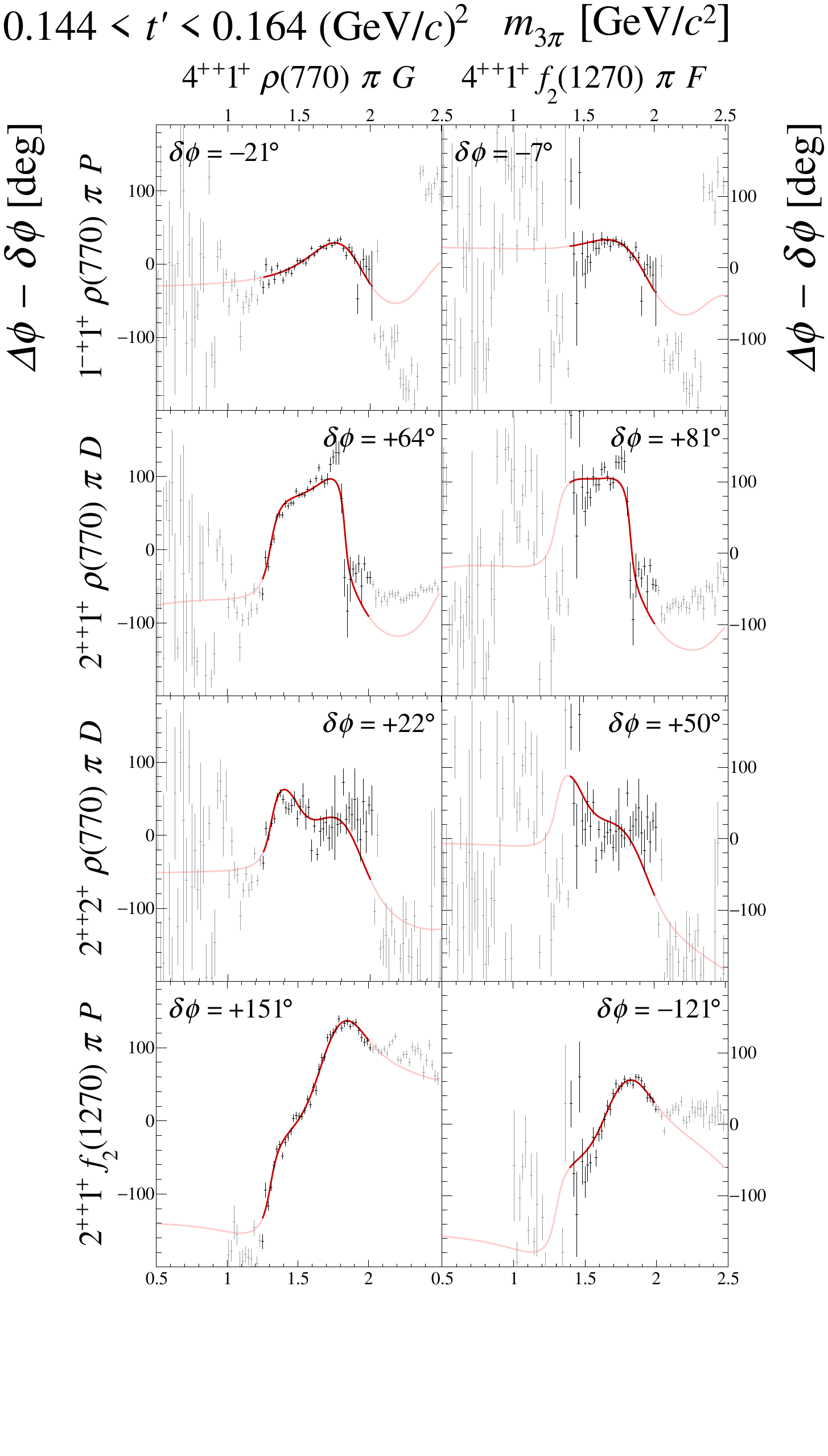}%
   \caption{Submatrix~G of the $14 \times 14$ matrix of graphs that
     represents the spin-density matrix (see
     \cref{tab:spin-dens_matrix_overview}).}
   \label{fig:spin-dens_submatrix_7_tbin_4}
 \end{minipage}
\end{textblock*}

\newpage\null
\begin{textblock*}{\textwidth}[0.5,0](0.5\paperwidth,\blockDistanceToTop)
 \begin{minipage}{\textwidth}
   \makeatletter
   \def\@captype{figure}
   \makeatother
   \centering
   \includegraphics[height=\matrixHeight]{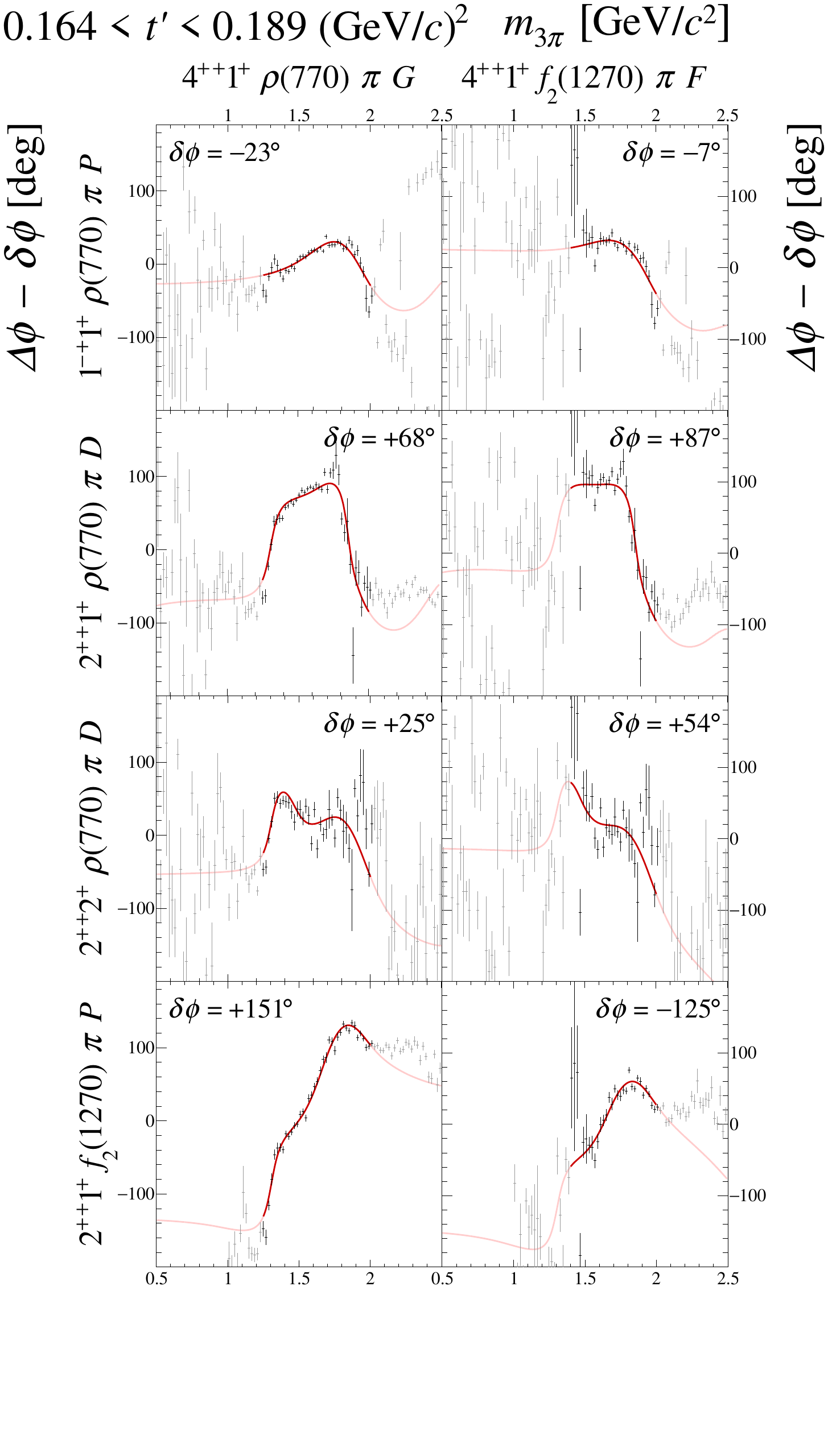}%
   \caption{Submatrix~G of the $14 \times 14$ matrix of graphs that
     represents the spin-density matrix (see
     \cref{tab:spin-dens_matrix_overview}).}
   \label{fig:spin-dens_submatrix_7_tbin_5}
 \end{minipage}
\end{textblock*}

\newpage\null
\begin{textblock*}{\textwidth}[0.5,0](0.5\paperwidth,\blockDistanceToTop)
 \begin{minipage}{\textwidth}
   \makeatletter
   \def\@captype{figure}
   \makeatother
   \centering
   \includegraphics[height=\matrixHeight]{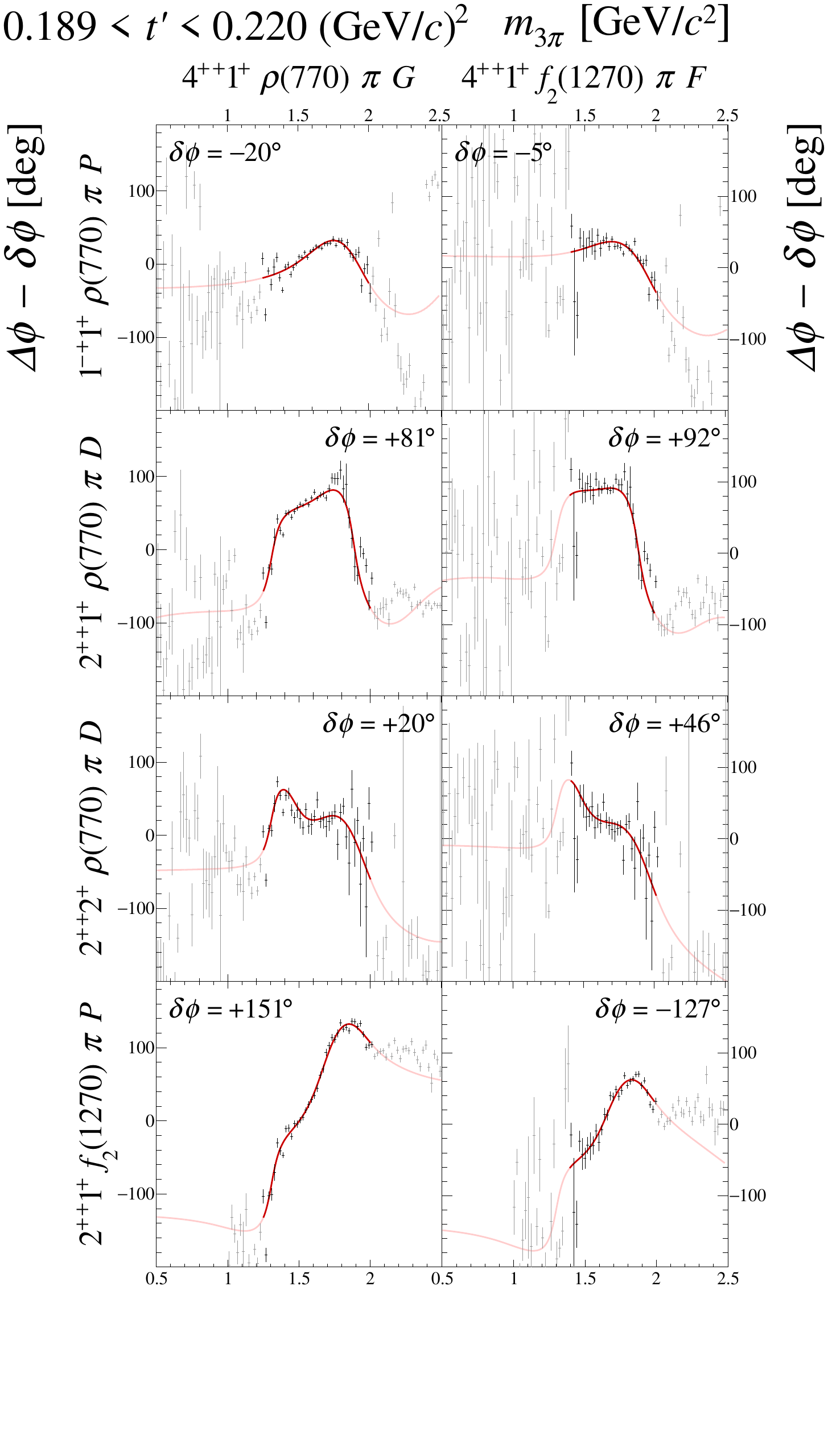}%
   \caption{Submatrix~G of the $14 \times 14$ matrix of graphs that
     represents the spin-density matrix (see
     \cref{tab:spin-dens_matrix_overview}).}
   \label{fig:spin-dens_submatrix_7_tbin_6}
 \end{minipage}
\end{textblock*}

\newpage\null
\begin{textblock*}{\textwidth}[0.5,0](0.5\paperwidth,\blockDistanceToTop)
 \begin{minipage}{\textwidth}
   \makeatletter
   \def\@captype{figure}
   \makeatother
   \centering
   \includegraphics[height=\matrixHeight]{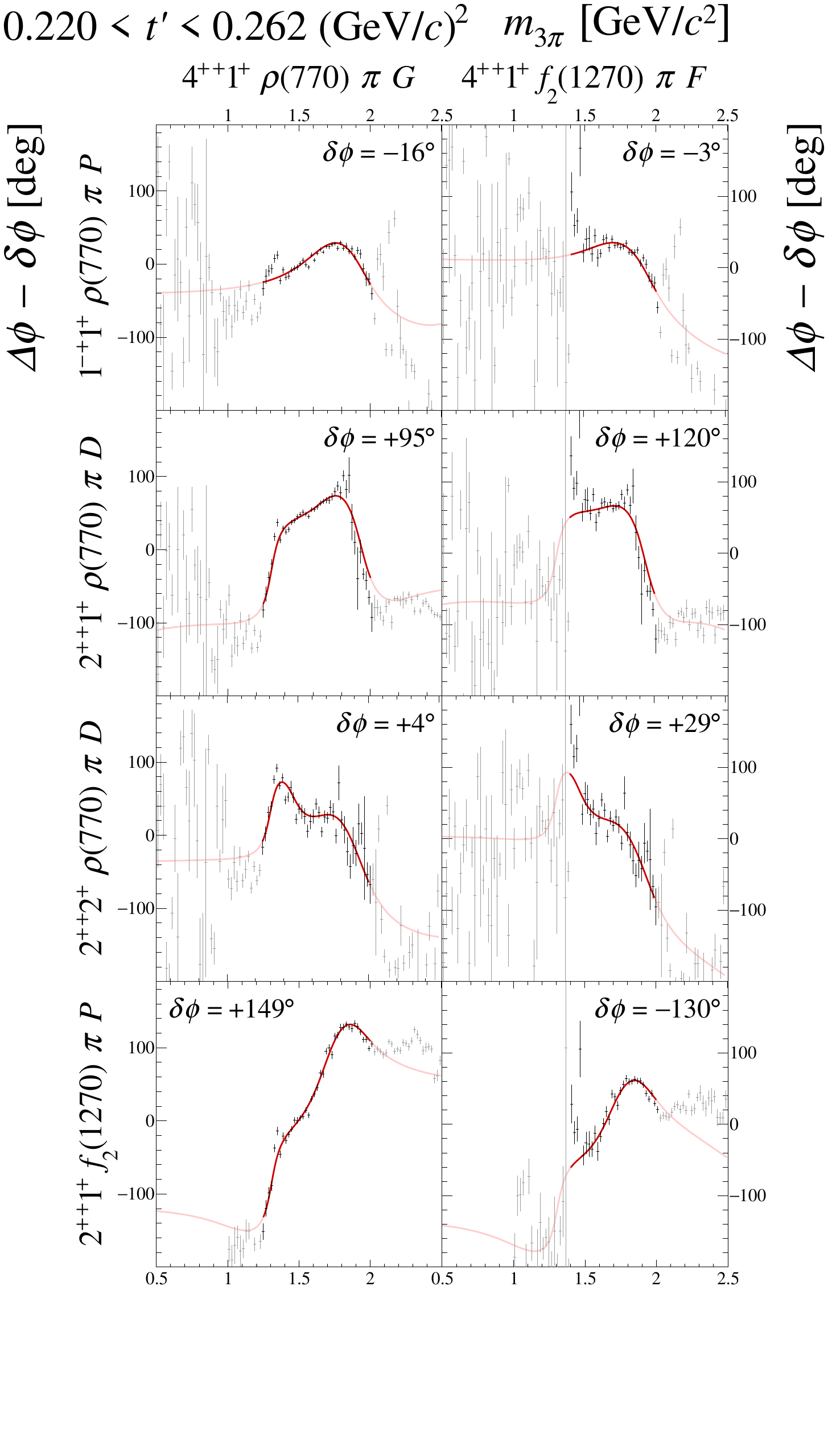}%
   \caption{Submatrix~G of the $14 \times 14$ matrix of graphs that
     represents the spin-density matrix (see
     \cref{tab:spin-dens_matrix_overview}).}
   \label{fig:spin-dens_submatrix_7_tbin_7}
 \end{minipage}
\end{textblock*}

\newpage\null
\begin{textblock*}{\textwidth}[0.5,0](0.5\paperwidth,\blockDistanceToTop)
 \begin{minipage}{\textwidth}
   \makeatletter
   \def\@captype{figure}
   \makeatother
   \centering
   \includegraphics[height=\matrixHeight]{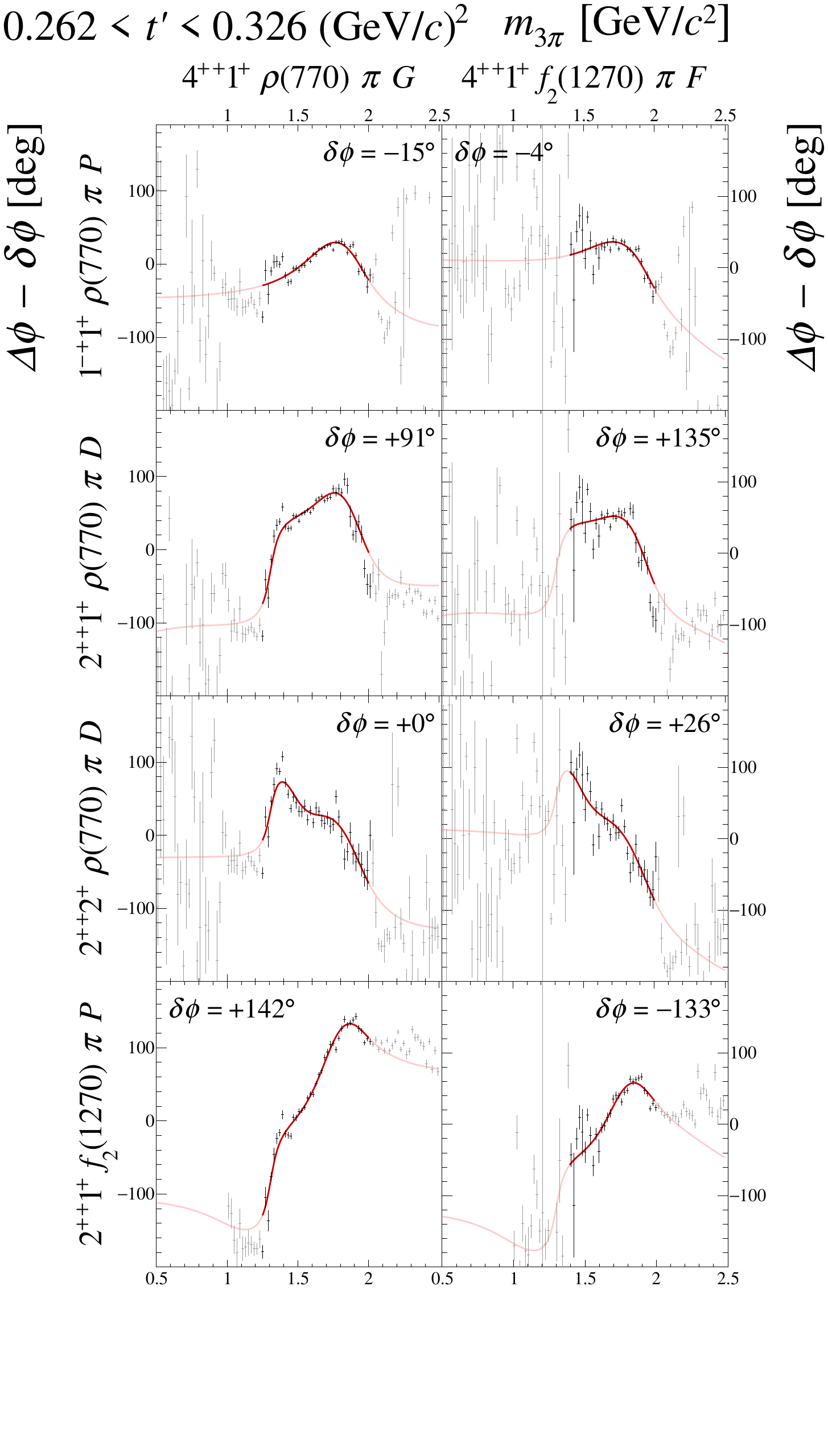}%
   \caption{Submatrix~G of the $14 \times 14$ matrix of graphs that
     represents the spin-density matrix (see
     \cref{tab:spin-dens_matrix_overview}).}
   \label{fig:spin-dens_submatrix_7_tbin_8}
 \end{minipage}
\end{textblock*}

\newpage\null
\begin{textblock*}{\textwidth}[0.5,0](0.5\paperwidth,\blockDistanceToTop)
 \begin{minipage}{\textwidth}
   \makeatletter
   \def\@captype{figure}
   \makeatother
   \centering
   \includegraphics[height=\matrixHeight]{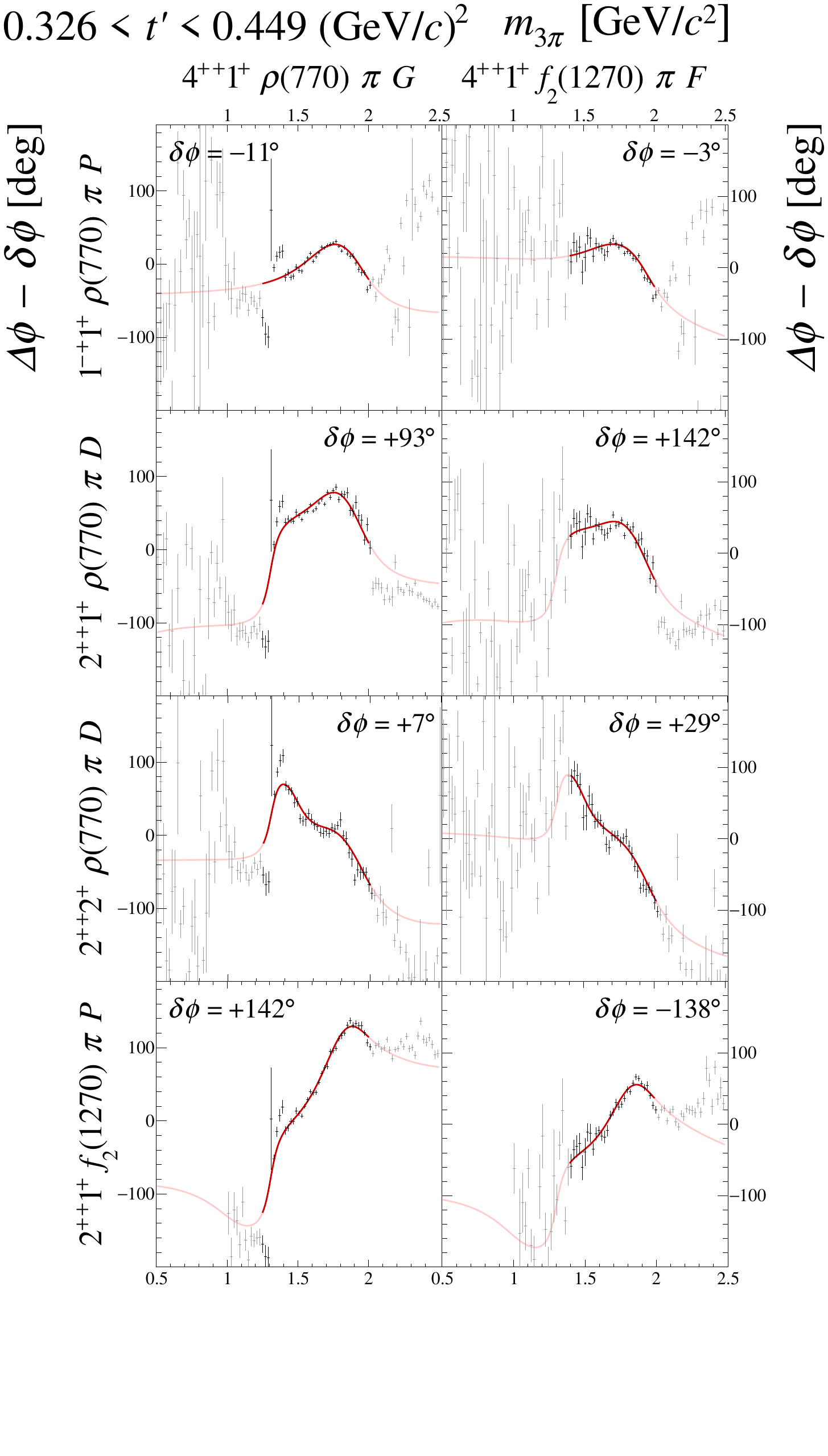}%
   \caption{Submatrix~G of the $14 \times 14$ matrix of graphs that
     represents the spin-density matrix (see
     \cref{tab:spin-dens_matrix_overview}).}
   \label{fig:spin-dens_submatrix_7_tbin_9}
 \end{minipage}
\end{textblock*}

\newpage\null
\begin{textblock*}{\textwidth}[0.5,0](0.5\paperwidth,\blockDistanceToTop)
 \begin{minipage}{\textwidth}
   \makeatletter
   \def\@captype{figure}
   \makeatother
   \centering
   \includegraphics[height=\matrixHeight]{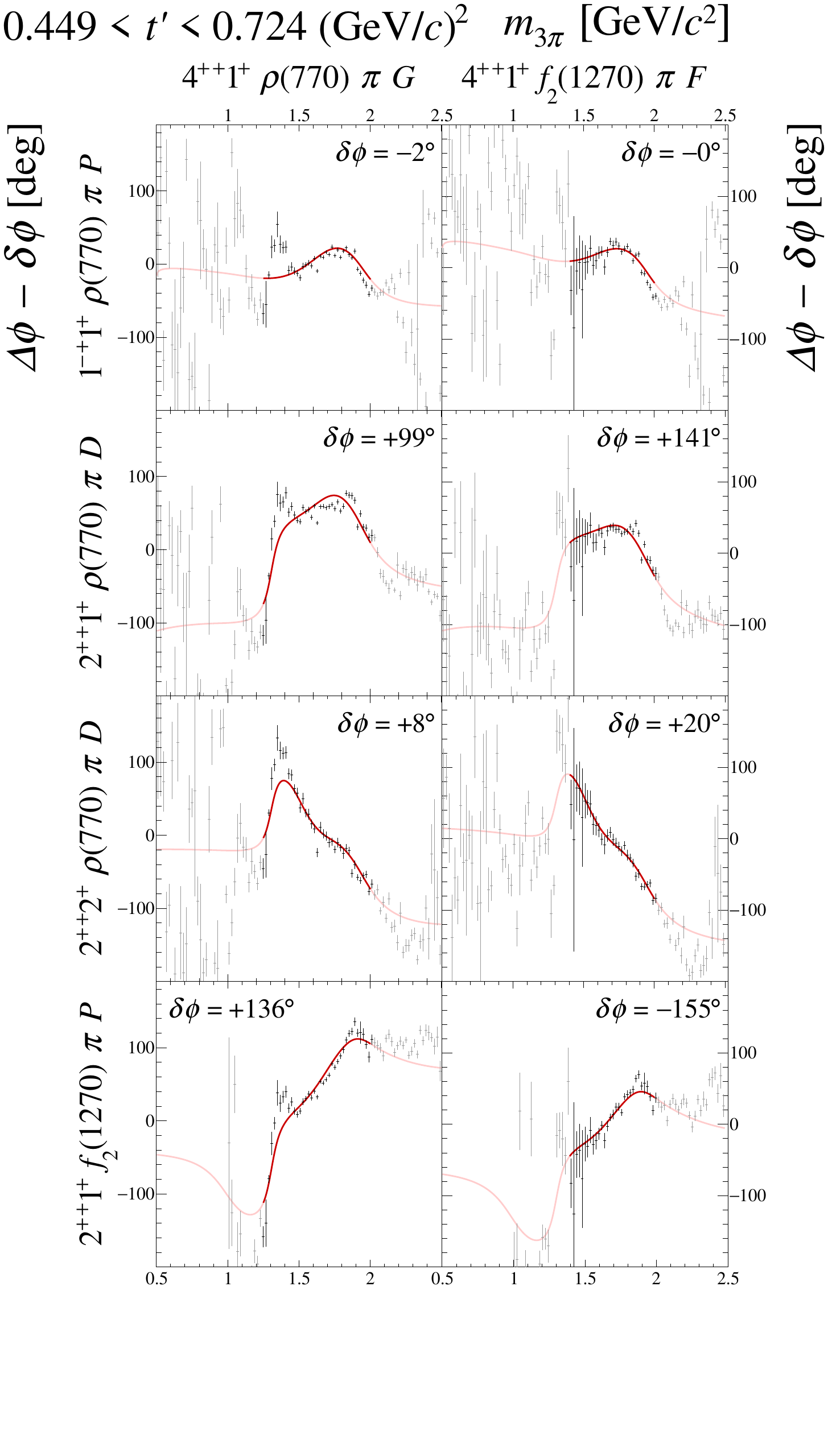}%
   \caption{Submatrix~G of the $14 \times 14$ matrix of graphs that
     represents the spin-density matrix (see
     \cref{tab:spin-dens_matrix_overview}).}
   \label{fig:spin-dens_submatrix_7_tbin_10}
 \end{minipage}
\end{textblock*}

\newpage\null
\begin{textblock*}{\textwidth}[0.5,0](0.5\paperwidth,\blockDistanceToTop)
 \begin{minipage}{\textwidth}
   \makeatletter
   \def\@captype{figure}
   \makeatother
   \centering
   \includegraphics[height=\matrixHeight]{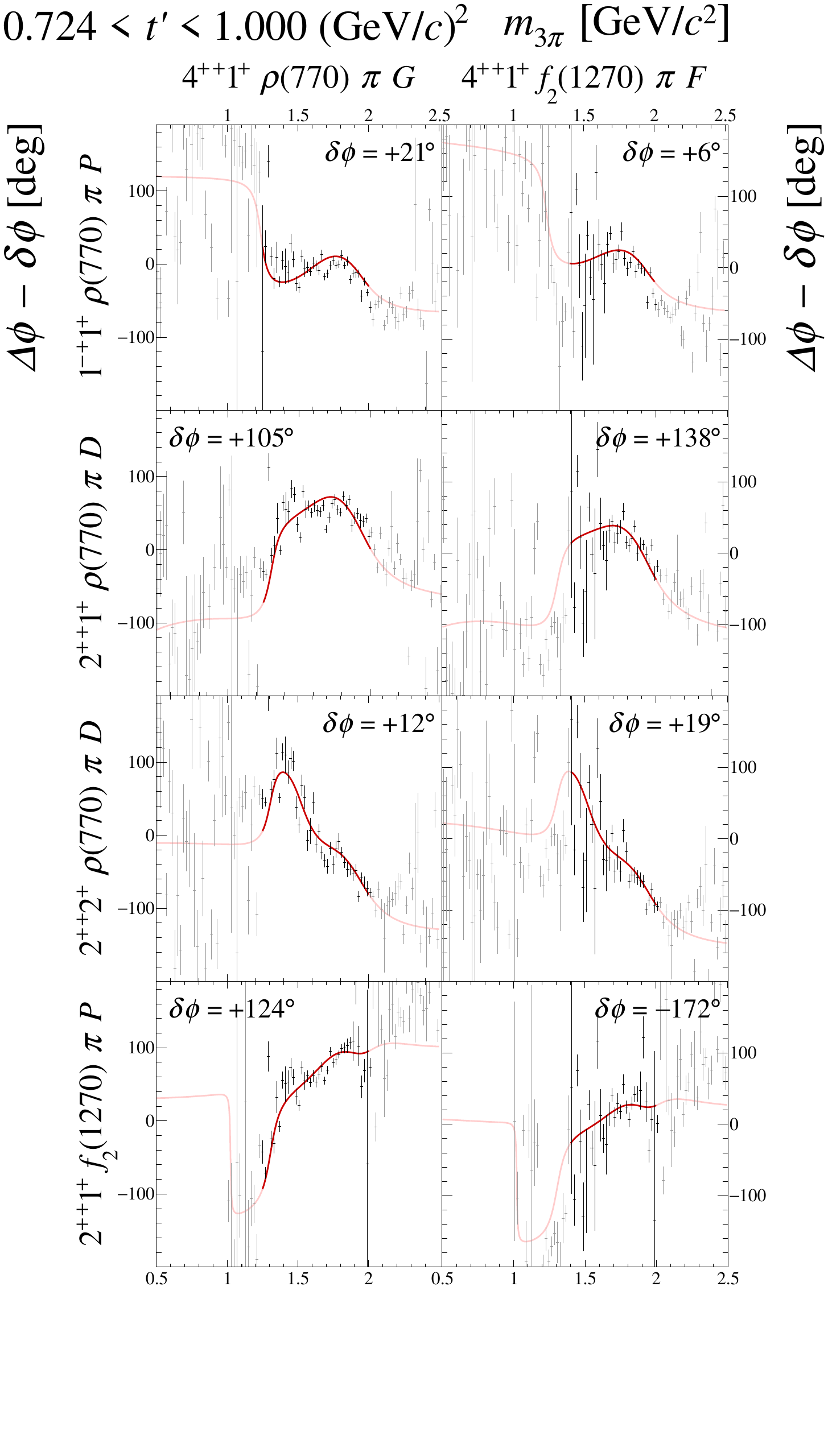}%
   \caption{Submatrix~G of the $14 \times 14$ matrix of graphs that
     represents the spin-density matrix (see
     \cref{tab:spin-dens_matrix_overview}).}
   \label{fig:spin-dens_submatrix_7_tbin_11}
 \end{minipage}
\end{textblock*}

\clearpage
\subsection{Submatrix H}
\label{sec:spin-dens_submatrix_8}

\begin{textblock*}{\textwidth}[0.5,0](0.5\paperwidth,\blockDistanceToTop)
 \begin{minipage}{\textwidth}
   \makeatletter
   \def\@captype{figure}
   \makeatother
   \centering
   \includegraphics[height=\matrixHeight]{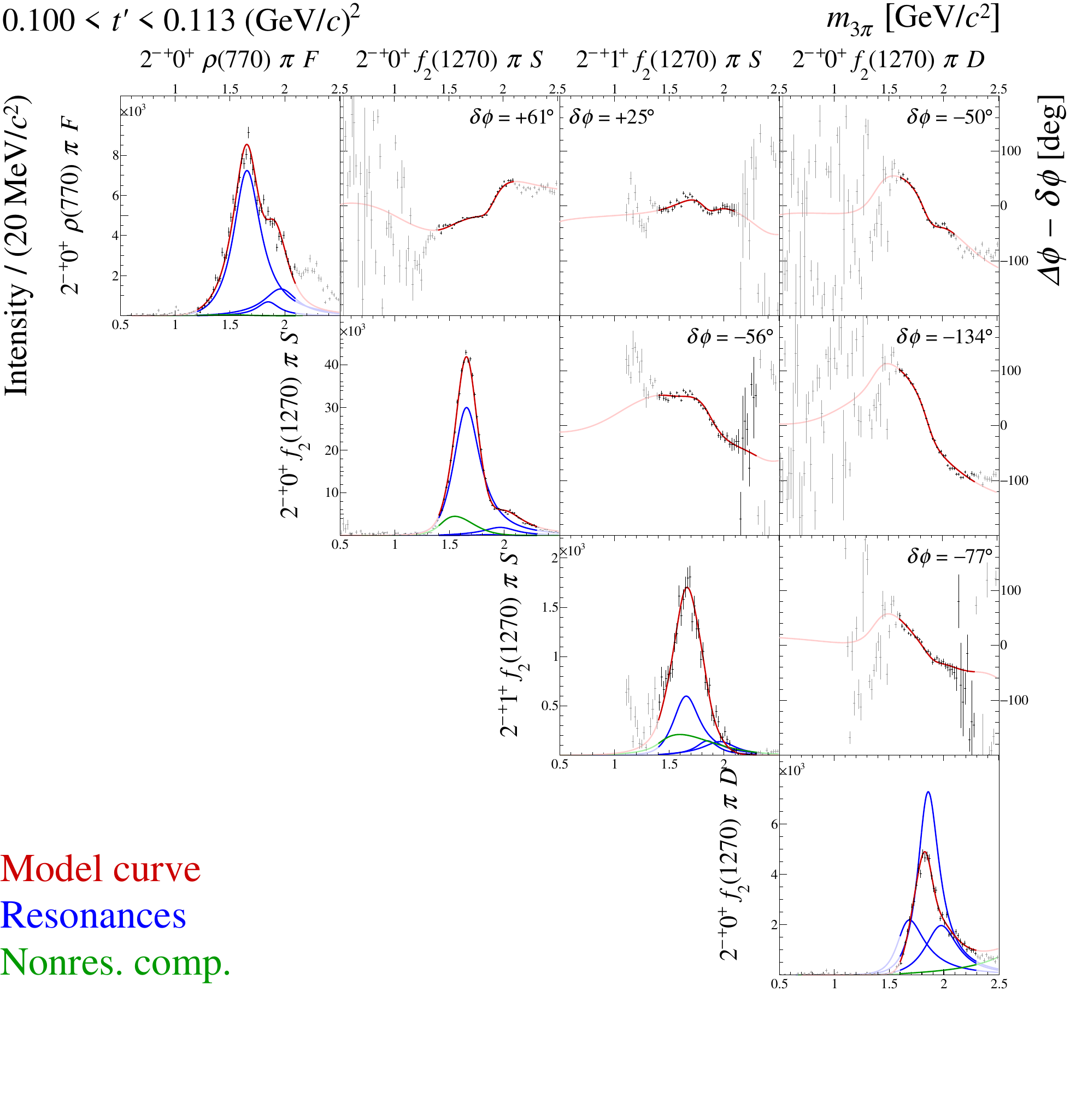}%
   \caption{Submatrix~H of the $14 \times 14$ matrix of graphs that
     represents the spin-density matrix (see
     \cref{tab:spin-dens_matrix_overview}).}
   \label{fig:spin-dens_submatrix_8_tbin_1}
 \end{minipage}
\end{textblock*}

\newpage\null
\begin{textblock*}{\textwidth}[0.5,0](0.5\paperwidth,\blockDistanceToTop)
 \begin{minipage}{\textwidth}
   \makeatletter
   \def\@captype{figure}
   \makeatother
   \centering
   \includegraphics[height=\matrixHeight]{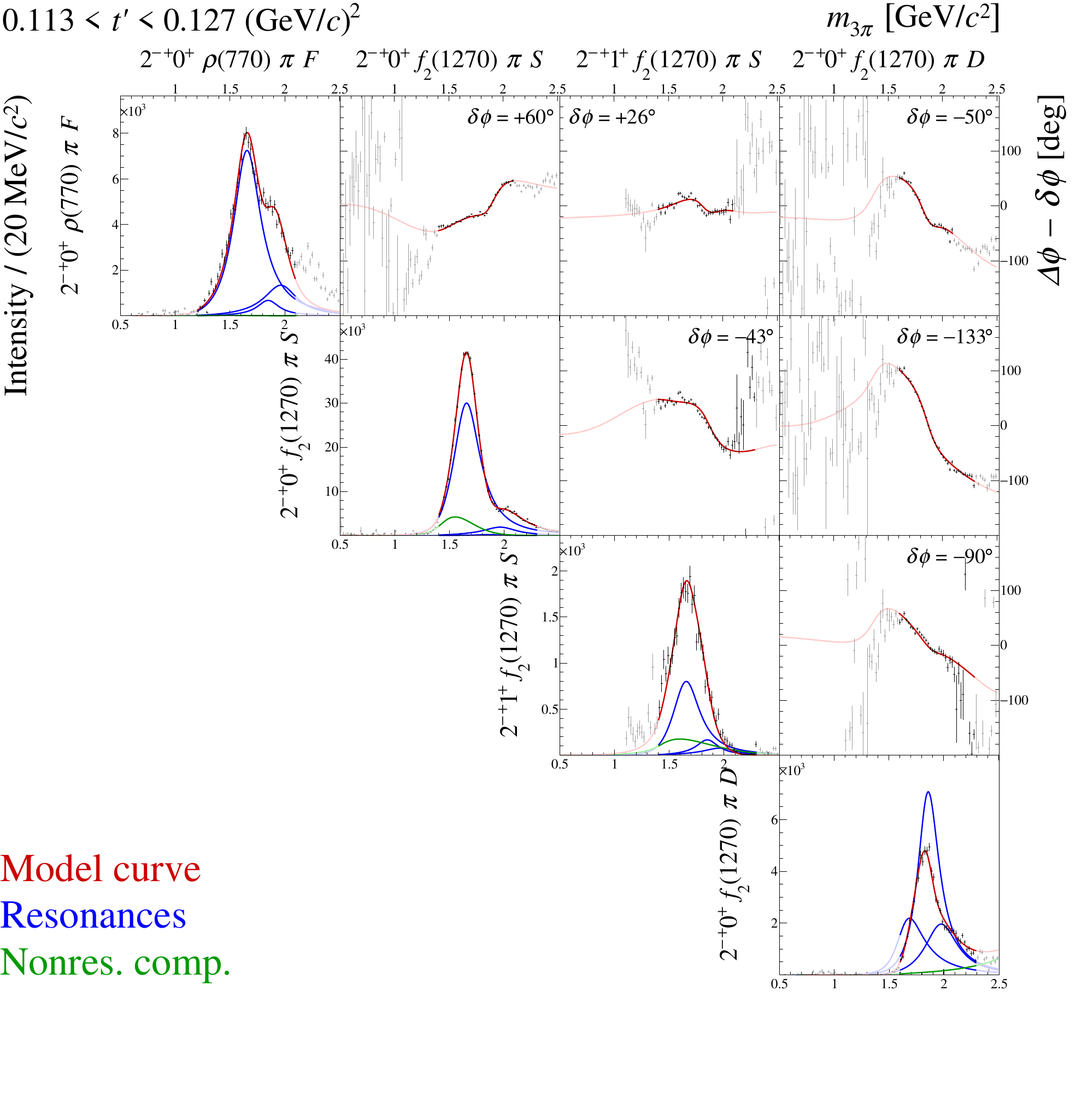}%
   \caption{Submatrix~H of the $14 \times 14$ matrix of graphs that
     represents the spin-density matrix (see
     \cref{tab:spin-dens_matrix_overview}).}
   \label{fig:spin-dens_submatrix_8_tbin_2}
 \end{minipage}
\end{textblock*}

\newpage\null
\begin{textblock*}{\textwidth}[0.5,0](0.5\paperwidth,\blockDistanceToTop)
 \begin{minipage}{\textwidth}
   \makeatletter
   \def\@captype{figure}
   \makeatother
   \centering
   \includegraphics[height=\matrixHeight]{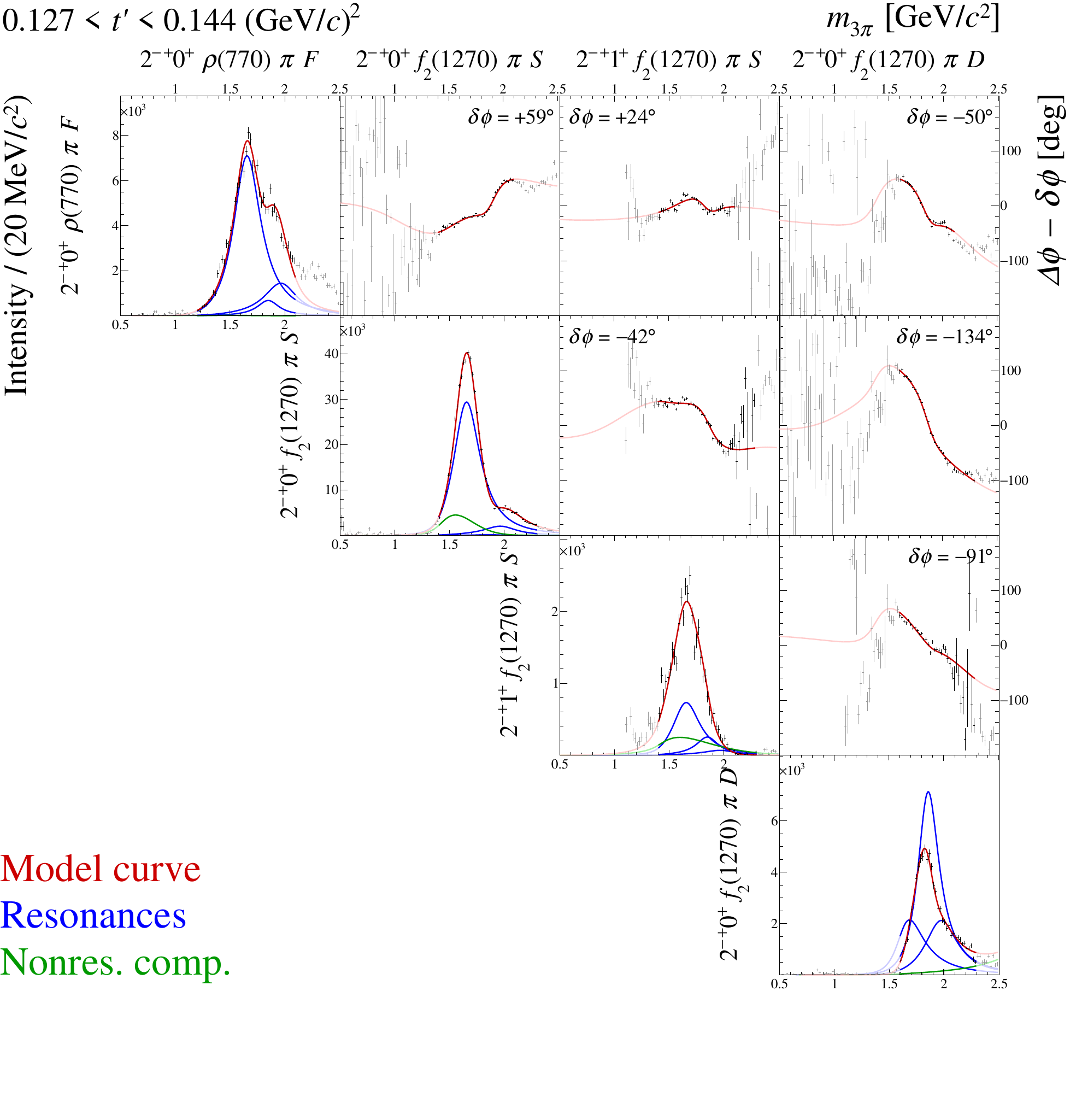}%
   \caption{Submatrix~H of the $14 \times 14$ matrix of graphs that
     represents the spin-density matrix (see
     \cref{tab:spin-dens_matrix_overview}).}
   \label{fig:spin-dens_submatrix_8_tbin_3}
 \end{minipage}
\end{textblock*}

\newpage\null
\begin{textblock*}{\textwidth}[0.5,0](0.5\paperwidth,\blockDistanceToTop)
 \begin{minipage}{\textwidth}
   \makeatletter
   \def\@captype{figure}
   \makeatother
   \centering
   \includegraphics[height=\matrixHeight]{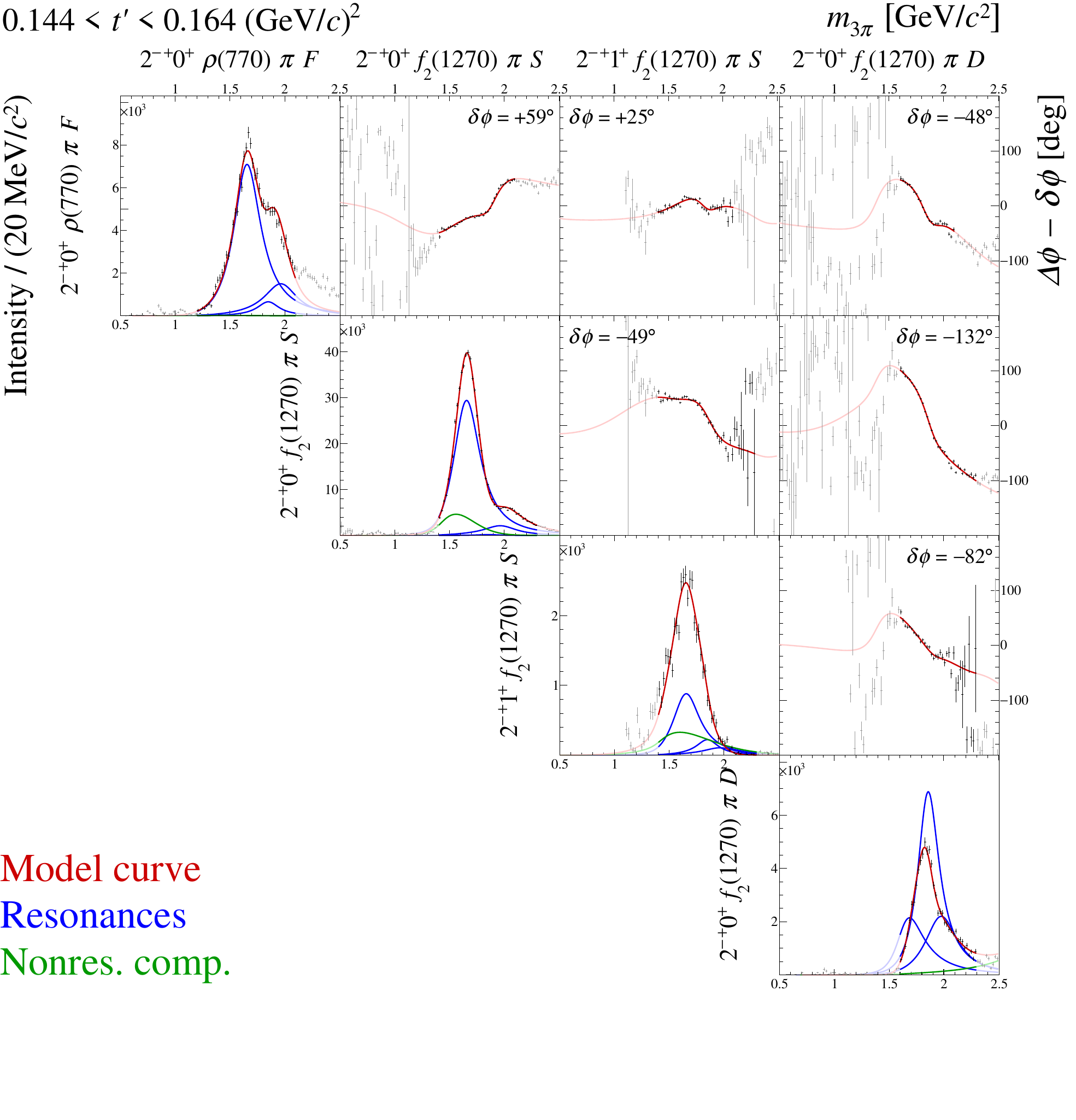}%
   \caption{Submatrix~H of the $14 \times 14$ matrix of graphs that
     represents the spin-density matrix (see
     \cref{tab:spin-dens_matrix_overview}).}
   \label{fig:spin-dens_submatrix_8_tbin_4}
 \end{minipage}
\end{textblock*}

\newpage\null
\begin{textblock*}{\textwidth}[0.5,0](0.5\paperwidth,\blockDistanceToTop)
 \begin{minipage}{\textwidth}
   \makeatletter
   \def\@captype{figure}
   \makeatother
   \centering
   \includegraphics[height=\matrixHeight]{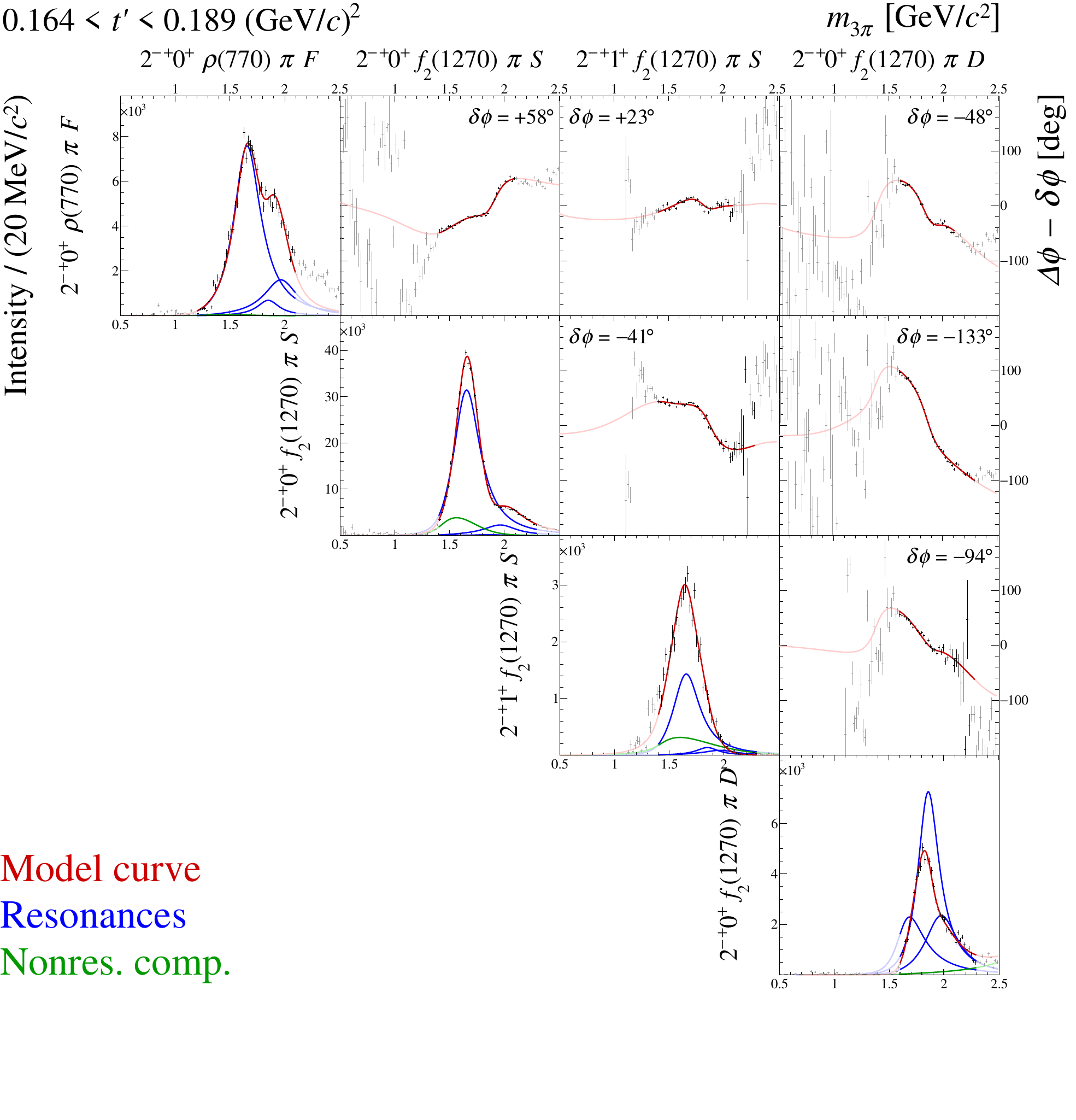}%
   \caption{Submatrix~H of the $14 \times 14$ matrix of graphs that
     represents the spin-density matrix (see
     \cref{tab:spin-dens_matrix_overview}).}
   \label{fig:spin-dens_submatrix_8_tbin_5}
 \end{minipage}
\end{textblock*}

\newpage\null
\begin{textblock*}{\textwidth}[0.5,0](0.5\paperwidth,\blockDistanceToTop)
 \begin{minipage}{\textwidth}
   \makeatletter
   \def\@captype{figure}
   \makeatother
   \centering
   \includegraphics[height=\matrixHeight]{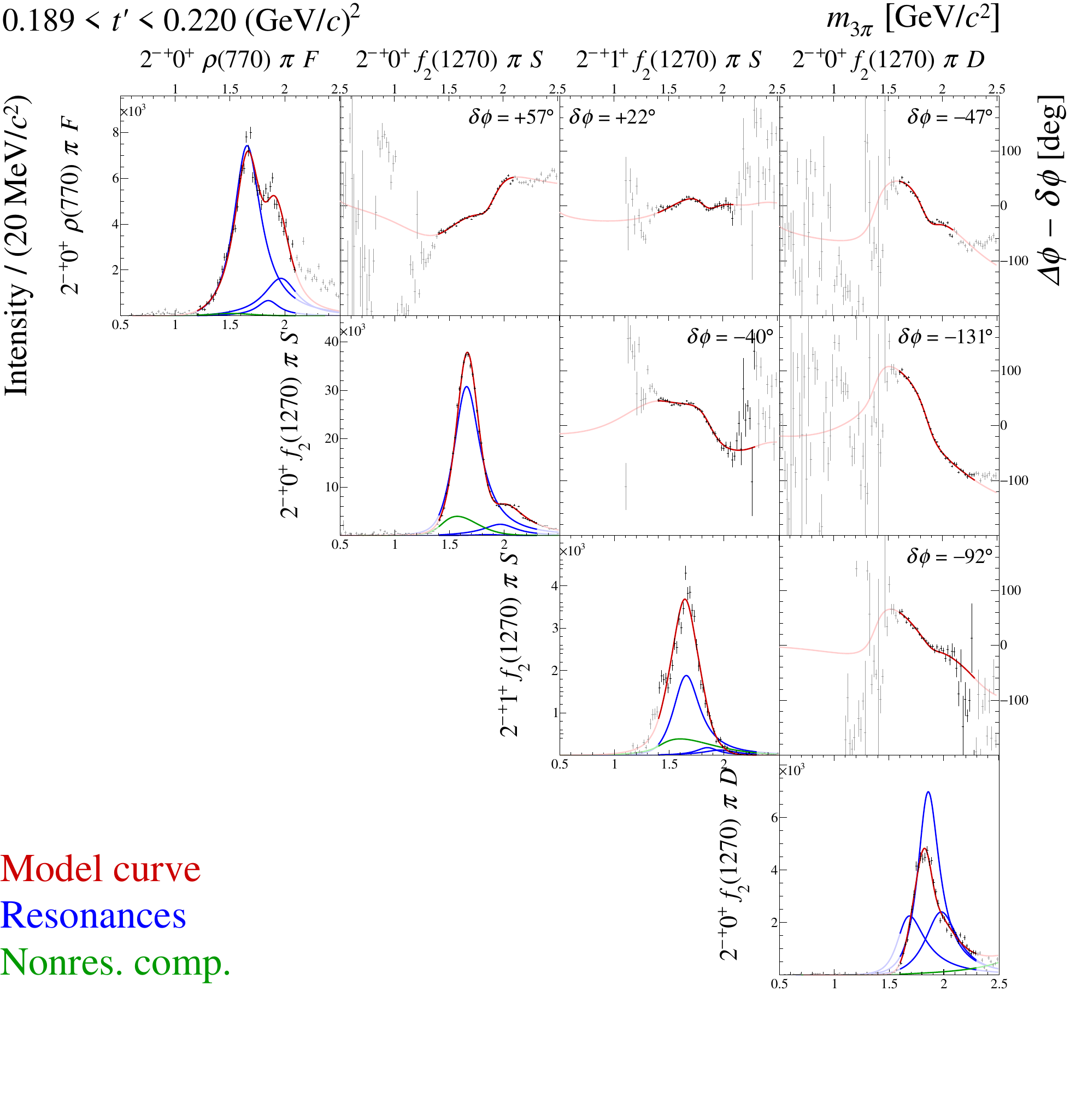}%
   \caption{Submatrix~H of the $14 \times 14$ matrix of graphs that
     represents the spin-density matrix (see
     \cref{tab:spin-dens_matrix_overview}).}
   \label{fig:spin-dens_submatrix_8_tbin_6}
 \end{minipage}
\end{textblock*}

\newpage\null
\begin{textblock*}{\textwidth}[0.5,0](0.5\paperwidth,\blockDistanceToTop)
 \begin{minipage}{\textwidth}
   \makeatletter
   \def\@captype{figure}
   \makeatother
   \centering
   \includegraphics[height=\matrixHeight]{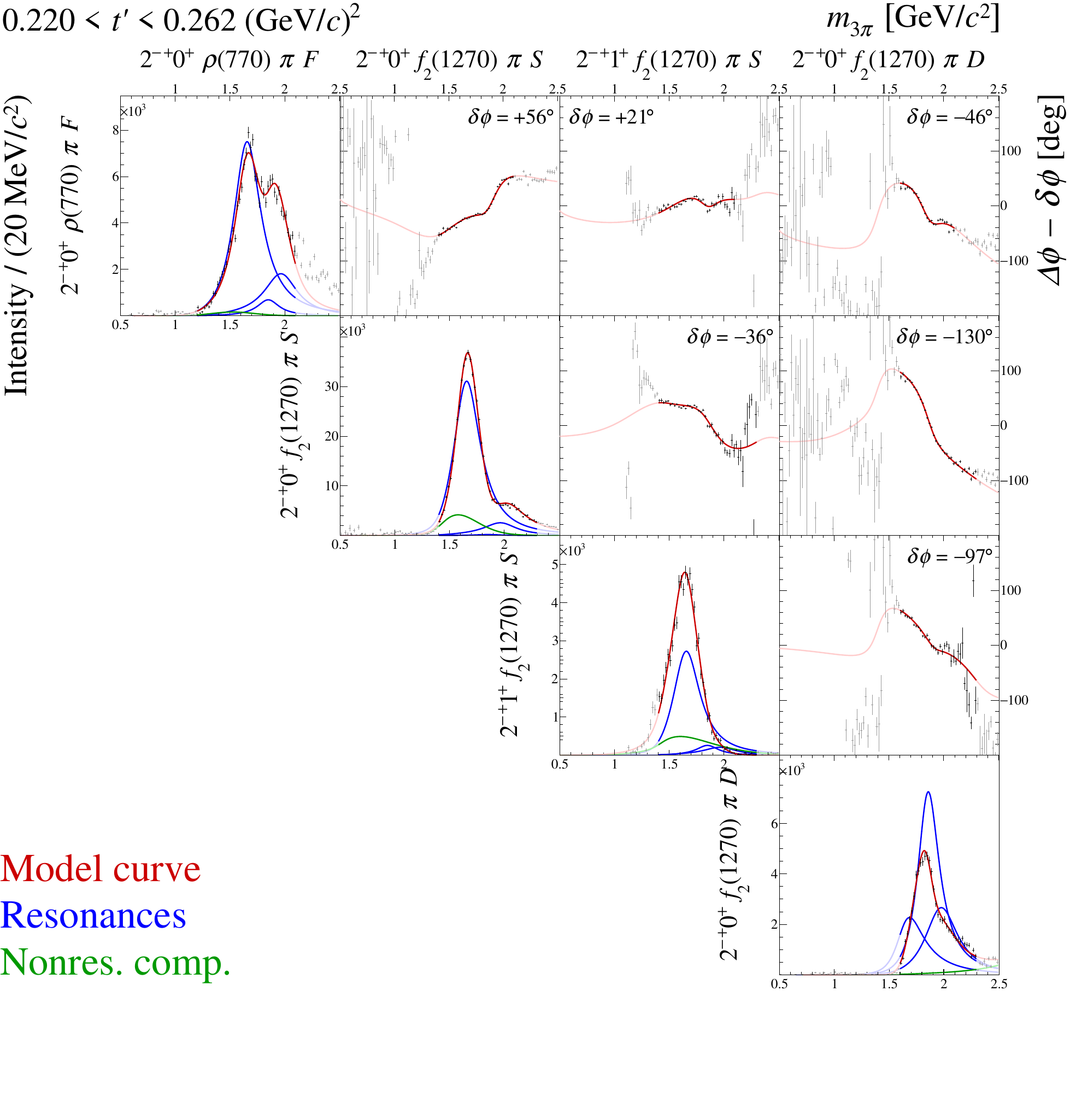}%
   \caption{Submatrix~H of the $14 \times 14$ matrix of graphs that
     represents the spin-density matrix (see
     \cref{tab:spin-dens_matrix_overview}).}
   \label{fig:spin-dens_submatrix_8_tbin_7}
 \end{minipage}
\end{textblock*}

\newpage\null
\begin{textblock*}{\textwidth}[0.5,0](0.5\paperwidth,\blockDistanceToTop)
 \begin{minipage}{\textwidth}
   \makeatletter
   \def\@captype{figure}
   \makeatother
   \centering
   \includegraphics[height=\matrixHeight]{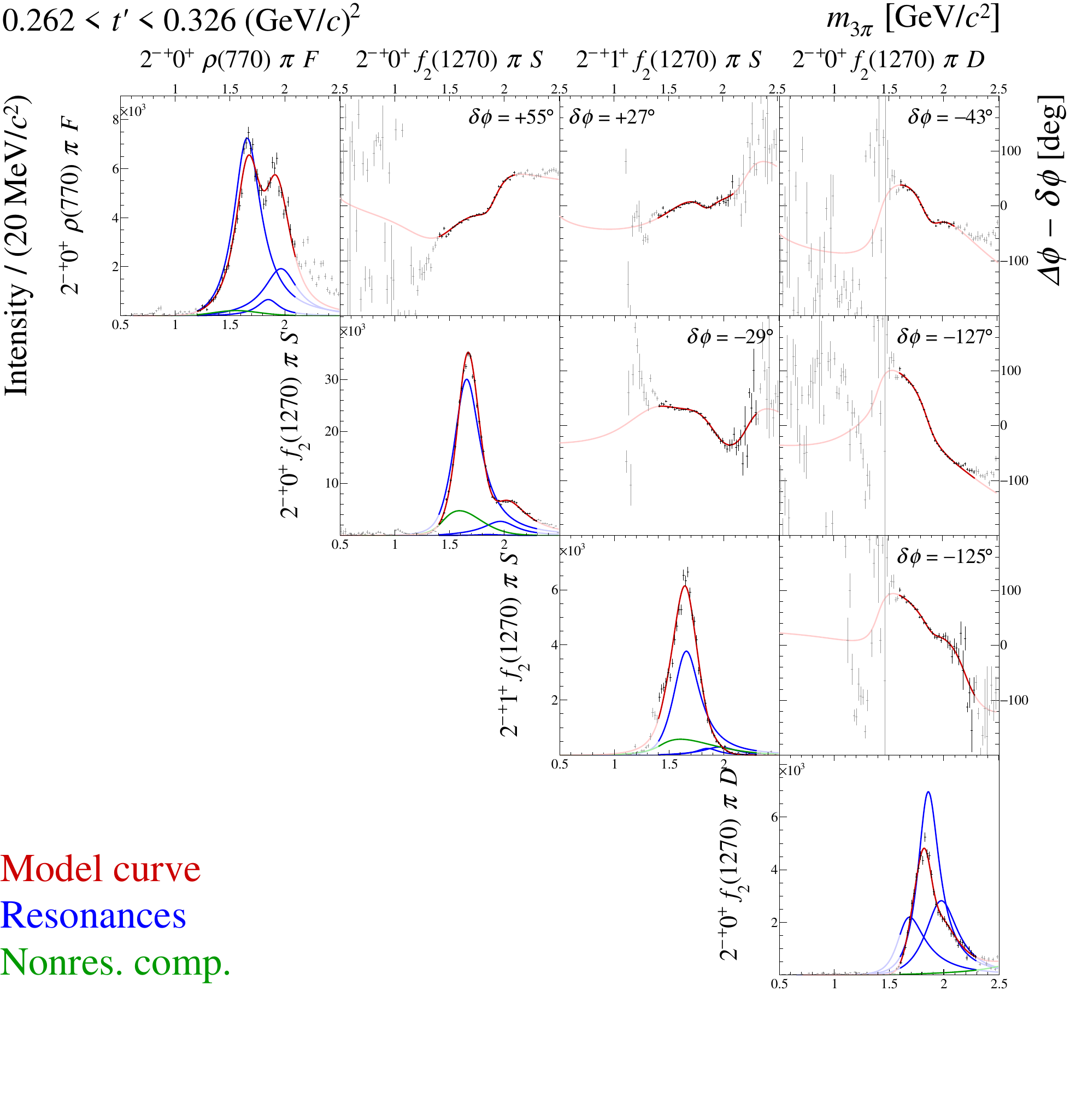}%
   \caption{Submatrix~H of the $14 \times 14$ matrix of graphs that
     represents the spin-density matrix (see
     \cref{tab:spin-dens_matrix_overview}).}
   \label{fig:spin-dens_submatrix_8_tbin_8}
 \end{minipage}
\end{textblock*}

\newpage\null
\begin{textblock*}{\textwidth}[0.5,0](0.5\paperwidth,\blockDistanceToTop)
 \begin{minipage}{\textwidth}
   \makeatletter
   \def\@captype{figure}
   \makeatother
   \centering
   \includegraphics[height=\matrixHeight]{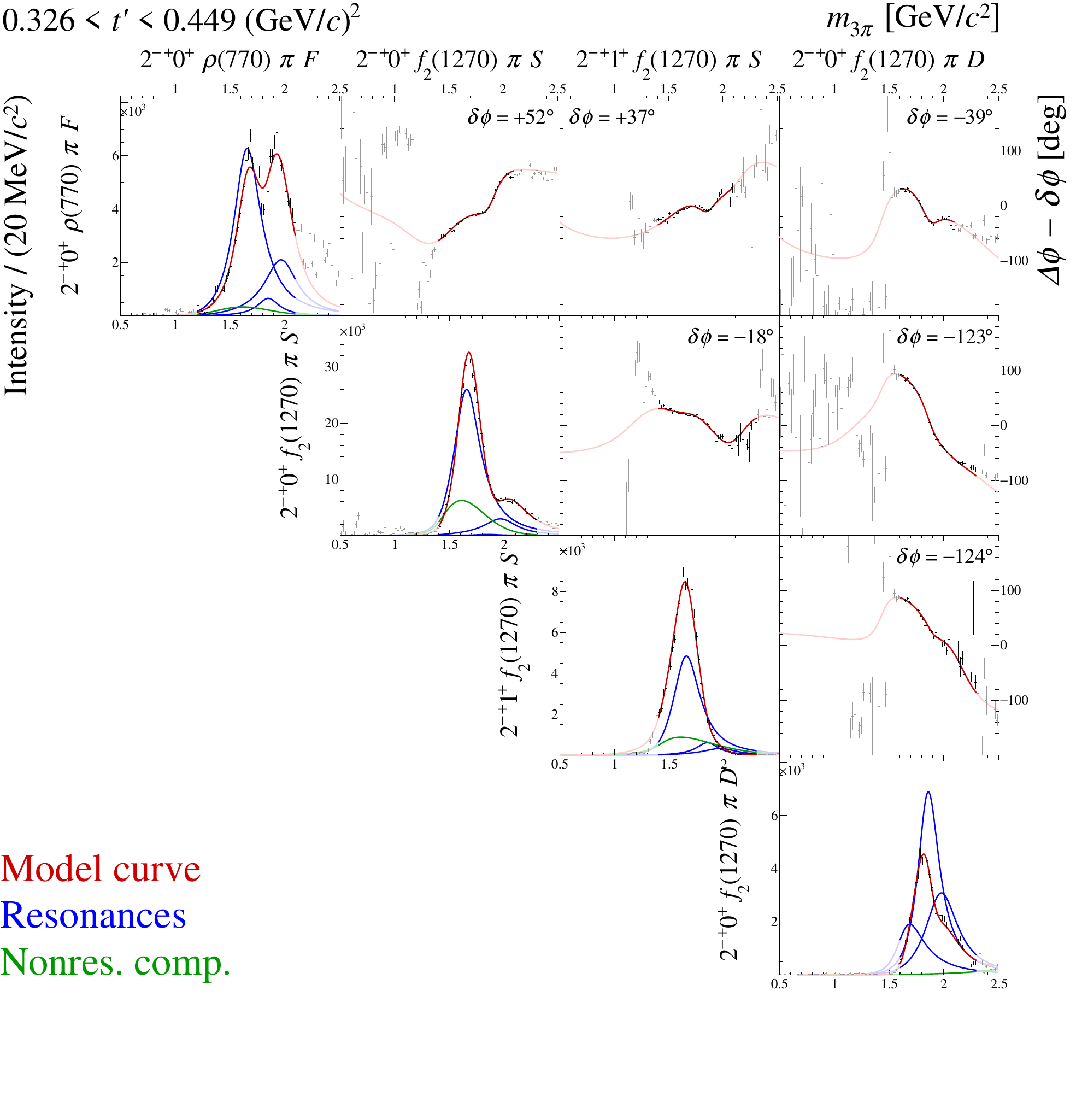}%
   \caption{Submatrix~H of the $14 \times 14$ matrix of graphs that
     represents the spin-density matrix (see
     \cref{tab:spin-dens_matrix_overview}).}
   \label{fig:spin-dens_submatrix_8_tbin_9}
 \end{minipage}
\end{textblock*}

\newpage\null
\begin{textblock*}{\textwidth}[0.5,0](0.5\paperwidth,\blockDistanceToTop)
 \begin{minipage}{\textwidth}
   \makeatletter
   \def\@captype{figure}
   \makeatother
   \centering
   \includegraphics[height=\matrixHeight]{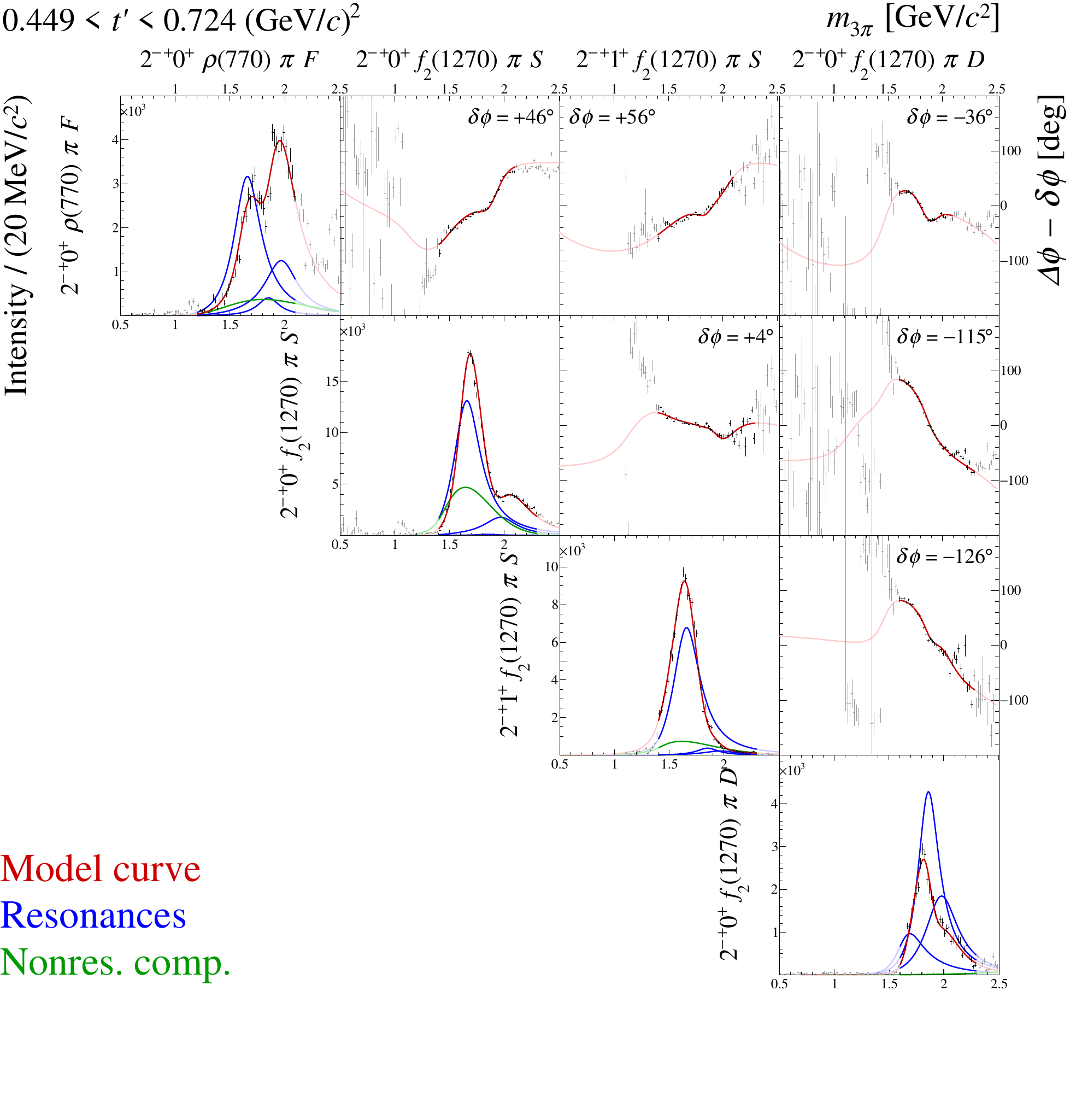}%
   \caption{Submatrix~H of the $14 \times 14$ matrix of graphs that
     represents the spin-density matrix (see
     \cref{tab:spin-dens_matrix_overview}).}
   \label{fig:spin-dens_submatrix_8_tbin_10}
 \end{minipage}
\end{textblock*}

\newpage\null
\begin{textblock*}{\textwidth}[0.5,0](0.5\paperwidth,\blockDistanceToTop)
 \begin{minipage}{\textwidth}
   \makeatletter
   \def\@captype{figure}
   \makeatother
   \centering
   \includegraphics[height=\matrixHeight]{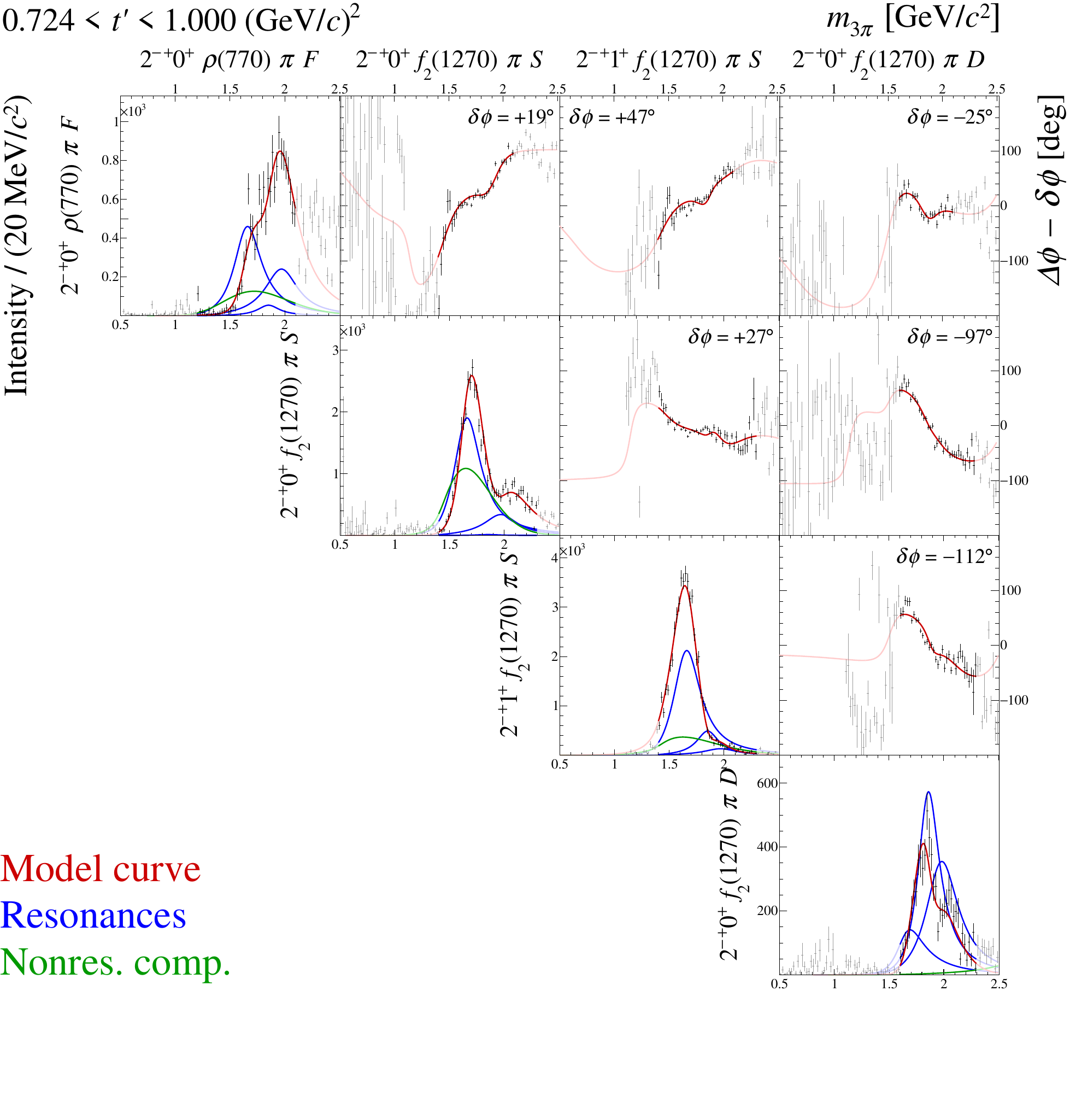}%
   \caption{Submatrix~H of the $14 \times 14$ matrix of graphs that
     represents the spin-density matrix (see
     \cref{tab:spin-dens_matrix_overview}).}
   \label{fig:spin-dens_submatrix_8_tbin_11}
 \end{minipage}
\end{textblock*}

\clearpage
\subsection{Submatrix I}
\label{sec:spin-dens_submatrix_9}

\begin{textblock*}{\textwidth}[0.5,0](0.5\paperwidth,\blockDistanceToTop)
 \begin{minipage}{\textwidth}
   \makeatletter
   \def\@captype{figure}
   \makeatother
   \centering
   \includegraphics[height=\matrixHeight]{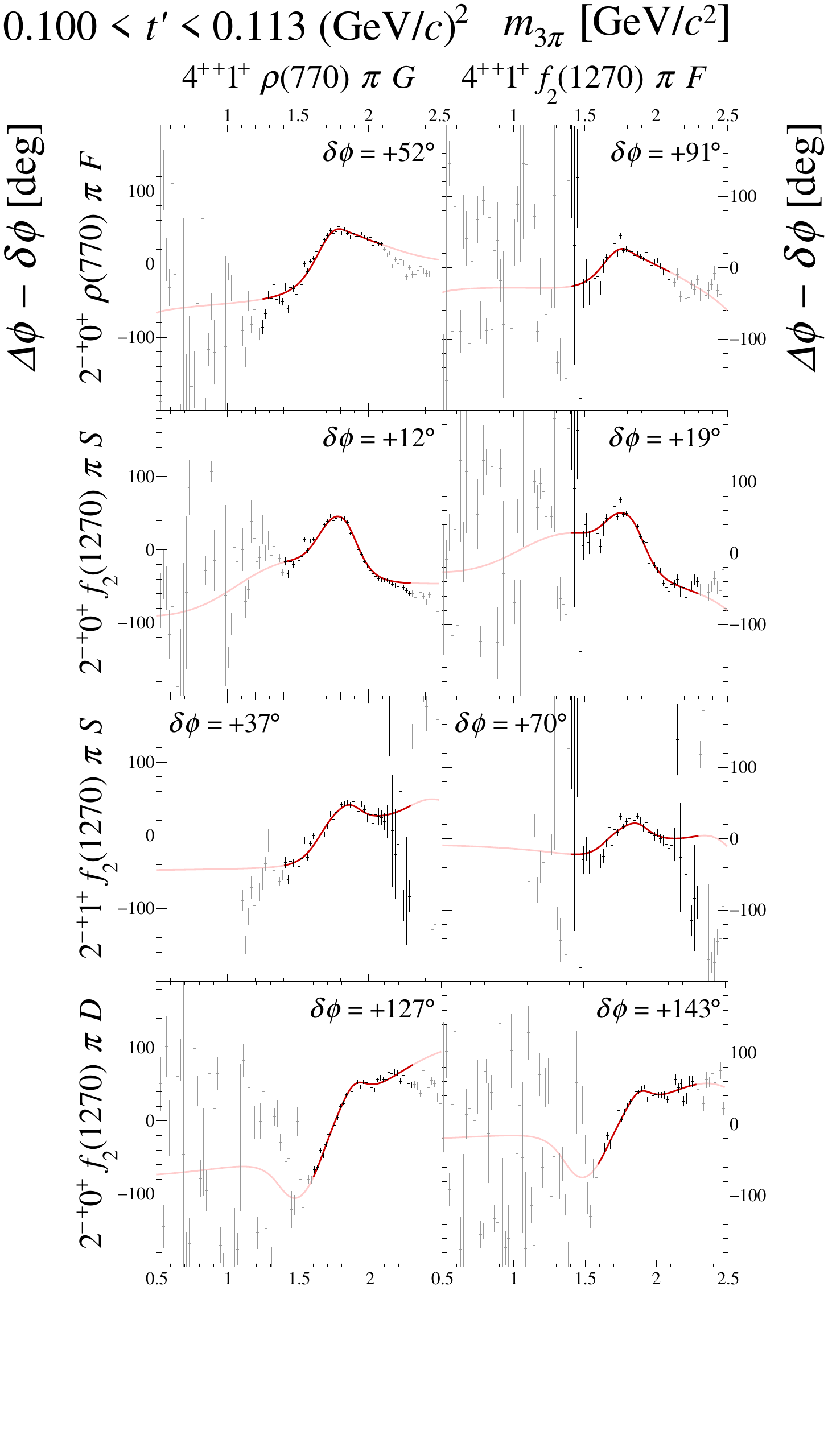}%
   \caption{Submatrix~I of the $14 \times 14$ matrix of graphs that
     represents the spin-density matrix (see
     \cref{tab:spin-dens_matrix_overview}).}
   \label{fig:spin-dens_submatrix_9_tbin_1}
 \end{minipage}
\end{textblock*}

\newpage\null
\begin{textblock*}{\textwidth}[0.5,0](0.5\paperwidth,\blockDistanceToTop)
 \begin{minipage}{\textwidth}
   \makeatletter
   \def\@captype{figure}
   \makeatother
   \centering
   \includegraphics[height=\matrixHeight]{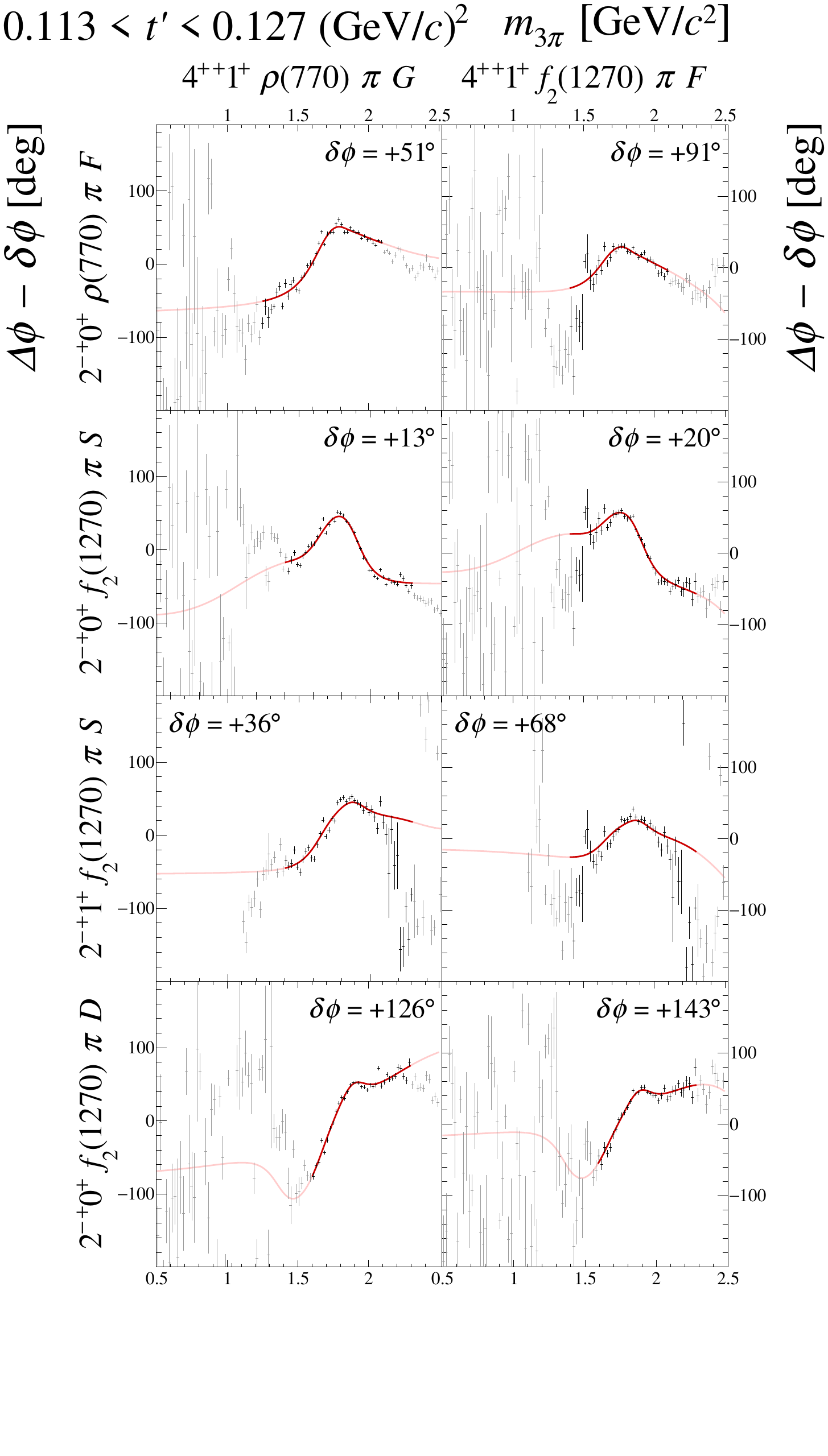}%
   \caption{Submatrix~I of the $14 \times 14$ matrix of graphs that
     represents the spin-density matrix (see
     \cref{tab:spin-dens_matrix_overview}).}
   \label{fig:spin-dens_submatrix_9_tbin_2}
 \end{minipage}
\end{textblock*}

\newpage\null
\begin{textblock*}{\textwidth}[0.5,0](0.5\paperwidth,\blockDistanceToTop)
 \begin{minipage}{\textwidth}
   \makeatletter
   \def\@captype{figure}
   \makeatother
   \centering
   \includegraphics[height=\matrixHeight]{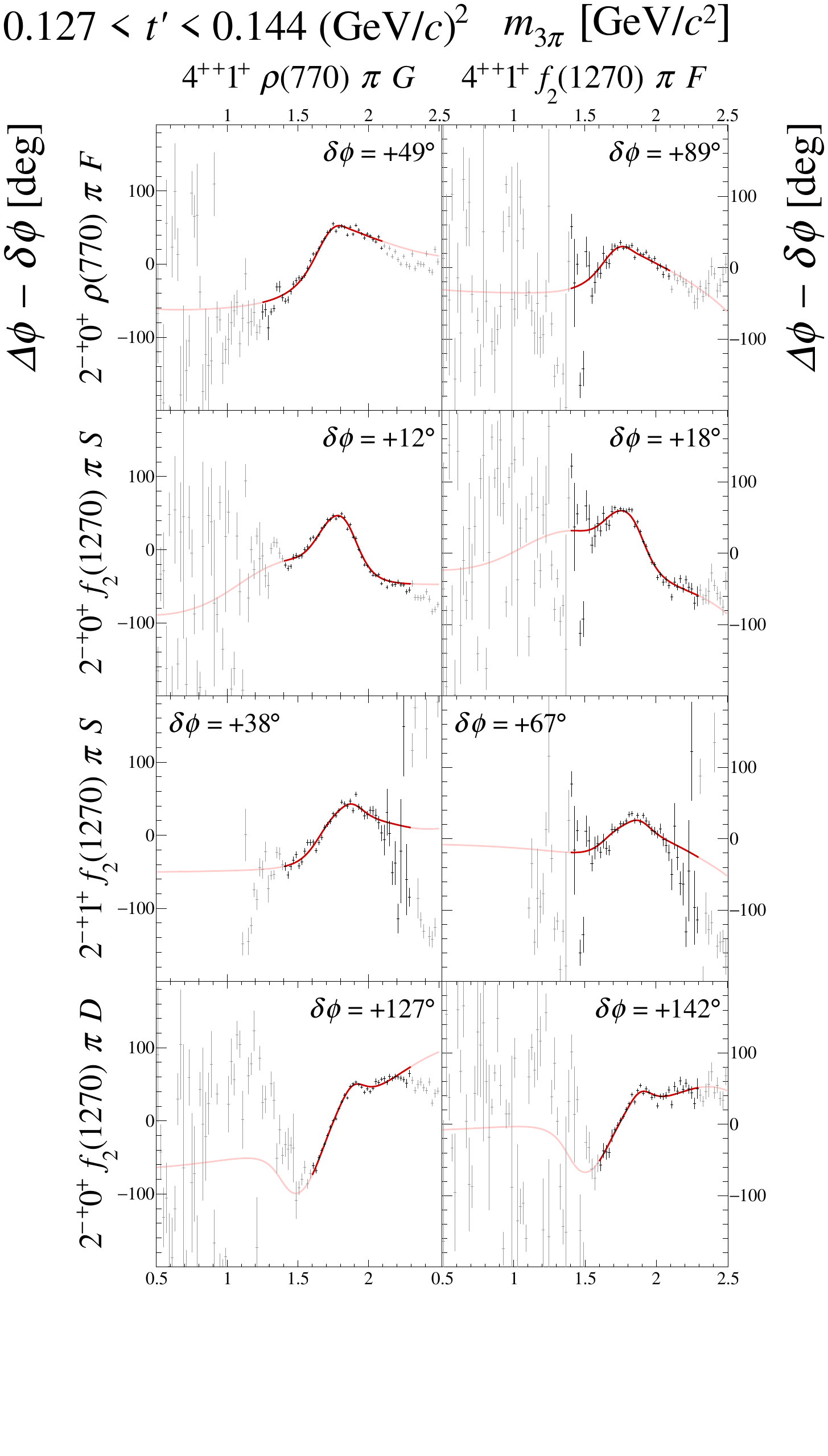}%
   \caption{Submatrix~I of the $14 \times 14$ matrix of graphs that
     represents the spin-density matrix (see
     \cref{tab:spin-dens_matrix_overview}).}
   \label{fig:spin-dens_submatrix_9_tbin_3}
 \end{minipage}
\end{textblock*}

\newpage\null
\begin{textblock*}{\textwidth}[0.5,0](0.5\paperwidth,\blockDistanceToTop)
 \begin{minipage}{\textwidth}
   \makeatletter
   \def\@captype{figure}
   \makeatother
   \centering
   \includegraphics[height=\matrixHeight]{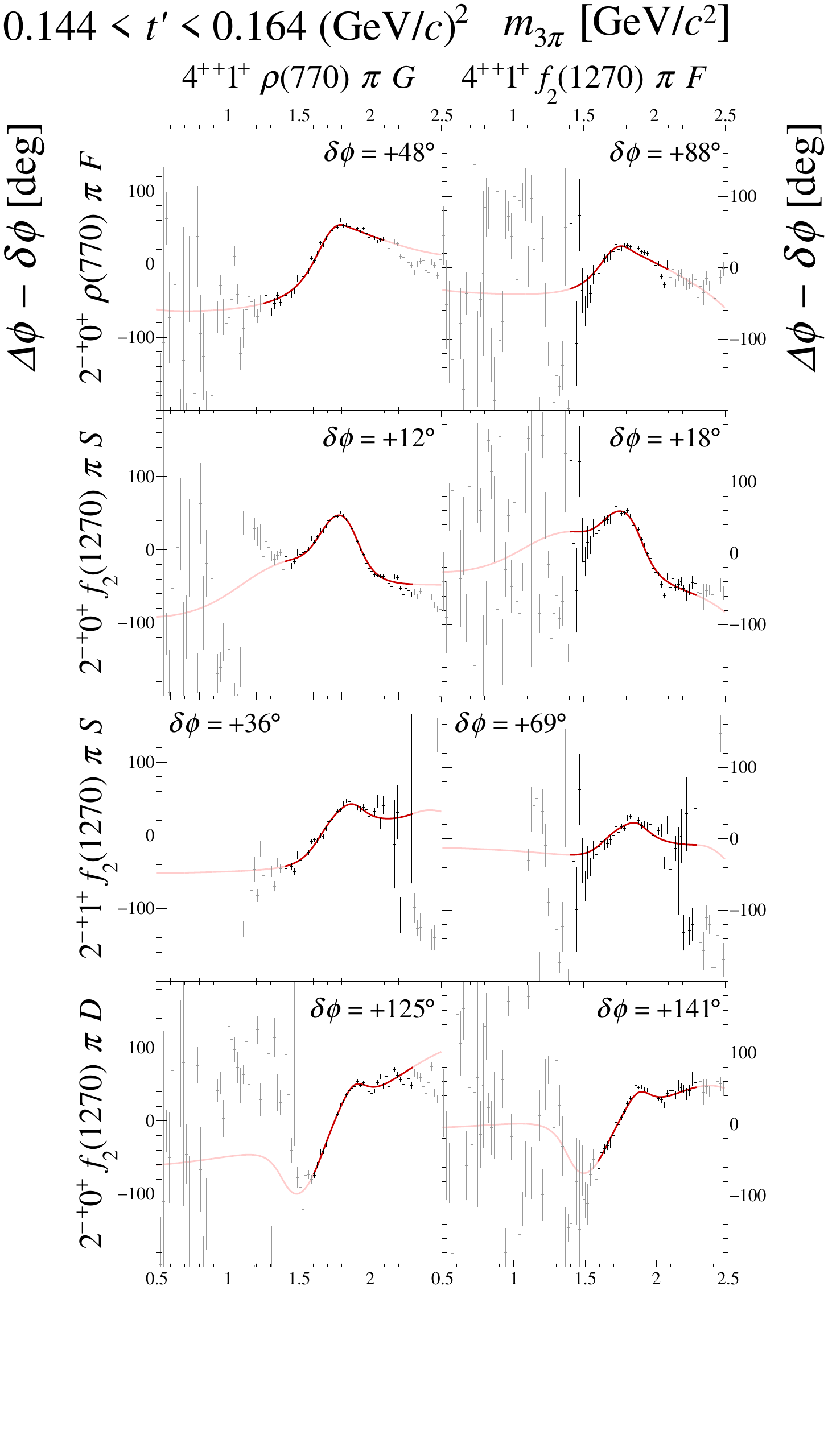}%
   \caption{Submatrix~I of the $14 \times 14$ matrix of graphs that
     represents the spin-density matrix (see
     \cref{tab:spin-dens_matrix_overview}).}
   \label{fig:spin-dens_submatrix_9_tbin_4}
 \end{minipage}
\end{textblock*}

\newpage\null
\begin{textblock*}{\textwidth}[0.5,0](0.5\paperwidth,\blockDistanceToTop)
 \begin{minipage}{\textwidth}
   \makeatletter
   \def\@captype{figure}
   \makeatother
   \centering
   \includegraphics[height=\matrixHeight]{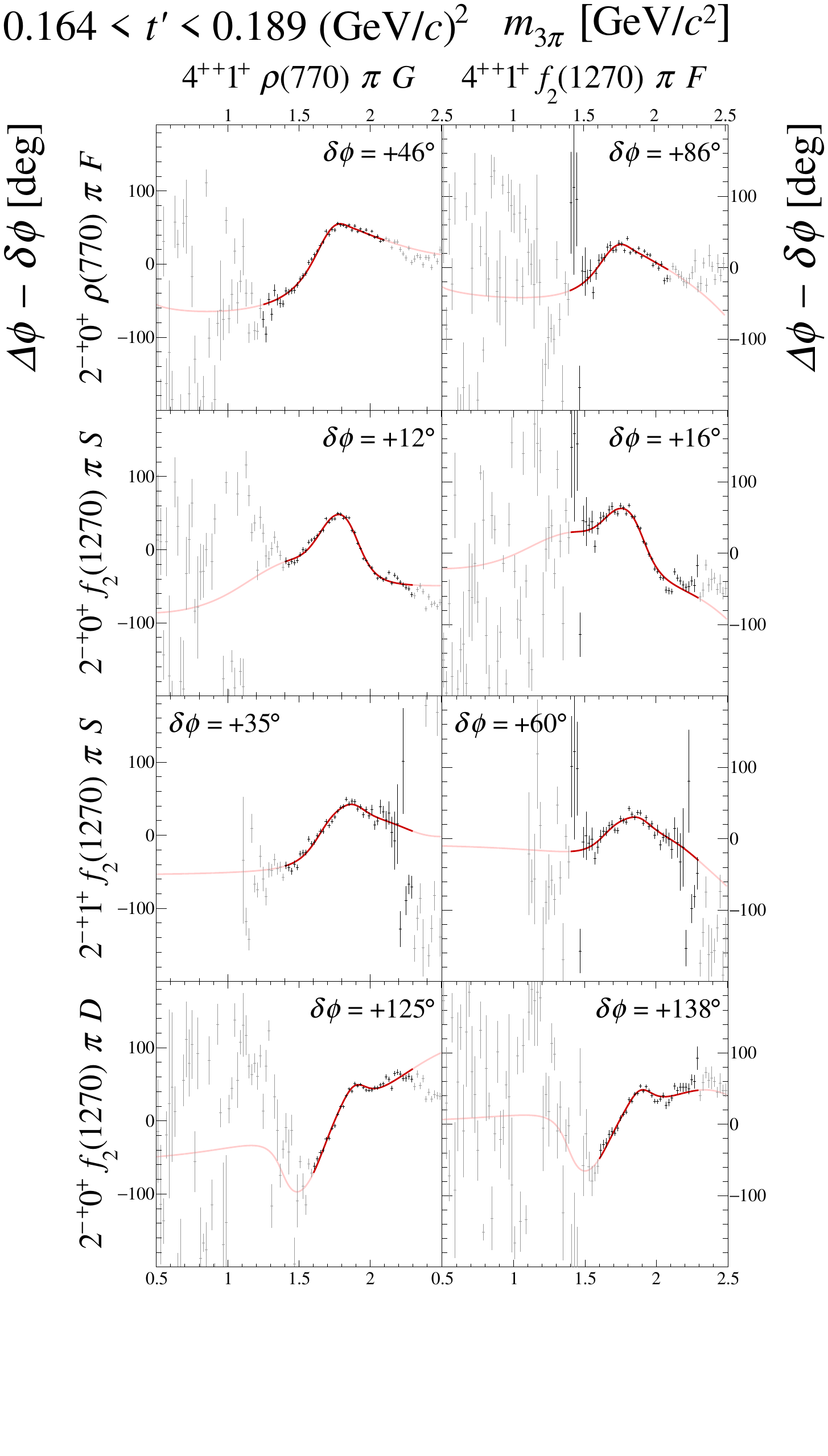}%
   \caption{Submatrix~I of the $14 \times 14$ matrix of graphs that
     represents the spin-density matrix (see
     \cref{tab:spin-dens_matrix_overview}).}
   \label{fig:spin-dens_submatrix_9_tbin_5}
 \end{minipage}
\end{textblock*}

\newpage\null
\begin{textblock*}{\textwidth}[0.5,0](0.5\paperwidth,\blockDistanceToTop)
 \begin{minipage}{\textwidth}
   \makeatletter
   \def\@captype{figure}
   \makeatother
   \centering
   \includegraphics[height=\matrixHeight]{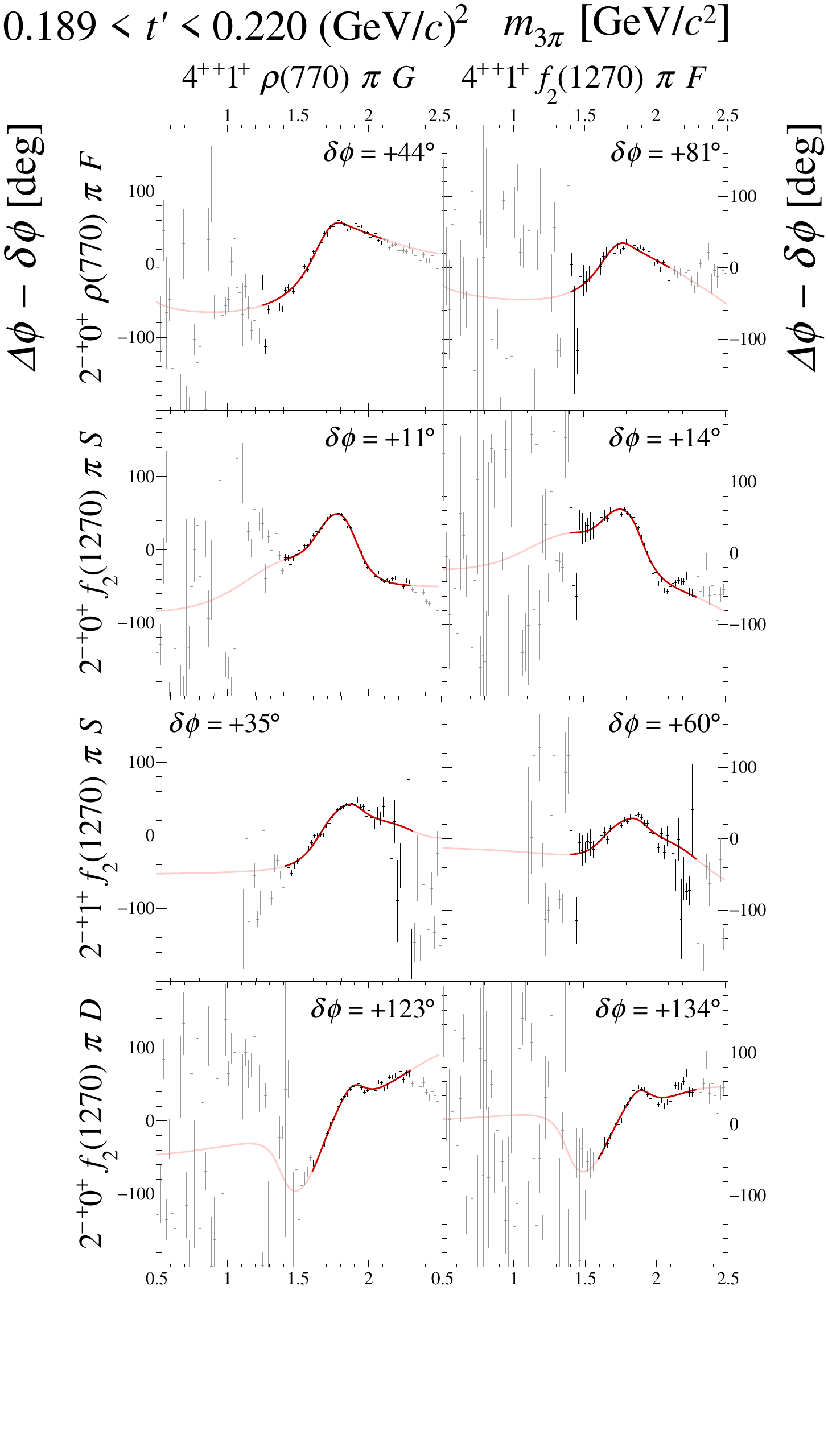}%
   \caption{Submatrix~I of the $14 \times 14$ matrix of graphs that
     represents the spin-density matrix (see
     \cref{tab:spin-dens_matrix_overview}).}
   \label{fig:spin-dens_submatrix_9_tbin_6}
 \end{minipage}
\end{textblock*}

\newpage\null
\begin{textblock*}{\textwidth}[0.5,0](0.5\paperwidth,\blockDistanceToTop)
 \begin{minipage}{\textwidth}
   \makeatletter
   \def\@captype{figure}
   \makeatother
   \centering
   \includegraphics[height=\matrixHeight]{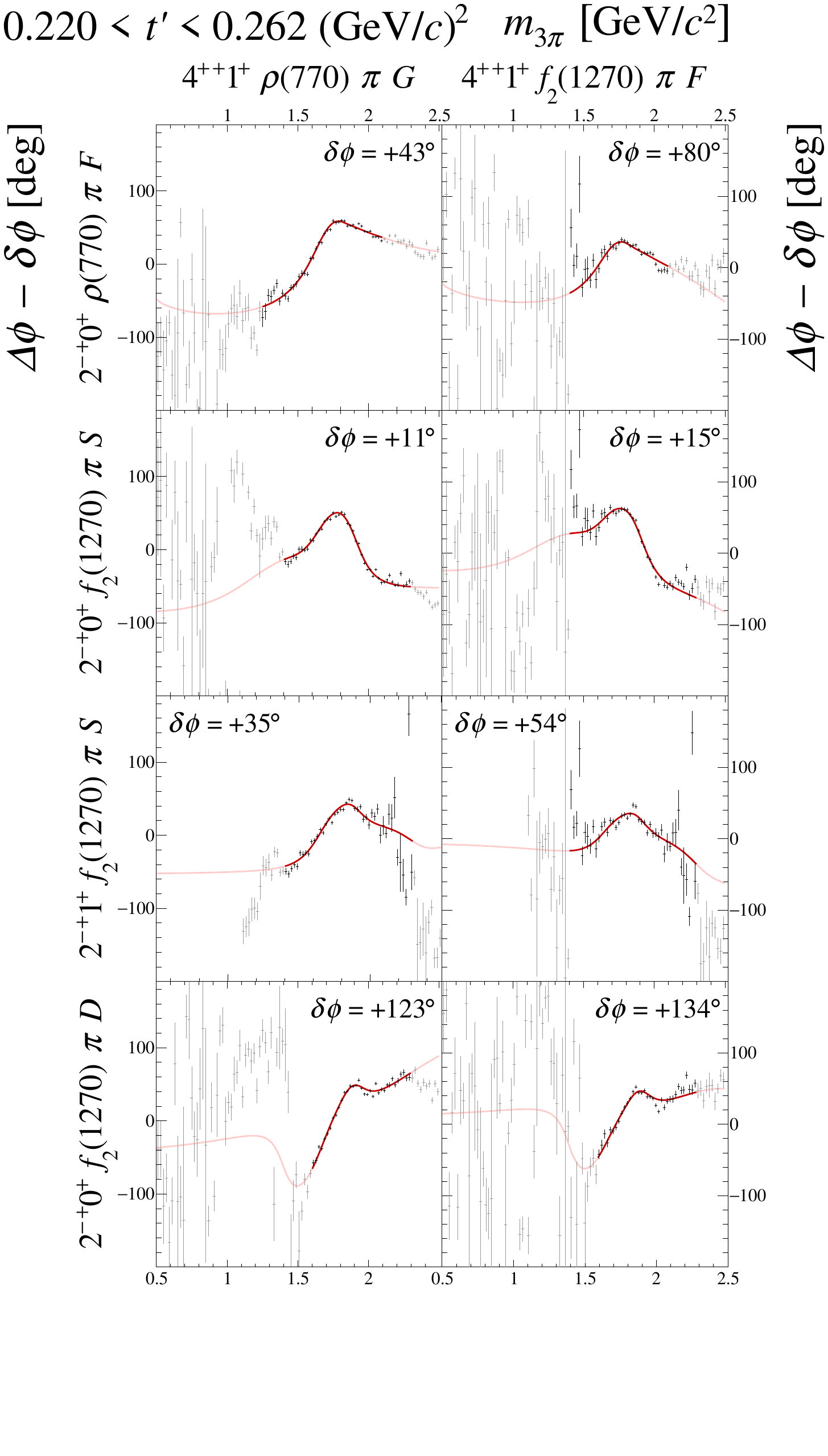}%
   \caption{Submatrix~I of the $14 \times 14$ matrix of graphs that
     represents the spin-density matrix (see
     \cref{tab:spin-dens_matrix_overview}).}
   \label{fig:spin-dens_submatrix_9_tbin_7}
 \end{minipage}
\end{textblock*}

\newpage\null
\begin{textblock*}{\textwidth}[0.5,0](0.5\paperwidth,\blockDistanceToTop)
 \begin{minipage}{\textwidth}
   \makeatletter
   \def\@captype{figure}
   \makeatother
   \centering
   \includegraphics[height=\matrixHeight]{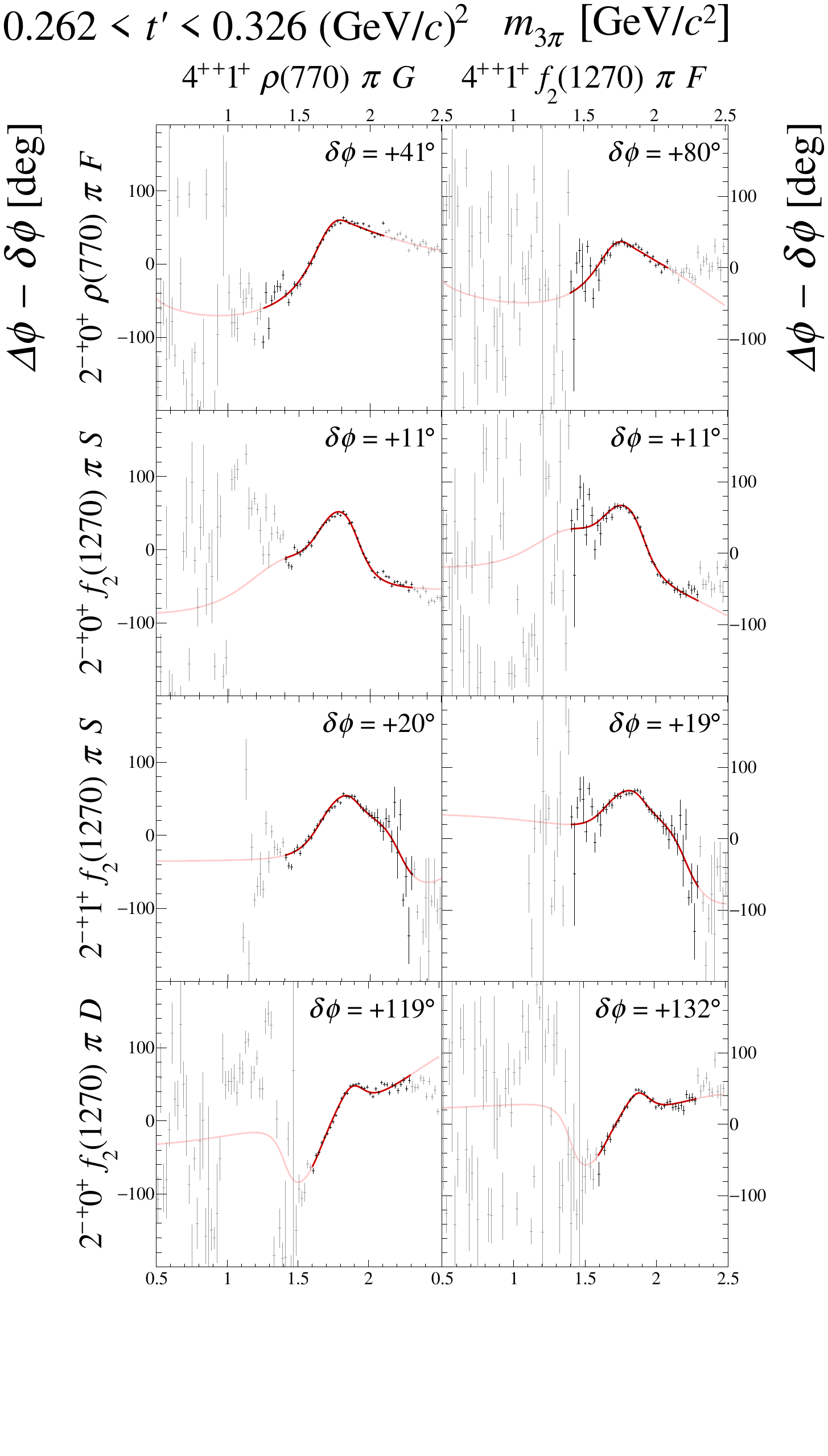}%
   \caption{Submatrix~I of the $14 \times 14$ matrix of graphs that
     represents the spin-density matrix (see
     \cref{tab:spin-dens_matrix_overview}).}
   \label{fig:spin-dens_submatrix_9_tbin_8}
 \end{minipage}
\end{textblock*}

\newpage\null
\begin{textblock*}{\textwidth}[0.5,0](0.5\paperwidth,\blockDistanceToTop)
 \begin{minipage}{\textwidth}
   \makeatletter
   \def\@captype{figure}
   \makeatother
   \centering
   \includegraphics[height=\matrixHeight]{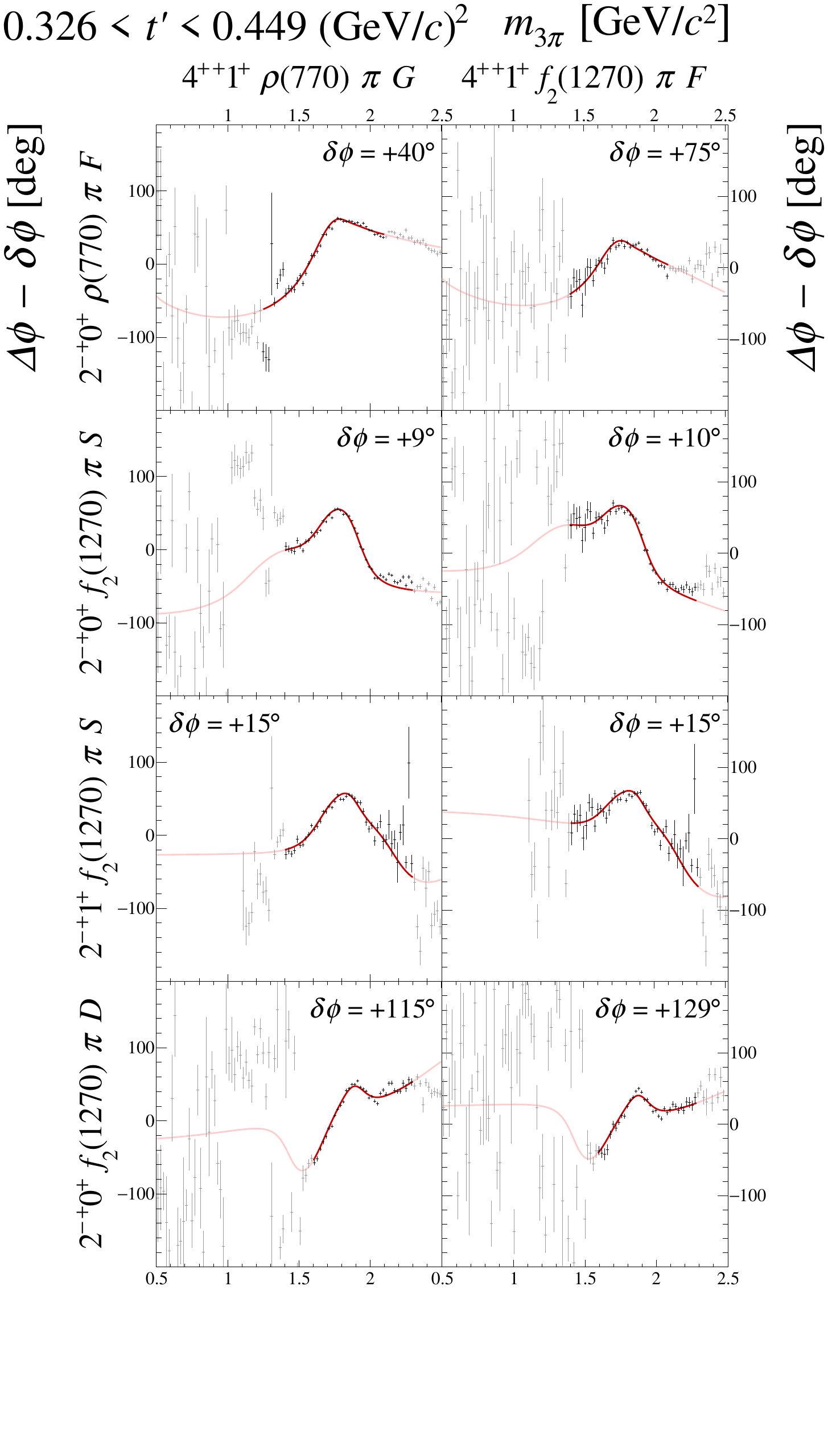}%
   \caption{Submatrix~I of the $14 \times 14$ matrix of graphs that
     represents the spin-density matrix (see
     \cref{tab:spin-dens_matrix_overview}).}
   \label{fig:spin-dens_submatrix_9_tbin_9}
 \end{minipage}
\end{textblock*}

\newpage\null
\begin{textblock*}{\textwidth}[0.5,0](0.5\paperwidth,\blockDistanceToTop)
 \begin{minipage}{\textwidth}
   \makeatletter
   \def\@captype{figure}
   \makeatother
   \centering
   \includegraphics[height=\matrixHeight]{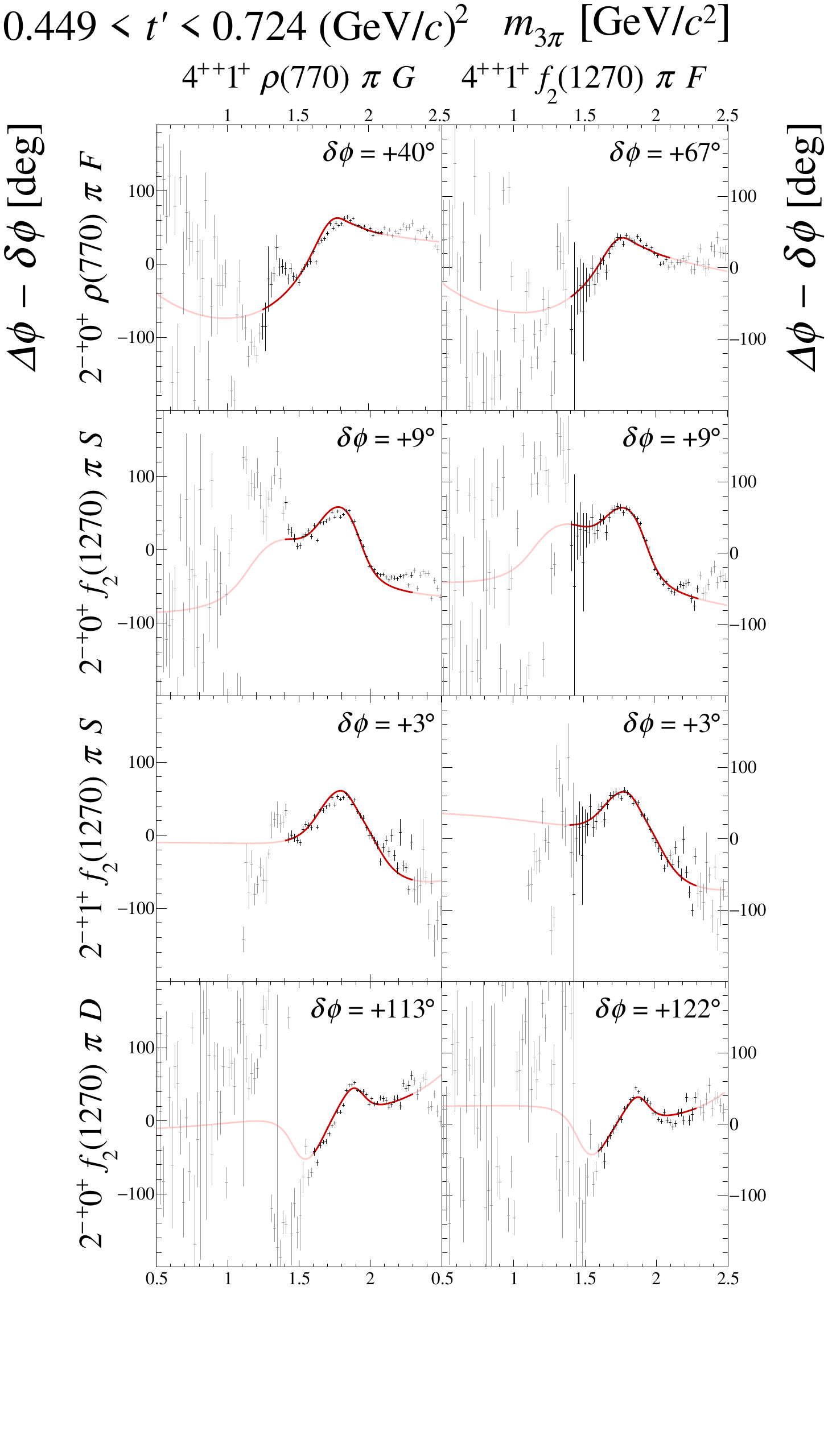}%
   \caption{Submatrix~I of the $14 \times 14$ matrix of graphs that
     represents the spin-density matrix (see
     \cref{tab:spin-dens_matrix_overview}).}
   \label{fig:spin-dens_submatrix_9_tbin_10}
 \end{minipage}
\end{textblock*}

\newpage\null
\begin{textblock*}{\textwidth}[0.5,0](0.5\paperwidth,\blockDistanceToTop)
 \begin{minipage}{\textwidth}
   \makeatletter
   \def\@captype{figure}
   \makeatother
   \centering
   \includegraphics[height=\matrixHeight]{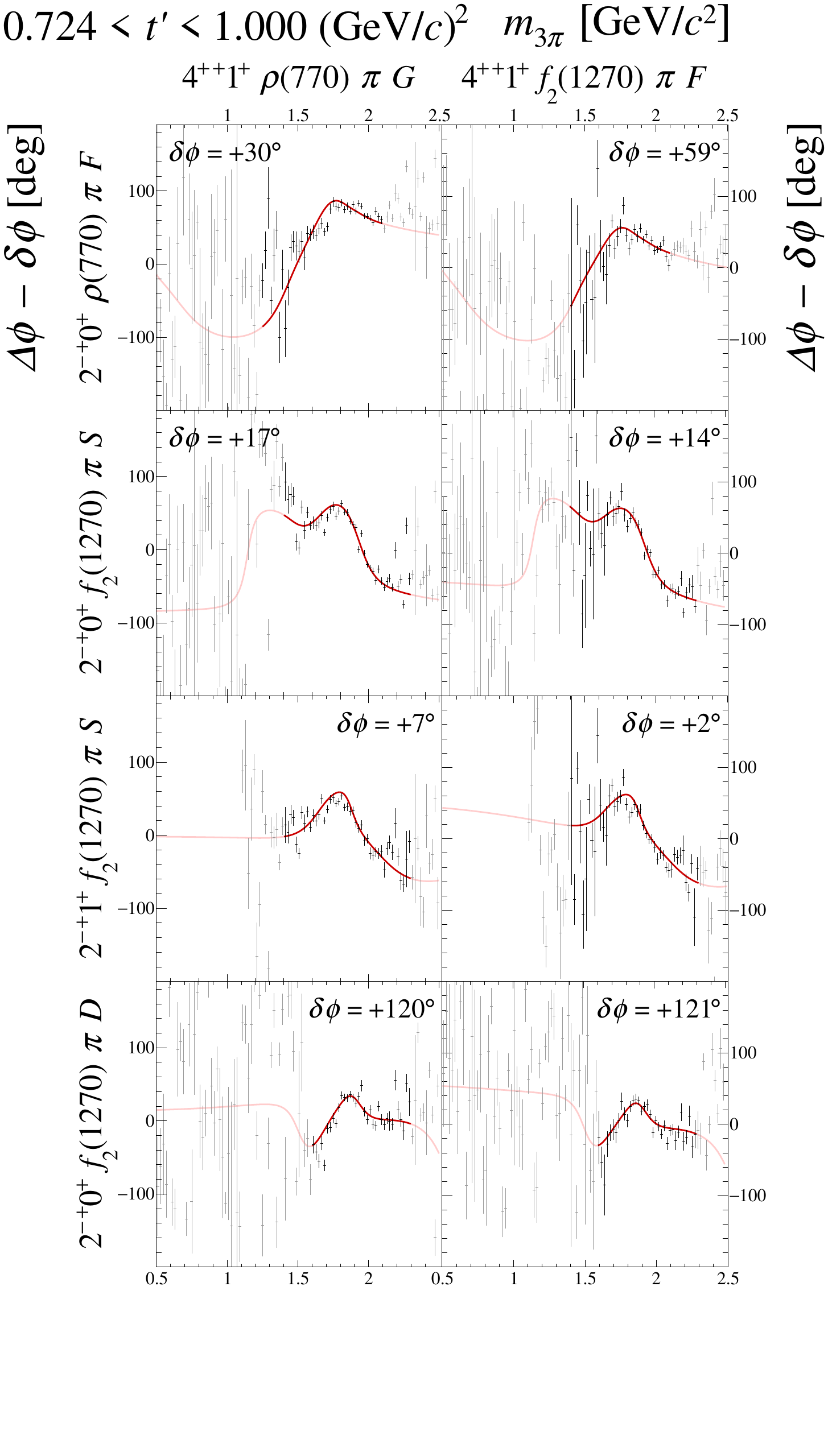}%
   \caption{Submatrix~I of the $14 \times 14$ matrix of graphs that
     represents the spin-density matrix (see
     \cref{tab:spin-dens_matrix_overview}).}
   \label{fig:spin-dens_submatrix_9_tbin_11}
 \end{minipage}
\end{textblock*}

\clearpage
\subsection{Submatrix J}
\label{sec:spin-dens_submatrix_10}

\begin{textblock*}{\textwidth}[0.5,0](0.5\paperwidth,\blockDistanceToTop)
 \begin{minipage}{\textwidth}
   \makeatletter
   \def\@captype{figure}
   \makeatother
   \centering
   \includegraphics[height=\matrixHeightHalf]{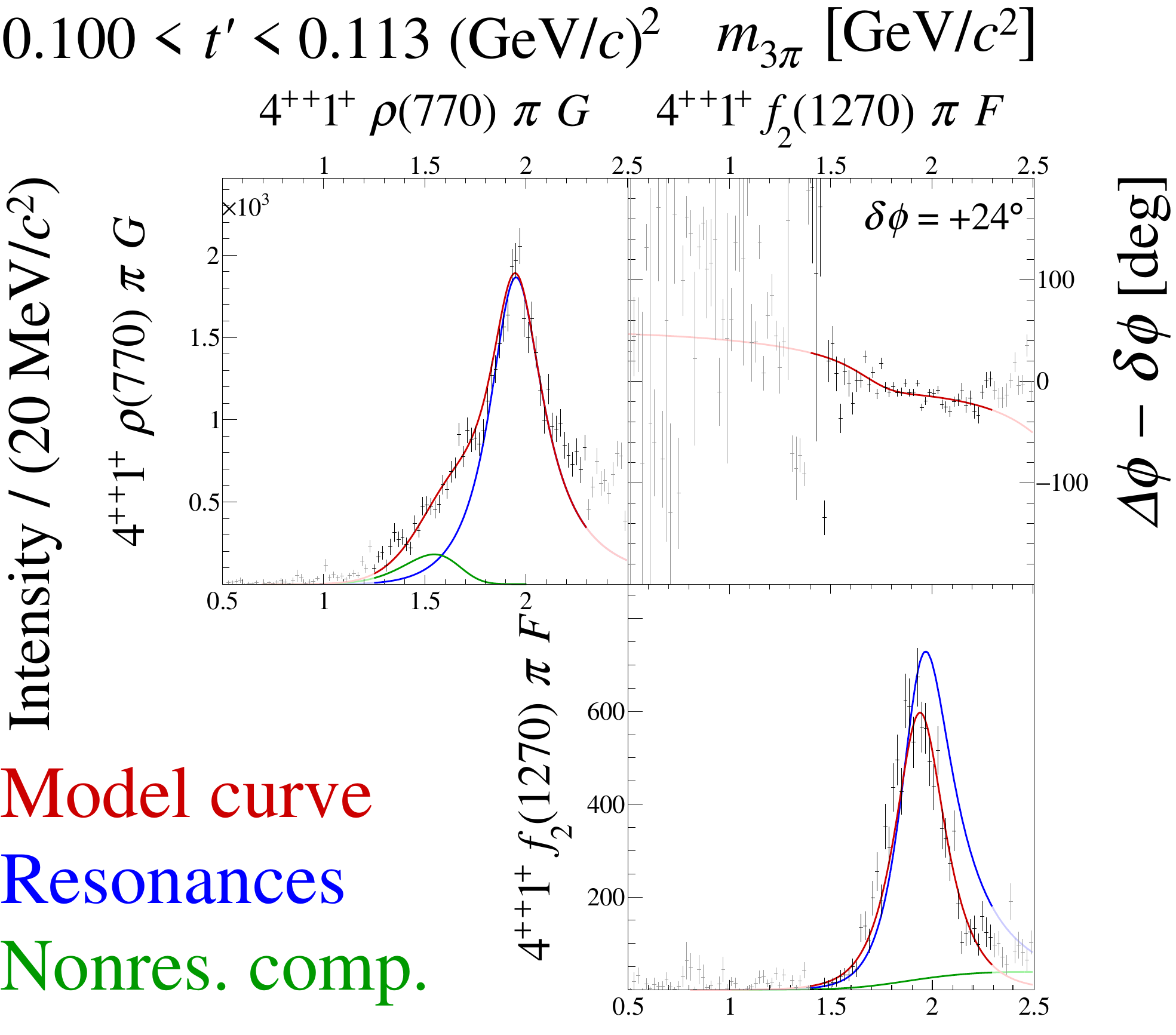}%
   \caption{Submatrix~J of the $14 \times 14$ matrix of graphs that
     represents the spin-density matrix (see
     \cref{tab:spin-dens_matrix_overview}).}
   \label{fig:spin-dens_submatrix_10_tbin_1}
 \end{minipage}
\end{textblock*}

\newpage\null
\begin{textblock*}{\textwidth}[0.5,0](0.5\paperwidth,\blockDistanceToTop)
 \begin{minipage}{\textwidth}
   \makeatletter
   \def\@captype{figure}
   \makeatother
   \centering
   \includegraphics[height=\matrixHeightHalf]{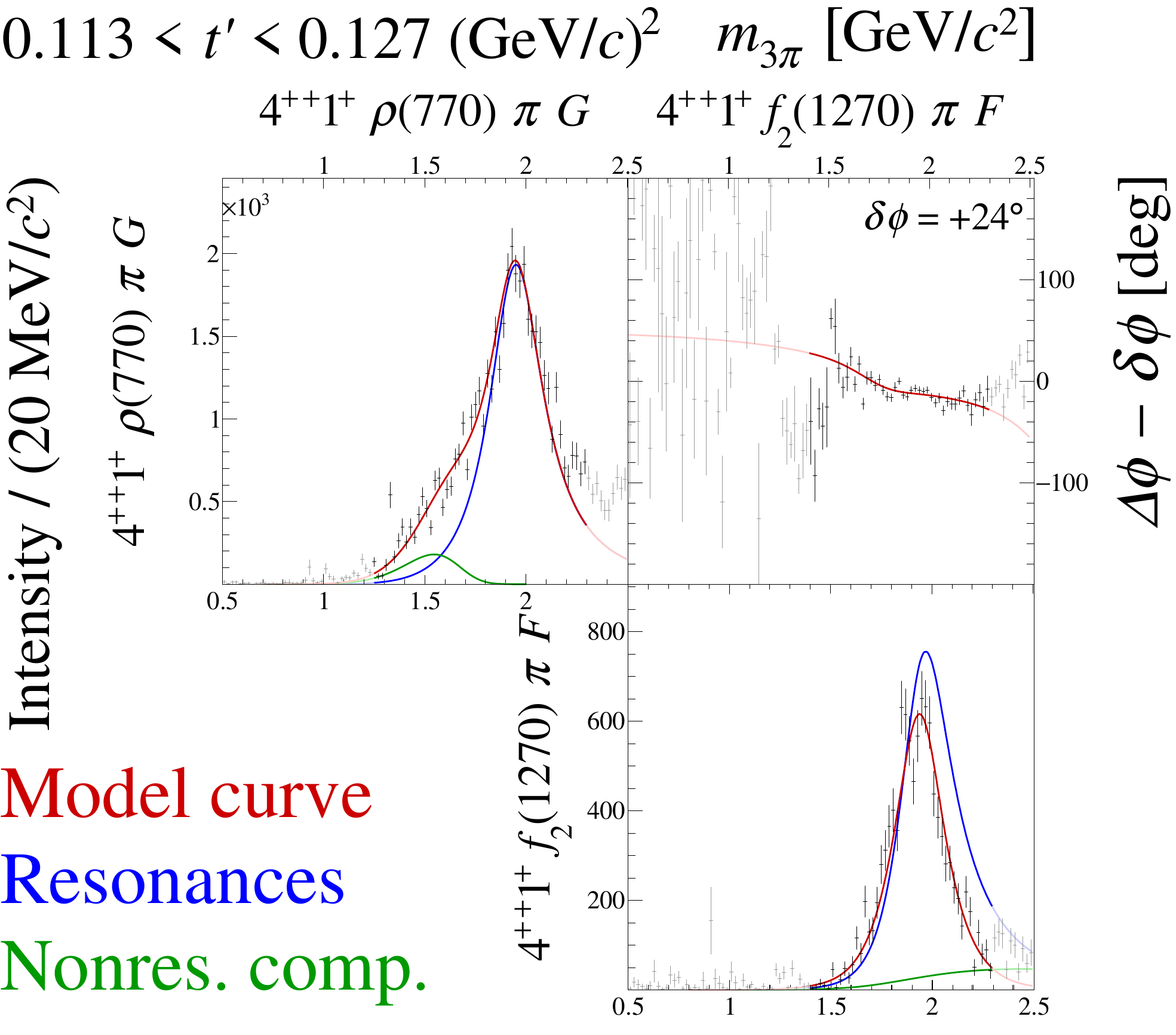}%
   \caption{Submatrix~J of the $14 \times 14$ matrix of graphs that
     represents the spin-density matrix (see
     \cref{tab:spin-dens_matrix_overview}).}
   \label{fig:spin-dens_submatrix_10_tbin_2}
 \end{minipage}
\end{textblock*}

\newpage\null
\begin{textblock*}{\textwidth}[0.5,0](0.5\paperwidth,\blockDistanceToTop)
 \begin{minipage}{\textwidth}
   \makeatletter
   \def\@captype{figure}
   \makeatother
   \centering
   \includegraphics[height=\matrixHeightHalf]{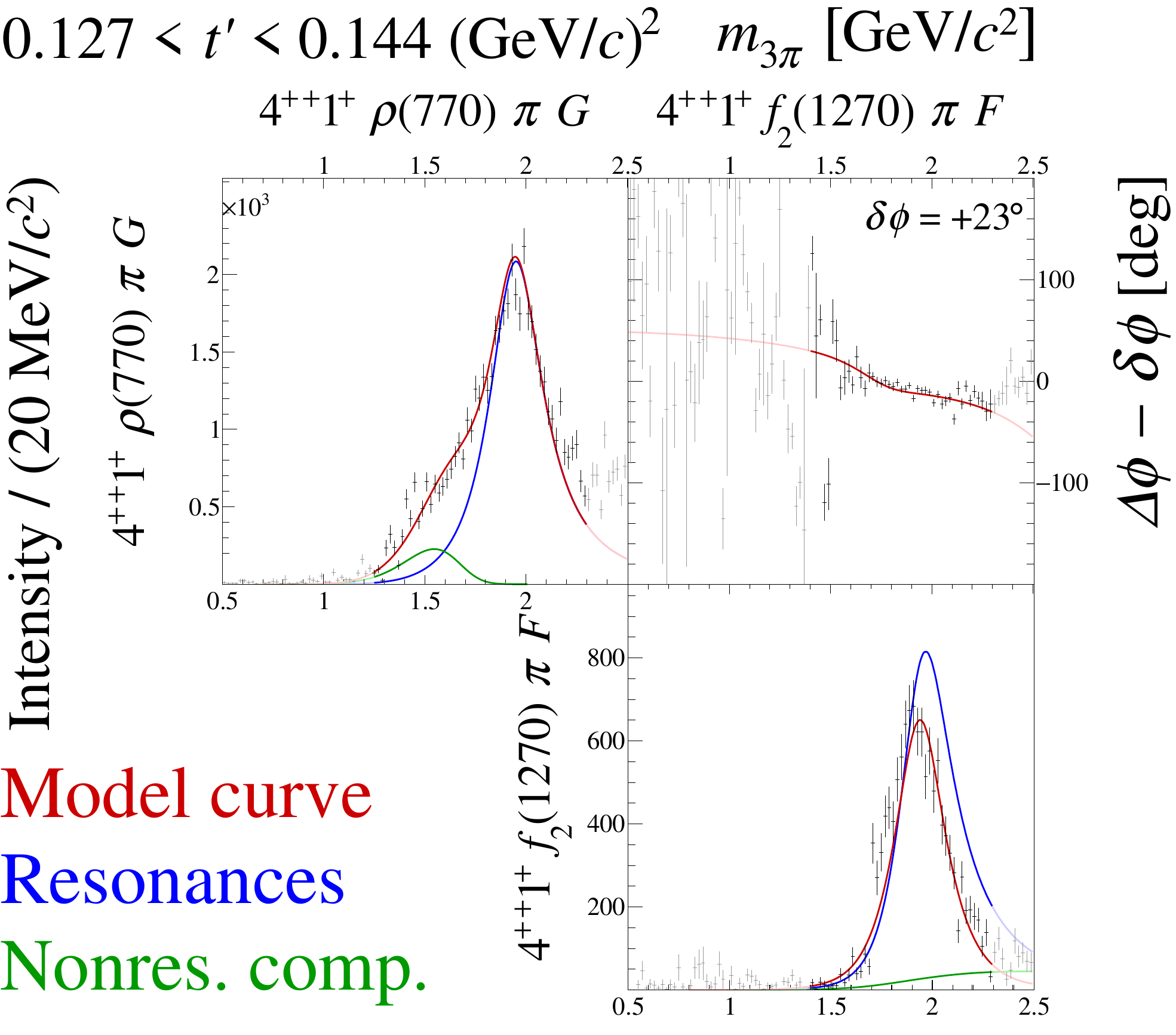}%
   \caption{Submatrix~J of the $14 \times 14$ matrix of graphs that
     represents the spin-density matrix (see
     \cref{tab:spin-dens_matrix_overview}).}
   \label{fig:spin-dens_submatrix_10_tbin_3}
 \end{minipage}
\end{textblock*}

\newpage\null
\begin{textblock*}{\textwidth}[0.5,0](0.5\paperwidth,\blockDistanceToTop)
 \begin{minipage}{\textwidth}
   \makeatletter
   \def\@captype{figure}
   \makeatother
   \centering
   \includegraphics[height=\matrixHeightHalf]{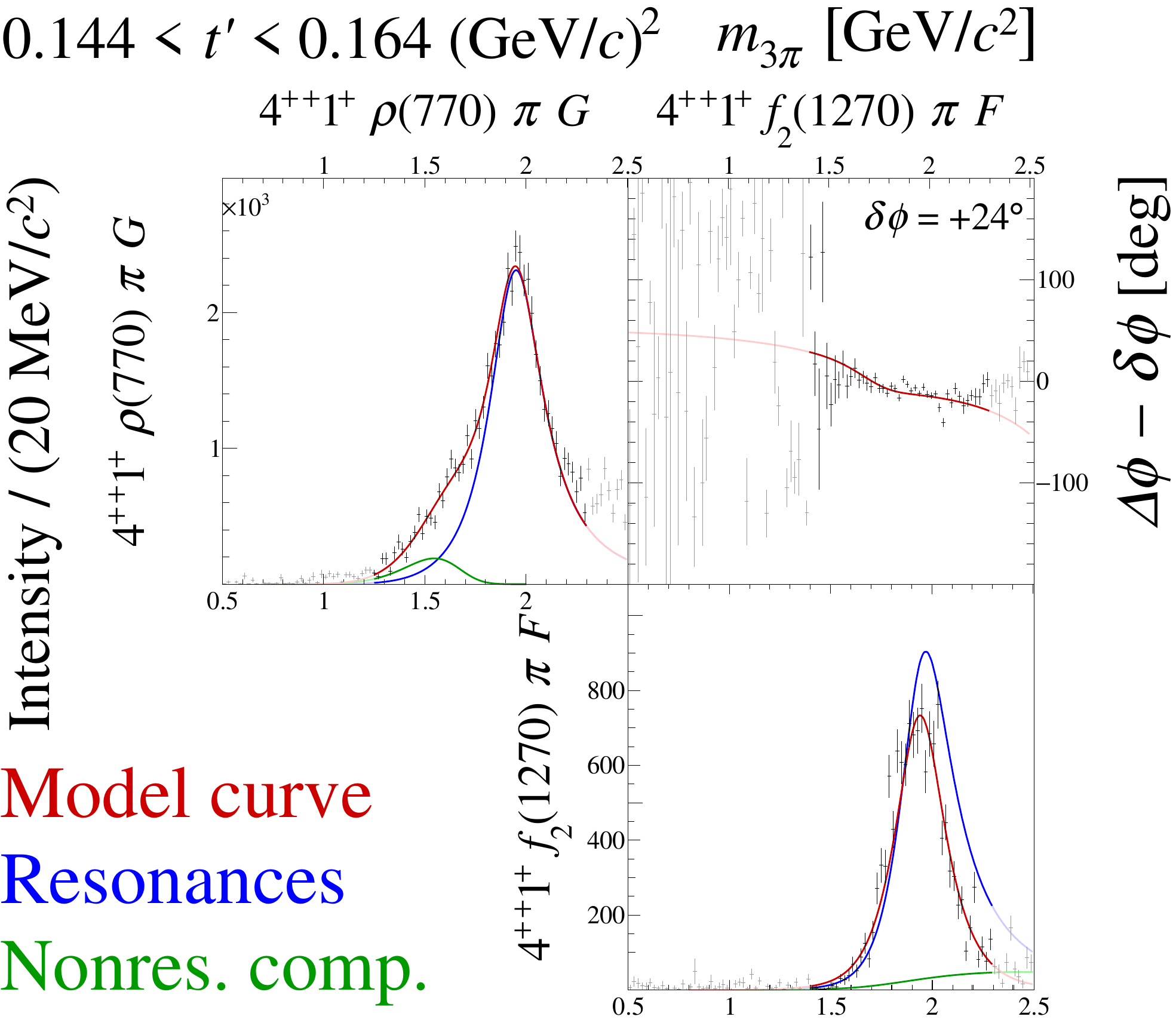}%
   \caption{Submatrix~J of the $14 \times 14$ matrix of graphs that
     represents the spin-density matrix (see
     \cref{tab:spin-dens_matrix_overview}).}
   \label{fig:spin-dens_submatrix_10_tbin_4}
 \end{minipage}
\end{textblock*}

\newpage\null
\begin{textblock*}{\textwidth}[0.5,0](0.5\paperwidth,\blockDistanceToTop)
 \begin{minipage}{\textwidth}
   \makeatletter
   \def\@captype{figure}
   \makeatother
   \centering
   \includegraphics[height=\matrixHeightHalf]{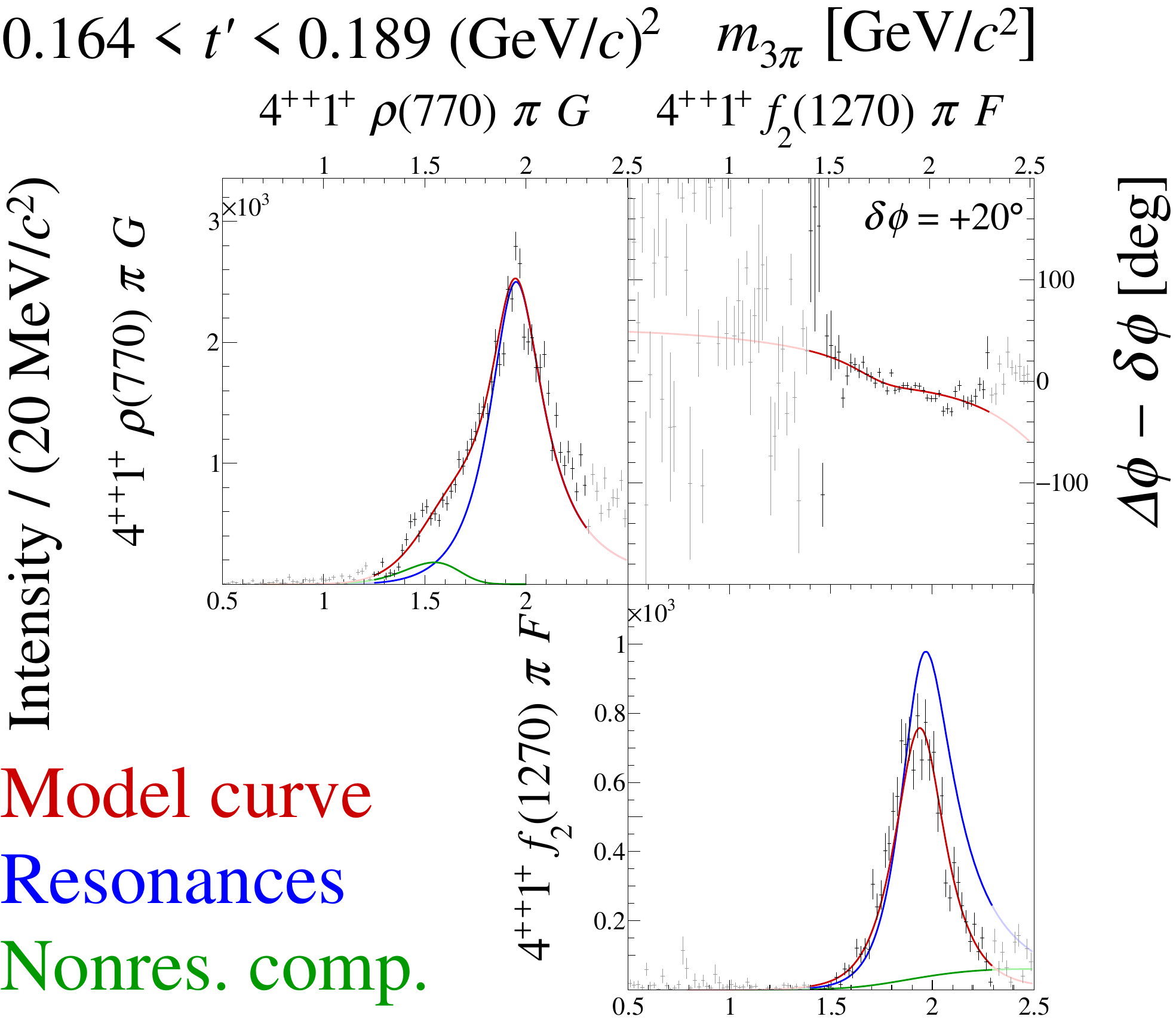}%
   \caption{Submatrix~J of the $14 \times 14$ matrix of graphs that
     represents the spin-density matrix (see
     \cref{tab:spin-dens_matrix_overview}).}
   \label{fig:spin-dens_submatrix_10_tbin_5}
 \end{minipage}
\end{textblock*}

\newpage\null
\begin{textblock*}{\textwidth}[0.5,0](0.5\paperwidth,\blockDistanceToTop)
 \begin{minipage}{\textwidth}
   \makeatletter
   \def\@captype{figure}
   \makeatother
   \centering
   \includegraphics[height=\matrixHeightHalf]{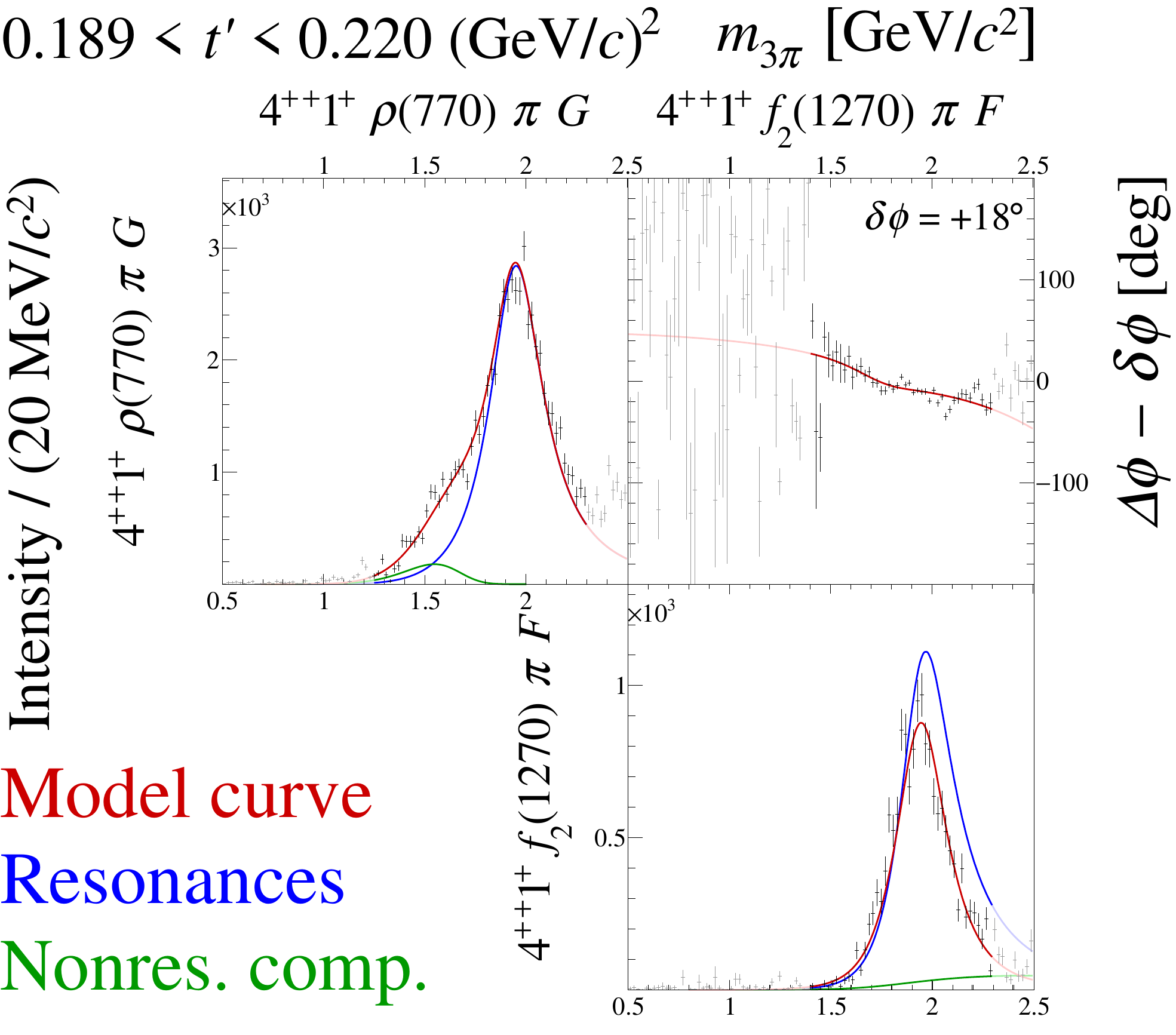}%
   \caption{Submatrix~J of the $14 \times 14$ matrix of graphs that
     represents the spin-density matrix (see
     \cref{tab:spin-dens_matrix_overview}).}
   \label{fig:spin-dens_submatrix_10_tbin_6}
 \end{minipage}
\end{textblock*}

\newpage\null
\begin{textblock*}{\textwidth}[0.5,0](0.5\paperwidth,\blockDistanceToTop)
 \begin{minipage}{\textwidth}
   \makeatletter
   \def\@captype{figure}
   \makeatother
   \centering
   \includegraphics[height=\matrixHeightHalf]{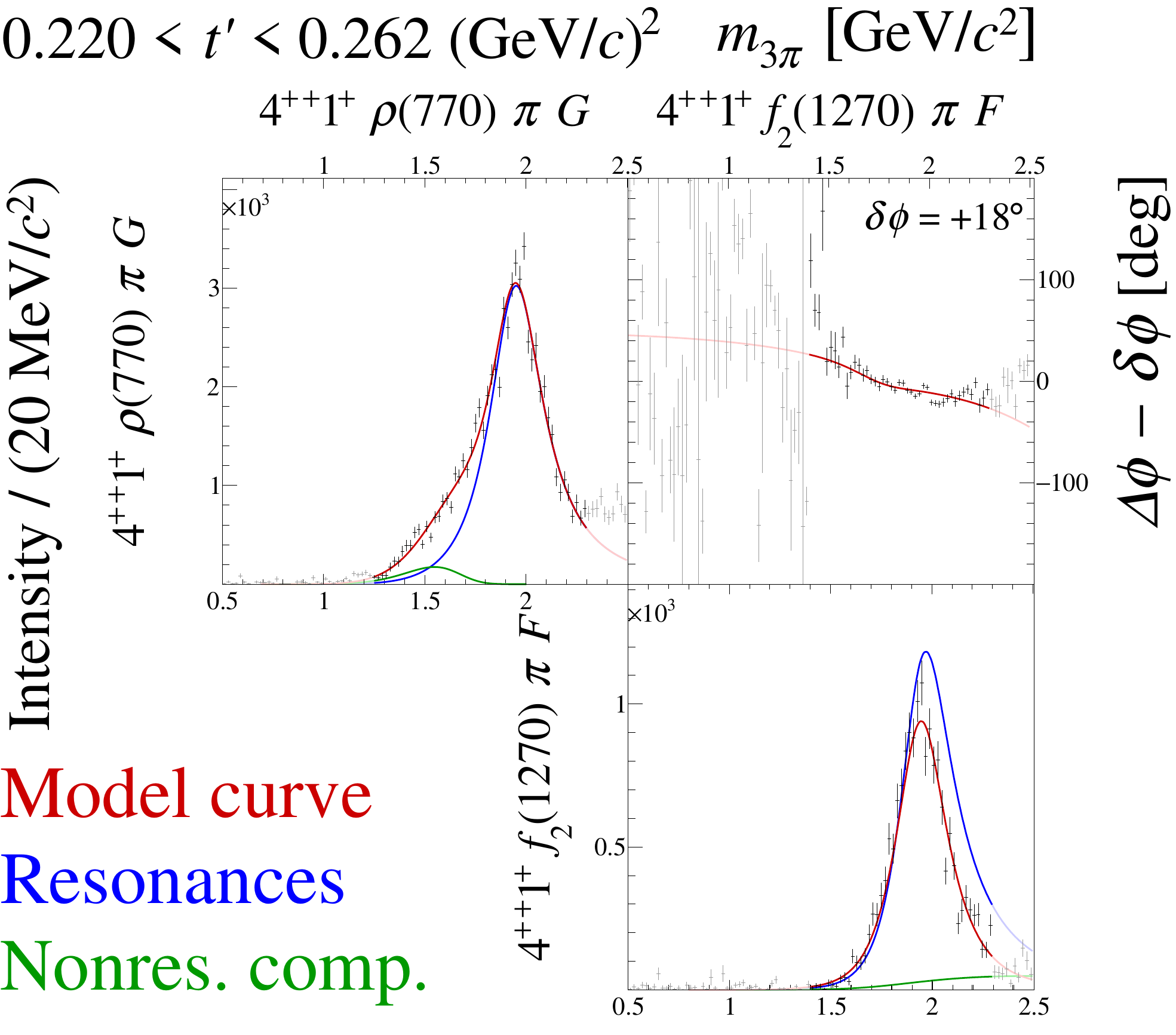}%
   \caption{Submatrix~J of the $14 \times 14$ matrix of graphs that
     represents the spin-density matrix (see
     \cref{tab:spin-dens_matrix_overview}).}
   \label{fig:spin-dens_submatrix_10_tbin_7}
 \end{minipage}
\end{textblock*}

\newpage\null
\begin{textblock*}{\textwidth}[0.5,0](0.5\paperwidth,\blockDistanceToTop)
 \begin{minipage}{\textwidth}
   \makeatletter
   \def\@captype{figure}
   \makeatother
   \centering
   \includegraphics[height=\matrixHeightHalf]{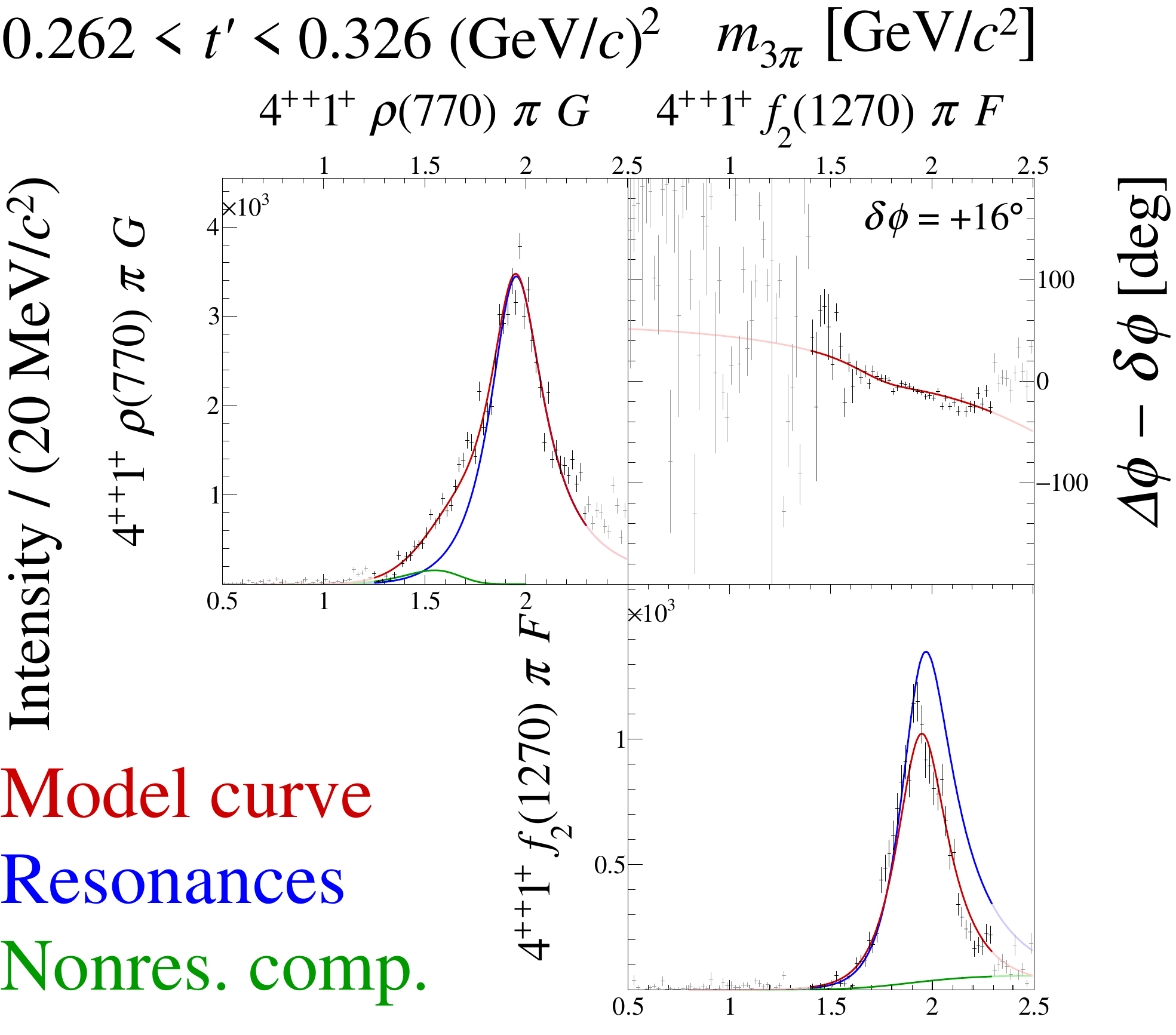}%
   \caption{Submatrix~J of the $14 \times 14$ matrix of graphs that
     represents the spin-density matrix (see
     \cref{tab:spin-dens_matrix_overview}).}
   \label{fig:spin-dens_submatrix_10_tbin_8}
 \end{minipage}
\end{textblock*}

\newpage\null
\begin{textblock*}{\textwidth}[0.5,0](0.5\paperwidth,\blockDistanceToTop)
 \begin{minipage}{\textwidth}
   \makeatletter
   \def\@captype{figure}
   \makeatother
   \centering
   \includegraphics[height=\matrixHeightHalf]{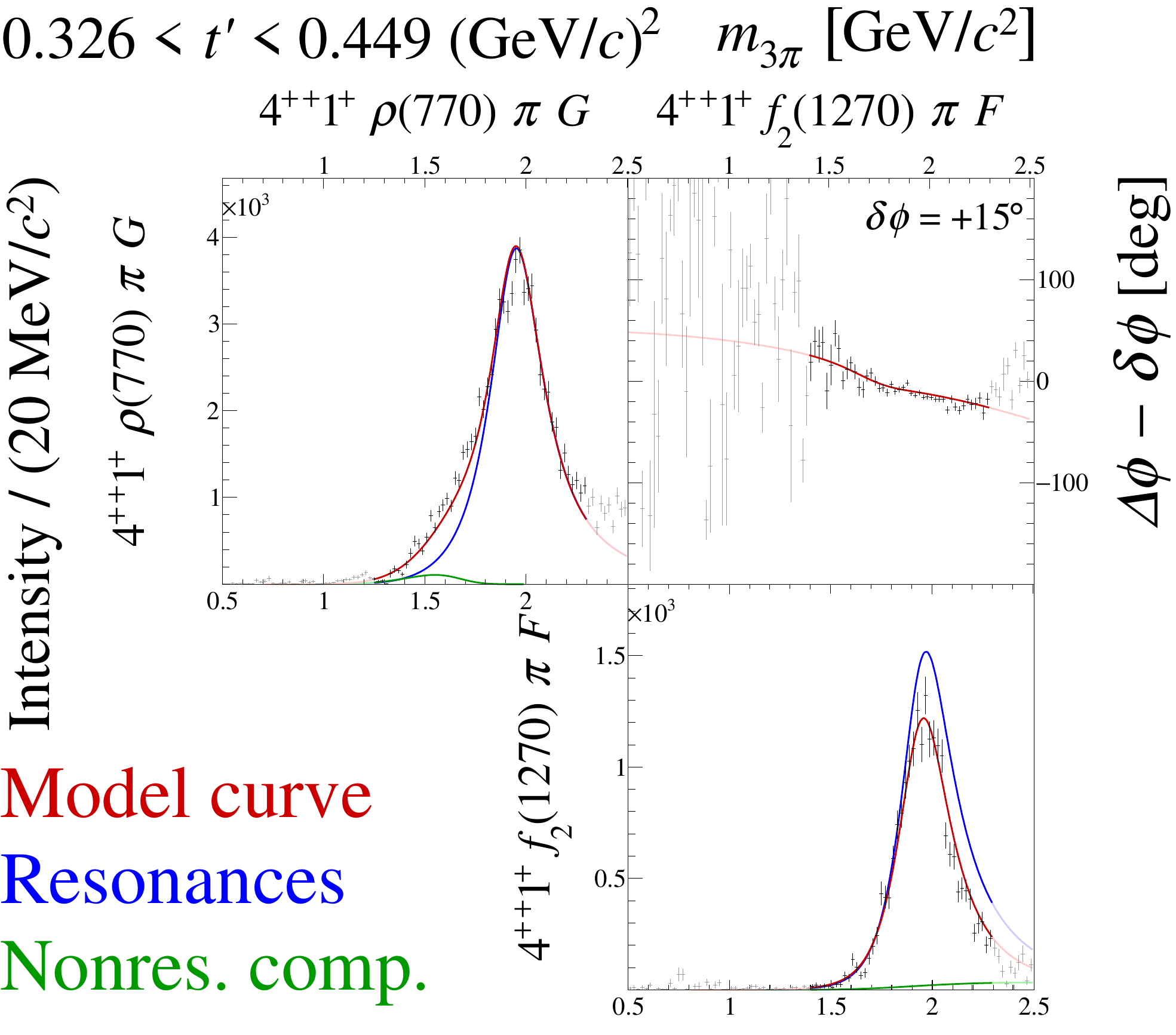}%
   \caption{Submatrix~J of the $14 \times 14$ matrix of graphs that
     represents the spin-density matrix (see
     \cref{tab:spin-dens_matrix_overview}).}
   \label{fig:spin-dens_submatrix_10_tbin_9}
 \end{minipage}
\end{textblock*}

\newpage\null
\begin{textblock*}{\textwidth}[0.5,0](0.5\paperwidth,\blockDistanceToTop)
 \begin{minipage}{\textwidth}
   \makeatletter
   \def\@captype{figure}
   \makeatother
   \centering
   \includegraphics[height=\matrixHeightHalf]{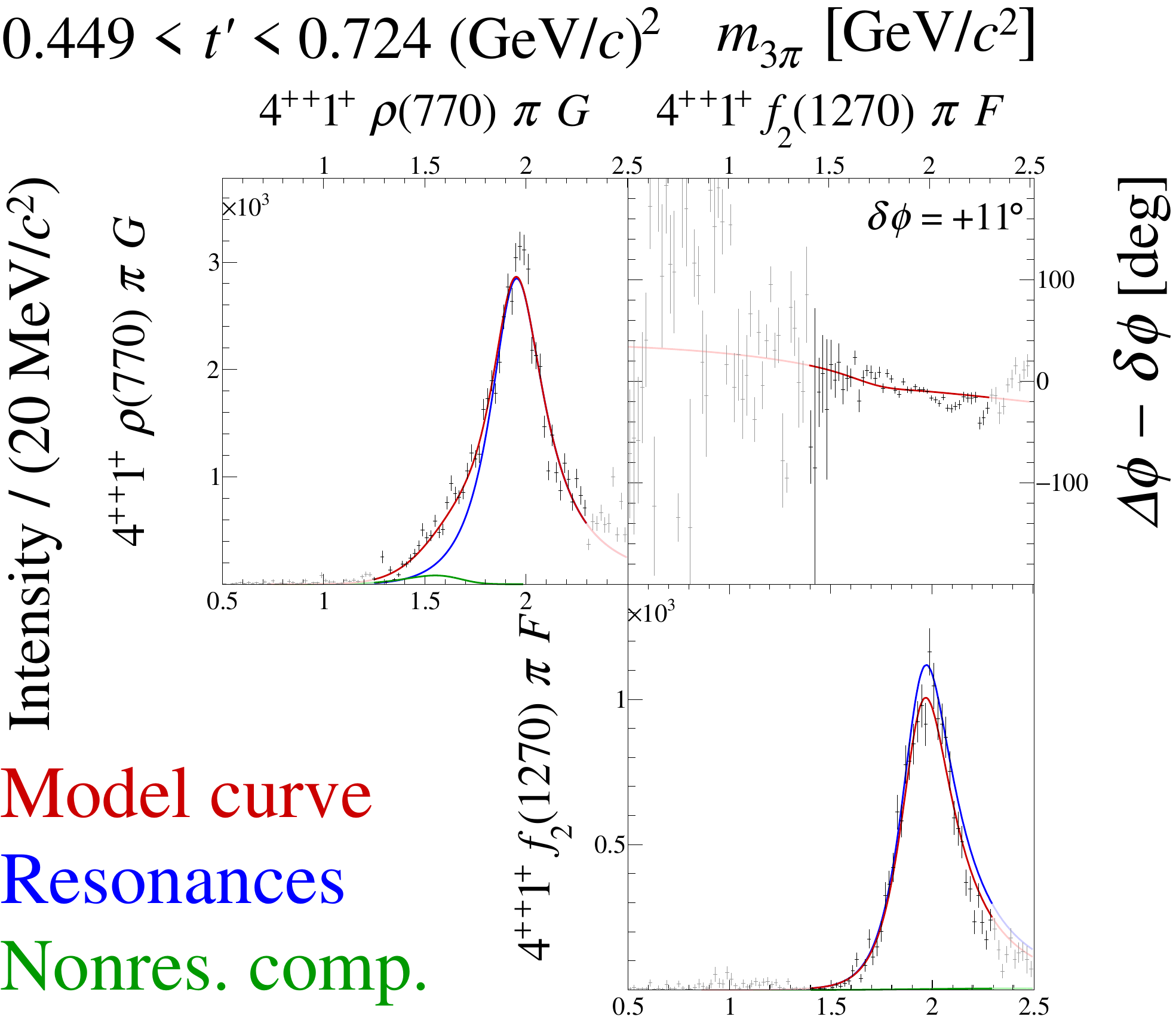}%
   \caption{Submatrix~J of the $14 \times 14$ matrix of graphs that
     represents the spin-density matrix (see
     \cref{tab:spin-dens_matrix_overview}).}
   \label{fig:spin-dens_submatrix_10_tbin_10}
 \end{minipage}
\end{textblock*}

\newpage\null
\begin{textblock*}{\textwidth}[0.5,0](0.5\paperwidth,\blockDistanceToTop)
 \begin{minipage}{\textwidth}
   \makeatletter
   \def\@captype{figure}
   \makeatother
   \centering
   \includegraphics[height=\matrixHeightHalf]{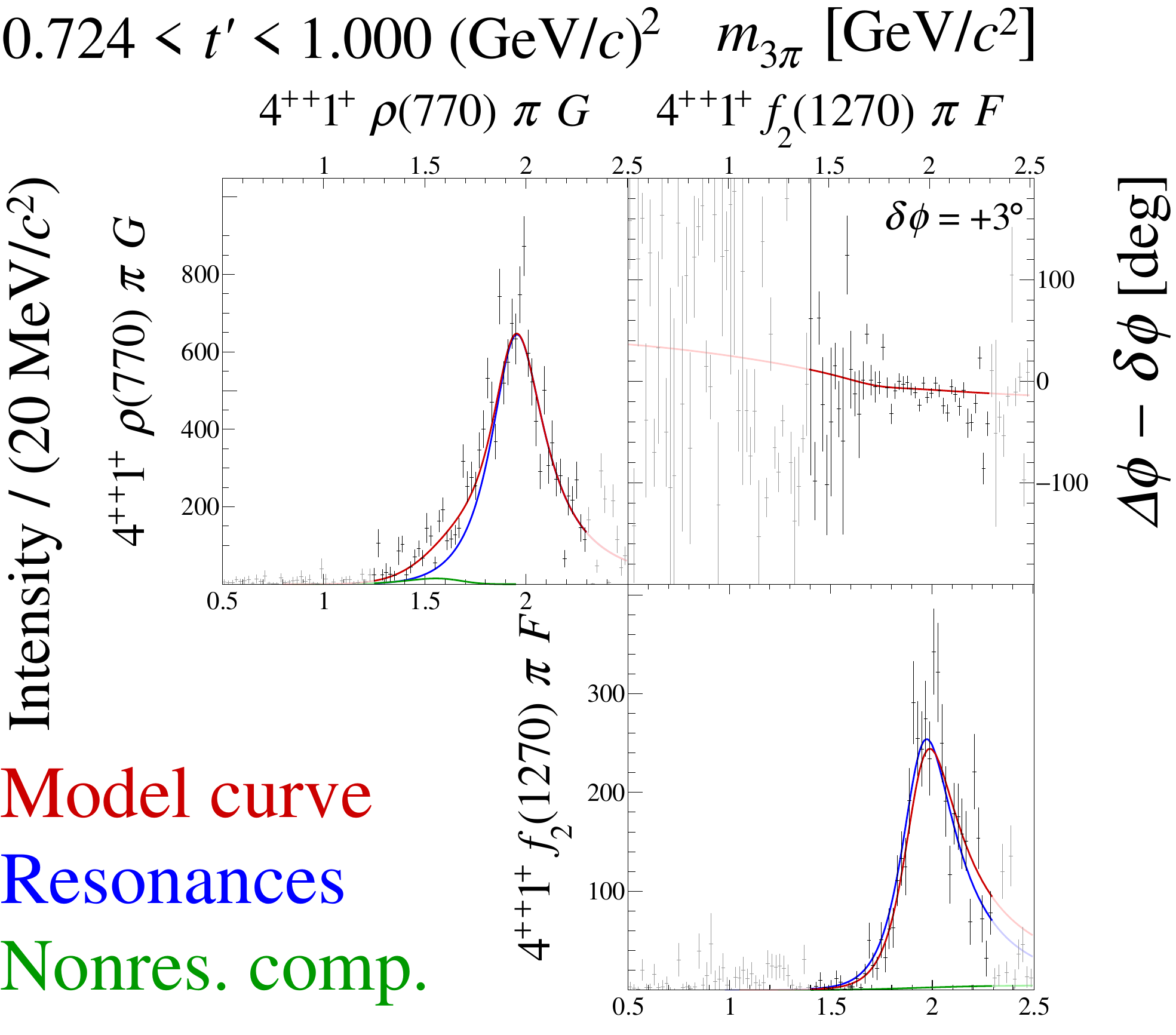}%
   \caption{Submatrix~J of the $14 \times 14$ matrix of graphs that
     represents the spin-density matrix (see
     \cref{tab:spin-dens_matrix_overview}).}
   \label{fig:spin-dens_submatrix_10_tbin_11}
 \end{minipage}
\end{textblock*}

\clearpage
\ifMultiColumnLayout{\twocolumngrid}{}
\section{Decay phase-space integrals for partial waves}
\label{sec:phase-space_vol}

\Crefrange{fig:phase-space_vol_0mp}{fig:phase-space_vol_4pp} show for
each of the 14~waves in the resonance-model fit the \mThreePi
dependence of the phase-space integrals $I_{a a}$
as defined in \ifMultiColumnLayout{Eq.~(6) in Sec.~III of
  \refCite{paper3}}{\cref{eq:integral_matrix_def}}.  The phase-space
integrals are normalized to their maximum value in the shown mass
range \SIvalRange{0.5}{\mThreePi}{2.5}{\GeVcc}.

\ifMultiColumnLayout{\onecolumngrid}{}

\begin{figure}[htbp]
  \centering
  \includegraphics[width=\threePlotWidth]{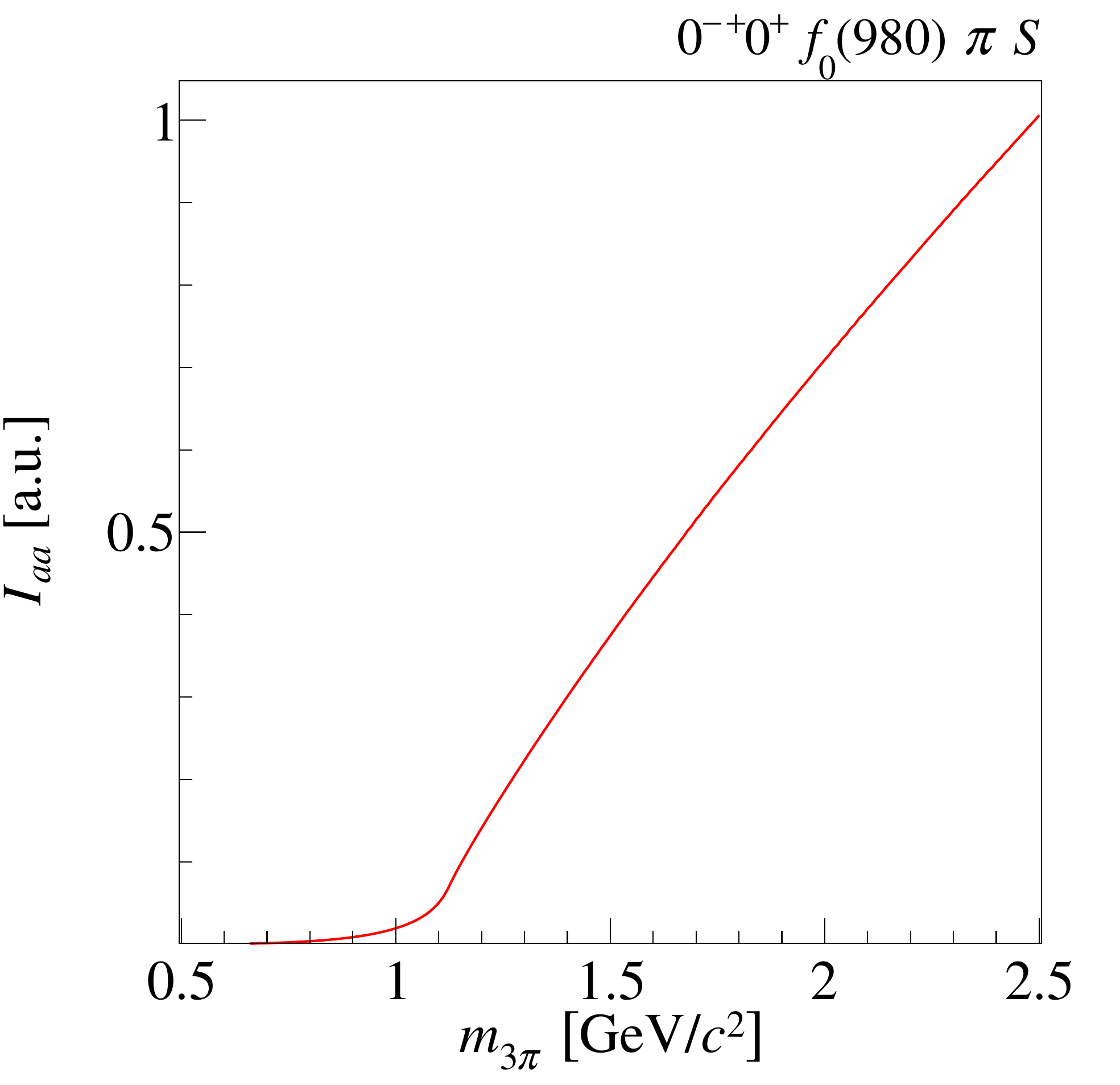}
  \caption{Phase-space integral $I_{a a}$ for the
    \wave{0}{-+}{0}{+}{\PfZero[980]}{S} wave in arbitrary units as a
    function of \mThreePi [see \ifMultiColumnLayout{Eq.~(6) in
      Sec.~III of \refCite{paper3}}{\cref{eq:integral_matrix_def}}].}
  \label{fig:phase-space_vol_0mp}
\end{figure}

\begin{figure}[htbp]
  \centering
  \subfloat[][]{%
    \includegraphics[width=\threePlotWidth]{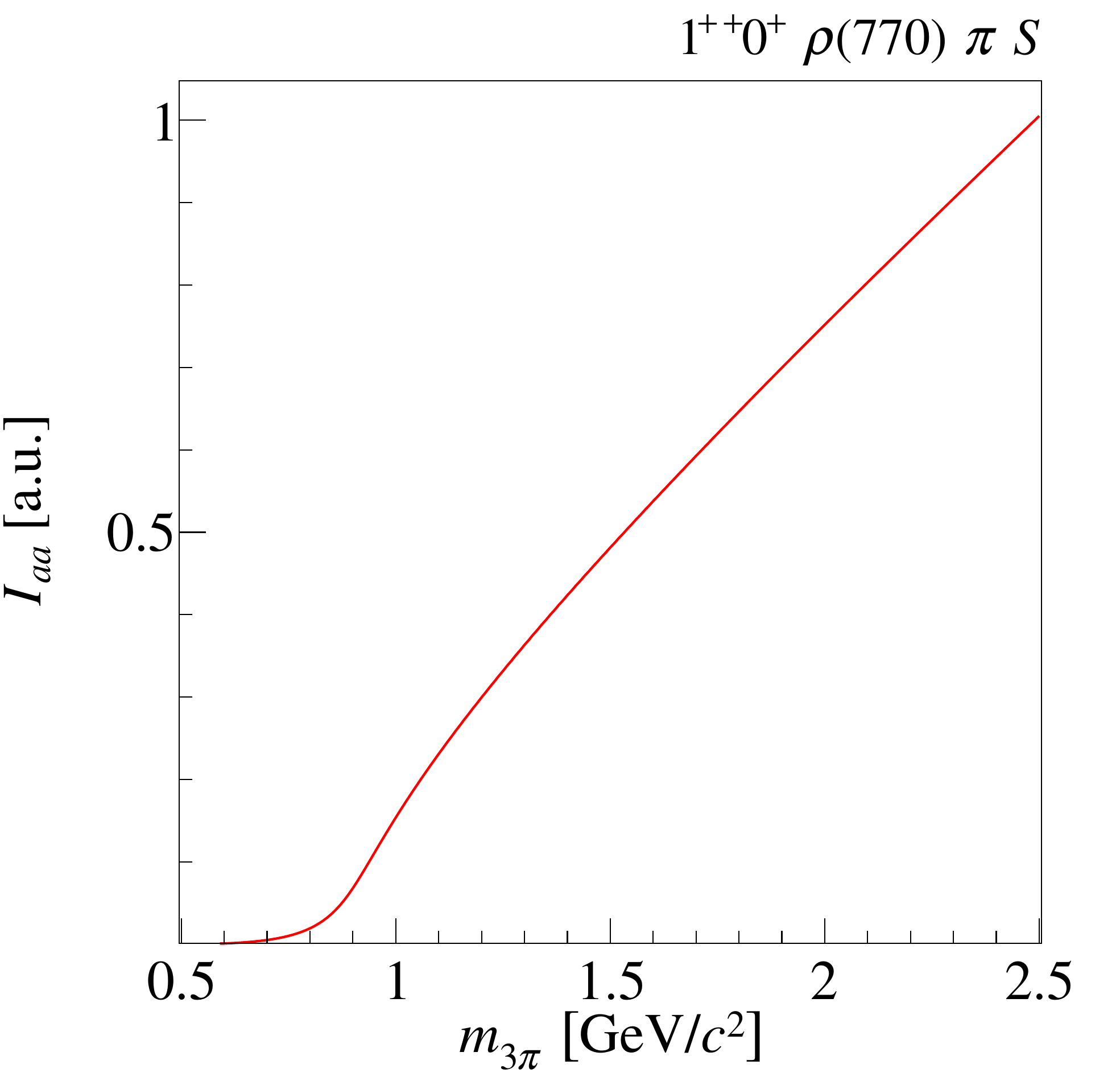}%
    \label{fig:phase-space_vol_1pp_rho}%
  }%
  \hspace*{\threePlotSpacing}%
  \subfloat[][]{%
    \includegraphics[width=\threePlotWidth]{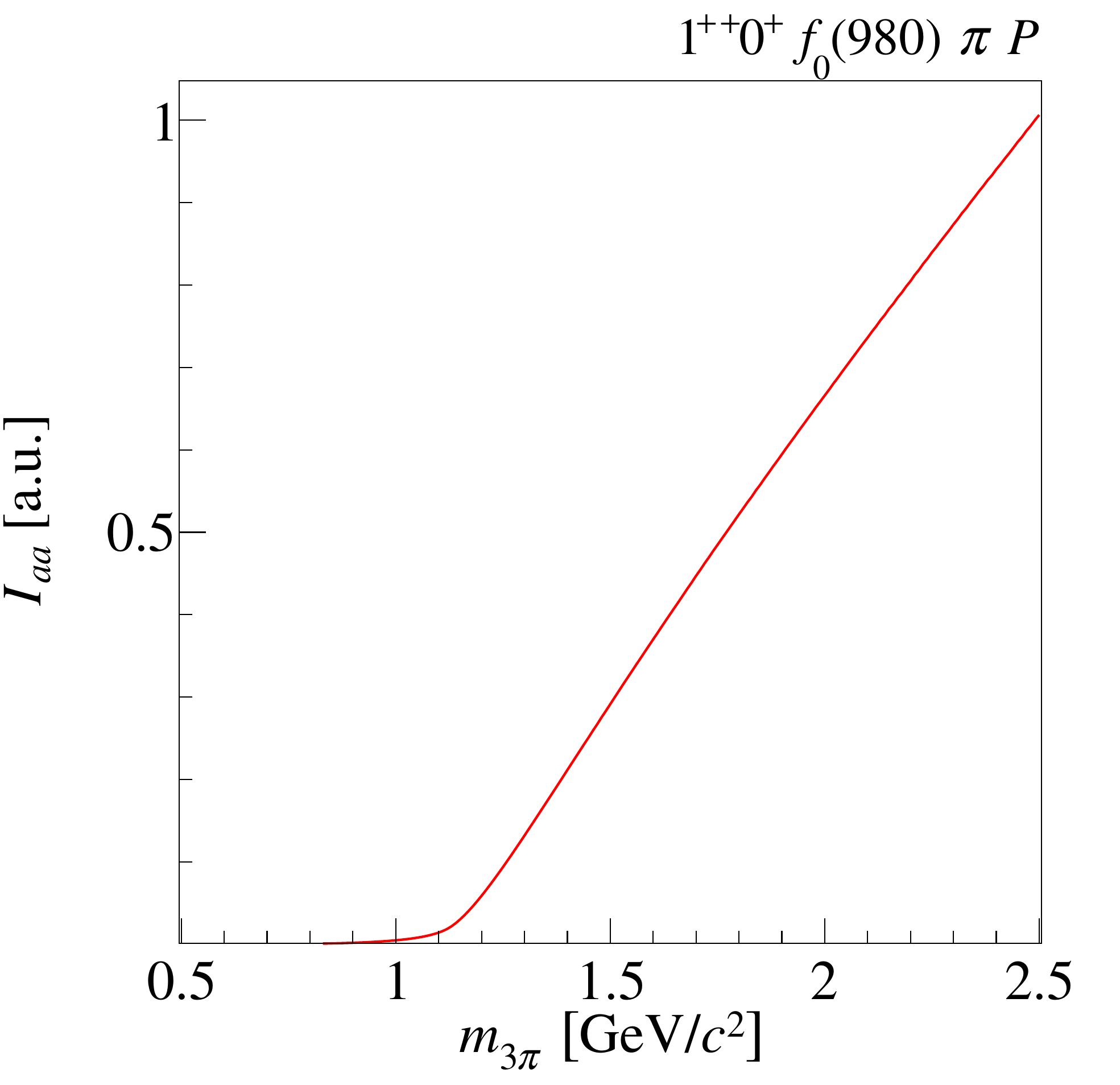}%
    \label{fig:phase-space_vol_1pp_f0}%
  }%
  \hspace*{\threePlotSpacing}%
  \subfloat[][]{%
    \includegraphics[width=\threePlotWidth]{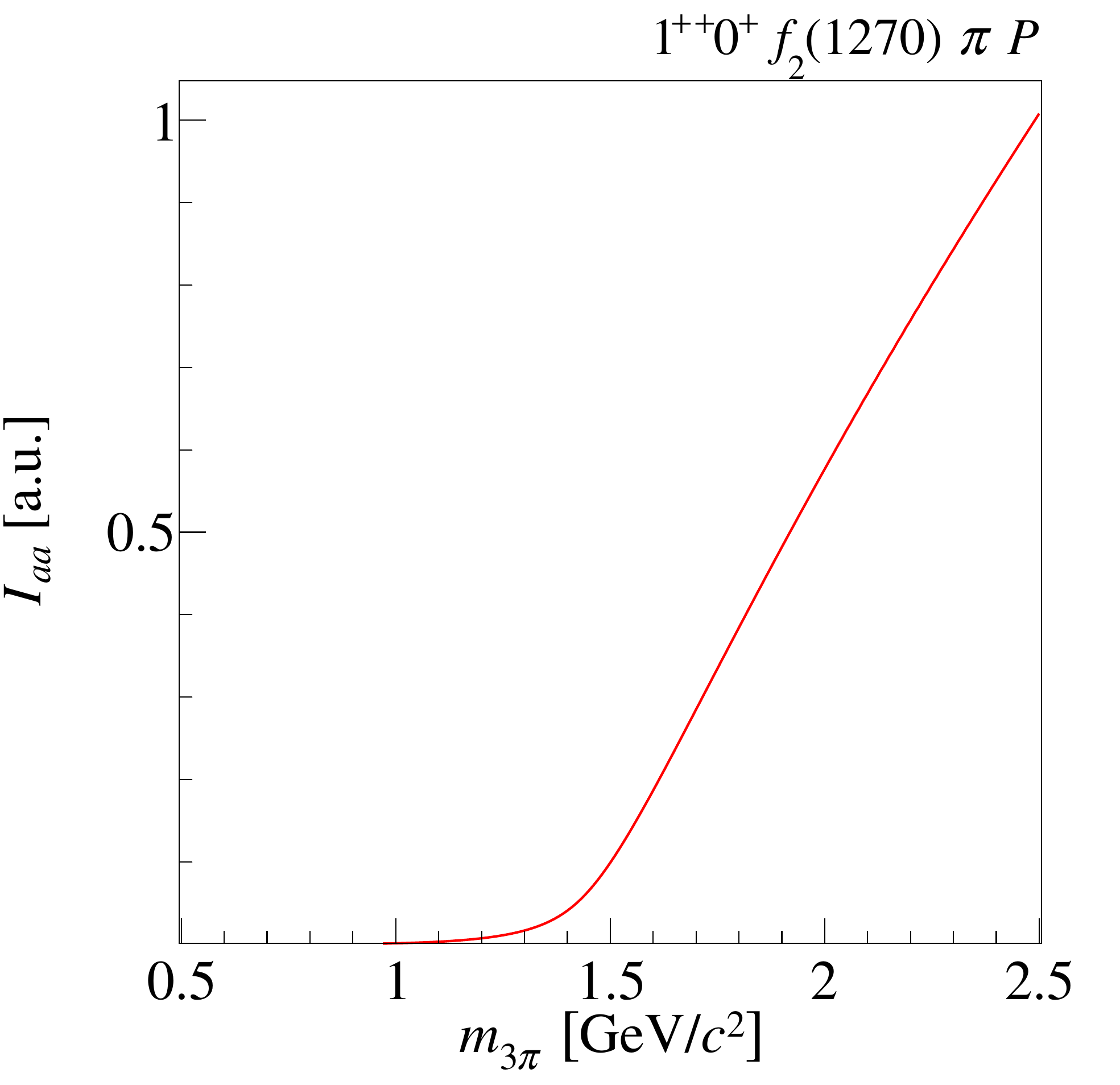}%
    \label{fig:phase-space_vol_1pp_f2}%
  }%
  \caption{Phase-space integrals $I_{a a}$ for
    \subfloatLabel{fig:phase-space_vol_1pp_rho}~the
    \wave{1}{++}{0}{+}{\Prho}{S},
    \subfloatLabel{fig:phase-space_vol_1pp_f0}~the
    \wave{1}{++}{0}{+}{\PfZero[980]}{P}, and
    \subfloatLabel{fig:phase-space_vol_1pp_f0}~the
    \wave{1}{++}{0}{+}{\PfTwo}{P} wave in arbitrary units as a
    function of \mThreePi [see \ifMultiColumnLayout{Eq.~(6) in
      Sec.~III of \refCite{paper3}}{\cref{eq:integral_matrix_def}}].}
  \label{fig:phase-space_vol_1pp}
\end{figure}

\begin{figure}[htbp]
  \centering
  \includegraphics[width=\threePlotWidth]{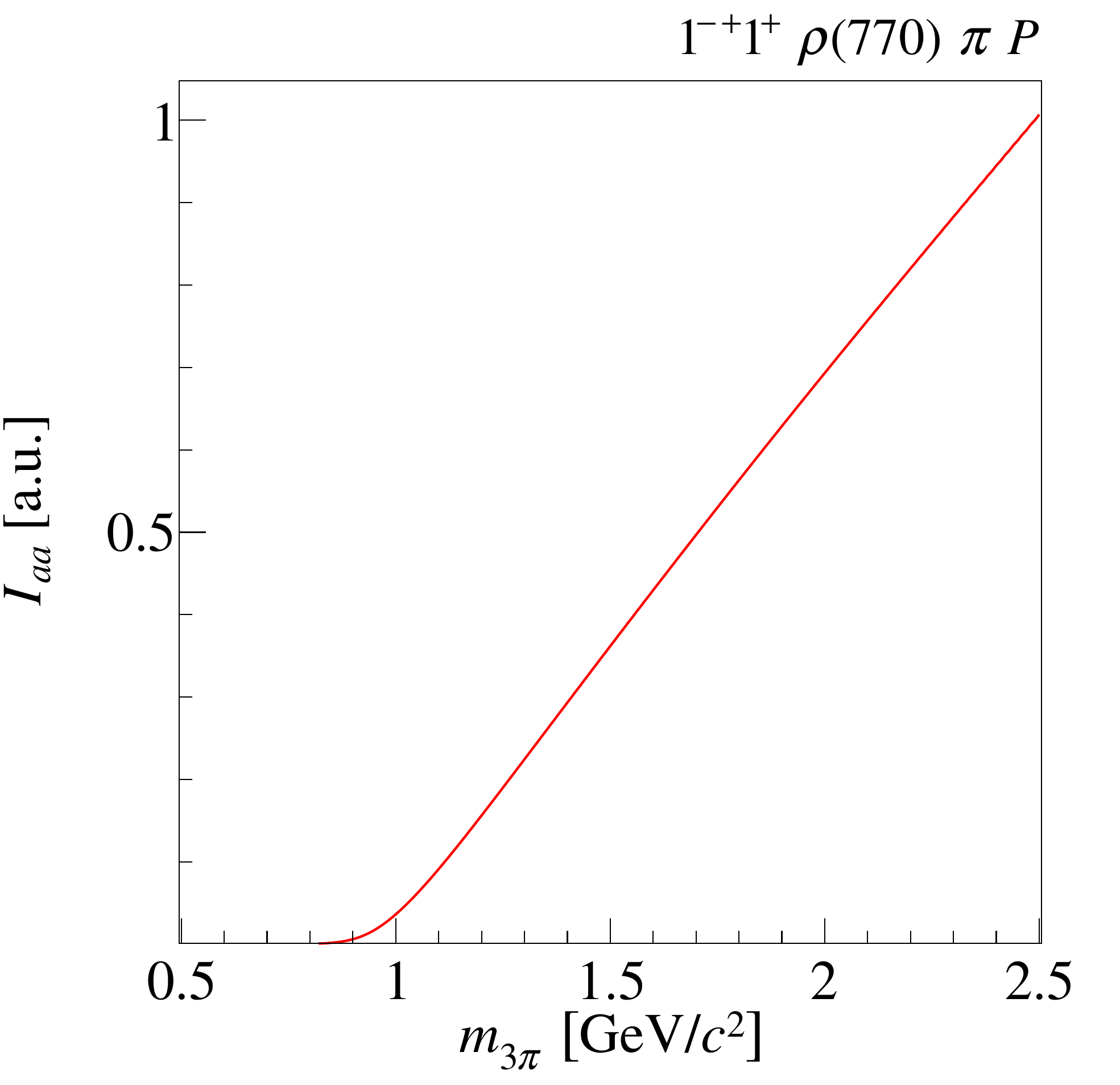}
  \caption{Phase-space integral $I_{a a}$ for the
    \wave{1}{-+}{1}{+}{\Prho}{P} wave in arbitrary units as a function
    of \mThreePi [see \ifMultiColumnLayout{Eq.~(6) in Sec.~III of
      \refCite{paper3}}{\cref{eq:integral_matrix_def}}].}
  \label{fig:phase-space_vol_1mp}
\end{figure}

\begin{figure}[htbp]
  \centering
  \subfloat[][]{%
    \includegraphics[width=\threePlotWidth]{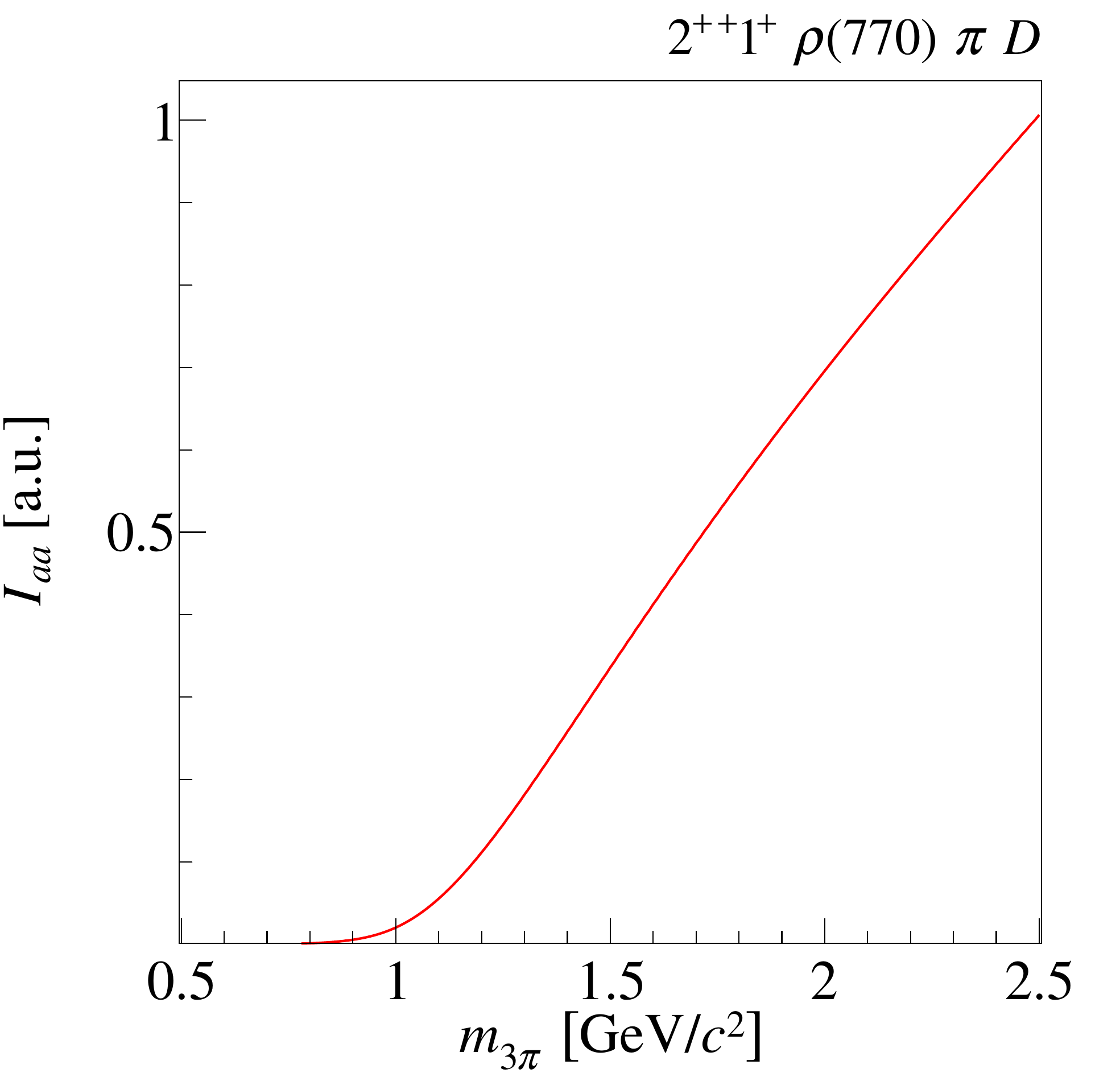}%
    \label{fig:phase-space_vol_2pp_rho_M1}%
  }%
  \hspace*{\threePlotSpacing}%
  \subfloat[][]{%
    \includegraphics[width=\threePlotWidth]{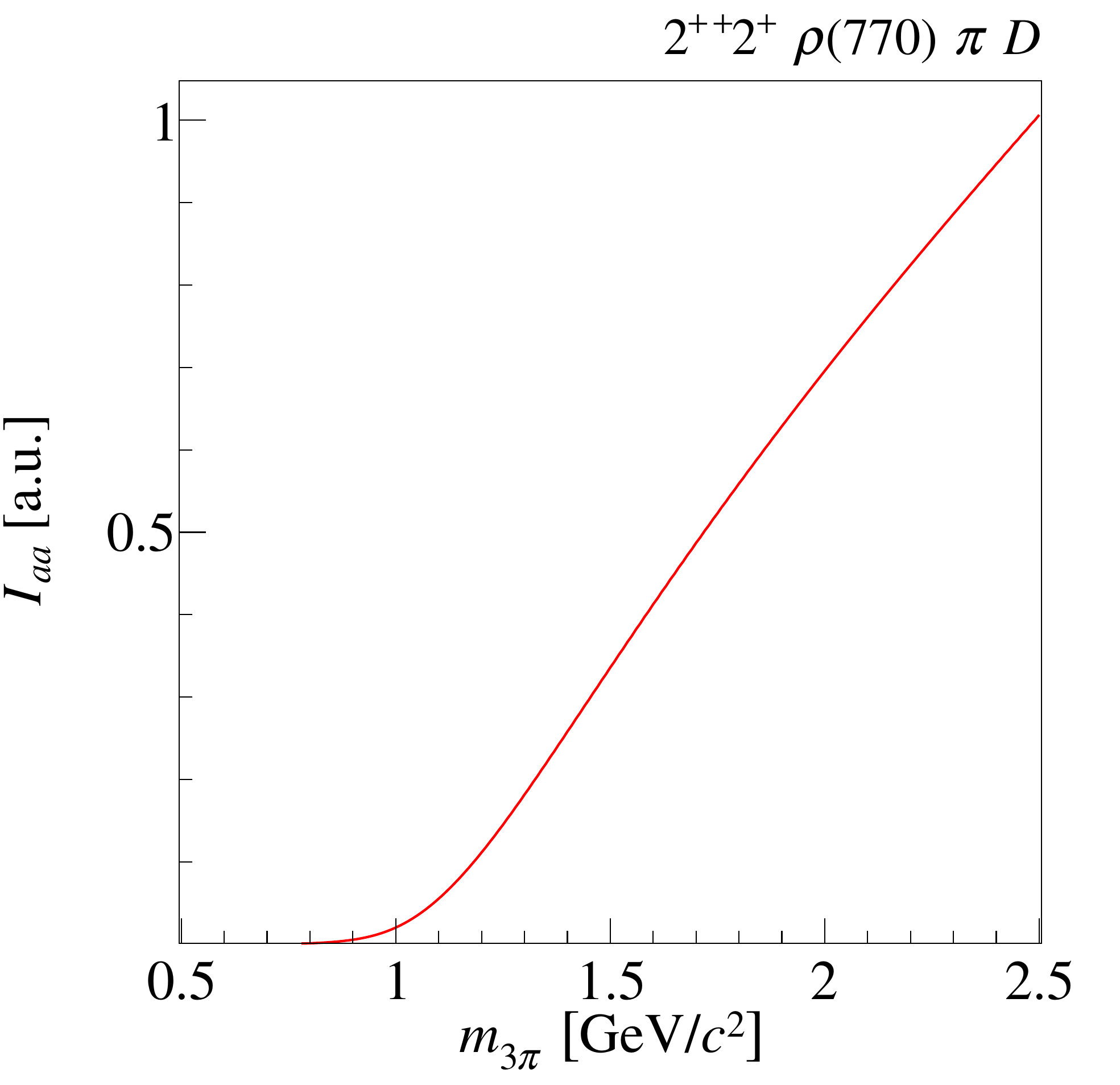}%
    \label{fig:phase-space_vol_2pp_rho_M2}%
  }%
  \hspace*{\threePlotSpacing}%
  \subfloat[][]{%
    \includegraphics[width=\threePlotWidth]{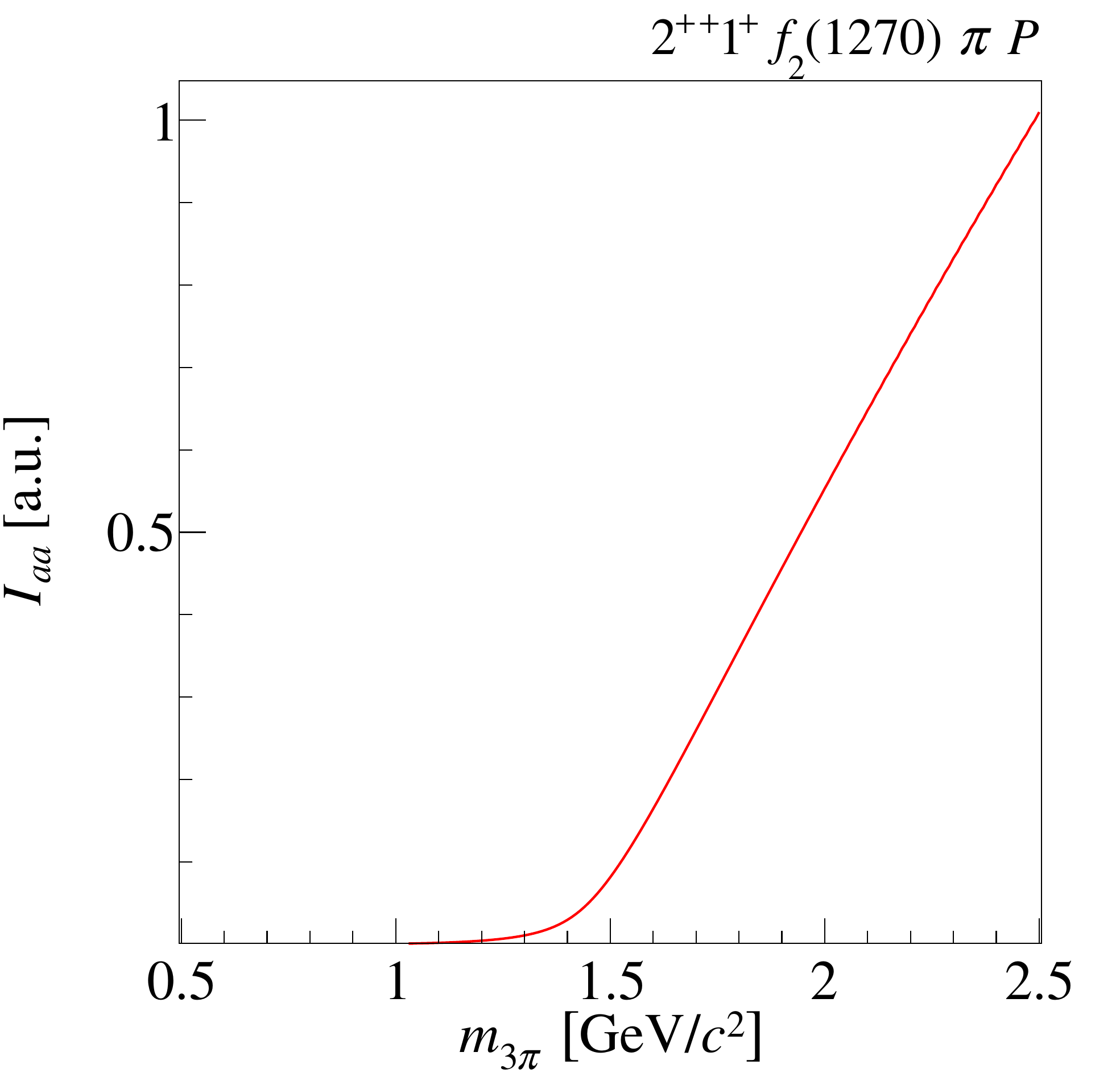}%
    \label{fig:phase-space_vol_2pp_f2}%
  }%
  \caption{Phase-space integrals $I_{a a}$ for
    \subfloatLabel{fig:phase-space_vol_2pp_rho_M1}~the
    \wave{2}{++}{1}{+}{\Prho}{D},
    \subfloatLabel{fig:phase-space_vol_2pp_rho_M2}~the
    \wave{2}{++}{2}{+}{\Prho}{D}, and
    \subfloatLabel{fig:phase-space_vol_2pp_f2}~the
    \wave{2}{++}{1}{+}{\PfTwo}{P} wave in arbitrary units as a
    function of \mThreePi [see \ifMultiColumnLayout{Eq.~(6) in
      Sec.~III of \refCite{paper3}}{\cref{eq:integral_matrix_def}}].}
  \label{fig:phase-space_vol_2pp}
\end{figure}

\begin{figure}[htbp]
  \centering
  \subfloat[][]{%
    \includegraphics[width=\threePlotWidth]{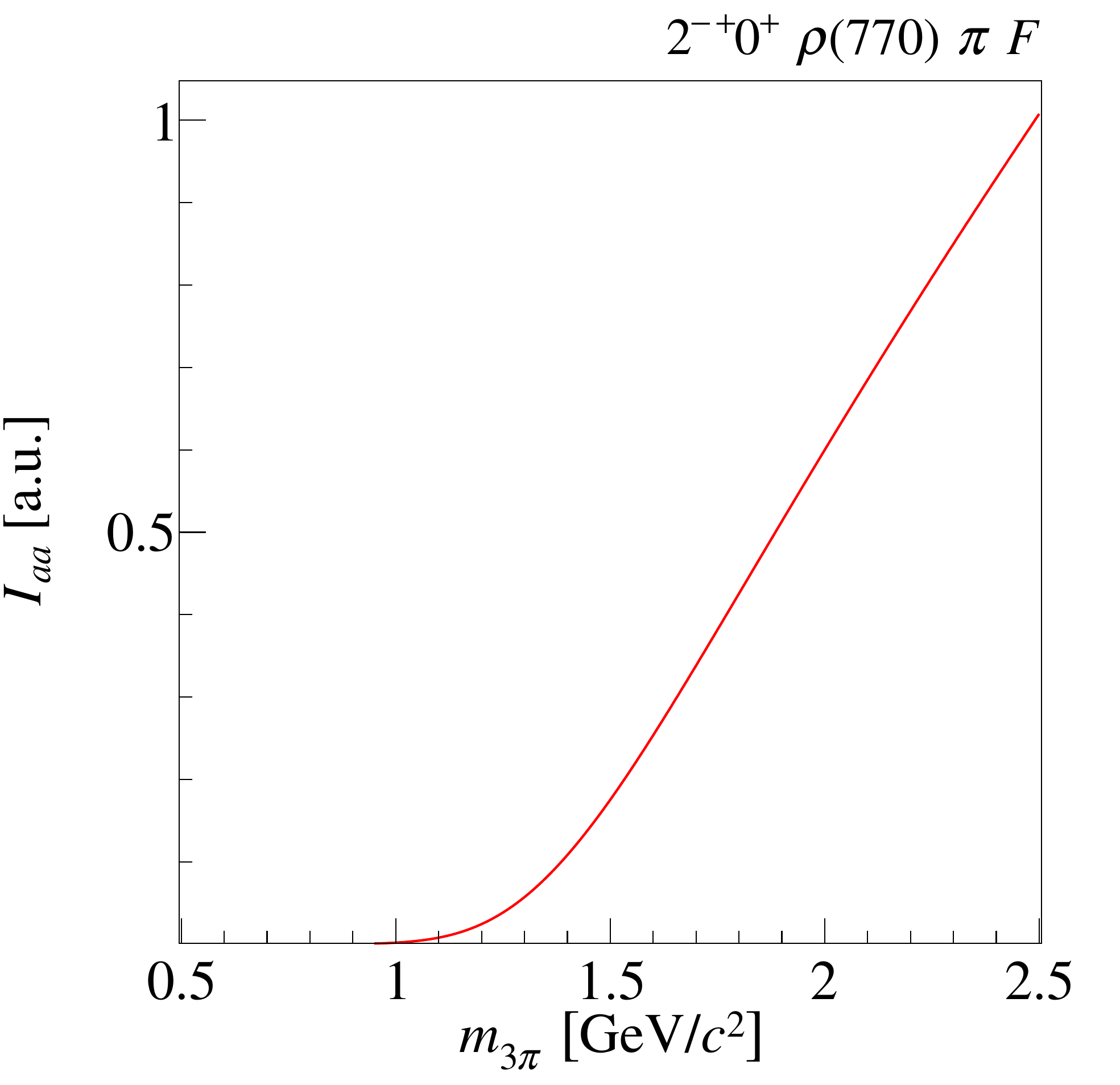}%
    \label{fig:phase-space_vol_2mp_rho}%
  }%
  \hspace*{\twoPlotSpacing}%
  \subfloat[][]{%
    \includegraphics[width=\threePlotWidth]{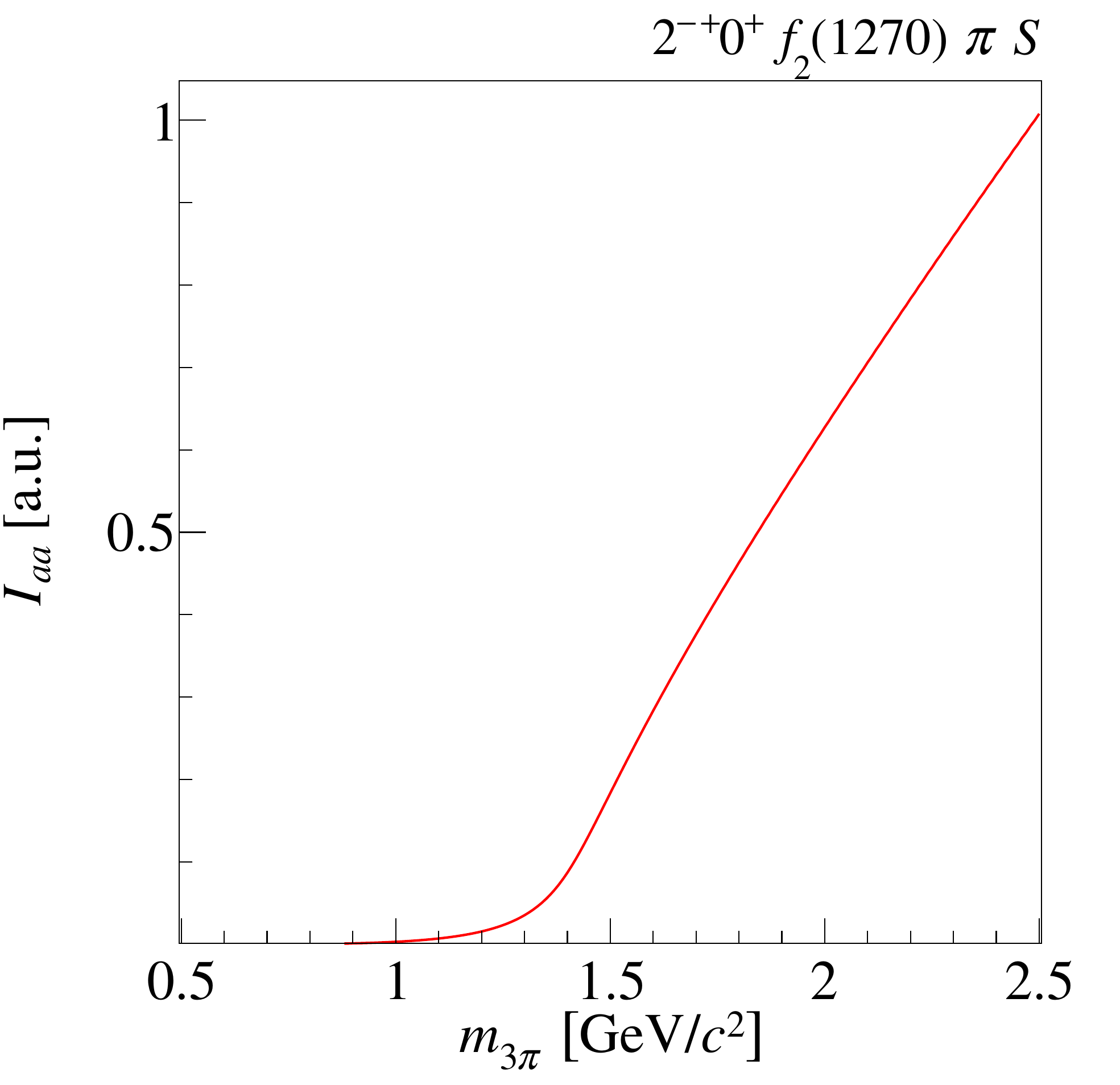}%
    \label{fig:phase-space_vol_2mp_f2_S_M0}%
  }%
  \\
  \subfloat[][]{%
    \includegraphics[width=\threePlotWidth]{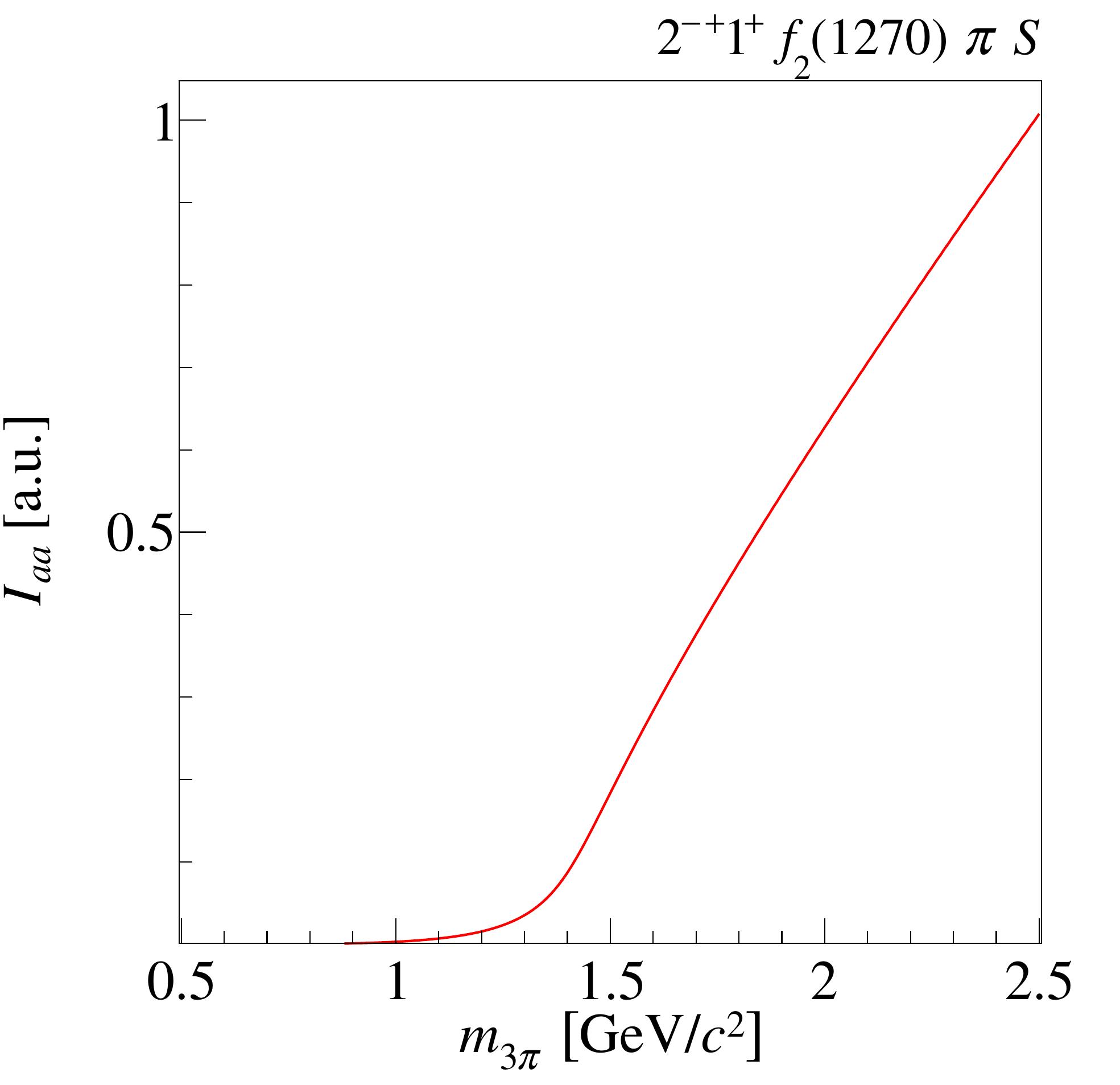}%
    \label{fig:phase-space_vol_2mp_f2_S_M1}%
  }%
  \hspace*{\twoPlotSpacing}%
  \subfloat[][]{%
    \includegraphics[width=\threePlotWidth]{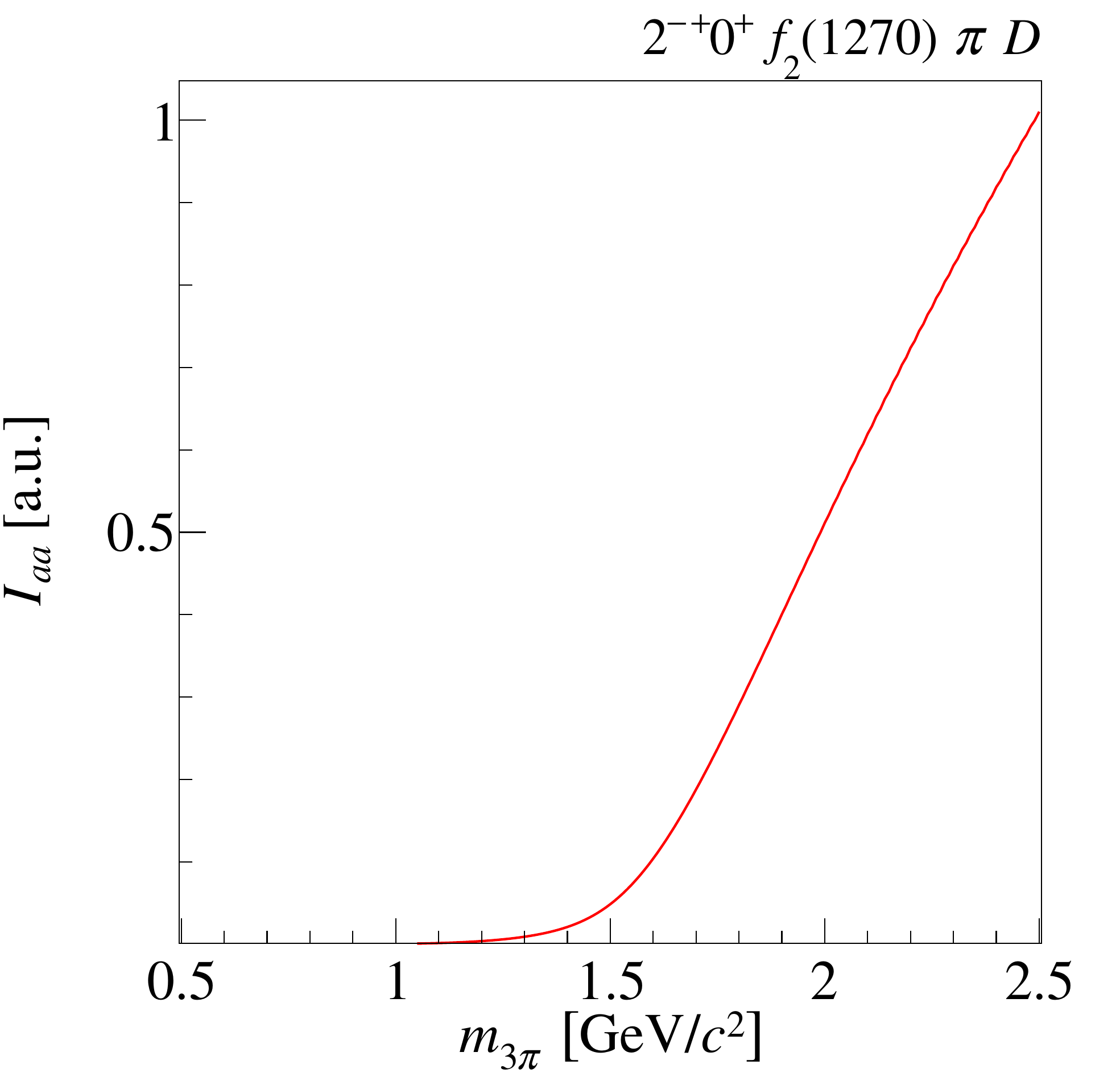}%
    \label{fig:phase-space_vol_2mp_f2_D}%
  }%
  \caption{Phase-space integrals $I_{a a}$ for
    \subfloatLabel{fig:phase-space_vol_2mp_rho}~the
    \wave{2}{-+}{0}{+}{\Prho}{F},
    \subfloatLabel{fig:phase-space_vol_2mp_f2_S_M0}~the
    \wave{2}{-+}{0}{+}{\PfTwo}{S},
    \subfloatLabel{fig:phase-space_vol_2mp_f2_S_M1}~the
    \wave{2}{-+}{1}{+}{\PfTwo}{S}, and
    \subfloatLabel{fig:phase-space_vol_2mp_f2_D}~the
    \wave{2}{-+}{0}{+}{\PfTwo}{D} wave in arbitrary units as a
    function of \mThreePi [see \ifMultiColumnLayout{Eq.~(6) in
      Sec.~III of \refCite{paper3}}{\cref{eq:integral_matrix_def}}].}
  \label{fig:phase-space_vol_2mp}
\end{figure}

\begin{figure}[htbp]
  \centering
  \subfloat[][]{%
    \includegraphics[width=\threePlotWidth]{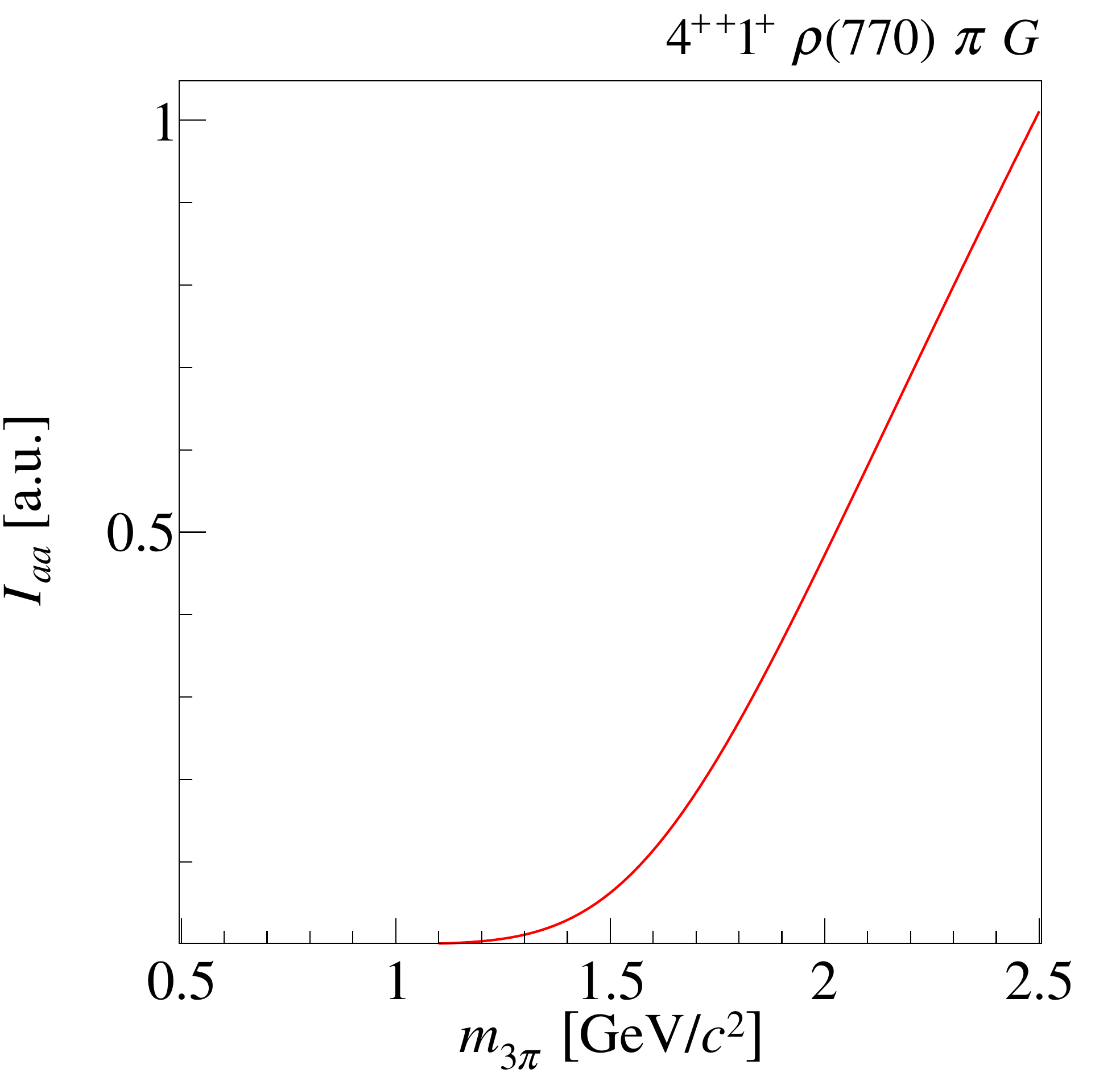}%
    \label{fig:phase-space_vol_4pp_rho}%
  }%
  \hspace*{\twoPlotSpacing}%
  \subfloat[][]{%
    \includegraphics[width=\threePlotWidth]{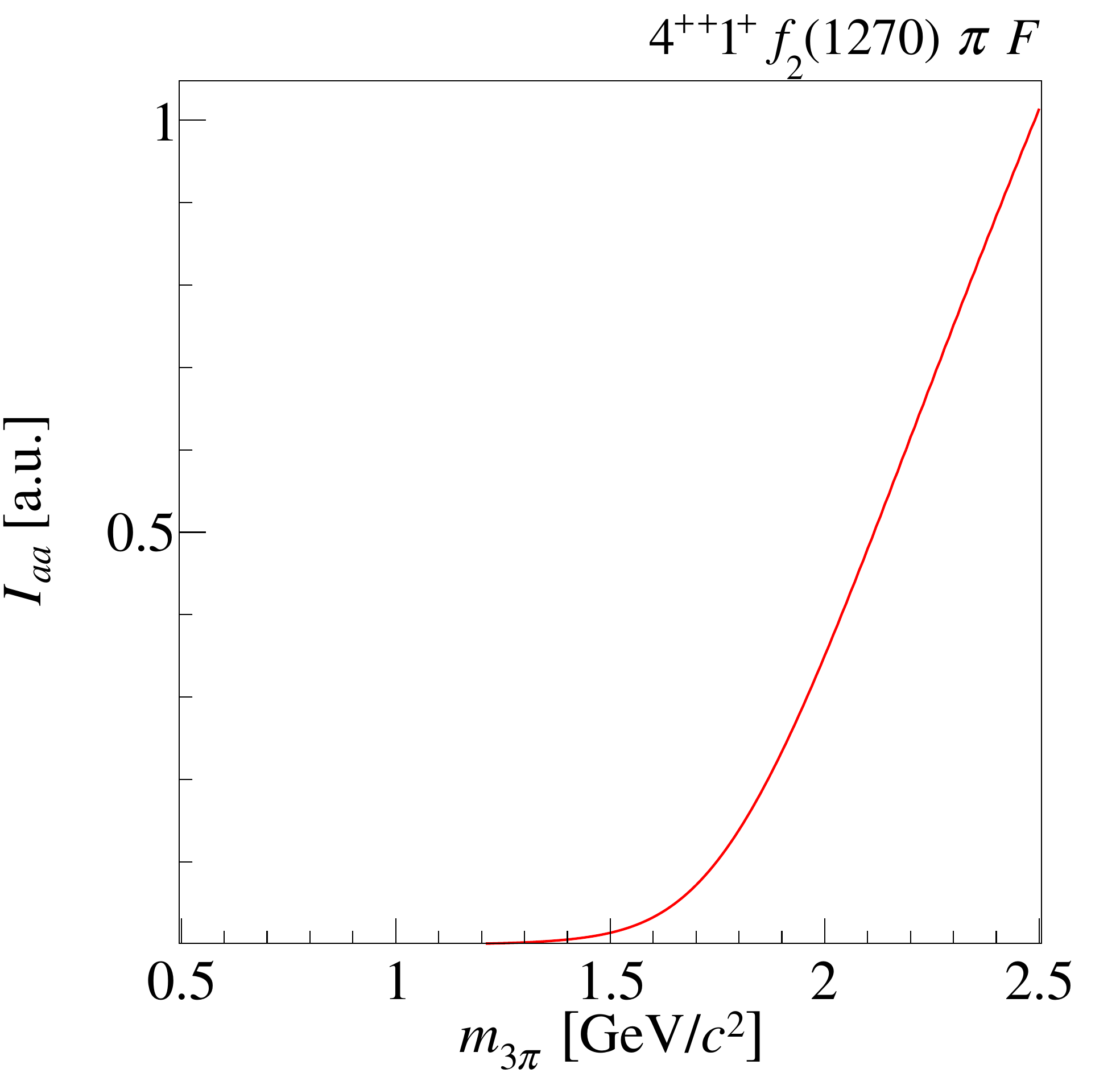}%
    \label{fig:phase-space_vol_4pp_f2}%
  }%
  \caption{Phase-space integrals $I_{a a}$ for
    \subfloatLabel{fig:phase-space_vol_4pp_rho}~the
    \wave{4}{++}{1}{+}{\Prho}{G} and
    \subfloatLabel{fig:phase-space_vol_4pp_f2}~the
    \wave{4}{++}{1}{+}{\PfTwo}{F} wave in arbitrary units as a
    function of \mThreePi [see \ifMultiColumnLayout{Eq.~(6) in
      Sec.~III of \refCite{paper3}}{\cref{eq:integral_matrix_def}}].}
  \label{fig:phase-space_vol_4pp}
\end{figure}
\clearpage{}%

\end{document}